\documentclass{article}
\usepackage[margin=0.6in]{geometry}
\usepackage{graphicx} 
\usepackage{titletoc}

\usepackage{adjustbox}
\usepackage{float}
\usepackage{subfig}

\usepackage{graphicx}
\usepackage{dcolumn}
\usepackage{bm}
\usepackage{tabularx}
\usepackage{longtable}
\usepackage{booktabs}
\usepackage{longtable}
\usepackage{hyperref}
\usepackage{multirow}

\usepackage{mathtools}
\usepackage[parfill]{parskip}
\usepackage{siunitx}
\usepackage{lipsum}
\usepackage{microtype}
\usepackage{booktabs}
\usepackage{bm}
\usepackage{xcolor}
\usepackage{graphicx}
\usepackage{longtable}
\usepackage{multirow}
\usepackage{setspace}

\usepackage[numbers,sort&compress]{natbib}
\usepackage{notoccite}
\usepackage[english]{babel}
\usepackage{amsmath}
\usepackage{amssymb}
\usepackage{microtype}
\usepackage{bm}
\usepackage{graphicx}
\usepackage{xcolor}
\usepackage{amsfonts}
\usepackage{mathtools}
\usepackage{dcolumn}
\usepackage{dsfont}
\usepackage{longtable}
\usepackage{extarrows}
\usepackage{booktabs}
\usepackage{multirow}
\usepackage{enumitem,adjustbox}
\usepackage{setspace}
\usepackage{authblk}
\usepackage{svg}
\usepackage{authblk} 
\usepackage{amsmath, amssymb, amsfonts}

\usepackage[titletoc,title]{appendix}
\usepackage[utf8]{inputenc}
\usepackage{csquotes}

\usepackage{arydshln}
\usepackage[utf8]{inputenc}
\usepackage{textcomp}
\usepackage{amsmath}

\newcommand{\equalcontribution}{\textsuperscript{*}}
\newcommand{\Correspondingg}{\textsuperscript{\dag}}

\title{High-throughput search for topological magnon materials}
\author[1]{Mohammed J. Karaki\equalcontribution}
\author[1]{Ahmed E. Fahmy\equalcontribution\Correspondingg}
\author[2]{Archibald J. Williams}
\author[3,4]{Sara Haravifard}
\author[2]{Joshua E. Goldberger}
\author[1]{Yuan-Ming Lu \Correspondingg}
\affil[1]{\small \textit{Department of Physics, The Ohio State University, Columbus, OH 43210, USA}}
\affil[2]{\textit{Department of Chemistry and Biochemistry, The Ohio State University, Columbus, OH 43210, USA}}
\affil[3]{\textit{Department of Physics, Duke University, Durham, North Carolina 27708, USA}}
\affil[4]{\textit{Department of Mechanical Engineering and Materials Science, Duke University, Durham, North Carolina 27708, USA}}
\date{\today}

\DeclareUnicodeCharacter{2212}{-}
\usepackage{newunicodechar}

\newunicodechar{≤}{\ensuremath{\leq}}
\begin{document}

\onehalfspacing

\maketitle

\begingroup
\renewcommand\thefootnote{}
\footnotetext{\equalcontribution These authors contributed equally to this work.}
\footnotetext{\Correspondingg Corresponding authors: abdelazim.2@osu.edu \& lu.1435@osu.edu}
\endgroup

\begin{abstract}

Topological magnons give rise to possibilities for engineering novel spintronics devices with critical applications in quantum information and computation, due to its symmetry-protected robustness and low dissipation. However, to make reliable and systematic predictions about material realization of topological magnons has been a major challenge, due to the lack of neutron scattering data for most materials. In this work, we significantly advance the symmetry-based approach for identifying topological magnons through developing a fully automated algorithm, utilizing the theory of symmetry indicators, that enables a highly efficient and large-scale search for candidate materials hosting field-induced topological magnons. This progress not only streamlines the discovery process but also expands the scope of materials exploration beyond previous manual or traditional methods, offering a powerful tool for uncovering novel topological phases in magnetic systems. Performing a large-scale search over all $1649$ magnetic materials in Bilbao Crystallographic Server with a commensurate magnetic order, we discover $387$ candidate materials for topological magnons, significantly expanding the pool of topological magnon materials. We further discuss examples and experimental accessibility of the candidate materials, shedding light on future experimental realizations of topological magnons in magnetic materials. 
\end{abstract}

\section{INTRODUCTION}

The discovery of topological insulators \cite{Hasan2010} revealed that in electronic materials, symmetry protected topological surface states can arise from nontrivial Berry phases of electronic bands in the bulk. Beyond electronic band structures, in strongly correlated magnetic materials, similar phenonema can also happen to spin wave excitations of ordered magnets, giving rise to so-called topological magnons \cite{annurev}. As collective excitations of spin degrees of freedom in magnetically ordered materials, magnons carry energy and spin through the crystal lattice without transporting electric charge, making them potential candidates for low-dissipation information processing \cite{Chumak2015}. The concept of topology introduces a robustness to these excitations, allowing for the existence of magnon modes that are protected against scattering by impurities or defects, making topological magnonics a robust route towards spintronic devices \cite{Zhuo2024}. It is therefore highly desirable to identify a list of material candidates that realize topological magnons. 



At the microscopic level, characterizing and understanding the magnon excitations in magnetically ordered structures, usually depends on a model spin Hamiltonian that captures spin-spin (and possible multi-spin) interactions, possible anisotropies, and couplings to external fields. Theoretically solving the spin model can determine the ground state magnetic order and the magnon spectrum, identify the possible nontrivial topology of the magnon bands, and predict topological surface states therein. 
In contrast to weakly correlated topological electronic \cite{Zhang2019,Tang2019,Vergniory2019,Cata4444Xu_2020,collinearmagnetictopologicalmaterialsSu2022,newmagnetictopologicalmaterials2024} and phonon \cite{Xu2024} materials, where \textit{ab initio} calculations have been instrumental in their discovery, strongly correlated magnetic materials present greater challenges for \textit{ab initio} methods to reliably predict their magnetic properties. Obtaining such a microscopic spin model has become possible thanks to the increasing development and accessibility of magnetic spectroscopies, specifically Inelastic Neutron Scattering (INS) spectroscopy for bulk structures \cite{Number1INSPhysRevB.7.336}, and other techniques such as Spin-Polarized Electron Energy Loss Spectroscopy (SPEELS) for thin-film structures \cite{Number2Qin2015LonglivingTM,Number3doi:10.1126/science.1125398}. Resonant Inelastic X-ray Scattering (RIXS) \cite{RIXSmitrano2024exploringquantummaterialsresonant} was able to probe magnetic excitations in both thin films \cite{Number8Pelliciari_2021,Number7Pelliciari_2021} and bulk structures \cite{Number6Lebert_2020,Number5PhysRevB.102.064412,Number4PhysRevB.97.155144}. 

Due to the difficulty for first principles methods to accurately predict spin models in strongly correlated magnetic materials, a good microscopic spin model is typically established by fitting experimentally-measured magnon dispersion data obtained by the aforementioned spectroscopic measurements. The topology of the magnon bands can also be diagnosed by calculations of the magnon spectra using this model. Identifying materials that can host topological magnons with this approach is extremely limiting as it relies on the existence of magnon dispersion data which is not available for the overwhelming majority of magnetic materials. Moreover, fitting the measured magnon dispersion with spin wave theory results can also encounter many challenges such as overfitting and local minima. These difficulties make this approach not applicable for a large-scale search for topological magnons materials, calling for a different methodology that does not depend on the specific spin model of magnetic materials. 

A symmetry-based approach was developed recently by \cite{KarakiSciAdvdoi:10.1126/sciadv.ade7731}, which searches topological magnons induced by external perturbations utilizing either constructed or literature-sourced microscopic spin Hamiltonians. This method starts with identifying the materials which host symmetry-enforced degeneracies in their magnon band structure. By applying external perturbations including electric field, magnetic field and/or mechanical strains, the original magnetic symmetries of the ordered magnet is broken down to a subgroup and the protected degeneracy can be lifted, leading to topological gaps in the magnon spectrum. The topological nature of the consequent magnon bands can be diagnosed using the theory of symmetry indicators \cite{MagneticSIPaperPeng_2022,MTQCElcoro_2021,Po2017} and topological quantum chemistry \cite{TQC1Bradlyn2017TopologicalQC}. Candidate materials for hosting topological magnons are then selected by filtering out those with trivial symmetry indicators for all of their magnetic subgroups. Finally, symmetry-allowed spin models are constructed for the candidate materials to examine the potential emergence of topological magnons induced by external perturbations. Applying this approach to all of the $23$ magnetic insulators in Bilbao Crystallographic Server (BCS) which remain ordered at room temperature, $12$ materials were predicted to host field-induced topological magnons \cite{KarakiSciAdvdoi:10.1126/sciadv.ade7731}. 

\begin{figure*}[h!]
 \hspace{-0.5cm}
 \includegraphics[width=1.13 \textwidth]{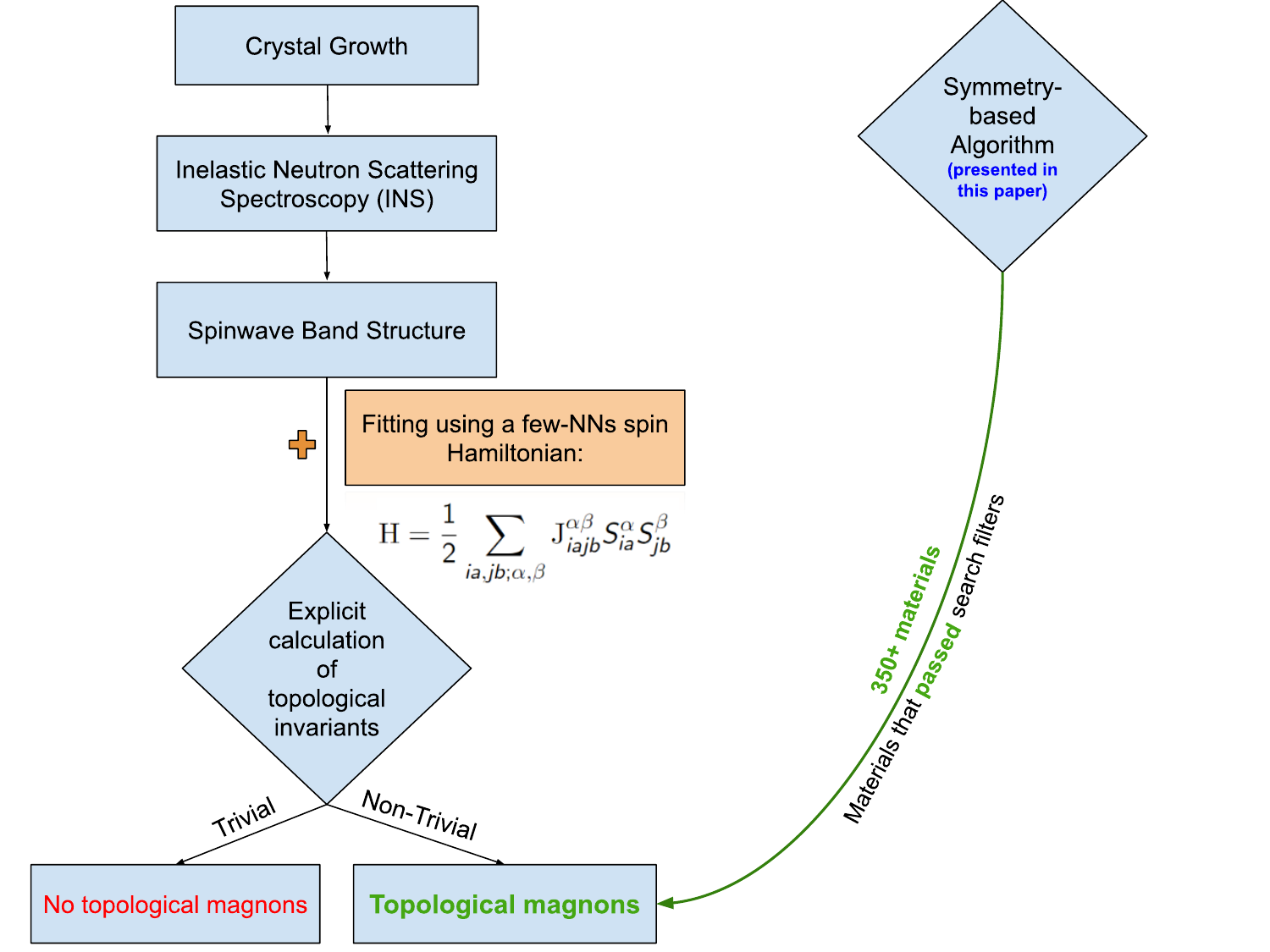}
 \caption{Flowchart summarizing the overall goal of this paper. On the left is the conventional approach used to identify topological magnon phases (including non-type-I topological magnons identified in \cite{KarakiSciAdvdoi:10.1126/sciadv.ade7731} where spin models were based on INS previous fittings or theoretically constructed and band topology was diagnosed through non-trivial $\mathcal{SI}$ values). Alternative approach is introduced through the symmetry-based analysis presented in this work, where more than $350$ materials, out of the $1649$ commensurate magnetic materials on the BCS MAGNDATA database \cite{MAGNDATACommensurateGallego:ks5532}, are found to host field-induced topological magnons.}
 \label{Overall Idea}
 \end{figure*}

However, this symmetry-based approach necessitates case-by-case examinations and detailed knowledge of spin Hamiltonians for potential candidate materials which limits its automation and makes large-scale, high-throughput searches for topological magnon materials challenging. This raises the natural question:  \textbf{Can the symmetry-based approach be generalized to facilitate a large scale material search for topological magnons?} Here, we demonstrate that not only is an efficient and automated search algorithm for topological magnons possible, but it also represents a significant advancement in the field. By eliminating the need for detailed spin models and leveraging symmetry-based approaches, our algorithm enables large-scale, high-throughput discovery of topological magnon materials, far surpassing traditional methods in both speed and scope. Building on the ideas of the previous work \cite{KarakiSciAdvdoi:10.1126/sciadv.ade7731}, we develop a fully automated search algorithm for field-induced topological magnons, and carry it out for all materials with commensurate magnetic orders in BCS. Among all $1649$ commensurate magnetic materials in BCS, we identify 387 candidates ($\sim 23.4 \%$) that host field-induced topological magnons. Flowchart \ref{Overall Idea} summarizes the conventional approach to identify topological magnons and the shortcut enabled by our algorithm presented in workflow diagram \ref{SearchAlgorithm}. 

This paper is organized as follows. In the main text, we start by outlining the theory framework, and then describe our automated search algorithm in detail. We highlight two examples out of our materials predictions, and then discuss the synthetic accessibility of the candidate materials. Finally we discuss the future directions to push forward based on this work. The details of the material search process and outcome are tabulated in the supplemental materials.
 

\section{\label{sec:level1}THEORY FRAMEWORK\protect   }

In this work, we are interested in crystalline solids with a commensurate long-range magnetic order formed by localized magnetic moments that is described by a local order parameter $\langle S_{i}^{\alpha} \rangle$ where $i$ is the lattice site index and $\alpha$ is the component of spin $\vec S_i$. A general bilinear spin Hamiltonian that captures the interactions between the localized moments takes the following form:
\begin{equation}
    \hat{\mathcal{H}}= \frac{1}{2} \sum_{i,j;a,b;\alpha,\beta} J_{ia,jb}^{\alpha,\beta} \hat{S}_{i,a}^{\alpha} \hat{S}_{j,b}^{\beta} - g \mu_{B} \sum_{i,a;\alpha} B^{\alpha} \hat{S}_{i;a}^{\alpha}
\end{equation}
where $i,j$ run over the primitive magnetic cells, $a,b$ run over the magnetic sublattices, and $\alpha,\beta$ run over the spin components. Such a Hamiltonian can describe exchange interactions, single-ion anisotropy and couplings to an external magnetic field. The magnetic order breaks the time reserval and space group $\boldsymbol{G}_H$ symmetry of this Hamiltonian down to a magnetic space group $\boldsymbol{G}_M$, whose elements leave the ordered magnetic moments $\langle S_i^\alpha\rangle$ invariant, therefore $\boldsymbol{G}_M$ is the relevant symmetry group for magnon bands.
Here, we restrict our attention to spin-orbit-coupled magnets where spin rotational symmetries are absent. This makes the relevant magnetic space groups for magnons \textit{single-valued} due to the bosonic nature of the excitations.\\

Spin wave excitations arise from the transverse fluctuations around the ordered moments and they give rise to a spectrum of dispersive, coherently propagating magnons, which is the focus of this study.
~In the framework of topological quantum chemistry (TQC) \cite{TQC1Bradlyn2017TopologicalQC,TQC2PhysRevB.97.035139,TQC3PhysRevB.97.035138,TQCcanOannurev:/content/journals/10.1146/annurev-conmatphys-041720-124134}, the band representations \cite{ZakBandRepresentationsPhysRevB.23.2824} can be constructed based on three symmetry ingredients: the space group ($\mathcal{SG}$) of lattice, the Wyckoff positions ($\mathcal{WP}$), and the nature of the atomic orbitals in the material \cite{TQC1Bradlyn2017TopologicalQC}\cite{TQC2PhysRevB.97.035139}. The band representation of the space group ($\mathcal{SG}$) is obtained by inducing the representation of the site symmetry group $\mathcal{SSG}$ of the $\mathcal{WP}$ to the full space group of the lattice:
\begin{equation}
    \rho_{G} = \rho \uparrow G
\end{equation}

where $\rho$ is a representation of the $\mathcal{SSG}$ which is related to the full space group through the coset decomposition:
\begin{equation}
    G = \cup_{\alpha} g_{\alpha} G_{\boldsymbol{q}}
\end{equation}
where $G_{\boldsymbol{q}}$ is the $\mathcal{SSG}$ of the $\mathcal{WP}$s $\textbf{q}$ of interest, and $g_{\alpha}$ are the coset representatives of $G_{\boldsymbol{q}}$ which map the points $\boldsymbol{q}$ to other points of the $\mathcal{WP}$.
Then this band representation is subduced to representation of the little co-groups of the points of interest in the BZ, usually high-symmetry momenta:
\begin{equation}
     ( \rho \uparrow G ) \downarrow G_{K} \approx \rho_{G_K}
\end{equation}
where $G_{K}$ is the little group of the momentum $K$ of interest. In analogy, to build a magnon band representation \cite{McClartyIdentigyingPhysRevLett.130.206702}, we start with building a representation of the magnetic $\mathcal{SSG}$ $\boldsymbol{G}_{\boldsymbol{q_1}}$ where $\boldsymbol{q}_1$ belongs to Wyckoff position of the magnetic moments,  which is a subgroup of $\boldsymbol{G}_M$ that leaves both the site (up to a primitive lattice translation) and the magnetic order invariant. This translates into the requirement that $\boldsymbol{S}^z$ transforms trivially under elements of $\boldsymbol{G}_{\boldsymbol{q_1}}$:
\begin{equation}
    g_i \boldsymbol{S}_{\boldsymbol{q}_1}^z = \boldsymbol{S}_{\boldsymbol{q}_1}^z \; \; \forall g_i \in \boldsymbol{G}_{\boldsymbol{q_1}}
\end{equation}
where $z$ refers to the local orientation of the magnetic moment. 
This reduces the number of allowed magnetic $\mathcal{SSG}$s to $31$ magnetic point groups that preserve the magnetic order out of the $122$ magnetic point groups \cite{McClartyIdentigyingPhysRevLett.130.206702}. Since magnons are transverse modes constructed from the $\hat{S}^{\pm}$ components, the allowed representations of the $31$ magnetic point groups compatible with the magnetic order are determined by how the transverse spin components transform under $\mathcal{SSG}$. The magnetic SSGs compatible with the magnetic order along with the induced pair of elementary band representations (reflecting the particle-hole symmetry of the spin wave Hamiltonian) were worked out and tabulated \cite{McClartyIdentigyingPhysRevLett.130.206702}. The band co-representations induced from a chosen $\mathcal{WP}$ of a $\mathcal{MSG}$ have been worked out \cite{MTQCElcoro_2021,Cata4444Xu_2020} and accessible via the BCS tool MBANDREP \cite{MTQCElcoro_2021,Cata4444Xu_2020}.
\\

TQC builds on enumerating all of the topologically-trivial band structures which are smoothly connected to the atomic insulator limit, constructed from a Wannier basis respecting all symmetries of the structure \cite{TQC1Bradlyn2017TopologicalQC,TQC2PhysRevB.97.035139,TQCcanOannurev:/content/journals/10.1146/annurev-conmatphys-041720-124134}. For any given space group, the band representations corresponding to a trivial atomic insulator can be identified by checking if a specific band representation belongs to one of the Elementary Band Representations (EBRs) or their composites. If not, the corresponding band is topological. The associated band representations at the high-symmetry points are referred to as Symmetry Indicators ($\mathcal{SI}s$) for band topology \cite{SI111Po_2017,SIElsamKhalafPhysRevX.8.031070}.\\

Technically, an isolated set of bands is characterized by a \textit{symmetry vector} $\boldsymbol{b}=\{\boldsymbol{n_k^{\alpha}}\}$ composed of $\boldsymbol{n_k^{\alpha}}$ multiples of the little groups' irreducible representations $\boldsymbol{\alpha}$ of the distinct high-symmetry momenta $\boldsymbol{k}$ appearing in the set of bands \cite{SI111Po_2017,MagneticSIPaperPeng_2022}. Such a vector is named a \textit{Band Structure} and its components are constrained by the \textit{compatibility relations} that determine how the \textit{irreps} at the high-symmetry points are connected to those at another \textbf{K} point while maintaining the gap condition. The set of all allowed band structures are denoted by $\{\boldsymbol{BS}\}$. A subgroup of $\{\boldsymbol{BS}\}$ is the set of atomic insulators (denoted by $\{\boldsymbol{AI}\}$) which are induced from symmetry-respecting localized Wannier orbitals. Therefore, given a band structure $\boldsymbol{b}$, we can diagnose its topology as follows. If $\boldsymbol{b} \in \{\boldsymbol{BS}\}$ and $\boldsymbol{b}\notin \{\boldsymbol{AI}\}$, these bands are not smoothly connected to the atomic limit and therefore are of non-trivial topology. If $\boldsymbol{b} \in \{\boldsymbol{AI}\}$, then symmetry alone is inconclusive regarding the existence of nontrivial topology. Last, $\boldsymbol{b} \notin \{\boldsymbol{BS}\}$ indicates a violation of the compatibility relations, and therefore a full gap cannot exist at all high-symmetry points in the BZ.\\

By allowing components of $\boldsymbol{b}$ to be negative integers, we can define an inverse operation to the direct sum of two representations, and therefore the sets $\{\boldsymbol{BS}\}$ and $\{\boldsymbol{AI}\}$ are now abelian groups of the same rank. The $SI$ group is defined as the quotient group \cite{SI111Po_2017}\cite{SIin230SGsPo_2017}:
\begin{equation}
    X_{BS}=\frac{\{\boldsymbol{BS}\}}{\{\boldsymbol{AI}\}}=\mathbb{Z}_{n_1} \times \mathbb{Z}_{n_2} \times ...
\end{equation}
The $\mathcal{SI}s$ for $\mathcal{MSG}$s have been tabulated in \cite{MagneticSIPaperPeng_2022,
MTQCElcoro_2021}. This forms a key ingredient to our search algorithm \ref{SearchAlgorithm} whose main strategy is to break the $\mathcal{MSG}$ down into one subgroup with nontrivial $\mathcal{SI}$ group through an external perturbation. The $\mathcal{SI}$ group of the perturbation-induced magnon band gaps are then calculated, while considering all different orders of the band \textit{irreps} both before and after applying the perturbation.\\

Although the induced magnon band representation encodes connectivity of the magnon bands through \textit{compatibility relations}, and symmetry-protected degeneracies encoded in the dimension of the \textit{irreps}, it does not determine energetics of the magnon bands. This leaves each high symmetry momentum with a number of possibilities originating from the different orderings of the \textit{irreps}. Therefore, combining these different orders of the \textit{irreps} at the high symmetry momenta, as long as the \textit{irreps} are connected according the \textit{compatibility relations}, gives a number of possibilities for the magnon band structure of a given magnetic material. This is why in our search algorithm shown in Fig. \ref{SearchAlgorithm}, it is necessary to sweep over all of the different orders of \textit{irreps}. This originates from the lack of spin models and reliable \textit{ab initio} calculations for magnon band structures which, if available, would collapse these possibilities to one specific order of \textit{irreps}.

\section{\label{sec:level2}SEARCH ALGORITHM\protect   }
\begin{figure*}[ht!]
 \hspace{-1.22cm}
 \includegraphics[width=1.12 \textwidth ]{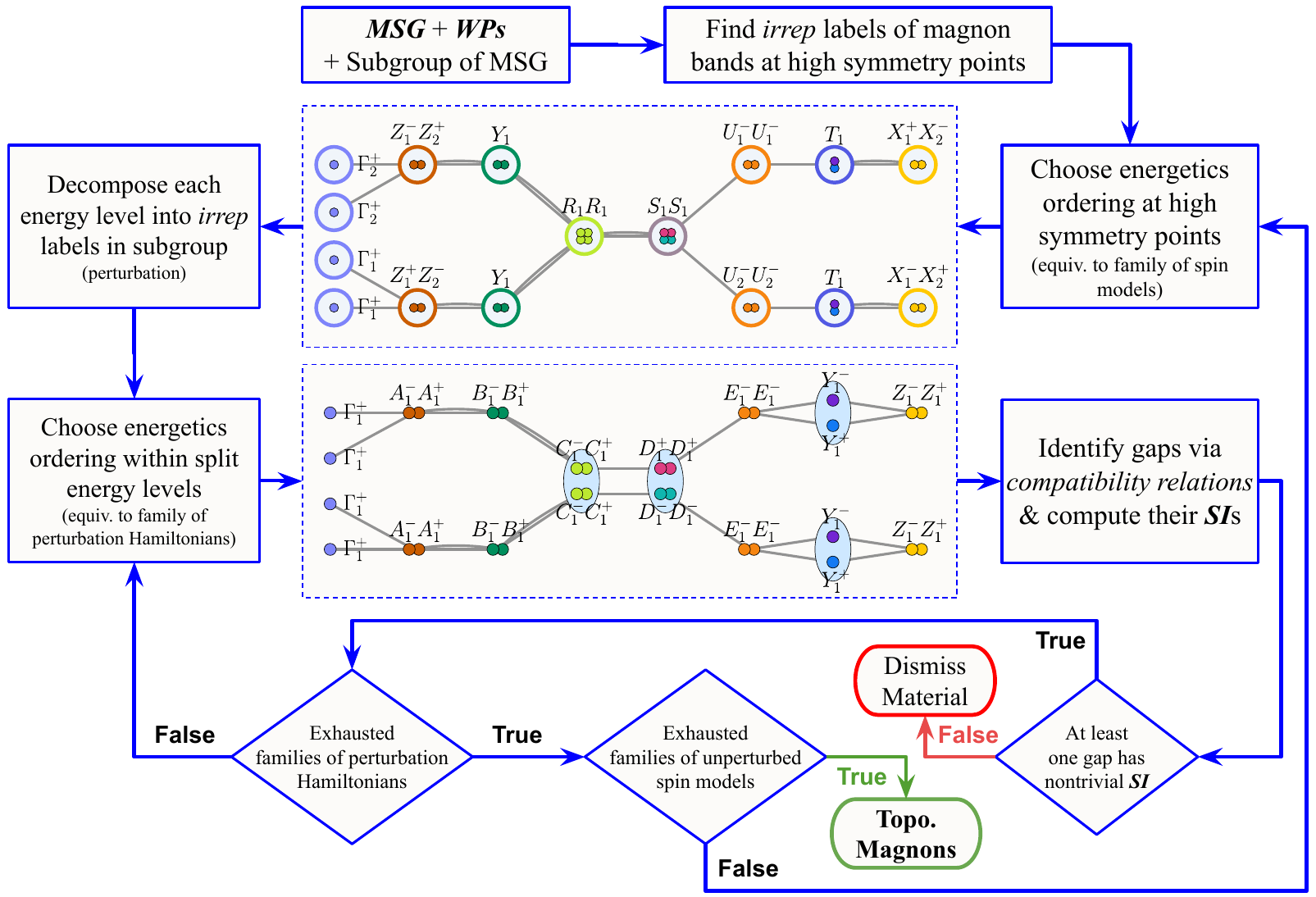}
 \caption{\textbf{Workflow diagram}. Flowchart summarizing the automated search algorithm for finding induced nontrivial magnon band topology given a generic magnetically-ordered structure.}
 \label{SearchAlgorithm}
 \end{figure*}
In this section, we describe the automated search algorithm, which is summarized as a flowchart diagram in Fig.~\ref{SearchAlgorithm}. Firstly, we classify all commensurate magnetic materials from BCS MAGNDATA tool \cite{MAGNDATACommensurateGallego:ks5532} according to their magnetic structure (i.e. their $\mathcal{MSG}$ and $\mathcal{WP}$). Secondly we filter out $\mathcal{MSG}s$ whose subgroups all have a trivial ($\mathbb{Z}_1$) SI group. This constraint excludes $345$ out of the $1651$ $\mathcal{MSG}s$. Among the $1649$ commensurate materials on BCS MAGNDATA \cite{MAGNDATACommensurateGallego:ks5532}, $1263$ pass this filter. Thirdly, we restrict to $\mathcal{MSG}s$ and $\mathcal{WP}s$ that host symmetry-enforced magnon band degeneracy that can be split upon the application of a symmetry-breaking external perturbation, such as a magnetic field $\boldsymbol{B}$, an electric field $\boldsymbol{E}$, or a mechanical strain $\boldsymbol{\sigma}$. This translates to searching for structures that give rise to at least one multi-dimensional \textit{irrep} that can be decomposed into lower dimensional \textit{irreps} upon the symmetry-lowering process. Out of the $1263$ materials that passes the first filter, $1171$ of them passes the second filter and proceed as following. For a given magnetic structure, the input data consists of two parts:
\begin{enumerate}
    \item  $\mathcal{MSG}$ of the material and the $\mathcal{WP}$s of its magnetic moments.
    \item a symmetry-indicated subgroup of the $\mathcal{MSG}$ corresponding to the perturbation of interest.
\end{enumerate}
The output of the algorithm is a binary result, with two possible outcomes:
\begin{itemize}
    \item[] {\bf Positive:} Based solely on the symmetry of the magnetic phase, the material will host topological magnon bands upon the application of the specified perturbation.
    \item[] {\bf Negative:} Symmetry is not sufficient to predict topological magnons under the specified perturbation. Topological magnons are not ruled out, but their presence depends on the spin interaction details in the material.
\end{itemize}
The algorithm proceeds in three main steps:
\begin{enumerate}
    \item\textbf{Iterate over different orderings of the original $\mathcal{MSG}$ \textit{irreps}.} This step aims to exhaust all equivalence classes of unpertrubed magnon Hamiltonians, where two Hamiltonians are said to belong to the same equivalence class if they share the same energetic ordering of irreps at high-symemtry momenta.
    \item\textbf{Iterate over different orderings of \textit{irreps} splitting upon small perturbations.} In this step, a perturbation is said to be small if the irrep splitting does not, at any momentum $\boldsymbol{k}$, swap the ordering of two subgroup irreps. This step is intended to exhaust all equivalence classes of small perturbation Hamitlonians.
    \item\textbf{Identify all gaps and evaluate their SIs.} For each energetic ordering associated with a perturbed Hamiltonian, the band gaps are verified by checking whether the collection of bands below a potential gap satisfy the compatibility relations of the subgroup. Concretely, this is decided based on whether the symmetry vector $\vec{b}$ of these bands belongs to the null space of the compatibility relation matrix $\mathcal{C}$.
    For each identified gap, we compute its $\mathcal{SI}$ \cite{MagneticSIPaperPeng_2022} to determine if it is topologically non-trivial.
\end{enumerate}

The algorithm terminates in two cases. First, if no gap with non-trivial topology is identified in step (3) for \textit{any} (equivalence class of) perturbed Hamiltonian, the algorithm terminates with a negative result. On the other hand, once all equivalence classes of perturbation Hamiltonians are exhausted (steps 1 and 2) without encountering a termination, the algorithm terminates with the positive result. This exhaustion ensures the existence of at least one topological gap upon applying the relevant perturbation.\\

\begin{figure}[H]
\centering
\includegraphics[scale=0.6]{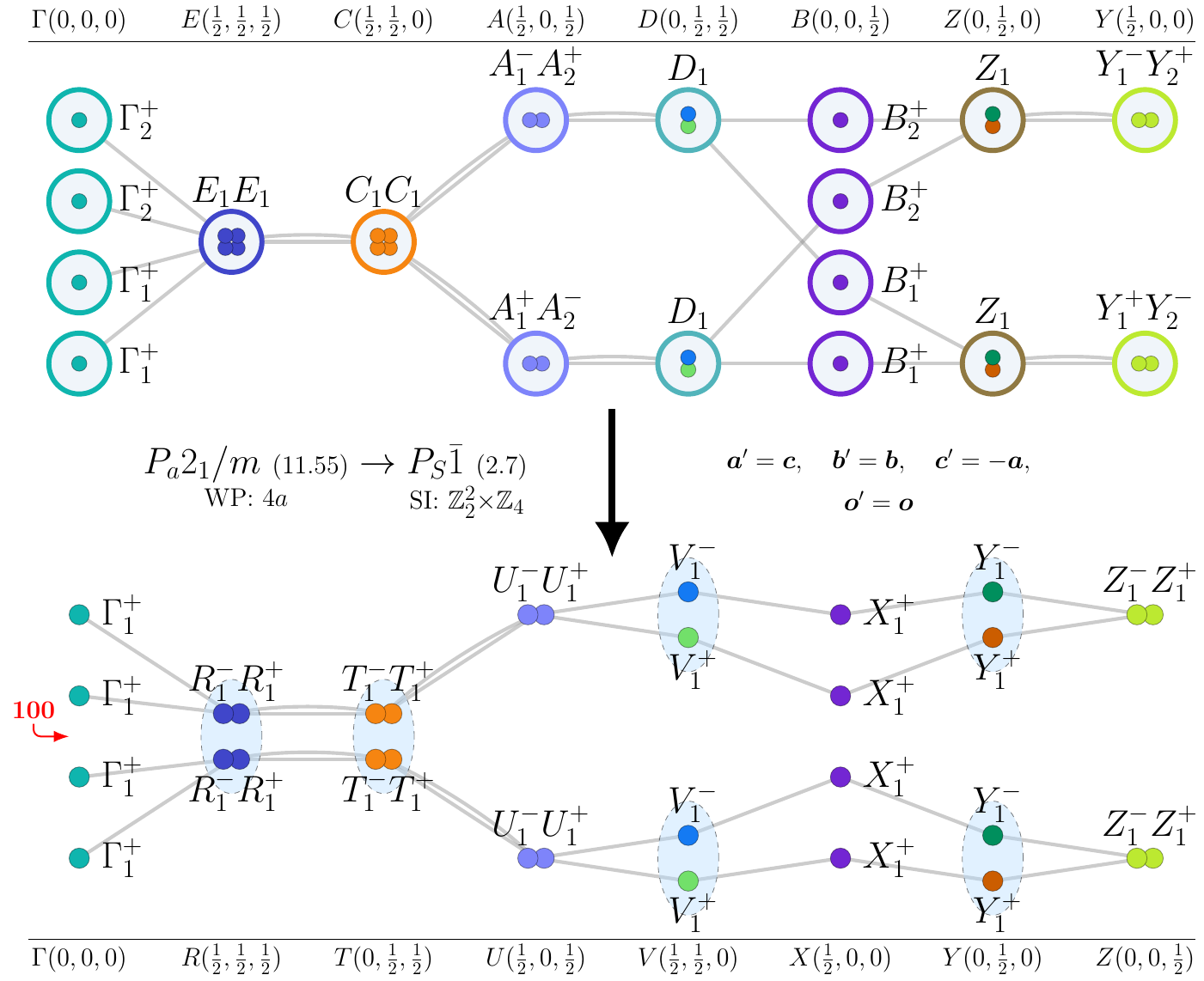}
\caption{Topological magnon bands in subgroup $P_{S}\bar{1}~(2.7)$ for magnetic moments on Wyckoff position $4a$ of supergroup $P_{a}2_{1}/m~(11.55)$.\label{fig_11.55_2.7_strainingenericdirection_4a}}
\end{figure}

Out of the $1171$ magnetic materials in BCS that passes the aforementioned second filter, $387$ successful candidates pass our search algorithm and therefore they are guaranteed to host topological magnons induced by the relevant perturbation. We present a supplementary section \ref{TopoMag2} where we summarize all topological magnon candidate materials along with the required perturbation to realize a guaranteed topological magnon phase. In subsequent supplementary sections (starting from section \ref{SuppSection1Label}), we provide detailed information of the magnetic structures that passed our search filters. Each section contains the BCS materials that belongs to this magnetic structure and subsections of the successful subgroups that met our filtering criteria. Each of those subsections has a list of the possible perturbations required to realize that subgroup, a table of the number of times a given $\mathcal{SI}$ can appear in the magnon spectrum, a table of the possible $\mathcal{SI}$ values for each band gap and one possible schematic arrangement of the magnon band structure. Each schematic picture (e.g. figure \ref{fig_11.55_2.7_strainingenericdirection_4a}) encodes the unperturbed magnon band representation, \textit{compatibility relations} between \textit{irreps}, the symmetry-lowering process, the induced band gaps and the $\mathcal{SI}$ values for this particular order of \textit{irreps}.

\begin{figure*}[h]
 \hspace{5.2cm}
 \includegraphics[scale=0.6]{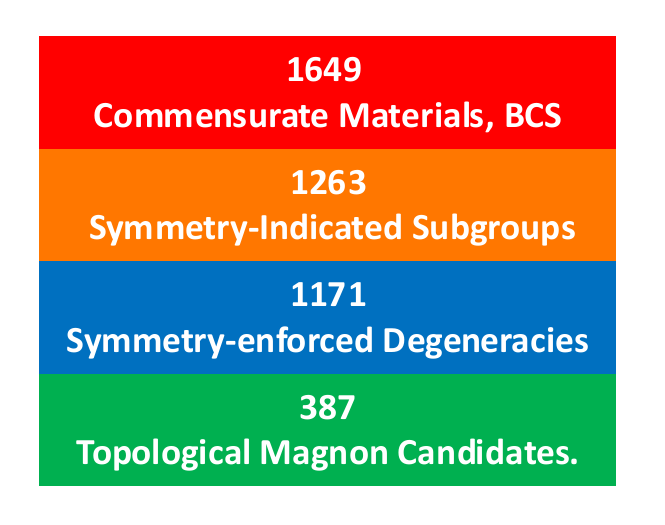}
 \caption{A summary of the search program. Out of the 1649 commensurate magnetic materials in the BCS database, 387 admit perturbation-induced type-I topological magnons. By type-I \cite{KarakiSciAdvdoi:10.1126/sciadv.ade7731}, we refer to those which admit the symmetry indicated topological magnons regardless the form of the perturbation as long as it breaks the $\mathcal{MSG}$ into the relevant magnetic subgroup. A first filter excludes $386$ materials that belong to $\mathcal{MSG}s$ without symmetry-indicated subgroups. A second filter excludes $92$ materials that do not host symmetry-enforced degeneracies in their magnon spectrum as a band gap cannot be induced by perturbations at all high symmetry momenta. This results in $1171$ materials that, after passing through our search algorithm \ref{SearchAlgorithm}, ended up with $387$ successful candidates. }
 \label{2}
 \end{figure*}

\section{\label{ExamplesSection}Examples\protect   }
To demonstrate our search results, we discuss two examples in detail, by working out the linear spin wave spectra of their spin Hamiltonians and identifying the magnon band representations therein to validate the predictions of automated search through calculating the $\mathcal{SI}s$ of the induced magnon band gaps. We focus on two candidate materials, both of which host Weyl points in their magnon spectra, known as Weyl magnons \cite{Li2016,WeylMagnonsMookPhysRevLett.117.157204}. As tabulated in the supplemental file, it turns out that many candidate materials hosting Weyl magnons are characterized by a topological invariant $\nu$, associated with Weyl-$\mathcal{SI}$ groups whose nontrivial values indicate the existence of bulk Weyl points \cite{MagneticSIPaperPeng_2022} that can either lie on a high-symmetry plane in the Brillouin zone or at generic momenta depending on the $\mathcal{MSG}$, along with surface magnon arcs \cite{KarakiSciAdvdoi:10.1126/sciadv.ade7731,WeylMagnonsMookPhysRevLett.117.157204} resembling Fermi arcs in electronic Weyl semimetals. In the first example, NdMnO$_3$, we analyze a case where Weyl magnons are $\mathcal{SI}$-signaled to lie on a high-symmetry momentum plane. To validate our prediction based on the tabulated values of the $\mathcal{SI}s$, we build a symmetric spin model, calculate its magnon spectrum, identify Weyl points on that high-symmetry plane, and identify their associated surface magnon arcs after terminating the periodicity of the lattice on one surface. In our second example, Ca$_2$RuO$_4$, we encounter a different situation where Weyl points exist at generic momenta, causing different Chern numbers on different planes in the momentum space. For this end, we check that the tabulated $\mathcal{SI}$ of the gap of interest is consistent with explicit calculations of the magnon spectrum. We then identify the Weyl magnons and validate the variance of the Chern numbers between the different high-symmetry planes.
\subsection{NdMnO$_3$}


\begin{figure}[ht!]
  \hspace{-1.22cm}
  \subfloat[]{\includegraphics[scale=0.16]{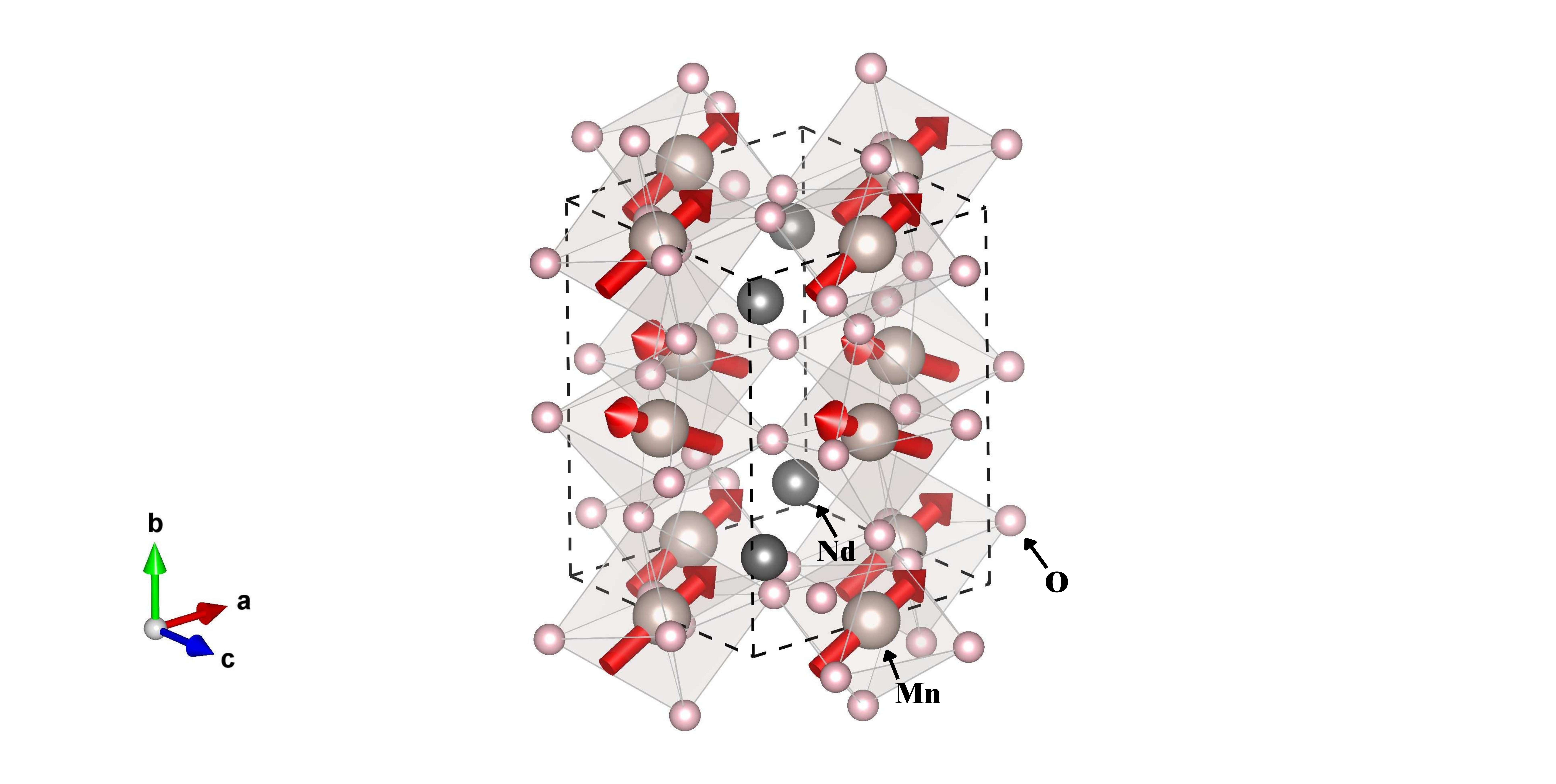}\label{Example1UnitCell}}
    
  \hspace*{-0.37cm}
  \subfloat[]{
    \raisebox{0cm}{\includegraphics[width=1.00\textwidth]{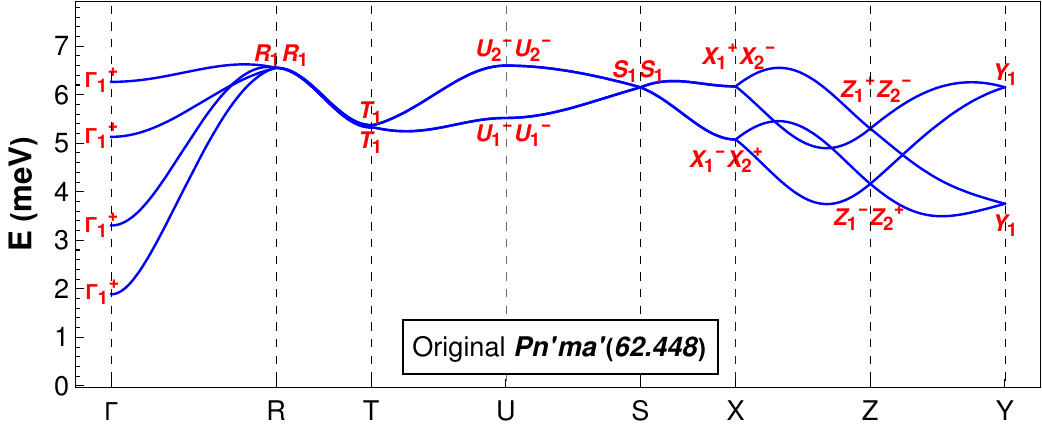}}
\label{Example1UnperturbedSpectrum}}  
  \caption{a) \textbf{Magnetic unit cell of NdMnO$_3$}. Light pink spheres indicate the Oxygen sites which form corner-sharing networks around the Mn rose spheres while black are Nd sites. Size of Nd spheres is reduced relative to those of Mn for a better visualization of the magnetic order. Red arrows indicate the magnetic moment direction of each of the four Mn magnetic sublattices. Spins mostly align along the $a$-axis with a small canting towards the $b$-axis. Dashed lines indicate the boundaries of the magnetic unit cell with the inset highlighting~\textit{a,b,c} axes. b) \textbf{Unperturbed magnon spectrum for NdMnO$_3$}. Red labels are the magnon band representations at high symmetry points ($\mathcal{HSP}s$). Band degeneracies are consistent with the dimensionality of the induced band representation discussed in the text which enforces four-fold degeneracy at $\mathcal{R}$,$\mathcal{S}$ and two-fold at all other $\mathcal{HSP}s$ except for the $\Gamma$ point that has four seperate bands (i.e. four 1D \textit{irreps}).}
  \label{Example1Unperturbedspectrum}
\end{figure}
Our first example is the orthorhombic perovskite NdMnO$_3$ in which Mn$^{2+}$ moments become magnetically ordered below a transition temperature $T_N = 78K$ \cite{NdMnO3Paper2000}. The magnetic structure belongs to $\mathcal{MSG}$ $\mathit{Pn'ma'}$ with the Mn$^{2+}$ ions located at $\mathcal{WP}$ 4b. The magnetic structure is such that the magnetic moments align antiferromagnetically along the $a$-axis and they acquire a small ferromagnetic component along the $b$-axis (see figure \ref{Example1UnitCell}). It turns out that below a second transition temperature of around $T_N^\prime = 15K$, the Nd$^{4+}$ moments located at $\mathcal{WP}$ 4c become magnetically ordered as well \cite{NdMnO3SecondPhasePhysRevB.96.014427}. In the low-temperature phase, both Mn$^{2+}$ moments at $\mathcal{WP}$s $4b$ and Nd$^{4+}$ moments $4c$ exhibit long-range magnetic orders. It turns out that both phases pass our filters, hosting topological magnons as detailed in the supplementary material.



Here we focus on the first phase in which only the $4b$ $\mathcal{WP}$ moments are magnetically ordered. The magnon band representation induced from this $\mathcal{WP}$ to the full $\mathcal{MSG}$ is:
\begin{align}
    (A_g)_{4b} \uparrow Pn'ma' \: (62.448) &= 2 \Gamma_1^+ (1) \: \oplus 
2 \Gamma_2^+ (1) \: \oplus R_1 R_1 (4) \: \oplus S_1 S_1 (4) \: \oplus 
2 T_1 (2) \nonumber\\
&\:\:\:\:\:\oplus U_1^- U_1^- (2) \: \oplus X_1^+ X_2^- (2) \: 
\oplus 2 Y_1^- (2) \: \oplus Z_1^+ Z_2^- (2) \: \oplus Z_1^- Z_2^+ (2).
\end{align}
therefore there is a symmetry-enforced four-fold degeneracy at $\mathcal{R}$ and $\mathcal{S}$, two-fold degeneracy at the $\mathcal{T},\mathcal{U},\mathcal{X},\mathcal{Y},\mathcal{Z}$ and no degeneracy at the $\Gamma$. Upon applying a magnetic field along the $a$-axis, the $\mathcal{MSG}$ is broken down into the subgroup $P2^\prime_1/c^\prime$ which can also be achieved using a uniaxial strain perpendicular to the $[001]$ direction. This perturbation results in the decomposition of the little groups \textit{irreps} of the $\mathit{Pn'ma'}$ $\mathcal{MSG}$ into \textit{irreps} of the $P2^\prime_1/c^\prime$ subgroup. To see this we firstly note that the standard axes and origin of the magnetic subgroup are related to those of the original $\mathcal{MSG}$ (unmarked) through the transformation:
\begin{equation}
\begin{aligned}
& a \rightarrow a^{\prime}=-b \\
& b \rightarrow b^{\prime}=-c \\
& c \rightarrow c^{\prime}=a \\
& o \rightarrow o^{\prime}=o
\end{aligned}    
\end{equation}
Then, the \textit{irreps} decomposition proceeds as following. At $A\left(\frac{1}{2}, 0, \frac{1}{2}\right)$, equivalent to $S\left(\frac{1}{2}, \frac{1}{2}, 0\right)$ in the original $\mathcal{MSG}$,

\begin{equation}
    S_1 S_1(4) \rightarrow 2 A_1^{-} A_1^{+}(2)
\end{equation}

At $B\left(0,0, \frac{1}{2}\right)$, equivalent to $X\left(\frac{1}{2}, 0,0\right)$ in the original $\mathcal{MSG}$,

\begin{equation}
   \begin{aligned}
& X_1^{+} X_2^{-}(2) \rightarrow B_1^{-} B_1^{+}(2) \\
& X_1^{-} X_2^{+}(2) \rightarrow B_1^{-} B_1^{+}(2)
\end{aligned} 
\end{equation}

At $C\left(\frac{1}{2}, \frac{1}{2}, 0\right)$, equivalent to $T\left(0, \frac{1}{2}, \frac{1}{2}\right)$ in the original $\mathcal{MSG}$,

\begin{equation}
    T_1(2) \rightarrow C_1^{-} C_1^{+}(2)
\end{equation}

At $D\left(0, \frac{1}{2}, \frac{1}{2}\right)$, equivalent to $U\left(\frac{1}{2}, 0, \frac{1}{2}\right)$ in the original $\mathcal{MSG}$,

\begin{equation}
   \begin{aligned}
& U_1^{+} U_1^{+}(2) \rightarrow D_1^{+} D_1^{+}(2) \\
& U_1^{-} U_1^{-}(2) \rightarrow D_1^{-} D_1^{-}(2) \\
& U_2^{+} U_2^{+}(2) \rightarrow D_1^{+} D_1^{+}(2) \\
& U_2^{-} U_2^{-}(2) \rightarrow D_1^{-} D_1^{-}(2)
\end{aligned} 
\end{equation}

At $E\left(\frac{1}{2}, \frac{1}{2}, \frac{1}{2}\right)$, equivalent to $R\left(\frac{1}{2}, \frac{1}{2}, \frac{1}{2}\right)$ in the original $\mathcal{MSG}$,

\begin{equation}
    R_1 R_1(4) \rightarrow E_1^{+} E_1^{+}(2) \oplus E_1^{-} E_1^{-}(2)
\end{equation}

$\operatorname{At} \Gamma(0,0,0)$,

\begin{equation}
    \begin{aligned}
& \Gamma_1^{+}(1) \rightarrow \Gamma_1^{+}(1) \\
& \Gamma_1^{-}(1) \rightarrow \Gamma_1^{-}(1) \\
& \Gamma_2^{+}(1) \rightarrow \Gamma_1^{+}(1) \\
& \Gamma_2^{-}(1) \rightarrow \Gamma_1^{-}(1)
\end{aligned}
\end{equation}

At $Y\left(\frac{1}{2}, 0,0\right)$, equivalent to $Y\left(0, \frac{1}{2}, 0\right)$ in the original $\mathcal{MSG}$,
\begin{equation}
    Y_1(2) \rightarrow Y_1^{+}(1) \oplus Y_1^{-}(1)
\end{equation}

Finally, at $Z\left(0, \frac{1}{2}, 0\right)$, equivalent to $Z\left(0,0, \frac{1}{2}\right)$ in the original $\mathcal{MSG}$,

\begin{equation}
    \begin{aligned}
& Z_1^{+} Z_2^{-}(2) \rightarrow Z_1^{-} Z_1^{+}(2) \\
& Z_1^{-} Z_2^{+}(2) \rightarrow Z_1^{-} Z_1^{+}(2)
\end{aligned}
\end{equation}

therefore the four-dimensional \textit{irreps} of the original $\mathcal{MSG}$ at the $\mathcal{R,S}$ points decompose into two two-dimensional \textit{irreps} of the magnetic subgroup $\mathit{P2^\prime_1/c^\prime}$, therefore inducing a band gap in the spectrum. \\



\begin{figure}[ht!]
  \hspace{-0.45cm}
  \subfloat[]{
    \includegraphics[width=1.0\textwidth]{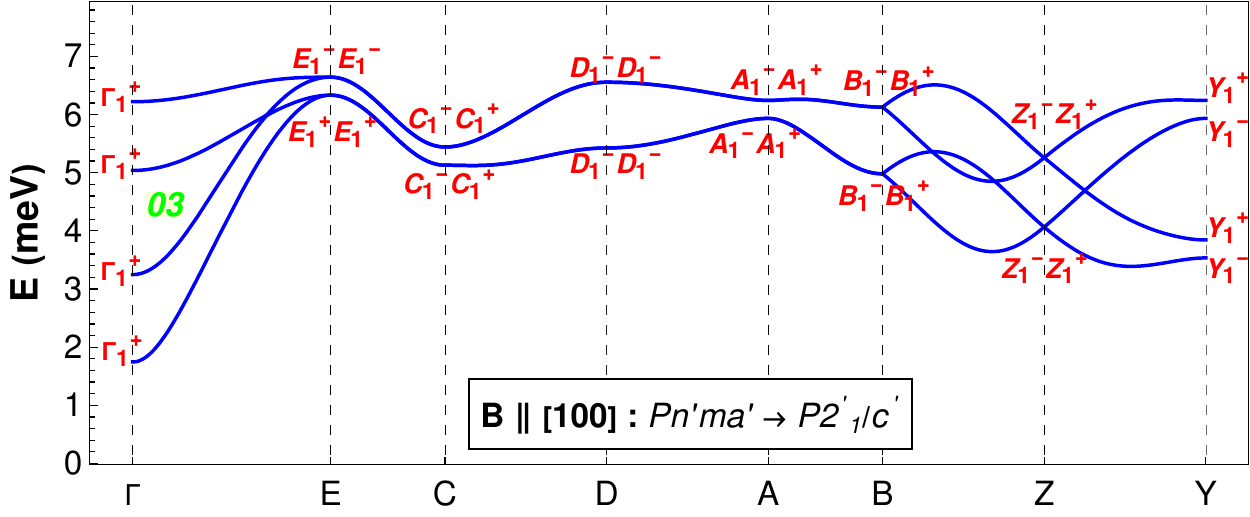}
    \label{Example1Perturbed}}
  \newline
  \hspace*{3.5cm}
  \subfloat[]{
    \raisebox{0cm}{\includegraphics[width=.55\textwidth]{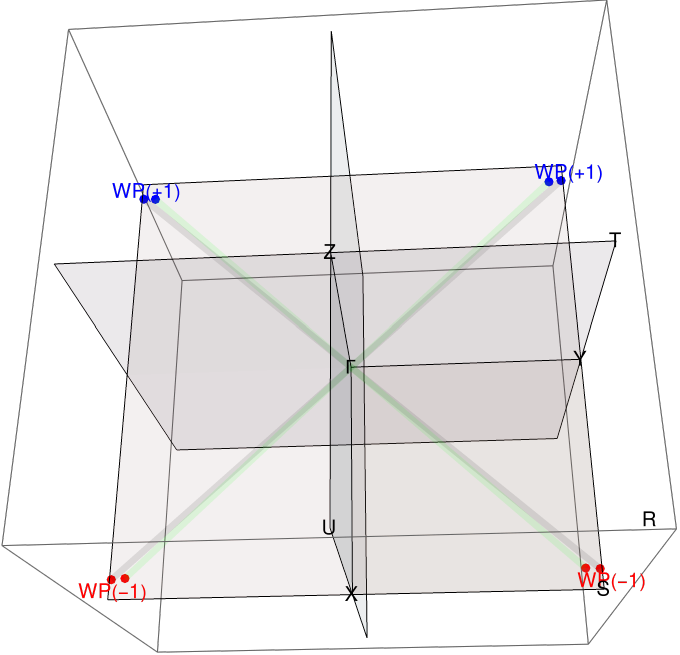}}
    \label{Example1WeylPoints}
  }  
  \caption{Magnetic field-induced Weyl points in the magnon band structure of \textbf{NdMnO}$_3$. a) \textbf{Perturbed magnon spectrum for NdMnO$_3$.} By breaking the $\mathcal{MSG}$ into the $P2^\prime_1/c^\prime$ subgroup through applying a magnetic field parallel to the $a$ axis, the four-fold degeneracies at $R,S$ points break into two two-folds, and therefore a band gap is opened at all of $\mathcal{HSP}s$. Calculated value of the band gap $\mathcal{SI}$ is labelled in green. Red labels are the magnon representations at the $\mathcal{HSP}s$. b) \textbf{Induced Weyl magnons in NdMnO$_3$} with blue and red labels indicate locations of Weyl points (labelled WP) of positive and negative charge respectively. Each Weyl point on the graph is connected to its inversion symmetric partner as  indicated by the dark lines. Due to the symmetries discussed in the main text, for each Weyl point $(k_x,k_y,0)$ there is a weyl point of the same charge located at $(k_x,-k_y,0)$, in addition to an opposite charge located at $(-k_x,-k_y,0)$ related by inversion, giving rise to a pair of magnon arcs between projections of Weyl points as shown in \ref{Example1FermiArc}. Solid green and black lines in the picture connect inversion-related weyl partners.}
  \label{Example1Figure2}
\end{figure}

To illustrate the guaranteed topological magnons in our materials search predictions, we explicitly construct a symmetric spin model for this material based on couplings up to third nearest neighbors (see methods \ref{SpinModelOfNdMnO3}). These values were chosen such that in the classical ground state of the Hamiltonian, ordered moments align along the experimentally-reported ground state directions \cite{NdMnO3Paper2000} shown in \ref{Example1UnitCell}. By performing a linear spinwave calculation around that ground state, the magnon band structure is obtained (see figure \ref{Example1UnperturbedSpectrum}). Now, we apply a small magnetic field, $3$ T, along the $a$-axis to achieve the symmetry breaking into the $P2^\prime_1/c^\prime$ magnetic subgroup. Upon turning the field on, a gap opens between the first and second pair of magnon bands at the High Symmetry Points ($\mathcal{HSP}s$) of the BZ. This allows us to calculate the $\mathcal{SI}$ of the induced band gap, using the subgroup $P2^\prime_1/c^\prime$ $\mathcal{SI}$ group $\mathbb{Z}_2 \times \mathbb{Z}_4$ which can be calculated as a function of the \textit{irreps} of the \textit{maximal} \textbf{k}-points:\begin{align}
z_{1} &=  n^{\Gamma^+_1}\mod \,2 \\
z_{2} &= 2 n^{D^+_1 \, D^+_1} + 2 n^{E^+_1 \, E^+_1} + n^{\Gamma^+_1} + 3 n^{Y^+_1}  \mod \,4 
   \end{align}
which we calculate using \textit{irreps} of the perturbed spectrum (shown in \ref{Example1Perturbed}) to be $03$ signaling Weyl points pinned to the high-symmetry plane $k_z=0$. This $\mathcal{SI}$ matches our automated search predictions (see supplementary material \ref{NdMnO3SuppSection}) for a nontrivial topology associated with the magnon spectrum gap. 

\begin{figure}[h]
\hspace{-2cm}
  \subfloat[]{
    \includegraphics[width=0.77\textwidth]{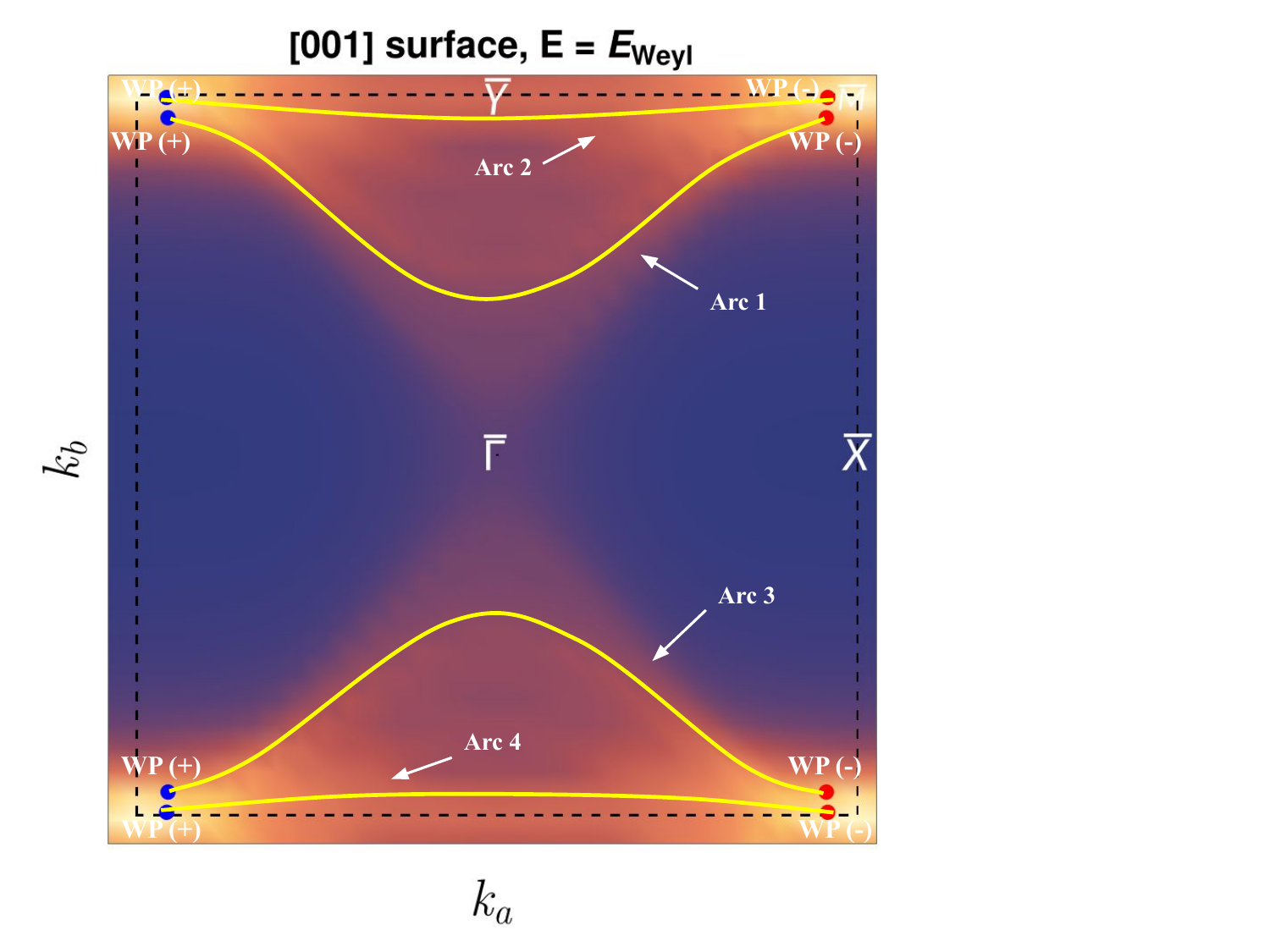}
    \label{Example1FermiArc}
  }
  \hspace{-4.5cm} 
  \vspace{0cm}
  \subfloat[]{
    \includegraphics[width=0.6\textwidth]{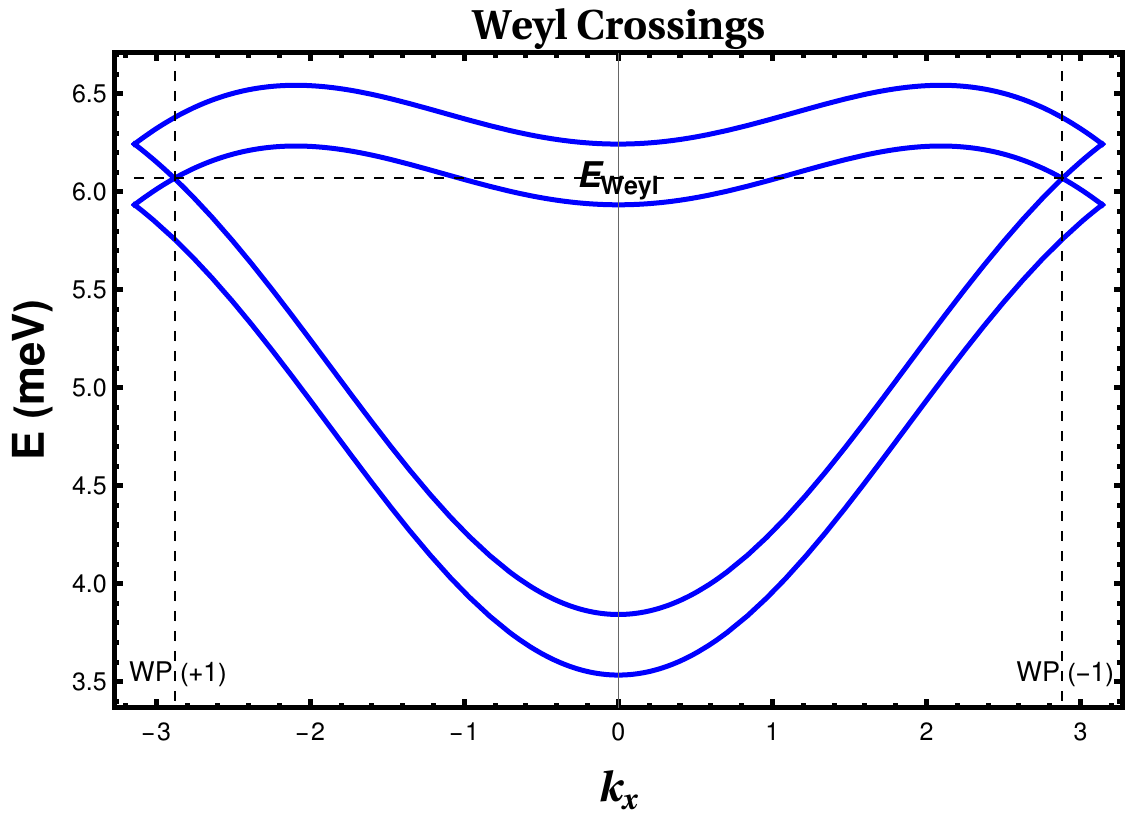}
    \label{Example1BulkWeylCrossings}
  }
  \caption{(a) \textbf{Surface spectral density} of magnons in NdMnO$_3$, in a semi-infinite system along the $c$-axis as a function of $(k_x,k_y)$. \textbf{WP} refer to the projections of the bulk Weyl points over the surface Brillouin zone with the sign indicating the chirality. Curves of highest (i.e., brightest) spectral density connecting the WPs are marked with yellow representing the magnon surface arcs. Projections of Weyl points of opposite chirality are connected with magnon surface arcs analogous to Fermi arcs in electronic Weyl semi-metals. (b) \textbf{Weyl points} in the magnon spectrum of NdMnO$_3$ between the middle two bulk bands. This is plotted on a line connecting two Weyl points with $(k_y=3.117,k_z=0)$. These two Weyl points are related by two-fold screw $\{2_{100} | \frac{1}{2} \frac{1}{2} \frac{1}{2} \}$.}
  \label{Example1figure3fermiarcplusWPs}
\end{figure}

\begin{figure*}[h!]
 \hspace{-0.3cm}
 \includegraphics[scale=0.7]{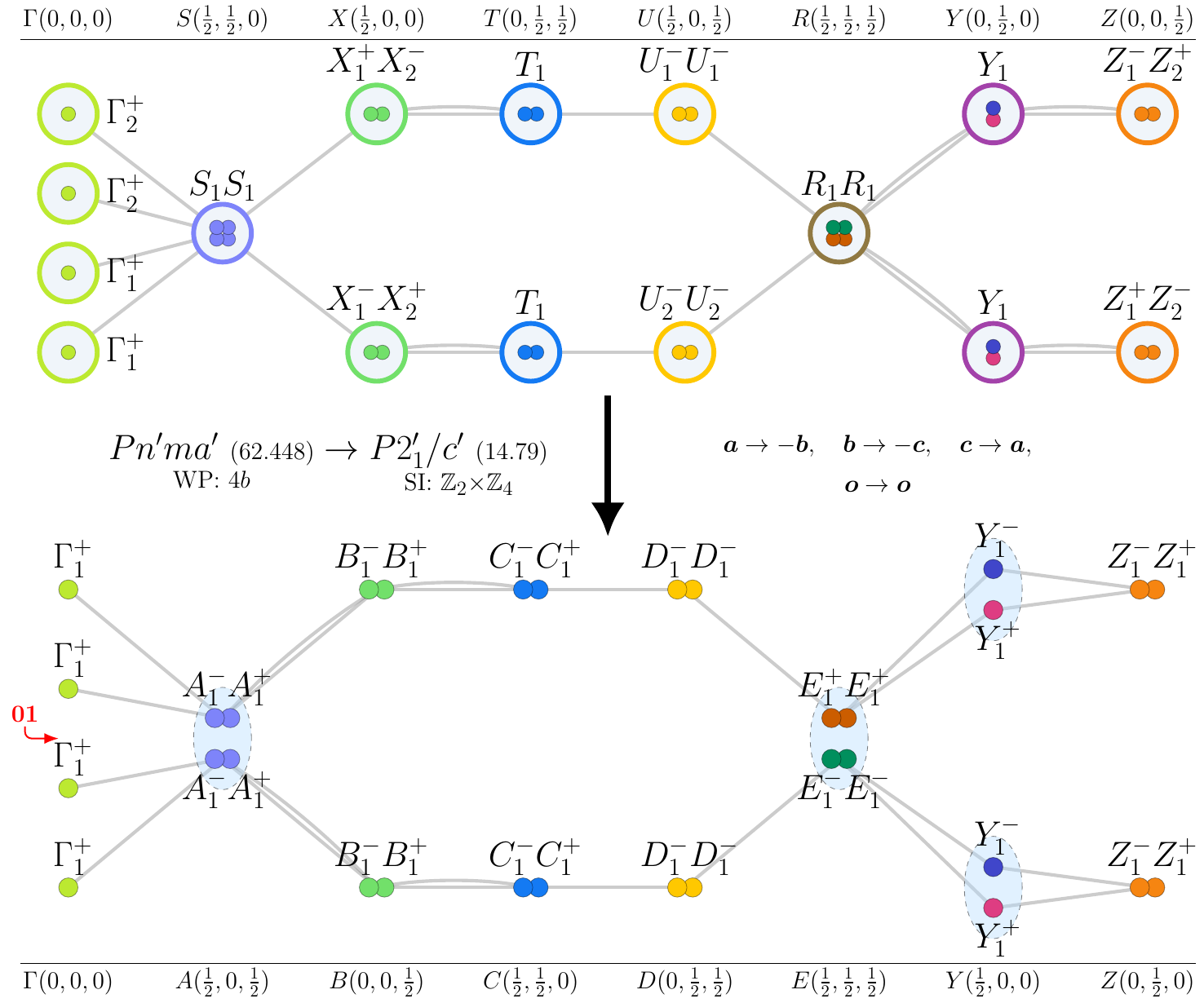}
 \caption{\textbf{Alternative schematic representation of the magnon band structure in NdMnO$_3$.} Upon the symmetry breaking discussed in the main text, the resulting order of the \textit{irreps} at the $\mathcal{HSP}s$ is not unique. This is one possibility the band structure might fall into. Unlike the constructed spin model that produced a $03$ $\mathcal{SI}$, a different spin model can give rise to a different possibility like the one presented here in which the $\mathcal{SI}$ is $03$. Regardless of the spin Hamiltonian details, the $\mathcal{SI}$ belongs to the same topological class. }
 \label{Example1OtherSchematicPicture}
 \end{figure*}


 In fact, any spin model, as long as it respects the space group symmetries and hosts the actual magnetic order in the ground state, will always feature an odd $\mathbb{Z}_4$ index after broken into $P2^\prime_1/c^\prime$ subgroup. It turns out, as tabulated in the supplemental file, that many other magnetic materials with different magnetic structures also have such an odd $Z_4$ index in their symmetry indicators. For example, $\mathcal{MSG}$ ${P_c4_2/ncm}$ \& $\mathcal{WP}$ 4b has $16$ materials that are guaranteed to host symmetry indicated Weyl magnons, if broken down to the subgroup $C2^{\prime}/c^{\prime}$ $(15.89)$ by perturbations. The same conclusion applies to $3$ materials belonging to the $\mathcal{MSG}$ $P_cmna$ $(53.335)$ with $\mathcal{WP}$ $4a$ after borken down to $P2^\prime_1/c^\prime$ subgroup. Here, we find four pairs of Weyl magnons around $E=E_{Weyl} \sim 6.08$ meV pinned to the $k_z=0$ plane at momenta $(2.882,3.117,0), (2.870,2.939,0)$ with $+1$ chirality. Other Weyl points are related to these two by symmetries as follows. After applying the field (along $[100]$), only two-fold screw $\{2_{100 }| \frac{1}{2} \frac{1}{2} \frac{1}{2}  \}'$, glide $ \{m_{100 }| \frac{1}{2} \frac{1}{2} \frac{1}{2} \}'$ (where $'$ means combined with time reversal) and inversion $\mathbf{ \mathcal{I} }$ survive. The bulk Weyl points (see Figs. \ref{Example1WeylPoints} and \ref{Example1FermiArc}) are related as follows: \\
 \begin{itemize}
     \item $ \mathcal{I}$ relates points $(k_x,k_y,k_z)$ to $(-k_x,-k_y,-k_z)$, which have opposite chiralities.
     \item $\{2_{100 }| \frac{1}{2} \frac{1}{2} \frac{1}{2}  \}'$ relates $(k_x,k_y,k_z)$ to $(k_x,-k_y,-k_z)$ with the same chirality. 
      \item $\{m_{100 }| \frac{1}{2} \frac{1}{2} \frac{1}{2}  \}'$ relates $(k_x,k_y,k_z)$ to $(-k_x,k_y,k_z)$ with opposite chiralities.
 \end{itemize}
   where $\{m_{100 }| \frac{1}{2} \frac{1}{2} \frac{1}{2}  \}'$ is nothing but inversion combined with $\{2_{100 }| \frac{1}{2} \frac{1}{2} \frac{1}{2}  \}'$. The topological charge of the Weyl points were calculated as the Chern number of the upper two bands on a small sphere enclosing the Weyl point using the method developed by \cite{ChernNumberCalculationFukui_2005} and the fermionic dual of the magnon Hamiltonian \cite{YML2018magnon}. To compute the surface states, we consider a semi-infinite system along the $c$-axis and calculate the surface spectral density $N_s(E,k_x,k_y)$ using the Green's function renormalization technique \cite{GreenFunctionMethodHENK199369}:
  \begin{equation}
      N_s(E, \boldsymbol{k})=-\frac{1}{\pi} \lim _{\eta \rightarrow 0^{+}} \operatorname{Im} \operatorname{Tr} G_{0 0}(E+\mathrm{i} \eta, \boldsymbol{k})
  \end{equation}
  where $G_{0 0}$ is the block of the Green's function that corresponds to the principal layer containing the surface \cite{GreenFunctionMethodHENK199369}. In the surface spectral density shown in Fig. \ref{Example1FermiArc}, we plot the paths of highest intensity in surface DOS that connect the projections of Weyl points with opposite chirality.  In Fig. \ref{Example1FermiArc} one can see that each pair of Weyl magnons is connected by a surface magnon arc, analogous to fermi arcs in Weyl semimetals. This results in a total of four magnon arcs due to the symmetry-related bulk Weyl magnons discussed earlier. One other possible arrangement of \textit{irreps} is shown in Fig. \ref{Example1OtherSchematicPicture} in which \textit{irreps} of $D$ point are swapped compared to Fig. \ref{Example1UnperturbedSpectrum}, resulting in a band gap with a $\mathcal{SI}$ value of $01$, which belongs to the same topological class as discussed here.

\subsection{Ca$_2$RuO$_4$}
\begin{figure*}[h!]
 \hspace{-1.33cm}
 \includegraphics[scale=0.16]{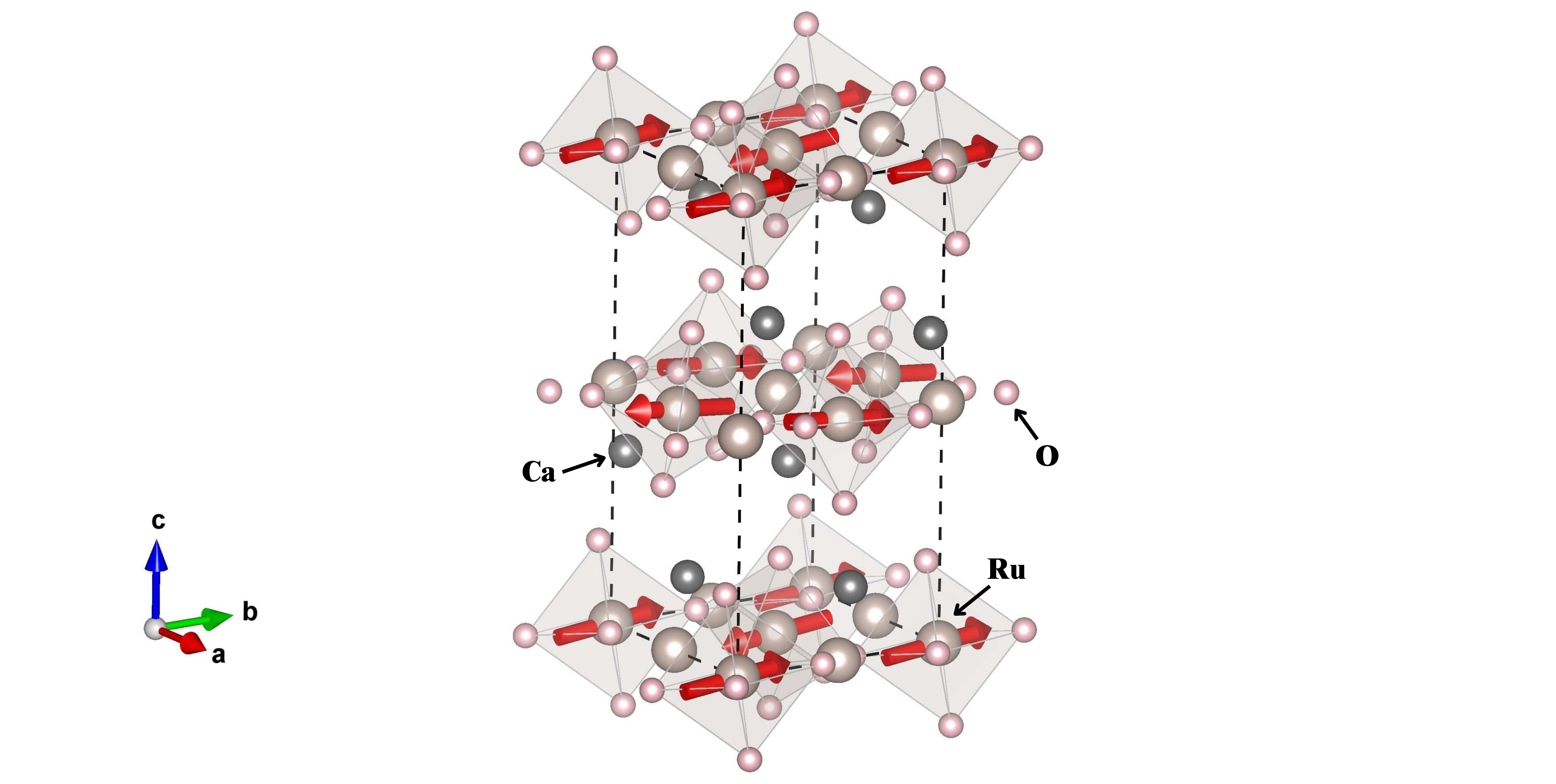}
 \caption{\textbf{Magnetic unit cell of Ca$_2$RuO$_4$}. Light pink spheres indicate the Oxygen sites which form corner-sharing networks around the $Ru$ rose spheres while black are Calcium sites. Size of \textit{Ca} spheres is reduced relative to those of $Ru$ for a better vision of the magnetic order. Red arrows indicate the magnetic moment direction of each of the four $Ru$ magnetic sublattices. Spins mostly align along the $b$-axis with a small canting towards the $c$-axis. Dashed lines indicate the boundaries of the magnetic unit cell with the \textit{a,b,c} axes are highlighted in the inset.}
 \label{Example2Structure}
 \end{figure*}

Now we discuss another example for which preliminary INS data exist in the literature\cite{Ca2RuO4SpinModelINSPhysRevLett.115.247201}. Ca$_2$RuO$_4$ is a Mott insulator that undergoes an antiferromagnetic ordering transition at $T_N=110$~K\cite{Ca2RuO4PaperPhysRevB.98.125142}. In this magnetically ordered phase, the magnetic moments of the $Ru^{4+}$ ions align mostly along the orthorhombic $b$ axis with a $\Vec{Q}=0$ propagation vector. It was shown also recently\cite{Ca2RuO4PaperPhysRevB.98.125142} that the magnetic moments have a canting angle from the $c$ axis. This magnetically-ordered structure is such that the moments are located at the $4a$ $\mathcal{WP}$ of the magnetic space group $Pbca$ $(61.433)$, as shown in Fig. \ref{Example2Structure}. These inputs indicate symmetry-enforced two-fold degeneracy at all high symmetry points ($\mathcal{HSP}s$) except $\Gamma$. 

\begin{figure*}[h!]
 \hspace{-0.78cm}
 \includegraphics[width=1.06\textwidth]{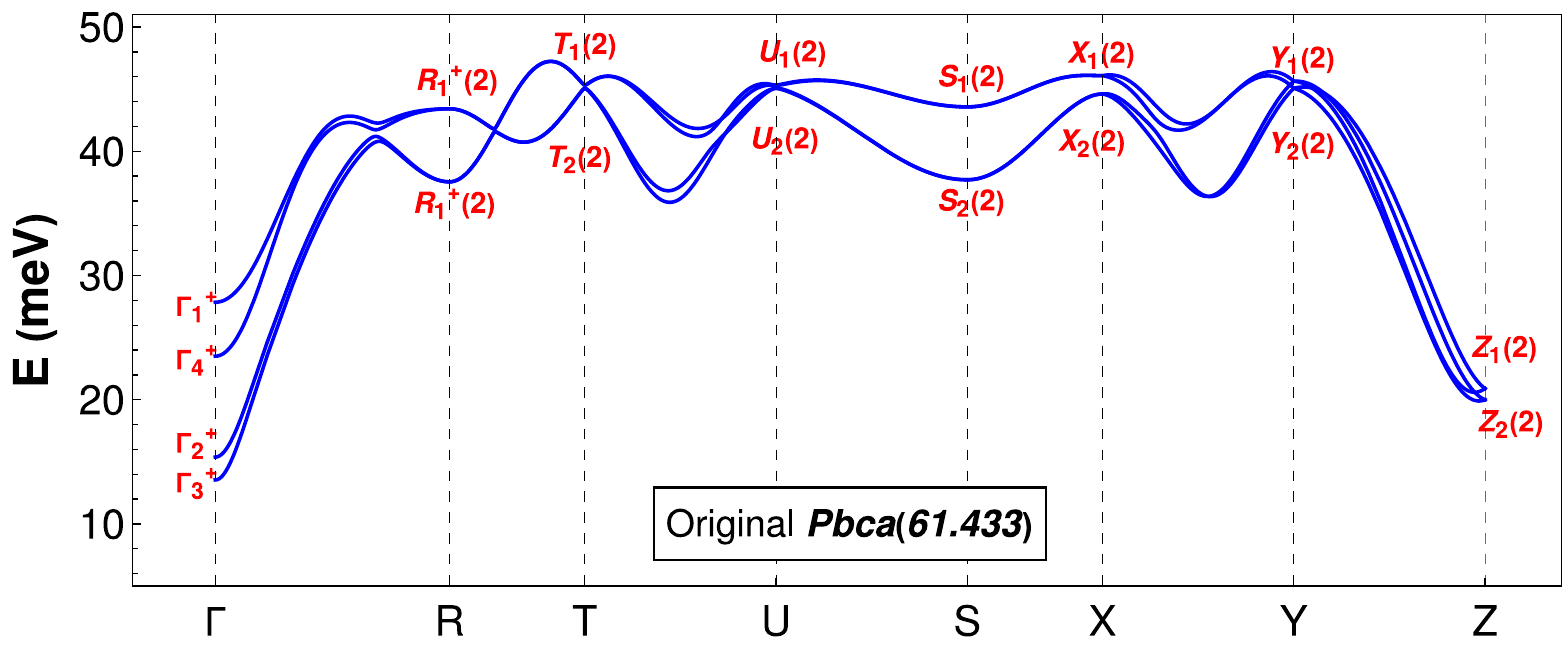}
 \caption{\textbf{Unperturbed magnon spectrum for Ca$_2$RuO$_4$}. Red labels are the magnon band representations at $\mathcal{HSP}s$. Band degeneracies are consistent with the dimensionality of the induced band representation of this magnetic structure.}
 \label{Example2UnperturbedSpectrum}
 \end{figure*}



According to our tabulated results (see supplementary section \ref{Ca2RuO4SuppSection}), this structure will feature at least one induced topological gap upon lowering the symmetry group down to the $P\Bar{1}$ $(2.4)$ subgroup. This subgroup has only inversion symmetry which can be realised by applying a magnetic field in a low-symmetry direction that breaks all $\mathcal{MSG}$ symmetries but inversion. Unlike the previous example, here we perform a linear spin wave (LSW) calculation based on the magnetic interactions reported in the INS study \cite{Ca2RuO4SpinModelINSPhysRevLett.115.247201}, including up to third nearest neighbor intraplane Heisenberg interaction, single-ion anisotropy along the $b$ axis, and an antiferromagnetic interlayer coupling. We also add extra small symmetry-allowed terms to remove accidental degeneracies, by requiring that our LSW spectrum closely matches the lowest pair of bands observed in the INS study at zero field. Our purpose is to show that upon the specified symmetry breaking, three band gaps are induced with nontrivial $\mathcal{SI}s$. The LSW spectrum based on our spin model (for more details, see Methods \ref{Ca2RuO4SpinModelParameters}) is shown in Fig. \ref{Example2UnperturbedSpectrum} where the four Ru$^{4+}$ magnetic sublattices give rise to four magnon bands. The two lower energy modes are very close in energy of magnons observed in INS study\cite{Ca2RuO4SpinModelINSPhysRevLett.115.247201}.\\

\begin{figure}[h!]
  \centering
  \subfloat[\textbf{Inversion $\mathcal{I}$ eigenvalues} of the upper two bands at the eight $\mathcal{HSP}s$ of the Brillouin zone.]{
    \includegraphics[width=0.4\textwidth]{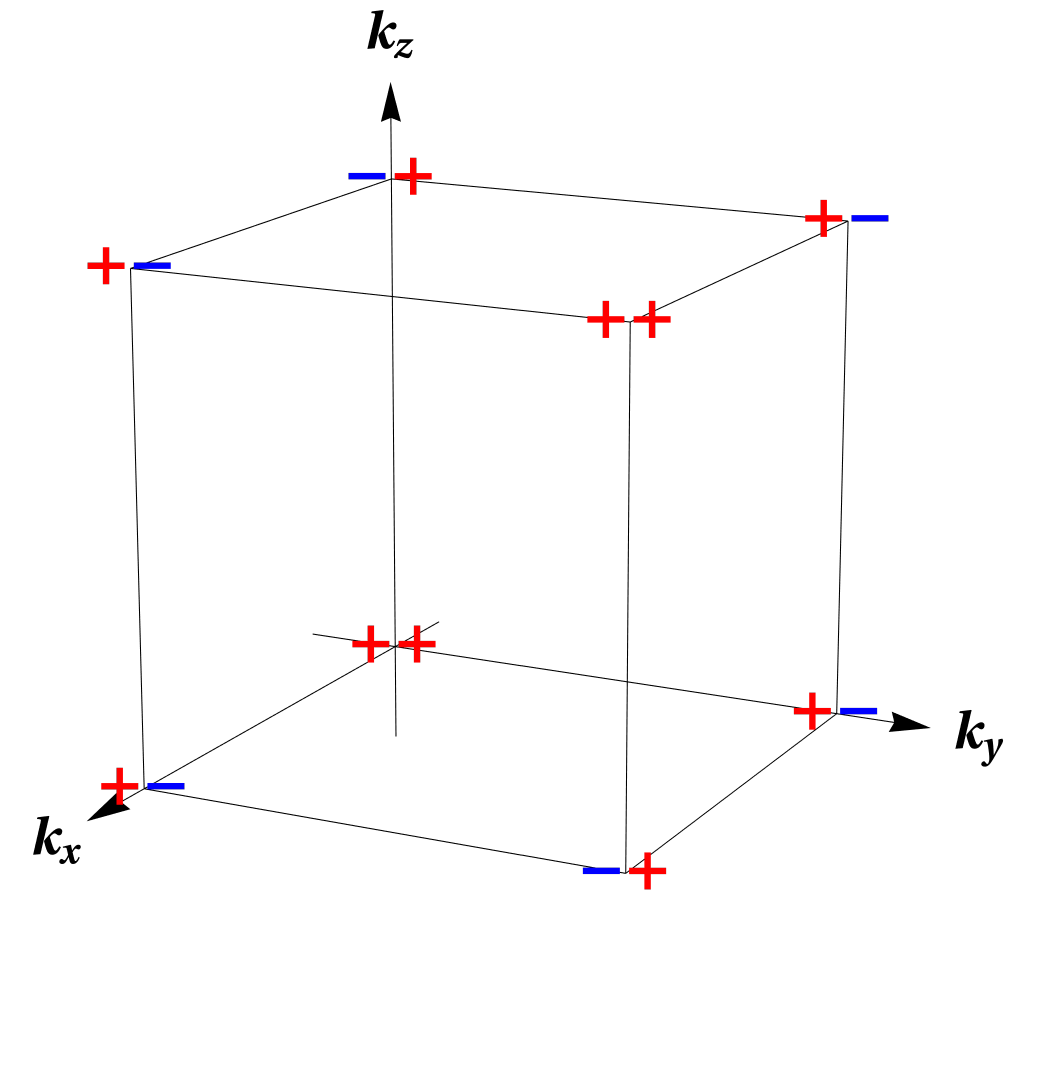}
    }
  \hspace{1cm}
  \subfloat[\textbf{Calculated Chern number} of the upper two bands on the $k_x = 0, \pi$ planes, in agreement with the $\mathcal{Z}_{2i,x} = 1$ SI.]{
    \includegraphics[width=0.4\textwidth]{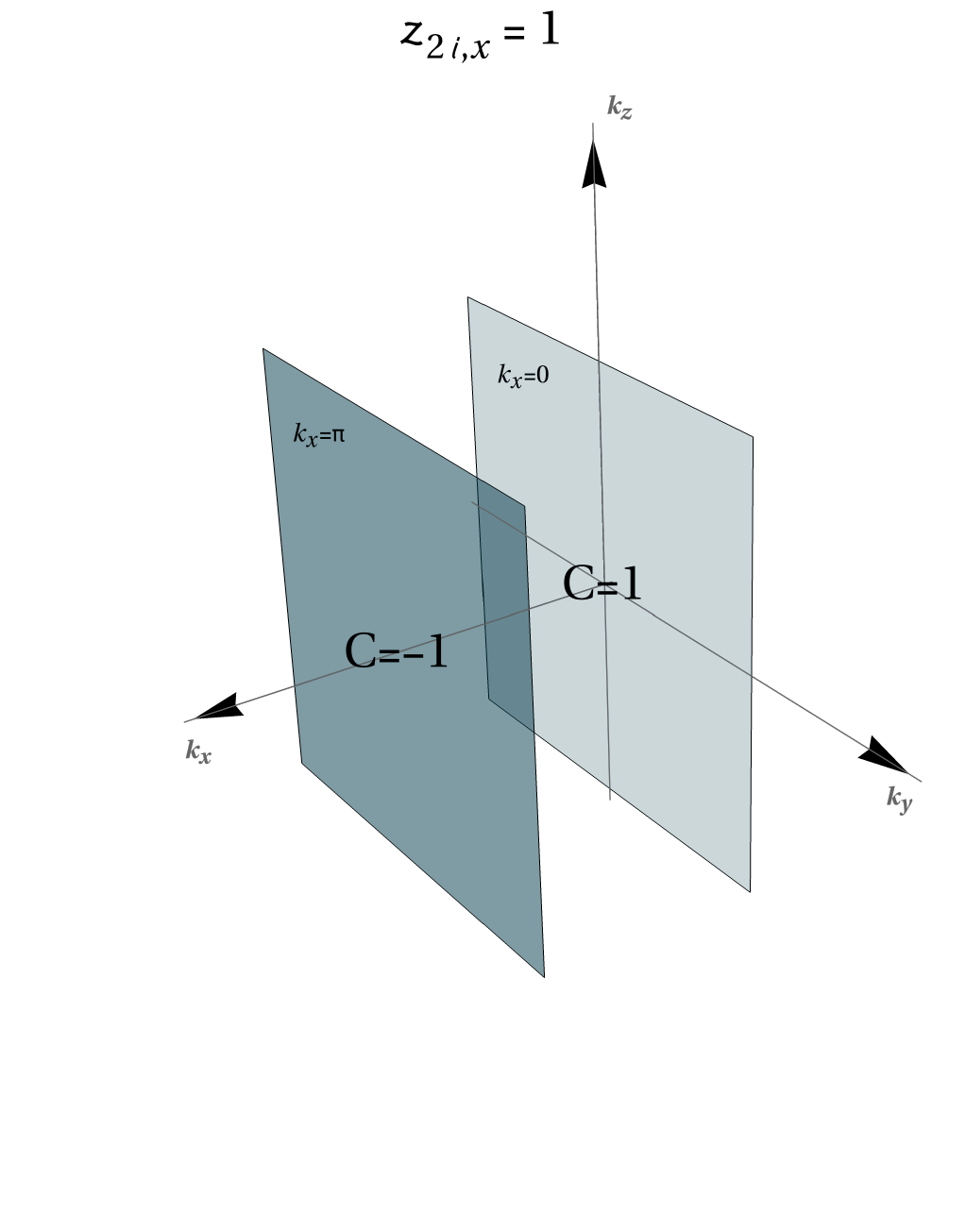}
    }
  \label{fig:subfigure_example}
  \hspace{1cm}
  \subfloat[\textbf{Calculated Chern number} of the upper two bands on the $k_y = 0, \pi$ planes, in agreement with the $\mathcal{Z}_{2i,y} = 1$ $\mathcal{SI}$.]{
    \includegraphics[width=0.4\textwidth]{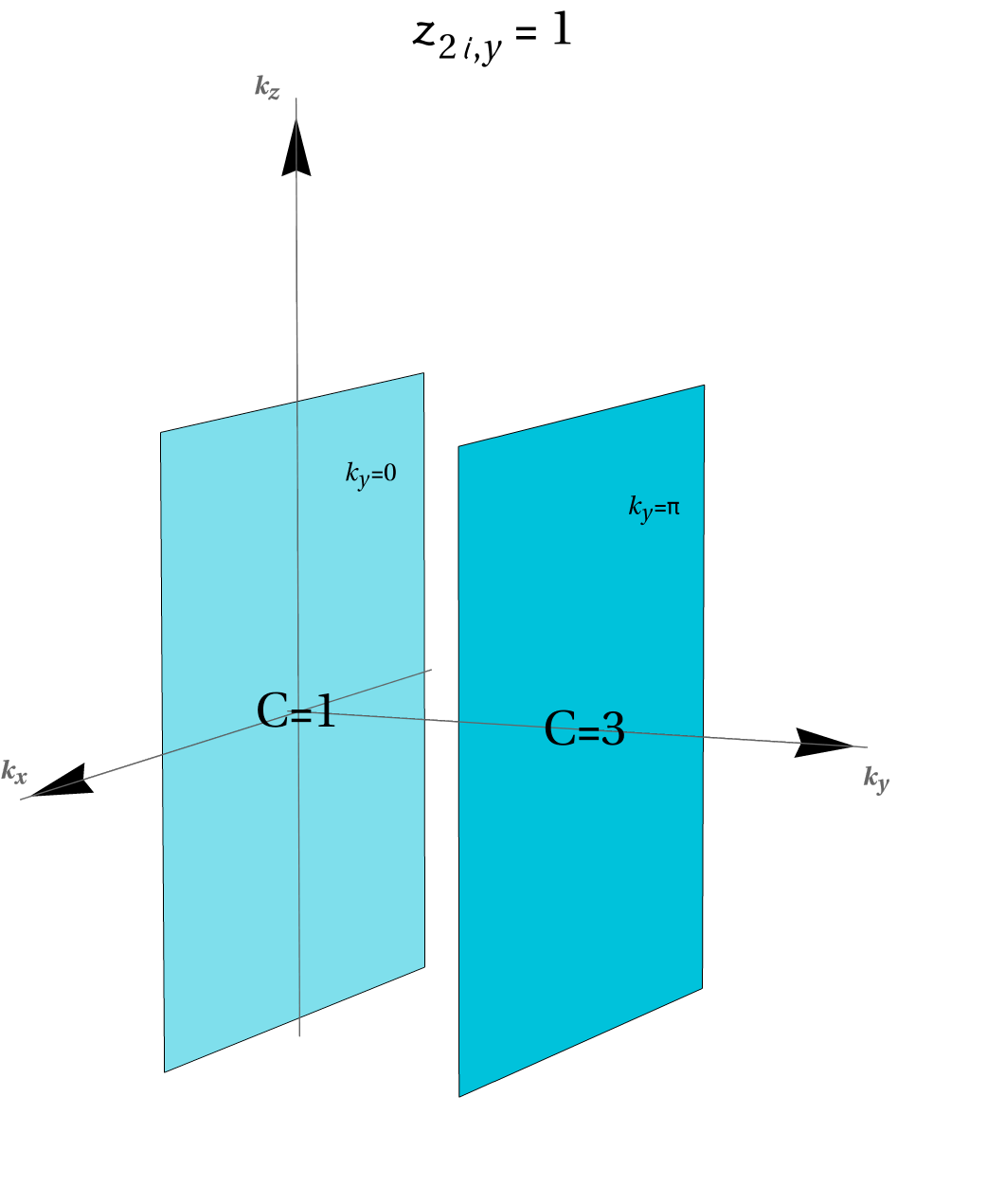}
    }
  \label{fig:subfigure_example2}
  \hspace{1cm}
  \subfloat[\textbf{Calculated Chern number} of the upper two bands on the $k_z = 0, \pi$ planes, in agreement with the $\mathcal{Z}_{2i,z} = 1$ $\mathcal{SI}$.]{
    \includegraphics[width=0.49\textwidth]{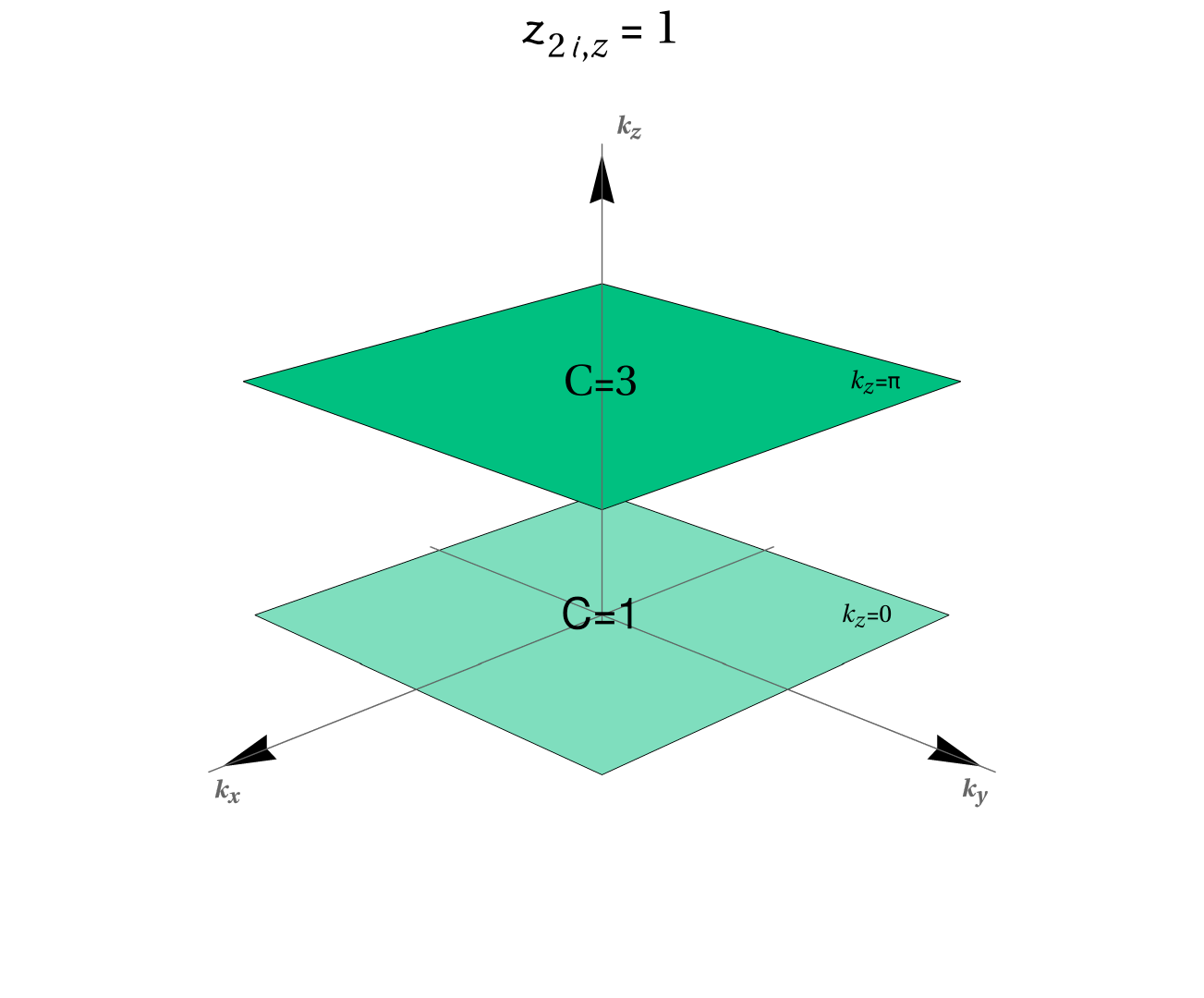}
    }
  \caption{\textbf{Chern numbers on high-symmetry planes in perturbed Ca$_2$RuO$_4$}. We check consistency between the $\mathcal{SI}$ value of the second magnon band gap enforced by symmetries and calculated using our spin model, and the Chern numbers over the $k_i=0,\pi$ planes of the BZ. The $\mathcal{Z}_4=2$ value reflects the difference between the Chern number over the $k_z=0$ and $k_z=\pi$ planes.}
  \label{Example2ChernNumbers}
\end{figure}

By applying a $10$ T magnetic field along a low-symmetry direction, we achieve a symmetry breaking down to the subgroup $P\Bar{1}$ $(2.4)$, which has a $\mathcal{SI}$ group $\mathbb{Z}^3_2 \times \mathbb{Z}_4$. In the symmetry breaking process, all 2-dimensional \textit{irreps} decompose into one-dimensional \textit{irreps} of $P\Bar{1}$. Therefore such a perturbation lifts all symmetry-protected degeneracies in both the higher and lower pairs of magnon bands at all $\mathcal{HSP}s$. This gives a total of three band gaps with the middle gap acquiring a nontrivial inversion $\mathcal{SI}$ $\textit{1112}$. 
This value indicates that the Chern number of the two lower (and accordingly the two higher) magnon bands on the $k_i = 0, \pi$ planes (where $i = x, y, z$) is an odd integer, i.e. it is $1$ mod $2$. This is confirmed by explicit Chern number calculations shown in Fig. \ref{Example2ChernNumbers}. The $\mathbb{Z}_4 = 2$ index diagnoses difference between the Chern numbers of $k_z = 0$ and $k_z=\pi$ planes, which is confirmed as shown in Fig. \ref{Example2ChernNumbers}. This signals an even number of Weyl points between the $k_i=0, \pi$ planes at generic momenta. This is consistent with our spin model that gives 14 Weyl points in half of the 1st BZ $k_i \in (0,\pi)$, with the sum of their topological charges in half of the Brillouin zone equal to 2. The presence of an even number of Weyl points accounts for the even disparity in Chern numbers between the high-symmetry planes. Their opposite-charge partners in $k_i \in (-\pi, 0)$ are related by inversion symmetry. Explicit locations of the Weyl points and their topological charges are summarized in Table \ref{Ca2RuO4WeylPointsTable}.\\


\begin{figure*}[h!]
 \hspace{-0.6cm}
 \includegraphics[width=1\textwidth]{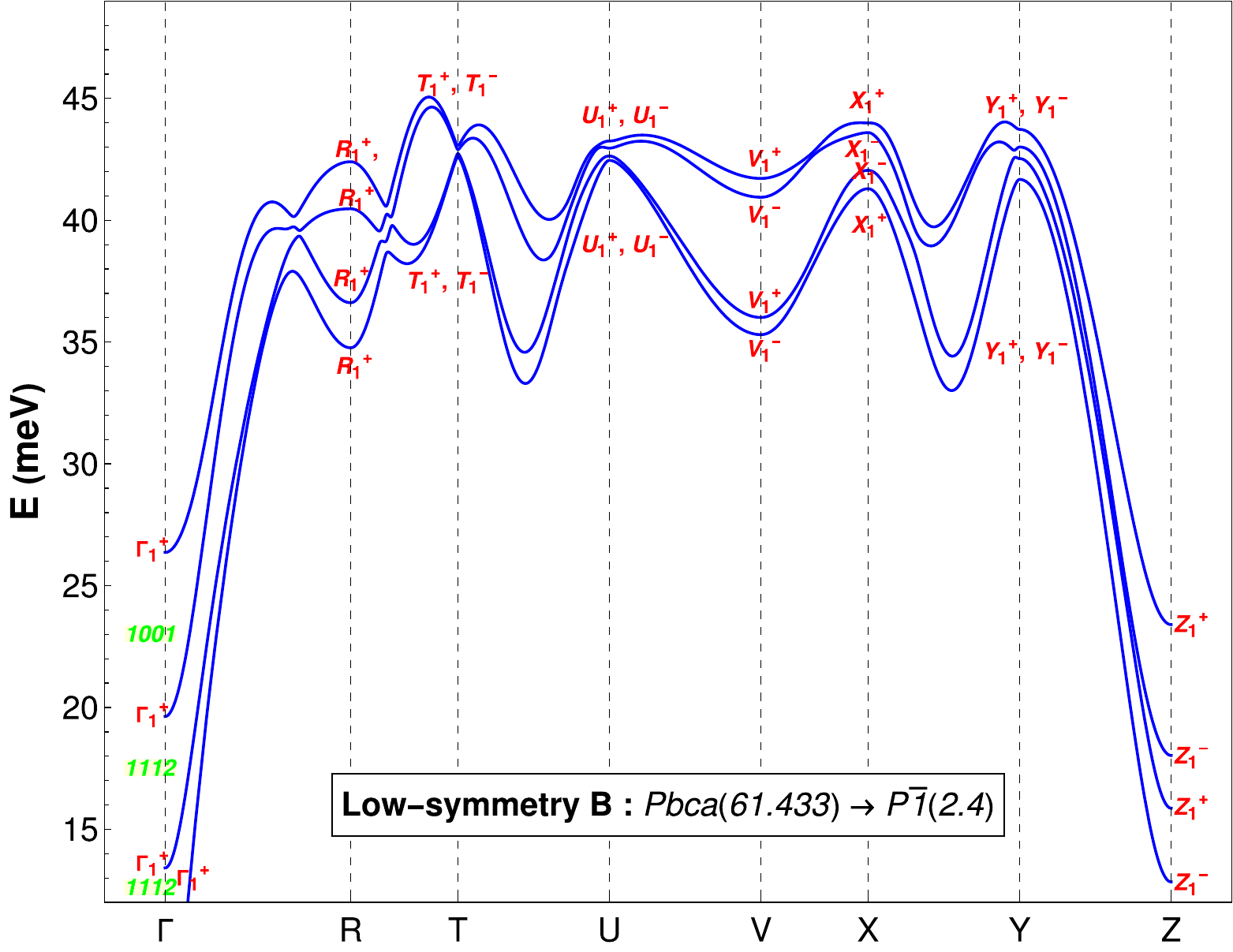}
 \caption{\textbf{Perturbed magnon spectrum for Ca$_2$RuO$_4$.} By breaking the $\mathcal{MSG}$ into the $\mathit{P\Bar{1}}$ subgroup through applying a magnetic field in a low-symmetry direction, the two-fold degeneracies at all $\mathcal{HSP}s$ break, and therefore three band gaps are opened at all $\mathcal{HSP}s$. Calculated values of the band gaps $\mathcal{SI}$ are labelled in green. Red labels are the magnon representations at the $\mathcal{HSP}s$. Since at some points the band gap is small, we write the band \textit{irreps} of two bands with a ($\boldsymbol{,}$) in between to be understood that the first label is for the lower energy band.}
 \label{Example2PerturbedSprctrum}
 \end{figure*}
 

\begin{figure*}[h!]
 \hspace{0cm}
 \includegraphics[scale=0.75]{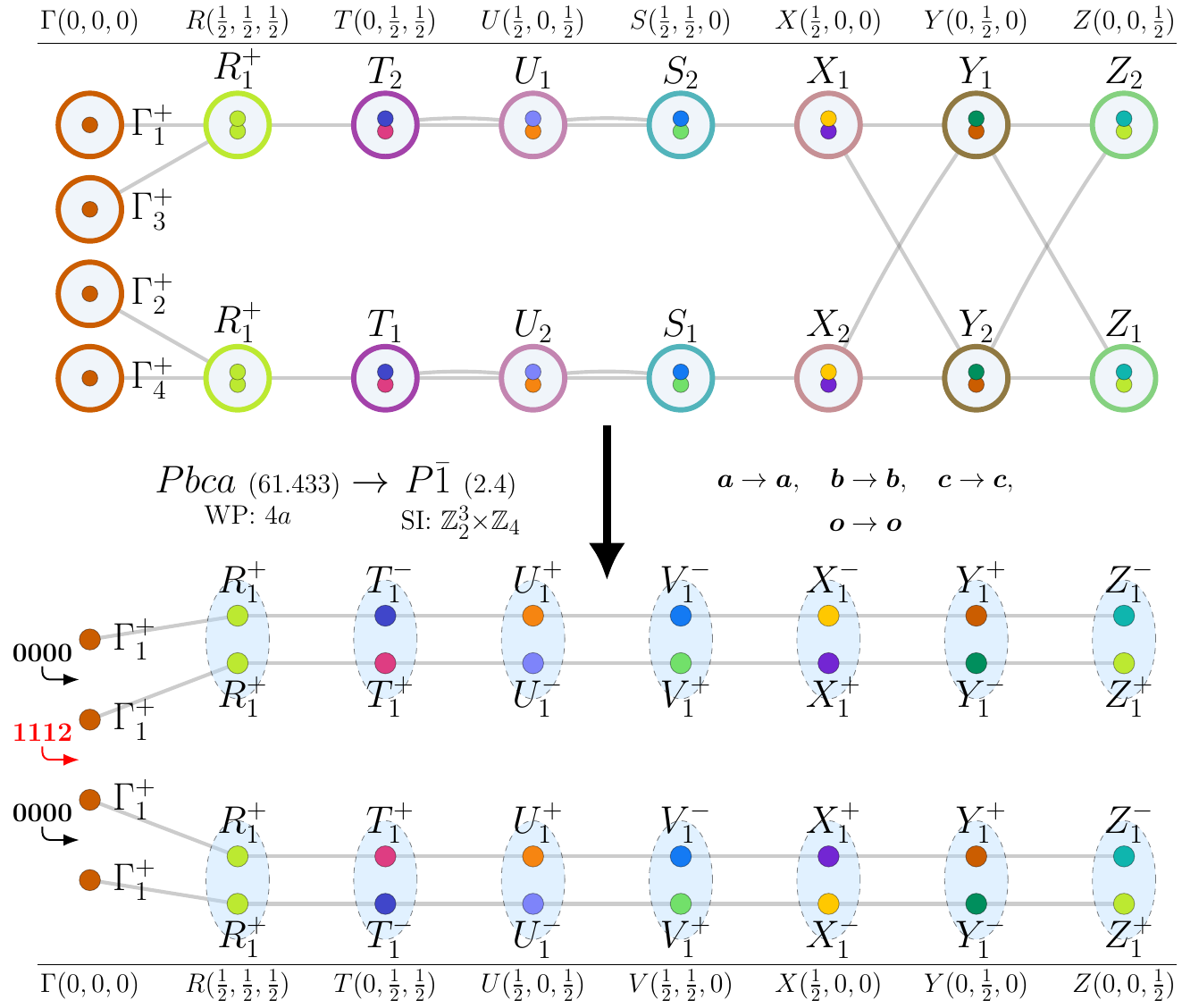}
 \caption{\textbf{Alternative schematic representation of the magnon band structure in Ca$_2$RuO$_4$ .} Upon the symmetry breaking discussed in the main text, the resulting order of the \textit{irreps} at the $\mathcal{HSP}s$ is not unique. This is one possibility the band structure might fall into. Unlike the constructed spin model that produced a $1112$ and $ 1001$ $\mathcal{SI}s$ for the first and third gaps, a different spin model can give rise to a different possibility like the one presented here in which the $\mathcal{SI}$ is $0000$. Regardless of the spin Hamiltonian details, the $\mathcal{SI}$ of the middle band gap belongs to the same topological class. }
 \label{Example2OtherSchematicPicture}
 \end{figure*}

The $\mathcal{SI}$ of $P\Bar{1}$ can be calculated using parities of magnon Bloch eigenstates at the 8 $\mathcal{HSP}s$ as follows: 
\begin{equation}
    \begin{aligned}
z_{2 \mathcal{I},x} &= \sum_{\boldsymbol{k} \in \text{HSPs},\; k_x = \pi }
\frac{1}{2}(N^-_{\boldsymbol{k}} - N^+_{\boldsymbol{k}}) \mod \, 2 \\
z_{2 \mathcal{I},y} &= \sum_{\boldsymbol{k} \in \text{HSPs},\; k_y = \pi }
\frac{1}{2}(N^-_{\boldsymbol{k}} - N^+_{\boldsymbol{k}}) \mod \, 2 \\ 
z_{2 \mathcal{I},z} &= \sum_{\boldsymbol{k} \in \text{HSPs},\; k_z = \pi }
\frac{1}{2}(N^-_{\boldsymbol{k}} - N^+_{\boldsymbol{k}}) \mod \, 2 \\
z_{4 \mathcal{I}} &= \sum_{\boldsymbol{k} \in \text{HSPs}}
\frac{1}{2}(N^-_{\boldsymbol{k}} - N^+_{\boldsymbol{k}}) \mod \, 4 
  \end{aligned}
\end{equation}
where $N^{\pm}_{\boldsymbol{k}}$ is the number of bands with $\pm$ inversion eigenvalue below a given gap. The spectra for the unpertubed and perturbed LSW Hamiltonian are shown in Fig. \ref{Example2UnperturbedSpectrum} and \ref{Example2PerturbedSprctrum} along with the band \textit{irreps} and the calculated values of the $\mathcal{SI}s$ of the band gaps. Note that as long as the \textit{irreps} of the $Pbca$ $(61.433)$ decompose into \textit{irreps} of the subgroup $P\Bar{1}$ $(2.4)$ via a small perturbation, the $\mathcal{SI}$ of the middle gap is robust with the same value as shown in Fig. \ref{Example2PerturbedSprctrum}. This is regardless of the form of the symmetry-allowed spin model or the external perturbation chosen to break the symmetry. The $\mathcal{SI}$ for the other two band gaps does depend on the spin model and the used perturbation. For example, a different spin model can lead to a different arrangement of \textit{irreps} (see Fig. \ref{Example2OtherSchematicPicture}) that has only one nontrivial $\mathcal{SI}$ in the middle gap. Although the focus so far was on the middle gap due to its robust $\mathcal{SI}$, according to our calculation of LSW spectrum, the first and third band gaps also exhibit nontrivial topology in the perturbed spin Hamiltonian reported in INS study\cite{Ca2RuO4SpinModelINSPhysRevLett.115.247201}. However, the sizes of these gaps are smaller, at least in our spin model, compared to the middle gap. Overall, all calculated values of $\mathcal{SI}s$ are consistent with our automated search results in \ref{Ca2RuO4SuppSection}.

\section{Experimental Accessibility of Candidates}
The experimental confirmation of the predicted topological magnons in this work often requires the application of mechanical strains and/or magnetic/electric fields in controlled directions. This requires the growth of thin films and crystals of macroscopic dimensions to realize these effects. Synthetic routes to these high quality and macroscopic sized crystals of many of the identified families of materials have been well established in the literature and for which a few demonstrative examples will be provided here. In the case of transition metal insulators, single crystals of NiFe$_2$O$_4$ have been grown at up to cm scale by Czochralski growth using a Na$_2$CO$_3$ flux \cite{EXP110.1063/1.1729463}. Meanwhile, doping of MnV$_2$O$_4$ has been achieved using Cr and Zn to adjust the p-type doping and modulate the magnetic transition temperature \cite{Exp222Shahi2014TransportMA}. High quality crystals of metal to insulator transition compound NiS$_2$ has been grown at the multiple mm scale by both Te flux and chemical vapor transport growths \cite{EXP333PhysRevMaterials.5.115003,EXP4444Yao1994GrowthON}. Thin films of perovskite candidate BiCrO$_3$ have been grown on many different substrates including SrTiO$_3$ $(001)$ and NdGdO$_3$ $(110)$, allowing for control of strains applied from the substrate \cite{EXP5555Murakami2006FabricationOM}. $f$-block insulator DyOCl is a 2D material and has been grown at the mm scale and exfoliated allowing for similar substrate control for measurement \cite{EXP666PhysRevB.104.214410}. Intermetallic Mn$_5$Si$_3$ growth has been demonstrated as both mm-sized single crystals and as thin films grown on Si $(111)$ \cite{Exp777Okada2001CrystalGB,EXP8888PhysRevMaterials.7.024416}. Additionally, large single crystals of NdMnO$_3$, highlighted in section \ref{ExamplesSection}, have been grown by a floating zone growth in air \cite{NdMnO3SynthesisReferenceBALBASHOV1996365}. For Ca$_2$RuO$_4$, single crystals of multi-millimeter size have been grown via floating growth using Ruthenium-self flux and an Ar/O$_2$ atmosphere to precisely control the oxygen concentration \cite{Ca2RuO4SythesisReferenceNAKATSUJI200126}. Further tuning of the structure of Ca$_2$RuO$_4$ can be obtained by alloying Sr in both powders and single crystal form. We list some interesting candidate materials, that host either gapless Weyl states or magnon axion insulating states upon the relevant perturbations detailed in the supplemental file, alongside their chemical categories in Table $1$.
\begin{longtable}
{|p{0.15\textwidth}|p{0.85\textwidth}|}
\hline
\textbf{Category} & \textbf{Materials} \\
\hline
\endfirsthead

\hline
\textbf{Category} & \textbf{Materials} \\
\hline
\endhead

\hline
\endfoot

\hline
\caption{Chemical categories of some synthesis-\textit{relevant} topological magnon materials identified in this work with a nontrivial $\mathbb{Z}_4$ inversion index upon breaking into one of the magnetic subgroups (\textit{e.g.} $P\bar{1}$, $P2^\prime_1/c^\prime$, $C2^\prime/m^\prime$, $C2^\prime/c^\prime$, ..\textit{etc}) of their respective $\mathcal{MSG}s$. Such values diagnose either gapless Weyl magnons or gapped magnon axion insulator states depending on the $\mathcal{MSG}$ of the chosen material and possibly the details of the magnetic interactions between spins therein.}
\endlastfoot

\multicolumn{2}{|l|}{\textbf{I- Transition Metal Magnets}} \\
\textit{IA- Transition Metal Insulators} & $\alpha$-Fe$_2$O$_3$, CoO, MnTe$_2$, NiFe$_2$O$_4$, NiCr$_2$O$_4$, MnV$_2$O$_4$, CdYb$_2$S$_4$, CdYb$_2$Se$_4$, MnPb$_4$Sb$_6$S$_{14}$, FePb$_4$Sb$_6$S$_{14}$, Ba$_2$CoO$_2$Ag$_2$Se$_2$, Sr$_2$CoO$_2$Ag$_2$Se$_2$, Sr$_2$Fe$_3$S$_2$O$_3$, BaCoSO, Sr$_3$ZnIrO$_6$, Ca$_4$IrO$_6$, Ca$_3$LiRuO$_6$, Sr$_3$LiRuO$_6$, Sr$_3$NaRuO$_6$, SrFeO$_2$, Ca$_3$LiOsO$_6$, Ca$_2$MnGaO$_5$, CuSb$_2$O$_6$, Cr$_2$ReO$_6$, Ca$_2$Fe$_2$O$_5$, Mn$_3$TeO$_6$, FeTa$_2$O$_6$, GeNi$_2$O$_4$, LiMnAs, CsMnP, CsMnBi, RbMnAs, RbMn, RbMnP, KMnP, KMnAs, RbMnAs, RbMnBi \\
\hdashline
\textit{IB- Metallic TM Pnictogens and Chalcogens} & CrN, Li$_{0.5}$FeCr$_{1.5}$S$_4$, FeCr$_2$S$_4$, SrMnSb$_2$, CoNb$_3$S$_6$, CsCo$_2$Se$_2$, CaCo$_{1.86}$As$_2$, Sr$_2$Cr$_3$As$_2$O$_2$, CuSe$_2$O$_5$ \\
\hdashline
\textit{IC- Metal to Insulator Transition Compounds} & NiS$_2$, KCuMnS$_2$, USb \\
\hdashline
\textit{ID- Transition Metal Intermetallics} & MnPt$_{0.5}$Pd$_{0.5}$, Mn$_3$Pt, MnCoGe, Mn$_5$Si$_3$, Fe$_{0.7}$Mn$_{0.3}$, Ni$_{1.64}$Co$_{0.36}$Mn$_{1.28}$Ga$_{0.72}$, Mn$_3$Sn$_2$, PrMnSi$_2$, CaFe$_4$Al$_8$, FeGe, FeSn, FeSn$_2$, FeGe$_2$, Mn$_3$Cu$_{0.5}$Ge$_{0.5}$N, Mn$_3$GaN, Mn$_3$ZnN, Mn$_3$GaC, Mn$_3$CuN, Mn$_2$As, Fe$_2$As, Cr$_2$As, EuFe$_2$As$_2$ \\
\hline
\multicolumn{2}{|l|}{\textbf{II- f-block Magnets}} \\
\textit{IIA- Insulating f-block Magnets} & UO$_2$, NpSe, NpS, DyOCl, HoP, La$_3$OsO$_7$, La$_{2.8}$Ca$_{0.2}$OsO$_7$, Cd$_2$Os$_2$O$_7$, Ho$_2$Ru$_2$O$_7$, Er$_2$Ru$_2$O$_7$, Dy$_3$Al$_5$O$_{12}$, Tb$_3$Al$_5$O$_{12}$, Ho$_3$Al$_5$O$_{12}$, Er$_3$Ga$_5$O$_{12}$, Dy$_3$Ga$_5$O$_{12}$, Tb$_3$Ga$_5$O$_{12}$, Ho$_3$Ga$_5$O$_{12}$, Ln$_3$Tr$_5$O$_{12}$, Tm$_2$Mn$_2$O$_7$, Ho$_2$CrSbO$_7$, Ho$_2$Ru$_2$O$_7$, Er$_2$Ti$_2$O$_7$, Gd$_2$Sn$_2$O$_7$, Er$_2$Pt$_2$O$_7$, Er$_2$Sn$_2$O$_7$, Gd$_2$Ti$_2$O$_7$, Nd$_2$Sn$_2$O$_7$, Nd$_2$Hf$_2$O$_7$, Nd$_2$Zr$_2$O$_7$, Nd$_2$ScNbO$_7$, Sm$_2$Ti$_2$O$_7$, Tb$_2$Sn$_2$O$_7$, Yb$_2$Ti$_2$O$_7$, Yb$_2$Sn$_2$O$_7$ \\
\hdashline
\textit{IIB- Metallic f-block Pnictogens and Chalcogens} & UP, USb$_2$, UP$_2$, UGeS, UGeTe, NpBi, NpTe, NdCoAsO, LaCrAsO, CeCo$_2$P$_2$, CeSbTe \\
\hdashline
\textit{IIC- f-block Intermetallics} & Ho(Co$_{0.66}$Ga$_{0.33}$)$_2$, CeIr(In$_{0.97}$Cd$_{0.03}$)$_5$, ErNiGe, NdCo$_2$, GdMg, Pr$_2$Pd$_2$In, NdZn, Ho$_3$NiGe$_2$, Pr$_3$CoGe$_2$, Tb$_{0.6}$Y$_{0.4}$RhIn$_5$, NdMg, NdPt, TbPd$_{2.05}$Sn$_{0.95}$, Ho$_3$Ge$_4$, NdNiMg$_{15}$, Er$_3$Ge$_4$, Ce$_2$Ni$_3$Ge$_5$, HoRh, Yb$_2$Pd(In$_{0.4}$Sn$_{0.6}$), NdPd$_5$Al$_2$, ErFe$_2$Si$_2$, NdScSiC$_{0.5}$H$_{0.2}$, UNiGa$_5$, UPd$_2$Ge$_2$, URh$_3$Si$_5$, NpNiGa$_5$, U$_2$Pd$_{2.35}$Sn$_{0.65}$, ErMn$_2$Ge$_2$, YMn$_2$Si$_2$, YMn$_2$Ge$_2$, EuMn$_2$Si$_2$, CeMn$_2$Si$_2$, NdMn$_2$Si$_2$, CeMn$_2$Si$_2$, PrMn$_2$Si$_2$, PrMn$_2$Si$_2$, ErMn$_2$Ge$_2$, ErMn$_2$Si$_2$, TbMn$_2$Si$_2$, DyCu, TbPt, TbPt$_{0.8}$Cu$_{0.2}$, NdSi, DyPt, TmNi, PrSi, NdNi$_{0.6}$Cu$_{0.4}$, HoNi, HoPt, ErPt, TmPt, Tb$_2$CoGa$_8$, Dy$_2$CoGa$_8$, Nd$_2$RhIn$_8$, U$_2$Rh$_2$Sn, U$_2$Ni$_2$Sn, CeRh$_2$Si$_2$, CeRu$_2$Al$_{10}$, U$_2$Ni$_2$In, YbCo$_2$Si$_2$ \\
\hline
\multicolumn{2}{|l|}{\textbf{III- Structurally Distinct}} \\
\textit{IIIA- Perovskite oxides and their derivatives} & BiCrO$_3$, TbCr$_{0.5}$Mn$_{0.5}$O$_3$, Lu$_{0.6}$Mn$_{0.4}$MnO$_3$, Pb$_{0.7}$Bi$_{0.3}$Fe$_{0.762}$W$_{0.231}$O$_3$, Pb$_{0.8}$Bi$_{0.2}$Fe$_{0.728}$W$_{0.264}$O$_3$, Nd$_2$CuO$_4$, Sr$_2$IrO$_4$, Nd$_2$NiO$_4$, Ca$_2$RuO$_4$, La$_2$NiO$_4$, LaSr$_3$Fe$_3$O$_9$, CaCu$_3$Fe$_2$Sb$_2$O$_{12}$, Sr$_2$CoOsO$_6$, Sr$_2$FeOsO$_6$, Cu$_3$Ni$_2$SbO$_6$, Bi$_2$RuMnO$_7$, NdFeO$_3$, CeFeO$_3$, TbFeO$_3$, SmFeO$_3$, NaOsO$_3$, LaCrO$_3$, YCrO$_3$, LaMnO$_3$, ErCrO$_3$, TmCrO$_3$, YRuO$_3$, PrMnO$_3$, Pr$_{0.95}$K$_{0.05}$MnO$_3$, NdMnO$_3$, KMnF$_3$, NdMnO$_3$, (CH$_3$NH$_3$)(Co(COOH)$_3$), Ho$_{0.2}$Bi$_{0.8}$FeO$_3$, Ho$_{0.15}$Bi$_{0.85}$FeO$_3$, La$_{0.875}$Ba$_{0.125}$Mn$_{0.875}$Ti$_{0.125}$O$_3$, La$_{0.90}$Ba$_{0.10}$Mn$_{0.90}$Ti$_{0.10}$O$_3$, La$_{0.95}$Ba$_{0.05}$Mn$_{0.95}$Ti$_{0.05}$O$_3$, La$_{0.95}$Ba$_{0.05}$MnO$_3$, LaMnO$_3$, YCr$_{0.5}$Mn$_{0.5}$O$_3$, SmFeO$_3$, SrRuO$_3$, TbCrO$_3$, DyCrO$_3$, TeNiO$_3$, YVO$_3$, LaSrFeO$_4$, Nd$_2$CuO$_4$, Pr$_2$CuO$_4$, LaSrFeO$_4$, LaCaFeO$_4$, LaBaFeO$_4$, La$_{0.75}$Bi$_{0.25}$Fe$_{0.5}$Cr$_{0.5}$O$_3$, LaCrO$_3$, CeFeO$_3$, InCrO$_3$, TlCrO$_3$, ScCrO$_3$, La$_2$NiO$_4$, Nd$_2$NiO$_4$, La$_2$CoO$_4$, Gd$_2$CuO$_4$, Sm$_2$CuO$_4$, Eu$_2$CuO$_4$, Bi$_{0.8}$La$_{0.2}$Fe$_{0.5}$Mn$_{0.5}$O$_3$, PrCrO$_3$, SmCrO$_3$, NdMn$_{0.8}$Fe$_{0.2}$O$_3$, Bi$_{0.8}$La$_{0.2}$Fe$_{0.5}$Mn$_{0.5}$O$_3$, PrCrO$_3$, SmCrO$_3$, NdMn$_{0.8}$Fe$_{0.2}$O$_3$, Rb$_2$Fe$_2$O(AsO$_4$)$_2$, NdMnO$_3$, NdMn$_{0.8}$Fe$_{0.2}$O$_3$, Pr$_{0.5}$Sr$_{0.5}$CoO$_3$, Pr$_{0.5}$Sr$_{0.4}$Ba$_{0.1}$CoO$_3$, Pr$_{0.5}$Sr$_{0.5}$MnO$_3$, (Tm$_{0.7}$Mn$_{0.3}$)MnO$_3$, (Ho$_{0.8}$Mn$_{0.2}$)MnO$_3$, Tb$_{0.55}$Sr$_{0.45}$MnO$_3$, Tb$_{0.55}$Sr$_{0.45}$MnO$_3$, SmCrO$_3$, TbFeO$_3$, Sc$_2$NiMnO$_6$, La$_2$CoPtO$_6$, Ca$_2$Fe$_{0.875}$Cr$_{0.125}$GaO$_5$ \\
\hdashline
\textit{IIIB- Halides and oxyhalides} & KNiF$_3$, KMnF$_3$, FeF$_3$, LiCoF$_4$, NaMnF$_4$, Na$_2$NiFeF$_7$, CsMn$_2$F$_6$, Fe$_2$F$_5$(H$_2$O)$_2$, Cr$_2$F$_5$, CsMnF$_4$, TlMnF$_4$, Na$_2$NiCrF$_7$, CsCoCl$_3$(D$_2$O)$_2$, CsNiF$_3$, SrFeO$_2$F, La$_2$NiO$_3$F$_2$, La$_{0.5}$Sr$_{0.5}$FeO$_{0.5}$F$_{0.5}$, Mn$_2$SeO$_3$F$_2$, Cu$_3$Mg(OD)$_6$Br$_2$, La$_2$NiO$_3$F$_{1.93}$ \\
\hdashline
\textit{IIIC- Polyatomic oxides (Sulfates, Phosphates, Carbonates, Vanadates, germanates and silicates)} & NiSO$_4$, FeSO$_4$, CoSO$_4$, FeSO$_4$F, FeOH$_4$, Li$_2$Fe(SO$_4$)$_2$, NaFeSO$_4$F, NaCoSO$_4$F, Co$_3$(PO$_4$)$_2$, CuFe$_2$(P$_2$O$_7$)$_2$, Fe$_2$MnBO$_5$, AgMnVO$_4$, VPO$_4$, Na$_3$Co(CO$_3$)$_2$Cl, Na$_2$BaMn(VO$_4$)$_2$, MnCO$_3$, NiCO$_3$, CoCO$_3$, Cu$_3$Y(SeO$_3$)$_2$O$_2$Cl, Cu$_3$Bi(SeO$_3$)$_2$O$_2$Br, (NH$_2$(CH$_3$)$_2$)(FeMn(HCOO)$_6$), (CH$_3$NH$_3$)Co(COOH)$_3$, (NH$_2$(CH$_3$)$_2$)(FeCo(HCOO)$_6$), NH$_4$FeCl$_2$(HCOO), [C(ND$_2$)$_3$]Mn(DCOO)$_3$, KFe$_3$(SO$_4$)$_2$(OH)$_6$, NaFe$_3$(SO$_4$)$_2$(OH)$_6$, AgFe$_3$(SO$_4$)$_2$(OD)$_6$, FeBO$_3$, CoFePO$_5$, NaFePO$_4$, Na$_2$BaFe(VO$_4$)$_2$, LaMn$_3$V$_4$O$_{12}$, ZrCo$_2$Ge$_4$O$_{12}$, CeCo$_2$Ge$_4$O$_{12}$, CeMnCoGe$_4$O$_{12}$, ZrMn$_2$Ge$_4$O$_{12}$, GdFeZnGe$_4$O$_{12}$, ErFeCuGe$_4$O$_{12}$, NaCeGe$_2$O$_6$, Fe$_4$Si$_2$Sn$_7$O$_{16}$, Mn$_2$SiO$_4$, Mn$_2$GeO$_4$, CuCl(C$_4$H$_4$N$_2$)$_2$(BF$_4$), CuBr(C$_4$H$_4$N$_2$)$_2$(BF$_4$) \\

\end{longtable}

\section{Discussion}

In this work, we present a major development in the symmetry-based approach for identifying topological magnons through a fully automated algorithm that enables an efficient, large-scale search for candidate materials hosting field-induced topological magnons. This innovation significantly broadens the scope of materials exploration, surpassing the limitations of manual or limited methods. It provides a powerful tool for uncovering novel topological phases in magnetic systems, opening new avenues for exploration in this area. Applying this algorithm to all $1649$ magnetic materials in BCS with a commensurate magnetic order, we ran a large scale search and discovered $387$ candidate materials for topological magnons. We further discussed examples and experimental accessibilities of these candidate materials. 

In the symmetry-based search algorithm discussed in this work, spin-orbit couplings (SOCs) are assumed in all magnetic materials, which explicitly break the $SO(3)$ spin rotational symmetries. Magnets of negligible (or weak) SOCs can exhibit distinct properties compared to spin-orbit-coupled ones discussed in this work, such as different topological properties \cite{Corticelli2022}, and different representations associated with spin-group symmetries in the case of negligible SOC \cite{Yang2021,Schiff2023,Chen2024}. New physical phenomena can emerge in magnets with negligible SOCs, such as recently discovered altermagnetism in collinear magnets \cite{Smejkel2024,Krempasky2024}. One future direction is to extend the current search algorithm to altermagnets with weak SOCs, making use of the representation theory of spin space groups. This can lead to new types of topological magnons in altermagnetic materials, protected by spin group symmetries. 

Another future direction is to generalize the current approach to magnon polarons, which arise from the hybridization of magnons and phonons in magnetically ordered materials. It is well established that this hybridization can lead to the emergence of topological magnon polarons \cite{Takahashi2016,Go2019,Zhang2019a}; however, only a few materials have been identified to host these topological bosonic modes \cite{Bao2023}. The current theoretical framework can be expanded to encompass not only pure topological magnons \cite{KarakiSciAdvdoi:10.1126/sciadv.ade7731} and topological phonons \cite{Xu2024}, but also to enable a systematic search for materials hosting topological magnon polarons. These intriguing possibilities open the door to future research directions, which we aim to explore in upcoming projects.

\section*{METHODS}
\textbf{Linear Spin Wave Theory}\\

For a generic bilinear spin Hamiltonian,
\begin{equation}
    \hat{\mathcal{H}}= \frac{1}{2} \sum_{i,j;a,b;\alpha,\beta} J_{ia,jb}^{\alpha,\beta} \hat{S}_{i,a}^{\alpha} \hat{S}_{j,b}^{\beta} 
\end{equation}
the standard method for calculating the spinwave excitations is linear spin wave theory which relies on an expansion in the spin fluctuations around the classical ground state order. In LSWT, transverse spin components are transformed to magnon creation $\hat{a}^{\dagger}$ and annihilation operators $\hat{a}$ using the \textit{Holestein-Primakoff} transformation:
\begin{equation}
    \begin{aligned}
& \hat{S}^z=S-\hat{a}^{\dagger} \hat{a} \\
& \hat{S}^{+}=\sqrt{2 S} \sqrt{1-\frac{\hat{a}^{\dagger} \hat{a}}{2 S}} \hat{a}=\sqrt{2 S}\left(1-\frac{\hat{a}^{\dagger} \hat{a}}{4 S}\right) \hat{a}+\ldots \\
& \hat{S}^{-}=\sqrt{2 S} \hat{a}^{\dagger} \sqrt{1-\frac{\hat{a}^{\dagger} \hat{a}}{2 S}}=\sqrt{2 S} \hat{a}^{\dagger}\left(1-\frac{\hat{a}^{\dagger} \hat{a}}{4 S}\right)+\ldots
\end{aligned}
\end{equation}
with the magnon operators satisfy the bosonic commutation relations $\left[\hat{a}_i, \hat{a}_j^{\dagger}\right]=\delta_{i j} \text { and }\left[\hat{a}_i, \hat{a}_j\right]=\left[\hat{a}_i^{\dagger}, \hat{a}_j^{\dagger}\right]=0$. ($S^x$, $S^y$) are the transverse spin components in its local frame that is locally rotated such that the z-axis is aligned along the polarization direction $\langle \boldsymbol{S}_i \rangle = S_i \hat{z}_i = (0,0,S_i)$. The square root is expanded in power of $\frac{1}{S}$ and by truncating it to a linear order in $\frac{1}{S}$, a Hamiltonian of the form:
\begin{equation}
    \hat{H}_{SW}= \hat{H}^{(0)} + \hat{H}^{(2)} + \hat{H}^{\prime}
\end{equation}
where $\hat{H}^{(0)}$ is the classical ground state energy, $\hat{H}^{(2)}$ is a quadratic Bogoliubov-de Genes (BdG) bosonic Hamiltonian and $\hat{H}^{\prime}$ represents higher order correction terms that describe magnon-magnon interactions. Finally, this BdG Hamiltonian is Fourier-transformed and diagonalized via a paraunitary matrix through a Bogoliuobov transformation to find the magnon wavefunctions and spectrum.\\

An alternative yet equivalent approach is the so-called equation of motion (EOM) approach \cite{YML2018magnon,KarakiSciAdvdoi:10.1126/sciadv.ade7731} in which the low-energy dynamics captured by the small deviations of spins from their ordered directions :
\begin{equation}
    \mathbf{s}_i \equiv \mathbf{S}_i-\left\langle\mathbf{S}_i\right\rangle \Rightarrow \mathbf{s}_i=\left(s_i^x, s_i^y, 0\right) \label{eq:LSW EOM Expansion}
\end{equation}
with $\left|\mathbf{s}_i\right| \ll \bar{S}_i $. By expanding the Hamiltonian in terms of the deviations, then considering their Heisenberg EOM, we reach a dynamical EOM that governs the spinwave dynamics:
\begin{equation}
    -\mathrm{i} \frac{\mathrm{d} s_i^\alpha}{\mathrm{d} t}=\sum_{j, \beta}\left(\delta_{i, j}\left(\sigma_y\right)^{\alpha, \beta} \cdot \mathbf{R}\right)_{i \alpha, j \beta} s_j^\beta = \sum_{j, \beta}\left(\mathbf{M} \cdot \mathbf{R}\right)_{i \alpha, j \beta} s_j^\beta  \label{eq:LSW EOM R Matrix}
\end{equation}
where $\mathbf{M}_{i \alpha, j \beta} = \delta_{i, j}\left(\sigma_y\right)^{\alpha, \beta}$ with $\sigma_y$ is y-Pauli matrix and $\mathbf{R}$ is a $2N \times 2N$, where N is the number of spins in the primitive magnetic unit cell, semi-positive definite symmetric matrix that describes the spin-spin interactions and is referred to as the "magnon" Hamiltonian. The spectrum is then calculated as the eigenvalues of $\mathbf{M}\cdot\mathbf{R}$.\\

\textbf{Fermionization map of LSW systems}\\
For the calculation of the topological charges of the Weyl magnons, we used the numerical technique \cite{ChernNumberCalculationFukui_2005} after mapping the LSW Hamiltonian into its fermionic counterpart. A LSW problem can be mapped into a free-fermion system through the similarity transformation \cite{YML2018magnon}:
\begin{equation}
    \mathbf{H}_f = \sqrt{\mathbf{R}} \cdot \mathbf{M} \cdot \sqrt{\mathbf{R}}
\end{equation}
where $\mathbf{R}$ is the magnon Hamiltonian in Eq. \ref{eq:LSW EOM R Matrix} of a generic spin-orbit coupled magnet and $\mathbf{H}_f$ is the free fermion Hamiltonian. Since $\mathbf{R}$ is a positive-definite matrix for a gapped LSW system, its square root is well-defined. The key feature is that $\mathbf{H}_f$ shares the same spectrum and band topology as the spinwave Hamiltonian. It also preserves the same symmetries of the magnon Hamiltonian, therefore if a unitary symmetry $g$ of the magnetic order exists, its implementation on the magnon Hamiltonian takes the form $[O_g,\mathbf{R}]=0$ and $[O_g, M]=0$ where $O_g \in SO(2N)$ and the fermionic counterpart preserves it $O_g \mathbf{H}_f O_g^{\dagger} = \mathbf{H}_f $. For an anti-unitary symmetry $p$, $[O_p,\mathbf{R}]=0$, $\{O_p, M\}=0$ and the fermionic Hamiltonian $\mathbf{H}_f$ satisfy $O_p \mathbf{H}_f O_p^{\dagger} = -\mathbf{H}_f $.  \\

Therefore this mapping can be used as an additional route to apply the results of TQC and SI theories to the spinwave problem. In this mapping, a spinless electronic counterpart replaces the spinwave problem, the spinwave variables $S^{\pm}$ takes the role of the atomic orbitals. Therefore the band representation is induced from the site symmetry group of the spinwave variables. Thus a tabulation \cite{KarakiSciAdvdoi:10.1126/sciadv.ade7731,McClartyIdentigyingPhysRevLett.130.206702} of the magnetic site symmetry groups compatible with the magnetic order and their representations provides a direct way to study the induced magnon band representation for a given $\mathcal{MSG}$ using MBANDREP \cite{MBANDREP2Xu2020HighthroughputCO,MTQCElcoro_2021} BCS tool as discussed in the text. 

\textbf{Spin wave spectrum of NdMnO$_3$} \\
\label{SpinModelOfNdMnO3}The crystal structure of NdMnO$_3$ belongs to the space group \textit{Pnma} (62.448) which has four Mn sublattices at the $\mathcal{WP}$ 4b:
\begin{equation}
    \begin{aligned}
        & r_A = (0, 0, c/2) \\
        & r_B = (a/2,0,0) \\
        & r_C = (0, b/2, c/2) \\
        & r_D = (a/2, b/2, 0)
    \end{aligned}
\end{equation}
where Cartesian coordinates are used. In the magnetic phase below $T_N \sim 78$ K, the Mn atoms become magnetized giving rise to the magnetic order \cite{NdMnO3Paper2000}:
\begin{equation}
    \begin{aligned}
        & \langle \boldsymbol{S}_{A,B} \rangle \propto (0.933, -0.359, 0) \\
        &\langle \boldsymbol{S}_{C,D} \rangle \propto (-0.933, -0.359, 0).
    \end{aligned}
\end{equation}
We build our spin model based on the interaction bonds defined between pairs of sites as following:
\begin{equation}
\begin{aligned}
    & \mathbf{J_1} : r_B \iff r_D \\
    & \mathbf{J_2} : r_B \iff r_A \\
    & \mathbf{J_3} : r_B \iff r_B + (0, 0, c) \\
    & \mathbf{J_4} : r_B \iff r_B + (a/2, b/2, c/2) \\
    & \mathbf{J_5} : r_B \iff r_B + (a, 0, 0) \\
\end{aligned}
\end{equation}
where $\mathbf{J_1}$ is constrained by $\{m_{010}|0, \frac{1}{2}, 0 \}$ and therefore have the anti-symmetric form:
\begin{equation}
    \mathbf{J_1} =
\begin{pmatrix}
  J_1^{xx} & D_1^z & \Gamma_1^{xz}\\
  -D_1^z & J_1^{yy} & D_1^x\\
  \Gamma_1^{xz} & -D_1^x & J_1^{zz}
\end{pmatrix}
\end{equation}
while $\mathbf{J_2}$ is not symmetry-constrained and therefore take the general form:
\begin{equation}
    \mathbf{J_2} = 
\begin{pmatrix}
  J_2^{xx} & D_2^z + \Gamma_2^{xy} & -D_2^y + \Gamma_2^{xz}\\
  -D_2^z + \Gamma_2^{xy} & J_2^{yy} & D_2^x + \Gamma_2^{yz}\\
  D_2^y + \Gamma_2^{xz} & -D_2^x + \Gamma_2^{yz} & J_2^{zz}
\end{pmatrix}
\end{equation}
$\mathbf{J_3}$ is constrained by inversion $\{-1|0\}$ and therefore take the general symmetric form:
\begin{equation}
    \mathbf{J_3} =
\begin{pmatrix}
  J_3^{xx} & \Gamma_3^{xy} & \Gamma_3^{xz}\\
  \Gamma_3^{xy} & J_3^{yy} & \Gamma_3^{yz}\\
  \Gamma_3^{xz} & \Gamma_3^{yz} & J_3^{zz}
\end{pmatrix}
\end{equation}
and $\mathbf{J_4}$ is not symmetry-constrained, so it takes the general form:
\begin{equation}
    \mathbf{J_4} = 
\begin{pmatrix}
  J_4^{xx} & D_4^z + \Gamma_4^{xy} & -D_4^y + \Gamma_4^{xz}\\
  -D_4^z + \Gamma_4^{xy} & J_4^{yy} & D_4^x + \Gamma_4^{yz}\\
  D_4^y + \Gamma_4^{xz} & -D_4^x + \Gamma_4^{yz} & J_4^{zz}
\end{pmatrix}
\end{equation}
and $\mathbf{J}_5$ is constrained by inversion $\{-1|0\}$ and therefore take the general symmetric form:
\begin{equation}
    \mathbf{J_5} =
\begin{pmatrix}
  J_5^{xx} & \Gamma_5^{xy} & \Gamma_5^{xz}\\
  \Gamma_5^{xy} & J_5^{yy} & \Gamma_5^{yz}\\
  \Gamma_5^{xz} & \Gamma_5^{yz} & J_5^{zz}
\end{pmatrix}
\end{equation}
and we include an extra single-ion anisotropy term of the form $-\sum_i A^{xx} (S_i^x)^2$ that helps in stabilizing the ground state along the experimentally-reported directions. We perform the the linear spinwave expansion as in Eq. \ref{eq:LSW EOM Expansion} using the Heisenberg exchange values $J_1 = 1$ meV, $J_2 = -0.3$ meV, $J_3 = -0.1$ meV, $J_4 = 0.05$ meV, $J_5 = -0.3$ meV and supplement them with a 
Dzyaloshinskii–Moriya interaction (DMI) on the first nearest neighbor (NN) bond $D_1^z = 1.1$ meV and an anisotropy $A^{xx}=0.4$ meV. we reach the magnon Hamiltonian $\mathbf{R}(\boldsymbol{k})$ in Eq. \ref{eq:LSW EOM R Matrix} which we use to diagonalize $\mathbf{M}\cdot\mathbf{R}(\boldsymbol{k})$ and find the magnon spectrum  as shown in Fig. \ref{Example1UnperturbedSpectrum} and the magnon wavefunctions which we use to calculate the \textit{irreps} of magnon bands.\\
\textbf{Spinwave spectrum of Ca$_2$RuO$_4$.} \label{SpinModelOfCa2RuO4}\\
The crystal structure of Ca$_2$RuO$_4$ belongs to the space group \textit{Pbca} (61.433) which has four Ru sublattices at the 4a $\mathcal{WP}$:
\begin{equation}
    \begin{aligned}
        & r_A = (0, 0, 0) \\
        & r_B = (a/2, b/2,0) \\
        & r_C = (0, b/2, c/2) \\
        & r_D = (a/2, 0, c/2)
    \end{aligned}
\end{equation}
where Cartesian coordinates are used. In the magnetic phase below $T_N \sim 110$ K, the Ru atoms become magnetized giving rise to the magnetic order \cite{Ca2RuO4PaperPhysRevB.98.125142}:
\begin{equation}
    \begin{aligned}
        & \langle \boldsymbol{S}_{A} \rangle \propto (0, 1, 0.1) \\
        &\langle \boldsymbol{S}_{B} \rangle \propto (0, -1, -0.1).\\
        &\langle \boldsymbol{S}_{C} \rangle \propto (0, 1, -0.1).\\
        &\langle \boldsymbol{S}_{D} \rangle \propto (0, -1, 0.1).
    \end{aligned}
\end{equation}
We build our spin model based on the interaction bonds defined between pairs of sites as following:
\begin{equation}
\begin{aligned}
    & \mathbf{J_1} : r_A \iff r_B \\
    & \mathbf{J_2} : r_A \iff r_A + (a, 0, 0) \\
    & \mathbf{J_3} : r_A \iff r_A + (0, b, 0) \\
    & \mathbf{J_4} : r_A \iff r_B + (a, 0, 0) \\
    & \mathbf{J_5} : r_A \iff r_A + (a, b, 0) \\
    & \mathbf{J_6} : r_A \iff r_D  \\
    & \mathbf{J_7} : r_A \iff r_A + (0, b/2, c/2) \\
    & \mathbf{J_8} : r_A \iff r_A + (a/2, b, c/2) \\
    & \mathbf{J_9} : r_A \iff r_A + (0, 0, C) \\
    & \mathbf{J_{10}} : r_A \iff r_A + (a, b/2, c/2) \\
    & \mathbf{J_{11}} : r_A \iff r_A + (a, b, 0) \\  
\end{aligned}
\end{equation}
where Inversion $\{-1|0\}$ constrain $\mathbf{J}_i$ where $i=2,3,5,9$ and therefore they take the general symmetric form:
\begin{equation}
    \mathbf{J_i} =
\begin{pmatrix}
  J_i^{xx} & \Gamma_i^{xy} & \Gamma_i^{xz}\\
  \Gamma_i^{xy} & J_i^{yy} & \Gamma_i^{yz}\\
  \Gamma_i^{xz} & \Gamma_i^{yz} & J_i^{zz}
\end{pmatrix}
\end{equation}
and the rest of bonds are not symmetry-constrained and therefore take the generic form:
\begin{equation}
    \mathbf{J_k} = 
\begin{pmatrix}
  J_k^{xx} & D_k^z + \Gamma_k^{xy} & -D_k^y + \Gamma_k^{xz}\\
  -D_k^z + \Gamma_k^{xy} & J_k^{yy} & D_k^x + \Gamma_k^{yz}\\
  D_k^y + \Gamma_k^{xz} & -D_k^x + \Gamma_k^{yz} & J_k^{zz}
\end{pmatrix}
\end{equation}
with $k=1,4,5,6,7,10$, and we include an extra single-ion anisotropy term of the form $-\sum_i A^{yy} (S_i^y)^2$ that helps in stabilizing the ground state along the experimentally-reported directions \cite{Ca2RuO4SpinModelINSPhysRevLett.115.247201}. We perform the linear spinwave calculation using the interaction parameters \label{Ca2RuO4SpinModelParameters}of the Heisenberg type to the first three NNs \cite{Ca2RuO4SpinModelINSPhysRevLett.115.247201} $J_1 = 8$ meV, $J_2 = 0.7$ meV, $J_3 = 0.7$ meV and we set $S=0.67$ similar to \cite{Ca2RuO4ValueofSPhysRevB.58.847}\cite{Ca2RuO4SpinModelINSPhysRevLett.115.247201}. For further couplings, we allow for off-diagonal exchange interactions to remove accidental degeneracies in the magnon spectrum and their used values are $J_4^{xx} = 0.2$ meV, $J_4^{xz} = 0.2$ meV, $J_4^{yz} = 0.09$ meV, $J_4^{zx} = 0.2$ meV, $J_4^{zy} = 0.06$ meV, $J_4^{yy} = 0.0002$ meV, $J_5^{xx} = J_5^{yy} = J_5^{zz} = -0.1$ meV, $J_5^{xz} = -0.6$ meV, $J_5^{yz} = 0.8$ meV, $J_6^{xx} = J_6^{yy} = J_6^{zz} = 0.03$ meV, $J_7^{xx} = J_7^{zz} = -0.25$ meV, $J_7^{yy} = -0.5$ meV, $J_8^{xx} = 0.01$ meV, $J_8^{xy} = 0.5$ meV, $J_8^{yy} = J_8^{zz} = 0.1$ meV, $J_8^{zy} = 0.2$ meV, $J_9^{xx} = J_9^{yy} = J_9^{zz} = -0.02$ meV, $J_9^{yz} = -0.5$ meV, $J_{10}^{xx} = J_{10}^{yy} = -0.02$ meV, $J_{10}^{zz} = -0.2$ meV, $J_{10}^{xy}= 0.5$ meV, $J_{10}^{yz}= -0.02$ meV and anisotropy $A^{yy}= 0.735$ meV. Based on this spin model, the magnon Hamiltonian $\mathbf{R}(\boldsymbol{k})$ is calculated and the magnon spectrum and band \textit{irreps} are shown in figure \ref{Example2UnperturbedSpectrum}.\\

\textbf{Weyl Magnons in Ca$_2$RuO$_4$}

\begin{table}[h!]
\centering
\begin{tabular}{|c|c|}
\hline
\textbf{($k_x$, $k_y$, $k_z$)} & \textbf{Topological Charge} \\
\hline
    \texttt{( 1.62561,\;-0.46326,\; 3.05048)}  & \texttt{+1} \\
    \texttt{(-1.60916,\; 0.45319,\; 2.99134)}  & \texttt{-1} \\
    \texttt{(-0.80420,\;-0.22994,\; 2.59234)}  & \texttt{-1} \\
    \texttt{(-0.34714,\;-0.09112,\; 2.35056)}  & \texttt{+1} \\
    \texttt{( 1.78258,\;-0.36203,\; 2.32929)}  & \texttt{-1} \\
    \texttt{( 2.10685,\;-1.06362,\; 2.15571)}  & \texttt{+1} \\
    \texttt{( 2.31297,\; 1.63148,\; 2.13981)}  & \texttt{+1} \\
    \texttt{( 0.09406,\;-0.89653,\; 2.00765)}  & \texttt{-1} \\
    \texttt{( 0.08200,\;-0.91530,\; 2.00562)}  & \texttt{-1} \\
    \texttt{(-0.01273,\;-3.07374,\; 1.99766)}  & \texttt{-1} \\
    \texttt{(-0.00477,\; 1.79253,\; 1.99405)}  & \texttt{-1} \\
    \texttt{(-2.58046,\; 0.10317,\; 1.82548)}  & \texttt{-1} \\
    \texttt{(-2.18417,\; 2.94025,\; 1.21087)}  & \texttt{+1} \\
    \texttt{(-2.02496,\; 3.03443,\; 0.93751)}  & \texttt{+1} \\
    \texttt{( 2.02496,\;-3.03443,\;-0.93751)}  & \texttt{-1} \\
    \texttt{( 2.18417,\;-2.94025,\;-1.21087)}  & \texttt{-1} \\
    \texttt{( 2.58046,\;-0.10317,\;-1.82548)}  & \texttt{+1} \\
    \texttt{( 0.00477,\;-1.79253,\;-1.99405)}  & \texttt{+1} \\
    \texttt{( 0.01273,\; 3.07374,\;-1.99766)}  & \texttt{+1} \\
    \texttt{(-0.08200,\; 0.91530,\;-2.00562)}  & \texttt{+1} \\
    \texttt{(-0.09406,\; 0.89653,\;-2.00765)}  & \texttt{+1} \\
    \texttt{(-2.31297,\;-1.63148,\;-2.13981)}  & \texttt{-1} \\
    \texttt{(-2.10685,\; 1.06362,\;-2.15571)}  & \texttt{-1} \\
    \texttt{(-1.78258,\; 0.36203,\;-2.32929)}  & \texttt{+1} \\
    \texttt{( 0.34714,\; 0.09112,\;-2.35056)}  & \texttt{-1} \\
    \texttt{( 0.80420,\; 0.22994,\;-2.59234)}  & \texttt{+1} \\
    \texttt{( 1.60916,\;-0.45319,\;-2.99134)}  & \texttt{+1} \\
    \texttt{(-1.62561,\; 0.46326,\;-3.05048)}  & \texttt{-1} \\
\hline
\end{tabular}
\caption{Explicit locations of Weyl Points between bands 2 and 3 in the perturbed LSW spectrum of Ca$_2$RuO$_4$ and their charges. Calculation is based on the spin model given above in Methods \ref{Ca2RuO4SpinModelParameters}.}
\label{Ca2RuO4WeylPointsTable}
\end{table}

\section*{Acknowledgments}
Authors would like to thank Prof. Rolando Valdés Aguilar for helpful discussions. AEF acknowledges Dr. Xu Yang (OSU), Dr. Yufei Li (OSU) and Mr. Mohamed Nawwar (OSU) for technical discussions. \\
\textbf{Funding:} This research was primarily supported by the Center for Emergent Materials, an NSF MRSEC, under award number DMR$-2011876$.

\section*{Authors Contribution}
MJK designed, implemented the search algorithm and prepared the supplementary materials. AEF thoroughly analyzed the algorithm results with the help of AJW, SH, JG, MJK and YML. Bulk and surface calculations along with band topology analyses were carried out by AEF. Manuscript was prepared by AEF with input from AJW, MJK, YML, SH and JG. Manuscript was edited by AEF and YML. All authors contributed to the scientific discussion. YML supervised the project.
\bibliographystyle{unsrtnat}
\bibliography{main}



\startcontents[sections]
\printcontents[sections]{}{1}{\section*{Supplementary materials: Details of type-I topological magnon materials}}

\setcounter{section}{7}

\label{TopoMagMaterials}
\section{Topological Magnon Materials\label{TopoMag2}}
\begin{center}
\begin{longtable}{cccl}
\caption{Summary of topological magnon materials.}\\
\toprule
&Material($T_\textrm{c}$[K])&MSG \& WPs&Perturbation \& Subgroup\\
\midrule
\endfirsthead
\multicolumn{4}{@{}l}{\ldots continued}\\
\toprule
&Material($T_\textrm{c}$[K])&MSG \& WPs&Perturbation \& Subgroup\\
\midrule
\endhead
\multicolumn{4}{r}{continued on next page\ldots}
\endfoot
\bottomrule
\endlastfoot
\multirow{2}{*}{1}&\multirow{2}{*}{\href{http://webbdcrista1.ehu.es/magndata/index.php?this_label=0.65}{$\alpha$-Fe\textsubscript{2}O\textsubscript{3}(955)}}&\multirow{1}{*}{$C2'/c'$}&\multirow{3}{*}{B in generic direction $P\bar{1}~(2.4)$}\\*\multirow{2}{*}{2}&\multirow{2}{*}{\href{http://webbdcrista1.ehu.es/magndata/index.php?this_label=0.493}{Ho(Co\textsubscript{0.667}Ga\textsubscript{0.333})\textsubscript{2}(31)}}&\multirow{1}{*}{$(15.89)$}&\\*&&\multirow{1}{*}{$8f$}&\\\midrule
\multirow{3}{*}{3}&\multirow{3}{*}{\href{http://webbdcrista1.ehu.es/magndata/index.php?this_label=1.466}{MnPt\textsubscript{0.5}Pd\textsubscript{0.5}(866)}}&\multirow{1}{*}{$C_{a}mma$}&\multirow{3}{*}{B $\perp$ [010] $C2'/m'~(12.62)$}\\*&&\multirow{1}{*}{$(67.509)$}&\\*&&\multirow{1}{*}{$4g$}&\\\midrule
\multirow{5}{*}{4}&\multirow{5}{*}{\href{http://webbdcrista1.ehu.es/magndata/index.php?this_label=0.713}{NiFe\textsubscript{2}O\textsubscript{4}(858)}}&\multirow{3}{*}{$I4_{1}/am'd'$}&\multirow{1}{*}{B in generic direction $P\bar{1}~(2.4)$}\\*&&\multirow{3}{*}{$(141.557)$}&\multirow{1}{*}{B $\parallel$ [110] $C2'/c'~(15.89)$}\\
\end{longtable}
\end{center}

\section{MSG $P_{a}2_{1}/m~(11.55)$\label{SuppSection1Label}}
\textbf{Nontrivial-SI Subgroups:} $P\bar{1}~(2.4)$, $P2_{1}'/c'~(14.79)$, $P_{S}\bar{1}~(2.7)$, $P2_{1}/m~(11.50)$.\\

\textbf{Trivial-SI Subgroups:} $Pc'~(7.26)$, $P2_{1}'~(4.9)$, $P_{S}1~(1.3)$, $Pm~(6.18)$, $P_{a}m~(6.21)$, $P2_{1}~(4.7)$, $P_{a}2_{1}~(4.10)$.\\

\subsection{WP: $4a$}
\textbf{BCS Materials:} {La\textsubscript{3}OsO\textsubscript{7}~(45 K)}\footnote{BCS web page: \texttt{\href{http://webbdcrista1.ehu.es/magndata/index.php?this\_label=1.571} {http://webbdcrista1.ehu.es/magndata/index.php?this\_label=1.571}}}, {La\textsubscript{3}OsO\textsubscript{7}~(45 K)}\footnote{BCS web page: \texttt{\href{http://webbdcrista1.ehu.es/magndata/index.php?this\_label=1.570} {http://webbdcrista1.ehu.es/magndata/index.php?this\_label=1.570}}}, {La\textsubscript{2.8}Ca\textsubscript{0.2}OsO\textsubscript{7}~(33 K)}\footnote{BCS web page: \texttt{\href{http://webbdcrista1.ehu.es/magndata/index.php?this\_label=1.572} {http://webbdcrista1.ehu.es/magndata/index.php?this\_label=1.572}}}.\\
\subsubsection{Topological bands in subgroup $P_{S}\bar{1}~(2.7)$}
\textbf{Perturbation:}
\begin{itemize}
\item strain in generic direction.
\end{itemize}
\begin{center}
\begin{longtable}{cc}
\caption{Number of times a given SI can appear in the band structure. Supergroup: $P_{a}2_{1}/m~(11.55)$, WP: $4a$, Subgroup: $P_{S}\bar{1}~(2.7)$.\label{gap_table_11.55_2.7_4a_strainingenericdirection}}\\
\toprule
SI&Possible \#Gaps\\
\midrule
\endfirsthead
\multicolumn{2}{@{}l}{\ldots continued}\\
\toprule
SI&Possible \#Gaps\\
\midrule
\endhead
\midrule\multicolumn{2}{r}{continued on next page\ldots}
\endfoot
\bottomrule
\endlastfoot
000&0\\
100&1\\
gapless&2\\
nontrivial&1\\
\end{longtable}
\end{center}
\begin{center}
\begin{longtable}{cc}
\caption{Possible SI values in each gap. Supergroup: $P_{a}2_{1}/m~(11.55)$, WP: $4a$, Subgroup: $P_{S}\bar{1}~(2.7)$.\label{si_table_11.55_2.7_4a_strainingenericdirection}}\\
\toprule
Gap\#& Possible SI Values\\
\midrule
\endfirsthead
\multicolumn{2}{@{}l}{\ldots continued}\\
\toprule
Gap\#& Possible SI Values\\
\midrule
\endhead
\multicolumn{2}{r}{continued on next page\ldots}
\endfoot
\bottomrule
\endlastfoot
\multirow{1}{*}{1}&\multirow{1}{*}{G}\\\midrule
\multirow{1}{*}{2}&\multirow{1}{*}{100}\\\midrule
\multirow{1}{*}{3}&\multirow{1}{*}{G}\\\midrule
\multirow{1}{*}{4}&\multirow{1}{*}{000}\\
\end{longtable}
\end{center}
\subsection{WP: $4a+4c$}
\textbf{BCS Materials:} {AgMnVO\textsubscript{4}~(12.1 K)}\footnote{BCS web page: \texttt{\href{http://webbdcrista1.ehu.es/magndata/index.php?this\_label=1.116} {http://webbdcrista1.ehu.es/magndata/index.php?this\_label=1.116}}}.\\
\subsubsection{Topological bands in subgroup $P_{S}\bar{1}~(2.7)$}
\textbf{Perturbation:}
\begin{itemize}
\item strain in generic direction.
\end{itemize}
\begin{figure}[H]
\centering
\includegraphics[scale=0.6]{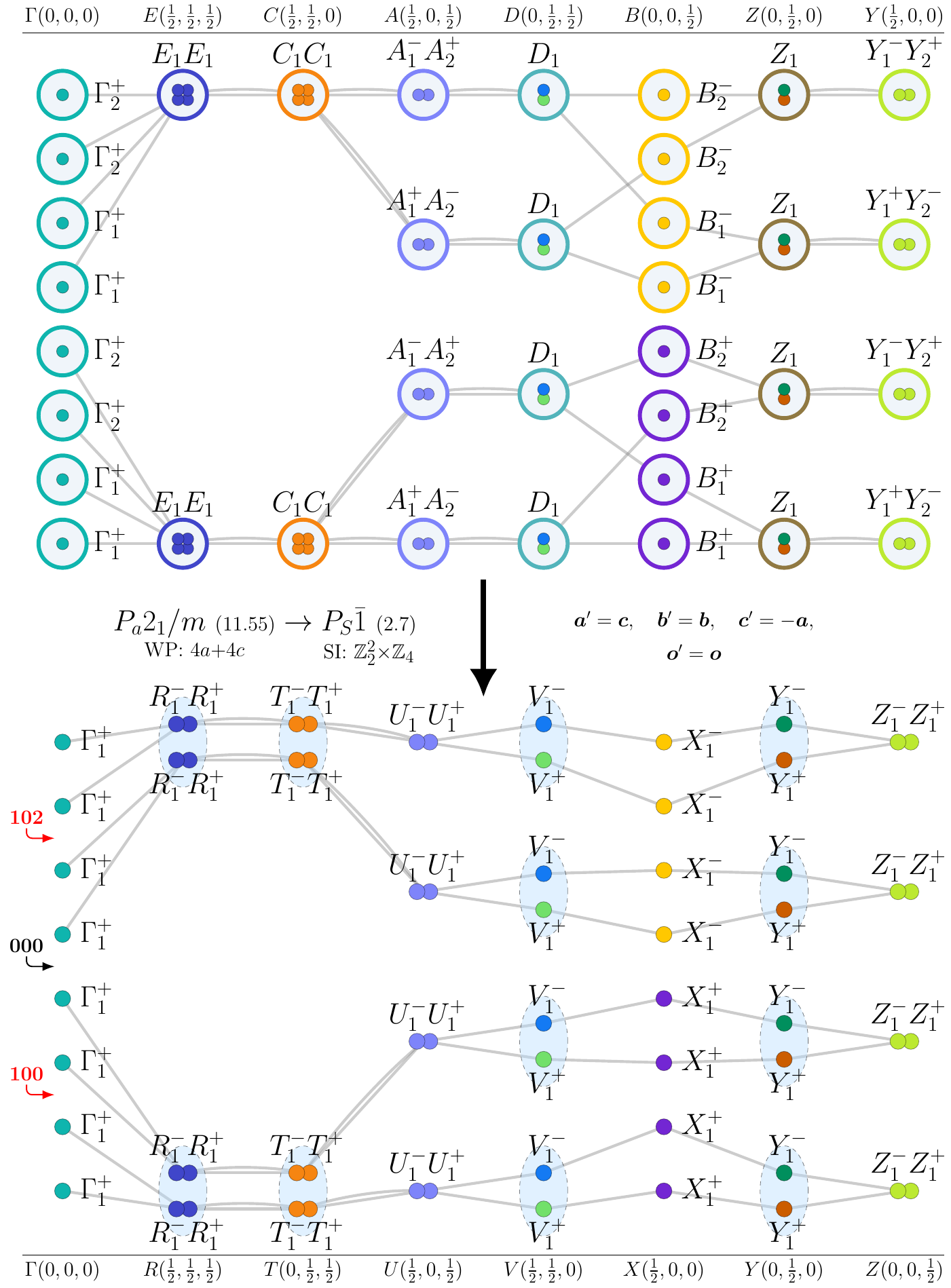}
\caption{Topological magnon bands in subgroup $P_{S}\bar{1}~(2.7)$ for magnetic moments on Wyckoff positions $4a+4c$ of supergroup $P_{a}2_{1}/m~(11.55)$.\label{fig_11.55_2.7_strainingenericdirection_4a+4c}}
\end{figure}
\begin{center}
\begin{longtable}{cc}
\caption{Number of times a given SI can appear in the band structure. Supergroup: $P_{a}2_{1}/m~(11.55)$, WP: $4a$+$4c$, Subgroup: $P_{S}\bar{1}~(2.7)$.\label{gap_table_11.55_2.7_4a+4c_strainingenericdirection}}\\
\toprule
SI&Possible \#Gaps\\
\midrule
\endfirsthead
\multicolumn{2}{@{}l}{\ldots continued}\\
\toprule
SI&Possible \#Gaps\\
\midrule
\endhead
\midrule\multicolumn{2}{r}{continued on next page\ldots}
\endfoot
\bottomrule
\endlastfoot
000&0, 1\\
002&0, 1\\
011&0, 1\\
013&0, 1\\
100&0, 1, 2\\
102&0, 1, 2\\
111&0, 1\\
113&0, 1\\
gapless&4\\
nontrivial&2, 3\\
\end{longtable}
\end{center}
\begin{center}
\begin{longtable}{cc}
\caption{Possible SI values in each gap. Supergroup: $P_{a}2_{1}/m~(11.55)$, WP: $4a$+$4c$, Subgroup: $P_{S}\bar{1}~(2.7)$.\label{si_table_11.55_2.7_4a+4c_strainingenericdirection}}\\
\toprule
Gap\#& Possible SI Values\\
\midrule
\endfirsthead
\multicolumn{2}{@{}l}{\ldots continued}\\
\toprule
Gap\#& Possible SI Values\\
\midrule
\endhead
\multicolumn{2}{r}{continued on next page\ldots}
\endfoot
\bottomrule
\endlastfoot
\multirow{1}{*}{1}&\multirow{1}{*}{G}\\\midrule
\multirow{1}{*}{2}&\multirow{1}{*}{100, 102, 111}\\\midrule
\multirow{1}{*}{3}&\multirow{1}{*}{G}\\\midrule
\multirow{1}{*}{4}&\multirow{1}{*}{000, 002, 011, 013}\\\midrule
\multirow{1}{*}{5}&\multirow{1}{*}{G}\\\midrule
\multirow{1}{*}{6}&\multirow{1}{*}{100, 102, 113}\\\midrule
\multirow{1}{*}{7}&\multirow{1}{*}{G}\\\midrule
\multirow{1}{*}{8}&\multirow{1}{*}{000}\\
\end{longtable}
\end{center}

\section{MSG $P_{C}2_{1}/m~(11.57)$}
\textbf{Nontrivial-SI Subgroups:} $P\bar{1}~(2.4)$, $P2'/c'~(13.69)$, $P_{S}\bar{1}~(2.7)$, $P2_{1}/m~(11.50)$.\\

\textbf{Trivial-SI Subgroups:} $Pc'~(7.26)$, $P2'~(3.3)$, $P_{S}1~(1.3)$, $Pm~(6.18)$, $P_{C}m~(6.23)$, $P2_{1}~(4.7)$, $P_{C}2_{1}~(4.12)$.\\

\subsection{WP: $4e$}
\textbf{BCS Materials:} {HoRh~(3.2 K)}\footnote{BCS web page: \texttt{\href{http://webbdcrista1.ehu.es/magndata/index.php?this\_label=2.71} {http://webbdcrista1.ehu.es/magndata/index.php?this\_label=2.71}}}.\\
\subsubsection{Topological bands in subgroup $P\bar{1}~(2.4)$}
\textbf{Perturbations:}
\begin{itemize}
\item B $\parallel$ [010] and strain in generic direction,
\item B $\perp$ [010] and strain in generic direction,
\item B in generic direction.
\end{itemize}
\begin{figure}[H]
\centering
\includegraphics[scale=0.6]{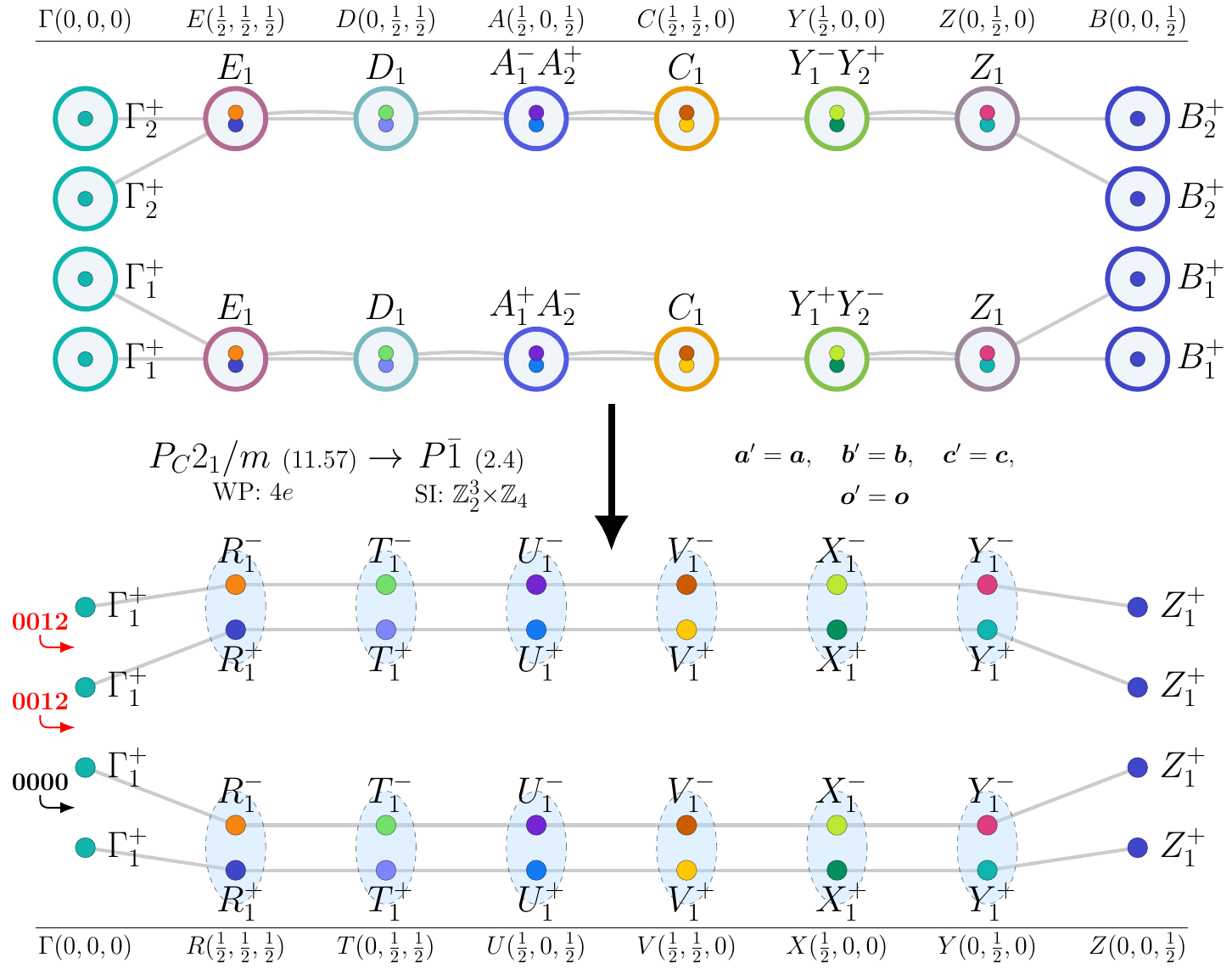}
\caption{Topological magnon bands in subgroup $P\bar{1}~(2.4)$ for magnetic moments on Wyckoff position $4e$ of supergroup $P_{C}2_{1}/m~(11.57)$.\label{fig_11.57_2.4_Bparallel010andstrainingenericdirection_4e}}
\end{figure}
\begin{center}
\begin{longtable}{cc}
\caption{Number of times a given SI can appear in the band structure. Supergroup: $P_{C}2_{1}/m~(11.57)$, WP: $4e$, Subgroup: $P\bar{1}~(2.4)$.\label{gap_table_11.57_2.4_4e_Bparallel010andstrainingenericdirection}}\\
\toprule
SI&Possible \#Gaps\\
\midrule
\endfirsthead
\multicolumn{2}{@{}l}{\ldots continued}\\
\toprule
SI&Possible \#Gaps\\
\midrule
\endhead
\midrule\multicolumn{2}{r}{continued on next page\ldots}
\endfoot
\bottomrule
\endlastfoot
0000&0, 1, 2\\
0001&0, 1, 2\\
0003&0, 1, 2\\
0011&0, 1, 2\\
0012&1, 2, 3\\
0013&0, 1, 2\\
0100&0, 1, 2\\
0101&0, 1, 2\\
0102&0, 1, 2\\
0103&0, 1, 2\\
0110&0, 1, 2\\
0111&0, 1, 2\\
0112&0, 1, 2\\
0113&0, 1, 2\\
1000&0, 1, 2\\
1001&0, 1, 2\\
1002&0, 1, 2\\
1003&0, 1, 2\\
1010&0, 1, 2\\
1011&0, 1, 2\\
1012&0, 1, 2\\
1013&0, 1, 2\\
1100&0, 1, 2\\
1101&0, 1\\
1102&0, 1, 2\\
1103&0, 1\\
1110&0, 1, 2\\
1111&0, 1\\
1112&0, 1, 2\\
1113&0, 1\\
gapless&0\\
nontrivial&1, 2, 3\\
\end{longtable}
\end{center}
\begin{center}
\begin{longtable}{cc}
\caption{Possible SI values in each gap. Supergroup: $P_{C}2_{1}/m~(11.57)$, WP: $4e$, Subgroup: $P\bar{1}~(2.4)$.\label{si_table_11.57_2.4_4e_Bparallel010andstrainingenericdirection}}\\
\toprule
Gap\#& Possible SI Values\\
\midrule
\endfirsthead
\multicolumn{2}{@{}l}{\ldots continued}\\
\toprule
Gap\#& Possible SI Values\\
\midrule
\endhead
\multicolumn{2}{r}{continued on next page\ldots}
\endfoot
\bottomrule
\endlastfoot
\multirow{6}{*}{1}&\multirow{1}{*}{0000, 0001, 0003, 0011, 0012}\\*
&\multirow{1}{*}{0013, 0100, 0101, 0102, 0103}\\*
&\multirow{1}{*}{0110, 0111, 0112, 0113, 1000}\\*
&\multirow{1}{*}{1001, 1002, 1003, 1010, 1011}\\*
&\multirow{1}{*}{1012, 1013, 1100, 1102, 1103}\\*
&\multirow{1}{*}{1110, 1112, 1113}\\\midrule
\multirow{1}{*}{2}&\multirow{1}{*}{0012}\\\midrule
\multirow{6}{*}{3}&\multirow{1}{*}{0000, 0001, 0003, 0011, 0012}\\*
&\multirow{1}{*}{0013, 0100, 0101, 0102, 0103}\\*
&\multirow{1}{*}{0110, 0111, 0112, 0113, 1000}\\*
&\multirow{1}{*}{1001, 1002, 1003, 1010, 1011}\\*
&\multirow{1}{*}{1012, 1013, 1100, 1101, 1102}\\*
&\multirow{1}{*}{1110, 1111, 1112}\\\midrule
\multirow{1}{*}{4}&\multirow{1}{*}{0000}\\
\end{longtable}
\end{center}
\subsubsection{Topological bands in subgroup $P2'/c'~(13.69)$}
\textbf{Perturbation:}
\begin{itemize}
\item B $\perp$ [010].
\end{itemize}
\begin{figure}[H]
\centering
\includegraphics[scale=0.6]{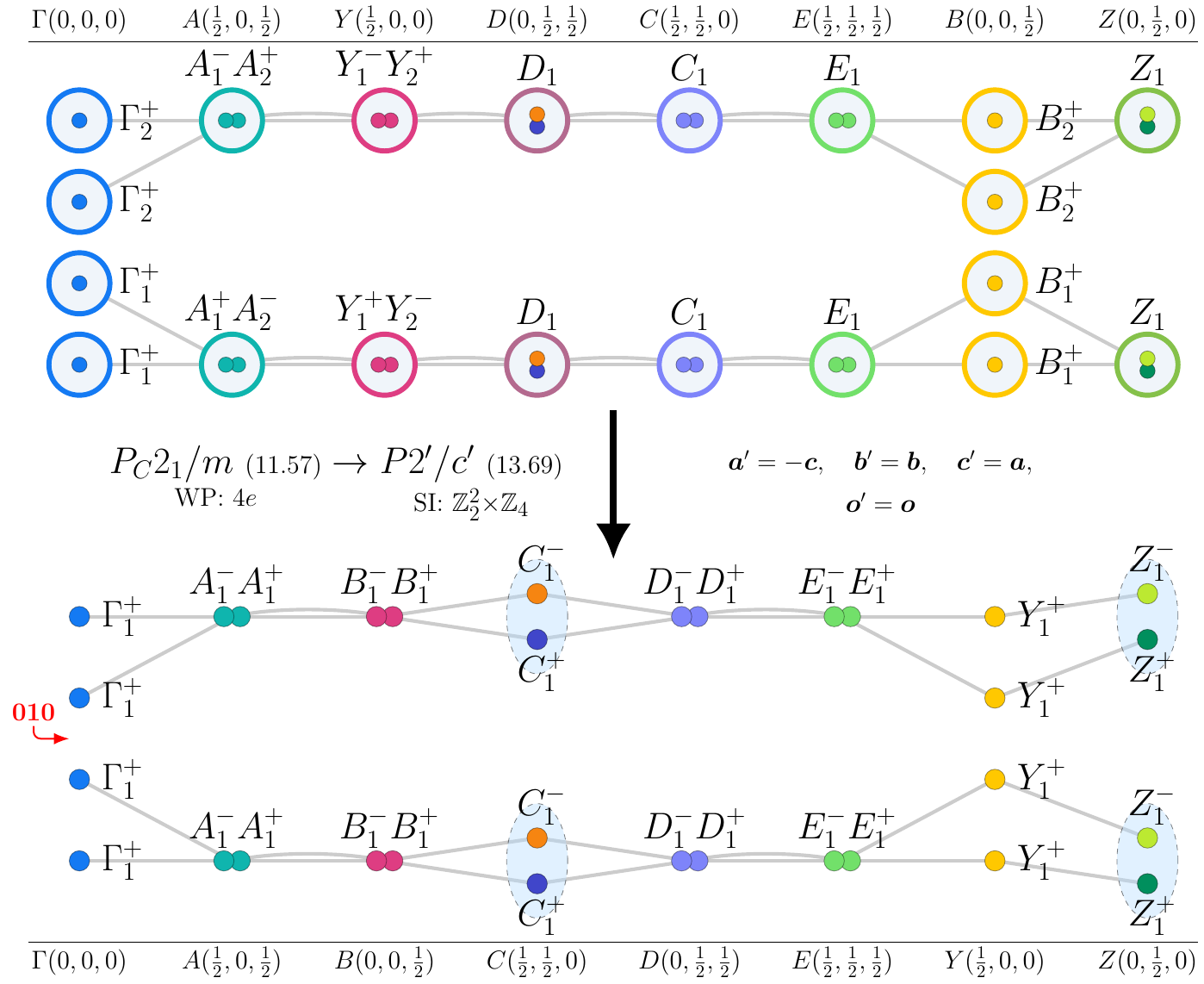}
\caption{Topological magnon bands in subgroup $P2'/c'~(13.69)$ for magnetic moments on Wyckoff position $4e$ of supergroup $P_{C}2_{1}/m~(11.57)$.\label{fig_11.57_13.69_Bperp010_4e}}
\end{figure}
\begin{center}
\begin{longtable}{cc}
\caption{Number of times a given SI can appear in the band structure. Supergroup: $P_{C}2_{1}/m~(11.57)$, WP: $4e$, Subgroup: $P2'/c'~(13.69)$.\label{gap_table_11.57_13.69_4e_Bperp010}}\\
\toprule
SI&Possible \#Gaps\\
\midrule
\endfirsthead
\multicolumn{2}{@{}l}{\ldots continued}\\
\toprule
SI&Possible \#Gaps\\
\midrule
\endhead
\midrule\multicolumn{2}{r}{continued on next page\ldots}
\endfoot
\bottomrule
\endlastfoot
000&0\\
010&1\\
gapless&2\\
nontrivial&1\\
\end{longtable}
\end{center}
\begin{center}
\begin{longtable}{cc}
\caption{Possible SI values in each gap. Supergroup: $P_{C}2_{1}/m~(11.57)$, WP: $4e$, Subgroup: $P2'/c'~(13.69)$.\label{si_table_11.57_13.69_4e_Bperp010}}\\
\toprule
Gap\#& Possible SI Values\\
\midrule
\endfirsthead
\multicolumn{2}{@{}l}{\ldots continued}\\
\toprule
Gap\#& Possible SI Values\\
\midrule
\endhead
\multicolumn{2}{r}{continued on next page\ldots}
\endfoot
\bottomrule
\endlastfoot
\multirow{1}{*}{1}&\multirow{1}{*}{G}\\\midrule
\multirow{1}{*}{2}&\multirow{1}{*}{010}\\\midrule
\multirow{1}{*}{3}&\multirow{1}{*}{G}\\\midrule
\multirow{1}{*}{4}&\multirow{1}{*}{000}\\
\end{longtable}
\end{center}
\subsubsection{Topological bands in subgroup $P_{S}\bar{1}~(2.7)$}
\textbf{Perturbation:}
\begin{itemize}
\item strain in generic direction.
\end{itemize}
\begin{figure}[H]
\centering
\includegraphics[scale=0.6]{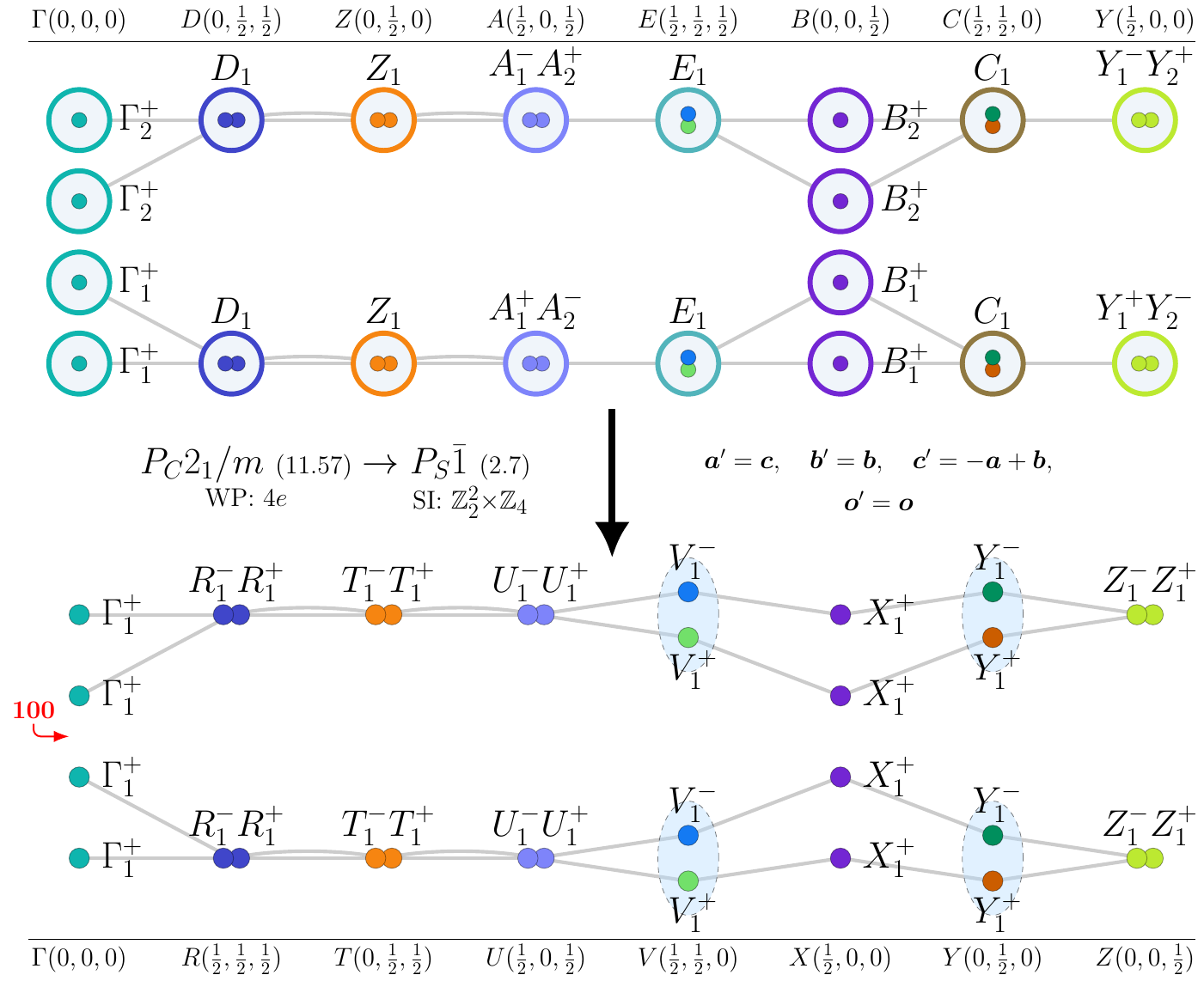}
\caption{Topological magnon bands in subgroup $P_{S}\bar{1}~(2.7)$ for magnetic moments on Wyckoff position $4e$ of supergroup $P_{C}2_{1}/m~(11.57)$.\label{fig_11.57_2.7_strainingenericdirection_4e}}
\end{figure}
\begin{center}
\begin{longtable}{cc}
\caption{Number of times a given SI can appear in the band structure. Supergroup: $P_{C}2_{1}/m~(11.57)$, WP: $4e$, Subgroup: $P_{S}\bar{1}~(2.7)$.\label{gap_table_11.57_2.7_4e_strainingenericdirection}}\\
\toprule
SI&Possible \#Gaps\\
\midrule
\endfirsthead
\multicolumn{2}{@{}l}{\ldots continued}\\
\toprule
SI&Possible \#Gaps\\
\midrule
\endhead
\midrule\multicolumn{2}{r}{continued on next page\ldots}
\endfoot
\bottomrule
\endlastfoot
000&0\\
100&1\\
gapless&2\\
nontrivial&1\\
\end{longtable}
\end{center}
\begin{center}
\begin{longtable}{cc}
\caption{Possible SI values in each gap. Supergroup: $P_{C}2_{1}/m~(11.57)$, WP: $4e$, Subgroup: $P_{S}\bar{1}~(2.7)$.\label{si_table_11.57_2.7_4e_strainingenericdirection}}\\
\toprule
Gap\#& Possible SI Values\\
\midrule
\endfirsthead
\multicolumn{2}{@{}l}{\ldots continued}\\
\toprule
Gap\#& Possible SI Values\\
\midrule
\endhead
\multicolumn{2}{r}{continued on next page\ldots}
\endfoot
\bottomrule
\endlastfoot
\multirow{1}{*}{1}&\multirow{1}{*}{G}\\\midrule
\multirow{1}{*}{2}&\multirow{1}{*}{100}\\\midrule
\multirow{1}{*}{3}&\multirow{1}{*}{G}\\\midrule
\multirow{1}{*}{4}&\multirow{1}{*}{000}\\
\end{longtable}
\end{center}

\section{MSG $P4'/nbm'~(125.367)$}
\textbf{Nontrivial-SI Subgroups:} $P\bar{1}~(2.4)$, $C2'/m'~(12.62)$, $C2'/m'~(12.62)$, $P2~(3.1)$, $Cm'm'2~(35.168)$, $P2/c~(13.65)$, $Cm'm'a~(67.505)$, $P2~(3.1)$, $P2/c~(13.65)$, $Pban~(50.277)$.\\

\textbf{Trivial-SI Subgroups:} $Cm'~(8.34)$, $Cm'~(8.34)$, $C2'~(5.15)$, $Pc~(7.24)$, $Abm'2'~(39.198)$, $Pc~(7.24)$, $Pba2~(32.135)$, $P4'bm'~(100.174)$, $Pnc2~(30.111)$.\\

\subsection{WP: $4f$}
\textbf{BCS Materials:} {ZrMn\textsubscript{2}Ge\textsubscript{4}O\textsubscript{12}~(8 K)}\footnote{BCS web page: \texttt{\href{http://webbdcrista1.ehu.es/magndata/index.php?this\_label=0.315} {http://webbdcrista1.ehu.es/magndata/index.php?this\_label=0.315}}}.\\
\subsubsection{Topological bands in subgroup $P\bar{1}~(2.4)$}
\textbf{Perturbations:}
\begin{itemize}
\item strain in generic direction,
\item B $\parallel$ [001] and strain $\perp$ [100],
\item B $\parallel$ [100] and strain $\parallel$ [110],
\item B $\parallel$ [100] and strain $\perp$ [001],
\item B $\parallel$ [100] and strain $\perp$ [110],
\item B $\parallel$ [110] and strain $\parallel$ [100],
\item B $\parallel$ [110] and strain $\perp$ [001],
\item B $\parallel$ [110] and strain $\perp$ [100],
\item B $\parallel$ [110] and strain $\perp$ [110],
\item B in generic direction,
\item B $\perp$ [110] and strain $\parallel$ [100],
\item B $\perp$ [110] and strain $\perp$ [001],
\item B $\perp$ [110] and strain $\perp$ [100].
\end{itemize}
\begin{figure}[H]
\centering
\includegraphics[scale=0.6]{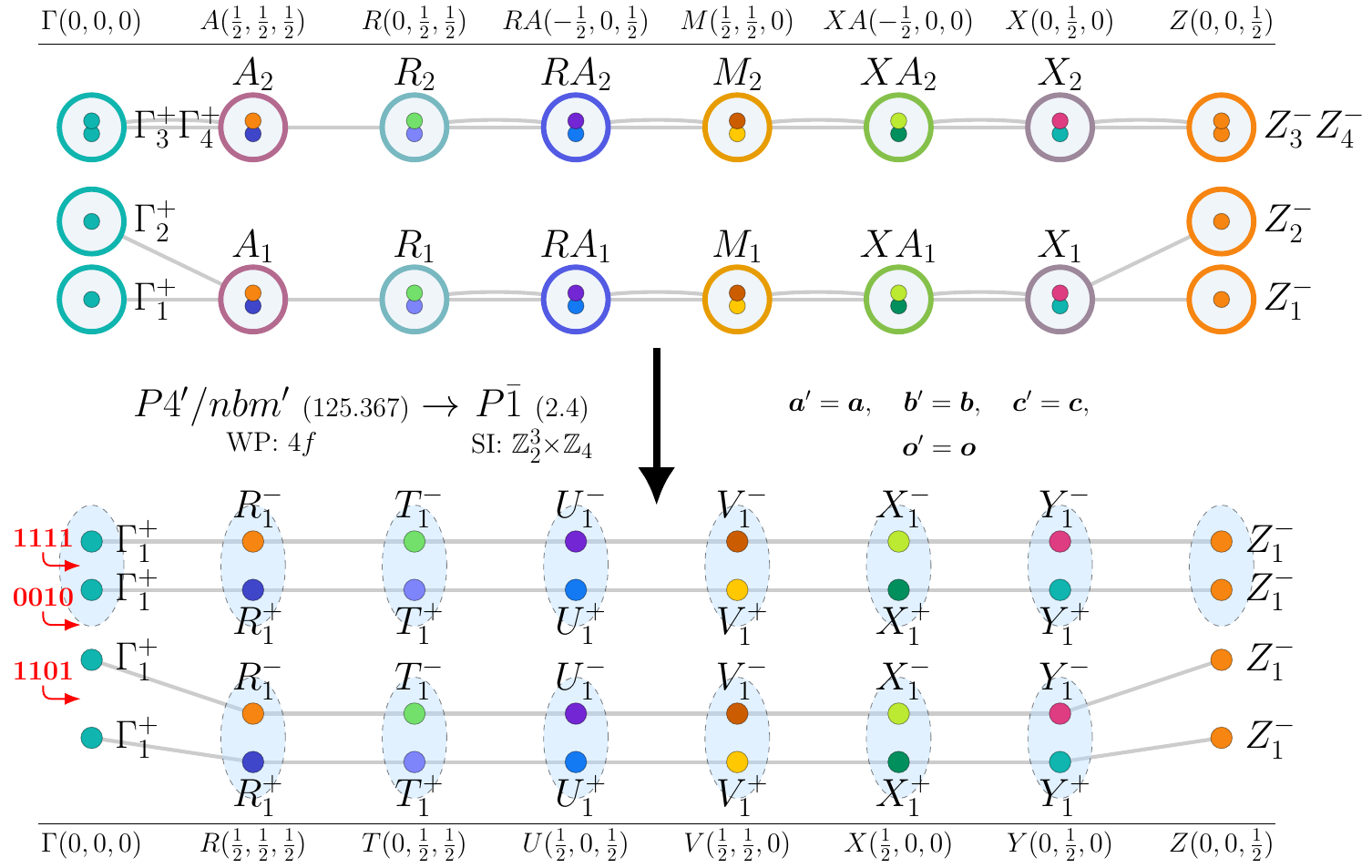}
\caption{Topological magnon bands in subgroup $P\bar{1}~(2.4)$ for magnetic moments on Wyckoff position $4f$ of supergroup $P4'/nbm'~(125.367)$.\label{fig_125.367_2.4_strainingenericdirection_4f}}
\end{figure}
\begin{center}
\begin{longtable}{cc}
\caption{Number of times a given SI can appear in the band structure. Supergroup: $P4'/nbm'~(125.367)$, WP: $4f$, Subgroup: $P\bar{1}~(2.4)$.\label{gap_table_125.367_2.4_4f_strainingenericdirection}}\\
\toprule
SI&Possible \#Gaps\\
\midrule
\endfirsthead
\multicolumn{2}{@{}l}{\ldots continued}\\
\toprule
SI&Possible \#Gaps\\
\midrule
\endhead
\midrule\multicolumn{2}{r}{continued on next page\ldots}
\endfoot
\bottomrule
\endlastfoot
0000&0, 1, 2\\
0001&0, 1, 2\\
0003&0, 1, 2\\
0010&1, 2, 3\\
0011&0, 1, 2\\
0013&0, 1, 2\\
0100&0, 1, 2\\
0101&0, 1, 2\\
0102&0, 1, 2\\
0103&0, 1, 2\\
0110&0, 1, 2\\
0111&0, 1, 2\\
0112&0, 1, 2\\
0113&0, 1, 2\\
1000&0, 1, 2\\
1001&0, 1, 2\\
1002&0, 1, 2\\
1003&0, 1, 2\\
1010&0, 1, 2\\
1011&0, 1, 2\\
1012&0, 1, 2\\
1013&0, 1, 2\\
1100&0, 1, 2\\
1101&0, 1\\
1102&0, 1, 2\\
1103&0, 1\\
1110&0, 1, 2\\
1111&0, 1\\
1112&0, 1, 2\\
1113&0, 1\\
gapless&0\\
nontrivial&1, 2, 3\\
\end{longtable}
\end{center}
\begin{center}
\begin{longtable}{cc}
\caption{Possible SI values in each gap. Supergroup: $P4'/nbm'~(125.367)$, WP: $4f$, Subgroup: $P\bar{1}~(2.4)$.\label{si_table_125.367_2.4_4f_strainingenericdirection}}\\
\toprule
Gap\#& Possible SI Values\\
\midrule
\endfirsthead
\multicolumn{2}{@{}l}{\ldots continued}\\
\toprule
Gap\#& Possible SI Values\\
\midrule
\endhead
\multicolumn{2}{r}{continued on next page\ldots}
\endfoot
\bottomrule
\endlastfoot
\multirow{6}{*}{1}&\multirow{1}{*}{0000, 0001, 0003, 0010, 0011}\\*
&\multirow{1}{*}{0013, 0100, 0101, 0102, 0103}\\*
&\multirow{1}{*}{0110, 0111, 0112, 0113, 1000}\\*
&\multirow{1}{*}{1001, 1002, 1003, 1010, 1011}\\*
&\multirow{1}{*}{1012, 1013, 1100, 1101, 1102}\\*
&\multirow{1}{*}{1110, 1112, 1113}\\\midrule
\multirow{1}{*}{2}&\multirow{1}{*}{0010}\\\midrule
\multirow{6}{*}{3}&\multirow{1}{*}{0000, 0001, 0003, 0010, 0011}\\*
&\multirow{1}{*}{0013, 0100, 0101, 0102, 0103}\\*
&\multirow{1}{*}{0110, 0111, 0112, 0113, 1000}\\*
&\multirow{1}{*}{1001, 1002, 1003, 1010, 1011}\\*
&\multirow{1}{*}{1012, 1013, 1100, 1102, 1103}\\*
&\multirow{1}{*}{1110, 1111, 1112}\\\midrule
\multirow{1}{*}{4}&\multirow{1}{*}{0000}\\
\end{longtable}
\end{center}
\subsubsection{Topological bands in subgroup $C2'/m'~(12.62)$}
\textbf{Perturbation:}
\begin{itemize}
\item B $\parallel$ [110].
\end{itemize}
\begin{figure}[H]
\centering
\includegraphics[scale=0.6]{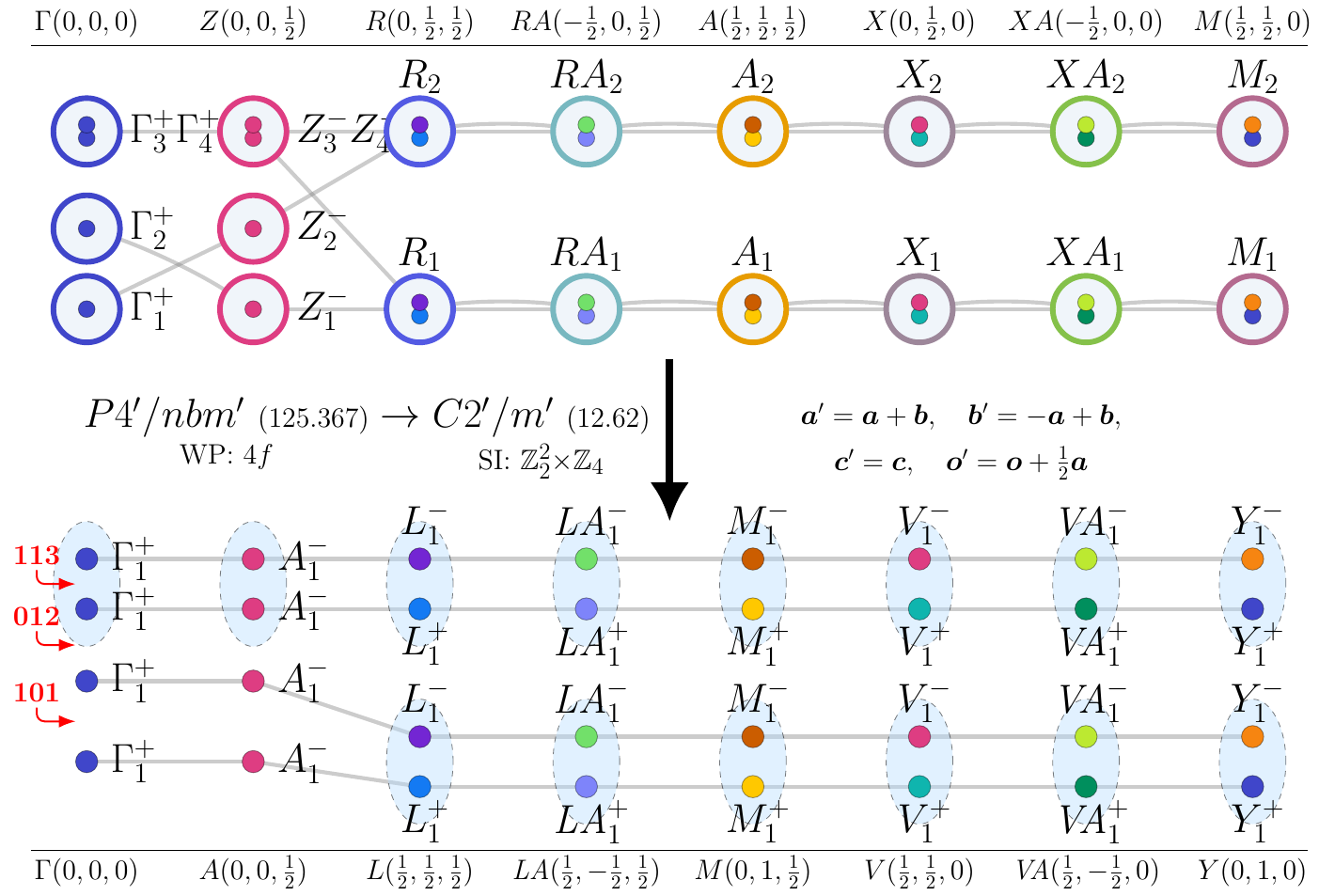}
\caption{Topological magnon bands in subgroup $C2'/m'~(12.62)$ for magnetic moments on Wyckoff position $4f$ of supergroup $P4'/nbm'~(125.367)$.\label{fig_125.367_12.62_Bparallel110_4f}}
\end{figure}
\begin{center}
\begin{longtable}{cc}
\caption{Number of times a given SI can appear in the band structure. Supergroup: $P4'/nbm'~(125.367)$, WP: $4f$, Subgroup: $C2'/m'~(12.62)$.\label{gap_table_125.367_12.62_4f_Bparallel110}}\\
\toprule
SI&Possible \#Gaps\\
\midrule
\endfirsthead
\multicolumn{2}{@{}l}{\ldots continued}\\
\toprule
SI&Possible \#Gaps\\
\midrule
\endhead
\midrule\multicolumn{2}{r}{continued on next page\ldots}
\endfoot
\bottomrule
\endlastfoot
000&0, 1, 2\\
001&0, 1\\
003&0, 1\\
011&0, 1\\
012&1, 2, 3\\
013&0, 1\\
100&0, 1, 2\\
101&0, 1\\
103&0, 1\\
111&0, 1\\
112&0, 1, 2\\
113&0, 1\\
gapless&0\\
nontrivial&1, 2, 3\\
\end{longtable}
\end{center}
\begin{center}
\begin{longtable}{cc}
\caption{Possible SI values in each gap. Supergroup: $P4'/nbm'~(125.367)$, WP: $4f$, Subgroup: $C2'/m'~(12.62)$.\label{si_table_125.367_12.62_4f_Bparallel110}}\\
\toprule
Gap\#& Possible SI Values\\
\midrule
\endfirsthead
\multicolumn{2}{@{}l}{\ldots continued}\\
\toprule
Gap\#& Possible SI Values\\
\midrule
\endhead
\multicolumn{2}{r}{continued on next page\ldots}
\endfoot
\bottomrule
\endlastfoot
\multirow{2}{*}{1}&\multirow{1}{*}{000, 001, 011, 012, 100}\\*
&\multirow{1}{*}{101, 111, 112}\\\midrule
\multirow{1}{*}{2}&\multirow{1}{*}{012}\\\midrule
\multirow{2}{*}{3}&\multirow{1}{*}{000, 003, 012, 013, 100}\\*
&\multirow{1}{*}{103, 112, 113}\\\midrule
\multirow{1}{*}{4}&\multirow{1}{*}{000}\\
\end{longtable}
\end{center}
\subsubsection{Topological bands in subgroup $C2'/m'~(12.62)$}
\textbf{Perturbations:}
\begin{itemize}
\item strain $\perp$ [110],
\item B $\perp$ [110].
\end{itemize}
\begin{figure}[H]
\centering
\includegraphics[scale=0.6]{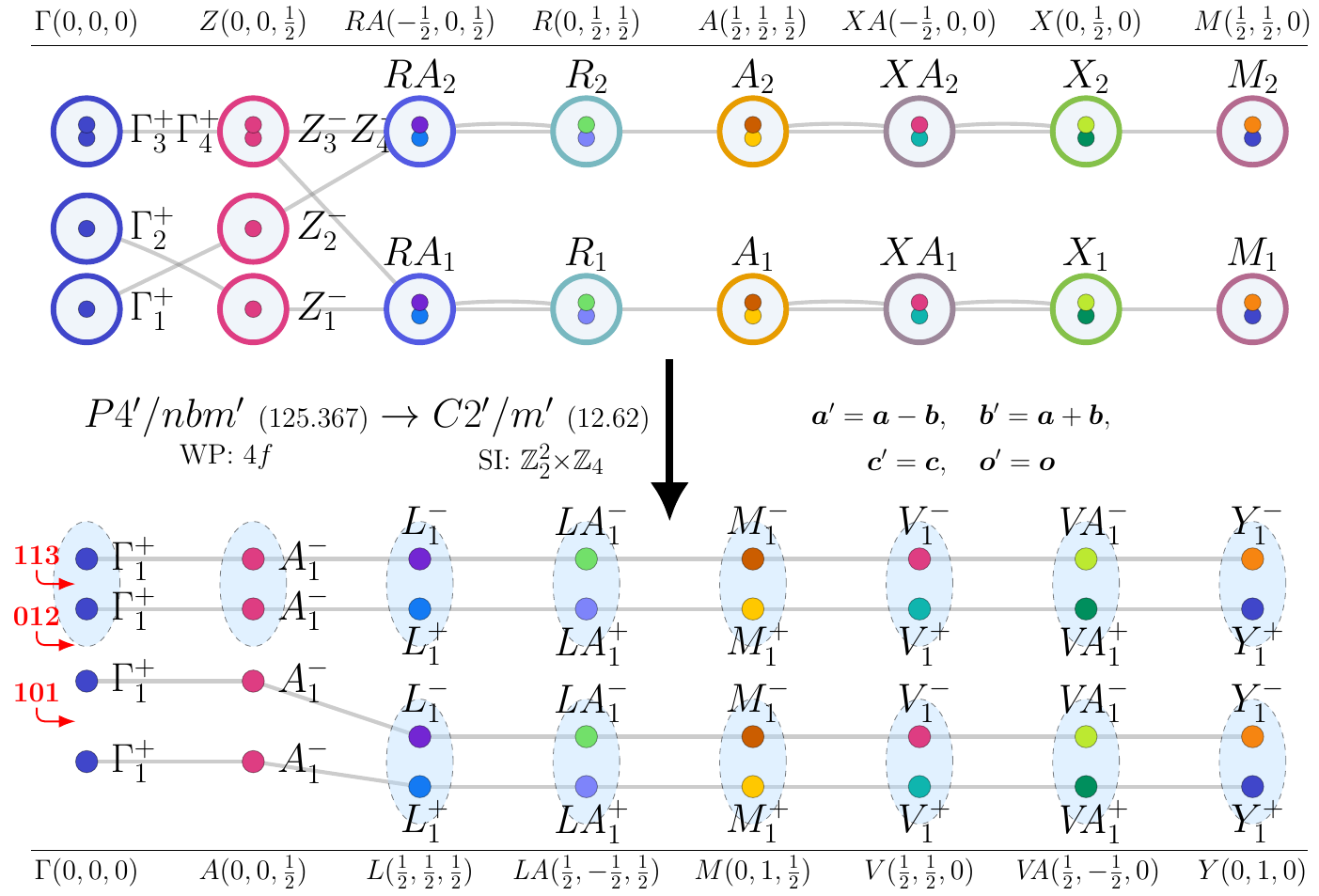}
\caption{Topological magnon bands in subgroup $C2'/m'~(12.62)$ for magnetic moments on Wyckoff position $4f$ of supergroup $P4'/nbm'~(125.367)$.\label{fig_125.367_12.62_strainperp110_4f}}
\end{figure}
\begin{center}
\begin{longtable}{cc}
\caption{Number of times a given SI can appear in the band structure. Supergroup: $P4'/nbm'~(125.367)$, WP: $4f$, Subgroup: $C2'/m'~(12.62)$.\label{gap_table_125.367_12.62_4f_strainperp110}}\\
\toprule
SI&Possible \#Gaps\\
\midrule
\endfirsthead
\multicolumn{2}{@{}l}{\ldots continued}\\
\toprule
SI&Possible \#Gaps\\
\midrule
\endhead
\midrule\multicolumn{2}{r}{continued on next page\ldots}
\endfoot
\bottomrule
\endlastfoot
000&0, 1, 2\\
001&0, 1\\
003&0, 1\\
011&0, 1\\
012&1, 2, 3\\
013&0, 1\\
100&0, 1, 2\\
101&0, 1\\
103&0, 1\\
111&0, 1\\
112&0, 1, 2\\
113&0, 1\\
gapless&0\\
nontrivial&1, 2, 3\\
\end{longtable}
\end{center}
\begin{center}
\begin{longtable}{cc}
\caption{Possible SI values in each gap. Supergroup: $P4'/nbm'~(125.367)$, WP: $4f$, Subgroup: $C2'/m'~(12.62)$.\label{si_table_125.367_12.62_4f_strainperp110}}\\
\toprule
Gap\#& Possible SI Values\\
\midrule
\endfirsthead
\multicolumn{2}{@{}l}{\ldots continued}\\
\toprule
Gap\#& Possible SI Values\\
\midrule
\endhead
\multicolumn{2}{r}{continued on next page\ldots}
\endfoot
\bottomrule
\endlastfoot
\multirow{2}{*}{1}&\multirow{1}{*}{000, 001, 011, 012, 100}\\*
&\multirow{1}{*}{101, 111, 112}\\\midrule
\multirow{1}{*}{2}&\multirow{1}{*}{012}\\\midrule
\multirow{2}{*}{3}&\multirow{1}{*}{000, 003, 012, 013, 100}\\*
&\multirow{1}{*}{103, 112, 113}\\\midrule
\multirow{1}{*}{4}&\multirow{1}{*}{000}\\
\end{longtable}
\end{center}

\section{MSG $P_{C}4/nbm~(125.373)$}
\textbf{Nontrivial-SI Subgroups:} $P\bar{1}~(2.4)$, $C2'/m'~(12.62)$, $P2'/m'~(10.46)$, $P2_{1}'/m'~(11.54)$, $P_{S}\bar{1}~(2.7)$, $C_{a}2~(5.17)$, $C2/m~(12.58)$, $Cmm'm'~(65.486)$, $C_{a}2/m~(12.64)$, $P2~(3.1)$, $Cm'm'2~(35.168)$, $Pm'm'2~(25.60)$, $P_{a}2~(3.4)$, $P2/c~(13.65)$, $Cm'm'a~(67.505)$, $Pm'm'n~(59.409)$, $P_{c}2/c~(13.72)$, $P2~(3.1)$, $Pm'm'2~(25.60)$, $P2/c~(13.65)$, $Pm'm'a~(51.294)$, $P_{A}2/c~(13.73)$, $P_{C}ban~(50.287)$, $P4m'm'~(99.167)$, $P4/nm'm'~(129.417)$.\\

\textbf{Trivial-SI Subgroups:} $Cm'~(8.34)$, $Pm'~(6.20)$, $Pm'~(6.20)$, $C2'~(5.15)$, $P2'~(3.3)$, $P2_{1}'~(4.9)$, $P_{S}1~(1.3)$, $Cm~(8.32)$, $Cm'm2'~(35.167)$, $C_{a}m~(8.36)$, $Pc~(7.24)$, $Abm'2'~(39.198)$, $Pm'n2_{1}'~(31.125)$, $P_{c}c~(7.28)$, $Pc~(7.24)$, $Pm'a2'~(28.89)$, $P_{A}c~(7.31)$, $C2~(5.13)$, $Am'm'2~(38.191)$, $A_{b}bm2~(39.201)$, $C_{a}mm2~(35.170)$, $P_{C}ba2~(32.141)$, $C_{a}mma~(67.509)$, $P_{C}2~(3.6)$, $P_{A}nc2~(30.119)$, $P_{C}4bm~(100.177)$.\\

\subsection{WP: $4f$}
\textbf{BCS Materials:} {NdMg~(35 K)}\footnote{BCS web page: \texttt{\href{http://webbdcrista1.ehu.es/magndata/index.php?this\_label=2.14} {http://webbdcrista1.ehu.es/magndata/index.php?this\_label=2.14}}}.\\
\subsubsection{Topological bands in subgroup $P\bar{1}~(2.4)$}
\textbf{Perturbations:}
\begin{itemize}
\item B $\parallel$ [001] and strain in generic direction,
\item B $\parallel$ [100] and strain $\perp$ [110],
\item B $\parallel$ [100] and strain in generic direction,
\item B $\parallel$ [110] and strain $\perp$ [100],
\item B $\parallel$ [110] and strain in generic direction,
\item B $\perp$ [001] and strain $\perp$ [100],
\item B $\perp$ [001] and strain $\perp$ [110],
\item B $\perp$ [001] and strain in generic direction,
\item B $\perp$ [100] and strain $\parallel$ [110],
\item B $\perp$ [100] and strain $\perp$ [001],
\item B $\perp$ [100] and strain $\perp$ [110],
\item B $\perp$ [100] and strain in generic direction,
\item B $\perp$ [110] and strain $\parallel$ [100],
\item B $\perp$ [110] and strain $\perp$ [001],
\item B $\perp$ [110] and strain $\perp$ [100],
\item B $\perp$ [110] and strain in generic direction,
\item B in generic direction.
\end{itemize}
\begin{figure}[H]
\centering
\includegraphics[scale=0.6]{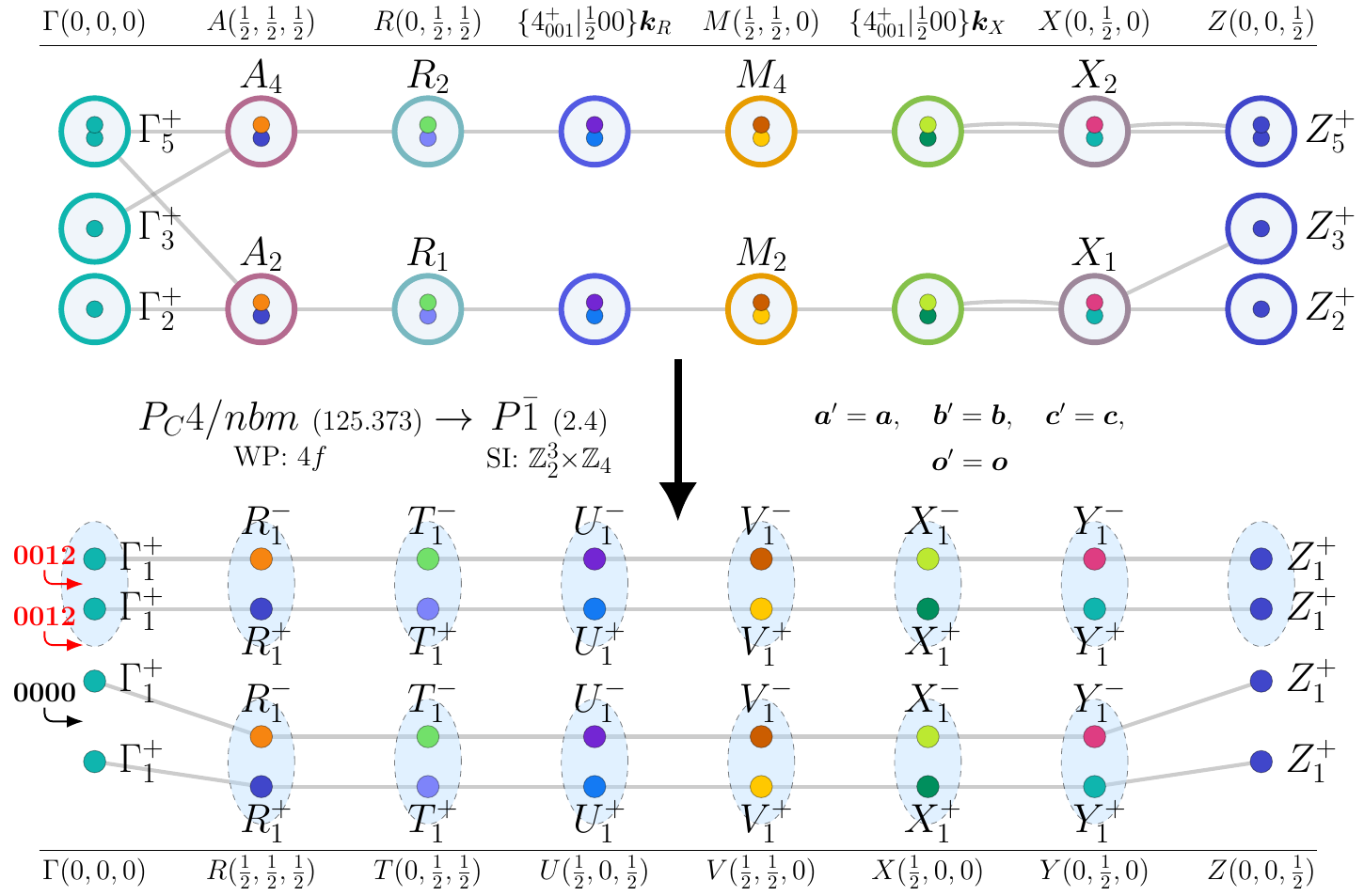}
\caption{Topological magnon bands in subgroup $P\bar{1}~(2.4)$ for magnetic moments on Wyckoff position $4f$ of supergroup $P_{C}4/nbm~(125.373)$.\label{fig_125.373_2.4_Bparallel001andstrainingenericdirection_4f}}
\end{figure}
\begin{center}
\begin{longtable}{cc}
\caption{Number of times a given SI can appear in the band structure. Supergroup: $P_{C}4/nbm~(125.373)$, WP: $4f$, Subgroup: $P\bar{1}~(2.4)$.\label{gap_table_125.373_2.4_4f_Bparallel001andstrainingenericdirection}}\\
\toprule
SI&Possible \#Gaps\\
\midrule
\endfirsthead
\multicolumn{2}{@{}l}{\ldots continued}\\
\toprule
SI&Possible \#Gaps\\
\midrule
\endhead
\midrule\multicolumn{2}{r}{continued on next page\ldots}
\endfoot
\bottomrule
\endlastfoot
0000&0, 1, 2\\
0001&0, 1, 2\\
0003&0, 1, 2\\
0011&0, 1, 2\\
0012&1, 2, 3\\
0013&0, 1, 2\\
0100&0, 1, 2\\
0101&0, 1, 2\\
0102&0, 1, 2\\
0103&0, 1, 2\\
0110&0, 1, 2\\
0111&0, 1, 2\\
0112&0, 1, 2\\
0113&0, 1, 2\\
1000&0, 1, 2\\
1001&0, 1, 2\\
1002&0, 1, 2\\
1003&0, 1, 2\\
1010&0, 1, 2\\
1011&0, 1, 2\\
1012&0, 1, 2\\
1013&0, 1, 2\\
1100&0, 1, 2\\
1101&0, 1\\
1102&0, 1, 2\\
1103&0, 1\\
1110&0, 1, 2\\
1111&0, 1\\
1112&0, 1, 2\\
1113&0, 1\\
gapless&0\\
nontrivial&1, 2, 3\\
\end{longtable}
\end{center}
\begin{center}
\begin{longtable}{cc}
\caption{Possible SI values in each gap. Supergroup: $P_{C}4/nbm~(125.373)$, WP: $4f$, Subgroup: $P\bar{1}~(2.4)$.\label{si_table_125.373_2.4_4f_Bparallel001andstrainingenericdirection}}\\
\toprule
Gap\#& Possible SI Values\\
\midrule
\endfirsthead
\multicolumn{2}{@{}l}{\ldots continued}\\
\toprule
Gap\#& Possible SI Values\\
\midrule
\endhead
\multicolumn{2}{r}{continued on next page\ldots}
\endfoot
\bottomrule
\endlastfoot
\multirow{6}{*}{1}&\multirow{1}{*}{0000, 0001, 0003, 0011, 0012}\\*
&\multirow{1}{*}{0013, 0100, 0101, 0102, 0103}\\*
&\multirow{1}{*}{0110, 0111, 0112, 0113, 1000}\\*
&\multirow{1}{*}{1001, 1002, 1003, 1010, 1011}\\*
&\multirow{1}{*}{1012, 1013, 1100, 1102, 1103}\\*
&\multirow{1}{*}{1110, 1112, 1113}\\\midrule
\multirow{1}{*}{2}&\multirow{1}{*}{0012}\\\midrule
\multirow{6}{*}{3}&\multirow{1}{*}{0000, 0001, 0003, 0011, 0012}\\*
&\multirow{1}{*}{0013, 0100, 0101, 0102, 0103}\\*
&\multirow{1}{*}{0110, 0111, 0112, 0113, 1000}\\*
&\multirow{1}{*}{1001, 1002, 1003, 1010, 1011}\\*
&\multirow{1}{*}{1012, 1013, 1100, 1101, 1102}\\*
&\multirow{1}{*}{1110, 1111, 1112}\\\midrule
\multirow{1}{*}{4}&\multirow{1}{*}{0000}\\
\end{longtable}
\end{center}
\subsubsection{Topological bands in subgroup $C2'/m'~(12.62)$}
\textbf{Perturbations:}
\begin{itemize}
\item B $\parallel$ [001] and strain $\perp$ [110],
\item B $\perp$ [110].
\end{itemize}
\begin{figure}[H]
\centering
\includegraphics[scale=0.6]{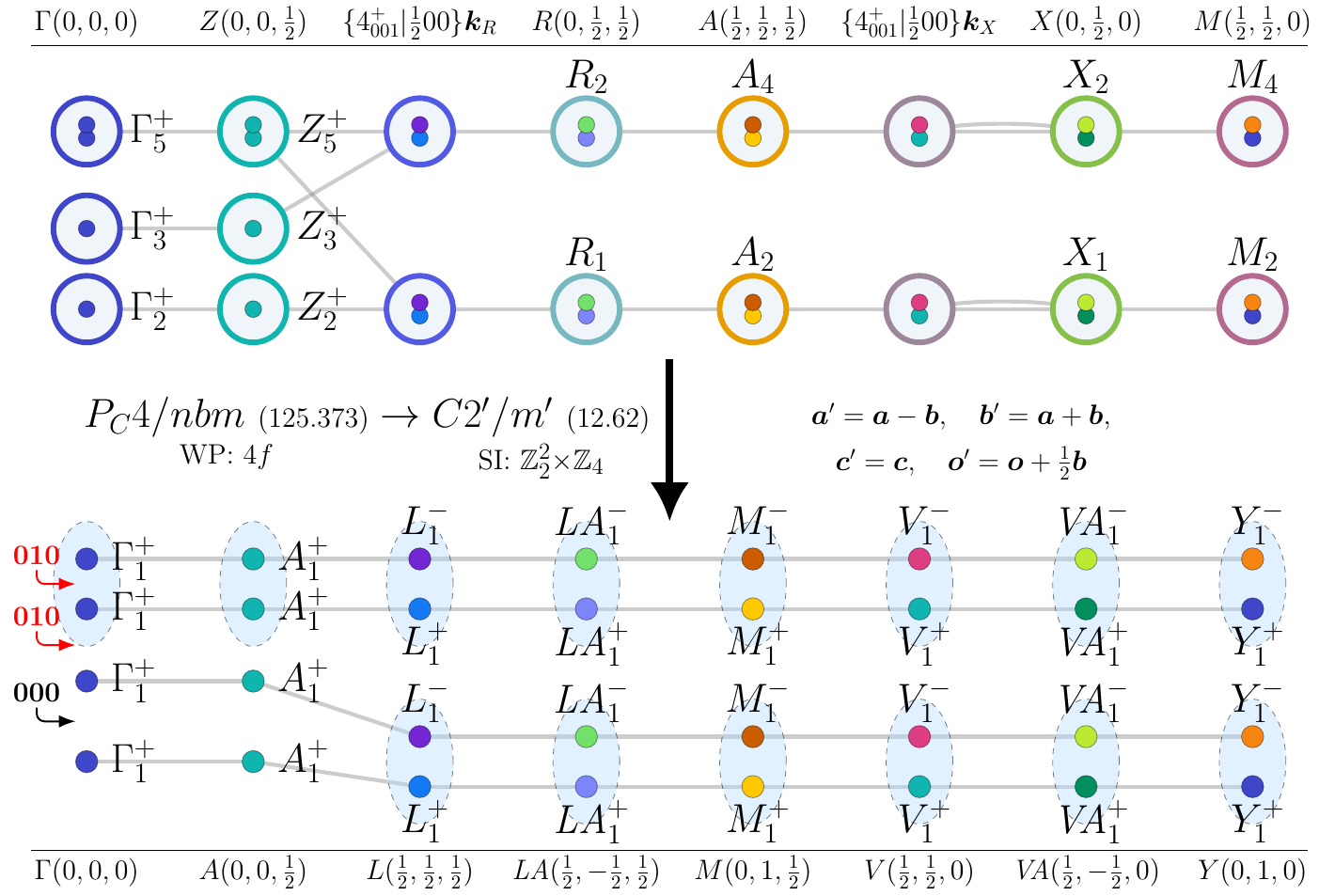}
\caption{Topological magnon bands in subgroup $C2'/m'~(12.62)$ for magnetic moments on Wyckoff position $4f$ of supergroup $P_{C}4/nbm~(125.373)$.\label{fig_125.373_12.62_Bparallel001andstrainperp110_4f}}
\end{figure}
\begin{center}
\begin{longtable}{cc}
\caption{Number of times a given SI can appear in the band structure. Supergroup: $P_{C}4/nbm~(125.373)$, WP: $4f$, Subgroup: $C2'/m'~(12.62)$.\label{gap_table_125.373_12.62_4f_Bparallel001andstrainperp110}}\\
\toprule
SI&Possible \#Gaps\\
\midrule
\endfirsthead
\multicolumn{2}{@{}l}{\ldots continued}\\
\toprule
SI&Possible \#Gaps\\
\midrule
\endhead
\midrule\multicolumn{2}{r}{continued on next page\ldots}
\endfoot
\bottomrule
\endlastfoot
000&0, 1, 2\\
001&0, 1\\
003&0, 1\\
010&1, 2, 3\\
011&0, 1\\
013&0, 1\\
100&0, 1, 2\\
101&0, 1\\
103&0, 1\\
110&0, 1, 2\\
111&0, 1\\
113&0, 1\\
gapless&0\\
nontrivial&1, 2, 3\\
\end{longtable}
\end{center}
\begin{center}
\begin{longtable}{cc}
\caption{Possible SI values in each gap. Supergroup: $P_{C}4/nbm~(125.373)$, WP: $4f$, Subgroup: $C2'/m'~(12.62)$.\label{si_table_125.373_12.62_4f_Bparallel001andstrainperp110}}\\
\toprule
Gap\#& Possible SI Values\\
\midrule
\endfirsthead
\multicolumn{2}{@{}l}{\ldots continued}\\
\toprule
Gap\#& Possible SI Values\\
\midrule
\endhead
\multicolumn{2}{r}{continued on next page\ldots}
\endfoot
\bottomrule
\endlastfoot
\multirow{2}{*}{1}&\multirow{1}{*}{000, 003, 010, 011, 100}\\*
&\multirow{1}{*}{103, 110, 111}\\\midrule
\multirow{1}{*}{2}&\multirow{1}{*}{010}\\\midrule
\multirow{2}{*}{3}&\multirow{1}{*}{000, 001, 010, 013, 100}\\*
&\multirow{1}{*}{101, 110, 113}\\\midrule
\multirow{1}{*}{4}&\multirow{1}{*}{000}\\
\end{longtable}
\end{center}
\subsubsection{Topological bands in subgroup $P2'/m'~(10.46)$}
\textbf{Perturbations:}
\begin{itemize}
\item B $\parallel$ [100] and strain $\parallel$ [110],
\item B $\parallel$ [100] and strain $\perp$ [001],
\item B $\parallel$ [110] and strain $\parallel$ [100],
\item B $\parallel$ [110] and strain $\perp$ [001],
\item B $\perp$ [001].
\end{itemize}
\begin{figure}[H]
\centering
\includegraphics[scale=0.6]{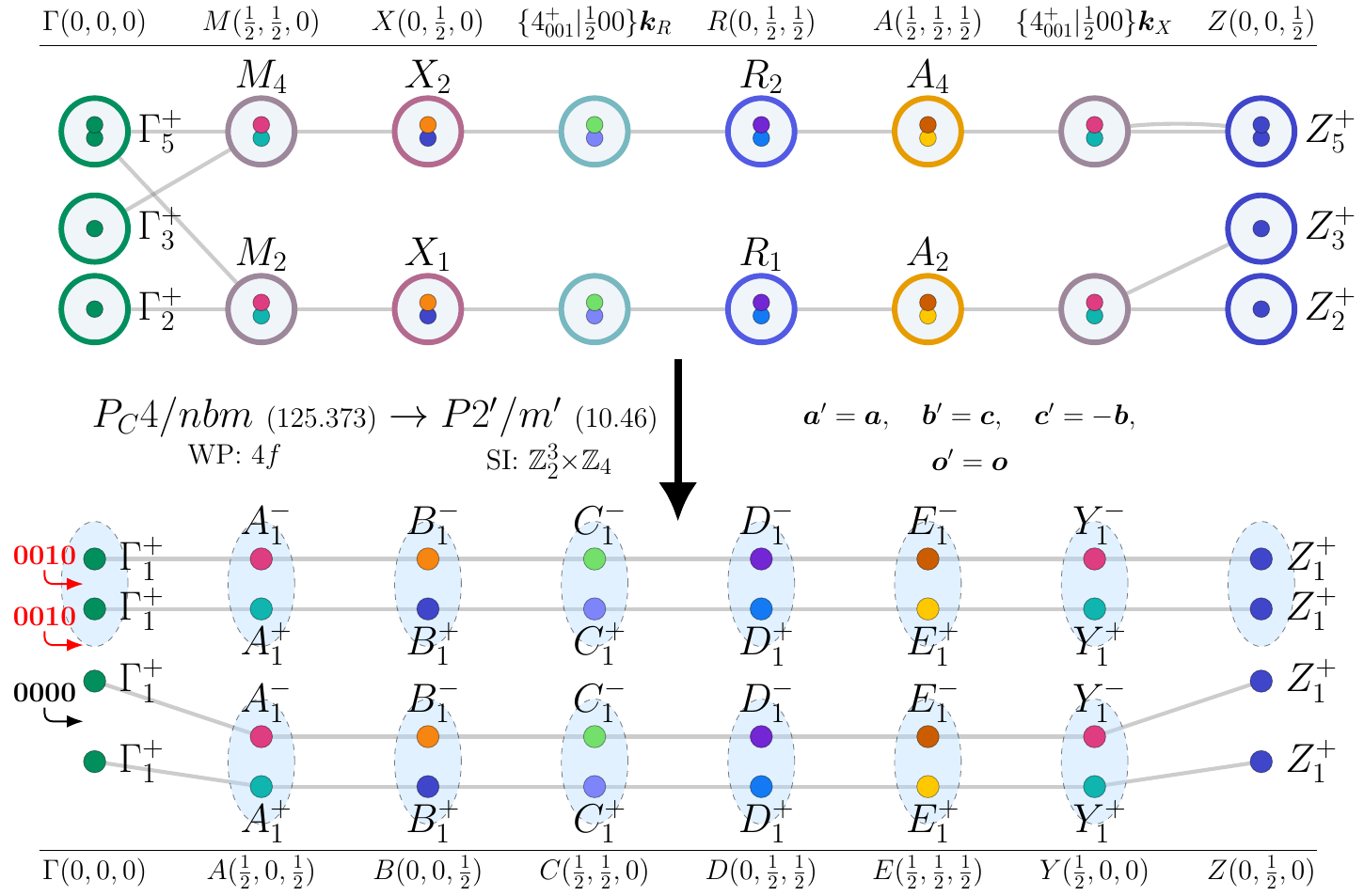}
\caption{Topological magnon bands in subgroup $P2'/m'~(10.46)$ for magnetic moments on Wyckoff position $4f$ of supergroup $P_{C}4/nbm~(125.373)$.\label{fig_125.373_10.46_Bparallel100andstrainparallel110_4f}}
\end{figure}
\begin{center}
\begin{longtable}{cc}
\caption{Number of times a given SI can appear in the band structure. Supergroup: $P_{C}4/nbm~(125.373)$, WP: $4f$, Subgroup: $P2'/m'~(10.46)$.\label{gap_table_125.373_10.46_4f_Bparallel100andstrainparallel110}}\\
\toprule
SI&Possible \#Gaps\\
\midrule
\endfirsthead
\multicolumn{2}{@{}l}{\ldots continued}\\
\toprule
SI&Possible \#Gaps\\
\midrule
\endhead
\midrule\multicolumn{2}{r}{continued on next page\ldots}
\endfoot
\bottomrule
\endlastfoot
0000&0, 1, 2\\
0001&0, 1, 2\\
0003&0, 1, 2\\
0010&1, 2, 3\\
0011&0, 1, 2\\
0013&0, 1, 2\\
0100&0, 1, 2\\
0101&0, 1, 2\\
0102&0, 1, 2\\
0103&0, 1, 2\\
0110&0, 1, 2\\
0111&0, 1, 2\\
0112&0, 1, 2\\
0113&0, 1, 2\\
1000&0, 1, 2\\
1001&0, 1, 2\\
1002&0, 1, 2\\
1003&0, 1, 2\\
1010&0, 1, 2\\
1011&0, 1, 2\\
1012&0, 1, 2\\
1013&0, 1, 2\\
1100&0, 1, 2\\
1101&0, 1\\
1102&0, 1, 2\\
1103&0, 1\\
1110&0, 1, 2\\
1111&0, 1\\
1112&0, 1, 2\\
1113&0, 1\\
gapless&0\\
nontrivial&1, 2, 3\\
\end{longtable}
\end{center}
\begin{center}
\begin{longtable}{cc}
\caption{Possible SI values in each gap. Supergroup: $P_{C}4/nbm~(125.373)$, WP: $4f$, Subgroup: $P2'/m'~(10.46)$.\label{si_table_125.373_10.46_4f_Bparallel100andstrainparallel110}}\\
\toprule
Gap\#& Possible SI Values\\
\midrule
\endfirsthead
\multicolumn{2}{@{}l}{\ldots continued}\\
\toprule
Gap\#& Possible SI Values\\
\midrule
\endhead
\multicolumn{2}{r}{continued on next page\ldots}
\endfoot
\bottomrule
\endlastfoot
\multirow{6}{*}{1}&\multirow{1}{*}{0000, 0001, 0003, 0010, 0011}\\*
&\multirow{1}{*}{0013, 0100, 0101, 0102, 0103}\\*
&\multirow{1}{*}{0110, 0111, 0112, 0113, 1000}\\*
&\multirow{1}{*}{1001, 1002, 1003, 1010, 1011}\\*
&\multirow{1}{*}{1012, 1013, 1100, 1102, 1103}\\*
&\multirow{1}{*}{1110, 1111, 1112}\\\midrule
\multirow{1}{*}{2}&\multirow{1}{*}{0010}\\\midrule
\multirow{6}{*}{3}&\multirow{1}{*}{0000, 0001, 0003, 0010, 0011}\\*
&\multirow{1}{*}{0013, 0100, 0101, 0102, 0103}\\*
&\multirow{1}{*}{0110, 0111, 0112, 0113, 1000}\\*
&\multirow{1}{*}{1001, 1002, 1003, 1010, 1011}\\*
&\multirow{1}{*}{1012, 1013, 1100, 1101, 1102}\\*
&\multirow{1}{*}{1110, 1112, 1113}\\\midrule
\multirow{1}{*}{4}&\multirow{1}{*}{0000}\\
\end{longtable}
\end{center}
\subsubsection{Topological bands in subgroup $P2_{1}'/m'~(11.54)$}
\textbf{Perturbations:}
\begin{itemize}
\item B $\parallel$ [001] and strain $\perp$ [100],
\item B $\perp$ [100].
\end{itemize}
\begin{figure}[H]
\centering
\includegraphics[scale=0.6]{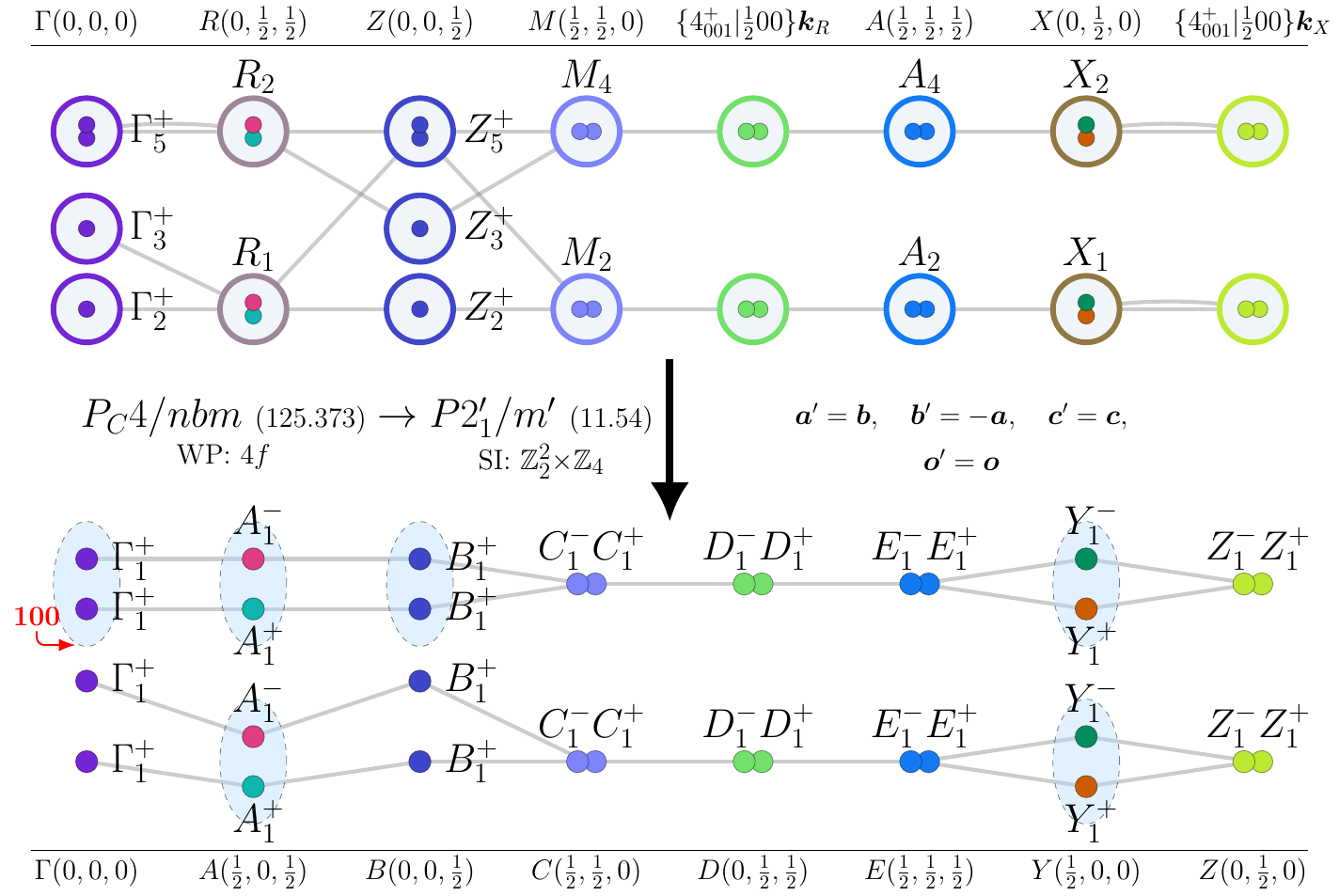}
\caption{Topological magnon bands in subgroup $P2_{1}'/m'~(11.54)$ for magnetic moments on Wyckoff position $4f$ of supergroup $P_{C}4/nbm~(125.373)$.\label{fig_125.373_11.54_Bparallel001andstrainperp100_4f}}
\end{figure}
\begin{center}
\begin{longtable}{cc}
\caption{Number of times a given SI can appear in the band structure. Supergroup: $P_{C}4/nbm~(125.373)$, WP: $4f$, Subgroup: $P2_{1}'/m'~(11.54)$.\label{gap_table_125.373_11.54_4f_Bparallel001andstrainperp100}}\\
\toprule
SI&Possible \#Gaps\\
\midrule
\endfirsthead
\multicolumn{2}{@{}l}{\ldots continued}\\
\toprule
SI&Possible \#Gaps\\
\midrule
\endhead
\midrule\multicolumn{2}{r}{continued on next page\ldots}
\endfoot
\bottomrule
\endlastfoot
000&0\\
100&1\\
gapless&2\\
nontrivial&1\\
\end{longtable}
\end{center}
\begin{center}
\begin{longtable}{cc}
\caption{Possible SI values in each gap. Supergroup: $P_{C}4/nbm~(125.373)$, WP: $4f$, Subgroup: $P2_{1}'/m'~(11.54)$.\label{si_table_125.373_11.54_4f_Bparallel001andstrainperp100}}\\
\toprule
Gap\#& Possible SI Values\\
\midrule
\endfirsthead
\multicolumn{2}{@{}l}{\ldots continued}\\
\toprule
Gap\#& Possible SI Values\\
\midrule
\endhead
\multicolumn{2}{r}{continued on next page\ldots}
\endfoot
\bottomrule
\endlastfoot
\multirow{1}{*}{1}&\multirow{1}{*}{G}\\\midrule
\multirow{1}{*}{2}&\multirow{1}{*}{100}\\\midrule
\multirow{1}{*}{3}&\multirow{1}{*}{G}\\\midrule
\multirow{1}{*}{4}&\multirow{1}{*}{000}\\
\end{longtable}
\end{center}
\subsubsection{Topological bands in subgroup $P_{S}\bar{1}~(2.7)$}
\textbf{Perturbation:}
\begin{itemize}
\item strain in generic direction.
\end{itemize}
\begin{figure}[H]
\centering
\includegraphics[scale=0.6]{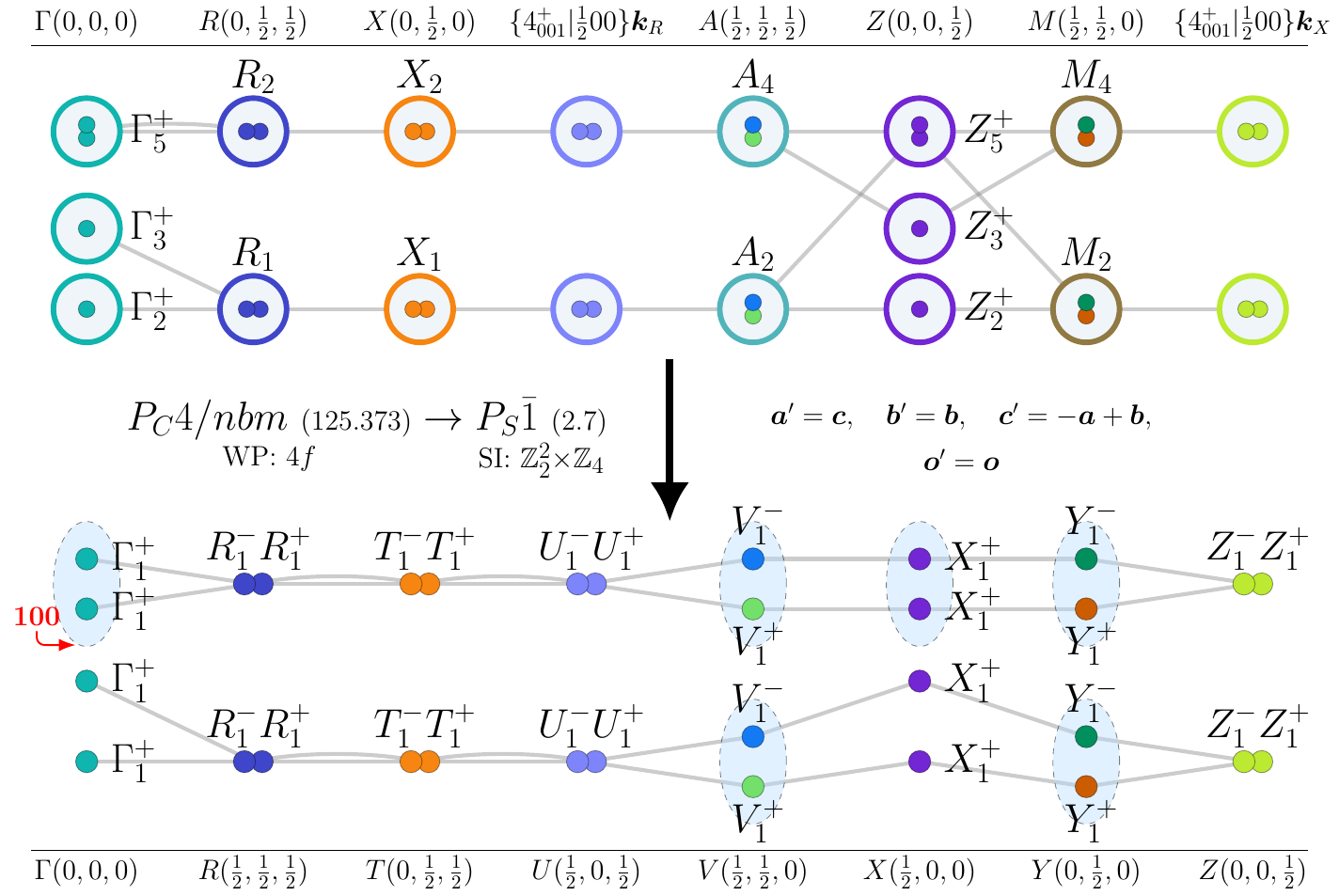}
\caption{Topological magnon bands in subgroup $P_{S}\bar{1}~(2.7)$ for magnetic moments on Wyckoff position $4f$ of supergroup $P_{C}4/nbm~(125.373)$.\label{fig_125.373_2.7_strainingenericdirection_4f}}
\end{figure}
\begin{center}
\begin{longtable}{cc}
\caption{Number of times a given SI can appear in the band structure. Supergroup: $P_{C}4/nbm~(125.373)$, WP: $4f$, Subgroup: $P_{S}\bar{1}~(2.7)$.\label{gap_table_125.373_2.7_4f_strainingenericdirection}}\\
\toprule
SI&Possible \#Gaps\\
\midrule
\endfirsthead
\multicolumn{2}{@{}l}{\ldots continued}\\
\toprule
SI&Possible \#Gaps\\
\midrule
\endhead
\midrule\multicolumn{2}{r}{continued on next page\ldots}
\endfoot
\bottomrule
\endlastfoot
000&0\\
100&1\\
gapless&2\\
nontrivial&1\\
\end{longtable}
\end{center}
\begin{center}
\begin{longtable}{cc}
\caption{Possible SI values in each gap. Supergroup: $P_{C}4/nbm~(125.373)$, WP: $4f$, Subgroup: $P_{S}\bar{1}~(2.7)$.\label{si_table_125.373_2.7_4f_strainingenericdirection}}\\
\toprule
Gap\#& Possible SI Values\\
\midrule
\endfirsthead
\multicolumn{2}{@{}l}{\ldots continued}\\
\toprule
Gap\#& Possible SI Values\\
\midrule
\endhead
\multicolumn{2}{r}{continued on next page\ldots}
\endfoot
\bottomrule
\endlastfoot
\multirow{1}{*}{1}&\multirow{1}{*}{G}\\\midrule
\multirow{1}{*}{2}&\multirow{1}{*}{100}\\\midrule
\multirow{1}{*}{3}&\multirow{1}{*}{G}\\\midrule
\multirow{1}{*}{4}&\multirow{1}{*}{000}\\
\end{longtable}
\end{center}

\section{MSG $P_{c}4/nnc~(126.384)$}
\textbf{Nontrivial-SI Subgroups:} $P\bar{1}~(2.4)$, $C2'/m'~(12.62)$, $P2_{1}'/c'~(14.79)$, $P2'/c'~(13.69)$, $P_{S}\bar{1}~(2.7)$, $Ab'm'2~(39.199)$, $C2/c~(15.85)$, $Cm'ca'~(64.476)$, $C_{c}2/c~(15.90)$, $P2~(3.1)$, $Cm'm'2~(35.168)$, $Pb'a'2~(32.138)$, $P_{b}2~(3.5)$, $C_{c}cc2~(37.184)$, $P_{c}nn2~(34.161)$, $P2/c~(13.65)$, $Cm'm'a~(67.505)$, $Pb'a'n~(50.281)$, $P_{b}2/c~(13.71)$, $C_{c}cca~(68.518)$, $P2~(3.1)$, $P_{a}2~(3.4)$, $P2/c~(13.65)$, $Pnn'a'~(52.311)$, $P_{a}2/c~(13.70)$, $P_{c}nnn~(48.262)$, $P4b'm'~(100.175)$, $P_{c}4nc~(104.208)$, $P4/nb'm'~(125.369)$.\\

\textbf{Trivial-SI Subgroups:} $Cm'~(8.34)$, $Pc'~(7.26)$, $Pc'~(7.26)$, $C2'~(5.15)$, $P2_{1}'~(4.9)$, $P2'~(3.3)$, $P_{S}1~(1.3)$, $Cc~(9.37)$, $Cm'c2_{1}'~(36.174)$, $C_{c}c~(9.40)$, $Pc~(7.24)$, $Abm'2'~(39.198)$, $Pnc'2'~(30.114)$, $P_{b}c~(7.29)$, $Pc~(7.24)$, $Pna'2_{1}'~(33.147)$, $P_{a}c~(7.27)$, $C2~(5.13)$, $C_{c}2~(5.16)$, $A_{a}ba2~(41.216)$, $Pn'c'2~(30.115)$, $P_{a}nn2~(34.160)$.\\

\subsection{WP: $4a+8e$}
\textbf{BCS Materials:} {ErFeCuGe\textsubscript{4}O\textsubscript{12}~(20 K)}\footnote{BCS web page: \texttt{\href{http://webbdcrista1.ehu.es/magndata/index.php?this\_label=1.236} {http://webbdcrista1.ehu.es/magndata/index.php?this\_label=1.236}}}.\\
\subsubsection{Topological bands in subgroup $P2_{1}'/c'~(14.79)$}
\textbf{Perturbations:}
\begin{itemize}
\item B $\parallel$ [100] and strain $\parallel$ [110],
\item B $\parallel$ [100] and strain $\perp$ [001],
\item B $\parallel$ [110] and strain $\parallel$ [100],
\item B $\parallel$ [110] and strain $\perp$ [001],
\item B $\perp$ [001].
\end{itemize}
\begin{figure}[H]
\centering
\includegraphics[scale=0.6]{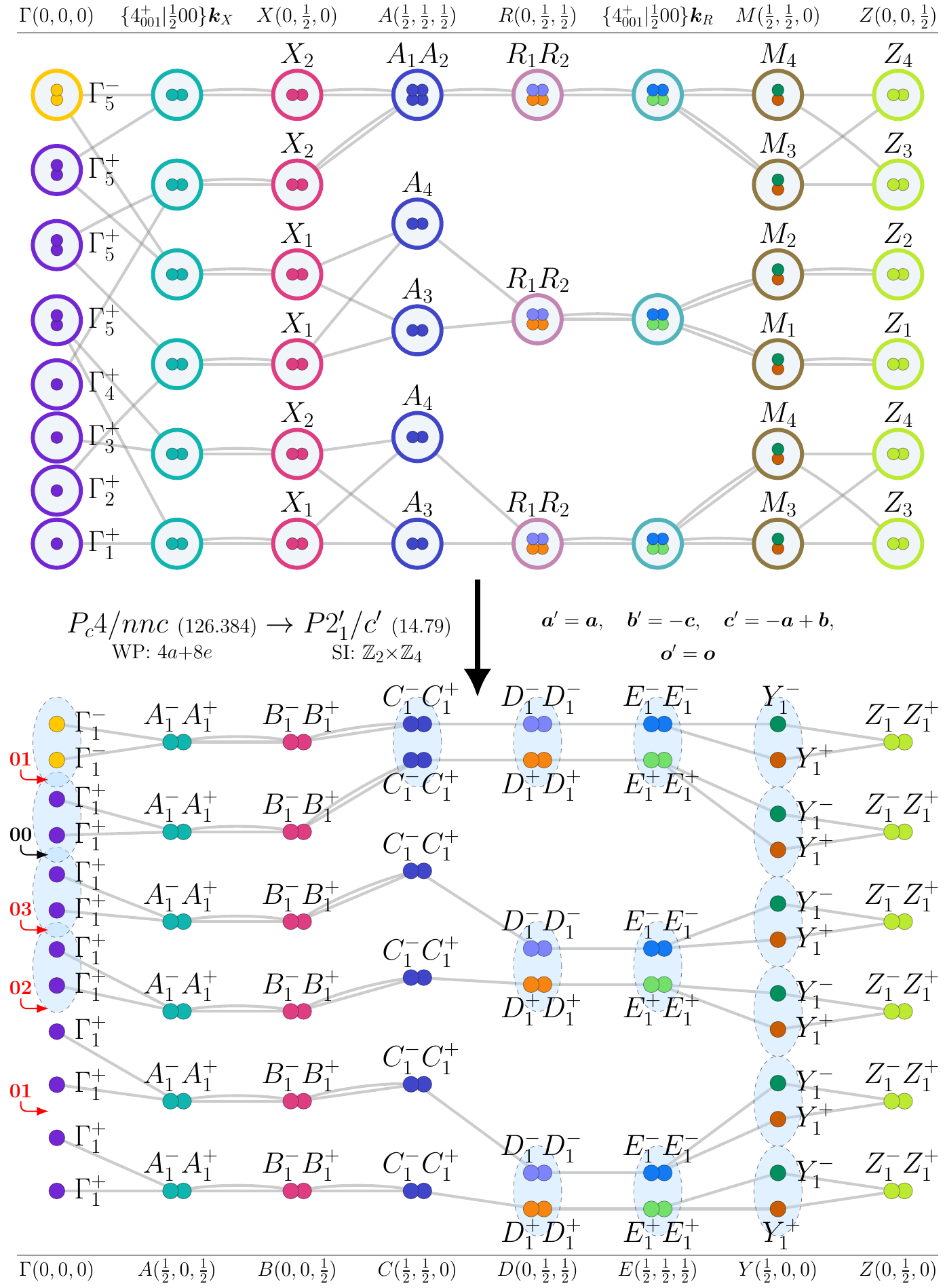}
\caption{Topological magnon bands in subgroup $P2_{1}'/c'~(14.79)$ for magnetic moments on Wyckoff positions $4a+8e$ of supergroup $P_{c}4/nnc~(126.384)$.\label{fig_126.384_14.79_Bparallel100andstrainparallel110_4a+8e}}
\end{figure}
\begin{center}
\begin{longtable}{cc}
\caption{Number of times a given SI can appear in the band structure. Supergroup: $P_{c}4/nnc~(126.384)$, WP: $4a$+$8e$, Subgroup: $P2_{1}'/c'~(14.79)$.\label{gap_table_126.384_14.79_4a+8e_Bparallel100andstrainparallel110}}\\
\toprule
SI&Possible \#Gaps\\
\midrule
\endfirsthead
\multicolumn{2}{@{}l}{\ldots continued}\\
\toprule
SI&Possible \#Gaps\\
\midrule
\endhead
\midrule\multicolumn{2}{r}{continued on next page\ldots}
\endfoot
\bottomrule
\endlastfoot
00&0, 1\\
01&0, 1, 2, 3\\
02&1, 2\\
03&0, 1, 2, 3\\
10&0, 1\\
11&0, 1\\
12&0, 1\\
13&0, 1\\
gapless&6\\
nontrivial&4, 5\\
\end{longtable}
\end{center}
\begin{center}
\begin{longtable}{cc}
\caption{Possible SI values in each gap. Supergroup: $P_{c}4/nnc~(126.384)$, WP: $4a$+$8e$, Subgroup: $P2_{1}'/c'~(14.79)$.\label{si_table_126.384_14.79_4a+8e_Bparallel100andstrainparallel110}}\\
\toprule
Gap\#& Possible SI Values\\
\midrule
\endfirsthead
\multicolumn{2}{@{}l}{\ldots continued}\\
\toprule
Gap\#& Possible SI Values\\
\midrule
\endhead
\multicolumn{2}{r}{continued on next page\ldots}
\endfoot
\bottomrule
\endlastfoot
\multirow{1}{*}{1}&\multirow{1}{*}{G}\\\midrule
\multirow{1}{*}{2}&\multirow{1}{*}{01, 03, 10, 12}\\\midrule
\multirow{1}{*}{3}&\multirow{1}{*}{G}\\\midrule
\multirow{1}{*}{4}&\multirow{1}{*}{00, 02, 11}\\\midrule
\multirow{1}{*}{5}&\multirow{1}{*}{G}\\\midrule
\multirow{1}{*}{6}&\multirow{1}{*}{01, 03, 10, 12}\\\midrule
\multirow{1}{*}{7}&\multirow{1}{*}{G}\\\midrule
\multirow{1}{*}{8}&\multirow{1}{*}{00, 02, 13}\\\midrule
\multirow{1}{*}{9}&\multirow{1}{*}{G}\\\midrule
\multirow{1}{*}{10}&\multirow{1}{*}{01, 03, 10, 12}\\\midrule
\multirow{1}{*}{11}&\multirow{1}{*}{G}\\\midrule
\multirow{1}{*}{12}&\multirow{1}{*}{00}\\
\end{longtable}
\end{center}
\subsubsection{Topological bands in subgroup $P2'/c'~(13.69)$}
\textbf{Perturbations:}
\begin{itemize}
\item B $\parallel$ [001] and strain $\perp$ [100],
\item B $\perp$ [100].
\end{itemize}
\begin{figure}[H]
\centering
\includegraphics[scale=0.6]{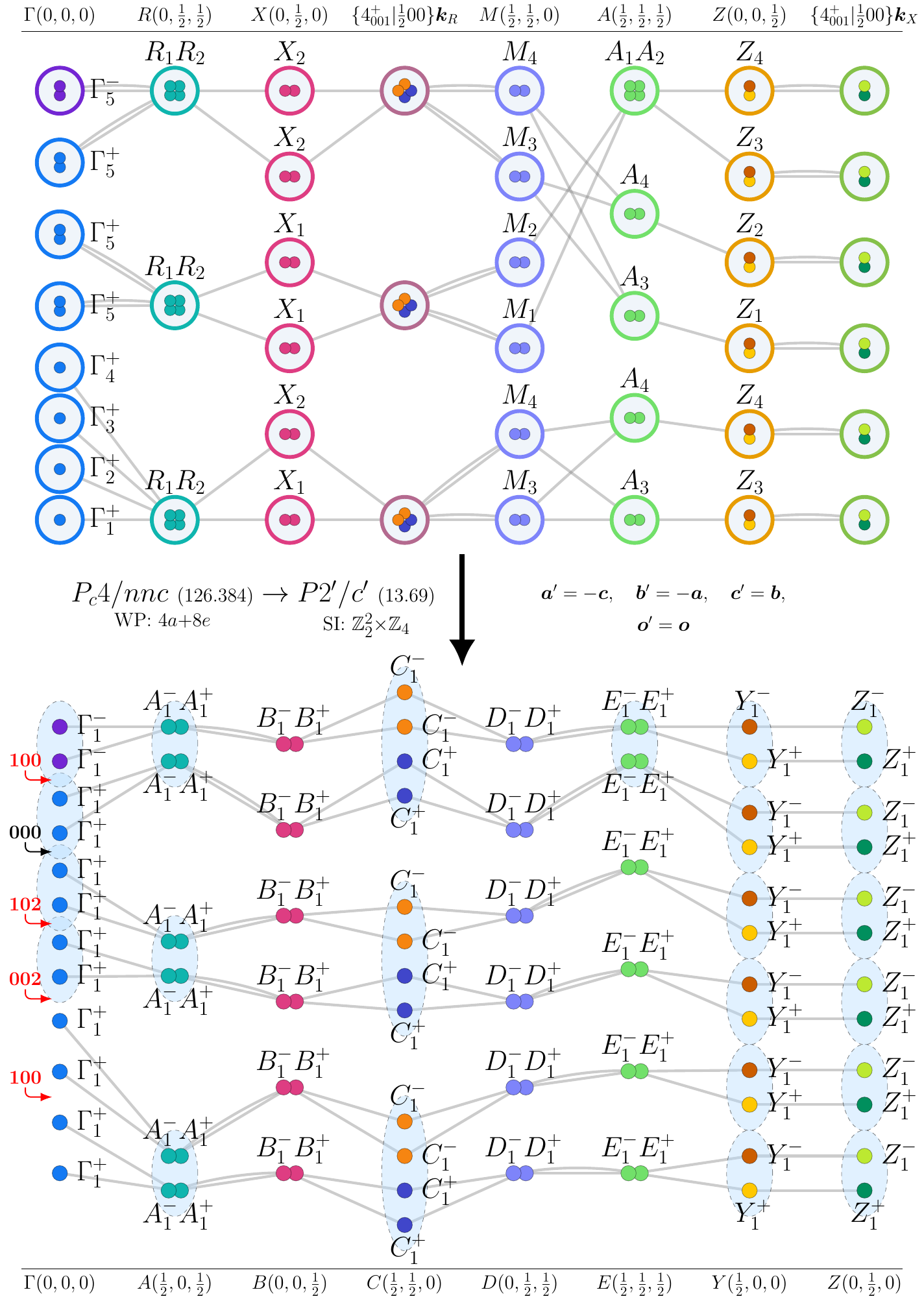}
\caption{Topological magnon bands in subgroup $P2'/c'~(13.69)$ for magnetic moments on Wyckoff positions $4a+8e$ of supergroup $P_{c}4/nnc~(126.384)$.\label{fig_126.384_13.69_Bparallel001andstrainperp100_4a+8e}}
\end{figure}
\begin{center}
\begin{longtable}{cc}
\caption{Number of times a given SI can appear in the band structure. Supergroup: $P_{c}4/nnc~(126.384)$, WP: $4a$+$8e$, Subgroup: $P2'/c'~(13.69)$.\label{gap_table_126.384_13.69_4a+8e_Bparallel001andstrainperp100}}\\
\toprule
SI&Possible \#Gaps\\
\midrule
\endfirsthead
\multicolumn{2}{@{}l}{\ldots continued}\\
\toprule
SI&Possible \#Gaps\\
\midrule
\endhead
\midrule\multicolumn{2}{r}{continued on next page\ldots}
\endfoot
\bottomrule
\endlastfoot
000&0, 1, 2\\
002&1, 2, 3\\
011&0, 1\\
013&0, 1\\
100&0, 1, 2, 3\\
102&0, 1, 2, 3\\
111&0, 1, 2, 3\\
113&0, 1, 2, 3\\
gapless&6\\
nontrivial&3, 4, 5\\
\end{longtable}
\end{center}
\begin{center}
\begin{longtable}{cc}
\caption{Possible SI values in each gap. Supergroup: $P_{c}4/nnc~(126.384)$, WP: $4a$+$8e$, Subgroup: $P2'/c'~(13.69)$.\label{si_table_126.384_13.69_4a+8e_Bparallel001andstrainperp100}}\\
\toprule
Gap\#& Possible SI Values\\
\midrule
\endfirsthead
\multicolumn{2}{@{}l}{\ldots continued}\\
\toprule
Gap\#& Possible SI Values\\
\midrule
\endhead
\multicolumn{2}{r}{continued on next page\ldots}
\endfoot
\bottomrule
\endlastfoot
\multirow{1}{*}{1}&\multirow{1}{*}{G}\\\midrule
\multirow{2}{*}{2}&\multirow{1}{*}{000, 011, 013, 100, 102}\\*
&\multirow{1}{*}{111, 113}\\\midrule
\multirow{1}{*}{3}&\multirow{1}{*}{G}\\\midrule
\multirow{1}{*}{4}&\multirow{1}{*}{000, 002, 111}\\\midrule
\multirow{1}{*}{5}&\multirow{1}{*}{G}\\\midrule
\multirow{2}{*}{6}&\multirow{1}{*}{002, 011, 013, 100, 102}\\*
&\multirow{1}{*}{111, 113}\\\midrule
\multirow{1}{*}{7}&\multirow{1}{*}{G}\\\midrule
\multirow{1}{*}{8}&\multirow{1}{*}{000, 002, 113}\\\midrule
\multirow{1}{*}{9}&\multirow{1}{*}{G}\\\midrule
\multirow{2}{*}{10}&\multirow{1}{*}{000, 011, 013, 100, 102}\\*
&\multirow{1}{*}{111, 113}\\\midrule
\multirow{1}{*}{11}&\multirow{1}{*}{G}\\\midrule
\multirow{1}{*}{12}&\multirow{1}{*}{000}\\
\end{longtable}
\end{center}
\subsubsection{Topological bands in subgroup $P_{S}\bar{1}~(2.7)$}
\textbf{Perturbation:}
\begin{itemize}
\item strain in generic direction.
\end{itemize}
\begin{figure}[H]
\centering
\includegraphics[scale=0.6]{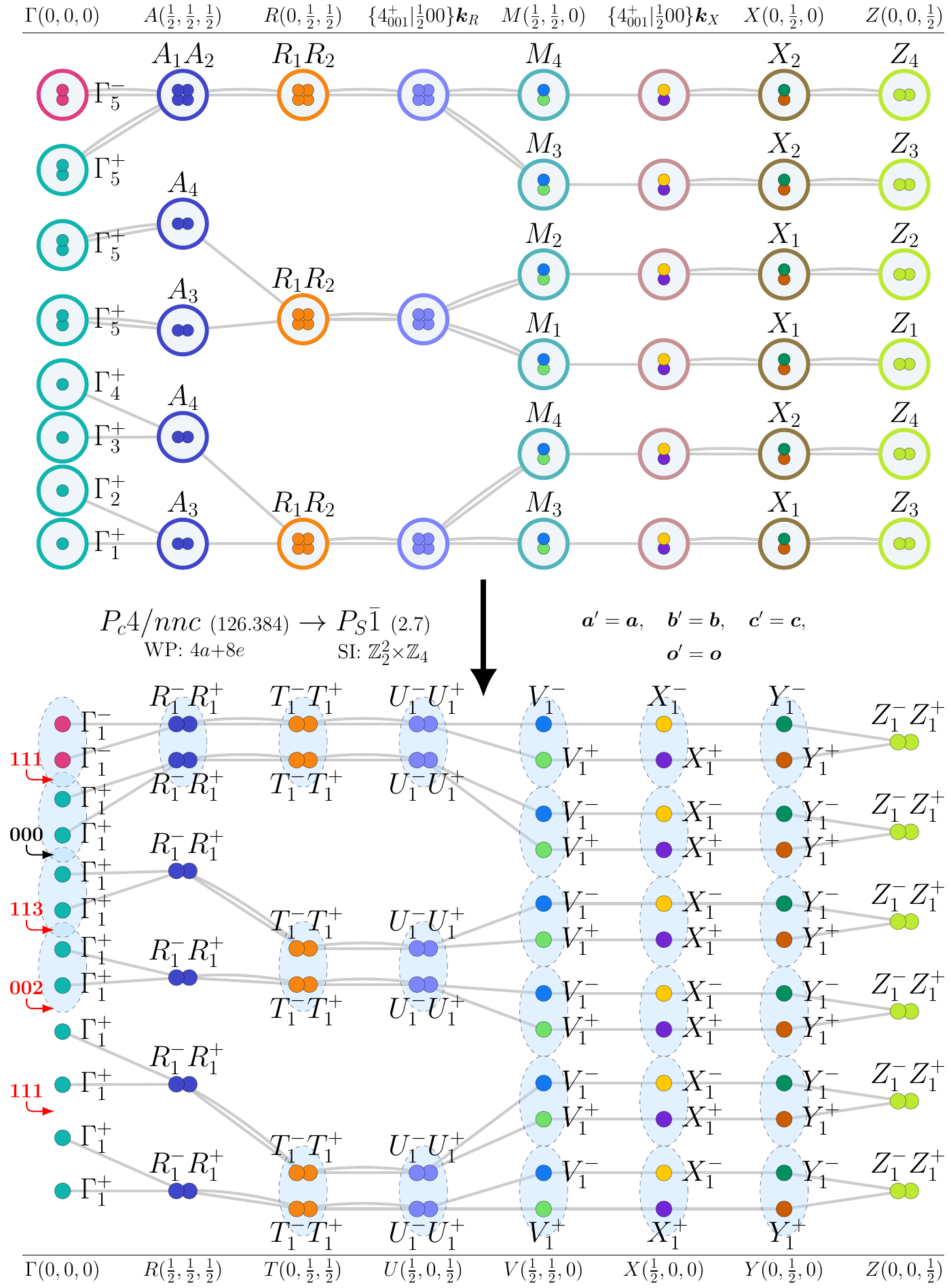}
\caption{Topological magnon bands in subgroup $P_{S}\bar{1}~(2.7)$ for magnetic moments on Wyckoff positions $4a+8e$ of supergroup $P_{c}4/nnc~(126.384)$.\label{fig_126.384_2.7_strainingenericdirection_4a+8e}}
\end{figure}
\begin{center}
\begin{longtable}{cc}
\caption{Number of times a given SI can appear in the band structure. Supergroup: $P_{c}4/nnc~(126.384)$, WP: $4a$+$8e$, Subgroup: $P_{S}\bar{1}~(2.7)$.\label{gap_table_126.384_2.7_4a+8e_strainingenericdirection}}\\
\toprule
SI&Possible \#Gaps\\
\midrule
\endfirsthead
\multicolumn{2}{@{}l}{\ldots continued}\\
\toprule
SI&Possible \#Gaps\\
\midrule
\endhead
\midrule\multicolumn{2}{r}{continued on next page\ldots}
\endfoot
\bottomrule
\endlastfoot
000&0, 1, 2\\
002&1, 2, 3\\
111&1, 2, 3\\
113&1, 2, 3\\
gapless&6\\
nontrivial&3, 4, 5\\
\end{longtable}
\end{center}
\begin{center}
\begin{longtable}{cc}
\caption{Possible SI values in each gap. Supergroup: $P_{c}4/nnc~(126.384)$, WP: $4a$+$8e$, Subgroup: $P_{S}\bar{1}~(2.7)$.\label{si_table_126.384_2.7_4a+8e_strainingenericdirection}}\\
\toprule
Gap\#& Possible SI Values\\
\midrule
\endfirsthead
\multicolumn{2}{@{}l}{\ldots continued}\\
\toprule
Gap\#& Possible SI Values\\
\midrule
\endhead
\multicolumn{2}{r}{continued on next page\ldots}
\endfoot
\bottomrule
\endlastfoot
\multirow{1}{*}{1}&\multirow{1}{*}{G}\\\midrule
\multirow{1}{*}{2}&\multirow{1}{*}{000, 111, 113}\\\midrule
\multirow{1}{*}{3}&\multirow{1}{*}{G}\\\midrule
\multirow{1}{*}{4}&\multirow{1}{*}{000, 002, 111}\\\midrule
\multirow{1}{*}{5}&\multirow{1}{*}{G}\\\midrule
\multirow{1}{*}{6}&\multirow{1}{*}{002, 111, 113}\\\midrule
\multirow{1}{*}{7}&\multirow{1}{*}{G}\\\midrule
\multirow{1}{*}{8}&\multirow{1}{*}{000, 002, 113}\\\midrule
\multirow{1}{*}{9}&\multirow{1}{*}{G}\\\midrule
\multirow{1}{*}{10}&\multirow{1}{*}{000, 111, 113}\\\midrule
\multirow{1}{*}{11}&\multirow{1}{*}{G}\\\midrule
\multirow{1}{*}{12}&\multirow{1}{*}{000}\\
\end{longtable}
\end{center}

\section{MSG $P_{I}4/nnc~(126.386)$}
\textbf{Nontrivial-SI Subgroups:} $P\bar{1}~(2.4)$, $C2'/m'~(12.62)$, $P2_{1}'/m'~(11.54)$, $P2_{1}'/m'~(11.54)$, $P_{S}\bar{1}~(2.7)$, $C2/c~(15.85)$, $Cm'cm'~(63.464)$, $C_{c}2/c~(15.90)$, $P2~(3.1)$, $Cm'm'2~(35.168)$, $Pm'm'2~(25.60)$, $C_{A}cc2~(37.186)$, $P2/c~(13.65)$, $Cm'm'a~(67.505)$, $Pm'm'n~(59.409)$, $P_{A}2/c~(13.73)$, $C_{A}cca~(68.520)$, $P2~(3.1)$, $Pm'm'2~(25.60)$, $P2/c~(13.65)$, $Pm'm'n~(59.409)$, $P_{A}2/c~(13.73)$, $P_{I}nnn~(48.264)$, $P4m'm'~(99.167)$, $P_{I}4nc~(104.210)$, $P4/nm'm'~(129.417)$.\\

\textbf{Trivial-SI Subgroups:} $Cm'~(8.34)$, $Pm'~(6.20)$, $Pm'~(6.20)$, $C2'~(5.15)$, $P2_{1}'~(4.9)$, $P2_{1}'~(4.9)$, $P_{S}1~(1.3)$, $Cc~(9.37)$, $Cm'c2_{1}'~(36.174)$, $C_{c}c~(9.40)$, $Pc~(7.24)$, $Abm'2'~(39.198)$, $Pm'n2_{1}'~(31.125)$, $P_{A}c~(7.31)$, $Pc~(7.24)$, $Pm'n2_{1}'~(31.125)$, $P_{A}c~(7.31)$, $C2~(5.13)$, $Am'm'2~(38.191)$, $C_{c}2~(5.16)$, $A_{B}ba2~(41.218)$, $P_{C}2~(3.6)$, $P_{I}nn2~(34.164)$, $P_{C}2~(3.6)$, $P_{I}nn2~(34.164)$.\\

\subsection{WP: $4d$}
\textbf{BCS Materials:} {ErMn\textsubscript{2}Ge\textsubscript{2}~(475 K)}\footnote{BCS web page: \texttt{\href{http://webbdcrista1.ehu.es/magndata/index.php?this\_label=1.640} {http://webbdcrista1.ehu.es/magndata/index.php?this\_label=1.640}}}, {YMn\textsubscript{2}Si\textsubscript{2}~(460 K)}\footnote{BCS web page: \texttt{\href{http://webbdcrista1.ehu.es/magndata/index.php?this\_label=1.495} {http://webbdcrista1.ehu.es/magndata/index.php?this\_label=1.495}}}, {CeCo\textsubscript{2}P\textsubscript{2}~(440 K)}\footnote{BCS web page: \texttt{\href{http://webbdcrista1.ehu.es/magndata/index.php?this\_label=1.253} {http://webbdcrista1.ehu.es/magndata/index.php?this\_label=1.253}}}, {YMn\textsubscript{2}Si\textsubscript{2}~(410 K)}\footnote{BCS web page: \texttt{\href{http://webbdcrista1.ehu.es/magndata/index.php?this\_label=1.469} {http://webbdcrista1.ehu.es/magndata/index.php?this\_label=1.469}}}, {YMn\textsubscript{2}Ge\textsubscript{2}~(395 K)}\footnote{BCS web page: \texttt{\href{http://webbdcrista1.ehu.es/magndata/index.php?this\_label=1.496} {http://webbdcrista1.ehu.es/magndata/index.php?this\_label=1.496}}}, {EuMn\textsubscript{2}Si\textsubscript{2}~(391 K)}\footnote{BCS web page: \texttt{\href{http://webbdcrista1.ehu.es/magndata/index.php?this\_label=1.453} {http://webbdcrista1.ehu.es/magndata/index.php?this\_label=1.453}}}, {CeMn\textsubscript{2}Si\textsubscript{2}~(384 K)}\footnote{BCS web page: \texttt{\href{http://webbdcrista1.ehu.es/magndata/index.php?this\_label=1.489} {http://webbdcrista1.ehu.es/magndata/index.php?this\_label=1.489}}}, {NdMn\textsubscript{2}Si\textsubscript{2}~(382 K)}\footnote{BCS web page: \texttt{\href{http://webbdcrista1.ehu.es/magndata/index.php?this\_label=1.493} {http://webbdcrista1.ehu.es/magndata/index.php?this\_label=1.493}}}, {NdMn\textsubscript{2}Si\textsubscript{2}~(380 K)}\footnote{BCS web page: \texttt{\href{http://webbdcrista1.ehu.es/magndata/index.php?this\_label=1.494} {http://webbdcrista1.ehu.es/magndata/index.php?this\_label=1.494}}}, {CeMn\textsubscript{2}Si\textsubscript{2}~(380 K)}\footnote{BCS web page: \texttt{\href{http://webbdcrista1.ehu.es/magndata/index.php?this\_label=1.490} {http://webbdcrista1.ehu.es/magndata/index.php?this\_label=1.490}}}, {CeMn\textsubscript{2}Si\textsubscript{2}~(379 K)}\footnote{BCS web page: \texttt{\href{http://webbdcrista1.ehu.es/magndata/index.php?this\_label=1.488} {http://webbdcrista1.ehu.es/magndata/index.php?this\_label=1.488}}}, {PrMn\textsubscript{2}Si\textsubscript{2}~(368 K)}\footnote{BCS web page: \texttt{\href{http://webbdcrista1.ehu.es/magndata/index.php?this\_label=1.492} {http://webbdcrista1.ehu.es/magndata/index.php?this\_label=1.492}}}, {PrMn\textsubscript{2}Si\textsubscript{2}~(348 K)}\footnote{BCS web page: \texttt{\href{http://webbdcrista1.ehu.es/magndata/index.php?this\_label=1.491} {http://webbdcrista1.ehu.es/magndata/index.php?this\_label=1.491}}}, {CaCo\textsubscript{1.86}As\textsubscript{2}~(52 K)}\footnote{BCS web page: \texttt{\href{http://webbdcrista1.ehu.es/magndata/index.php?this\_label=0.691} {http://webbdcrista1.ehu.es/magndata/index.php?this\_label=0.691}}}, {ErMn\textsubscript{2}Ge\textsubscript{2}}\footnote{BCS web page: \texttt{\href{http://webbdcrista1.ehu.es/magndata/index.php?this\_label=1.639} {http://webbdcrista1.ehu.es/magndata/index.php?this\_label=1.639}}}, {ErMn\textsubscript{2}Ge\textsubscript{2}}\footnote{BCS web page: \texttt{\href{http://webbdcrista1.ehu.es/magndata/index.php?this\_label=1.638} {http://webbdcrista1.ehu.es/magndata/index.php?this\_label=1.638}}}, {ErMn\textsubscript{2}Si\textsubscript{2}}\footnote{BCS web page: \texttt{\href{http://webbdcrista1.ehu.es/magndata/index.php?this\_label=1.637} {http://webbdcrista1.ehu.es/magndata/index.php?this\_label=1.637}}}, {ErMn\textsubscript{2}Si\textsubscript{2}}\footnote{BCS web page: \texttt{\href{http://webbdcrista1.ehu.es/magndata/index.php?this\_label=1.636} {http://webbdcrista1.ehu.es/magndata/index.php?this\_label=1.636}}}, {TbMn\textsubscript{2}Si\textsubscript{2}}\footnote{BCS web page: \texttt{\href{http://webbdcrista1.ehu.es/magndata/index.php?this\_label=1.468} {http://webbdcrista1.ehu.es/magndata/index.php?this\_label=1.468}}}.\\
\subsubsection{Topological bands in subgroup $P\bar{1}~(2.4)$}
\textbf{Perturbations:}
\begin{itemize}
\item B $\parallel$ [001] and strain in generic direction,
\item B $\parallel$ [100] and strain $\perp$ [110],
\item B $\parallel$ [100] and strain in generic direction,
\item B $\parallel$ [110] and strain $\perp$ [100],
\item B $\parallel$ [110] and strain in generic direction,
\item B $\perp$ [001] and strain $\perp$ [100],
\item B $\perp$ [001] and strain $\perp$ [110],
\item B $\perp$ [001] and strain in generic direction,
\item B $\perp$ [100] and strain $\parallel$ [110],
\item B $\perp$ [100] and strain $\perp$ [001],
\item B $\perp$ [100] and strain $\perp$ [110],
\item B $\perp$ [100] and strain in generic direction,
\item B $\perp$ [110] and strain $\parallel$ [100],
\item B $\perp$ [110] and strain $\perp$ [001],
\item B $\perp$ [110] and strain $\perp$ [100],
\item B $\perp$ [110] and strain in generic direction,
\item B in generic direction.
\end{itemize}
\begin{figure}[H]
\centering
\includegraphics[scale=0.6]{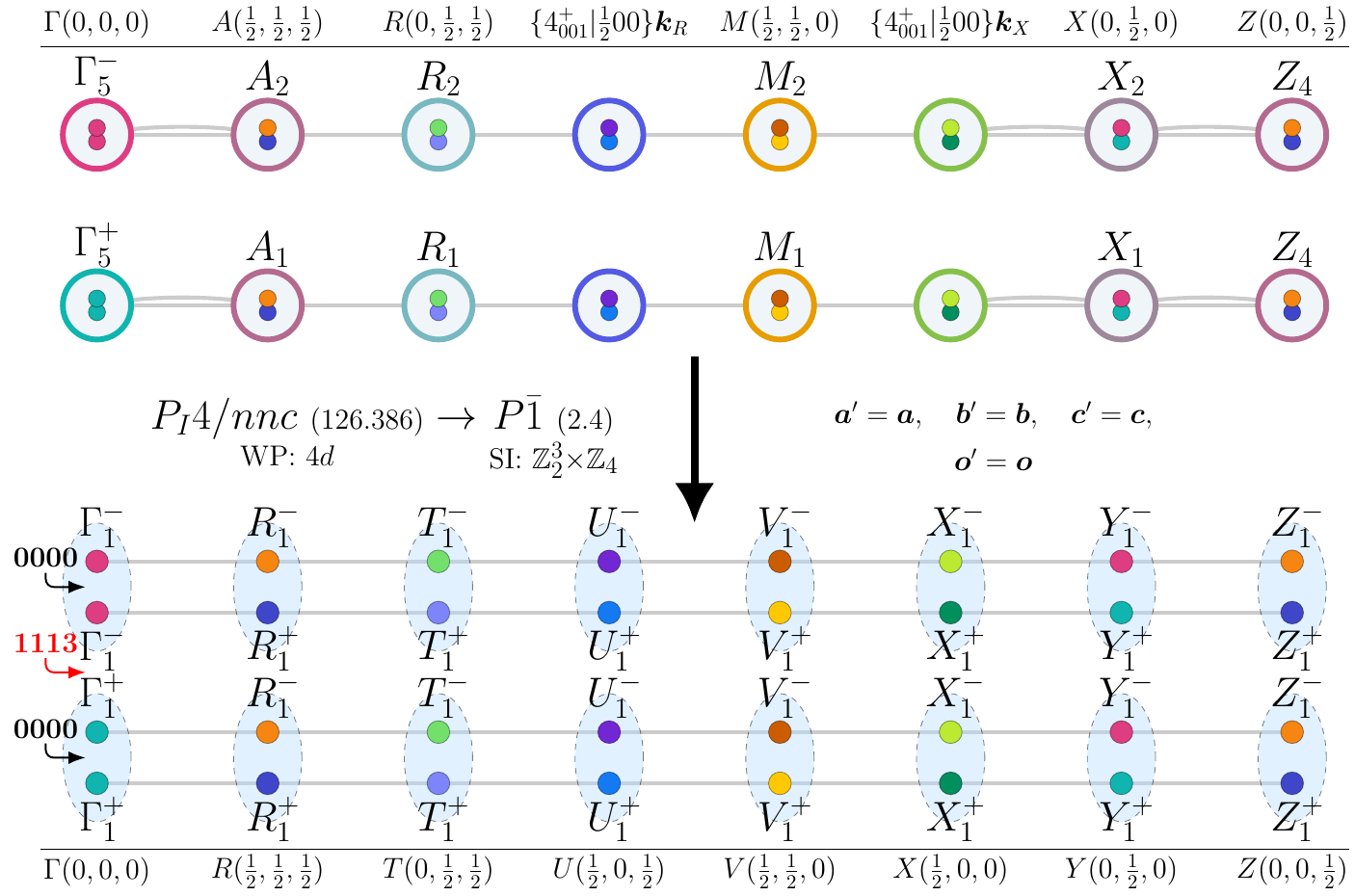}
\caption{Topological magnon bands in subgroup $P\bar{1}~(2.4)$ for magnetic moments on Wyckoff position $4d$ of supergroup $P_{I}4/nnc~(126.386)$.\label{fig_126.386_2.4_Bparallel001andstrainingenericdirection_4d}}
\end{figure}
\begin{center}
\begin{longtable}{cc}
\caption{Number of times a given SI can appear in the band structure. Supergroup: $P_{I}4/nnc~(126.386)$, WP: $4d$, Subgroup: $P\bar{1}~(2.4)$.\label{gap_table_126.386_2.4_4d_Bparallel001andstrainingenericdirection}}\\
\toprule
SI&Possible \#Gaps\\
\midrule
\endfirsthead
\multicolumn{2}{@{}l}{\ldots continued}\\
\toprule
SI&Possible \#Gaps\\
\midrule
\endhead
\midrule\multicolumn{2}{r}{continued on next page\ldots}
\endfoot
\bottomrule
\endlastfoot
0000&0, 1, 2\\
0001&0, 1, 2\\
0003&0, 1, 2\\
0010&0, 1, 2\\
0011&0, 1, 2\\
0012&0, 1, 2\\
0013&0, 1, 2\\
0100&0, 1, 2\\
0101&0, 1, 2\\
0102&0, 1, 2\\
0103&0, 1, 2\\
0110&0, 1, 2\\
0111&0, 1, 2\\
0112&0, 1, 2\\
0113&0, 1, 2\\
1000&0, 1, 2\\
1001&0, 1, 2\\
1002&0, 1, 2\\
1003&0, 1, 2\\
1010&0, 1, 2\\
1011&0, 1, 2\\
1012&0, 1, 2\\
1013&0, 1, 2\\
1100&0, 1, 2\\
1101&0, 1, 2\\
1102&0, 1, 2\\
1103&0, 1, 2\\
1110&0, 1, 2\\
1111&0, 1, 2, 3\\
1112&0, 1, 2\\
1113&0, 1, 2, 3\\
gapless&0\\
nontrivial&1, 2, 3\\
\end{longtable}
\end{center}
\begin{center}
\begin{longtable}{cc}
\caption{Possible SI values in each gap. Supergroup: $P_{I}4/nnc~(126.386)$, WP: $4d$, Subgroup: $P\bar{1}~(2.4)$.\label{si_table_126.386_2.4_4d_Bparallel001andstrainingenericdirection}}\\
\toprule
Gap\#& Possible SI Values\\
\midrule
\endfirsthead
\multicolumn{2}{@{}l}{\ldots continued}\\
\toprule
Gap\#& Possible SI Values\\
\midrule
\endhead
\multicolumn{2}{r}{continued on next page\ldots}
\endfoot
\bottomrule
\endlastfoot
\multirow{7}{*}{1}&\multirow{1}{*}{0000, 0001, 0003, 0010, 0011}\\*
&\multirow{1}{*}{0012, 0013, 0100, 0101, 0102}\\*
&\multirow{1}{*}{0103, 0110, 0111, 0112, 0113}\\*
&\multirow{1}{*}{1000, 1001, 1002, 1003, 1010}\\*
&\multirow{1}{*}{1011, 1012, 1013, 1100, 1101}\\*
&\multirow{1}{*}{1102, 1103, 1110, 1111, 1112}\\*
&\multirow{1}{*}{1113}\\\midrule
\multirow{1}{*}{2}&\multirow{1}{*}{1111, 1113}\\\midrule
\multirow{7}{*}{3}&\multirow{1}{*}{0000, 0001, 0003, 0010, 0011}\\*
&\multirow{1}{*}{0012, 0013, 0100, 0101, 0102}\\*
&\multirow{1}{*}{0103, 0110, 0111, 0112, 0113}\\*
&\multirow{1}{*}{1000, 1001, 1002, 1003, 1010}\\*
&\multirow{1}{*}{1011, 1012, 1013, 1100, 1101}\\*
&\multirow{1}{*}{1102, 1103, 1110, 1111, 1112}\\*
&\multirow{1}{*}{1113}\\\midrule
\multirow{1}{*}{4}&\multirow{1}{*}{0000}\\
\end{longtable}
\end{center}
\subsubsection{Topological bands in subgroup $C2'/m'~(12.62)$}
\textbf{Perturbations:}
\begin{itemize}
\item B $\parallel$ [001] and strain $\perp$ [110],
\item B $\perp$ [110].
\end{itemize}
\begin{figure}[H]
\centering
\includegraphics[scale=0.6]{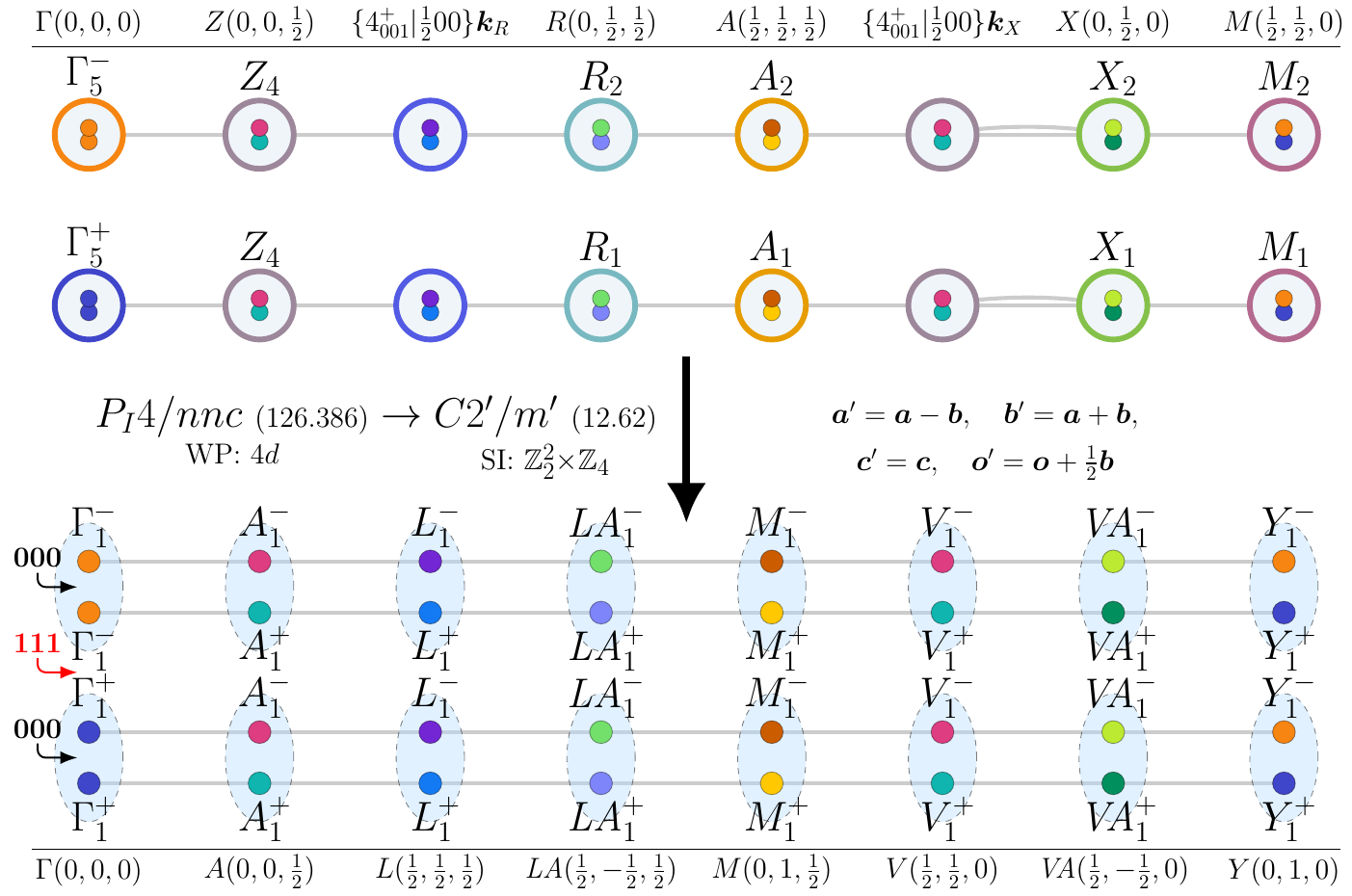}
\caption{Topological magnon bands in subgroup $C2'/m'~(12.62)$ for magnetic moments on Wyckoff position $4d$ of supergroup $P_{I}4/nnc~(126.386)$.\label{fig_126.386_12.62_Bparallel001andstrainperp110_4d}}
\end{figure}
\begin{center}
\begin{longtable}{cc}
\caption{Number of times a given SI can appear in the band structure. Supergroup: $P_{I}4/nnc~(126.386)$, WP: $4d$, Subgroup: $C2'/m'~(12.62)$.\label{gap_table_126.386_12.62_4d_Bparallel001andstrainperp110}}\\
\toprule
SI&Possible \#Gaps\\
\midrule
\endfirsthead
\multicolumn{2}{@{}l}{\ldots continued}\\
\toprule
SI&Possible \#Gaps\\
\midrule
\endhead
\midrule\multicolumn{2}{r}{continued on next page\ldots}
\endfoot
\bottomrule
\endlastfoot
000&0, 1, 2\\
001&0, 1, 2\\
003&0, 1, 2\\
010&0, 1, 2\\
011&0, 1, 2\\
012&0, 1, 2\\
013&0, 1, 2\\
100&0, 1, 2\\
101&0, 1, 2\\
103&0, 1, 2\\
110&0, 1, 2\\
111&0, 1, 2, 3\\
112&0, 1, 2\\
113&0, 1, 2, 3\\
gapless&0\\
nontrivial&1, 2, 3\\
\end{longtable}
\end{center}
\begin{center}
\begin{longtable}{cc}
\caption{Possible SI values in each gap. Supergroup: $P_{I}4/nnc~(126.386)$, WP: $4d$, Subgroup: $C2'/m'~(12.62)$.\label{si_table_126.386_12.62_4d_Bparallel001andstrainperp110}}\\
\toprule
Gap\#& Possible SI Values\\
\midrule
\endfirsthead
\multicolumn{2}{@{}l}{\ldots continued}\\
\toprule
Gap\#& Possible SI Values\\
\midrule
\endhead
\multicolumn{2}{r}{continued on next page\ldots}
\endfoot
\bottomrule
\endlastfoot
\multirow{3}{*}{1}&\multirow{1}{*}{000, 001, 003, 010, 011}\\*
&\multirow{1}{*}{012, 013, 100, 101, 103}\\*
&\multirow{1}{*}{110, 111, 112, 113}\\\midrule
\multirow{1}{*}{2}&\multirow{1}{*}{111, 113}\\\midrule
\multirow{3}{*}{3}&\multirow{1}{*}{000, 001, 003, 010, 011}\\*
&\multirow{1}{*}{012, 013, 100, 101, 103}\\*
&\multirow{1}{*}{110, 111, 112, 113}\\\midrule
\multirow{1}{*}{4}&\multirow{1}{*}{000}\\
\end{longtable}
\end{center}
\subsubsection{Topological bands in subgroup $P2_{1}'/m'~(11.54)$}
\textbf{Perturbations:}
\begin{itemize}
\item B $\parallel$ [100] and strain $\parallel$ [110],
\item B $\parallel$ [100] and strain $\perp$ [001],
\item B $\parallel$ [110] and strain $\parallel$ [100],
\item B $\parallel$ [110] and strain $\perp$ [001],
\item B $\perp$ [001].
\end{itemize}
\begin{figure}[H]
\centering
\includegraphics[scale=0.6]{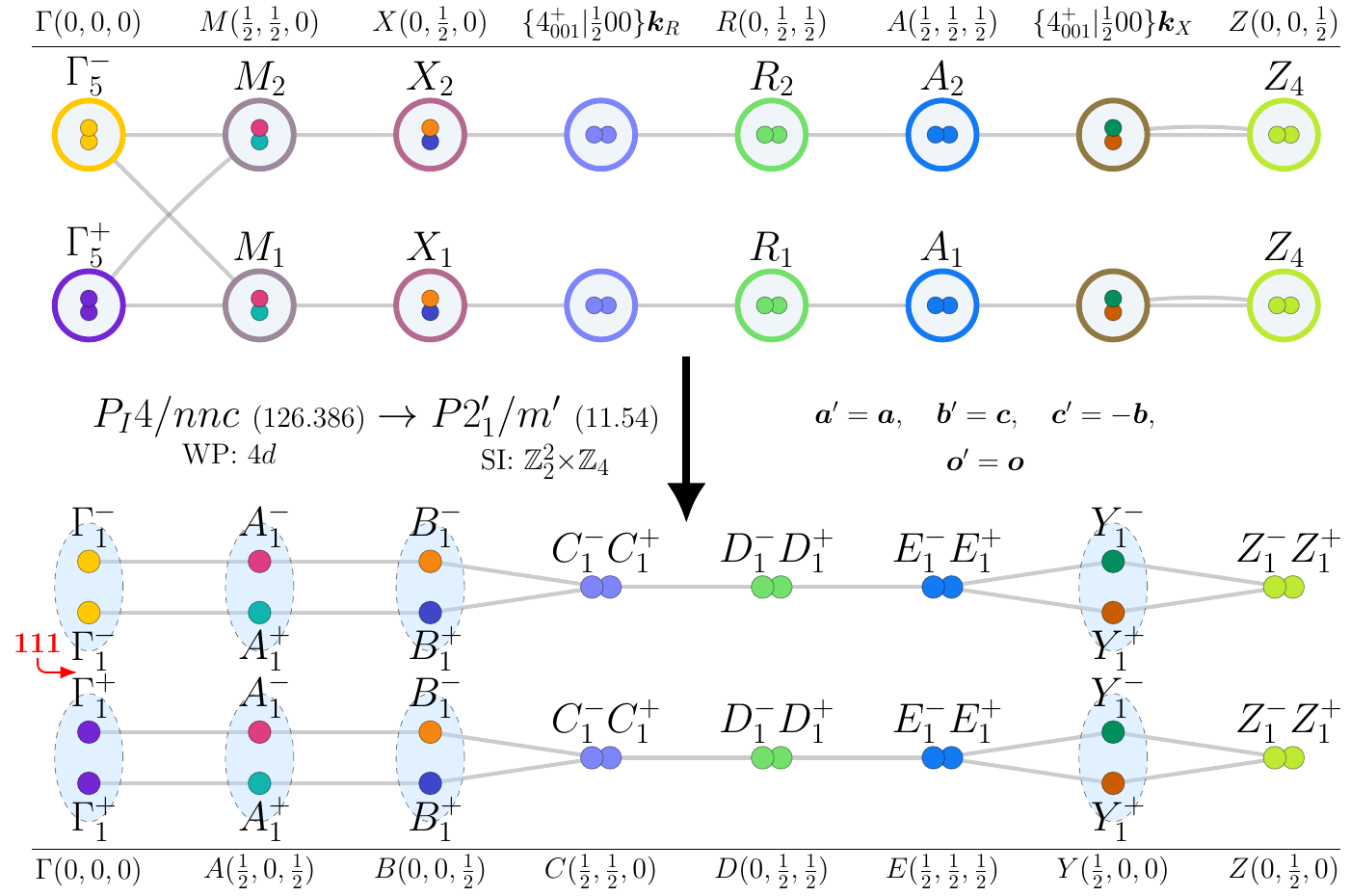}
\caption{Topological magnon bands in subgroup $P2_{1}'/m'~(11.54)$ for magnetic moments on Wyckoff position $4d$ of supergroup $P_{I}4/nnc~(126.386)$.\label{fig_126.386_11.54_Bparallel100andstrainparallel110_4d}}
\end{figure}
\begin{center}
\begin{longtable}{cc}
\caption{Number of times a given SI can appear in the band structure. Supergroup: $P_{I}4/nnc~(126.386)$, WP: $4d$, Subgroup: $P2_{1}'/m'~(11.54)$.\label{gap_table_126.386_11.54_4d_Bparallel100andstrainparallel110}}\\
\toprule
SI&Possible \#Gaps\\
\midrule
\endfirsthead
\multicolumn{2}{@{}l}{\ldots continued}\\
\toprule
SI&Possible \#Gaps\\
\midrule
\endhead
\midrule\multicolumn{2}{r}{continued on next page\ldots}
\endfoot
\bottomrule
\endlastfoot
000&0\\
111&0, 1\\
113&0, 1\\
gapless&2\\
nontrivial&1\\
\end{longtable}
\end{center}
\begin{center}
\begin{longtable}{cc}
\caption{Possible SI values in each gap. Supergroup: $P_{I}4/nnc~(126.386)$, WP: $4d$, Subgroup: $P2_{1}'/m'~(11.54)$.\label{si_table_126.386_11.54_4d_Bparallel100andstrainparallel110}}\\
\toprule
Gap\#& Possible SI Values\\
\midrule
\endfirsthead
\multicolumn{2}{@{}l}{\ldots continued}\\
\toprule
Gap\#& Possible SI Values\\
\midrule
\endhead
\multicolumn{2}{r}{continued on next page\ldots}
\endfoot
\bottomrule
\endlastfoot
\multirow{1}{*}{1}&\multirow{1}{*}{G}\\\midrule
\multirow{1}{*}{2}&\multirow{1}{*}{111, 113}\\\midrule
\multirow{1}{*}{3}&\multirow{1}{*}{G}\\\midrule
\multirow{1}{*}{4}&\multirow{1}{*}{000}\\
\end{longtable}
\end{center}
\subsubsection{Topological bands in subgroup $P2_{1}'/m'~(11.54)$}
\textbf{Perturbations:}
\begin{itemize}
\item B $\parallel$ [001] and strain $\perp$ [100],
\item B $\perp$ [100].
\end{itemize}
\begin{figure}[H]
\centering
\includegraphics[scale=0.6]{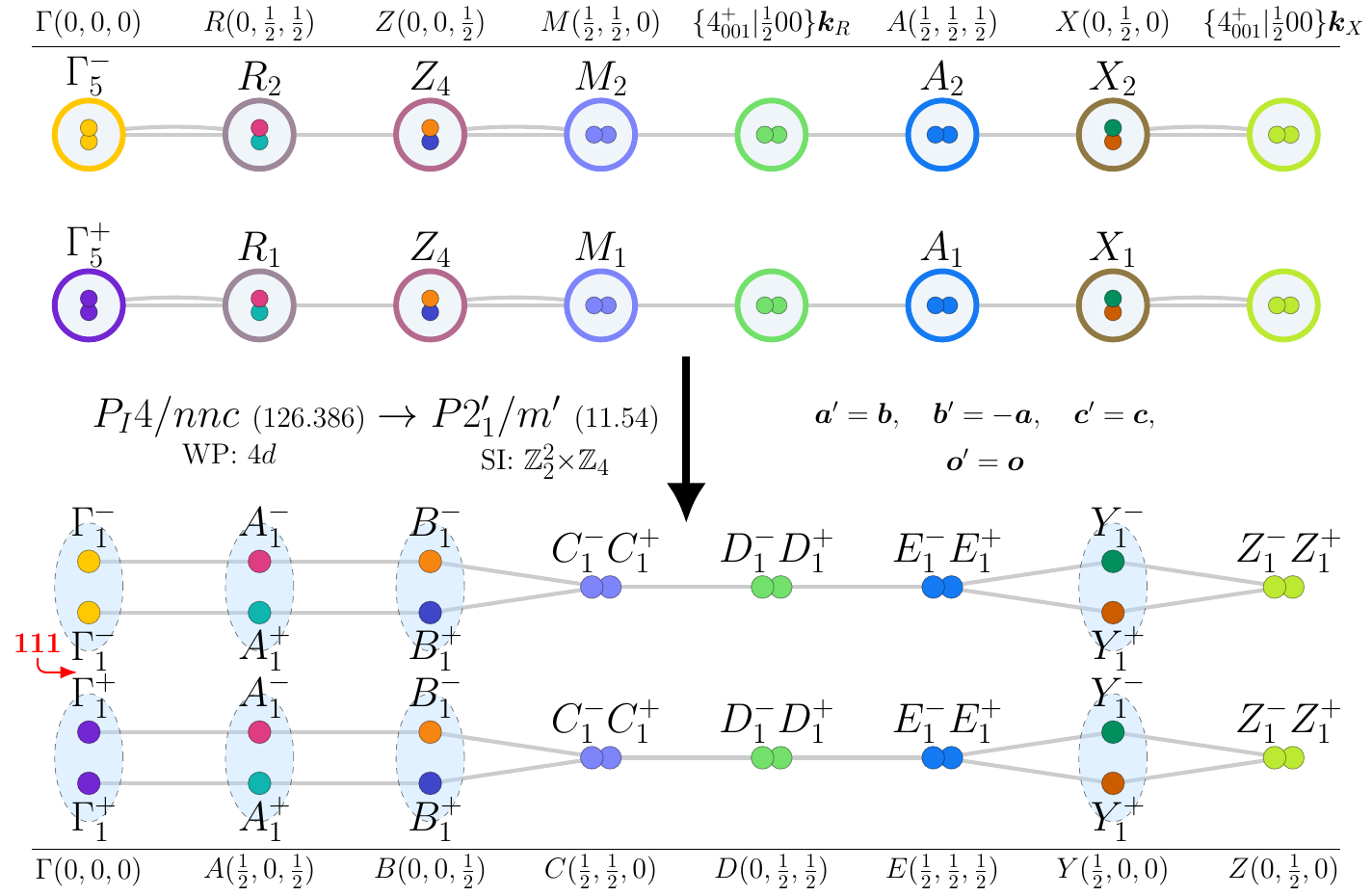}
\caption{Topological magnon bands in subgroup $P2_{1}'/m'~(11.54)$ for magnetic moments on Wyckoff position $4d$ of supergroup $P_{I}4/nnc~(126.386)$.\label{fig_126.386_11.54_Bparallel001andstrainperp100_4d}}
\end{figure}
\begin{center}
\begin{longtable}{cc}
\caption{Number of times a given SI can appear in the band structure. Supergroup: $P_{I}4/nnc~(126.386)$, WP: $4d$, Subgroup: $P2_{1}'/m'~(11.54)$.\label{gap_table_126.386_11.54_4d_Bparallel001andstrainperp100}}\\
\toprule
SI&Possible \#Gaps\\
\midrule
\endfirsthead
\multicolumn{2}{@{}l}{\ldots continued}\\
\toprule
SI&Possible \#Gaps\\
\midrule
\endhead
\midrule\multicolumn{2}{r}{continued on next page\ldots}
\endfoot
\bottomrule
\endlastfoot
000&0\\
111&0, 1\\
113&0, 1\\
gapless&2\\
nontrivial&1\\
\end{longtable}
\end{center}
\begin{center}
\begin{longtable}{cc}
\caption{Possible SI values in each gap. Supergroup: $P_{I}4/nnc~(126.386)$, WP: $4d$, Subgroup: $P2_{1}'/m'~(11.54)$.\label{si_table_126.386_11.54_4d_Bparallel001andstrainperp100}}\\
\toprule
Gap\#& Possible SI Values\\
\midrule
\endfirsthead
\multicolumn{2}{@{}l}{\ldots continued}\\
\toprule
Gap\#& Possible SI Values\\
\midrule
\endhead
\multicolumn{2}{r}{continued on next page\ldots}
\endfoot
\bottomrule
\endlastfoot
\multirow{1}{*}{1}&\multirow{1}{*}{G}\\\midrule
\multirow{1}{*}{2}&\multirow{1}{*}{111, 113}\\\midrule
\multirow{1}{*}{3}&\multirow{1}{*}{G}\\\midrule
\multirow{1}{*}{4}&\multirow{1}{*}{000}\\
\end{longtable}
\end{center}
\subsubsection{Topological bands in subgroup $P_{S}\bar{1}~(2.7)$}
\textbf{Perturbation:}
\begin{itemize}
\item strain in generic direction.
\end{itemize}
\begin{figure}[H]
\centering
\includegraphics[scale=0.6]{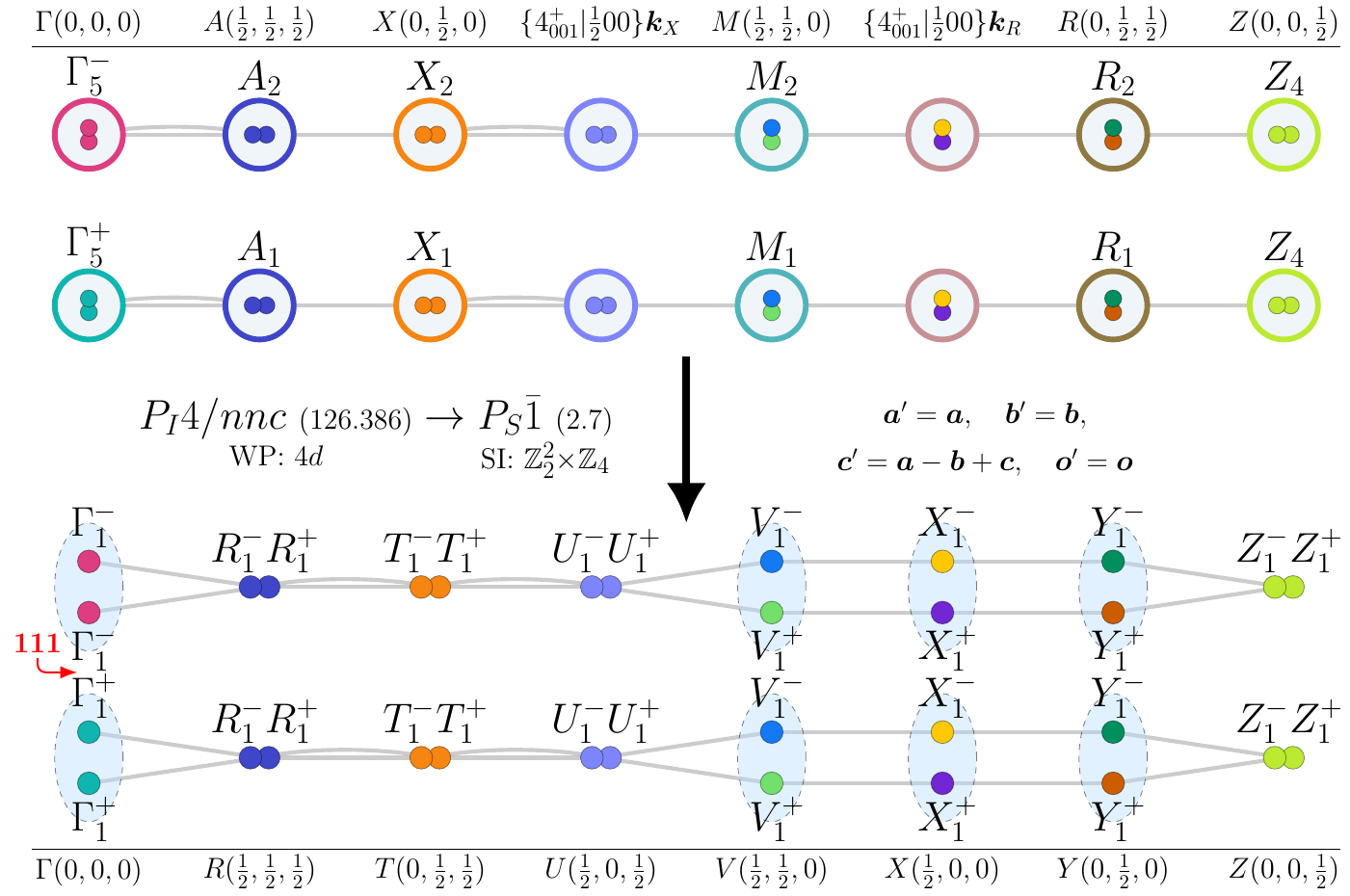}
\caption{Topological magnon bands in subgroup $P_{S}\bar{1}~(2.7)$ for magnetic moments on Wyckoff position $4d$ of supergroup $P_{I}4/nnc~(126.386)$.\label{fig_126.386_2.7_strainingenericdirection_4d}}
\end{figure}
\begin{center}
\begin{longtable}{cc}
\caption{Number of times a given SI can appear in the band structure. Supergroup: $P_{I}4/nnc~(126.386)$, WP: $4d$, Subgroup: $P_{S}\bar{1}~(2.7)$.\label{gap_table_126.386_2.7_4d_strainingenericdirection}}\\
\toprule
SI&Possible \#Gaps\\
\midrule
\endfirsthead
\multicolumn{2}{@{}l}{\ldots continued}\\
\toprule
SI&Possible \#Gaps\\
\midrule
\endhead
\midrule\multicolumn{2}{r}{continued on next page\ldots}
\endfoot
\bottomrule
\endlastfoot
000&0\\
111&0, 1\\
113&0, 1\\
gapless&2\\
nontrivial&1\\
\end{longtable}
\end{center}
\begin{center}
\begin{longtable}{cc}
\caption{Possible SI values in each gap. Supergroup: $P_{I}4/nnc~(126.386)$, WP: $4d$, Subgroup: $P_{S}\bar{1}~(2.7)$.\label{si_table_126.386_2.7_4d_strainingenericdirection}}\\
\toprule
Gap\#& Possible SI Values\\
\midrule
\endfirsthead
\multicolumn{2}{@{}l}{\ldots continued}\\
\toprule
Gap\#& Possible SI Values\\
\midrule
\endhead
\multicolumn{2}{r}{continued on next page\ldots}
\endfoot
\bottomrule
\endlastfoot
\multirow{1}{*}{1}&\multirow{1}{*}{G}\\\midrule
\multirow{1}{*}{2}&\multirow{1}{*}{111, 113}\\\midrule
\multirow{1}{*}{3}&\multirow{1}{*}{G}\\\midrule
\multirow{1}{*}{4}&\multirow{1}{*}{000}\\
\end{longtable}
\end{center}
\subsubsection{Topological bands in subgroup $C_{A}cc2~(37.186)$}
\textbf{Perturbation:}
\begin{itemize}
\item E $\parallel$ [001] and strain $\parallel$ [110].
\end{itemize}
\begin{figure}[H]
\centering
\includegraphics[scale=0.6]{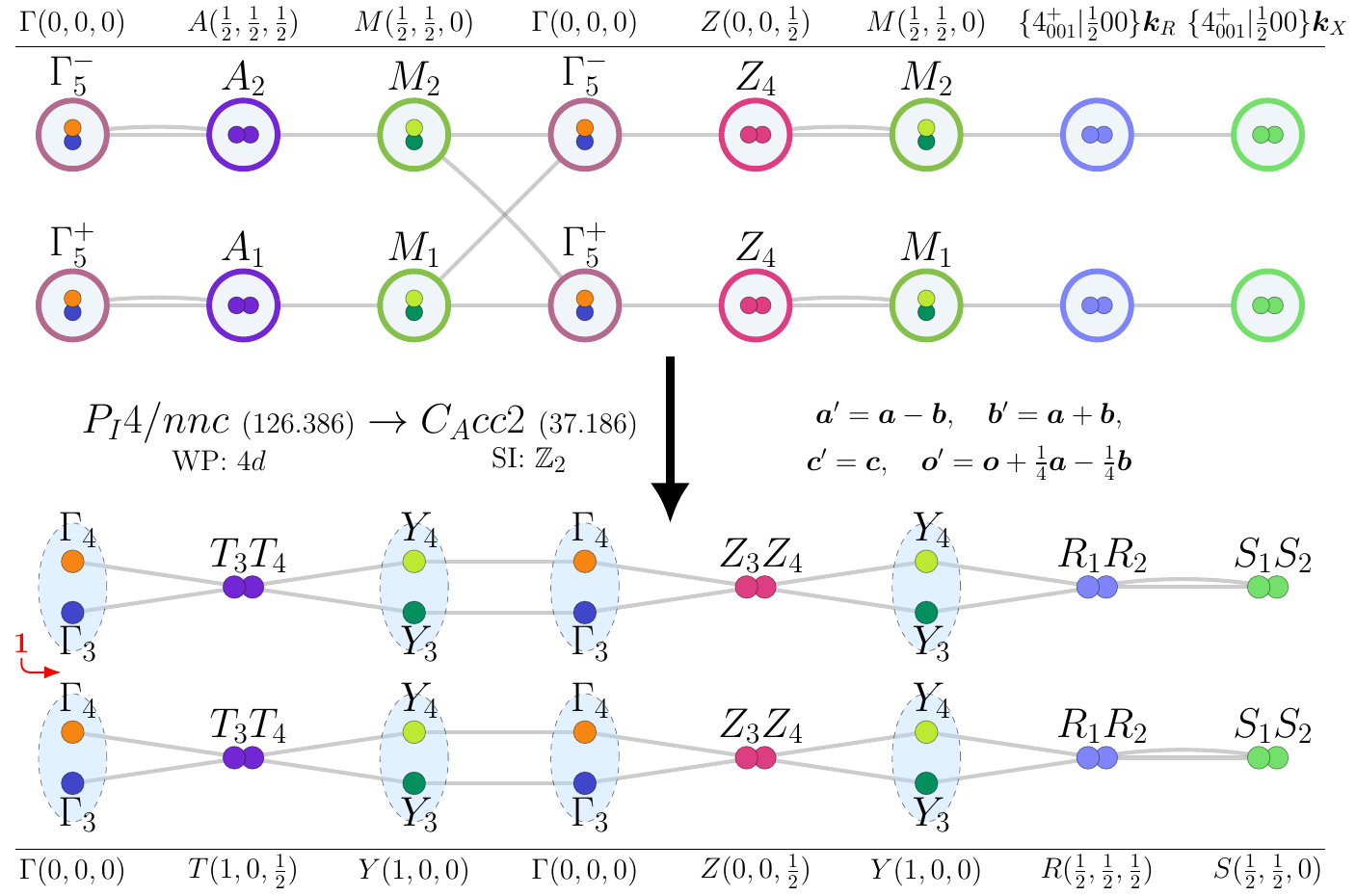}
\caption{Topological magnon bands in subgroup $C_{A}cc2~(37.186)$ for magnetic moments on Wyckoff position $4d$ of supergroup $P_{I}4/nnc~(126.386)$.\label{fig_126.386_37.186_Eparallel001andstrainparallel110_4d}}
\end{figure}
\begin{center}
\begin{longtable}{cc}
\caption{Number of times a given SI can appear in the band structure. Supergroup: $P_{I}4/nnc~(126.386)$, WP: $4d$, Subgroup: $C_{A}cc2~(37.186)$.\label{gap_table_126.386_37.186_4d_Eparallel001andstrainparallel110}}\\
\toprule
SI&Possible \#Gaps\\
\midrule
\endfirsthead
\multicolumn{2}{@{}l}{\ldots continued}\\
\toprule
SI&Possible \#Gaps\\
\midrule
\endhead
\midrule\multicolumn{2}{r}{continued on next page\ldots}
\endfoot
\bottomrule
\endlastfoot
0&0\\
1&1\\
gapless&2\\
nontrivial&1\\
\end{longtable}
\end{center}
\begin{center}
\begin{longtable}{cc}
\caption{Possible SI values in each gap. Supergroup: $P_{I}4/nnc~(126.386)$, WP: $4d$, Subgroup: $C_{A}cc2~(37.186)$.\label{si_table_126.386_37.186_4d_Eparallel001andstrainparallel110}}\\
\toprule
Gap\#& Possible SI Values\\
\midrule
\endfirsthead
\multicolumn{2}{@{}l}{\ldots continued}\\
\toprule
Gap\#& Possible SI Values\\
\midrule
\endhead
\multicolumn{2}{r}{continued on next page\ldots}
\endfoot
\bottomrule
\endlastfoot
\multirow{1}{*}{1}&\multirow{1}{*}{G}\\\midrule
\multirow{1}{*}{2}&\multirow{1}{*}{1}\\\midrule
\multirow{1}{*}{3}&\multirow{1}{*}{G}\\\midrule
\multirow{1}{*}{4}&\multirow{1}{*}{0}\\
\end{longtable}
\end{center}
\subsubsection{Topological bands in subgroup $P_{I}4nc~(104.210)$}
\textbf{Perturbation:}
\begin{itemize}
\item E $\parallel$ [001].
\end{itemize}
\begin{figure}[H]
\centering
\includegraphics[scale=0.6]{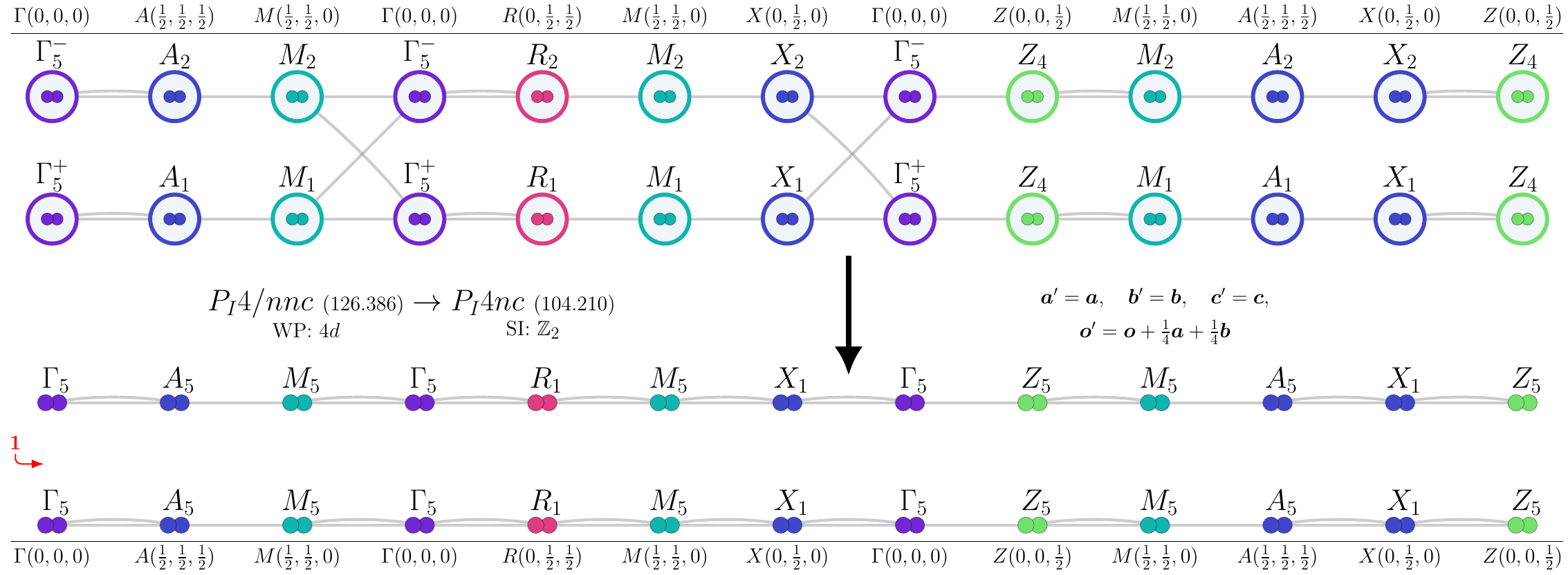}
\caption{Topological magnon bands in subgroup $P_{I}4nc~(104.210)$ for magnetic moments on Wyckoff position $4d$ of supergroup $P_{I}4/nnc~(126.386)$.\label{fig_126.386_104.210_Eparallel001_4d}}
\end{figure}
\begin{center}
\begin{longtable}{cc}
\caption{Number of times a given SI can appear in the band structure. Supergroup: $P_{I}4/nnc~(126.386)$, WP: $4d$, Subgroup: $P_{I}4nc~(104.210)$.\label{gap_table_126.386_104.210_4d_Eparallel001}}\\
\toprule
SI&Possible \#Gaps\\
\midrule
\endfirsthead
\multicolumn{2}{@{}l}{\ldots continued}\\
\toprule
SI&Possible \#Gaps\\
\midrule
\endhead
\midrule\multicolumn{2}{r}{continued on next page\ldots}
\endfoot
\bottomrule
\endlastfoot
0&0\\
1&1\\
gapless&2\\
nontrivial&1\\
\end{longtable}
\end{center}
\begin{center}
\begin{longtable}{cc}
\caption{Possible SI values in each gap. Supergroup: $P_{I}4/nnc~(126.386)$, WP: $4d$, Subgroup: $P_{I}4nc~(104.210)$.\label{si_table_126.386_104.210_4d_Eparallel001}}\\
\toprule
Gap\#& Possible SI Values\\
\midrule
\endfirsthead
\multicolumn{2}{@{}l}{\ldots continued}\\
\toprule
Gap\#& Possible SI Values\\
\midrule
\endhead
\multicolumn{2}{r}{continued on next page\ldots}
\endfoot
\bottomrule
\endlastfoot
\multirow{1}{*}{1}&\multirow{1}{*}{G}\\\midrule
\multirow{1}{*}{2}&\multirow{1}{*}{1}\\\midrule
\multirow{1}{*}{3}&\multirow{1}{*}{G}\\\midrule
\multirow{1}{*}{4}&\multirow{1}{*}{0}\\
\end{longtable}
\end{center}
\subsection{WP: $4e$}
\textbf{BCS Materials:} {UGeTe~(73 K)}\footnote{BCS web page: \texttt{\href{http://webbdcrista1.ehu.es/magndata/index.php?this\_label=1.425} {http://webbdcrista1.ehu.es/magndata/index.php?this\_label=1.425}}}, {NdScSiC\textsubscript{0.5}H\textsubscript{0.2}~(10 K)}\footnote{BCS web page: \texttt{\href{http://webbdcrista1.ehu.es/magndata/index.php?this\_label=1.503} {http://webbdcrista1.ehu.es/magndata/index.php?this\_label=1.503}}}.\\
\subsubsection{Topological bands in subgroup $P\bar{1}~(2.4)$}
\textbf{Perturbations:}
\begin{itemize}
\item B $\parallel$ [001] and strain in generic direction,
\item B $\parallel$ [100] and strain $\perp$ [110],
\item B $\parallel$ [100] and strain in generic direction,
\item B $\parallel$ [110] and strain $\perp$ [100],
\item B $\parallel$ [110] and strain in generic direction,
\item B $\perp$ [001] and strain $\perp$ [100],
\item B $\perp$ [001] and strain $\perp$ [110],
\item B $\perp$ [001] and strain in generic direction,
\item B $\perp$ [100] and strain $\parallel$ [110],
\item B $\perp$ [100] and strain $\perp$ [001],
\item B $\perp$ [100] and strain $\perp$ [110],
\item B $\perp$ [100] and strain in generic direction,
\item B $\perp$ [110] and strain $\parallel$ [100],
\item B $\perp$ [110] and strain $\perp$ [001],
\item B $\perp$ [110] and strain $\perp$ [100],
\item B $\perp$ [110] and strain in generic direction,
\item B in generic direction.
\end{itemize}
\begin{figure}[H]
\centering
\includegraphics[scale=0.6]{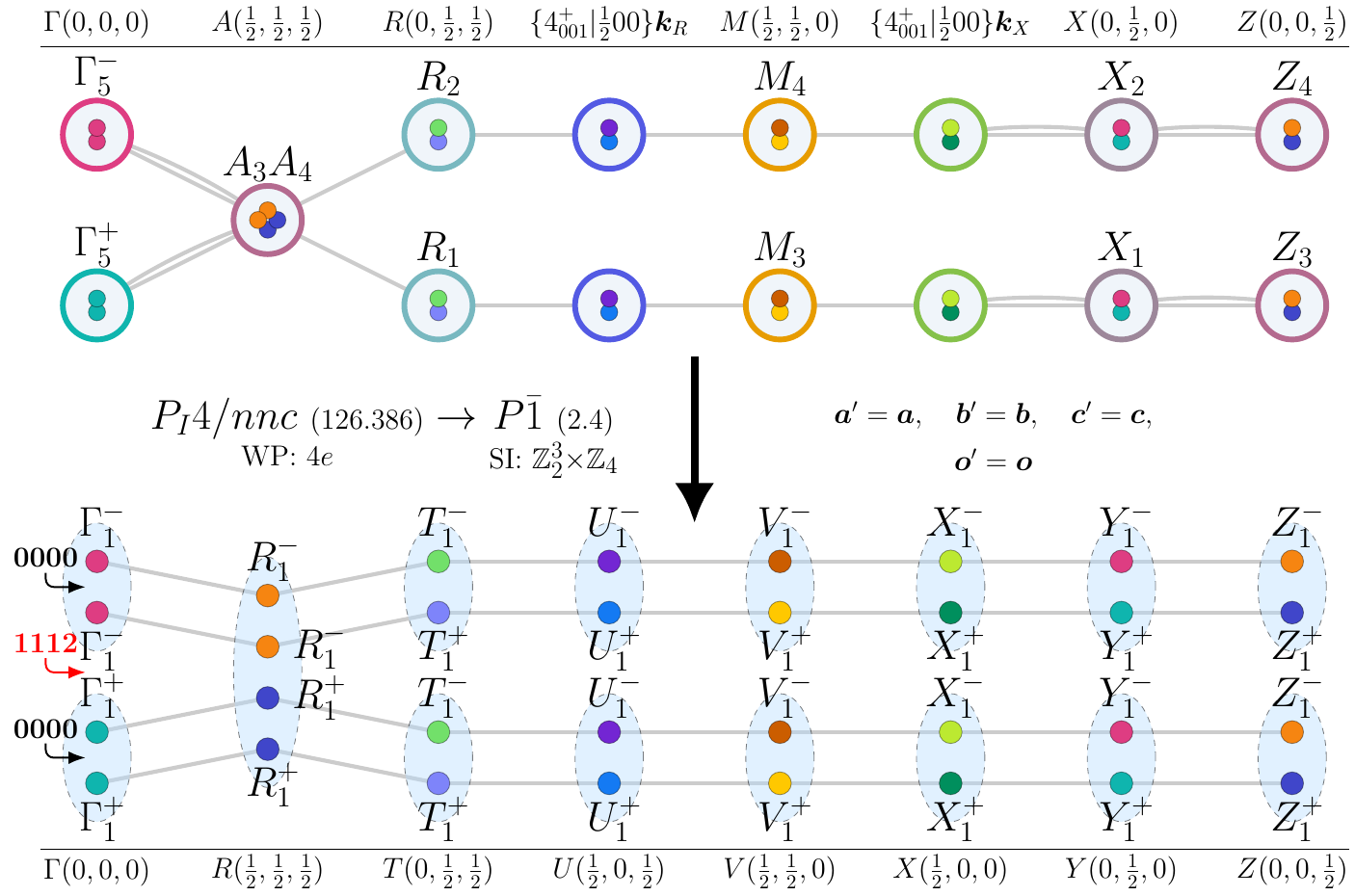}
\caption{Topological magnon bands in subgroup $P\bar{1}~(2.4)$ for magnetic moments on Wyckoff position $4e$ of supergroup $P_{I}4/nnc~(126.386)$.\label{fig_126.386_2.4_Bparallel001andstrainingenericdirection_4e}}
\end{figure}
\begin{center}
\begin{longtable}{cc}
\caption{Number of times a given SI can appear in the band structure. Supergroup: $P_{I}4/nnc~(126.386)$, WP: $4e$, Subgroup: $P\bar{1}~(2.4)$.\label{gap_table_126.386_2.4_4e_Bparallel001andstrainingenericdirection}}\\
\toprule
SI&Possible \#Gaps\\
\midrule
\endfirsthead
\multicolumn{2}{@{}l}{\ldots continued}\\
\toprule
SI&Possible \#Gaps\\
\midrule
\endhead
\midrule\multicolumn{2}{r}{continued on next page\ldots}
\endfoot
\bottomrule
\endlastfoot
0000&0, 1, 2\\
0001&0, 1, 2\\
0003&0, 1, 2\\
0010&0, 1, 2\\
0011&0, 1, 2\\
0012&0, 1, 2\\
0013&0, 1, 2\\
0100&0, 1, 2\\
0101&0, 1, 2\\
0102&0, 1, 2\\
0103&0, 1, 2\\
0110&0, 1, 2\\
0111&0, 1, 2\\
0112&0, 1, 2\\
0113&0, 1, 2\\
1000&0, 1, 2\\
1001&0, 1, 2\\
1002&0, 1, 2\\
1003&0, 1, 2\\
1010&0, 1, 2\\
1011&0, 1, 2\\
1012&0, 1, 2\\
1013&0, 1, 2\\
1100&0, 1, 2\\
1101&0, 1, 2\\
1102&0, 1, 2\\
1103&0, 1, 2\\
1110&0, 1, 2, 3\\
1111&0, 1, 2, 3\\
1112&0, 1, 2, 3\\
1113&0, 1, 2, 3\\
gapless&0\\
nontrivial&1, 2, 3\\
\end{longtable}
\end{center}
\begin{center}
\begin{longtable}{cc}
\caption{Possible SI values in each gap. Supergroup: $P_{I}4/nnc~(126.386)$, WP: $4e$, Subgroup: $P\bar{1}~(2.4)$.\label{si_table_126.386_2.4_4e_Bparallel001andstrainingenericdirection}}\\
\toprule
Gap\#& Possible SI Values\\
\midrule
\endfirsthead
\multicolumn{2}{@{}l}{\ldots continued}\\
\toprule
Gap\#& Possible SI Values\\
\midrule
\endhead
\multicolumn{2}{r}{continued on next page\ldots}
\endfoot
\bottomrule
\endlastfoot
\multirow{7}{*}{1}&\multirow{1}{*}{0000, 0001, 0003, 0010, 0011}\\*
&\multirow{1}{*}{0012, 0013, 0100, 0101, 0102}\\*
&\multirow{1}{*}{0103, 0110, 0111, 0112, 0113}\\*
&\multirow{1}{*}{1000, 1001, 1002, 1003, 1010}\\*
&\multirow{1}{*}{1011, 1012, 1013, 1100, 1101}\\*
&\multirow{1}{*}{1102, 1103, 1110, 1111, 1112}\\*
&\multirow{1}{*}{1113}\\\midrule
\multirow{1}{*}{2}&\multirow{1}{*}{1110, 1111, 1112, 1113}\\\midrule
\multirow{7}{*}{3}&\multirow{1}{*}{0000, 0001, 0003, 0010, 0011}\\*
&\multirow{1}{*}{0012, 0013, 0100, 0101, 0102}\\*
&\multirow{1}{*}{0103, 0110, 0111, 0112, 0113}\\*
&\multirow{1}{*}{1000, 1001, 1002, 1003, 1010}\\*
&\multirow{1}{*}{1011, 1012, 1013, 1100, 1101}\\*
&\multirow{1}{*}{1102, 1103, 1110, 1111, 1112}\\*
&\multirow{1}{*}{1113}\\\midrule
\multirow{1}{*}{4}&\multirow{1}{*}{0000}\\
\end{longtable}
\end{center}
\subsubsection{Topological bands in subgroup $C2'/m'~(12.62)$}
\textbf{Perturbations:}
\begin{itemize}
\item B $\parallel$ [001] and strain $\perp$ [110],
\item B $\perp$ [110].
\end{itemize}
\begin{figure}[H]
\centering
\includegraphics[scale=0.6]{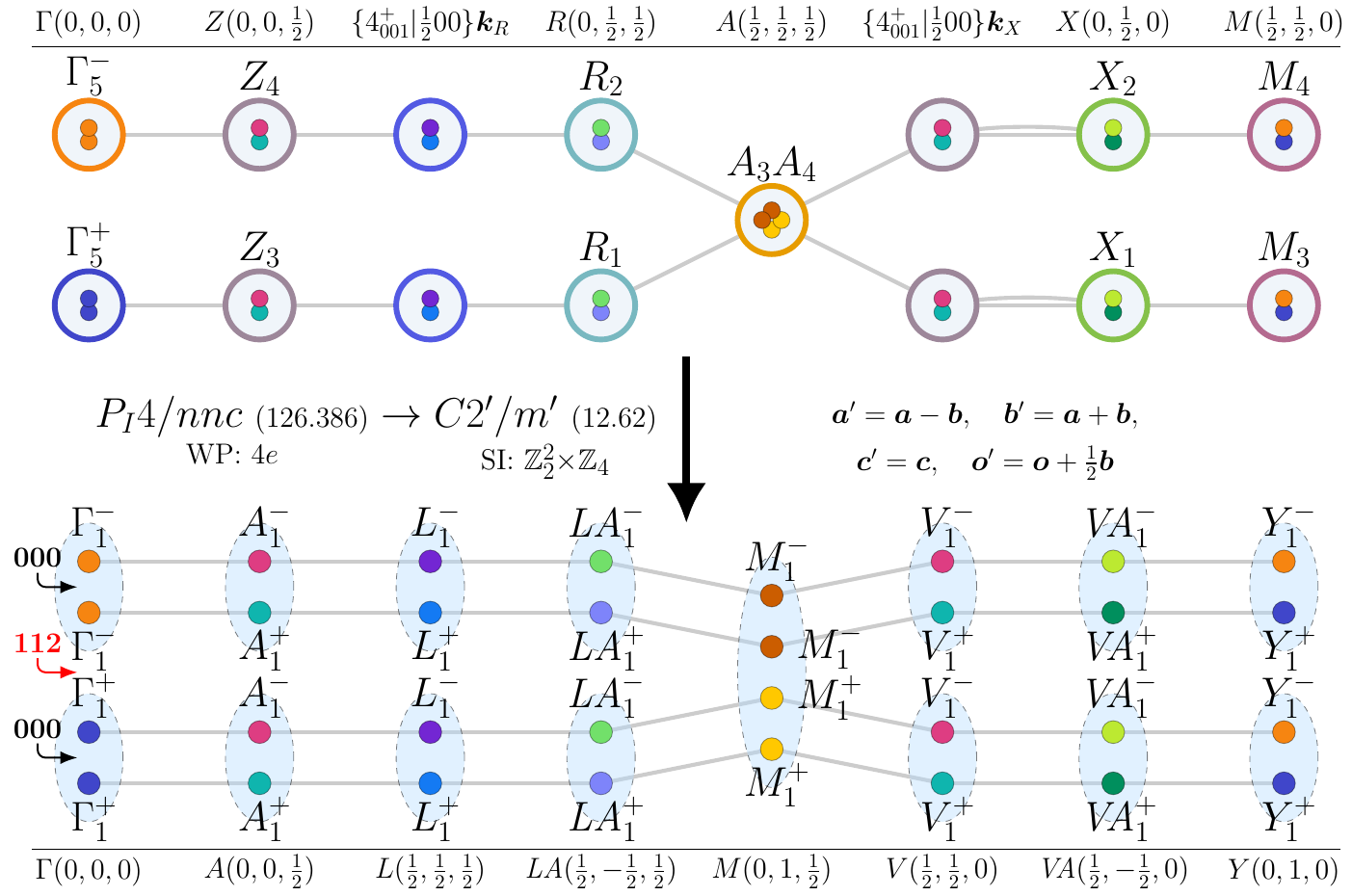}
\caption{Topological magnon bands in subgroup $C2'/m'~(12.62)$ for magnetic moments on Wyckoff position $4e$ of supergroup $P_{I}4/nnc~(126.386)$.\label{fig_126.386_12.62_Bparallel001andstrainperp110_4e}}
\end{figure}
\begin{center}
\begin{longtable}{cc}
\caption{Number of times a given SI can appear in the band structure. Supergroup: $P_{I}4/nnc~(126.386)$, WP: $4e$, Subgroup: $C2'/m'~(12.62)$.\label{gap_table_126.386_12.62_4e_Bparallel001andstrainperp110}}\\
\toprule
SI&Possible \#Gaps\\
\midrule
\endfirsthead
\multicolumn{2}{@{}l}{\ldots continued}\\
\toprule
SI&Possible \#Gaps\\
\midrule
\endhead
\midrule\multicolumn{2}{r}{continued on next page\ldots}
\endfoot
\bottomrule
\endlastfoot
000&0, 1, 2\\
001&0, 1, 2\\
003&0, 1, 2\\
010&0, 1, 2\\
011&0, 1, 2\\
012&0, 1, 2\\
013&0, 1, 2\\
100&0, 1, 2\\
101&0, 1, 2\\
103&0, 1, 2\\
110&0, 1, 2, 3\\
111&0, 1, 2, 3\\
112&0, 1, 2, 3\\
113&0, 1, 2, 3\\
gapless&0\\
nontrivial&1, 2, 3\\
\end{longtable}
\end{center}
\begin{center}
\begin{longtable}{cc}
\caption{Possible SI values in each gap. Supergroup: $P_{I}4/nnc~(126.386)$, WP: $4e$, Subgroup: $C2'/m'~(12.62)$.\label{si_table_126.386_12.62_4e_Bparallel001andstrainperp110}}\\
\toprule
Gap\#& Possible SI Values\\
\midrule
\endfirsthead
\multicolumn{2}{@{}l}{\ldots continued}\\
\toprule
Gap\#& Possible SI Values\\
\midrule
\endhead
\multicolumn{2}{r}{continued on next page\ldots}
\endfoot
\bottomrule
\endlastfoot
\multirow{3}{*}{1}&\multirow{1}{*}{000, 001, 003, 010, 011}\\*
&\multirow{1}{*}{012, 013, 100, 101, 103}\\*
&\multirow{1}{*}{110, 111, 112, 113}\\\midrule
\multirow{1}{*}{2}&\multirow{1}{*}{110, 111, 112, 113}\\\midrule
\multirow{3}{*}{3}&\multirow{1}{*}{000, 001, 003, 010, 011}\\*
&\multirow{1}{*}{012, 013, 100, 101, 103}\\*
&\multirow{1}{*}{110, 111, 112, 113}\\\midrule
\multirow{1}{*}{4}&\multirow{1}{*}{000}\\
\end{longtable}
\end{center}
\subsubsection{Topological bands in subgroup $P2_{1}'/m'~(11.54)$}
\textbf{Perturbations:}
\begin{itemize}
\item B $\parallel$ [100] and strain $\parallel$ [110],
\item B $\parallel$ [100] and strain $\perp$ [001],
\item B $\parallel$ [110] and strain $\parallel$ [100],
\item B $\parallel$ [110] and strain $\perp$ [001],
\item B $\perp$ [001].
\end{itemize}
\begin{figure}[H]
\centering
\includegraphics[scale=0.6]{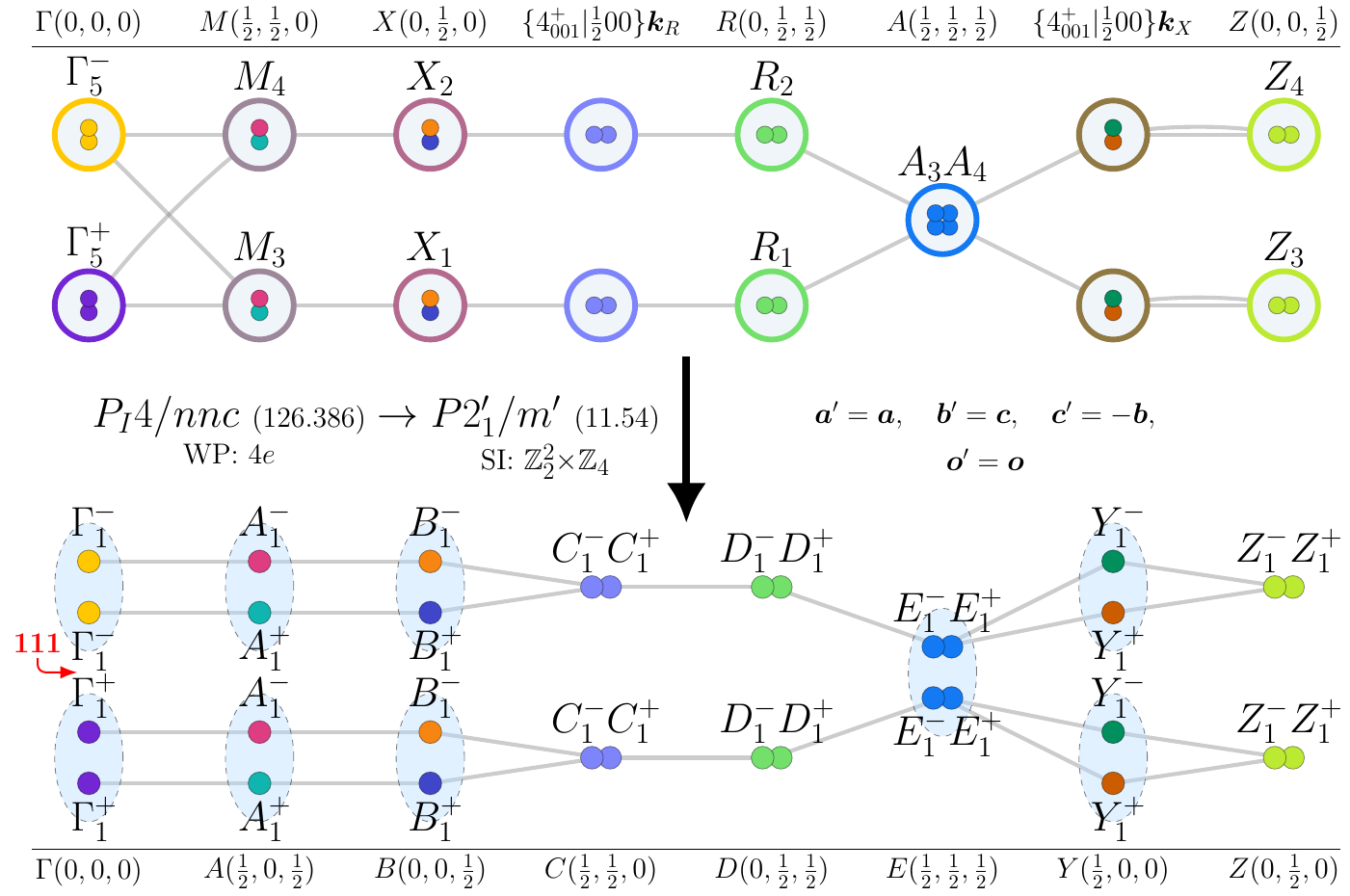}
\caption{Topological magnon bands in subgroup $P2_{1}'/m'~(11.54)$ for magnetic moments on Wyckoff position $4e$ of supergroup $P_{I}4/nnc~(126.386)$.\label{fig_126.386_11.54_Bparallel100andstrainparallel110_4e}}
\end{figure}
\begin{center}
\begin{longtable}{cc}
\caption{Number of times a given SI can appear in the band structure. Supergroup: $P_{I}4/nnc~(126.386)$, WP: $4e$, Subgroup: $P2_{1}'/m'~(11.54)$.\label{gap_table_126.386_11.54_4e_Bparallel100andstrainparallel110}}\\
\toprule
SI&Possible \#Gaps\\
\midrule
\endfirsthead
\multicolumn{2}{@{}l}{\ldots continued}\\
\toprule
SI&Possible \#Gaps\\
\midrule
\endhead
\midrule\multicolumn{2}{r}{continued on next page\ldots}
\endfoot
\bottomrule
\endlastfoot
000&0\\
111&0, 1\\
113&0, 1\\
gapless&2\\
nontrivial&1\\
\end{longtable}
\end{center}
\begin{center}
\begin{longtable}{cc}
\caption{Possible SI values in each gap. Supergroup: $P_{I}4/nnc~(126.386)$, WP: $4e$, Subgroup: $P2_{1}'/m'~(11.54)$.\label{si_table_126.386_11.54_4e_Bparallel100andstrainparallel110}}\\
\toprule
Gap\#& Possible SI Values\\
\midrule
\endfirsthead
\multicolumn{2}{@{}l}{\ldots continued}\\
\toprule
Gap\#& Possible SI Values\\
\midrule
\endhead
\multicolumn{2}{r}{continued on next page\ldots}
\endfoot
\bottomrule
\endlastfoot
\multirow{1}{*}{1}&\multirow{1}{*}{G}\\\midrule
\multirow{1}{*}{2}&\multirow{1}{*}{111, 113}\\\midrule
\multirow{1}{*}{3}&\multirow{1}{*}{G}\\\midrule
\multirow{1}{*}{4}&\multirow{1}{*}{000}\\
\end{longtable}
\end{center}
\subsubsection{Topological bands in subgroup $P2_{1}'/m'~(11.54)$}
\textbf{Perturbations:}
\begin{itemize}
\item B $\parallel$ [001] and strain $\perp$ [100],
\item B $\perp$ [100].
\end{itemize}
\begin{figure}[H]
\centering
\includegraphics[scale=0.6]{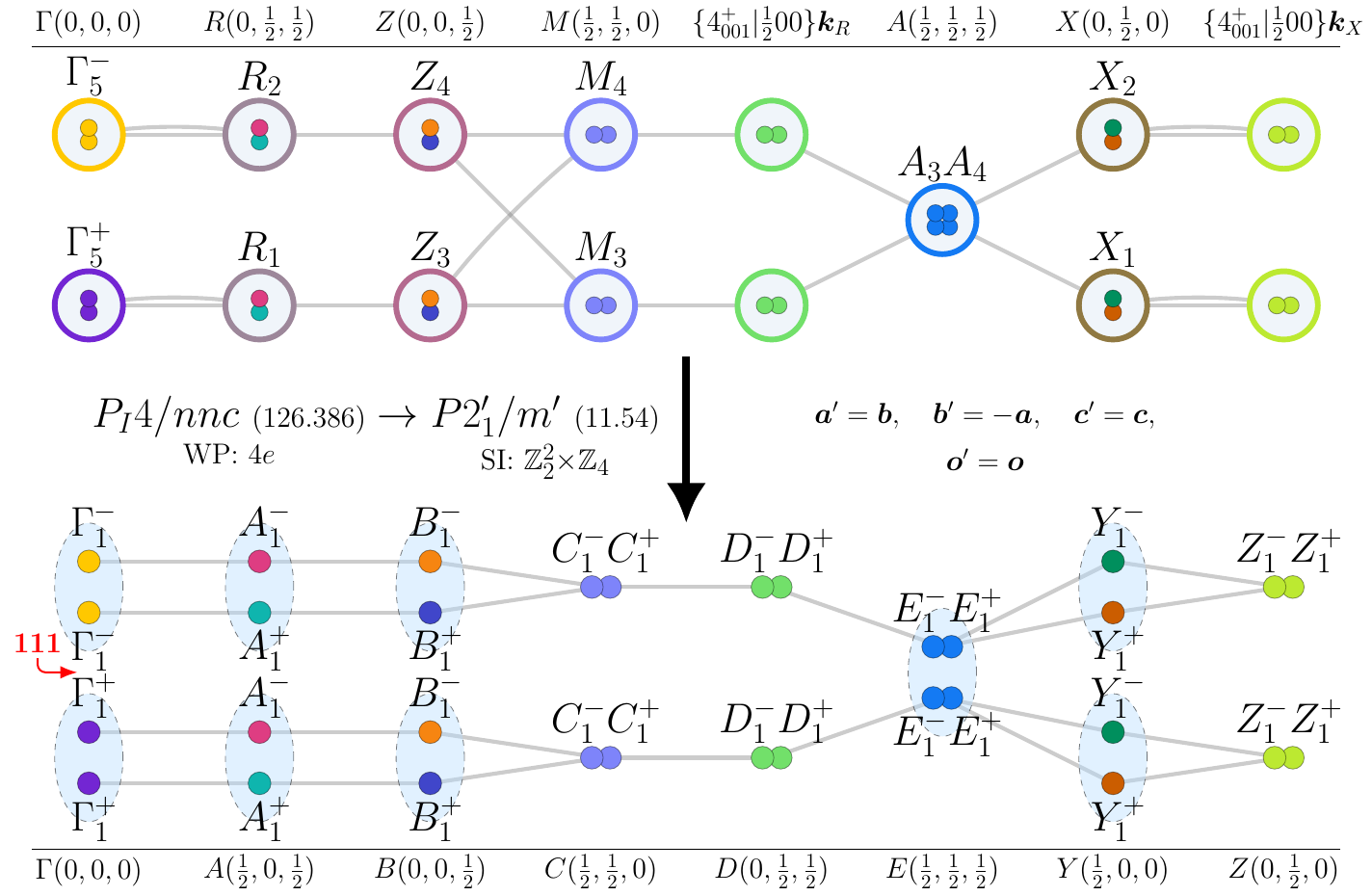}
\caption{Topological magnon bands in subgroup $P2_{1}'/m'~(11.54)$ for magnetic moments on Wyckoff position $4e$ of supergroup $P_{I}4/nnc~(126.386)$.\label{fig_126.386_11.54_Bparallel001andstrainperp100_4e}}
\end{figure}
\begin{center}
\begin{longtable}{cc}
\caption{Number of times a given SI can appear in the band structure. Supergroup: $P_{I}4/nnc~(126.386)$, WP: $4e$, Subgroup: $P2_{1}'/m'~(11.54)$.\label{gap_table_126.386_11.54_4e_Bparallel001andstrainperp100}}\\
\toprule
SI&Possible \#Gaps\\
\midrule
\endfirsthead
\multicolumn{2}{@{}l}{\ldots continued}\\
\toprule
SI&Possible \#Gaps\\
\midrule
\endhead
\midrule\multicolumn{2}{r}{continued on next page\ldots}
\endfoot
\bottomrule
\endlastfoot
000&0\\
111&0, 1\\
113&0, 1\\
gapless&2\\
nontrivial&1\\
\end{longtable}
\end{center}
\begin{center}
\begin{longtable}{cc}
\caption{Possible SI values in each gap. Supergroup: $P_{I}4/nnc~(126.386)$, WP: $4e$, Subgroup: $P2_{1}'/m'~(11.54)$.\label{si_table_126.386_11.54_4e_Bparallel001andstrainperp100}}\\
\toprule
Gap\#& Possible SI Values\\
\midrule
\endfirsthead
\multicolumn{2}{@{}l}{\ldots continued}\\
\toprule
Gap\#& Possible SI Values\\
\midrule
\endhead
\multicolumn{2}{r}{continued on next page\ldots}
\endfoot
\bottomrule
\endlastfoot
\multirow{1}{*}{1}&\multirow{1}{*}{G}\\\midrule
\multirow{1}{*}{2}&\multirow{1}{*}{111, 113}\\\midrule
\multirow{1}{*}{3}&\multirow{1}{*}{G}\\\midrule
\multirow{1}{*}{4}&\multirow{1}{*}{000}\\
\end{longtable}
\end{center}
\subsubsection{Topological bands in subgroup $P_{S}\bar{1}~(2.7)$}
\textbf{Perturbation:}
\begin{itemize}
\item strain in generic direction.
\end{itemize}
\begin{figure}[H]
\centering
\includegraphics[scale=0.6]{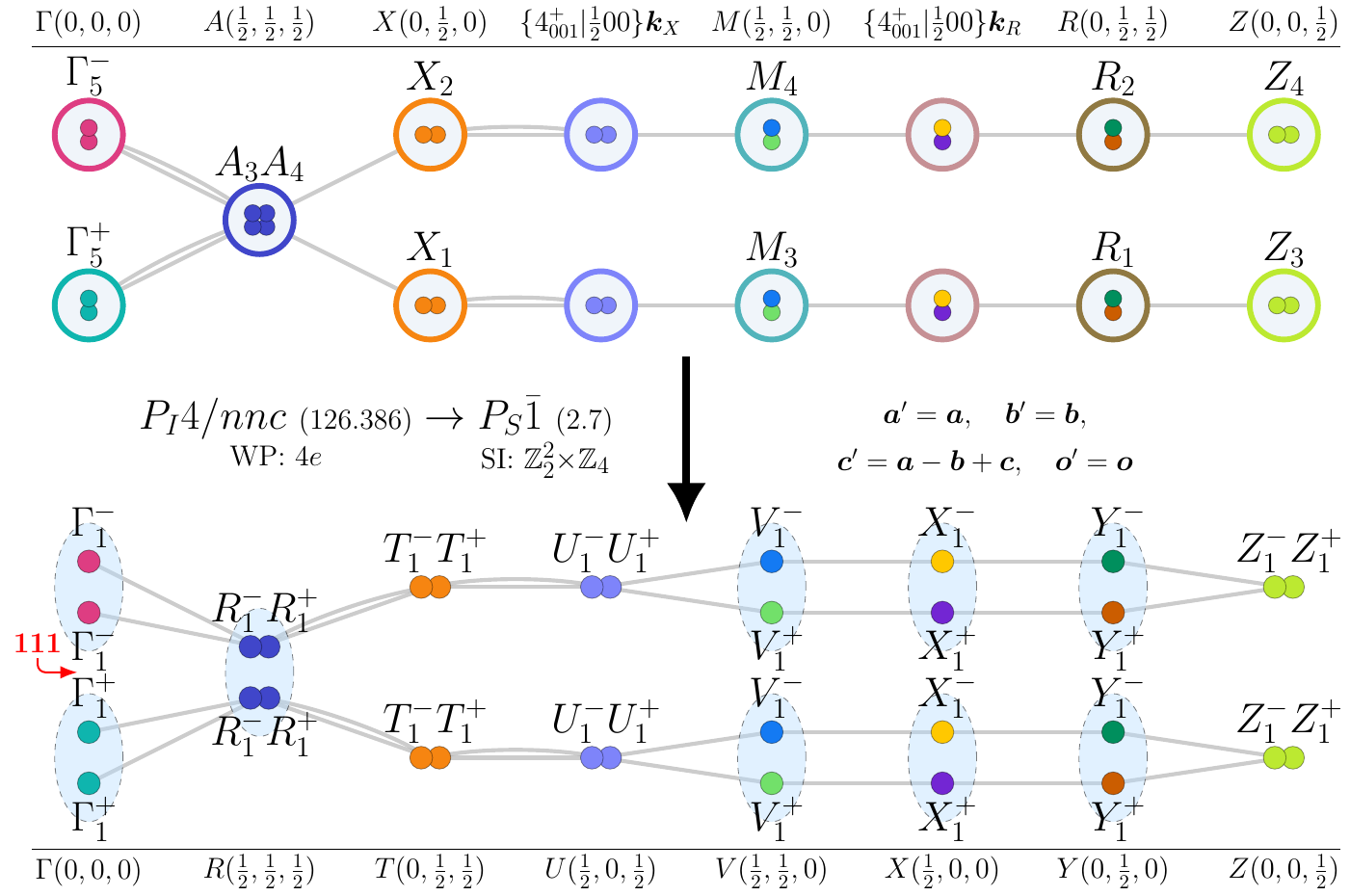}
\caption{Topological magnon bands in subgroup $P_{S}\bar{1}~(2.7)$ for magnetic moments on Wyckoff position $4e$ of supergroup $P_{I}4/nnc~(126.386)$.\label{fig_126.386_2.7_strainingenericdirection_4e}}
\end{figure}
\begin{center}
\begin{longtable}{cc}
\caption{Number of times a given SI can appear in the band structure. Supergroup: $P_{I}4/nnc~(126.386)$, WP: $4e$, Subgroup: $P_{S}\bar{1}~(2.7)$.\label{gap_table_126.386_2.7_4e_strainingenericdirection}}\\
\toprule
SI&Possible \#Gaps\\
\midrule
\endfirsthead
\multicolumn{2}{@{}l}{\ldots continued}\\
\toprule
SI&Possible \#Gaps\\
\midrule
\endhead
\midrule\multicolumn{2}{r}{continued on next page\ldots}
\endfoot
\bottomrule
\endlastfoot
000&0\\
111&0, 1\\
113&0, 1\\
gapless&2\\
nontrivial&1\\
\end{longtable}
\end{center}
\begin{center}
\begin{longtable}{cc}
\caption{Possible SI values in each gap. Supergroup: $P_{I}4/nnc~(126.386)$, WP: $4e$, Subgroup: $P_{S}\bar{1}~(2.7)$.\label{si_table_126.386_2.7_4e_strainingenericdirection}}\\
\toprule
Gap\#& Possible SI Values\\
\midrule
\endfirsthead
\multicolumn{2}{@{}l}{\ldots continued}\\
\toprule
Gap\#& Possible SI Values\\
\midrule
\endhead
\multicolumn{2}{r}{continued on next page\ldots}
\endfoot
\bottomrule
\endlastfoot
\multirow{1}{*}{1}&\multirow{1}{*}{G}\\\midrule
\multirow{1}{*}{2}&\multirow{1}{*}{111, 113}\\\midrule
\multirow{1}{*}{3}&\multirow{1}{*}{G}\\\midrule
\multirow{1}{*}{4}&\multirow{1}{*}{000}\\
\end{longtable}
\end{center}

\section{MSG $C_{c}2/m~(12.63)$}
\textbf{Nontrivial-SI Subgroups:} $P\bar{1}~(2.4)$, $C2'/c'~(15.89)$, $P_{S}\bar{1}~(2.7)$, $C2/m~(12.58)$.\\

\textbf{Trivial-SI Subgroups:} $Cc'~(9.39)$, $C2'~(5.15)$, $P_{S}1~(1.3)$, $Cm~(8.32)$, $C_{c}m~(8.35)$, $C2~(5.13)$, $C_{c}2~(5.16)$.\\

\subsection{WP: $4b+8e$}
\textbf{BCS Materials:} {GeNi\textsubscript{2}O\textsubscript{4}~(12.1 K)}\footnote{BCS web page: \texttt{\href{http://webbdcrista1.ehu.es/magndata/index.php?this\_label=1.560} {http://webbdcrista1.ehu.es/magndata/index.php?this\_label=1.560}}}, {Cu\textsubscript{3}Mg(OD)\textsubscript{6}Br\textsubscript{2}~(5.4 K)}\footnote{BCS web page: \texttt{\href{http://webbdcrista1.ehu.es/magndata/index.php?this\_label=1.397} {http://webbdcrista1.ehu.es/magndata/index.php?this\_label=1.397}}}.\\
\subsubsection{Topological bands in subgroup $P\bar{1}~(2.4)$}
\textbf{Perturbations:}
\begin{itemize}
\item B $\parallel$ [010] and strain in generic direction,
\item B $\perp$ [010] and strain in generic direction,
\item B in generic direction.
\end{itemize}
\begin{figure}[H]
\centering
\includegraphics[scale=0.6]{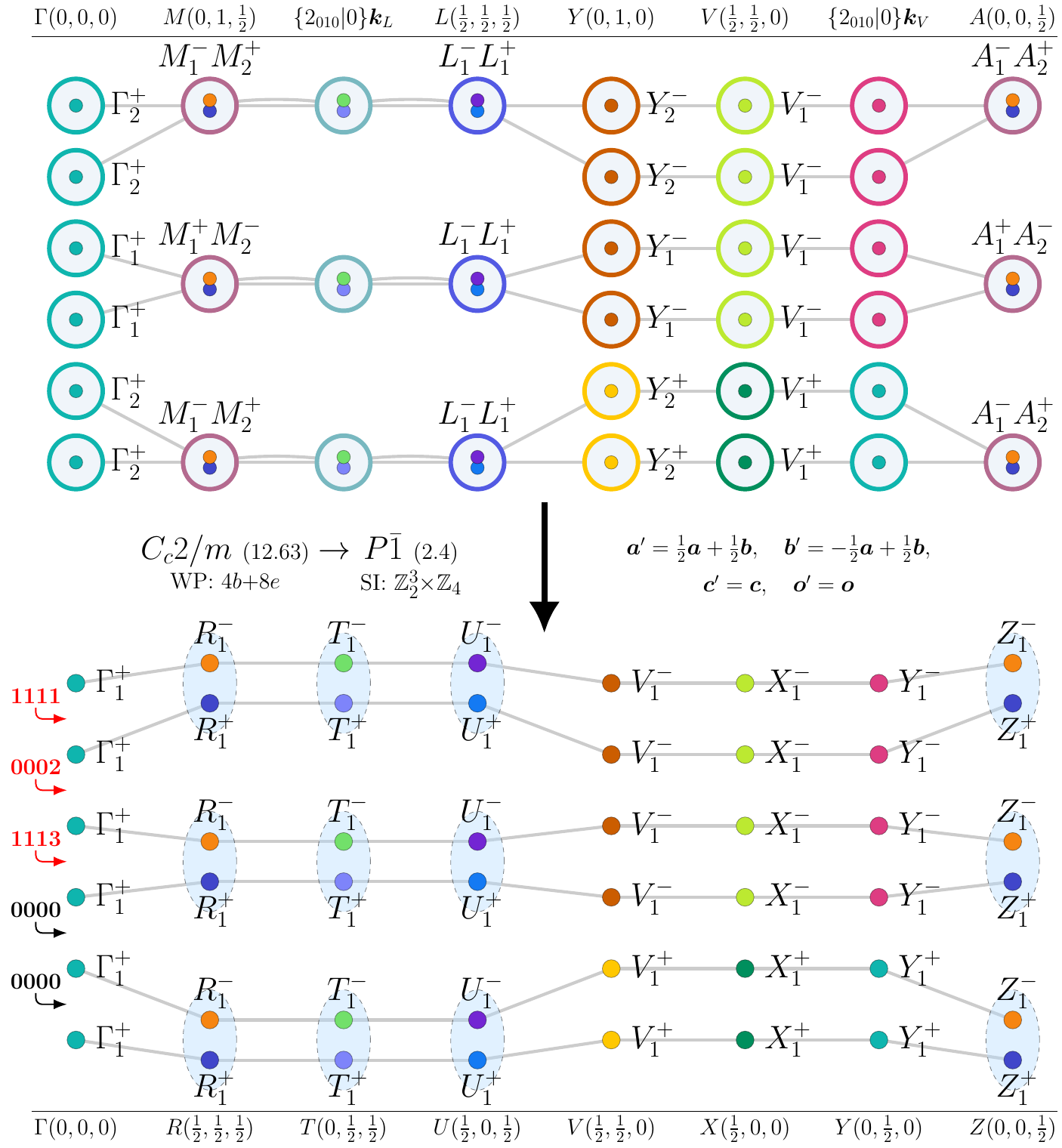}
\caption{Topological magnon bands in subgroup $P\bar{1}~(2.4)$ for magnetic moments on Wyckoff positions $4b+8e$ of supergroup $C_{c}2/m~(12.63)$.\label{fig_12.63_2.4_Bparallel010andstrainingenericdirection_4b+8e}}
\end{figure}
\begin{center}
\begin{longtable}{cc}
\caption{Number of times a given SI can appear in the band structure. Supergroup: $C_{c}2/m~(12.63)$, WP: $4b$+$8e$, Subgroup: $P\bar{1}~(2.4)$.\label{gap_table_12.63_2.4_4b+8e_Bparallel010andstrainingenericdirection}}\\
\toprule
SI&Possible \#Gaps\\
\midrule
\endfirsthead
\multicolumn{2}{@{}l}{\ldots continued}\\
\toprule
SI&Possible \#Gaps\\
\midrule
\endhead
\midrule\multicolumn{2}{r}{continued on next page\ldots}
\endfoot
\bottomrule
\endlastfoot
0000&0, 1, 2\\
0001&0, 1, 2, 3\\
0002&0, 1, 2, 3\\
0003&0, 1, 2, 3\\
0010&0, 1, 2, 3\\
0011&0, 1, 2, 3\\
0012&0, 1, 2, 3\\
0013&0, 1, 2, 3\\
0100&0, 1, 2, 3\\
0101&0, 1, 2, 3\\
0102&0, 1, 2\\
0103&0, 1, 2, 3\\
0110&0, 1, 2, 3\\
0111&0, 1, 2, 3\\
0112&0, 1, 2, 3\\
0113&0, 1, 2, 3\\
1000&0, 1, 2, 3\\
1001&0, 1, 2, 3\\
1002&0, 1, 2\\
1003&0, 1, 2, 3\\
1010&0, 1, 2, 3\\
1011&0, 1, 2, 3\\
1012&0, 1, 2, 3\\
1013&0, 1, 2, 3\\
1100&0, 1, 2, 3\\
1101&0, 1, 2, 3\\
1102&0, 1, 2\\
1103&0, 1, 2, 3\\
1110&0, 1, 2, 3\\
1111&0, 1, 2, 3\\
1112&0, 1, 2, 3\\
1113&0, 1, 2, 3\\
gapless&0\\
nontrivial&3, 4, 5\\
\end{longtable}
\end{center}
\begin{center}
\begin{longtable}{cc}
\caption{Possible SI values in each gap. Supergroup: $C_{c}2/m~(12.63)$, WP: $4b$+$8e$, Subgroup: $P\bar{1}~(2.4)$.\label{si_table_12.63_2.4_4b+8e_Bparallel010andstrainingenericdirection}}\\
\toprule
Gap\#& Possible SI Values\\
\midrule
\endfirsthead
\multicolumn{2}{@{}l}{\ldots continued}\\
\toprule
Gap\#& Possible SI Values\\
\midrule
\endhead
\multicolumn{2}{r}{continued on next page\ldots}
\endfoot
\bottomrule
\endlastfoot
\multirow{6}{*}{1}&\multirow{1}{*}{0000, 0001, 0003, 0010, 0011}\\*
&\multirow{1}{*}{0012, 0100, 0101, 0102, 0103}\\*
&\multirow{1}{*}{0110, 0111, 0112, 0113, 1000}\\*
&\multirow{1}{*}{1001, 1002, 1003, 1010, 1011}\\*
&\multirow{1}{*}{1012, 1013, 1101, 1102, 1103}\\*
&\multirow{1}{*}{1110, 1112, 1113}\\\midrule
\multirow{2}{*}{2}&\multirow{1}{*}{0000, 0002, 0011, 1100, 1102}\\*
&\multirow{1}{*}{1113}\\\midrule
\multirow{6}{*}{3}&\multirow{1}{*}{0001, 0002, 0003, 0010, 0011}\\*
&\multirow{1}{*}{0012, 0013, 0100, 0101, 0102}\\*
&\multirow{1}{*}{0103, 0110, 0111, 0112, 0113}\\*
&\multirow{1}{*}{1000, 1001, 1002, 1003, 1010}\\*
&\multirow{1}{*}{1011, 1012, 1013, 1100, 1101}\\*
&\multirow{1}{*}{1103, 1110, 1111, 1112, 1113}\\\midrule
\multirow{2}{*}{4}&\multirow{1}{*}{0000, 0002, 0013, 1100, 1102}\\*
&\multirow{1}{*}{1111}\\\midrule
\multirow{6}{*}{5}&\multirow{1}{*}{0000, 0001, 0003, 0010, 0012}\\*
&\multirow{1}{*}{0013, 0100, 0101, 0102, 0103}\\*
&\multirow{1}{*}{0110, 0111, 0112, 0113, 1000}\\*
&\multirow{1}{*}{1001, 1002, 1003, 1010, 1011}\\*
&\multirow{1}{*}{1012, 1013, 1101, 1102, 1103}\\*
&\multirow{1}{*}{1110, 1111, 1112}\\\midrule
\multirow{1}{*}{6}&\multirow{1}{*}{0000}\\
\end{longtable}
\end{center}
\subsubsection{Topological bands in subgroup $C2'/c'~(15.89)$}
\textbf{Perturbation:}
\begin{itemize}
\item B $\perp$ [010].
\end{itemize}
\begin{figure}[H]
\centering
\includegraphics[scale=0.6]{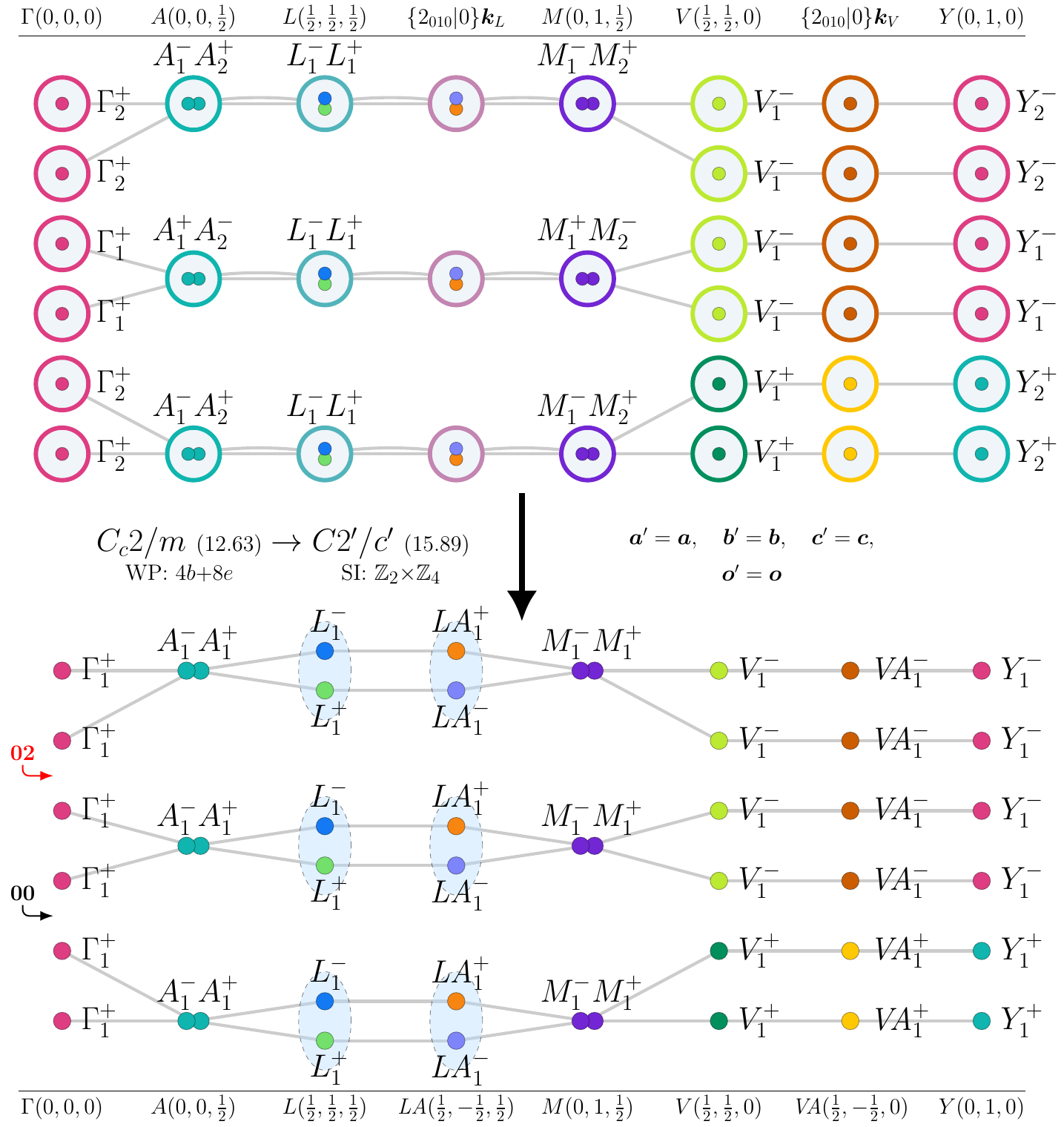}
\caption{Topological magnon bands in subgroup $C2'/c'~(15.89)$ for magnetic moments on Wyckoff positions $4b+8e$ of supergroup $C_{c}2/m~(12.63)$.\label{fig_12.63_15.89_Bperp010_4b+8e}}
\end{figure}
\begin{center}
\begin{longtable}{cc}
\caption{Number of times a given SI can appear in the band structure. Supergroup: $C_{c}2/m~(12.63)$, WP: $4b$+$8e$, Subgroup: $C2'/c'~(15.89)$.\label{gap_table_12.63_15.89_4b+8e_Bperp010}}\\
\toprule
SI&Possible \#Gaps\\
\midrule
\endfirsthead
\multicolumn{2}{@{}l}{\ldots continued}\\
\toprule
SI&Possible \#Gaps\\
\midrule
\endhead
\midrule\multicolumn{2}{r}{continued on next page\ldots}
\endfoot
\bottomrule
\endlastfoot
00&0, 1\\
01&0, 1\\
02&0, 1, 2\\
03&0, 1\\
10&0, 1, 2\\
11&0, 1\\
12&0, 1\\
13&0, 1\\
gapless&3\\
nontrivial&1, 2\\
\end{longtable}
\end{center}
\begin{center}
\begin{longtable}{cc}
\caption{Possible SI values in each gap. Supergroup: $C_{c}2/m~(12.63)$, WP: $4b$+$8e$, Subgroup: $C2'/c'~(15.89)$.\label{si_table_12.63_15.89_4b+8e_Bperp010}}\\
\toprule
Gap\#& Possible SI Values\\
\midrule
\endfirsthead
\multicolumn{2}{@{}l}{\ldots continued}\\
\toprule
Gap\#& Possible SI Values\\
\midrule
\endhead
\multicolumn{2}{r}{continued on next page\ldots}
\endfoot
\bottomrule
\endlastfoot
\multirow{1}{*}{1}&\multirow{1}{*}{G}\\\midrule
\multirow{2}{*}{2}&\multirow{1}{*}{00, 01, 02, 10, 12}\\*
&\multirow{1}{*}{13}\\\midrule
\multirow{1}{*}{3}&\multirow{1}{*}{G}\\\midrule
\multirow{2}{*}{4}&\multirow{1}{*}{00, 02, 03, 10, 11}\\*
&\multirow{1}{*}{12}\\\midrule
\multirow{1}{*}{5}&\multirow{1}{*}{G}\\\midrule
\multirow{1}{*}{6}&\multirow{1}{*}{00}\\
\end{longtable}
\end{center}
\subsubsection{Topological bands in subgroup $P_{S}\bar{1}~(2.7)$}
\textbf{Perturbation:}
\begin{itemize}
\item strain in generic direction.
\end{itemize}
\begin{figure}[H]
\centering
\includegraphics[scale=0.6]{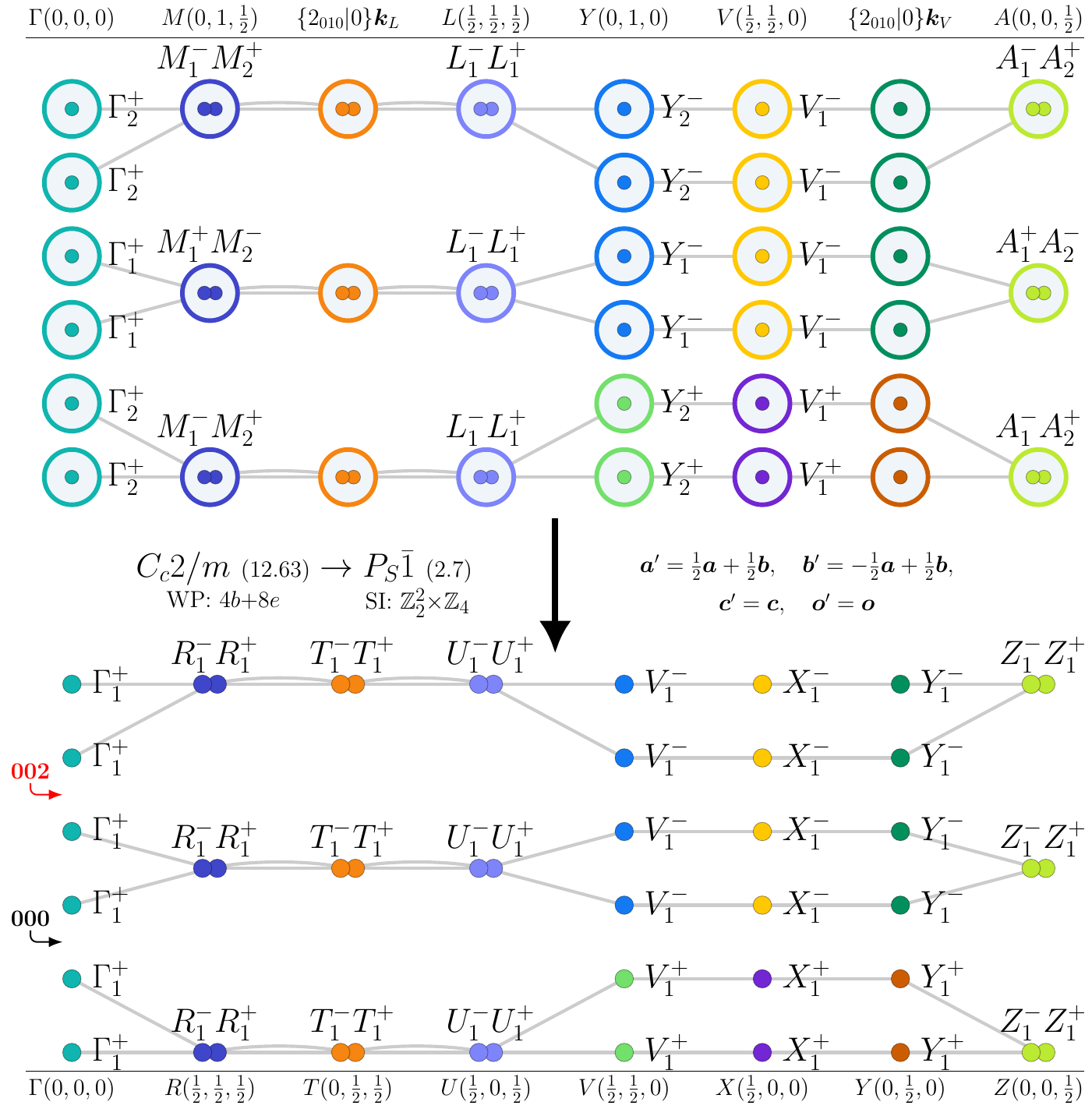}
\caption{Topological magnon bands in subgroup $P_{S}\bar{1}~(2.7)$ for magnetic moments on Wyckoff positions $4b+8e$ of supergroup $C_{c}2/m~(12.63)$.\label{fig_12.63_2.7_strainingenericdirection_4b+8e}}
\end{figure}
\begin{center}
\begin{longtable}{cc}
\caption{Number of times a given SI can appear in the band structure. Supergroup: $C_{c}2/m~(12.63)$, WP: $4b$+$8e$, Subgroup: $P_{S}\bar{1}~(2.7)$.\label{gap_table_12.63_2.7_4b+8e_strainingenericdirection}}\\
\toprule
SI&Possible \#Gaps\\
\midrule
\endfirsthead
\multicolumn{2}{@{}l}{\ldots continued}\\
\toprule
SI&Possible \#Gaps\\
\midrule
\endhead
\midrule\multicolumn{2}{r}{continued on next page\ldots}
\endfoot
\bottomrule
\endlastfoot
000&0, 1\\
001&0, 1\\
002&0, 1, 2\\
003&0, 1\\
110&0, 1, 2\\
111&0, 1\\
112&0, 1\\
113&0, 1\\
gapless&3\\
nontrivial&1, 2\\
\end{longtable}
\end{center}
\begin{center}
\begin{longtable}{cc}
\caption{Possible SI values in each gap. Supergroup: $C_{c}2/m~(12.63)$, WP: $4b$+$8e$, Subgroup: $P_{S}\bar{1}~(2.7)$.\label{si_table_12.63_2.7_4b+8e_strainingenericdirection}}\\
\toprule
Gap\#& Possible SI Values\\
\midrule
\endfirsthead
\multicolumn{2}{@{}l}{\ldots continued}\\
\toprule
Gap\#& Possible SI Values\\
\midrule
\endhead
\multicolumn{2}{r}{continued on next page\ldots}
\endfoot
\bottomrule
\endlastfoot
\multirow{1}{*}{1}&\multirow{1}{*}{G}\\\midrule
\multirow{2}{*}{2}&\multirow{1}{*}{000, 002, 003, 110, 111}\\*
&\multirow{1}{*}{112}\\\midrule
\multirow{1}{*}{3}&\multirow{1}{*}{G}\\\midrule
\multirow{2}{*}{4}&\multirow{1}{*}{000, 001, 002, 110, 112}\\*
&\multirow{1}{*}{113}\\\midrule
\multirow{1}{*}{5}&\multirow{1}{*}{G}\\\midrule
\multirow{1}{*}{6}&\multirow{1}{*}{000}\\
\end{longtable}
\end{center}
\subsection{WP: $8e$}
\textbf{BCS Materials:} {Fe\textsubscript{4}Si\textsubscript{2}Sn\textsubscript{7}O\textsubscript{16}~(3.5 K)}\footnote{BCS web page: \texttt{\href{http://webbdcrista1.ehu.es/magndata/index.php?this\_label=1.197} {http://webbdcrista1.ehu.es/magndata/index.php?this\_label=1.197}}}.\\
\subsubsection{Topological bands in subgroup $P\bar{1}~(2.4)$}
\textbf{Perturbations:}
\begin{itemize}
\item B $\parallel$ [010] and strain in generic direction,
\item B $\perp$ [010] and strain in generic direction,
\item B in generic direction.
\end{itemize}
\begin{figure}[H]
\centering
\includegraphics[scale=0.6]{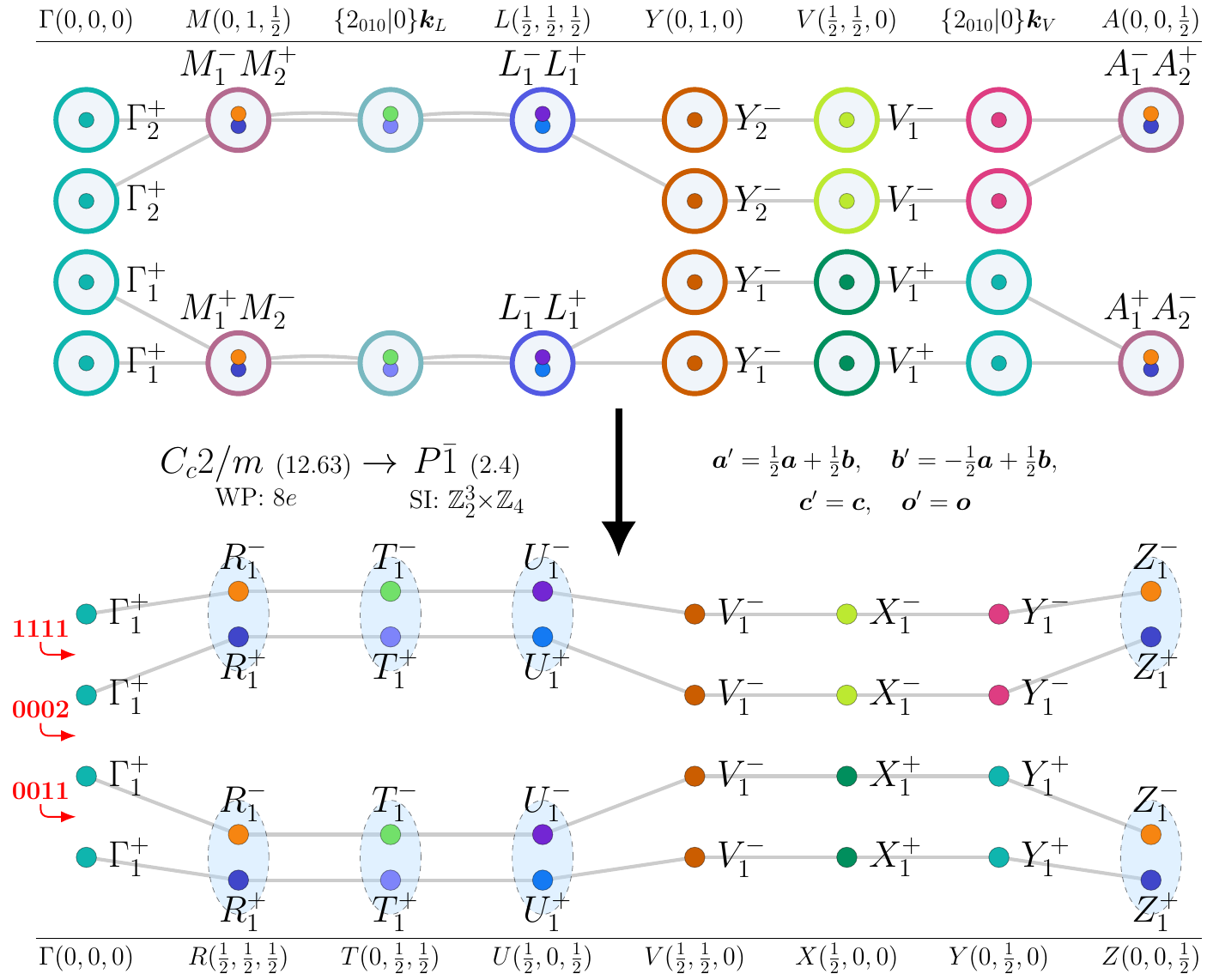}
\caption{Topological magnon bands in subgroup $P\bar{1}~(2.4)$ for magnetic moments on Wyckoff position $8e$ of supergroup $C_{c}2/m~(12.63)$.\label{fig_12.63_2.4_Bparallel010andstrainingenericdirection_8e}}
\end{figure}
\input{gap_tables_tex/12.63_2.4_Bparallel010andstrainingenericdirection_8e_table.tex}
\input{si_tables_tex/12.63_2.4_Bparallel010andstrainingenericdirection_8e_table.tex}
\subsubsection{Topological bands in subgroup $C2'/c'~(15.89)$}
\textbf{Perturbation:}
\begin{itemize}
\item B $\perp$ [010].
\end{itemize}
\begin{figure}[H]
\centering
\includegraphics[scale=0.6]{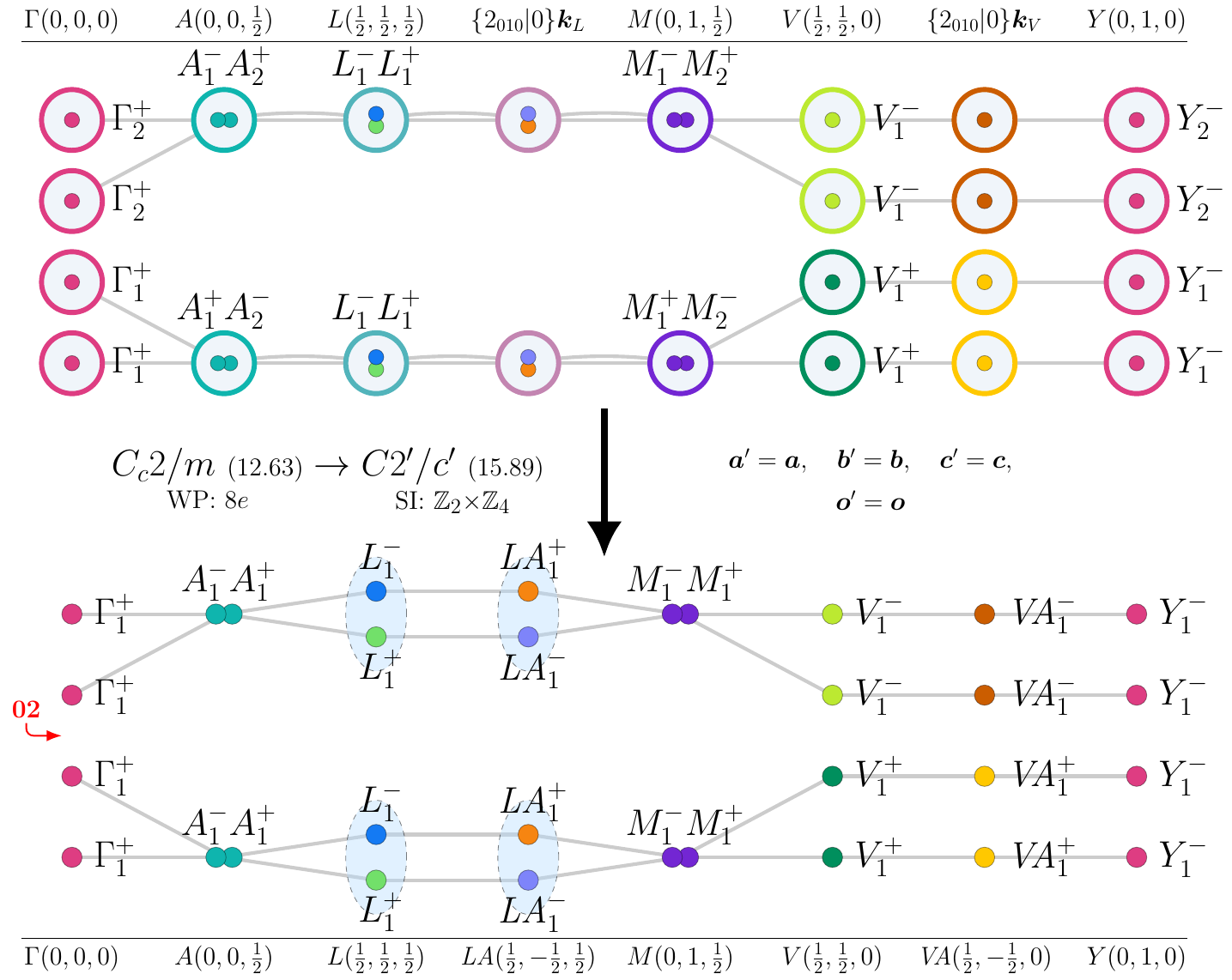}
\caption{Topological magnon bands in subgroup $C2'/c'~(15.89)$ for magnetic moments on Wyckoff position $8e$ of supergroup $C_{c}2/m~(12.63)$.\label{fig_12.63_15.89_Bperp010_8e}}
\end{figure}
\input{gap_tables_tex/12.63_15.89_Bperp010_8e_table.tex}
\input{si_tables_tex/12.63_15.89_Bperp010_8e_table.tex}
\subsubsection{Topological bands in subgroup $P_{S}\bar{1}~(2.7)$}
\textbf{Perturbation:}
\begin{itemize}
\item strain in generic direction.
\end{itemize}
\begin{figure}[H]
\centering
\includegraphics[scale=0.6]{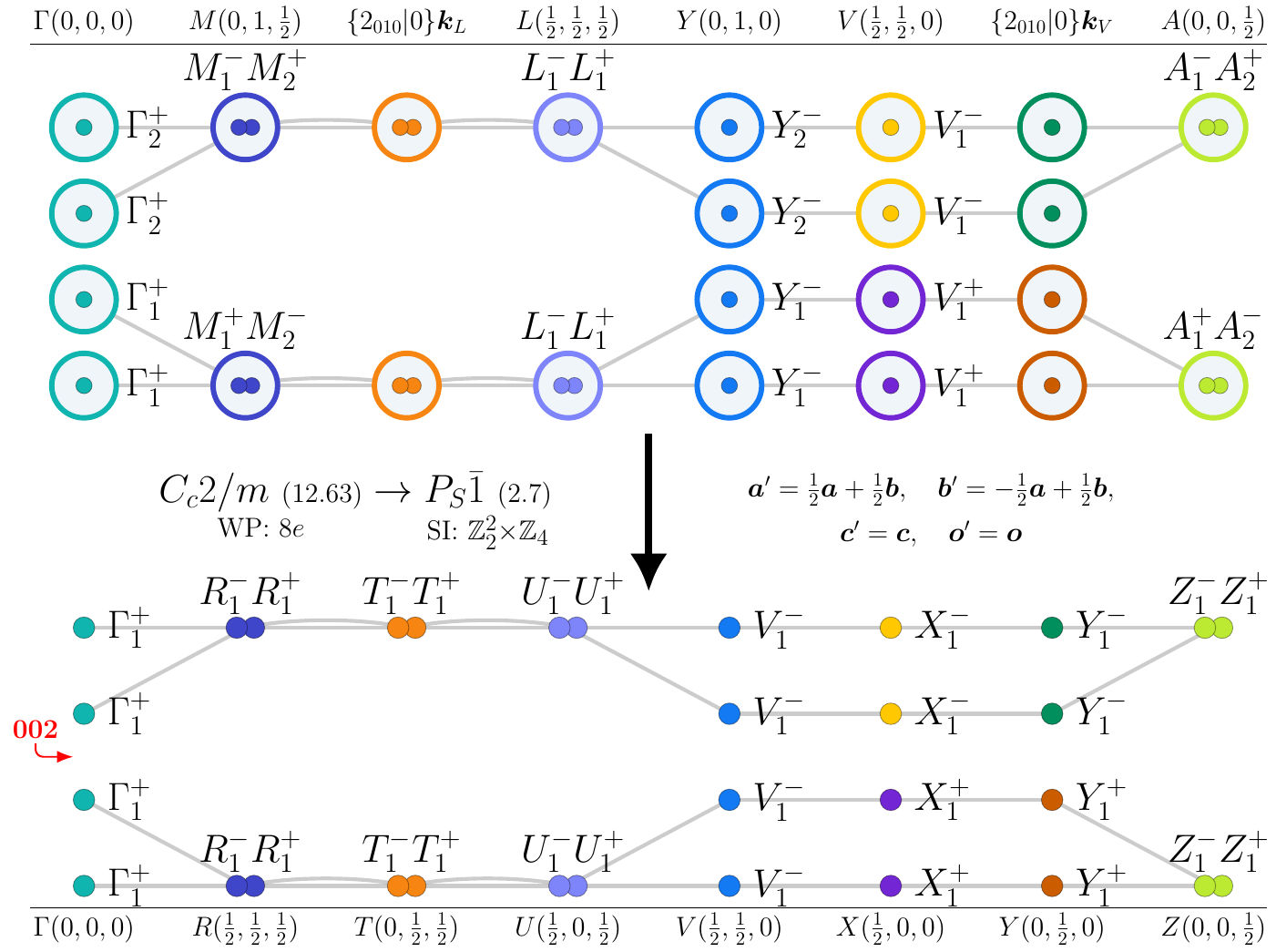}
\caption{Topological magnon bands in subgroup $P_{S}\bar{1}~(2.7)$ for magnetic moments on Wyckoff position $8e$ of supergroup $C_{c}2/m~(12.63)$.\label{fig_12.63_2.7_strainingenericdirection_8e}}
\end{figure}
\input{gap_tables_tex/12.63_2.7_strainingenericdirection_8e_table.tex}
\input{si_tables_tex/12.63_2.7_strainingenericdirection_8e_table.tex}

\section{MSG $P_{c}4/mbm~(127.396)$}
\textbf{Nontrivial-SI Subgroups:} $P\bar{1}~(2.4)$, $C2'/c'~(15.89)$, $P2_{1}'/m'~(11.54)$, $P2_{1}'/c'~(14.79)$, $P_{S}\bar{1}~(2.7)$, $C2/m~(12.58)$, $Cmc'm'~(63.463)$, $C_{c}2/m~(12.63)$, $P2~(3.1)$, $Cc'c'2~(37.183)$, $P_{b}2~(3.5)$, $P_{c}ba2~(32.139)$, $P2/m~(10.42)$, $Cc'c'm~(66.495)$, $Pn'n'm~(58.397)$, $P_{b}2/m~(10.48)$, $P2_{1}/c~(14.75)$, $Pn'm'a~(62.446)$, $P_{a}2_{1}/c~(14.80)$, $P4n'c'~(104.207)$, $P4/mn'c'~(128.405)$.\\

\textbf{Trivial-SI Subgroups:} $Cc'~(9.39)$, $Pm'~(6.20)$, $Pc'~(7.26)$, $C2'~(5.15)$, $P2_{1}'~(4.9)$, $P2_{1}'~(4.9)$, $P_{S}1~(1.3)$, $Cm~(8.32)$, $Cmc'2_{1}'~(36.175)$, $C_{c}m~(8.35)$, $Pm~(6.18)$, $Ama'2'~(40.206)$, $Pmn'2_{1}'~(31.126)$, $P_{b}m~(6.22)$, $Pc~(7.24)$, $Pn'a2_{1}'~(33.146)$, $P_{a}c~(7.27)$, $C2~(5.13)$, $Am'a'2~(40.207)$, $C_{c}2~(5.16)$, $A_{a}mm2~(38.192)$, $Pn'n'2~(34.159)$, $C_{c}mm2~(35.169)$, $C_{c}mmm~(65.488)$, $P2_{1}~(4.7)$, $Pm'n'2_{1}~(31.127)$, $P_{a}2_{1}~(4.10)$, $P_{a}mc2_{1}~(26.71)$, $P_{c}bam~(55.361)$, $P_{c}4bm~(100.176)$.\\

\subsection{WP: $8h$}
\textbf{BCS Materials:} {Yb\textsubscript{2}Pd\textsubscript{2}(In\textsubscript{0.4}Sn\textsubscript{0.6})~(3 K)}\footnote{BCS web page: \texttt{\href{http://webbdcrista1.ehu.es/magndata/index.php?this\_label=1.333} {http://webbdcrista1.ehu.es/magndata/index.php?this\_label=1.333}}}.\\
\subsubsection{Topological bands in subgroup $P2_{1}'/m'~(11.54)$}
\textbf{Perturbations:}
\begin{itemize}
\item B $\parallel$ [100] and strain $\parallel$ [110],
\item B $\parallel$ [100] and strain $\perp$ [001],
\item B $\parallel$ [110] and strain $\parallel$ [100],
\item B $\parallel$ [110] and strain $\perp$ [001],
\item B $\perp$ [001].
\end{itemize}
\begin{figure}[H]
\centering
\includegraphics[scale=0.6]{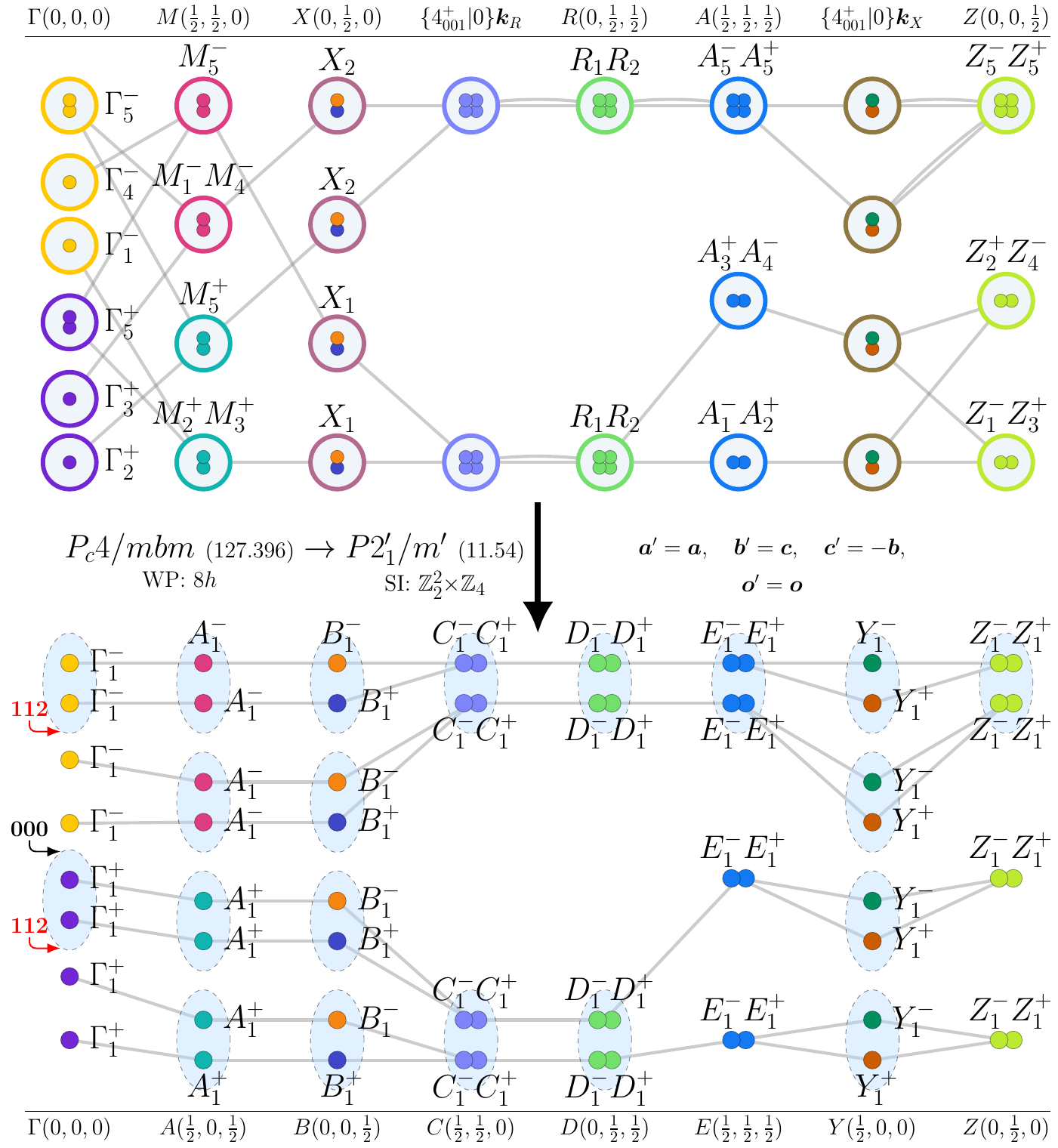}
\caption{Topological magnon bands in subgroup $P2_{1}'/m'~(11.54)$ for magnetic moments on Wyckoff position $8h$ of supergroup $P_{c}4/mbm~(127.396)$.\label{fig_127.396_11.54_Bparallel100andstrainparallel110_8h}}
\end{figure}
\input{gap_tables_tex/127.396_11.54_Bparallel100andstrainparallel110_8h_table.tex}
\input{si_tables_tex/127.396_11.54_Bparallel100andstrainparallel110_8h_table.tex}
\subsubsection{Topological bands in subgroup $P_{S}\bar{1}~(2.7)$}
\textbf{Perturbation:}
\begin{itemize}
\item strain in generic direction.
\end{itemize}
\begin{figure}[H]
\centering
\includegraphics[scale=0.6]{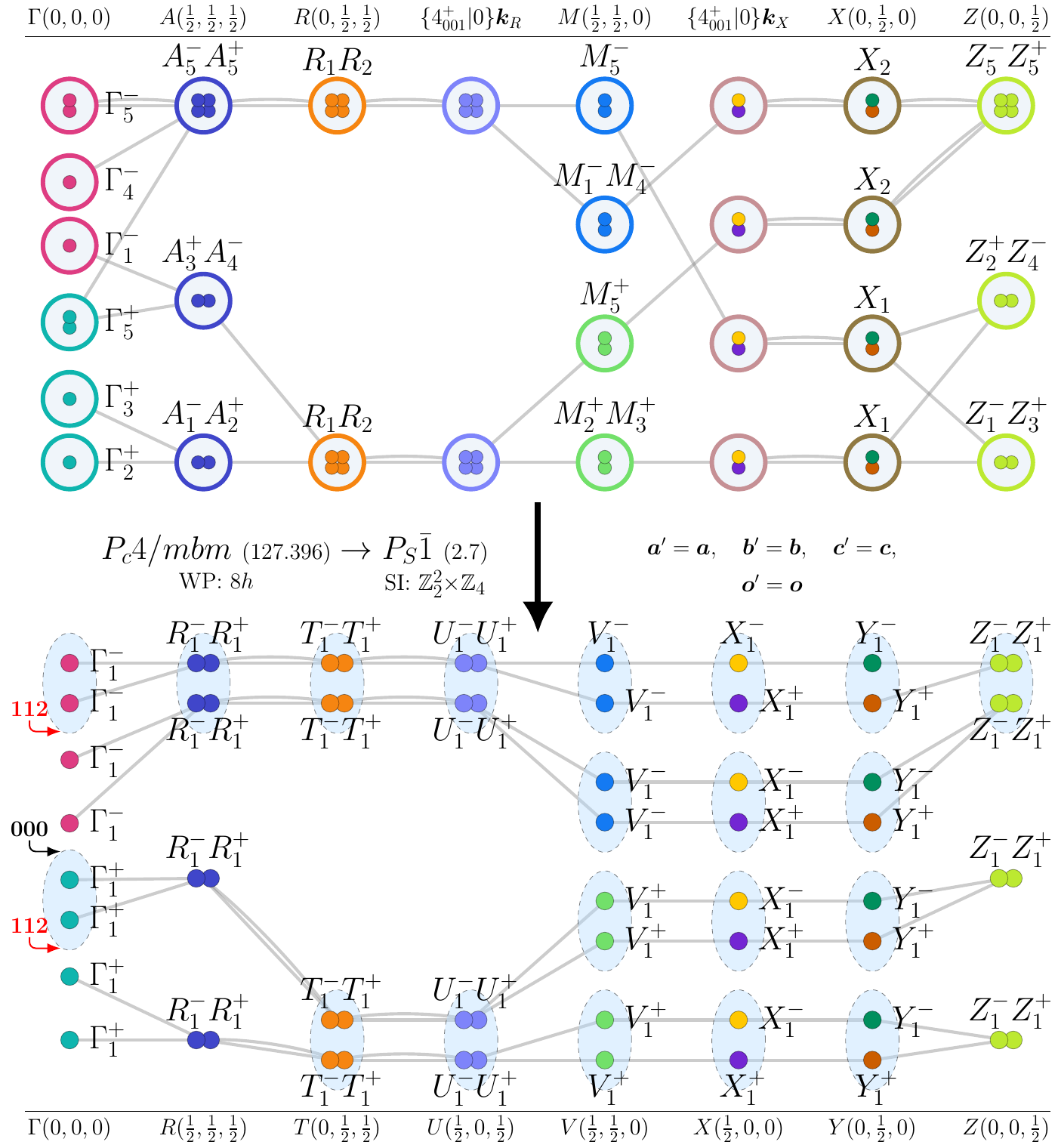}
\caption{Topological magnon bands in subgroup $P_{S}\bar{1}~(2.7)$ for magnetic moments on Wyckoff position $8h$ of supergroup $P_{c}4/mbm~(127.396)$.\label{fig_127.396_2.7_strainingenericdirection_8h}}
\end{figure}
\input{gap_tables_tex/127.396_2.7_strainingenericdirection_8h_table.tex}
\input{si_tables_tex/127.396_2.7_strainingenericdirection_8h_table.tex}

\section{MSG $P_{c}4/mnc~(128.408)$}
\textbf{Nontrivial-SI Subgroups:} $P\bar{1}~(2.4)$, $C2'/m'~(12.62)$, $P2_{1}'/m'~(11.54)$, $P2_{1}'/c'~(14.79)$, $P_{S}\bar{1}~(2.7)$, $C2/c~(15.85)$, $Cm'cm'~(63.464)$, $C_{c}2/c~(15.90)$, $P2~(3.1)$, $Cm'm'2~(35.168)$, $Pb'a'2~(32.138)$, $P_{b}2~(3.5)$, $C_{c}cc2~(37.184)$, $P_{c}nn2~(34.161)$, $P2/m~(10.42)$, $Cm'm'm~(65.485)$, $Pb'a'm~(55.357)$, $P_{b}2/m~(10.48)$, $C_{c}ccm~(66.498)$, $P2_{1}/c~(14.75)$, $Pnm'a'~(62.447)$, $P_{a}2_{1}/c~(14.80)$, $P_{c}nnm~(58.401)$, $P4b'm'~(100.175)$, $P_{c}4nc~(104.208)$, $P4/mb'm'~(127.393)$.\\

\textbf{Trivial-SI Subgroups:} $Cm'~(8.34)$, $Pm'~(6.20)$, $Pc'~(7.26)$, $C2'~(5.15)$, $P2_{1}'~(4.9)$, $P2_{1}'~(4.9)$, $P_{S}1~(1.3)$, $Cc~(9.37)$, $Cm'c2_{1}'~(36.174)$, $C_{c}c~(9.40)$, $Pm~(6.18)$, $Amm'2'~(38.190)$, $Pmc'2_{1}'~(26.69)$, $P_{b}m~(6.22)$, $Pc~(7.24)$, $Pna'2_{1}'~(33.147)$, $P_{a}c~(7.27)$, $C2~(5.13)$, $Am'm'2~(38.191)$, $C_{c}2~(5.16)$, $A_{a}ma2~(40.208)$, $P2_{1}~(4.7)$, $Pm'c'2_{1}~(26.70)$, $P_{a}2_{1}~(4.10)$, $P_{a}mn2_{1}~(31.128)$.\\

\subsection{WP: $8g$}
\textbf{BCS Materials:} {U\textsubscript{2}Pd\textsubscript{2.35}Sn\textsubscript{0.65}~(17 K)}\footnote{BCS web page: \texttt{\href{http://webbdcrista1.ehu.es/magndata/index.php?this\_label=1.337} {http://webbdcrista1.ehu.es/magndata/index.php?this\_label=1.337}}}.\\
\subsubsection{Topological bands in subgroup $P2_{1}'/m'~(11.54)$}
\textbf{Perturbations:}
\begin{itemize}
\item B $\parallel$ [100] and strain $\parallel$ [110],
\item B $\parallel$ [100] and strain $\perp$ [001],
\item B $\parallel$ [110] and strain $\parallel$ [100],
\item B $\parallel$ [110] and strain $\perp$ [001],
\item B $\perp$ [001].
\end{itemize}
\begin{figure}[H]
\centering
\includegraphics[scale=0.6]{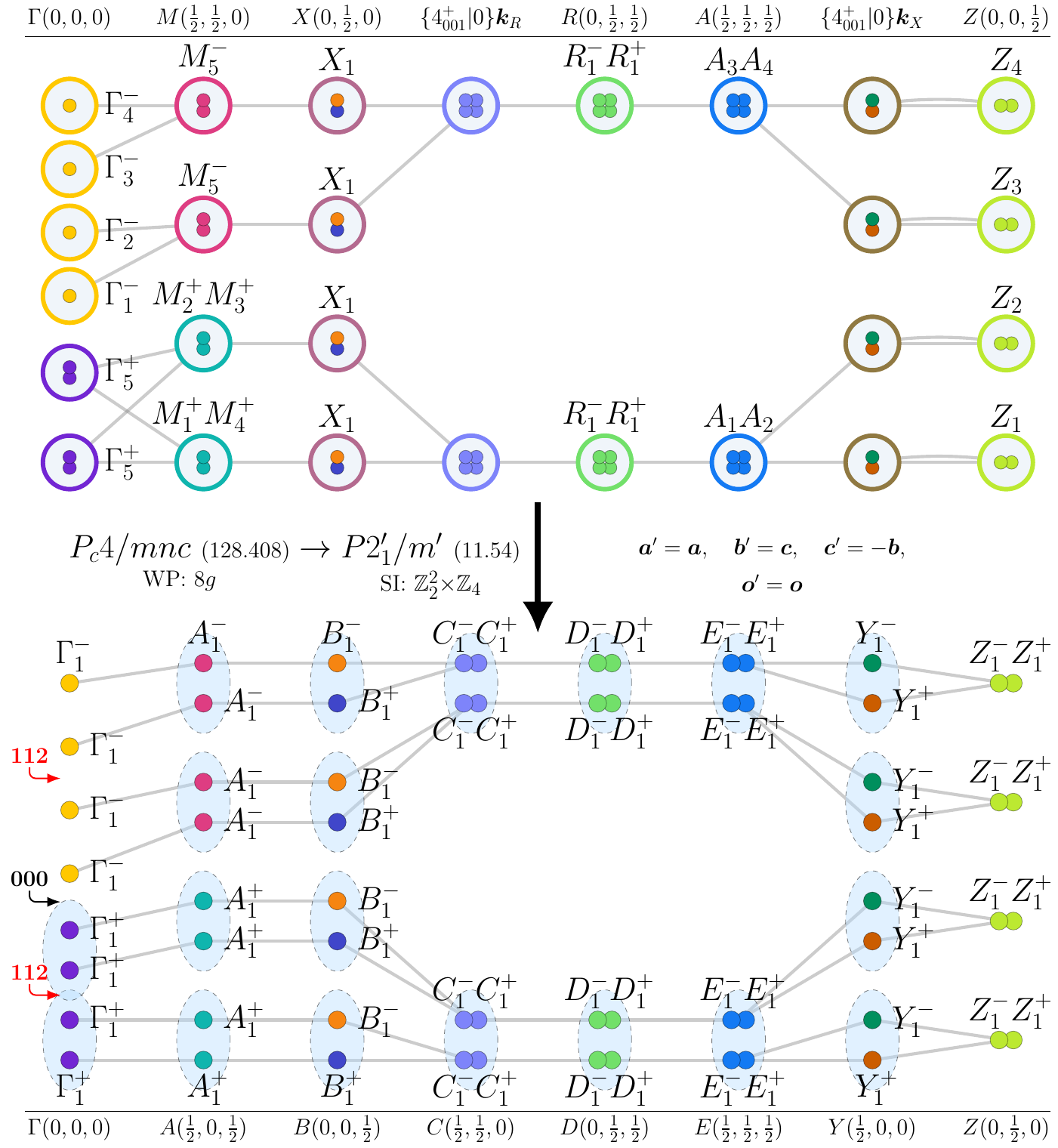}
\caption{Topological magnon bands in subgroup $P2_{1}'/m'~(11.54)$ for magnetic moments on Wyckoff position $8g$ of supergroup $P_{c}4/mnc~(128.408)$.\label{fig_128.408_11.54_Bparallel100andstrainparallel110_8g}}
\end{figure}
\input{gap_tables_tex/128.408_11.54_Bparallel100andstrainparallel110_8g_table.tex}
\input{si_tables_tex/128.408_11.54_Bparallel100andstrainparallel110_8g_table.tex}
\subsubsection{Topological bands in subgroup $P2_{1}'/c'~(14.79)$}
\textbf{Perturbations:}
\begin{itemize}
\item B $\parallel$ [001] and strain $\perp$ [100],
\item B $\perp$ [100].
\end{itemize}
\begin{figure}[H]
\centering
\includegraphics[scale=0.6]{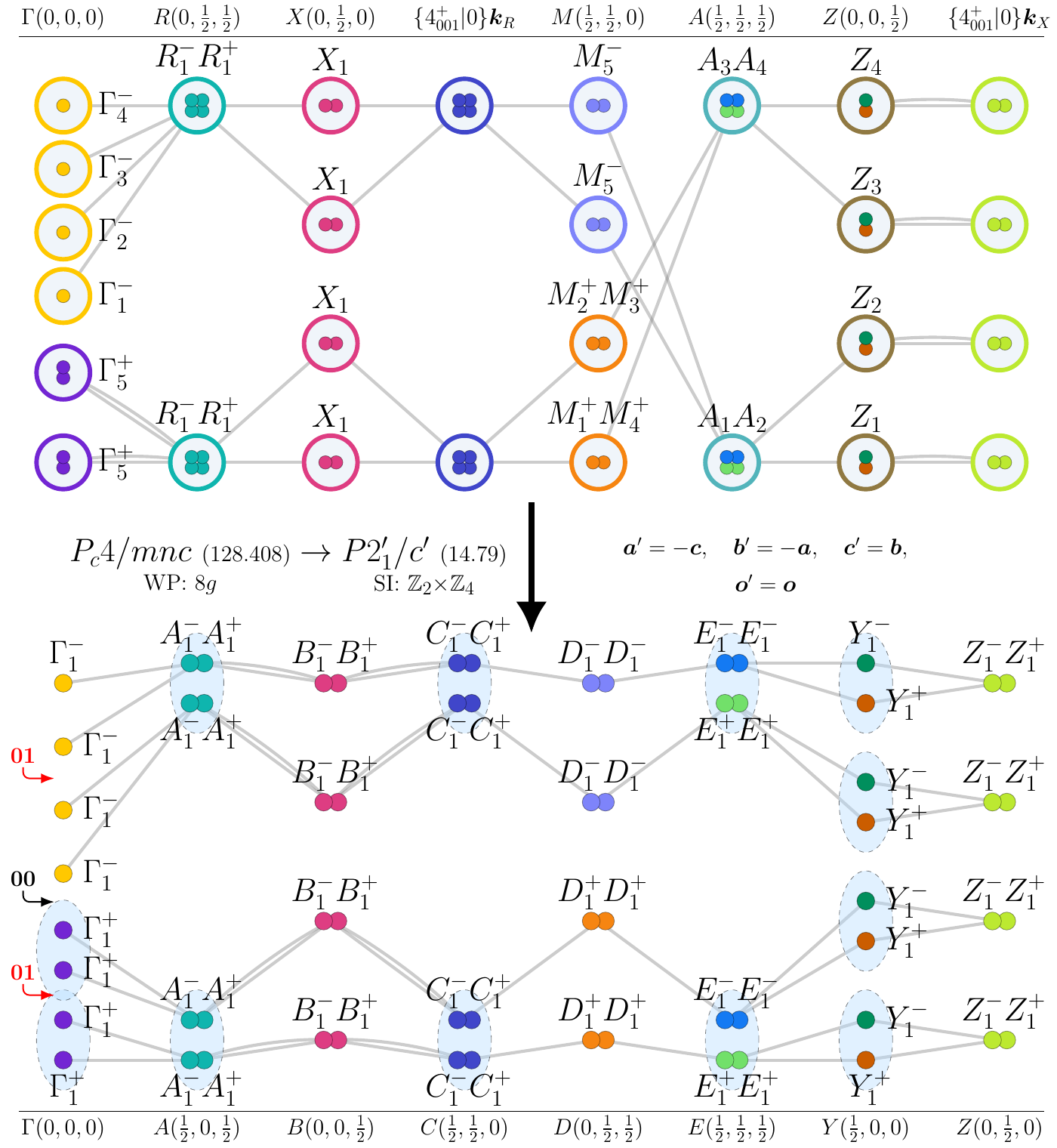}
\caption{Topological magnon bands in subgroup $P2_{1}'/c'~(14.79)$ for magnetic moments on Wyckoff position $8g$ of supergroup $P_{c}4/mnc~(128.408)$.\label{fig_128.408_14.79_Bparallel001andstrainperp100_8g}}
\end{figure}
\input{gap_tables_tex/128.408_14.79_Bparallel001andstrainperp100_8g_table.tex}
\input{si_tables_tex/128.408_14.79_Bparallel001andstrainperp100_8g_table.tex}
\subsubsection{Topological bands in subgroup $P_{S}\bar{1}~(2.7)$}
\textbf{Perturbation:}
\begin{itemize}
\item strain in generic direction.
\end{itemize}
\begin{figure}[H]
\centering
\includegraphics[scale=0.6]{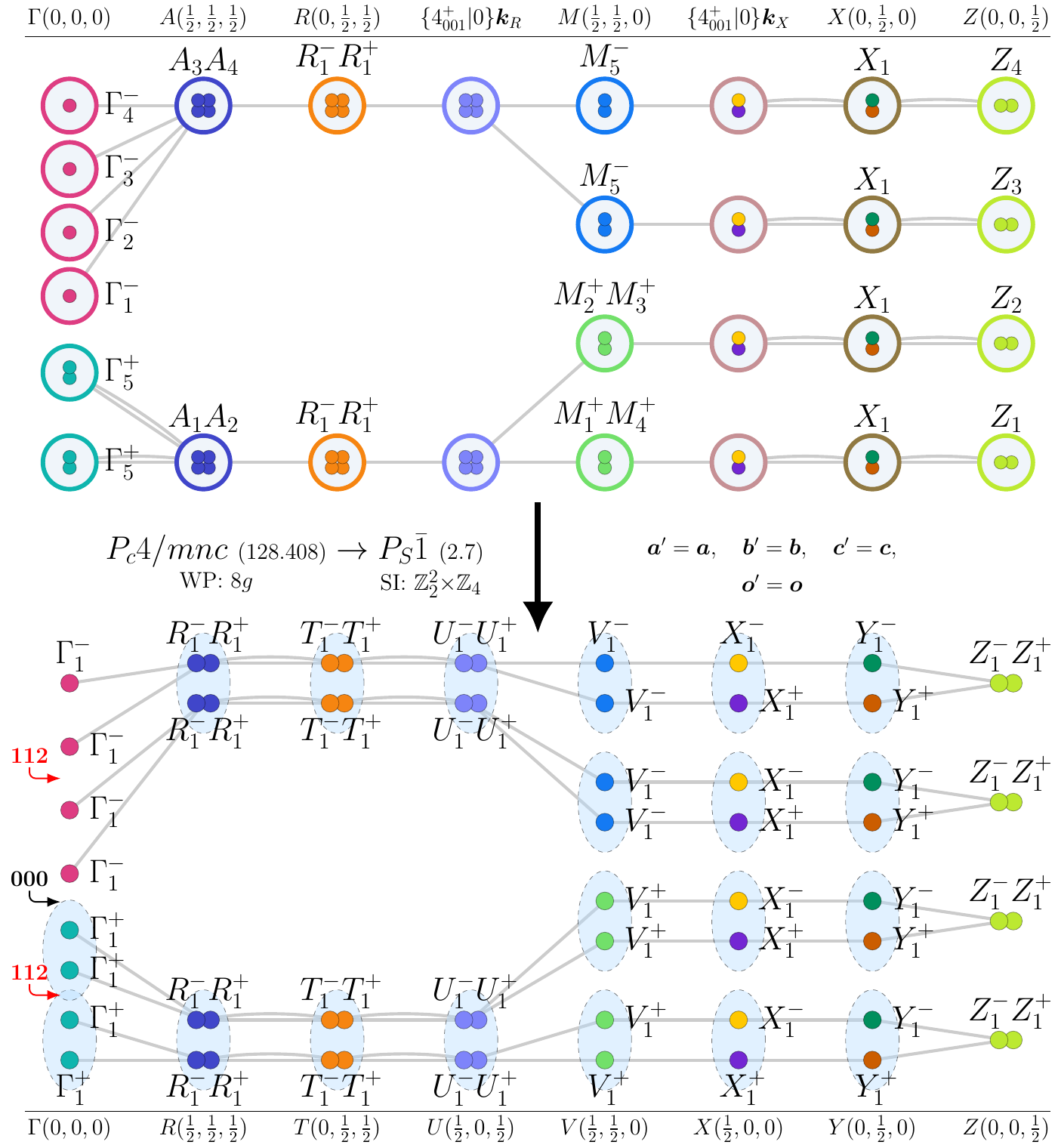}
\caption{Topological magnon bands in subgroup $P_{S}\bar{1}~(2.7)$ for magnetic moments on Wyckoff position $8g$ of supergroup $P_{c}4/mnc~(128.408)$.\label{fig_128.408_2.7_strainingenericdirection_8g}}
\end{figure}
\input{gap_tables_tex/128.408_2.7_strainingenericdirection_8g_table.tex}
\input{si_tables_tex/128.408_2.7_strainingenericdirection_8g_table.tex}
\subsection{WP: $8h$}
\textbf{BCS Materials:} {U\textsubscript{2}Ni\textsubscript{2}In~(15 K)}\footnote{BCS web page: \texttt{\href{http://webbdcrista1.ehu.es/magndata/index.php?this\_label=1.338} {http://webbdcrista1.ehu.es/magndata/index.php?this\_label=1.338}}}, {U\textsubscript{2}Ni\textsubscript{2}In~(14.0 K)}\footnote{BCS web page: \texttt{\href{http://webbdcrista1.ehu.es/magndata/index.php?this\_label=1.549} {http://webbdcrista1.ehu.es/magndata/index.php?this\_label=1.549}}}.\\
\subsubsection{Topological bands in subgroup $P2_{1}'/m'~(11.54)$}
\textbf{Perturbations:}
\begin{itemize}
\item B $\parallel$ [100] and strain $\parallel$ [110],
\item B $\parallel$ [100] and strain $\perp$ [001],
\item B $\parallel$ [110] and strain $\parallel$ [100],
\item B $\parallel$ [110] and strain $\perp$ [001],
\item B $\perp$ [001].
\end{itemize}
\begin{figure}[H]
\centering
\includegraphics[scale=0.6]{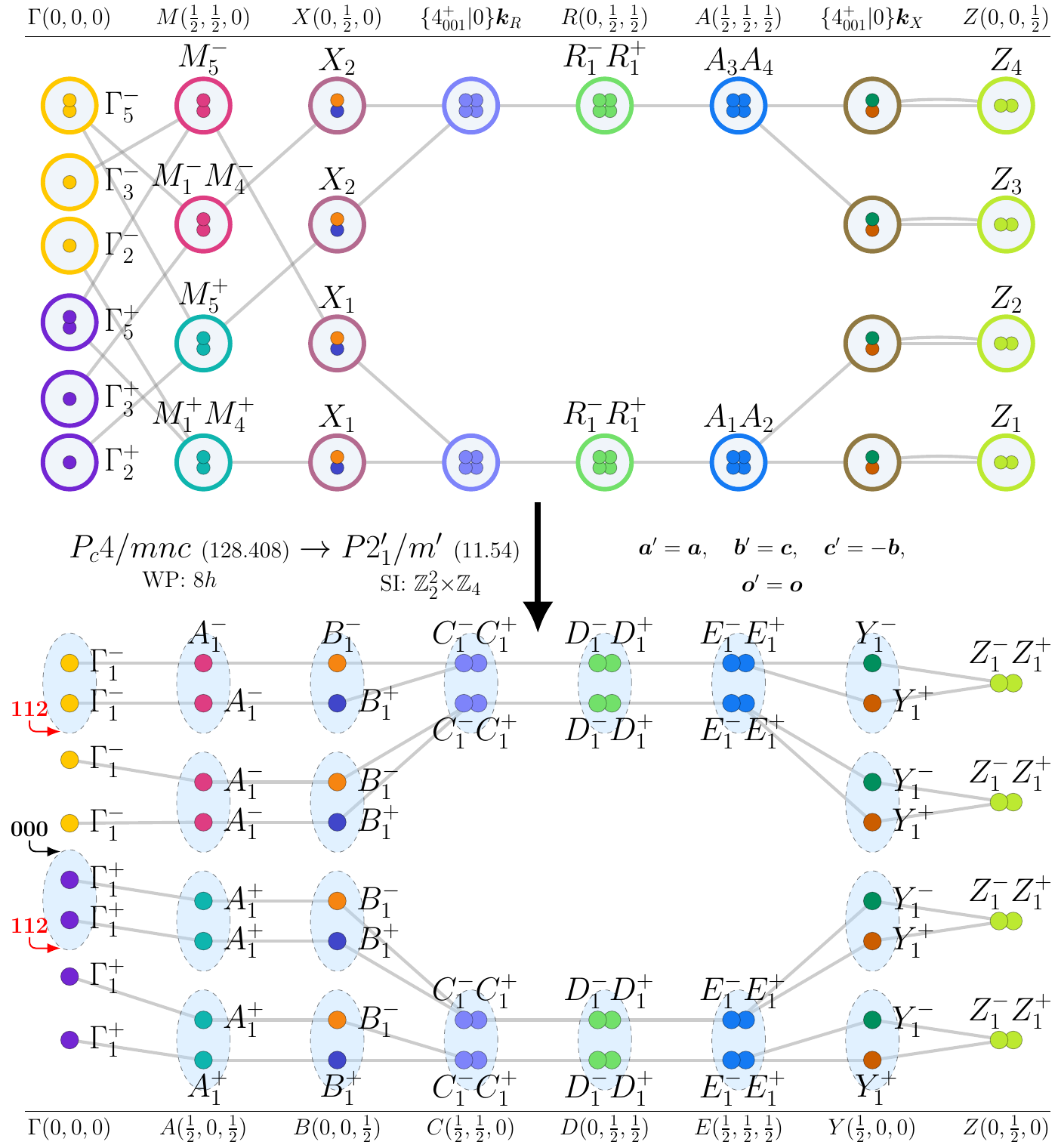}
\caption{Topological magnon bands in subgroup $P2_{1}'/m'~(11.54)$ for magnetic moments on Wyckoff position $8h$ of supergroup $P_{c}4/mnc~(128.408)$.\label{fig_128.408_11.54_Bparallel100andstrainparallel110_8h}}
\end{figure}
\input{gap_tables_tex/128.408_11.54_Bparallel100andstrainparallel110_8h_table.tex}
\input{si_tables_tex/128.408_11.54_Bparallel100andstrainparallel110_8h_table.tex}
\subsubsection{Topological bands in subgroup $P2_{1}'/c'~(14.79)$}
\textbf{Perturbations:}
\begin{itemize}
\item B $\parallel$ [001] and strain $\perp$ [100],
\item B $\perp$ [100].
\end{itemize}
\begin{figure}[H]
\centering
\includegraphics[scale=0.6]{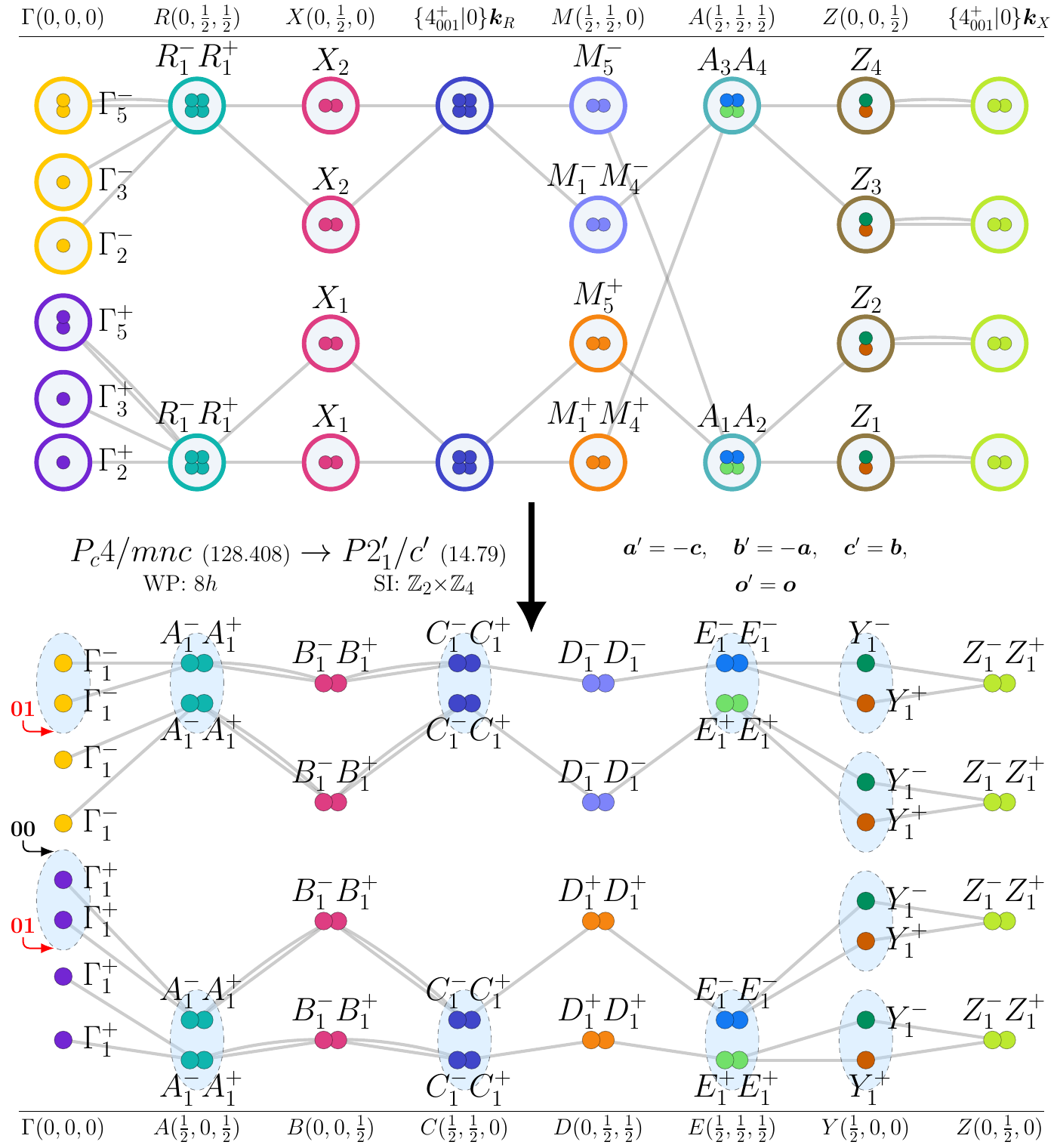}
\caption{Topological magnon bands in subgroup $P2_{1}'/c'~(14.79)$ for magnetic moments on Wyckoff position $8h$ of supergroup $P_{c}4/mnc~(128.408)$.\label{fig_128.408_14.79_Bparallel001andstrainperp100_8h}}
\end{figure}
\input{gap_tables_tex/128.408_14.79_Bparallel001andstrainperp100_8h_table.tex}
\input{si_tables_tex/128.408_14.79_Bparallel001andstrainperp100_8h_table.tex}
\subsubsection{Topological bands in subgroup $P_{S}\bar{1}~(2.7)$}
\textbf{Perturbation:}
\begin{itemize}
\item strain in generic direction.
\end{itemize}
\begin{figure}[H]
\centering
\includegraphics[scale=0.6]{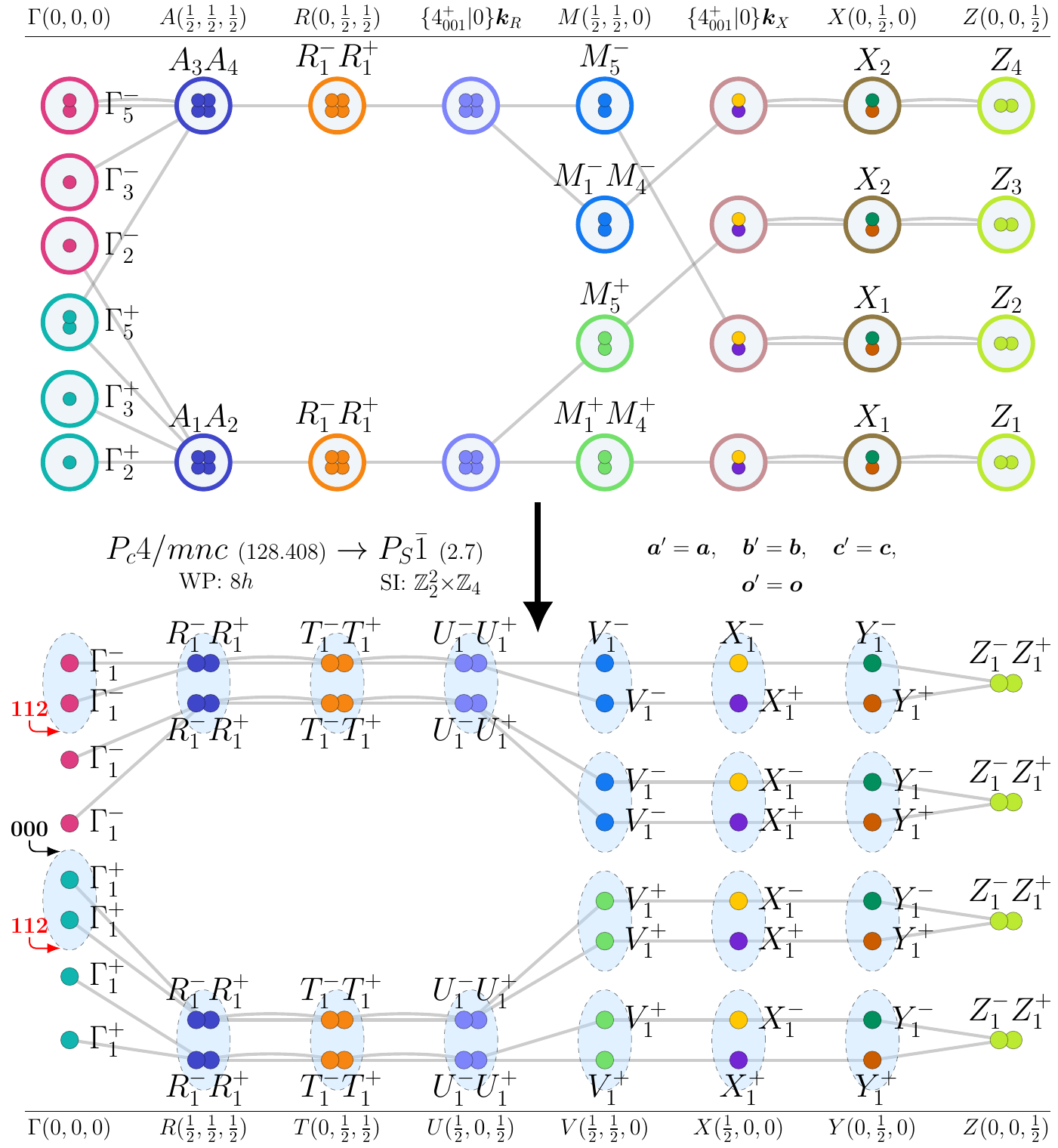}
\caption{Topological magnon bands in subgroup $P_{S}\bar{1}~(2.7)$ for magnetic moments on Wyckoff position $8h$ of supergroup $P_{c}4/mnc~(128.408)$.\label{fig_128.408_2.7_strainingenericdirection_8h}}
\end{figure}
\input{gap_tables_tex/128.408_2.7_strainingenericdirection_8h_table.tex}
\input{si_tables_tex/128.408_2.7_strainingenericdirection_8h_table.tex}

\section{MSG $P_{c}4/ncc~(130.432)$}
\textbf{Nontrivial-SI Subgroups:} $P\bar{1}~(2.4)$, $C2'/m'~(12.62)$, $P2_{1}'/c'~(14.79)$, $P2_{1}'/m'~(11.54)$, $P_{S}\bar{1}~(2.7)$, $Ab'm'2~(39.199)$, $C2/c~(15.85)$, $Cm'ca'~(64.476)$, $C_{c}2/c~(15.90)$, $P2~(3.1)$, $Cm'm'2~(35.168)$, $Pm'm'2~(25.60)$, $P_{b}2~(3.5)$, $C_{c}cc2~(37.184)$, $P_{c}cc2~(27.82)$, $P2/c~(13.65)$, $Cm'm'a~(67.505)$, $Pm'm'n~(59.409)$, $P_{b}2/c~(13.71)$, $C_{c}cca~(68.518)$, $P2_{1}/c~(14.75)$, $Pn'm'a~(62.446)$, $P_{c}2_{1}/c~(14.82)$, $P_{c}ccn~(56.373)$, $P4m'm'~(99.167)$, $P_{c}4cc~(103.200)$, $P4/nm'm'~(129.417)$.\\

\textbf{Trivial-SI Subgroups:} $Cm'~(8.34)$, $Pc'~(7.26)$, $Pm'~(6.20)$, $C2'~(5.15)$, $P2_{1}'~(4.9)$, $P2_{1}'~(4.9)$, $P_{S}1~(1.3)$, $Cc~(9.37)$, $Cm'c2_{1}'~(36.174)$, $C_{c}c~(9.40)$, $Pc~(7.24)$, $Abm'2'~(39.198)$, $Pm'n2_{1}'~(31.125)$, $P_{b}c~(7.29)$, $Pc~(7.24)$, $Pm'c2_{1}'~(26.68)$, $P_{c}c~(7.28)$, $C2~(5.13)$, $C_{c}2~(5.16)$, $A_{a}ba2~(41.216)$, $P2_{1}~(4.7)$, $Pm'n'2_{1}~(31.127)$, $P_{a}2_{1}~(4.10)$, $P_{a}na2_{1}~(33.149)$.\\

\subsection{WP: $4c$}
\textbf{BCS Materials:} {USb\textsubscript{2}~(206 K)}\footnote{BCS web page: \texttt{\href{http://webbdcrista1.ehu.es/magndata/index.php?this\_label=1.384} {http://webbdcrista1.ehu.es/magndata/index.php?this\_label=1.384}}}, {UP\textsubscript{2}~(203 K)}\footnote{BCS web page: \texttt{\href{http://webbdcrista1.ehu.es/magndata/index.php?this\_label=1.215} {http://webbdcrista1.ehu.es/magndata/index.php?this\_label=1.215}}}, {UGeS~(90 K)}\footnote{BCS web page: \texttt{\href{http://webbdcrista1.ehu.es/magndata/index.php?this\_label=1.426} {http://webbdcrista1.ehu.es/magndata/index.php?this\_label=1.426}}}, {CeSbTe~(2.3 K)}\footnote{BCS web page: \texttt{\href{http://webbdcrista1.ehu.es/magndata/index.php?this\_label=1.271} {http://webbdcrista1.ehu.es/magndata/index.php?this\_label=1.271}}}.\\
\subsubsection{Topological bands in subgroup $P\bar{1}~(2.4)$}
\textbf{Perturbations:}
\begin{itemize}
\item B $\parallel$ [001] and strain in generic direction,
\item B $\parallel$ [100] and strain $\perp$ [110],
\item B $\parallel$ [100] and strain in generic direction,
\item B $\parallel$ [110] and strain $\perp$ [100],
\item B $\parallel$ [110] and strain in generic direction,
\item B $\perp$ [001] and strain $\perp$ [100],
\item B $\perp$ [001] and strain $\perp$ [110],
\item B $\perp$ [001] and strain in generic direction,
\item B $\perp$ [100] and strain $\parallel$ [110],
\item B $\perp$ [100] and strain $\perp$ [001],
\item B $\perp$ [100] and strain $\perp$ [110],
\item B $\perp$ [100] and strain in generic direction,
\item B $\perp$ [110] and strain $\parallel$ [100],
\item B $\perp$ [110] and strain $\perp$ [001],
\item B $\perp$ [110] and strain $\perp$ [100],
\item B $\perp$ [110] and strain in generic direction,
\item B in generic direction.
\end{itemize}
\begin{figure}[H]
\centering
\includegraphics[scale=0.6]{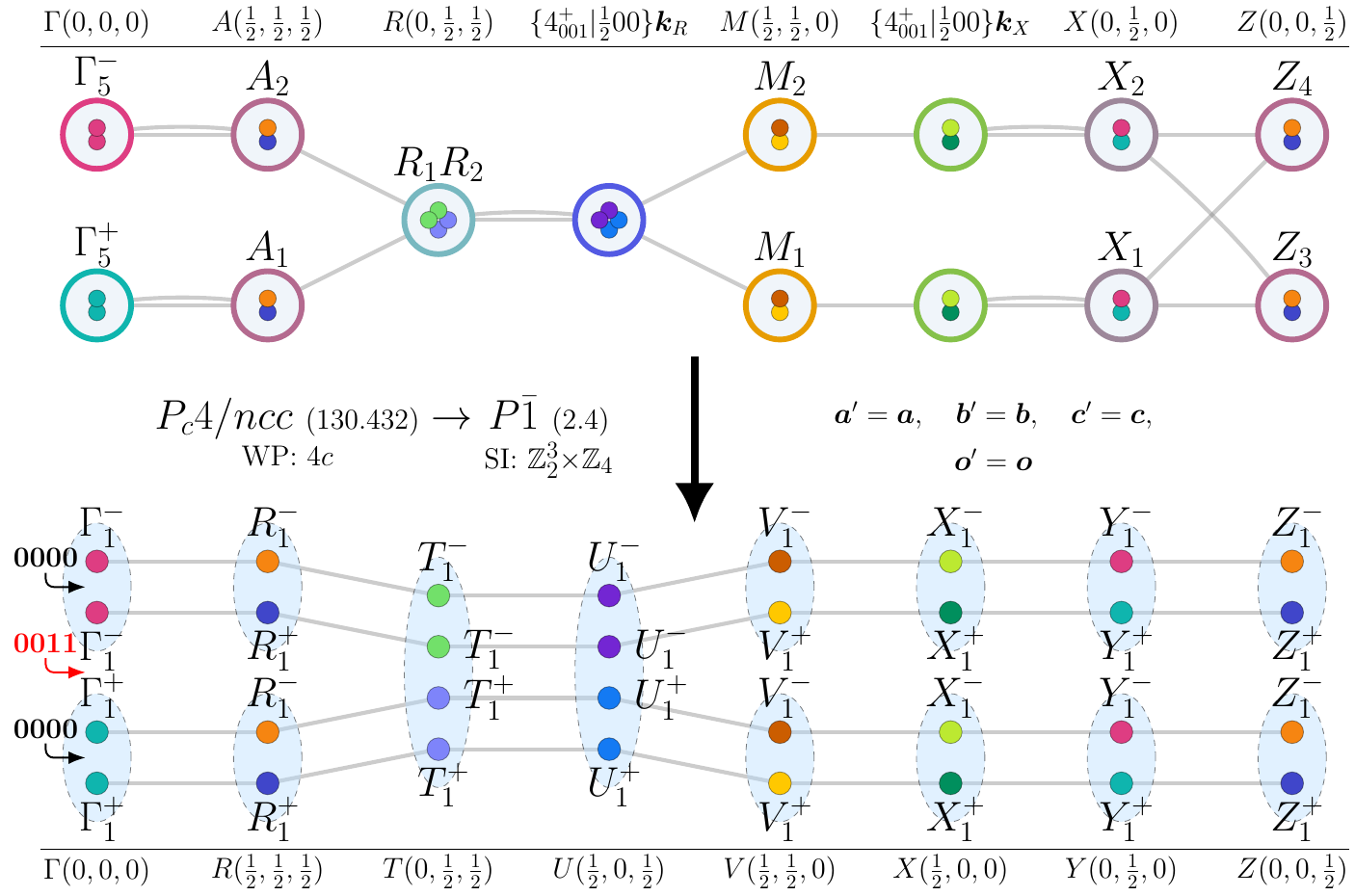}
\caption{Topological magnon bands in subgroup $P\bar{1}~(2.4)$ for magnetic moments on Wyckoff position $4c$ of supergroup $P_{c}4/ncc~(130.432)$.\label{fig_130.432_2.4_Bparallel001andstrainingenericdirection_4c}}
\end{figure}
\input{gap_tables_tex/130.432_2.4_Bparallel001andstrainingenericdirection_4c_table.tex}
\input{si_tables_tex/130.432_2.4_Bparallel001andstrainingenericdirection_4c_table.tex}
\subsubsection{Topological bands in subgroup $C2'/m'~(12.62)$}
\textbf{Perturbations:}
\begin{itemize}
\item B $\parallel$ [001] and strain $\perp$ [110],
\item B $\perp$ [110].
\end{itemize}
\begin{figure}[H]
\centering
\includegraphics[scale=0.6]{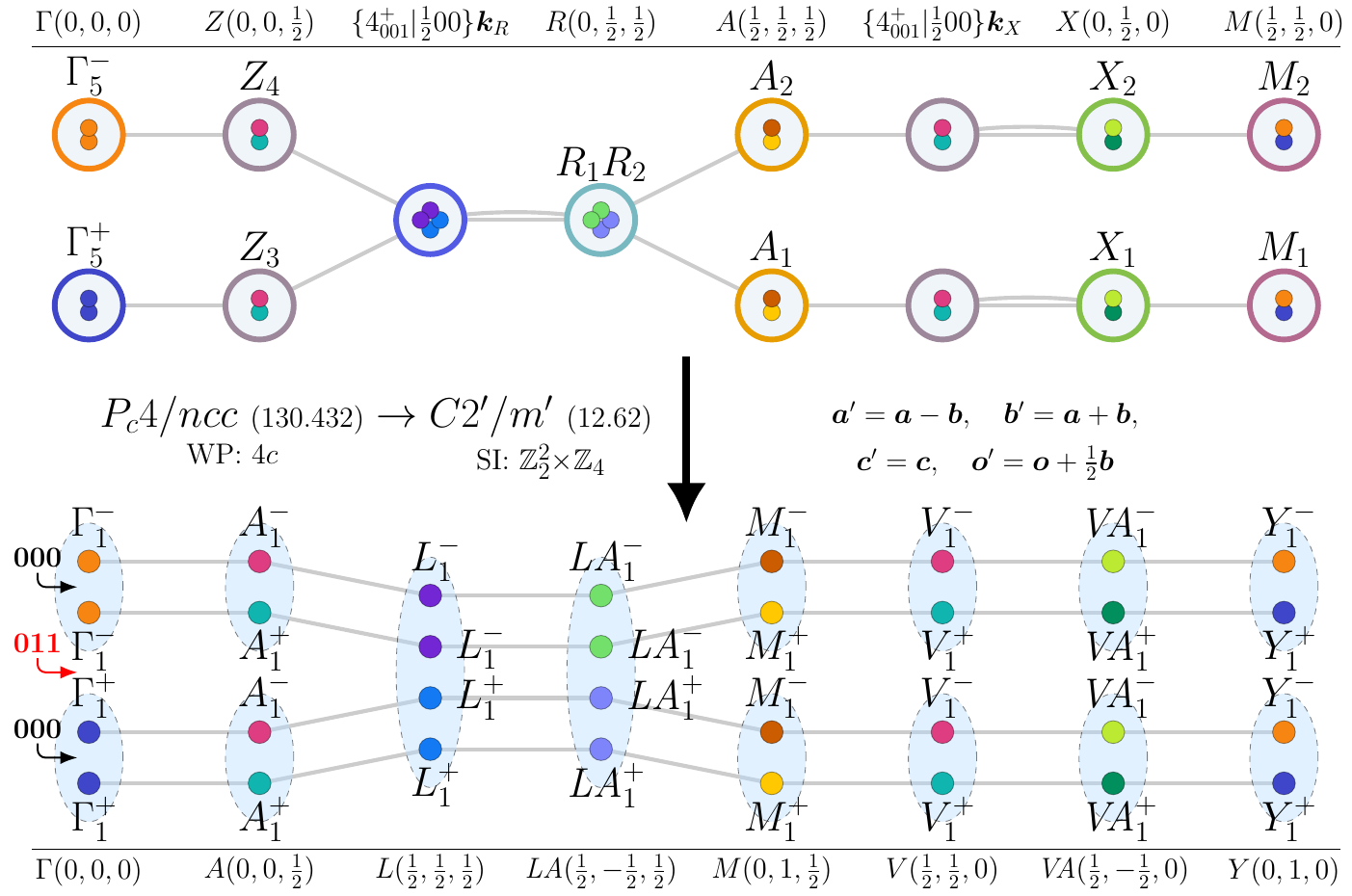}
\caption{Topological magnon bands in subgroup $C2'/m'~(12.62)$ for magnetic moments on Wyckoff position $4c$ of supergroup $P_{c}4/ncc~(130.432)$.\label{fig_130.432_12.62_Bparallel001andstrainperp110_4c}}
\end{figure}
\input{gap_tables_tex/130.432_12.62_Bparallel001andstrainperp110_4c_table.tex}
\input{si_tables_tex/130.432_12.62_Bparallel001andstrainperp110_4c_table.tex}
\subsubsection{Topological bands in subgroup $P2_{1}'/c'~(14.79)$}
\textbf{Perturbations:}
\begin{itemize}
\item B $\parallel$ [100] and strain $\parallel$ [110],
\item B $\parallel$ [100] and strain $\perp$ [001],
\item B $\parallel$ [110] and strain $\parallel$ [100],
\item B $\parallel$ [110] and strain $\perp$ [001],
\item B $\perp$ [001].
\end{itemize}
\begin{figure}[H]
\centering
\includegraphics[scale=0.6]{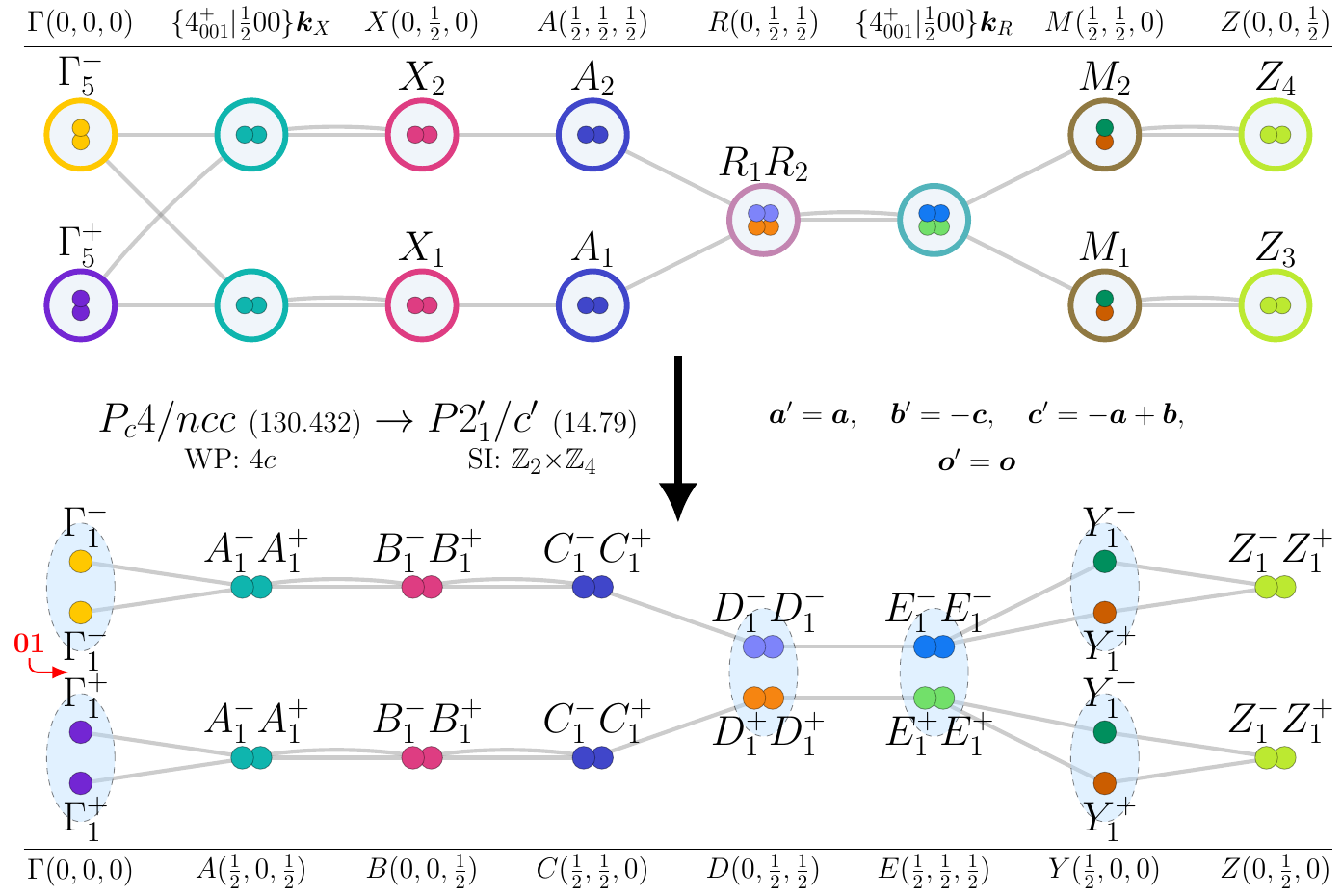}
\caption{Topological magnon bands in subgroup $P2_{1}'/c'~(14.79)$ for magnetic moments on Wyckoff position $4c$ of supergroup $P_{c}4/ncc~(130.432)$.\label{fig_130.432_14.79_Bparallel100andstrainparallel110_4c}}
\end{figure}
\input{gap_tables_tex/130.432_14.79_Bparallel100andstrainparallel110_4c_table.tex}
\input{si_tables_tex/130.432_14.79_Bparallel100andstrainparallel110_4c_table.tex}
\subsubsection{Topological bands in subgroup $P2_{1}'/m'~(11.54)$}
\textbf{Perturbations:}
\begin{itemize}
\item B $\parallel$ [001] and strain $\perp$ [100],
\item B $\perp$ [100].
\end{itemize}
\begin{figure}[H]
\centering
\includegraphics[scale=0.6]{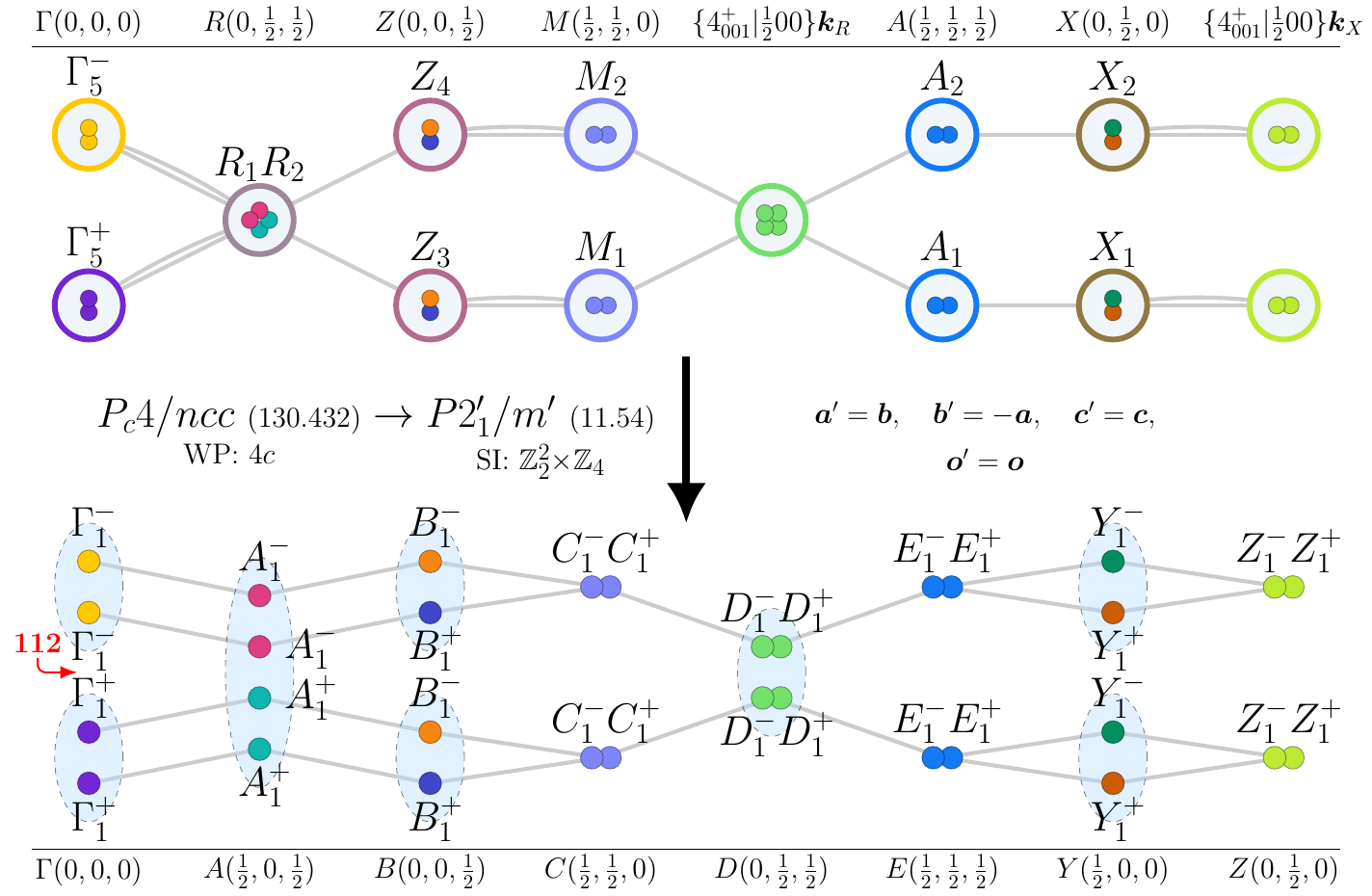}
\caption{Topological magnon bands in subgroup $P2_{1}'/m'~(11.54)$ for magnetic moments on Wyckoff position $4c$ of supergroup $P_{c}4/ncc~(130.432)$.\label{fig_130.432_11.54_Bparallel001andstrainperp100_4c}}
\end{figure}
\input{gap_tables_tex/130.432_11.54_Bparallel001andstrainperp100_4c_table.tex}
\input{si_tables_tex/130.432_11.54_Bparallel001andstrainperp100_4c_table.tex}
\subsubsection{Topological bands in subgroup $P_{S}\bar{1}~(2.7)$}
\textbf{Perturbation:}
\begin{itemize}
\item strain in generic direction.
\end{itemize}
\begin{figure}[H]
\centering
\includegraphics[scale=0.6]{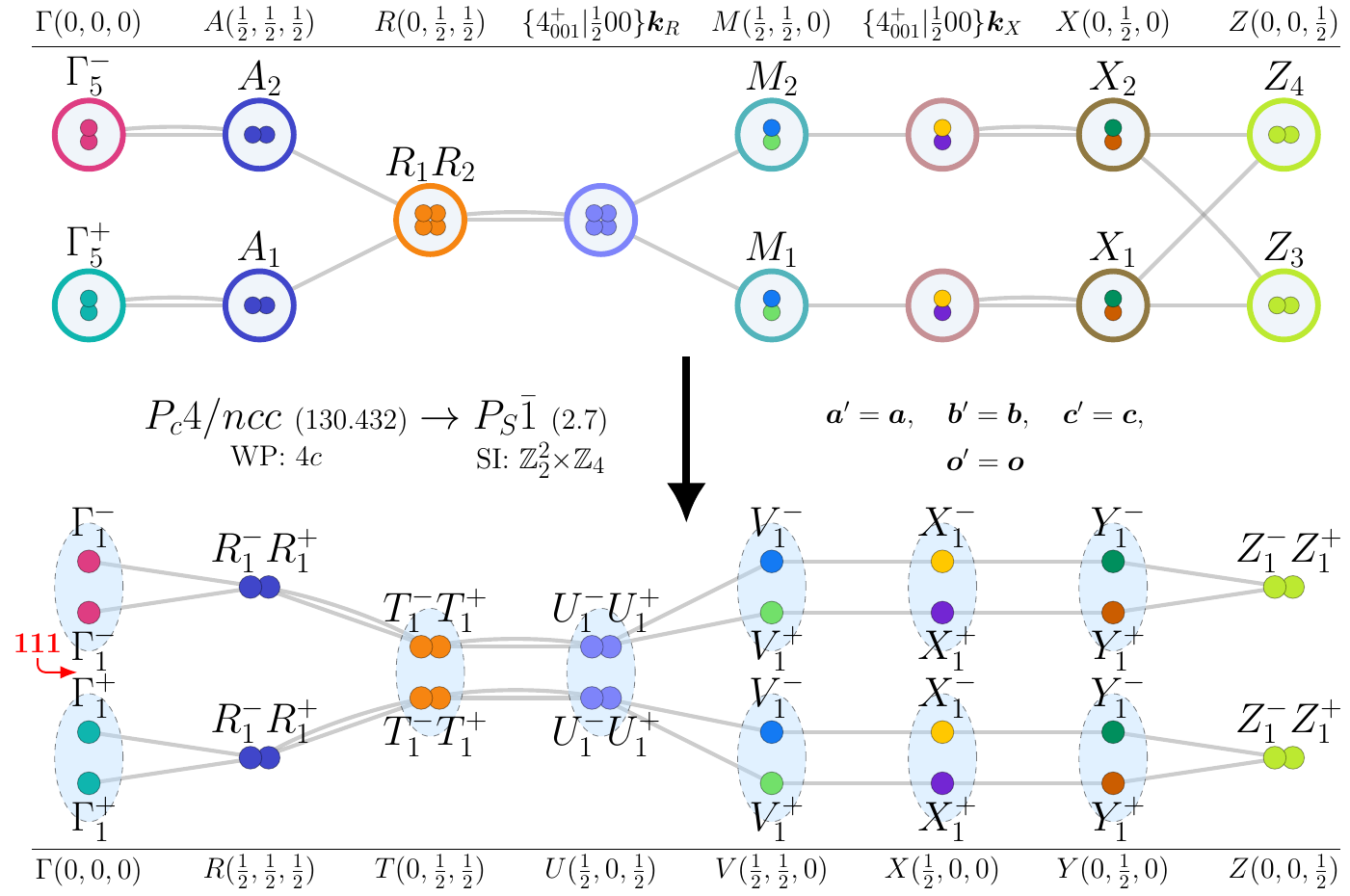}
\caption{Topological magnon bands in subgroup $P_{S}\bar{1}~(2.7)$ for magnetic moments on Wyckoff position $4c$ of supergroup $P_{c}4/ncc~(130.432)$.\label{fig_130.432_2.7_strainingenericdirection_4c}}
\end{figure}
\input{gap_tables_tex/130.432_2.7_strainingenericdirection_4c_table.tex}
\input{si_tables_tex/130.432_2.7_strainingenericdirection_4c_table.tex}
\subsection{WP: $4c+4c$}
\textbf{BCS Materials:} {UPd\textsubscript{2}Ge\textsubscript{2}~(80 K)}\footnote{BCS web page: \texttt{\href{http://webbdcrista1.ehu.es/magndata/index.php?this\_label=1.535} {http://webbdcrista1.ehu.es/magndata/index.php?this\_label=1.535}}}.\\
\subsubsection{Topological bands in subgroup $C2'/m'~(12.62)$}
\textbf{Perturbations:}
\begin{itemize}
\item B $\parallel$ [001] and strain $\perp$ [110],
\item B $\perp$ [110].
\end{itemize}
\begin{figure}[H]
\centering
\includegraphics[scale=0.6]{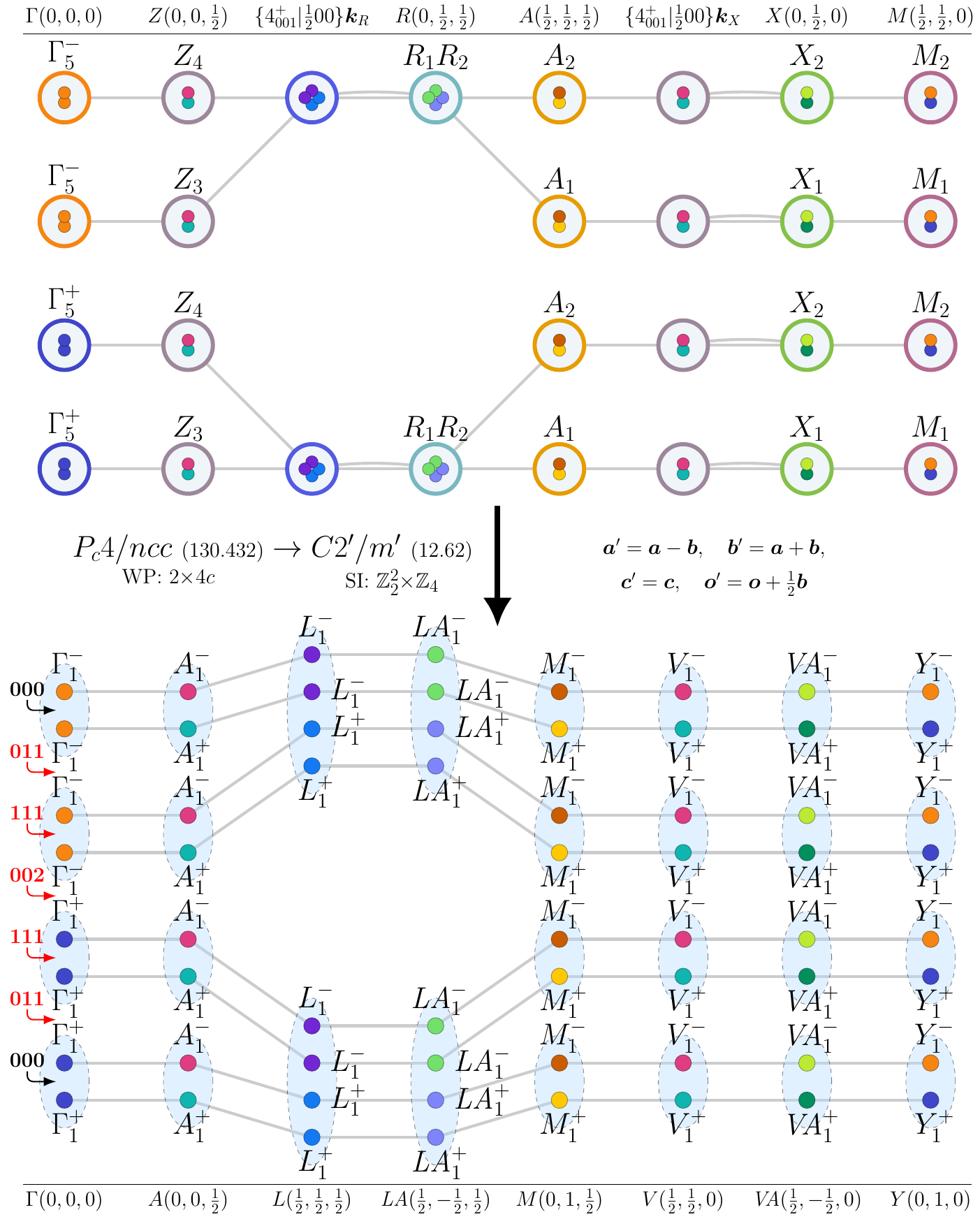}
\caption{Topological magnon bands in subgroup $C2'/m'~(12.62)$ for magnetic moments on Wyckoff positions $4c+4c$ of supergroup $P_{c}4/ncc~(130.432)$.\label{fig_130.432_12.62_Bparallel001andstrainperp110_4c+4c}}
\end{figure}
\input{gap_tables_tex/130.432_12.62_Bparallel001andstrainperp110_4c+4c_table.tex}
\input{si_tables_tex/130.432_12.62_Bparallel001andstrainperp110_4c+4c_table.tex}
\subsubsection{Topological bands in subgroup $P2_{1}'/c'~(14.79)$}
\textbf{Perturbations:}
\begin{itemize}
\item B $\parallel$ [100] and strain $\parallel$ [110],
\item B $\parallel$ [100] and strain $\perp$ [001],
\item B $\parallel$ [110] and strain $\parallel$ [100],
\item B $\parallel$ [110] and strain $\perp$ [001],
\item B $\perp$ [001].
\end{itemize}
\begin{figure}[H]
\centering
\includegraphics[scale=0.6]{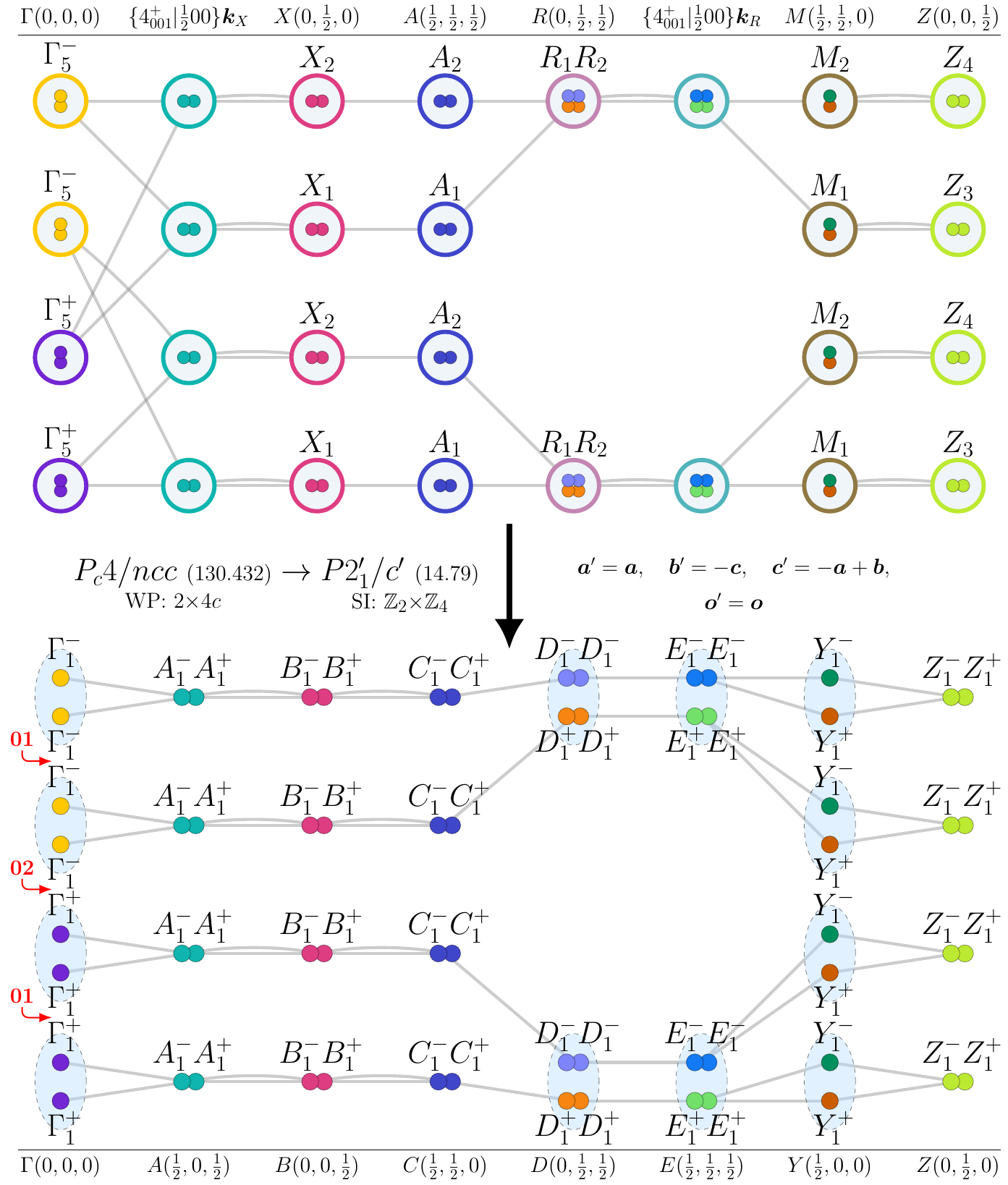}
\caption{Topological magnon bands in subgroup $P2_{1}'/c'~(14.79)$ for magnetic moments on Wyckoff positions $4c+4c$ of supergroup $P_{c}4/ncc~(130.432)$.\label{fig_130.432_14.79_Bparallel100andstrainparallel110_4c+4c}}
\end{figure}
\input{gap_tables_tex/130.432_14.79_Bparallel100andstrainparallel110_4c+4c_table.tex}
\input{si_tables_tex/130.432_14.79_Bparallel100andstrainparallel110_4c+4c_table.tex}
\subsubsection{Topological bands in subgroup $P2_{1}'/m'~(11.54)$}
\textbf{Perturbations:}
\begin{itemize}
\item B $\parallel$ [001] and strain $\perp$ [100],
\item B $\perp$ [100].
\end{itemize}
\begin{figure}[H]
\centering
\includegraphics[scale=0.6]{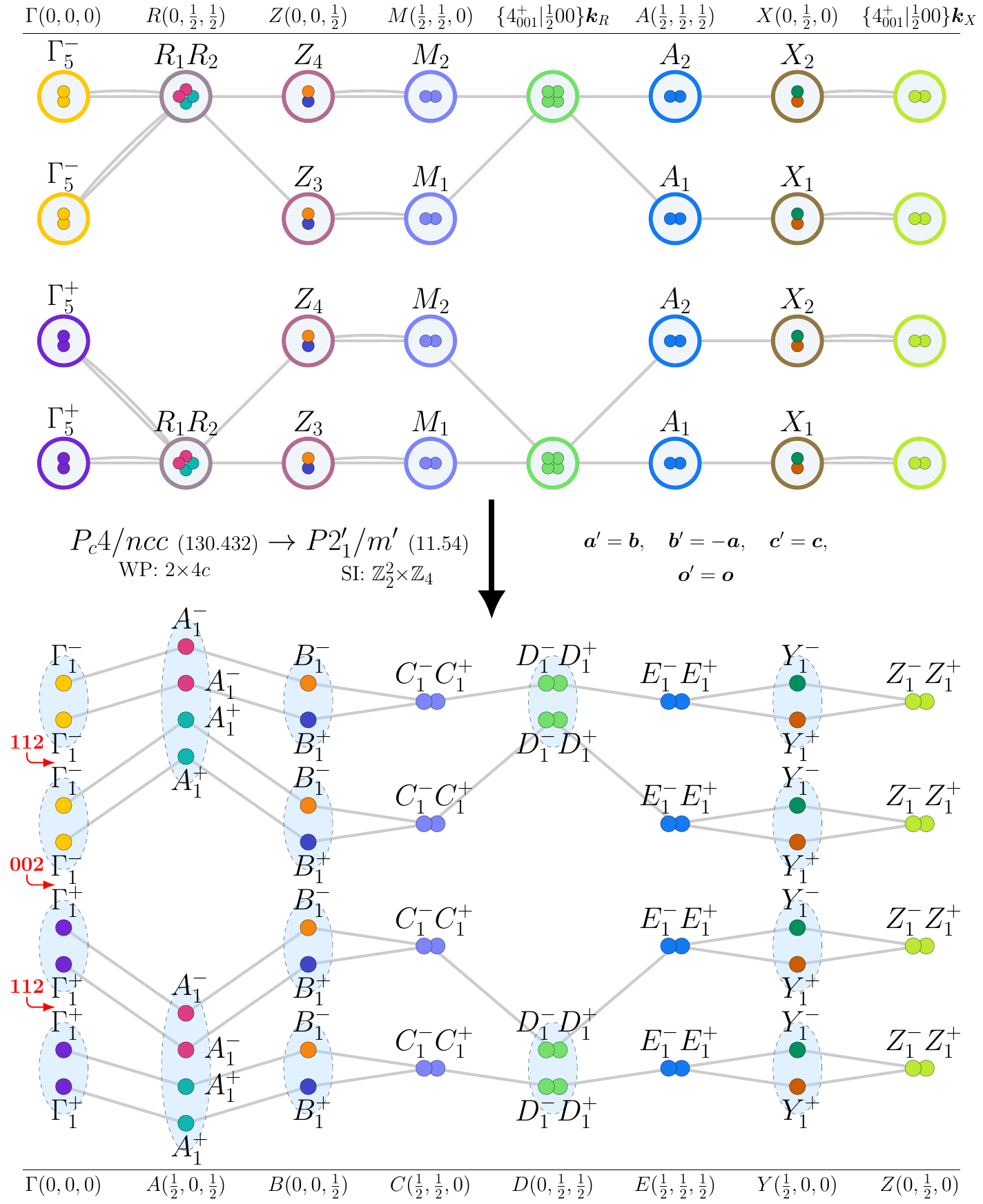}
\caption{Topological magnon bands in subgroup $P2_{1}'/m'~(11.54)$ for magnetic moments on Wyckoff positions $4c+4c$ of supergroup $P_{c}4/ncc~(130.432)$.\label{fig_130.432_11.54_Bparallel001andstrainperp100_4c+4c}}
\end{figure}
\input{gap_tables_tex/130.432_11.54_Bparallel001andstrainperp100_4c+4c_table.tex}
\input{si_tables_tex/130.432_11.54_Bparallel001andstrainperp100_4c+4c_table.tex}
\subsubsection{Topological bands in subgroup $P_{S}\bar{1}~(2.7)$}
\textbf{Perturbation:}
\begin{itemize}
\item strain in generic direction.
\end{itemize}
\begin{figure}[H]
\centering
\includegraphics[scale=0.6]{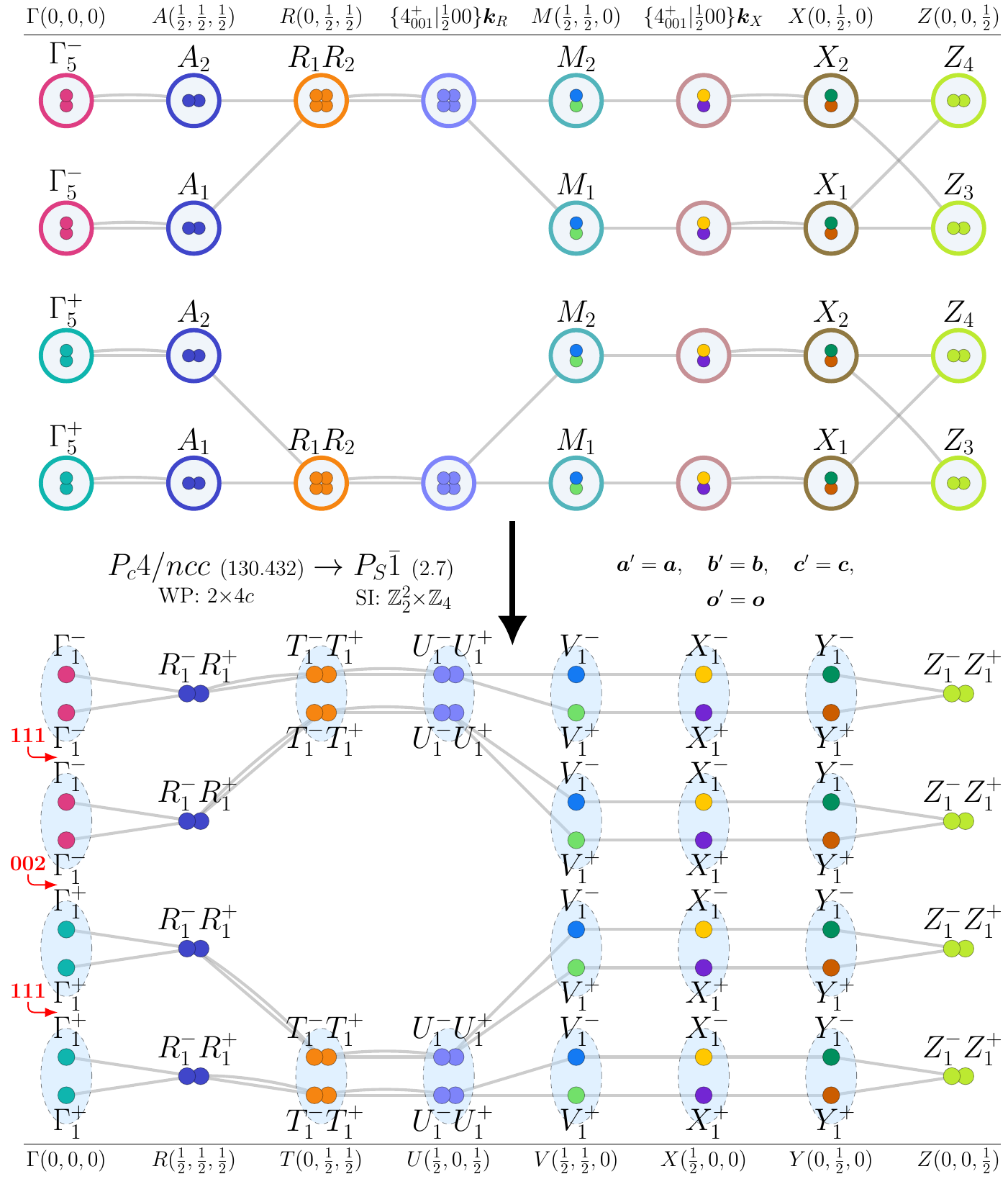}
\caption{Topological magnon bands in subgroup $P_{S}\bar{1}~(2.7)$ for magnetic moments on Wyckoff positions $4c+4c$ of supergroup $P_{c}4/ncc~(130.432)$.\label{fig_130.432_2.7_strainingenericdirection_4c+4c}}
\end{figure}
\input{gap_tables_tex/130.432_2.7_strainingenericdirection_4c+4c_table.tex}
\input{si_tables_tex/130.432_2.7_strainingenericdirection_4c+4c_table.tex}

\section{MSG $P_{c}4_{2}/mcm~(132.456)$}
\textbf{Nontrivial-SI Subgroups:} $P\bar{1}~(2.4)$, $C2'/c'~(15.89)$, $P2_{1}'/m'~(11.54)$, $P2'/m'~(10.46)$, $P_{S}\bar{1}~(2.7)$, $C2/m~(12.58)$, $Cmc'm'~(63.463)$, $C_{c}2/m~(12.63)$, $P2~(3.1)$, $Cc'c'2~(37.183)$, $Pm'm'2~(25.60)$, $P_{b}2~(3.5)$, $P_{c}cc2~(27.82)$, $P2/m~(10.42)$, $Cc'c'm~(66.495)$, $Pm'm'm~(47.252)$, $P_{b}2/m~(10.48)$, $P2~(3.1)$, $Pm'm'2~(25.60)$, $P_{a}2~(3.4)$, $P2/c~(13.65)$, $Pm'm'a~(51.294)$, $P_{c}2/c~(13.72)$, $P_{c}ccm~(49.273)$, $P4_{2}m'c'~(105.215)$, $P4_{2}/mm'c'~(131.441)$.\\

\textbf{Trivial-SI Subgroups:} $Cc'~(9.39)$, $Pm'~(6.20)$, $Pm'~(6.20)$, $C2'~(5.15)$, $P2_{1}'~(4.9)$, $P2'~(3.3)$, $P_{S}1~(1.3)$, $Cm~(8.32)$, $Cmc'2_{1}'~(36.175)$, $C_{c}m~(8.35)$, $Pm~(6.18)$, $Ama'2'~(40.206)$, $Pm'm2'~(25.59)$, $P_{b}m~(6.22)$, $Pc~(7.24)$, $Pm'c2_{1}'~(26.68)$, $P_{c}c~(7.28)$, $C2~(5.13)$, $Am'a'2~(40.207)$, $C_{c}2~(5.16)$, $A_{a}mm2~(38.192)$, $C_{c}mm2~(35.169)$, $C_{c}mmm~(65.488)$, $P_{a}ma2~(28.92)$, $P_{c}4_{2}cm~(101.184)$.\\

\subsection{WP: $4f$}
\textbf{BCS Materials:} {Mn\textsubscript{3}Pt~(475 K)}\footnote{BCS web page: \texttt{\href{http://webbdcrista1.ehu.es/magndata/index.php?this\_label=1.143} {http://webbdcrista1.ehu.es/magndata/index.php?this\_label=1.143}}}.\\
\subsubsection{Topological bands in subgroup $P\bar{1}~(2.4)$}
\textbf{Perturbations:}
\begin{itemize}
\item B $\parallel$ [001] and strain in generic direction,
\item B $\parallel$ [100] and strain $\perp$ [110],
\item B $\parallel$ [100] and strain in generic direction,
\item B $\parallel$ [110] and strain $\perp$ [100],
\item B $\parallel$ [110] and strain in generic direction,
\item B $\perp$ [001] and strain $\perp$ [100],
\item B $\perp$ [001] and strain $\perp$ [110],
\item B $\perp$ [001] and strain in generic direction,
\item B $\perp$ [100] and strain $\parallel$ [110],
\item B $\perp$ [100] and strain $\perp$ [001],
\item B $\perp$ [100] and strain $\perp$ [110],
\item B $\perp$ [100] and strain in generic direction,
\item B $\perp$ [110] and strain $\parallel$ [100],
\item B $\perp$ [110] and strain $\perp$ [001],
\item B $\perp$ [110] and strain $\perp$ [100],
\item B $\perp$ [110] and strain in generic direction,
\item B in generic direction.
\end{itemize}
\begin{figure}[H]
\centering
\includegraphics[scale=0.6]{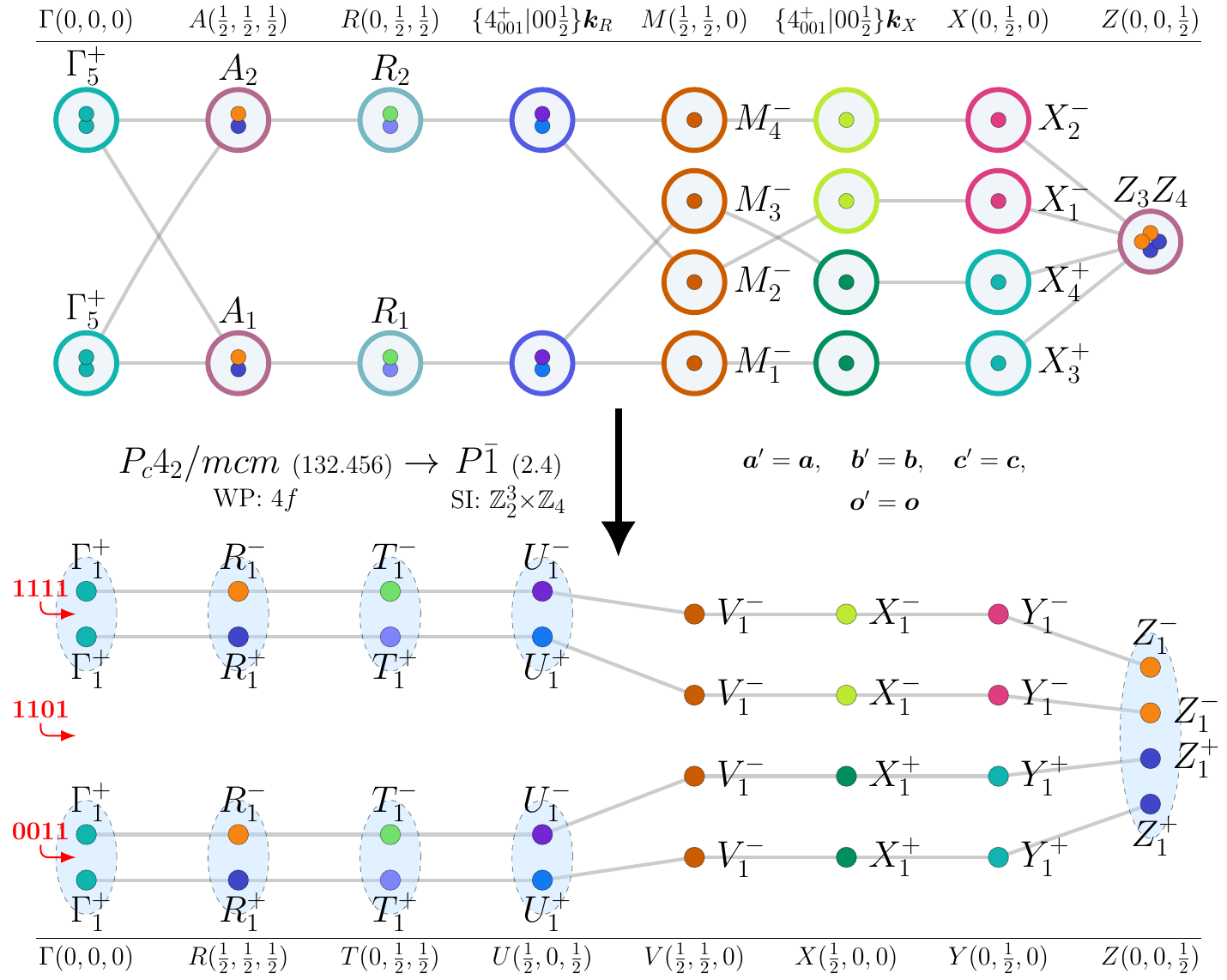}
\caption{Topological magnon bands in subgroup $P\bar{1}~(2.4)$ for magnetic moments on Wyckoff position $4f$ of supergroup $P_{c}4_{2}/mcm~(132.456)$.\label{fig_132.456_2.4_Bparallel001andstrainingenericdirection_4f}}
\end{figure}
\input{gap_tables_tex/132.456_2.4_Bparallel001andstrainingenericdirection_4f_table.tex}
\input{si_tables_tex/132.456_2.4_Bparallel001andstrainingenericdirection_4f_table.tex}
\subsubsection{Topological bands in subgroup $C2'/c'~(15.89)$}
\textbf{Perturbations:}
\begin{itemize}
\item B $\parallel$ [001] and strain $\perp$ [110],
\item B $\perp$ [110].
\end{itemize}
\begin{figure}[H]
\centering
\includegraphics[scale=0.6]{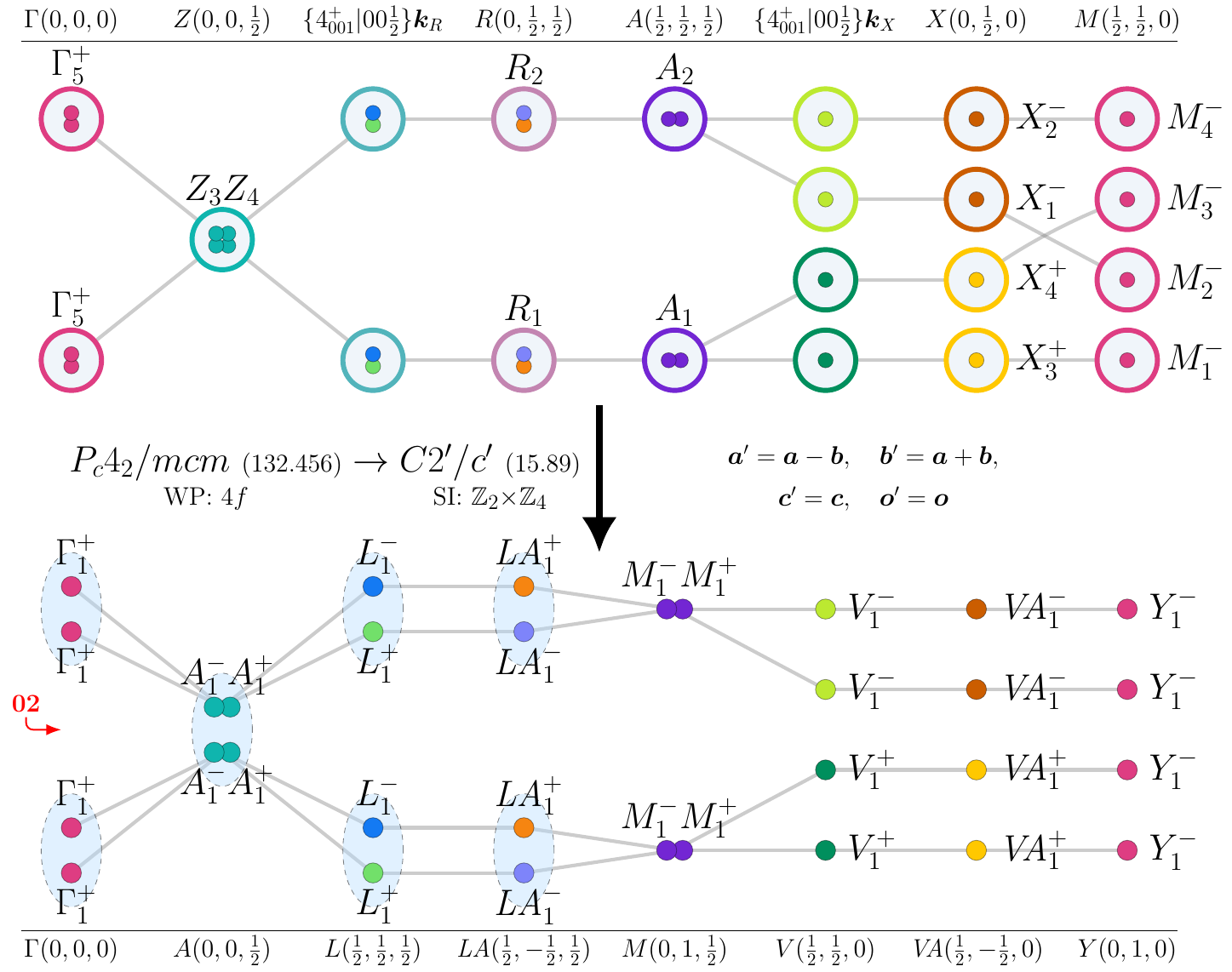}
\caption{Topological magnon bands in subgroup $C2'/c'~(15.89)$ for magnetic moments on Wyckoff position $4f$ of supergroup $P_{c}4_{2}/mcm~(132.456)$.\label{fig_132.456_15.89_Bparallel001andstrainperp110_4f}}
\end{figure}
\input{gap_tables_tex/132.456_15.89_Bparallel001andstrainperp110_4f_table.tex}
\input{si_tables_tex/132.456_15.89_Bparallel001andstrainperp110_4f_table.tex}
\subsubsection{Topological bands in subgroup $P2_{1}'/m'~(11.54)$}
\textbf{Perturbations:}
\begin{itemize}
\item B $\parallel$ [100] and strain $\parallel$ [110],
\item B $\parallel$ [100] and strain $\perp$ [001],
\item B $\parallel$ [110] and strain $\parallel$ [100],
\item B $\parallel$ [110] and strain $\perp$ [001],
\item B $\perp$ [001].
\end{itemize}
\begin{figure}[H]
\centering
\includegraphics[scale=0.6]{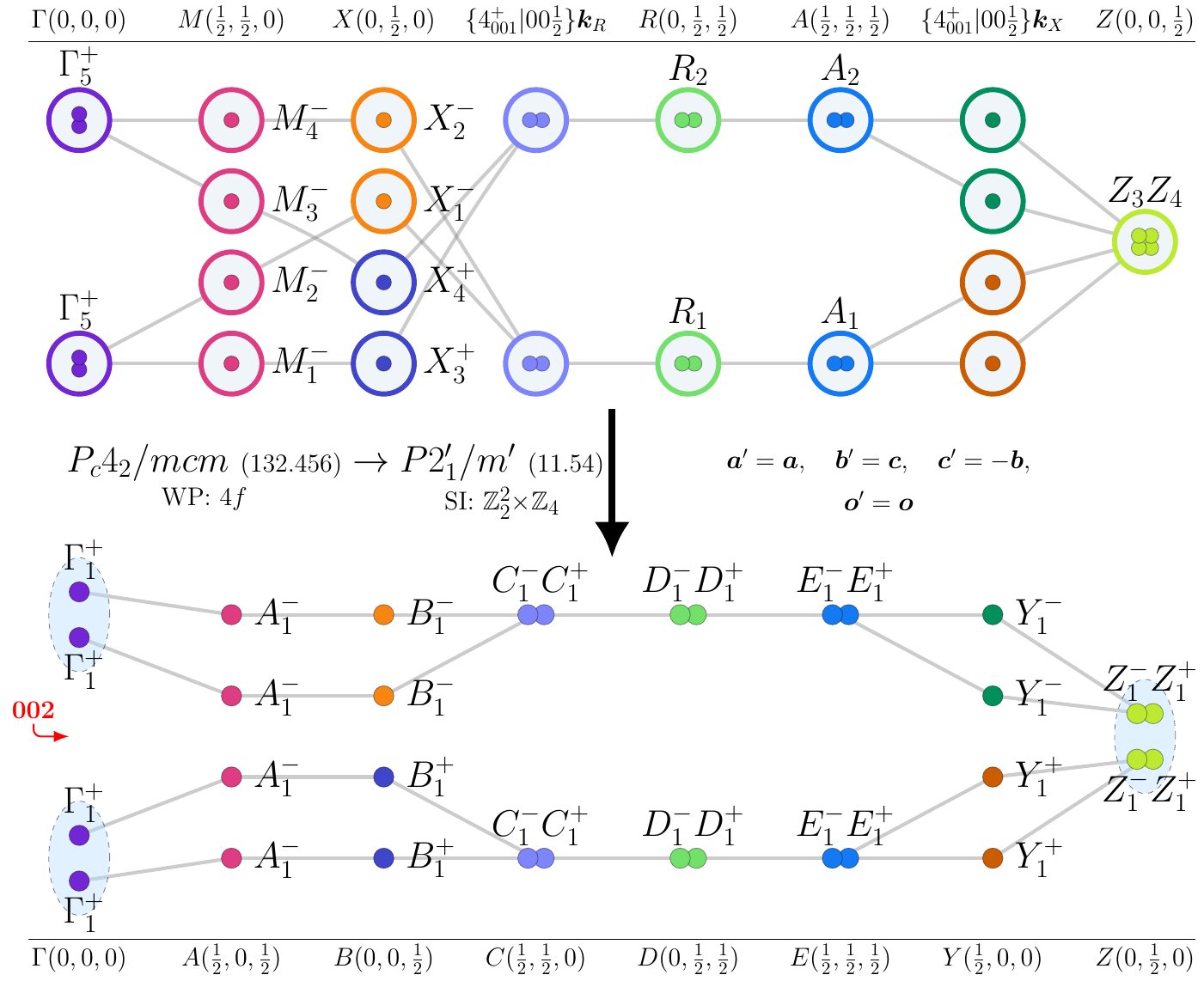}
\caption{Topological magnon bands in subgroup $P2_{1}'/m'~(11.54)$ for magnetic moments on Wyckoff position $4f$ of supergroup $P_{c}4_{2}/mcm~(132.456)$.\label{fig_132.456_11.54_Bparallel100andstrainparallel110_4f}}
\end{figure}
\input{gap_tables_tex/132.456_11.54_Bparallel100andstrainparallel110_4f_table.tex}
\input{si_tables_tex/132.456_11.54_Bparallel100andstrainparallel110_4f_table.tex}
\subsubsection{Topological bands in subgroup $P2'/m'~(10.46)$}
\textbf{Perturbations:}
\begin{itemize}
\item B $\parallel$ [001] and strain $\perp$ [100],
\item B $\perp$ [100].
\end{itemize}
\begin{figure}[H]
\centering
\includegraphics[scale=0.6]{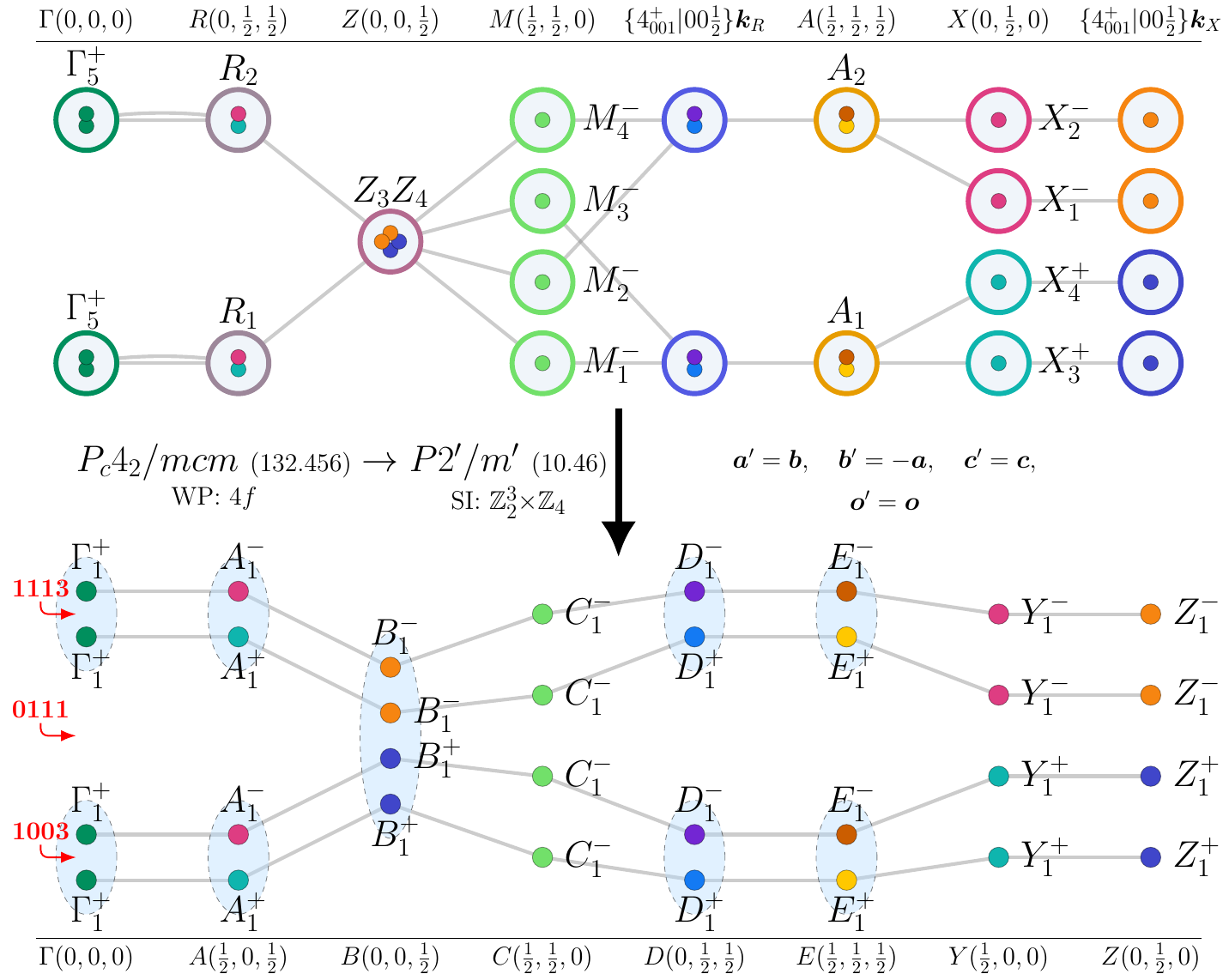}
\caption{Topological magnon bands in subgroup $P2'/m'~(10.46)$ for magnetic moments on Wyckoff position $4f$ of supergroup $P_{c}4_{2}/mcm~(132.456)$.\label{fig_132.456_10.46_Bparallel001andstrainperp100_4f}}
\end{figure}
\input{gap_tables_tex/132.456_10.46_Bparallel001andstrainperp100_4f_table.tex}
\input{si_tables_tex/132.456_10.46_Bparallel001andstrainperp100_4f_table.tex}
\subsubsection{Topological bands in subgroup $P_{S}\bar{1}~(2.7)$}
\textbf{Perturbation:}
\begin{itemize}
\item strain in generic direction.
\end{itemize}
\begin{figure}[H]
\centering
\includegraphics[scale=0.6]{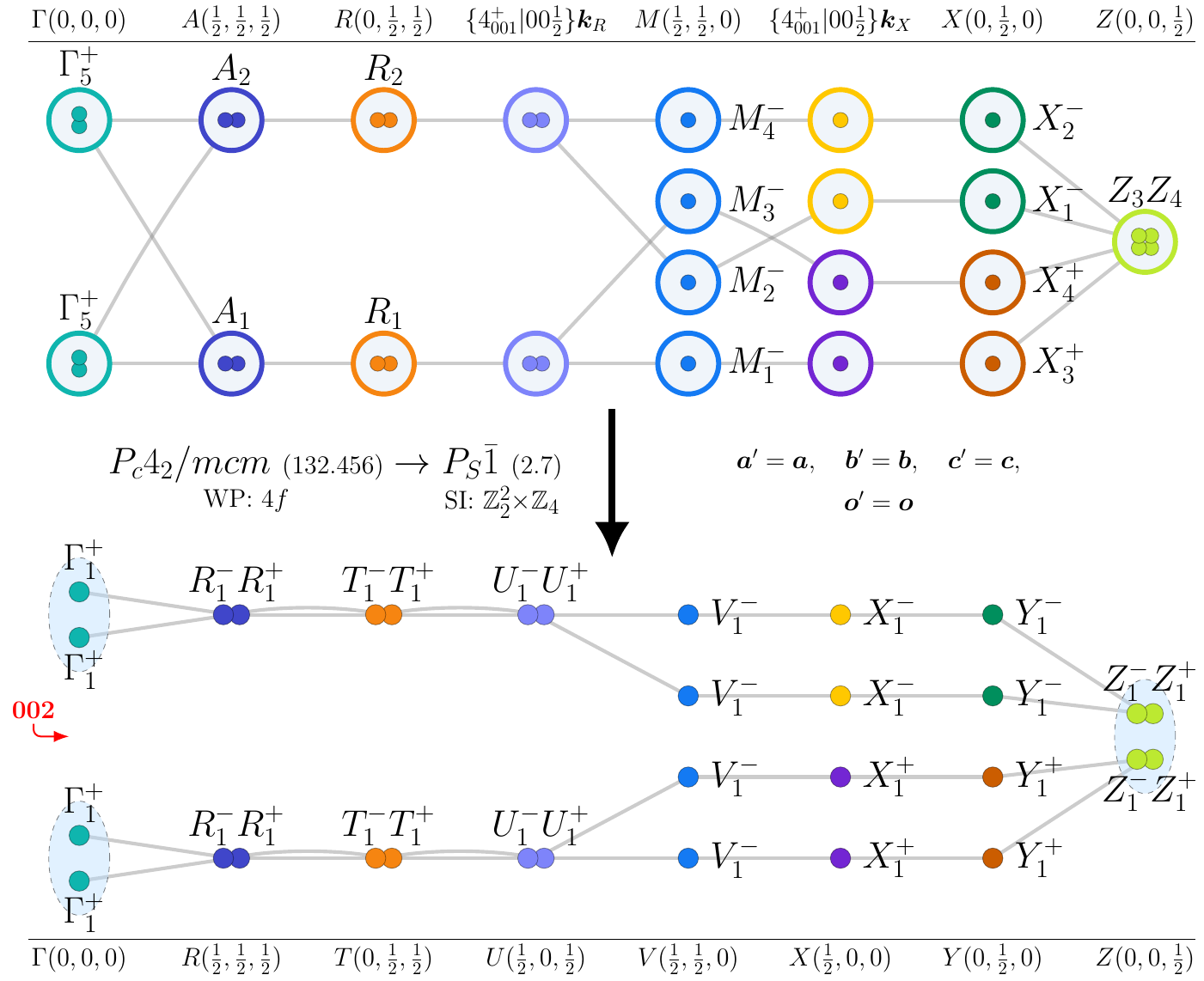}
\caption{Topological magnon bands in subgroup $P_{S}\bar{1}~(2.7)$ for magnetic moments on Wyckoff position $4f$ of supergroup $P_{c}4_{2}/mcm~(132.456)$.\label{fig_132.456_2.7_strainingenericdirection_4f}}
\end{figure}
\input{gap_tables_tex/132.456_2.7_strainingenericdirection_4f_table.tex}
\input{si_tables_tex/132.456_2.7_strainingenericdirection_4f_table.tex}

\section{MSG $P_{C}4_{2}/nnm~(134.481)$}
\textbf{Nontrivial-SI Subgroups:} $P\bar{1}~(2.4)$, $C2'/m'~(12.62)$, $P2'/m'~(10.46)$, $P2_{1}'/c'~(14.79)$, $P_{S}\bar{1}~(2.7)$, $C_{a}2~(5.17)$, $C2/m~(12.58)$, $Cmm'm'~(65.486)$, $C_{a}2/m~(12.64)$, $P2~(3.1)$, $Cm'm'2~(35.168)$, $Pc'c'2~(27.81)$, $P_{a}2~(3.4)$, $P_{C}nn2~(34.163)$, $P2/c~(13.65)$, $Cm'm'a~(67.505)$, $Pc'c'n~(56.369)$, $P_{c}2/c~(13.72)$, $P2~(3.1)$, $Pm'a'2~(28.91)$, $P2/c~(13.65)$, $Pm'na'~(53.328)$, $P_{C}2/c~(13.74)$, $P_{C}nnn~(48.263)$, $P4_{2}c'm'~(101.183)$, $P4_{2}/nc'm'~(138.525)$.\\

\textbf{Trivial-SI Subgroups:} $Cm'~(8.34)$, $Pm'~(6.20)$, $Pc'~(7.26)$, $C2'~(5.15)$, $P2'~(3.3)$, $P2_{1}'~(4.9)$, $P_{S}1~(1.3)$, $Cm~(8.32)$, $Cm'm2'~(35.167)$, $C_{a}m~(8.36)$, $Pc~(7.24)$, $Abm'2'~(39.198)$, $Pna'2_{1}'~(33.147)$, $P_{c}c~(7.28)$, $Pc~(7.24)$, $Pnc'2'~(30.114)$, $P_{C}c~(7.30)$, $C2~(5.13)$, $Am'm'2~(38.191)$, $A_{b}bm2~(39.201)$, $C_{a}mm2~(35.170)$, $C_{a}mma~(67.509)$, $P_{C}2~(3.6)$, $P_{A}nn2~(34.162)$, $P_{C}4_{2}nm~(102.193)$.\\

\subsection{WP: $4d$}
\textbf{BCS Materials:} {LaSrFeO\textsubscript{4}~(380 K)}\footnote{BCS web page: \texttt{\href{http://webbdcrista1.ehu.es/magndata/index.php?this\_label=2.42} {http://webbdcrista1.ehu.es/magndata/index.php?this\_label=2.42}}}, {Nd\textsubscript{2}CuO\textsubscript{4}~(276 K)}\footnote{BCS web page: \texttt{\href{http://webbdcrista1.ehu.es/magndata/index.php?this\_label=2.6} {http://webbdcrista1.ehu.es/magndata/index.php?this\_label=2.6}}}, {Pr\textsubscript{2}CuO\textsubscript{4}~(270 K)}\footnote{BCS web page: \texttt{\href{http://webbdcrista1.ehu.es/magndata/index.php?this\_label=2.48} {http://webbdcrista1.ehu.es/magndata/index.php?this\_label=2.48}}}.\\
\subsubsection{Topological bands in subgroup $P\bar{1}~(2.4)$}
\textbf{Perturbations:}
\begin{itemize}
\item B $\parallel$ [001] and strain in generic direction,
\item B $\parallel$ [100] and strain $\perp$ [110],
\item B $\parallel$ [100] and strain in generic direction,
\item B $\parallel$ [110] and strain $\perp$ [100],
\item B $\parallel$ [110] and strain in generic direction,
\item B $\perp$ [001] and strain $\perp$ [100],
\item B $\perp$ [001] and strain $\perp$ [110],
\item B $\perp$ [001] and strain in generic direction,
\item B $\perp$ [100] and strain $\parallel$ [110],
\item B $\perp$ [100] and strain $\perp$ [001],
\item B $\perp$ [100] and strain $\perp$ [110],
\item B $\perp$ [100] and strain in generic direction,
\item B $\perp$ [110] and strain $\parallel$ [100],
\item B $\perp$ [110] and strain $\perp$ [001],
\item B $\perp$ [110] and strain $\perp$ [100],
\item B $\perp$ [110] and strain in generic direction,
\item B in generic direction.
\end{itemize}
\begin{figure}[H]
\centering
\includegraphics[scale=0.6]{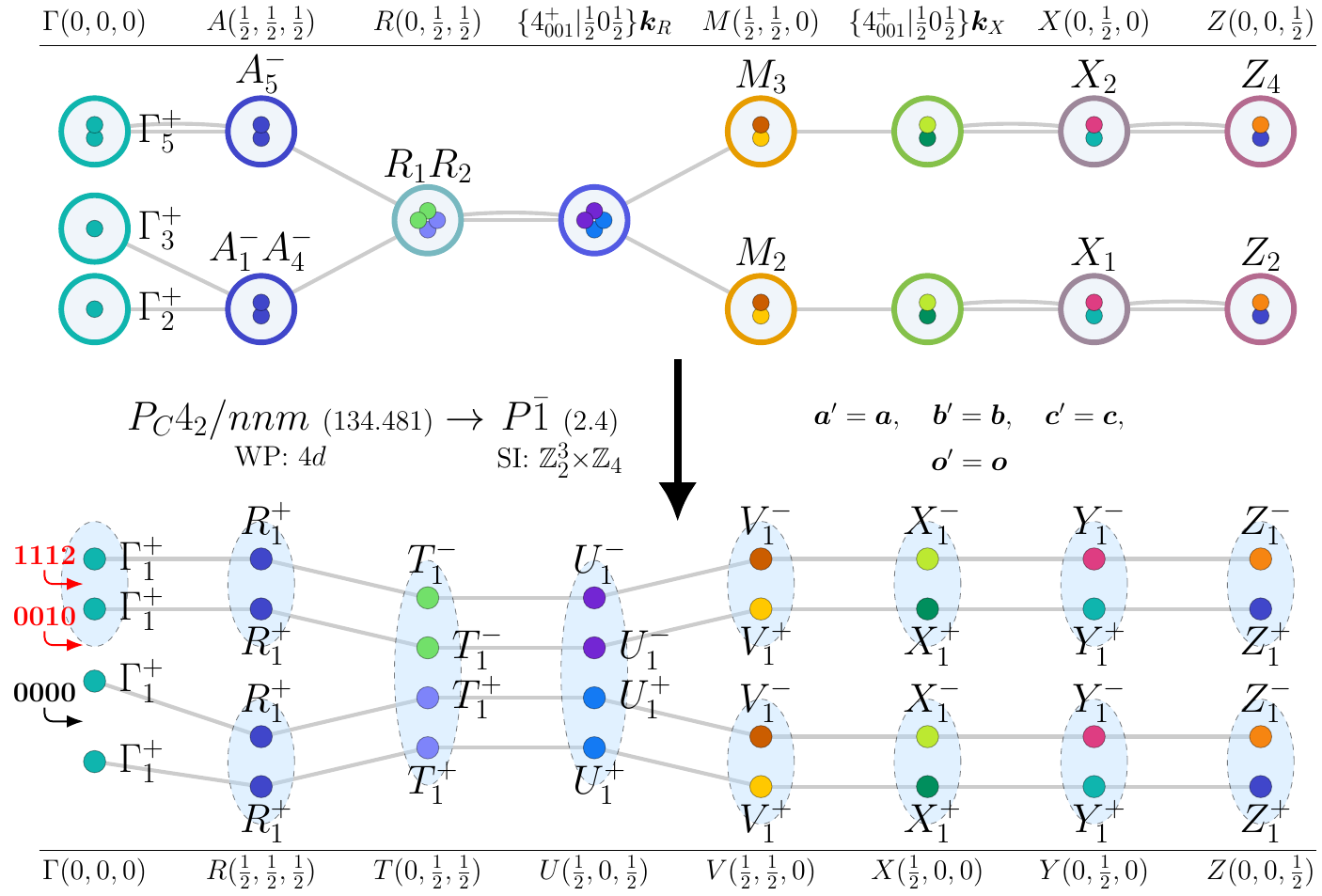}
\caption{Topological magnon bands in subgroup $P\bar{1}~(2.4)$ for magnetic moments on Wyckoff position $4d$ of supergroup $P_{C}4_{2}/nnm~(134.481)$.\label{fig_134.481_2.4_Bparallel001andstrainingenericdirection_4d}}
\end{figure}
\input{gap_tables_tex/134.481_2.4_Bparallel001andstrainingenericdirection_4d_table.tex}
\input{si_tables_tex/134.481_2.4_Bparallel001andstrainingenericdirection_4d_table.tex}
\subsubsection{Topological bands in subgroup $C2'/m'~(12.62)$}
\textbf{Perturbations:}
\begin{itemize}
\item B $\parallel$ [001] and strain $\perp$ [110],
\item B $\perp$ [110].
\end{itemize}
\begin{figure}[H]
\centering
\includegraphics[scale=0.6]{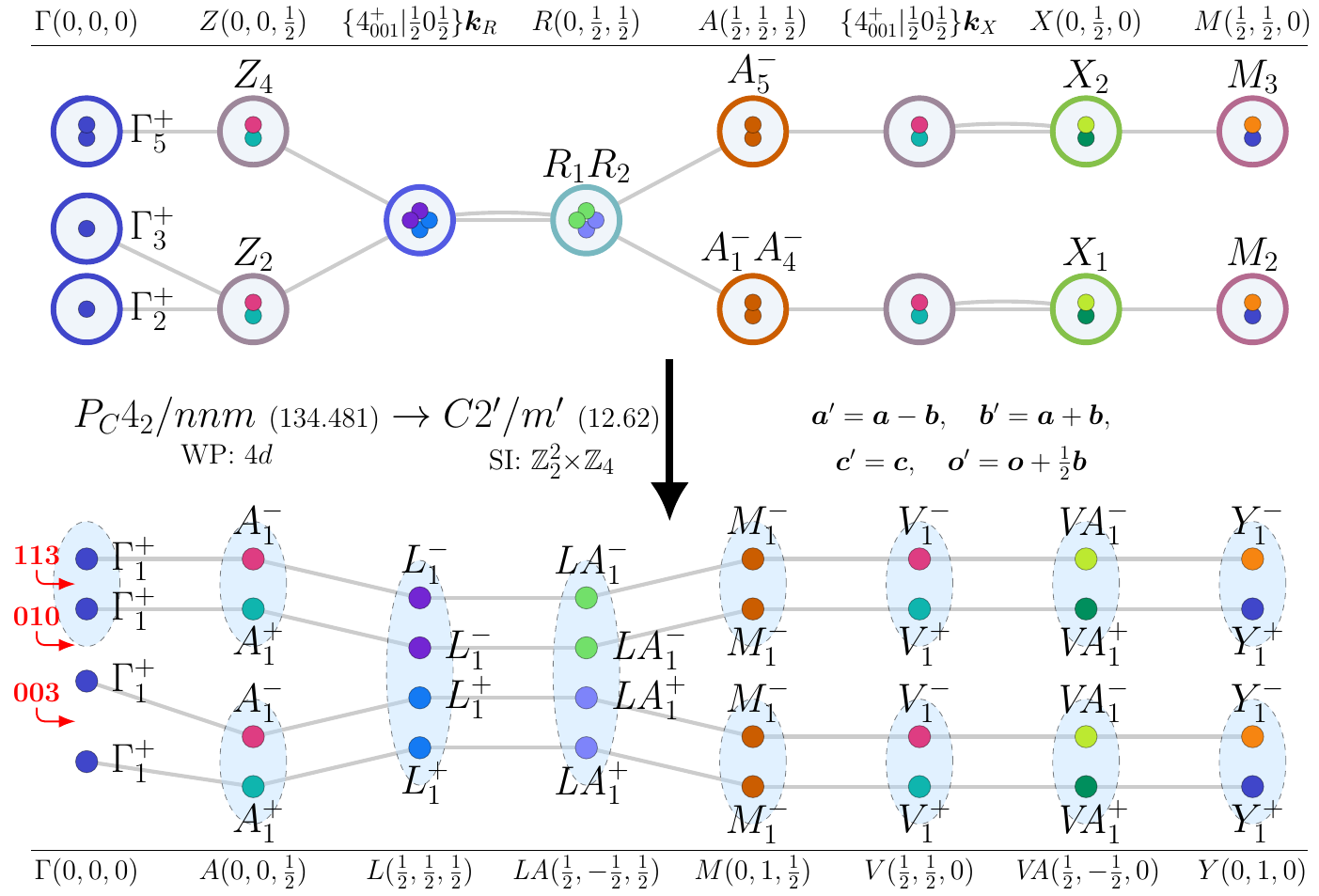}
\caption{Topological magnon bands in subgroup $C2'/m'~(12.62)$ for magnetic moments on Wyckoff position $4d$ of supergroup $P_{C}4_{2}/nnm~(134.481)$.\label{fig_134.481_12.62_Bparallel001andstrainperp110_4d}}
\end{figure}
\input{gap_tables_tex/134.481_12.62_Bparallel001andstrainperp110_4d_table.tex}
\input{si_tables_tex/134.481_12.62_Bparallel001andstrainperp110_4d_table.tex}
\subsubsection{Topological bands in subgroup $P2'/m'~(10.46)$}
\textbf{Perturbations:}
\begin{itemize}
\item B $\parallel$ [100] and strain $\parallel$ [110],
\item B $\parallel$ [100] and strain $\perp$ [001],
\item B $\parallel$ [110] and strain $\parallel$ [100],
\item B $\parallel$ [110] and strain $\perp$ [001],
\item B $\perp$ [001].
\end{itemize}
\begin{figure}[H]
\centering
\includegraphics[scale=0.6]{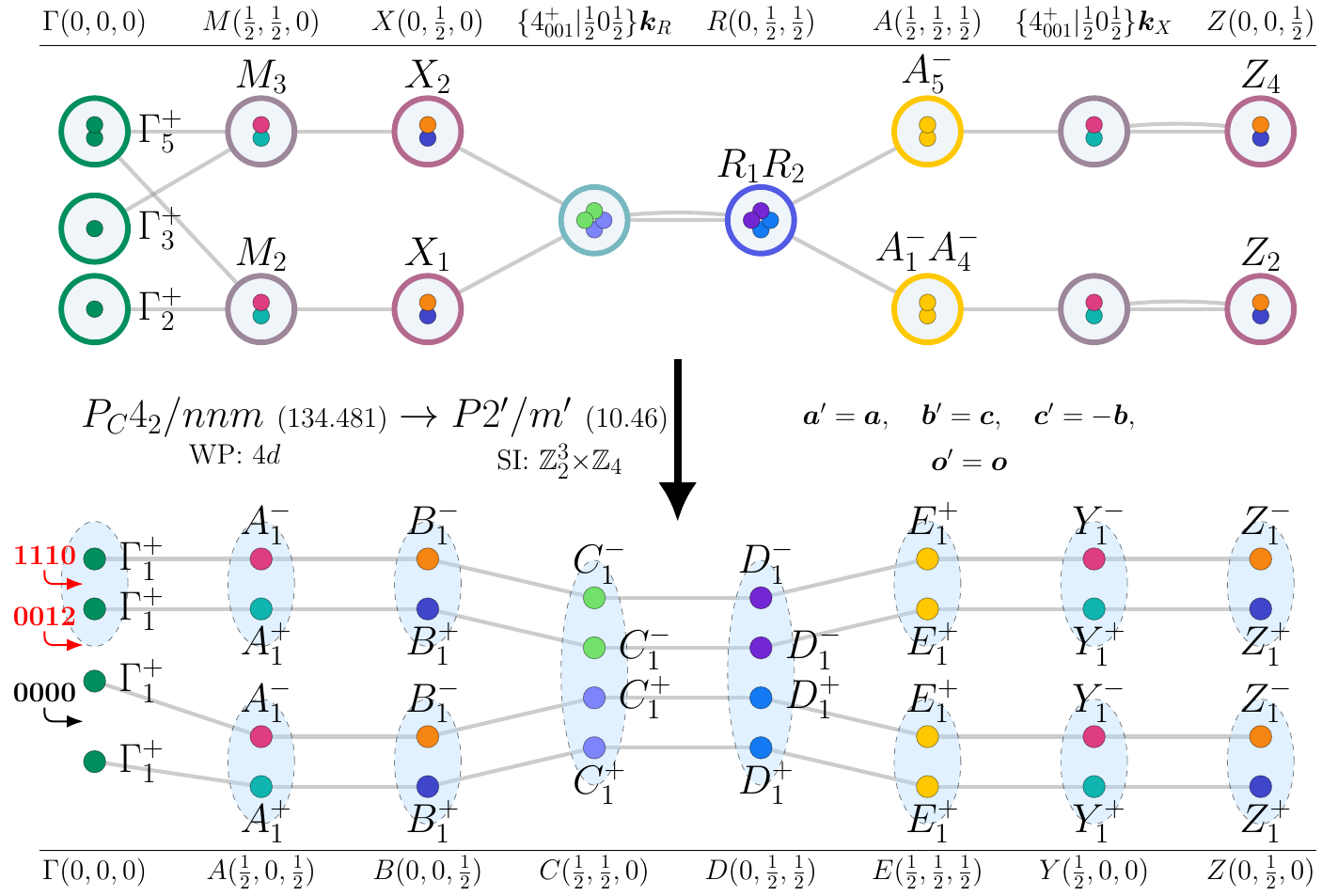}
\caption{Topological magnon bands in subgroup $P2'/m'~(10.46)$ for magnetic moments on Wyckoff position $4d$ of supergroup $P_{C}4_{2}/nnm~(134.481)$.\label{fig_134.481_10.46_Bparallel100andstrainparallel110_4d}}
\end{figure}
\input{gap_tables_tex/134.481_10.46_Bparallel100andstrainparallel110_4d_table.tex}
\input{si_tables_tex/134.481_10.46_Bparallel100andstrainparallel110_4d_table.tex}
\subsubsection{Topological bands in subgroup $P2_{1}'/c'~(14.79)$}
\textbf{Perturbations:}
\begin{itemize}
\item B $\parallel$ [001] and strain $\perp$ [100],
\item B $\perp$ [100].
\end{itemize}
\begin{figure}[H]
\centering
\includegraphics[scale=0.6]{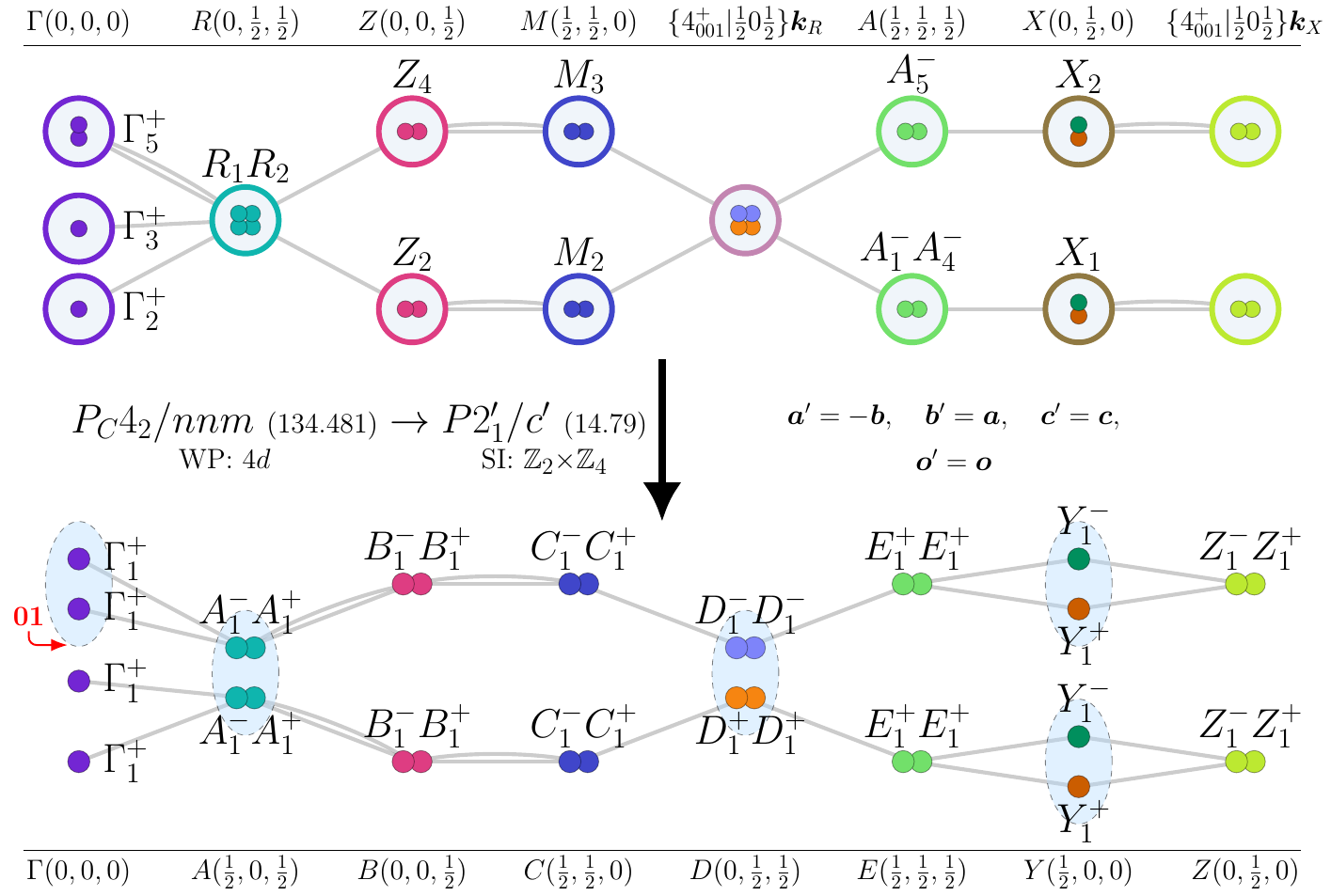}
\caption{Topological magnon bands in subgroup $P2_{1}'/c'~(14.79)$ for magnetic moments on Wyckoff position $4d$ of supergroup $P_{C}4_{2}/nnm~(134.481)$.\label{fig_134.481_14.79_Bparallel001andstrainperp100_4d}}
\end{figure}
\input{gap_tables_tex/134.481_14.79_Bparallel001andstrainperp100_4d_table.tex}
\input{si_tables_tex/134.481_14.79_Bparallel001andstrainperp100_4d_table.tex}
\subsubsection{Topological bands in subgroup $P_{S}\bar{1}~(2.7)$}
\textbf{Perturbation:}
\begin{itemize}
\item strain in generic direction.
\end{itemize}
\begin{figure}[H]
\centering
\includegraphics[scale=0.6]{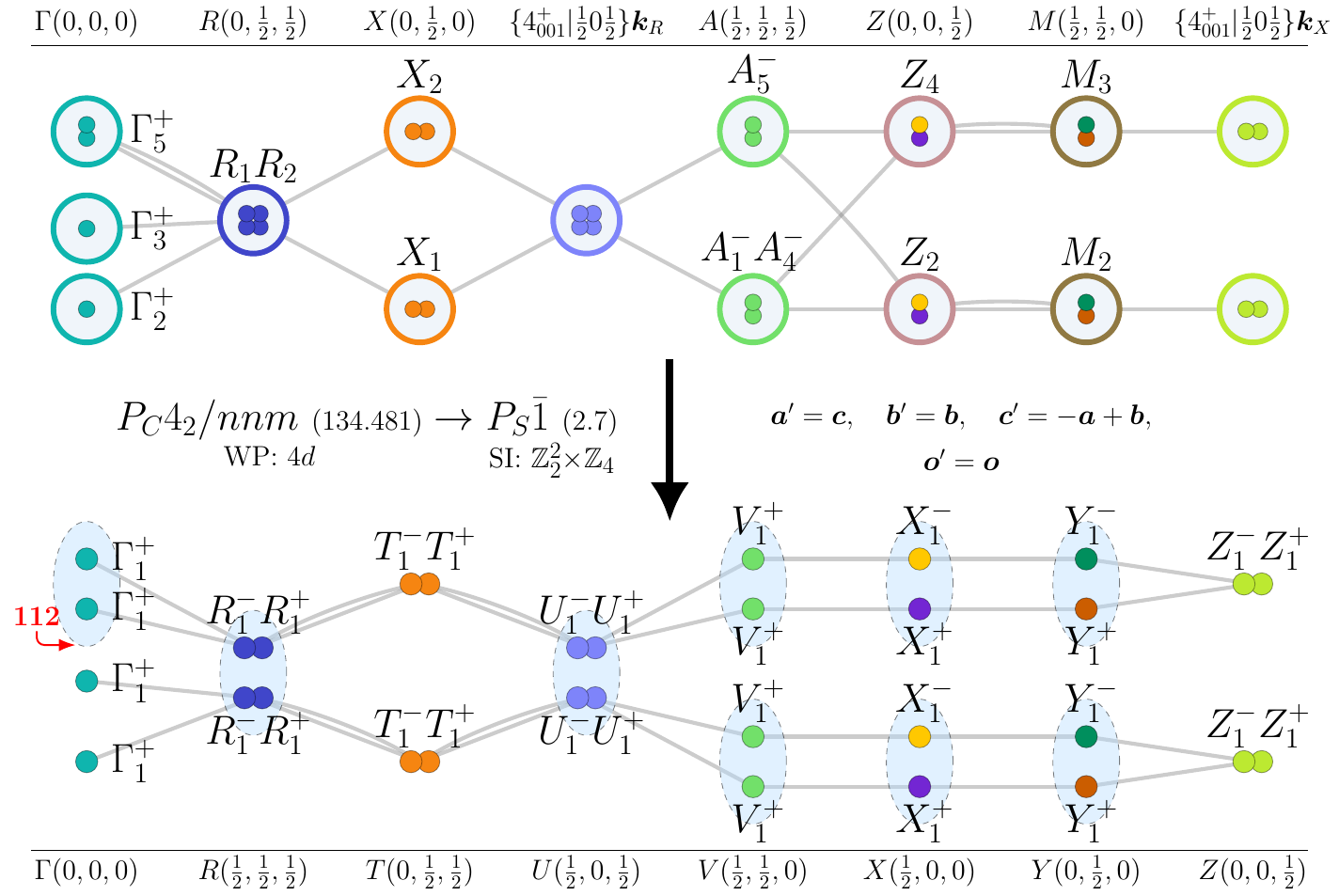}
\caption{Topological magnon bands in subgroup $P_{S}\bar{1}~(2.7)$ for magnetic moments on Wyckoff position $4d$ of supergroup $P_{C}4_{2}/nnm~(134.481)$.\label{fig_134.481_2.7_strainingenericdirection_4d}}
\end{figure}
\input{gap_tables_tex/134.481_2.7_strainingenericdirection_4d_table.tex}
\input{si_tables_tex/134.481_2.7_strainingenericdirection_4d_table.tex}
\subsection{WP: $4d+8i$}
\textbf{BCS Materials:} {Pr\textsubscript{2}CuO\textsubscript{4}~(284 K)}\footnote{BCS web page: \texttt{\href{http://webbdcrista1.ehu.es/magndata/index.php?this\_label=2.79} {http://webbdcrista1.ehu.es/magndata/index.php?this\_label=2.79}}}.\\
\subsubsection{Topological bands in subgroup $P2_{1}'/c'~(14.79)$}
\textbf{Perturbations:}
\begin{itemize}
\item B $\parallel$ [001] and strain $\perp$ [100],
\item B $\perp$ [100].
\end{itemize}
\begin{figure}[H]
\centering
\includegraphics[scale=0.6]{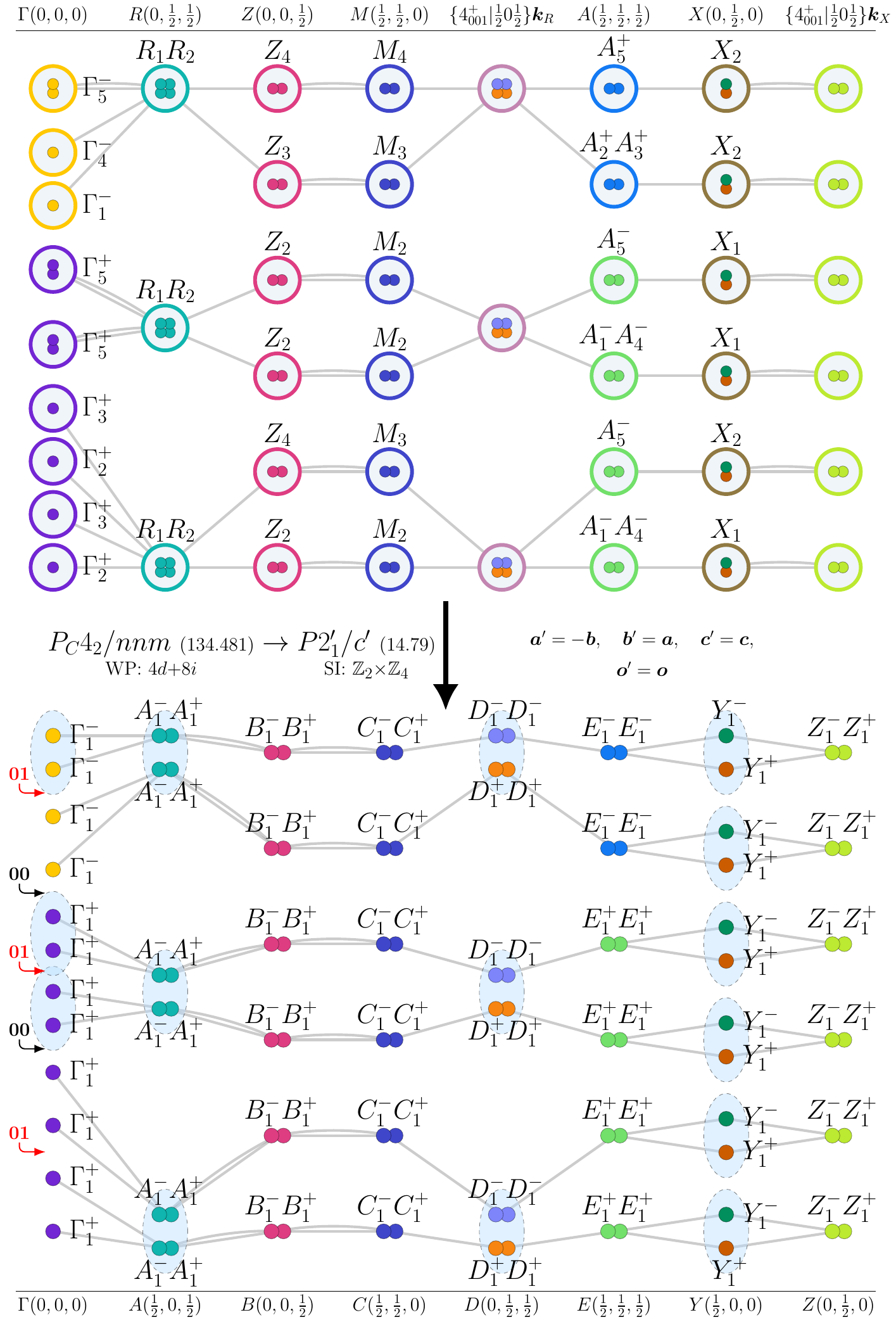}
\caption{Topological magnon bands in subgroup $P2_{1}'/c'~(14.79)$ for magnetic moments on Wyckoff positions $4d+8i$ of supergroup $P_{C}4_{2}/nnm~(134.481)$.\label{fig_134.481_14.79_Bparallel001andstrainperp100_4d+8i}}
\end{figure}
\input{gap_tables_tex/134.481_14.79_Bparallel001andstrainperp100_4d+8i_table.tex}
\input{si_tables_tex/134.481_14.79_Bparallel001andstrainperp100_4d+8i_table.tex}
\subsubsection{Topological bands in subgroup $P_{S}\bar{1}~(2.7)$}
\textbf{Perturbation:}
\begin{itemize}
\item strain in generic direction.
\end{itemize}
\begin{figure}[H]
\centering
\includegraphics[scale=0.6]{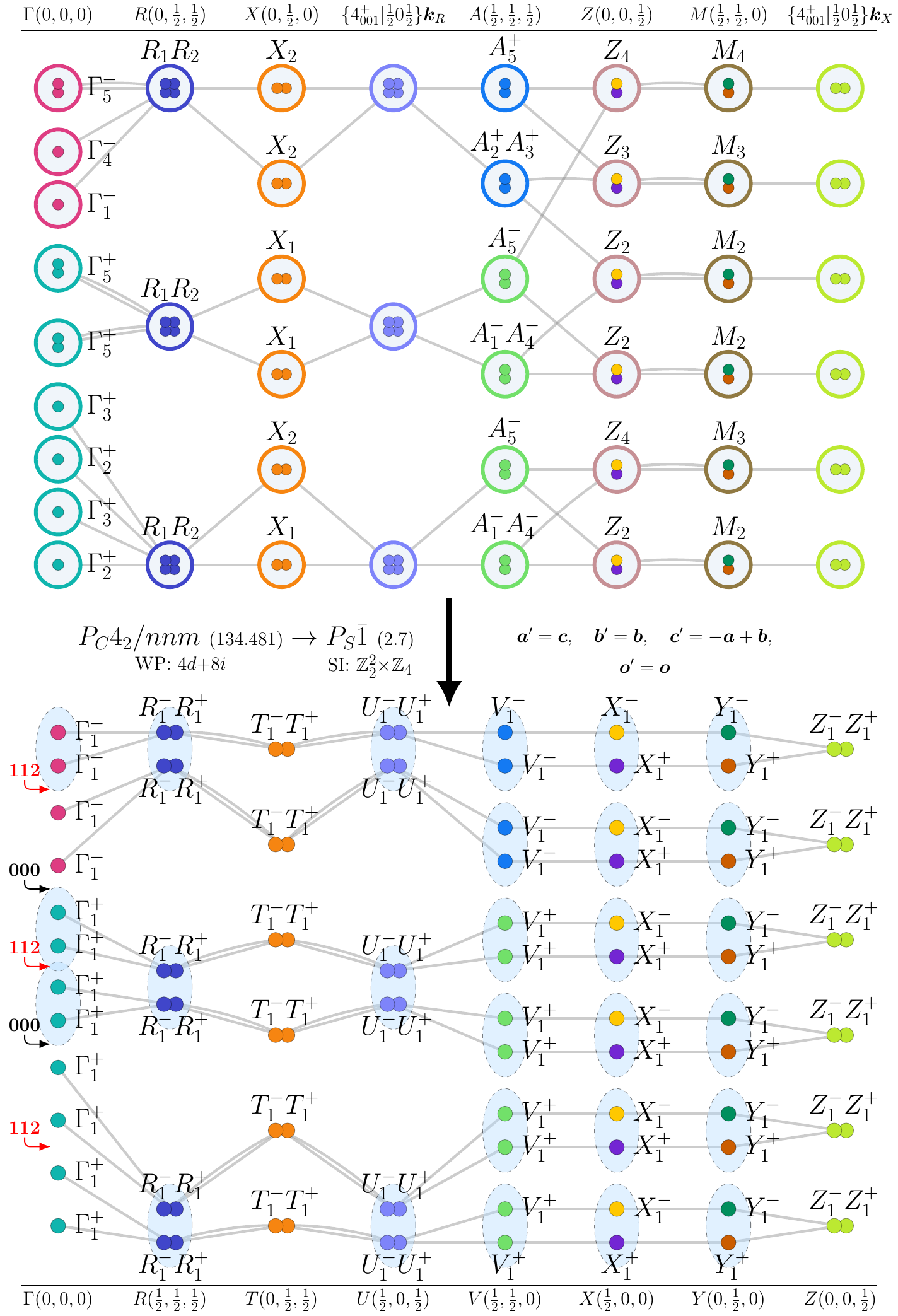}
\caption{Topological magnon bands in subgroup $P_{S}\bar{1}~(2.7)$ for magnetic moments on Wyckoff positions $4d+8i$ of supergroup $P_{C}4_{2}/nnm~(134.481)$.\label{fig_134.481_2.7_strainingenericdirection_4d+8i}}
\end{figure}
\input{gap_tables_tex/134.481_2.7_strainingenericdirection_4d+8i_table.tex}
\input{si_tables_tex/134.481_2.7_strainingenericdirection_4d+8i_table.tex}
\subsection{WP: $4c$}
\textbf{BCS Materials:} {Nd\textsubscript{2}CuO\textsubscript{4}~(245 K)}\footnote{BCS web page: \texttt{\href{http://webbdcrista1.ehu.es/magndata/index.php?this\_label=2.78} {http://webbdcrista1.ehu.es/magndata/index.php?this\_label=2.78}}}, {UP~(22.5 K)}\footnote{BCS web page: \texttt{\href{http://webbdcrista1.ehu.es/magndata/index.php?this\_label=2.13} {http://webbdcrista1.ehu.es/magndata/index.php?this\_label=2.13}}}.\\
\subsubsection{Topological bands in subgroup $P\bar{1}~(2.4)$}
\textbf{Perturbations:}
\begin{itemize}
\item B $\parallel$ [001] and strain in generic direction,
\item B $\parallel$ [100] and strain $\perp$ [110],
\item B $\parallel$ [100] and strain in generic direction,
\item B $\parallel$ [110] and strain $\perp$ [100],
\item B $\parallel$ [110] and strain in generic direction,
\item B $\perp$ [001] and strain $\perp$ [100],
\item B $\perp$ [001] and strain $\perp$ [110],
\item B $\perp$ [001] and strain in generic direction,
\item B $\perp$ [100] and strain $\parallel$ [110],
\item B $\perp$ [100] and strain $\perp$ [001],
\item B $\perp$ [100] and strain $\perp$ [110],
\item B $\perp$ [100] and strain in generic direction,
\item B $\perp$ [110] and strain $\parallel$ [100],
\item B $\perp$ [110] and strain $\perp$ [001],
\item B $\perp$ [110] and strain $\perp$ [100],
\item B $\perp$ [110] and strain in generic direction,
\item B in generic direction.
\end{itemize}
\begin{figure}[H]
\centering
\includegraphics[scale=0.6]{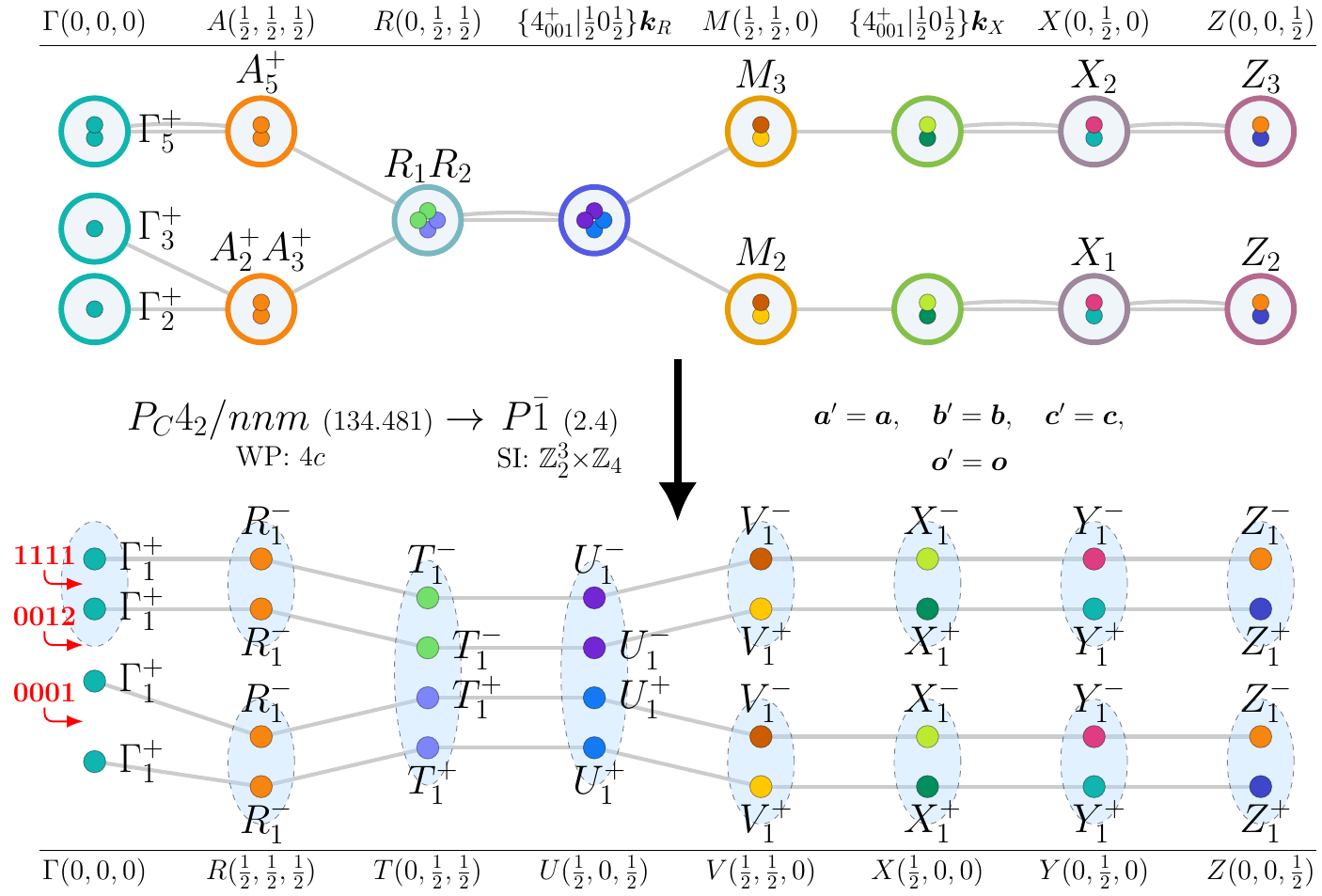}
\caption{Topological magnon bands in subgroup $P\bar{1}~(2.4)$ for magnetic moments on Wyckoff position $4c$ of supergroup $P_{C}4_{2}/nnm~(134.481)$.\label{fig_134.481_2.4_Bparallel001andstrainingenericdirection_4c}}
\end{figure}
\input{gap_tables_tex/134.481_2.4_Bparallel001andstrainingenericdirection_4c_table.tex}
\input{si_tables_tex/134.481_2.4_Bparallel001andstrainingenericdirection_4c_table.tex}
\subsubsection{Topological bands in subgroup $C2'/m'~(12.62)$}
\textbf{Perturbations:}
\begin{itemize}
\item B $\parallel$ [001] and strain $\perp$ [110],
\item B $\perp$ [110].
\end{itemize}
\begin{figure}[H]
\centering
\includegraphics[scale=0.6]{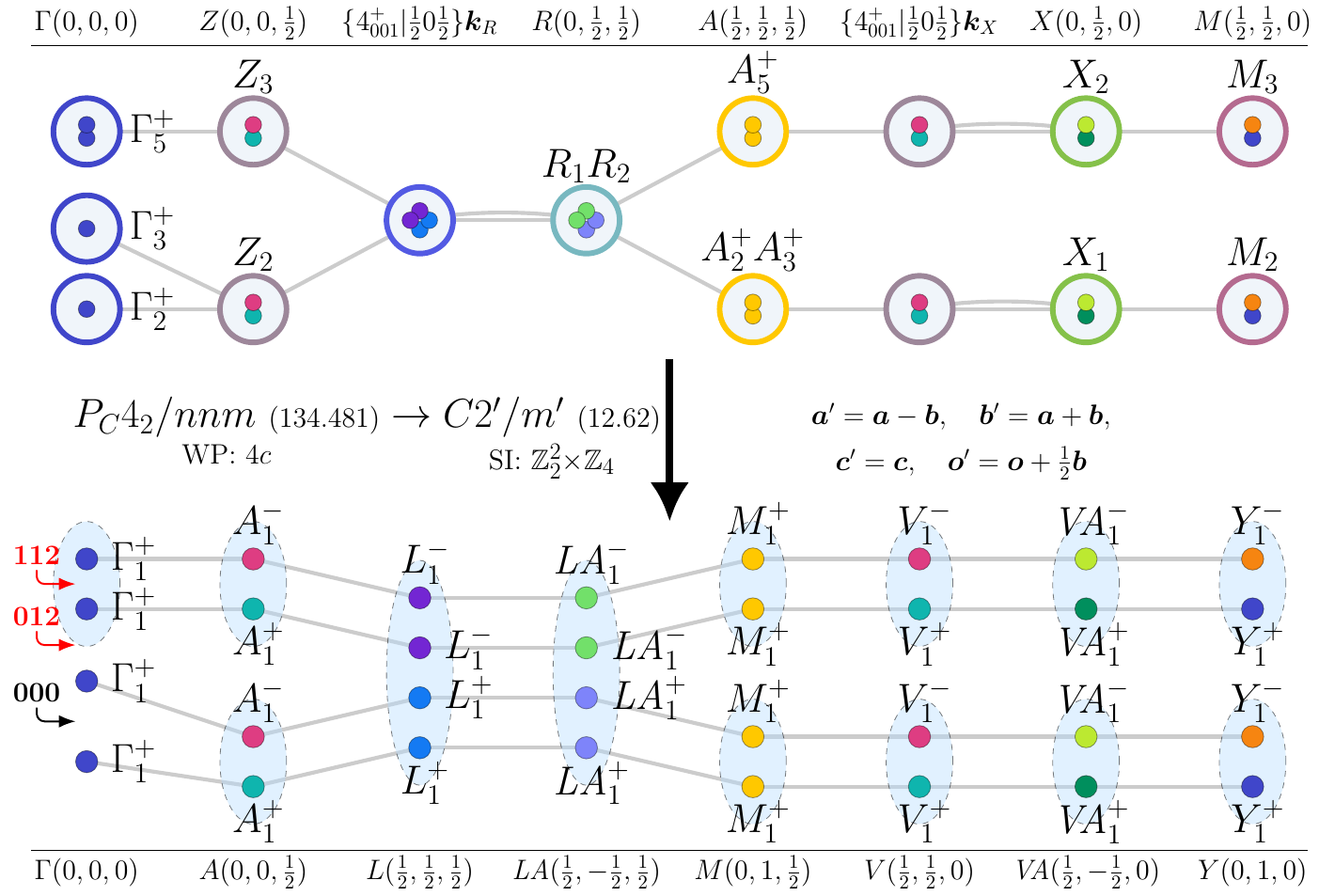}
\caption{Topological magnon bands in subgroup $C2'/m'~(12.62)$ for magnetic moments on Wyckoff position $4c$ of supergroup $P_{C}4_{2}/nnm~(134.481)$.\label{fig_134.481_12.62_Bparallel001andstrainperp110_4c}}
\end{figure}
\input{gap_tables_tex/134.481_12.62_Bparallel001andstrainperp110_4c_table.tex}
\input{si_tables_tex/134.481_12.62_Bparallel001andstrainperp110_4c_table.tex}
\subsubsection{Topological bands in subgroup $P2'/m'~(10.46)$}
\textbf{Perturbations:}
\begin{itemize}
\item B $\parallel$ [100] and strain $\parallel$ [110],
\item B $\parallel$ [100] and strain $\perp$ [001],
\item B $\parallel$ [110] and strain $\parallel$ [100],
\item B $\parallel$ [110] and strain $\perp$ [001],
\item B $\perp$ [001].
\end{itemize}
\begin{figure}[H]
\centering
\includegraphics[scale=0.6]{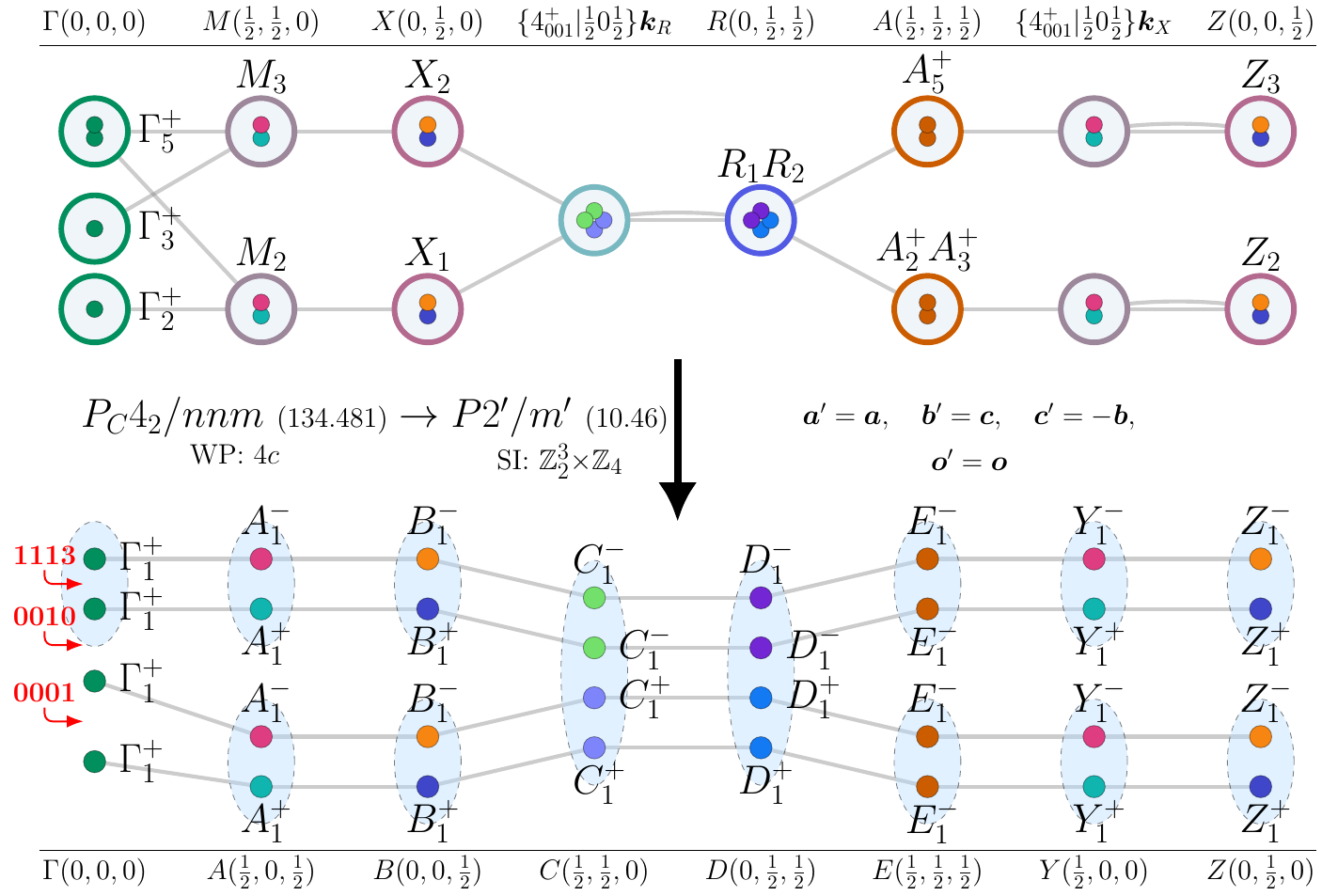}
\caption{Topological magnon bands in subgroup $P2'/m'~(10.46)$ for magnetic moments on Wyckoff position $4c$ of supergroup $P_{C}4_{2}/nnm~(134.481)$.\label{fig_134.481_10.46_Bparallel100andstrainparallel110_4c}}
\end{figure}
\input{gap_tables_tex/134.481_10.46_Bparallel100andstrainparallel110_4c_table.tex}
\input{si_tables_tex/134.481_10.46_Bparallel100andstrainparallel110_4c_table.tex}
\subsubsection{Topological bands in subgroup $P2_{1}'/c'~(14.79)$}
\textbf{Perturbations:}
\begin{itemize}
\item B $\parallel$ [001] and strain $\perp$ [100],
\item B $\perp$ [100].
\end{itemize}
\begin{figure}[H]
\centering
\includegraphics[scale=0.6]{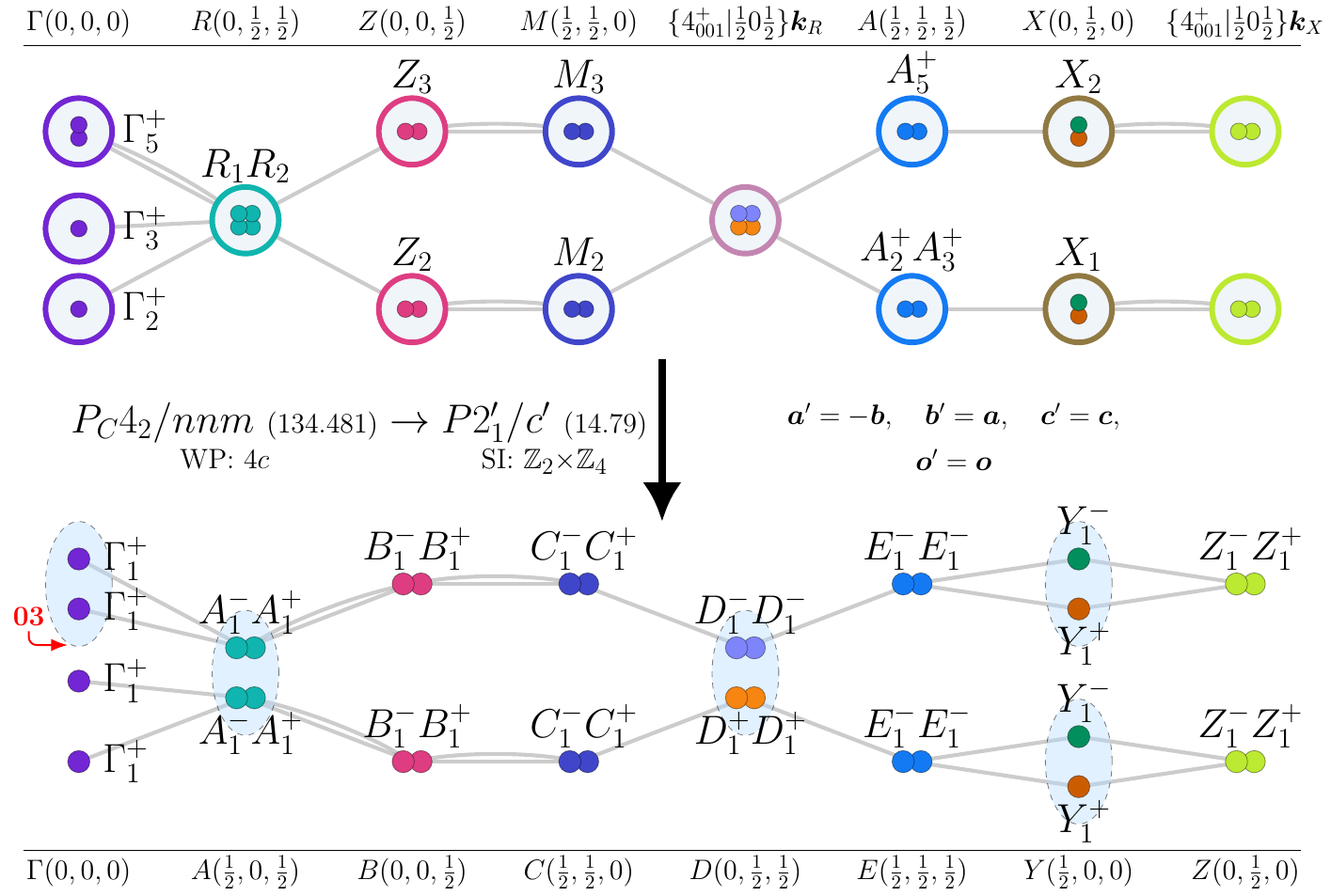}
\caption{Topological magnon bands in subgroup $P2_{1}'/c'~(14.79)$ for magnetic moments on Wyckoff position $4c$ of supergroup $P_{C}4_{2}/nnm~(134.481)$.\label{fig_134.481_14.79_Bparallel001andstrainperp100_4c}}
\end{figure}
\input{gap_tables_tex/134.481_14.79_Bparallel001andstrainperp100_4c_table.tex}
\input{si_tables_tex/134.481_14.79_Bparallel001andstrainperp100_4c_table.tex}
\subsubsection{Topological bands in subgroup $P_{S}\bar{1}~(2.7)$}
\textbf{Perturbation:}
\begin{itemize}
\item strain in generic direction.
\end{itemize}
\begin{figure}[H]
\centering
\includegraphics[scale=0.6]{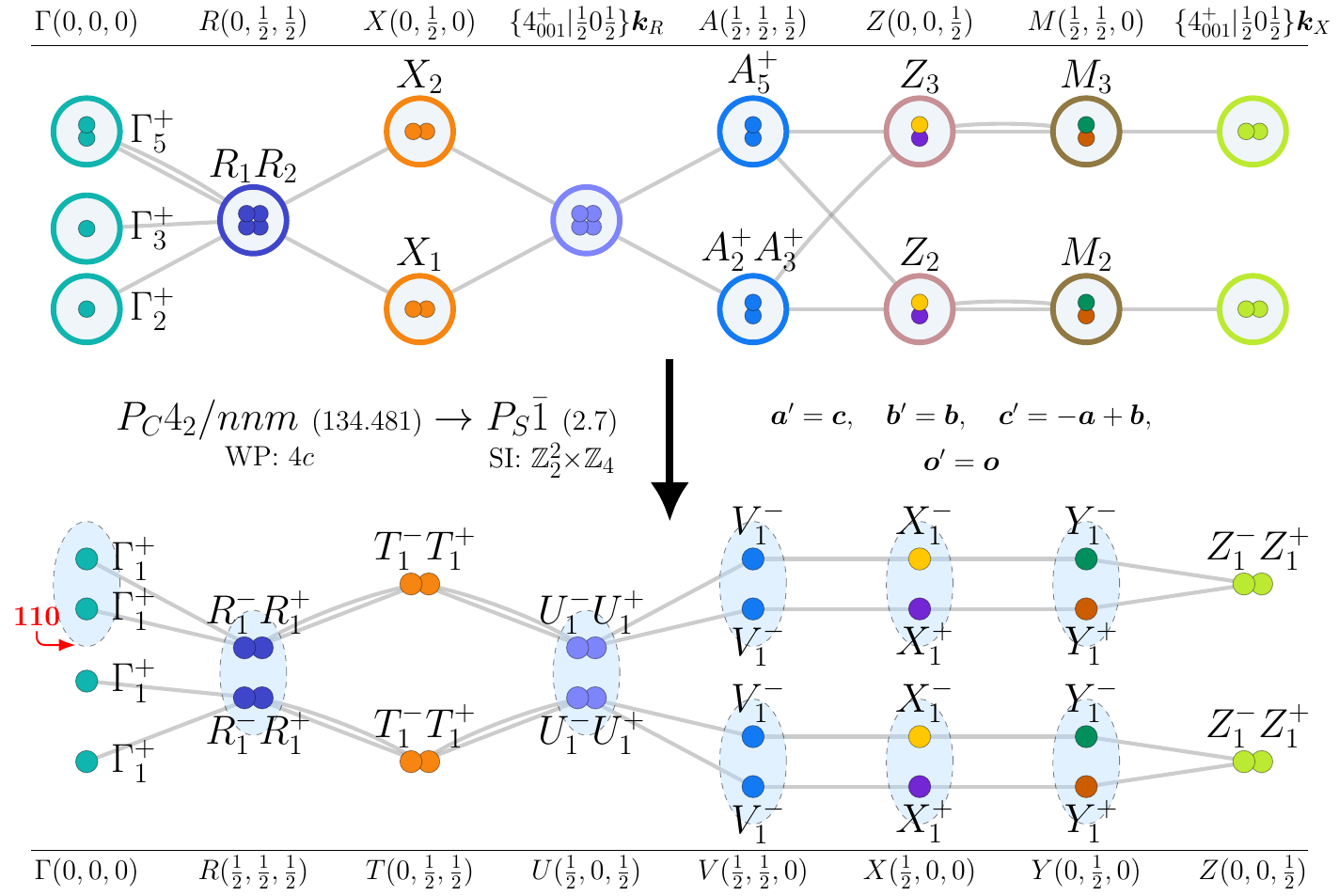}
\caption{Topological magnon bands in subgroup $P_{S}\bar{1}~(2.7)$ for magnetic moments on Wyckoff position $4c$ of supergroup $P_{C}4_{2}/nnm~(134.481)$.\label{fig_134.481_2.7_strainingenericdirection_4c}}
\end{figure}
\input{gap_tables_tex/134.481_2.7_strainingenericdirection_4c_table.tex}
\input{si_tables_tex/134.481_2.7_strainingenericdirection_4c_table.tex}

\section{MSG $P_{c}4_{2}/mbc~(135.492)$}
\textbf{Nontrivial-SI Subgroups:} $P\bar{1}~(2.4)$, $C2'/m'~(12.62)$, $P2_{1}'/m'~(11.54)$, $P2_{1}'/c'~(14.79)$, $P_{S}\bar{1}~(2.7)$, $C2/c~(15.85)$, $Cm'cm'~(63.464)$, $C_{c}2/c~(15.90)$, $P2~(3.1)$, $Cm'm'2~(35.168)$, $P_{b}2~(3.5)$, $C_{c}cc2~(37.184)$, $P_{c}ba2~(32.139)$, $P2/m~(10.42)$, $Cm'm'm~(65.485)$, $Pn'n'm~(58.397)$, $P_{b}2/m~(10.48)$, $C_{c}ccm~(66.498)$, $P2_{1}/c~(14.75)$, $Pn'm'a~(62.446)$, $P_{a}2_{1}/c~(14.80)$, $P4_{2}n'm'~(102.191)$, $P4_{2}/mn'm'~(136.501)$.\\

\textbf{Trivial-SI Subgroups:} $Cm'~(8.34)$, $Pm'~(6.20)$, $Pc'~(7.26)$, $C2'~(5.15)$, $P2_{1}'~(4.9)$, $P2_{1}'~(4.9)$, $P_{S}1~(1.3)$, $Cc~(9.37)$, $Cm'c2_{1}'~(36.174)$, $C_{c}c~(9.40)$, $Pm~(6.18)$, $Amm'2'~(38.190)$, $Pmn'2_{1}'~(31.126)$, $P_{b}m~(6.22)$, $Pc~(7.24)$, $Pn'a2_{1}'~(33.146)$, $P_{a}c~(7.27)$, $C2~(5.13)$, $Am'm'2~(38.191)$, $C_{c}2~(5.16)$, $A_{a}ma2~(40.208)$, $Pn'n'2~(34.159)$, $P2_{1}~(4.7)$, $Pm'n'2_{1}~(31.127)$, $P_{a}2_{1}~(4.10)$, $P_{a}mc2_{1}~(26.71)$, $P_{c}bam~(55.361)$, $P_{c}4_{2}bc~(106.224)$.\\

\subsection{WP: $8g$}
\textbf{BCS Materials:} {U\textsubscript{2}Rh\textsubscript{2}Sn~(28 K)}\footnote{BCS web page: \texttt{\href{http://webbdcrista1.ehu.es/magndata/index.php?this\_label=1.103} {http://webbdcrista1.ehu.es/magndata/index.php?this\_label=1.103}}}, {U\textsubscript{2}Ni\textsubscript{2}Sn~(25 K)}\footnote{BCS web page: \texttt{\href{http://webbdcrista1.ehu.es/magndata/index.php?this\_label=1.479} {http://webbdcrista1.ehu.es/magndata/index.php?this\_label=1.479}}}.\\
\subsubsection{Topological bands in subgroup $P2_{1}'/m'~(11.54)$}
\textbf{Perturbations:}
\begin{itemize}
\item B $\parallel$ [100] and strain $\parallel$ [110],
\item B $\parallel$ [100] and strain $\perp$ [001],
\item B $\parallel$ [110] and strain $\parallel$ [100],
\item B $\parallel$ [110] and strain $\perp$ [001],
\item B $\perp$ [001].
\end{itemize}
\begin{figure}[H]
\centering
\includegraphics[scale=0.6]{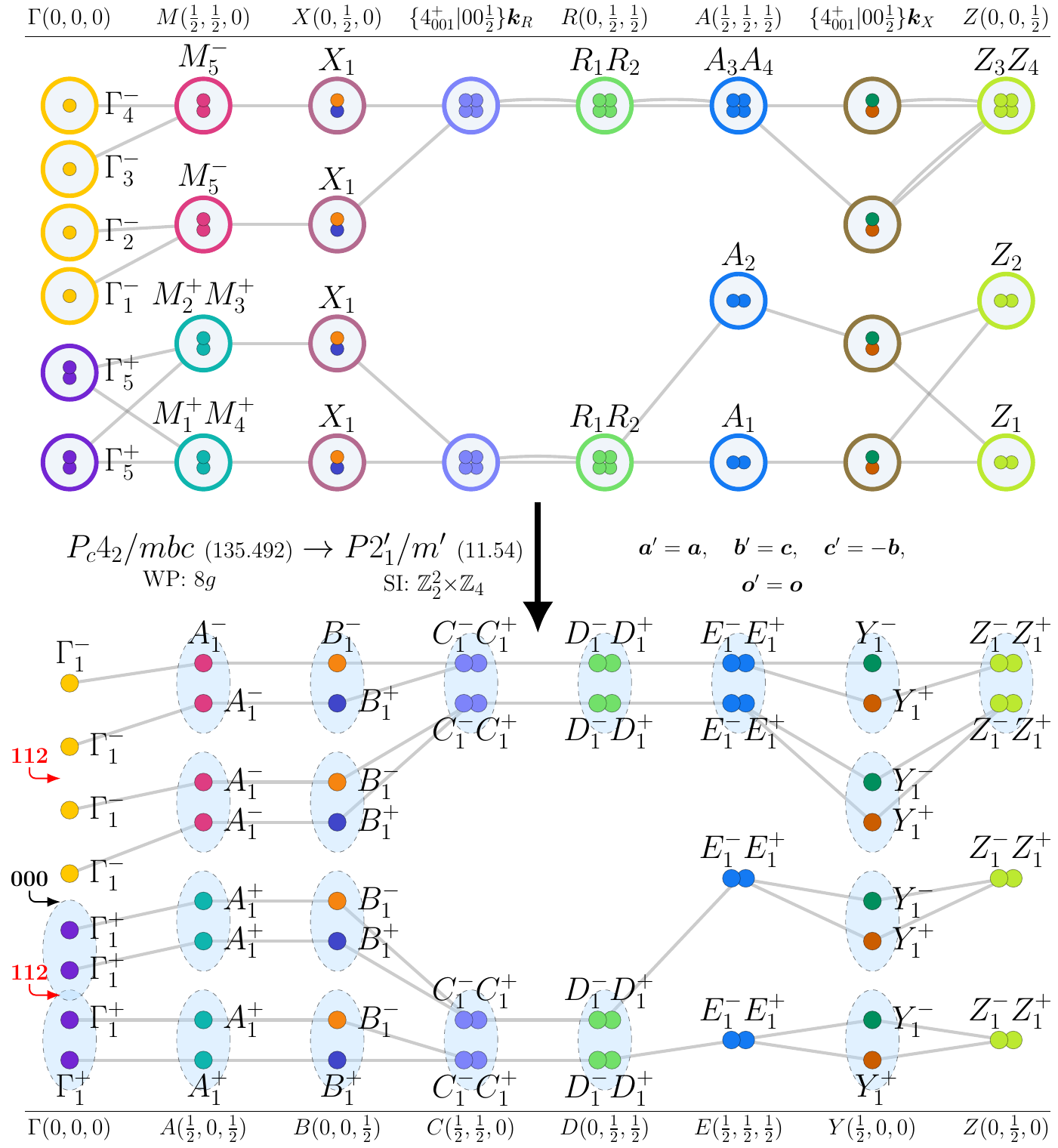}
\caption{Topological magnon bands in subgroup $P2_{1}'/m'~(11.54)$ for magnetic moments on Wyckoff position $8g$ of supergroup $P_{c}4_{2}/mbc~(135.492)$.\label{fig_135.492_11.54_Bparallel100andstrainparallel110_8g}}
\end{figure}
\input{gap_tables_tex/135.492_11.54_Bparallel100andstrainparallel110_8g_table.tex}
\input{si_tables_tex/135.492_11.54_Bparallel100andstrainparallel110_8g_table.tex}
\subsubsection{Topological bands in subgroup $P_{S}\bar{1}~(2.7)$}
\textbf{Perturbation:}
\begin{itemize}
\item strain in generic direction.
\end{itemize}
\begin{figure}[H]
\centering
\includegraphics[scale=0.6]{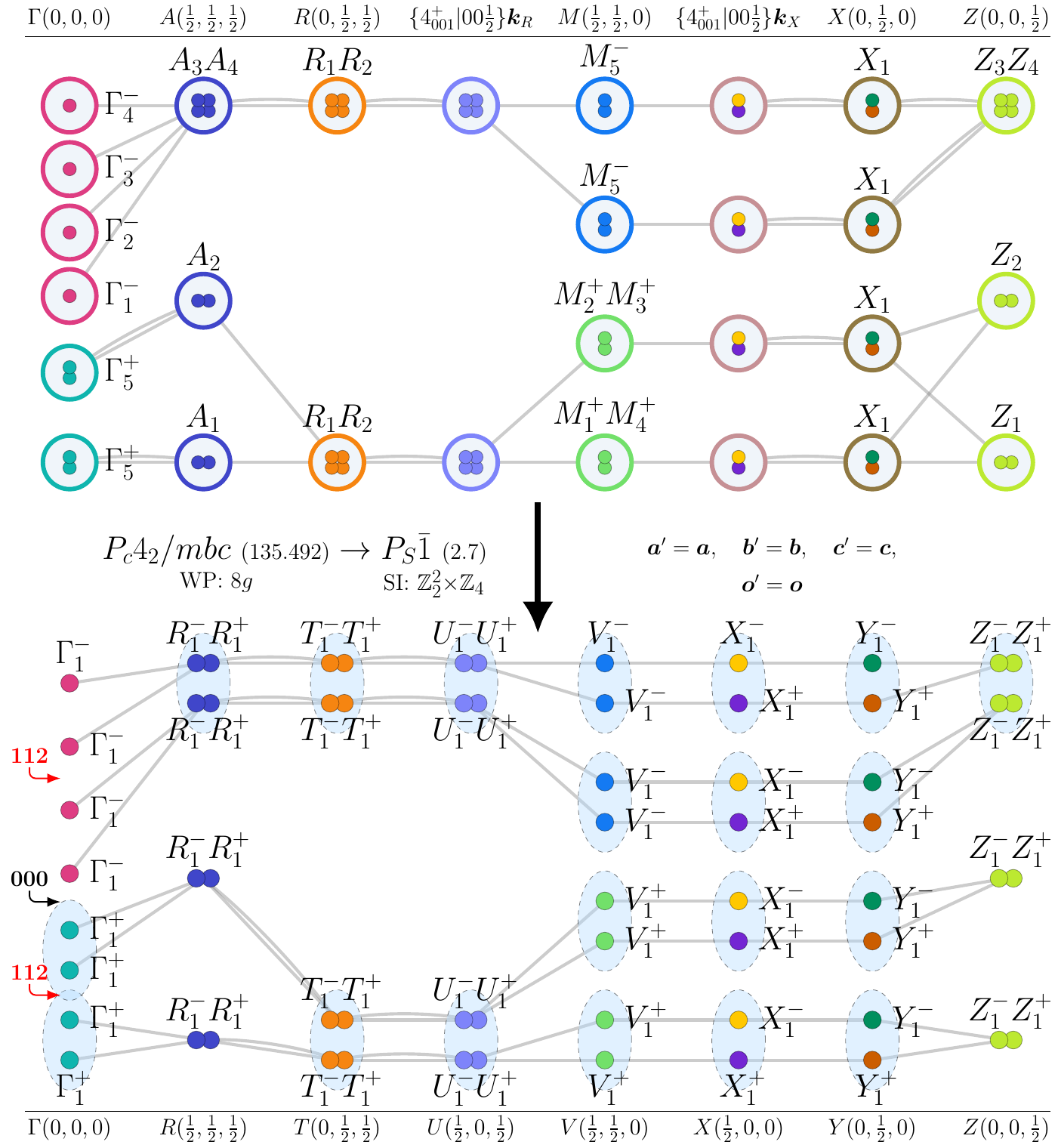}
\caption{Topological magnon bands in subgroup $P_{S}\bar{1}~(2.7)$ for magnetic moments on Wyckoff position $8g$ of supergroup $P_{c}4_{2}/mbc~(135.492)$.\label{fig_135.492_2.7_strainingenericdirection_8g}}
\end{figure}
\input{gap_tables_tex/135.492_2.7_strainingenericdirection_8g_table.tex}
\input{si_tables_tex/135.492_2.7_strainingenericdirection_8g_table.tex}

\section{MSG $P_{I}4_{2}/mnm~(136.506)$}
\textbf{Nontrivial-SI Subgroups:} $P\bar{1}~(2.4)$, $C2'/c'~(15.89)$, $P2_{1}'/c'~(14.79)$, $P2'/m'~(10.46)$, $P_{S}\bar{1}~(2.7)$, $Ab'a'2~(41.215)$, $C2/m~(12.58)$, $Cmc'a'~(64.475)$, $C_{c}2/m~(12.63)$, $P2~(3.1)$, $Cc'c'2~(37.183)$, $Pm'm'2~(25.60)$, $P2/m~(10.42)$, $Cc'c'm~(66.495)$, $Pm'm'm~(47.252)$, $P_{C}2/m~(10.49)$, $P2_{1}/c~(14.75)$, $Pnn'm'~(58.398)$, $P_{A}2_{1}/c~(14.83)$, $P_{I}nnm~(58.404)$, $P4_{2}m'c'~(105.215)$, $P4_{2}/mm'c'~(131.441)$.\\

\textbf{Trivial-SI Subgroups:} $Cc'~(9.39)$, $Pc'~(7.26)$, $Pm'~(6.20)$, $C2'~(5.15)$, $P2_{1}'~(4.9)$, $P2'~(3.3)$, $P_{S}1~(1.3)$, $Cm~(8.32)$, $Cmc'2_{1}'~(36.175)$, $C_{c}m~(8.35)$, $Pm~(6.18)$, $Ama'2'~(40.206)$, $Pm'm2'~(25.59)$, $P_{C}m~(6.23)$, $Pc~(7.24)$, $Pm'n2_{1}'~(31.125)$, $P_{A}c~(7.31)$, $C2~(5.13)$, $C_{c}2~(5.16)$, $A_{B}mm2~(38.194)$, $P_{C}2~(3.6)$, $C_{A}mm2~(35.171)$, $P_{I}nn2~(34.164)$, $C_{A}mmm~(65.490)$, $P2_{1}~(4.7)$, $Pm'n'2_{1}~(31.127)$, $P_{C}2_{1}~(4.12)$, $P_{I}mn2_{1}~(31.134)$, $P_{I}4_{2}nm~(102.194)$.\\

\subsection{WP: $4d$}
\textbf{BCS Materials:} {Sr\textsubscript{2}Cr\textsubscript{3}As\textsubscript{2}O\textsubscript{2}~(590 K)}\footnote{BCS web page: \texttt{\href{http://webbdcrista1.ehu.es/magndata/index.php?this\_label=1.461} {http://webbdcrista1.ehu.es/magndata/index.php?this\_label=1.461}}}.\\
\subsubsection{Topological bands in subgroup $Cc'c'2~(37.183)$}
\textbf{Perturbation:}
\begin{itemize}
\item E $\parallel$ [001] and B $\parallel$ [001] and strain $\parallel$ [110].
\end{itemize}
\begin{figure}[H]
\centering
\includegraphics[scale=0.6]{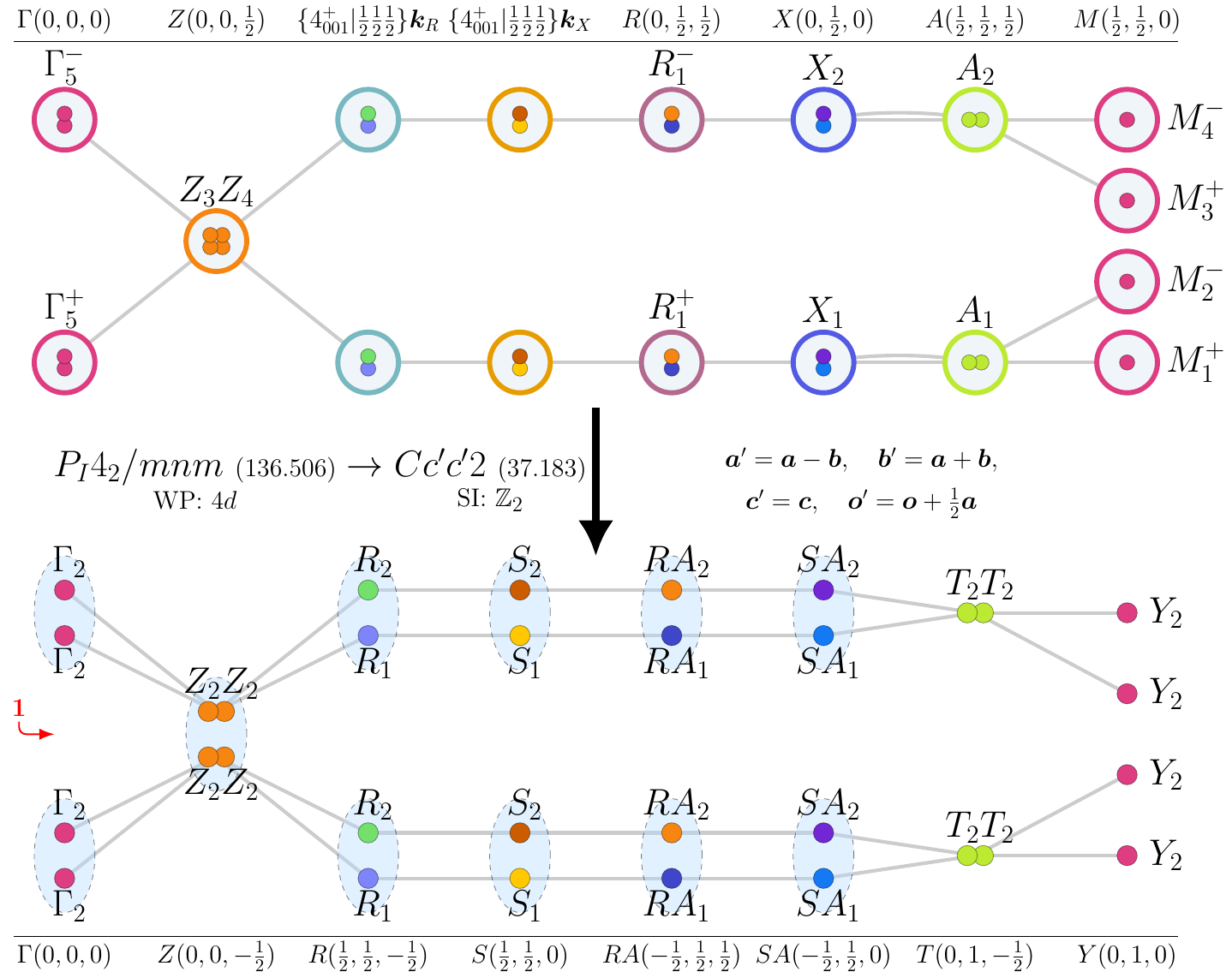}
\caption{Topological magnon bands in subgroup $Cc'c'2~(37.183)$ for magnetic moments on Wyckoff position $4d$ of supergroup $P_{I}4_{2}/mnm~(136.506)$.\label{fig_136.506_37.183_Eparallel001andBparallel001andstrainparallel110_4d}}
\end{figure}
\input{gap_tables_tex/136.506_37.183_Eparallel001andBparallel001andstrainparallel110_4d_table.tex}
\input{si_tables_tex/136.506_37.183_Eparallel001andBparallel001andstrainparallel110_4d_table.tex}
\subsection{WP: $4d+4d$}
\textbf{BCS Materials:} {Ni\textsubscript{1.64}Co\textsubscript{0.36}Mn\textsubscript{1.28}Ga\textsubscript{0.72}~(450 K)}\footnote{BCS web page: \texttt{\href{http://webbdcrista1.ehu.es/magndata/index.php?this\_label=1.198} {http://webbdcrista1.ehu.es/magndata/index.php?this\_label=1.198}}}.\\
\subsubsection{Topological bands in subgroup $Cc'c'2~(37.183)$}
\textbf{Perturbation:}
\begin{itemize}
\item E $\parallel$ [001] and B $\parallel$ [001] and strain $\parallel$ [110].
\end{itemize}
\begin{figure}[H]
\centering
\includegraphics[scale=0.6]{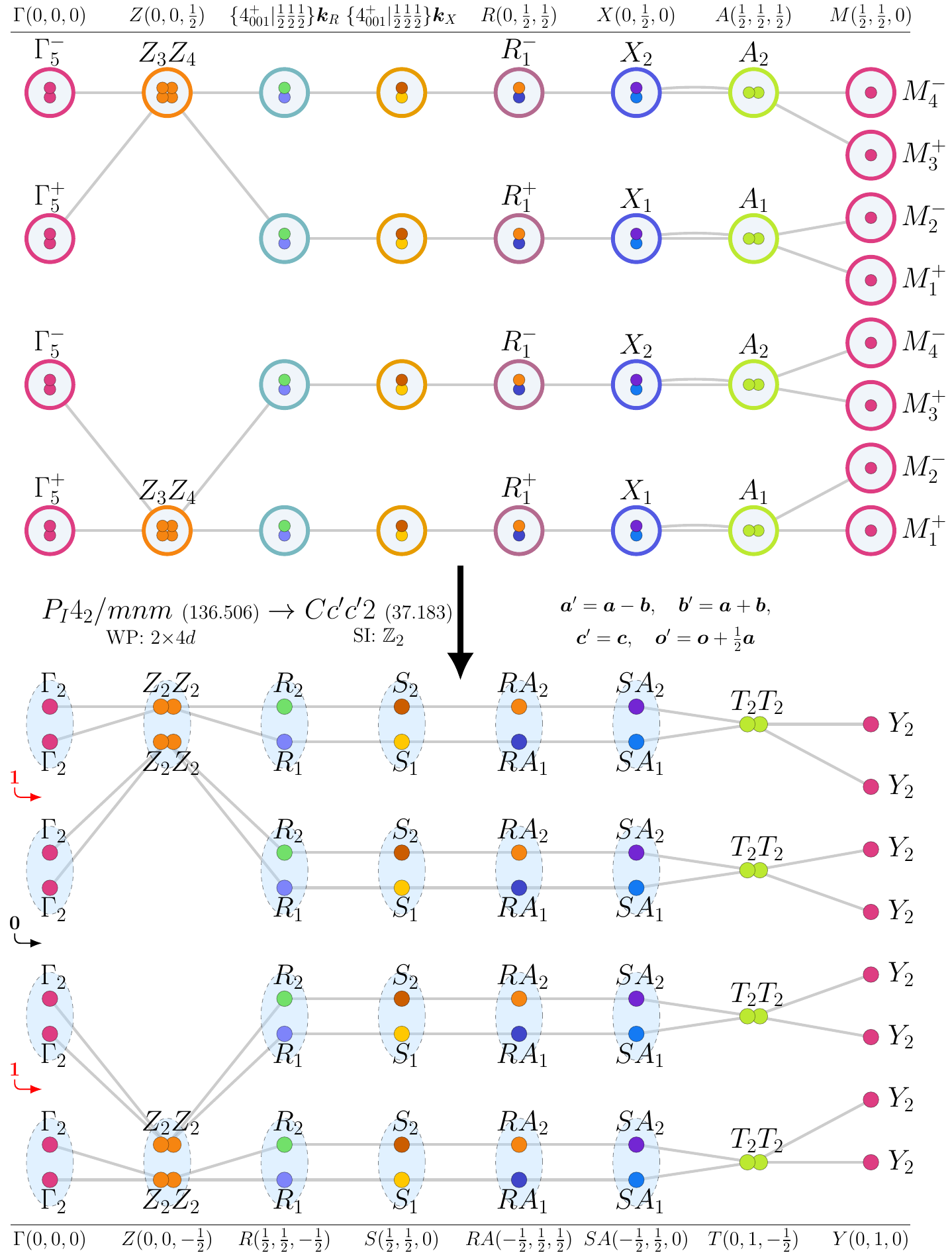}
\caption{Topological magnon bands in subgroup $Cc'c'2~(37.183)$ for magnetic moments on Wyckoff positions $4d+4d$ of supergroup $P_{I}4_{2}/mnm~(136.506)$.\label{fig_136.506_37.183_Eparallel001andBparallel001andstrainparallel110_4d+4d}}
\end{figure}
\input{gap_tables_tex/136.506_37.183_Eparallel001andBparallel001andstrainparallel110_4d+4d_table.tex}
\input{si_tables_tex/136.506_37.183_Eparallel001andBparallel001andstrainparallel110_4d+4d_table.tex}

\section{MSG $P_{C}2/c~(13.74)$}
\textbf{Nontrivial-SI Subgroups:} $P\bar{1}~(2.4)$, $P2_{1}'/c'~(14.79)$, $P_{S}\bar{1}~(2.7)$, $P2~(3.1)$, $P2/c~(13.65)$.\\

\textbf{Trivial-SI Subgroups:} $Pc'~(7.26)$, $P2_{1}'~(4.9)$, $P_{S}1~(1.3)$, $Pc~(7.24)$, $P_{C}c~(7.30)$, $P_{C}2~(3.6)$.\\

\subsection{WP: $4a$}
\textbf{BCS Materials:} {NaFeSO\textsubscript{4}F~(36 K)}\footnote{BCS web page: \texttt{\href{http://webbdcrista1.ehu.es/magndata/index.php?this\_label=1.121} {http://webbdcrista1.ehu.es/magndata/index.php?this\_label=1.121}}}, {NaCoSO\textsubscript{4}F~(29 K)}\footnote{BCS web page: \texttt{\href{http://webbdcrista1.ehu.es/magndata/index.php?this\_label=1.126} {http://webbdcrista1.ehu.es/magndata/index.php?this\_label=1.126}}}, {Sr\textsubscript{3}ZnIrO\textsubscript{6}~(17 K)}\footnote{BCS web page: \texttt{\href{http://webbdcrista1.ehu.es/magndata/index.php?this\_label=1.649} {http://webbdcrista1.ehu.es/magndata/index.php?this\_label=1.649}}}, {Ca\textsubscript{4}IrO\textsubscript{6}~(12 K)}\footnote{BCS web page: \texttt{\href{http://webbdcrista1.ehu.es/magndata/index.php?this\_label=1.114} {http://webbdcrista1.ehu.es/magndata/index.php?this\_label=1.114}}}.\\
\subsubsection{Topological bands in subgroup $P\bar{1}~(2.4)$}
\textbf{Perturbations:}
\begin{itemize}
\item B $\parallel$ [010] and strain in generic direction,
\item B $\perp$ [010] and strain in generic direction,
\item B in generic direction.
\end{itemize}
\begin{figure}[H]
\centering
\includegraphics[scale=0.6]{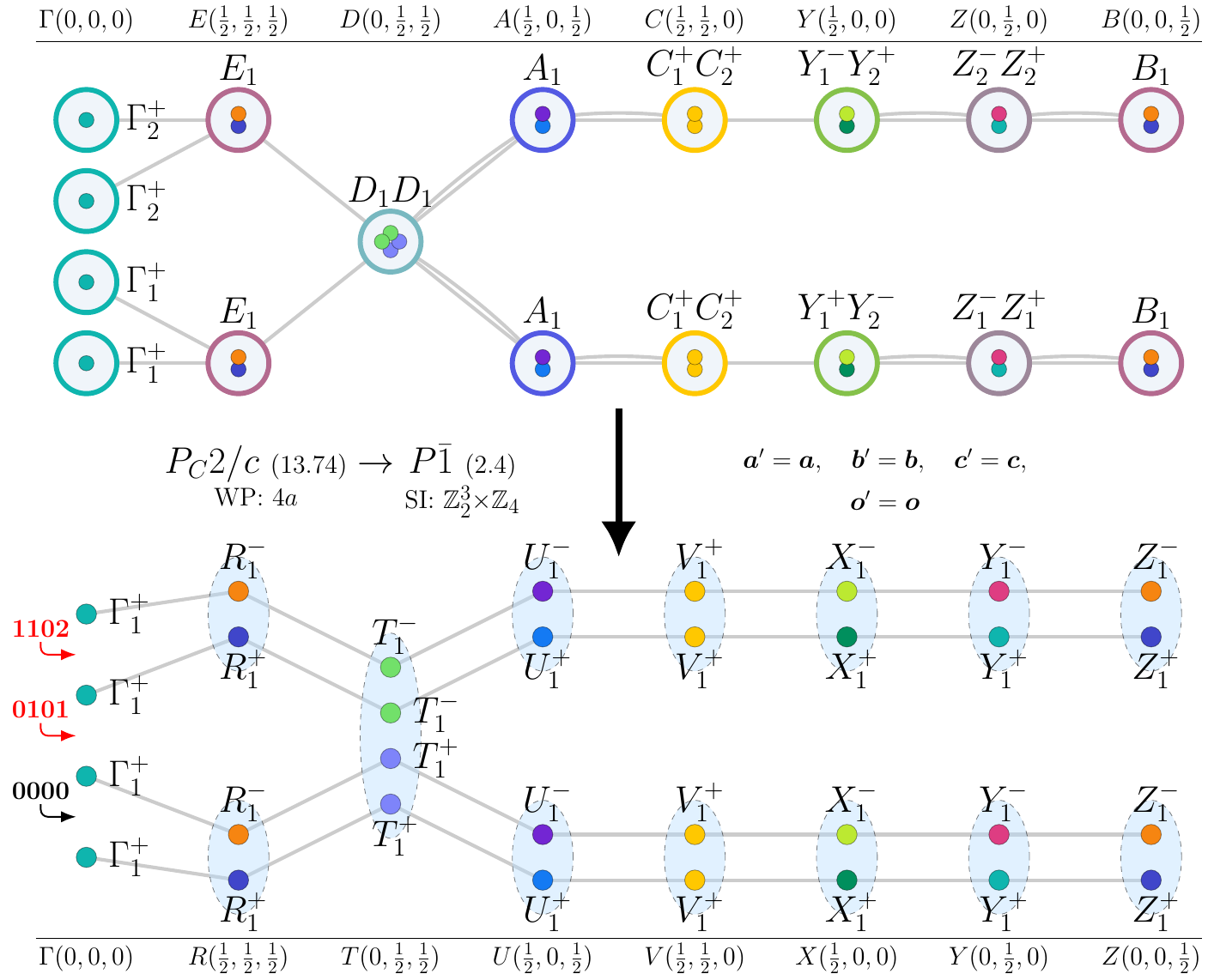}
\caption{Topological magnon bands in subgroup $P\bar{1}~(2.4)$ for magnetic moments on Wyckoff position $4a$ of supergroup $P_{C}2/c~(13.74)$.\label{fig_13.74_2.4_Bparallel010andstrainingenericdirection_4a}}
\end{figure}
\input{gap_tables_tex/13.74_2.4_Bparallel010andstrainingenericdirection_4a_table.tex}
\input{si_tables_tex/13.74_2.4_Bparallel010andstrainingenericdirection_4a_table.tex}
\subsubsection{Topological bands in subgroup $P2_{1}'/c'~(14.79)$}
\textbf{Perturbation:}
\begin{itemize}
\item B $\perp$ [010].
\end{itemize}
\begin{figure}[H]
\centering
\includegraphics[scale=0.6]{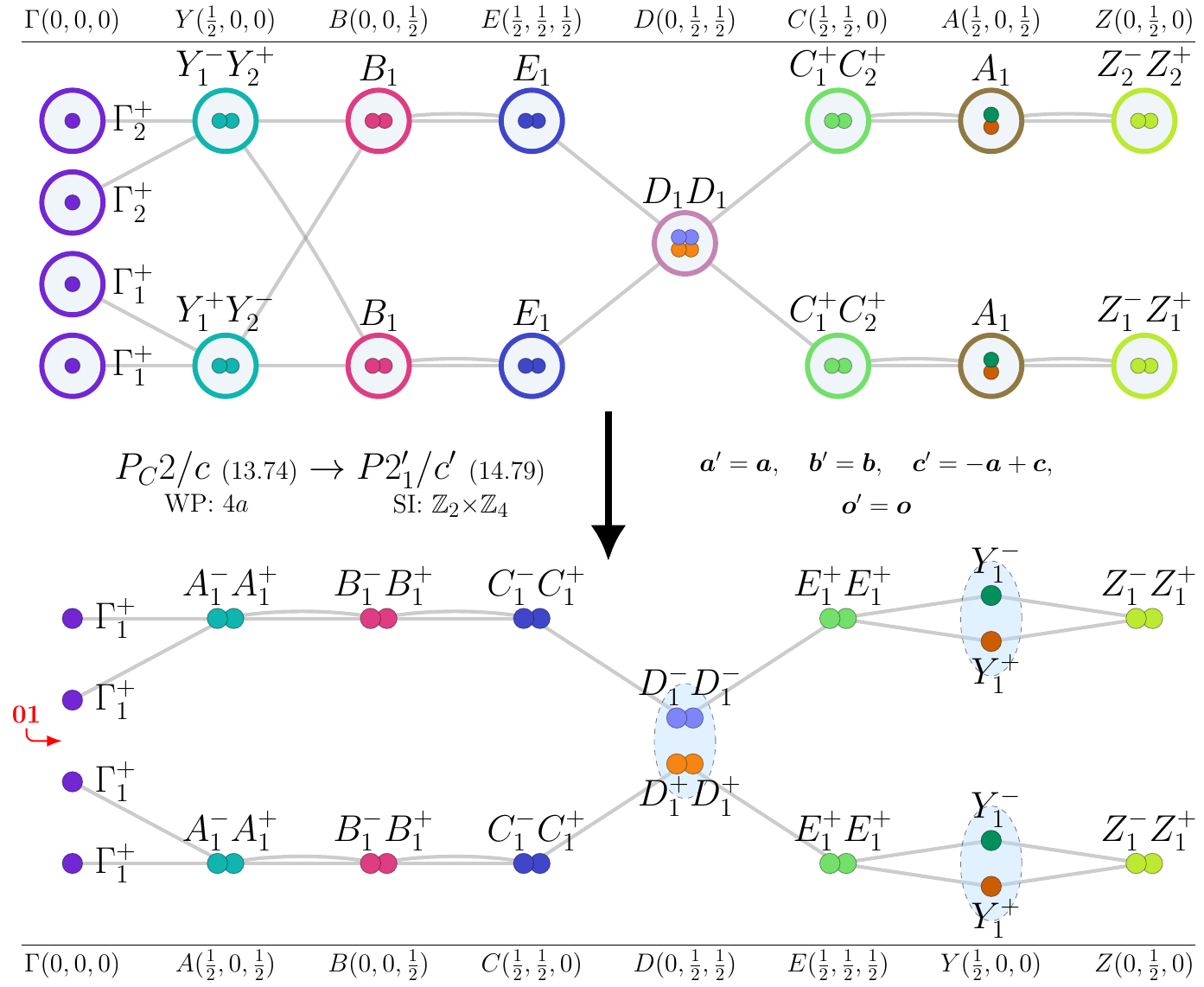}
\caption{Topological magnon bands in subgroup $P2_{1}'/c'~(14.79)$ for magnetic moments on Wyckoff position $4a$ of supergroup $P_{C}2/c~(13.74)$.\label{fig_13.74_14.79_Bperp010_4a}}
\end{figure}
\input{gap_tables_tex/13.74_14.79_Bperp010_4a_table.tex}
\input{si_tables_tex/13.74_14.79_Bperp010_4a_table.tex}
\subsubsection{Topological bands in subgroup $P_{S}\bar{1}~(2.7)$}
\textbf{Perturbation:}
\begin{itemize}
\item strain in generic direction.
\end{itemize}
\begin{figure}[H]
\centering
\includegraphics[scale=0.6]{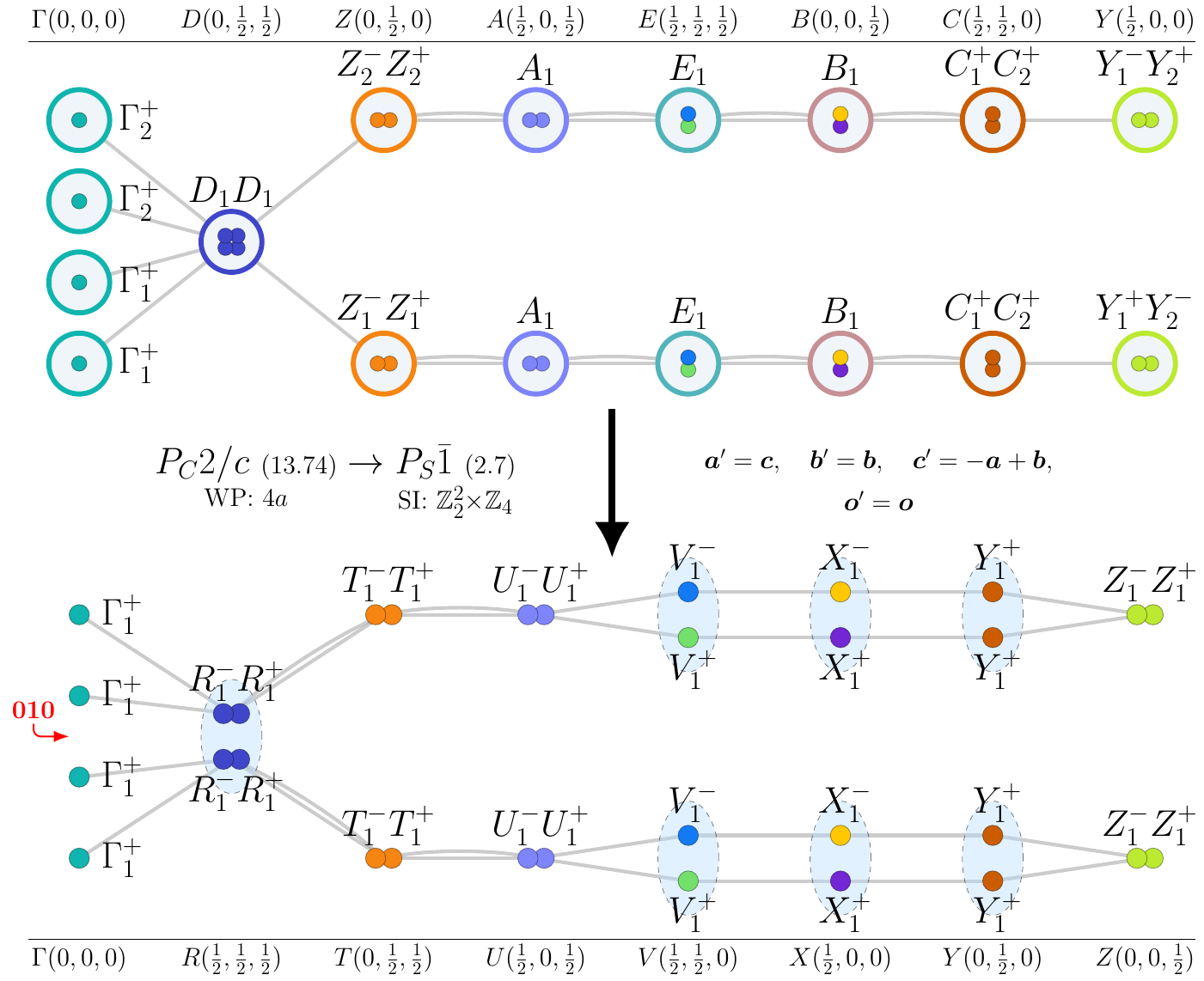}
\caption{Topological magnon bands in subgroup $P_{S}\bar{1}~(2.7)$ for magnetic moments on Wyckoff position $4a$ of supergroup $P_{C}2/c~(13.74)$.\label{fig_13.74_2.7_strainingenericdirection_4a}}
\end{figure}
\input{gap_tables_tex/13.74_2.7_strainingenericdirection_4a_table.tex}
\input{si_tables_tex/13.74_2.7_strainingenericdirection_4a_table.tex}
\subsection{WP: $8f$}
\textbf{BCS Materials:} {U\textsubscript{2}Rh\textsubscript{3}Si\textsubscript{5}~(26 K)}\footnote{BCS web page: \texttt{\href{http://webbdcrista1.ehu.es/magndata/index.php?this\_label=0.564} {http://webbdcrista1.ehu.es/magndata/index.php?this\_label=0.564}}}, {Cu\textsubscript{3}Ni\textsubscript{2}SbO\textsubscript{6}~(22.3 K)}\footnote{BCS web page: \texttt{\href{http://webbdcrista1.ehu.es/magndata/index.php?this\_label=1.259} {http://webbdcrista1.ehu.es/magndata/index.php?this\_label=1.259}}}.\\
\subsubsection{Topological bands in subgroup $P2_{1}'/c'~(14.79)$}
\textbf{Perturbation:}
\begin{itemize}
\item B $\perp$ [010].
\end{itemize}
\begin{figure}[H]
\centering
\includegraphics[scale=0.6]{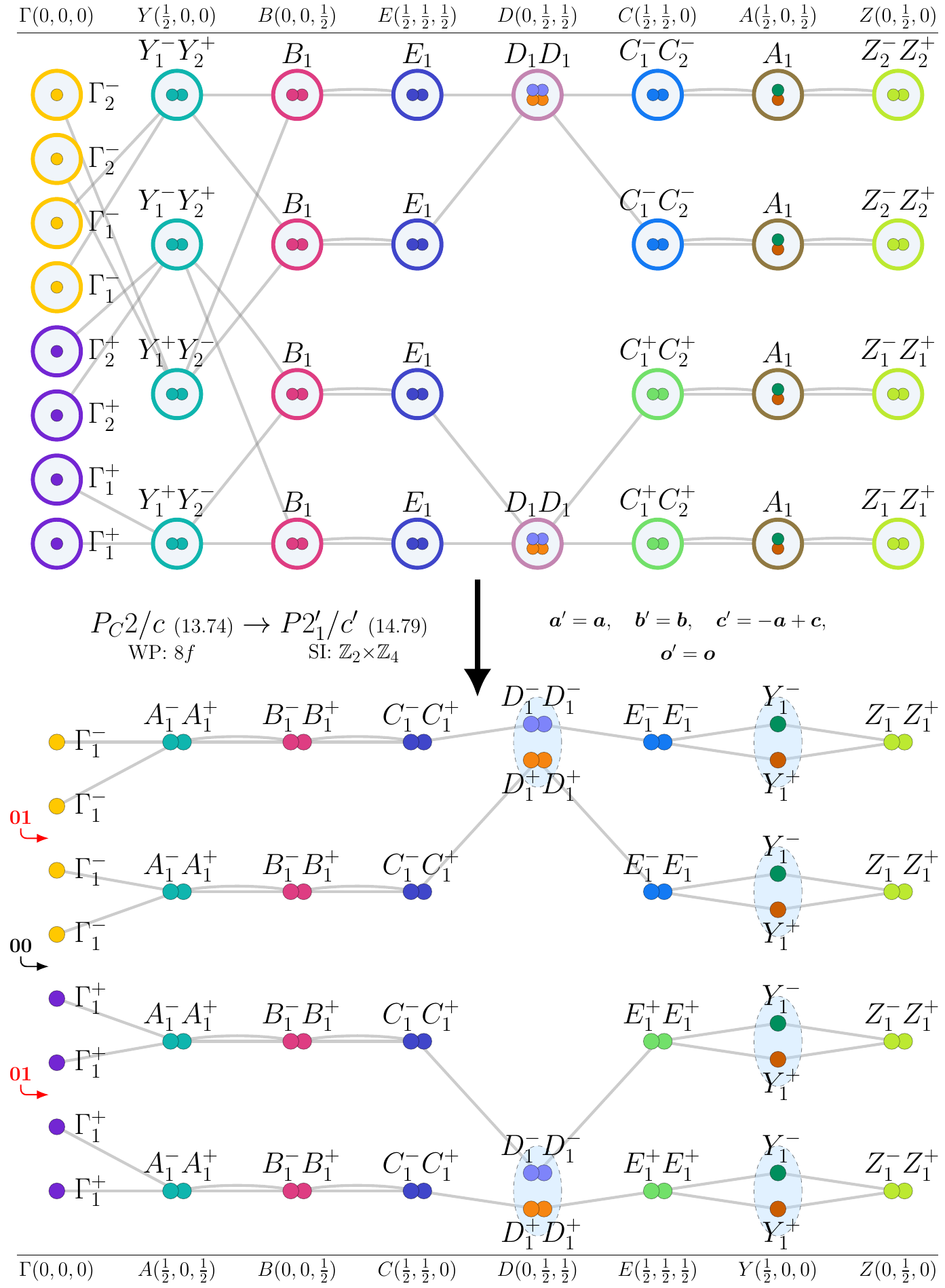}
\caption{Topological magnon bands in subgroup $P2_{1}'/c'~(14.79)$ for magnetic moments on Wyckoff position $8f$ of supergroup $P_{C}2/c~(13.74)$.\label{fig_13.74_14.79_Bperp010_8f}}
\end{figure}
\input{gap_tables_tex/13.74_14.79_Bperp010_8f_table.tex}
\input{si_tables_tex/13.74_14.79_Bperp010_8f_table.tex}
\subsubsection{Topological bands in subgroup $P_{S}\bar{1}~(2.7)$}
\textbf{Perturbation:}
\begin{itemize}
\item strain in generic direction.
\end{itemize}
\begin{figure}[H]
\centering
\includegraphics[scale=0.6]{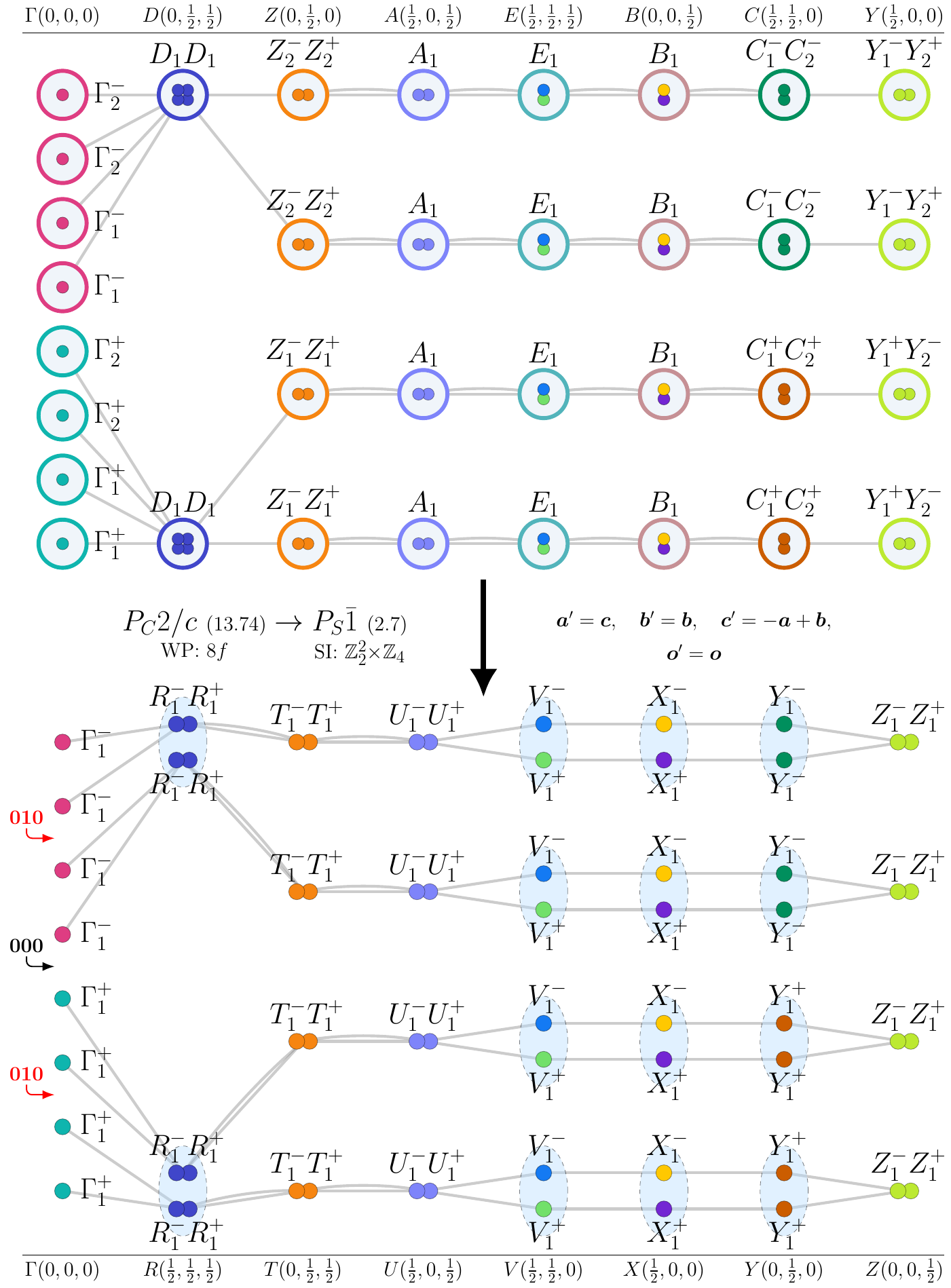}
\caption{Topological magnon bands in subgroup $P_{S}\bar{1}~(2.7)$ for magnetic moments on Wyckoff position $8f$ of supergroup $P_{C}2/c~(13.74)$.\label{fig_13.74_2.7_strainingenericdirection_8f}}
\end{figure}
\input{gap_tables_tex/13.74_2.7_strainingenericdirection_8f_table.tex}
\input{si_tables_tex/13.74_2.7_strainingenericdirection_8f_table.tex}
\subsection{WP: $4b$}
\textbf{BCS Materials:} {TlMnF\textsubscript{4}~(4.2 K)}\footnote{BCS web page: \texttt{\href{http://webbdcrista1.ehu.es/magndata/index.php?this\_label=1.346} {http://webbdcrista1.ehu.es/magndata/index.php?this\_label=1.346}}}.\\
\subsubsection{Topological bands in subgroup $P\bar{1}~(2.4)$}
\textbf{Perturbations:}
\begin{itemize}
\item B $\parallel$ [010] and strain in generic direction,
\item B $\perp$ [010] and strain in generic direction,
\item B in generic direction.
\end{itemize}
\begin{figure}[H]
\centering
\includegraphics[scale=0.6]{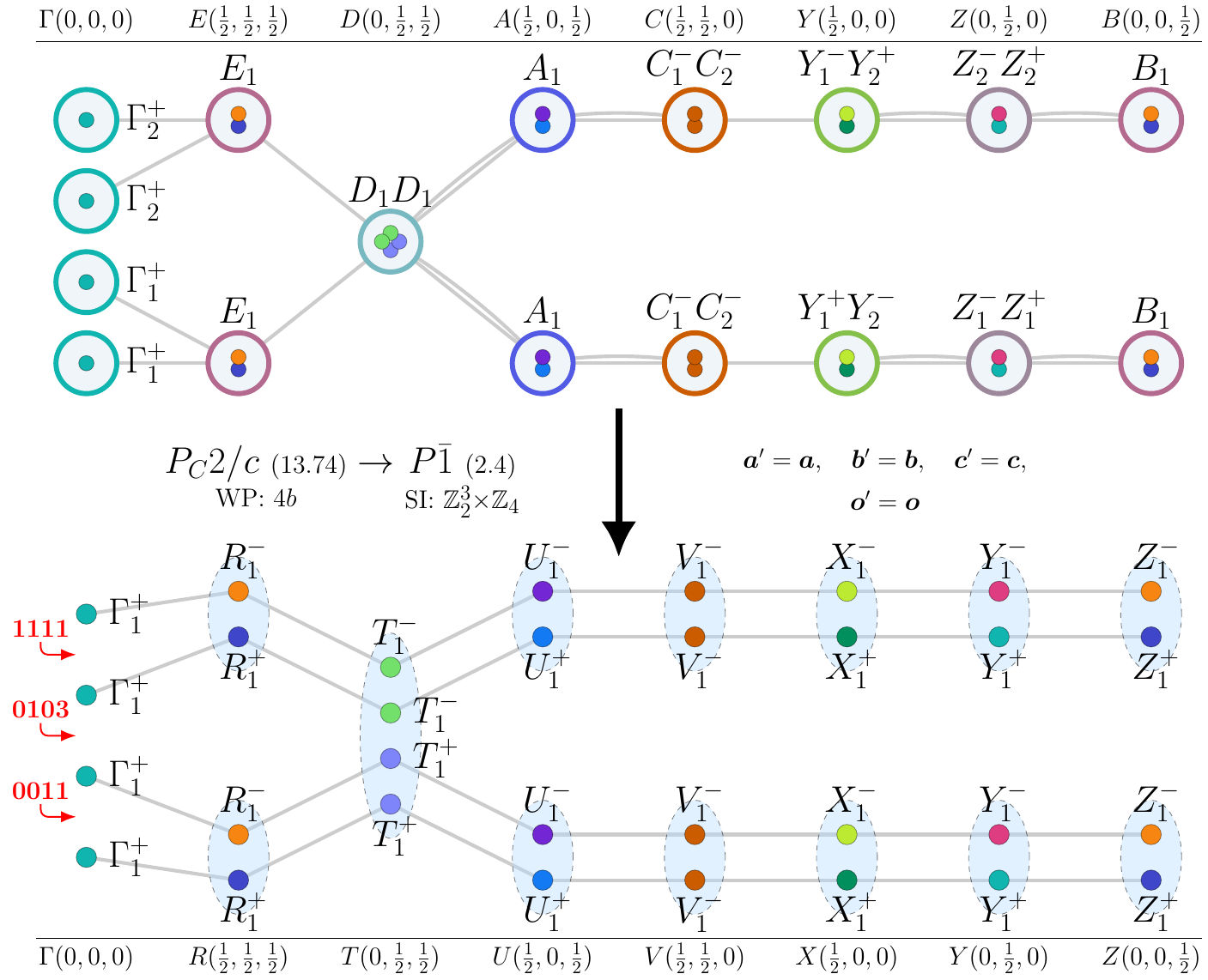}
\caption{Topological magnon bands in subgroup $P\bar{1}~(2.4)$ for magnetic moments on Wyckoff position $4b$ of supergroup $P_{C}2/c~(13.74)$.\label{fig_13.74_2.4_Bparallel010andstrainingenericdirection_4b}}
\end{figure}
\input{gap_tables_tex/13.74_2.4_Bparallel010andstrainingenericdirection_4b_table.tex}
\input{si_tables_tex/13.74_2.4_Bparallel010andstrainingenericdirection_4b_table.tex}
\subsubsection{Topological bands in subgroup $P2_{1}'/c'~(14.79)$}
\textbf{Perturbation:}
\begin{itemize}
\item B $\perp$ [010].
\end{itemize}
\begin{figure}[H]
\centering
\includegraphics[scale=0.6]{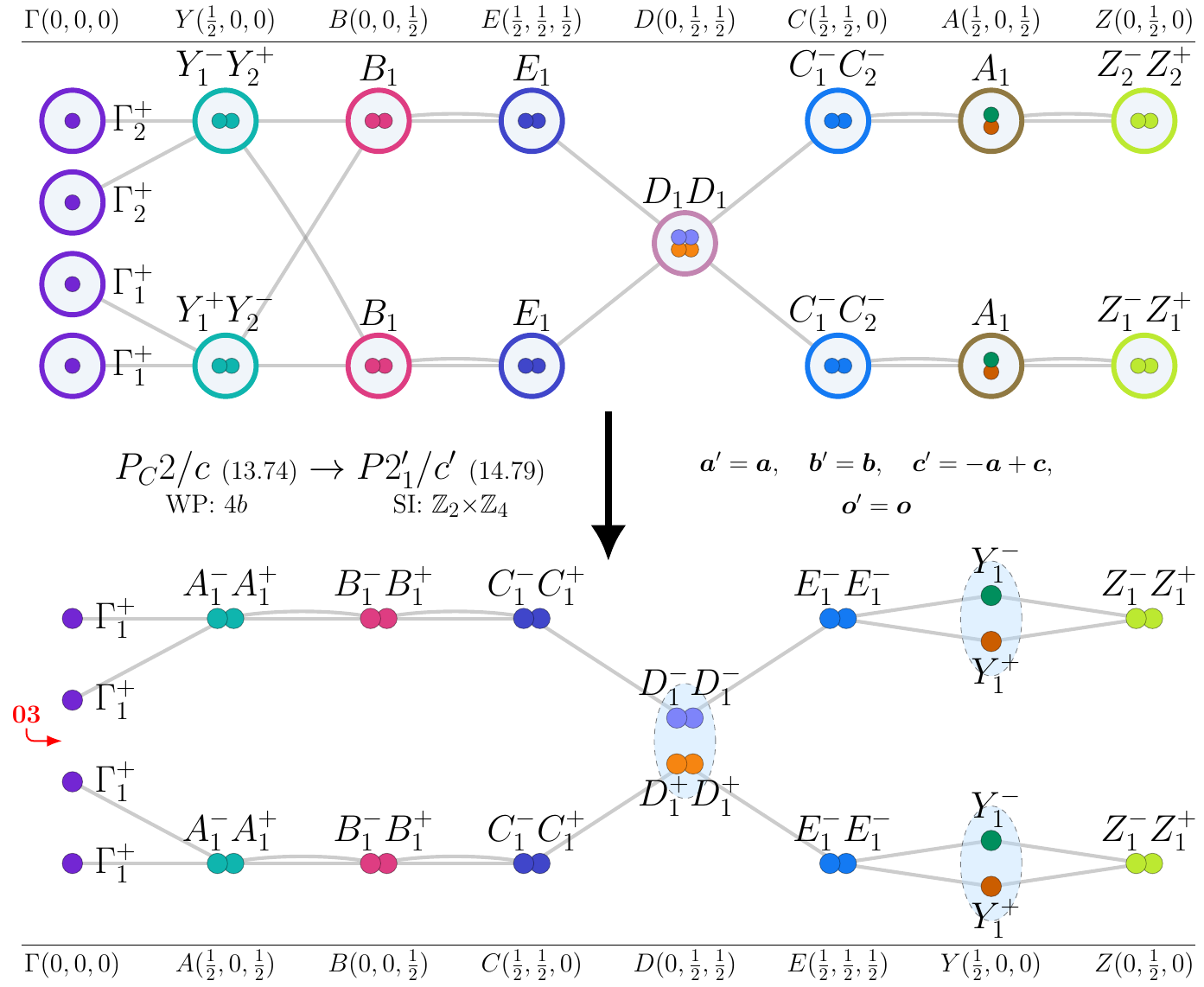}
\caption{Topological magnon bands in subgroup $P2_{1}'/c'~(14.79)$ for magnetic moments on Wyckoff position $4b$ of supergroup $P_{C}2/c~(13.74)$.\label{fig_13.74_14.79_Bperp010_4b}}
\end{figure}
\input{gap_tables_tex/13.74_14.79_Bperp010_4b_table.tex}
\input{si_tables_tex/13.74_14.79_Bperp010_4b_table.tex}
\subsubsection{Topological bands in subgroup $P_{S}\bar{1}~(2.7)$}
\textbf{Perturbation:}
\begin{itemize}
\item strain in generic direction.
\end{itemize}
\begin{figure}[H]
\centering
\includegraphics[scale=0.6]{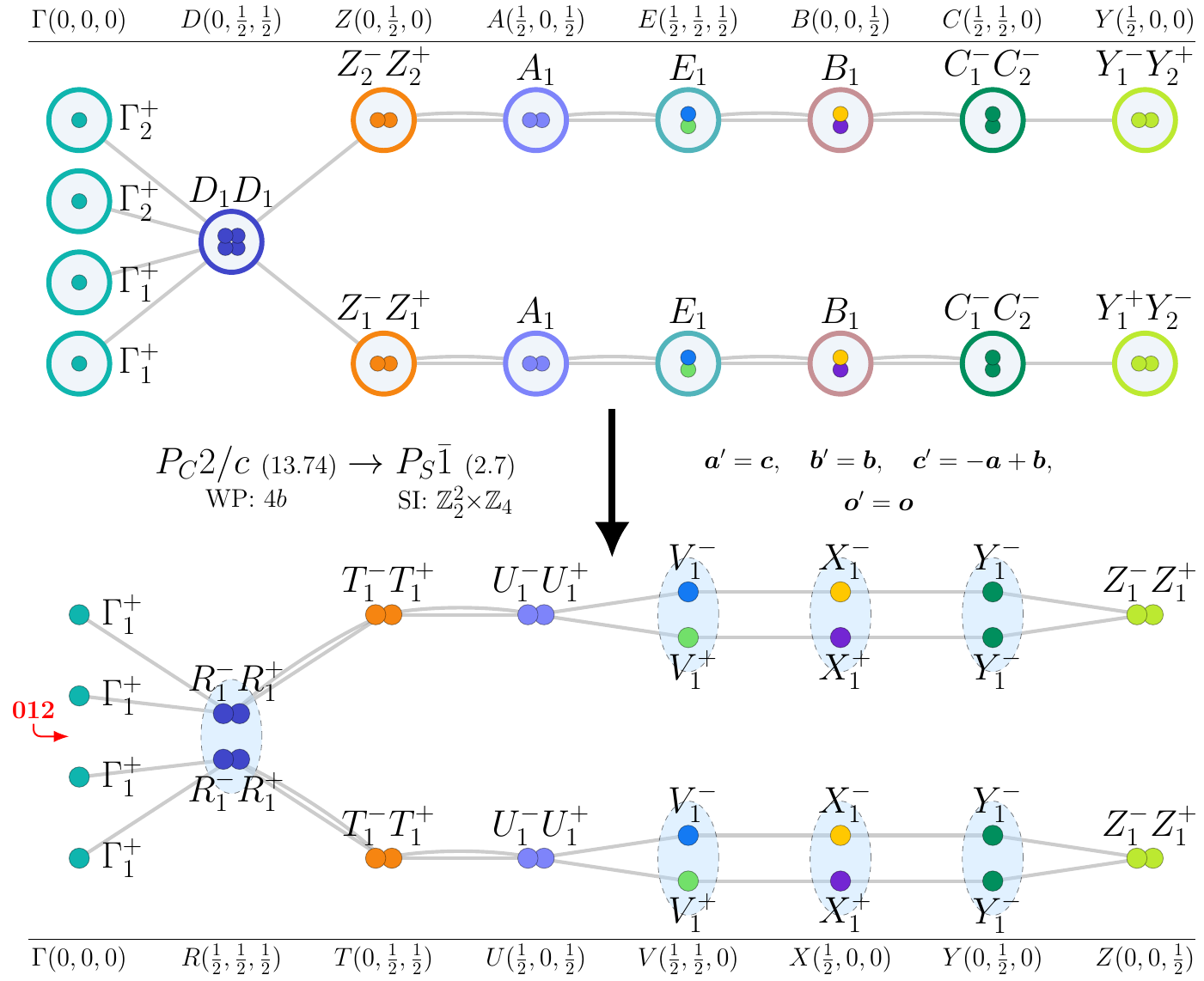}
\caption{Topological magnon bands in subgroup $P_{S}\bar{1}~(2.7)$ for magnetic moments on Wyckoff position $4b$ of supergroup $P_{C}2/c~(13.74)$.\label{fig_13.74_2.7_strainingenericdirection_4b}}
\end{figure}
\input{gap_tables_tex/13.74_2.7_strainingenericdirection_4b_table.tex}
\input{si_tables_tex/13.74_2.7_strainingenericdirection_4b_table.tex}

\section{MSG $P4_{2}/nc'm'~(138.525)$}
\textbf{Nontrivial-SI Subgroups:} $P\bar{1}~(2.4)$, $C2'/m'~(12.62)$, $C2'/m'~(12.62)$, $P2_{1}'/c'~(14.79)$, $P2_{1}'/c'~(14.79)$, $P2~(3.1)$, $Cm'm'2~(35.168)$, $Pc'c'2~(27.81)$, $P2/c~(13.65)$, $Cm'm'a~(67.505)$, $Pc'c'n~(56.369)$, $P4_{2}c'm'~(101.183)$.\\

\textbf{Trivial-SI Subgroups:} $Cm'~(8.34)$, $Cm'~(8.34)$, $Pc'~(7.26)$, $Pc'~(7.26)$, $C2'~(5.15)$, $P2_{1}'~(4.9)$, $Pc~(7.24)$, $Abm'2'~(39.198)$, $Pna'2_{1}'~(33.147)$.\\

\subsection{WP: $4d+8i$}
\textbf{BCS Materials:} {Nd\textsubscript{2}NiO\textsubscript{4}~(130 K)}\footnote{BCS web page: \texttt{\href{http://webbdcrista1.ehu.es/magndata/index.php?this\_label=0.349} {http://webbdcrista1.ehu.es/magndata/index.php?this\_label=0.349}}}, {Nd\textsubscript{2}NiO\textsubscript{4.11}~(53 K)}\footnote{BCS web page: \texttt{\href{http://webbdcrista1.ehu.es/magndata/index.php?this\_label=0.247} {http://webbdcrista1.ehu.es/magndata/index.php?this\_label=0.247}}}.\\
\subsubsection{Topological bands in subgroup $P2_{1}'/c'~(14.79)$}
\textbf{Perturbation:}
\begin{itemize}
\item B $\parallel$ [100].
\end{itemize}
\begin{figure}[H]
\centering
\includegraphics[scale=0.6]{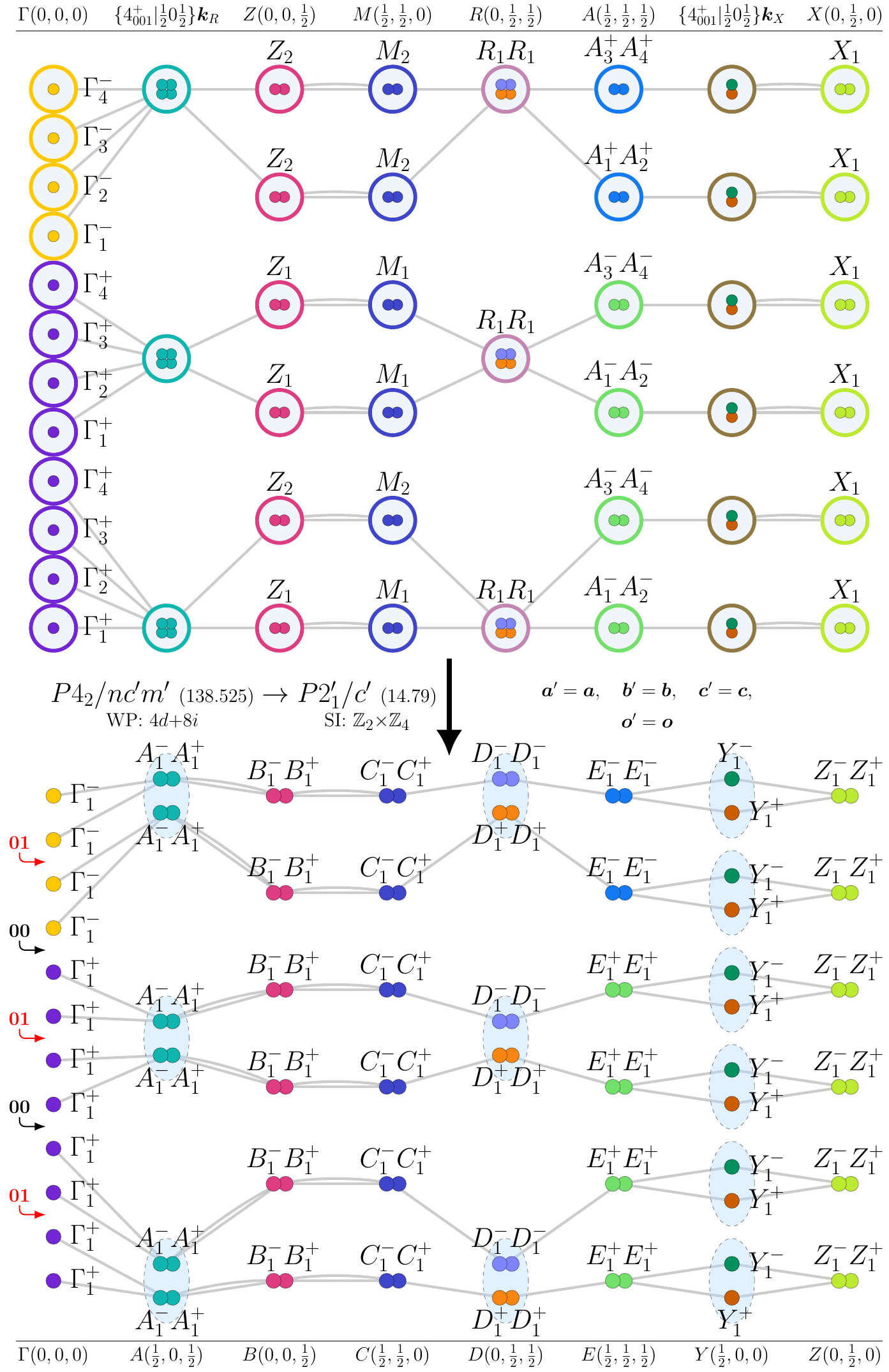}
\caption{Topological magnon bands in subgroup $P2_{1}'/c'~(14.79)$ for magnetic moments on Wyckoff positions $4d+8i$ of supergroup $P4_{2}/nc'm'~(138.525)$.\label{fig_138.525_14.79_Bparallel100_4d+8i}}
\end{figure}
\input{gap_tables_tex/138.525_14.79_Bparallel100_4d+8i_table.tex}
\input{si_tables_tex/138.525_14.79_Bparallel100_4d+8i_table.tex}
\subsubsection{Topological bands in subgroup $P2_{1}'/c'~(14.79)$}
\textbf{Perturbations:}
\begin{itemize}
\item strain $\perp$ [100],
\item B $\perp$ [100].
\end{itemize}
\begin{figure}[H]
\centering
\includegraphics[scale=0.6]{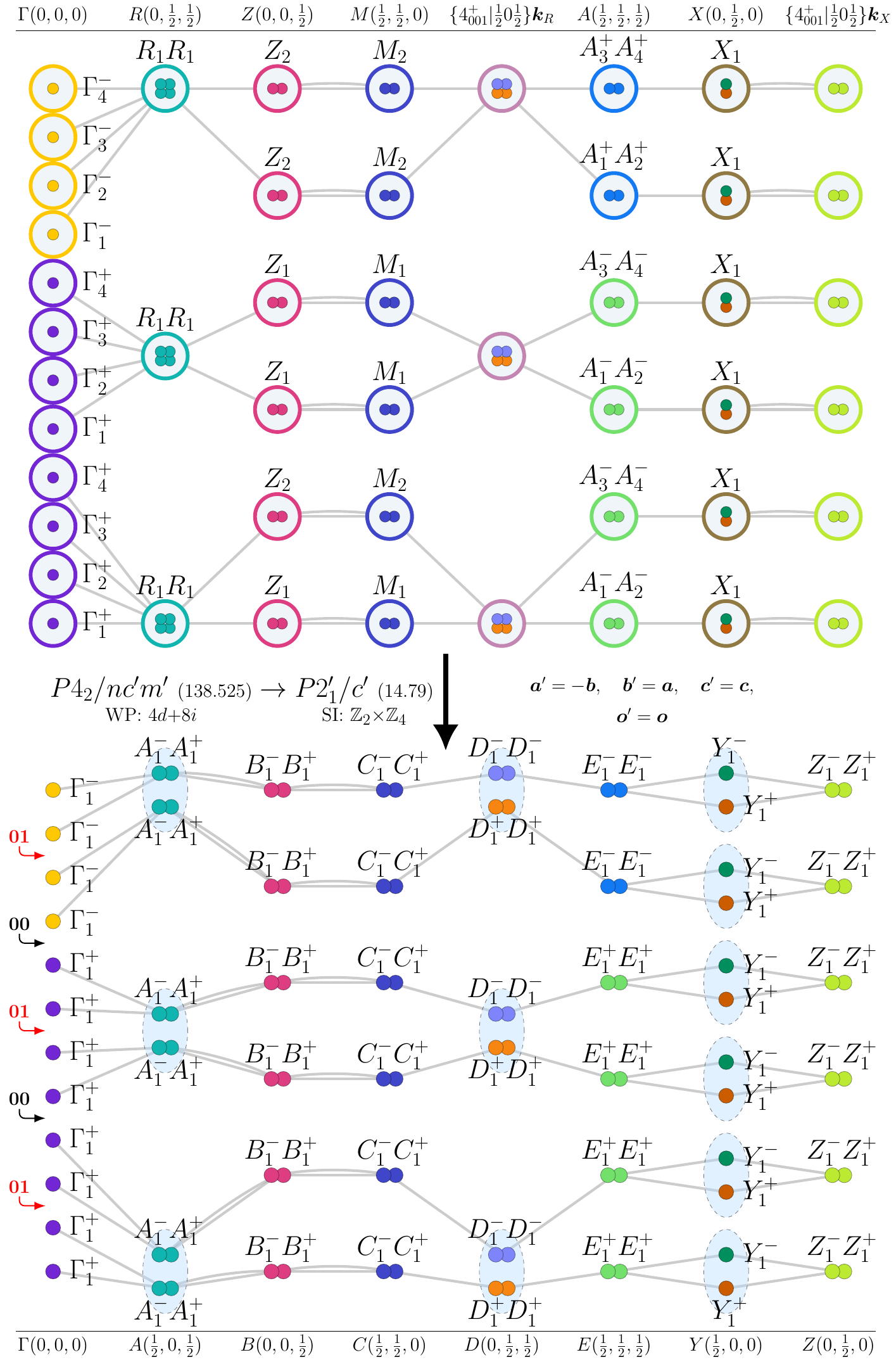}
\caption{Topological magnon bands in subgroup $P2_{1}'/c'~(14.79)$ for magnetic moments on Wyckoff positions $4d+8i$ of supergroup $P4_{2}/nc'm'~(138.525)$.\label{fig_138.525_14.79_strainperp100_4d+8i}}
\end{figure}
\input{gap_tables_tex/138.525_14.79_strainperp100_4d+8i_table.tex}
\input{si_tables_tex/138.525_14.79_strainperp100_4d+8i_table.tex}

\section{MSG $P_{c}4_{2}/ncm~(138.528)$}
\textbf{Nontrivial-SI Subgroups:} $P\bar{1}~(2.4)$, $C2'/c'~(15.89)$, $P2_{1}'/c'~(14.79)$, $P2_{1}'/m'~(11.54)$, $P_{S}\bar{1}~(2.7)$, $Ab'a'2~(41.215)$, $C2/m~(12.58)$, $Cmc'a'~(64.475)$, $C_{c}2/m~(12.63)$, $P2~(3.1)$, $Cc'c'2~(37.183)$, $Pm'm'2~(25.60)$, $P_{b}2~(3.5)$, $P_{c}cc2~(27.82)$, $P2/c~(13.65)$, $Cc'c'a~(68.515)$, $Pm'm'n~(59.409)$, $P_{b}2/c~(13.71)$, $P2_{1}/c~(14.75)$, $Pn'm'a~(62.446)$, $P_{c}2_{1}/c~(14.82)$, $P_{c}ccn~(56.373)$, $P4_{2}m'c'~(105.215)$, $P4_{2}/nm'c'~(137.513)$.\\

\textbf{Trivial-SI Subgroups:} $Cc'~(9.39)$, $Pc'~(7.26)$, $Pm'~(6.20)$, $C2'~(5.15)$, $P2_{1}'~(4.9)$, $P2_{1}'~(4.9)$, $P_{S}1~(1.3)$, $Cm~(8.32)$, $Cmc'2_{1}'~(36.175)$, $C_{c}m~(8.35)$, $Pc~(7.24)$, $Aba'2'~(41.214)$, $Pm'n2_{1}'~(31.125)$, $P_{b}c~(7.29)$, $Pc~(7.24)$, $Pm'c2_{1}'~(26.68)$, $P_{c}c~(7.28)$, $C2~(5.13)$, $C_{c}2~(5.16)$, $A_{a}bm2~(39.200)$, $C_{c}mm2~(35.169)$, $C_{c}mma~(67.508)$, $P2_{1}~(4.7)$, $Pm'n'2_{1}~(31.127)$, $P_{a}2_{1}~(4.10)$, $P_{a}na2_{1}~(33.149)$, $P_{c}4_{2}cm~(101.184)$.\\

\subsection{WP: $4b$}
\textbf{BCS Materials:} {LiMnAs~(393 K)}\footnote{BCS web page: \texttt{\href{http://webbdcrista1.ehu.es/magndata/index.php?this\_label=1.552} {http://webbdcrista1.ehu.es/magndata/index.php?this\_label=1.552}}}, {LiMnAs~(393 K)}\footnote{BCS web page: \texttt{\href{http://webbdcrista1.ehu.es/magndata/index.php?this\_label=1.551} {http://webbdcrista1.ehu.es/magndata/index.php?this\_label=1.551}}}, {LiMnAs~(393 K)}\footnote{BCS web page: \texttt{\href{http://webbdcrista1.ehu.es/magndata/index.php?this\_label=1.550} {http://webbdcrista1.ehu.es/magndata/index.php?this\_label=1.550}}}, {CsMnP~(295 K)}\footnote{BCS web page: \texttt{\href{http://webbdcrista1.ehu.es/magndata/index.php?this\_label=1.548} {http://webbdcrista1.ehu.es/magndata/index.php?this\_label=1.548}}}, {CsMnP~(295 K)}\footnote{BCS web page: \texttt{\href{http://webbdcrista1.ehu.es/magndata/index.php?this\_label=1.547} {http://webbdcrista1.ehu.es/magndata/index.php?this\_label=1.547}}}, {CsMnBi~(295 K)}\footnote{BCS web page: \texttt{\href{http://webbdcrista1.ehu.es/magndata/index.php?this\_label=1.546} {http://webbdcrista1.ehu.es/magndata/index.php?this\_label=1.546}}}, {RbMnAs~(295 K)}\footnote{BCS web page: \texttt{\href{http://webbdcrista1.ehu.es/magndata/index.php?this\_label=1.543} {http://webbdcrista1.ehu.es/magndata/index.php?this\_label=1.543}}}, {RbMnP~(295 K)}\footnote{BCS web page: \texttt{\href{http://webbdcrista1.ehu.es/magndata/index.php?this\_label=1.542} {http://webbdcrista1.ehu.es/magndata/index.php?this\_label=1.542}}}, {RbMnP~(295 K)}\footnote{BCS web page: \texttt{\href{http://webbdcrista1.ehu.es/magndata/index.php?this\_label=1.541} {http://webbdcrista1.ehu.es/magndata/index.php?this\_label=1.541}}}, {KMnP~(295 K)}\footnote{BCS web page: \texttt{\href{http://webbdcrista1.ehu.es/magndata/index.php?this\_label=1.540} {http://webbdcrista1.ehu.es/magndata/index.php?this\_label=1.540}}}, {KMnP~(295 K)}\footnote{BCS web page: \texttt{\href{http://webbdcrista1.ehu.es/magndata/index.php?this\_label=1.539} {http://webbdcrista1.ehu.es/magndata/index.php?this\_label=1.539}}}, {KMnAs~(293 K)}\footnote{BCS web page: \texttt{\href{http://webbdcrista1.ehu.es/magndata/index.php?this\_label=1.554} {http://webbdcrista1.ehu.es/magndata/index.php?this\_label=1.554}}}, {KMnAs~(293 K)}\footnote{BCS web page: \texttt{\href{http://webbdcrista1.ehu.es/magndata/index.php?this\_label=1.553} {http://webbdcrista1.ehu.es/magndata/index.php?this\_label=1.553}}}, {RbMnAs~(293 K)}\footnote{BCS web page: \texttt{\href{http://webbdcrista1.ehu.es/magndata/index.php?this\_label=1.544} {http://webbdcrista1.ehu.es/magndata/index.php?this\_label=1.544}}}, {RbMnBi}\footnote{BCS web page: \texttt{\href{http://webbdcrista1.ehu.es/magndata/index.php?this\_label=1.545} {http://webbdcrista1.ehu.es/magndata/index.php?this\_label=1.545}}}, {LaCrAsO}\footnote{BCS web page: \texttt{\href{http://webbdcrista1.ehu.es/magndata/index.php?this\_label=1.146} {http://webbdcrista1.ehu.es/magndata/index.php?this\_label=1.146}}}.\\
\subsubsection{Topological bands in subgroup $P\bar{1}~(2.4)$}
\textbf{Perturbations:}
\begin{itemize}
\item B $\parallel$ [001] and strain in generic direction,
\item B $\parallel$ [100] and strain $\perp$ [110],
\item B $\parallel$ [100] and strain in generic direction,
\item B $\parallel$ [110] and strain $\perp$ [100],
\item B $\parallel$ [110] and strain in generic direction,
\item B $\perp$ [001] and strain $\perp$ [100],
\item B $\perp$ [001] and strain $\perp$ [110],
\item B $\perp$ [001] and strain in generic direction,
\item B $\perp$ [100] and strain $\parallel$ [110],
\item B $\perp$ [100] and strain $\perp$ [001],
\item B $\perp$ [100] and strain $\perp$ [110],
\item B $\perp$ [100] and strain in generic direction,
\item B $\perp$ [110] and strain $\parallel$ [100],
\item B $\perp$ [110] and strain $\perp$ [001],
\item B $\perp$ [110] and strain $\perp$ [100],
\item B $\perp$ [110] and strain in generic direction,
\item B in generic direction.
\end{itemize}
\begin{figure}[H]
\centering
\includegraphics[scale=0.6]{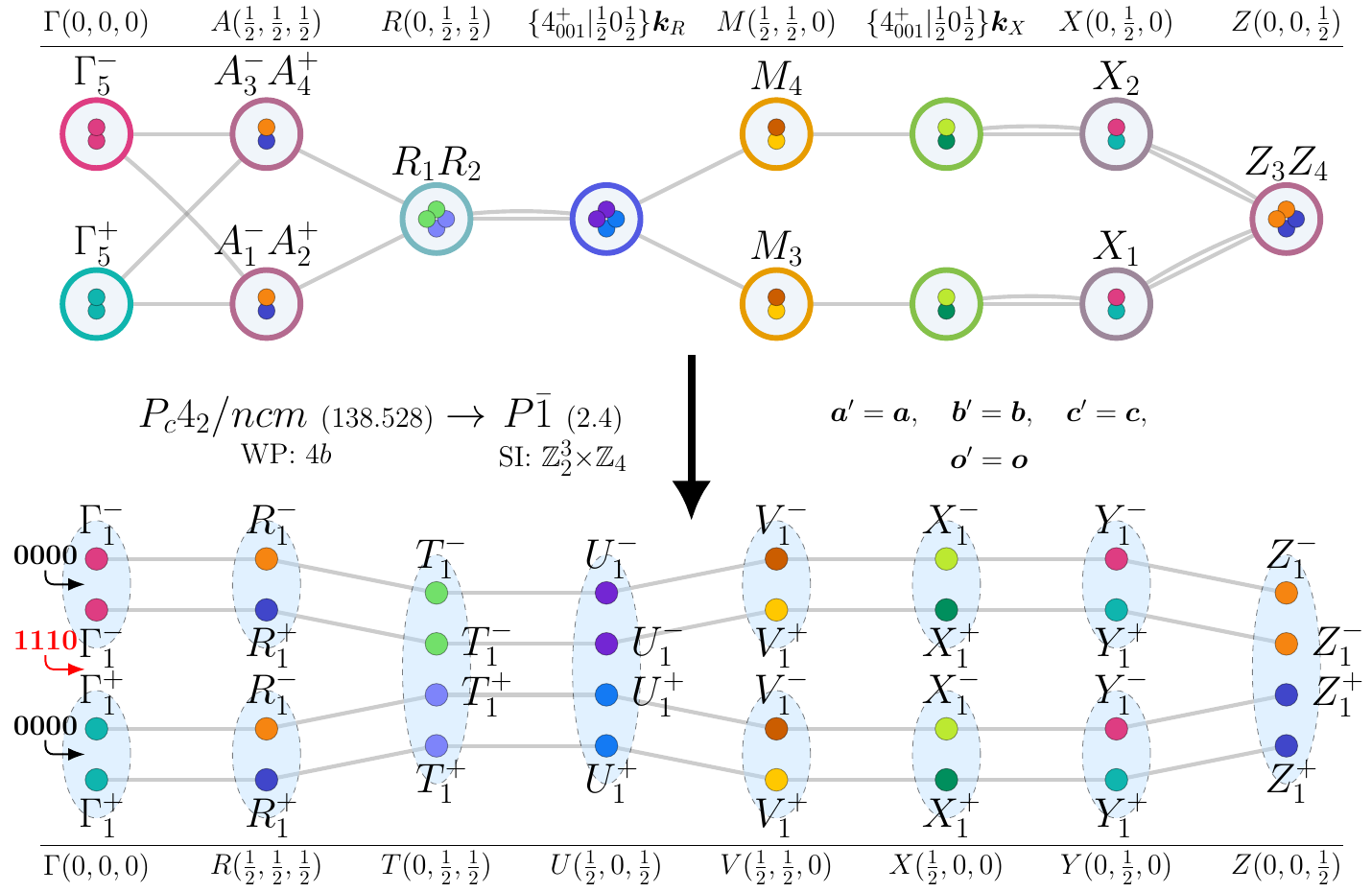}
\caption{Topological magnon bands in subgroup $P\bar{1}~(2.4)$ for magnetic moments on Wyckoff position $4b$ of supergroup $P_{c}4_{2}/ncm~(138.528)$.\label{fig_138.528_2.4_Bparallel001andstrainingenericdirection_4b}}
\end{figure}
\input{gap_tables_tex/138.528_2.4_Bparallel001andstrainingenericdirection_4b_table.tex}
\input{si_tables_tex/138.528_2.4_Bparallel001andstrainingenericdirection_4b_table.tex}
\subsubsection{Topological bands in subgroup $C2'/c'~(15.89)$}
\textbf{Perturbations:}
\begin{itemize}
\item B $\parallel$ [001] and strain $\perp$ [110],
\item B $\perp$ [110].
\end{itemize}
\begin{figure}[H]
\centering
\includegraphics[scale=0.6]{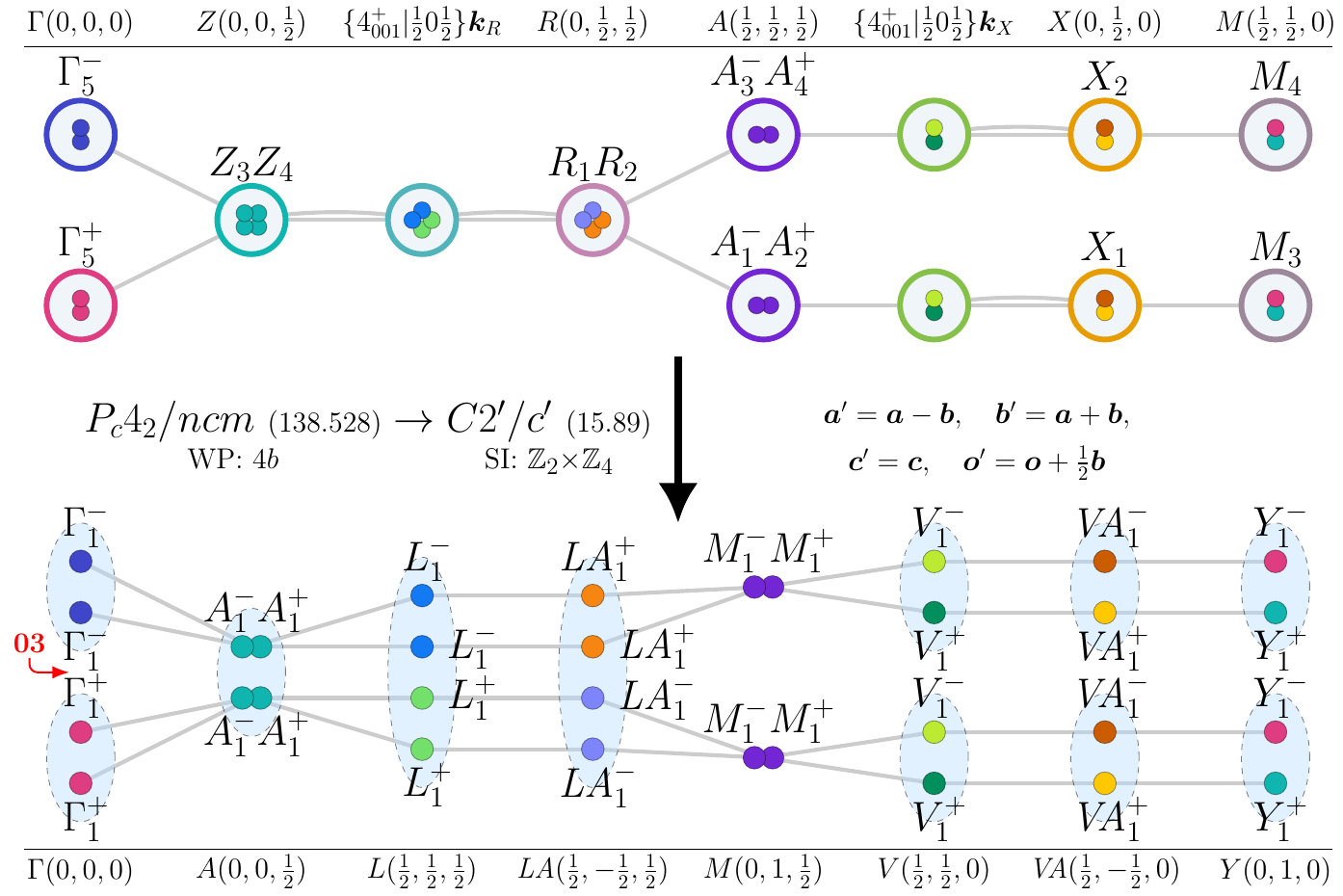}
\caption{Topological magnon bands in subgroup $C2'/c'~(15.89)$ for magnetic moments on Wyckoff position $4b$ of supergroup $P_{c}4_{2}/ncm~(138.528)$.\label{fig_138.528_15.89_Bparallel001andstrainperp110_4b}}
\end{figure}
\input{gap_tables_tex/138.528_15.89_Bparallel001andstrainperp110_4b_table.tex}
\input{si_tables_tex/138.528_15.89_Bparallel001andstrainperp110_4b_table.tex}
\subsubsection{Topological bands in subgroup $P2_{1}'/c'~(14.79)$}
\textbf{Perturbations:}
\begin{itemize}
\item B $\parallel$ [100] and strain $\parallel$ [110],
\item B $\parallel$ [100] and strain $\perp$ [001],
\item B $\parallel$ [110] and strain $\parallel$ [100],
\item B $\parallel$ [110] and strain $\perp$ [001],
\item B $\perp$ [001].
\end{itemize}
\begin{figure}[H]
\centering
\includegraphics[scale=0.6]{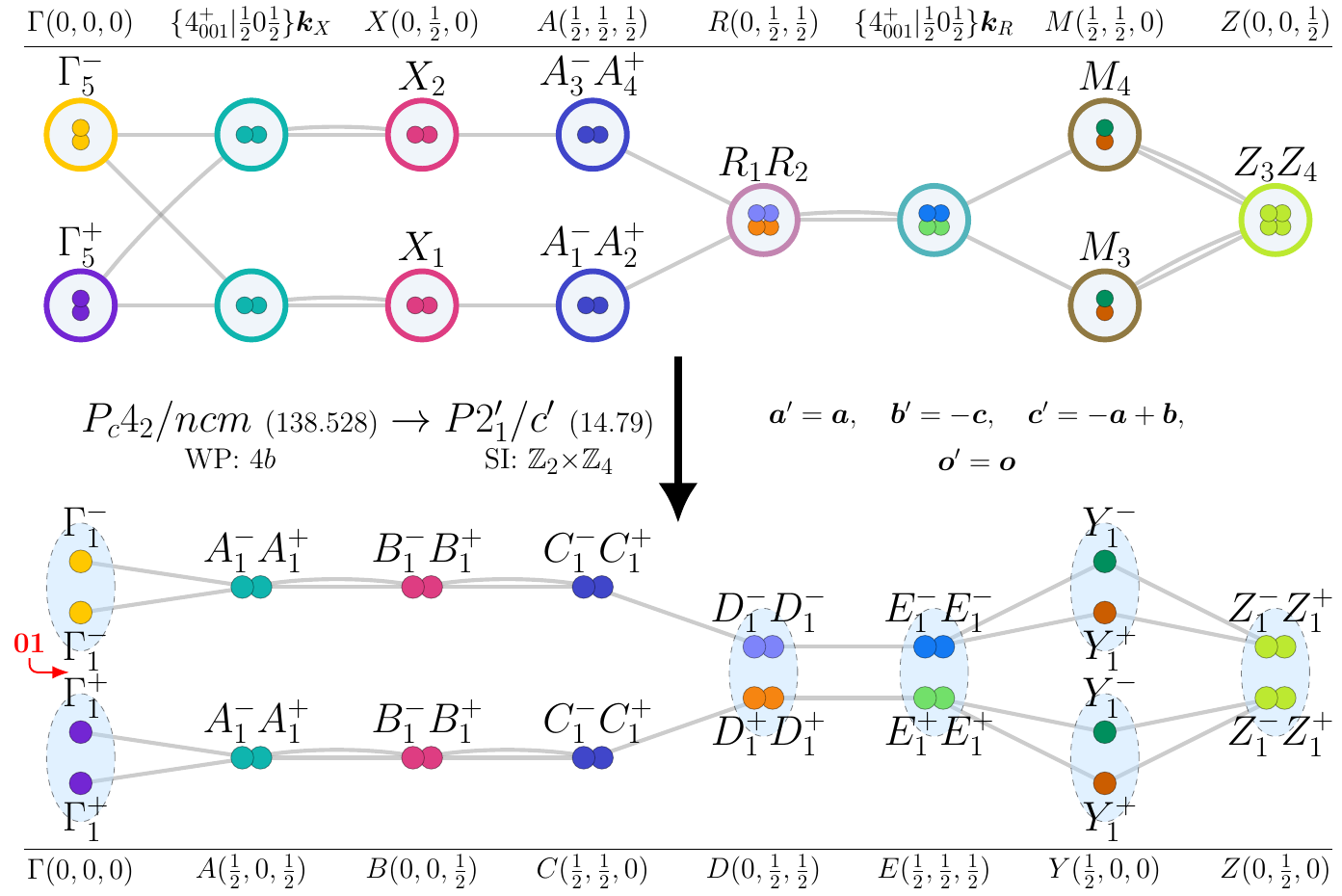}
\caption{Topological magnon bands in subgroup $P2_{1}'/c'~(14.79)$ for magnetic moments on Wyckoff position $4b$ of supergroup $P_{c}4_{2}/ncm~(138.528)$.\label{fig_138.528_14.79_Bparallel100andstrainparallel110_4b}}
\end{figure}
\input{gap_tables_tex/138.528_14.79_Bparallel100andstrainparallel110_4b_table.tex}
\input{si_tables_tex/138.528_14.79_Bparallel100andstrainparallel110_4b_table.tex}
\subsubsection{Topological bands in subgroup $P2_{1}'/m'~(11.54)$}
\textbf{Perturbations:}
\begin{itemize}
\item B $\parallel$ [001] and strain $\perp$ [100],
\item B $\perp$ [100].
\end{itemize}
\begin{figure}[H]
\centering
\includegraphics[scale=0.6]{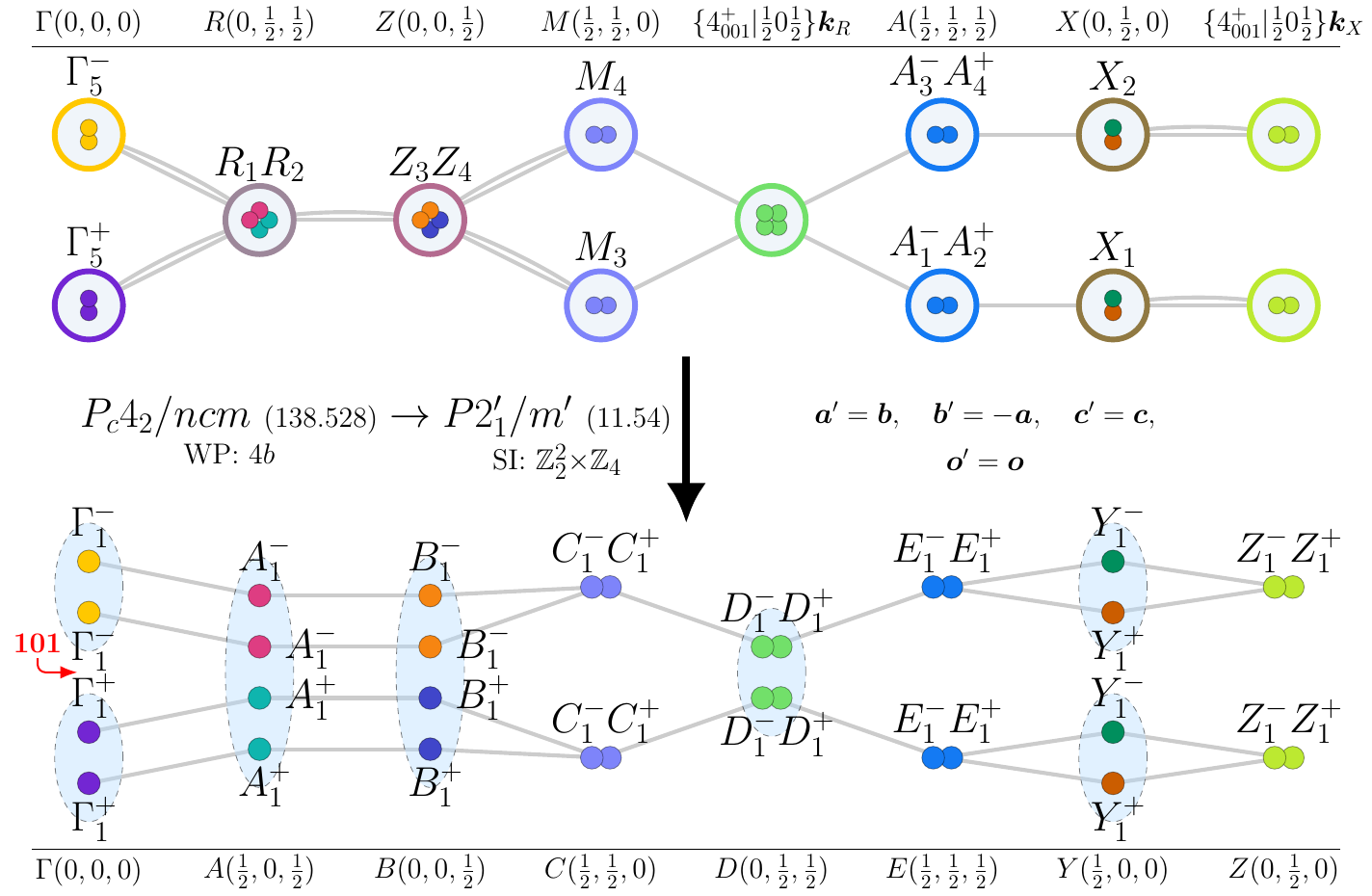}
\caption{Topological magnon bands in subgroup $P2_{1}'/m'~(11.54)$ for magnetic moments on Wyckoff position $4b$ of supergroup $P_{c}4_{2}/ncm~(138.528)$.\label{fig_138.528_11.54_Bparallel001andstrainperp100_4b}}
\end{figure}
\input{gap_tables_tex/138.528_11.54_Bparallel001andstrainperp100_4b_table.tex}
\input{si_tables_tex/138.528_11.54_Bparallel001andstrainperp100_4b_table.tex}
\subsubsection{Topological bands in subgroup $P_{S}\bar{1}~(2.7)$}
\textbf{Perturbation:}
\begin{itemize}
\item strain in generic direction.
\end{itemize}
\begin{figure}[H]
\centering
\includegraphics[scale=0.6]{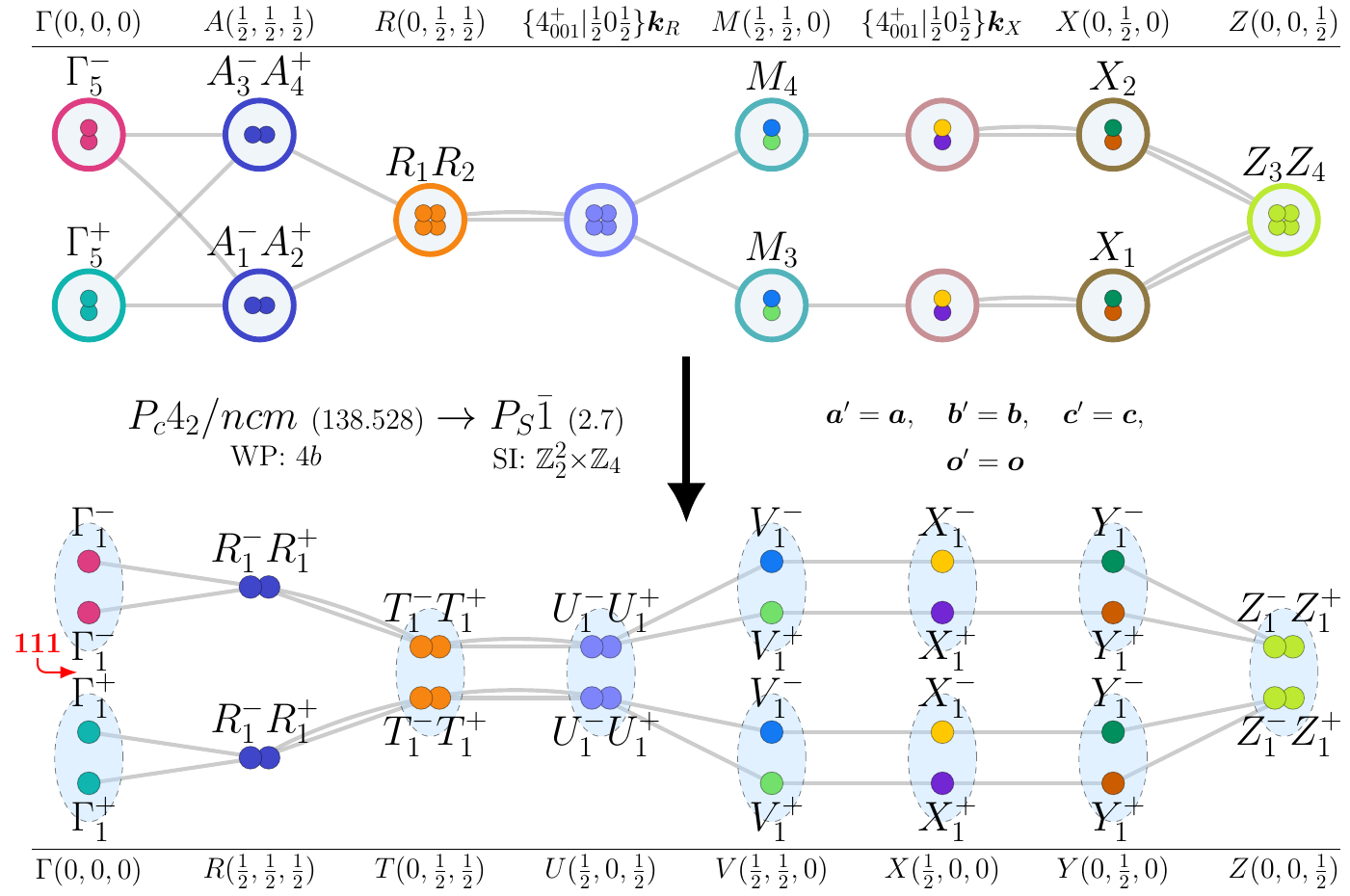}
\caption{Topological magnon bands in subgroup $P_{S}\bar{1}~(2.7)$ for magnetic moments on Wyckoff position $4b$ of supergroup $P_{c}4_{2}/ncm~(138.528)$.\label{fig_138.528_2.7_strainingenericdirection_4b}}
\end{figure}
\input{gap_tables_tex/138.528_2.7_strainingenericdirection_4b_table.tex}
\input{si_tables_tex/138.528_2.7_strainingenericdirection_4b_table.tex}
\subsubsection{Topological bands in subgroup $Ab'a'2~(41.215)$}
\textbf{Perturbation:}
\begin{itemize}
\item E $\parallel$ [110] and B $\parallel$ [110].
\end{itemize}
\begin{figure}[H]
\centering
\includegraphics[scale=0.6]{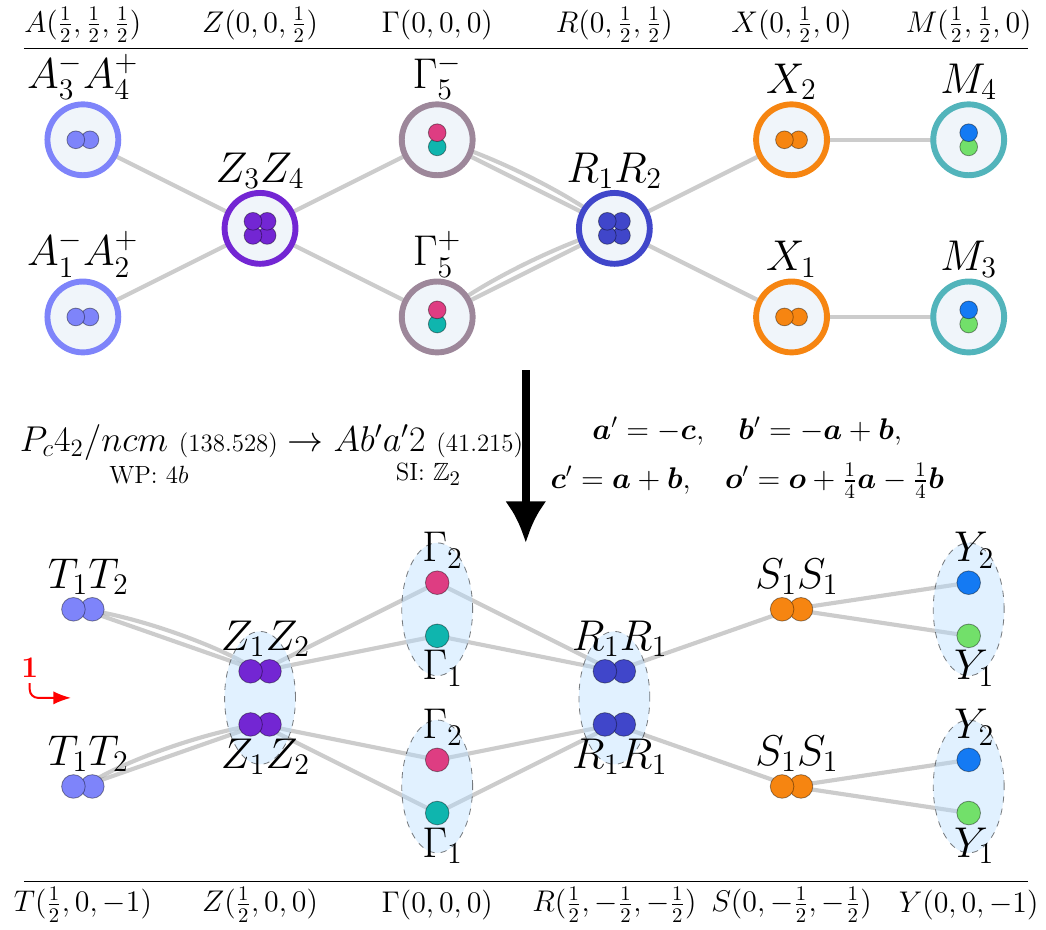}
\caption{Topological magnon bands in subgroup $Ab'a'2~(41.215)$ for magnetic moments on Wyckoff position $4b$ of supergroup $P_{c}4_{2}/ncm~(138.528)$.\label{fig_138.528_41.215_Eparallel110andBparallel110_4b}}
\end{figure}
\input{gap_tables_tex/138.528_41.215_Eparallel110andBparallel110_4b_table.tex}
\input{si_tables_tex/138.528_41.215_Eparallel110andBparallel110_4b_table.tex}

\section{MSG $P_{C}4_{2}/ncm~(138.529)$}
\textbf{Nontrivial-SI Subgroups:} $P\bar{1}~(2.4)$, $C2'/m'~(12.62)$, $P2'/m'~(10.46)$, $P2'/c'~(13.69)$, $P_{S}\bar{1}~(2.7)$, $C_{a}2~(5.17)$, $C2/m~(12.58)$, $Cmm'm'~(65.486)$, $C_{a}2/m~(12.64)$, $P2~(3.1)$, $Cm'm'2~(35.168)$, $P_{a}2~(3.4)$, $P_{C}cc2~(27.84)$, $P2/c~(13.65)$, $Cm'm'a~(67.505)$, $Pn'n'n~(48.260)$, $P_{c}2/c~(13.72)$, $P2_{1}/c~(14.75)$, $Pm'n'a~(53.326)$, $P_{C}2_{1}/c~(14.84)$, $P_{C}ccn~(56.375)$, $P4_{2}n'm'~(102.191)$, $P4_{2}/nn'm'~(134.477)$.\\

\textbf{Trivial-SI Subgroups:} $Cm'~(8.34)$, $Pm'~(6.20)$, $Pc'~(7.26)$, $C2'~(5.15)$, $P2'~(3.3)$, $P2'~(3.3)$, $P_{S}1~(1.3)$, $Cm~(8.32)$, $Cm'm2'~(35.167)$, $C_{a}m~(8.36)$, $Pc~(7.24)$, $Abm'2'~(39.198)$, $Pn'n2'~(34.158)$, $P_{c}c~(7.28)$, $Pc~(7.24)$, $Pn'c2'~(30.113)$, $P_{C}c~(7.30)$, $C2~(5.13)$, $Am'm'2~(38.191)$, $A_{b}bm2~(39.201)$, $Pn'n'2~(34.159)$, $C_{a}mm2~(35.170)$, $C_{a}mma~(67.509)$, $P2_{1}~(4.7)$, $Pm'n'2_{1}~(31.127)$, $P_{C}2_{1}~(4.12)$, $P_{A}na2_{1}~(33.152)$, $P_{C}4_{2}cm~(101.185)$.\\

\subsection{WP: $4d$}
\textbf{BCS Materials:} {LaSrFeO\textsubscript{4}~(380 K)}\footnote{BCS web page: \texttt{\href{http://webbdcrista1.ehu.es/magndata/index.php?this\_label=2.41} {http://webbdcrista1.ehu.es/magndata/index.php?this\_label=2.41}}}, {LaCaFeO\textsubscript{4}~(373 K)}\footnote{BCS web page: \texttt{\href{http://webbdcrista1.ehu.es/magndata/index.php?this\_label=2.39} {http://webbdcrista1.ehu.es/magndata/index.php?this\_label=2.39}}}, {LaBaFeO\textsubscript{4}}\footnote{BCS web page: \texttt{\href{http://webbdcrista1.ehu.es/magndata/index.php?this\_label=2.40} {http://webbdcrista1.ehu.es/magndata/index.php?this\_label=2.40}}}.\\
\subsubsection{Topological bands in subgroup $P\bar{1}~(2.4)$}
\textbf{Perturbations:}
\begin{itemize}
\item B $\parallel$ [001] and strain in generic direction,
\item B $\parallel$ [100] and strain $\perp$ [110],
\item B $\parallel$ [100] and strain in generic direction,
\item B $\parallel$ [110] and strain $\perp$ [100],
\item B $\parallel$ [110] and strain in generic direction,
\item B $\perp$ [001] and strain $\perp$ [100],
\item B $\perp$ [001] and strain $\perp$ [110],
\item B $\perp$ [001] and strain in generic direction,
\item B $\perp$ [100] and strain $\parallel$ [110],
\item B $\perp$ [100] and strain $\perp$ [001],
\item B $\perp$ [100] and strain $\perp$ [110],
\item B $\perp$ [100] and strain in generic direction,
\item B $\perp$ [110] and strain $\parallel$ [100],
\item B $\perp$ [110] and strain $\perp$ [001],
\item B $\perp$ [110] and strain $\perp$ [100],
\item B $\perp$ [110] and strain in generic direction,
\item B in generic direction.
\end{itemize}
\begin{figure}[H]
\centering
\includegraphics[scale=0.6]{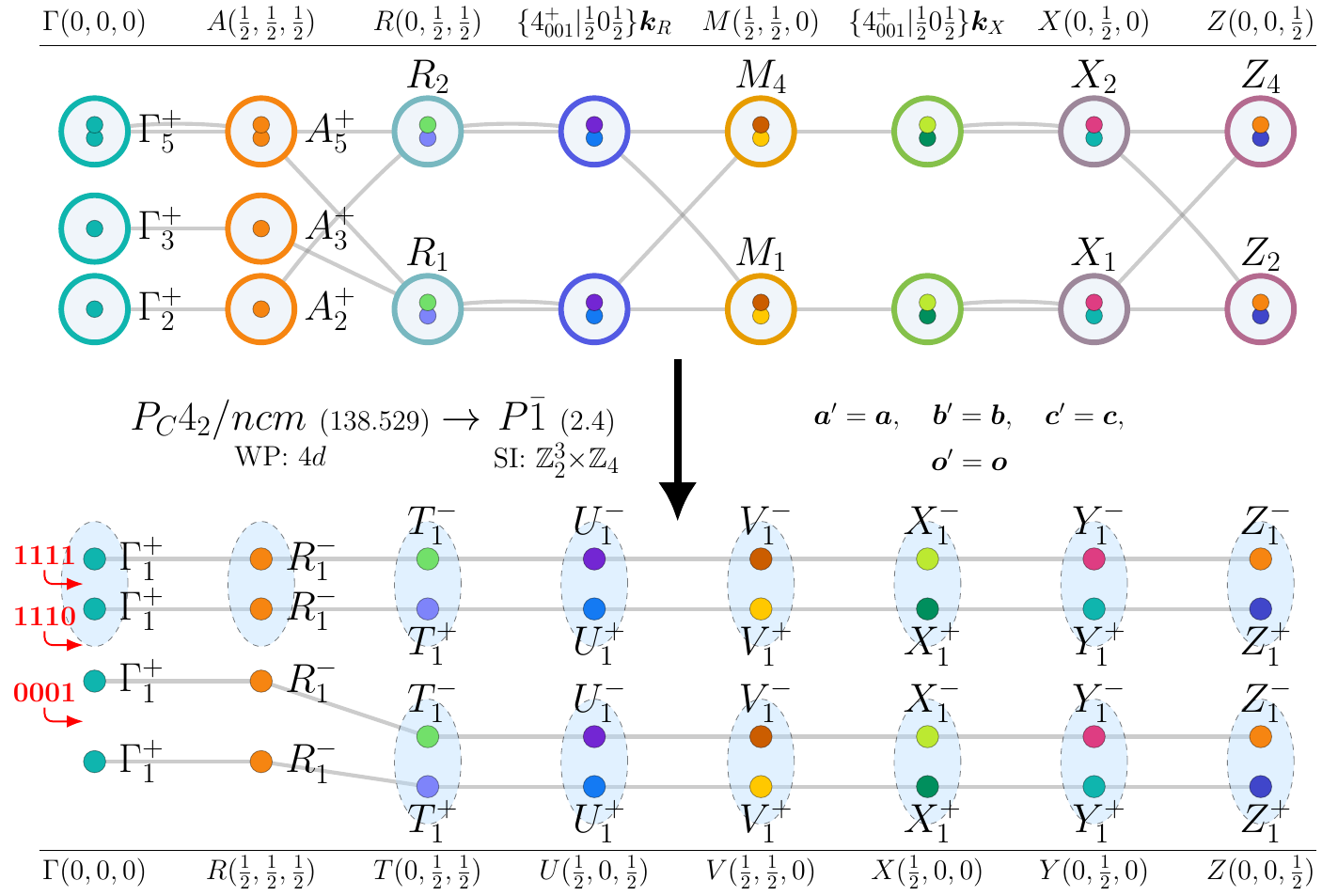}
\caption{Topological magnon bands in subgroup $P\bar{1}~(2.4)$ for magnetic moments on Wyckoff position $4d$ of supergroup $P_{C}4_{2}/ncm~(138.529)$.\label{fig_138.529_2.4_Bparallel001andstrainingenericdirection_4d}}
\end{figure}
\input{gap_tables_tex/138.529_2.4_Bparallel001andstrainingenericdirection_4d_table.tex}
\input{si_tables_tex/138.529_2.4_Bparallel001andstrainingenericdirection_4d_table.tex}
\subsubsection{Topological bands in subgroup $C2'/m'~(12.62)$}
\textbf{Perturbations:}
\begin{itemize}
\item B $\parallel$ [001] and strain $\perp$ [110],
\item B $\perp$ [110].
\end{itemize}
\begin{figure}[H]
\centering
\includegraphics[scale=0.6]{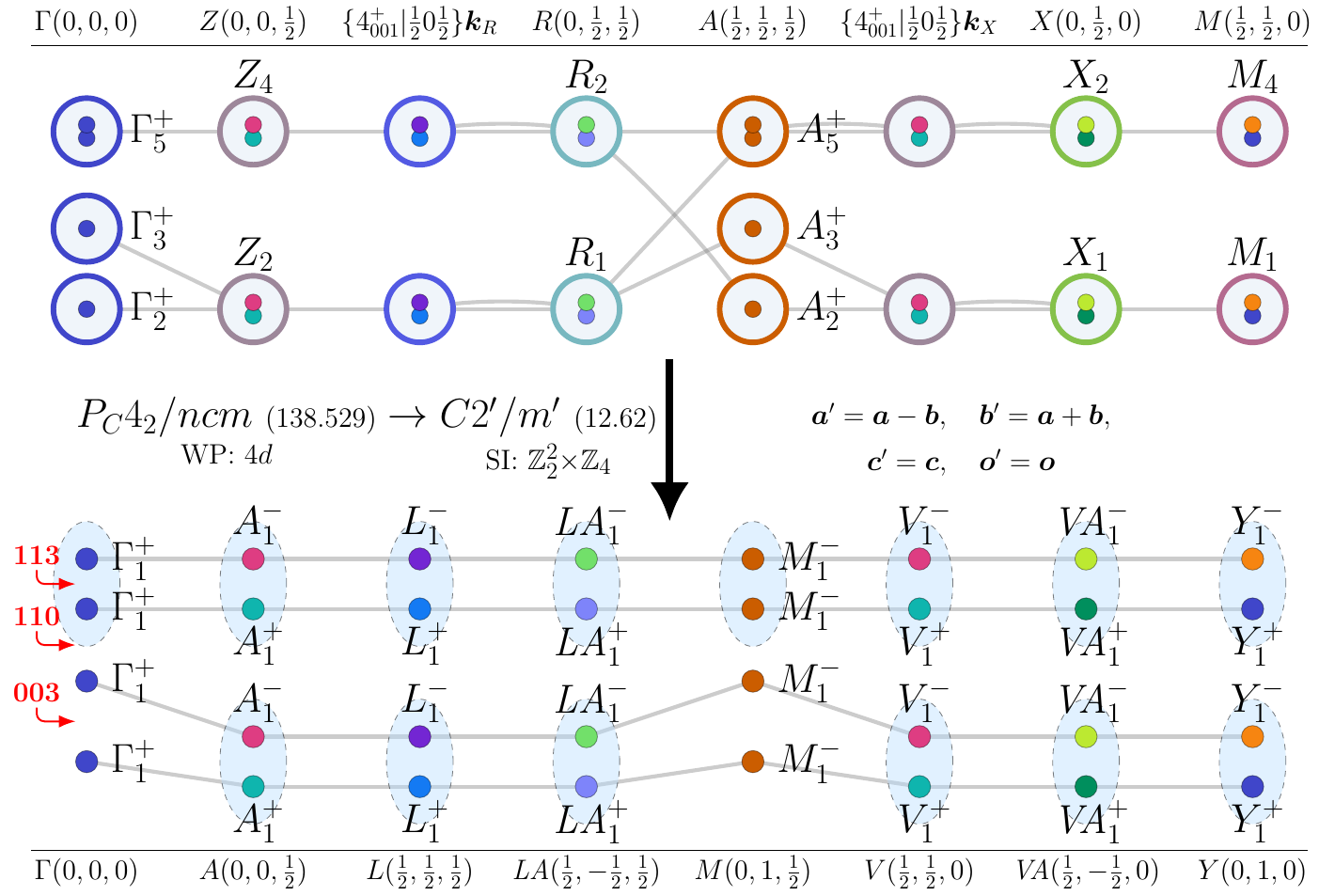}
\caption{Topological magnon bands in subgroup $C2'/m'~(12.62)$ for magnetic moments on Wyckoff position $4d$ of supergroup $P_{C}4_{2}/ncm~(138.529)$.\label{fig_138.529_12.62_Bparallel001andstrainperp110_4d}}
\end{figure}
\input{gap_tables_tex/138.529_12.62_Bparallel001andstrainperp110_4d_table.tex}
\input{si_tables_tex/138.529_12.62_Bparallel001andstrainperp110_4d_table.tex}
\subsubsection{Topological bands in subgroup $P2'/m'~(10.46)$}
\textbf{Perturbations:}
\begin{itemize}
\item B $\parallel$ [100] and strain $\parallel$ [110],
\item B $\parallel$ [100] and strain $\perp$ [001],
\item B $\parallel$ [110] and strain $\parallel$ [100],
\item B $\parallel$ [110] and strain $\perp$ [001],
\item B $\perp$ [001].
\end{itemize}
\begin{figure}[H]
\centering
\includegraphics[scale=0.6]{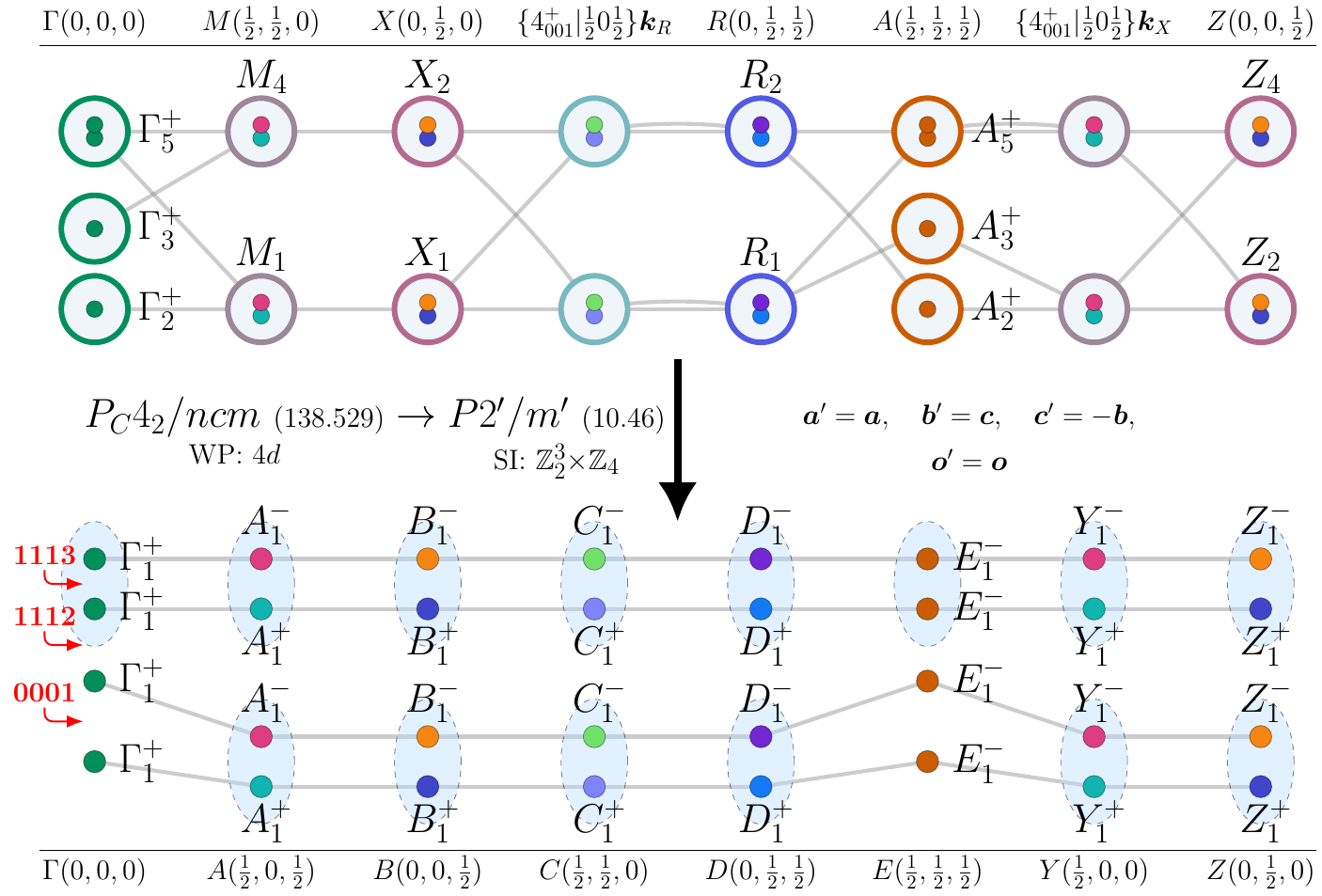}
\caption{Topological magnon bands in subgroup $P2'/m'~(10.46)$ for magnetic moments on Wyckoff position $4d$ of supergroup $P_{C}4_{2}/ncm~(138.529)$.\label{fig_138.529_10.46_Bparallel100andstrainparallel110_4d}}
\end{figure}
\input{gap_tables_tex/138.529_10.46_Bparallel100andstrainparallel110_4d_table.tex}
\input{si_tables_tex/138.529_10.46_Bparallel100andstrainparallel110_4d_table.tex}
\subsubsection{Topological bands in subgroup $P2'/c'~(13.69)$}
\textbf{Perturbations:}
\begin{itemize}
\item B $\parallel$ [001] and strain $\perp$ [100],
\item B $\perp$ [100].
\end{itemize}
\begin{figure}[H]
\centering
\includegraphics[scale=0.6]{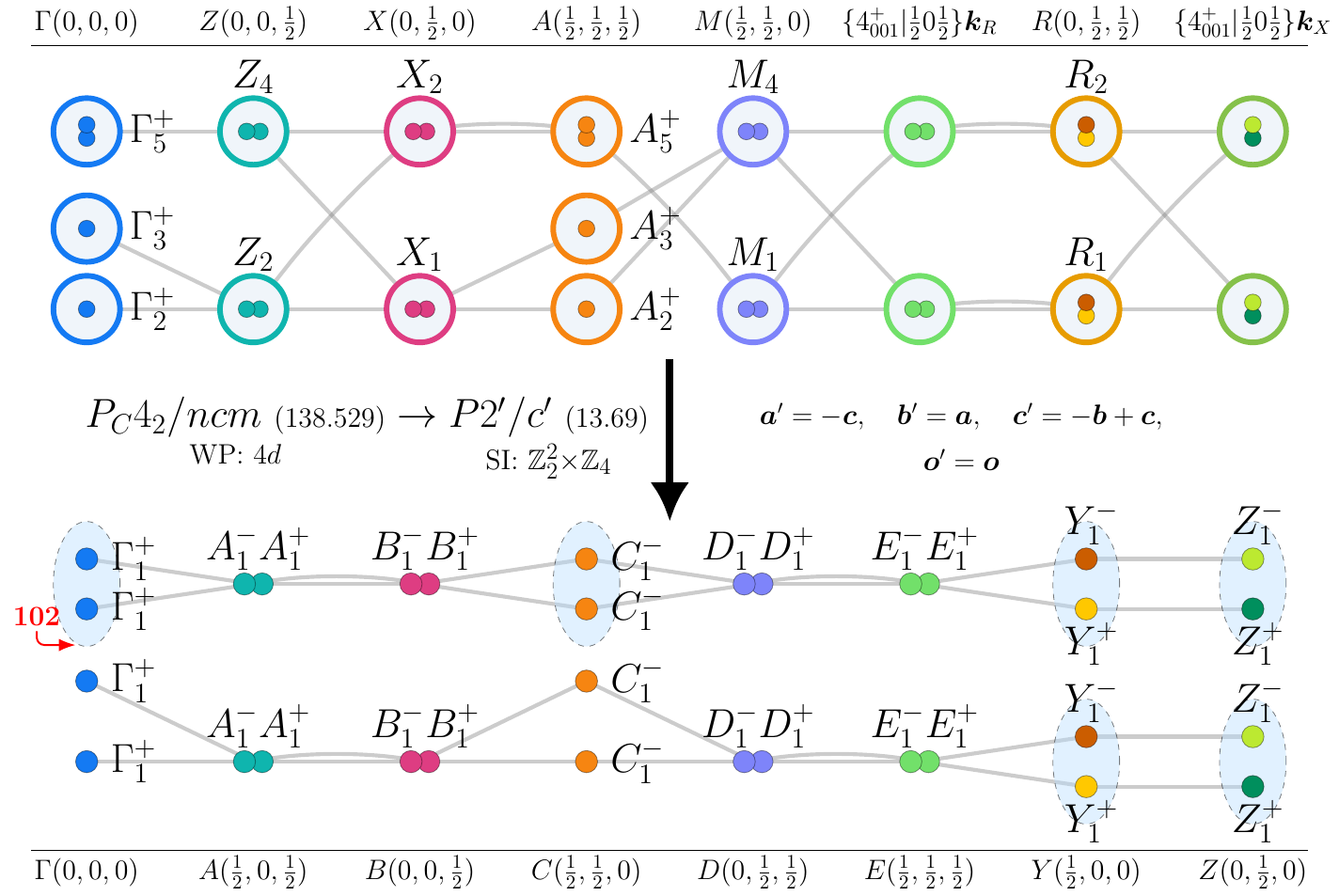}
\caption{Topological magnon bands in subgroup $P2'/c'~(13.69)$ for magnetic moments on Wyckoff position $4d$ of supergroup $P_{C}4_{2}/ncm~(138.529)$.\label{fig_138.529_13.69_Bparallel001andstrainperp100_4d}}
\end{figure}
\input{gap_tables_tex/138.529_13.69_Bparallel001andstrainperp100_4d_table.tex}
\input{si_tables_tex/138.529_13.69_Bparallel001andstrainperp100_4d_table.tex}
\subsubsection{Topological bands in subgroup $P_{S}\bar{1}~(2.7)$}
\textbf{Perturbation:}
\begin{itemize}
\item strain in generic direction.
\end{itemize}
\begin{figure}[H]
\centering
\includegraphics[scale=0.6]{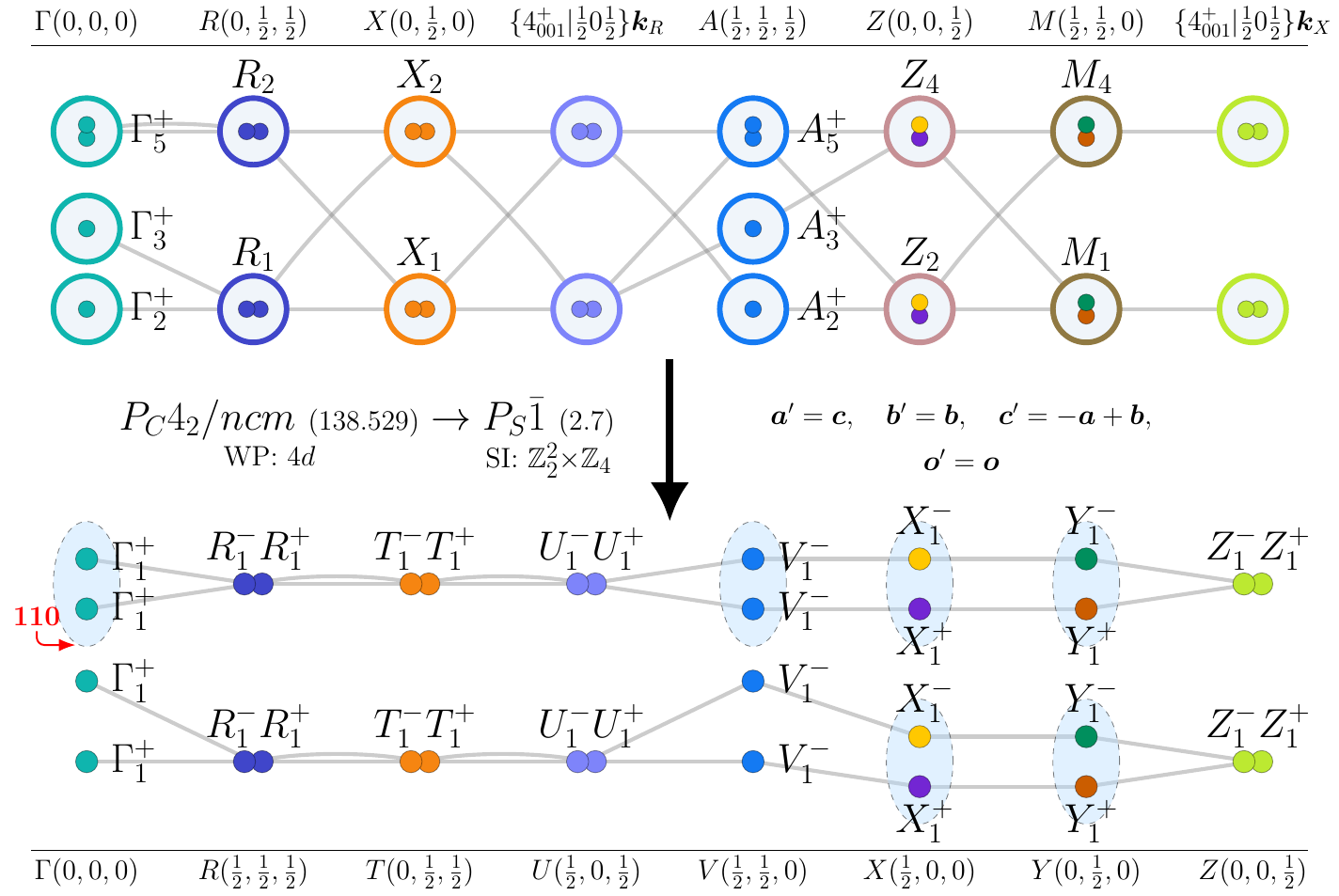}
\caption{Topological magnon bands in subgroup $P_{S}\bar{1}~(2.7)$ for magnetic moments on Wyckoff position $4d$ of supergroup $P_{C}4_{2}/ncm~(138.529)$.\label{fig_138.529_2.7_strainingenericdirection_4d}}
\end{figure}
\input{gap_tables_tex/138.529_2.7_strainingenericdirection_4d_table.tex}
\input{si_tables_tex/138.529_2.7_strainingenericdirection_4d_table.tex}
\subsection{WP: $4c$}
\textbf{BCS Materials:} {Sm\textsubscript{2}CuO\textsubscript{4}~(280 K)}\footnote{BCS web page: \texttt{\href{http://webbdcrista1.ehu.es/magndata/index.php?this\_label=2.7} {http://webbdcrista1.ehu.es/magndata/index.php?this\_label=2.7}}}, {Eu\textsubscript{2}CuO\textsubscript{4}~(265 K)}\footnote{BCS web page: \texttt{\href{http://webbdcrista1.ehu.es/magndata/index.php?this\_label=2.77} {http://webbdcrista1.ehu.es/magndata/index.php?this\_label=2.77}}}.\\
\subsubsection{Topological bands in subgroup $P\bar{1}~(2.4)$}
\textbf{Perturbations:}
\begin{itemize}
\item B $\parallel$ [001] and strain in generic direction,
\item B $\parallel$ [100] and strain $\perp$ [110],
\item B $\parallel$ [100] and strain in generic direction,
\item B $\parallel$ [110] and strain $\perp$ [100],
\item B $\parallel$ [110] and strain in generic direction,
\item B $\perp$ [001] and strain $\perp$ [100],
\item B $\perp$ [001] and strain $\perp$ [110],
\item B $\perp$ [001] and strain in generic direction,
\item B $\perp$ [100] and strain $\parallel$ [110],
\item B $\perp$ [100] and strain $\perp$ [001],
\item B $\perp$ [100] and strain $\perp$ [110],
\item B $\perp$ [100] and strain in generic direction,
\item B $\perp$ [110] and strain $\parallel$ [100],
\item B $\perp$ [110] and strain $\perp$ [001],
\item B $\perp$ [110] and strain $\perp$ [100],
\item B $\perp$ [110] and strain in generic direction,
\item B in generic direction.
\end{itemize}
\begin{figure}[H]
\centering
\includegraphics[scale=0.6]{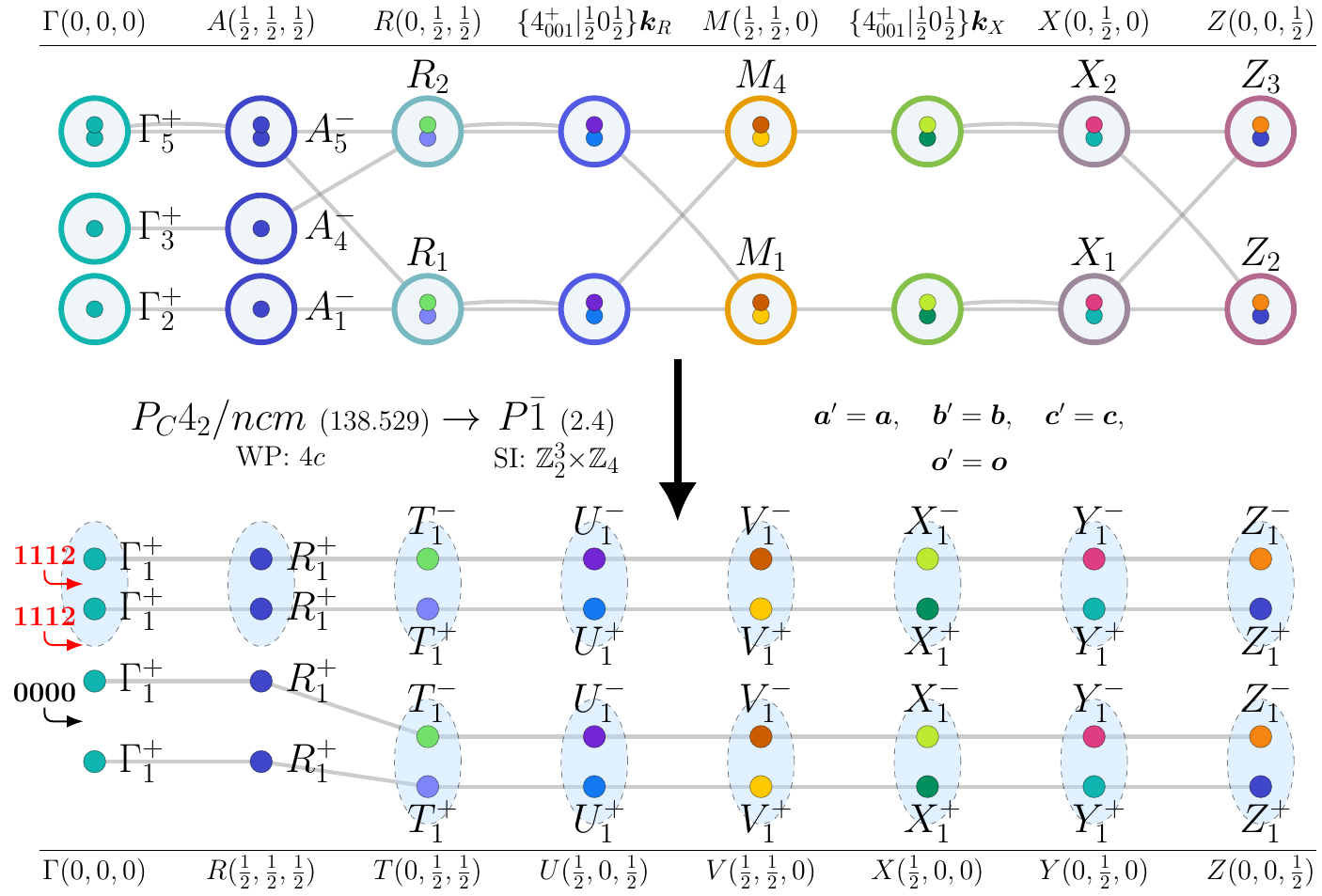}
\caption{Topological magnon bands in subgroup $P\bar{1}~(2.4)$ for magnetic moments on Wyckoff position $4c$ of supergroup $P_{C}4_{2}/ncm~(138.529)$.\label{fig_138.529_2.4_Bparallel001andstrainingenericdirection_4c}}
\end{figure}
\input{gap_tables_tex/138.529_2.4_Bparallel001andstrainingenericdirection_4c_table.tex}
\input{si_tables_tex/138.529_2.4_Bparallel001andstrainingenericdirection_4c_table.tex}
\subsubsection{Topological bands in subgroup $C2'/m'~(12.62)$}
\textbf{Perturbations:}
\begin{itemize}
\item B $\parallel$ [001] and strain $\perp$ [110],
\item B $\perp$ [110].
\end{itemize}
\begin{figure}[H]
\centering
\includegraphics[scale=0.6]{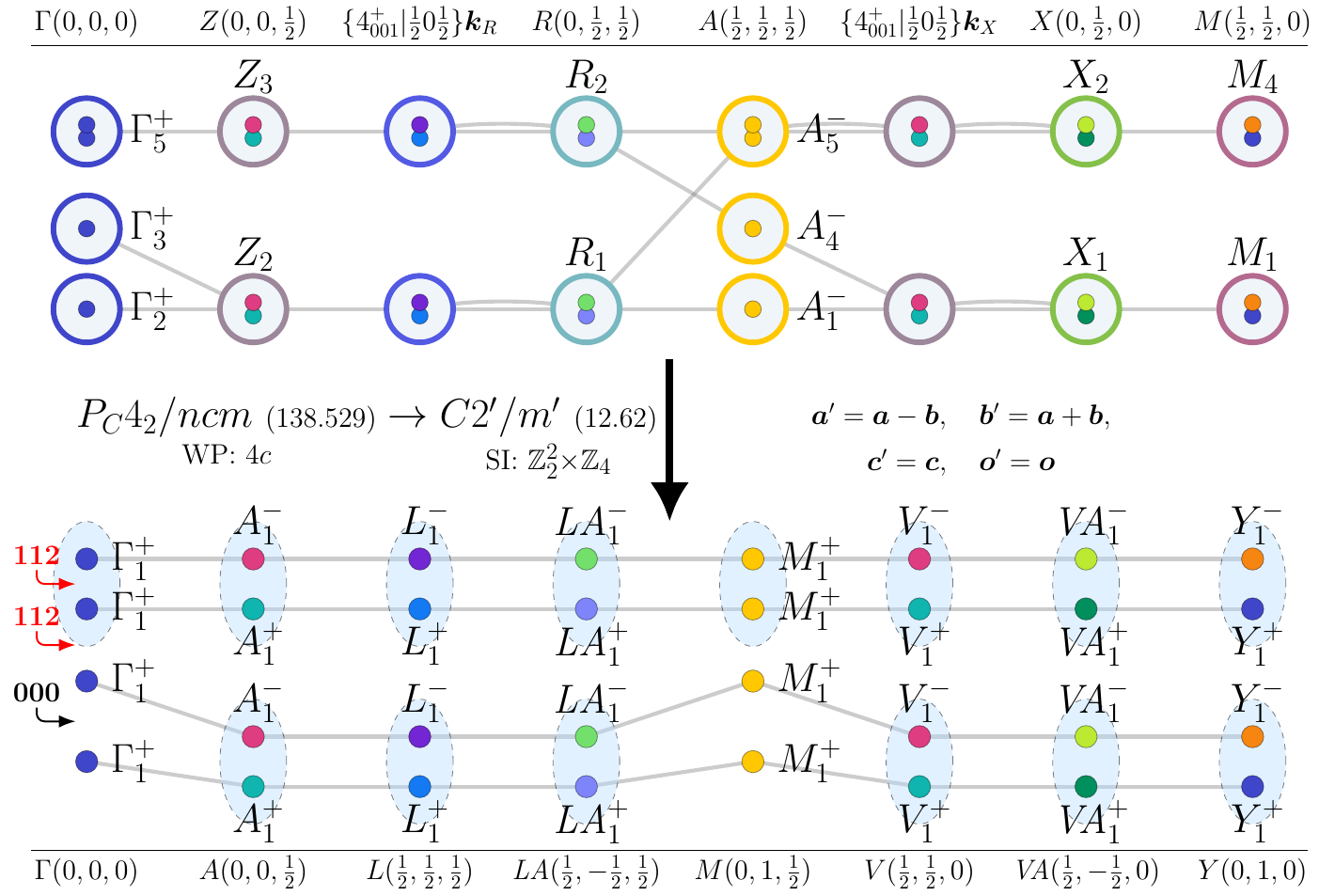}
\caption{Topological magnon bands in subgroup $C2'/m'~(12.62)$ for magnetic moments on Wyckoff position $4c$ of supergroup $P_{C}4_{2}/ncm~(138.529)$.\label{fig_138.529_12.62_Bparallel001andstrainperp110_4c}}
\end{figure}
\input{gap_tables_tex/138.529_12.62_Bparallel001andstrainperp110_4c_table.tex}
\input{si_tables_tex/138.529_12.62_Bparallel001andstrainperp110_4c_table.tex}
\subsubsection{Topological bands in subgroup $P2'/m'~(10.46)$}
\textbf{Perturbations:}
\begin{itemize}
\item B $\parallel$ [100] and strain $\parallel$ [110],
\item B $\parallel$ [100] and strain $\perp$ [001],
\item B $\parallel$ [110] and strain $\parallel$ [100],
\item B $\parallel$ [110] and strain $\perp$ [001],
\item B $\perp$ [001].
\end{itemize}
\begin{figure}[H]
\centering
\includegraphics[scale=0.6]{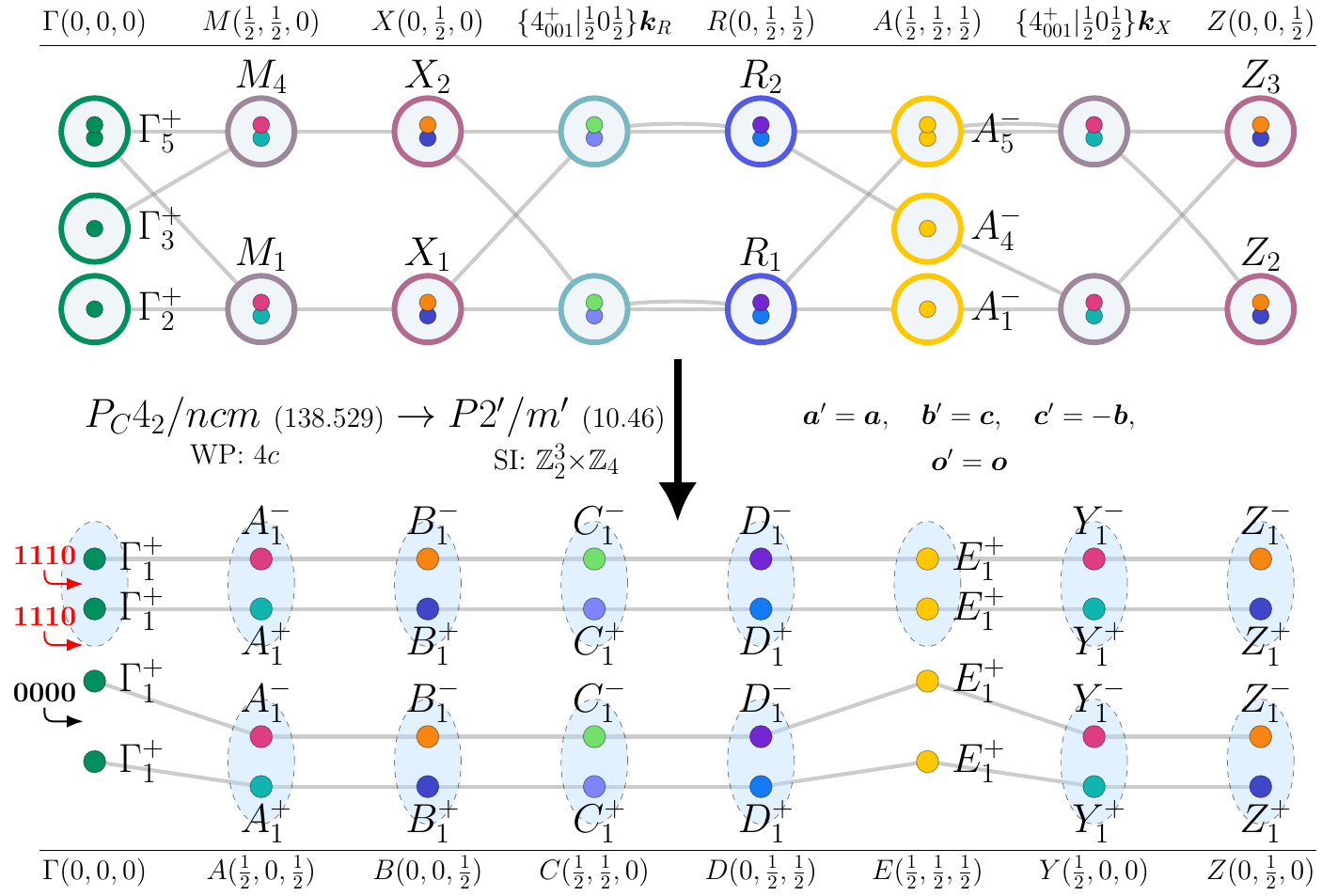}
\caption{Topological magnon bands in subgroup $P2'/m'~(10.46)$ for magnetic moments on Wyckoff position $4c$ of supergroup $P_{C}4_{2}/ncm~(138.529)$.\label{fig_138.529_10.46_Bparallel100andstrainparallel110_4c}}
\end{figure}
\input{gap_tables_tex/138.529_10.46_Bparallel100andstrainparallel110_4c_table.tex}
\input{si_tables_tex/138.529_10.46_Bparallel100andstrainparallel110_4c_table.tex}
\subsubsection{Topological bands in subgroup $P2'/c'~(13.69)$}
\textbf{Perturbations:}
\begin{itemize}
\item B $\parallel$ [001] and strain $\perp$ [100],
\item B $\perp$ [100].
\end{itemize}
\begin{figure}[H]
\centering
\includegraphics[scale=0.6]{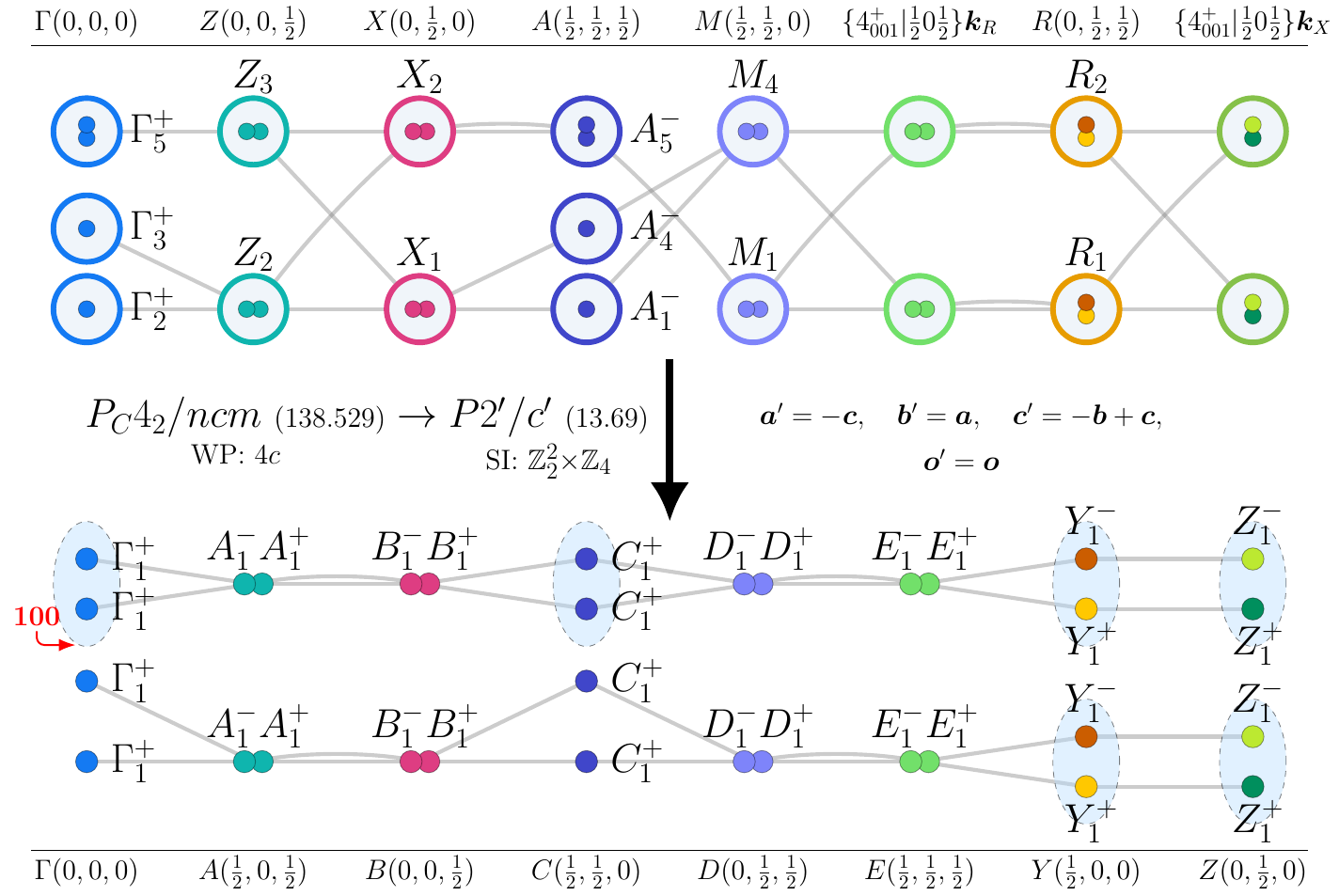}
\caption{Topological magnon bands in subgroup $P2'/c'~(13.69)$ for magnetic moments on Wyckoff position $4c$ of supergroup $P_{C}4_{2}/ncm~(138.529)$.\label{fig_138.529_13.69_Bparallel001andstrainperp100_4c}}
\end{figure}
\input{gap_tables_tex/138.529_13.69_Bparallel001andstrainperp100_4c_table.tex}
\input{si_tables_tex/138.529_13.69_Bparallel001andstrainperp100_4c_table.tex}
\subsubsection{Topological bands in subgroup $P_{S}\bar{1}~(2.7)$}
\textbf{Perturbation:}
\begin{itemize}
\item strain in generic direction.
\end{itemize}
\begin{figure}[H]
\centering
\includegraphics[scale=0.6]{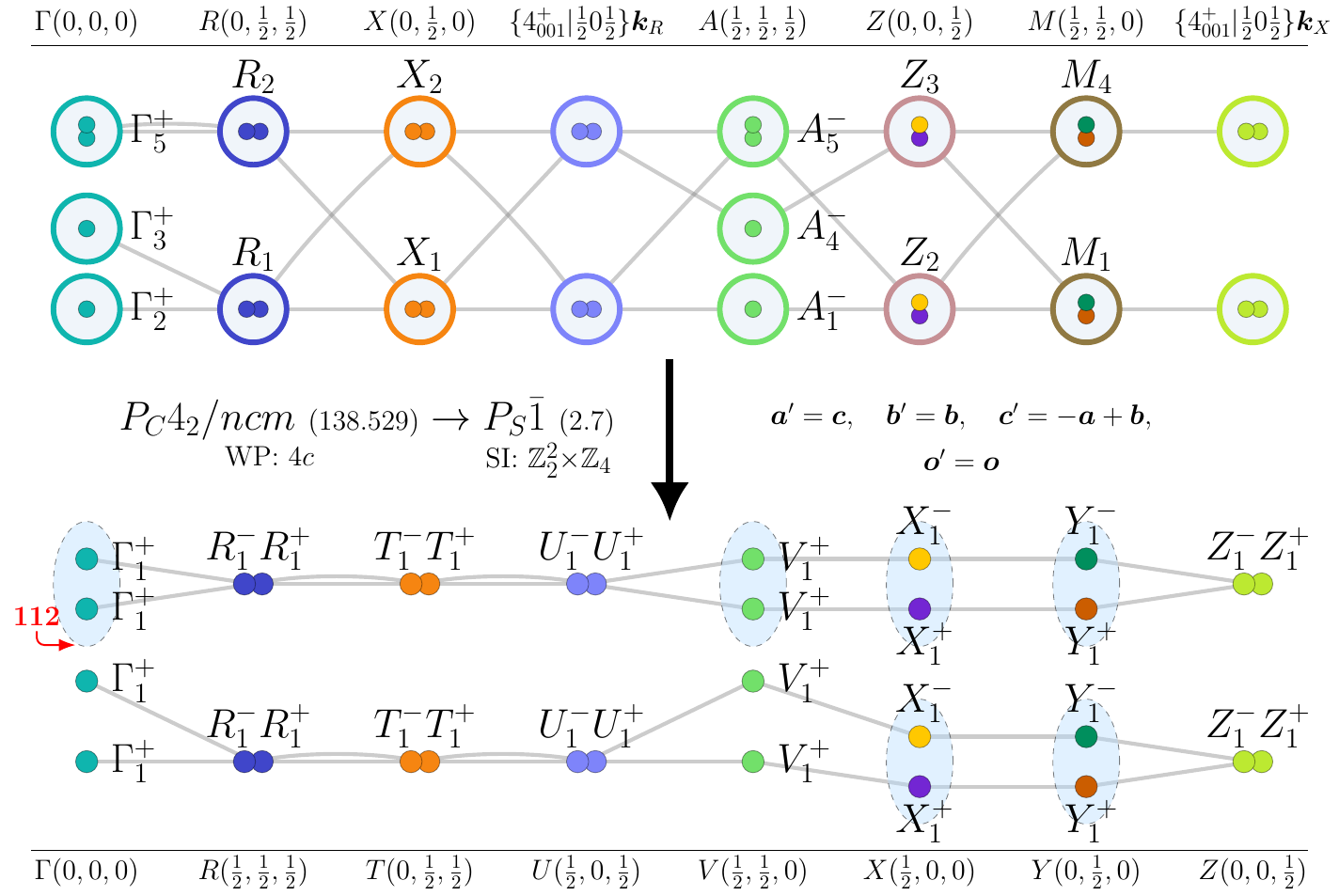}
\caption{Topological magnon bands in subgroup $P_{S}\bar{1}~(2.7)$ for magnetic moments on Wyckoff position $4c$ of supergroup $P_{C}4_{2}/ncm~(138.529)$.\label{fig_138.529_2.7_strainingenericdirection_4c}}
\end{figure}
\input{gap_tables_tex/138.529_2.7_strainingenericdirection_4c_table.tex}
\input{si_tables_tex/138.529_2.7_strainingenericdirection_4c_table.tex}

\section{MSG $I4'/mmm'~(139.535)$}
\textbf{Nontrivial-SI Subgroups:} $P\bar{1}~(2.4)$, $C2'/m'~(12.62)$, $C2'/m'~(12.62)$, $Fm'm'2~(42.222)$, $C2/m~(12.58)$, $Fm'm'm~(69.524)$, $C2/m~(12.58)$.\\

\textbf{Trivial-SI Subgroups:} $Cm'~(8.34)$, $Cm'~(8.34)$, $C2'~(5.15)$, $Cm~(8.32)$, $Fm'm2'~(42.221)$, $Cm~(8.32)$, $C2~(5.13)$, $Imm2~(44.229)$, $I4'mm'~(107.230)$, $C2~(5.13)$, $Imm2~(44.229)$, $Immm~(71.533)$.\\

\subsection{WP: $8f$}
\textbf{BCS Materials:} {CaFe\textsubscript{4}Al\textsubscript{8}~(180 K)}\footnote{BCS web page: \texttt{\href{http://webbdcrista1.ehu.es/magndata/index.php?this\_label=0.236} {http://webbdcrista1.ehu.es/magndata/index.php?this\_label=0.236}}}.\\
\subsubsection{Topological bands in subgroup $P\bar{1}~(2.4)$}
\textbf{Perturbations:}
\begin{itemize}
\item strain in generic direction,
\item B $\parallel$ [001] and strain $\perp$ [100],
\item B $\parallel$ [100] and strain $\parallel$ [110],
\item B $\parallel$ [100] and strain $\perp$ [001],
\item B $\parallel$ [100] and strain $\perp$ [110],
\item B $\parallel$ [110] and strain $\parallel$ [100],
\item B $\parallel$ [110] and strain $\perp$ [001],
\item B $\parallel$ [110] and strain $\perp$ [100],
\item B $\parallel$ [110] and strain $\perp$ [110],
\item B in generic direction,
\item B $\perp$ [110] and strain $\parallel$ [100],
\item B $\perp$ [110] and strain $\perp$ [001],
\item B $\perp$ [110] and strain $\perp$ [100].
\end{itemize}
\begin{figure}[H]
\centering
\includegraphics[scale=0.6]{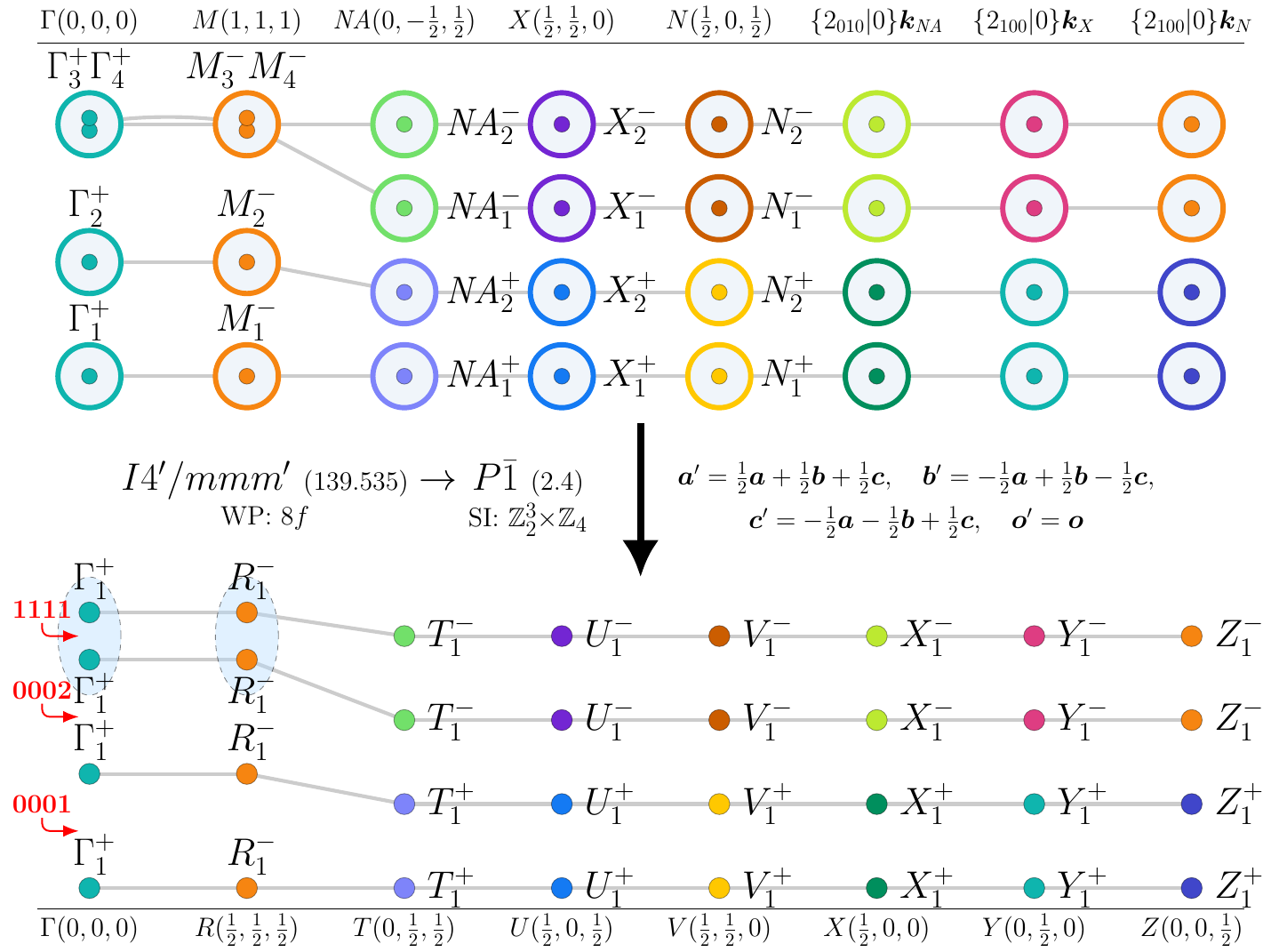}
\caption{Topological magnon bands in subgroup $P\bar{1}~(2.4)$ for magnetic moments on Wyckoff position $8f$ of supergroup $I4'/mmm'~(139.535)$.\label{fig_139.535_2.4_strainingenericdirection_8f}}
\end{figure}
\input{gap_tables_tex/139.535_2.4_strainingenericdirection_8f_table.tex}
\input{si_tables_tex/139.535_2.4_strainingenericdirection_8f_table.tex}
\subsubsection{Topological bands in subgroup $C2'/m'~(12.62)$}
\textbf{Perturbation:}
\begin{itemize}
\item B $\parallel$ [110].
\end{itemize}
\begin{figure}[H]
\centering
\includegraphics[scale=0.6]{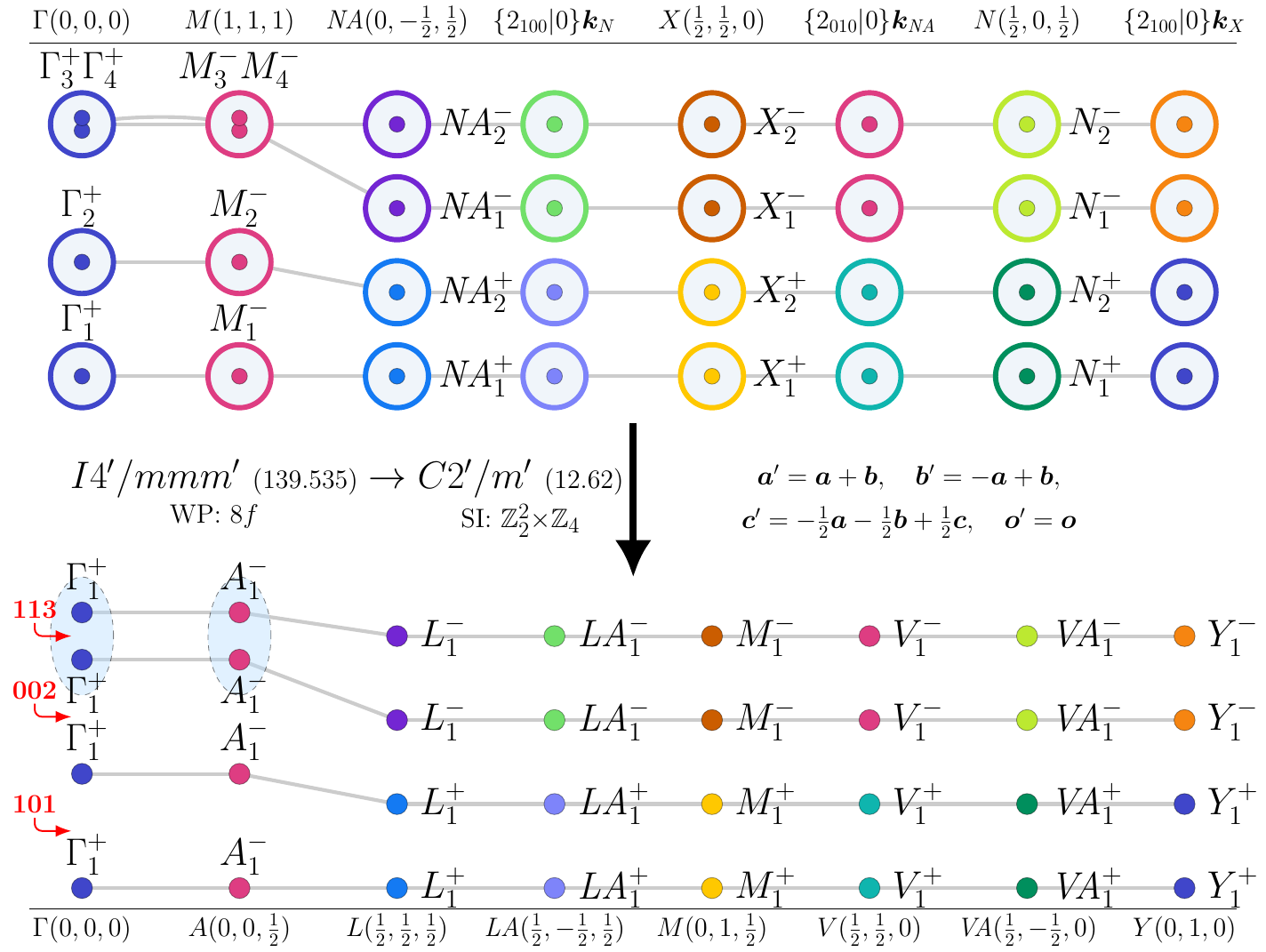}
\caption{Topological magnon bands in subgroup $C2'/m'~(12.62)$ for magnetic moments on Wyckoff position $8f$ of supergroup $I4'/mmm'~(139.535)$.\label{fig_139.535_12.62_Bparallel110_8f}}
\end{figure}
\input{gap_tables_tex/139.535_12.62_Bparallel110_8f_table.tex}
\input{si_tables_tex/139.535_12.62_Bparallel110_8f_table.tex}
\subsubsection{Topological bands in subgroup $C2'/m'~(12.62)$}
\textbf{Perturbations:}
\begin{itemize}
\item strain $\perp$ [110],
\item B $\perp$ [110].
\end{itemize}
\begin{figure}[H]
\centering
\includegraphics[scale=0.6]{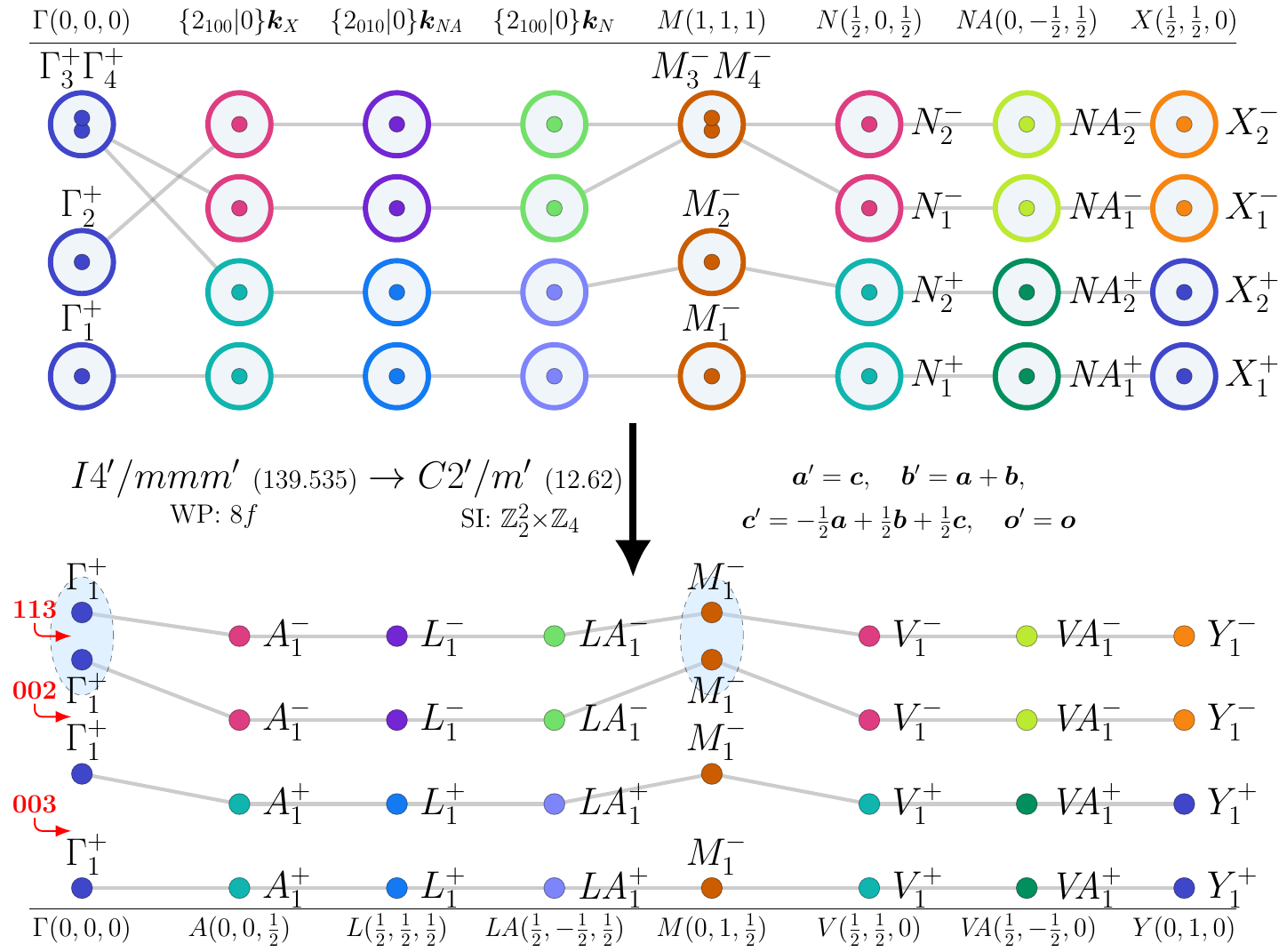}
\caption{Topological magnon bands in subgroup $C2'/m'~(12.62)$ for magnetic moments on Wyckoff position $8f$ of supergroup $I4'/mmm'~(139.535)$.\label{fig_139.535_12.62_strainperp110_8f}}
\end{figure}
\input{gap_tables_tex/139.535_12.62_strainperp110_8f_table.tex}
\input{si_tables_tex/139.535_12.62_strainperp110_8f_table.tex}

\section{MSG $I_{c}4/mcm~(140.550)$}
\textbf{Nontrivial-SI Subgroups:} $P\bar{1}~(2.4)$, $C2'/m'~(12.62)$, $C2'/m'~(12.62)$, $C2'/m'~(12.62)$, $P_{S}\bar{1}~(2.7)$, $Fm'm'2~(42.222)$, $C_{a}2~(5.17)$, $C2/m~(12.58)$, $Fm'm'm~(69.524)$, $C_{a}2/m~(12.64)$, $Fm'm'2~(42.222)$, $C_{a}2~(5.17)$, $I_{c}ba2~(45.239)$, $C2/m~(12.58)$, $Fm'm'm~(69.524)$, $Im'm'm~(71.536)$, $C_{a}2/m~(12.64)$, $C2/c~(15.85)$, $Im'm'a~(74.558)$, $C_{c}2/c~(15.90)$, $I4m'm'~(107.231)$, $I4/mm'm'~(139.537)$.\\

\textbf{Trivial-SI Subgroups:} $Cm'~(8.34)$, $Cm'~(8.34)$, $Cm'~(8.34)$, $C2'~(5.15)$, $C2'~(5.15)$, $C2'~(5.15)$, $P_{S}1~(1.3)$, $Cm~(8.32)$, $Fm'm2'~(42.221)$, $C_{a}m~(8.36)$, $Cm~(8.32)$, $Fm'm2'~(42.221)$, $Im'm2'~(44.231)$, $C_{a}m~(8.36)$, $Cc~(9.37)$, $Im'a2'~(46.243)$, $C_{c}c~(9.40)$, $C2~(5.13)$, $F_{S}mm2~(42.223)$, $C2~(5.13)$, $Im'm'2~(44.232)$, $F_{S}mm2~(42.223)$, $F_{S}mmm~(69.526)$, $C2~(5.13)$, $Im'm'2~(44.232)$, $C_{c}2~(5.16)$, $I_{a}ma2~(46.247)$, $I_{c}bam~(72.546)$, $I_{c}4cm~(108.238)$.\\

\subsection{WP: $4a$}
\textbf{BCS Materials:} {SrFeO\textsubscript{2}F~(710 K)}\footnote{BCS web page: \texttt{\href{http://webbdcrista1.ehu.es/magndata/index.php?this\_label=1.84} {http://webbdcrista1.ehu.es/magndata/index.php?this\_label=1.84}}}, {Pb\textsubscript{0.7}Bi\textsubscript{0.3}Fe\textsubscript{0.762}W\textsubscript{0.231}O\textsubscript{3}~(504 K)}\footnote{BCS web page: \texttt{\href{http://webbdcrista1.ehu.es/magndata/index.php?this\_label=1.591} {http://webbdcrista1.ehu.es/magndata/index.php?this\_label=1.591}}}, {Pb\textsubscript{0.8}Bi\textsubscript{0.2}Fe\textsubscript{0.728}W\textsubscript{0.264}O\textsubscript{3}~(435 K)}\footnote{BCS web page: \texttt{\href{http://webbdcrista1.ehu.es/magndata/index.php?this\_label=1.590} {http://webbdcrista1.ehu.es/magndata/index.php?this\_label=1.590}}}, {KNiF\textsubscript{3}~(275 K)}\footnote{BCS web page: \texttt{\href{http://webbdcrista1.ehu.es/magndata/index.php?this\_label=1.250} {http://webbdcrista1.ehu.es/magndata/index.php?this\_label=1.250}}}, {UNiGa\textsubscript{5}~(86 K)}\footnote{BCS web page: \texttt{\href{http://webbdcrista1.ehu.es/magndata/index.php?this\_label=1.254} {http://webbdcrista1.ehu.es/magndata/index.php?this\_label=1.254}}}, {CeIr(In\textsubscript{0.97}Cd\textsubscript{0.03})\textsubscript{5}~(3 K)}\footnote{BCS web page: \texttt{\href{http://webbdcrista1.ehu.es/magndata/index.php?this\_label=1.598} {http://webbdcrista1.ehu.es/magndata/index.php?this\_label=1.598}}}.\\
\subsubsection{Topological bands in subgroup $C2'/m'~(12.62)$}
\textbf{Perturbations:}
\begin{itemize}
\item B $\parallel$ [100] and strain $\parallel$ [110],
\item B $\parallel$ [100] and strain $\perp$ [001],
\item B $\parallel$ [110] and strain $\parallel$ [100],
\item B $\parallel$ [110] and strain $\perp$ [001],
\item B $\perp$ [001].
\end{itemize}
\begin{figure}[H]
\centering
\includegraphics[scale=0.6]{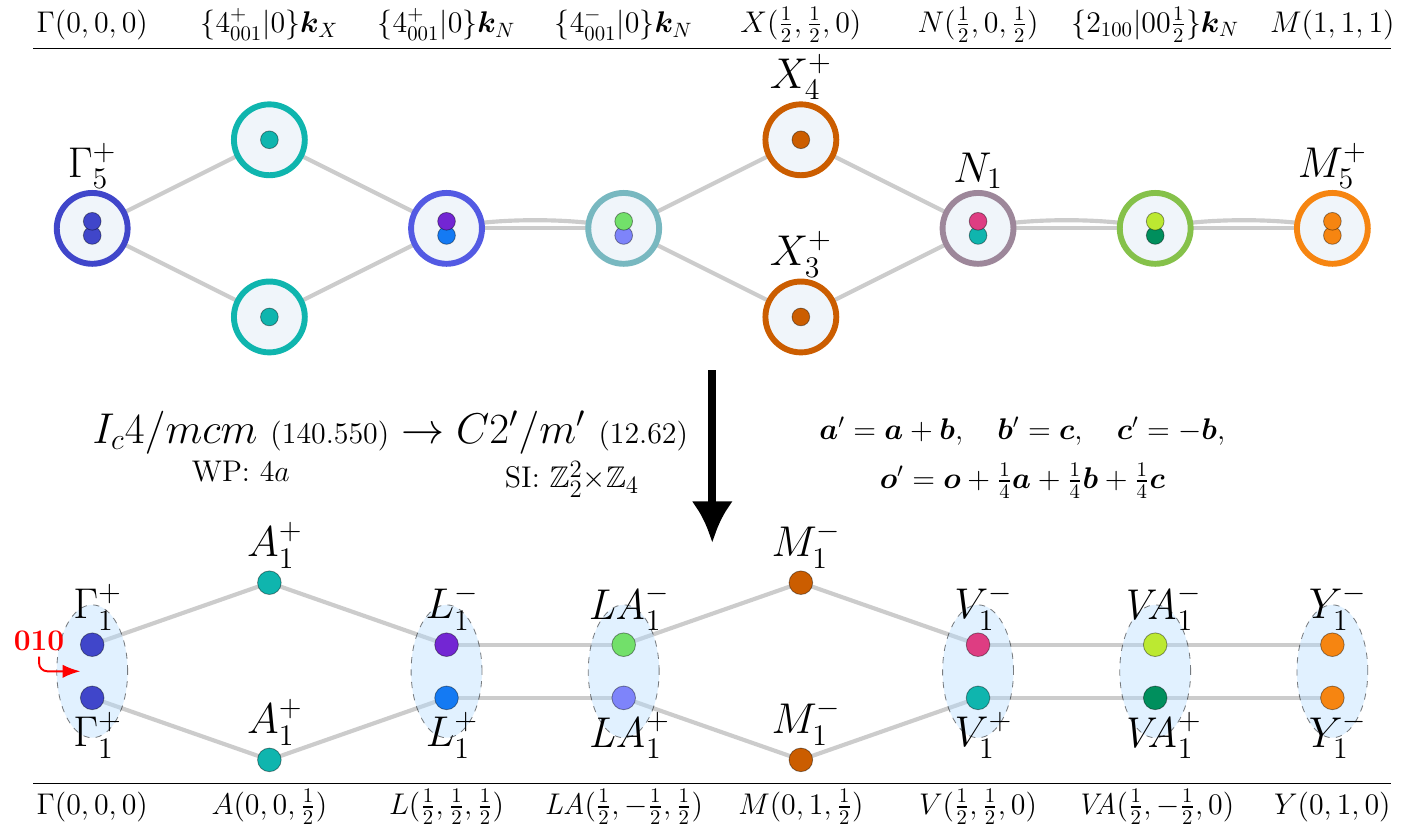}
\caption{Topological magnon bands in subgroup $C2'/m'~(12.62)$ for magnetic moments on Wyckoff position $4a$ of supergroup $I_{c}4/mcm~(140.550)$.\label{fig_140.550_12.62_Bparallel100andstrainparallel110_4a}}
\end{figure}
\input{gap_tables_tex/140.550_12.62_Bparallel100andstrainparallel110_4a_table.tex}
\input{si_tables_tex/140.550_12.62_Bparallel100andstrainparallel110_4a_table.tex}
\subsection{WP: $8g$}
\textbf{BCS Materials:} {Tb\textsubscript{2}CoGa\textsubscript{8}~(28.5 K)}\footnote{BCS web page: \texttt{\href{http://webbdcrista1.ehu.es/magndata/index.php?this\_label=1.87} {http://webbdcrista1.ehu.es/magndata/index.php?this\_label=1.87}}}, {Dy\textsubscript{2}CoGa\textsubscript{8}~(15.2 K)}\footnote{BCS web page: \texttt{\href{http://webbdcrista1.ehu.es/magndata/index.php?this\_label=1.80} {http://webbdcrista1.ehu.es/magndata/index.php?this\_label=1.80}}}, {Nd\textsubscript{2}RhIn\textsubscript{8}~(10.63 K)}\footnote{BCS web page: \texttt{\href{http://webbdcrista1.ehu.es/magndata/index.php?this\_label=1.82} {http://webbdcrista1.ehu.es/magndata/index.php?this\_label=1.82}}}.\\
\subsubsection{Topological bands in subgroup $C2'/m'~(12.62)$}
\textbf{Perturbations:}
\begin{itemize}
\item B $\parallel$ [100] and strain $\parallel$ [110],
\item B $\parallel$ [100] and strain $\perp$ [001],
\item B $\parallel$ [110] and strain $\parallel$ [100],
\item B $\parallel$ [110] and strain $\perp$ [001],
\item B $\perp$ [001].
\end{itemize}
\begin{figure}[H]
\centering
\includegraphics[scale=0.6]{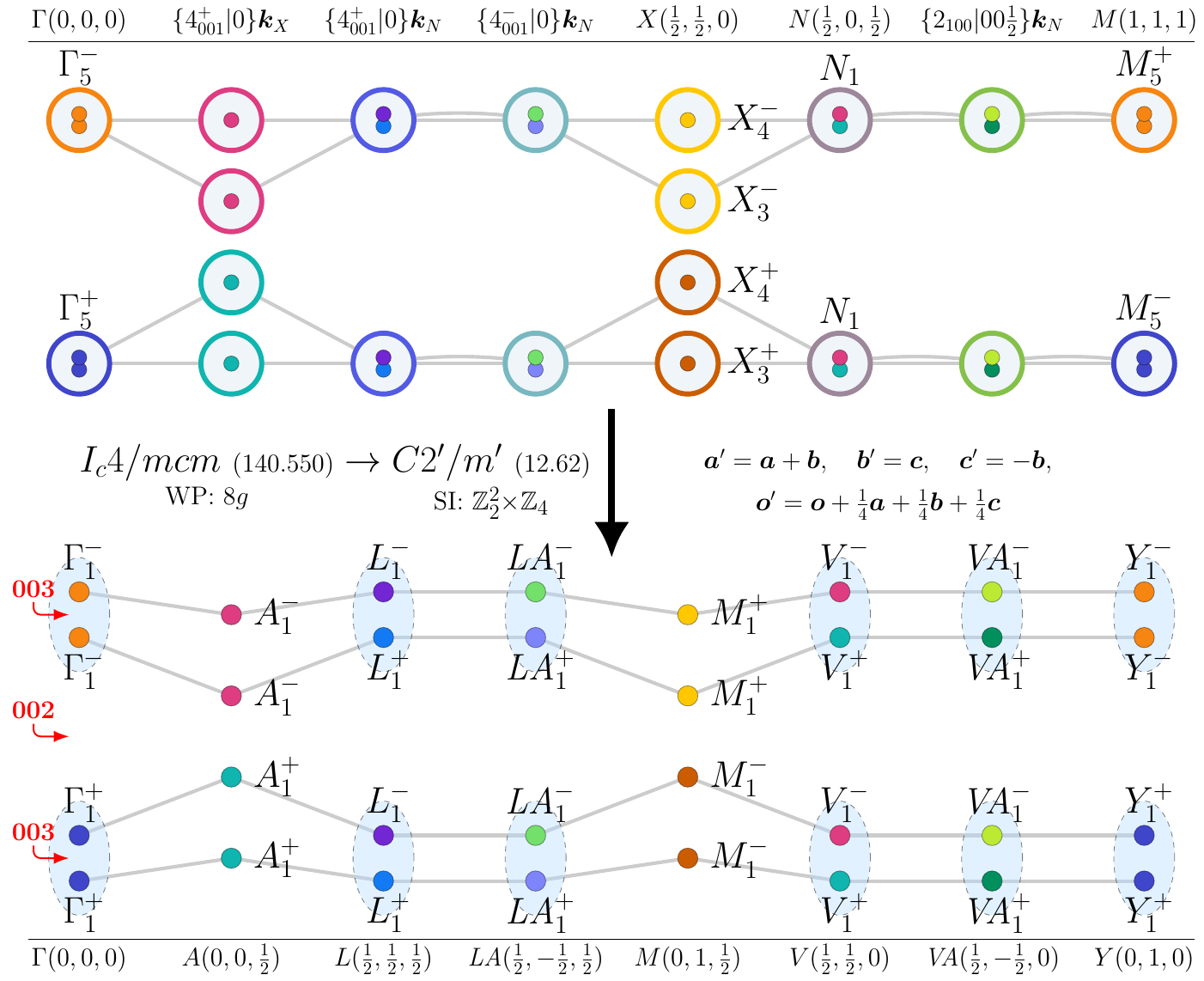}
\caption{Topological magnon bands in subgroup $C2'/m'~(12.62)$ for magnetic moments on Wyckoff position $8g$ of supergroup $I_{c}4/mcm~(140.550)$.\label{fig_140.550_12.62_Bparallel100andstrainparallel110_8g}}
\end{figure}
\input{gap_tables_tex/140.550_12.62_Bparallel100andstrainparallel110_8g_table.tex}
\input{si_tables_tex/140.550_12.62_Bparallel100andstrainparallel110_8g_table.tex}

\section{MSG $I4_{1}/amd~(141.551)$}
\textbf{Nontrivial-SI Subgroups:} $P\bar{1}~(2.4)$, $C2/c~(15.85)$, $C2/c~(15.85)$, $Fddd~(70.527)$, $C2/m~(12.58)$, $I4_{1}/a~(88.81)$.\\

\textbf{Trivial-SI Subgroups:} $Cc~(9.37)$, $Cc~(9.37)$, $Cm~(8.32)$, $C2~(5.13)$, $Fdd2~(43.224)$, $C2~(5.13)$, $Fdd2~(43.224)$, $Imm2~(44.229)$, $C2~(5.13)$, $Ima2~(46.241)$, $Imma~(74.554)$, $I4_{1}~(80.29)$, $I4_{1}md~(109.239)$.\\

\subsection{WP: $8d$}
\textbf{BCS Materials:} {CdYb\textsubscript{2}S\textsubscript{4}~(1.92 K)}\footnote{BCS web page: \texttt{\href{http://webbdcrista1.ehu.es/magndata/index.php?this\_label=0.324} {http://webbdcrista1.ehu.es/magndata/index.php?this\_label=0.324}}}, {CdYb\textsubscript{2}Se\textsubscript{4}~(1.75 K)}\footnote{BCS web page: \texttt{\href{http://webbdcrista1.ehu.es/magndata/index.php?this\_label=0.325} {http://webbdcrista1.ehu.es/magndata/index.php?this\_label=0.325}}}.\\
\subsubsection{Topological bands in subgroup $P\bar{1}~(2.4)$}
\textbf{Perturbations:}
\begin{itemize}
\item strain in generic direction,
\item B $\parallel$ [001] and strain $\perp$ [100],
\item B $\parallel$ [001] and strain $\perp$ [110],
\item B $\parallel$ [100] and strain $\parallel$ [110],
\item B $\parallel$ [100] and strain $\perp$ [001],
\item B $\parallel$ [100] and strain $\perp$ [110],
\item B $\parallel$ [110] and strain $\parallel$ [100],
\item B $\parallel$ [110] and strain $\perp$ [001],
\item B $\parallel$ [110] and strain $\perp$ [100],
\item B in generic direction.
\end{itemize}
\begin{figure}[H]
\centering
\includegraphics[scale=0.6]{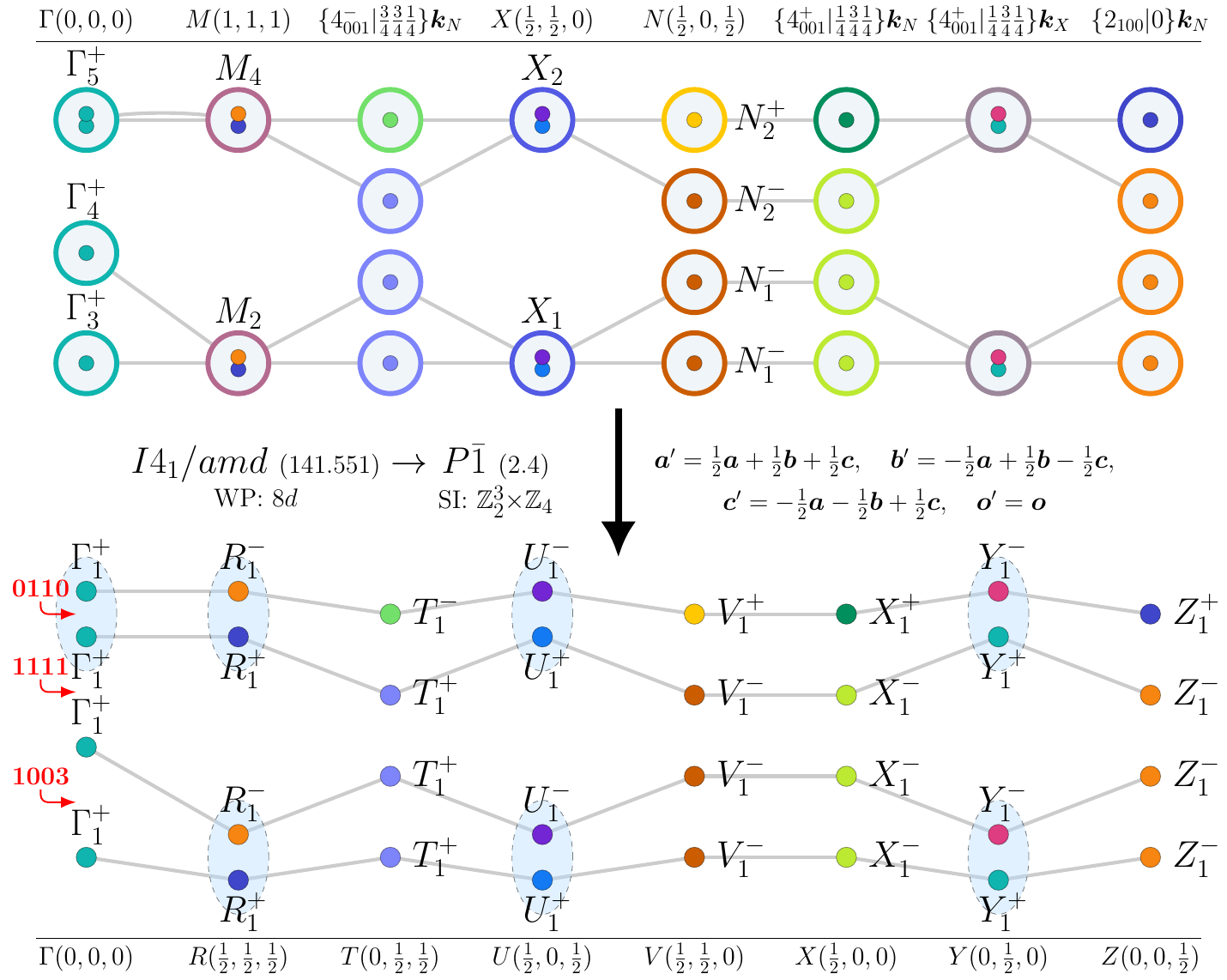}
\caption{Topological magnon bands in subgroup $P\bar{1}~(2.4)$ for magnetic moments on Wyckoff position $8d$ of supergroup $I4_{1}/amd~(141.551)$.\label{fig_141.551_2.4_strainingenericdirection_8d}}
\end{figure}
\input{gap_tables_tex/141.551_2.4_strainingenericdirection_8d_table.tex}
\input{si_tables_tex/141.551_2.4_strainingenericdirection_8d_table.tex}

\section{MSG $I4_{1}'/am'd~(141.554)$}
\textbf{Nontrivial-SI Subgroups:} $P\bar{1}~(2.4)$, $C2'/m'~(12.62)$, $C2'/m'~(12.62)$, $C2/c~(15.85)$, $C2/c~(15.85)$, $Im'm'a~(74.558)$, $Fddd~(70.527)$.\\

\textbf{Trivial-SI Subgroups:} $Cm'~(8.34)$, $Cm'~(8.34)$, $C2'~(5.15)$, $Cc~(9.37)$, $Cc~(9.37)$, $Im'a2'~(46.243)$, $C2~(5.13)$, $Fdd2~(43.224)$, $C2~(5.13)$, $Im'm'2~(44.232)$, $Fdd2~(43.224)$, $I4_{1}'m'd~(109.241)$.\\

\subsection{WP: $8c+8d$}
\textbf{BCS Materials:} {Er\textsubscript{2}Ru\textsubscript{2}O\textsubscript{7}~(90 K)}\footnote{BCS web page: \texttt{\href{http://webbdcrista1.ehu.es/magndata/index.php?this\_label=0.154} {http://webbdcrista1.ehu.es/magndata/index.php?this\_label=0.154}}}.\\
\subsubsection{Topological bands in subgroup $P\bar{1}~(2.4)$}
\textbf{Perturbations:}
\begin{itemize}
\item strain in generic direction,
\item B $\parallel$ [001] and strain $\perp$ [110],
\item B $\parallel$ [100] and strain $\parallel$ [110],
\item B $\parallel$ [100] and strain $\perp$ [001],
\item B $\parallel$ [100] and strain $\perp$ [100],
\item B $\parallel$ [100] and strain $\perp$ [110],
\item B $\parallel$ [110] and strain $\parallel$ [100],
\item B $\parallel$ [110] and strain $\perp$ [001],
\item B $\parallel$ [110] and strain $\perp$ [100],
\item B in generic direction,
\item B $\perp$ [100] and strain $\parallel$ [110],
\item B $\perp$ [100] and strain $\perp$ [001],
\item B $\perp$ [100] and strain $\perp$ [110].
\end{itemize}
\begin{figure}[H]
\centering
\includegraphics[scale=0.6]{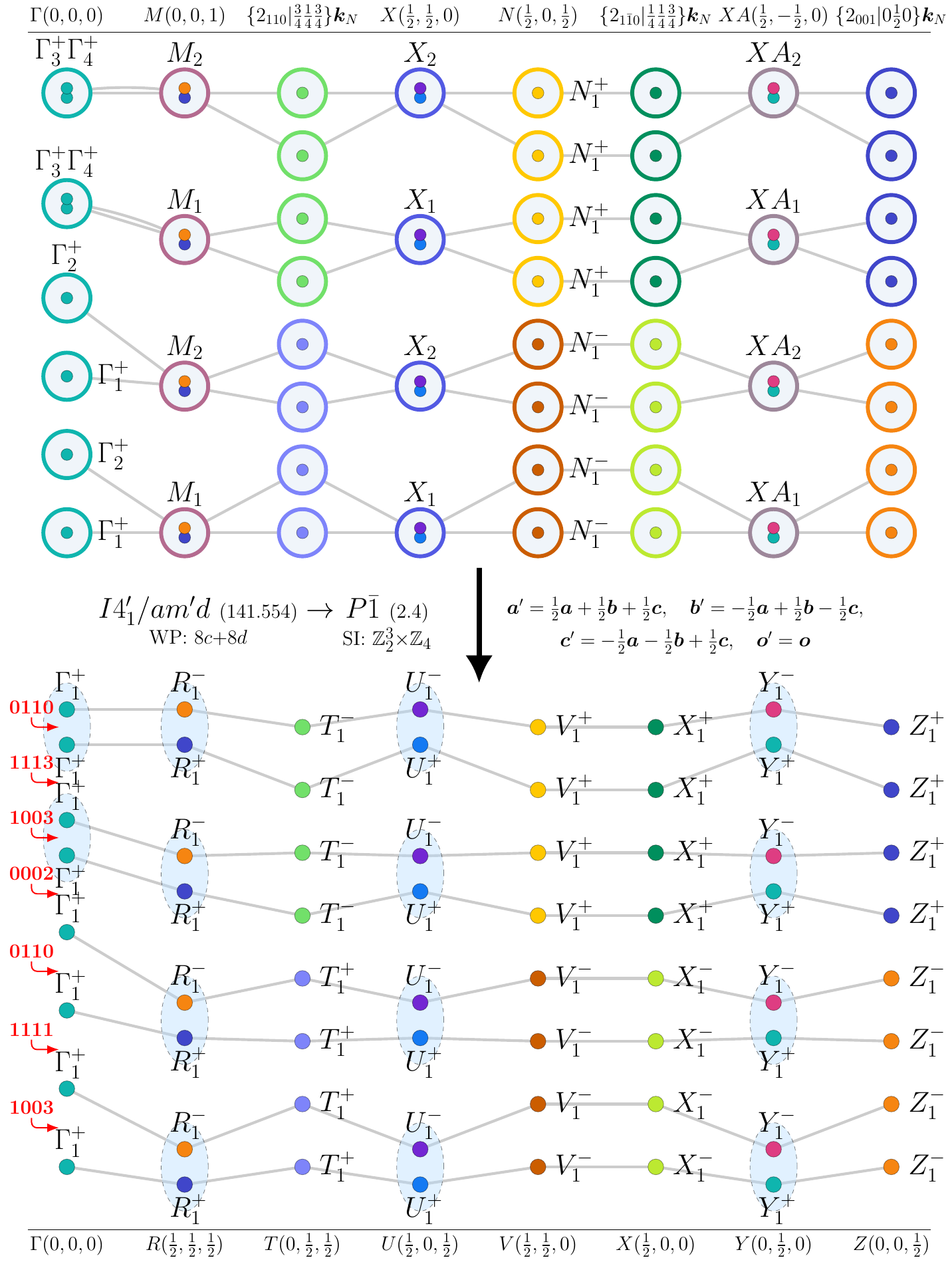}
\caption{Topological magnon bands in subgroup $P\bar{1}~(2.4)$ for magnetic moments on Wyckoff positions $8c+8d$ of supergroup $I4_{1}'/am'd~(141.554)$.\label{fig_141.554_2.4_strainingenericdirection_8c+8d}}
\end{figure}
\input{gap_tables_tex/141.554_2.4_strainingenericdirection_8c+8d_table.tex}
\input{si_tables_tex/141.554_2.4_strainingenericdirection_8c+8d_table.tex}
\subsubsection{Topological bands in subgroup $C2'/m'~(12.62)$}
\textbf{Perturbation:}
\begin{itemize}
\item B $\parallel$ [100].
\end{itemize}
\begin{figure}[H]
\centering
\includegraphics[scale=0.6]{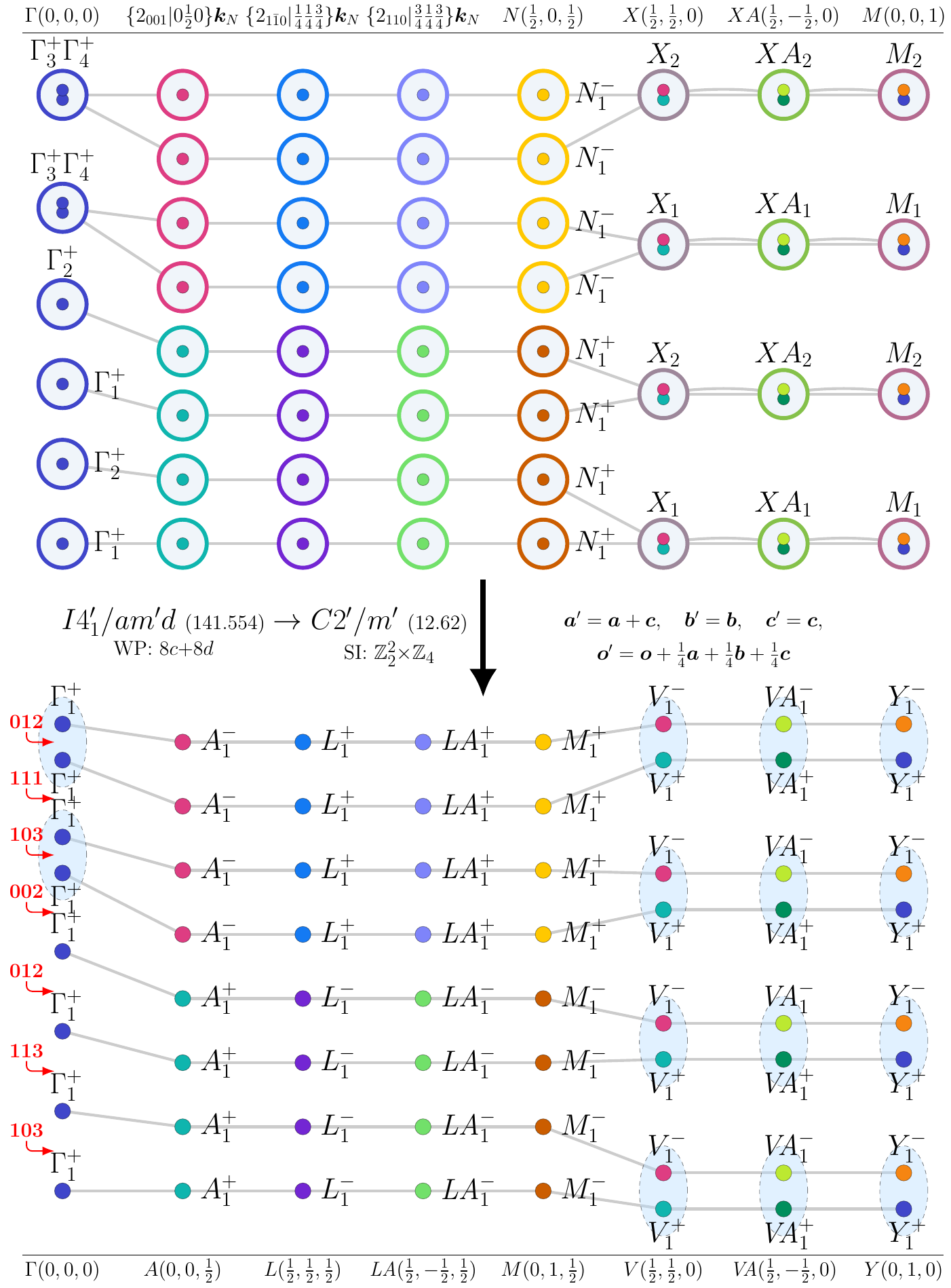}
\caption{Topological magnon bands in subgroup $C2'/m'~(12.62)$ for magnetic moments on Wyckoff positions $8c+8d$ of supergroup $I4_{1}'/am'd~(141.554)$.\label{fig_141.554_12.62_Bparallel100_8c+8d}}
\end{figure}
\input{gap_tables_tex/141.554_12.62_Bparallel100_8c+8d_table.tex}
\input{si_tables_tex/141.554_12.62_Bparallel100_8c+8d_table.tex}
\subsubsection{Topological bands in subgroup $C2'/m'~(12.62)$}
\textbf{Perturbations:}
\begin{itemize}
\item strain $\perp$ [100],
\item B $\perp$ [100].
\end{itemize}
\begin{figure}[H]
\centering
\includegraphics[scale=0.6]{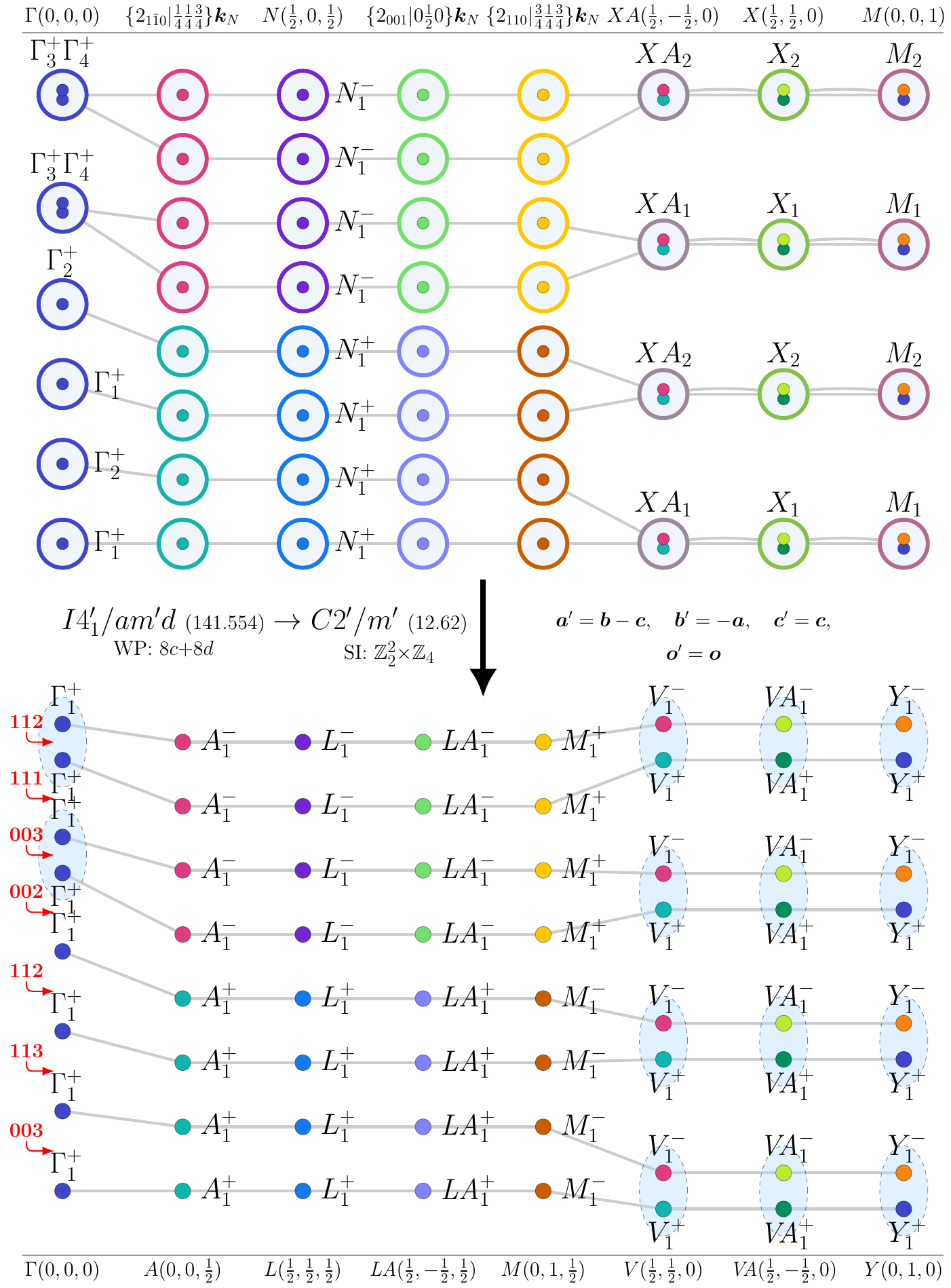}
\caption{Topological magnon bands in subgroup $C2'/m'~(12.62)$ for magnetic moments on Wyckoff positions $8c+8d$ of supergroup $I4_{1}'/am'd~(141.554)$.\label{fig_141.554_12.62_strainperp100_8c+8d}}
\end{figure}
\input{gap_tables_tex/141.554_12.62_strainperp100_8c+8d_table.tex}
\input{si_tables_tex/141.554_12.62_strainperp100_8c+8d_table.tex}
\subsection{WP: $8c$}
\textbf{BCS Materials:} {Er\textsubscript{2}Ti\textsubscript{2}O\textsubscript{7}~(1.173 K)}\footnote{BCS web page: \texttt{\href{http://webbdcrista1.ehu.es/magndata/index.php?this\_label=0.29} {http://webbdcrista1.ehu.es/magndata/index.php?this\_label=0.29}}}.\\
\subsubsection{Topological bands in subgroup $P\bar{1}~(2.4)$}
\textbf{Perturbations:}
\begin{itemize}
\item strain in generic direction,
\item B $\parallel$ [001] and strain $\perp$ [110],
\item B $\parallel$ [100] and strain $\parallel$ [110],
\item B $\parallel$ [100] and strain $\perp$ [001],
\item B $\parallel$ [100] and strain $\perp$ [100],
\item B $\parallel$ [100] and strain $\perp$ [110],
\item B $\parallel$ [110] and strain $\parallel$ [100],
\item B $\parallel$ [110] and strain $\perp$ [001],
\item B $\parallel$ [110] and strain $\perp$ [100],
\item B in generic direction,
\item B $\perp$ [100] and strain $\parallel$ [110],
\item B $\perp$ [100] and strain $\perp$ [001],
\item B $\perp$ [100] and strain $\perp$ [110].
\end{itemize}
\begin{figure}[H]
\centering
\includegraphics[scale=0.6]{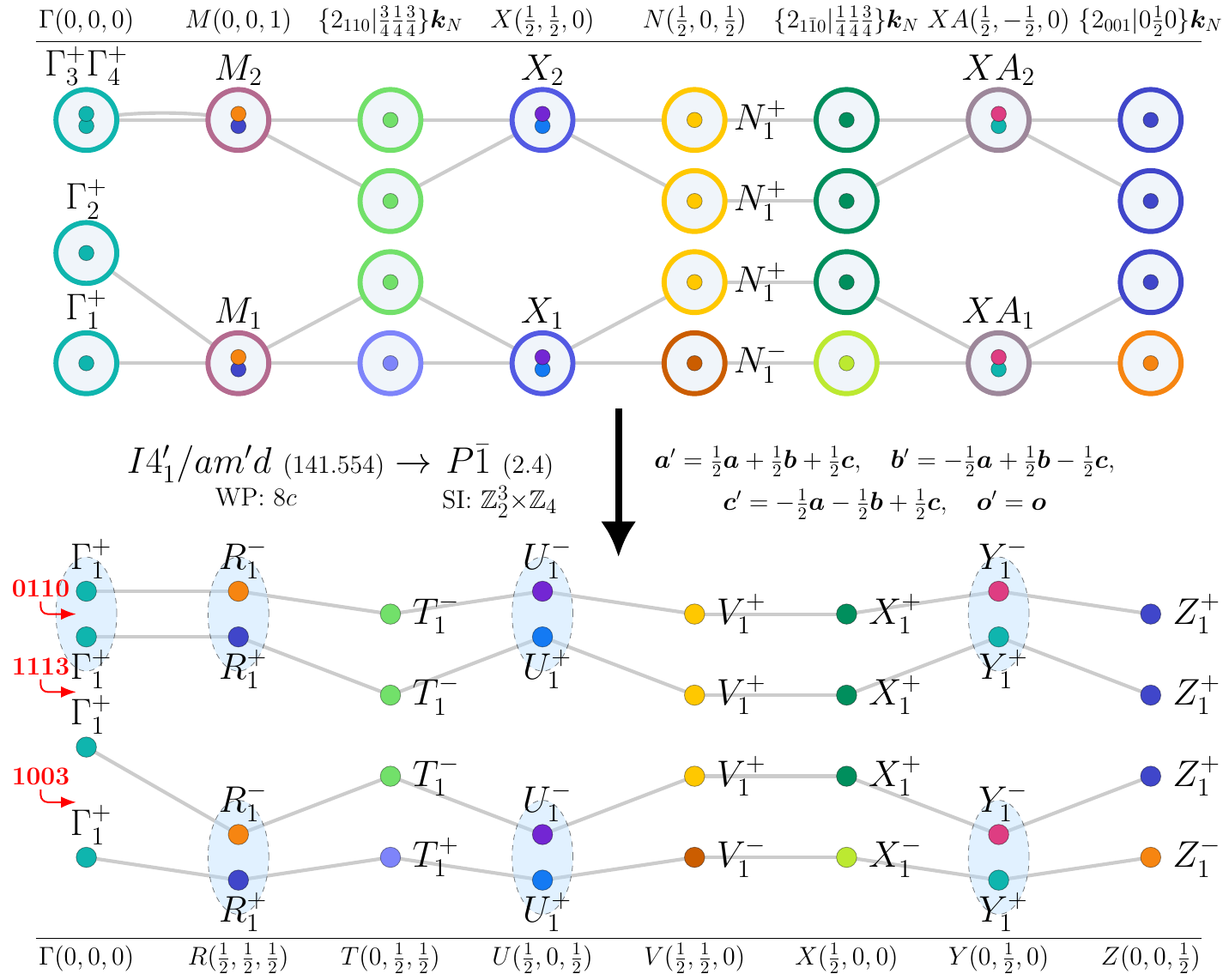}
\caption{Topological magnon bands in subgroup $P\bar{1}~(2.4)$ for magnetic moments on Wyckoff position $8c$ of supergroup $I4_{1}'/am'd~(141.554)$.\label{fig_141.554_2.4_strainingenericdirection_8c}}
\end{figure}
\input{gap_tables_tex/141.554_2.4_strainingenericdirection_8c_table.tex}
\input{si_tables_tex/141.554_2.4_strainingenericdirection_8c_table.tex}
\subsubsection{Topological bands in subgroup $C2'/m'~(12.62)$}
\textbf{Perturbation:}
\begin{itemize}
\item B $\parallel$ [100].
\end{itemize}
\begin{figure}[H]
\centering
\includegraphics[scale=0.6]{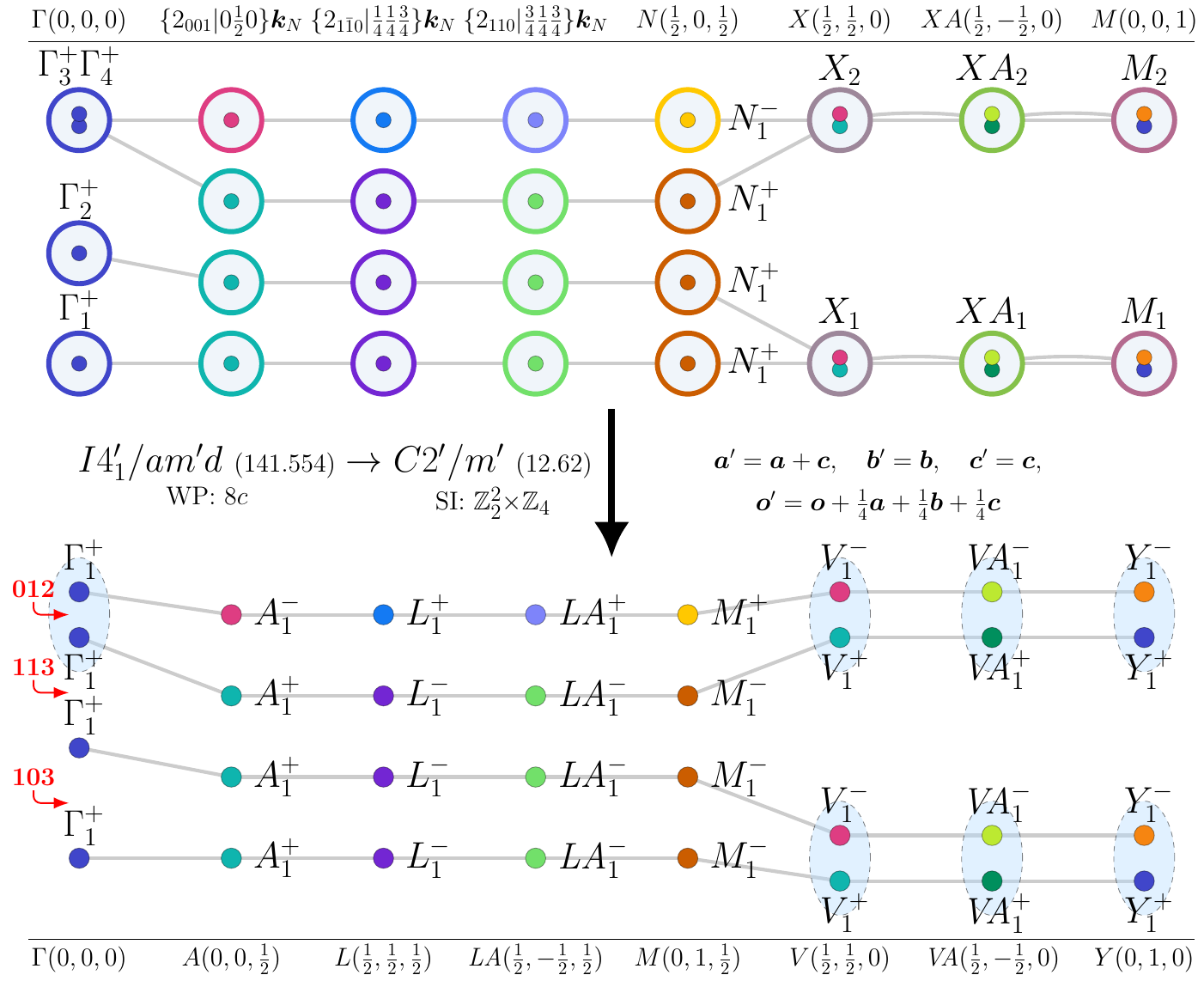}
\caption{Topological magnon bands in subgroup $C2'/m'~(12.62)$ for magnetic moments on Wyckoff position $8c$ of supergroup $I4_{1}'/am'd~(141.554)$.\label{fig_141.554_12.62_Bparallel100_8c}}
\end{figure}
\input{gap_tables_tex/141.554_12.62_Bparallel100_8c_table.tex}
\input{si_tables_tex/141.554_12.62_Bparallel100_8c_table.tex}
\subsubsection{Topological bands in subgroup $C2'/m'~(12.62)$}
\textbf{Perturbations:}
\begin{itemize}
\item strain $\perp$ [100],
\item B $\perp$ [100].
\end{itemize}
\begin{figure}[H]
\centering
\includegraphics[scale=0.6]{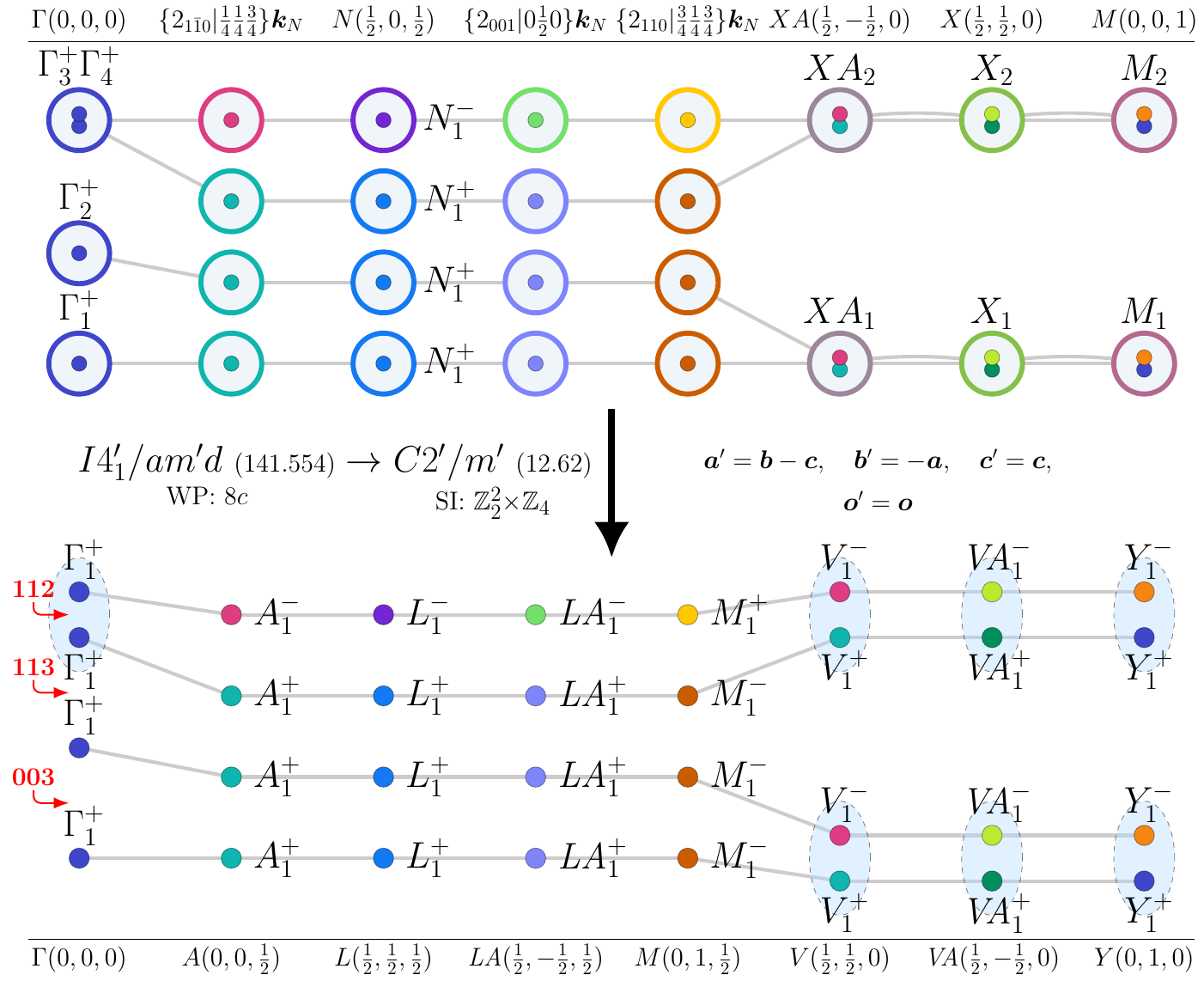}
\caption{Topological magnon bands in subgroup $C2'/m'~(12.62)$ for magnetic moments on Wyckoff position $8c$ of supergroup $I4_{1}'/am'd~(141.554)$.\label{fig_141.554_12.62_strainperp100_8c}}
\end{figure}
\input{gap_tables_tex/141.554_12.62_strainperp100_8c_table.tex}
\input{si_tables_tex/141.554_12.62_strainperp100_8c_table.tex}

\section{MSG $I4_{1}'/amd'~(141.555)$}
\textbf{Nontrivial-SI Subgroups:} $P\bar{1}~(2.4)$, $C2'/c'~(15.89)$, $C2'/c'~(15.89)$, $C2/c~(15.85)$, $Fd'd'd~(70.530)$, $C2/m~(12.58)$.\\

\textbf{Trivial-SI Subgroups:} $Cc'~(9.39)$, $Cc'~(9.39)$, $C2'~(5.15)$, $Cc~(9.37)$, $Fd'd2'~(43.226)$, $Cm~(8.32)$, $C2~(5.13)$, $Fd'd'2~(43.227)$, $Imm2~(44.229)$, $I4_{1}'md'~(109.242)$, $C2~(5.13)$, $Ima2~(46.241)$, $Imma~(74.554)$.\\

\subsection{WP: $8d$}
\textbf{BCS Materials:} {Gd\textsubscript{2}Sn\textsubscript{2}O\textsubscript{7}~(1 K)}\footnote{BCS web page: \texttt{\href{http://webbdcrista1.ehu.es/magndata/index.php?this\_label=0.47} {http://webbdcrista1.ehu.es/magndata/index.php?this\_label=0.47}}}, {Er\textsubscript{2}Pt\textsubscript{2}O\textsubscript{7}~(0.38 K)}\footnote{BCS web page: \texttt{\href{http://webbdcrista1.ehu.es/magndata/index.php?this\_label=0.238} {http://webbdcrista1.ehu.es/magndata/index.php?this\_label=0.238}}}, {Er\textsubscript{2}Sn\textsubscript{2}O\textsubscript{7}~(0.1 K)}\footnote{BCS web page: \texttt{\href{http://webbdcrista1.ehu.es/magndata/index.php?this\_label=0.237} {http://webbdcrista1.ehu.es/magndata/index.php?this\_label=0.237}}}.\\
\subsubsection{Topological bands in subgroup $P\bar{1}~(2.4)$}
\textbf{Perturbations:}
\begin{itemize}
\item strain in generic direction,
\item B $\parallel$ [001] and strain $\perp$ [100],
\item B $\parallel$ [100] and strain $\parallel$ [110],
\item B $\parallel$ [100] and strain $\perp$ [001],
\item B $\parallel$ [100] and strain $\perp$ [110],
\item B $\parallel$ [110] and strain $\parallel$ [100],
\item B $\parallel$ [110] and strain $\perp$ [001],
\item B $\parallel$ [110] and strain $\perp$ [100],
\item B $\parallel$ [110] and strain $\perp$ [110],
\item B in generic direction,
\item B $\perp$ [110] and strain $\parallel$ [100],
\item B $\perp$ [110] and strain $\perp$ [001],
\item B $\perp$ [110] and strain $\perp$ [100].
\end{itemize}
\begin{figure}[H]
\centering
\includegraphics[scale=0.6]{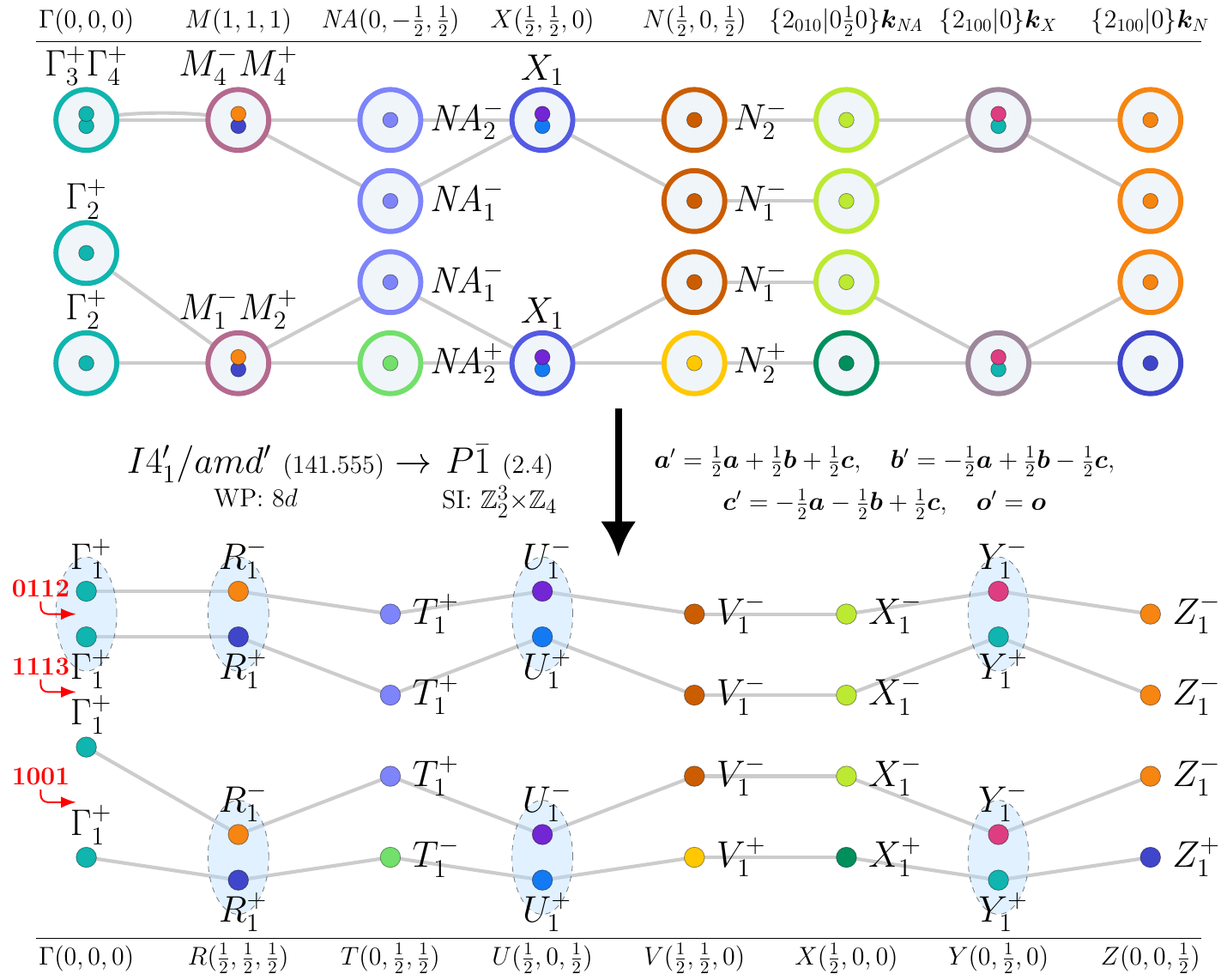}
\caption{Topological magnon bands in subgroup $P\bar{1}~(2.4)$ for magnetic moments on Wyckoff position $8d$ of supergroup $I4_{1}'/amd'~(141.555)$.\label{fig_141.555_2.4_strainingenericdirection_8d}}
\end{figure}
\input{gap_tables_tex/141.555_2.4_strainingenericdirection_8d_table.tex}
\input{si_tables_tex/141.555_2.4_strainingenericdirection_8d_table.tex}
\subsubsection{Topological bands in subgroup $C2'/c'~(15.89)$}
\textbf{Perturbation:}
\begin{itemize}
\item B $\parallel$ [110].
\end{itemize}
\begin{figure}[H]
\centering
\includegraphics[scale=0.6]{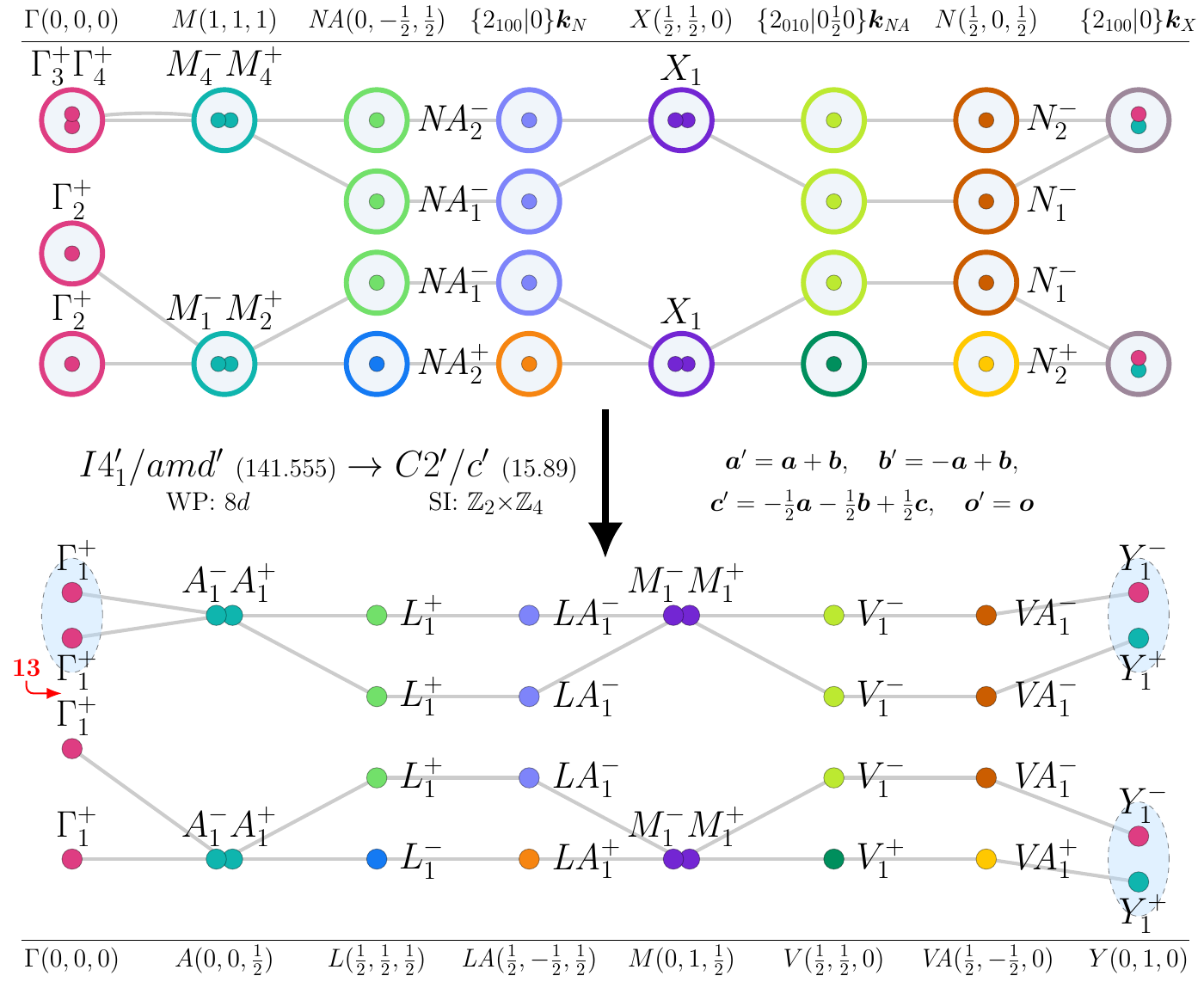}
\caption{Topological magnon bands in subgroup $C2'/c'~(15.89)$ for magnetic moments on Wyckoff position $8d$ of supergroup $I4_{1}'/amd'~(141.555)$.\label{fig_141.555_15.89_Bparallel110_8d}}
\end{figure}
\input{gap_tables_tex/141.555_15.89_Bparallel110_8d_table.tex}
\input{si_tables_tex/141.555_15.89_Bparallel110_8d_table.tex}
\subsubsection{Topological bands in subgroup $C2'/c'~(15.89)$}
\textbf{Perturbations:}
\begin{itemize}
\item strain $\perp$ [110],
\item B $\perp$ [110].
\end{itemize}
\begin{figure}[H]
\centering
\includegraphics[scale=0.6]{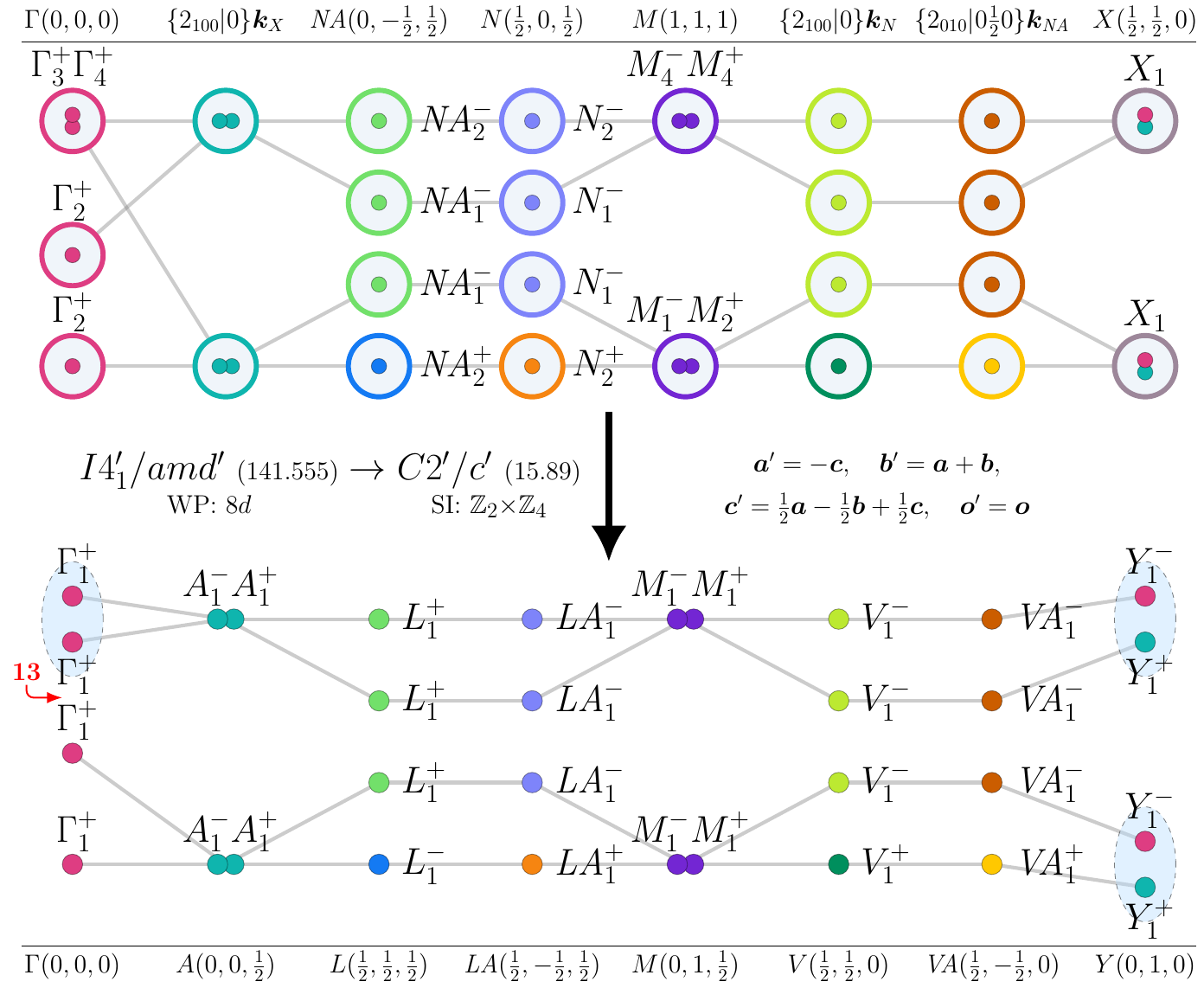}
\caption{Topological magnon bands in subgroup $C2'/c'~(15.89)$ for magnetic moments on Wyckoff position $8d$ of supergroup $I4_{1}'/amd'~(141.555)$.\label{fig_141.555_15.89_strainperp110_8d}}
\end{figure}
\input{gap_tables_tex/141.555_15.89_strainperp110_8d_table.tex}
\input{si_tables_tex/141.555_15.89_strainperp110_8d_table.tex}

\section{MSG $I4_{1}/am'd'~(141.557)$}
\textbf{Nontrivial-SI Subgroups:} $P\bar{1}~(2.4)$, $C2'/c'~(15.89)$, $C2'/c'~(15.89)$, $C2'/m'~(12.62)$, $C2'/m'~(12.62)$, $C2/c~(15.85)$, $Fd'd'd~(70.530)$, $Im'm'a~(74.558)$.\\

\textbf{Trivial-SI Subgroups:} $Cc'~(9.39)$, $Cc'~(9.39)$, $Cm'~(8.34)$, $Cm'~(8.34)$, $C2'~(5.15)$, $C2'~(5.15)$, $Cc~(9.37)$, $Fd'd2'~(43.226)$, $Im'a2'~(46.243)$, $C2~(5.13)$, $Fd'd'2~(43.227)$, $Im'm'2~(44.232)$, $I4_{1}m'd'~(109.243)$.\\

\subsection{WP: $4a+8d+8d$}
\textbf{BCS Materials:} {NiFe\textsubscript{2}O\textsubscript{4}~(858 K)}\footnote{BCS web page: \texttt{\href{http://webbdcrista1.ehu.es/magndata/index.php?this\_label=0.713} {http://webbdcrista1.ehu.es/magndata/index.php?this\_label=0.713}}}.\\
\subsubsection{Topological bands in subgroup $P\bar{1}~(2.4)$}
\textbf{Perturbations:}
\begin{itemize}
\item strain in generic direction,
\item B $\parallel$ [100] and strain $\parallel$ [110],
\item B $\parallel$ [100] and strain $\perp$ [001],
\item B $\parallel$ [100] and strain $\perp$ [100],
\item B $\parallel$ [100] and strain $\perp$ [110],
\item B $\parallel$ [110] and strain $\parallel$ [100],
\item B $\parallel$ [110] and strain $\perp$ [001],
\item B $\parallel$ [110] and strain $\perp$ [100],
\item B $\parallel$ [110] and strain $\perp$ [110],
\item B in generic direction,
\item B $\perp$ [100] and strain $\parallel$ [110],
\item B $\perp$ [100] and strain $\perp$ [001],
\item B $\perp$ [100] and strain $\perp$ [110],
\item B $\perp$ [110] and strain $\parallel$ [100],
\item B $\perp$ [110] and strain $\perp$ [001],
\item B $\perp$ [110] and strain $\perp$ [100].
\end{itemize}
\begin{figure}[H]
\centering
\includegraphics[scale=0.6]{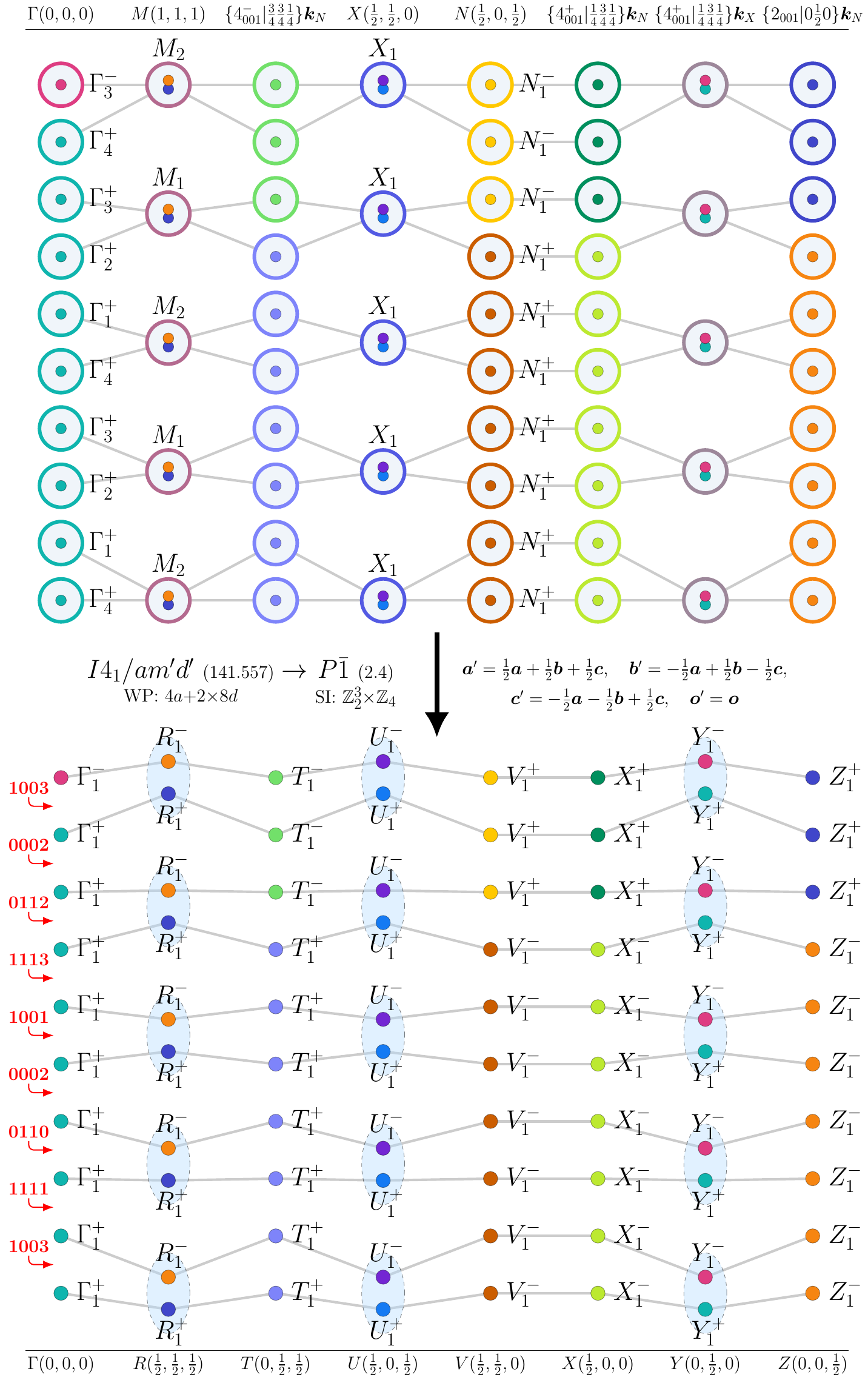}
\caption{Topological magnon bands in subgroup $P\bar{1}~(2.4)$ for magnetic moments on Wyckoff positions $4a+8d+8d$ of supergroup $I4_{1}/am'd'~(141.557)$.\label{fig_141.557_2.4_strainingenericdirection_4a+8d+8d}}
\end{figure}
\input{gap_tables_tex/141.557_2.4_strainingenericdirection_4a+8d+8d_table.tex}
\input{si_tables_tex/141.557_2.4_strainingenericdirection_4a+8d+8d_table.tex}
\subsubsection{Topological bands in subgroup $C2'/c'~(15.89)$}
\textbf{Perturbation:}
\begin{itemize}
\item B $\parallel$ [110].
\end{itemize}
\begin{figure}[H]
\centering
\includegraphics[scale=0.6]{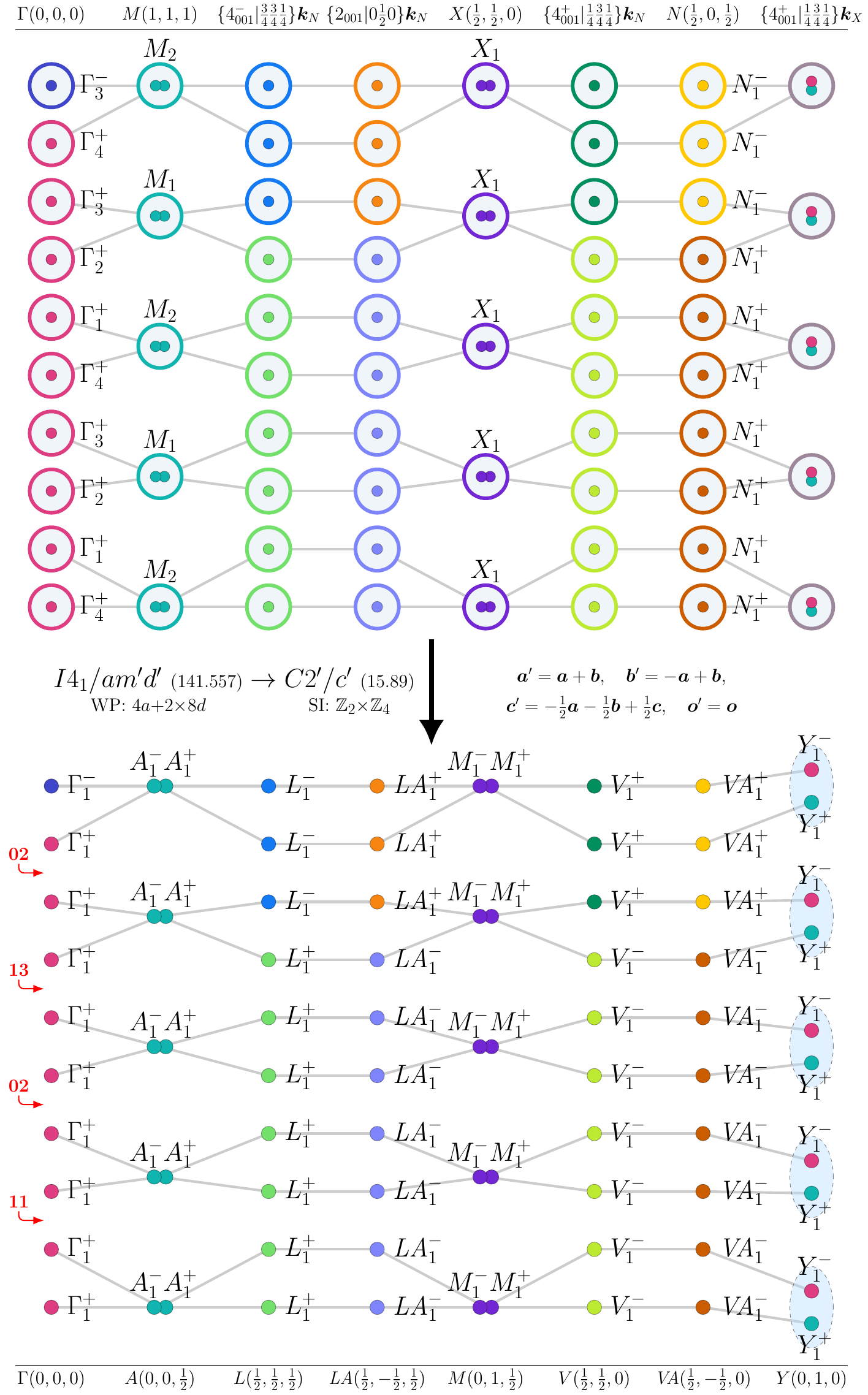}
\caption{Topological magnon bands in subgroup $C2'/c'~(15.89)$ for magnetic moments on Wyckoff positions $4a+8d+8d$ of supergroup $I4_{1}/am'd'~(141.557)$.\label{fig_141.557_15.89_Bparallel110_4a+8d+8d}}
\end{figure}
\input{gap_tables_tex/141.557_15.89_Bparallel110_4a+8d+8d_table.tex}
\input{si_tables_tex/141.557_15.89_Bparallel110_4a+8d+8d_table.tex}
\subsubsection{Topological bands in subgroup $C2'/c'~(15.89)$}
\textbf{Perturbations:}
\begin{itemize}
\item strain $\perp$ [110],
\item B $\perp$ [110].
\end{itemize}
\begin{figure}[H]
\centering
\includegraphics[scale=0.6]{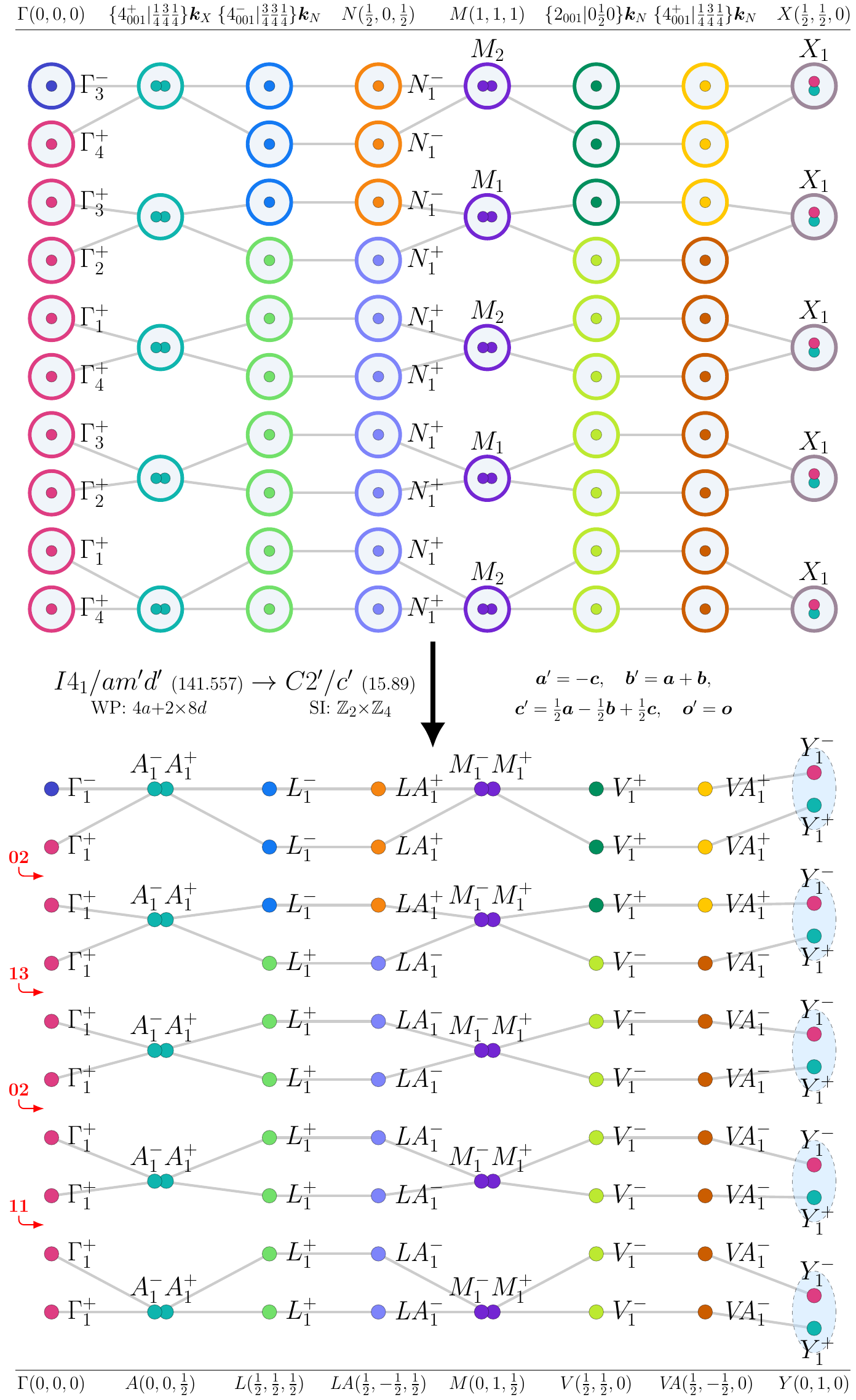}
\caption{Topological magnon bands in subgroup $C2'/c'~(15.89)$ for magnetic moments on Wyckoff positions $4a+8d+8d$ of supergroup $I4_{1}/am'd'~(141.557)$.\label{fig_141.557_15.89_strainperp110_4a+8d+8d}}
\end{figure}
\input{gap_tables_tex/141.557_15.89_strainperp110_4a+8d+8d_table.tex}
\input{si_tables_tex/141.557_15.89_strainperp110_4a+8d+8d_table.tex}
\subsubsection{Topological bands in subgroup $C2'/m'~(12.62)$}
\textbf{Perturbation:}
\begin{itemize}
\item B $\parallel$ [100].
\end{itemize}
\begin{figure}[H]
\centering
\includegraphics[scale=0.6]{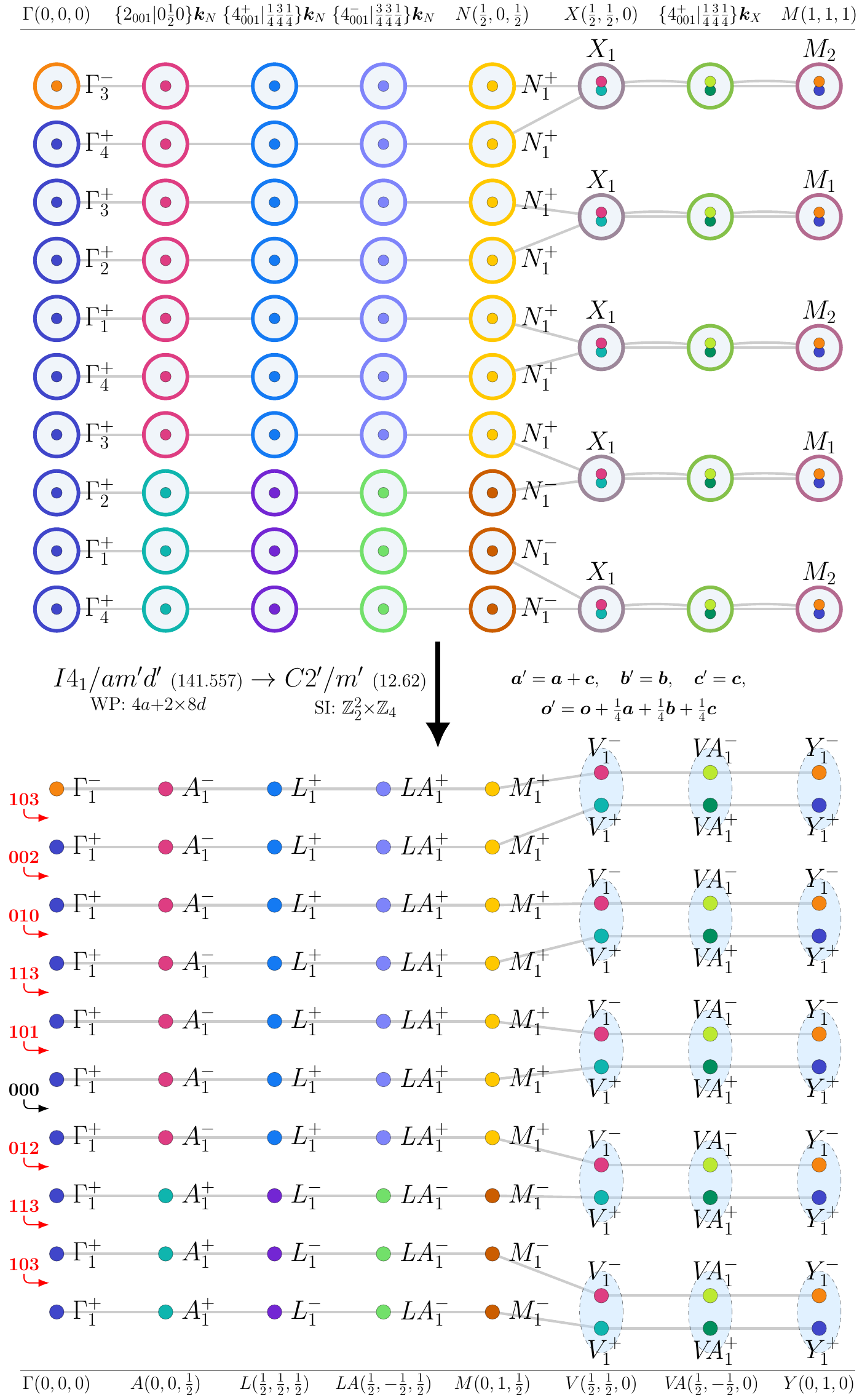}
\caption{Topological magnon bands in subgroup $C2'/m'~(12.62)$ for magnetic moments on Wyckoff positions $4a+8d+8d$ of supergroup $I4_{1}/am'd'~(141.557)$.\label{fig_141.557_12.62_Bparallel100_4a+8d+8d}}
\end{figure}
\input{gap_tables_tex/141.557_12.62_Bparallel100_4a+8d+8d_table.tex}
\input{si_tables_tex/141.557_12.62_Bparallel100_4a+8d+8d_table.tex}
\subsubsection{Topological bands in subgroup $C2'/m'~(12.62)$}
\textbf{Perturbations:}
\begin{itemize}
\item strain $\perp$ [100],
\item B $\perp$ [100].
\end{itemize}
\begin{figure}[H]
\centering
\includegraphics[scale=0.6]{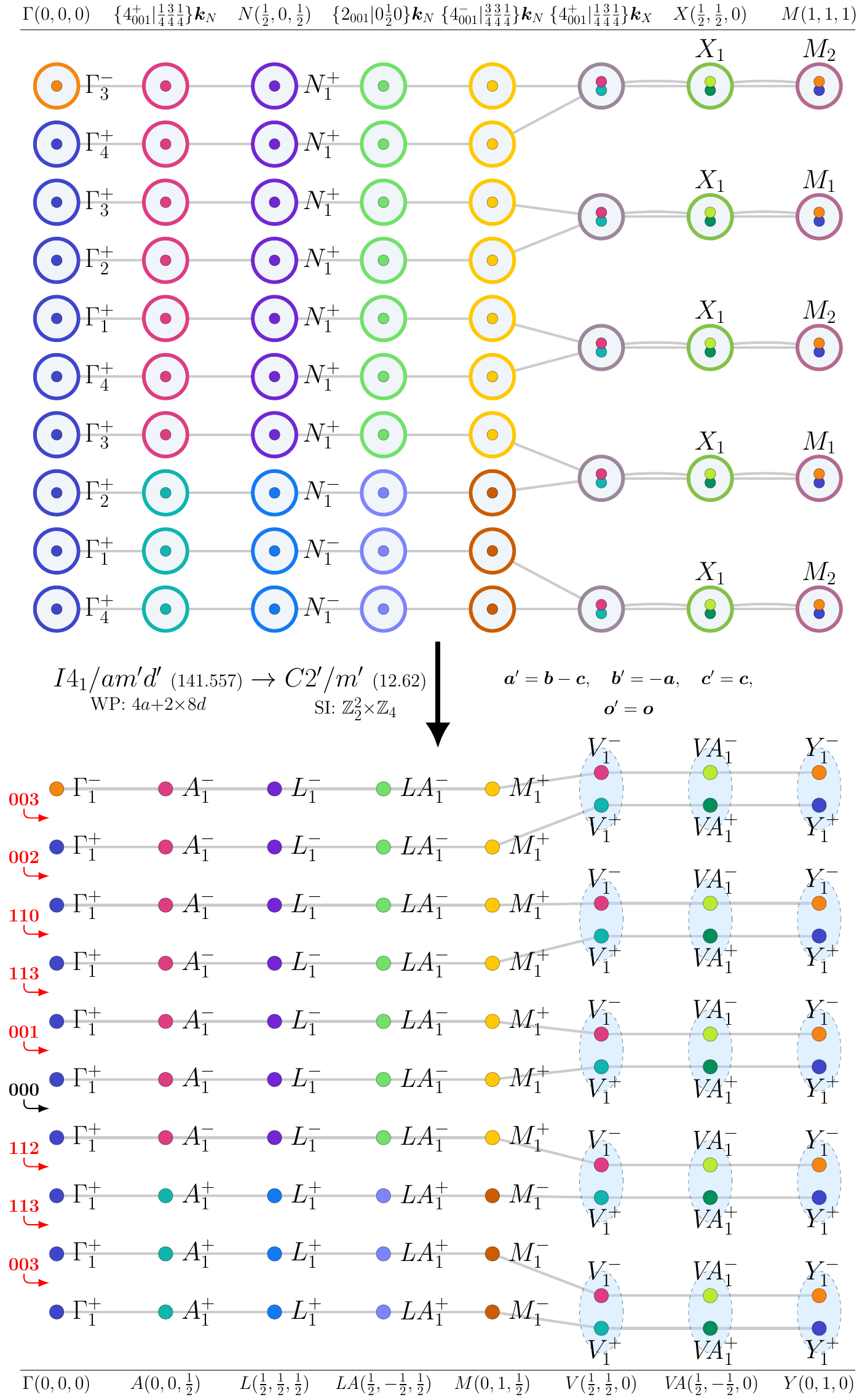}
\caption{Topological magnon bands in subgroup $C2'/m'~(12.62)$ for magnetic moments on Wyckoff positions $4a+8d+8d$ of supergroup $I4_{1}/am'd'~(141.557)$.\label{fig_141.557_12.62_strainperp100_4a+8d+8d}}
\end{figure}
\input{gap_tables_tex/141.557_12.62_strainperp100_4a+8d+8d_table.tex}
\input{si_tables_tex/141.557_12.62_strainperp100_4a+8d+8d_table.tex}
\subsection{WP: $4a+8d$}
\textbf{BCS Materials:} {Li\textsubscript{0.5}FeCr\textsubscript{1.5}O\textsubscript{4}~(417 K)}\footnote{BCS web page: \texttt{\href{http://webbdcrista1.ehu.es/magndata/index.php?this\_label=0.570} {http://webbdcrista1.ehu.es/magndata/index.php?this\_label=0.570}}}, {FeCr\textsubscript{2}S\textsubscript{4}~(172 K)}\footnote{BCS web page: \texttt{\href{http://webbdcrista1.ehu.es/magndata/index.php?this\_label=0.613} {http://webbdcrista1.ehu.es/magndata/index.php?this\_label=0.613}}}, {Ce\textsubscript{5}TeO\textsubscript{8}~(40 K)}\footnote{BCS web page: \texttt{\href{http://webbdcrista1.ehu.es/magndata/index.php?this\_label=0.725} {http://webbdcrista1.ehu.es/magndata/index.php?this\_label=0.725}}}.\\
\subsubsection{Topological bands in subgroup $P\bar{1}~(2.4)$}
\textbf{Perturbations:}
\begin{itemize}
\item strain in generic direction,
\item B $\parallel$ [100] and strain $\parallel$ [110],
\item B $\parallel$ [100] and strain $\perp$ [001],
\item B $\parallel$ [100] and strain $\perp$ [100],
\item B $\parallel$ [100] and strain $\perp$ [110],
\item B $\parallel$ [110] and strain $\parallel$ [100],
\item B $\parallel$ [110] and strain $\perp$ [001],
\item B $\parallel$ [110] and strain $\perp$ [100],
\item B $\parallel$ [110] and strain $\perp$ [110],
\item B in generic direction,
\item B $\perp$ [100] and strain $\parallel$ [110],
\item B $\perp$ [100] and strain $\perp$ [001],
\item B $\perp$ [100] and strain $\perp$ [110],
\item B $\perp$ [110] and strain $\parallel$ [100],
\item B $\perp$ [110] and strain $\perp$ [001],
\item B $\perp$ [110] and strain $\perp$ [100].
\end{itemize}
\begin{figure}[H]
\centering
\includegraphics[scale=0.6]{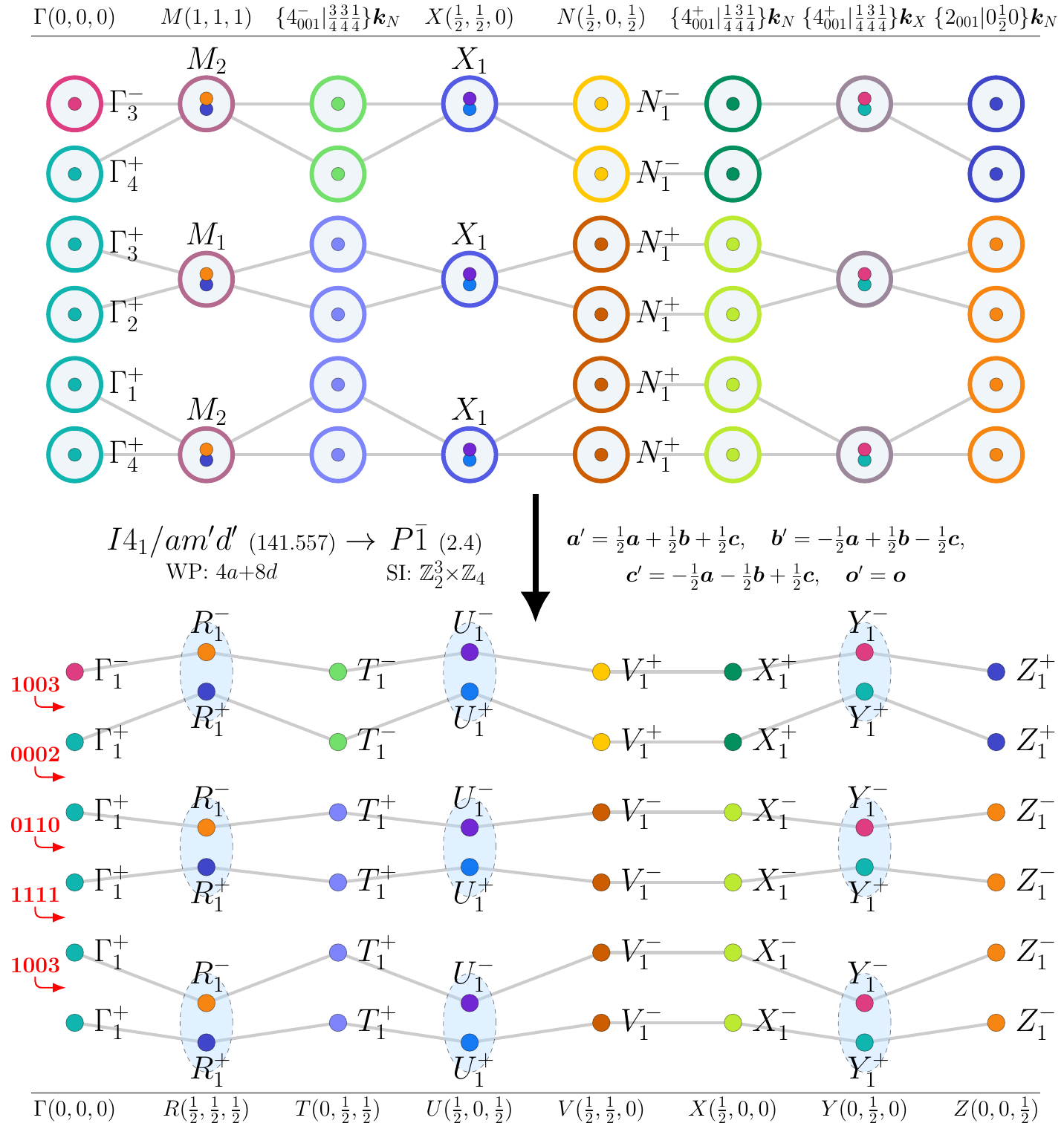}
\caption{Topological magnon bands in subgroup $P\bar{1}~(2.4)$ for magnetic moments on Wyckoff positions $4a+8d$ of supergroup $I4_{1}/am'd'~(141.557)$.\label{fig_141.557_2.4_strainingenericdirection_4a+8d}}
\end{figure}
\input{gap_tables_tex/141.557_2.4_strainingenericdirection_4a+8d_table.tex}
\input{si_tables_tex/141.557_2.4_strainingenericdirection_4a+8d_table.tex}
\subsubsection{Topological bands in subgroup $C2'/c'~(15.89)$}
\textbf{Perturbation:}
\begin{itemize}
\item B $\parallel$ [110].
\end{itemize}
\begin{figure}[H]
\centering
\includegraphics[scale=0.6]{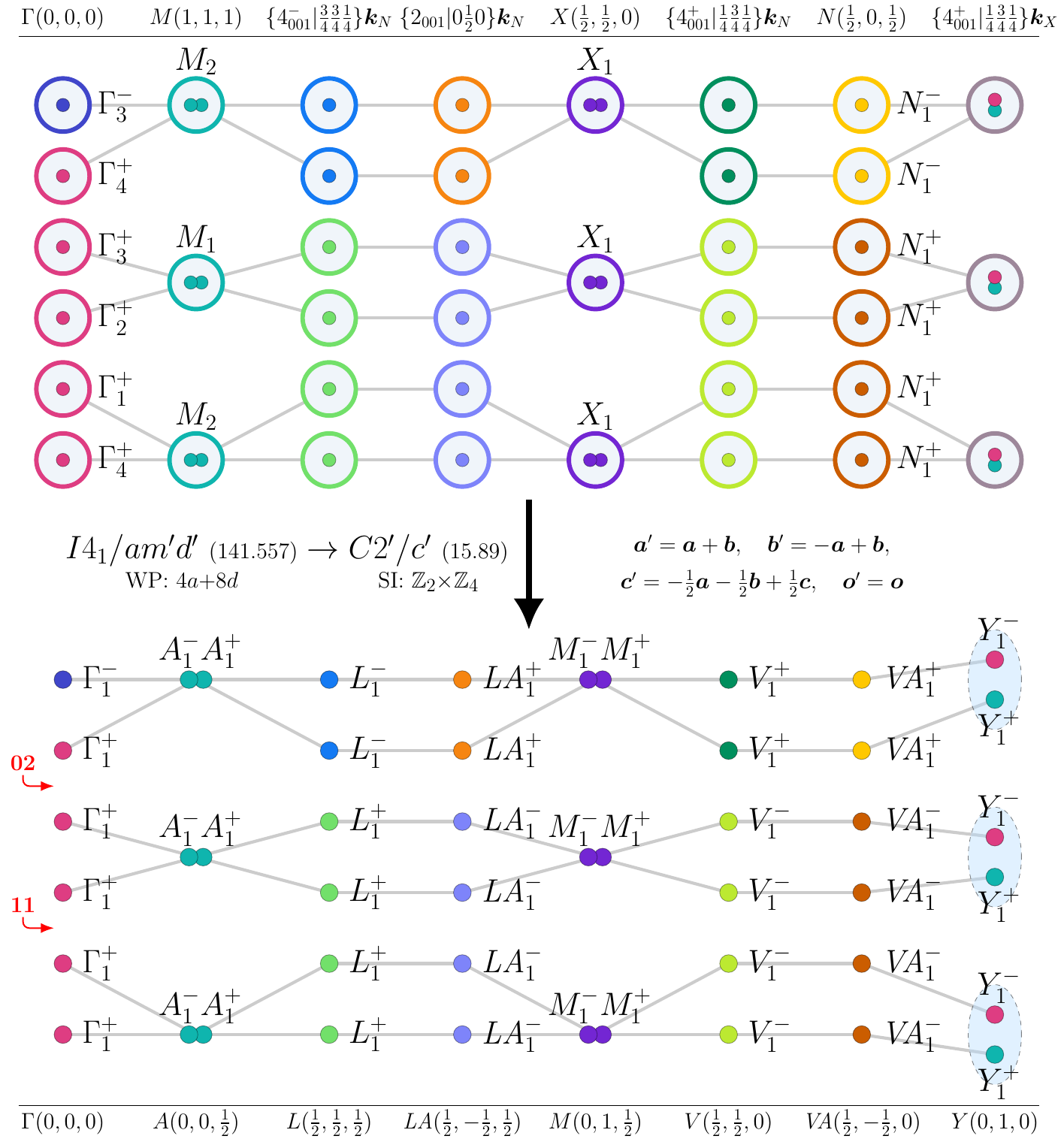}
\caption{Topological magnon bands in subgroup $C2'/c'~(15.89)$ for magnetic moments on Wyckoff positions $4a+8d$ of supergroup $I4_{1}/am'd'~(141.557)$.\label{fig_141.557_15.89_Bparallel110_4a+8d}}
\end{figure}
\input{gap_tables_tex/141.557_15.89_Bparallel110_4a+8d_table.tex}
\input{si_tables_tex/141.557_15.89_Bparallel110_4a+8d_table.tex}
\subsubsection{Topological bands in subgroup $C2'/c'~(15.89)$}
\textbf{Perturbations:}
\begin{itemize}
\item strain $\perp$ [110],
\item B $\perp$ [110].
\end{itemize}
\begin{figure}[H]
\centering
\includegraphics[scale=0.6]{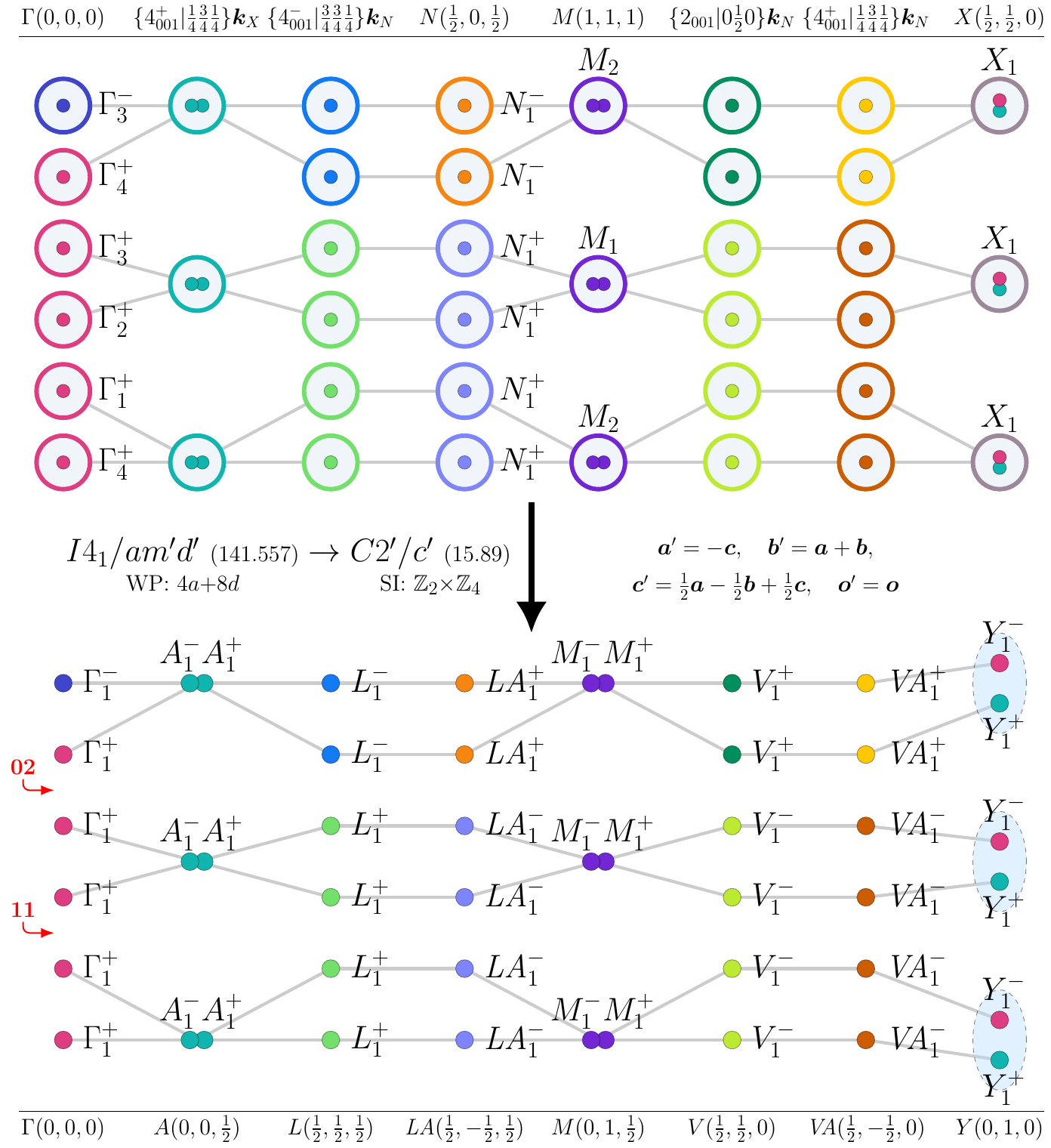}
\caption{Topological magnon bands in subgroup $C2'/c'~(15.89)$ for magnetic moments on Wyckoff positions $4a+8d$ of supergroup $I4_{1}/am'd'~(141.557)$.\label{fig_141.557_15.89_strainperp110_4a+8d}}
\end{figure}
\input{gap_tables_tex/141.557_15.89_strainperp110_4a+8d_table.tex}
\input{si_tables_tex/141.557_15.89_strainperp110_4a+8d_table.tex}
\subsubsection{Topological bands in subgroup $C2'/m'~(12.62)$}
\textbf{Perturbation:}
\begin{itemize}
\item B $\parallel$ [100].
\end{itemize}
\begin{figure}[H]
\centering
\includegraphics[scale=0.6]{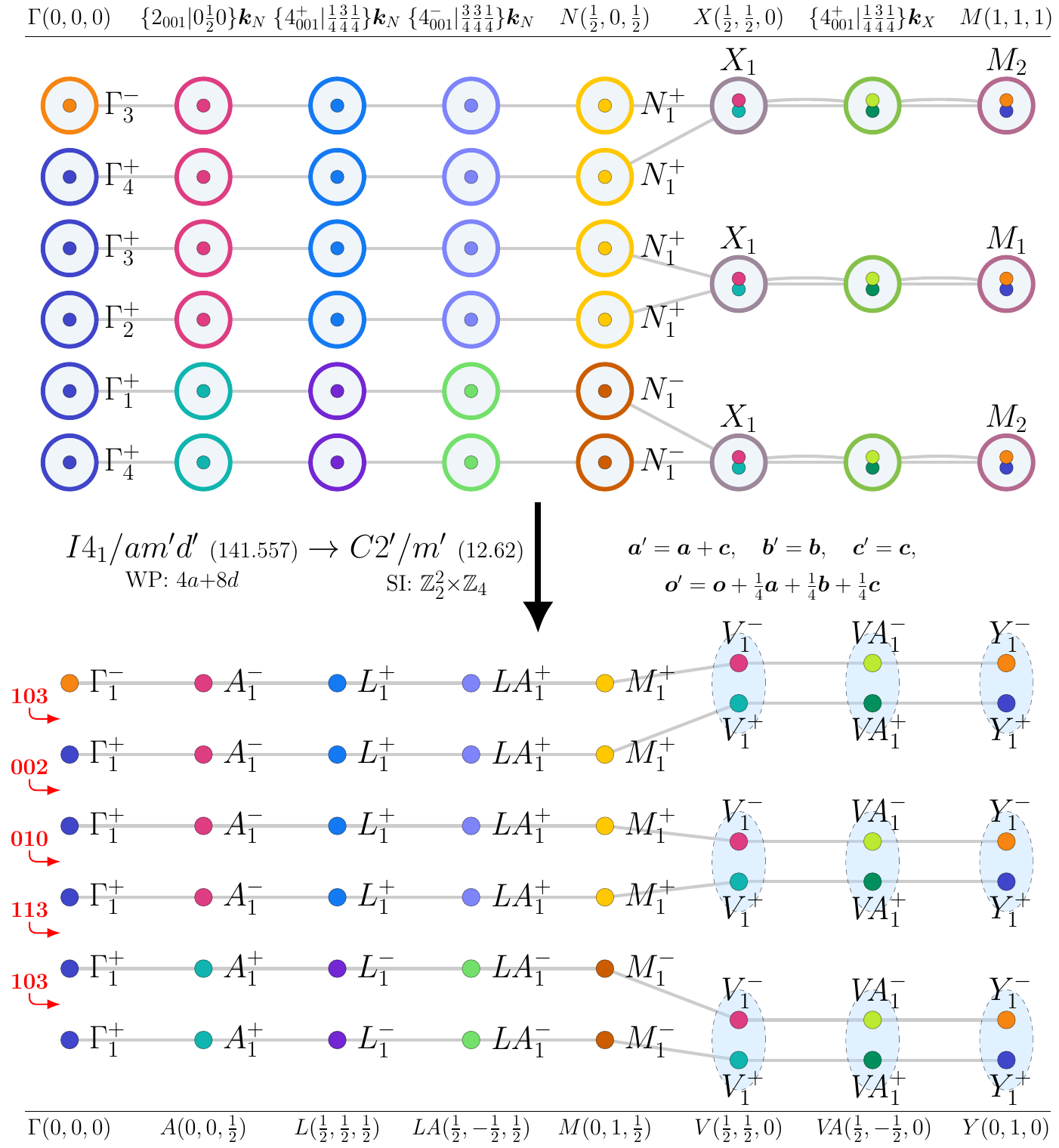}
\caption{Topological magnon bands in subgroup $C2'/m'~(12.62)$ for magnetic moments on Wyckoff positions $4a+8d$ of supergroup $I4_{1}/am'd'~(141.557)$.\label{fig_141.557_12.62_Bparallel100_4a+8d}}
\end{figure}
\input{gap_tables_tex/141.557_12.62_Bparallel100_4a+8d_table.tex}
\input{si_tables_tex/141.557_12.62_Bparallel100_4a+8d_table.tex}
\subsubsection{Topological bands in subgroup $C2'/m'~(12.62)$}
\textbf{Perturbations:}
\begin{itemize}
\item strain $\perp$ [100],
\item B $\perp$ [100].
\end{itemize}
\begin{figure}[H]
\centering
\includegraphics[scale=0.6]{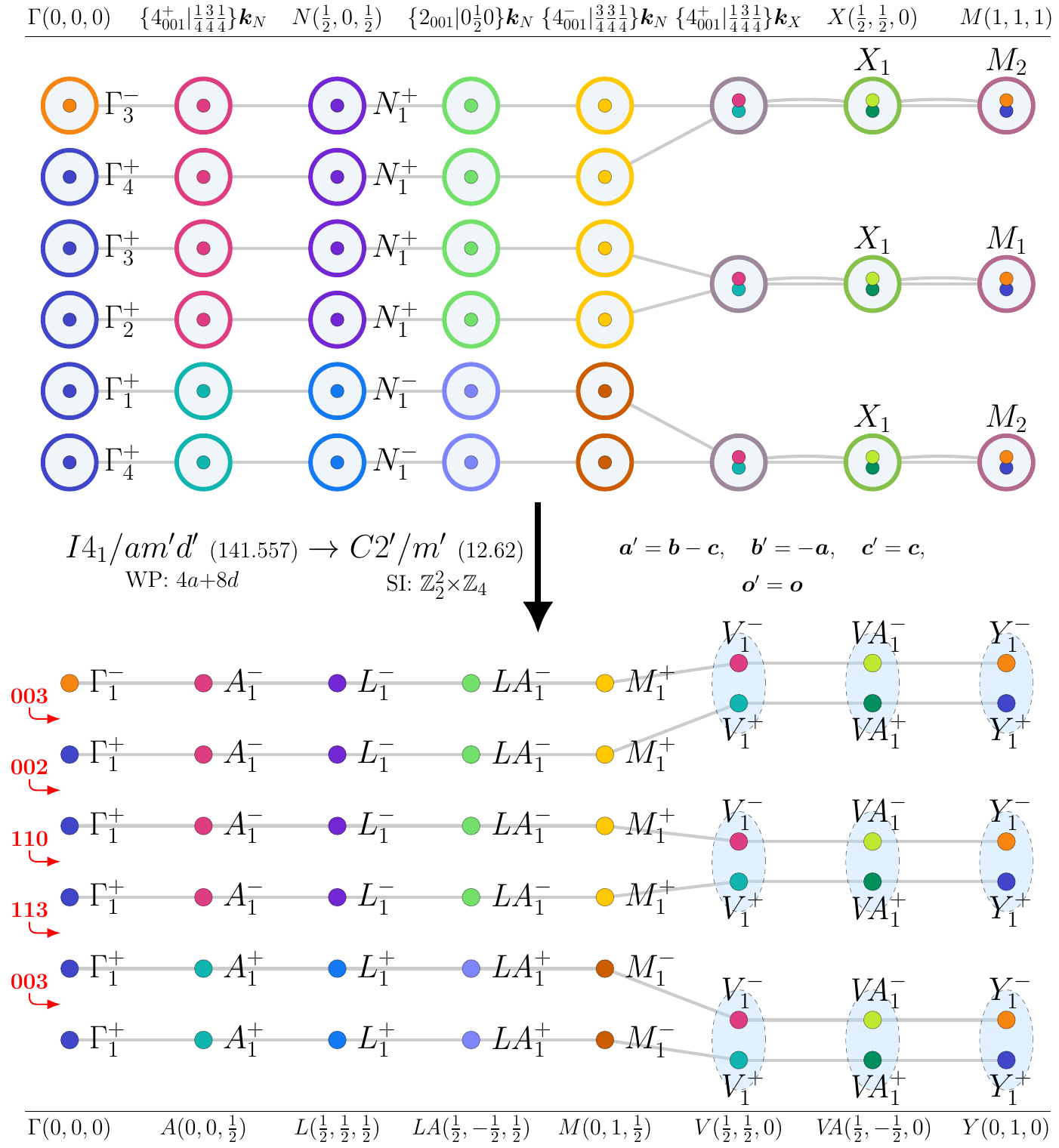}
\caption{Topological magnon bands in subgroup $C2'/m'~(12.62)$ for magnetic moments on Wyckoff positions $4a+8d$ of supergroup $I4_{1}/am'd'~(141.557)$.\label{fig_141.557_12.62_strainperp100_4a+8d}}
\end{figure}
\input{gap_tables_tex/141.557_12.62_strainperp100_4a+8d_table.tex}
\input{si_tables_tex/141.557_12.62_strainperp100_4a+8d_table.tex}
\subsection{WP: $4b+8c$}
\textbf{BCS Materials:} {FeCr\textsubscript{2}S\textsubscript{4}~(180 K)}\footnote{BCS web page: \texttt{\href{http://webbdcrista1.ehu.es/magndata/index.php?this\_label=0.615} {http://webbdcrista1.ehu.es/magndata/index.php?this\_label=0.615}}}, {FeCr\textsubscript{2}S\textsubscript{4}~(180 K)}\footnote{BCS web page: \texttt{\href{http://webbdcrista1.ehu.es/magndata/index.php?this\_label=0.614} {http://webbdcrista1.ehu.es/magndata/index.php?this\_label=0.614}}}, {NdCo\textsubscript{2}~(100 K)}\footnote{BCS web page: \texttt{\href{http://webbdcrista1.ehu.es/magndata/index.php?this\_label=0.227} {http://webbdcrista1.ehu.es/magndata/index.php?this\_label=0.227}}}.\\
\subsubsection{Topological bands in subgroup $P\bar{1}~(2.4)$}
\textbf{Perturbations:}
\begin{itemize}
\item strain in generic direction,
\item B $\parallel$ [100] and strain $\parallel$ [110],
\item B $\parallel$ [100] and strain $\perp$ [001],
\item B $\parallel$ [100] and strain $\perp$ [100],
\item B $\parallel$ [100] and strain $\perp$ [110],
\item B $\parallel$ [110] and strain $\parallel$ [100],
\item B $\parallel$ [110] and strain $\perp$ [001],
\item B $\parallel$ [110] and strain $\perp$ [100],
\item B $\parallel$ [110] and strain $\perp$ [110],
\item B in generic direction,
\item B $\perp$ [100] and strain $\parallel$ [110],
\item B $\perp$ [100] and strain $\perp$ [001],
\item B $\perp$ [100] and strain $\perp$ [110],
\item B $\perp$ [110] and strain $\parallel$ [100],
\item B $\perp$ [110] and strain $\perp$ [001],
\item B $\perp$ [110] and strain $\perp$ [100].
\end{itemize}
\begin{figure}[H]
\centering
\includegraphics[scale=0.6]{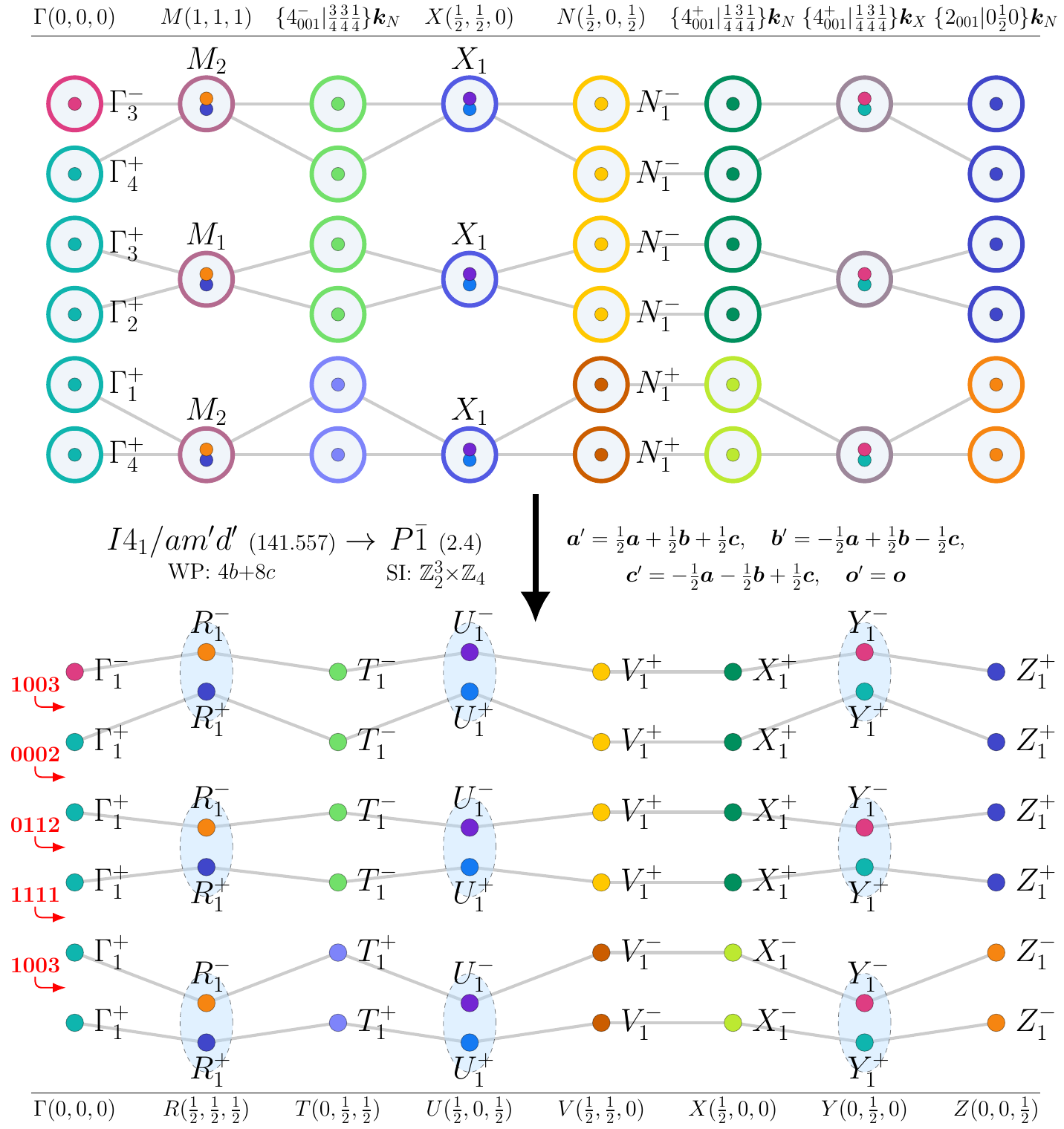}
\caption{Topological magnon bands in subgroup $P\bar{1}~(2.4)$ for magnetic moments on Wyckoff positions $4b+8c$ of supergroup $I4_{1}/am'd'~(141.557)$.\label{fig_141.557_2.4_strainingenericdirection_4b+8c}}
\end{figure}
\input{gap_tables_tex/141.557_2.4_strainingenericdirection_4b+8c_table.tex}
\input{si_tables_tex/141.557_2.4_strainingenericdirection_4b+8c_table.tex}
\subsubsection{Topological bands in subgroup $C2'/c'~(15.89)$}
\textbf{Perturbation:}
\begin{itemize}
\item B $\parallel$ [110].
\end{itemize}
\begin{figure}[H]
\centering
\includegraphics[scale=0.6]{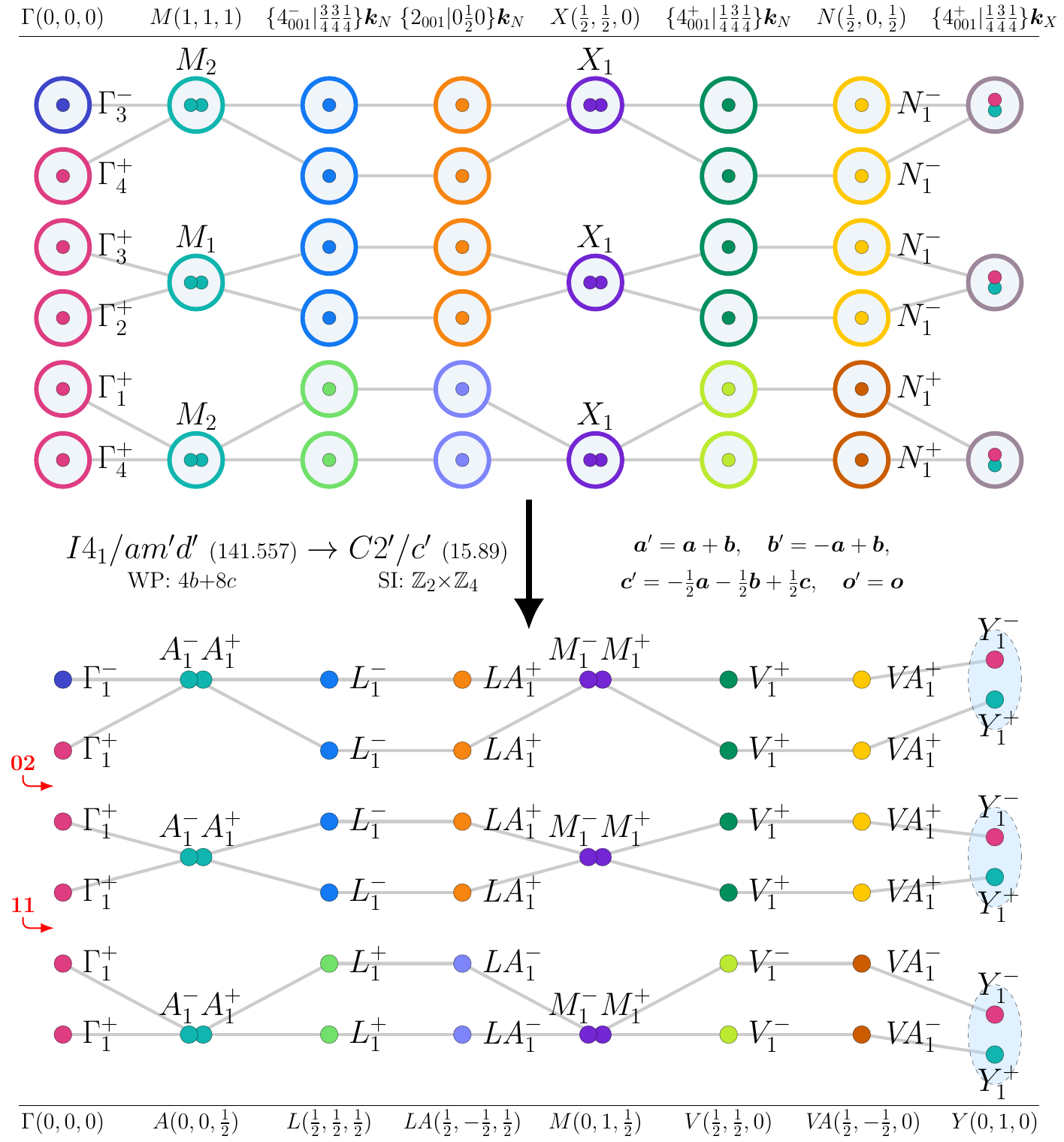}
\caption{Topological magnon bands in subgroup $C2'/c'~(15.89)$ for magnetic moments on Wyckoff positions $4b+8c$ of supergroup $I4_{1}/am'd'~(141.557)$.\label{fig_141.557_15.89_Bparallel110_4b+8c}}
\end{figure}
\input{gap_tables_tex/141.557_15.89_Bparallel110_4b+8c_table.tex}
\input{si_tables_tex/141.557_15.89_Bparallel110_4b+8c_table.tex}
\subsubsection{Topological bands in subgroup $C2'/c'~(15.89)$}
\textbf{Perturbations:}
\begin{itemize}
\item strain $\perp$ [110],
\item B $\perp$ [110].
\end{itemize}
\begin{figure}[H]
\centering
\includegraphics[scale=0.6]{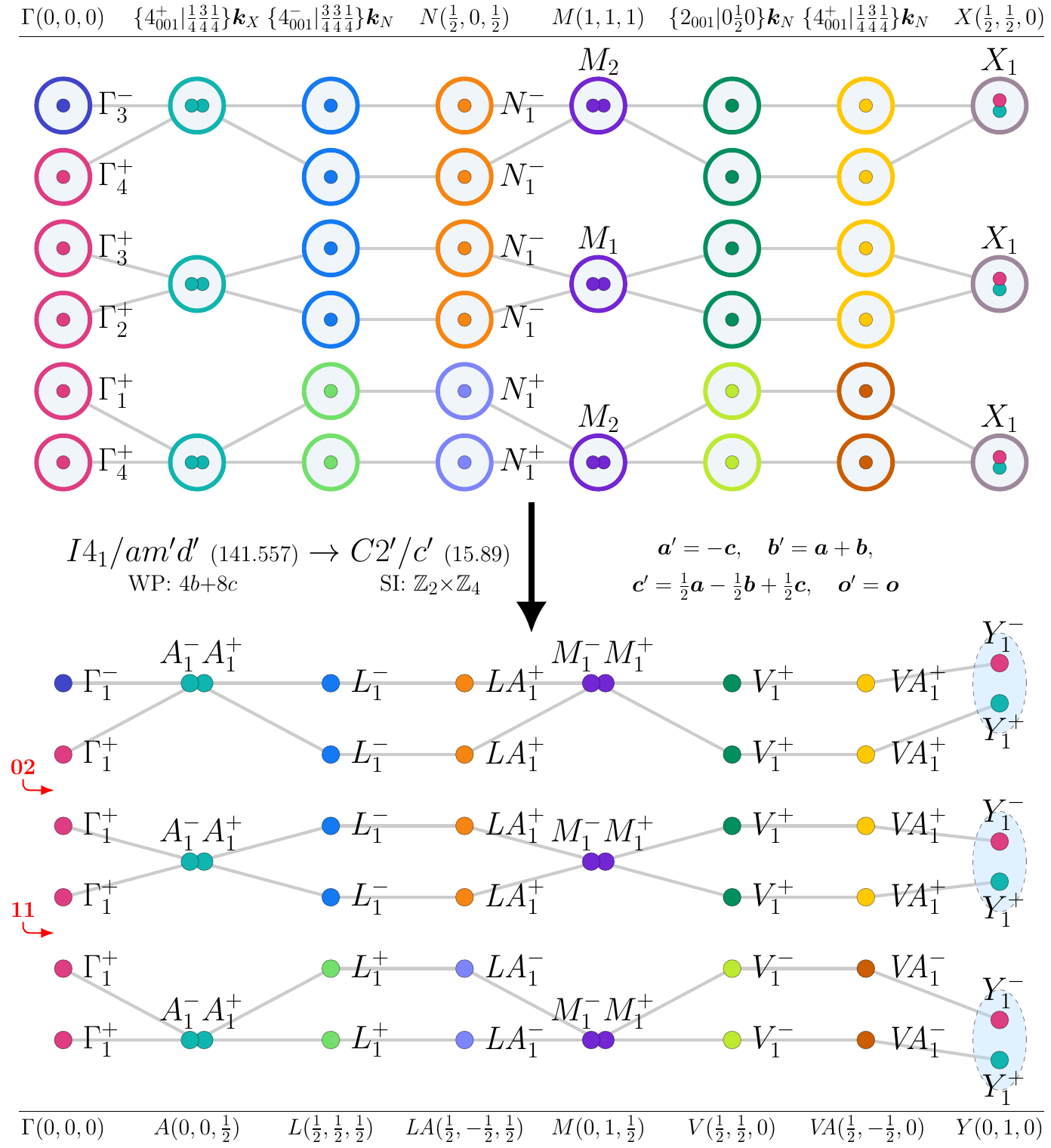}
\caption{Topological magnon bands in subgroup $C2'/c'~(15.89)$ for magnetic moments on Wyckoff positions $4b+8c$ of supergroup $I4_{1}/am'd'~(141.557)$.\label{fig_141.557_15.89_strainperp110_4b+8c}}
\end{figure}
\input{gap_tables_tex/141.557_15.89_strainperp110_4b+8c_table.tex}
\input{si_tables_tex/141.557_15.89_strainperp110_4b+8c_table.tex}
\subsubsection{Topological bands in subgroup $C2'/m'~(12.62)$}
\textbf{Perturbation:}
\begin{itemize}
\item B $\parallel$ [100].
\end{itemize}
\begin{figure}[H]
\centering
\includegraphics[scale=0.6]{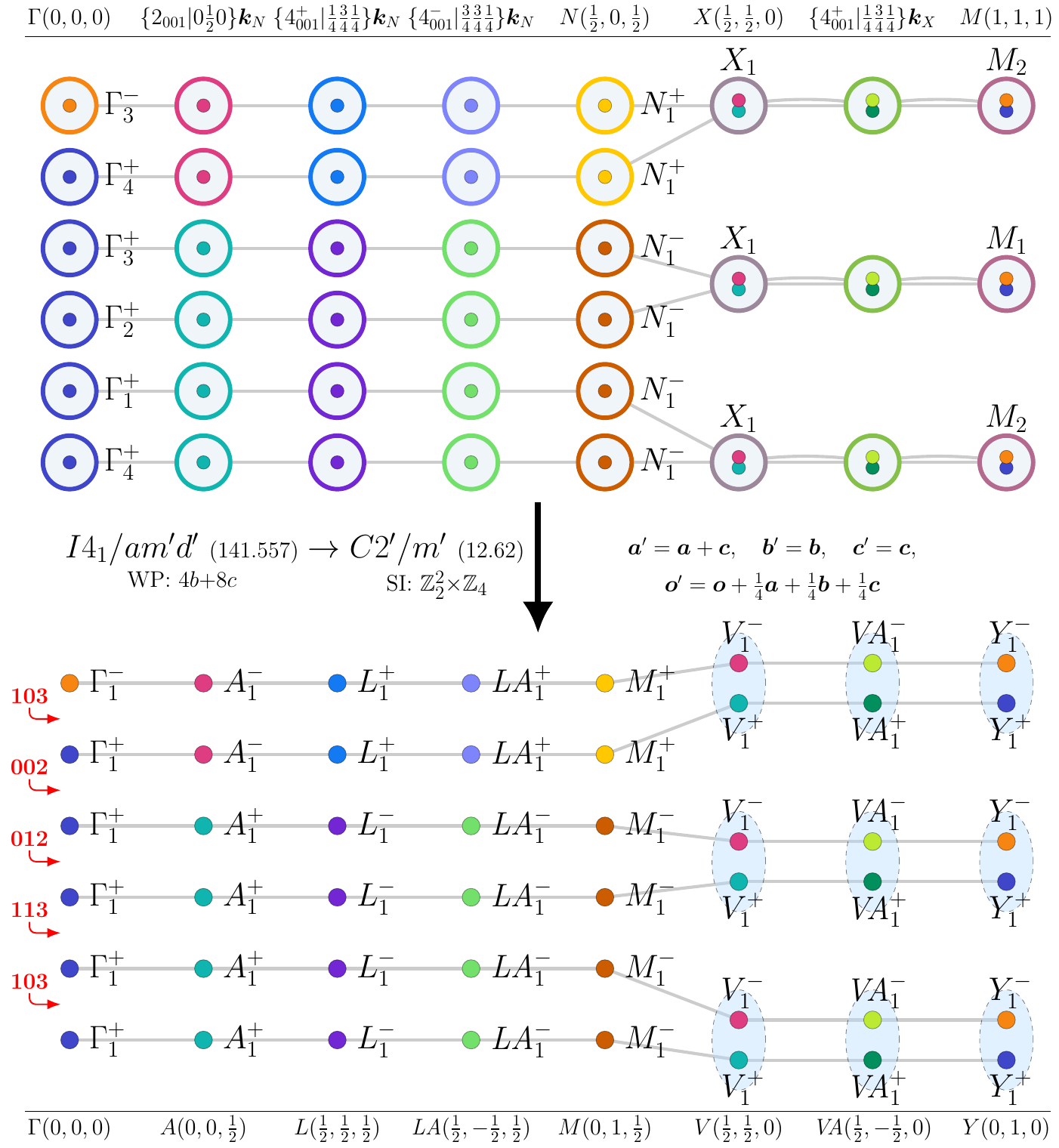}
\caption{Topological magnon bands in subgroup $C2'/m'~(12.62)$ for magnetic moments on Wyckoff positions $4b+8c$ of supergroup $I4_{1}/am'd'~(141.557)$.\label{fig_141.557_12.62_Bparallel100_4b+8c}}
\end{figure}
\input{gap_tables_tex/141.557_12.62_Bparallel100_4b+8c_table.tex}
\input{si_tables_tex/141.557_12.62_Bparallel100_4b+8c_table.tex}
\subsubsection{Topological bands in subgroup $C2'/m'~(12.62)$}
\textbf{Perturbations:}
\begin{itemize}
\item strain $\perp$ [100],
\item B $\perp$ [100].
\end{itemize}
\begin{figure}[H]
\centering
\includegraphics[scale=0.6]{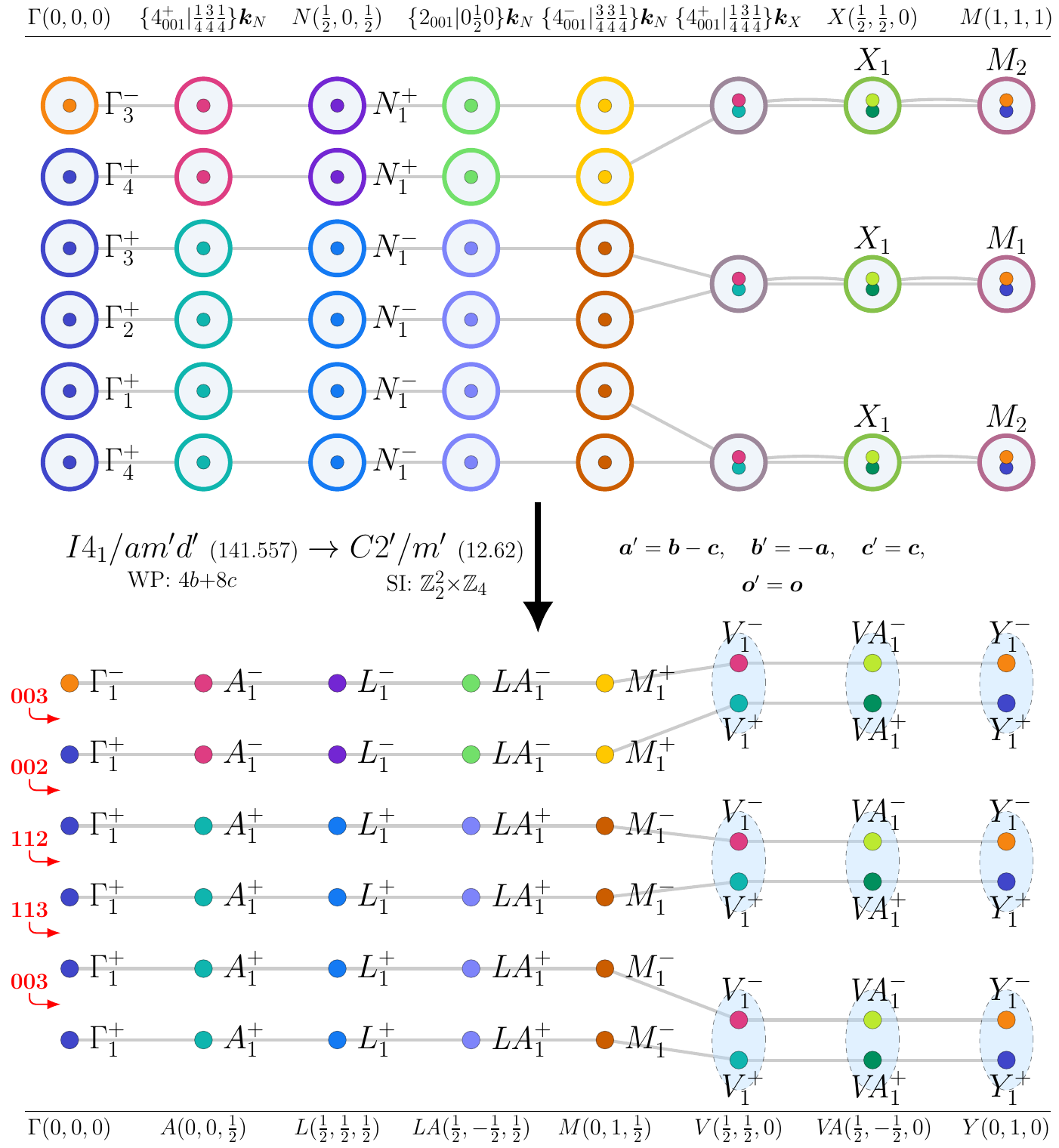}
\caption{Topological magnon bands in subgroup $C2'/m'~(12.62)$ for magnetic moments on Wyckoff positions $4b+8c$ of supergroup $I4_{1}/am'd'~(141.557)$.\label{fig_141.557_12.62_strainperp100_4b+8c}}
\end{figure}
\input{gap_tables_tex/141.557_12.62_strainperp100_4b+8c_table.tex}
\input{si_tables_tex/141.557_12.62_strainperp100_4b+8c_table.tex}
\subsection{WP: $8c$}
\textbf{BCS Materials:} {Ho\textsubscript{2}Ru\textsubscript{2}O\textsubscript{7}~(95 K)}\footnote{BCS web page: \texttt{\href{http://webbdcrista1.ehu.es/magndata/index.php?this\_label=0.49} {http://webbdcrista1.ehu.es/magndata/index.php?this\_label=0.49}}}.\\
\subsubsection{Topological bands in subgroup $P\bar{1}~(2.4)$}
\textbf{Perturbations:}
\begin{itemize}
\item strain in generic direction,
\item B $\parallel$ [100] and strain $\parallel$ [110],
\item B $\parallel$ [100] and strain $\perp$ [001],
\item B $\parallel$ [100] and strain $\perp$ [100],
\item B $\parallel$ [100] and strain $\perp$ [110],
\item B $\parallel$ [110] and strain $\parallel$ [100],
\item B $\parallel$ [110] and strain $\perp$ [001],
\item B $\parallel$ [110] and strain $\perp$ [100],
\item B $\parallel$ [110] and strain $\perp$ [110],
\item B in generic direction,
\item B $\perp$ [100] and strain $\parallel$ [110],
\item B $\perp$ [100] and strain $\perp$ [001],
\item B $\perp$ [100] and strain $\perp$ [110],
\item B $\perp$ [110] and strain $\parallel$ [100],
\item B $\perp$ [110] and strain $\perp$ [001],
\item B $\perp$ [110] and strain $\perp$ [100].
\end{itemize}
\begin{figure}[H]
\centering
\includegraphics[scale=0.6]{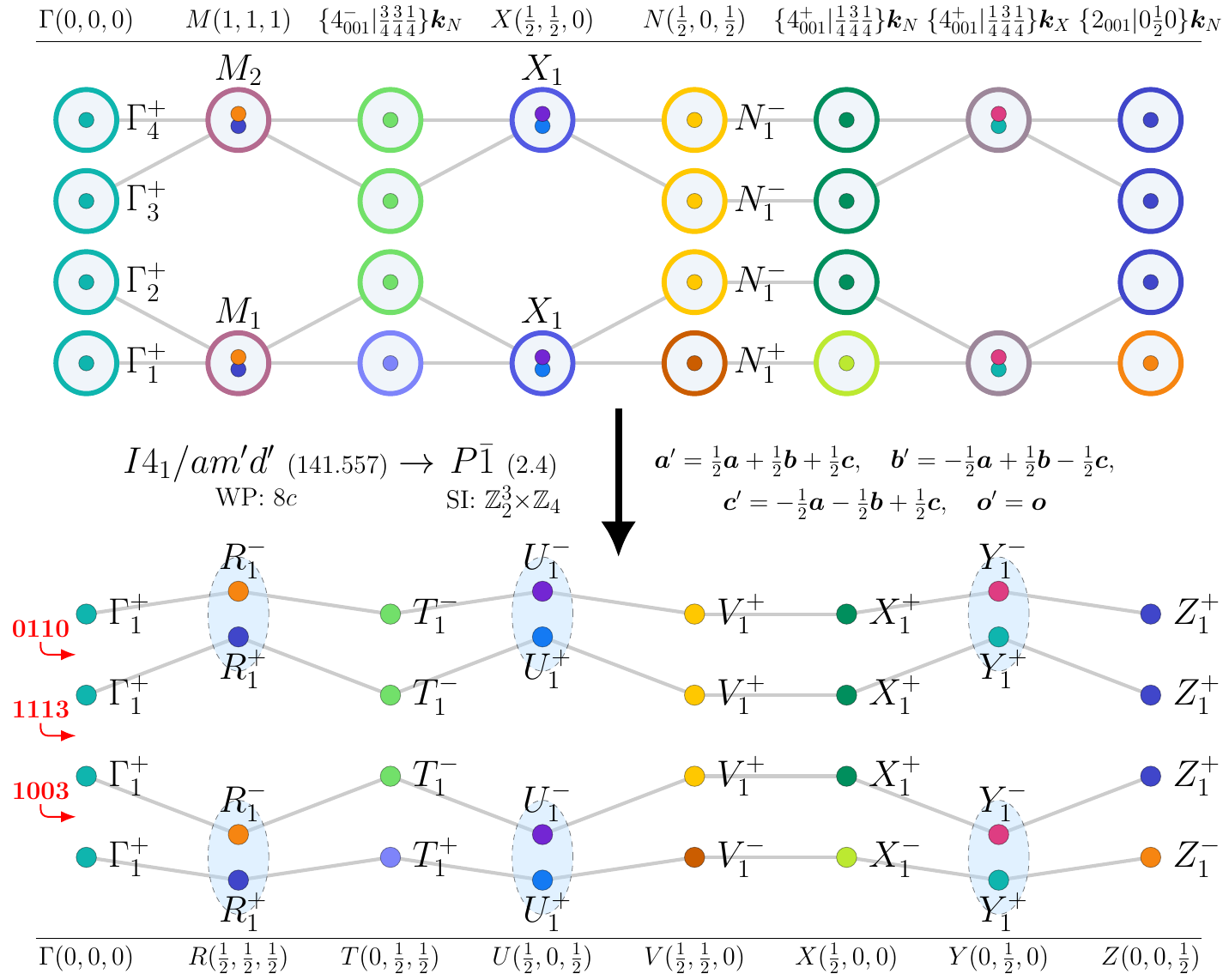}
\caption{Topological magnon bands in subgroup $P\bar{1}~(2.4)$ for magnetic moments on Wyckoff position $8c$ of supergroup $I4_{1}/am'd'~(141.557)$.\label{fig_141.557_2.4_strainingenericdirection_8c}}
\end{figure}
\input{gap_tables_tex/141.557_2.4_strainingenericdirection_8c_table.tex}
\input{si_tables_tex/141.557_2.4_strainingenericdirection_8c_table.tex}
\subsubsection{Topological bands in subgroup $C2'/c'~(15.89)$}
\textbf{Perturbation:}
\begin{itemize}
\item B $\parallel$ [110].
\end{itemize}
\begin{figure}[H]
\centering
\includegraphics[scale=0.6]{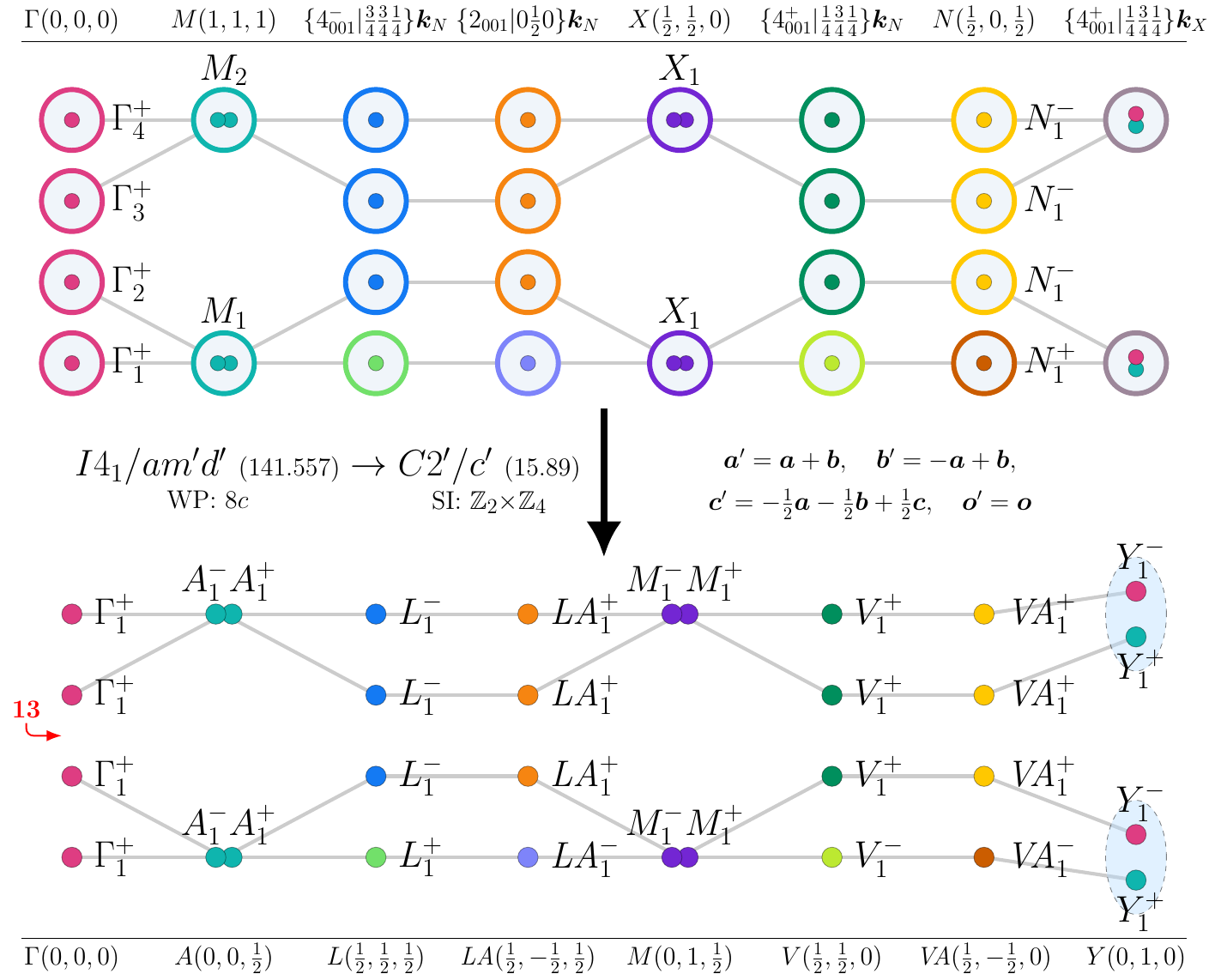}
\caption{Topological magnon bands in subgroup $C2'/c'~(15.89)$ for magnetic moments on Wyckoff position $8c$ of supergroup $I4_{1}/am'd'~(141.557)$.\label{fig_141.557_15.89_Bparallel110_8c}}
\end{figure}
\input{gap_tables_tex/141.557_15.89_Bparallel110_8c_table.tex}
\input{si_tables_tex/141.557_15.89_Bparallel110_8c_table.tex}
\subsubsection{Topological bands in subgroup $C2'/c'~(15.89)$}
\textbf{Perturbations:}
\begin{itemize}
\item strain $\perp$ [110],
\item B $\perp$ [110].
\end{itemize}
\begin{figure}[H]
\centering
\includegraphics[scale=0.6]{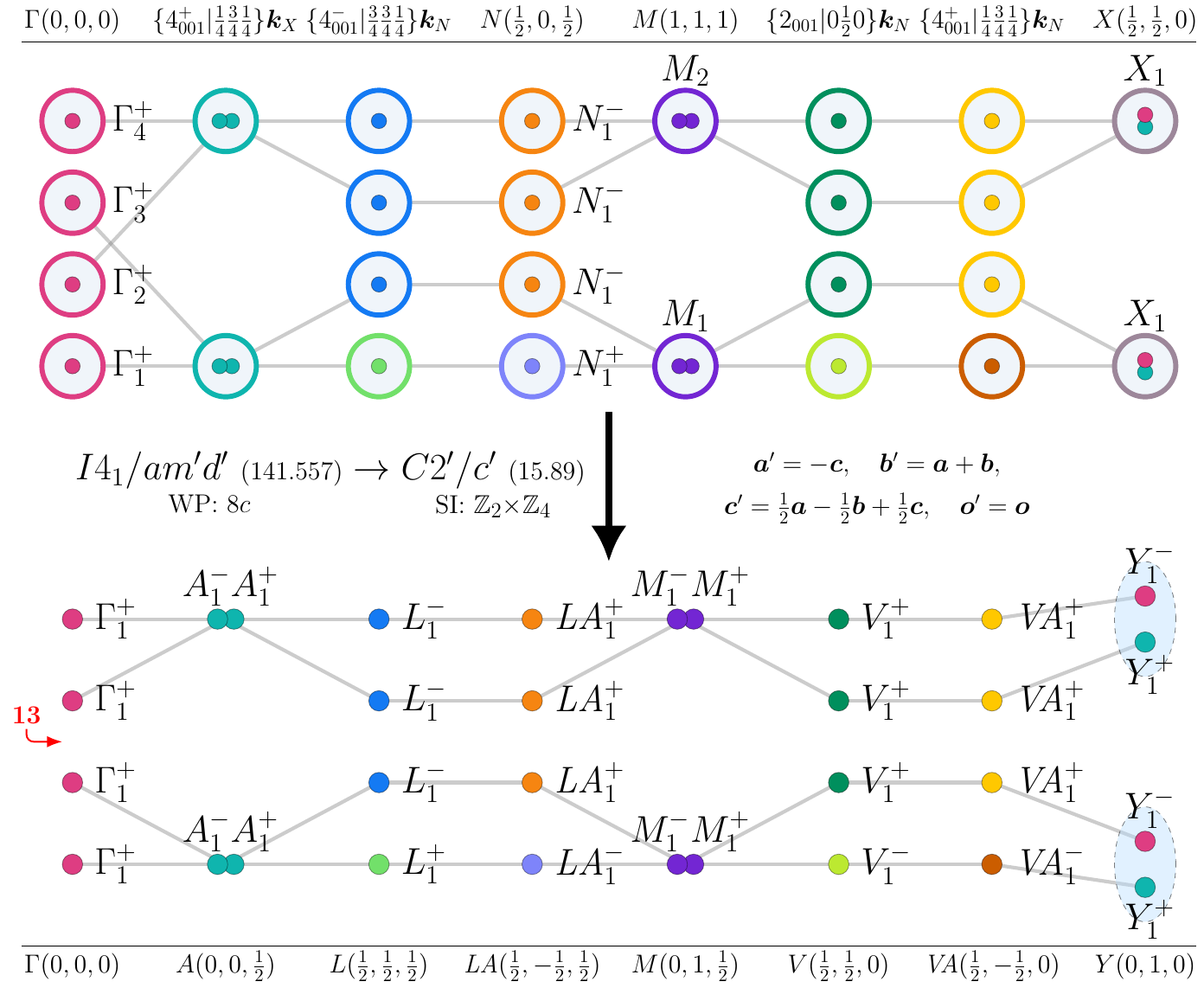}
\caption{Topological magnon bands in subgroup $C2'/c'~(15.89)$ for magnetic moments on Wyckoff position $8c$ of supergroup $I4_{1}/am'd'~(141.557)$.\label{fig_141.557_15.89_strainperp110_8c}}
\end{figure}
\input{gap_tables_tex/141.557_15.89_strainperp110_8c_table.tex}
\input{si_tables_tex/141.557_15.89_strainperp110_8c_table.tex}
\subsubsection{Topological bands in subgroup $C2'/m'~(12.62)$}
\textbf{Perturbation:}
\begin{itemize}
\item B $\parallel$ [100].
\end{itemize}
\begin{figure}[H]
\centering
\includegraphics[scale=0.6]{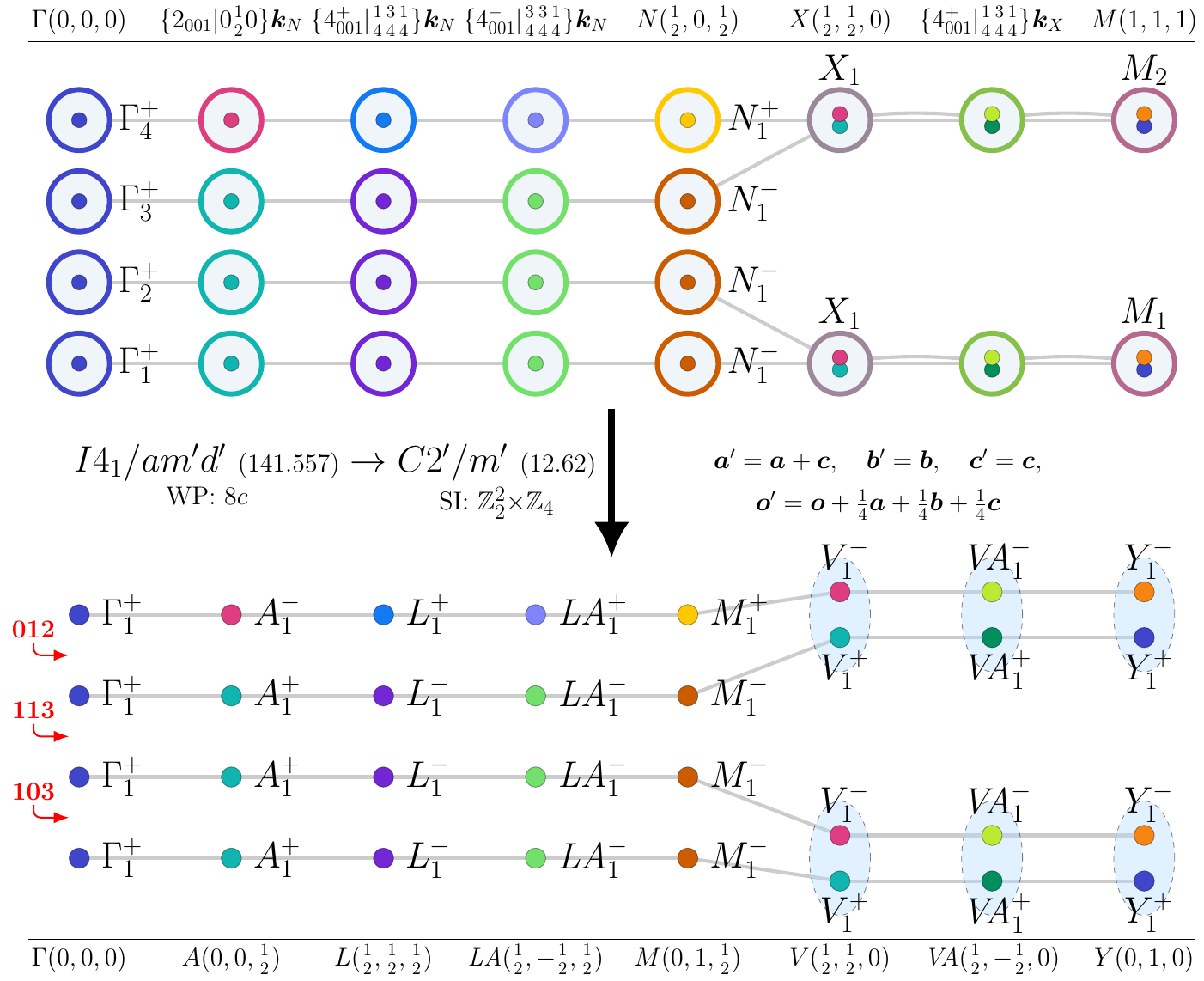}
\caption{Topological magnon bands in subgroup $C2'/m'~(12.62)$ for magnetic moments on Wyckoff position $8c$ of supergroup $I4_{1}/am'd'~(141.557)$.\label{fig_141.557_12.62_Bparallel100_8c}}
\end{figure}
\input{gap_tables_tex/141.557_12.62_Bparallel100_8c_table.tex}
\input{si_tables_tex/141.557_12.62_Bparallel100_8c_table.tex}
\subsubsection{Topological bands in subgroup $C2'/m'~(12.62)$}
\textbf{Perturbations:}
\begin{itemize}
\item strain $\perp$ [100],
\item B $\perp$ [100].
\end{itemize}
\begin{figure}[H]
\centering
\includegraphics[scale=0.6]{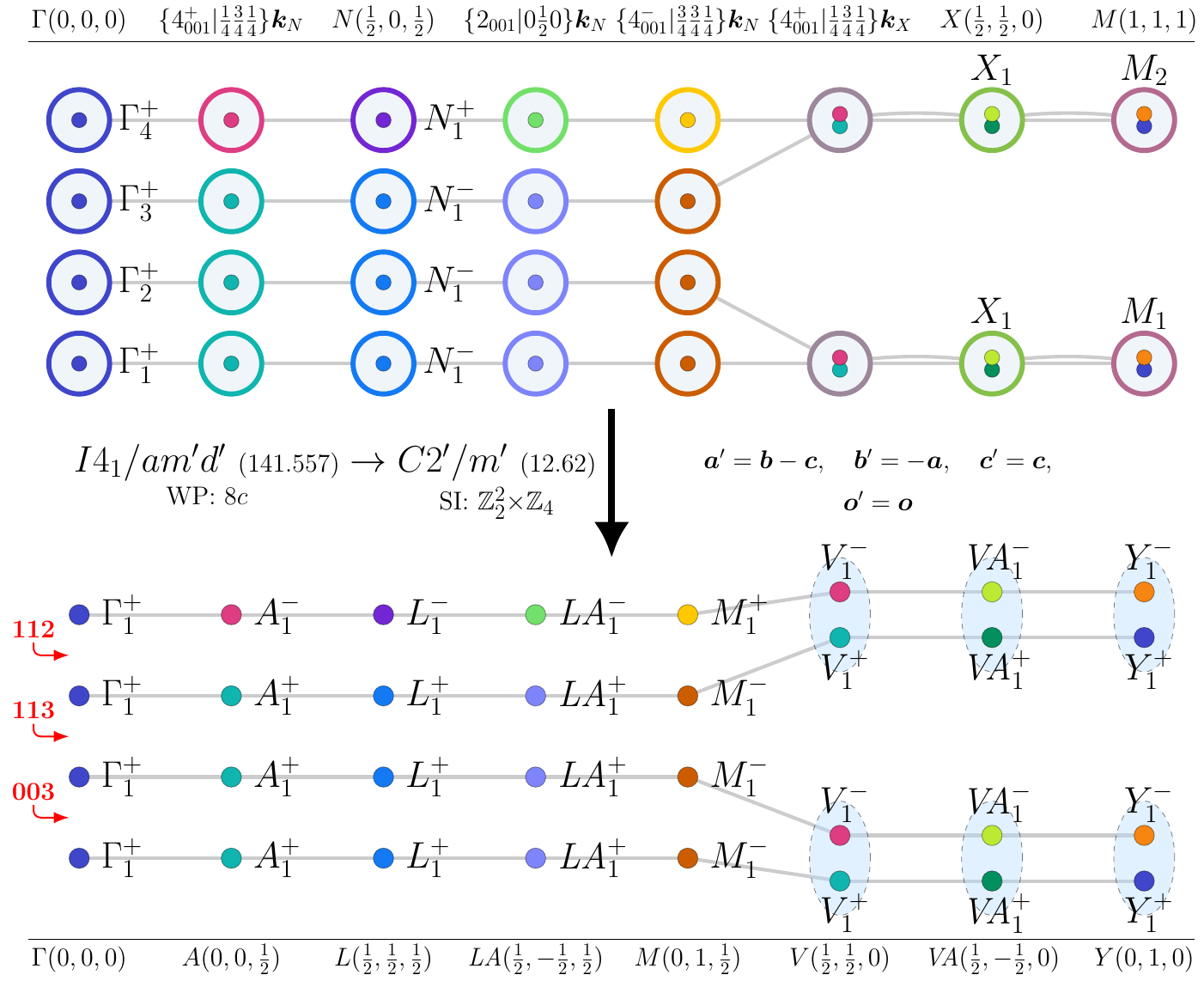}
\caption{Topological magnon bands in subgroup $C2'/m'~(12.62)$ for magnetic moments on Wyckoff position $8c$ of supergroup $I4_{1}/am'd'~(141.557)$.\label{fig_141.557_12.62_strainperp100_8c}}
\end{figure}
\input{gap_tables_tex/141.557_12.62_strainperp100_8c_table.tex}
\input{si_tables_tex/141.557_12.62_strainperp100_8c_table.tex}
\subsection{WP: $8c+8d$}
\textbf{BCS Materials:} {Tm\textsubscript{2}Mn\textsubscript{2}O\textsubscript{7}~(25 K)}\footnote{BCS web page: \texttt{\href{http://webbdcrista1.ehu.es/magndata/index.php?this\_label=0.151} {http://webbdcrista1.ehu.es/magndata/index.php?this\_label=0.151}}}, {Ho\textsubscript{2}CrSbO\textsubscript{7}~(13 K)}\footnote{BCS web page: \texttt{\href{http://webbdcrista1.ehu.es/magndata/index.php?this\_label=0.63} {http://webbdcrista1.ehu.es/magndata/index.php?this\_label=0.63}}}, {Ho\textsubscript{2}Ru\textsubscript{2}O\textsubscript{7}~(1.4 K)}\footnote{BCS web page: \texttt{\href{http://webbdcrista1.ehu.es/magndata/index.php?this\_label=0.51} {http://webbdcrista1.ehu.es/magndata/index.php?this\_label=0.51}}}.\\
\subsubsection{Topological bands in subgroup $P\bar{1}~(2.4)$}
\textbf{Perturbations:}
\begin{itemize}
\item strain in generic direction,
\item B $\parallel$ [100] and strain $\parallel$ [110],
\item B $\parallel$ [100] and strain $\perp$ [001],
\item B $\parallel$ [100] and strain $\perp$ [100],
\item B $\parallel$ [100] and strain $\perp$ [110],
\item B $\parallel$ [110] and strain $\parallel$ [100],
\item B $\parallel$ [110] and strain $\perp$ [001],
\item B $\parallel$ [110] and strain $\perp$ [100],
\item B $\parallel$ [110] and strain $\perp$ [110],
\item B in generic direction,
\item B $\perp$ [100] and strain $\parallel$ [110],
\item B $\perp$ [100] and strain $\perp$ [001],
\item B $\perp$ [100] and strain $\perp$ [110],
\item B $\perp$ [110] and strain $\parallel$ [100],
\item B $\perp$ [110] and strain $\perp$ [001],
\item B $\perp$ [110] and strain $\perp$ [100].
\end{itemize}
\begin{figure}[H]
\centering
\includegraphics[scale=0.6]{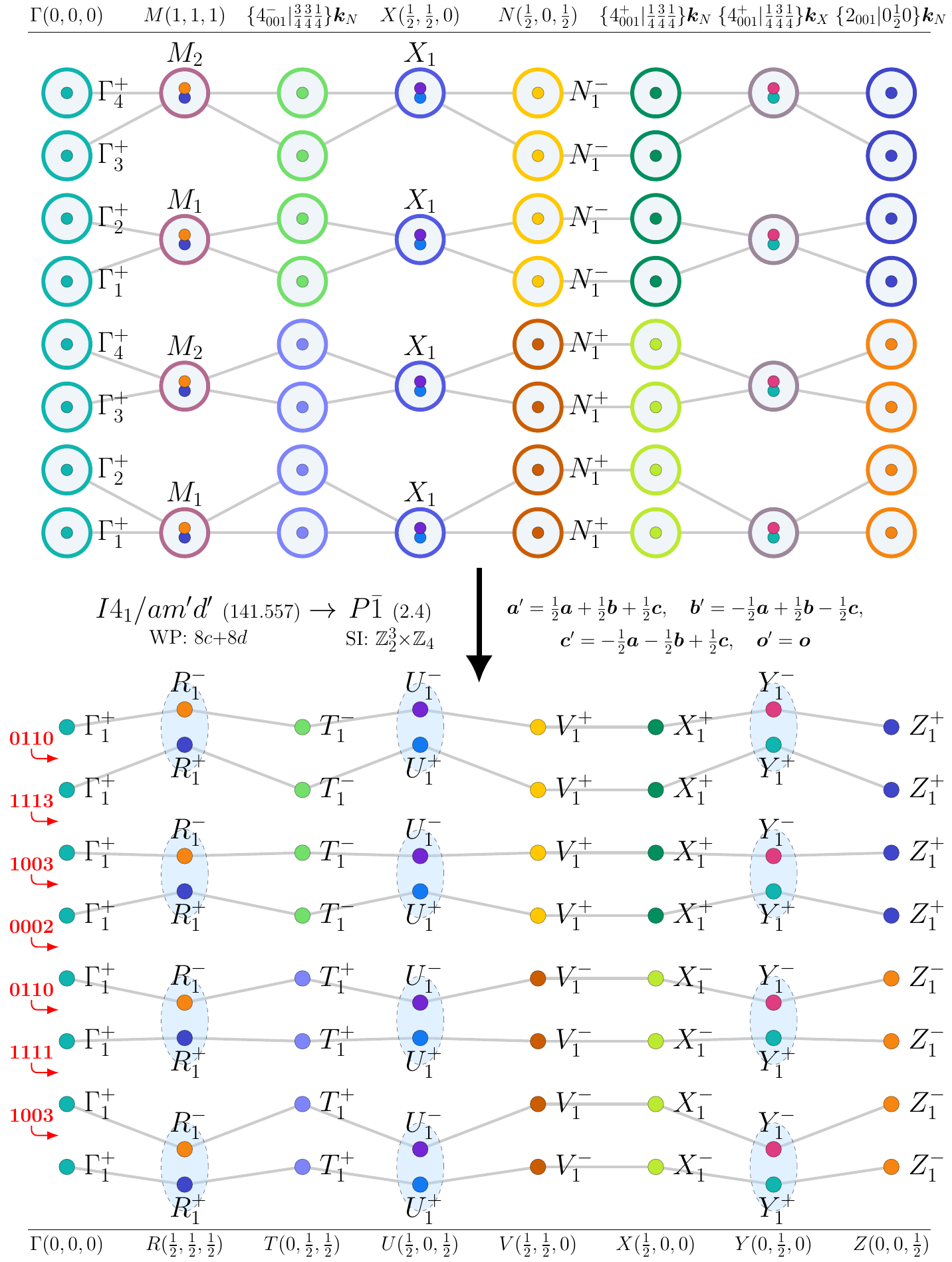}
\caption{Topological magnon bands in subgroup $P\bar{1}~(2.4)$ for magnetic moments on Wyckoff positions $8c+8d$ of supergroup $I4_{1}/am'd'~(141.557)$.\label{fig_141.557_2.4_strainingenericdirection_8c+8d}}
\end{figure}
\input{gap_tables_tex/141.557_2.4_strainingenericdirection_8c+8d_table.tex}
\input{si_tables_tex/141.557_2.4_strainingenericdirection_8c+8d_table.tex}
\subsubsection{Topological bands in subgroup $C2'/c'~(15.89)$}
\textbf{Perturbation:}
\begin{itemize}
\item B $\parallel$ [110].
\end{itemize}
\begin{figure}[H]
\centering
\includegraphics[scale=0.6]{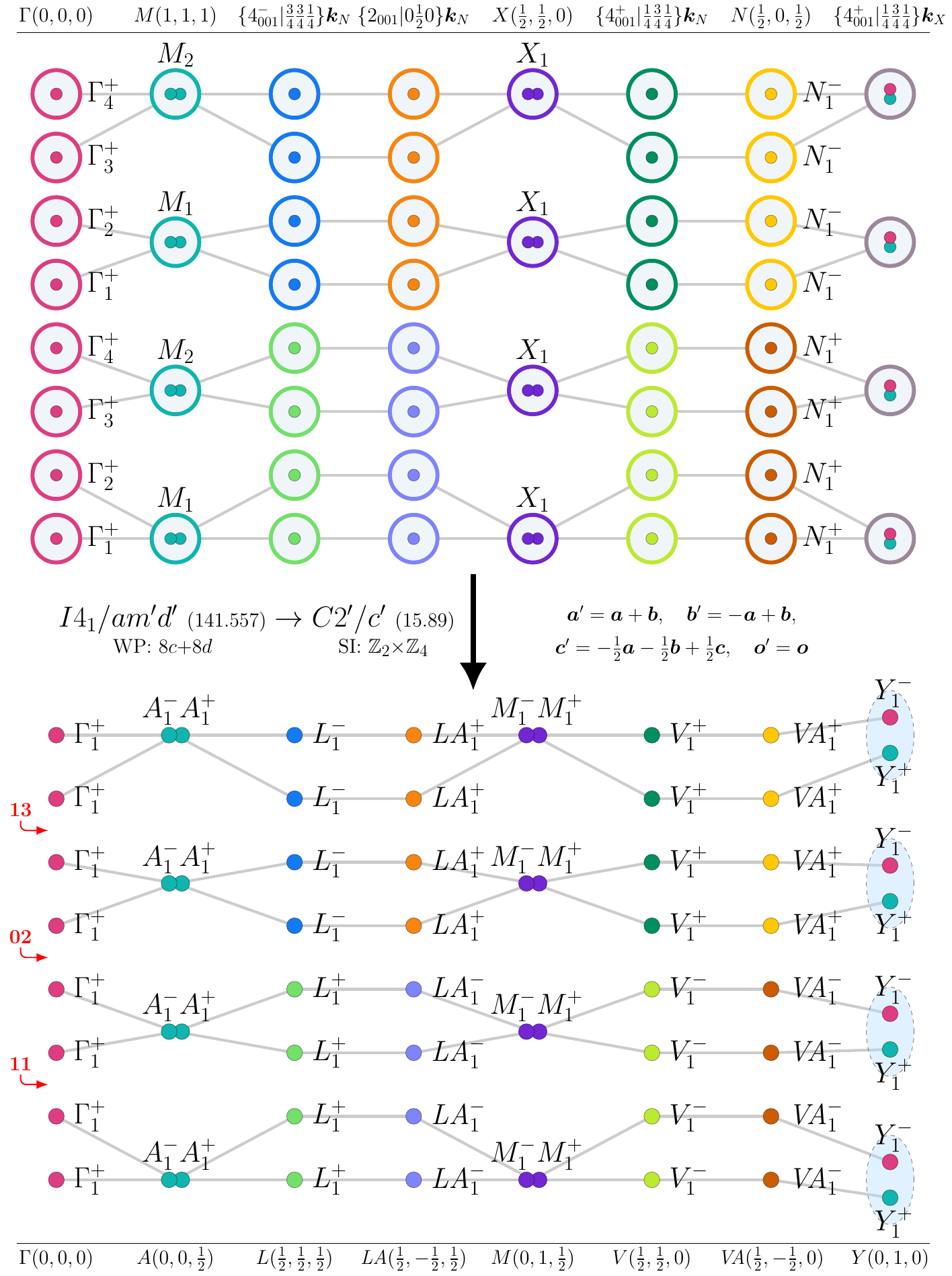}
\caption{Topological magnon bands in subgroup $C2'/c'~(15.89)$ for magnetic moments on Wyckoff positions $8c+8d$ of supergroup $I4_{1}/am'd'~(141.557)$.\label{fig_141.557_15.89_Bparallel110_8c+8d}}
\end{figure}
\input{gap_tables_tex/141.557_15.89_Bparallel110_8c+8d_table.tex}
\input{si_tables_tex/141.557_15.89_Bparallel110_8c+8d_table.tex}
\subsubsection{Topological bands in subgroup $C2'/c'~(15.89)$}
\textbf{Perturbations:}
\begin{itemize}
\item strain $\perp$ [110],
\item B $\perp$ [110].
\end{itemize}
\begin{figure}[H]
\centering
\includegraphics[scale=0.6]{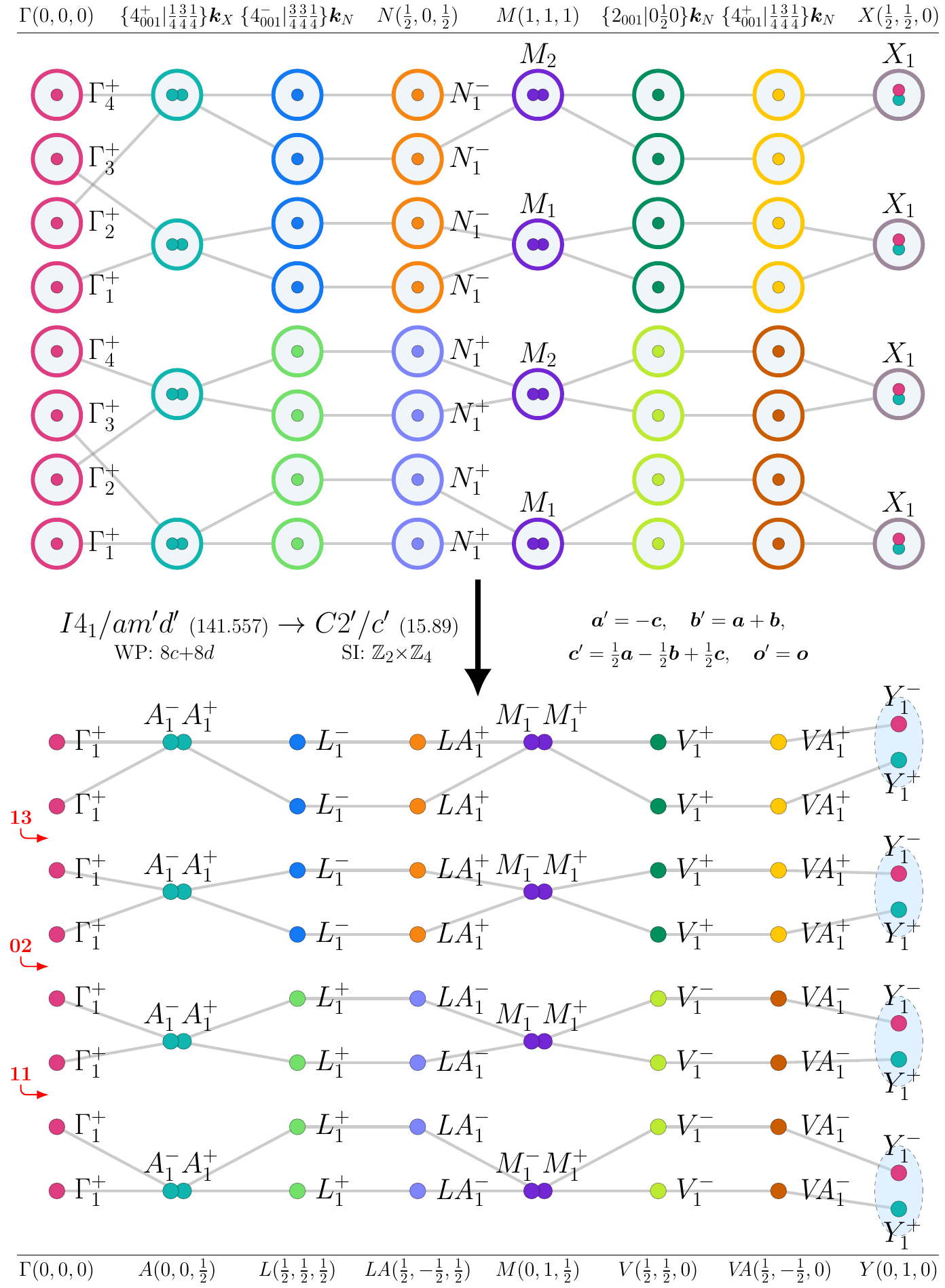}
\caption{Topological magnon bands in subgroup $C2'/c'~(15.89)$ for magnetic moments on Wyckoff positions $8c+8d$ of supergroup $I4_{1}/am'd'~(141.557)$.\label{fig_141.557_15.89_strainperp110_8c+8d}}
\end{figure}
\input{gap_tables_tex/141.557_15.89_strainperp110_8c+8d_table.tex}
\input{si_tables_tex/141.557_15.89_strainperp110_8c+8d_table.tex}
\subsubsection{Topological bands in subgroup $C2'/m'~(12.62)$}
\textbf{Perturbation:}
\begin{itemize}
\item B $\parallel$ [100].
\end{itemize}
\begin{figure}[H]
\centering
\includegraphics[scale=0.6]{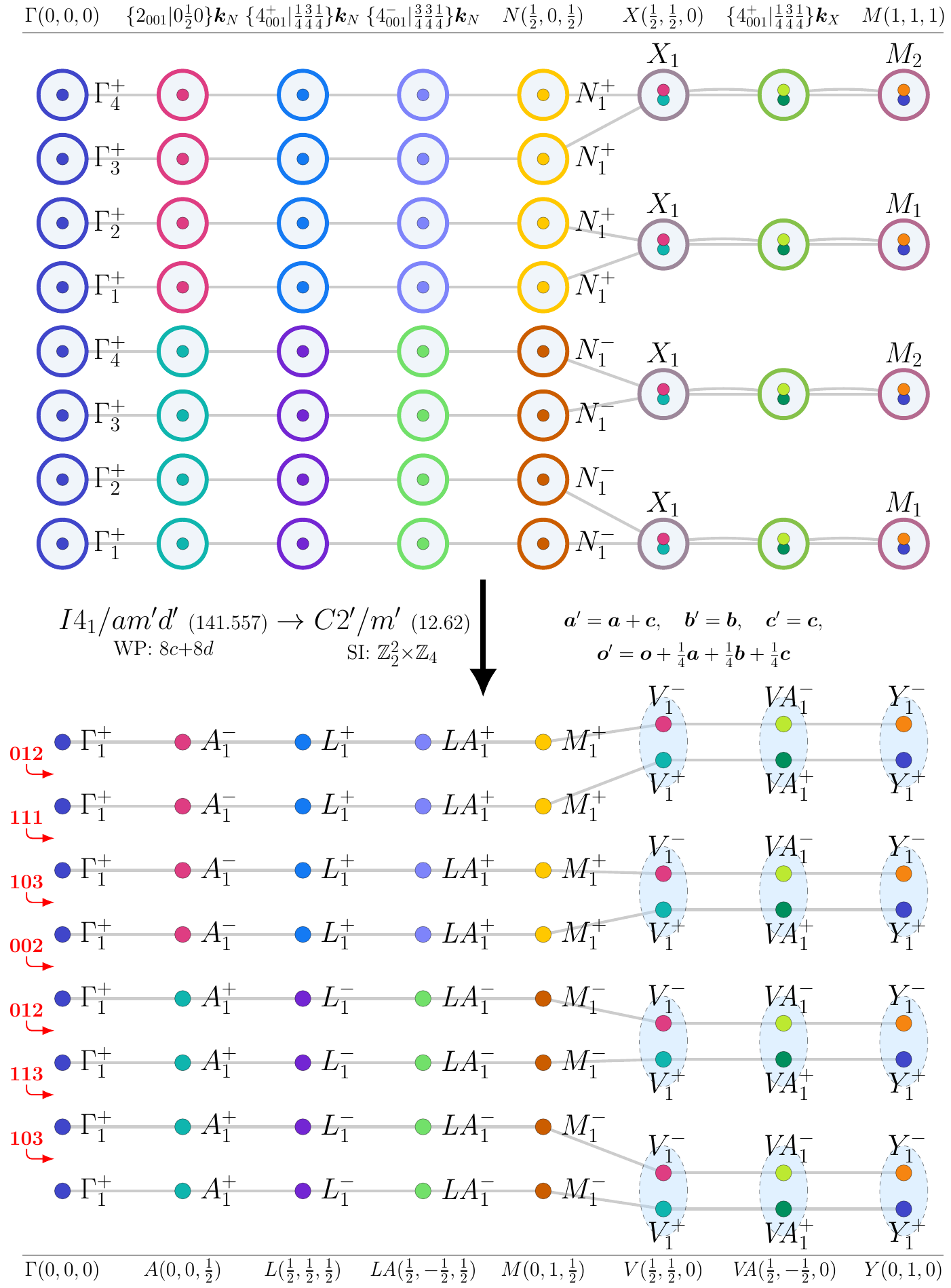}
\caption{Topological magnon bands in subgroup $C2'/m'~(12.62)$ for magnetic moments on Wyckoff positions $8c+8d$ of supergroup $I4_{1}/am'd'~(141.557)$.\label{fig_141.557_12.62_Bparallel100_8c+8d}}
\end{figure}
\input{gap_tables_tex/141.557_12.62_Bparallel100_8c+8d_table.tex}
\input{si_tables_tex/141.557_12.62_Bparallel100_8c+8d_table.tex}
\subsubsection{Topological bands in subgroup $C2'/m'~(12.62)$}
\textbf{Perturbations:}
\begin{itemize}
\item strain $\perp$ [100],
\item B $\perp$ [100].
\end{itemize}
\begin{figure}[H]
\centering
\includegraphics[scale=0.6]{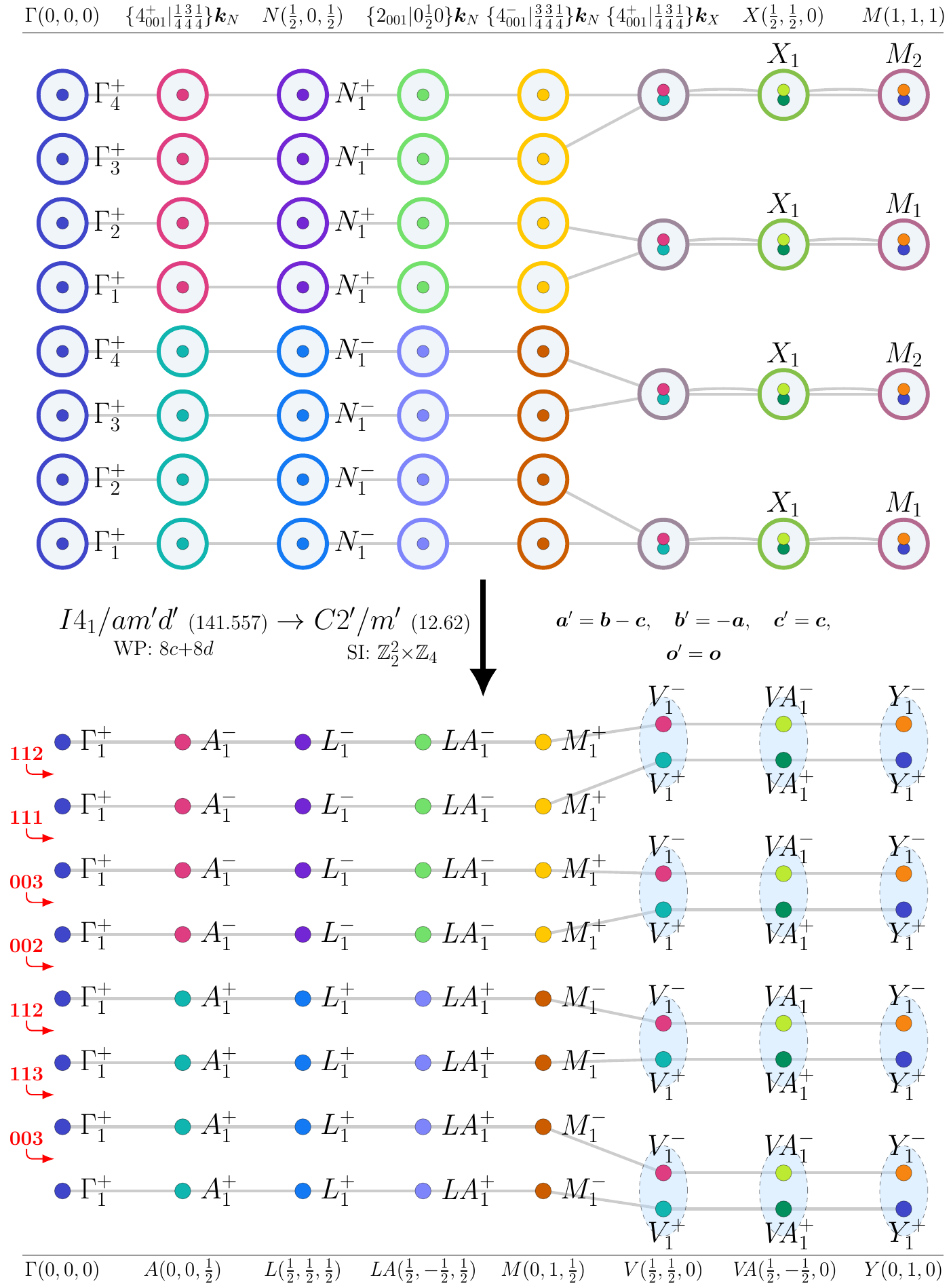}
\caption{Topological magnon bands in subgroup $C2'/m'~(12.62)$ for magnetic moments on Wyckoff positions $8c+8d$ of supergroup $I4_{1}/am'd'~(141.557)$.\label{fig_141.557_12.62_strainperp100_8c+8d}}
\end{figure}
\input{gap_tables_tex/141.557_12.62_strainperp100_8c+8d_table.tex}
\input{si_tables_tex/141.557_12.62_strainperp100_8c+8d_table.tex}
\subsection{WP: $8d$}
\textbf{BCS Materials:} {Tb\textsubscript{2}Sn\textsubscript{2}O\textsubscript{7}~(0.87 K)}\footnote{BCS web page: \texttt{\href{http://webbdcrista1.ehu.es/magndata/index.php?this\_label=0.48} {http://webbdcrista1.ehu.es/magndata/index.php?this\_label=0.48}}}, {Yb\textsubscript{2}Ti\textsubscript{2}O\textsubscript{7}~(0.26 K)}\footnote{BCS web page: \texttt{\href{http://webbdcrista1.ehu.es/magndata/index.php?this\_label=0.158} {http://webbdcrista1.ehu.es/magndata/index.php?this\_label=0.158}}}, {Yb\textsubscript{2}Sn\textsubscript{2}O\textsubscript{7}~(0.15 K)}\footnote{BCS web page: \texttt{\href{http://webbdcrista1.ehu.es/magndata/index.php?this\_label=0.157} {http://webbdcrista1.ehu.es/magndata/index.php?this\_label=0.157}}}.\\
\subsubsection{Topological bands in subgroup $P\bar{1}~(2.4)$}
\textbf{Perturbations:}
\begin{itemize}
\item strain in generic direction,
\item B $\parallel$ [100] and strain $\parallel$ [110],
\item B $\parallel$ [100] and strain $\perp$ [001],
\item B $\parallel$ [100] and strain $\perp$ [100],
\item B $\parallel$ [100] and strain $\perp$ [110],
\item B $\parallel$ [110] and strain $\parallel$ [100],
\item B $\parallel$ [110] and strain $\perp$ [001],
\item B $\parallel$ [110] and strain $\perp$ [100],
\item B $\parallel$ [110] and strain $\perp$ [110],
\item B in generic direction,
\item B $\perp$ [100] and strain $\parallel$ [110],
\item B $\perp$ [100] and strain $\perp$ [001],
\item B $\perp$ [100] and strain $\perp$ [110],
\item B $\perp$ [110] and strain $\parallel$ [100],
\item B $\perp$ [110] and strain $\perp$ [001],
\item B $\perp$ [110] and strain $\perp$ [100].
\end{itemize}
\begin{figure}[H]
\centering
\includegraphics[scale=0.6]{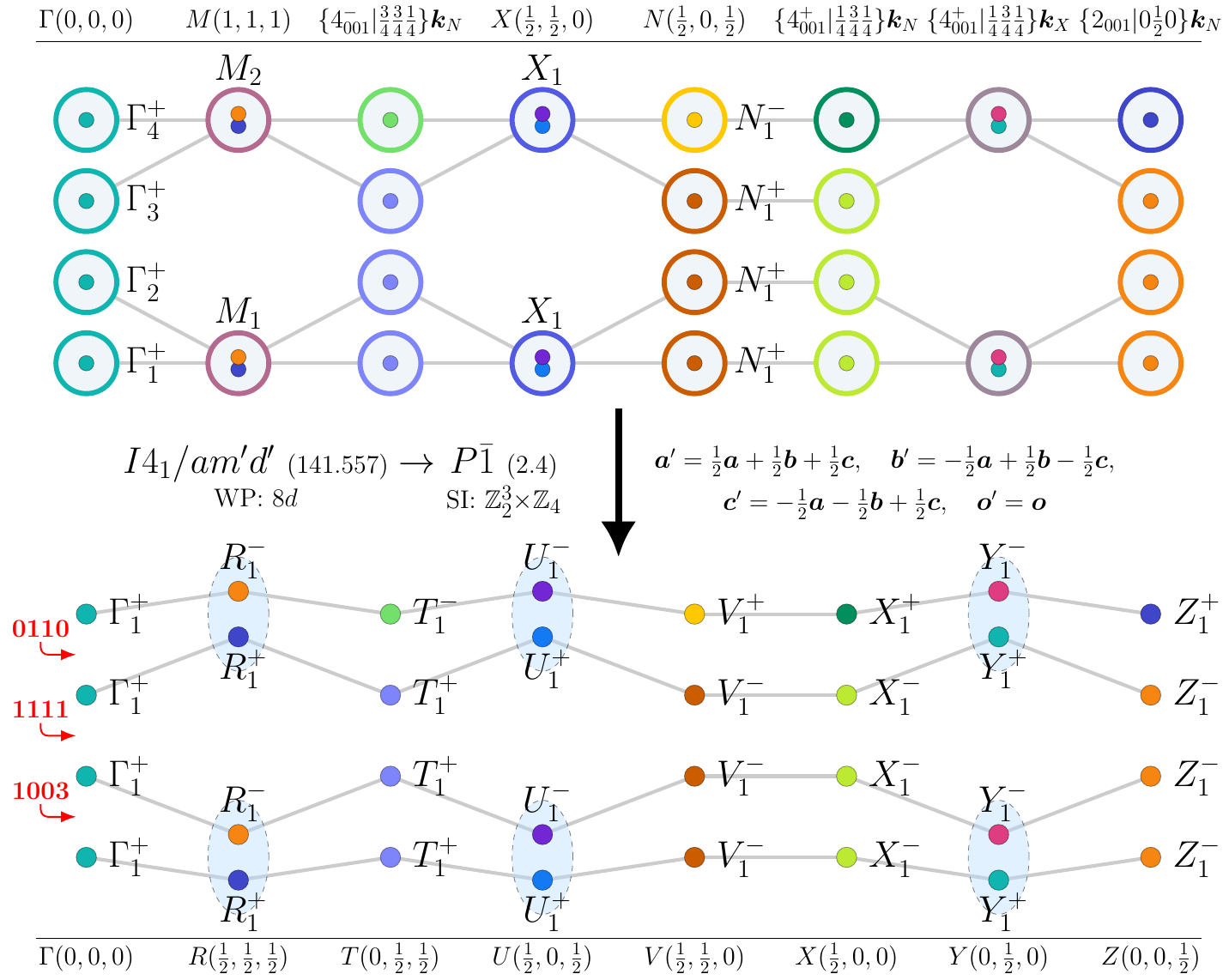}
\caption{Topological magnon bands in subgroup $P\bar{1}~(2.4)$ for magnetic moments on Wyckoff position $8d$ of supergroup $I4_{1}/am'd'~(141.557)$.\label{fig_141.557_2.4_strainingenericdirection_8d}}
\end{figure}
\input{gap_tables_tex/141.557_2.4_strainingenericdirection_8d_table.tex}
\input{si_tables_tex/141.557_2.4_strainingenericdirection_8d_table.tex}
\subsubsection{Topological bands in subgroup $C2'/c'~(15.89)$}
\textbf{Perturbation:}
\begin{itemize}
\item B $\parallel$ [110].
\end{itemize}
\begin{figure}[H]
\centering
\includegraphics[scale=0.6]{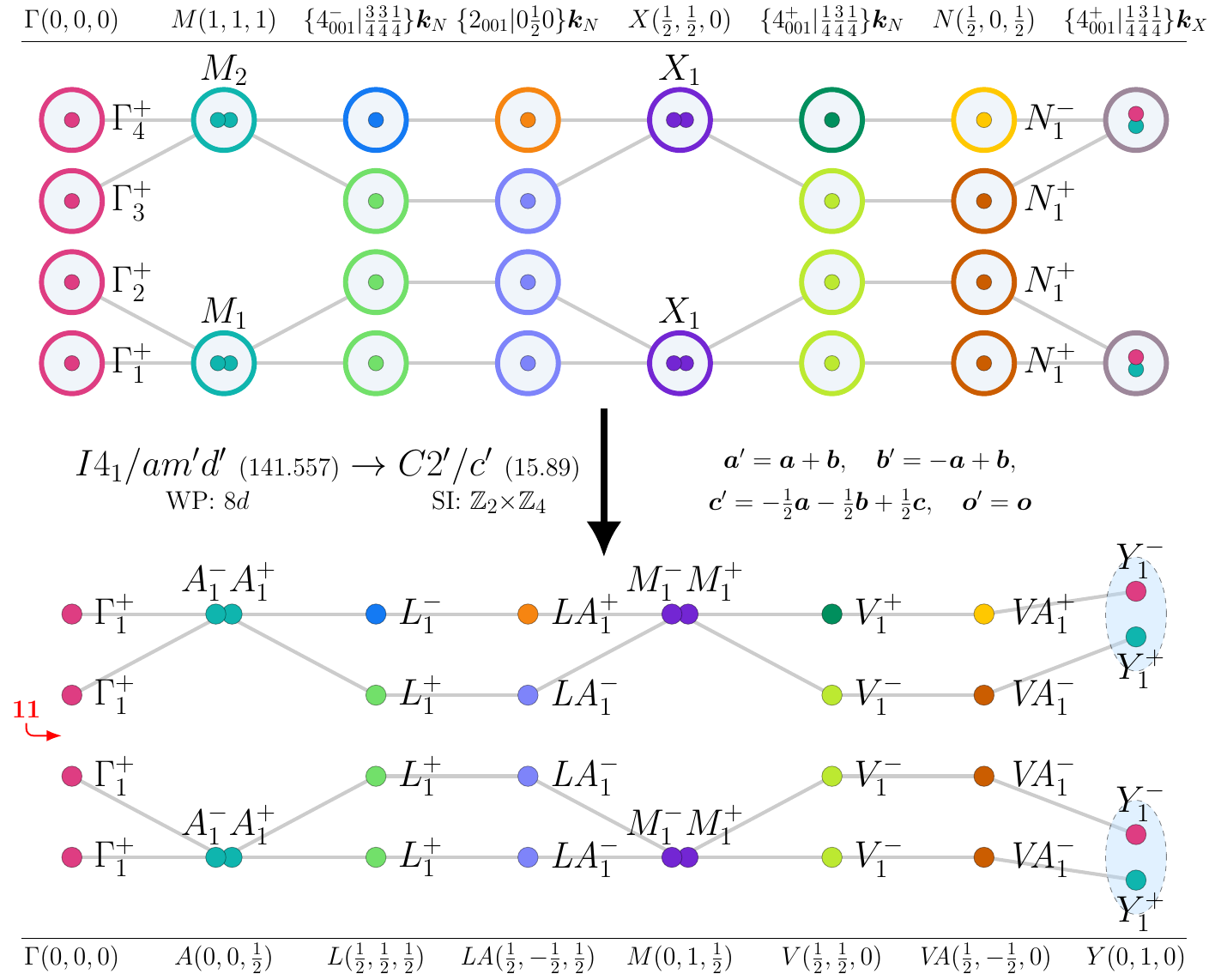}
\caption{Topological magnon bands in subgroup $C2'/c'~(15.89)$ for magnetic moments on Wyckoff position $8d$ of supergroup $I4_{1}/am'd'~(141.557)$.\label{fig_141.557_15.89_Bparallel110_8d}}
\end{figure}
\input{gap_tables_tex/141.557_15.89_Bparallel110_8d_table.tex}
\input{si_tables_tex/141.557_15.89_Bparallel110_8d_table.tex}
\subsubsection{Topological bands in subgroup $C2'/c'~(15.89)$}
\textbf{Perturbations:}
\begin{itemize}
\item strain $\perp$ [110],
\item B $\perp$ [110].
\end{itemize}
\begin{figure}[H]
\centering
\includegraphics[scale=0.6]{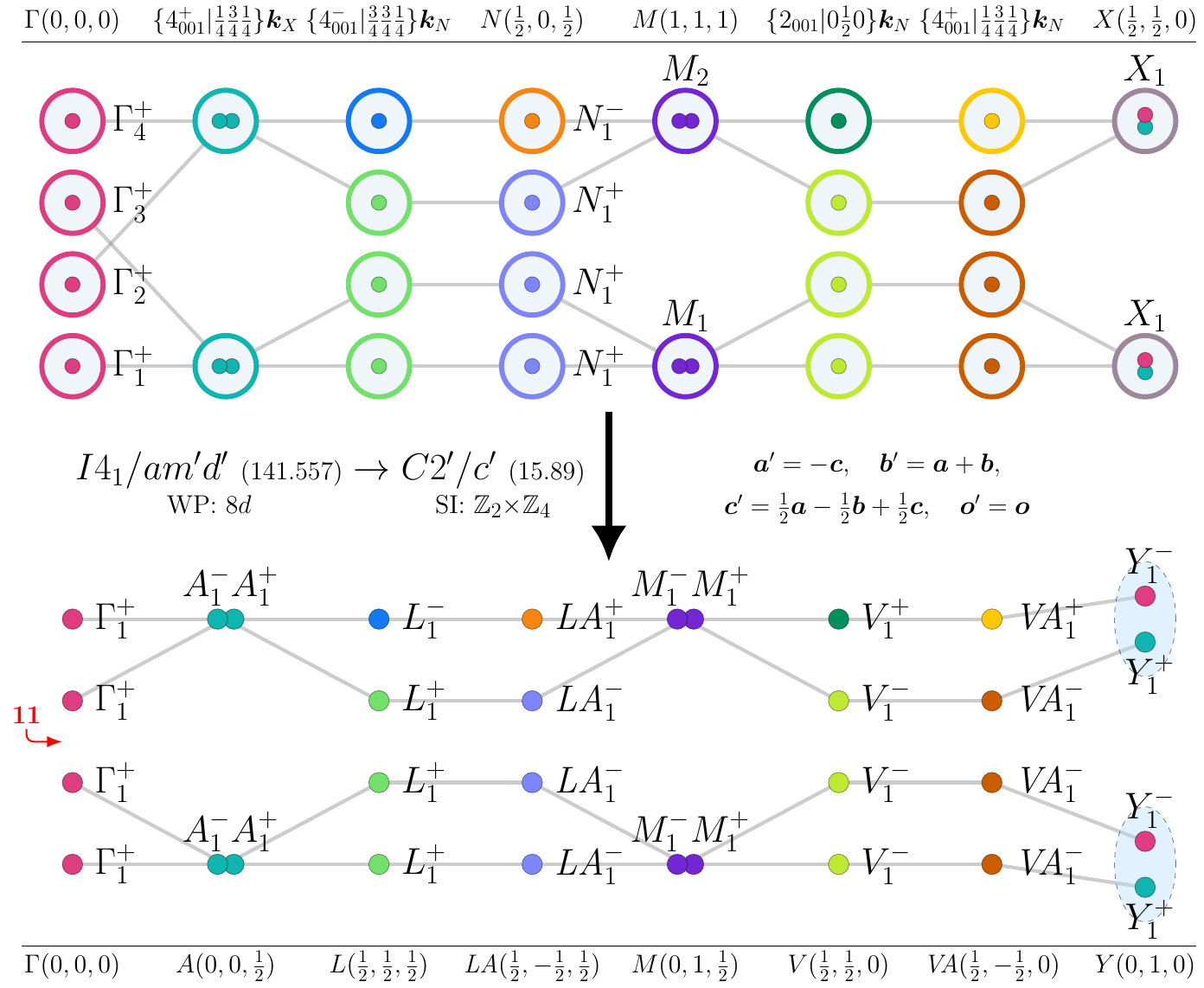}
\caption{Topological magnon bands in subgroup $C2'/c'~(15.89)$ for magnetic moments on Wyckoff position $8d$ of supergroup $I4_{1}/am'd'~(141.557)$.\label{fig_141.557_15.89_strainperp110_8d}}
\end{figure}
\input{gap_tables_tex/141.557_15.89_strainperp110_8d_table.tex}
\input{si_tables_tex/141.557_15.89_strainperp110_8d_table.tex}
\subsubsection{Topological bands in subgroup $C2'/m'~(12.62)$}
\textbf{Perturbation:}
\begin{itemize}
\item B $\parallel$ [100].
\end{itemize}
\begin{figure}[H]
\centering
\includegraphics[scale=0.6]{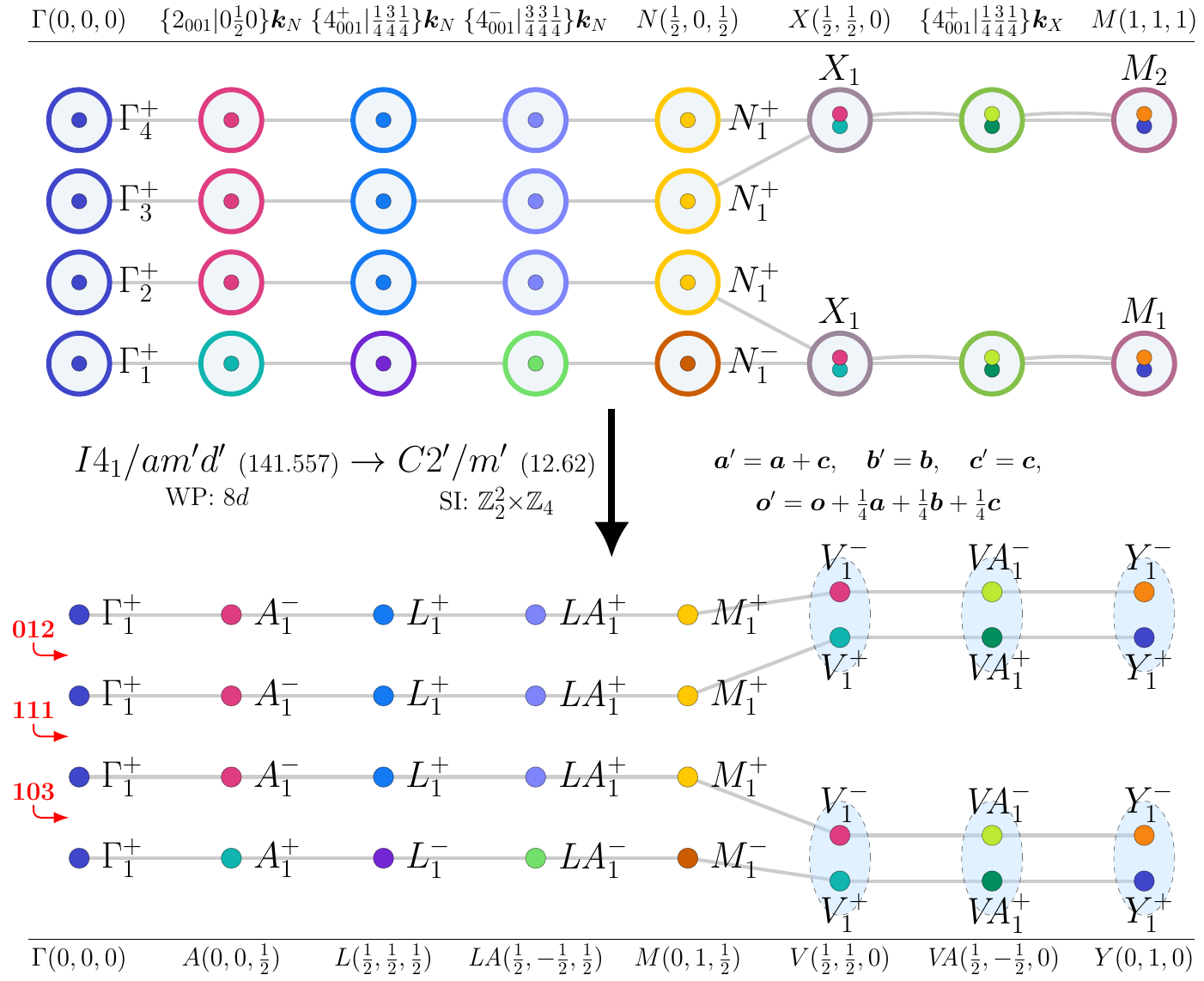}
\caption{Topological magnon bands in subgroup $C2'/m'~(12.62)$ for magnetic moments on Wyckoff position $8d$ of supergroup $I4_{1}/am'd'~(141.557)$.\label{fig_141.557_12.62_Bparallel100_8d}}
\end{figure}
\input{gap_tables_tex/141.557_12.62_Bparallel100_8d_table.tex}
\input{si_tables_tex/141.557_12.62_Bparallel100_8d_table.tex}
\subsubsection{Topological bands in subgroup $C2'/m'~(12.62)$}
\textbf{Perturbations:}
\begin{itemize}
\item strain $\perp$ [100],
\item B $\perp$ [100].
\end{itemize}
\begin{figure}[H]
\centering
\includegraphics[scale=0.6]{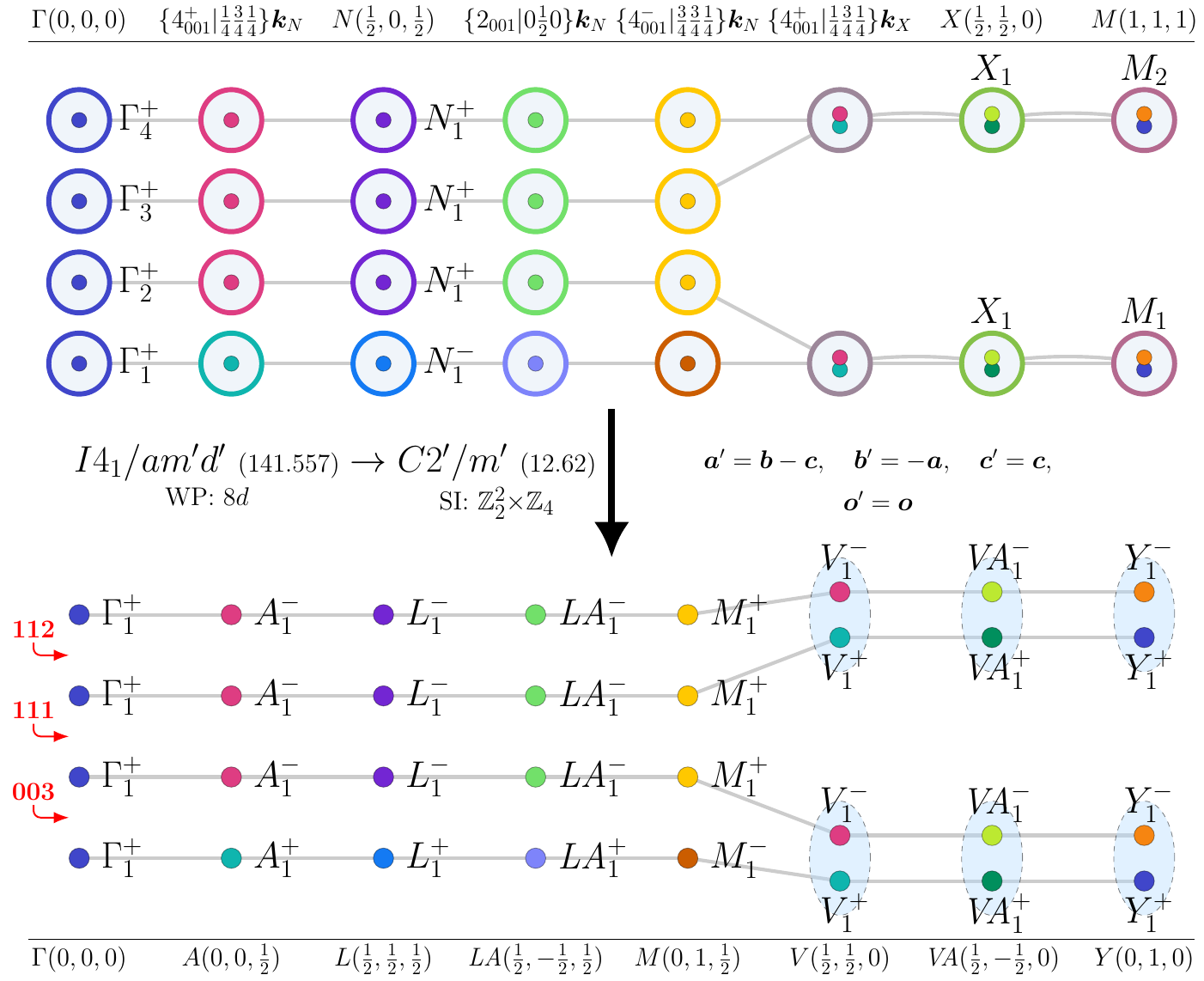}
\caption{Topological magnon bands in subgroup $C2'/m'~(12.62)$ for magnetic moments on Wyckoff position $8d$ of supergroup $I4_{1}/am'd'~(141.557)$.\label{fig_141.557_12.62_strainperp100_8d}}
\end{figure}
\input{gap_tables_tex/141.557_12.62_strainperp100_8d_table.tex}
\input{si_tables_tex/141.557_12.62_strainperp100_8d_table.tex}

\section{MSG $I4_{1}'/a'cd'~(142.568)$}
\textbf{Nontrivial-SI Subgroups:} $Iba2~(45.235)$, $Ib'ca~(73.550)$, $I4_{1}'cd'~(110.248)$, $I\bar{4}2'd'~(122.337)$.\\

\textbf{Trivial-SI Subgroups:} $Cc'~(9.39)$, $Cc'~(9.39)$, $Cc'~(9.39)$, $P\bar{1}'~(2.6)$, $C2'~(5.15)$, $Ib'a2'~(45.237)$, $Cc~(9.37)$, $Ib'a2'~(45.237)$, $C2'/c~(15.87)$, $C2~(5.13)$, $Fd'd'2~(43.227)$, $C2/c'~(15.88)$, $C2~(5.13)$, $Fd'd'2~(43.227)$, $C2/c'~(15.88)$, $I2_{1}'2_{1}'2_{1}~(24.55)$, $Fd'd'd'~(70.531)$.\\

\subsection{WP: $8a$}
\textbf{BCS Materials:} {Ca\textsubscript{2}MnO\textsubscript{4}~(110 K)}\footnote{BCS web page: \texttt{\href{http://webbdcrista1.ehu.es/magndata/index.php?this\_label=0.211} {http://webbdcrista1.ehu.es/magndata/index.php?this\_label=0.211}}}.\\
\subsubsection{Topological bands in subgroup $I\bar{4}2'd'~(122.337)$}
\textbf{Perturbation:}
\begin{itemize}
\item B $\parallel$ [001].
\end{itemize}
\begin{figure}[H]
\centering
\includegraphics[scale=0.6]{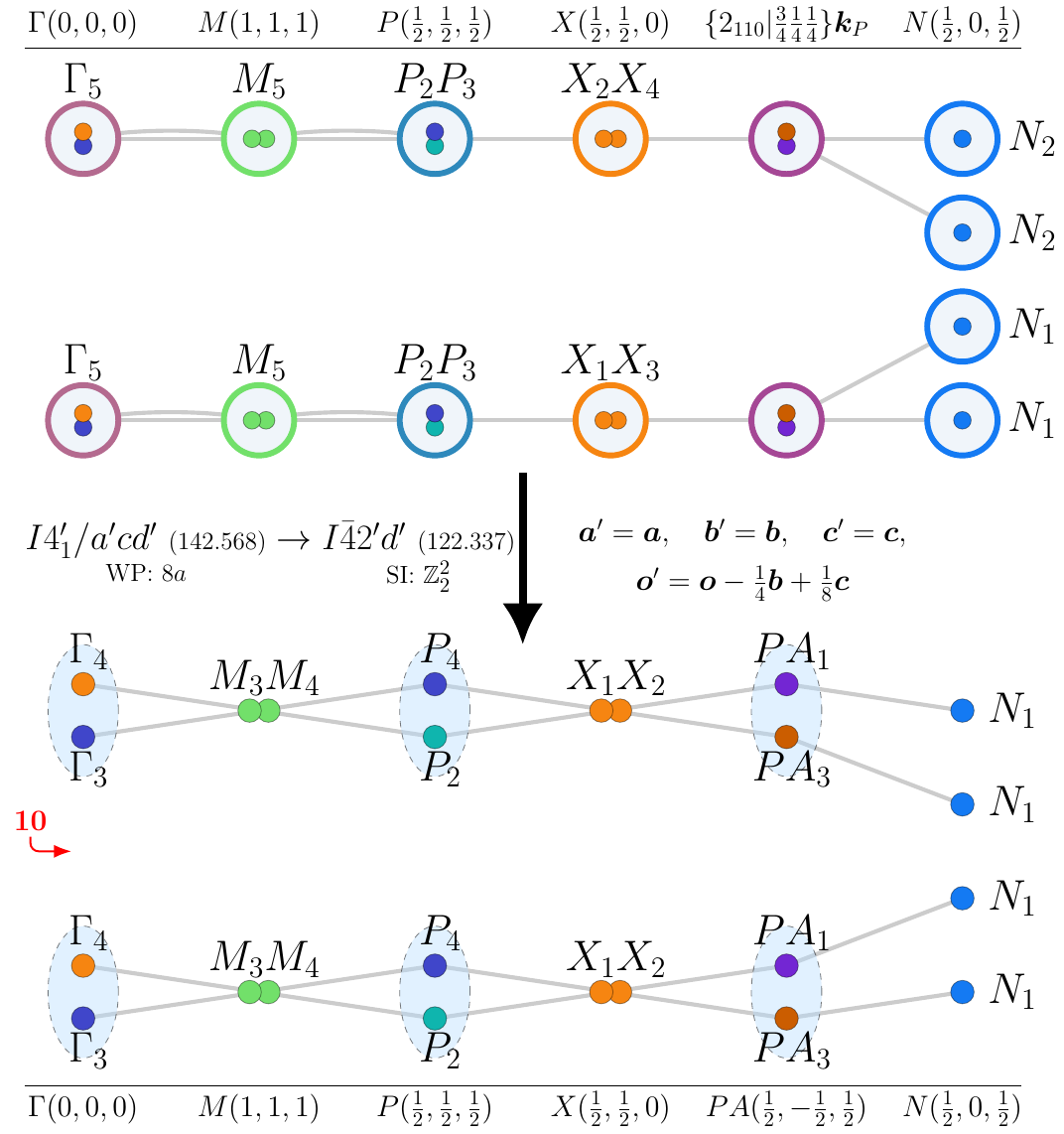}
\caption{Topological magnon bands in subgroup $I\bar{4}2'd'~(122.337)$ for magnetic moments on Wyckoff position $8a$ of supergroup $I4_{1}'/a'cd'~(142.568)$.\label{fig_142.568_122.337_Bparallel001_8a}}
\end{figure}
\input{gap_tables_tex/142.568_122.337_Bparallel001_8a_table.tex}
\input{si_tables_tex/142.568_122.337_Bparallel001_8a_table.tex}

\section{MSG $I_{c}4_{1}/acd~(142.570)$}
\textbf{Nontrivial-SI Subgroups:} $C_{a}c~(9.41)$, $C_{a}c~(9.41)$, $P\bar{1}~(2.4)$, $C2'/c'~(15.89)$, $C2'/c'~(15.89)$, $C2'/m'~(12.62)$, $P_{S}\bar{1}~(2.7)$, $C_{a}2~(5.17)$, $F_{S}dd2~(43.228)$, $C2/c~(15.85)$, $Fd'd'd~(70.530)$, $C_{a}2/c~(15.91)$, $C_{a}2~(5.17)$, $F_{S}dd2~(43.228)$, $I_{c}ba2~(45.239)$, $C2/c~(15.85)$, $Fd'd'd~(70.530)$, $Im'm'a~(74.558)$, $C_{a}2/c~(15.91)$, $F_{S}ddd~(70.532)$, $I_{a}ba2~(45.240)$, $C2/c~(15.85)$, $Iba'm'~(72.544)$, $C_{c}2/c~(15.90)$, $I_{c}bca~(73.553)$, $I_{c}4_{1}cd~(110.250)$, $I4_{1}/am'd'~(141.557)$.\\

\textbf{Trivial-SI Subgroups:} $Cc'~(9.39)$, $Cc'~(9.39)$, $Cm'~(8.34)$, $C2'~(5.15)$, $C2'~(5.15)$, $C2'~(5.15)$, $P_{S}1~(1.3)$, $Cc~(9.37)$, $Fd'd2'~(43.226)$, $Cc~(9.37)$, $Fd'd2'~(43.226)$, $Im'a2'~(46.243)$, $Cc~(9.37)$, $Im'a2'~(46.243)$, $C_{c}c~(9.40)$, $C2~(5.13)$, $Fd'd'2~(43.227)$, $C2~(5.13)$, $Fd'd'2~(43.227)$, $Im'm'2~(44.232)$, $C2~(5.13)$, $Im'a'2~(46.245)$, $C_{c}2~(5.16)$, $I4_{1}m'd'~(109.243)$.\\

\subsection{WP: $16e$}
\textbf{BCS Materials:} {CoO~(290 K)}\footnote{BCS web page: \texttt{\href{http://webbdcrista1.ehu.es/magndata/index.php?this\_label=3.19} {http://webbdcrista1.ehu.es/magndata/index.php?this\_label=3.19}}}.\\
\subsubsection{Topological bands in subgroup $P\bar{1}~(2.4)$}
\textbf{Perturbations:}
\begin{itemize}
\item B $\parallel$ [001] and strain in generic direction,
\item B $\parallel$ [100] and strain $\perp$ [110],
\item B $\parallel$ [100] and strain in generic direction,
\item B $\parallel$ [110] and strain $\perp$ [100],
\item B $\parallel$ [110] and strain in generic direction,
\item B $\perp$ [001] and strain $\perp$ [100],
\item B $\perp$ [001] and strain $\perp$ [110],
\item B $\perp$ [001] and strain in generic direction,
\item B $\perp$ [100] and strain $\parallel$ [110],
\item B $\perp$ [100] and strain $\perp$ [001],
\item B $\perp$ [100] and strain $\perp$ [110],
\item B $\perp$ [100] and strain in generic direction,
\item B $\perp$ [110] and strain $\parallel$ [100],
\item B $\perp$ [110] and strain $\perp$ [001],
\item B $\perp$ [110] and strain $\perp$ [100],
\item B $\perp$ [110] and strain in generic direction,
\item B in generic direction.
\end{itemize}
\begin{figure}[H]
\centering
\includegraphics[scale=0.6]{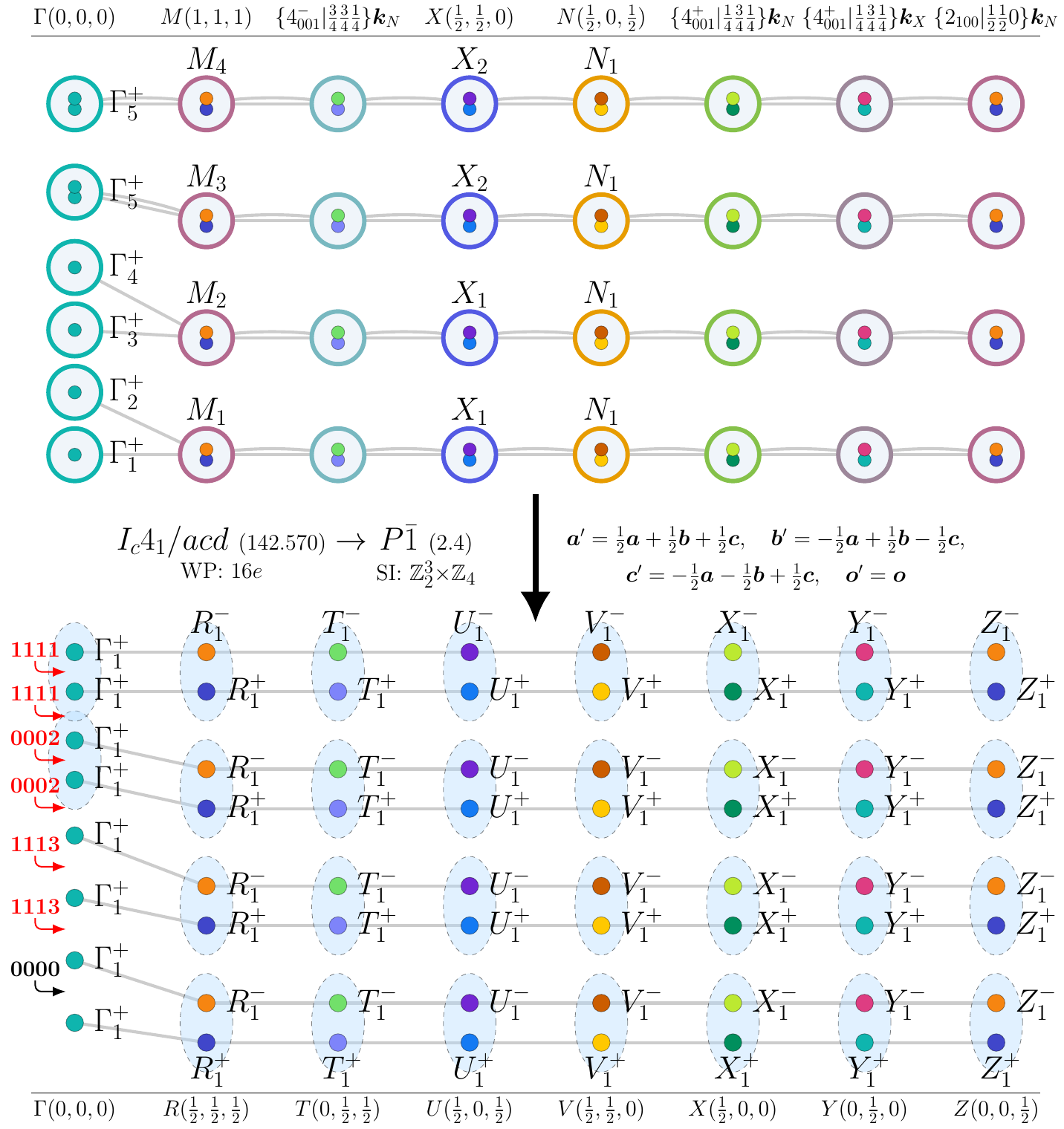}
\caption{Topological magnon bands in subgroup $P\bar{1}~(2.4)$ for magnetic moments on Wyckoff position $16e$ of supergroup $I_{c}4_{1}/acd~(142.570)$.\label{fig_142.570_2.4_Bparallel001andstrainingenericdirection_16e}}
\end{figure}
\input{gap_tables_tex/142.570_2.4_Bparallel001andstrainingenericdirection_16e_table.tex}
\input{si_tables_tex/142.570_2.4_Bparallel001andstrainingenericdirection_16e_table.tex}
\subsubsection{Topological bands in subgroup $C2'/c'~(15.89)$}
\textbf{Perturbations:}
\begin{itemize}
\item B $\parallel$ [001] and strain $\perp$ [110],
\item B $\perp$ [110].
\end{itemize}
\begin{figure}[H]
\centering
\includegraphics[scale=0.6]{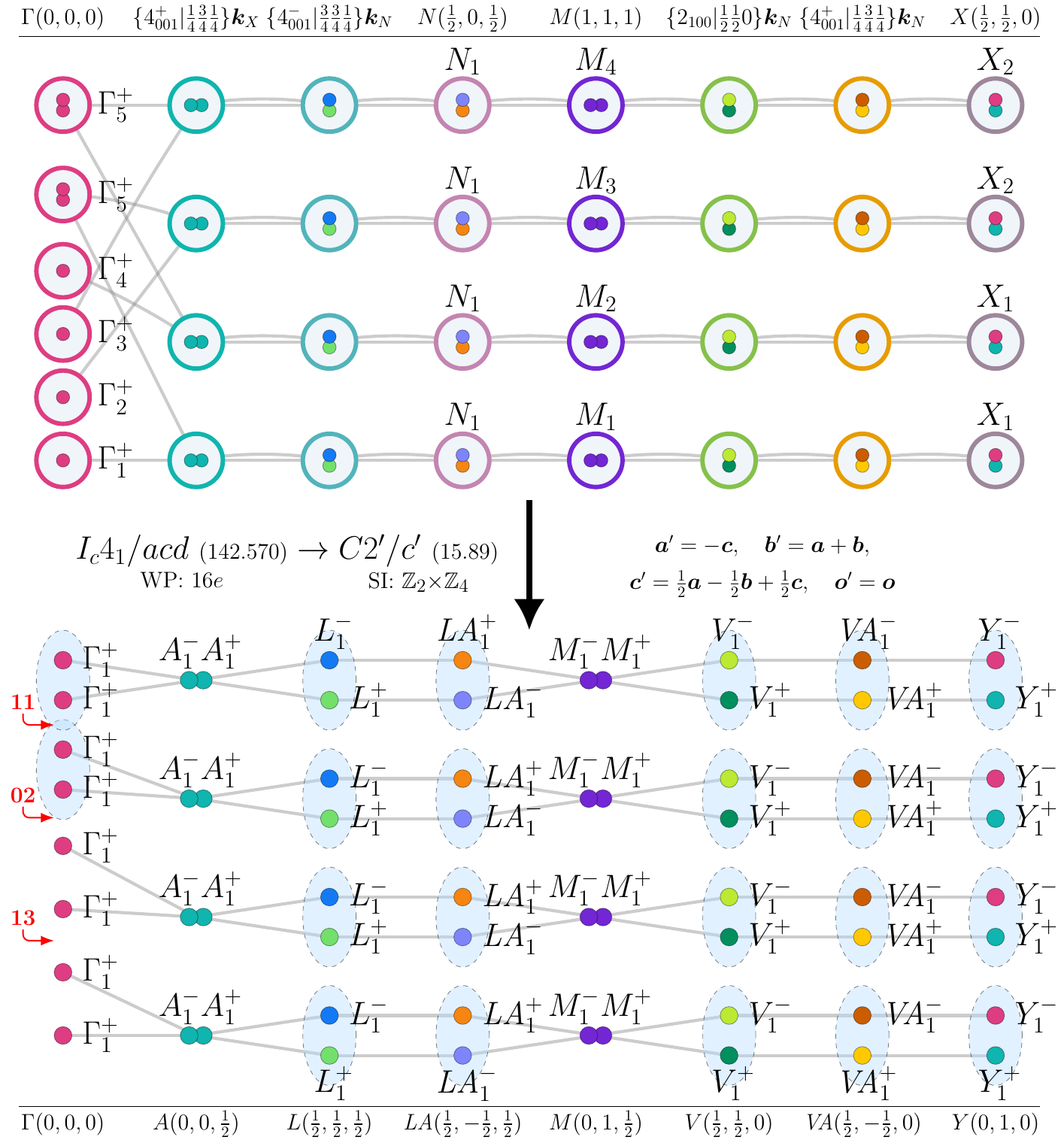}
\caption{Topological magnon bands in subgroup $C2'/c'~(15.89)$ for magnetic moments on Wyckoff position $16e$ of supergroup $I_{c}4_{1}/acd~(142.570)$.\label{fig_142.570_15.89_Bparallel001andstrainperp110_16e}}
\end{figure}
\input{gap_tables_tex/142.570_15.89_Bparallel001andstrainperp110_16e_table.tex}
\input{si_tables_tex/142.570_15.89_Bparallel001andstrainperp110_16e_table.tex}
\subsubsection{Topological bands in subgroup $C2'/c'~(15.89)$}
\textbf{Perturbations:}
\begin{itemize}
\item B $\parallel$ [100] and strain $\parallel$ [110],
\item B $\parallel$ [100] and strain $\perp$ [001],
\item B $\parallel$ [110] and strain $\parallel$ [100],
\item B $\parallel$ [110] and strain $\perp$ [001],
\item B $\perp$ [001].
\end{itemize}
\begin{figure}[H]
\centering
\includegraphics[scale=0.6]{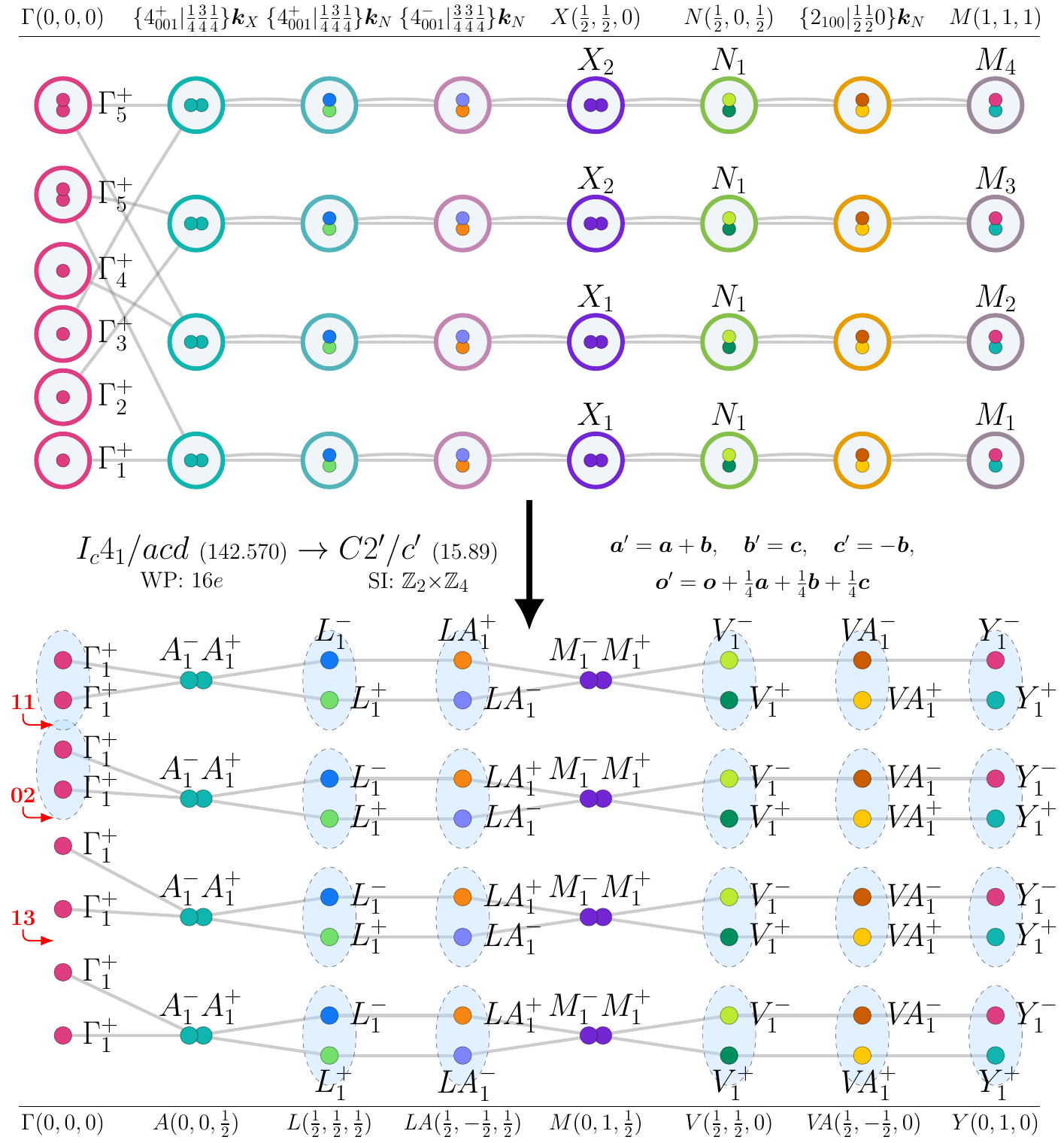}
\caption{Topological magnon bands in subgroup $C2'/c'~(15.89)$ for magnetic moments on Wyckoff position $16e$ of supergroup $I_{c}4_{1}/acd~(142.570)$.\label{fig_142.570_15.89_Bparallel100andstrainparallel110_16e}}
\end{figure}
\input{gap_tables_tex/142.570_15.89_Bparallel100andstrainparallel110_16e_table.tex}
\input{si_tables_tex/142.570_15.89_Bparallel100andstrainparallel110_16e_table.tex}
\subsubsection{Topological bands in subgroup $C2'/m'~(12.62)$}
\textbf{Perturbations:}
\begin{itemize}
\item B $\parallel$ [001] and strain $\perp$ [100],
\item B $\perp$ [100].
\end{itemize}
\begin{figure}[H]
\centering
\includegraphics[scale=0.6]{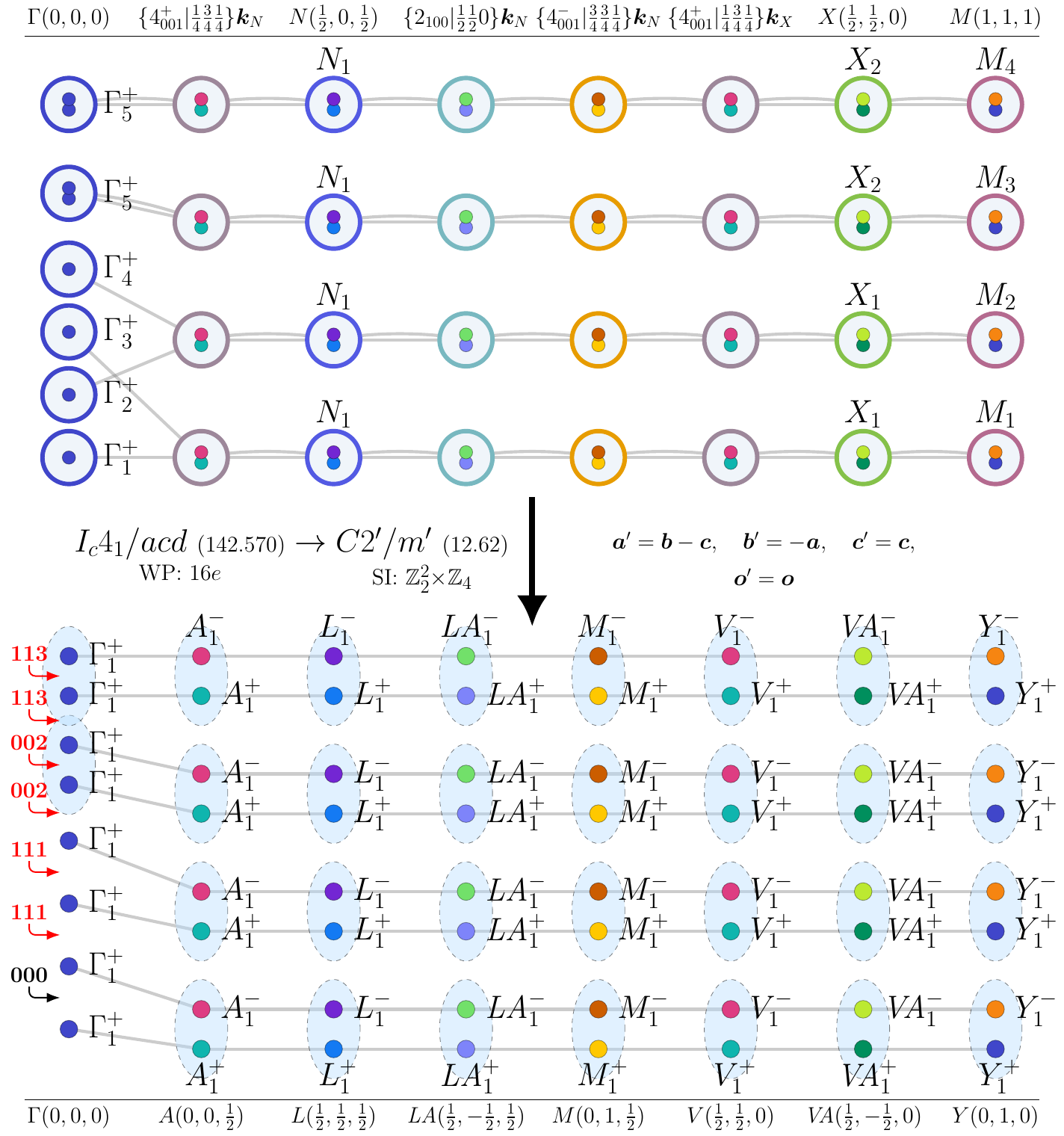}
\caption{Topological magnon bands in subgroup $C2'/m'~(12.62)$ for magnetic moments on Wyckoff position $16e$ of supergroup $I_{c}4_{1}/acd~(142.570)$.\label{fig_142.570_12.62_Bparallel001andstrainperp100_16e}}
\end{figure}
\input{gap_tables_tex/142.570_12.62_Bparallel001andstrainperp100_16e_table.tex}
\input{si_tables_tex/142.570_12.62_Bparallel001andstrainperp100_16e_table.tex}
\subsubsection{Topological bands in subgroup $P_{S}\bar{1}~(2.7)$}
\textbf{Perturbation:}
\begin{itemize}
\item strain in generic direction.
\end{itemize}
\begin{figure}[H]
\centering
\includegraphics[scale=0.6]{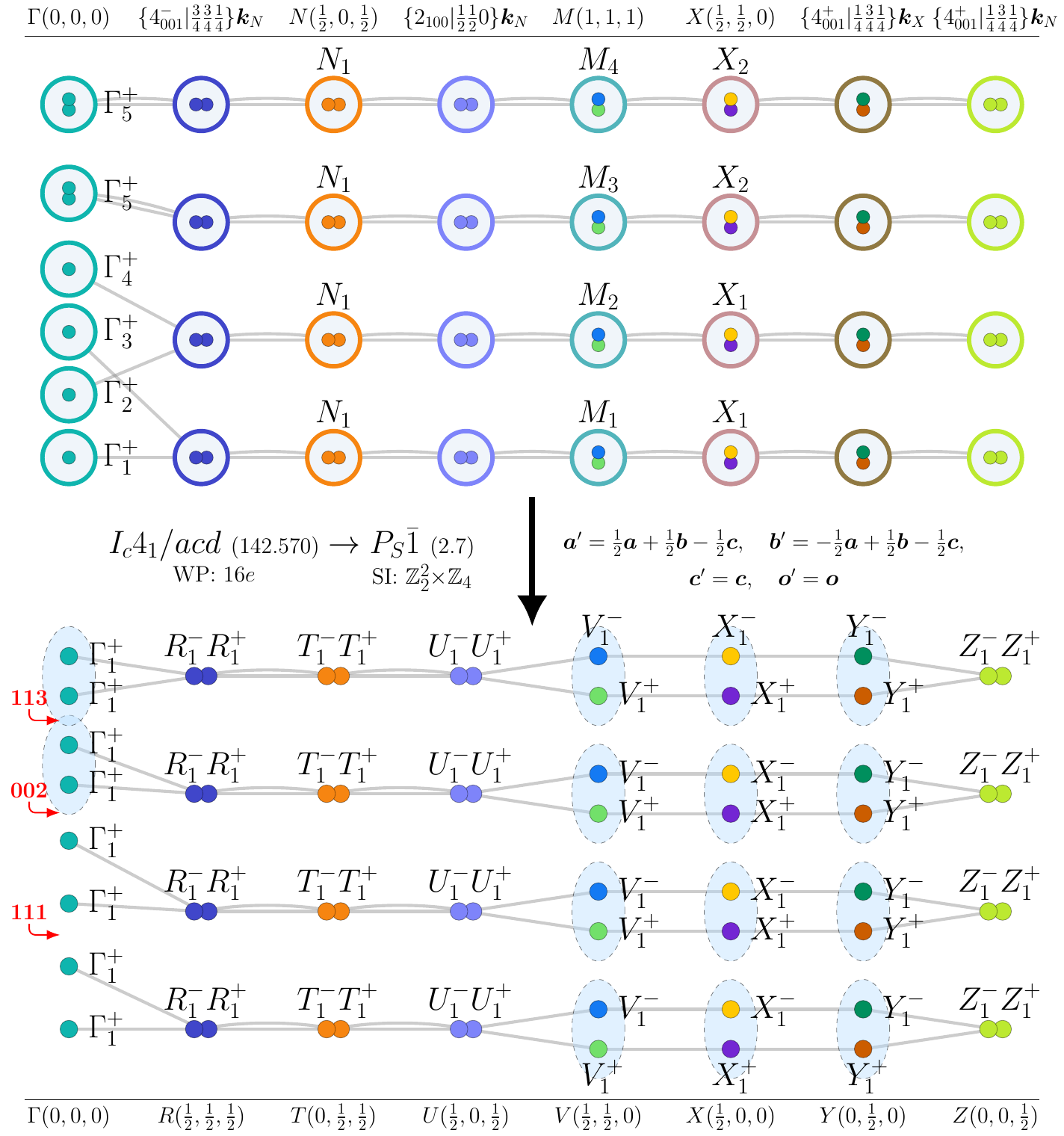}
\caption{Topological magnon bands in subgroup $P_{S}\bar{1}~(2.7)$ for magnetic moments on Wyckoff position $16e$ of supergroup $I_{c}4_{1}/acd~(142.570)$.\label{fig_142.570_2.7_strainingenericdirection_16e}}
\end{figure}
\input{gap_tables_tex/142.570_2.7_strainingenericdirection_16e_table.tex}
\input{si_tables_tex/142.570_2.7_strainingenericdirection_16e_table.tex}

\section{MSG $P_{a}2_{1}/c~(14.80)$}
\textbf{Nontrivial-SI Subgroups:} $P\bar{1}~(2.4)$, $P2_{1}'/c'~(14.79)$, $P_{S}\bar{1}~(2.7)$, $P2_{1}/c~(14.75)$.\\

\textbf{Trivial-SI Subgroups:} $Pc'~(7.26)$, $P2_{1}'~(4.9)$, $P_{S}1~(1.3)$, $Pc~(7.24)$, $P_{a}c~(7.27)$, $P2_{1}~(4.7)$, $P_{a}2_{1}~(4.10)$.\\

\subsection{WP: $8e$}
\textbf{BCS Materials:} {CuO~(213 K)}\footnote{BCS web page: \texttt{\href{http://webbdcrista1.ehu.es/magndata/index.php?this\_label=1.62} {http://webbdcrista1.ehu.es/magndata/index.php?this\_label=1.62}}}, {Ba\textsubscript{2}CoO\textsubscript{4}~(26 K)}\footnote{BCS web page: \texttt{\href{http://webbdcrista1.ehu.es/magndata/index.php?this\_label=1.302} {http://webbdcrista1.ehu.es/magndata/index.php?this\_label=1.302}}}, {Ba\textsubscript{2}CoO\textsubscript{4}~(25 K)}\footnote{BCS web page: \texttt{\href{http://webbdcrista1.ehu.es/magndata/index.php?this\_label=1.476} {http://webbdcrista1.ehu.es/magndata/index.php?this\_label=1.476}}}, {Ba\textsubscript{2}CoO\textsubscript{4}~(23 K)}\footnote{BCS web page: \texttt{\href{http://webbdcrista1.ehu.es/magndata/index.php?this\_label=1.477} {http://webbdcrista1.ehu.es/magndata/index.php?this\_label=1.477}}}, {LiFeGe\textsubscript{2}O\textsubscript{6}~(20 K)}\footnote{BCS web page: \texttt{\href{http://webbdcrista1.ehu.es/magndata/index.php?this\_label=1.39} {http://webbdcrista1.ehu.es/magndata/index.php?this\_label=1.39}}}, {Na\textsubscript{0.5}Li\textsubscript{0.5}FeGe\textsubscript{2}O\textsubscript{6}~(18 K)}\footnote{BCS web page: \texttt{\href{http://webbdcrista1.ehu.es/magndata/index.php?this\_label=1.276} {http://webbdcrista1.ehu.es/magndata/index.php?this\_label=1.276}}}, {BiNiO(PO\textsubscript{4})~(17.5 K)}\footnote{BCS web page: \texttt{\href{http://webbdcrista1.ehu.es/magndata/index.php?this\_label=1.127} {http://webbdcrista1.ehu.es/magndata/index.php?this\_label=1.127}}}, {Li\textsubscript{0.31}Na\textsubscript{0.69}FeGe\textsubscript{2}O\textsubscript{6}~(15 K)}\footnote{BCS web page: \texttt{\href{http://webbdcrista1.ehu.es/magndata/index.php?this\_label=1.331} {http://webbdcrista1.ehu.es/magndata/index.php?this\_label=1.331}}}, {BiCoO(PO\textsubscript{4})~(15 K)}\footnote{BCS web page: \texttt{\href{http://webbdcrista1.ehu.es/magndata/index.php?this\_label=1.128} {http://webbdcrista1.ehu.es/magndata/index.php?this\_label=1.128}}}, {Li\textsubscript{2}MnSiO\textsubscript{4}~(12 K)}\footnote{BCS web page: \texttt{\href{http://webbdcrista1.ehu.es/magndata/index.php?this\_label=1.78} {http://webbdcrista1.ehu.es/magndata/index.php?this\_label=1.78}}}, {BiMnTeO\textsubscript{6}~(10 K)}\footnote{BCS web page: \texttt{\href{http://webbdcrista1.ehu.es/magndata/index.php?this\_label=1.301} {http://webbdcrista1.ehu.es/magndata/index.php?this\_label=1.301}}}, {MnV\textsubscript{2}O\textsubscript{6}~(4.7 K)}\footnote{BCS web page: \texttt{\href{http://webbdcrista1.ehu.es/magndata/index.php?this\_label=1.196} {http://webbdcrista1.ehu.es/magndata/index.php?this\_label=1.196}}}, {ErNiGe~(2.9 K)}\footnote{BCS web page: \texttt{\href{http://webbdcrista1.ehu.es/magndata/index.php?this\_label=1.379} {http://webbdcrista1.ehu.es/magndata/index.php?this\_label=1.379}}}, {BaNd\textsubscript{2}O\textsubscript{4}~(1.7 K)}\footnote{BCS web page: \texttt{\href{http://webbdcrista1.ehu.es/magndata/index.php?this\_label=1.96} {http://webbdcrista1.ehu.es/magndata/index.php?this\_label=1.96}}}, {BaNd\textsubscript{2}O\textsubscript{4}~(1.7 K)}\footnote{BCS web page: \texttt{\href{http://webbdcrista1.ehu.es/magndata/index.php?this\_label=1.95} {http://webbdcrista1.ehu.es/magndata/index.php?this\_label=1.95}}}, {GdPO\textsubscript{4}~(0.8 K)}\footnote{BCS web page: \texttt{\href{http://webbdcrista1.ehu.es/magndata/index.php?this\_label=1.118} {http://webbdcrista1.ehu.es/magndata/index.php?this\_label=1.118}}}, {CoNb\textsubscript{2}O\textsubscript{6}}\footnote{BCS web page: \texttt{\href{http://webbdcrista1.ehu.es/magndata/index.php?this\_label=1.656} {http://webbdcrista1.ehu.es/magndata/index.php?this\_label=1.656}}}, {Y\textsubscript{2}BaCuO\textsubscript{5}}\footnote{BCS web page: \texttt{\href{http://webbdcrista1.ehu.es/magndata/index.php?this\_label=1.445} {http://webbdcrista1.ehu.es/magndata/index.php?this\_label=1.445}}}.\\
\subsubsection{Topological bands in subgroup $P2_{1}'/c'~(14.79)$}
\textbf{Perturbation:}
\begin{itemize}
\item B $\perp$ [010].
\end{itemize}
\begin{figure}[H]
\centering
\includegraphics[scale=0.6]{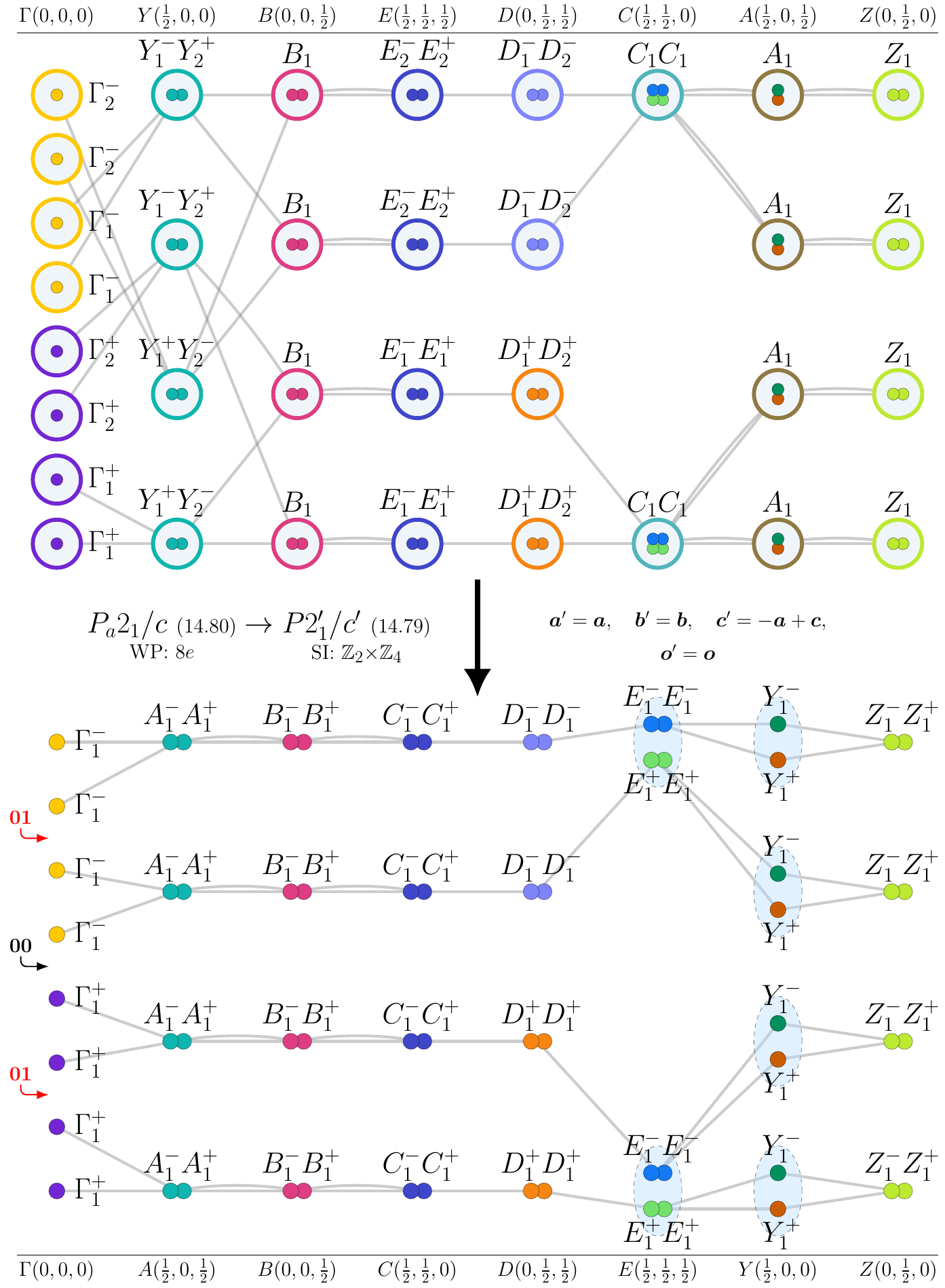}
\caption{Topological magnon bands in subgroup $P2_{1}'/c'~(14.79)$ for magnetic moments on Wyckoff position $8e$ of supergroup $P_{a}2_{1}/c~(14.80)$.\label{fig_14.80_14.79_Bperp010_8e}}
\end{figure}
\input{gap_tables_tex/14.80_14.79_Bperp010_8e_table.tex}
\input{si_tables_tex/14.80_14.79_Bperp010_8e_table.tex}
\subsubsection{Topological bands in subgroup $P_{S}\bar{1}~(2.7)$}
\textbf{Perturbation:}
\begin{itemize}
\item strain in generic direction.
\end{itemize}
\begin{figure}[H]
\centering
\includegraphics[scale=0.6]{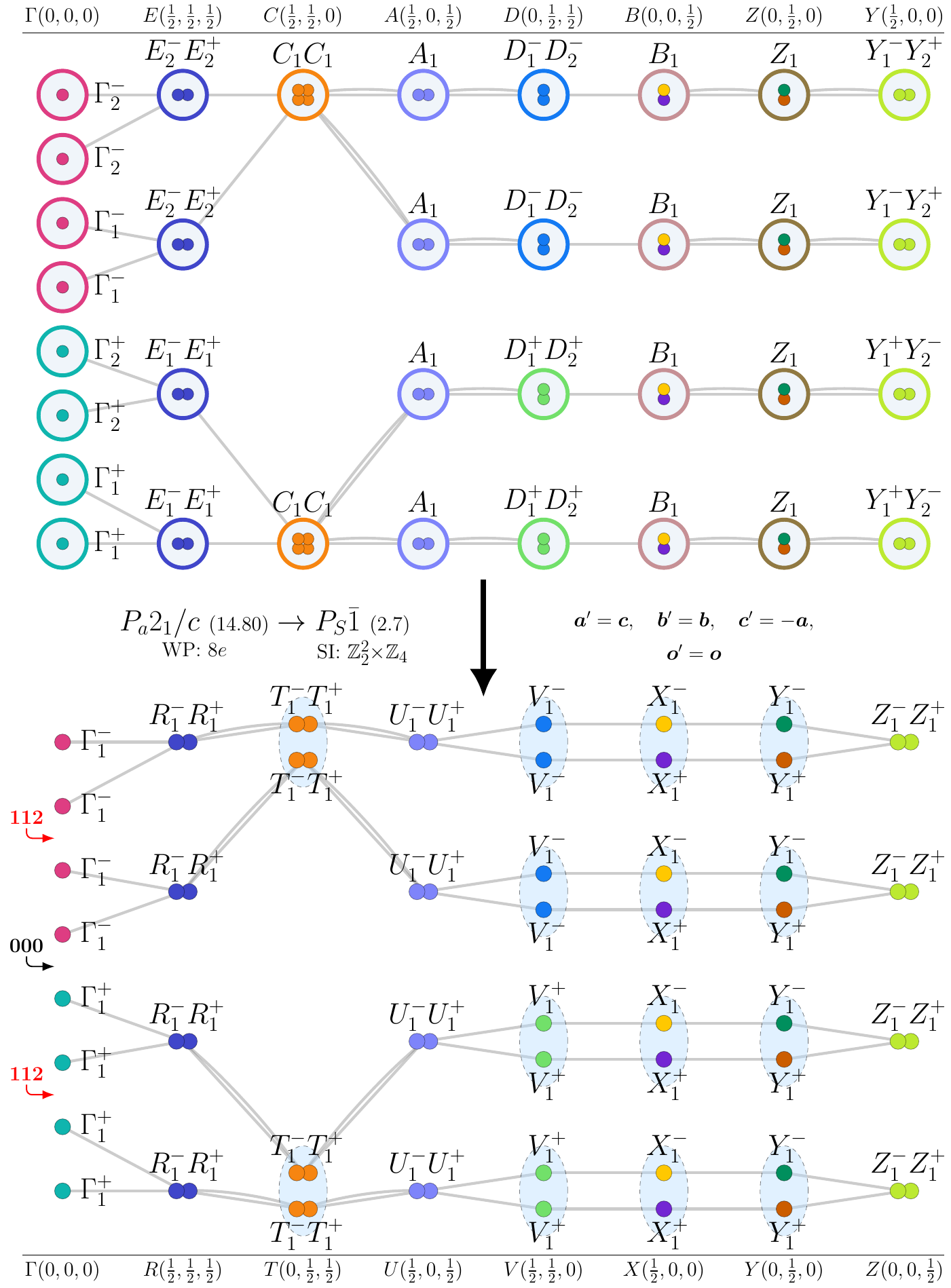}
\caption{Topological magnon bands in subgroup $P_{S}\bar{1}~(2.7)$ for magnetic moments on Wyckoff position $8e$ of supergroup $P_{a}2_{1}/c~(14.80)$.\label{fig_14.80_2.7_strainingenericdirection_8e}}
\end{figure}
\input{gap_tables_tex/14.80_2.7_strainingenericdirection_8e_table.tex}
\input{si_tables_tex/14.80_2.7_strainingenericdirection_8e_table.tex}
\subsection{WP: $4a$}
\textbf{BCS Materials:} {LiCoF\textsubscript{4}~(150 K)}\footnote{BCS web page: \texttt{\href{http://webbdcrista1.ehu.es/magndata/index.php?this\_label=1.526} {http://webbdcrista1.ehu.es/magndata/index.php?this\_label=1.526}}}, {NaMnF\textsubscript{4}~(13 K)}\footnote{BCS web page: \texttt{\href{http://webbdcrista1.ehu.es/magndata/index.php?this\_label=1.345} {http://webbdcrista1.ehu.es/magndata/index.php?this\_label=1.345}}}, {CuSb\textsubscript{2}O\textsubscript{6}~(8.7 K)}\footnote{BCS web page: \texttt{\href{http://webbdcrista1.ehu.es/magndata/index.php?this\_label=1.133} {http://webbdcrista1.ehu.es/magndata/index.php?this\_label=1.133}}}, {MnPb\textsubscript{4}Sb\textsubscript{6}S\textsubscript{14}~(6 K)}\footnote{BCS web page: \texttt{\href{http://webbdcrista1.ehu.es/magndata/index.php?this\_label=1.63} {http://webbdcrista1.ehu.es/magndata/index.php?this\_label=1.63}}}, {FePb\textsubscript{4}Sb\textsubscript{6}S\textsubscript{14}~(5 K)}\footnote{BCS web page: \texttt{\href{http://webbdcrista1.ehu.es/magndata/index.php?this\_label=1.660} {http://webbdcrista1.ehu.es/magndata/index.php?this\_label=1.660}}}, {Li\textsubscript{2}Fe(SO\textsubscript{4})\textsubscript{2}~(4 K)}\footnote{BCS web page: \texttt{\href{http://webbdcrista1.ehu.es/magndata/index.php?this\_label=1.147} {http://webbdcrista1.ehu.es/magndata/index.php?this\_label=1.147}}}.\\
\subsubsection{Topological bands in subgroup $P\bar{1}~(2.4)$}
\textbf{Perturbations:}
\begin{itemize}
\item B $\parallel$ [010] and strain in generic direction,
\item B $\perp$ [010] and strain in generic direction,
\item B in generic direction.
\end{itemize}
\begin{figure}[H]
\centering
\includegraphics[scale=0.6]{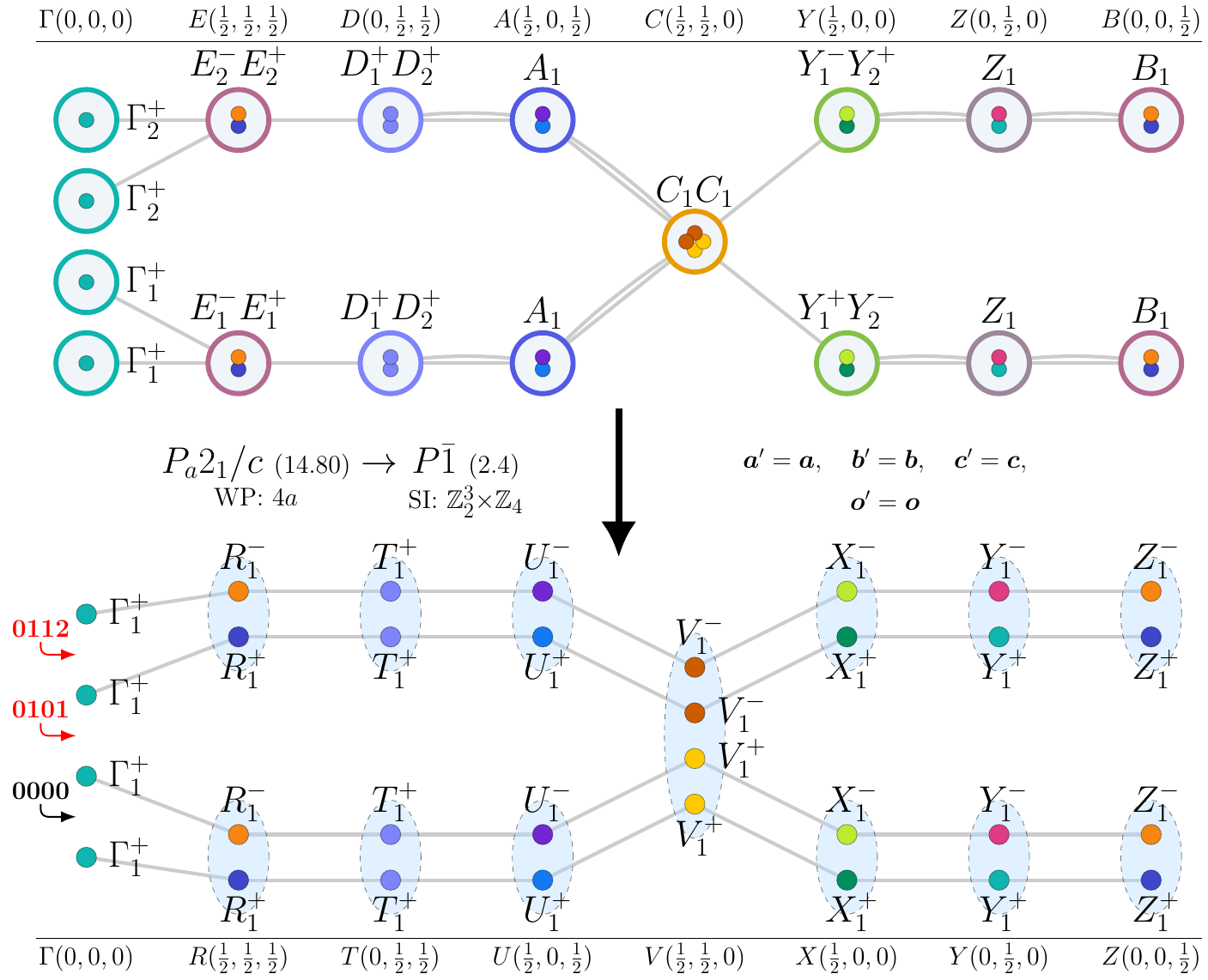}
\caption{Topological magnon bands in subgroup $P\bar{1}~(2.4)$ for magnetic moments on Wyckoff position $4a$ of supergroup $P_{a}2_{1}/c~(14.80)$.\label{fig_14.80_2.4_Bparallel010andstrainingenericdirection_4a}}
\end{figure}
\input{gap_tables_tex/14.80_2.4_Bparallel010andstrainingenericdirection_4a_table.tex}
\input{si_tables_tex/14.80_2.4_Bparallel010andstrainingenericdirection_4a_table.tex}
\subsubsection{Topological bands in subgroup $P2_{1}'/c'~(14.79)$}
\textbf{Perturbation:}
\begin{itemize}
\item B $\perp$ [010].
\end{itemize}
\begin{figure}[H]
\centering
\includegraphics[scale=0.6]{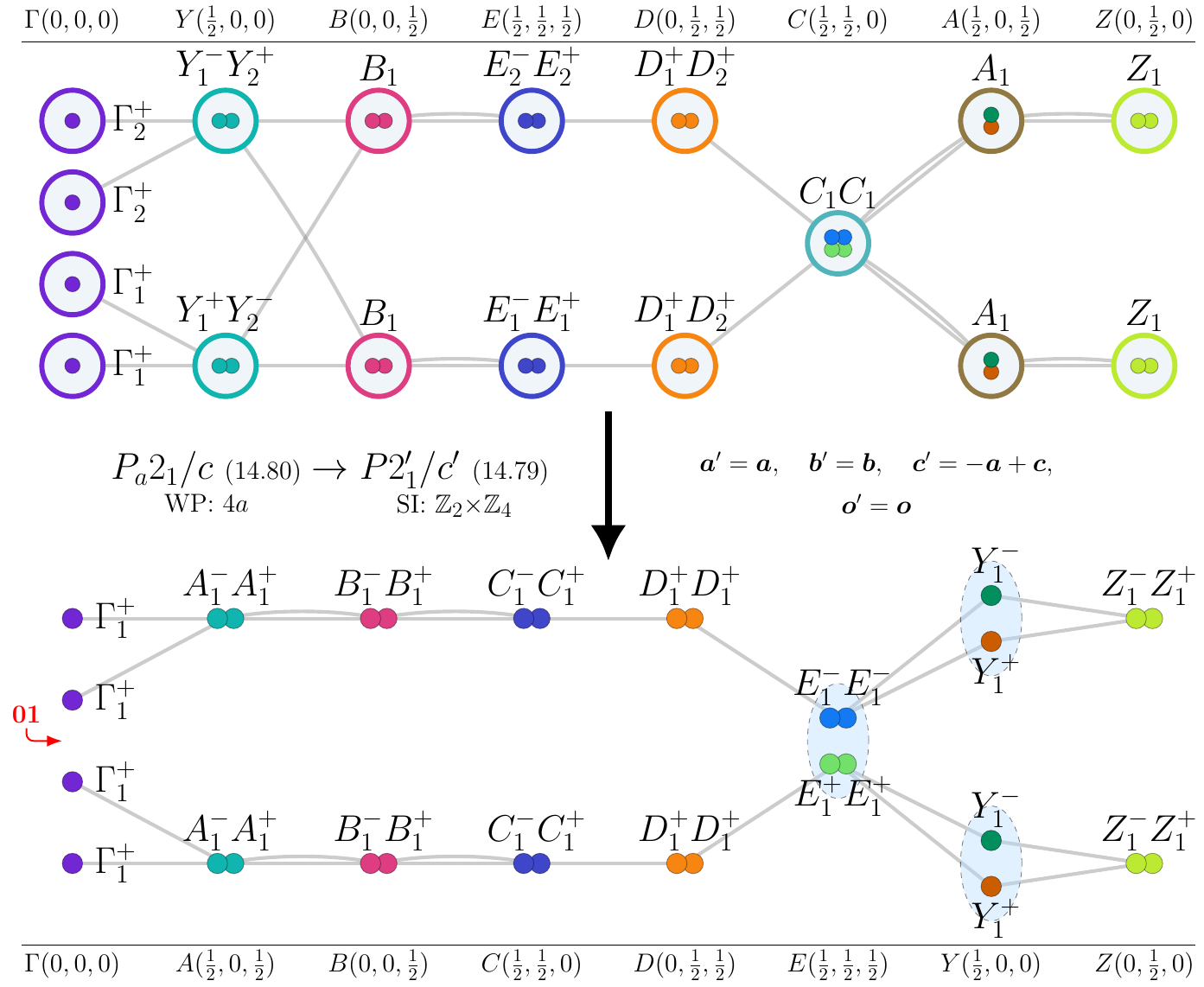}
\caption{Topological magnon bands in subgroup $P2_{1}'/c'~(14.79)$ for magnetic moments on Wyckoff position $4a$ of supergroup $P_{a}2_{1}/c~(14.80)$.\label{fig_14.80_14.79_Bperp010_4a}}
\end{figure}
\input{gap_tables_tex/14.80_14.79_Bperp010_4a_table.tex}
\input{si_tables_tex/14.80_14.79_Bperp010_4a_table.tex}
\subsubsection{Topological bands in subgroup $P_{S}\bar{1}~(2.7)$}
\textbf{Perturbation:}
\begin{itemize}
\item strain in generic direction.
\end{itemize}
\begin{figure}[H]
\centering
\includegraphics[scale=0.6]{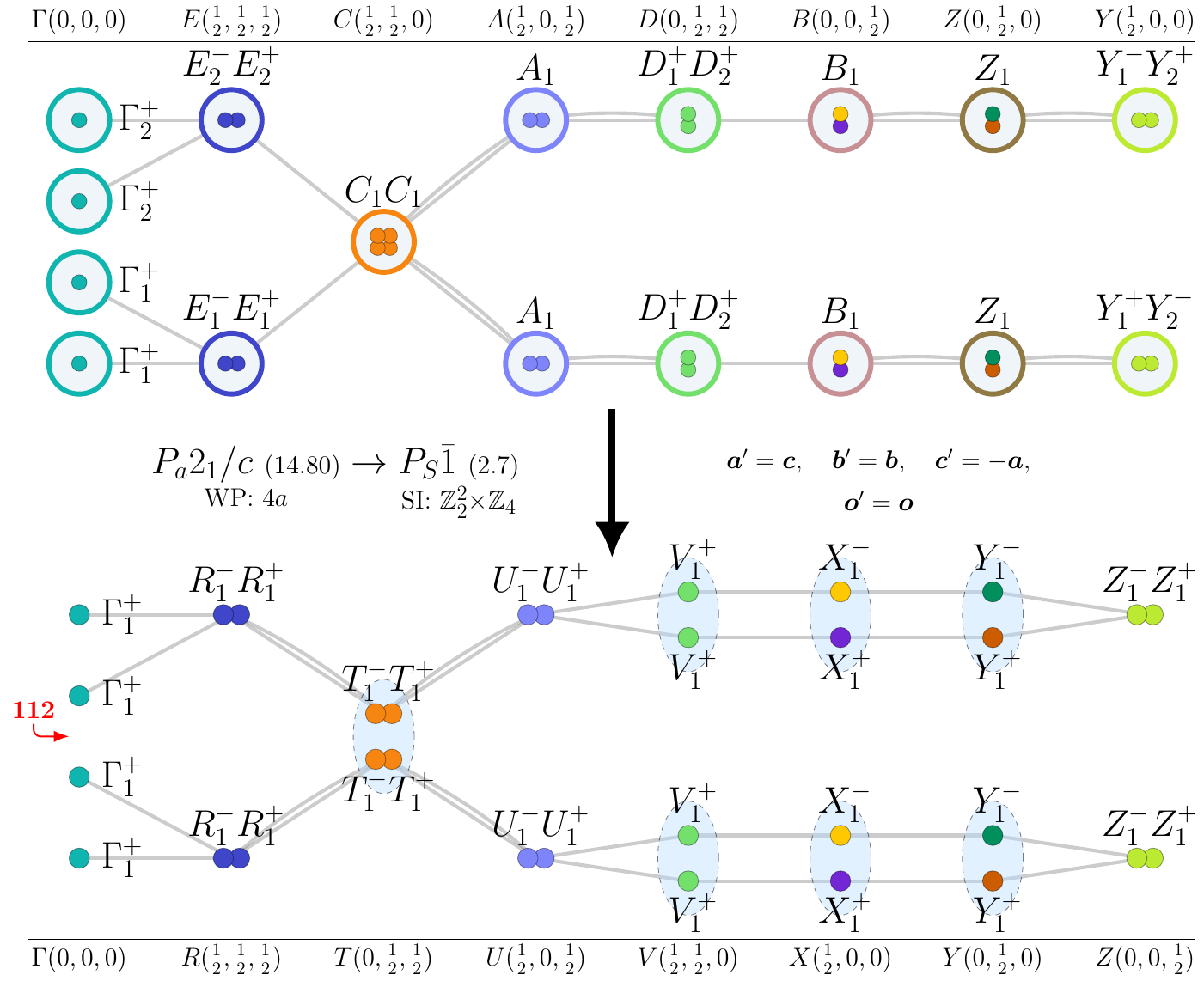}
\caption{Topological magnon bands in subgroup $P_{S}\bar{1}~(2.7)$ for magnetic moments on Wyckoff position $4a$ of supergroup $P_{a}2_{1}/c~(14.80)$.\label{fig_14.80_2.7_strainingenericdirection_4a}}
\end{figure}
\input{gap_tables_tex/14.80_2.7_strainingenericdirection_4a_table.tex}
\input{si_tables_tex/14.80_2.7_strainingenericdirection_4a_table.tex}
\subsection{WP: $4a+8e$}
\textbf{BCS Materials:} {Cr\textsubscript{2}ReO\textsubscript{6}~(67 K)}\footnote{BCS web page: \texttt{\href{http://webbdcrista1.ehu.es/magndata/index.php?this\_label=1.201} {http://webbdcrista1.ehu.es/magndata/index.php?this\_label=1.201}}}, {Co\textsubscript{3}(PO\textsubscript{4})\textsubscript{2}~(30 K)}\footnote{BCS web page: \texttt{\href{http://webbdcrista1.ehu.es/magndata/index.php?this\_label=1.342} {http://webbdcrista1.ehu.es/magndata/index.php?this\_label=1.342}}}, {CuFe\textsubscript{2}(P\textsubscript{2}O\textsubscript{7})\textsubscript{2}~(15.5 K)}\footnote{BCS web page: \texttt{\href{http://webbdcrista1.ehu.es/magndata/index.php?this\_label=1.297} {http://webbdcrista1.ehu.es/magndata/index.php?this\_label=1.297}}}.\\
\subsubsection{Topological bands in subgroup $P2_{1}'/c'~(14.79)$}
\textbf{Perturbation:}
\begin{itemize}
\item B $\perp$ [010].
\end{itemize}
\begin{figure}[H]
\centering
\includegraphics[scale=0.6]{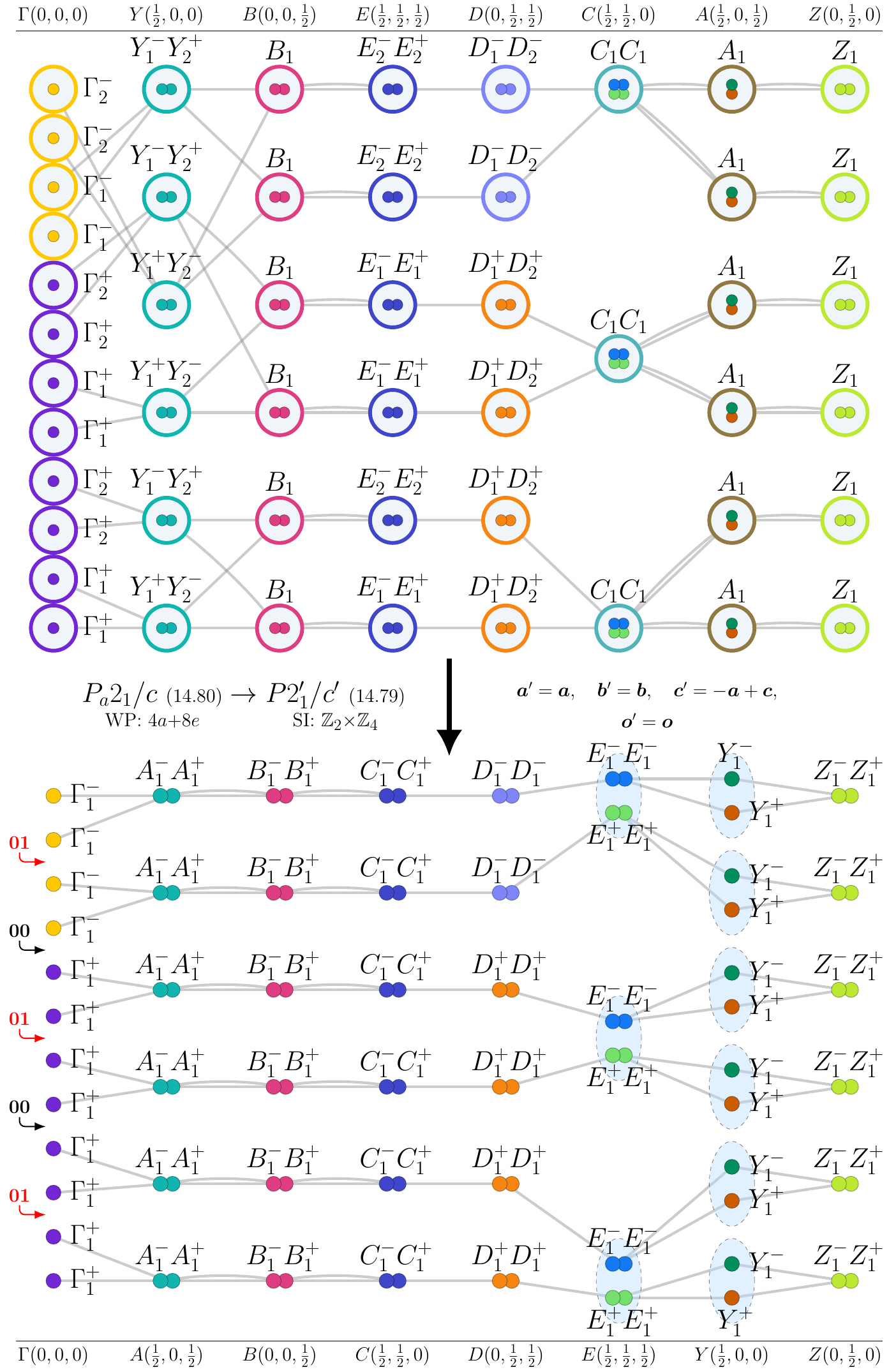}
\caption{Topological magnon bands in subgroup $P2_{1}'/c'~(14.79)$ for magnetic moments on Wyckoff positions $4a+8e$ of supergroup $P_{a}2_{1}/c~(14.80)$.\label{fig_14.80_14.79_Bperp010_4a+8e}}
\end{figure}
\input{gap_tables_tex/14.80_14.79_Bperp010_4a+8e_table.tex}
\input{si_tables_tex/14.80_14.79_Bperp010_4a+8e_table.tex}
\subsubsection{Topological bands in subgroup $P_{S}\bar{1}~(2.7)$}
\textbf{Perturbation:}
\begin{itemize}
\item strain in generic direction.
\end{itemize}
\begin{figure}[H]
\centering
\includegraphics[scale=0.6]{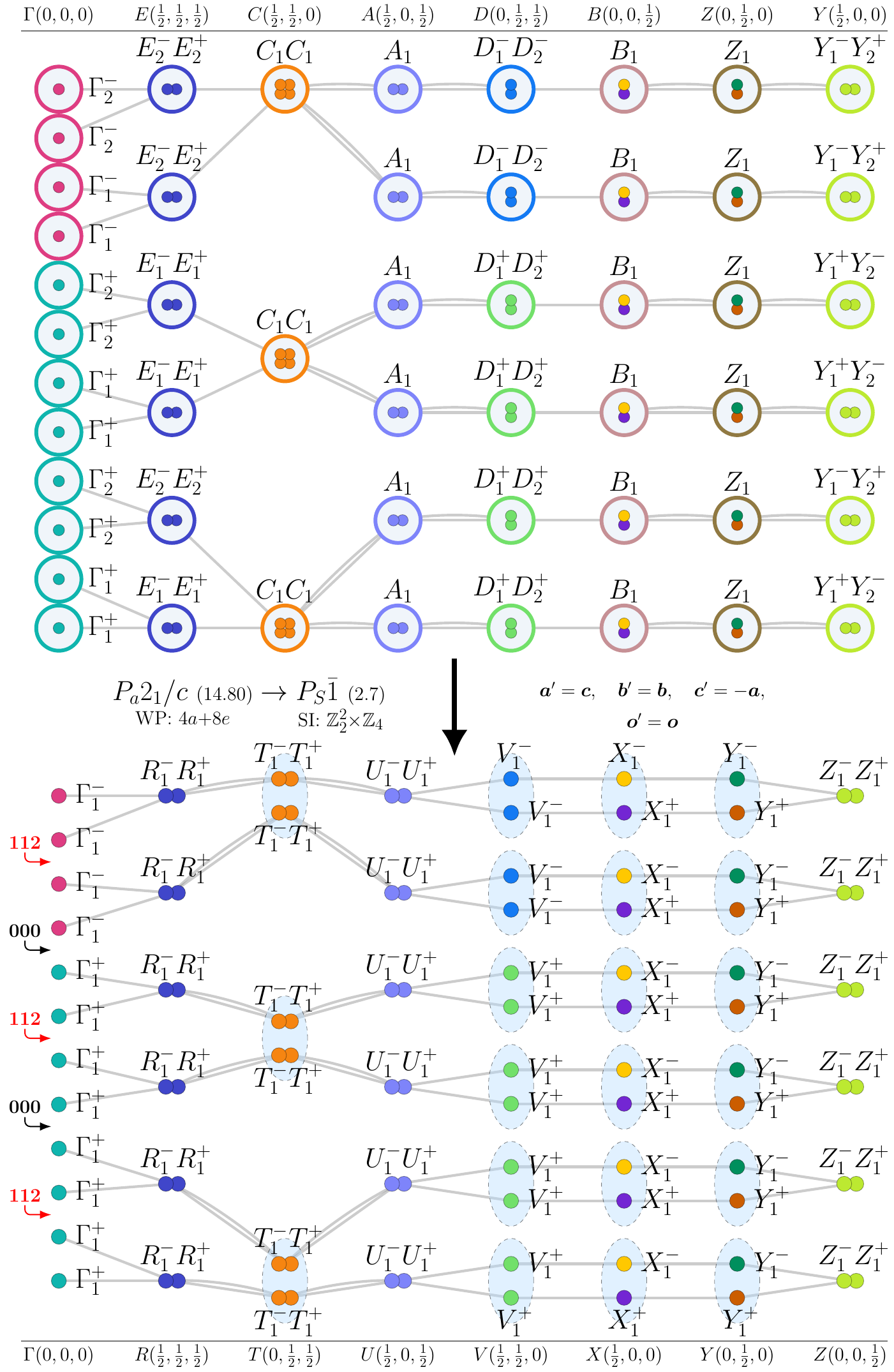}
\caption{Topological magnon bands in subgroup $P_{S}\bar{1}~(2.7)$ for magnetic moments on Wyckoff positions $4a+8e$ of supergroup $P_{a}2_{1}/c~(14.80)$.\label{fig_14.80_2.7_strainingenericdirection_4a+8e}}
\end{figure}
\input{gap_tables_tex/14.80_2.7_strainingenericdirection_4a+8e_table.tex}
\input{si_tables_tex/14.80_2.7_strainingenericdirection_4a+8e_table.tex}
\subsection{WP: $4c+8e$}
\textbf{BCS Materials:} {Mn\textsubscript{3}TeO\textsubscript{6}~(35.8 K)}\footnote{BCS web page: \texttt{\href{http://webbdcrista1.ehu.es/magndata/index.php?this\_label=1.485} {http://webbdcrista1.ehu.es/magndata/index.php?this\_label=1.485}}}.\\
\subsubsection{Topological bands in subgroup $P2_{1}'/c'~(14.79)$}
\textbf{Perturbation:}
\begin{itemize}
\item B $\perp$ [010].
\end{itemize}
\begin{figure}[H]
\centering
\includegraphics[scale=0.6]{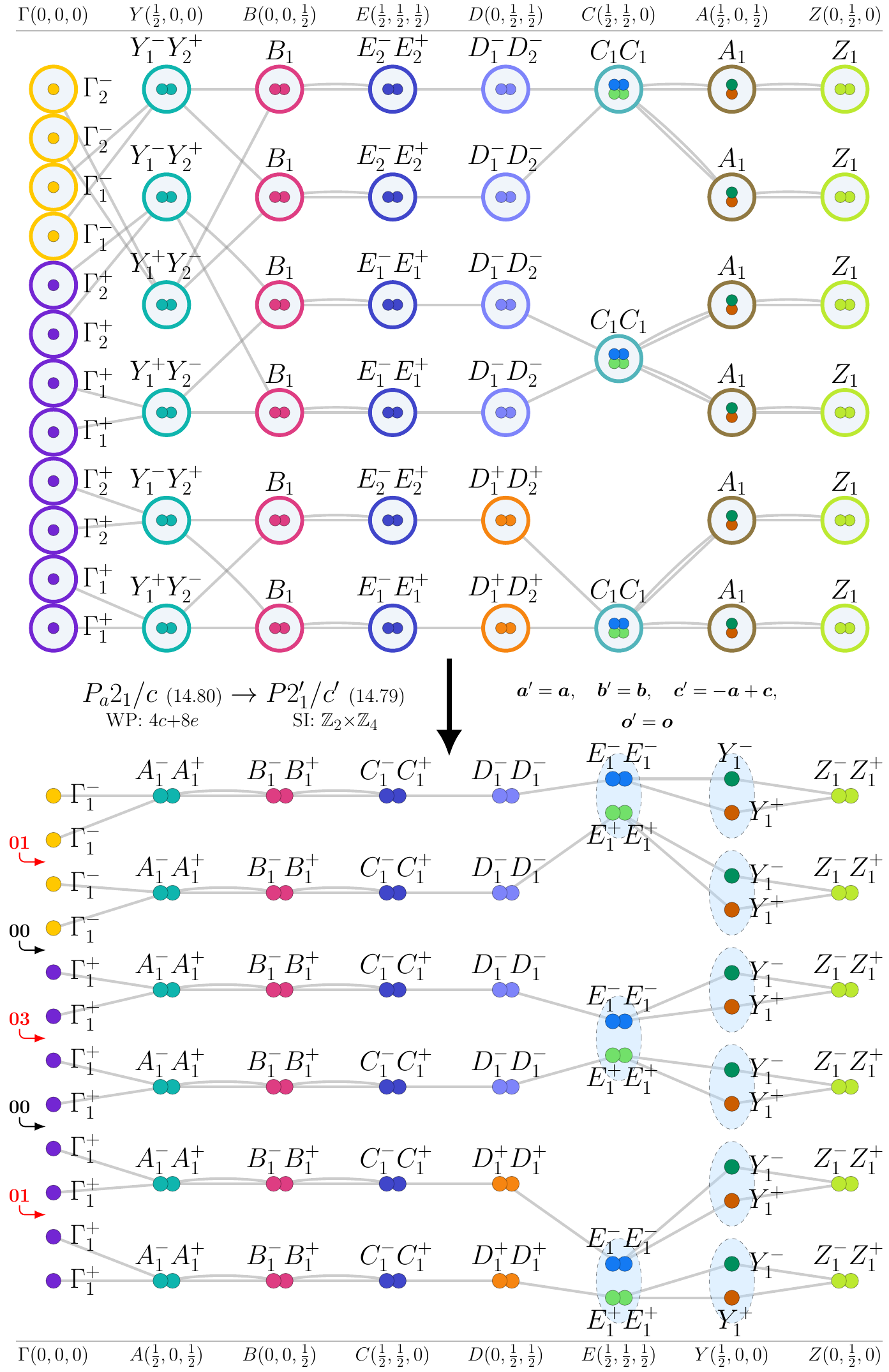}
\caption{Topological magnon bands in subgroup $P2_{1}'/c'~(14.79)$ for magnetic moments on Wyckoff positions $4c+8e$ of supergroup $P_{a}2_{1}/c~(14.80)$.\label{fig_14.80_14.79_Bperp010_4c+8e}}
\end{figure}
\input{gap_tables_tex/14.80_14.79_Bperp010_4c+8e_table.tex}
\input{si_tables_tex/14.80_14.79_Bperp010_4c+8e_table.tex}
\subsubsection{Topological bands in subgroup $P_{S}\bar{1}~(2.7)$}
\textbf{Perturbation:}
\begin{itemize}
\item strain in generic direction.
\end{itemize}
\begin{figure}[H]
\centering
\includegraphics[scale=0.6]{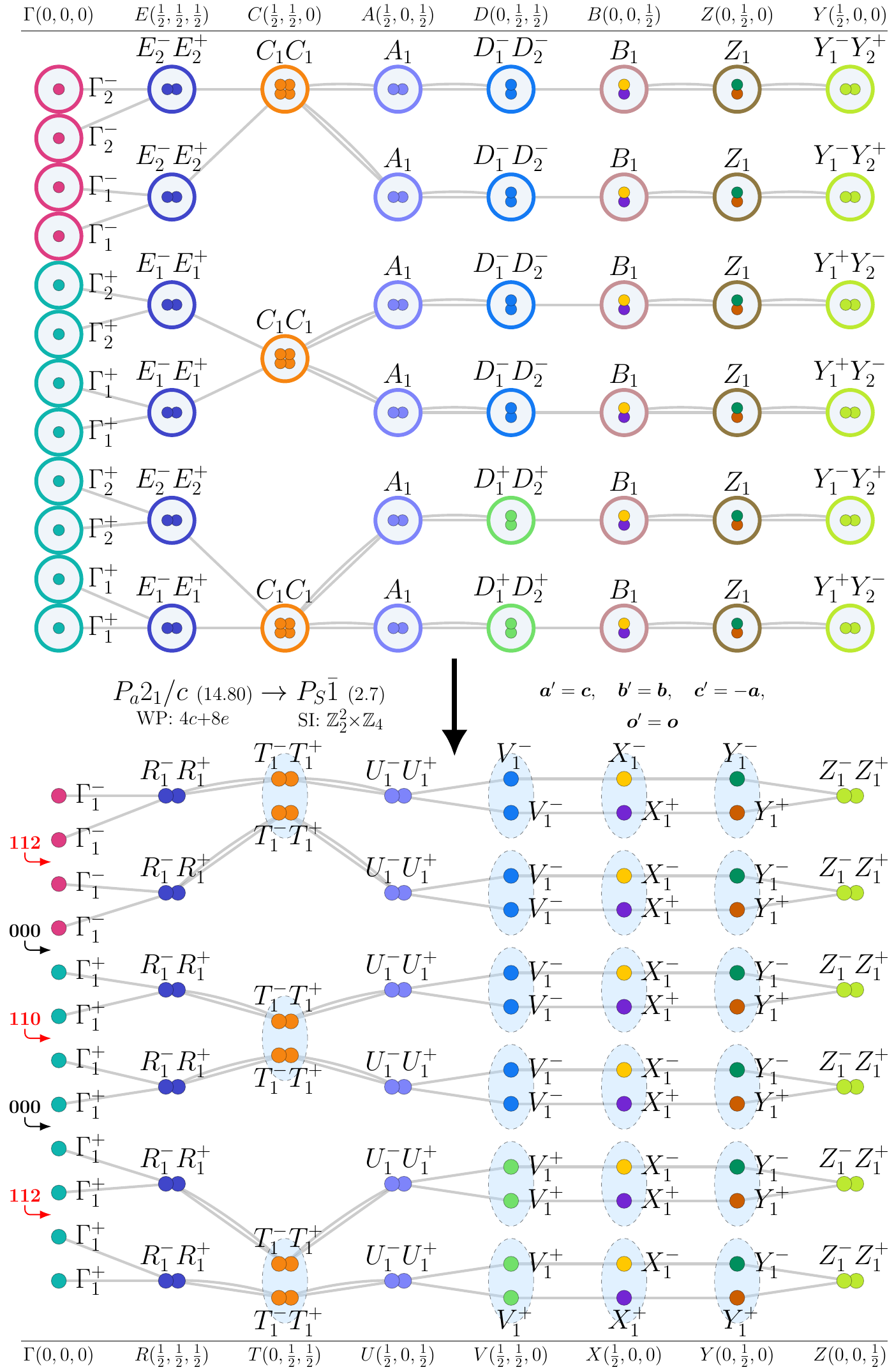}
\caption{Topological magnon bands in subgroup $P_{S}\bar{1}~(2.7)$ for magnetic moments on Wyckoff positions $4c+8e$ of supergroup $P_{a}2_{1}/c~(14.80)$.\label{fig_14.80_2.7_strainingenericdirection_4c+8e}}
\end{figure}
\input{gap_tables_tex/14.80_2.7_strainingenericdirection_4c+8e_table.tex}
\input{si_tables_tex/14.80_2.7_strainingenericdirection_4c+8e_table.tex}
\subsection{WP: $4c$}
\textbf{BCS Materials:} {Sc\textsubscript{2}NiMnO\textsubscript{6}~(35 K)}\footnote{BCS web page: \texttt{\href{http://webbdcrista1.ehu.es/magndata/index.php?this\_label=1.199} {http://webbdcrista1.ehu.es/magndata/index.php?this\_label=1.199}}}, {La\textsubscript{2}CoPtO\textsubscript{6}~(28 K)}\footnote{BCS web page: \texttt{\href{http://webbdcrista1.ehu.es/magndata/index.php?this\_label=1.462} {http://webbdcrista1.ehu.es/magndata/index.php?this\_label=1.462}}}.\\
\subsubsection{Topological bands in subgroup $P\bar{1}~(2.4)$}
\textbf{Perturbations:}
\begin{itemize}
\item B $\parallel$ [010] and strain in generic direction,
\item B $\perp$ [010] and strain in generic direction,
\item B in generic direction.
\end{itemize}
\begin{figure}[H]
\centering
\includegraphics[scale=0.6]{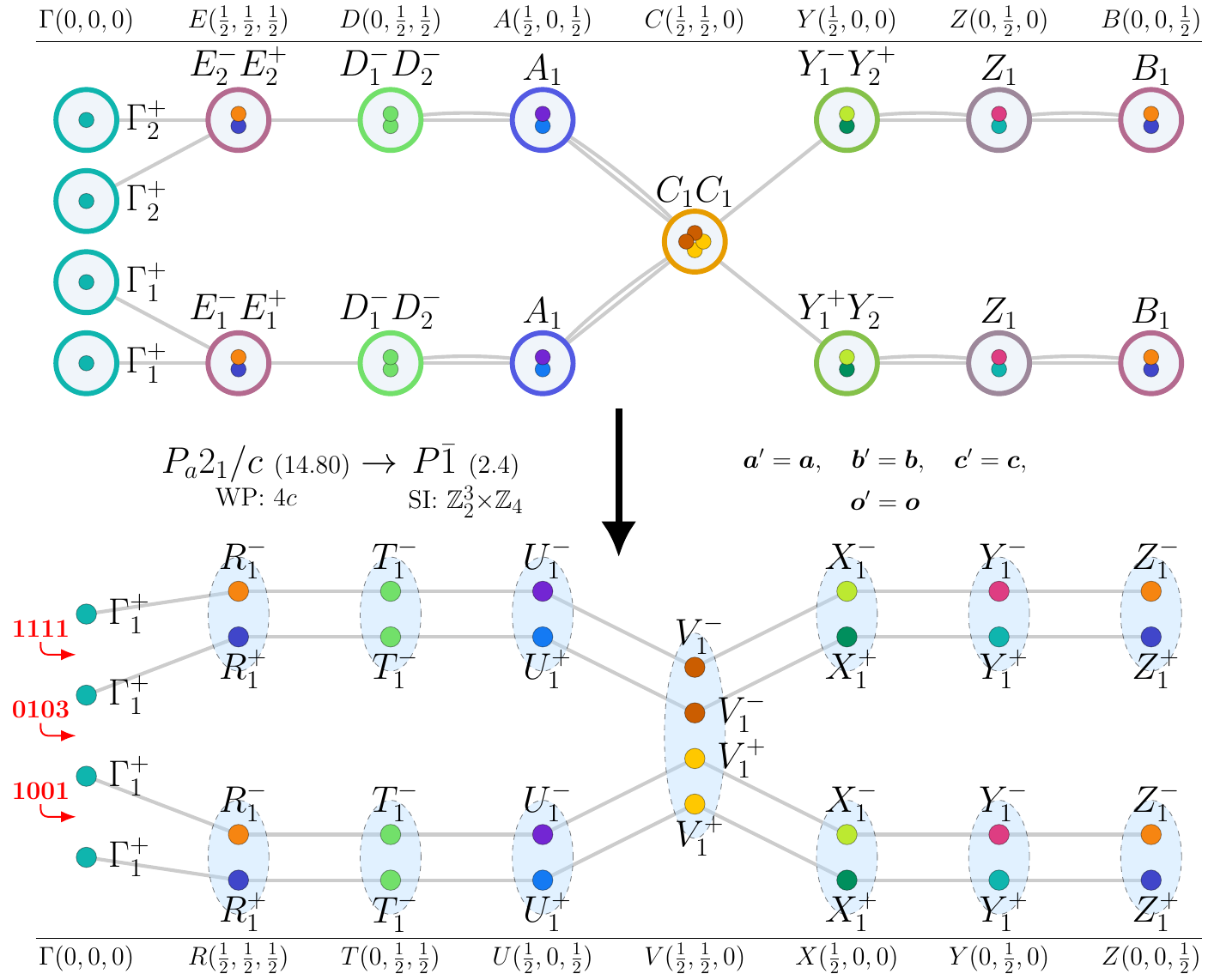}
\caption{Topological magnon bands in subgroup $P\bar{1}~(2.4)$ for magnetic moments on Wyckoff position $4c$ of supergroup $P_{a}2_{1}/c~(14.80)$.\label{fig_14.80_2.4_Bparallel010andstrainingenericdirection_4c}}
\end{figure}
\input{gap_tables_tex/14.80_2.4_Bparallel010andstrainingenericdirection_4c_table.tex}
\input{si_tables_tex/14.80_2.4_Bparallel010andstrainingenericdirection_4c_table.tex}
\subsubsection{Topological bands in subgroup $P2_{1}'/c'~(14.79)$}
\textbf{Perturbation:}
\begin{itemize}
\item B $\perp$ [010].
\end{itemize}
\begin{figure}[H]
\centering
\includegraphics[scale=0.6]{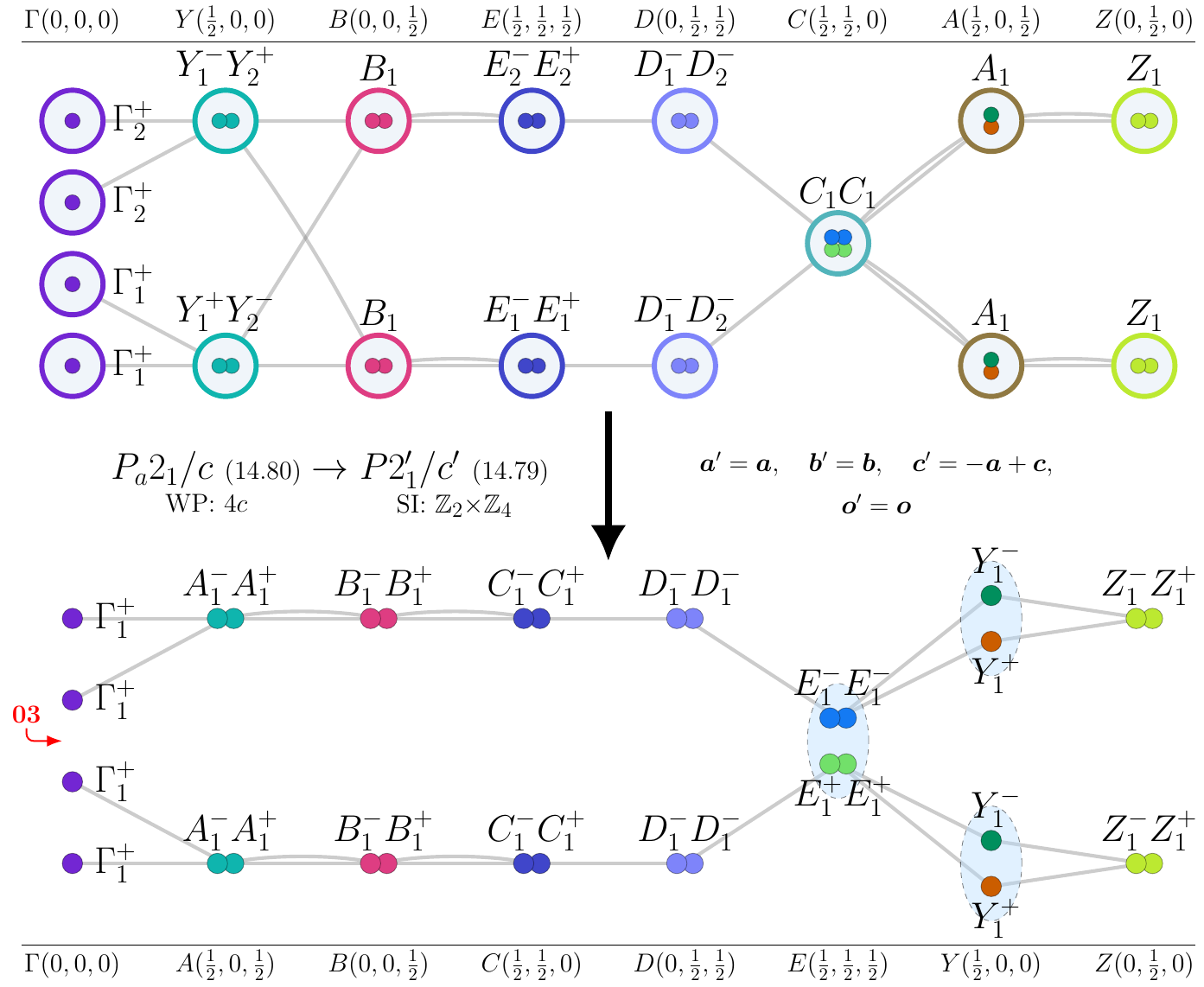}
\caption{Topological magnon bands in subgroup $P2_{1}'/c'~(14.79)$ for magnetic moments on Wyckoff position $4c$ of supergroup $P_{a}2_{1}/c~(14.80)$.\label{fig_14.80_14.79_Bperp010_4c}}
\end{figure}
\input{gap_tables_tex/14.80_14.79_Bperp010_4c_table.tex}
\input{si_tables_tex/14.80_14.79_Bperp010_4c_table.tex}
\subsubsection{Topological bands in subgroup $P_{S}\bar{1}~(2.7)$}
\textbf{Perturbation:}
\begin{itemize}
\item strain in generic direction.
\end{itemize}
\begin{figure}[H]
\centering
\includegraphics[scale=0.6]{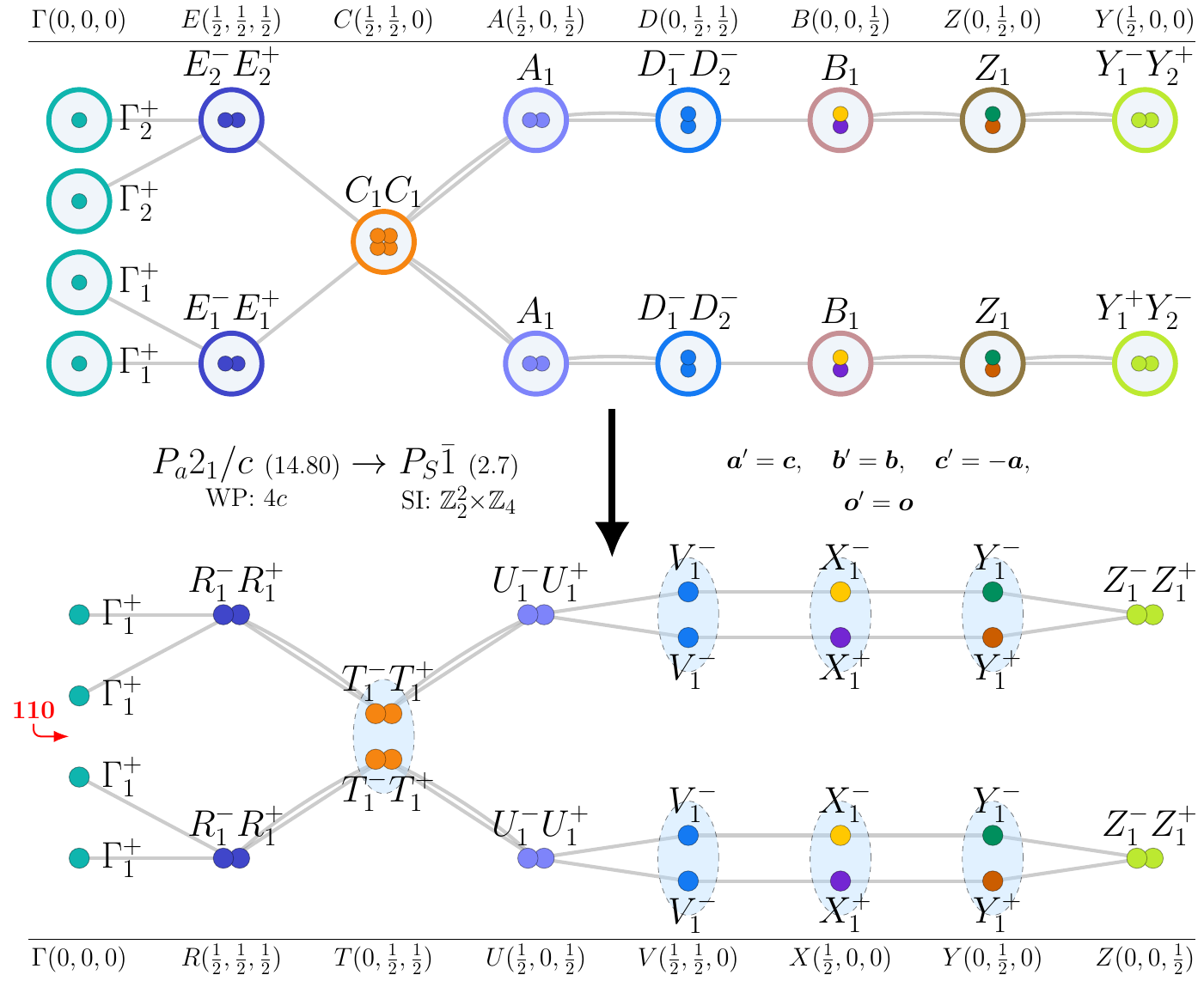}
\caption{Topological magnon bands in subgroup $P_{S}\bar{1}~(2.7)$ for magnetic moments on Wyckoff position $4c$ of supergroup $P_{a}2_{1}/c~(14.80)$.\label{fig_14.80_2.7_strainingenericdirection_4c}}
\end{figure}
\input{gap_tables_tex/14.80_2.7_strainingenericdirection_4c_table.tex}
\input{si_tables_tex/14.80_2.7_strainingenericdirection_4c_table.tex}

\section{MSG $R_{I}\bar{3}~(148.20)$}
\textbf{Nontrivial-SI Subgroups:} $P\bar{1}~(2.4)$, $P_{S}\bar{1}~(2.7)$, $R\bar{3}~(148.17)$.\\

\textbf{Trivial-SI Subgroups:} $P_{S}1~(1.3)$, $R3~(146.10)$, $R_{I}3~(146.12)$.\\

\subsection{WP: $18e$}
\textbf{BCS Materials:} {LaMn\textsubscript{3}V\textsubscript{4}O\textsubscript{12}~(44 K)}\footnote{BCS web page: \texttt{\href{http://webbdcrista1.ehu.es/magndata/index.php?this\_label=1.119} {http://webbdcrista1.ehu.es/magndata/index.php?this\_label=1.119}}}.\\
\subsubsection{Topological bands in subgroup $P\bar{1}~(2.4)$}
\textbf{Perturbations:}
\begin{itemize}
\item B $\parallel$ [001] and strain in generic direction,
\item B in generic direction.
\end{itemize}
\begin{figure}[H]
\centering
\includegraphics[scale=0.6]{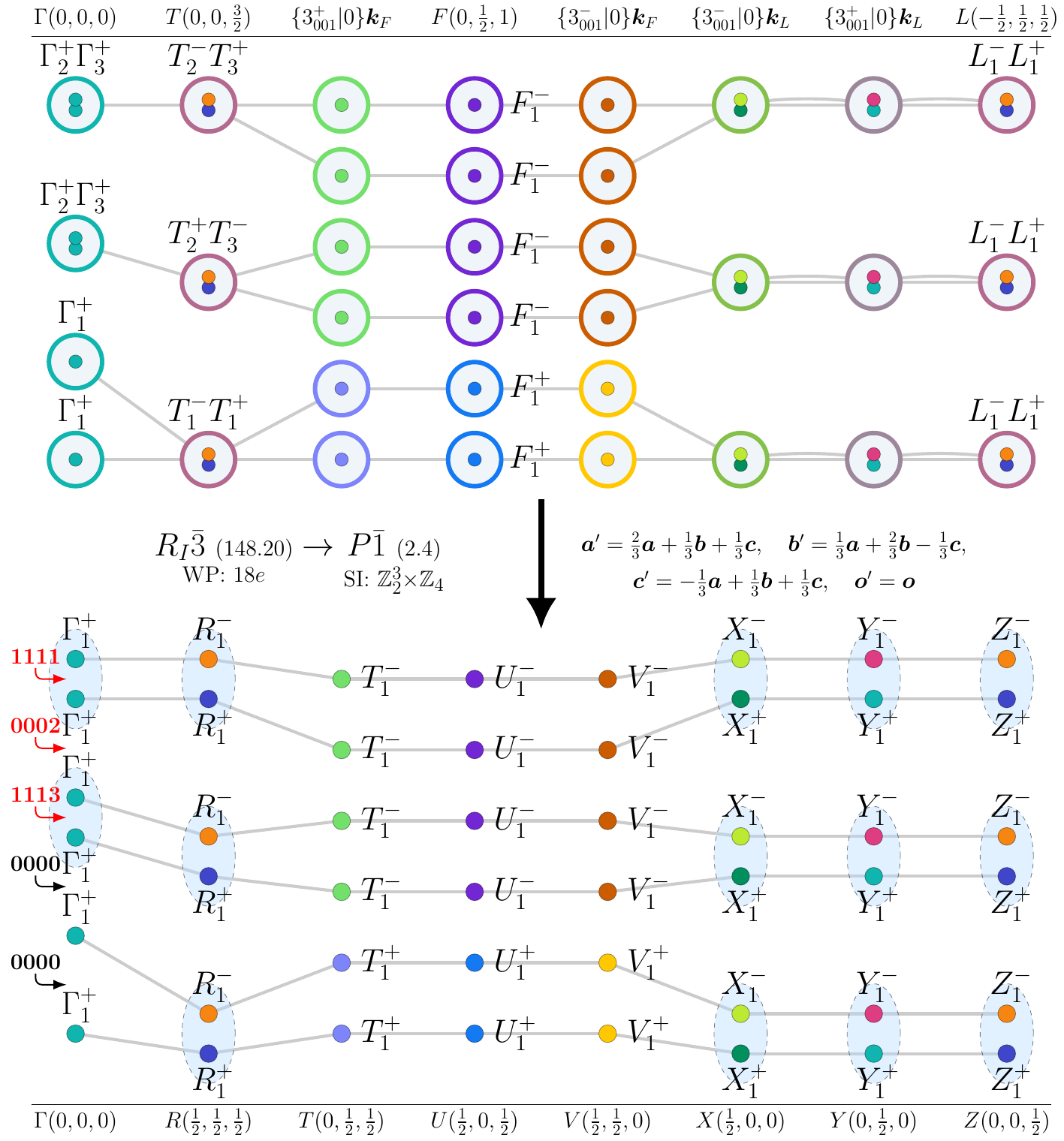}
\caption{Topological magnon bands in subgroup $P\bar{1}~(2.4)$ for magnetic moments on Wyckoff position $18e$ of supergroup $R_{I}\bar{3}~(148.20)$.\label{fig_148.20_2.4_Bparallel001andstrainingenericdirection_18e}}
\end{figure}
\input{gap_tables_tex/148.20_2.4_Bparallel001andstrainingenericdirection_18e_table.tex}
\input{si_tables_tex/148.20_2.4_Bparallel001andstrainingenericdirection_18e_table.tex}
\subsubsection{Topological bands in subgroup $P_{S}\bar{1}~(2.7)$}
\textbf{Perturbation:}
\begin{itemize}
\item strain in generic direction.
\end{itemize}
\begin{figure}[H]
\centering
\includegraphics[scale=0.6]{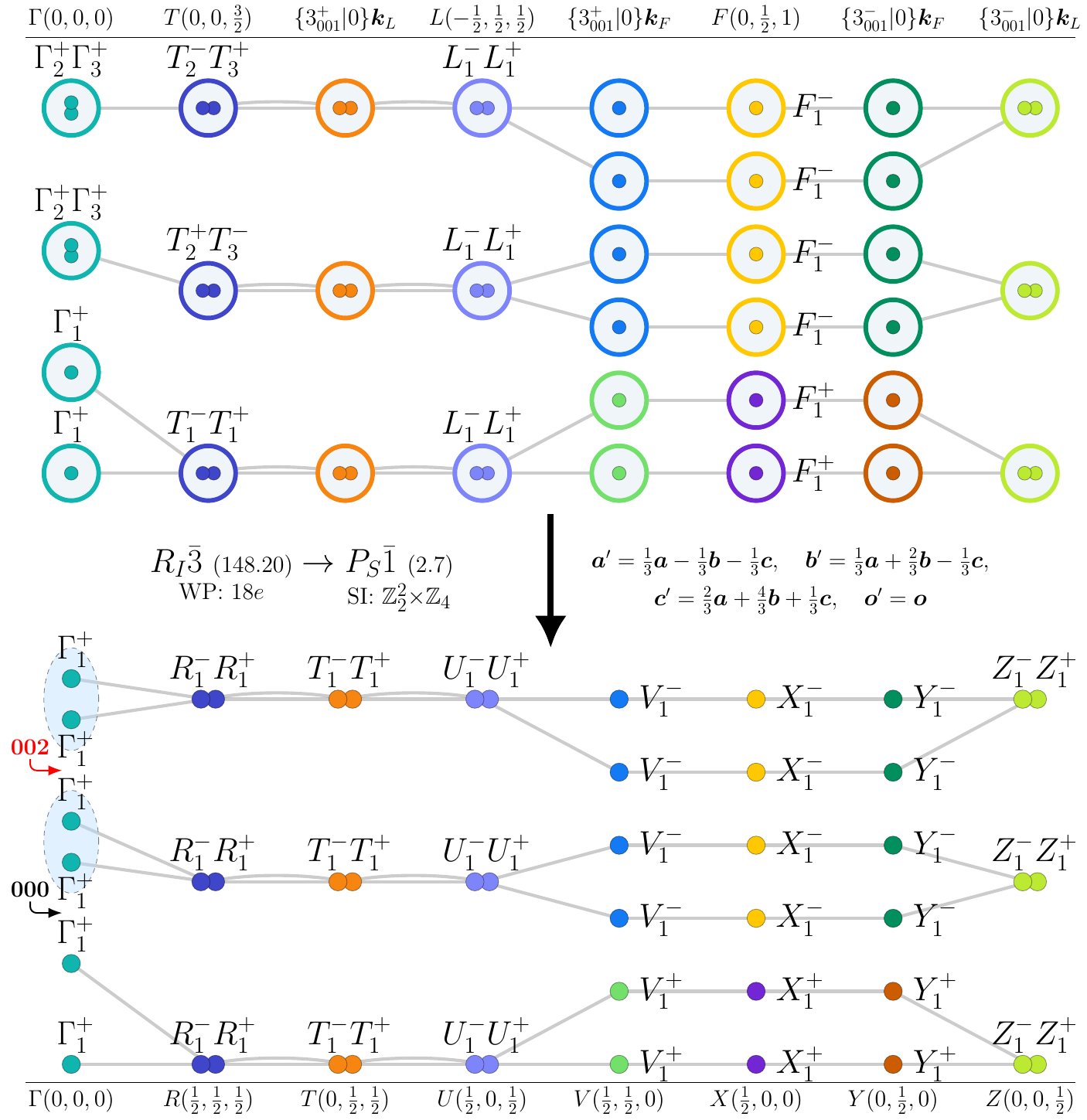}
\caption{Topological magnon bands in subgroup $P_{S}\bar{1}~(2.7)$ for magnetic moments on Wyckoff position $18e$ of supergroup $R_{I}\bar{3}~(148.20)$.\label{fig_148.20_2.7_strainingenericdirection_18e}}
\end{figure}
\input{gap_tables_tex/148.20_2.7_strainingenericdirection_18e_table.tex}
\input{si_tables_tex/148.20_2.7_strainingenericdirection_18e_table.tex}

\section{MSG $P_{c}2_{1}/c~(14.82)$}
\textbf{Nontrivial-SI Subgroups:} $P\bar{1}~(2.4)$, $P2_{1}'/m'~(11.54)$, $P_{S}\bar{1}~(2.7)$, $P2_{1}/c~(14.75)$.\\

\textbf{Trivial-SI Subgroups:} $Pm'~(6.20)$, $P2_{1}'~(4.9)$, $P_{S}1~(1.3)$, $Pc~(7.24)$, $P_{c}c~(7.28)$, $P2_{1}~(4.7)$, $P_{a}2_{1}~(4.10)$.\\

\subsection{WP: $4c+4d$}
\textbf{BCS Materials:} {FeSO\textsubscript{4}~(14 K)}\footnote{BCS web page: \texttt{\href{http://webbdcrista1.ehu.es/magndata/index.php?this\_label=1.573} {http://webbdcrista1.ehu.es/magndata/index.php?this\_label=1.573}}}, {NaFePO\textsubscript{4}~(13 K)}\footnote{BCS web page: \texttt{\href{http://webbdcrista1.ehu.es/magndata/index.php?this\_label=1.117} {http://webbdcrista1.ehu.es/magndata/index.php?this\_label=1.117}}}.\\
\subsubsection{Topological bands in subgroup $P\bar{1}~(2.4)$}
\textbf{Perturbations:}
\begin{itemize}
\item B $\parallel$ [010] and strain in generic direction,
\item B $\perp$ [010] and strain in generic direction,
\item B in generic direction.
\end{itemize}
\begin{figure}[H]
\centering
\includegraphics[scale=0.6]{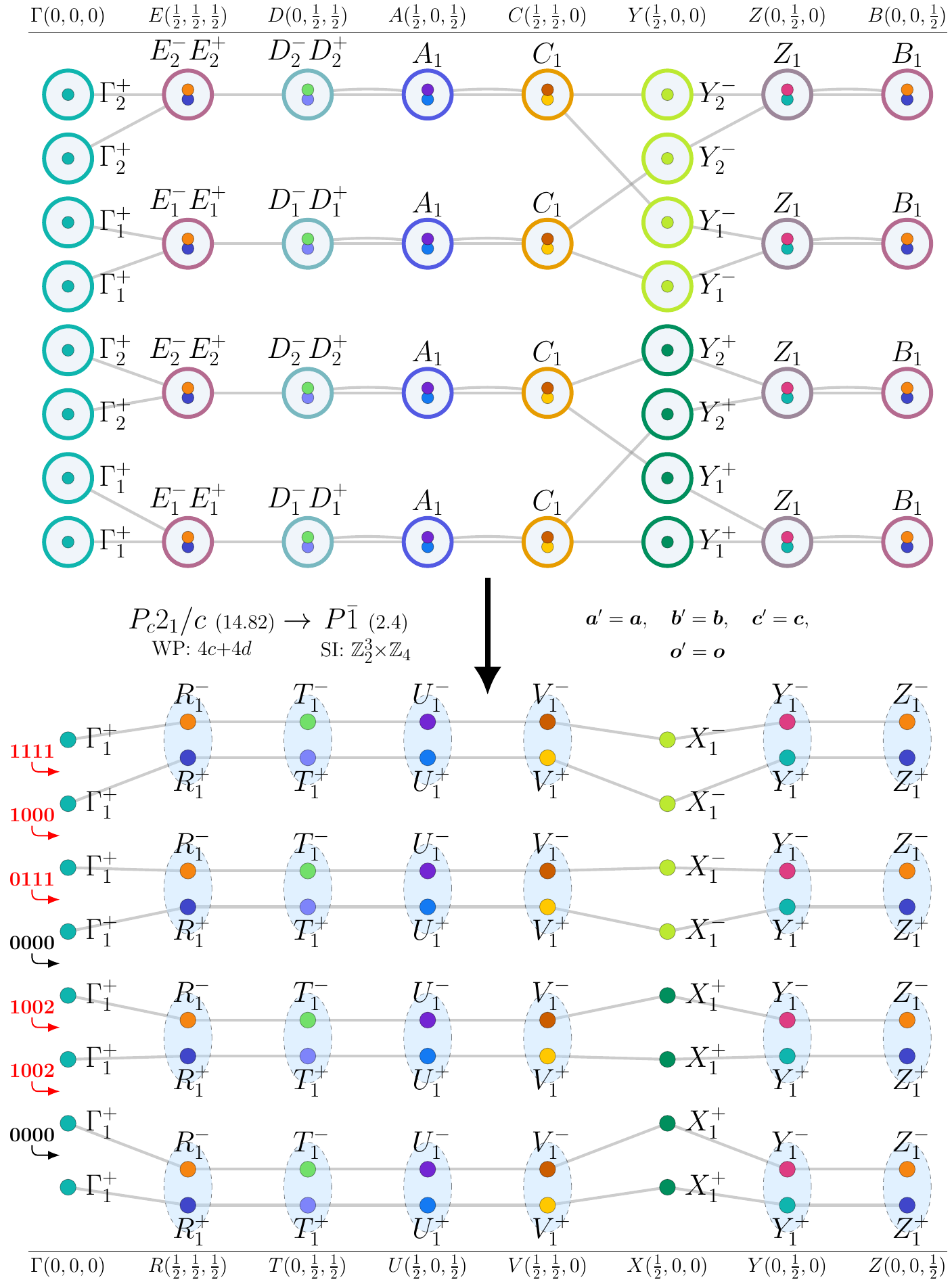}
\caption{Topological magnon bands in subgroup $P\bar{1}~(2.4)$ for magnetic moments on Wyckoff positions $4c+4d$ of supergroup $P_{c}2_{1}/c~(14.82)$.\label{fig_14.82_2.4_Bparallel010andstrainingenericdirection_4c+4d}}
\end{figure}
\input{gap_tables_tex/14.82_2.4_Bparallel010andstrainingenericdirection_4c+4d_table.tex}
\input{si_tables_tex/14.82_2.4_Bparallel010andstrainingenericdirection_4c+4d_table.tex}
\subsubsection{Topological bands in subgroup $P2_{1}'/m'~(11.54)$}
\textbf{Perturbation:}
\begin{itemize}
\item B $\perp$ [010].
\end{itemize}
\begin{figure}[H]
\centering
\includegraphics[scale=0.6]{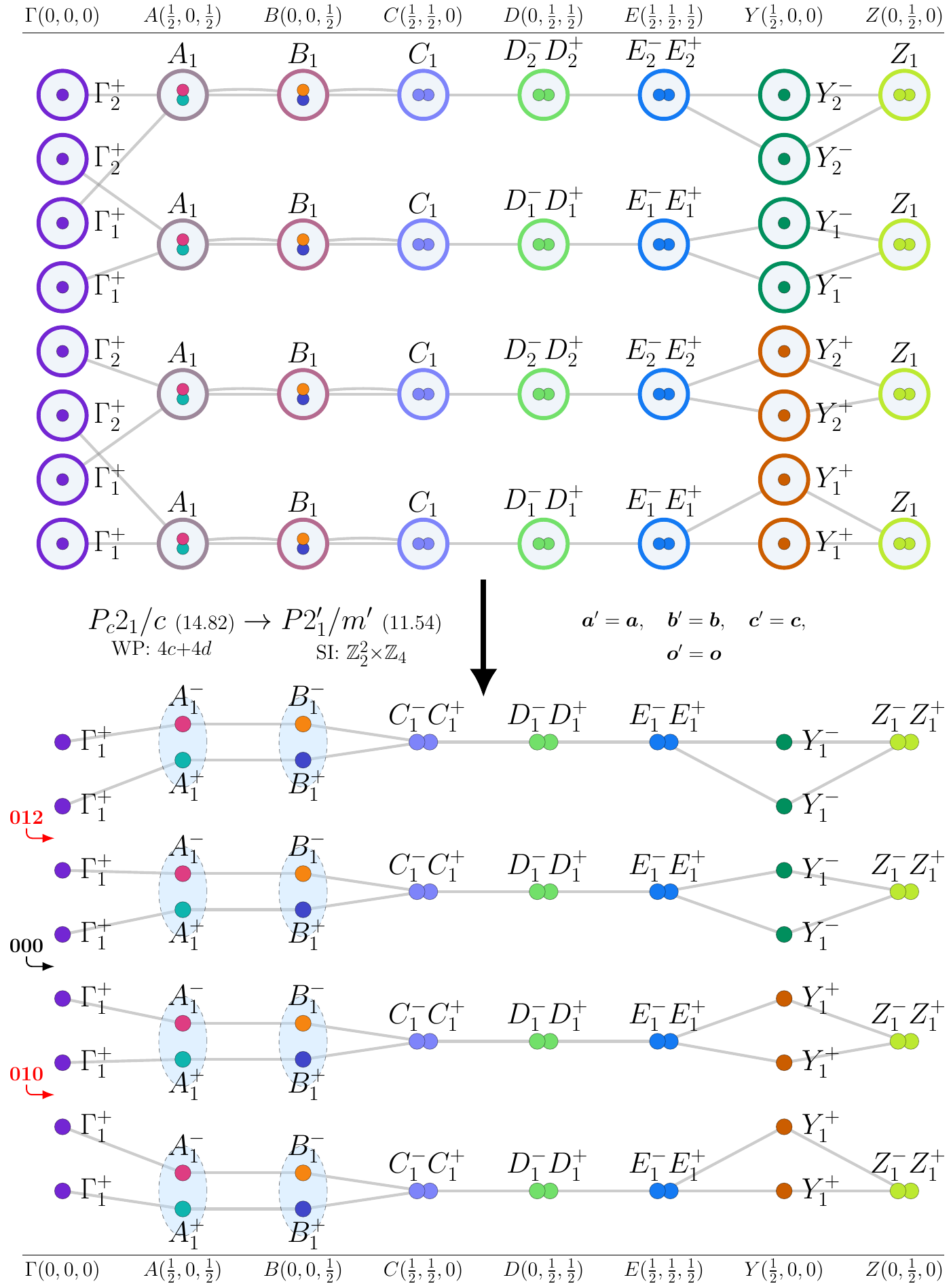}
\caption{Topological magnon bands in subgroup $P2_{1}'/m'~(11.54)$ for magnetic moments on Wyckoff positions $4c+4d$ of supergroup $P_{c}2_{1}/c~(14.82)$.\label{fig_14.82_11.54_Bperp010_4c+4d}}
\end{figure}
\input{gap_tables_tex/14.82_11.54_Bperp010_4c+4d_table.tex}
\input{si_tables_tex/14.82_11.54_Bperp010_4c+4d_table.tex}
\subsubsection{Topological bands in subgroup $P_{S}\bar{1}~(2.7)$}
\textbf{Perturbation:}
\begin{itemize}
\item strain in generic direction.
\end{itemize}
\begin{figure}[H]
\centering
\includegraphics[scale=0.6]{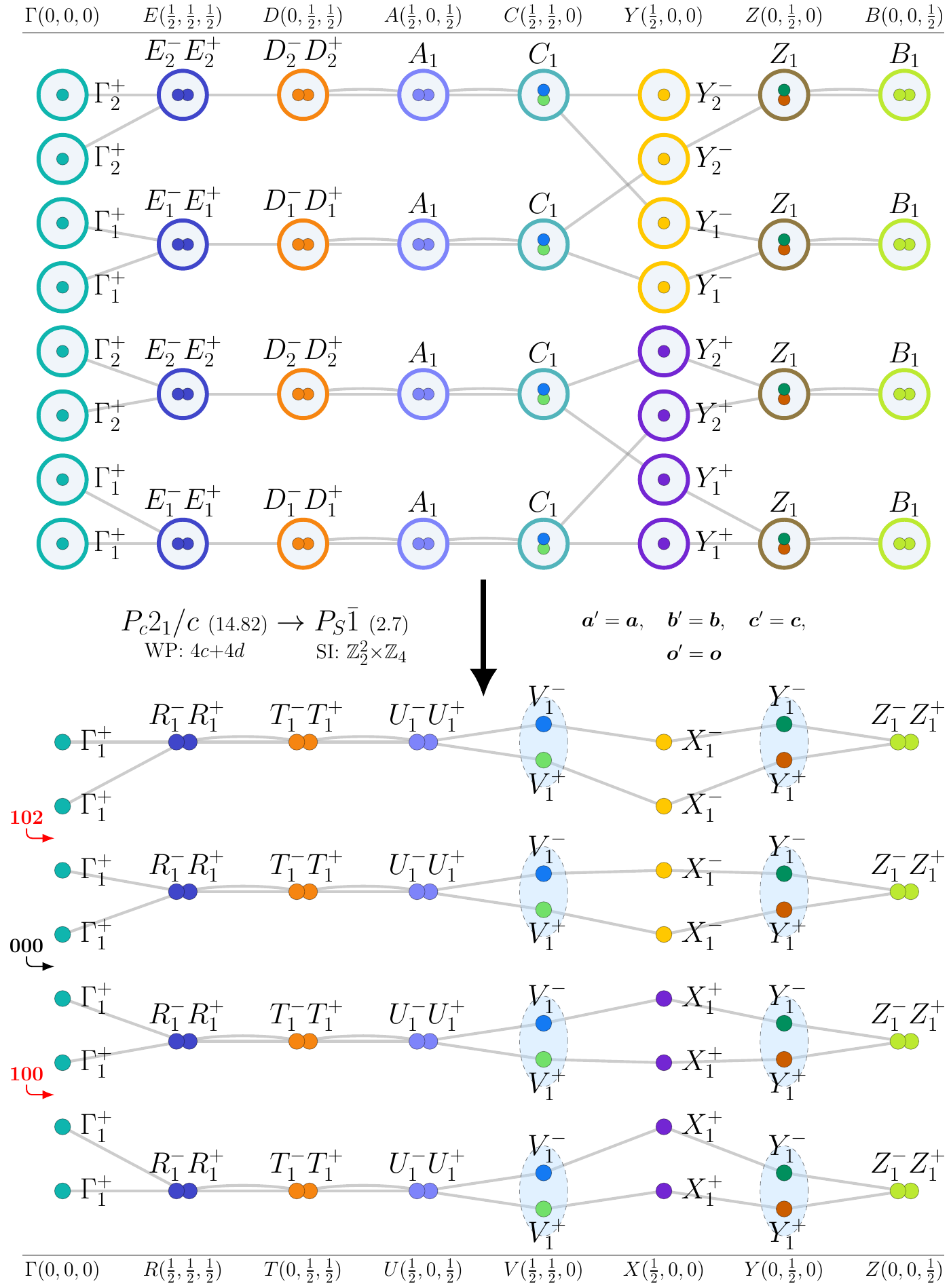}
\caption{Topological magnon bands in subgroup $P_{S}\bar{1}~(2.7)$ for magnetic moments on Wyckoff positions $4c+4d$ of supergroup $P_{c}2_{1}/c~(14.82)$.\label{fig_14.82_2.7_strainingenericdirection_4c+4d}}
\end{figure}
\input{gap_tables_tex/14.82_2.7_strainingenericdirection_4c+4d_table.tex}
\input{si_tables_tex/14.82_2.7_strainingenericdirection_4c+4d_table.tex}

\section{MSG $P_{C}2_{1}/c~(14.84)$}
\textbf{Nontrivial-SI Subgroups:} $P\bar{1}~(2.4)$, $P2'/c'~(13.69)$, $P_{S}\bar{1}~(2.7)$, $P2_{1}/c~(14.75)$.\\

\textbf{Trivial-SI Subgroups:} $Pc'~(7.26)$, $P2'~(3.3)$, $P_{S}1~(1.3)$, $Pc~(7.24)$, $P_{C}c~(7.30)$, $P2_{1}~(4.7)$, $P_{C}2_{1}~(4.12)$.\\

\subsection{WP: $4c$}
\textbf{BCS Materials:} {CuSe\textsubscript{2}O\textsubscript{5}~(17 K)}\footnote{BCS web page: \texttt{\href{http://webbdcrista1.ehu.es/magndata/index.php?this\_label=1.2} {http://webbdcrista1.ehu.es/magndata/index.php?this\_label=1.2}}}, {NH\textsubscript{4}FeCl\textsubscript{2}(HCOO)~(6 K)}\footnote{BCS web page: \texttt{\href{http://webbdcrista1.ehu.es/magndata/index.php?this\_label=1.144} {http://webbdcrista1.ehu.es/magndata/index.php?this\_label=1.144}}}, {La\textsubscript{2}NiO\textsubscript{3}F\textsubscript{1.93}}\footnote{BCS web page: \texttt{\href{http://webbdcrista1.ehu.es/magndata/index.php?this\_label=1.390} {http://webbdcrista1.ehu.es/magndata/index.php?this\_label=1.390}}}.\\
\subsubsection{Topological bands in subgroup $P\bar{1}~(2.4)$}
\textbf{Perturbations:}
\begin{itemize}
\item B $\parallel$ [010] and strain in generic direction,
\item B $\perp$ [010] and strain in generic direction,
\item B in generic direction.
\end{itemize}
\begin{figure}[H]
\centering
\includegraphics[scale=0.6]{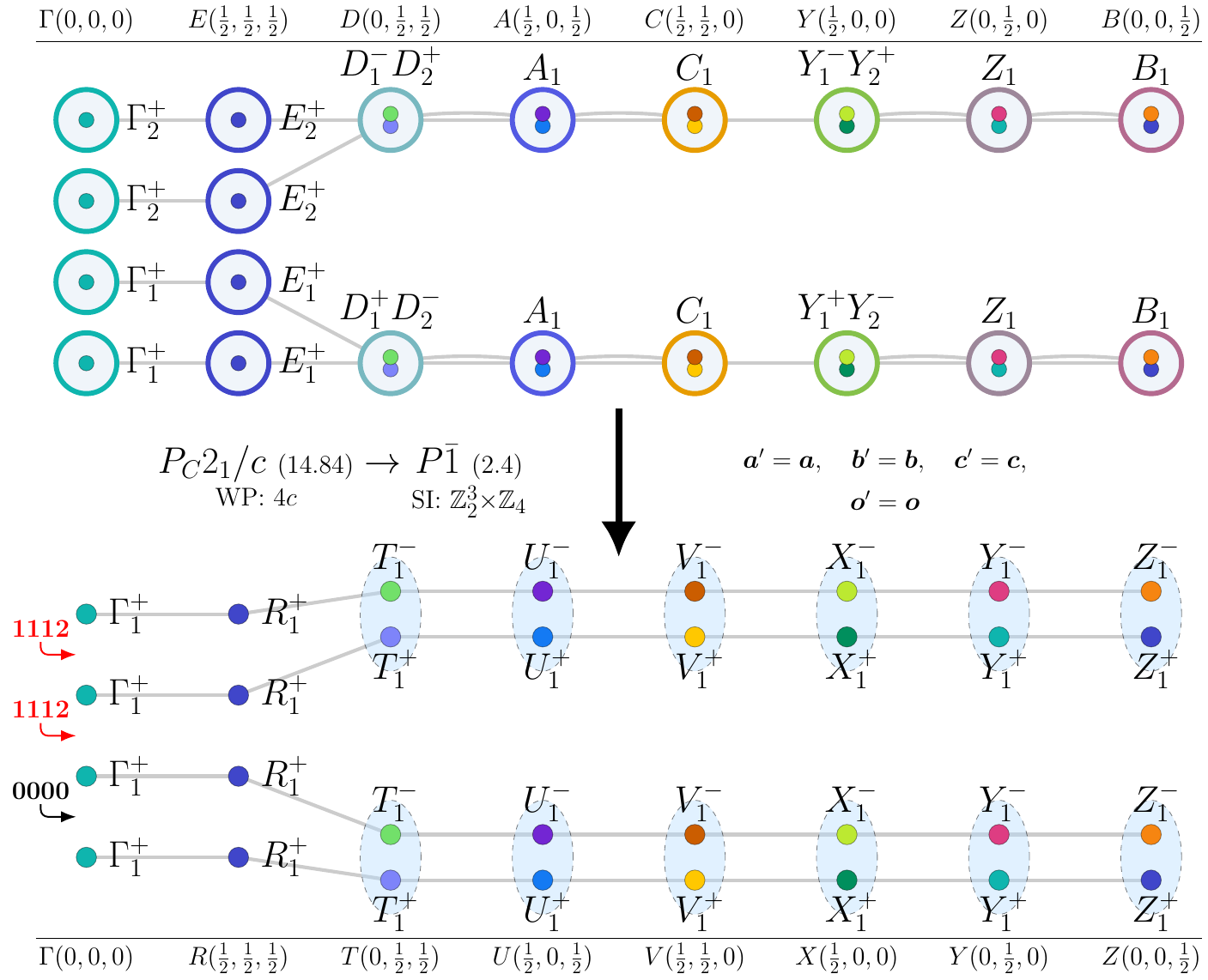}
\caption{Topological magnon bands in subgroup $P\bar{1}~(2.4)$ for magnetic moments on Wyckoff position $4c$ of supergroup $P_{C}2_{1}/c~(14.84)$.\label{fig_14.84_2.4_Bparallel010andstrainingenericdirection_4c}}
\end{figure}
\input{gap_tables_tex/14.84_2.4_Bparallel010andstrainingenericdirection_4c_table.tex}
\input{si_tables_tex/14.84_2.4_Bparallel010andstrainingenericdirection_4c_table.tex}
\subsubsection{Topological bands in subgroup $P2'/c'~(13.69)$}
\textbf{Perturbation:}
\begin{itemize}
\item B $\perp$ [010].
\end{itemize}
\begin{figure}[H]
\centering
\includegraphics[scale=0.6]{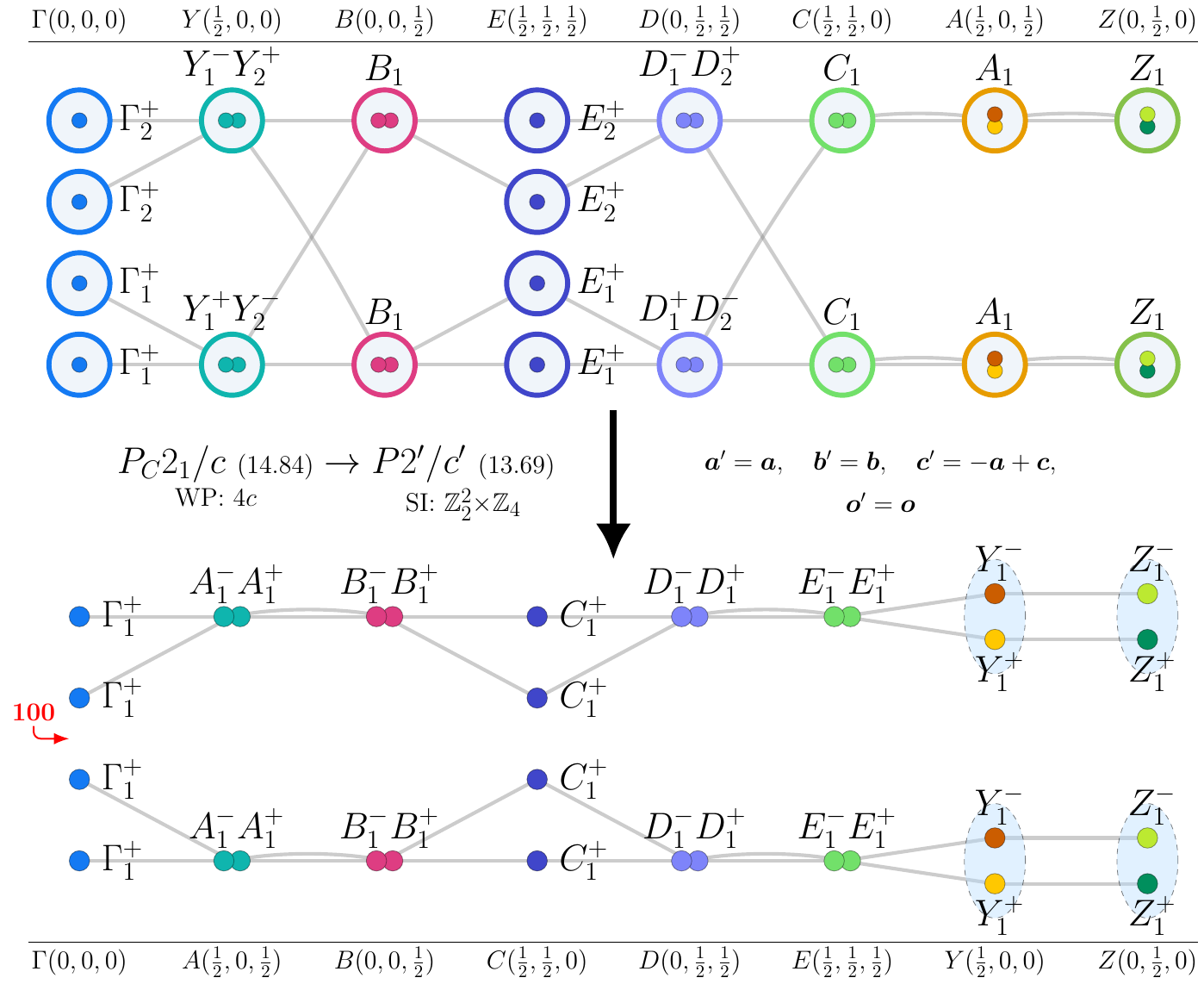}
\caption{Topological magnon bands in subgroup $P2'/c'~(13.69)$ for magnetic moments on Wyckoff position $4c$ of supergroup $P_{C}2_{1}/c~(14.84)$.\label{fig_14.84_13.69_Bperp010_4c}}
\end{figure}
\input{gap_tables_tex/14.84_13.69_Bperp010_4c_table.tex}
\input{si_tables_tex/14.84_13.69_Bperp010_4c_table.tex}
\subsubsection{Topological bands in subgroup $P_{S}\bar{1}~(2.7)$}
\textbf{Perturbation:}
\begin{itemize}
\item strain in generic direction.
\end{itemize}
\begin{figure}[H]
\centering
\includegraphics[scale=0.6]{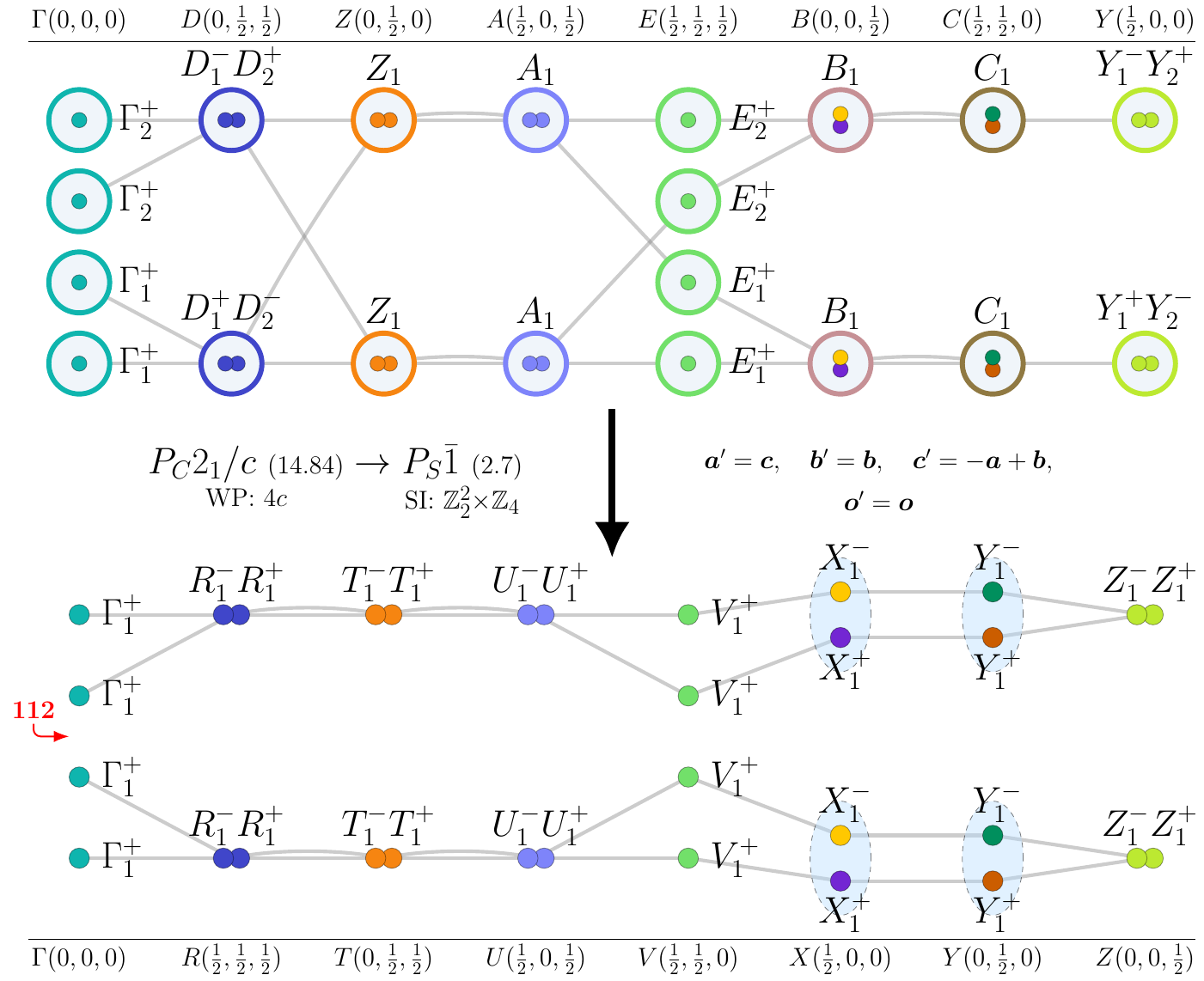}
\caption{Topological magnon bands in subgroup $P_{S}\bar{1}~(2.7)$ for magnetic moments on Wyckoff position $4c$ of supergroup $P_{C}2_{1}/c~(14.84)$.\label{fig_14.84_2.7_strainingenericdirection_4c}}
\end{figure}
\input{gap_tables_tex/14.84_2.7_strainingenericdirection_4c_table.tex}
\input{si_tables_tex/14.84_2.7_strainingenericdirection_4c_table.tex}
\subsection{WP: $4d$}
\textbf{BCS Materials:} {Na\textsubscript{2}BaMn(VO\textsubscript{4})\textsubscript{2}~(1.3 K)}\footnote{BCS web page: \texttt{\href{http://webbdcrista1.ehu.es/magndata/index.php?this\_label=1.306} {http://webbdcrista1.ehu.es/magndata/index.php?this\_label=1.306}}}.\\
\subsubsection{Topological bands in subgroup $P\bar{1}~(2.4)$}
\textbf{Perturbations:}
\begin{itemize}
\item B $\parallel$ [010] and strain in generic direction,
\item B $\perp$ [010] and strain in generic direction,
\item B in generic direction.
\end{itemize}
\begin{figure}[H]
\centering
\includegraphics[scale=0.6]{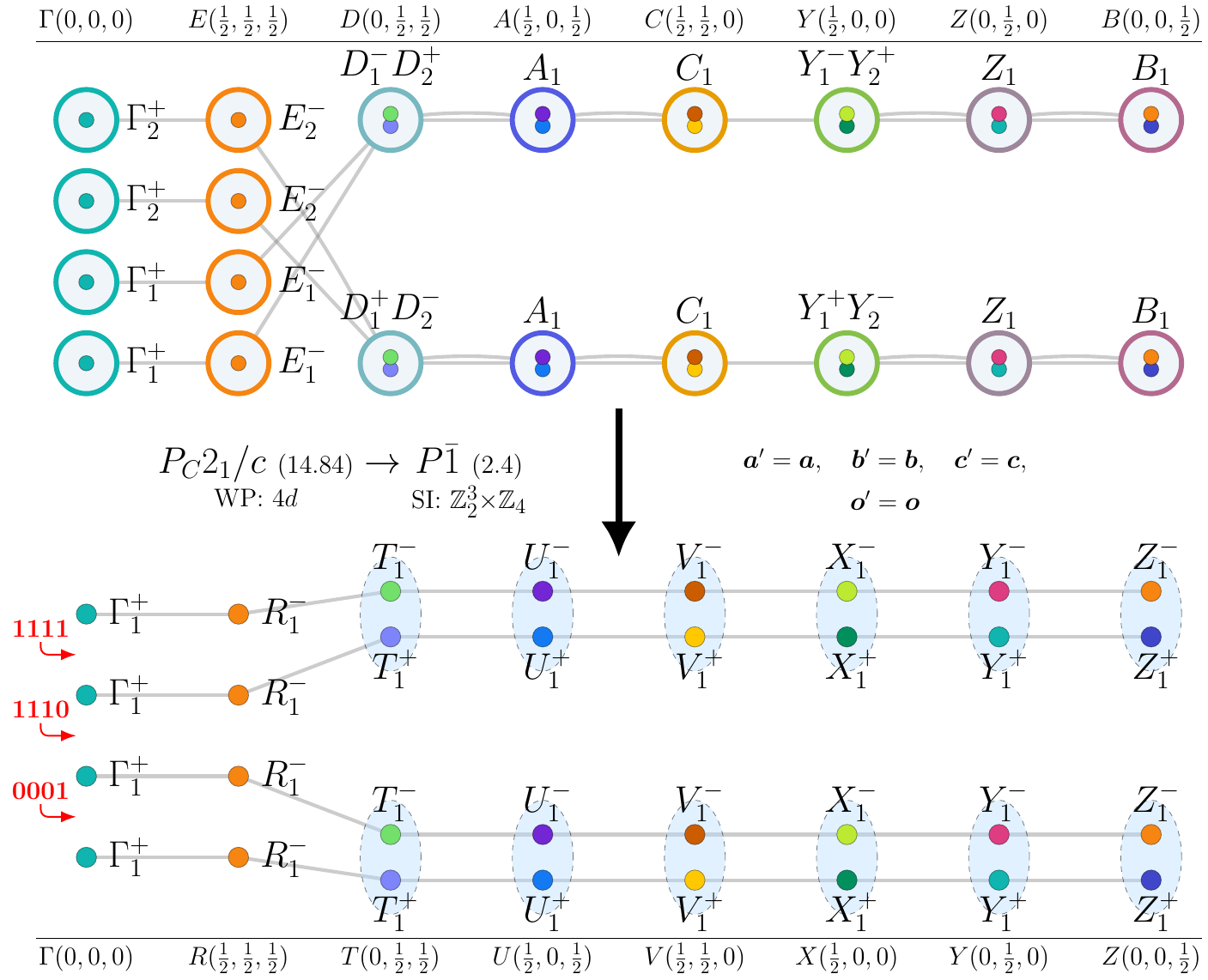}
\caption{Topological magnon bands in subgroup $P\bar{1}~(2.4)$ for magnetic moments on Wyckoff position $4d$ of supergroup $P_{C}2_{1}/c~(14.84)$.\label{fig_14.84_2.4_Bparallel010andstrainingenericdirection_4d}}
\end{figure}
\input{gap_tables_tex/14.84_2.4_Bparallel010andstrainingenericdirection_4d_table.tex}
\input{si_tables_tex/14.84_2.4_Bparallel010andstrainingenericdirection_4d_table.tex}
\subsubsection{Topological bands in subgroup $P2'/c'~(13.69)$}
\textbf{Perturbation:}
\begin{itemize}
\item B $\perp$ [010].
\end{itemize}
\begin{figure}[H]
\centering
\includegraphics[scale=0.6]{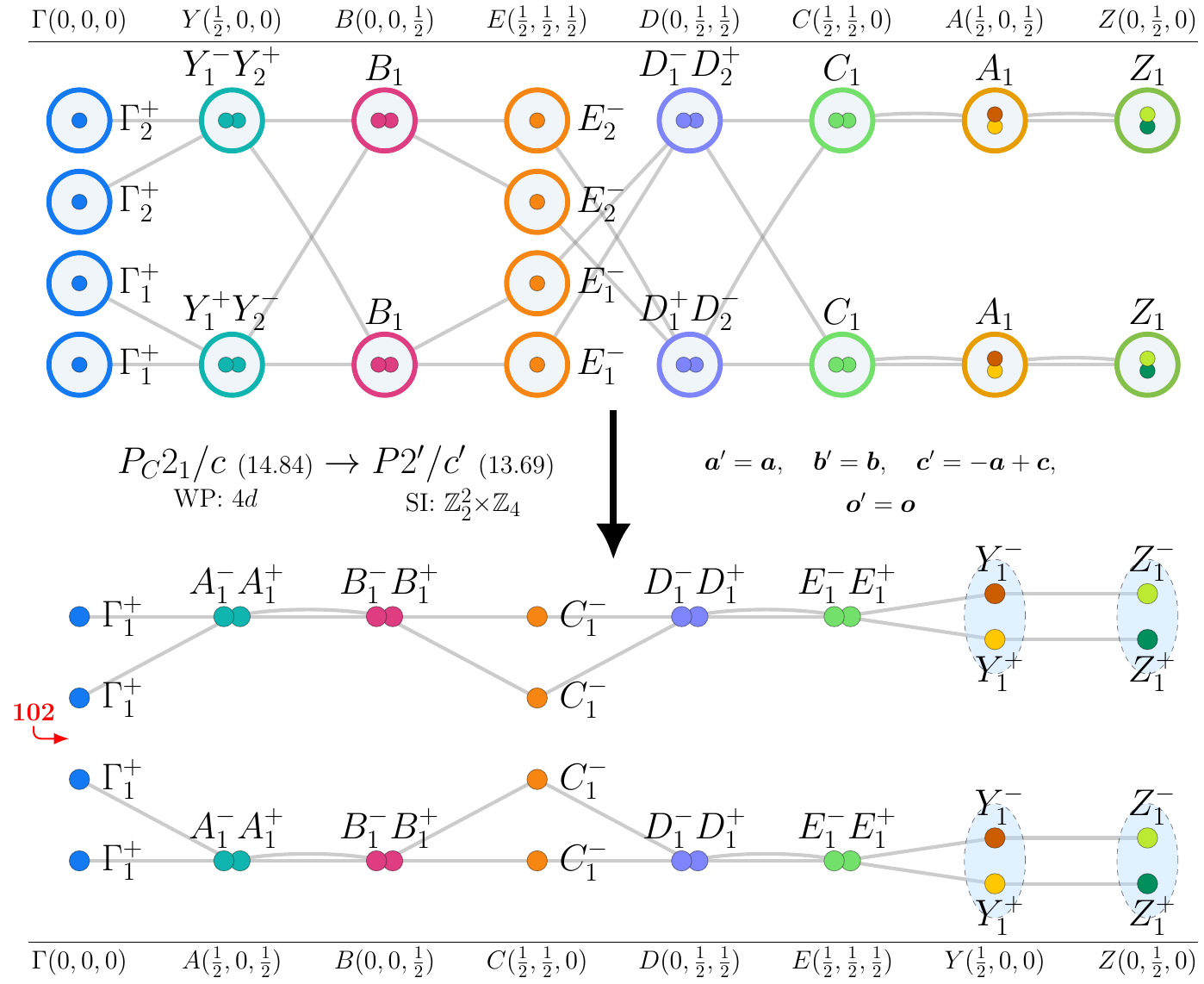}
\caption{Topological magnon bands in subgroup $P2'/c'~(13.69)$ for magnetic moments on Wyckoff position $4d$ of supergroup $P_{C}2_{1}/c~(14.84)$.\label{fig_14.84_13.69_Bperp010_4d}}
\end{figure}
\input{gap_tables_tex/14.84_13.69_Bperp010_4d_table.tex}
\input{si_tables_tex/14.84_13.69_Bperp010_4d_table.tex}
\subsubsection{Topological bands in subgroup $P_{S}\bar{1}~(2.7)$}
\textbf{Perturbation:}
\begin{itemize}
\item strain in generic direction.
\end{itemize}
\begin{figure}[H]
\centering
\includegraphics[scale=0.6]{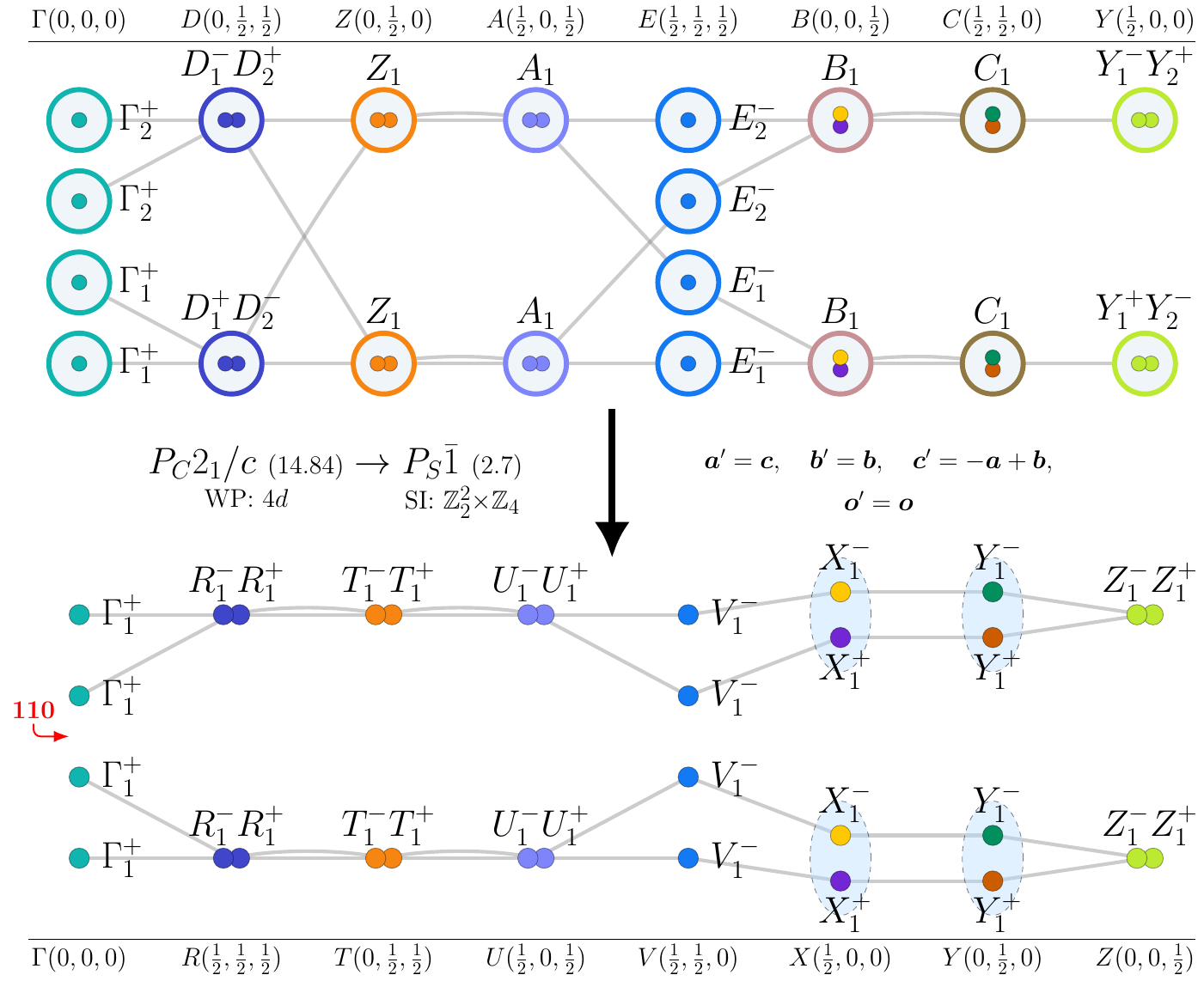}
\caption{Topological magnon bands in subgroup $P_{S}\bar{1}~(2.7)$ for magnetic moments on Wyckoff position $4d$ of supergroup $P_{C}2_{1}/c~(14.84)$.\label{fig_14.84_2.7_strainingenericdirection_4d}}
\end{figure}
\input{gap_tables_tex/14.84_2.7_strainingenericdirection_4d_table.tex}
\input{si_tables_tex/14.84_2.7_strainingenericdirection_4d_table.tex}

\section{MSG $C2/c~(15.85)$}
\textbf{Nontrivial-SI Subgroups:} $P\bar{1}~(2.4)$.\\

\textbf{Trivial-SI Subgroups:} $Cc~(9.37)$, $C2~(5.13)$.\\

\subsection{WP: $4d+4e$}
\textbf{BCS Materials:} {BiCrO\textsubscript{3}~(114 K)}\footnote{BCS web page: \texttt{\href{http://webbdcrista1.ehu.es/magndata/index.php?this\_label=0.138} {http://webbdcrista1.ehu.es/magndata/index.php?this\_label=0.138}}}.\\
\subsubsection{Topological bands in subgroup $P\bar{1}~(2.4)$}
\textbf{Perturbations:}
\begin{itemize}
\item strain in generic direction,
\item B in generic direction.
\end{itemize}
\begin{figure}[H]
\centering
\includegraphics[scale=0.6]{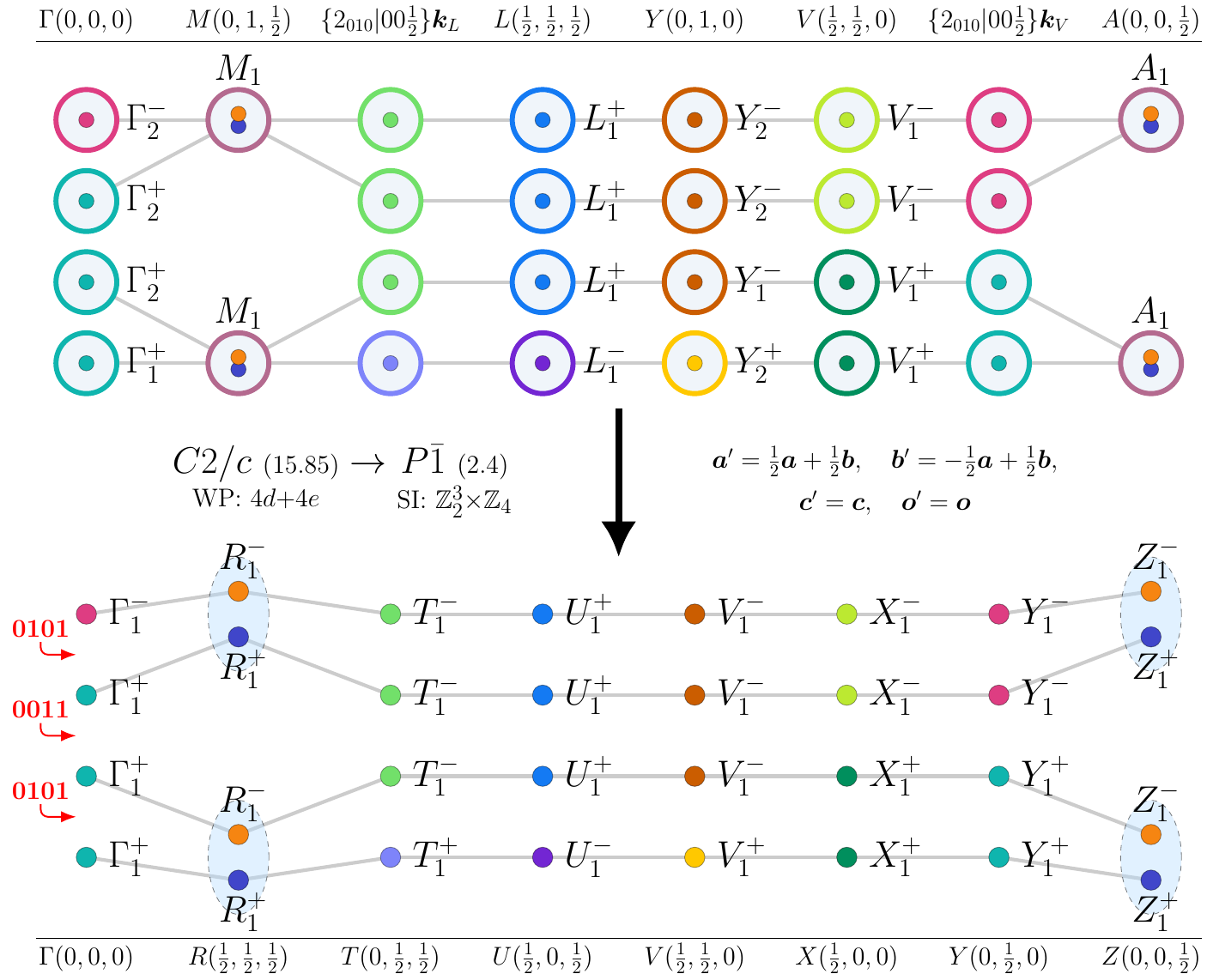}
\caption{Topological magnon bands in subgroup $P\bar{1}~(2.4)$ for magnetic moments on Wyckoff positions $4d+4e$ of supergroup $C2/c~(15.85)$.\label{fig_15.85_2.4_strainingenericdirection_4d+4e}}
\end{figure}
\input{gap_tables_tex/15.85_2.4_strainingenericdirection_4d+4e_table.tex}
\input{si_tables_tex/15.85_2.4_strainingenericdirection_4d+4e_table.tex}
\subsection{WP: $4c+4d$}
\textbf{BCS Materials:} {Sr\textsubscript{2}CoOsO\textsubscript{6}~(108 K)}\footnote{BCS web page: \texttt{\href{http://webbdcrista1.ehu.es/magndata/index.php?this\_label=0.210} {http://webbdcrista1.ehu.es/magndata/index.php?this\_label=0.210}}}.\\
\subsubsection{Topological bands in subgroup $P\bar{1}~(2.4)$}
\textbf{Perturbations:}
\begin{itemize}
\item strain in generic direction,
\item B in generic direction.
\end{itemize}
\begin{figure}[H]
\centering
\includegraphics[scale=0.6]{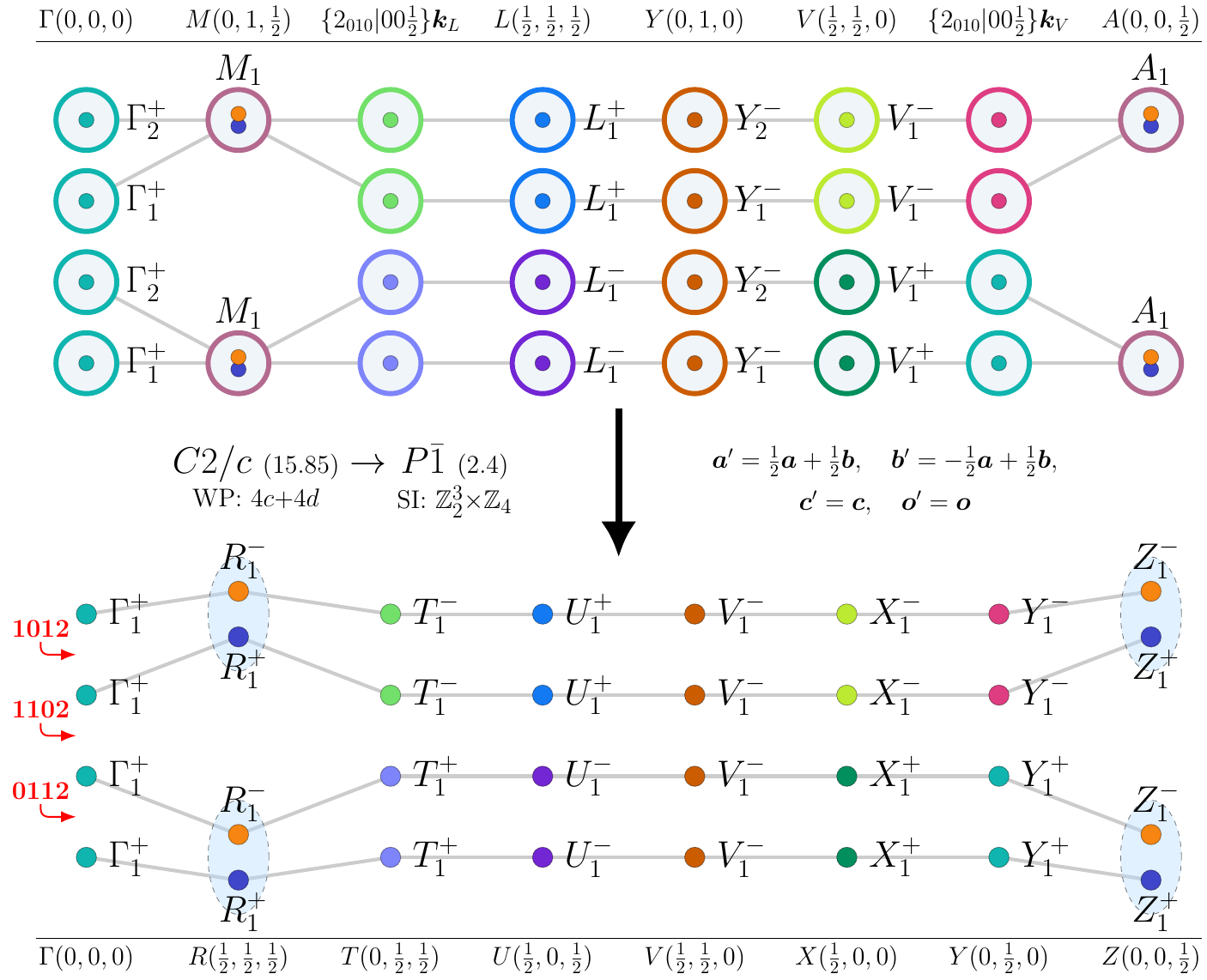}
\caption{Topological magnon bands in subgroup $P\bar{1}~(2.4)$ for magnetic moments on Wyckoff positions $4c+4d$ of supergroup $C2/c~(15.85)$.\label{fig_15.85_2.4_strainingenericdirection_4c+4d}}
\end{figure}
\input{gap_tables_tex/15.85_2.4_strainingenericdirection_4c+4d_table.tex}
\input{si_tables_tex/15.85_2.4_strainingenericdirection_4c+4d_table.tex}
\subsection{WP: $4a+4b$}
\textbf{BCS Materials:} {Cr\textsubscript{2}F\textsubscript{5}~(40 K)}\footnote{BCS web page: \texttt{\href{http://webbdcrista1.ehu.es/magndata/index.php?this\_label=0.576} {http://webbdcrista1.ehu.es/magndata/index.php?this\_label=0.576}}}.\\
\subsubsection{Topological bands in subgroup $P\bar{1}~(2.4)$}
\textbf{Perturbations:}
\begin{itemize}
\item strain in generic direction,
\item B in generic direction.
\end{itemize}
\begin{figure}[H]
\centering
\includegraphics[scale=0.6]{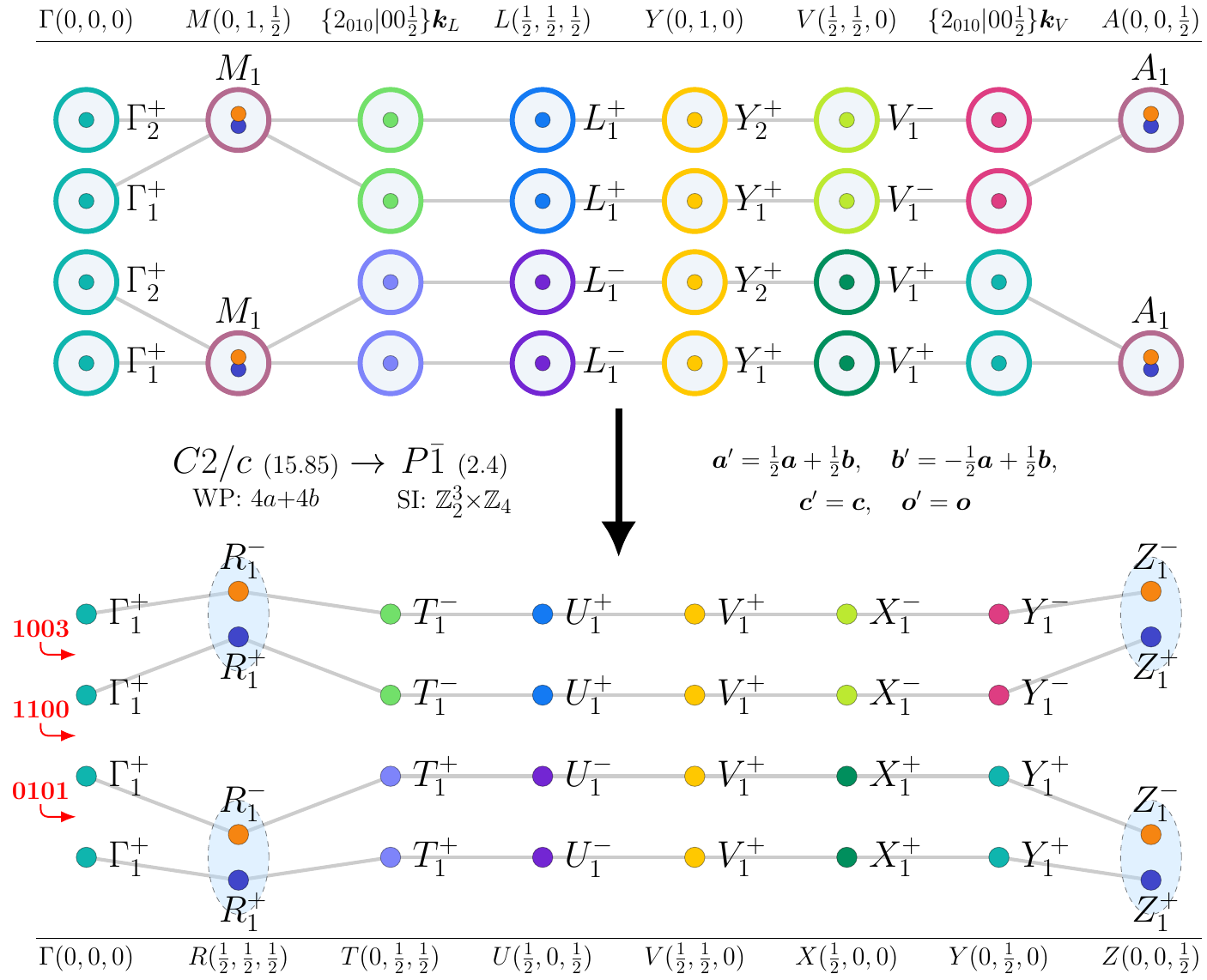}
\caption{Topological magnon bands in subgroup $P\bar{1}~(2.4)$ for magnetic moments on Wyckoff positions $4a+4b$ of supergroup $C2/c~(15.85)$.\label{fig_15.85_2.4_strainingenericdirection_4a+4b}}
\end{figure}
\input{gap_tables_tex/15.85_2.4_strainingenericdirection_4a+4b_table.tex}
\input{si_tables_tex/15.85_2.4_strainingenericdirection_4a+4b_table.tex}
\subsection{WP: $4c$}
\textbf{BCS Materials:} {MnCO\textsubscript{3}~(32 K)}\footnote{BCS web page: \texttt{\href{http://webbdcrista1.ehu.es/magndata/index.php?this\_label=0.115} {http://webbdcrista1.ehu.es/magndata/index.php?this\_label=0.115}}}, {NiCO\textsubscript{3}~(30 K)}\footnote{BCS web page: \texttt{\href{http://webbdcrista1.ehu.es/magndata/index.php?this\_label=0.113} {http://webbdcrista1.ehu.es/magndata/index.php?this\_label=0.113}}}, {CoCO\textsubscript{3}~(18 K)}\footnote{BCS web page: \texttt{\href{http://webbdcrista1.ehu.es/magndata/index.php?this\_label=0.114} {http://webbdcrista1.ehu.es/magndata/index.php?this\_label=0.114}}}.\\
\subsubsection{Topological bands in subgroup $P\bar{1}~(2.4)$}
\textbf{Perturbations:}
\begin{itemize}
\item strain in generic direction,
\item B in generic direction.
\end{itemize}
\begin{figure}[H]
\centering
\includegraphics[scale=0.6]{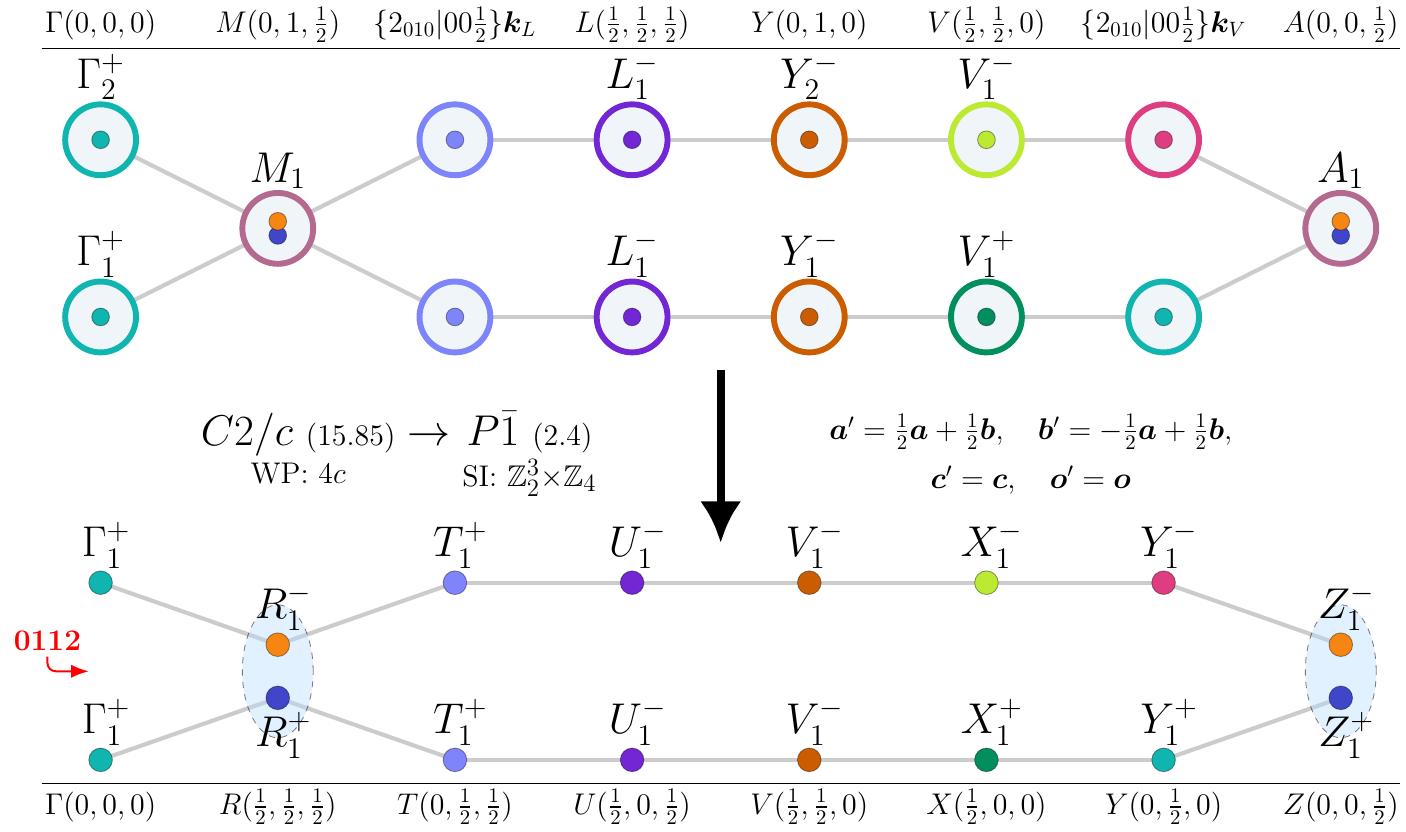}
\caption{Topological magnon bands in subgroup $P\bar{1}~(2.4)$ for magnetic moments on Wyckoff position $4c$ of supergroup $C2/c~(15.85)$.\label{fig_15.85_2.4_strainingenericdirection_4c}}
\end{figure}
\input{gap_tables_tex/15.85_2.4_strainingenericdirection_4c_table.tex}
\input{si_tables_tex/15.85_2.4_strainingenericdirection_4c_table.tex}

\section{MSG $C2'/c'~(15.89)$}
\textbf{Nontrivial-SI Subgroups:} $P\bar{1}~(2.4)$.\\

\textbf{Trivial-SI Subgroups:} $Cc'~(9.39)$, $C2'~(5.15)$.\\

\subsection{WP: $8f$}
\textbf{BCS Materials:} {$\alpha$-Fe\textsubscript{2}O\textsubscript{3}~(955 K)}\footnote{BCS web page: \texttt{\href{http://webbdcrista1.ehu.es/magndata/index.php?this\_label=0.65} {http://webbdcrista1.ehu.es/magndata/index.php?this\_label=0.65}}}, {Ho(Co\textsubscript{0.667}Ga\textsubscript{0.333})\textsubscript{2}~(31 K)}\footnote{BCS web page: \texttt{\href{http://webbdcrista1.ehu.es/magndata/index.php?this\_label=0.493} {http://webbdcrista1.ehu.es/magndata/index.php?this\_label=0.493}}}.\\
\subsubsection{Topological bands in subgroup $P\bar{1}~(2.4)$}
\textbf{Perturbations:}
\begin{itemize}
\item strain in generic direction,
\item B in generic direction.
\end{itemize}
\begin{figure}[H]
\centering
\includegraphics[scale=0.6]{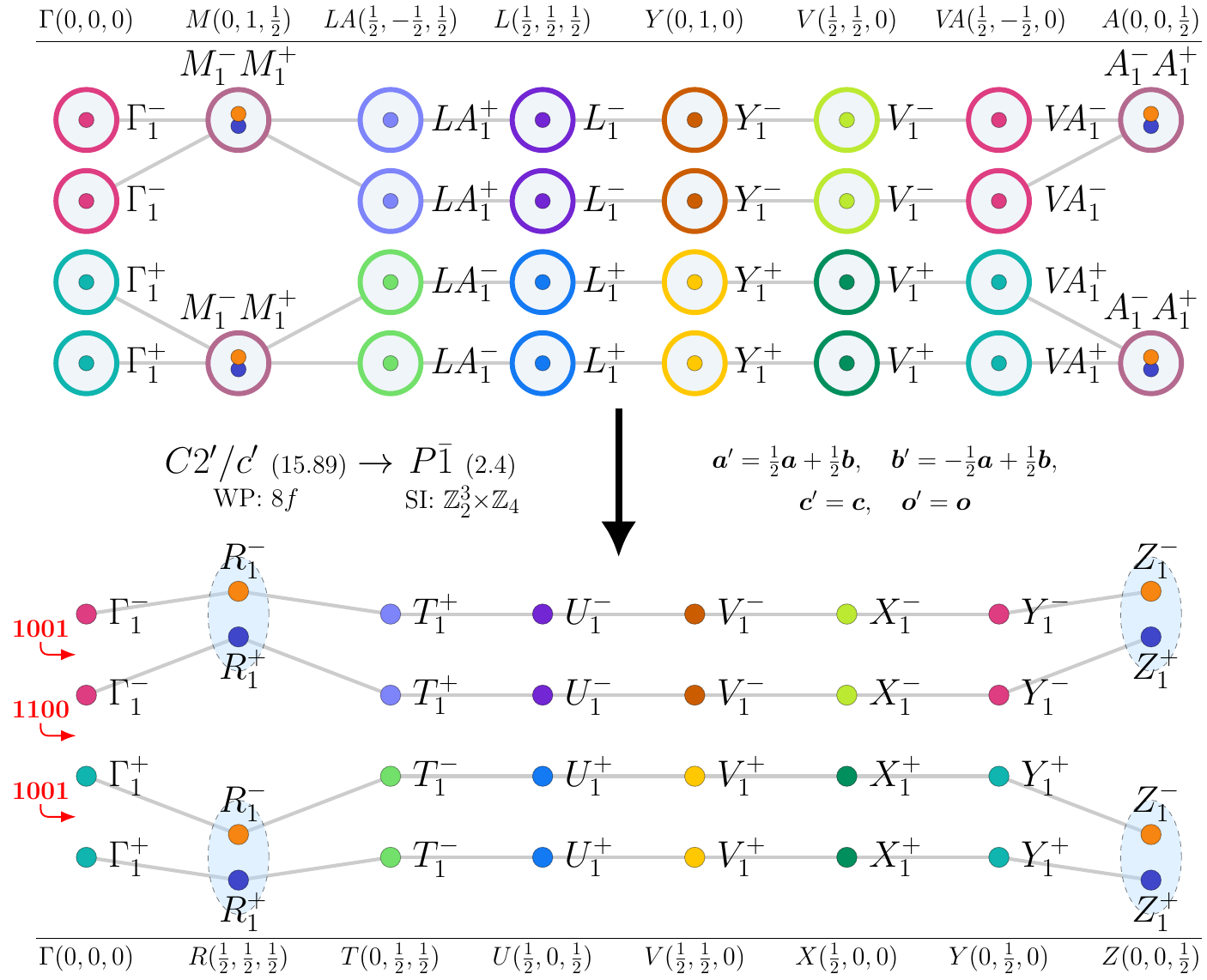}
\caption{Topological magnon bands in subgroup $P\bar{1}~(2.4)$ for magnetic moments on Wyckoff position $8f$ of supergroup $C2'/c'~(15.89)$.\label{fig_15.89_2.4_strainingenericdirection_8f}}
\end{figure}
\input{gap_tables_tex/15.89_2.4_strainingenericdirection_8f_table.tex}
\input{si_tables_tex/15.89_2.4_strainingenericdirection_8f_table.tex}
\subsection{WP: $4d$}
\textbf{BCS Materials:} {FeF\textsubscript{3}~(394 K)}\footnote{BCS web page: \texttt{\href{http://webbdcrista1.ehu.es/magndata/index.php?this\_label=0.581} {http://webbdcrista1.ehu.es/magndata/index.php?this\_label=0.581}}}, {Ca\textsubscript{3}LiRuO\textsubscript{6}~(117 K)}\footnote{BCS web page: \texttt{\href{http://webbdcrista1.ehu.es/magndata/index.php?this\_label=0.239} {http://webbdcrista1.ehu.es/magndata/index.php?this\_label=0.239}}}, {FeSO\textsubscript{4}F~(100 K)}\footnote{BCS web page: \texttt{\href{http://webbdcrista1.ehu.es/magndata/index.php?this\_label=0.128} {http://webbdcrista1.ehu.es/magndata/index.php?this\_label=0.128}}}, {Sr\textsubscript{3}LiRuO\textsubscript{6}~(90 K)}\footnote{BCS web page: \texttt{\href{http://webbdcrista1.ehu.es/magndata/index.php?this\_label=0.361} {http://webbdcrista1.ehu.es/magndata/index.php?this\_label=0.361}}}, {Sr\textsubscript{3}NaRuO\textsubscript{6}~(70 K)}\footnote{BCS web page: \texttt{\href{http://webbdcrista1.ehu.es/magndata/index.php?this\_label=0.404} {http://webbdcrista1.ehu.es/magndata/index.php?this\_label=0.404}}}, {HoP~(5.5 K)}\footnote{BCS web page: \texttt{\href{http://webbdcrista1.ehu.es/magndata/index.php?this\_label=2.10} {http://webbdcrista1.ehu.es/magndata/index.php?this\_label=2.10}}}, {FeF\textsubscript{3}}\footnote{BCS web page: \texttt{\href{http://webbdcrista1.ehu.es/magndata/index.php?this\_label=0.335} {http://webbdcrista1.ehu.es/magndata/index.php?this\_label=0.335}}}.\\
\subsubsection{Topological bands in subgroup $P\bar{1}~(2.4)$}
\textbf{Perturbations:}
\begin{itemize}
\item strain in generic direction,
\item B in generic direction.
\end{itemize}
\begin{figure}[H]
\centering
\includegraphics[scale=0.6]{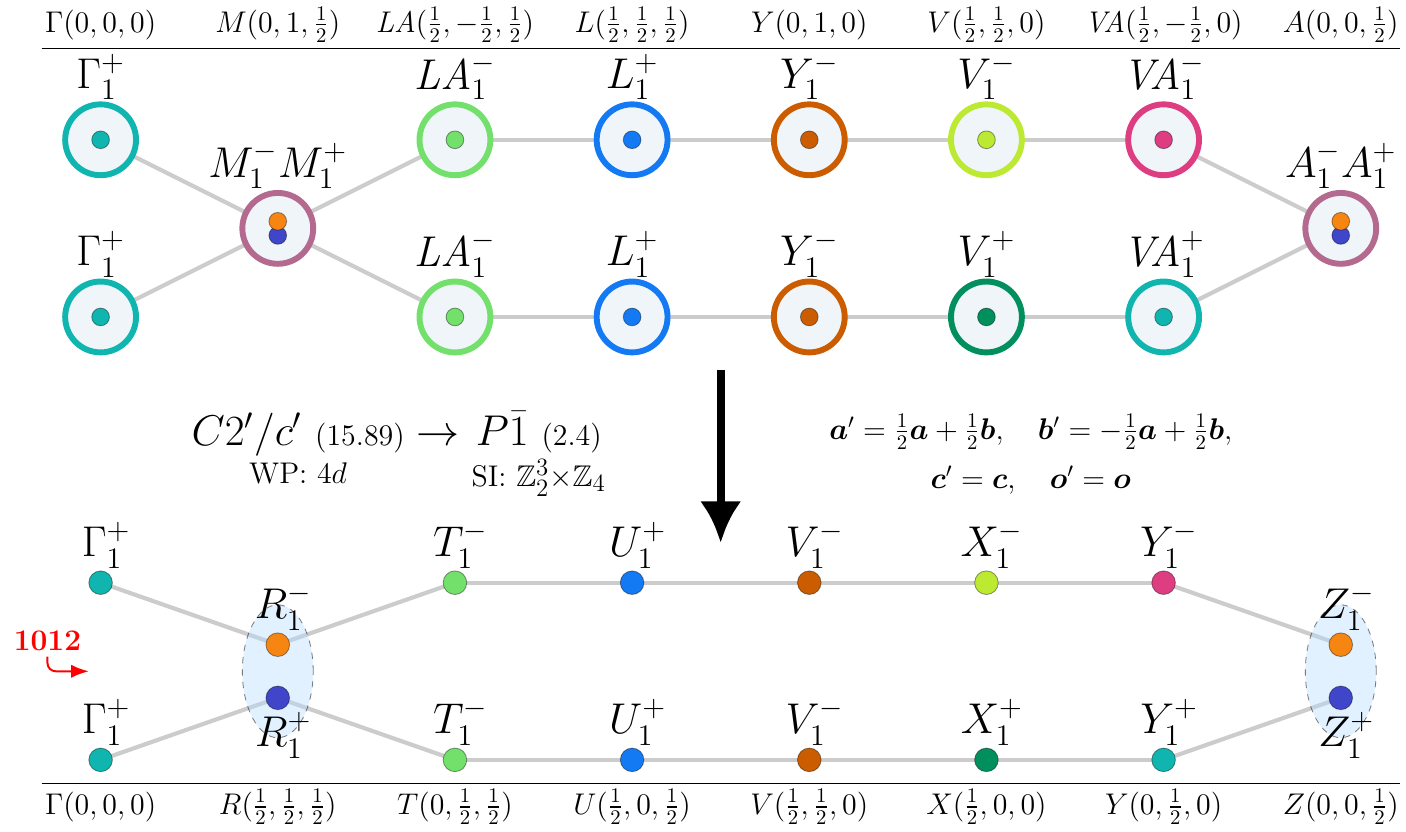}
\caption{Topological magnon bands in subgroup $P\bar{1}~(2.4)$ for magnetic moments on Wyckoff position $4d$ of supergroup $C2'/c'~(15.89)$.\label{fig_15.89_2.4_strainingenericdirection_4d}}
\end{figure}
\input{gap_tables_tex/15.89_2.4_strainingenericdirection_4d_table.tex}
\input{si_tables_tex/15.89_2.4_strainingenericdirection_4d_table.tex}
\subsection{WP: $4c$}
\textbf{BCS Materials:} {FeBO\textsubscript{3}~(348 K)}\footnote{BCS web page: \texttt{\href{http://webbdcrista1.ehu.es/magndata/index.php?this\_label=0.112} {http://webbdcrista1.ehu.es/magndata/index.php?this\_label=0.112}}}, {FeOHSO\textsubscript{4}~(125 K)}\footnote{BCS web page: \texttt{\href{http://webbdcrista1.ehu.es/magndata/index.php?this\_label=0.760} {http://webbdcrista1.ehu.es/magndata/index.php?this\_label=0.760}}}, {Ca\textsubscript{3}LiOsO\textsubscript{6}~(117.1 K)}\footnote{BCS web page: \texttt{\href{http://webbdcrista1.ehu.es/magndata/index.php?this\_label=0.3} {http://webbdcrista1.ehu.es/magndata/index.php?this\_label=0.3}}}.\\
\subsubsection{Topological bands in subgroup $P\bar{1}~(2.4)$}
\textbf{Perturbations:}
\begin{itemize}
\item strain in generic direction,
\item B in generic direction.
\end{itemize}
\begin{figure}[H]
\centering
\includegraphics[scale=0.6]{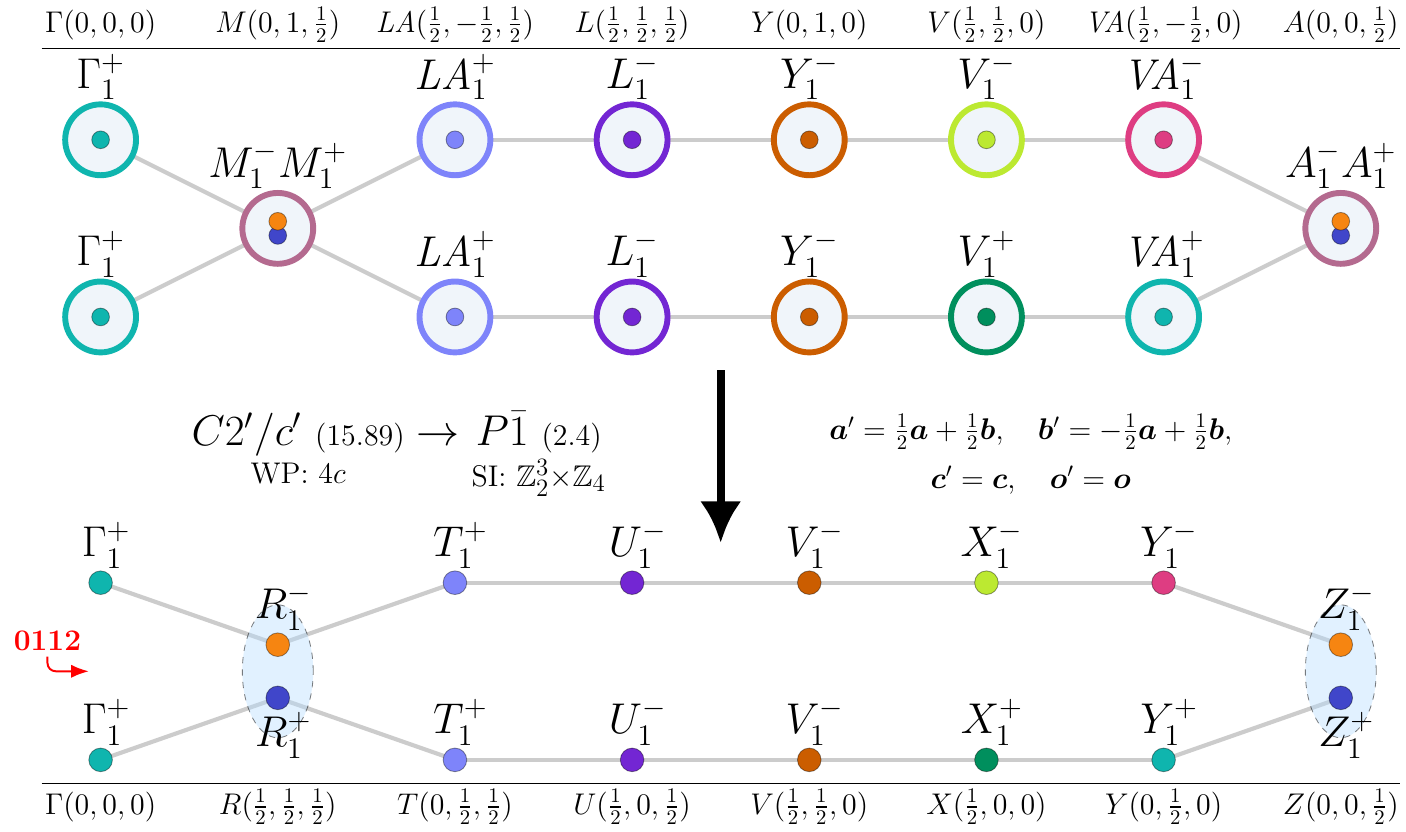}
\caption{Topological magnon bands in subgroup $P\bar{1}~(2.4)$ for magnetic moments on Wyckoff position $4c$ of supergroup $C2'/c'~(15.89)$.\label{fig_15.89_2.4_strainingenericdirection_4c}}
\end{figure}
\input{gap_tables_tex/15.89_2.4_strainingenericdirection_4c_table.tex}
\input{si_tables_tex/15.89_2.4_strainingenericdirection_4c_table.tex}
\subsection{WP: $4a$}
\textbf{BCS Materials:} {GdMg~(110 K)}\footnote{BCS web page: \texttt{\href{http://webbdcrista1.ehu.es/magndata/index.php?this\_label=2.70} {http://webbdcrista1.ehu.es/magndata/index.php?this\_label=2.70}}}.\\
\subsubsection{Topological bands in subgroup $P\bar{1}~(2.4)$}
\textbf{Perturbations:}
\begin{itemize}
\item strain in generic direction,
\item B in generic direction.
\end{itemize}
\begin{figure}[H]
\centering
\includegraphics[scale=0.6]{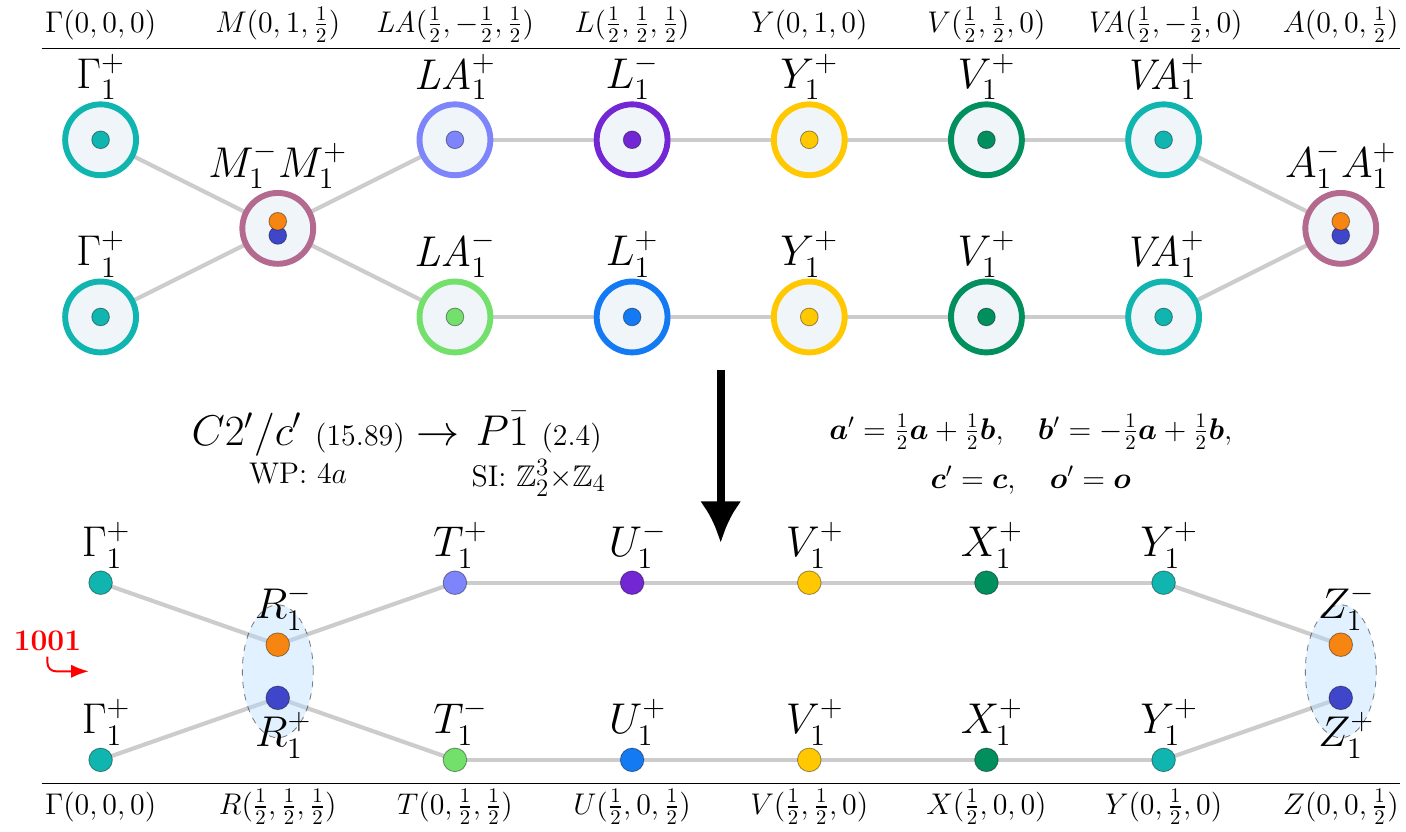}
\caption{Topological magnon bands in subgroup $P\bar{1}~(2.4)$ for magnetic moments on Wyckoff position $4a$ of supergroup $C2'/c'~(15.89)$.\label{fig_15.89_2.4_strainingenericdirection_4a}}
\end{figure}
\input{gap_tables_tex/15.89_2.4_strainingenericdirection_4a_table.tex}
\input{si_tables_tex/15.89_2.4_strainingenericdirection_4a_table.tex}
\subsection{WP: $4b+4d+4e$}
\textbf{BCS Materials:} {NdCo\textsubscript{2}~(42 K)}\footnote{BCS web page: \texttt{\href{http://webbdcrista1.ehu.es/magndata/index.php?this\_label=0.226} {http://webbdcrista1.ehu.es/magndata/index.php?this\_label=0.226}}}.\\
\subsubsection{Topological bands in subgroup $P\bar{1}~(2.4)$}
\textbf{Perturbations:}
\begin{itemize}
\item strain in generic direction,
\item B in generic direction.
\end{itemize}
\begin{figure}[H]
\centering
\includegraphics[scale=0.6]{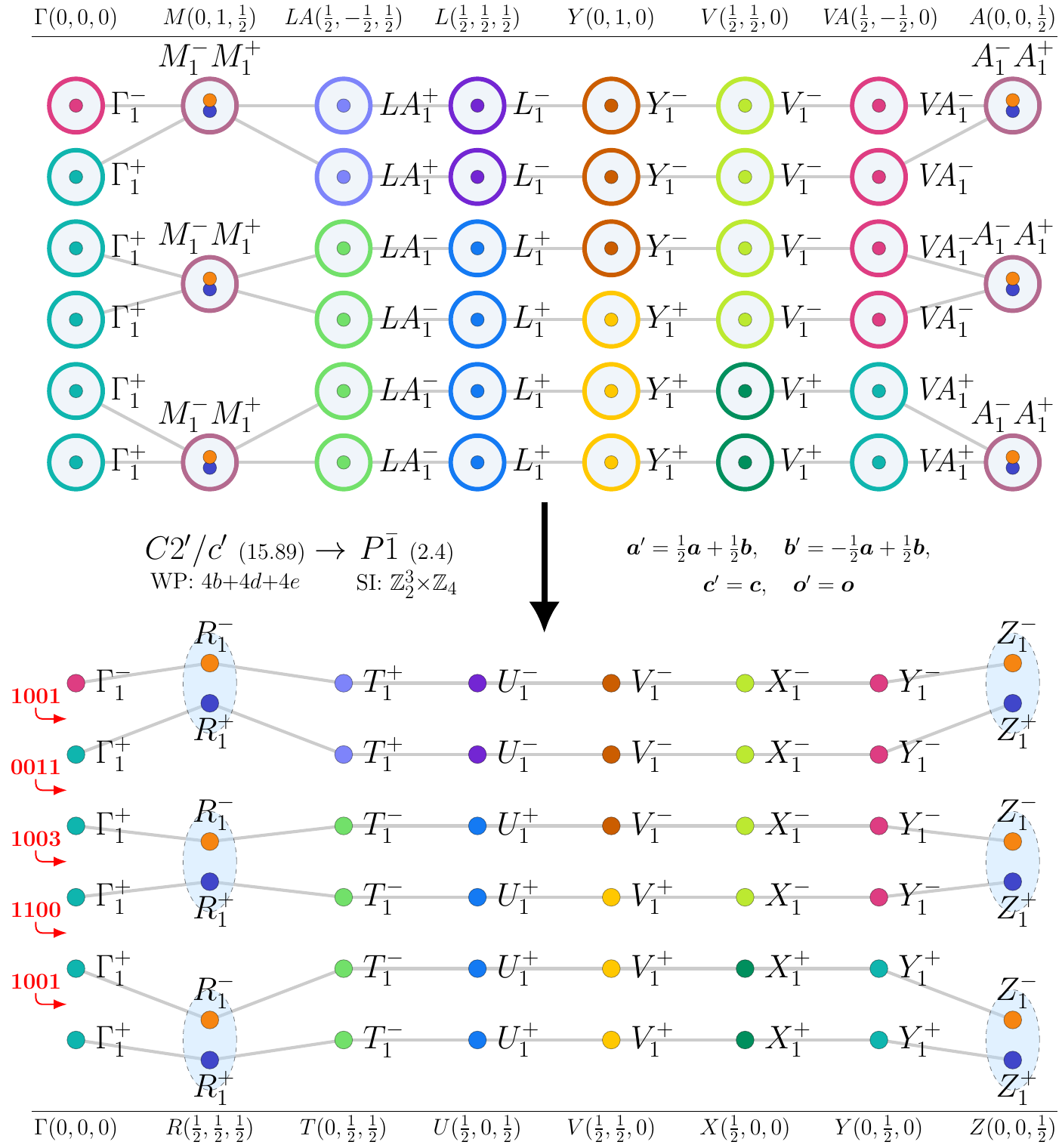}
\caption{Topological magnon bands in subgroup $P\bar{1}~(2.4)$ for magnetic moments on Wyckoff positions $4b+4d+4e$ of supergroup $C2'/c'~(15.89)$.\label{fig_15.89_2.4_strainingenericdirection_4b+4d+4e}}
\end{figure}
\input{gap_tables_tex/15.89_2.4_strainingenericdirection_4b+4d+4e_table.tex}
\input{si_tables_tex/15.89_2.4_strainingenericdirection_4b+4d+4e_table.tex}
\subsection{WP: $4a+4e$}
\textbf{BCS Materials:} {(NH\textsubscript{2}(CH\textsubscript{3})\textsubscript{2})(FeMn(HCOO)\textsubscript{6})~(35 K)}\footnote{BCS web page: \texttt{\href{http://webbdcrista1.ehu.es/magndata/index.php?this\_label=0.251} {http://webbdcrista1.ehu.es/magndata/index.php?this\_label=0.251}}}, {(NH\textsubscript{2}(CH\textsubscript{3})\textsubscript{2})(FeCo(HCOO)\textsubscript{6})~(32 K)}\footnote{BCS web page: \texttt{\href{http://webbdcrista1.ehu.es/magndata/index.php?this\_label=0.250} {http://webbdcrista1.ehu.es/magndata/index.php?this\_label=0.250}}}.\\
\subsubsection{Topological bands in subgroup $P\bar{1}~(2.4)$}
\textbf{Perturbations:}
\begin{itemize}
\item strain in generic direction,
\item B in generic direction.
\end{itemize}
\begin{figure}[H]
\centering
\includegraphics[scale=0.6]{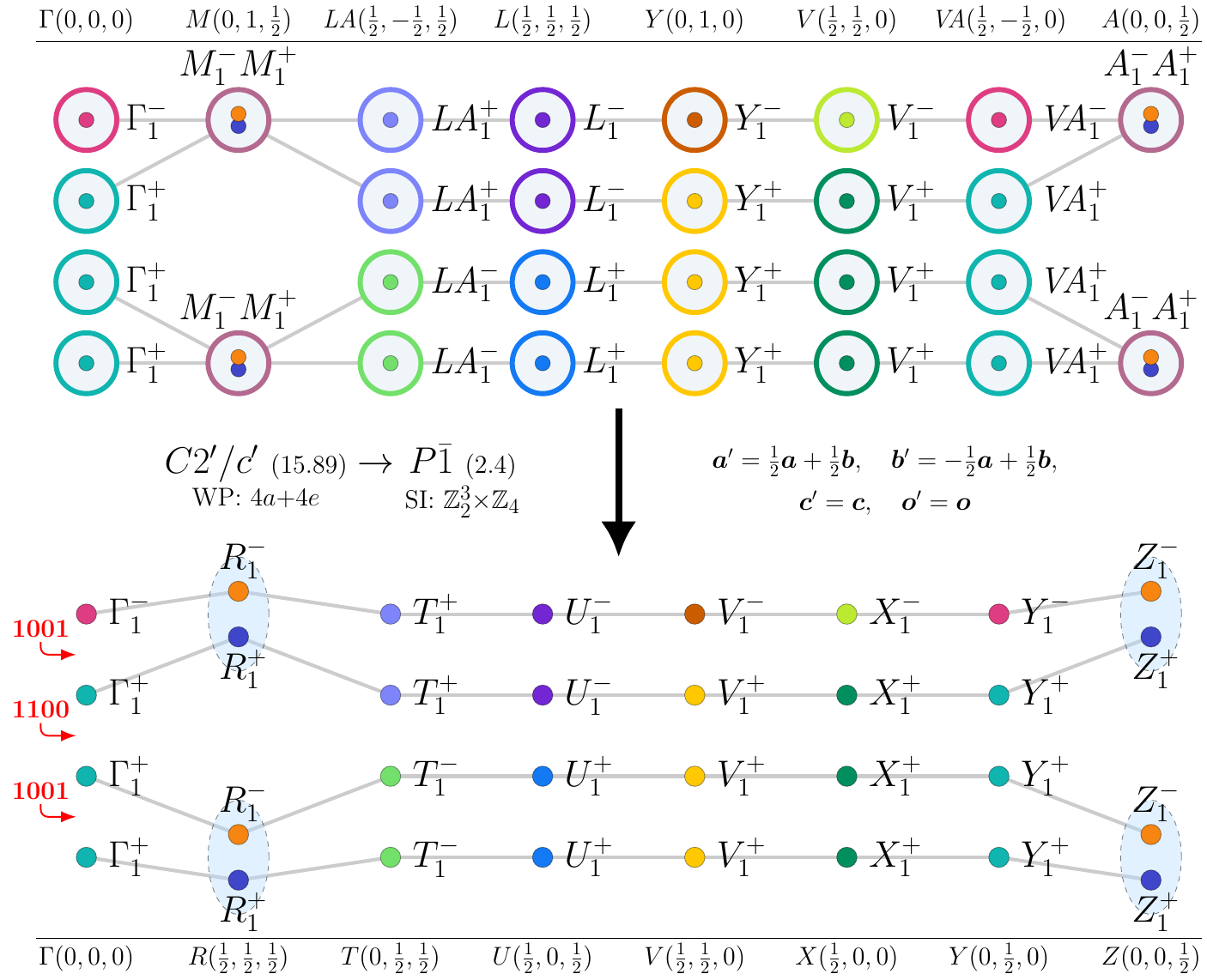}
\caption{Topological magnon bands in subgroup $P\bar{1}~(2.4)$ for magnetic moments on Wyckoff positions $4a+4e$ of supergroup $C2'/c'~(15.89)$.\label{fig_15.89_2.4_strainingenericdirection_4a+4e}}
\end{figure}
\input{gap_tables_tex/15.89_2.4_strainingenericdirection_4a+4e_table.tex}
\input{si_tables_tex/15.89_2.4_strainingenericdirection_4a+4e_table.tex}
\subsection{WP: $4a+4c$}
\textbf{BCS Materials:} {Fe\textsubscript{2}F\textsubscript{5}(H\textsubscript{2}O)\textsubscript{2}~(26 K)}\footnote{BCS web page: \texttt{\href{http://webbdcrista1.ehu.es/magndata/index.php?this\_label=0.584} {http://webbdcrista1.ehu.es/magndata/index.php?this\_label=0.584}}}.\\
\subsubsection{Topological bands in subgroup $P\bar{1}~(2.4)$}
\textbf{Perturbations:}
\begin{itemize}
\item strain in generic direction,
\item B in generic direction.
\end{itemize}
\begin{figure}[H]
\centering
\includegraphics[scale=0.6]{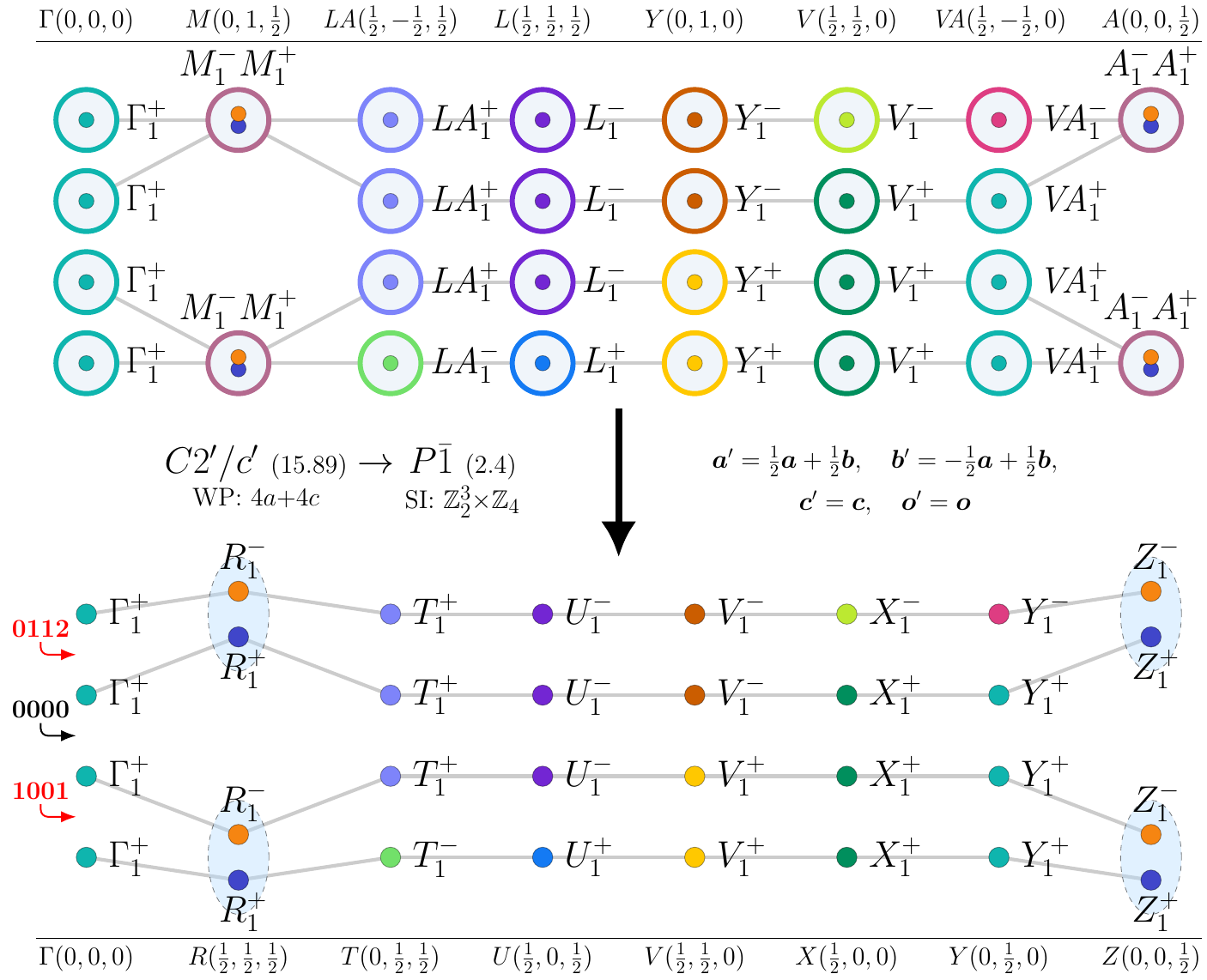}
\caption{Topological magnon bands in subgroup $P\bar{1}~(2.4)$ for magnetic moments on Wyckoff positions $4a+4c$ of supergroup $C2'/c'~(15.89)$.\label{fig_15.89_2.4_strainingenericdirection_4a+4c}}
\end{figure}
\input{gap_tables_tex/15.89_2.4_strainingenericdirection_4a+4c_table.tex}
\input{si_tables_tex/15.89_2.4_strainingenericdirection_4a+4c_table.tex}
\subsection{WP: $4e$}
\textbf{BCS Materials:} {NdPt~(23 K)}\footnote{BCS web page: \texttt{\href{http://webbdcrista1.ehu.es/magndata/index.php?this\_label=0.690} {http://webbdcrista1.ehu.es/magndata/index.php?this\_label=0.690}}}, {NaCrGe\textsubscript{2}O\textsubscript{6}~(6 K)}\footnote{BCS web page: \texttt{\href{http://webbdcrista1.ehu.es/magndata/index.php?this\_label=0.297} {http://webbdcrista1.ehu.es/magndata/index.php?this\_label=0.297}}}.\\
\subsubsection{Topological bands in subgroup $P\bar{1}~(2.4)$}
\textbf{Perturbations:}
\begin{itemize}
\item strain in generic direction,
\item B in generic direction.
\end{itemize}
\begin{figure}[H]
\centering
\includegraphics[scale=0.6]{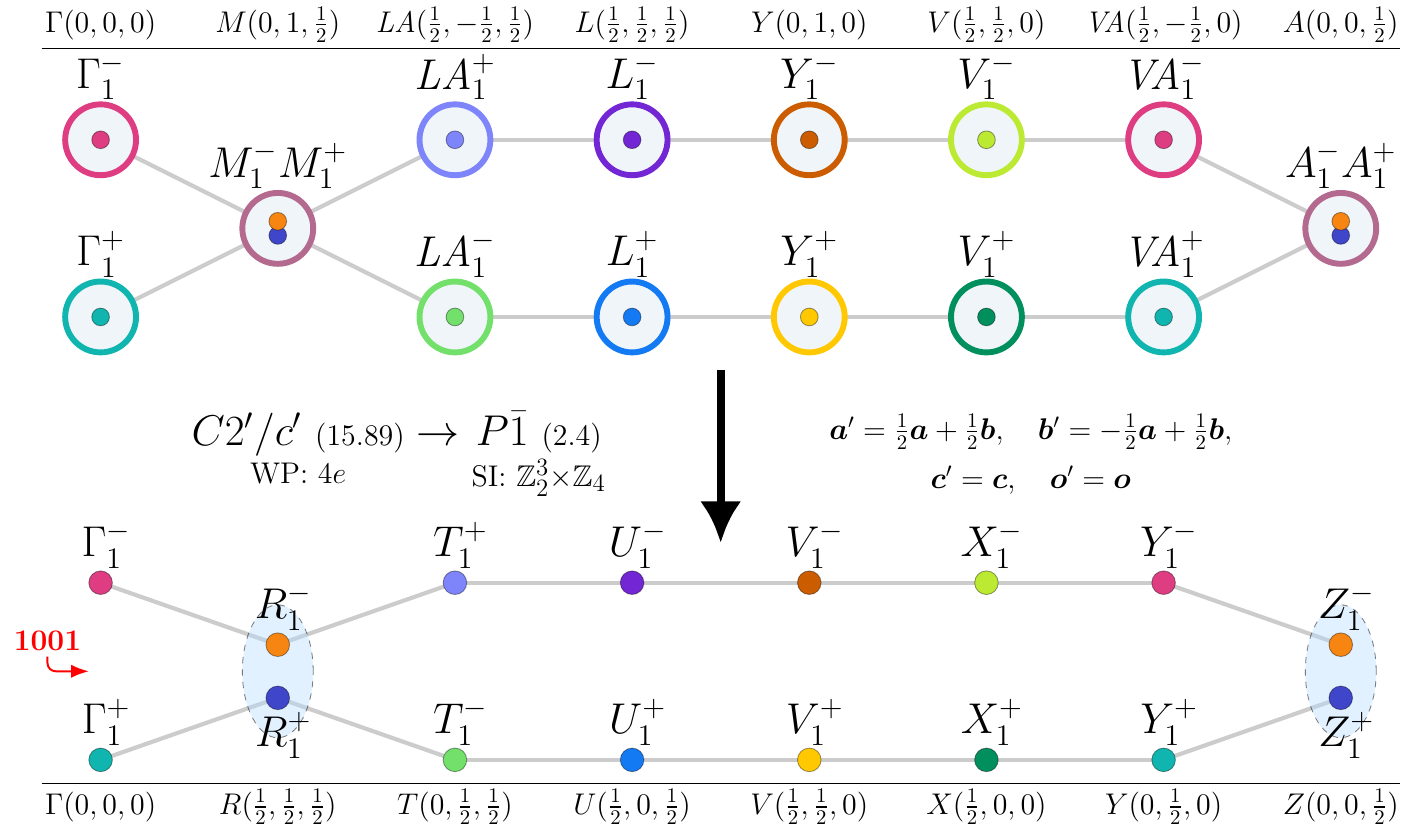}
\caption{Topological magnon bands in subgroup $P\bar{1}~(2.4)$ for magnetic moments on Wyckoff position $4e$ of supergroup $C2'/c'~(15.89)$.\label{fig_15.89_2.4_strainingenericdirection_4e}}
\end{figure}
\input{gap_tables_tex/15.89_2.4_strainingenericdirection_4e_table.tex}
\input{si_tables_tex/15.89_2.4_strainingenericdirection_4e_table.tex}
\subsection{WP: $4b$}
\textbf{BCS Materials:} {Na\textsubscript{2}BaFe(VO\textsubscript{4})\textsubscript{2}~(7 K)}\footnote{BCS web page: \texttt{\href{http://webbdcrista1.ehu.es/magndata/index.php?this\_label=0.298} {http://webbdcrista1.ehu.es/magndata/index.php?this\_label=0.298}}}.\\
\subsubsection{Topological bands in subgroup $P\bar{1}~(2.4)$}
\textbf{Perturbations:}
\begin{itemize}
\item strain in generic direction,
\item B in generic direction.
\end{itemize}
\begin{figure}[H]
\centering
\includegraphics[scale=0.6]{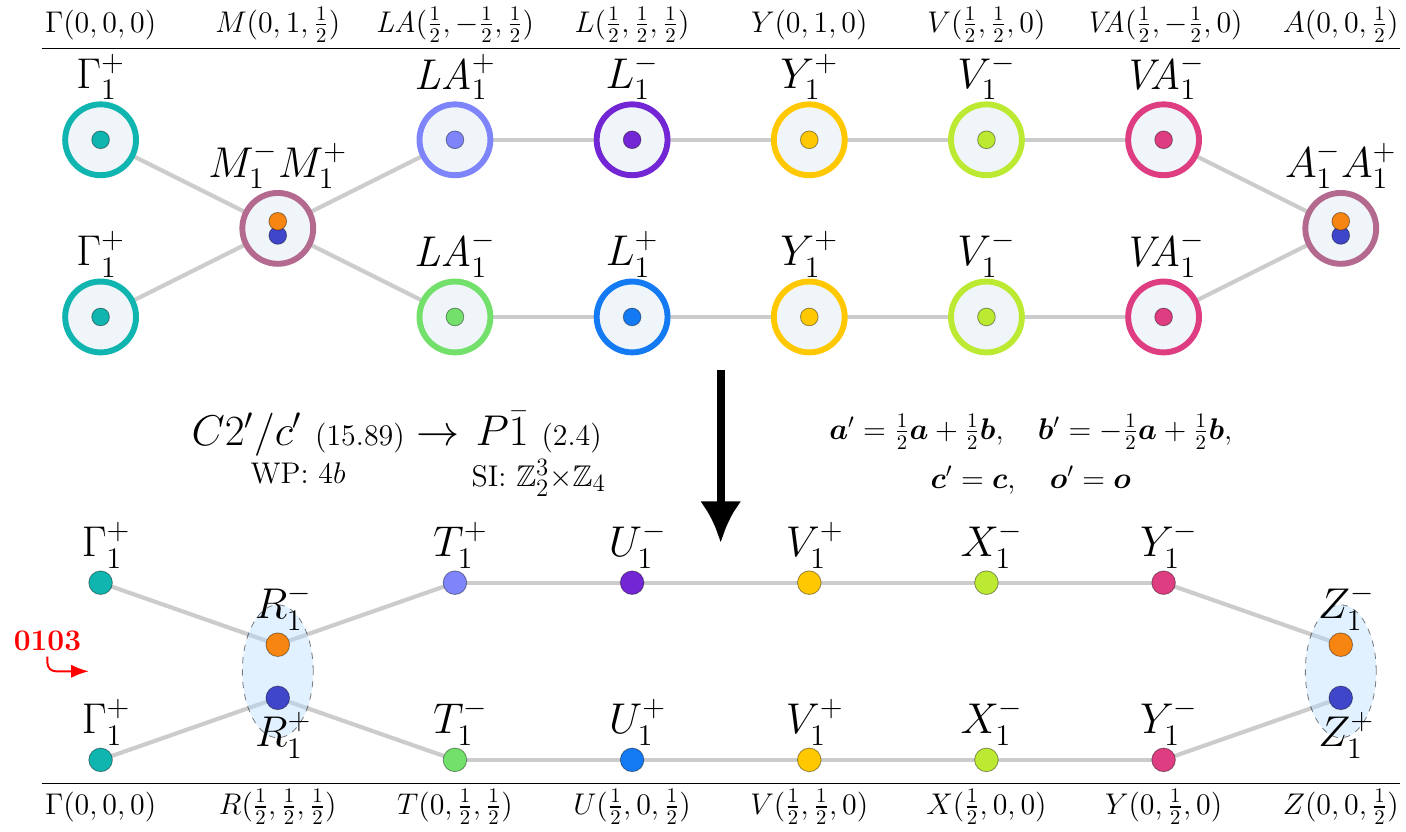}
\caption{Topological magnon bands in subgroup $P\bar{1}~(2.4)$ for magnetic moments on Wyckoff position $4b$ of supergroup $C2'/c'~(15.89)$.\label{fig_15.89_2.4_strainingenericdirection_4b}}
\end{figure}
\input{gap_tables_tex/15.89_2.4_strainingenericdirection_4b_table.tex}
\input{si_tables_tex/15.89_2.4_strainingenericdirection_4b_table.tex}

\section{MSG $R_{I}\bar{3}m~(166.102)$}
\textbf{Nontrivial-SI Subgroups:} $P\bar{1}~(2.4)$, $C2'/c'~(15.89)$, $C2'/c'~(15.89)$, $P_{S}\bar{1}~(2.7)$, $C2/m~(12.58)$, $C_{c}2/m~(12.63)$, $C2/m~(12.58)$, $C_{c}2/m~(12.63)$, $R\bar{3}c'~(167.107)$.\\

\textbf{Trivial-SI Subgroups:} $Cc'~(9.39)$, $Cc'~(9.39)$, $C2'~(5.15)$, $C2'~(5.15)$, $P_{S}1~(1.3)$, $Cm~(8.32)$, $C_{c}m~(8.35)$, $Cm~(8.32)$, $C_{c}m~(8.35)$, $C2~(5.13)$, $C_{c}2~(5.16)$, $C2~(5.13)$, $C_{c}2~(5.16)$, $R3c'~(161.71)$, $R_{I}3m~(160.68)$.\\

\subsection{WP: $18e$}
\textbf{BCS Materials:} {Gd\textsubscript{2}Ti\textsubscript{2}O\textsubscript{7}~(0.97 K)}\footnote{BCS web page: \texttt{\href{http://webbdcrista1.ehu.es/magndata/index.php?this\_label=1.56} {http://webbdcrista1.ehu.es/magndata/index.php?this\_label=1.56}}}.\\
\subsubsection{Topological bands in subgroup $P\bar{1}~(2.4)$}
\textbf{Perturbations:}
\begin{itemize}
\item B $\parallel$ [001] and strain in generic direction,
\item B $\parallel$ [100] and (strain $\parallel$ [110] or strain $\perp$ [110]),
\item B $\parallel$ [100] and strain in generic direction,
\item B $\parallel$ [110] and (strain $\parallel$ [100] or strain $\perp$ [100]),
\item B $\parallel$ [110] and strain in generic direction,
\item B in generic direction,
\item B $\perp$ [100] and (strain $\parallel$ [110] or strain $\perp$ [110]),
\item B $\perp$ [100] and strain in generic direction,
\item B $\perp$ [110] and (strain $\parallel$ [100] or strain $\perp$ [100]),
\item B $\perp$ [110] and strain in generic direction.
\end{itemize}
\begin{figure}[H]
\centering
\includegraphics[scale=0.6]{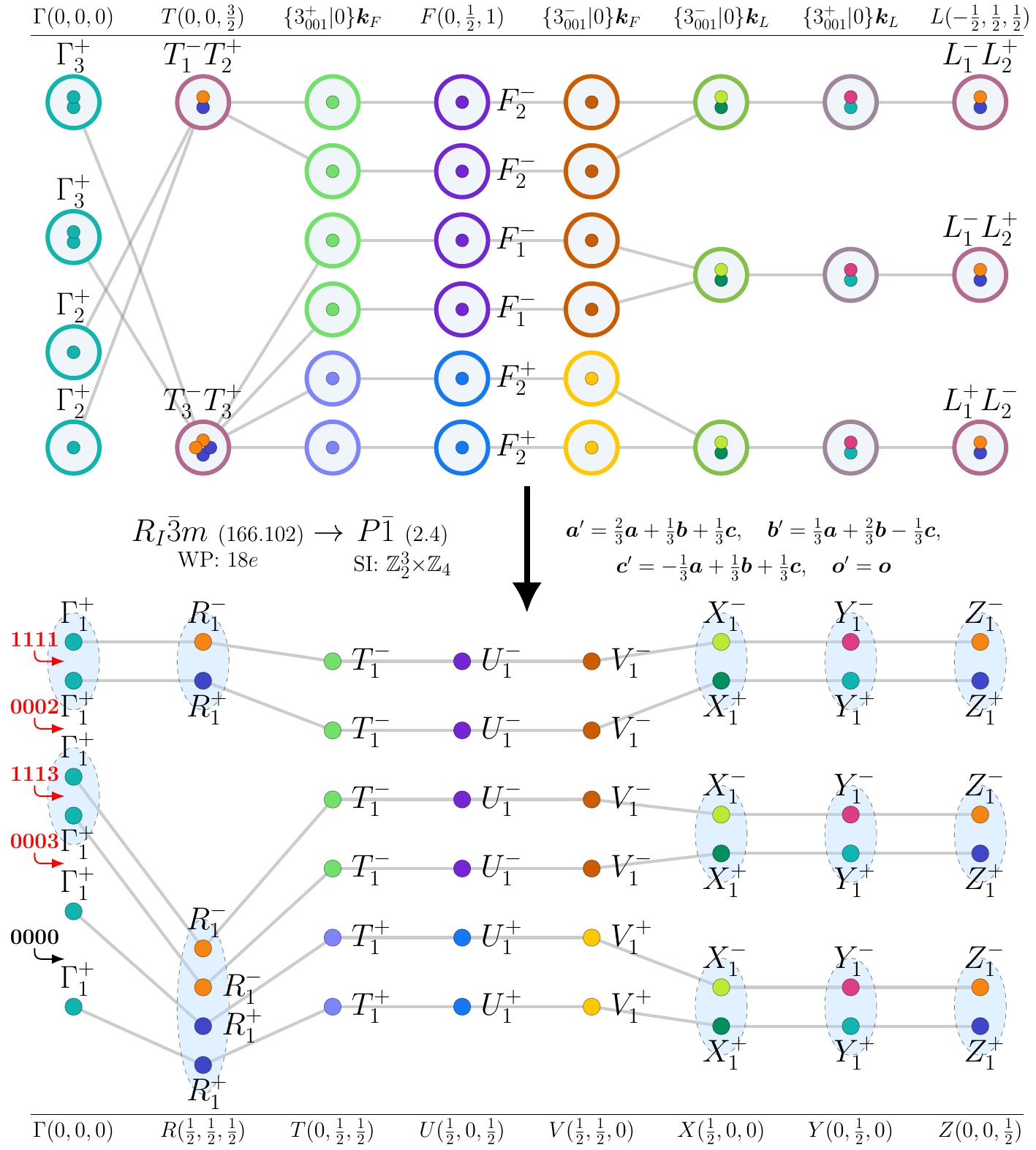}
\caption{Topological magnon bands in subgroup $P\bar{1}~(2.4)$ for magnetic moments on Wyckoff position $18e$ of supergroup $R_{I}\bar{3}m~(166.102)$.\label{fig_166.102_2.4_Bparallel001andstrainingenericdirection_18e}}
\end{figure}
\input{gap_tables_tex/166.102_2.4_Bparallel001andstrainingenericdirection_18e_table.tex}
\input{si_tables_tex/166.102_2.4_Bparallel001andstrainingenericdirection_18e_table.tex}
\subsubsection{Topological bands in subgroup $C2'/c'~(15.89)$}
\textbf{Perturbations:}
\begin{itemize}
\item B $\parallel$ [001] and (strain $\parallel$ [110] or strain $\perp$ [110]),
\item B $\perp$ [110].
\end{itemize}
\begin{figure}[H]
\centering
\includegraphics[scale=0.6]{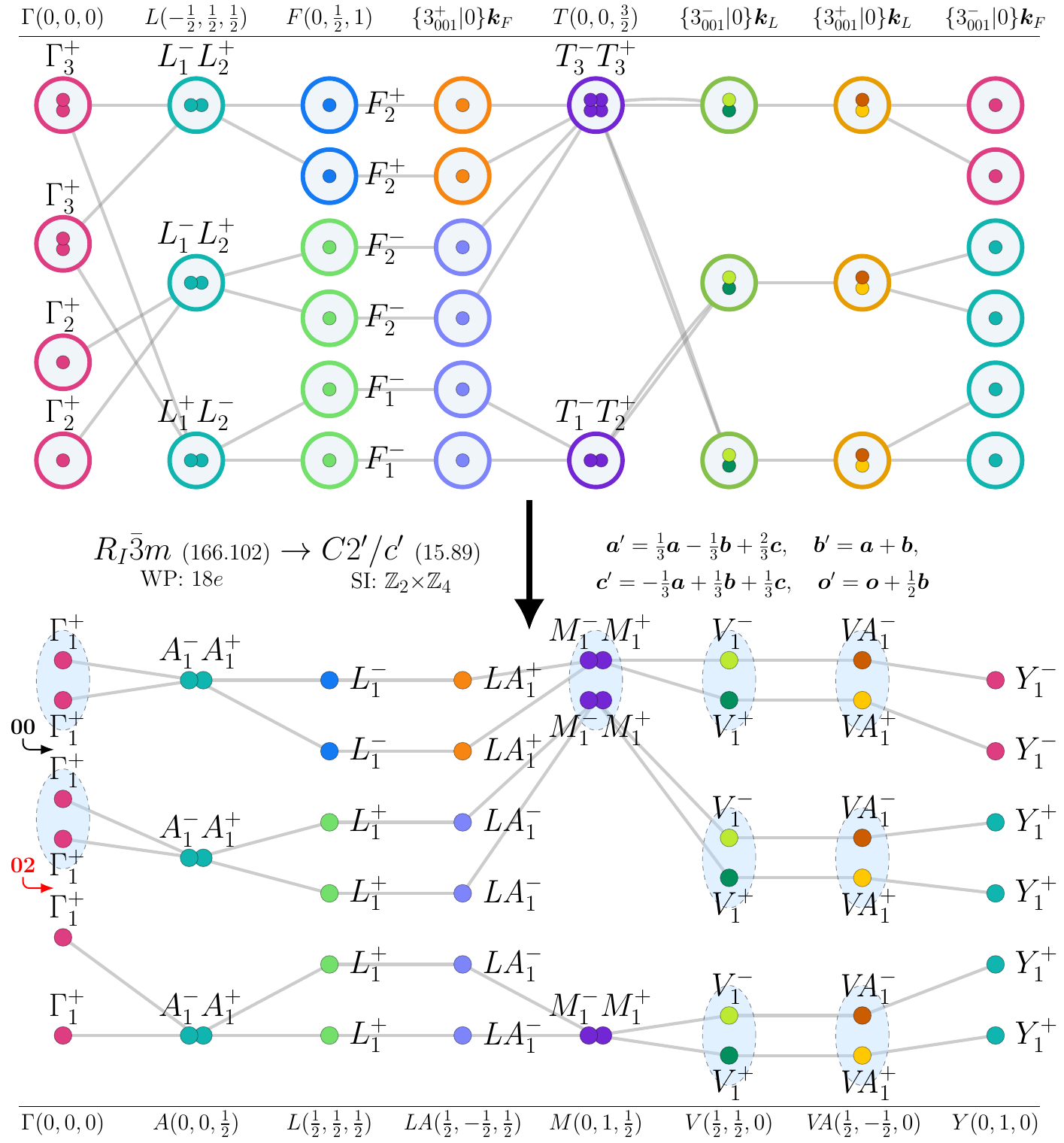}
\caption{Topological magnon bands in subgroup $C2'/c'~(15.89)$ for magnetic moments on Wyckoff position $18e$ of supergroup $R_{I}\bar{3}m~(166.102)$.\label{fig_166.102_15.89_Bparallel001andstrainparallel110orstrainperp110_18e}}
\end{figure}
\input{gap_tables_tex/166.102_15.89_Bparallel001andstrainparallel110orstrainperp110_18e_table.tex}
\input{si_tables_tex/166.102_15.89_Bparallel001andstrainparallel110orstrainperp110_18e_table.tex}
\subsubsection{Topological bands in subgroup $C2'/c'~(15.89)$}
\textbf{Perturbations:}
\begin{itemize}
\item B $\parallel$ [001] and (strain $\parallel$ [100] or strain $\perp$ [100]),
\item B $\perp$ [100].
\end{itemize}
\begin{figure}[H]
\centering
\includegraphics[scale=0.6]{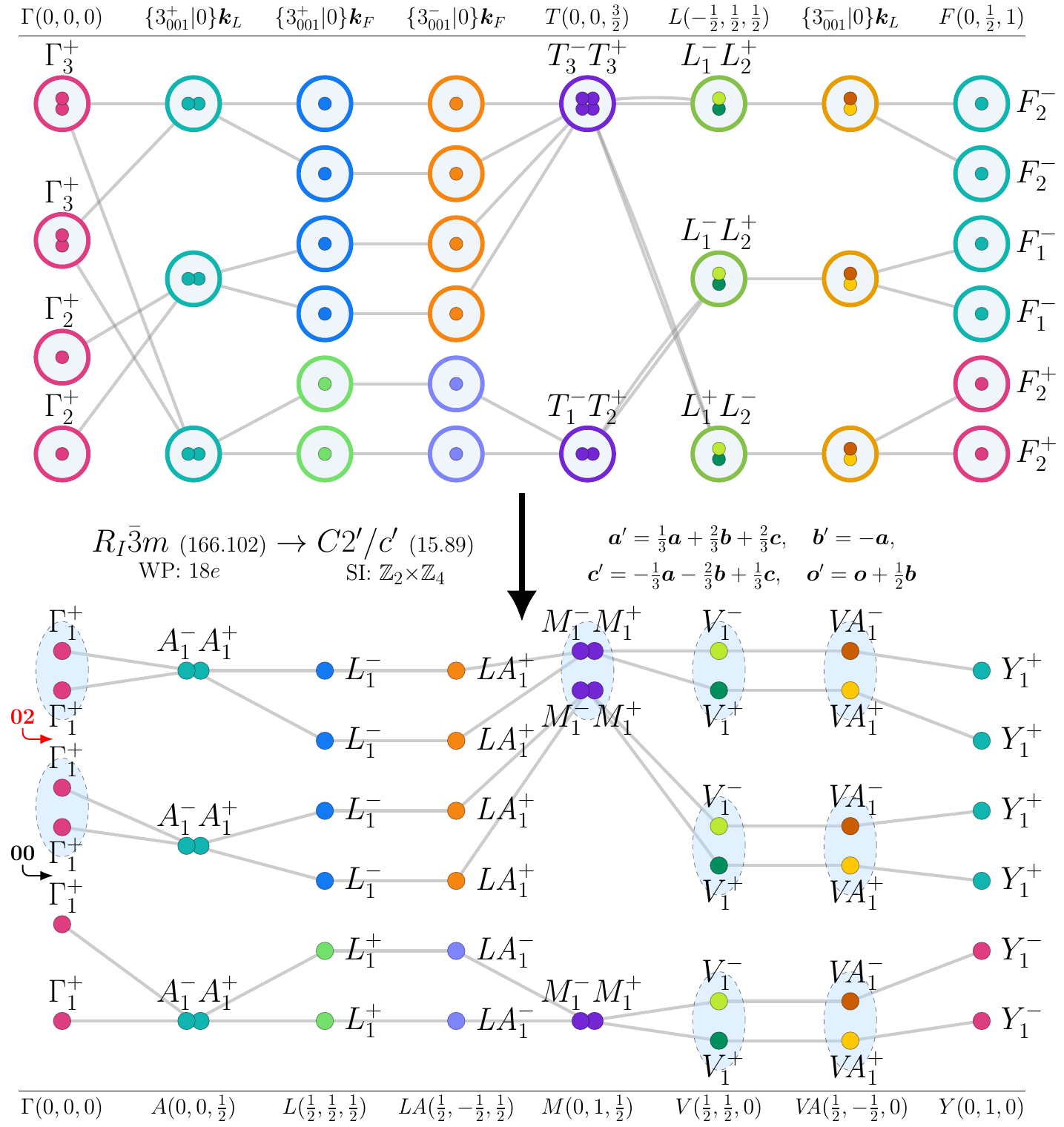}
\caption{Topological magnon bands in subgroup $C2'/c'~(15.89)$ for magnetic moments on Wyckoff position $18e$ of supergroup $R_{I}\bar{3}m~(166.102)$.\label{fig_166.102_15.89_Bparallel001andstrainparallel100orstrainperp100_18e}}
\end{figure}
\input{gap_tables_tex/166.102_15.89_Bparallel001andstrainparallel100orstrainperp100_18e_table.tex}
\input{si_tables_tex/166.102_15.89_Bparallel001andstrainparallel100orstrainperp100_18e_table.tex}
\subsubsection{Topological bands in subgroup $P_{S}\bar{1}~(2.7)$}
\textbf{Perturbation:}
\begin{itemize}
\item strain in generic direction.
\end{itemize}
\begin{figure}[H]
\centering
\includegraphics[scale=0.6]{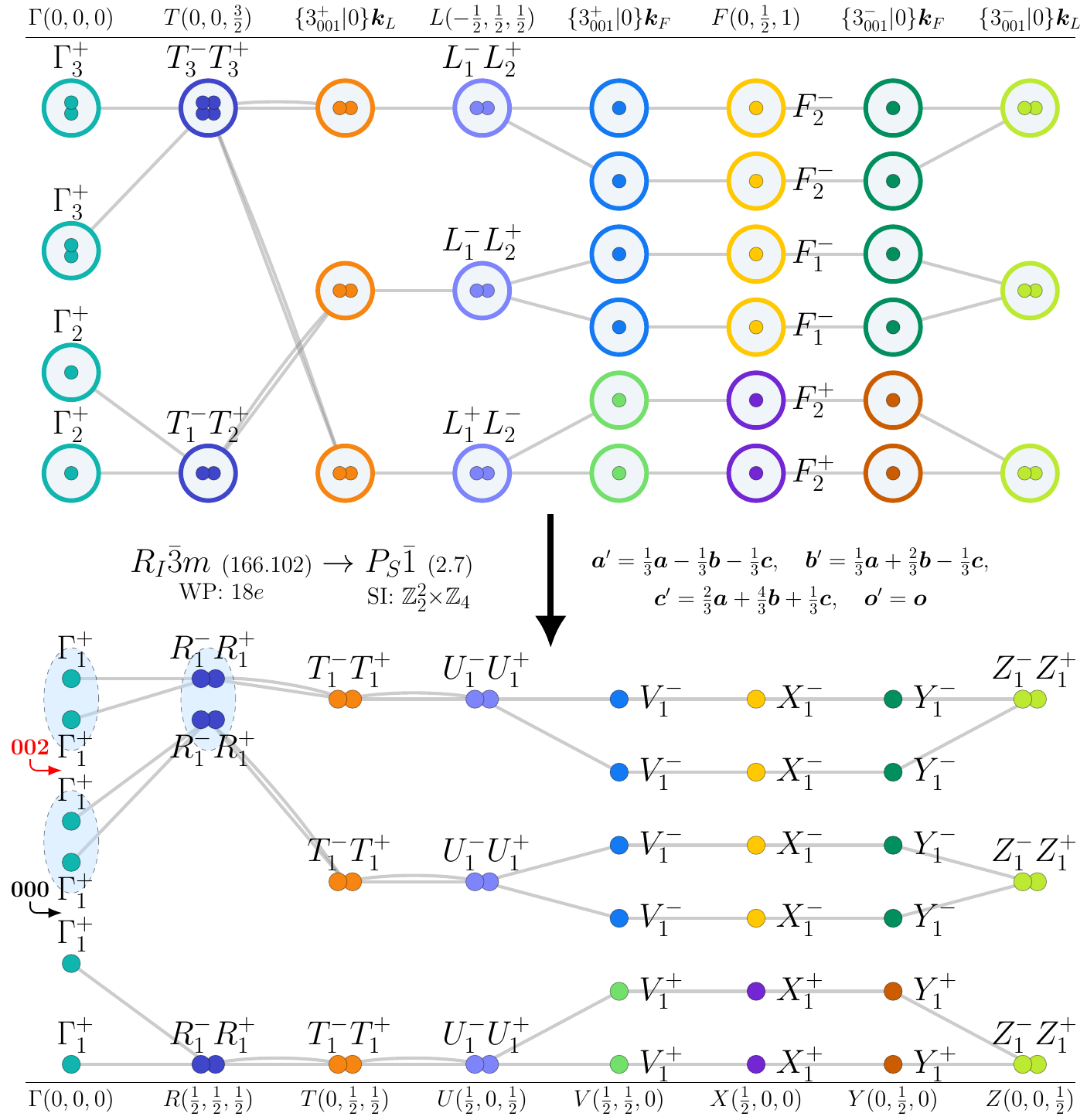}
\caption{Topological magnon bands in subgroup $P_{S}\bar{1}~(2.7)$ for magnetic moments on Wyckoff position $18e$ of supergroup $R_{I}\bar{3}m~(166.102)$.\label{fig_166.102_2.7_strainingenericdirection_18e}}
\end{figure}
\input{gap_tables_tex/166.102_2.7_strainingenericdirection_18e_table.tex}
\input{si_tables_tex/166.102_2.7_strainingenericdirection_18e_table.tex}

\section{MSG $R\bar{3}m~(166.97)$}
\textbf{Nontrivial-SI Subgroups:} $P\bar{1}~(2.4)$, $C2/m~(12.58)$, $C2/m~(12.58)$, $R\bar{3}~(148.17)$.\\

\textbf{Trivial-SI Subgroups:} $Cm~(8.32)$, $Cm~(8.32)$, $C2~(5.13)$, $C2~(5.13)$, $R3~(146.10)$, $R3m~(160.65)$.\\

\subsection{WP: $9e$}
\textbf{BCS Materials:} {Mn\textsubscript{3}Cu\textsubscript{0.5}Ge\textsubscript{0.5}N~(380 K)}\footnote{BCS web page: \texttt{\href{http://webbdcrista1.ehu.es/magndata/index.php?this\_label=0.74} {http://webbdcrista1.ehu.es/magndata/index.php?this\_label=0.74}}}, {Mn\textsubscript{3}GaN~(298 K)}\footnote{BCS web page: \texttt{\href{http://webbdcrista1.ehu.es/magndata/index.php?this\_label=0.177} {http://webbdcrista1.ehu.es/magndata/index.php?this\_label=0.177}}}, {Mn\textsubscript{3}ZnN~(183 K)}\footnote{BCS web page: \texttt{\href{http://webbdcrista1.ehu.es/magndata/index.php?this\_label=0.273} {http://webbdcrista1.ehu.es/magndata/index.php?this\_label=0.273}}}.\\
\subsubsection{Topological bands in subgroup $P\bar{1}~(2.4)$}
\textbf{Perturbations:}
\begin{itemize}
\item strain in generic direction,
\item B $\parallel$ [001] and (strain $\parallel$ [100] or strain $\perp$ [100]),
\item B $\parallel$ [001] and (strain $\parallel$ [110] or strain $\perp$ [110]),
\item B $\parallel$ [100] and (strain $\parallel$ [110] or strain $\perp$ [110]),
\item B $\parallel$ [110] and (strain $\parallel$ [100] or strain $\perp$ [100]),
\item B in generic direction.
\end{itemize}
\begin{figure}[H]
\centering
\includegraphics[scale=0.6]{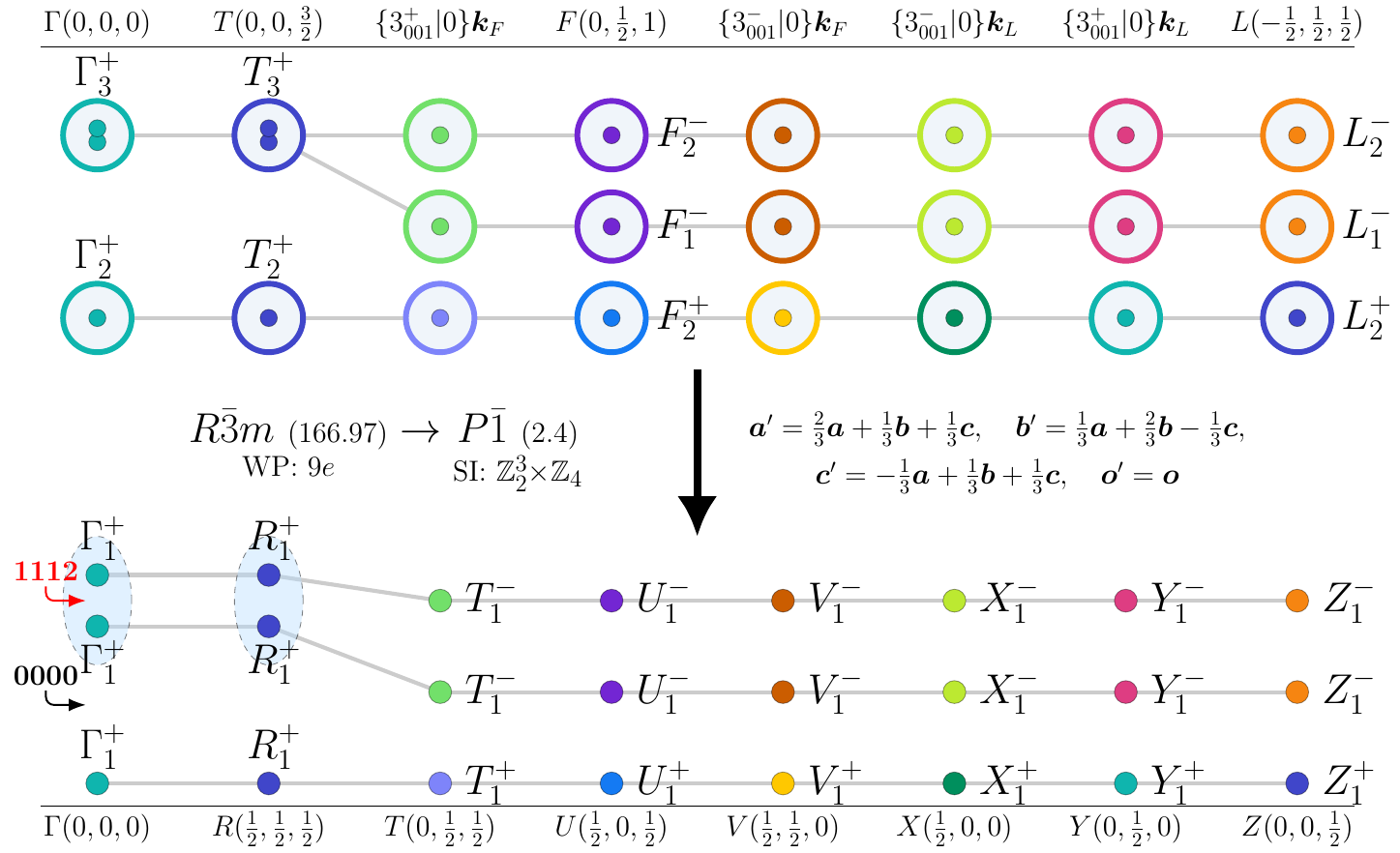}
\caption{Topological magnon bands in subgroup $P\bar{1}~(2.4)$ for magnetic moments on Wyckoff position $9e$ of supergroup $R\bar{3}m~(166.97)$.\label{fig_166.97_2.4_strainingenericdirection_9e}}
\end{figure}
\input{gap_tables_tex/166.97_2.4_strainingenericdirection_9e_table.tex}
\input{si_tables_tex/166.97_2.4_strainingenericdirection_9e_table.tex}

\section{MSG $R_{I}\bar{3}c~(167.108)$}
\textbf{Nontrivial-SI Subgroups:} $P\bar{1}~(2.4)$, $C2'/m'~(12.62)$, $C2'/m'~(12.62)$, $P_{S}\bar{1}~(2.7)$, $C2/c~(15.85)$, $C_{c}2/c~(15.90)$, $C2/c~(15.85)$, $C_{c}2/c~(15.90)$, $R\bar{3}m'~(166.101)$.\\

\textbf{Trivial-SI Subgroups:} $Cm'~(8.34)$, $Cm'~(8.34)$, $C2'~(5.15)$, $C2'~(5.15)$, $P_{S}1~(1.3)$, $Cc~(9.37)$, $C_{c}c~(9.40)$, $Cc~(9.37)$, $C_{c}c~(9.40)$, $C2~(5.13)$, $C_{c}2~(5.16)$, $C2~(5.13)$, $C_{c}2~(5.16)$, $R3m'~(160.67)$, $R_{I}3c~(161.72)$.\\

\subsection{WP: $18e$}
\textbf{BCS Materials:} {Mn\textsubscript{3}GaC~(164 K)}\footnote{BCS web page: \texttt{\href{http://webbdcrista1.ehu.es/magndata/index.php?this\_label=1.153} {http://webbdcrista1.ehu.es/magndata/index.php?this\_label=1.153}}}, {KFe\textsubscript{3}(OH)\textsubscript{6}(SO\textsubscript{4})\textsubscript{2}~(64 K)}\footnote{BCS web page: \texttt{\href{http://webbdcrista1.ehu.es/magndata/index.php?this\_label=1.25} {http://webbdcrista1.ehu.es/magndata/index.php?this\_label=1.25}}}, {NaFe\textsubscript{3}(SO\textsubscript{4})\textsubscript{2}(OH)\textsubscript{6}~(50 K)}\footnote{BCS web page: \texttt{\href{http://webbdcrista1.ehu.es/magndata/index.php?this\_label=1.441} {http://webbdcrista1.ehu.es/magndata/index.php?this\_label=1.441}}}, {AgFe\textsubscript{3}(SO\textsubscript{4})\textsubscript{2}(OD)\textsubscript{6}}\footnote{BCS web page: \texttt{\href{http://webbdcrista1.ehu.es/magndata/index.php?this\_label=1.129} {http://webbdcrista1.ehu.es/magndata/index.php?this\_label=1.129}}}.\\
\subsubsection{Topological bands in subgroup $P\bar{1}~(2.4)$}
\textbf{Perturbations:}
\begin{itemize}
\item B $\parallel$ [001] and strain in generic direction,
\item B $\parallel$ [100] and (strain $\parallel$ [110] or strain $\perp$ [110]),
\item B $\parallel$ [100] and strain in generic direction,
\item B $\parallel$ [110] and (strain $\parallel$ [100] or strain $\perp$ [100]),
\item B $\parallel$ [110] and strain in generic direction,
\item B in generic direction,
\item B $\perp$ [100] and (strain $\parallel$ [110] or strain $\perp$ [110]),
\item B $\perp$ [100] and strain in generic direction,
\item B $\perp$ [110] and (strain $\parallel$ [100] or strain $\perp$ [100]),
\item B $\perp$ [110] and strain in generic direction.
\end{itemize}
\begin{figure}[H]
\centering
\includegraphics[scale=0.6]{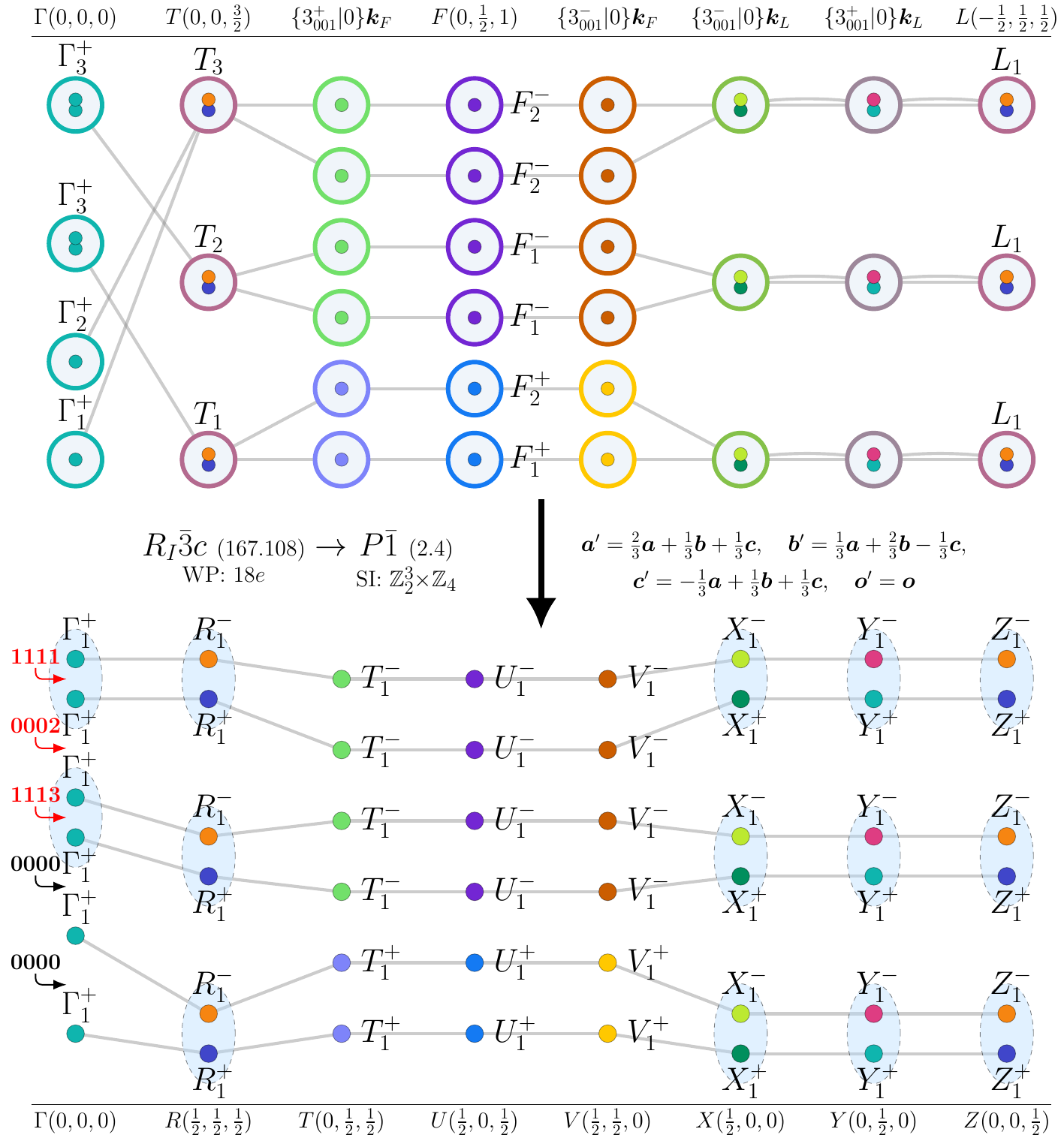}
\caption{Topological magnon bands in subgroup $P\bar{1}~(2.4)$ for magnetic moments on Wyckoff position $18e$ of supergroup $R_{I}\bar{3}c~(167.108)$.\label{fig_167.108_2.4_Bparallel001andstrainingenericdirection_18e}}
\end{figure}
\input{gap_tables_tex/167.108_2.4_Bparallel001andstrainingenericdirection_18e_table.tex}
\input{si_tables_tex/167.108_2.4_Bparallel001andstrainingenericdirection_18e_table.tex}
\subsubsection{Topological bands in subgroup $C2'/m'~(12.62)$}
\textbf{Perturbations:}
\begin{itemize}
\item B $\parallel$ [001] and (strain $\parallel$ [110] or strain $\perp$ [110]),
\item B $\perp$ [110].
\end{itemize}
\begin{figure}[H]
\centering
\includegraphics[scale=0.6]{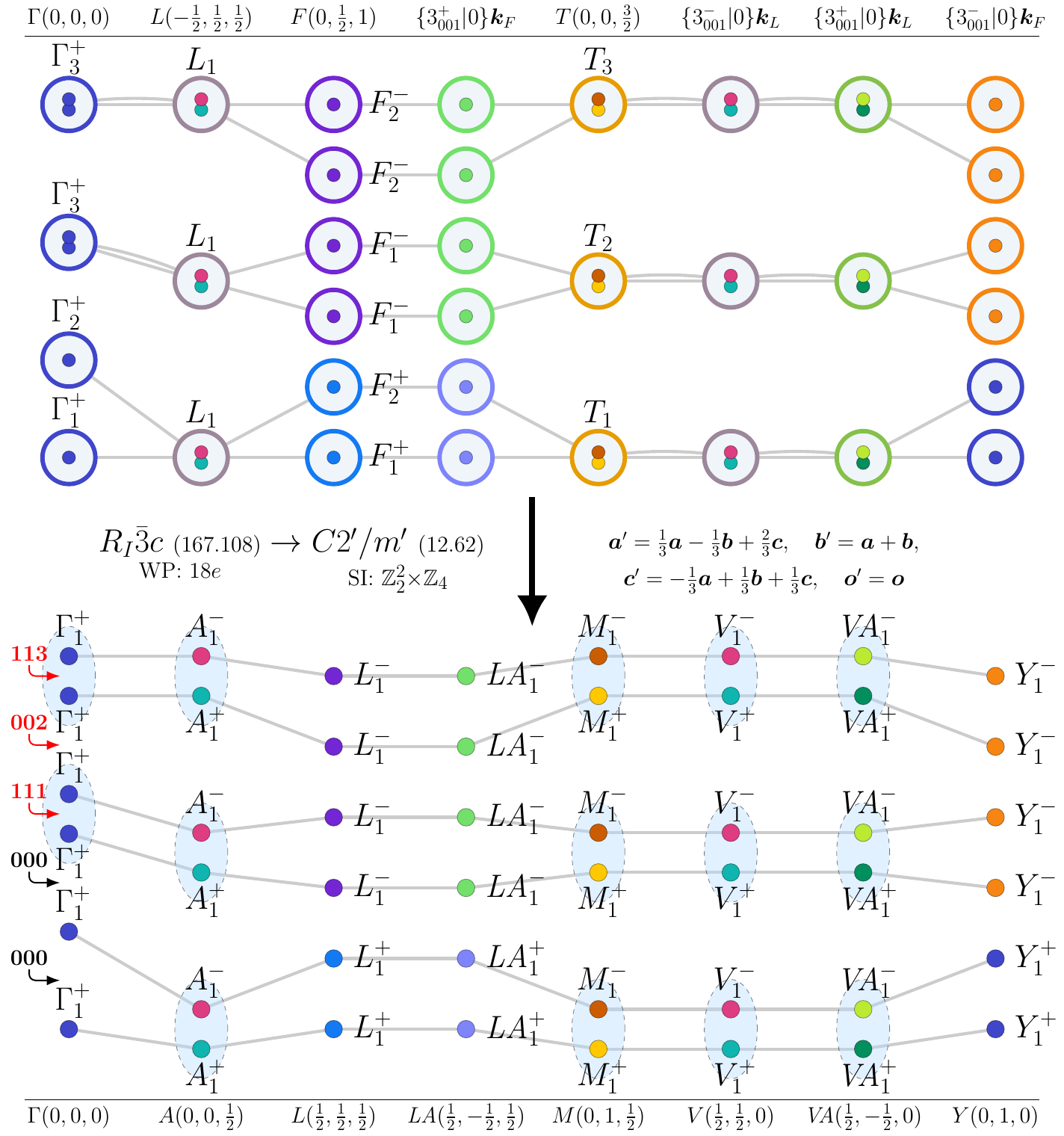}
\caption{Topological magnon bands in subgroup $C2'/m'~(12.62)$ for magnetic moments on Wyckoff position $18e$ of supergroup $R_{I}\bar{3}c~(167.108)$.\label{fig_167.108_12.62_Bparallel001andstrainparallel110orstrainperp110_18e}}
\end{figure}
\input{gap_tables_tex/167.108_12.62_Bparallel001andstrainparallel110orstrainperp110_18e_table.tex}
\input{si_tables_tex/167.108_12.62_Bparallel001andstrainparallel110orstrainperp110_18e_table.tex}
\subsubsection{Topological bands in subgroup $C2'/m'~(12.62)$}
\textbf{Perturbations:}
\begin{itemize}
\item B $\parallel$ [001] and (strain $\parallel$ [100] or strain $\perp$ [100]),
\item B $\perp$ [100].
\end{itemize}
\begin{figure}[H]
\centering
\includegraphics[scale=0.6]{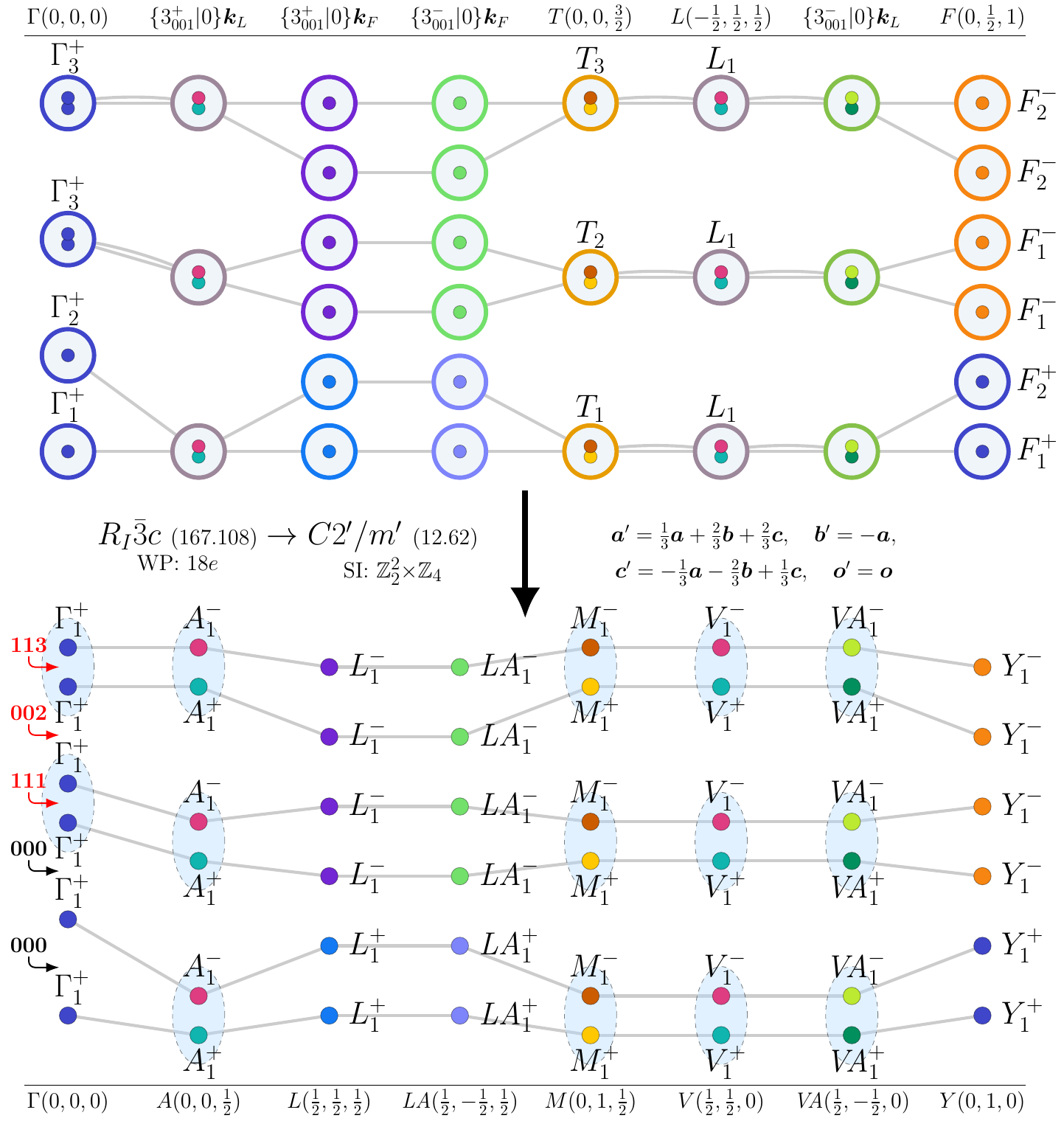}
\caption{Topological magnon bands in subgroup $C2'/m'~(12.62)$ for magnetic moments on Wyckoff position $18e$ of supergroup $R_{I}\bar{3}c~(167.108)$.\label{fig_167.108_12.62_Bparallel001andstrainparallel100orstrainperp100_18e}}
\end{figure}
\input{gap_tables_tex/167.108_12.62_Bparallel001andstrainparallel100orstrainperp100_18e_table.tex}
\input{si_tables_tex/167.108_12.62_Bparallel001andstrainparallel100orstrainperp100_18e_table.tex}
\subsubsection{Topological bands in subgroup $P_{S}\bar{1}~(2.7)$}
\textbf{Perturbation:}
\begin{itemize}
\item strain in generic direction.
\end{itemize}
\begin{figure}[H]
\centering
\includegraphics[scale=0.6]{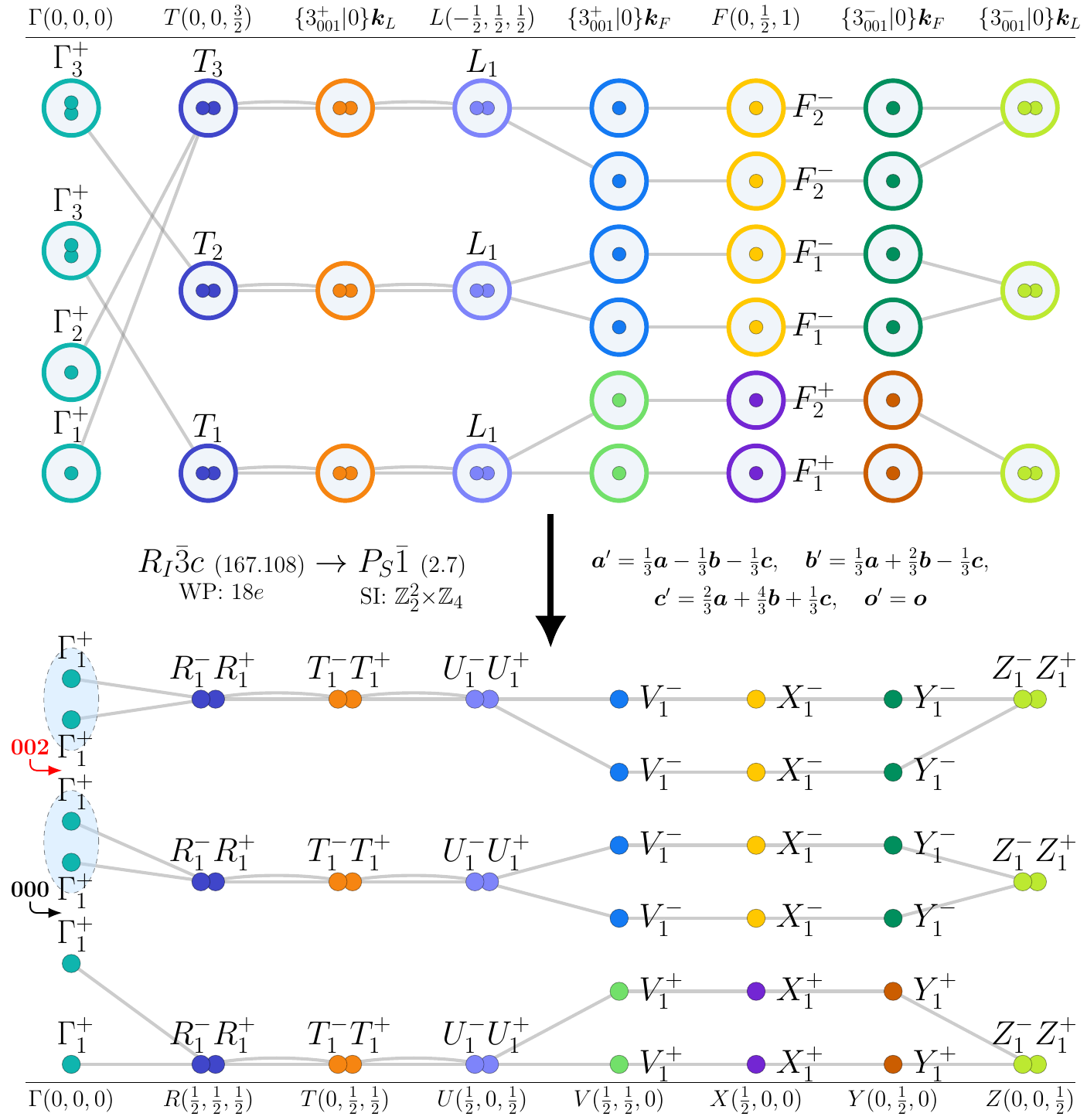}
\caption{Topological magnon bands in subgroup $P_{S}\bar{1}~(2.7)$ for magnetic moments on Wyckoff position $18e$ of supergroup $R_{I}\bar{3}c~(167.108)$.\label{fig_167.108_2.7_strainingenericdirection_18e}}
\end{figure}
\input{gap_tables_tex/167.108_2.7_strainingenericdirection_18e_table.tex}
\input{si_tables_tex/167.108_2.7_strainingenericdirection_18e_table.tex}

\section{MSG $P6_{3}/m~(176.143)$}
\textbf{Nontrivial-SI Subgroups:} $P\bar{1}~(2.4)$, $P2_{1}/m~(11.50)$, $P6_{3}~(173.129)$.\\

\textbf{Trivial-SI Subgroups:} $Pm~(6.18)$, $P2_{1}~(4.7)$.\\

\subsection{WP: $6g$}
\textbf{BCS Materials:} {FeF\textsubscript{3}~(97 K)}\footnote{BCS web page: \texttt{\href{http://webbdcrista1.ehu.es/magndata/index.php?this\_label=1.0.33} {http://webbdcrista1.ehu.es/magndata/index.php?this\_label=1.0.33}}}.\\
\subsubsection{Topological bands in subgroup $P\bar{1}~(2.4)$}
\textbf{Perturbations:}
\begin{itemize}
\item strain in generic direction,
\item B in generic direction.
\end{itemize}
\begin{figure}[H]
\centering
\includegraphics[scale=0.6]{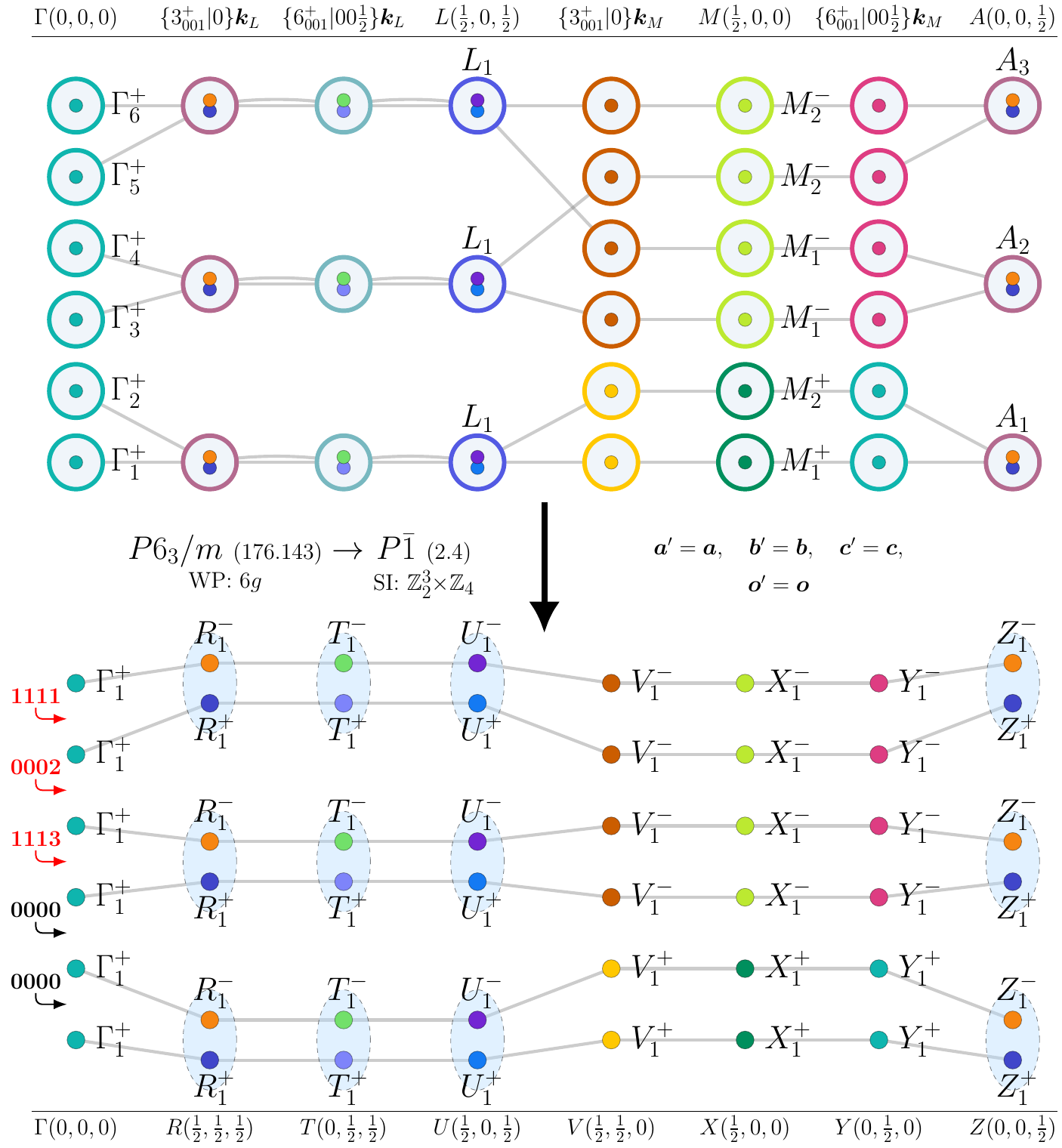}
\caption{Topological magnon bands in subgroup $P\bar{1}~(2.4)$ for magnetic moments on Wyckoff position $6g$ of supergroup $P6_{3}/m~(176.143)$.\label{fig_176.143_2.4_strainingenericdirection_6g}}
\end{figure}
\input{gap_tables_tex/176.143_2.4_strainingenericdirection_6g_table.tex}
\input{si_tables_tex/176.143_2.4_strainingenericdirection_6g_table.tex}

\section{MSG $P_{B}2_{1}2_{1}2~(18.22)$}
\textbf{Nontrivial-SI Subgroups:} $P2~(3.1)$, $P22'2_{1}'~(17.10)$.\\

\textbf{Trivial-SI Subgroups:} $P2_{1}'~(4.9)$, $P2_{1}'~(4.9)$, $P2'~(3.3)$, $P_{S}1~(1.3)$, $P_{C}2~(3.6)$, $P2_{1}~(4.7)$, $P2_{1}2_{1}'2'~(18.19)$, $P_{a}2_{1}~(4.10)$, $P2_{1}~(4.7)$, $P2_{1}'2_{1}'2_{1}~(19.27)$, $P_{C}2_{1}~(4.12)$.\\

\subsection{WP: $4a$}
\textbf{BCS Materials:} {CoNb\textsubscript{3}S\textsubscript{6}~(25 K)}\footnote{BCS web page: \texttt{\href{http://webbdcrista1.ehu.es/magndata/index.php?this\_label=1.349} {http://webbdcrista1.ehu.es/magndata/index.php?this\_label=1.349}}}.\\
\subsubsection{Topological bands in subgroup $P22'2_{1}'~(17.10)$}
\textbf{Perturbation:}
\begin{itemize}
\item B $\parallel$ [001].
\end{itemize}
\begin{figure}[H]
\centering
\includegraphics[scale=0.6]{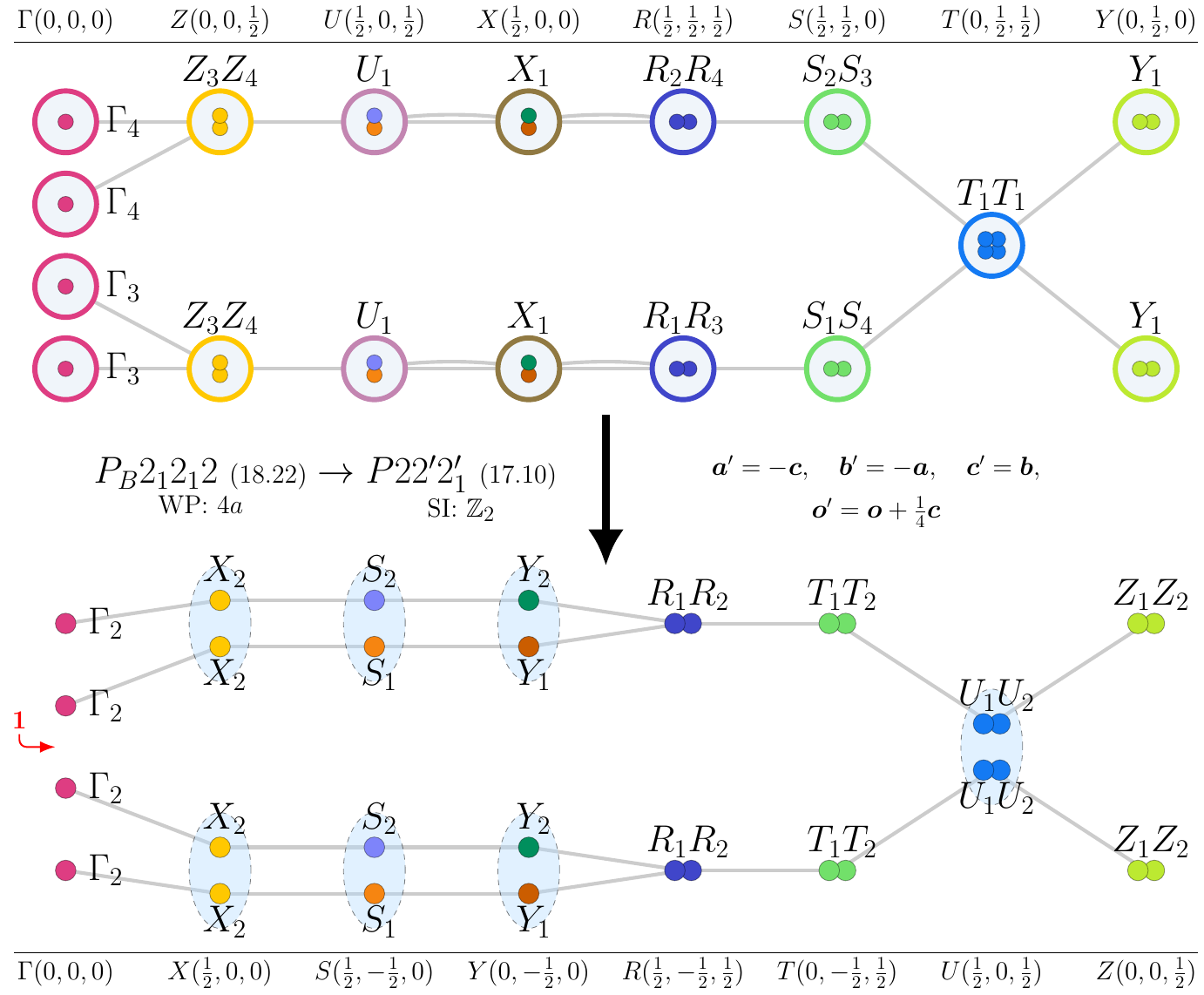}
\caption{Topological magnon bands in subgroup $P22'2_{1}'~(17.10)$ for magnetic moments on Wyckoff position $4a$ of supergroup $P_{B}2_{1}2_{1}2~(18.22)$.\label{fig_18.22_17.10_Bparallel001_4a}}
\end{figure}
\input{gap_tables_tex/18.22_17.10_Bparallel001_4a_table.tex}
\input{si_tables_tex/18.22_17.10_Bparallel001_4a_table.tex}

\section{MSG $P_{c}6/mcc~(192.252)$}
\textbf{Nontrivial-SI Subgroups:} $P\bar{1}~(2.4)$, $C2'/m'~(12.62)$, $C2'/m'~(12.62)$, $P2_{1}'/m'~(11.54)$, $P_{S}\bar{1}~(2.7)$, $C2/c~(15.85)$, $Cm'cm'~(63.464)$, $C_{c}2/c~(15.90)$, $C2/c~(15.85)$, $Cm'cm'~(63.464)$, $C_{c}2/c~(15.90)$, $P2~(3.1)$, $Cm'm'2~(35.168)$, $Cm'm'2~(35.168)$, $P_{b}2~(3.5)$, $C_{c}cc2~(37.184)$, $C_{c}cc2~(37.184)$, $P2/m~(10.42)$, $Cm'm'm~(65.485)$, $Cm'm'm~(65.485)$, $P_{b}2/m~(10.48)$, $C_{c}ccm~(66.498)$, $C_{c}ccm~(66.498)$, $P6m'm'~(183.189)$, $P_{c}6cc~(184.196)$, $P6/mm'm'~(191.240)$.\\

\textbf{Trivial-SI Subgroups:} $Cm'~(8.34)$, $Cm'~(8.34)$, $Pm'~(6.20)$, $C2'~(5.15)$, $C2'~(5.15)$, $P2_{1}'~(4.9)$, $P_{S}1~(1.3)$, $Cc~(9.37)$, $Cm'c2_{1}'~(36.174)$, $C_{c}c~(9.40)$, $Cc~(9.37)$, $Cm'c2_{1}'~(36.174)$, $C_{c}c~(9.40)$, $Pm~(6.18)$, $Amm'2'~(38.190)$, $Amm'2'~(38.190)$, $P_{b}m~(6.22)$, $C2~(5.13)$, $Am'm'2~(38.191)$, $C_{c}2~(5.16)$, $A_{a}ma2~(40.208)$, $C2~(5.13)$, $Am'm'2~(38.191)$, $C_{c}2~(5.16)$, $A_{a}ma2~(40.208)$.\\

\subsection{WP: $6f$}
\textbf{BCS Materials:} {FeGe~(410 K)}\footnote{BCS web page: \texttt{\href{http://webbdcrista1.ehu.es/magndata/index.php?this\_label=1.629} {http://webbdcrista1.ehu.es/magndata/index.php?this\_label=1.629}}}.\\
\subsubsection{Topological bands in subgroup $P\bar{1}~(2.4)$}
\textbf{Perturbations:}
\begin{itemize}
\item B $\parallel$ [001] and strain in generic direction,
\item B $\parallel$ [100] and strain $\perp$ [110],
\item B $\parallel$ [100] and strain in generic direction,
\item B $\parallel$ [110] and strain $\perp$ [100],
\item B $\parallel$ [110] and strain in generic direction,
\item B $\perp$ [001] and strain $\perp$ [100],
\item B $\perp$ [001] and strain $\perp$ [110],
\item B $\perp$ [001] and strain in generic direction,
\item B $\perp$ [100] and strain $\parallel$ [110],
\item B $\perp$ [100] and strain $\perp$ [001],
\item B $\perp$ [100] and strain $\perp$ [110],
\item B $\perp$ [100] and strain in generic direction,
\item B $\perp$ [110] and strain $\parallel$ [100],
\item B $\perp$ [110] and strain $\perp$ [001],
\item B $\perp$ [110] and strain $\perp$ [100],
\item B $\perp$ [110] and strain in generic direction,
\item B in generic direction.
\end{itemize}
\begin{figure}[H]
\centering
\includegraphics[scale=0.6]{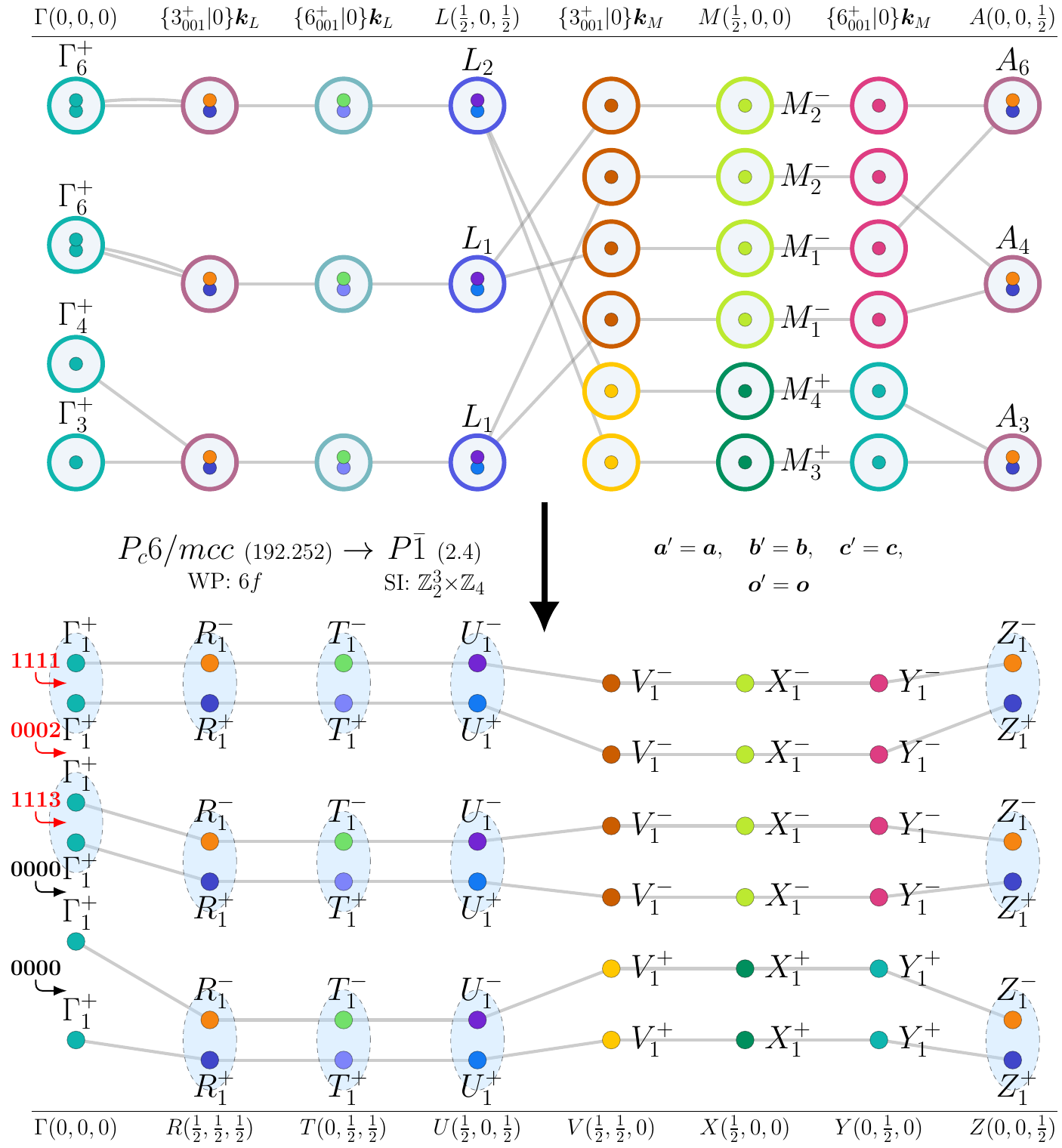}
\caption{Topological magnon bands in subgroup $P\bar{1}~(2.4)$ for magnetic moments on Wyckoff position $6f$ of supergroup $P_{c}6/mcc~(192.252)$.\label{fig_192.252_2.4_Bparallel001andstrainingenericdirection_6f}}
\end{figure}
\input{gap_tables_tex/192.252_2.4_Bparallel001andstrainingenericdirection_6f_table.tex}
\input{si_tables_tex/192.252_2.4_Bparallel001andstrainingenericdirection_6f_table.tex}
\subsubsection{Topological bands in subgroup $C2'/m'~(12.62)$}
\textbf{Perturbations:}
\begin{itemize}
\item B $\parallel$ [001] and strain $\perp$ [110],
\item B $\perp$ [110].
\end{itemize}
\begin{figure}[H]
\centering
\includegraphics[scale=0.6]{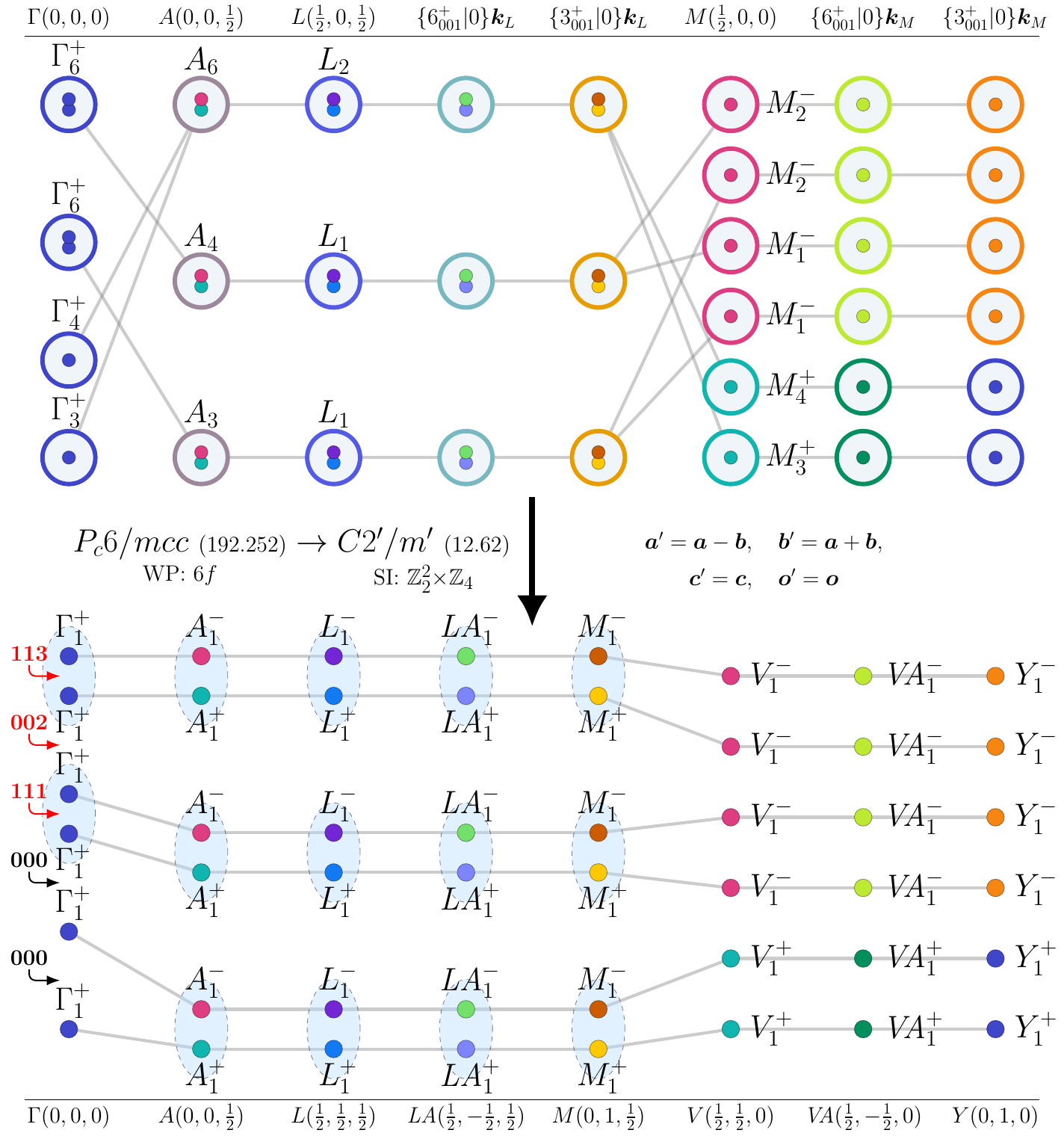}
\caption{Topological magnon bands in subgroup $C2'/m'~(12.62)$ for magnetic moments on Wyckoff position $6f$ of supergroup $P_{c}6/mcc~(192.252)$.\label{fig_192.252_12.62_Bparallel001andstrainperp110_6f}}
\end{figure}
\input{gap_tables_tex/192.252_12.62_Bparallel001andstrainperp110_6f_table.tex}
\input{si_tables_tex/192.252_12.62_Bparallel001andstrainperp110_6f_table.tex}
\subsubsection{Topological bands in subgroup $C2'/m'~(12.62)$}
\textbf{Perturbations:}
\begin{itemize}
\item B $\parallel$ [001] and strain $\perp$ [100],
\item B $\perp$ [100].
\end{itemize}
\begin{figure}[H]
\centering
\includegraphics[scale=0.6]{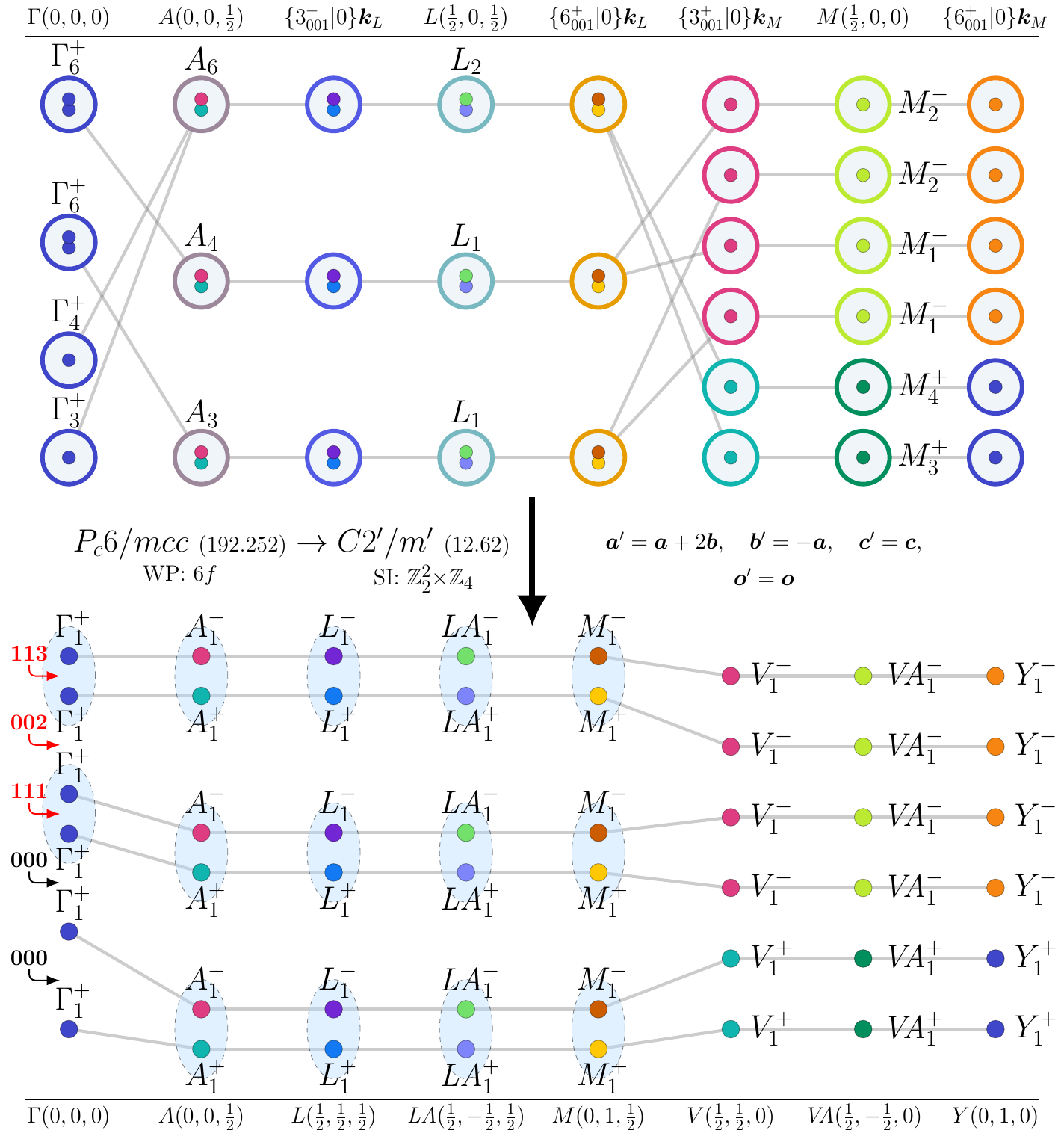}
\caption{Topological magnon bands in subgroup $C2'/m'~(12.62)$ for magnetic moments on Wyckoff position $6f$ of supergroup $P_{c}6/mcc~(192.252)$.\label{fig_192.252_12.62_Bparallel001andstrainperp100_6f}}
\end{figure}
\input{gap_tables_tex/192.252_12.62_Bparallel001andstrainperp100_6f_table.tex}
\input{si_tables_tex/192.252_12.62_Bparallel001andstrainperp100_6f_table.tex}
\subsubsection{Topological bands in subgroup $P2_{1}'/m'~(11.54)$}
\textbf{Perturbations:}
\begin{itemize}
\item B $\parallel$ [100] and strain $\parallel$ [110],
\item B $\parallel$ [100] and strain $\perp$ [001],
\item B $\parallel$ [110] and strain $\parallel$ [100],
\item B $\parallel$ [110] and strain $\perp$ [001],
\item B $\perp$ [001].
\end{itemize}
\begin{figure}[H]
\centering
\includegraphics[scale=0.6]{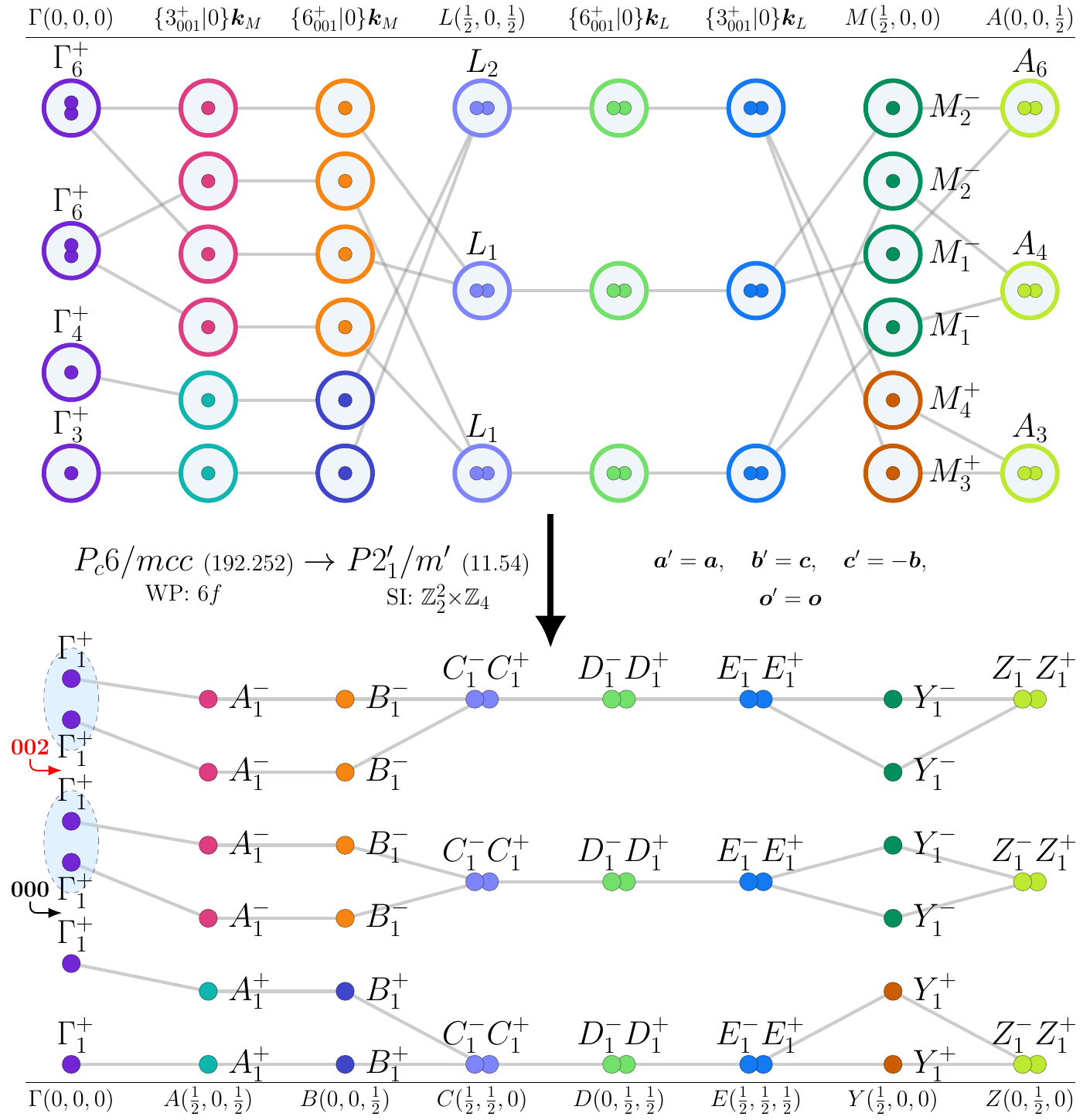}
\caption{Topological magnon bands in subgroup $P2_{1}'/m'~(11.54)$ for magnetic moments on Wyckoff position $6f$ of supergroup $P_{c}6/mcc~(192.252)$.\label{fig_192.252_11.54_Bparallel100andstrainparallel110_6f}}
\end{figure}
\input{gap_tables_tex/192.252_11.54_Bparallel100andstrainparallel110_6f_table.tex}
\input{si_tables_tex/192.252_11.54_Bparallel100andstrainparallel110_6f_table.tex}
\subsubsection{Topological bands in subgroup $P_{S}\bar{1}~(2.7)$}
\textbf{Perturbation:}
\begin{itemize}
\item strain in generic direction.
\end{itemize}
\begin{figure}[H]
\centering
\includegraphics[scale=0.6]{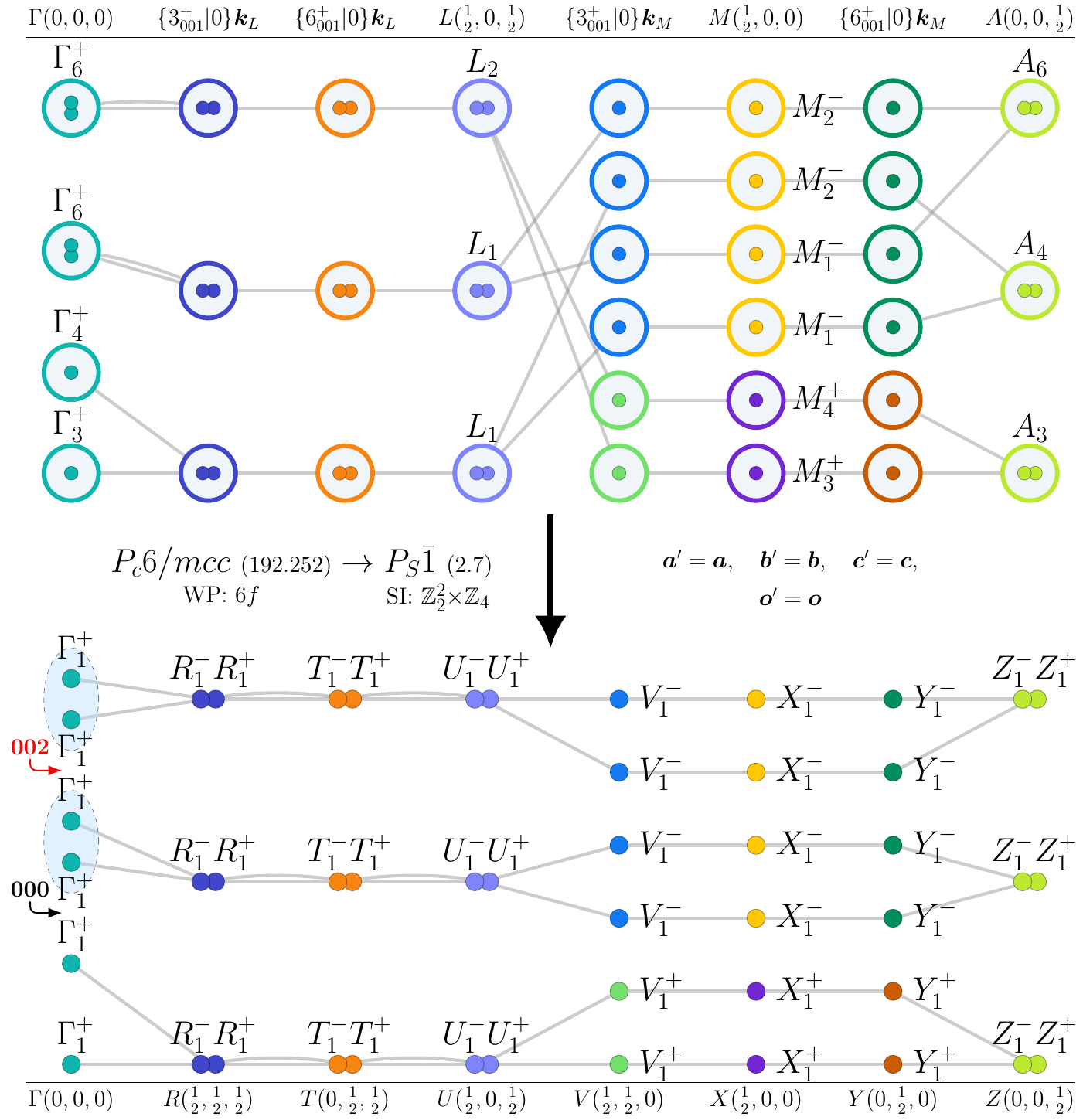}
\caption{Topological magnon bands in subgroup $P_{S}\bar{1}~(2.7)$ for magnetic moments on Wyckoff position $6f$ of supergroup $P_{c}6/mcc~(192.252)$.\label{fig_192.252_2.7_strainingenericdirection_6f}}
\end{figure}
\input{gap_tables_tex/192.252_2.7_strainingenericdirection_6f_table.tex}
\input{si_tables_tex/192.252_2.7_strainingenericdirection_6f_table.tex}

\section{MSG $Fd\bar{3}~(203.26)$}
\textbf{Nontrivial-SI Subgroups:} $P\bar{1}~(2.4)$, $R\bar{3}~(148.17)$, $C2/c~(15.85)$, $C2/c~(15.85)$, $Fddd~(70.527)$.\\

\textbf{Trivial-SI Subgroups:} $Cc~(9.37)$, $Cc~(9.37)$, $R3~(146.10)$, $C2~(5.13)$, $Fdd2~(43.224)$.\\

\subsection{WP: $16c$}
\textbf{BCS Materials:} {Na\textsubscript{3}Co(CO\textsubscript{3})\textsubscript{2}Cl~(1.5 K)}\footnote{BCS web page: \texttt{\href{http://webbdcrista1.ehu.es/magndata/index.php?this\_label=0.70} {http://webbdcrista1.ehu.es/magndata/index.php?this\_label=0.70}}}.\\
\subsubsection{Topological bands in subgroup $P\bar{1}~(2.4)$}
\textbf{Perturbations:}
\begin{itemize}
\item strain in generic direction,
\item B $\parallel$ [100] and strain $\parallel$ [110],
\item B $\parallel$ [100] and strain $\parallel$ [111],
\item B in generic direction,
\item B $\parallel$ [111] and strain $\parallel$ [100],
\item B $\parallel$ [111] and strain $\parallel$ [110],
\item B $\parallel$ [111] and strain $\perp$ [100].
\end{itemize}
\begin{figure}[H]
\centering
\includegraphics[scale=0.6]{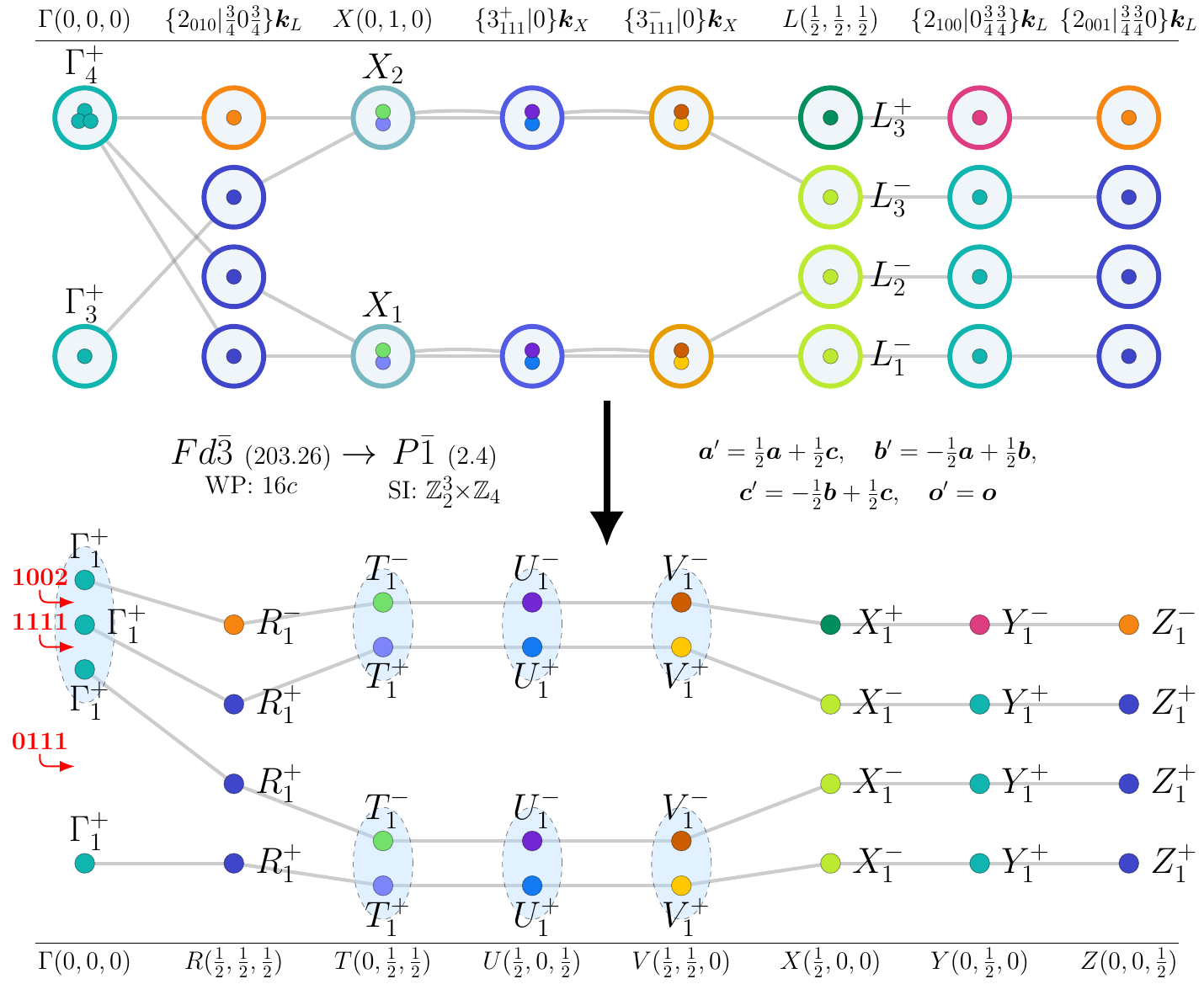}
\caption{Topological magnon bands in subgroup $P\bar{1}~(2.4)$ for magnetic moments on Wyckoff position $16c$ of supergroup $Fd\bar{3}~(203.26)$.\label{fig_203.26_2.4_strainingenericdirection_16c}}
\end{figure}
\input{gap_tables_tex/203.26_2.4_strainingenericdirection_16c_table.tex}
\input{si_tables_tex/203.26_2.4_strainingenericdirection_16c_table.tex}

\section{MSG $Pa\bar{3}~(205.33)$}
\textbf{Nontrivial-SI Subgroups:} $P\bar{1}~(2.4)$, $R\bar{3}~(148.17)$, $P2_{1}/c~(14.75)$, $P2_{1}/c~(14.75)$, $Pbca~(61.433)$.\\

\textbf{Trivial-SI Subgroups:} $Pc~(7.24)$, $Pc~(7.24)$, $R3~(146.10)$, $P2_{1}~(4.7)$, $Pca2_{1}~(29.99)$.\\

\subsection{WP: $4a$}
\textbf{BCS Materials:} {MnTe\textsubscript{2}~(86.5 K)}\footnote{BCS web page: \texttt{\href{http://webbdcrista1.ehu.es/magndata/index.php?this\_label=0.20} {http://webbdcrista1.ehu.es/magndata/index.php?this\_label=0.20}}}, {NiS\textsubscript{2}~(39 K)}\footnote{BCS web page: \texttt{\href{http://webbdcrista1.ehu.es/magndata/index.php?this\_label=0.150} {http://webbdcrista1.ehu.es/magndata/index.php?this\_label=0.150}}}.\\
\subsubsection{Topological bands in subgroup $P\bar{1}~(2.4)$}
\textbf{Perturbations:}
\begin{itemize}
\item strain in generic direction,
\item B $\parallel$ [100] and strain $\parallel$ [110],
\item B $\parallel$ [100] and strain $\parallel$ [111],
\item B in generic direction,
\item B $\parallel$ [111] and strain $\parallel$ [100],
\item B $\parallel$ [111] and strain $\parallel$ [110],
\item B $\parallel$ [111] and strain $\perp$ [100].
\end{itemize}
\begin{figure}[H]
\centering
\includegraphics[scale=0.6]{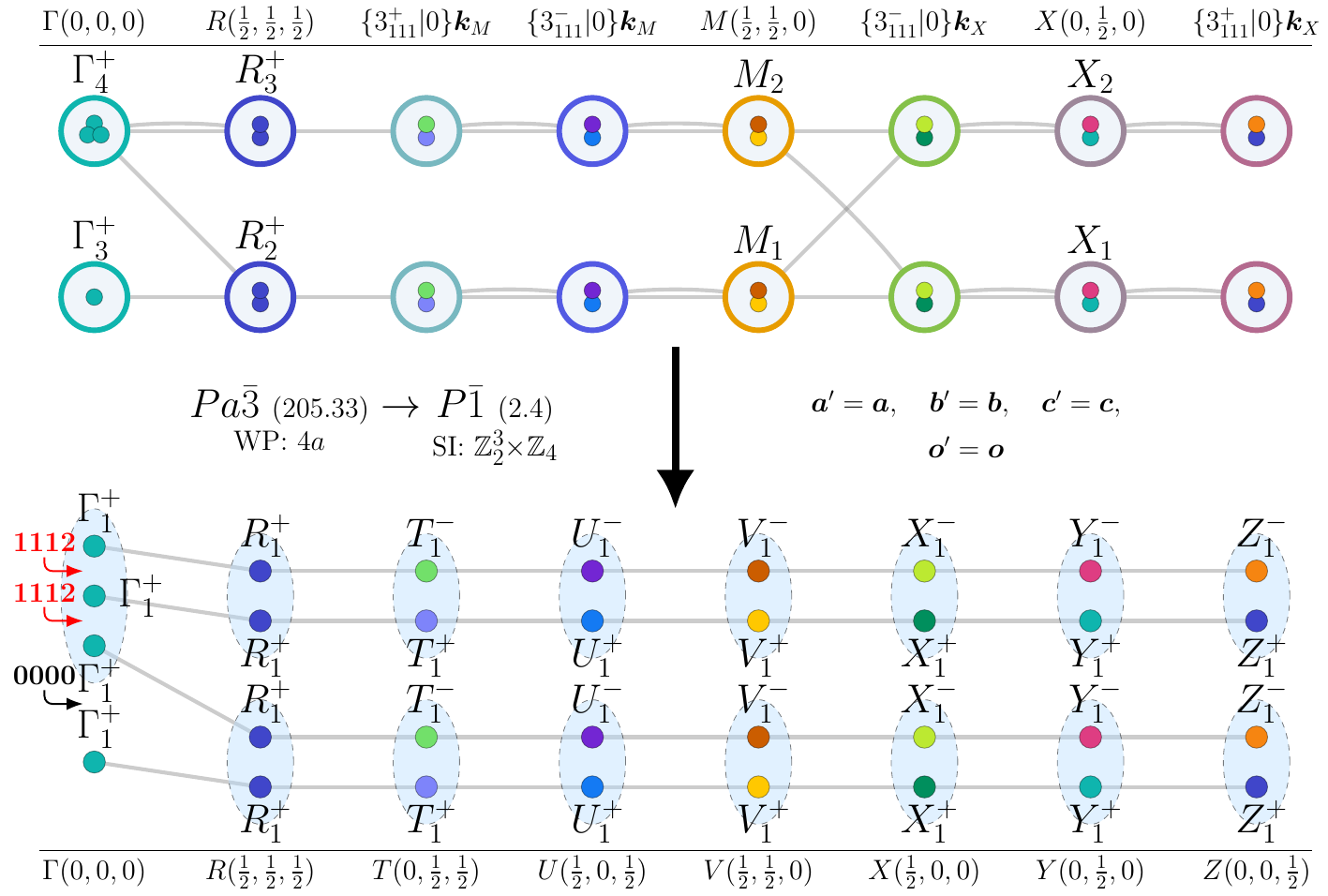}
\caption{Topological magnon bands in subgroup $P\bar{1}~(2.4)$ for magnetic moments on Wyckoff position $4a$ of supergroup $Pa\bar{3}~(205.33)$.\label{fig_205.33_2.4_strainingenericdirection_4a}}
\end{figure}
\input{gap_tables_tex/205.33_2.4_strainingenericdirection_4a_table.tex}
\input{si_tables_tex/205.33_2.4_strainingenericdirection_4a_table.tex}

\section{MSG $P_{I}a\bar{3}~(205.36)$}
\textbf{Nontrivial-SI Subgroups:} $P\bar{1}~(2.4)$, $P2'/c'~(13.69)$, $P2'/c'~(13.69)$, $P_{S}\bar{1}~(2.7)$, $R\bar{3}~(148.17)$, $R_{I}\bar{3}~(148.20)$, $P_{C}2_{1}/c~(14.84)$, $P2_{1}/c~(14.75)$, $Pcc'a'~(54.343)$, $P_{C}2_{1}/c~(14.84)$, $P_{I}bca~(61.440)$.\\

\textbf{Trivial-SI Subgroups:} $Pc'~(7.26)$, $Pc'~(7.26)$, $P2'~(3.3)$, $P_{S}1~(1.3)$, $P_{C}c~(7.30)$, $Pc~(7.24)$, $P_{C}c~(7.30)$, $R3~(146.10)$, $R_{I}3~(146.12)$, $P2_{1}~(4.7)$, $Pc'a'2_{1}~(29.103)$, $P_{C}2_{1}~(4.12)$, $P_{I}ca2_{1}~(29.110)$.\\

\subsection{WP: $8b$}
\textbf{BCS Materials:} {HoRh~(3.2 K)}\footnote{BCS web page: \texttt{\href{http://webbdcrista1.ehu.es/magndata/index.php?this\_label=3.18} {http://webbdcrista1.ehu.es/magndata/index.php?this\_label=3.18}}}.\\
\subsubsection{Topological bands in subgroup $P\bar{1}~(2.4)$}
\textbf{Perturbations:}
\begin{itemize}
\item B $\parallel$ [100] and strain $\parallel$ [111],
\item B $\parallel$ [100] and strain in generic direction,
\item B $\parallel$ [110] and strain $\parallel$ [111],
\item B $\parallel$ [110] and strain $\perp$ [100],
\item B $\parallel$ [110] and strain in generic direction,
\item B $\parallel$ [111] and strain $\parallel$ [100],
\item B $\parallel$ [111] and strain $\parallel$ [110],
\item B $\parallel$ [111] and strain $\perp$ [100],
\item B $\parallel$ [111] and strain in generic direction,
\item B $\perp$ [100] and strain $\parallel$ [110],
\item B $\perp$ [100] and strain $\parallel$ [111],
\item B $\perp$ [100] and strain in generic direction,
\item B in generic direction.
\end{itemize}
\begin{figure}[H]
\centering
\includegraphics[scale=0.6]{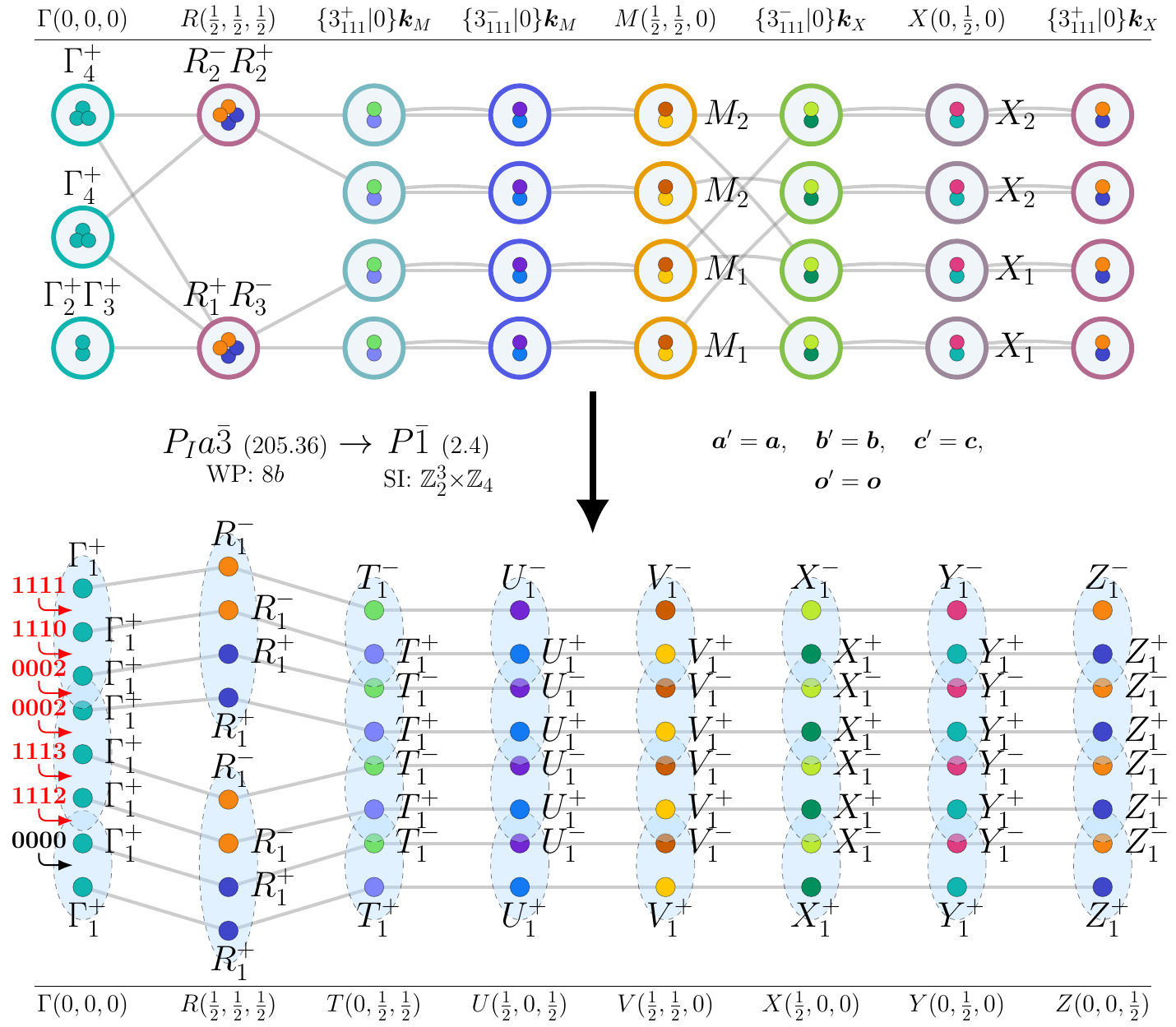}
\caption{Topological magnon bands in subgroup $P\bar{1}~(2.4)$ for magnetic moments on Wyckoff position $8b$ of supergroup $P_{I}a\bar{3}~(205.36)$.\label{fig_205.36_2.4_Bparallel100andstrainparallel111_8b}}
\end{figure}
\input{gap_tables_tex/205.36_2.4_Bparallel100andstrainparallel111_8b_table.tex}
\input{si_tables_tex/205.36_2.4_Bparallel100andstrainparallel111_8b_table.tex}
\subsubsection{Topological bands in subgroup $P2'/c'~(13.69)$}
\textbf{Perturbations:}
\begin{itemize}
\item B $\parallel$ [100] and strain $\parallel$ [110],
\item B $\parallel$ [110].
\end{itemize}
\begin{figure}[H]
\centering
\includegraphics[scale=0.6]{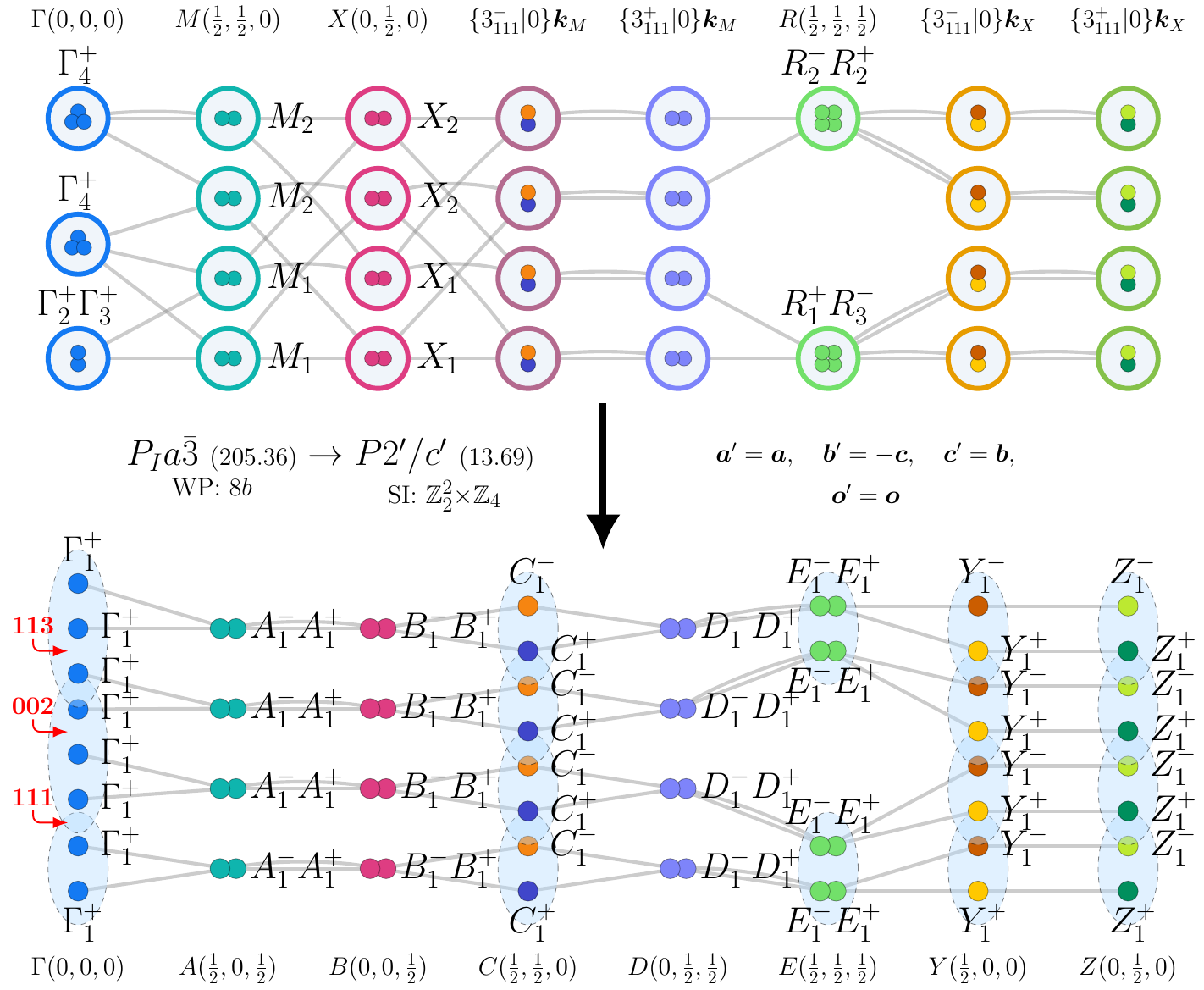}
\caption{Topological magnon bands in subgroup $P2'/c'~(13.69)$ for magnetic moments on Wyckoff position $8b$ of supergroup $P_{I}a\bar{3}~(205.36)$.\label{fig_205.36_13.69_Bparallel100andstrainparallel110_8b}}
\end{figure}
\input{gap_tables_tex/205.36_13.69_Bparallel100andstrainparallel110_8b_table.tex}
\input{si_tables_tex/205.36_13.69_Bparallel100andstrainparallel110_8b_table.tex}
\subsubsection{Topological bands in subgroup $P2'/c'~(13.69)$}
\textbf{Perturbation:}
\begin{itemize}
\item B $\perp$ [100].
\end{itemize}
\begin{figure}[H]
\centering
\includegraphics[scale=0.6]{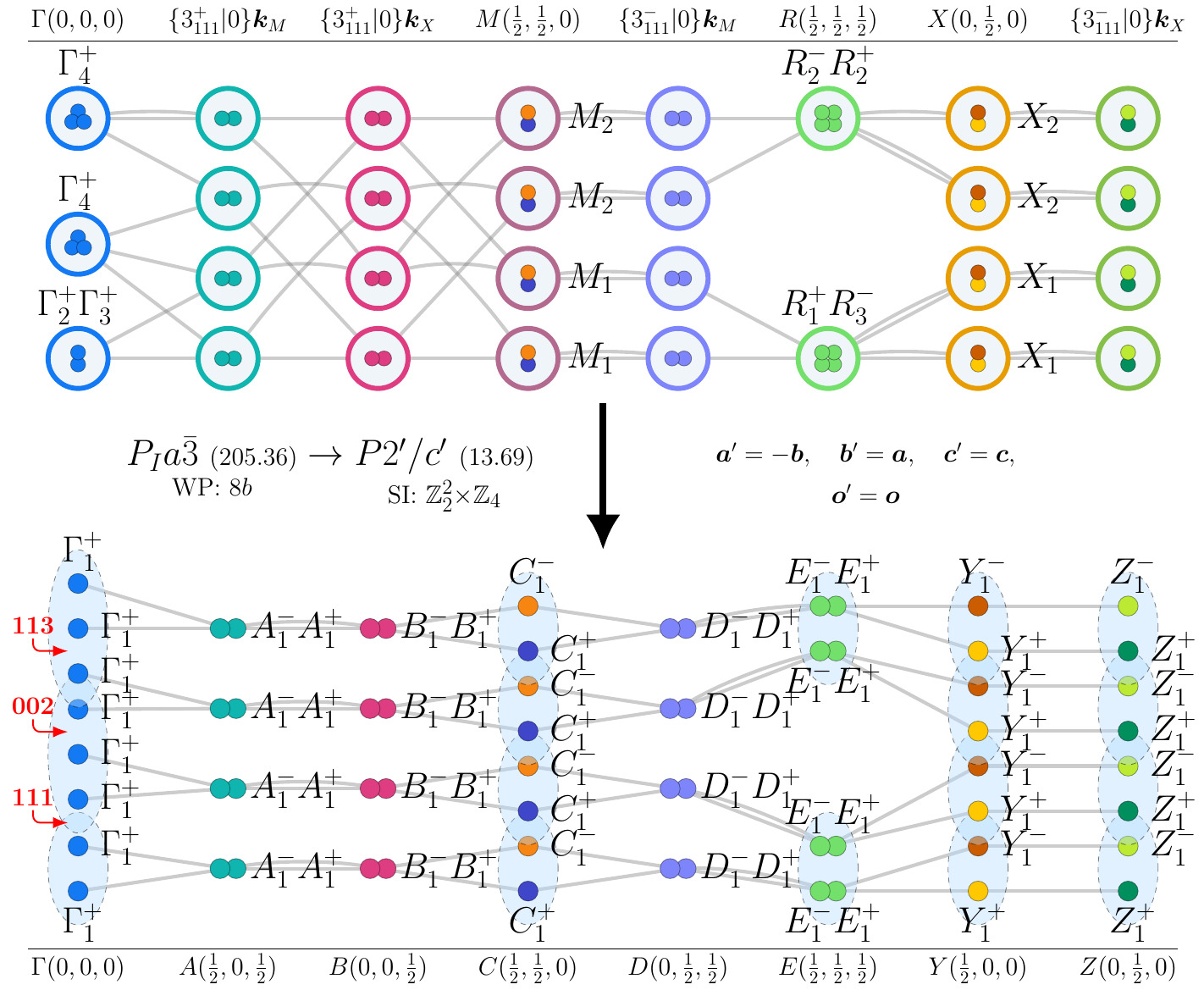}
\caption{Topological magnon bands in subgroup $P2'/c'~(13.69)$ for magnetic moments on Wyckoff position $8b$ of supergroup $P_{I}a\bar{3}~(205.36)$.\label{fig_205.36_13.69_Bperp100_8b}}
\end{figure}
\input{gap_tables_tex/205.36_13.69_Bperp100_8b_table.tex}
\input{si_tables_tex/205.36_13.69_Bperp100_8b_table.tex}
\subsubsection{Topological bands in subgroup $P_{S}\bar{1}~(2.7)$}
\textbf{Perturbation:}
\begin{itemize}
\item strain in generic direction.
\end{itemize}
\begin{figure}[H]
\centering
\includegraphics[scale=0.6]{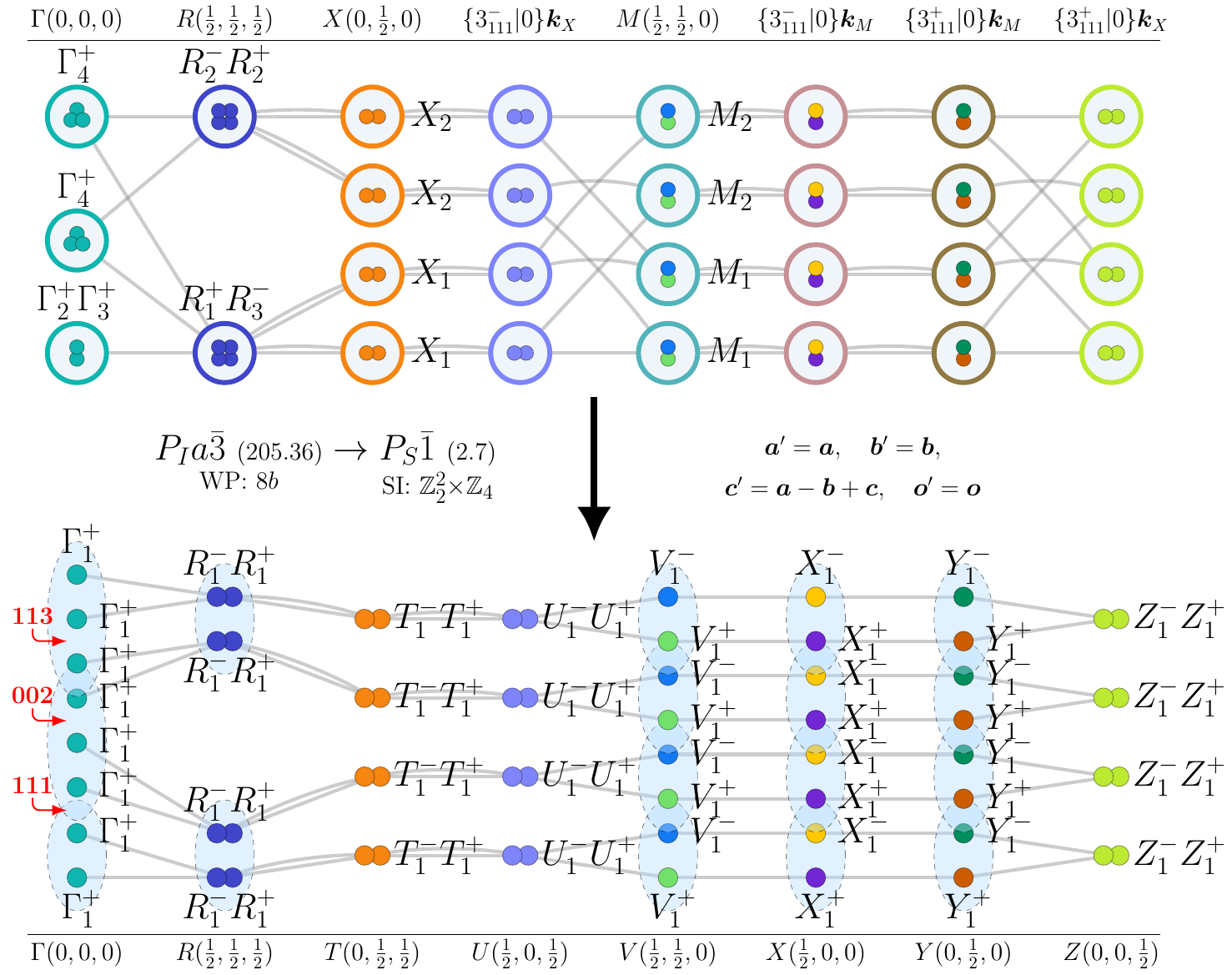}
\caption{Topological magnon bands in subgroup $P_{S}\bar{1}~(2.7)$ for magnetic moments on Wyckoff position $8b$ of supergroup $P_{I}a\bar{3}~(205.36)$.\label{fig_205.36_2.7_strainingenericdirection_8b}}
\end{figure}
\input{gap_tables_tex/205.36_2.7_strainingenericdirection_8b_table.tex}
\input{si_tables_tex/205.36_2.7_strainingenericdirection_8b_table.tex}

\section{MSG $P_{I}n\bar{3}n~(222.103)$}
\textbf{Nontrivial-SI Subgroups:} $P\bar{1}~(2.4)$, $C2'/m'~(12.62)$, $C2'/m'~(12.62)$, $C2'/m'~(12.62)$, $P2_{1}'/m'~(11.54)$, $P2_{1}'/m'~(11.54)$, $P_{S}\bar{1}~(2.7)$, $R\bar{3}m'~(166.101)$, $R_{I}\bar{3}c~(167.108)$, $C2/c~(15.85)$, $Cm'cm'~(63.464)$, $C_{c}2/c~(15.90)$, $C_{A}cca~(68.520)$, $P2~(3.1)$, $P2/c~(13.65)$, $P_{A}2/c~(13.73)$, $P4m'm'~(99.167)$, $P_{I}4nc~(104.210)$, $P4/nm'm'~(129.417)$, $P_{I}4/nnc~(126.386)$.\\

\textbf{Trivial-SI Subgroups:} $Cm'~(8.34)$, $Cm'~(8.34)$, $Cm'~(8.34)$, $Pm'~(6.20)$, $Pm'~(6.20)$, $C2'~(5.15)$, $P2_{1}'~(4.9)$, $P_{S}1~(1.3)$, $C_{c}c~(9.40)$, $C_{c}c~(9.40)$, $Cc~(9.37)$, $C_{c}c~(9.40)$, $P_{A}c~(7.31)$, $Pc~(7.24)$, $P_{A}c~(7.31)$, $R3m'~(160.67)$, $R_{I}3c~(161.72)$, $C2~(5.13)$, $Am'm'2~(38.191)$, $C_{c}2~(5.16)$, $A_{B}ba2~(41.218)$, $P_{C}2~(3.6)$.\\

\subsection{WP: $8c$}
\textbf{BCS Materials:} {NdZn~(70 K)}\footnote{BCS web page: \texttt{\href{http://webbdcrista1.ehu.es/magndata/index.php?this\_label=3.8} {http://webbdcrista1.ehu.es/magndata/index.php?this\_label=3.8}}}.\\
\subsubsection{Topological bands in subgroup $P2_{1}'/m'~(11.54)$}
\textbf{Perturbations:}
\begin{itemize}
\item B $\parallel$ [100] and strain $\parallel$ [110],
\item B $\parallel$ [110] and strain $\parallel$ [100].
\end{itemize}
\begin{figure}[H]
\centering
\includegraphics[scale=0.6]{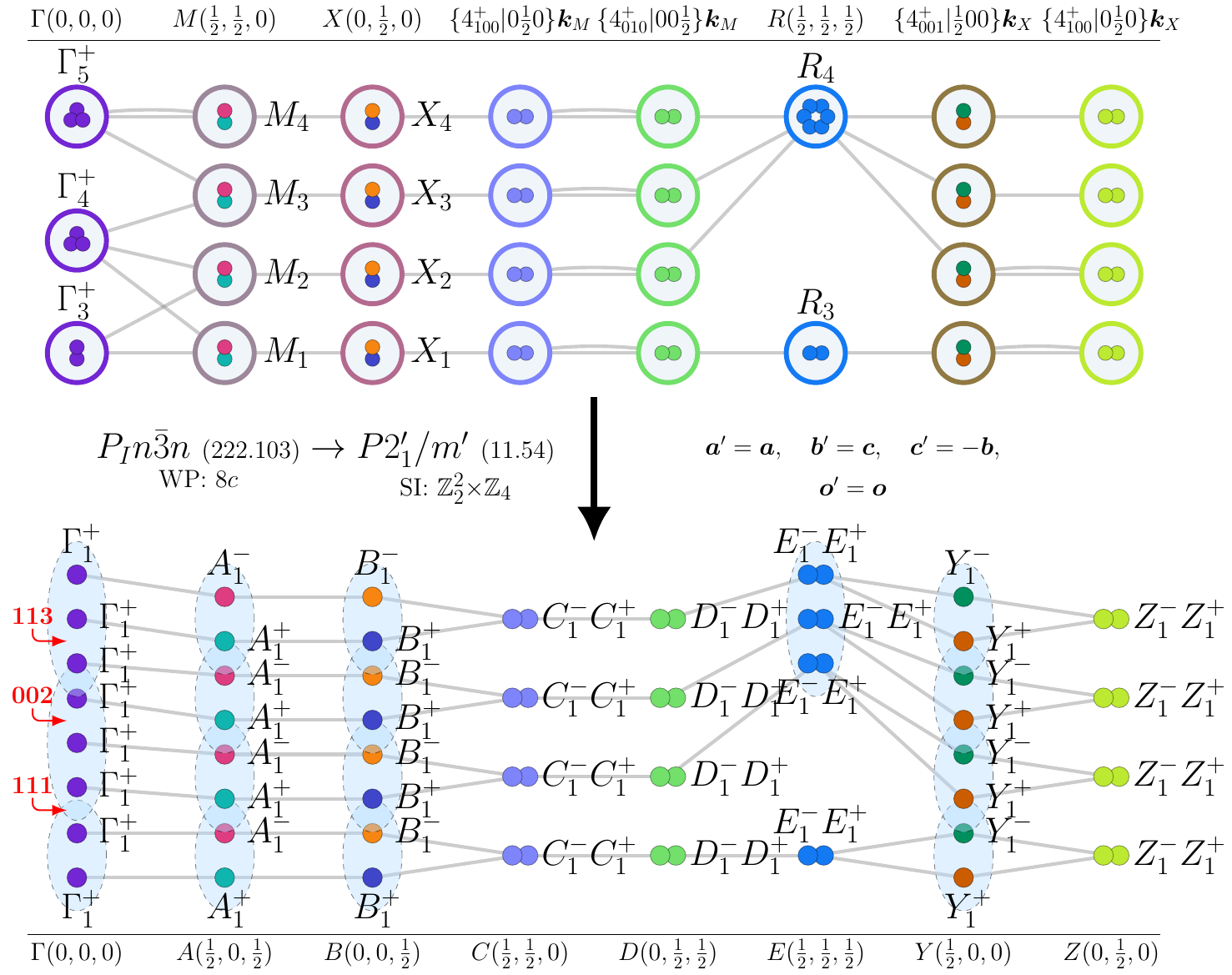}
\caption{Topological magnon bands in subgroup $P2_{1}'/m'~(11.54)$ for magnetic moments on Wyckoff position $8c$ of supergroup $P_{I}n\bar{3}n~(222.103)$.\label{fig_222.103_11.54_Bparallel100andstrainparallel110_8c}}
\end{figure}
\input{gap_tables_tex/222.103_11.54_Bparallel100andstrainparallel110_8c_table.tex}
\input{si_tables_tex/222.103_11.54_Bparallel100andstrainparallel110_8c_table.tex}
\subsubsection{Topological bands in subgroup $P2_{1}'/m'~(11.54)$}
\textbf{Perturbation:}
\begin{itemize}
\item B $\perp$ [100].
\end{itemize}
\begin{figure}[H]
\centering
\includegraphics[scale=0.6]{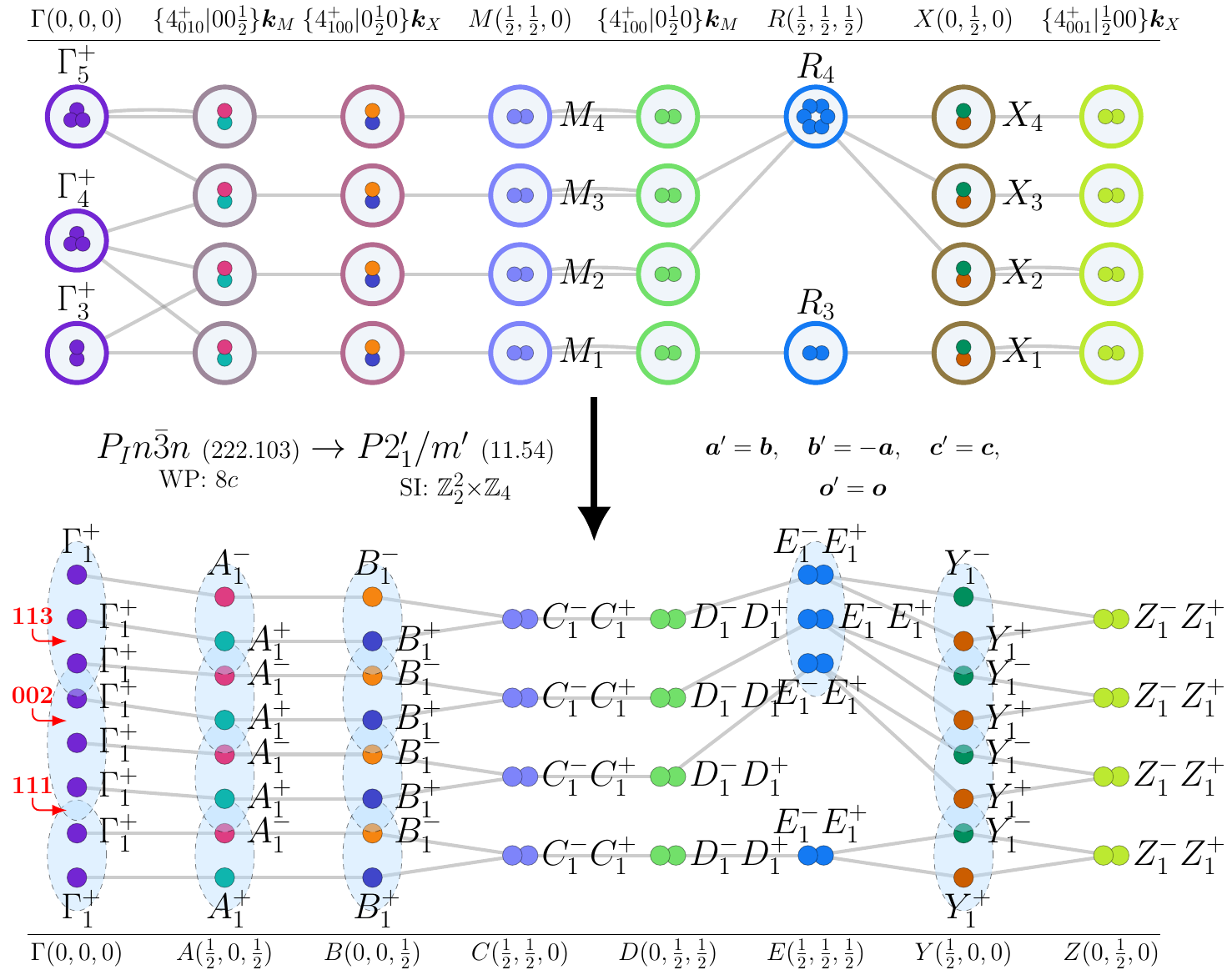}
\caption{Topological magnon bands in subgroup $P2_{1}'/m'~(11.54)$ for magnetic moments on Wyckoff position $8c$ of supergroup $P_{I}n\bar{3}n~(222.103)$.\label{fig_222.103_11.54_Bperp100_8c}}
\end{figure}
\input{gap_tables_tex/222.103_11.54_Bperp100_8c_table.tex}
\input{si_tables_tex/222.103_11.54_Bperp100_8c_table.tex}
\subsubsection{Topological bands in subgroup $P_{S}\bar{1}~(2.7)$}
\textbf{Perturbation:}
\begin{itemize}
\item strain in generic direction.
\end{itemize}
\begin{figure}[H]
\centering
\includegraphics[scale=0.6]{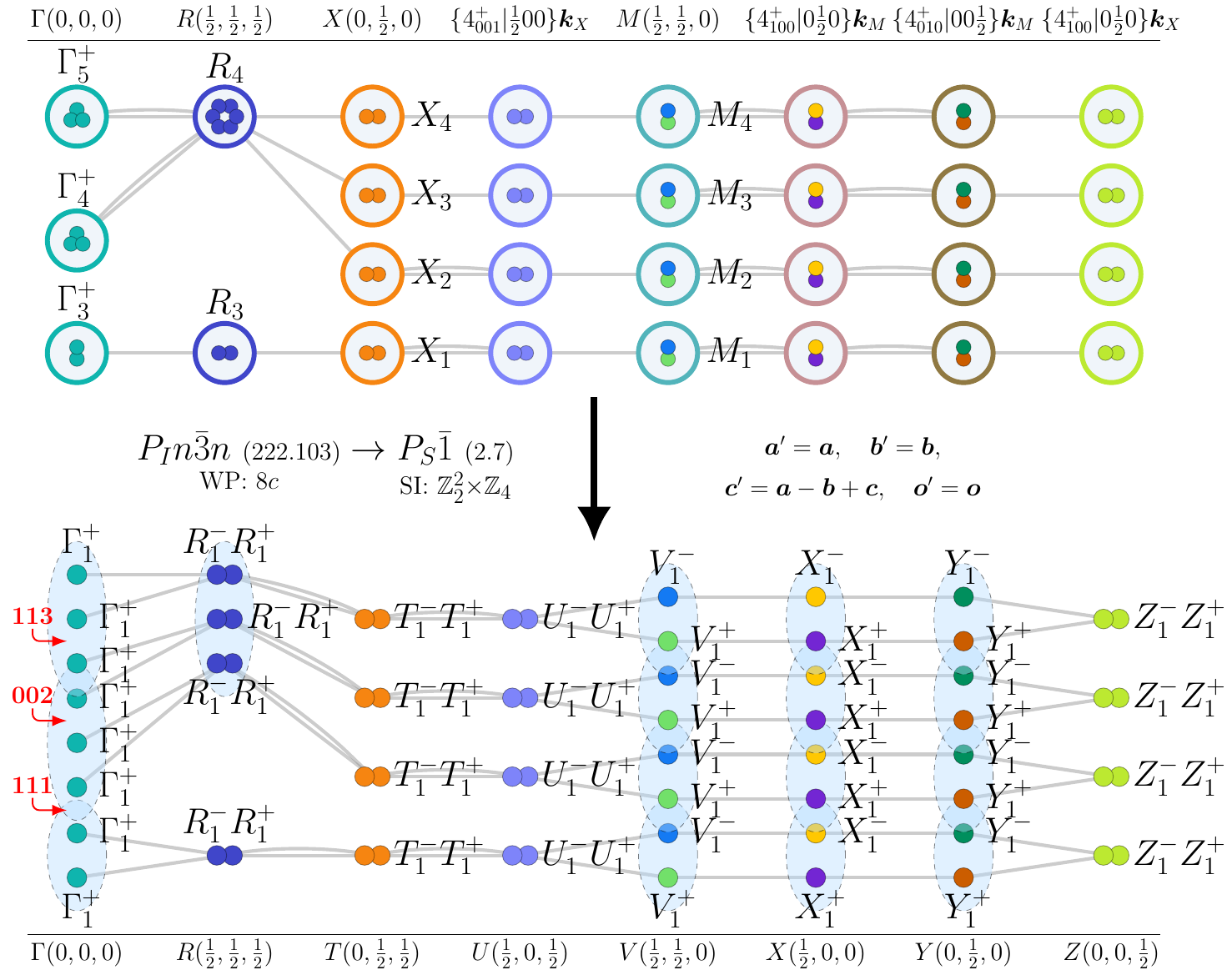}
\caption{Topological magnon bands in subgroup $P_{S}\bar{1}~(2.7)$ for magnetic moments on Wyckoff position $8c$ of supergroup $P_{I}n\bar{3}n~(222.103)$.\label{fig_222.103_2.7_strainingenericdirection_8c}}
\end{figure}
\input{gap_tables_tex/222.103_2.7_strainingenericdirection_8c_table.tex}
\input{si_tables_tex/222.103_2.7_strainingenericdirection_8c_table.tex}

\section{MSG $Pn\bar{3}m'~(224.113)$}
\textbf{Nontrivial-SI Subgroups:} $P\bar{1}~(2.4)$, $C2'/m'~(12.62)$, $C2'/m'~(12.62)$, $C2'/m'~(12.62)$, $R\bar{3}m'~(166.101)$, $Cm'm'a~(67.505)$, $P2~(3.1)$, $Cm'm'2~(35.168)$, $P4_{2}'nm'~(102.190)$, $P2/c~(13.65)$, $Cm'm'a~(67.505)$, $P4_{2}'/nnm'~(134.475)$.\\

\textbf{Trivial-SI Subgroups:} $Cm'~(8.34)$, $Cm'~(8.34)$, $Cm'~(8.34)$, $C2'~(5.15)$, $Pc~(7.24)$, $Abm'2'~(39.198)$, $Pc~(7.24)$, $R3m'~(160.67)$.\\

\subsection{WP: $4b$}
\textbf{BCS Materials:} {USb~(213 K)}\footnote{BCS web page: \texttt{\href{http://webbdcrista1.ehu.es/magndata/index.php?this\_label=3.12} {http://webbdcrista1.ehu.es/magndata/index.php?this\_label=3.12}}}, {NpBi~(192.5 K)}\footnote{BCS web page: \texttt{\href{http://webbdcrista1.ehu.es/magndata/index.php?this\_label=3.7} {http://webbdcrista1.ehu.es/magndata/index.php?this\_label=3.7}}}, {Fe\textsubscript{0.7}Mn\textsubscript{0.3}}\footnote{BCS web page: \texttt{\href{http://webbdcrista1.ehu.es/magndata/index.php?this\_label=3.5} {http://webbdcrista1.ehu.es/magndata/index.php?this\_label=3.5}}}.\\
\subsubsection{Topological bands in subgroup $P\bar{1}~(2.4)$}
\textbf{Perturbations:}
\begin{itemize}
\item strain in generic direction,
\item B $\parallel$ [100] and strain $\parallel$ [110],
\item B $\parallel$ [100] and strain $\perp$ [110],
\item B $\parallel$ [110] and strain $\parallel$ [100],
\item B $\parallel$ [110] and strain $\perp$ [100],
\item B $\parallel$ [110] and strain $\perp$ [110],
\item B $\parallel$ [111] and strain $\perp$ [100],
\item B $\parallel$ [111] and strain $\perp$ [110],
\item B in generic direction,
\item B $\perp$ [110] and strain $\parallel$ [100],
\item B $\perp$ [110] and strain $\parallel$ [111],
\item B $\perp$ [110] and strain $\perp$ [100].
\end{itemize}
\begin{figure}[H]
\centering
\includegraphics[scale=0.6]{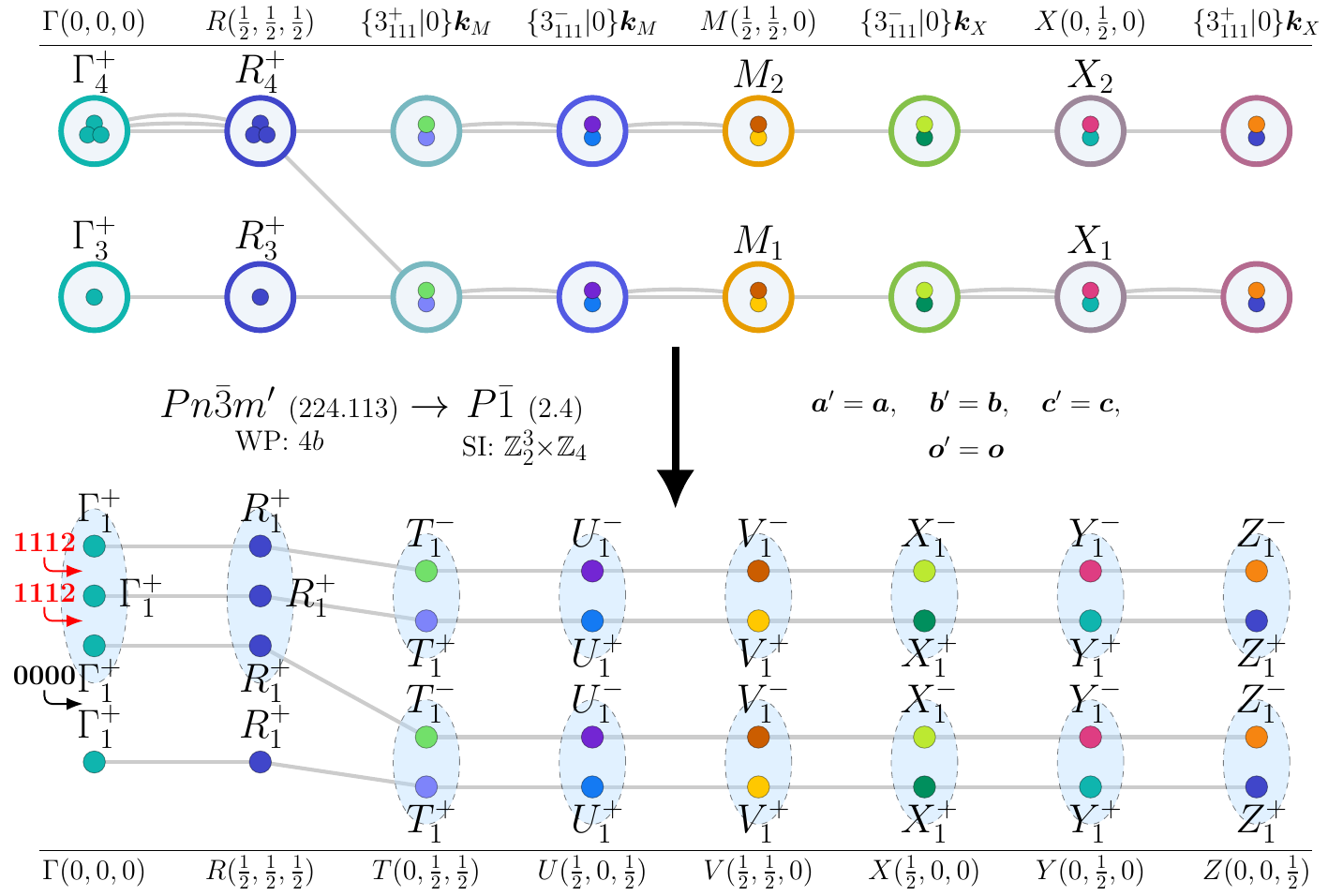}
\caption{Topological magnon bands in subgroup $P\bar{1}~(2.4)$ for magnetic moments on Wyckoff position $4b$ of supergroup $Pn\bar{3}m'~(224.113)$.\label{fig_224.113_2.4_strainingenericdirection_4b}}
\end{figure}
\input{gap_tables_tex/224.113_2.4_strainingenericdirection_4b_table.tex}
\input{si_tables_tex/224.113_2.4_strainingenericdirection_4b_table.tex}
\subsubsection{Topological bands in subgroup $C2'/m'~(12.62)$}
\textbf{Perturbations:}
\begin{itemize}
\item B $\parallel$ [100] and strain $\parallel$ [111],
\item B $\parallel$ [111] and strain $\parallel$ [100].
\end{itemize}
\begin{figure}[H]
\centering
\includegraphics[scale=0.6]{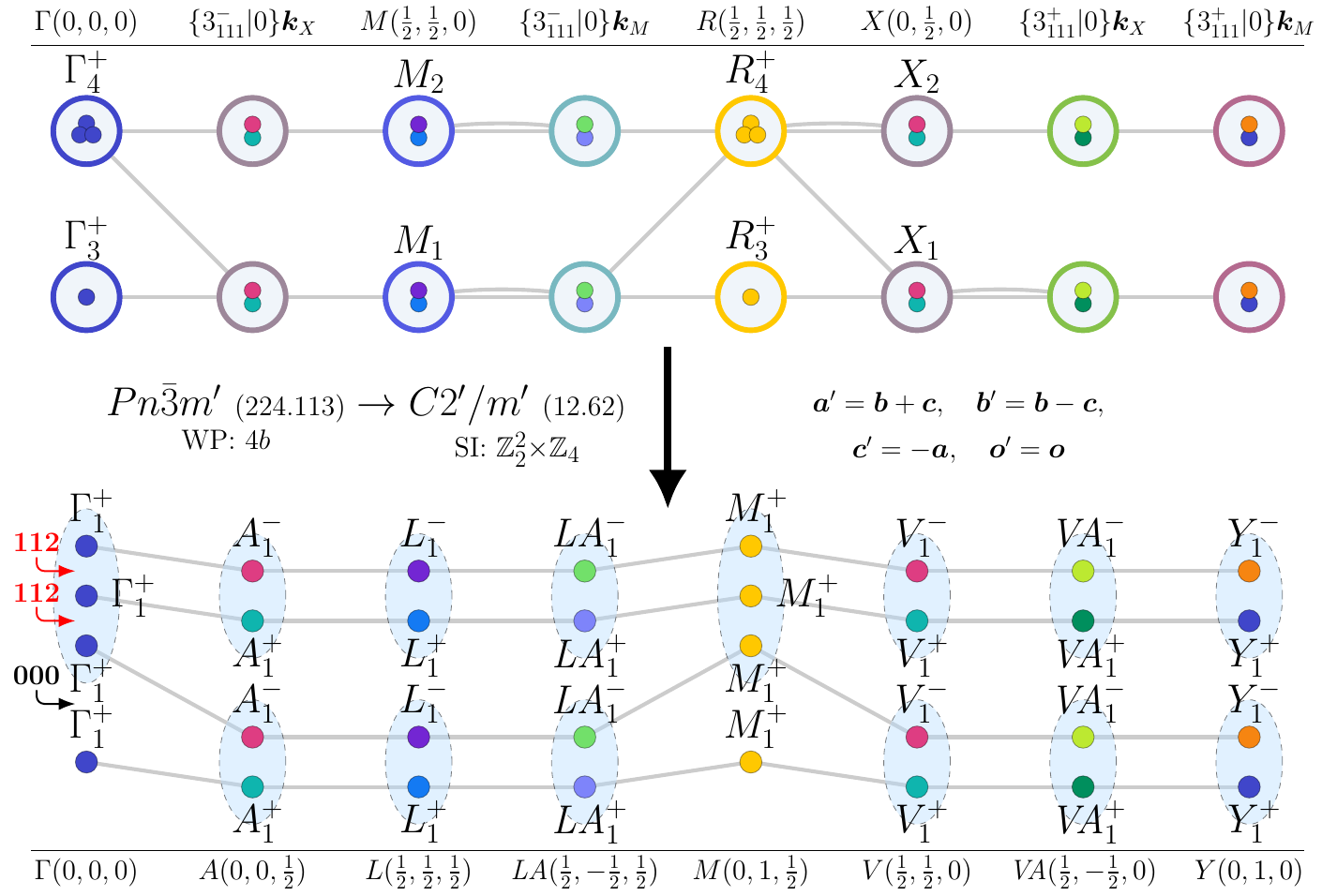}
\caption{Topological magnon bands in subgroup $C2'/m'~(12.62)$ for magnetic moments on Wyckoff position $4b$ of supergroup $Pn\bar{3}m'~(224.113)$.\label{fig_224.113_12.62_Bparallel100andstrainparallel111_4b}}
\end{figure}
\input{gap_tables_tex/224.113_12.62_Bparallel100andstrainparallel111_4b_table.tex}
\input{si_tables_tex/224.113_12.62_Bparallel100andstrainparallel111_4b_table.tex}
\subsubsection{Topological bands in subgroup $C2'/m'~(12.62)$}
\textbf{Perturbations:}
\begin{itemize}
\item B $\parallel$ [110],
\item B $\parallel$ [111] and strain $\parallel$ [110].
\end{itemize}
\begin{figure}[H]
\centering
\includegraphics[scale=0.6]{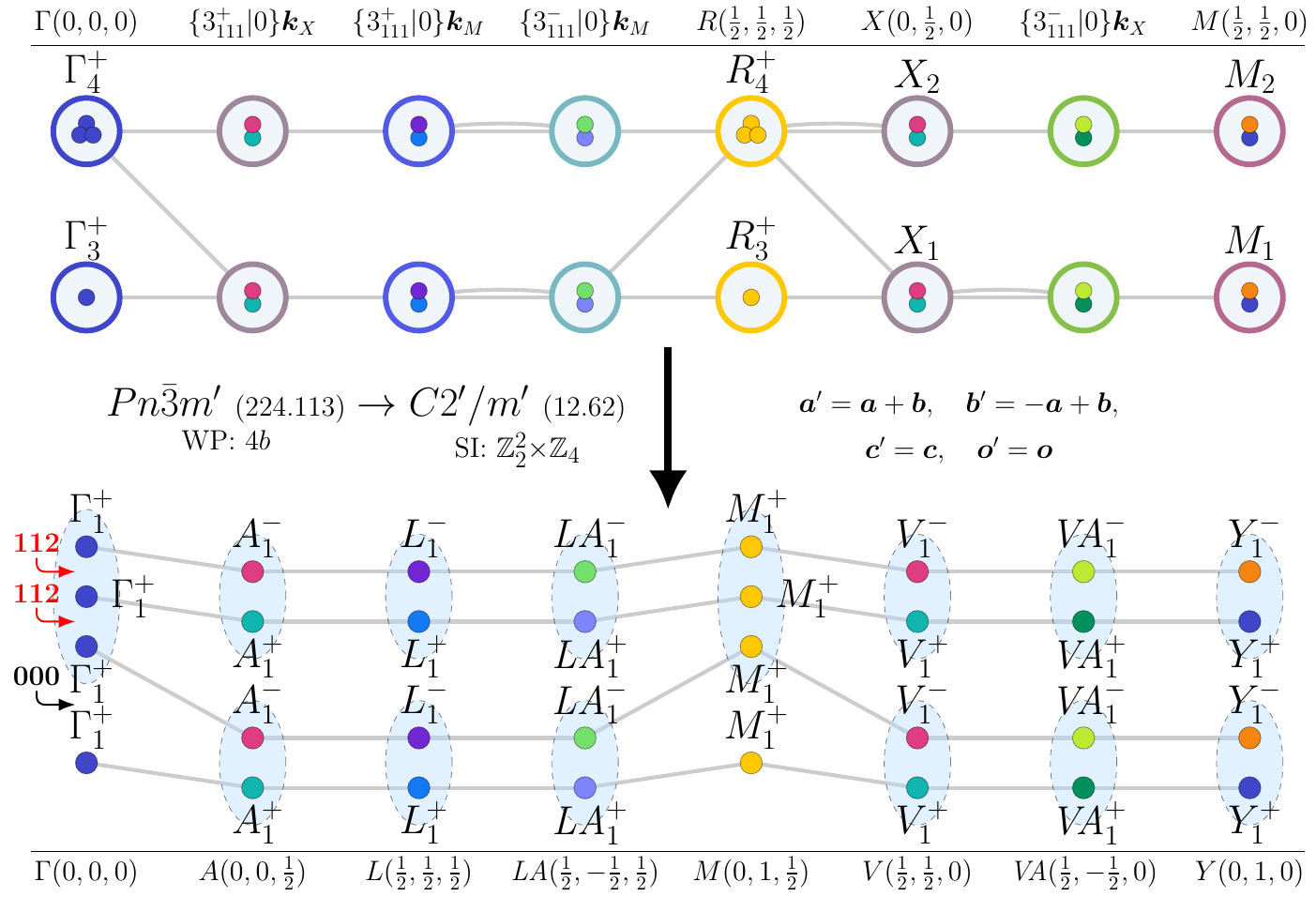}
\caption{Topological magnon bands in subgroup $C2'/m'~(12.62)$ for magnetic moments on Wyckoff position $4b$ of supergroup $Pn\bar{3}m'~(224.113)$.\label{fig_224.113_12.62_Bparallel110_4b}}
\end{figure}
\input{gap_tables_tex/224.113_12.62_Bparallel110_4b_table.tex}
\input{si_tables_tex/224.113_12.62_Bparallel110_4b_table.tex}
\subsubsection{Topological bands in subgroup $C2'/m'~(12.62)$}
\textbf{Perturbations:}
\begin{itemize}
\item strain $\perp$ [110],
\item B $\perp$ [110].
\end{itemize}
\begin{figure}[H]
\centering
\includegraphics[scale=0.6]{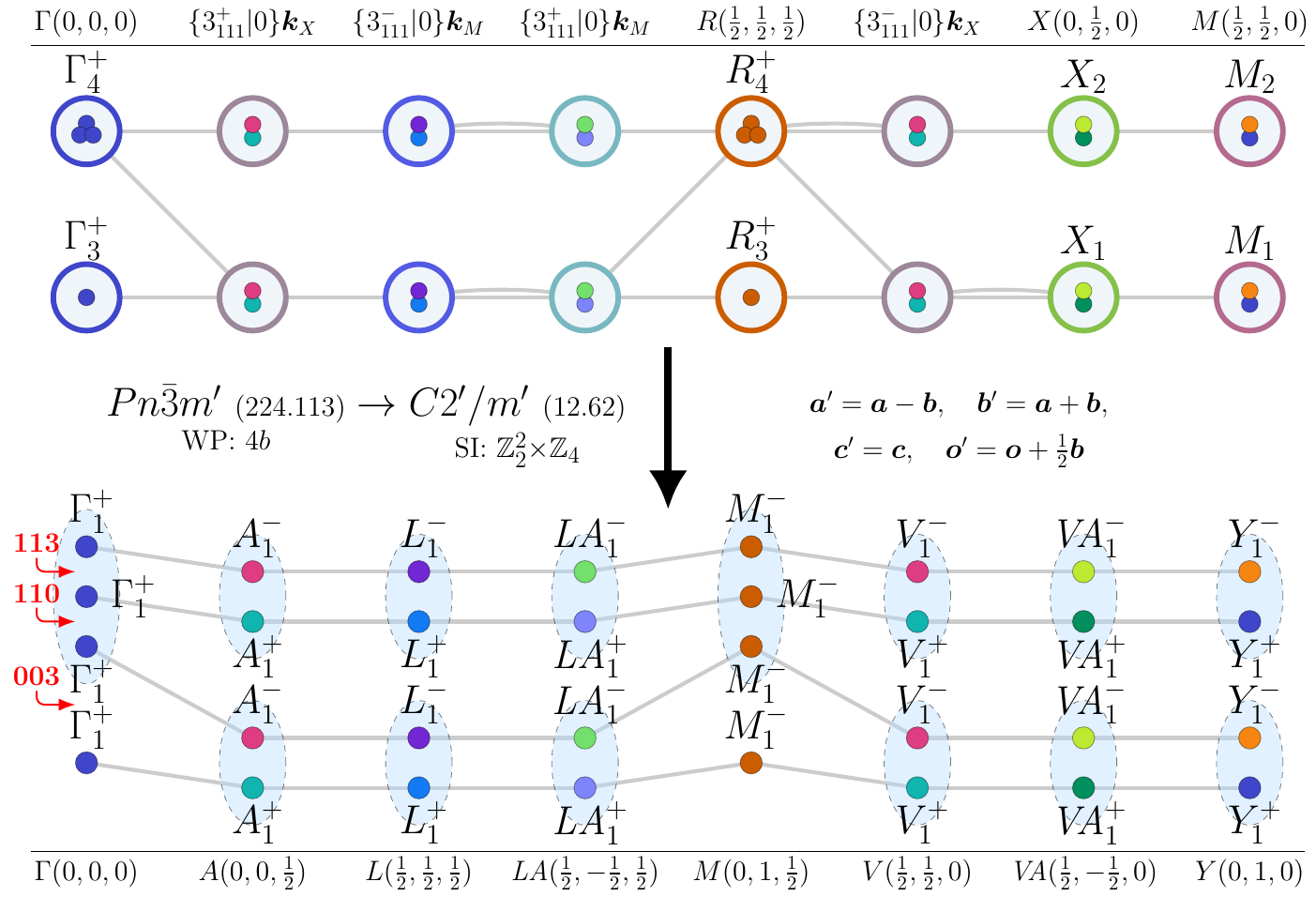}
\caption{Topological magnon bands in subgroup $C2'/m'~(12.62)$ for magnetic moments on Wyckoff position $4b$ of supergroup $Pn\bar{3}m'~(224.113)$.\label{fig_224.113_12.62_strainperp110_4b}}
\end{figure}
\input{gap_tables_tex/224.113_12.62_strainperp110_4b_table.tex}
\input{si_tables_tex/224.113_12.62_strainperp110_4b_table.tex}
\subsection{WP: $4c$}
\textbf{BCS Materials:} {UO\textsubscript{2}~(30.8 K)}\footnote{BCS web page: \texttt{\href{http://webbdcrista1.ehu.es/magndata/index.php?this\_label=3.2} {http://webbdcrista1.ehu.es/magndata/index.php?this\_label=3.2}}}.\\
\subsubsection{Topological bands in subgroup $P\bar{1}~(2.4)$}
\textbf{Perturbations:}
\begin{itemize}
\item strain in generic direction,
\item B $\parallel$ [100] and strain $\parallel$ [110],
\item B $\parallel$ [100] and strain $\perp$ [110],
\item B $\parallel$ [110] and strain $\parallel$ [100],
\item B $\parallel$ [110] and strain $\perp$ [100],
\item B $\parallel$ [110] and strain $\perp$ [110],
\item B $\parallel$ [111] and strain $\perp$ [100],
\item B $\parallel$ [111] and strain $\perp$ [110],
\item B in generic direction,
\item B $\perp$ [110] and strain $\parallel$ [100],
\item B $\perp$ [110] and strain $\parallel$ [111],
\item B $\perp$ [110] and strain $\perp$ [100].
\end{itemize}
\begin{figure}[H]
\centering
\includegraphics[scale=0.6]{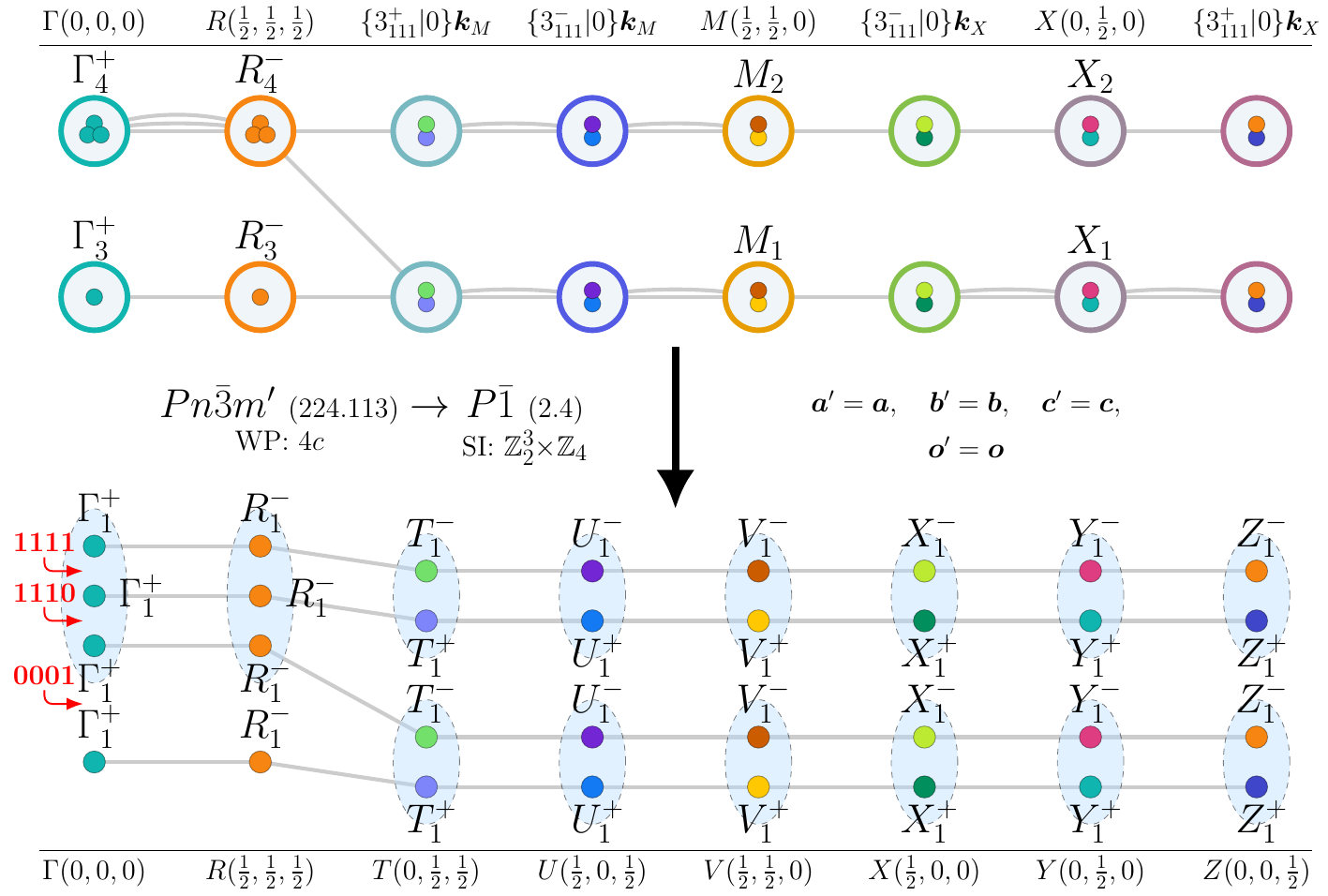}
\caption{Topological magnon bands in subgroup $P\bar{1}~(2.4)$ for magnetic moments on Wyckoff position $4c$ of supergroup $Pn\bar{3}m'~(224.113)$.\label{fig_224.113_2.4_strainingenericdirection_4c}}
\end{figure}
\input{gap_tables_tex/224.113_2.4_strainingenericdirection_4c_table.tex}
\input{si_tables_tex/224.113_2.4_strainingenericdirection_4c_table.tex}
\subsubsection{Topological bands in subgroup $C2'/m'~(12.62)$}
\textbf{Perturbations:}
\begin{itemize}
\item B $\parallel$ [100] and strain $\parallel$ [111],
\item B $\parallel$ [111] and strain $\parallel$ [100].
\end{itemize}
\begin{figure}[H]
\centering
\includegraphics[scale=0.6]{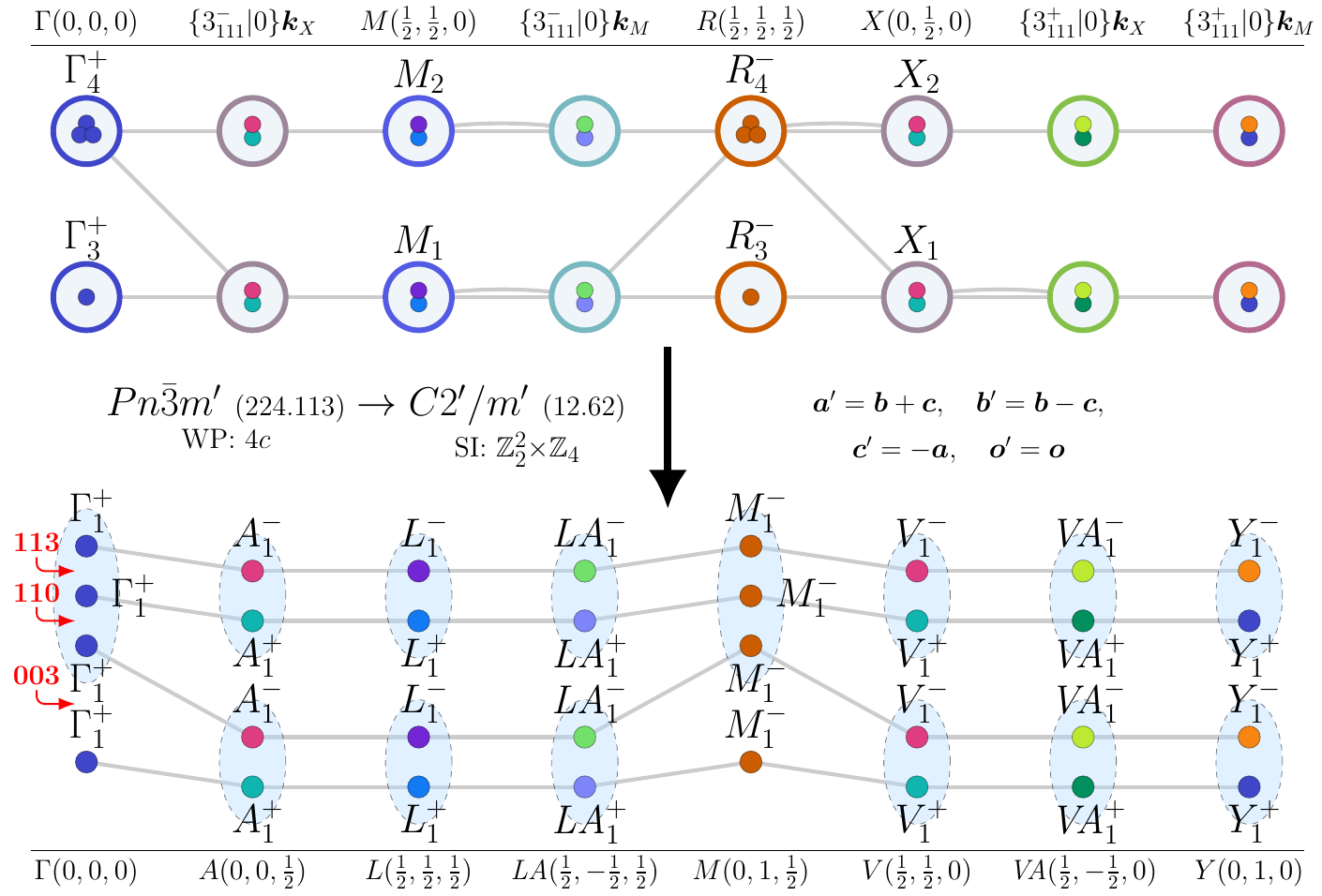}
\caption{Topological magnon bands in subgroup $C2'/m'~(12.62)$ for magnetic moments on Wyckoff position $4c$ of supergroup $Pn\bar{3}m'~(224.113)$.\label{fig_224.113_12.62_Bparallel100andstrainparallel111_4c}}
\end{figure}
\input{gap_tables_tex/224.113_12.62_Bparallel100andstrainparallel111_4c_table.tex}
\input{si_tables_tex/224.113_12.62_Bparallel100andstrainparallel111_4c_table.tex}
\subsubsection{Topological bands in subgroup $C2'/m'~(12.62)$}
\textbf{Perturbations:}
\begin{itemize}
\item B $\parallel$ [110],
\item B $\parallel$ [111] and strain $\parallel$ [110].
\end{itemize}
\begin{figure}[H]
\centering
\includegraphics[scale=0.6]{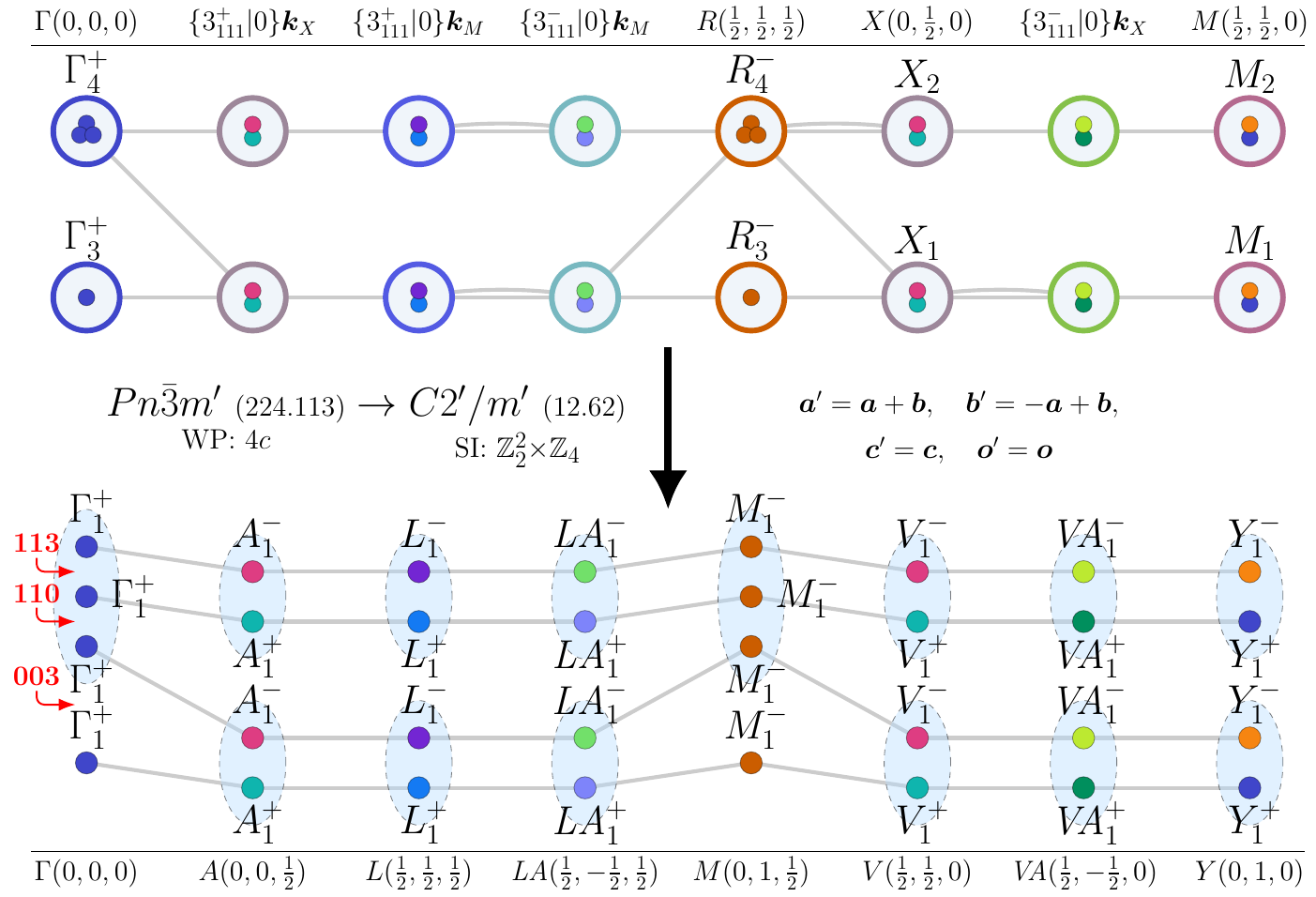}
\caption{Topological magnon bands in subgroup $C2'/m'~(12.62)$ for magnetic moments on Wyckoff position $4c$ of supergroup $Pn\bar{3}m'~(224.113)$.\label{fig_224.113_12.62_Bparallel110_4c}}
\end{figure}
\input{gap_tables_tex/224.113_12.62_Bparallel110_4c_table.tex}
\input{si_tables_tex/224.113_12.62_Bparallel110_4c_table.tex}
\subsubsection{Topological bands in subgroup $C2'/m'~(12.62)$}
\textbf{Perturbations:}
\begin{itemize}
\item strain $\perp$ [110],
\item B $\perp$ [110].
\end{itemize}
\begin{figure}[H]
\centering
\includegraphics[scale=0.6]{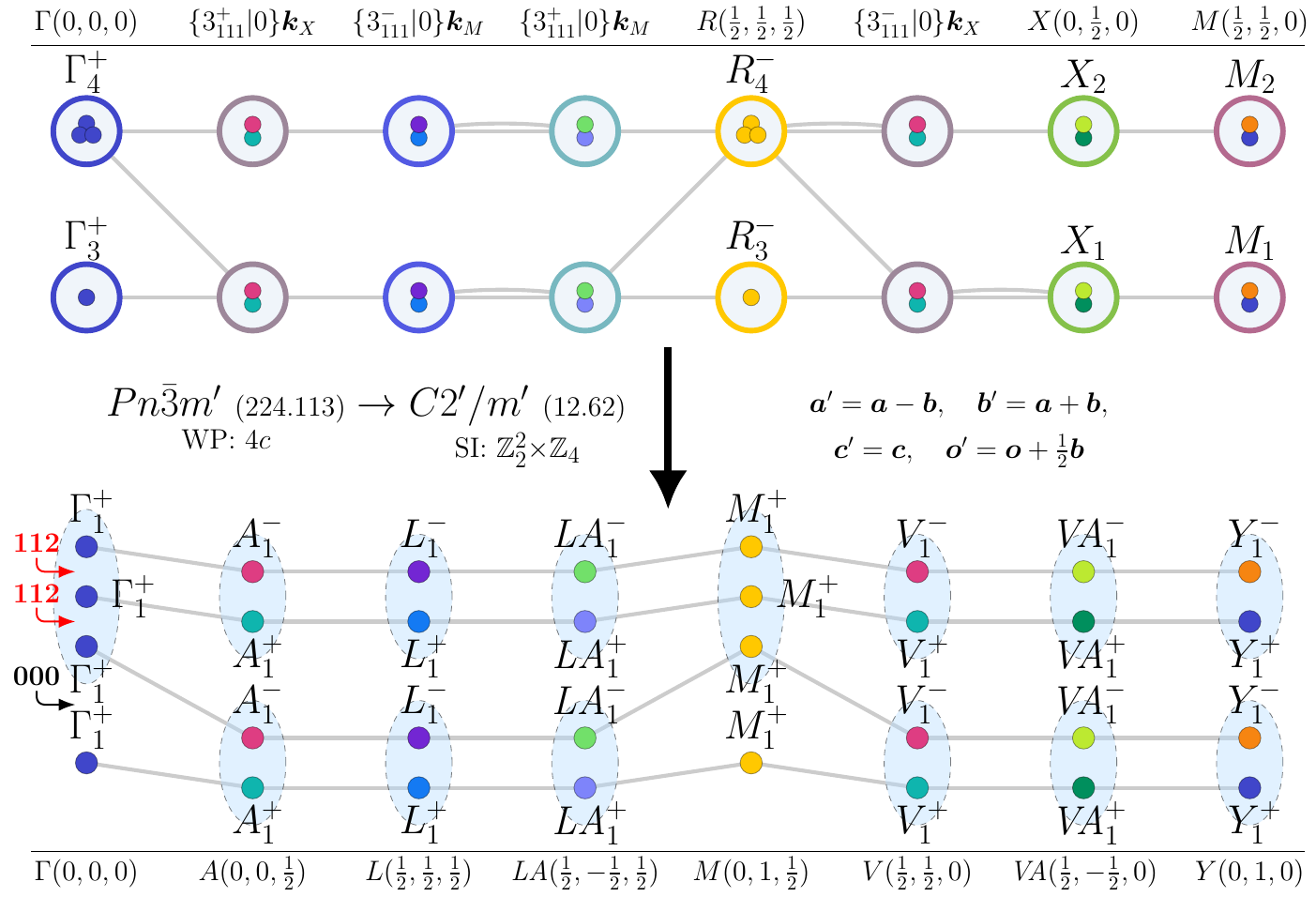}
\caption{Topological magnon bands in subgroup $C2'/m'~(12.62)$ for magnetic moments on Wyckoff position $4c$ of supergroup $Pn\bar{3}m'~(224.113)$.\label{fig_224.113_12.62_strainperp110_4c}}
\end{figure}
\input{gap_tables_tex/224.113_12.62_strainperp110_4c_table.tex}
\input{si_tables_tex/224.113_12.62_strainperp110_4c_table.tex}

\section{MSG $Fd\bar{3}m'~(227.131)$}
\textbf{Nontrivial-SI Subgroups:} $P\bar{1}~(2.4)$, $C2'/m'~(12.62)$, $C2'/m'~(12.62)$, $C2'/m'~(12.62)$, $R\bar{3}m'~(166.101)$, $Im'm'a~(74.558)$, $C2/c~(15.85)$, $Im'm'a~(74.558)$, $I4_{1}'/am'd~(141.554)$.\\

\textbf{Trivial-SI Subgroups:} $Cm'~(8.34)$, $Cm'~(8.34)$, $Cm'~(8.34)$, $C2'~(5.15)$, $Cc~(9.37)$, $Im'a2'~(46.243)$, $Cc~(9.37)$, $R3m'~(160.67)$, $C2~(5.13)$, $Im'm'2~(44.232)$, $I4_{1}'m'd~(109.241)$.\\

\subsection{WP: $16c$}
\textbf{BCS Materials:} {Cd\textsubscript{2}Os\textsubscript{2}O\textsubscript{7}~(227 K)}\footnote{BCS web page: \texttt{\href{http://webbdcrista1.ehu.es/magndata/index.php?this\_label=0.2} {http://webbdcrista1.ehu.es/magndata/index.php?this\_label=0.2}}}.\\
\subsubsection{Topological bands in subgroup $P\bar{1}~(2.4)$}
\textbf{Perturbations:}
\begin{itemize}
\item strain in generic direction,
\item B $\parallel$ [100] and strain $\parallel$ [110],
\item B $\parallel$ [100] and strain $\perp$ [110],
\item B $\parallel$ [110] and strain $\parallel$ [100],
\item B $\parallel$ [110] and strain $\perp$ [100],
\item B $\parallel$ [110] and strain $\perp$ [110],
\item B $\parallel$ [111] and strain $\perp$ [100],
\item B $\parallel$ [111] and strain $\perp$ [110],
\item B in generic direction,
\item B $\perp$ [110] and strain $\parallel$ [100],
\item B $\perp$ [110] and strain $\parallel$ [111],
\item B $\perp$ [110] and strain $\perp$ [100].
\end{itemize}
\begin{figure}[H]
\centering
\includegraphics[scale=0.6]{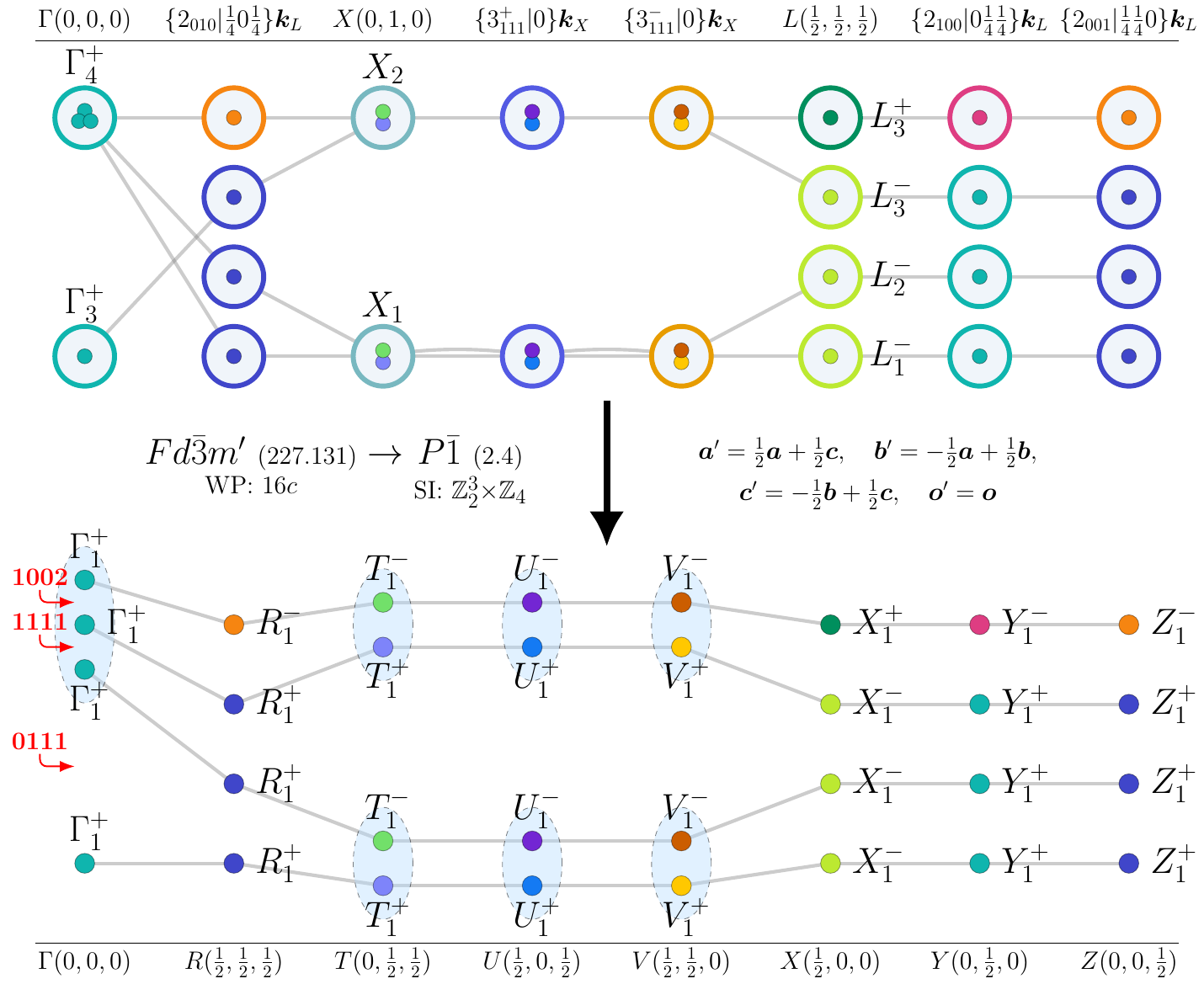}
\caption{Topological magnon bands in subgroup $P\bar{1}~(2.4)$ for magnetic moments on Wyckoff position $16c$ of supergroup $Fd\bar{3}m'~(227.131)$.\label{fig_227.131_2.4_strainingenericdirection_16c}}
\end{figure}
\input{gap_tables_tex/227.131_2.4_strainingenericdirection_16c_table.tex}
\input{si_tables_tex/227.131_2.4_strainingenericdirection_16c_table.tex}
\subsubsection{Topological bands in subgroup $C2'/m'~(12.62)$}
\textbf{Perturbations:}
\begin{itemize}
\item B $\parallel$ [100] and strain $\parallel$ [111],
\item B $\parallel$ [111] and strain $\parallel$ [100].
\end{itemize}
\begin{figure}[H]
\centering
\includegraphics[scale=0.6]{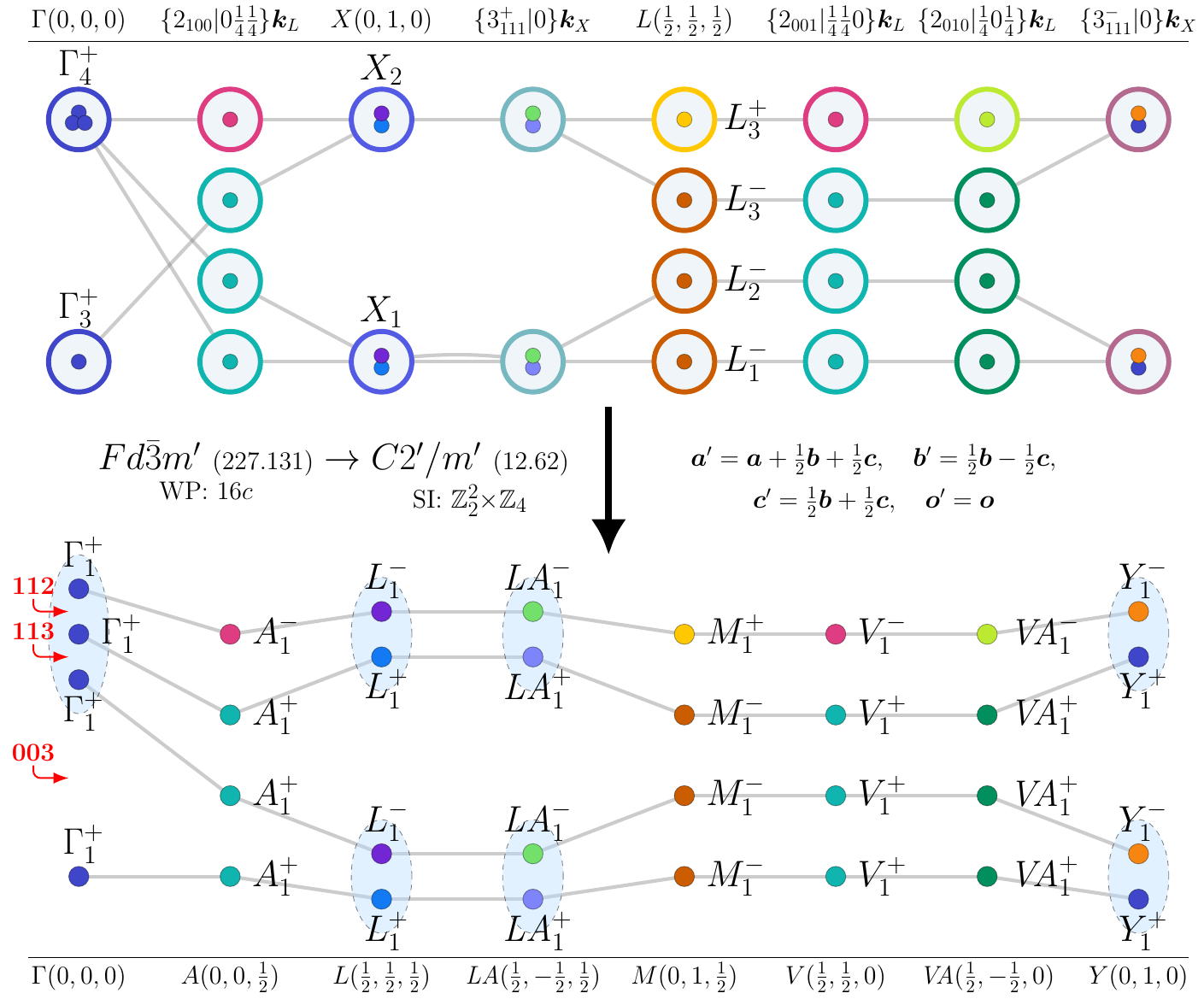}
\caption{Topological magnon bands in subgroup $C2'/m'~(12.62)$ for magnetic moments on Wyckoff position $16c$ of supergroup $Fd\bar{3}m'~(227.131)$.\label{fig_227.131_12.62_Bparallel100andstrainparallel111_16c}}
\end{figure}
\input{gap_tables_tex/227.131_12.62_Bparallel100andstrainparallel111_16c_table.tex}
\input{si_tables_tex/227.131_12.62_Bparallel100andstrainparallel111_16c_table.tex}
\subsubsection{Topological bands in subgroup $C2'/m'~(12.62)$}
\textbf{Perturbations:}
\begin{itemize}
\item B $\parallel$ [110],
\item B $\parallel$ [111] and strain $\parallel$ [110].
\end{itemize}
\begin{figure}[H]
\centering
\includegraphics[scale=0.6]{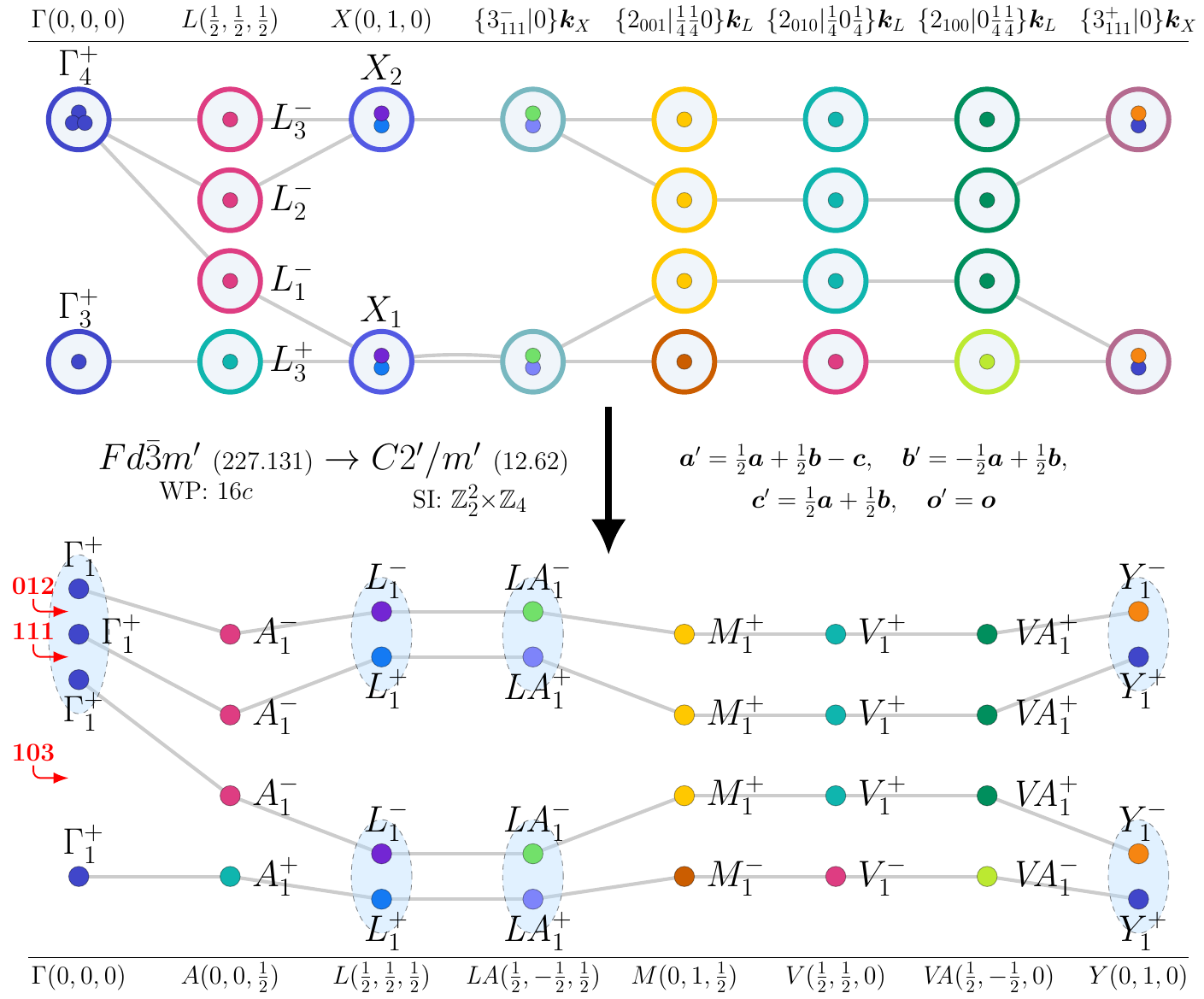}
\caption{Topological magnon bands in subgroup $C2'/m'~(12.62)$ for magnetic moments on Wyckoff position $16c$ of supergroup $Fd\bar{3}m'~(227.131)$.\label{fig_227.131_12.62_Bparallel110_16c}}
\end{figure}
\input{gap_tables_tex/227.131_12.62_Bparallel110_16c_table.tex}
\input{si_tables_tex/227.131_12.62_Bparallel110_16c_table.tex}
\subsubsection{Topological bands in subgroup $C2'/m'~(12.62)$}
\textbf{Perturbations:}
\begin{itemize}
\item strain $\perp$ [110],
\item B $\perp$ [110].
\end{itemize}
\begin{figure}[H]
\centering
\includegraphics[scale=0.6]{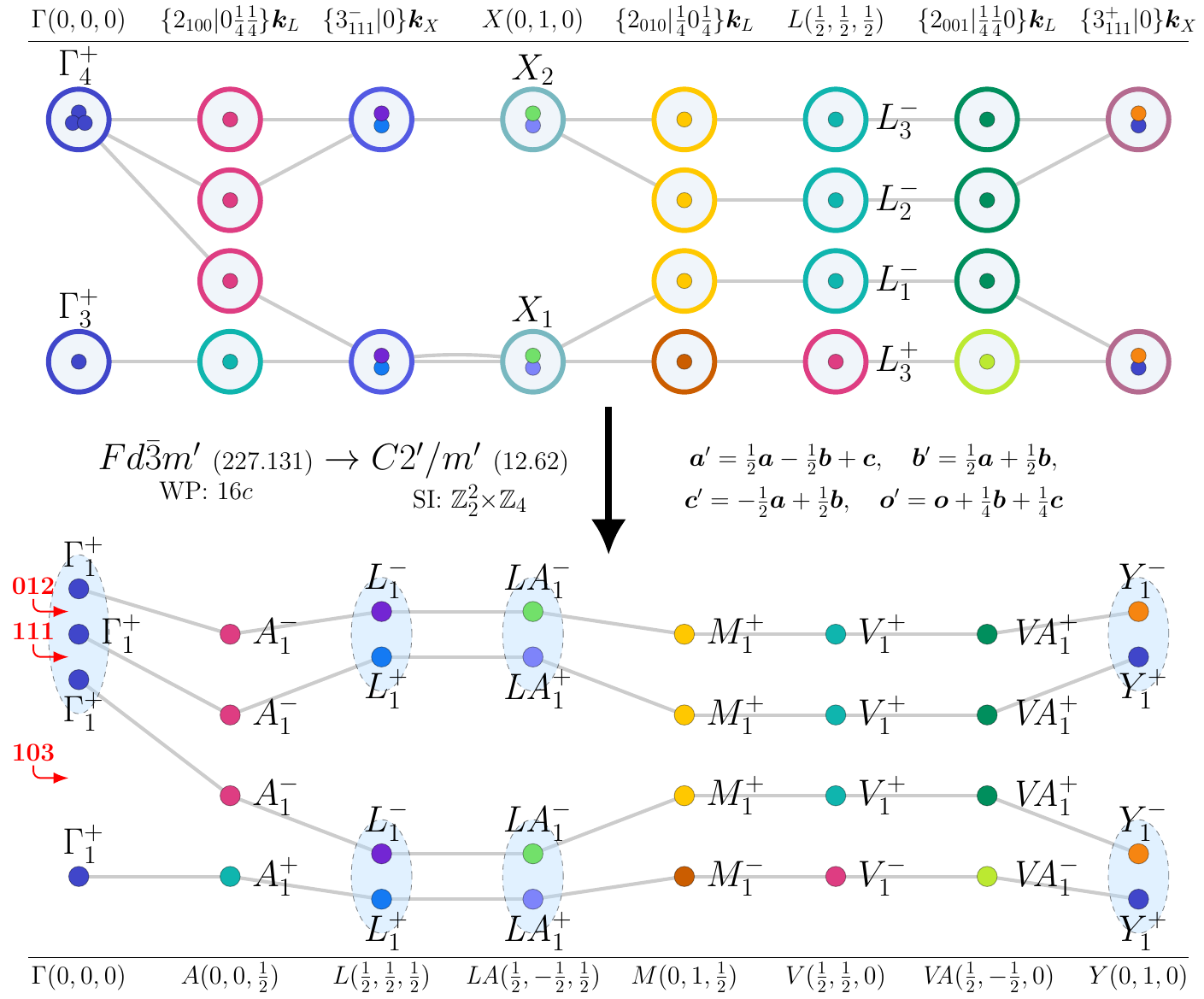}
\caption{Topological magnon bands in subgroup $C2'/m'~(12.62)$ for magnetic moments on Wyckoff position $16c$ of supergroup $Fd\bar{3}m'~(227.131)$.\label{fig_227.131_12.62_strainperp110_16c}}
\end{figure}
\input{gap_tables_tex/227.131_12.62_strainperp110_16c_table.tex}
\input{si_tables_tex/227.131_12.62_strainperp110_16c_table.tex}
\subsection{WP: $16d$}
\textbf{BCS Materials:} {Nd\textsubscript{2}Sn\textsubscript{2}O\textsubscript{7}~(0.91 K)}\footnote{BCS web page: \texttt{\href{http://webbdcrista1.ehu.es/magndata/index.php?this\_label=0.326} {http://webbdcrista1.ehu.es/magndata/index.php?this\_label=0.326}}}, {Nd\textsubscript{2}Hf\textsubscript{2}O\textsubscript{7}~(0.55 K)}\footnote{BCS web page: \texttt{\href{http://webbdcrista1.ehu.es/magndata/index.php?this\_label=0.339} {http://webbdcrista1.ehu.es/magndata/index.php?this\_label=0.339}}}, {Nd\textsubscript{2}Zr\textsubscript{2}O\textsubscript{7}~(0.4 K)}\footnote{BCS web page: \texttt{\href{http://webbdcrista1.ehu.es/magndata/index.php?this\_label=0.340} {http://webbdcrista1.ehu.es/magndata/index.php?this\_label=0.340}}}, {Nd\textsubscript{2}ScNbO\textsubscript{7}~(0.371 K)}\footnote{BCS web page: \texttt{\href{http://webbdcrista1.ehu.es/magndata/index.php?this\_label=0.822} {http://webbdcrista1.ehu.es/magndata/index.php?this\_label=0.822}}}, {Sm\textsubscript{2}Ti\textsubscript{2}O\textsubscript{7}~(0.35 K)}\footnote{BCS web page: \texttt{\href{http://webbdcrista1.ehu.es/magndata/index.php?this\_label=0.427} {http://webbdcrista1.ehu.es/magndata/index.php?this\_label=0.427}}}.\\
\subsubsection{Topological bands in subgroup $P\bar{1}~(2.4)$}
\textbf{Perturbations:}
\begin{itemize}
\item strain in generic direction,
\item B $\parallel$ [100] and strain $\parallel$ [110],
\item B $\parallel$ [100] and strain $\perp$ [110],
\item B $\parallel$ [110] and strain $\parallel$ [100],
\item B $\parallel$ [110] and strain $\perp$ [100],
\item B $\parallel$ [110] and strain $\perp$ [110],
\item B $\parallel$ [111] and strain $\perp$ [100],
\item B $\parallel$ [111] and strain $\perp$ [110],
\item B in generic direction,
\item B $\perp$ [110] and strain $\parallel$ [100],
\item B $\perp$ [110] and strain $\parallel$ [111],
\item B $\perp$ [110] and strain $\perp$ [100].
\end{itemize}
\begin{figure}[H]
\centering
\includegraphics[scale=0.6]{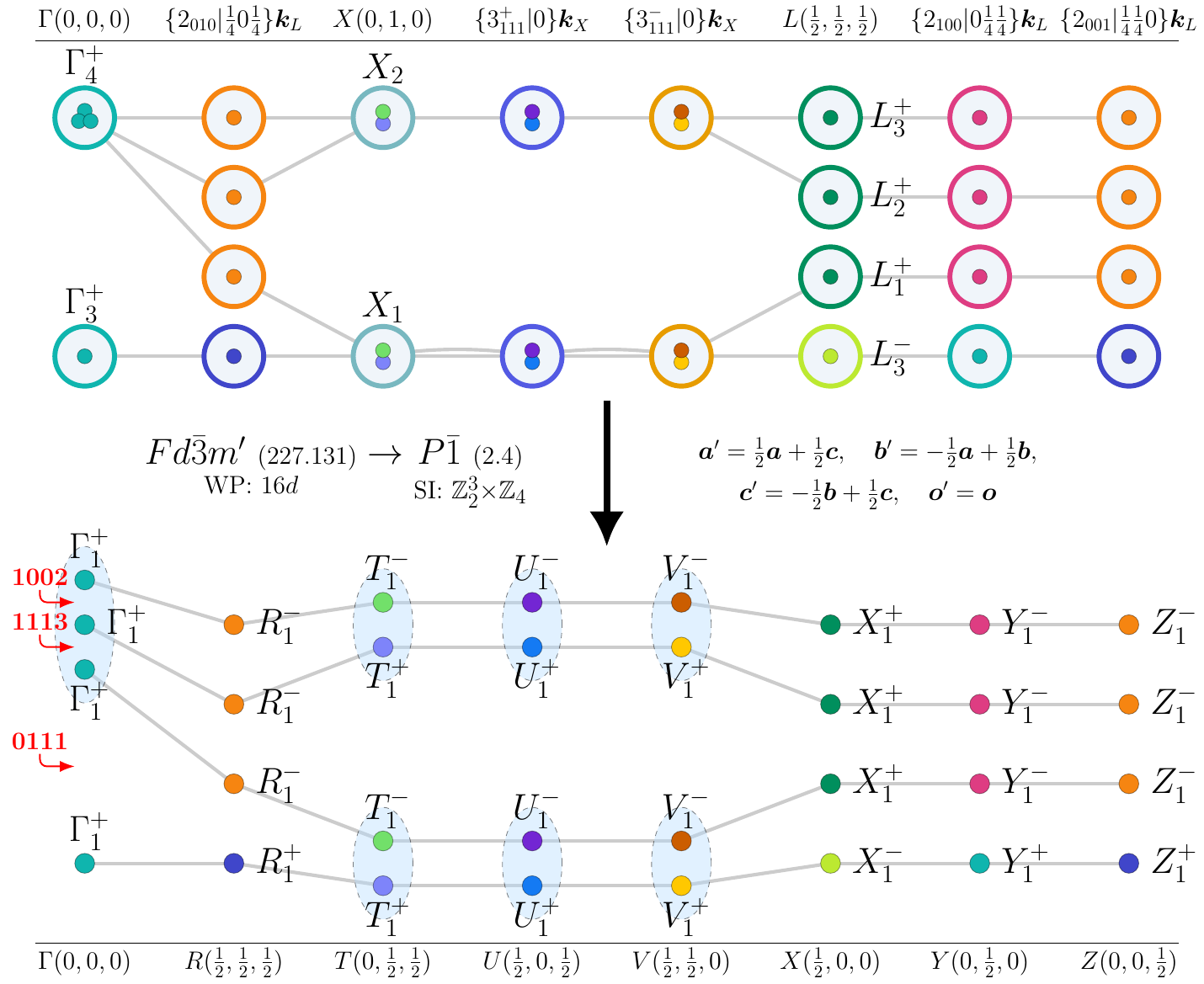}
\caption{Topological magnon bands in subgroup $P\bar{1}~(2.4)$ for magnetic moments on Wyckoff position $16d$ of supergroup $Fd\bar{3}m'~(227.131)$.\label{fig_227.131_2.4_strainingenericdirection_16d}}
\end{figure}
\input{gap_tables_tex/227.131_2.4_strainingenericdirection_16d_table.tex}
\input{si_tables_tex/227.131_2.4_strainingenericdirection_16d_table.tex}
\subsubsection{Topological bands in subgroup $C2'/m'~(12.62)$}
\textbf{Perturbations:}
\begin{itemize}
\item B $\parallel$ [100] and strain $\parallel$ [111],
\item B $\parallel$ [111] and strain $\parallel$ [100].
\end{itemize}
\begin{figure}[H]
\centering
\includegraphics[scale=0.6]{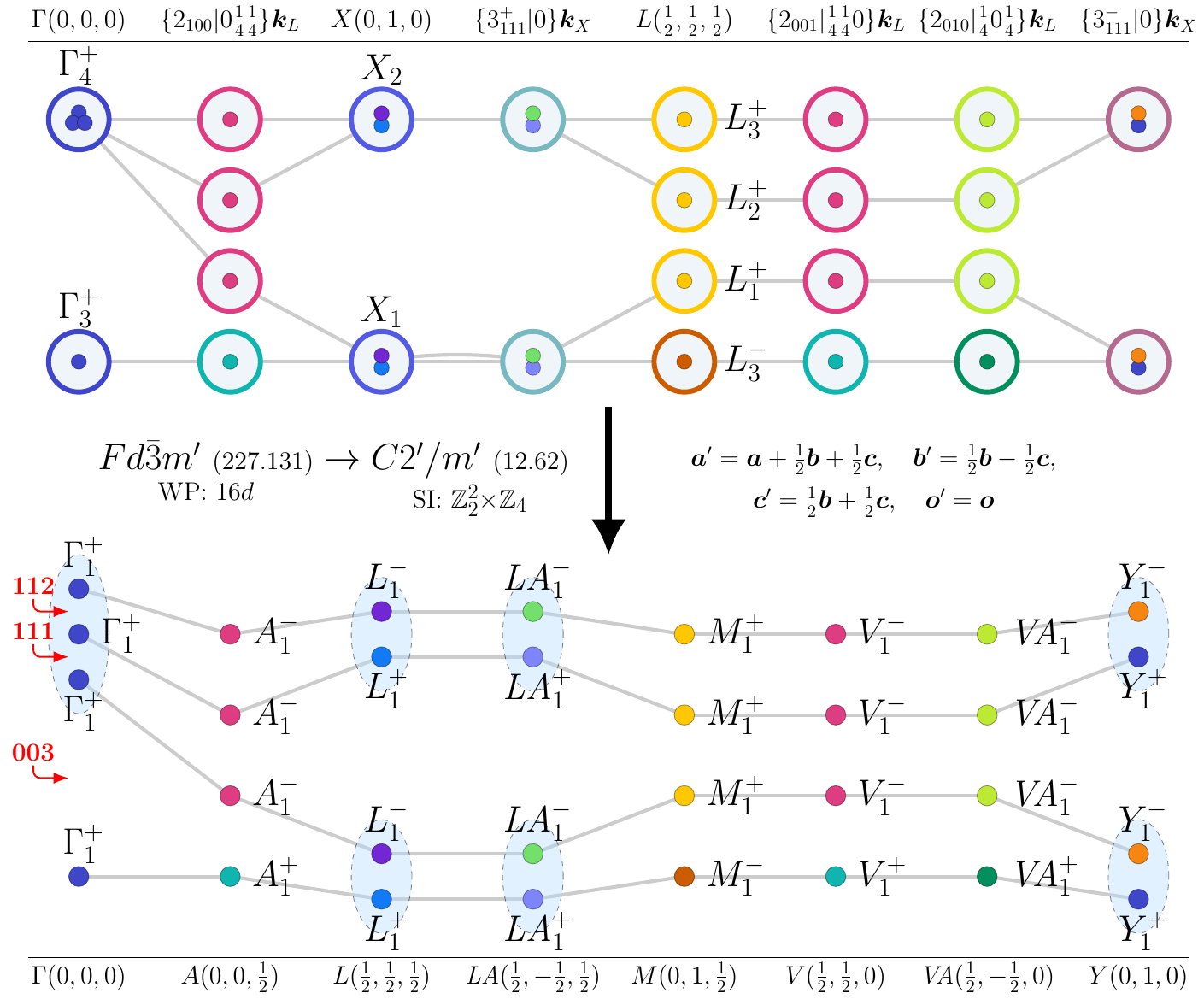}
\caption{Topological magnon bands in subgroup $C2'/m'~(12.62)$ for magnetic moments on Wyckoff position $16d$ of supergroup $Fd\bar{3}m'~(227.131)$.\label{fig_227.131_12.62_Bparallel100andstrainparallel111_16d}}
\end{figure}
\input{gap_tables_tex/227.131_12.62_Bparallel100andstrainparallel111_16d_table.tex}
\input{si_tables_tex/227.131_12.62_Bparallel100andstrainparallel111_16d_table.tex}
\subsubsection{Topological bands in subgroup $C2'/m'~(12.62)$}
\textbf{Perturbations:}
\begin{itemize}
\item B $\parallel$ [110],
\item B $\parallel$ [111] and strain $\parallel$ [110].
\end{itemize}
\begin{figure}[H]
\centering
\includegraphics[scale=0.6]{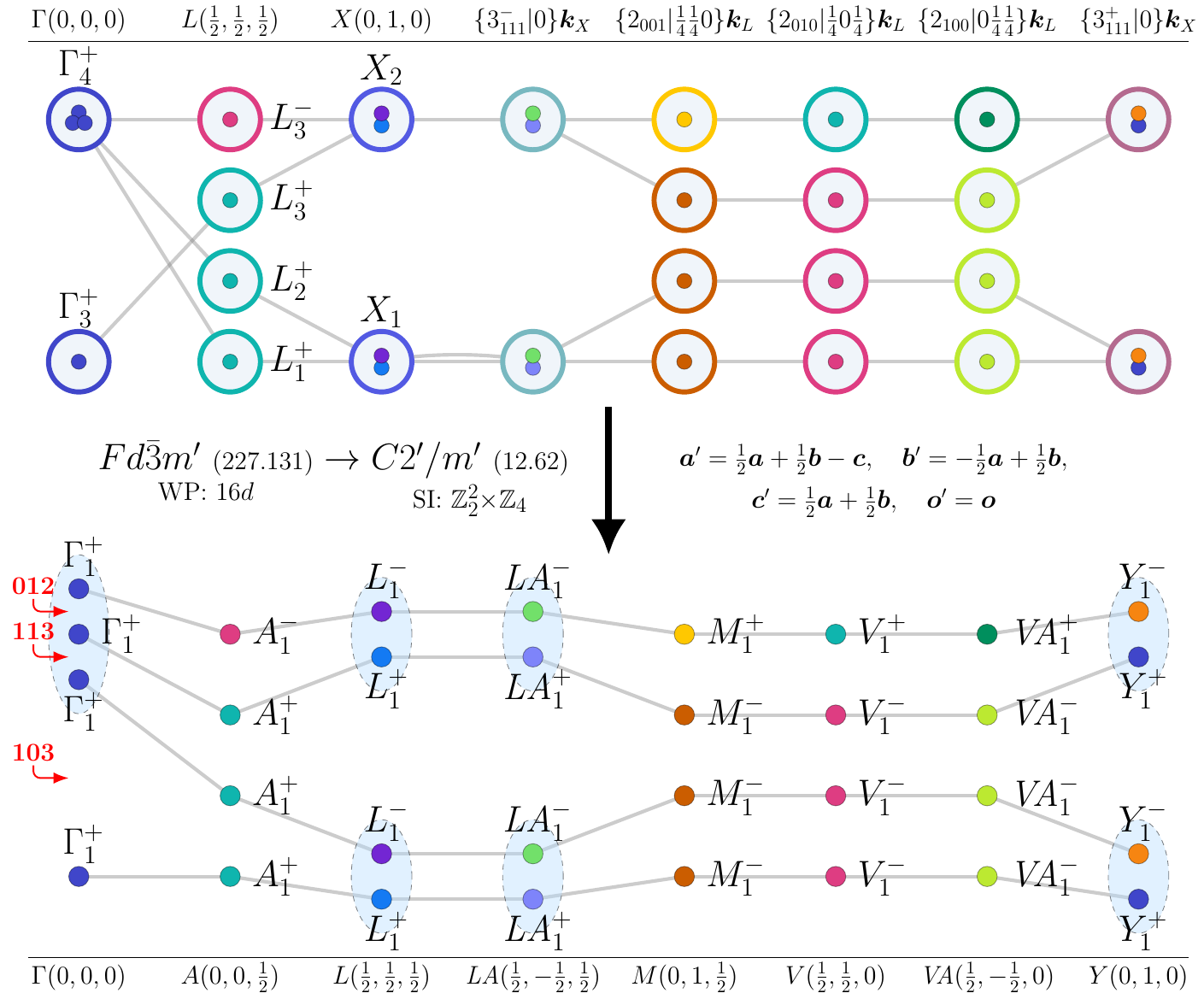}
\caption{Topological magnon bands in subgroup $C2'/m'~(12.62)$ for magnetic moments on Wyckoff position $16d$ of supergroup $Fd\bar{3}m'~(227.131)$.\label{fig_227.131_12.62_Bparallel110_16d}}
\end{figure}
\input{gap_tables_tex/227.131_12.62_Bparallel110_16d_table.tex}
\input{si_tables_tex/227.131_12.62_Bparallel110_16d_table.tex}
\subsubsection{Topological bands in subgroup $C2'/m'~(12.62)$}
\textbf{Perturbations:}
\begin{itemize}
\item strain $\perp$ [110],
\item B $\perp$ [110].
\end{itemize}
\begin{figure}[H]
\centering
\includegraphics[scale=0.6]{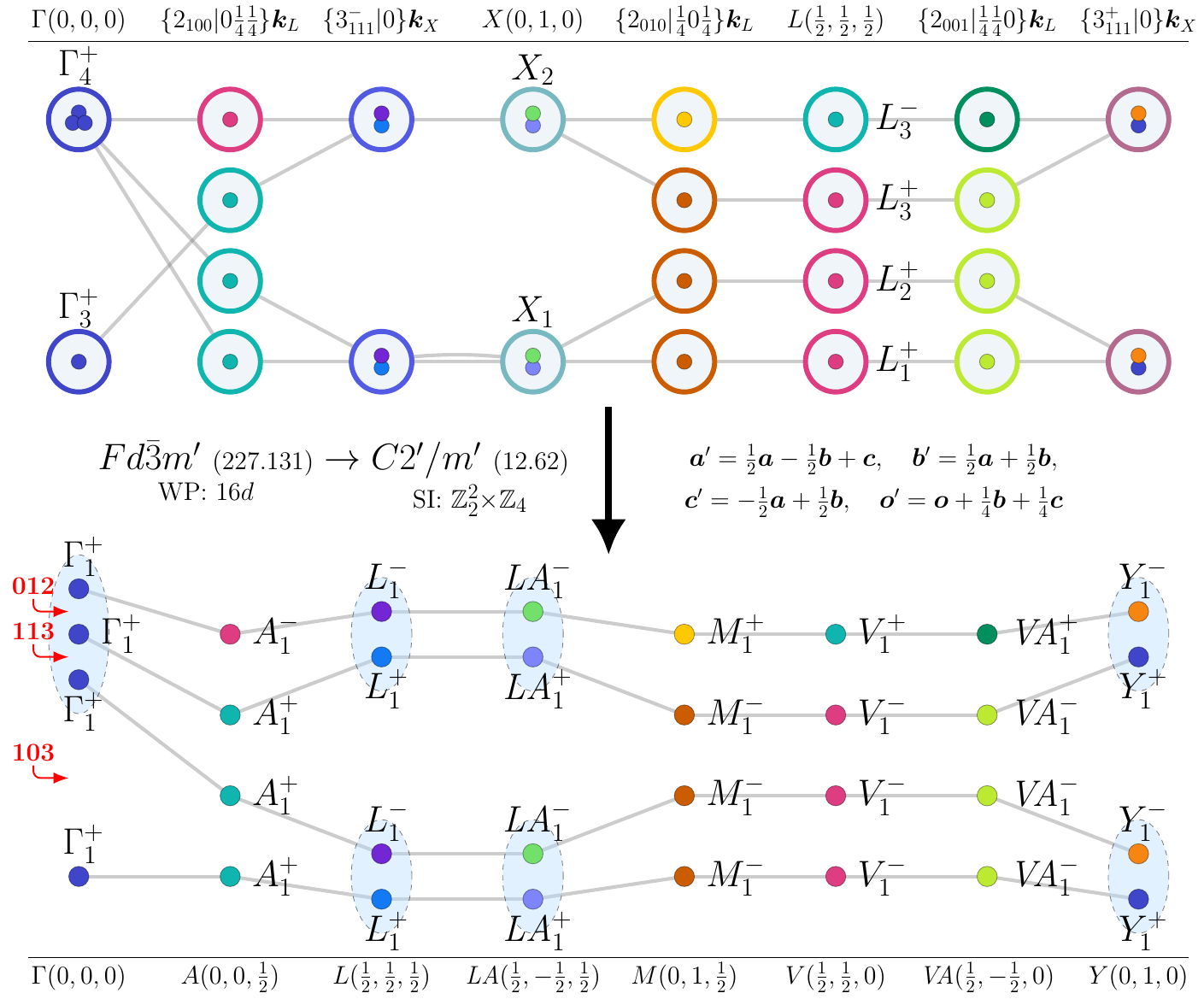}
\caption{Topological magnon bands in subgroup $C2'/m'~(12.62)$ for magnetic moments on Wyckoff position $16d$ of supergroup $Fd\bar{3}m'~(227.131)$.\label{fig_227.131_12.62_strainperp110_16d}}
\end{figure}
\input{gap_tables_tex/227.131_12.62_strainperp110_16d_table.tex}
\input{si_tables_tex/227.131_12.62_strainperp110_16d_table.tex}

\section{MSG $F_{S}d\bar{3}c~(228.139)$}
\textbf{Nontrivial-SI Subgroups:} $C_{a}c~(9.41)$, $C_{a}c~(9.41)$, $P\bar{1}~(2.4)$, $C2'/m'~(12.62)$, $C2'/m'~(12.62)$, $C2'/m'~(12.62)$, $C2'/c'~(15.89)$, $C2'/c'~(15.89)$, $P_{S}\bar{1}~(2.7)$, $R\bar{3}m'~(166.101)$, $R_{I}\bar{3}c~(167.108)$, $I_{a}ba2~(45.240)$, $C2/c~(15.85)$, $Iba'm'~(72.544)$, $C_{c}2/c~(15.90)$, $I_{c}bca~(73.553)$, $C_{a}2~(5.17)$, $C2/c~(15.85)$, $C_{a}2/c~(15.91)$, $I_{c}4_{1}cd~(110.250)$, $I4_{1}/am'd'~(141.557)$, $I_{c}4_{1}/acd~(142.570)$.\\

\textbf{Trivial-SI Subgroups:} $Cm'~(8.34)$, $Cm'~(8.34)$, $Cm'~(8.34)$, $Cc'~(9.39)$, $Cc'~(9.39)$, $C2'~(5.15)$, $C2'~(5.15)$, $P_{S}1~(1.3)$, $C_{c}c~(9.40)$, $C_{c}c~(9.40)$, $Cc~(9.37)$, $C_{c}c~(9.40)$, $Cc~(9.37)$, $R3m'~(160.67)$, $R_{I}3c~(161.72)$, $C2~(5.13)$, $Im'a'2~(46.245)$, $C_{c}2~(5.16)$, $C2~(5.13)$, $I4_{1}m'd'~(109.243)$.\\

\subsection{WP: $32b$}
\textbf{BCS Materials:} {NpTe~(40 K)}\footnote{BCS web page: \texttt{\href{http://webbdcrista1.ehu.es/magndata/index.php?this\_label=3.11} {http://webbdcrista1.ehu.es/magndata/index.php?this\_label=3.11}}}, {NpSe~(38 K)}\footnote{BCS web page: \texttt{\href{http://webbdcrista1.ehu.es/magndata/index.php?this\_label=3.10} {http://webbdcrista1.ehu.es/magndata/index.php?this\_label=3.10}}}, {NpS~(23 K)}\footnote{BCS web page: \texttt{\href{http://webbdcrista1.ehu.es/magndata/index.php?this\_label=3.9} {http://webbdcrista1.ehu.es/magndata/index.php?this\_label=3.9}}}.\\
\subsubsection{Topological bands in subgroup $P\bar{1}~(2.4)$}
\textbf{Perturbations:}
\begin{itemize}
\item B $\parallel$ [100] and strain $\perp$ [110],
\item B $\parallel$ [100] and strain in generic direction,
\item B $\parallel$ [110] and strain $\perp$ [100],
\item B $\parallel$ [110] and strain in generic direction,
\item B $\parallel$ [111] and strain $\perp$ [100],
\item B $\parallel$ [111] and strain $\perp$ [110],
\item B $\parallel$ [111] and strain in generic direction,
\item B $\perp$ [100] and strain $\parallel$ [110],
\item B $\perp$ [100] and strain $\parallel$ [111],
\item B $\perp$ [100] and strain $\perp$ [110],
\item B $\perp$ [100] and strain in generic direction,
\item B $\perp$ [110] and strain $\parallel$ [100],
\item B $\perp$ [110] and strain $\parallel$ [111],
\item B $\perp$ [110] and strain $\perp$ [100],
\item B $\perp$ [110] and strain in generic direction,
\item B in generic direction.
\end{itemize}
\begin{figure}[H]
\centering
\includegraphics[scale=0.6]{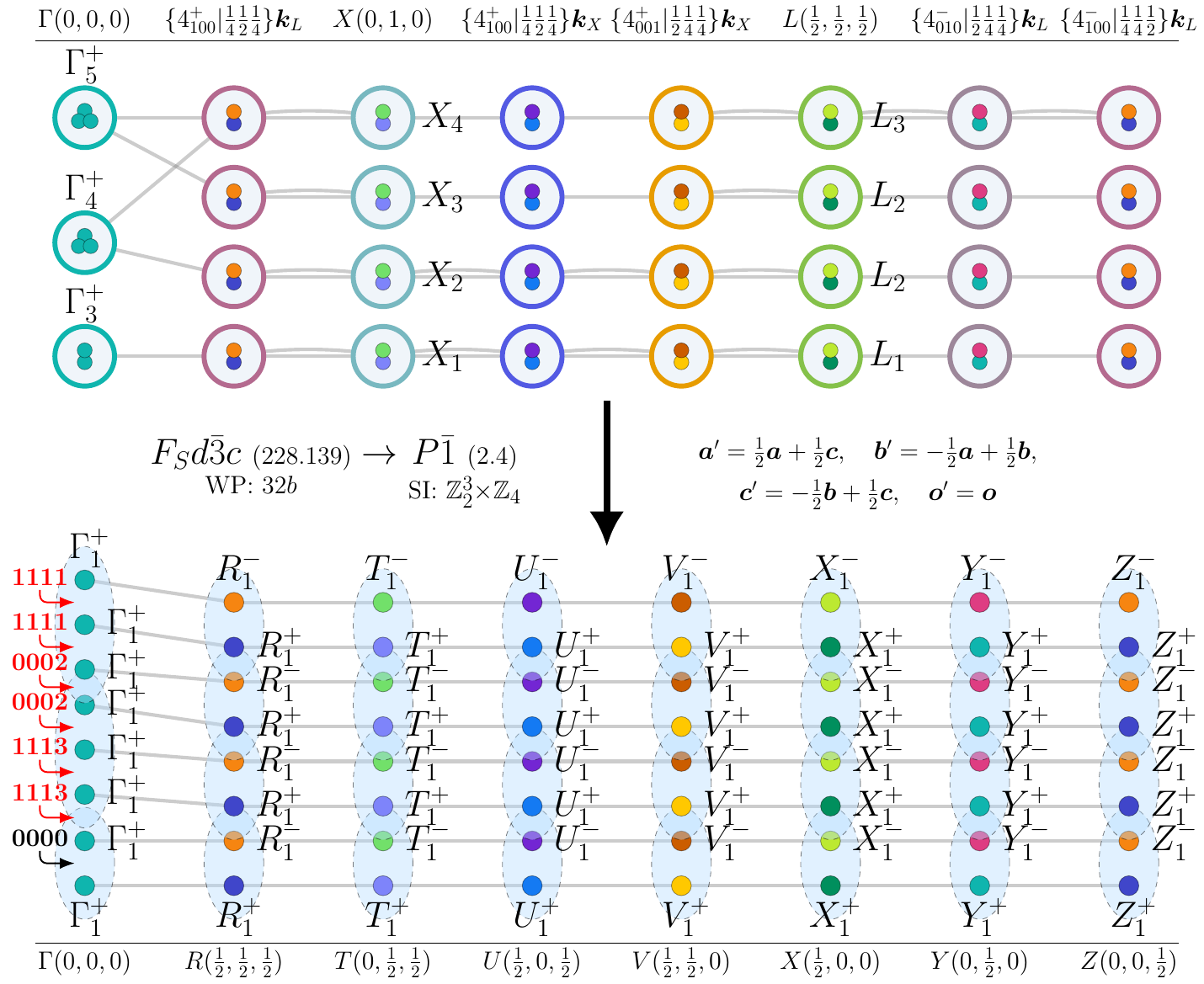}
\caption{Topological magnon bands in subgroup $P\bar{1}~(2.4)$ for magnetic moments on Wyckoff position $32b$ of supergroup $F_{S}d\bar{3}c~(228.139)$.\label{fig_228.139_2.4_Bparallel100andstrainperp110_32b}}
\end{figure}
\input{gap_tables_tex/228.139_2.4_Bparallel100andstrainperp110_32b_table.tex}
\input{si_tables_tex/228.139_2.4_Bparallel100andstrainperp110_32b_table.tex}
\subsubsection{Topological bands in subgroup $C2'/m'~(12.62)$}
\textbf{Perturbations:}
\begin{itemize}
\item B $\parallel$ [100] and strain $\parallel$ [111],
\item B $\parallel$ [111] and strain $\parallel$ [100].
\end{itemize}
\begin{figure}[H]
\centering
\includegraphics[scale=0.6]{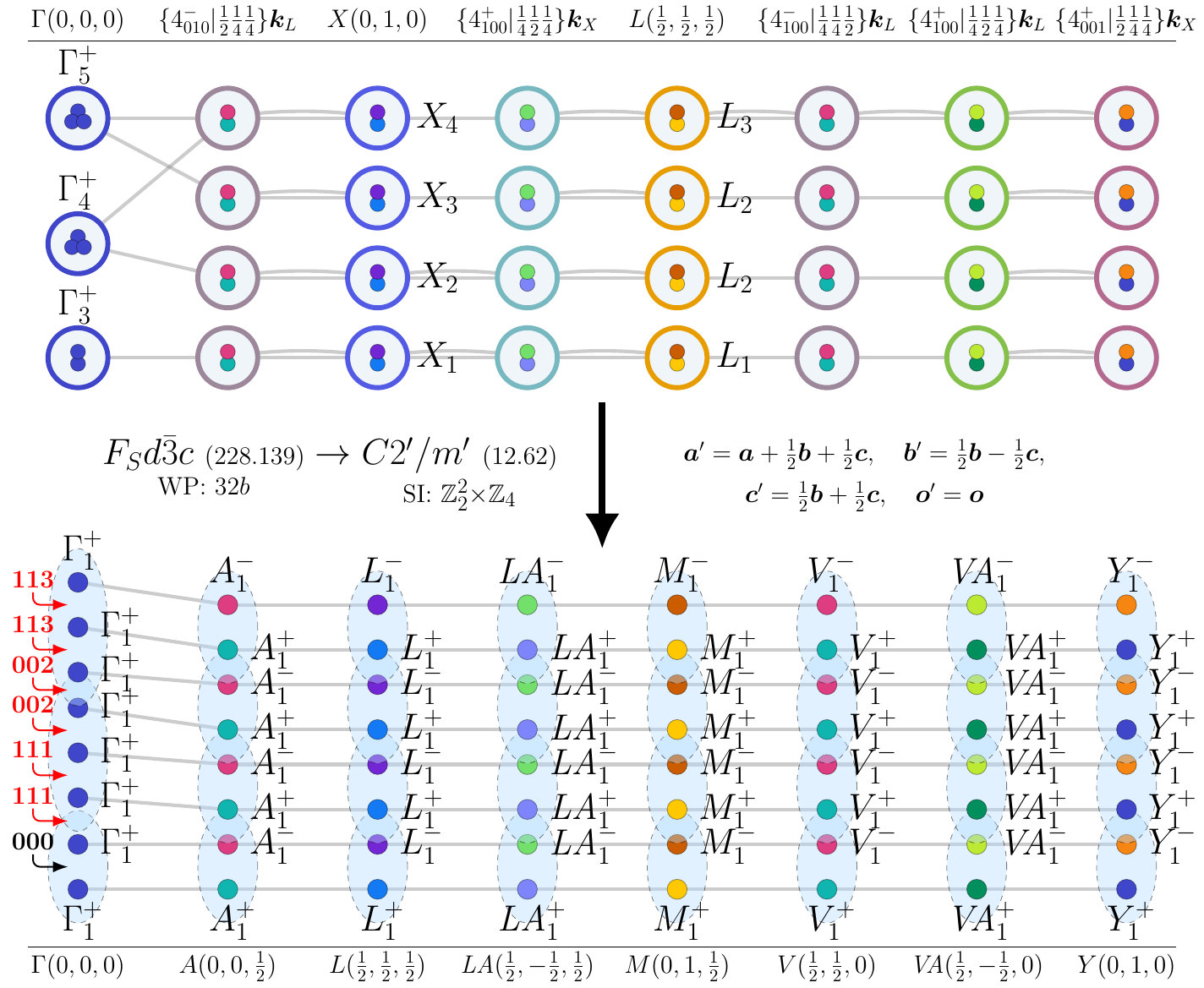}
\caption{Topological magnon bands in subgroup $C2'/m'~(12.62)$ for magnetic moments on Wyckoff position $32b$ of supergroup $F_{S}d\bar{3}c~(228.139)$.\label{fig_228.139_12.62_Bparallel100andstrainparallel111_32b}}
\end{figure}
\input{gap_tables_tex/228.139_12.62_Bparallel100andstrainparallel111_32b_table.tex}
\input{si_tables_tex/228.139_12.62_Bparallel100andstrainparallel111_32b_table.tex}
\subsubsection{Topological bands in subgroup $C2'/m'~(12.62)$}
\textbf{Perturbations:}
\begin{itemize}
\item B $\parallel$ [110] and strain $\parallel$ [111],
\item B $\parallel$ [111] and strain $\parallel$ [110].
\end{itemize}
\begin{figure}[H]
\centering
\includegraphics[scale=0.6]{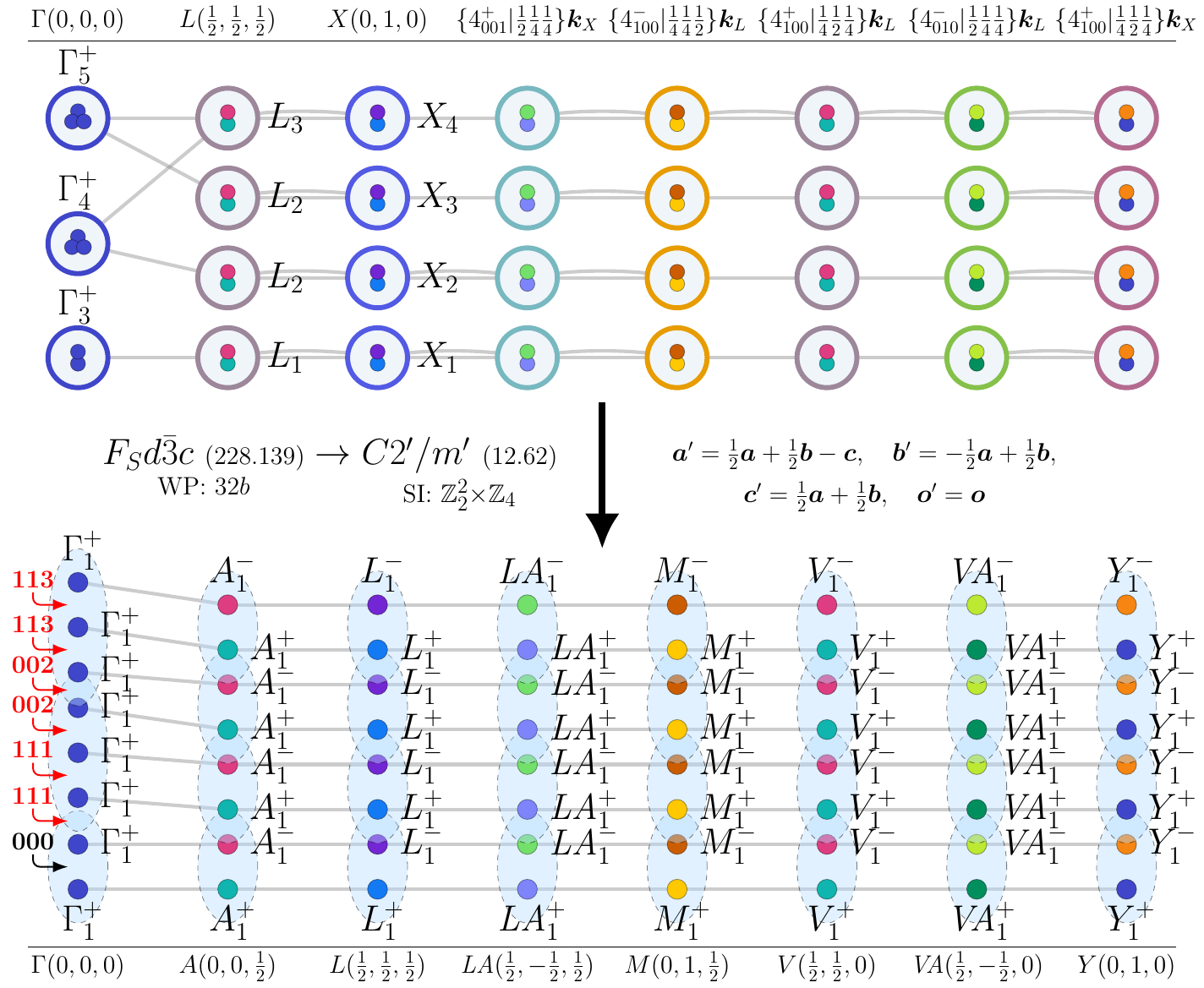}
\caption{Topological magnon bands in subgroup $C2'/m'~(12.62)$ for magnetic moments on Wyckoff position $32b$ of supergroup $F_{S}d\bar{3}c~(228.139)$.\label{fig_228.139_12.62_Bparallel110andstrainparallel111_32b}}
\end{figure}
\input{gap_tables_tex/228.139_12.62_Bparallel110andstrainparallel111_32b_table.tex}
\input{si_tables_tex/228.139_12.62_Bparallel110andstrainparallel111_32b_table.tex}
\subsubsection{Topological bands in subgroup $C2'/m'~(12.62)$}
\textbf{Perturbation:}
\begin{itemize}
\item B $\perp$ [110].
\end{itemize}
\begin{figure}[H]
\centering
\includegraphics[scale=0.6]{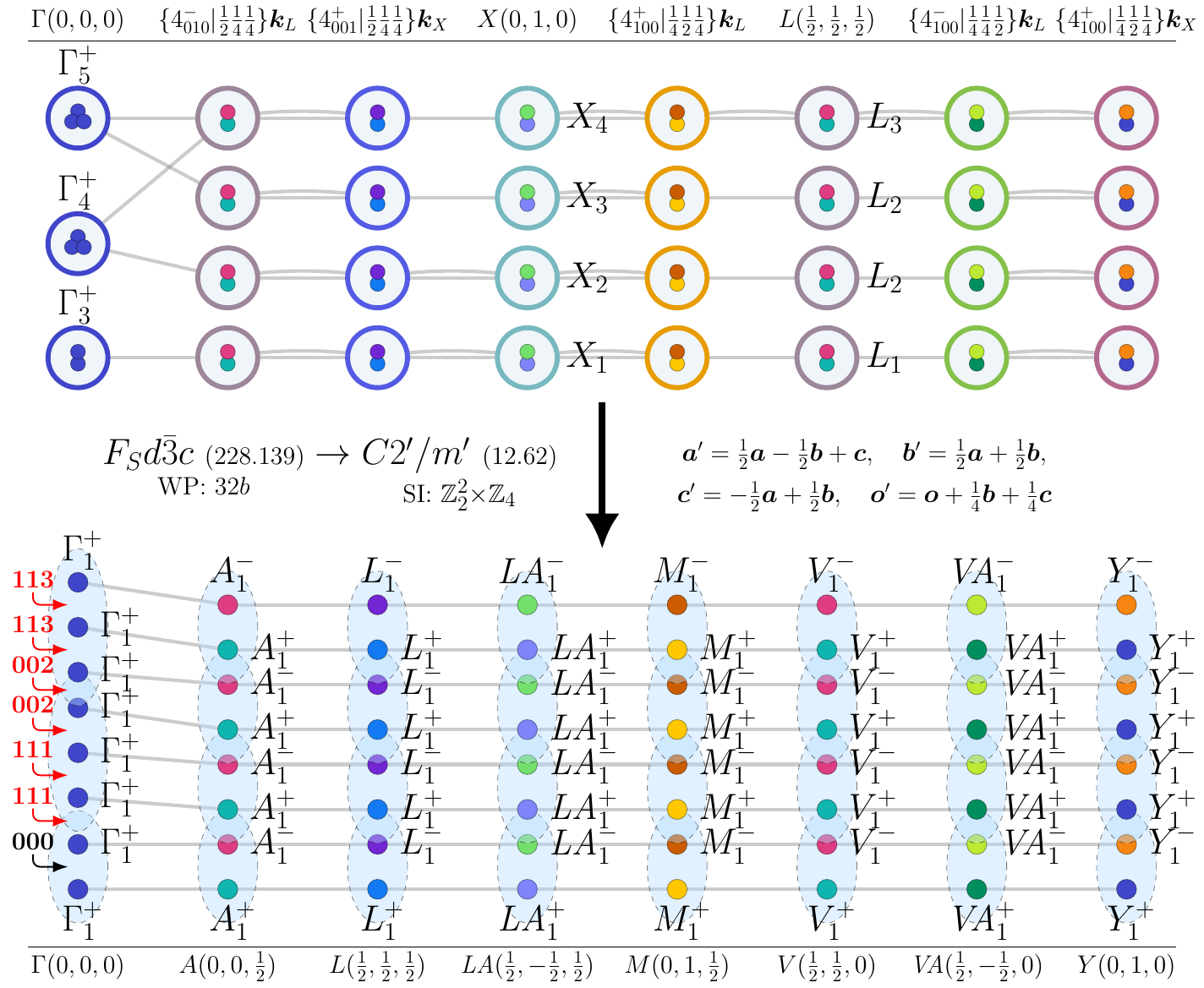}
\caption{Topological magnon bands in subgroup $C2'/m'~(12.62)$ for magnetic moments on Wyckoff position $32b$ of supergroup $F_{S}d\bar{3}c~(228.139)$.\label{fig_228.139_12.62_Bperp110_32b}}
\end{figure}
\input{gap_tables_tex/228.139_12.62_Bperp110_32b_table.tex}
\input{si_tables_tex/228.139_12.62_Bperp110_32b_table.tex}
\subsubsection{Topological bands in subgroup $C2'/c'~(15.89)$}
\textbf{Perturbations:}
\begin{itemize}
\item B $\parallel$ [100] and strain $\parallel$ [110],
\item B $\parallel$ [110] and strain $\parallel$ [100].
\end{itemize}
\begin{figure}[H]
\centering
\includegraphics[scale=0.6]{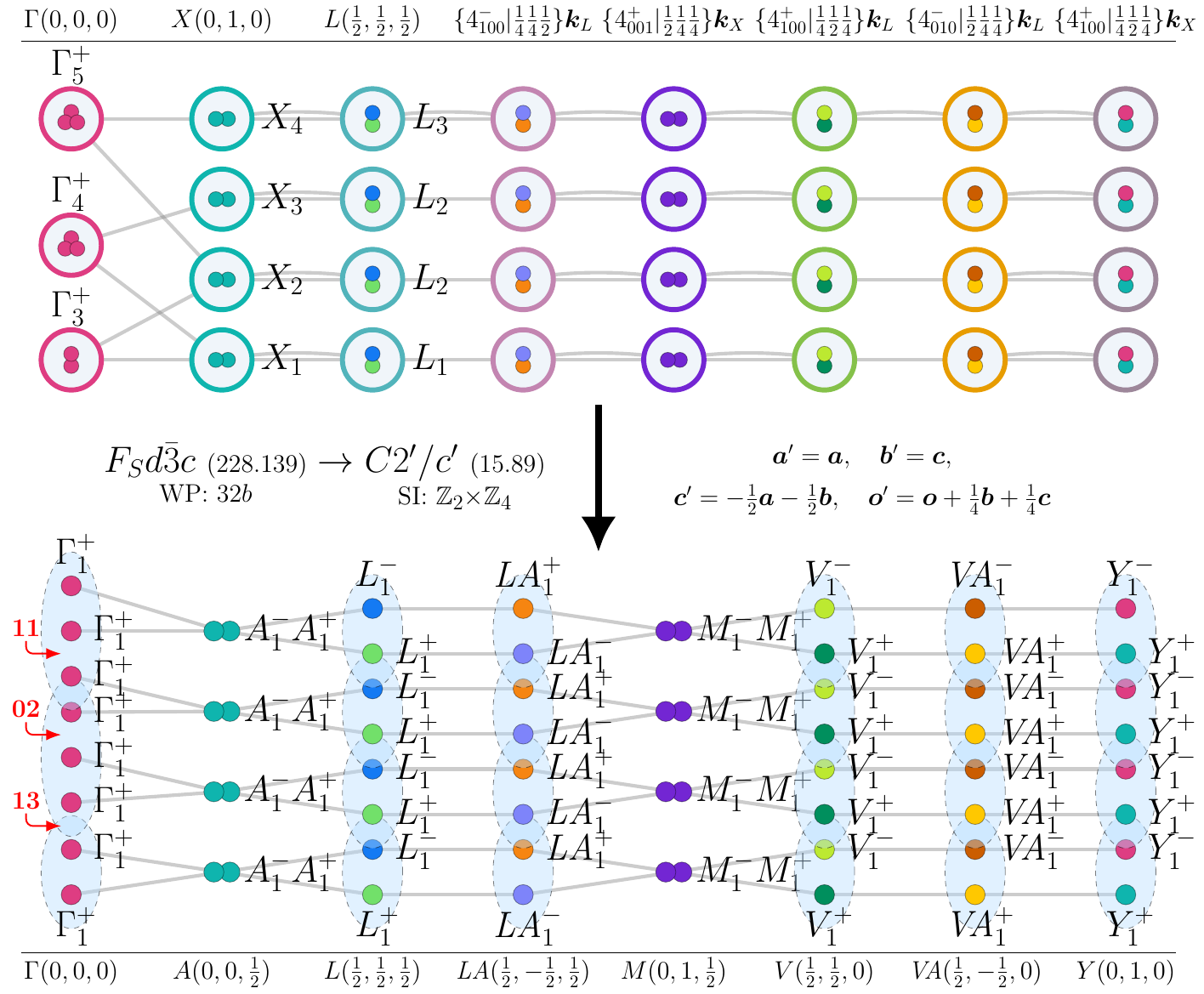}
\caption{Topological magnon bands in subgroup $C2'/c'~(15.89)$ for magnetic moments on Wyckoff position $32b$ of supergroup $F_{S}d\bar{3}c~(228.139)$.\label{fig_228.139_15.89_Bparallel100andstrainparallel110_32b}}
\end{figure}
\input{gap_tables_tex/228.139_15.89_Bparallel100andstrainparallel110_32b_table.tex}
\input{si_tables_tex/228.139_15.89_Bparallel100andstrainparallel110_32b_table.tex}
\subsubsection{Topological bands in subgroup $C2'/c'~(15.89)$}
\textbf{Perturbation:}
\begin{itemize}
\item B $\perp$ [100].
\end{itemize}
\begin{figure}[H]
\centering
\includegraphics[scale=0.6]{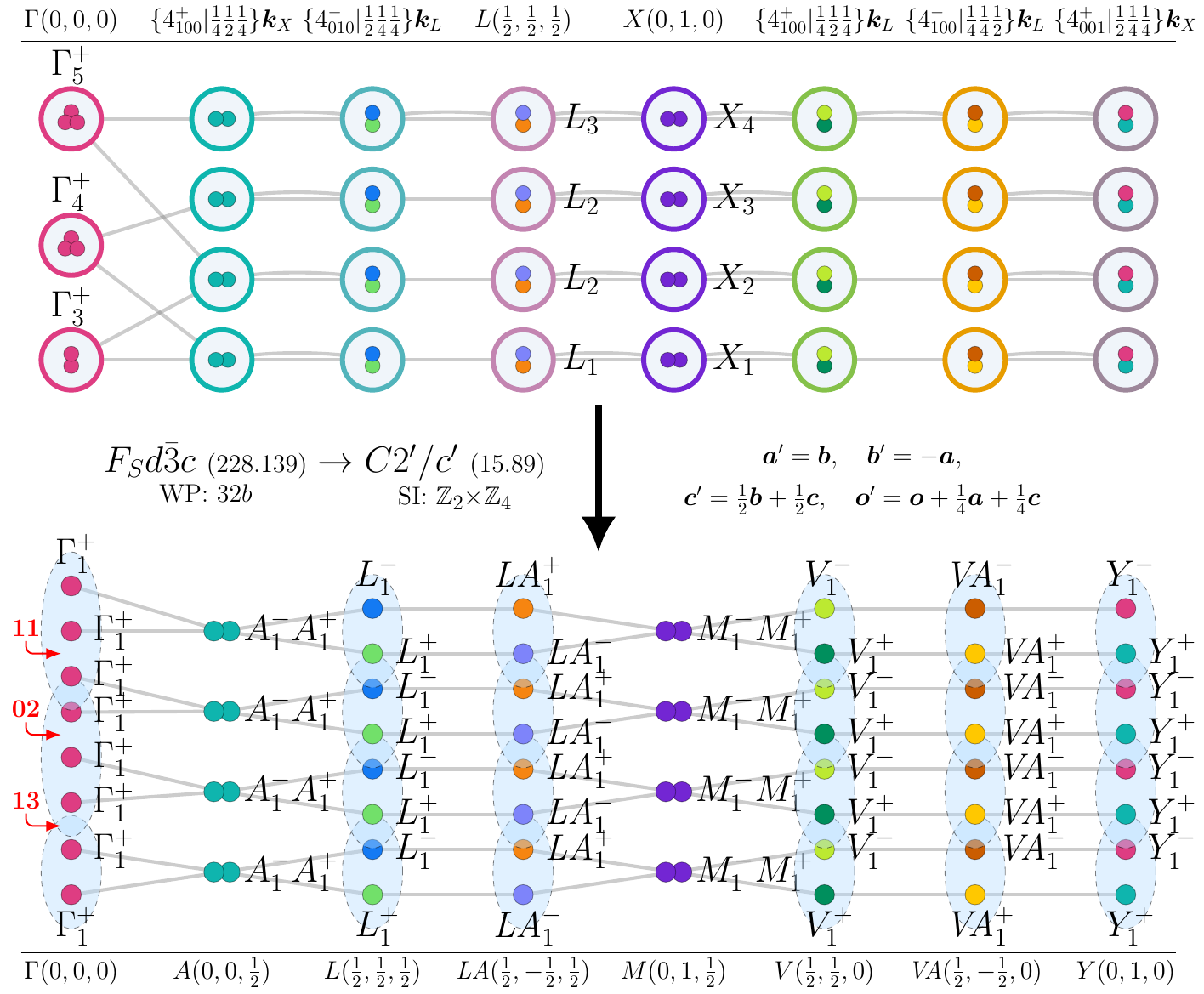}
\caption{Topological magnon bands in subgroup $C2'/c'~(15.89)$ for magnetic moments on Wyckoff position $32b$ of supergroup $F_{S}d\bar{3}c~(228.139)$.\label{fig_228.139_15.89_Bperp100_32b}}
\end{figure}
\input{gap_tables_tex/228.139_15.89_Bperp100_32b_table.tex}
\input{si_tables_tex/228.139_15.89_Bperp100_32b_table.tex}
\subsubsection{Topological bands in subgroup $P_{S}\bar{1}~(2.7)$}
\textbf{Perturbation:}
\begin{itemize}
\item strain in generic direction.
\end{itemize}
\begin{figure}[H]
\centering
\includegraphics[scale=0.6]{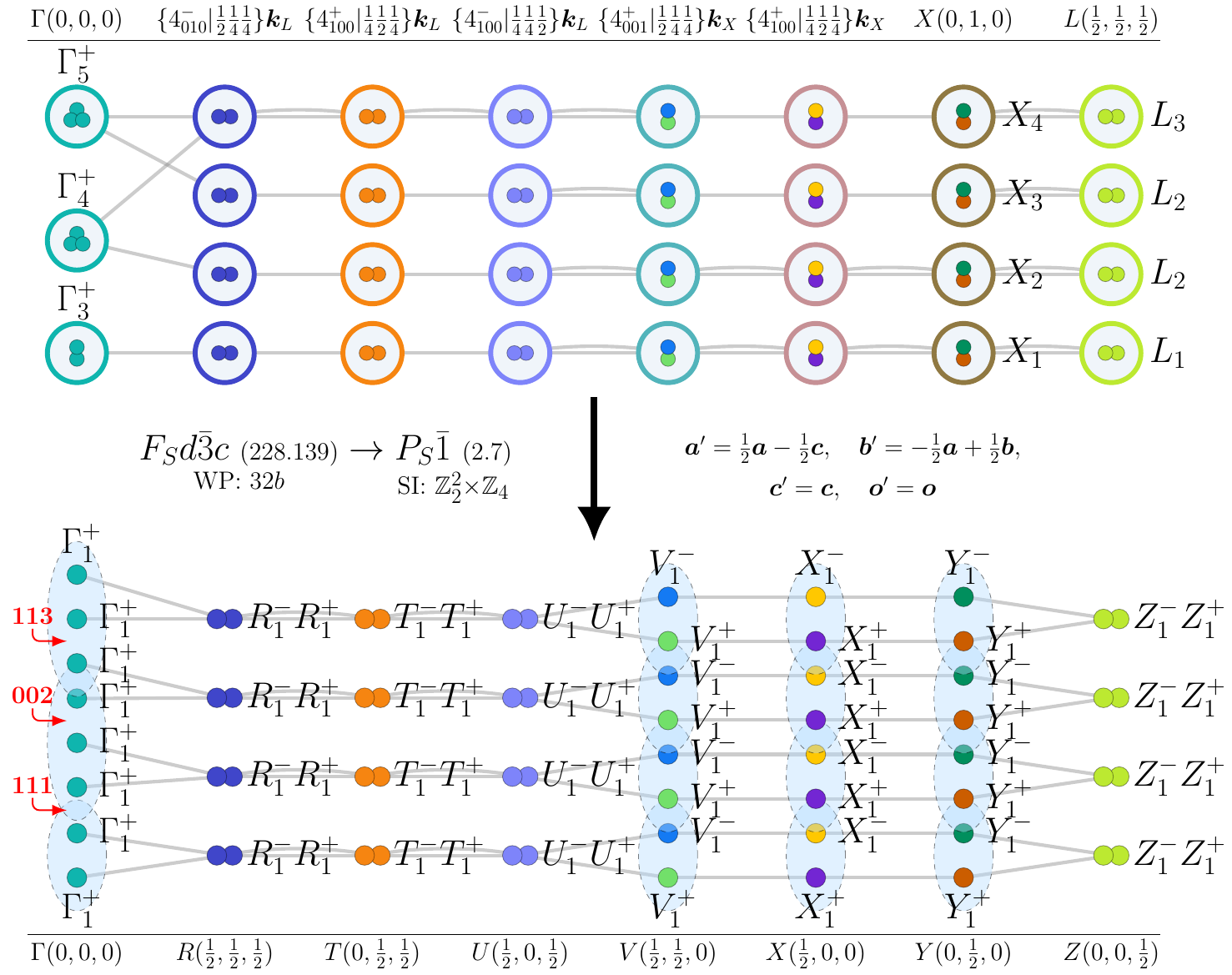}
\caption{Topological magnon bands in subgroup $P_{S}\bar{1}~(2.7)$ for magnetic moments on Wyckoff position $32b$ of supergroup $F_{S}d\bar{3}c~(228.139)$.\label{fig_228.139_2.7_strainingenericdirection_32b}}
\end{figure}
\input{gap_tables_tex/228.139_2.7_strainingenericdirection_32b_table.tex}
\input{si_tables_tex/228.139_2.7_strainingenericdirection_32b_table.tex}

\section{MSG $Im\bar{3}m'~(229.143)$}
\textbf{Nontrivial-SI Subgroups:} $P\bar{1}~(2.4)$, $C2'/m'~(12.62)$, $C2'/m'~(12.62)$, $C2'/m'~(12.62)$, $R\bar{3}m'~(166.101)$, $Fm'm'm~(69.524)$, $Fm'm'2~(42.222)$, $C2/m~(12.58)$, $Fm'm'm~(69.524)$.\\

\textbf{Trivial-SI Subgroups:} $Cm'~(8.34)$, $Cm'~(8.34)$, $Cm'~(8.34)$, $C2'~(5.15)$, $Cm~(8.32)$, $Fm'm2'~(42.221)$, $Cm~(8.32)$, $R3m'~(160.67)$, $C2~(5.13)$, $I4'mm'~(107.230)$, $I4'/mmm'~(139.535)$.\\

\subsection{WP: $8c$}
\textbf{BCS Materials:} {DyCu~(64 K)}\footnote{BCS web page: \texttt{\href{http://webbdcrista1.ehu.es/magndata/index.php?this\_label=3.6} {http://webbdcrista1.ehu.es/magndata/index.php?this\_label=3.6}}}.\\
\subsubsection{Topological bands in subgroup $P\bar{1}~(2.4)$}
\textbf{Perturbations:}
\begin{itemize}
\item strain in generic direction,
\item B $\parallel$ [100] and strain $\parallel$ [110],
\item B $\parallel$ [100] and strain $\perp$ [110],
\item B $\parallel$ [110] and strain $\parallel$ [100],
\item B $\parallel$ [110] and strain $\perp$ [100],
\item B $\parallel$ [110] and strain $\perp$ [110],
\item B $\parallel$ [111] and strain $\perp$ [100],
\item B $\parallel$ [111] and strain $\perp$ [110],
\item B in generic direction,
\item B $\perp$ [110] and strain $\parallel$ [100],
\item B $\perp$ [110] and strain $\parallel$ [111],
\item B $\perp$ [110] and strain $\perp$ [100].
\end{itemize}
\begin{figure}[H]
\centering
\includegraphics[scale=0.6]{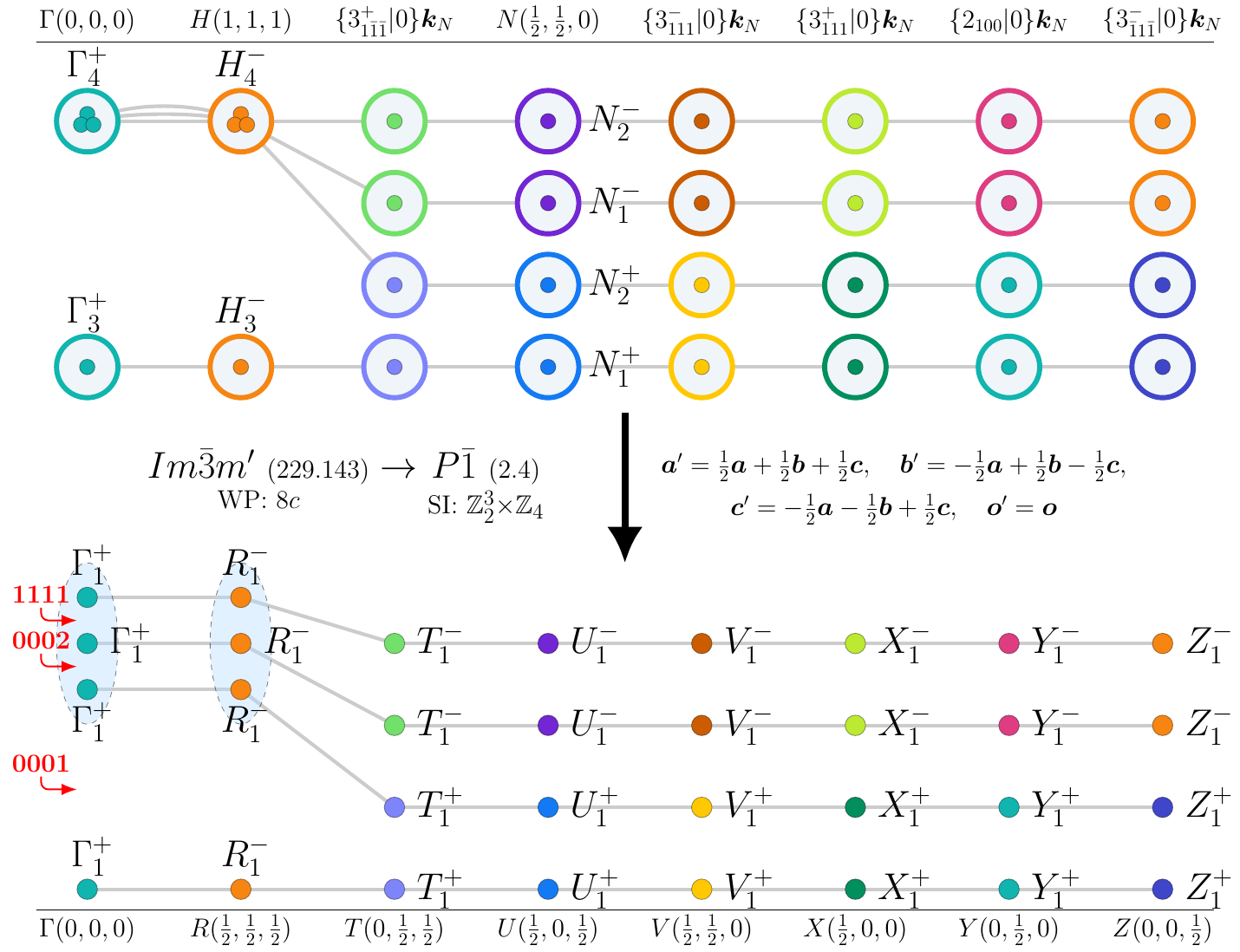}
\caption{Topological magnon bands in subgroup $P\bar{1}~(2.4)$ for magnetic moments on Wyckoff position $8c$ of supergroup $Im\bar{3}m'~(229.143)$.\label{fig_229.143_2.4_strainingenericdirection_8c}}
\end{figure}
\input{gap_tables_tex/229.143_2.4_strainingenericdirection_8c_table.tex}
\input{si_tables_tex/229.143_2.4_strainingenericdirection_8c_table.tex}
\subsubsection{Topological bands in subgroup $C2'/m'~(12.62)$}
\textbf{Perturbations:}
\begin{itemize}
\item B $\parallel$ [100] and strain $\parallel$ [111],
\item B $\parallel$ [111] and strain $\parallel$ [100].
\end{itemize}
\begin{figure}[H]
\centering
\includegraphics[scale=0.6]{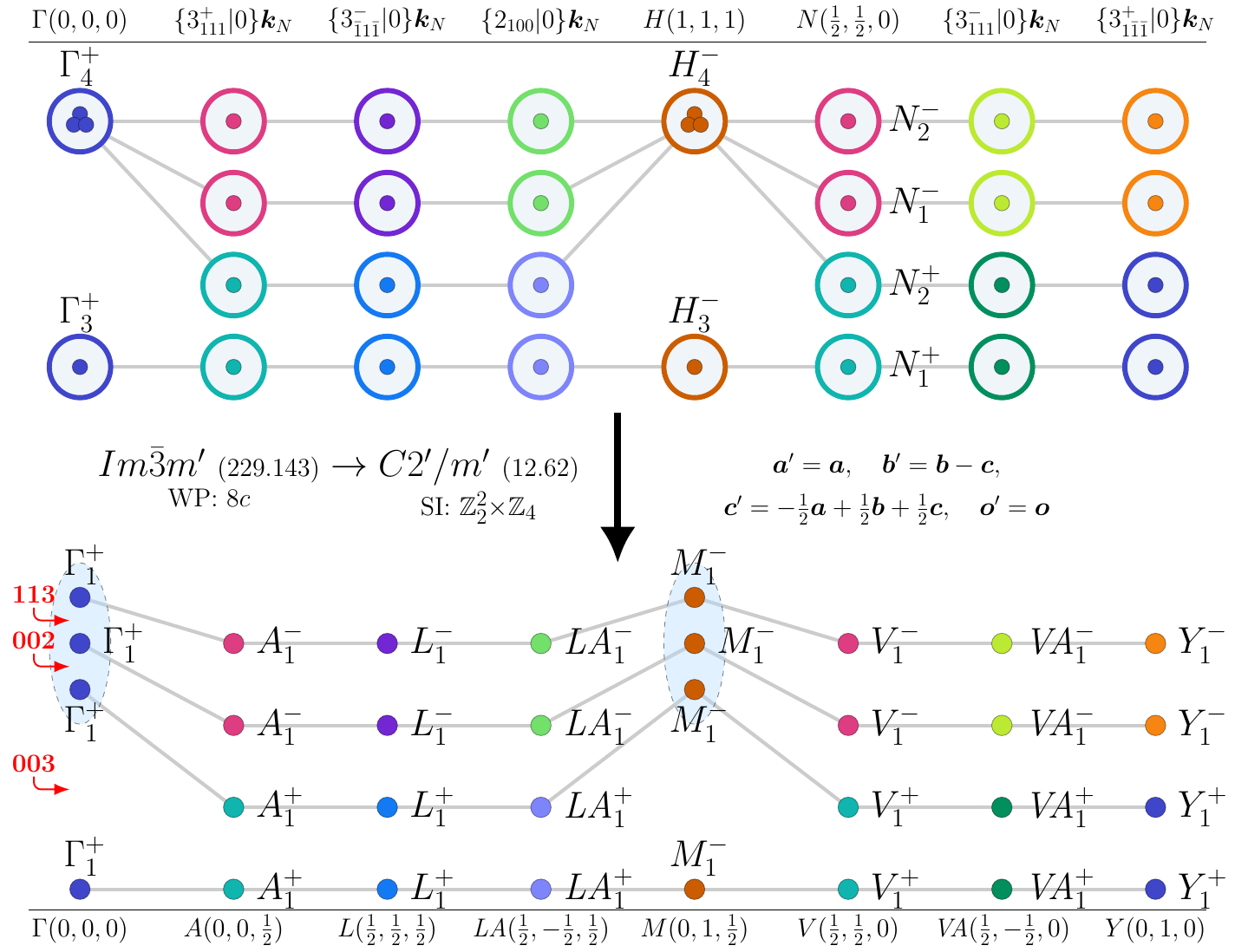}
\caption{Topological magnon bands in subgroup $C2'/m'~(12.62)$ for magnetic moments on Wyckoff position $8c$ of supergroup $Im\bar{3}m'~(229.143)$.\label{fig_229.143_12.62_Bparallel100andstrainparallel111_8c}}
\end{figure}
\input{gap_tables_tex/229.143_12.62_Bparallel100andstrainparallel111_8c_table.tex}
\input{si_tables_tex/229.143_12.62_Bparallel100andstrainparallel111_8c_table.tex}
\subsubsection{Topological bands in subgroup $C2'/m'~(12.62)$}
\textbf{Perturbations:}
\begin{itemize}
\item B $\parallel$ [110],
\item B $\parallel$ [111] and strain $\parallel$ [110].
\end{itemize}
\begin{figure}[H]
\centering
\includegraphics[scale=0.6]{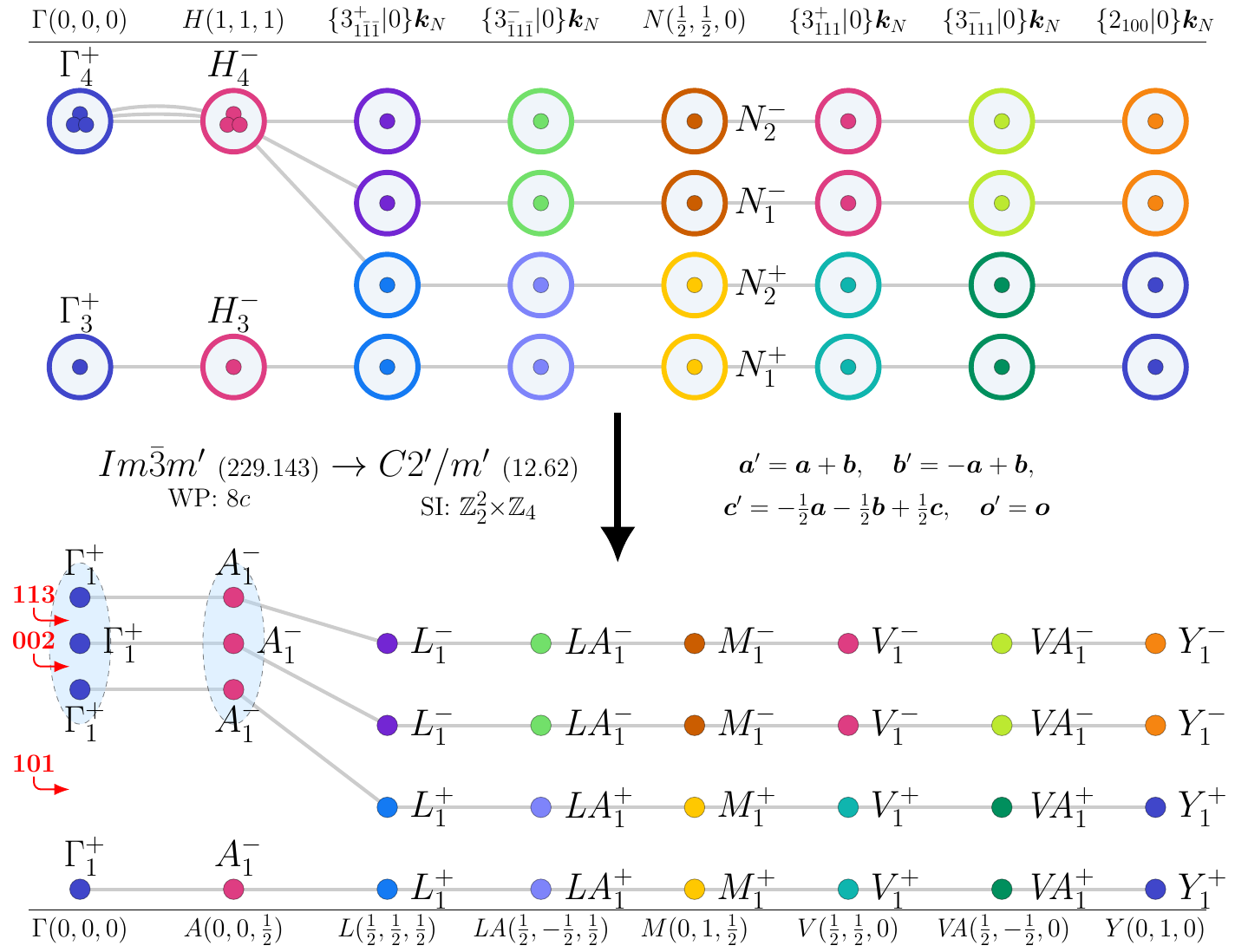}
\caption{Topological magnon bands in subgroup $C2'/m'~(12.62)$ for magnetic moments on Wyckoff position $8c$ of supergroup $Im\bar{3}m'~(229.143)$.\label{fig_229.143_12.62_Bparallel110_8c}}
\end{figure}
\input{gap_tables_tex/229.143_12.62_Bparallel110_8c_table.tex}
\input{si_tables_tex/229.143_12.62_Bparallel110_8c_table.tex}
\subsubsection{Topological bands in subgroup $C2'/m'~(12.62)$}
\textbf{Perturbations:}
\begin{itemize}
\item strain $\perp$ [110],
\item B $\perp$ [110].
\end{itemize}
\begin{figure}[H]
\centering
\includegraphics[scale=0.6]{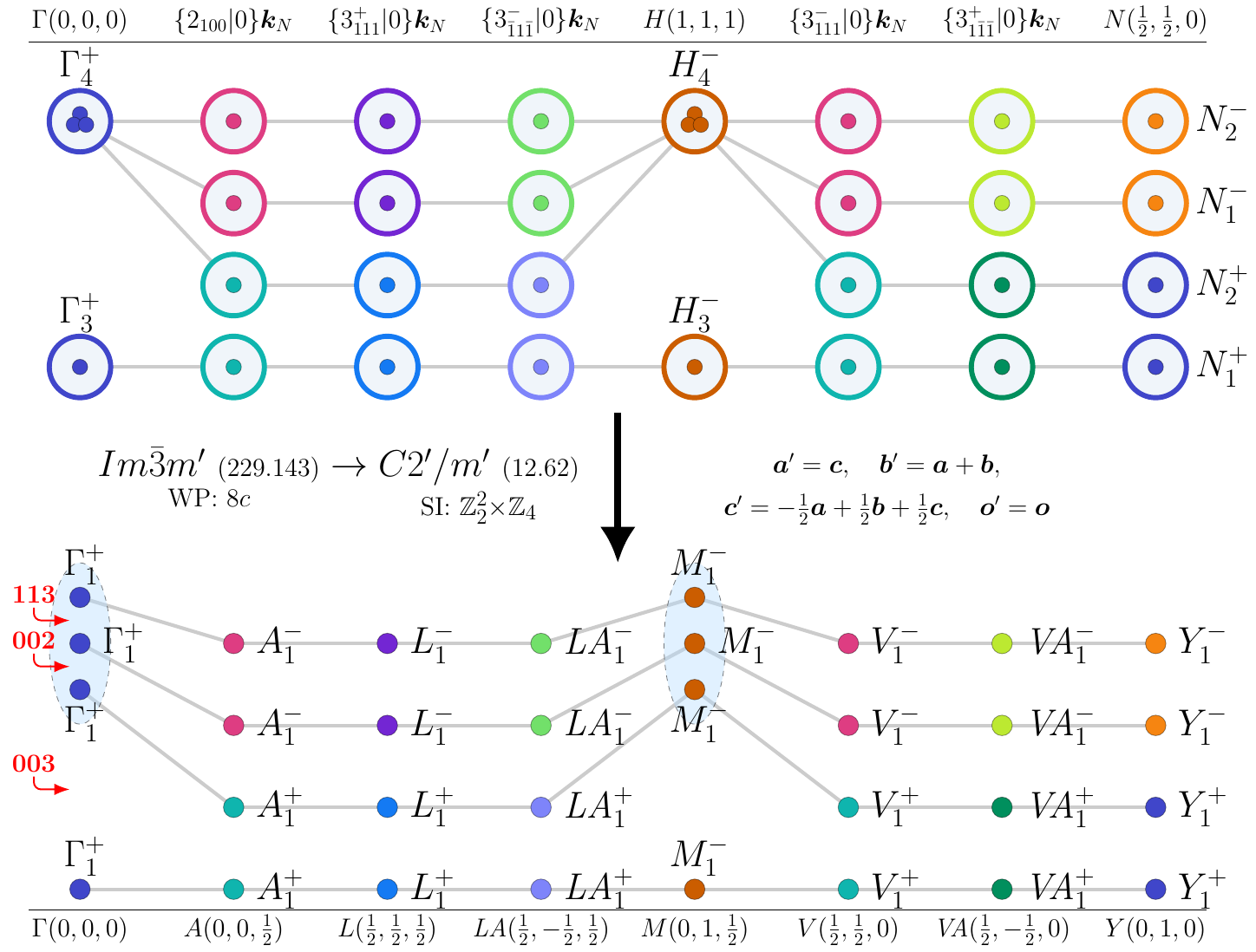}
\caption{Topological magnon bands in subgroup $C2'/m'~(12.62)$ for magnetic moments on Wyckoff position $8c$ of supergroup $Im\bar{3}m'~(229.143)$.\label{fig_229.143_12.62_strainperp110_8c}}
\end{figure}
\input{gap_tables_tex/229.143_12.62_strainperp110_8c_table.tex}
\input{si_tables_tex/229.143_12.62_strainperp110_8c_table.tex}

\section{MSG $Ia\bar{3}d'~(230.148)$}
\textbf{Nontrivial-SI Subgroups:} $P\bar{1}~(2.4)$, $C2'/c'~(15.89)$, $C2'/c'~(15.89)$, $C2'/c'~(15.89)$, $R\bar{3}c'~(167.107)$, $Fd'd'd~(70.530)$, $I4_{1}'cd'~(110.248)$, $C2/c~(15.85)$, $Fd'd'd~(70.530)$, $I4_{1}'/acd'~(142.565)$.\\

\textbf{Trivial-SI Subgroups:} $Cc'~(9.39)$, $Cc'~(9.39)$, $Cc'~(9.39)$, $C2'~(5.15)$, $Cc~(9.37)$, $Fd'd2'~(43.226)$, $Cc~(9.37)$, $R3c'~(161.71)$, $C2~(5.13)$, $Fd'd'2~(43.227)$.\\

\subsection{WP: $24c$}
\textbf{BCS Materials:} {Dy\textsubscript{3}Al\textsubscript{5}O\textsubscript{12}~(2.49 K)}\footnote{BCS web page: \texttt{\href{http://webbdcrista1.ehu.es/magndata/index.php?this\_label=0.127} {http://webbdcrista1.ehu.es/magndata/index.php?this\_label=0.127}}}, {Tb\textsubscript{3}Al\textsubscript{5}O\textsubscript{12}~(1.35 K)}\footnote{BCS web page: \texttt{\href{http://webbdcrista1.ehu.es/magndata/index.php?this\_label=0.744} {http://webbdcrista1.ehu.es/magndata/index.php?this\_label=0.744}}}, {Ho\textsubscript{3}Al\textsubscript{5}O\textsubscript{12}~(0.85 K)}\footnote{BCS web page: \texttt{\href{http://webbdcrista1.ehu.es/magndata/index.php?this\_label=0.743} {http://webbdcrista1.ehu.es/magndata/index.php?this\_label=0.743}}}, {Er\textsubscript{3}Ga\textsubscript{5}O\textsubscript{12}~(0.8 K)}\footnote{BCS web page: \texttt{\href{http://webbdcrista1.ehu.es/magndata/index.php?this\_label=0.741} {http://webbdcrista1.ehu.es/magndata/index.php?this\_label=0.741}}}, {Dy\textsubscript{3}Ga\textsubscript{5}O\textsubscript{12}~(0.37 K)}\footnote{BCS web page: \texttt{\href{http://webbdcrista1.ehu.es/magndata/index.php?this\_label=0.740} {http://webbdcrista1.ehu.es/magndata/index.php?this\_label=0.740}}}, {Tb\textsubscript{3}Ga\textsubscript{5}O\textsubscript{12}~(0.25 K)}\footnote{BCS web page: \texttt{\href{http://webbdcrista1.ehu.es/magndata/index.php?this\_label=0.742} {http://webbdcrista1.ehu.es/magndata/index.php?this\_label=0.742}}}, {Tb\textsubscript{3}Ga\textsubscript{5}O\textsubscript{12}~(0.24 K)}\footnote{BCS web page: \texttt{\href{http://webbdcrista1.ehu.es/magndata/index.php?this\_label=0.746} {http://webbdcrista1.ehu.es/magndata/index.php?this\_label=0.746}}}, {Ho\textsubscript{3}Ga\textsubscript{5}O\textsubscript{12}~(0.15 K)}\footnote{BCS web page: \texttt{\href{http://webbdcrista1.ehu.es/magndata/index.php?this\_label=0.745} {http://webbdcrista1.ehu.es/magndata/index.php?this\_label=0.745}}}.\\
\subsubsection{Topological bands in subgroup $C2'/c'~(15.89)$}
\textbf{Perturbations:}
\begin{itemize}
\item B $\parallel$ [100] and strain $\parallel$ [111],
\item B $\parallel$ [111] and strain $\parallel$ [100].
\end{itemize}
\begin{figure}[H]
\centering
\includegraphics[scale=0.6]{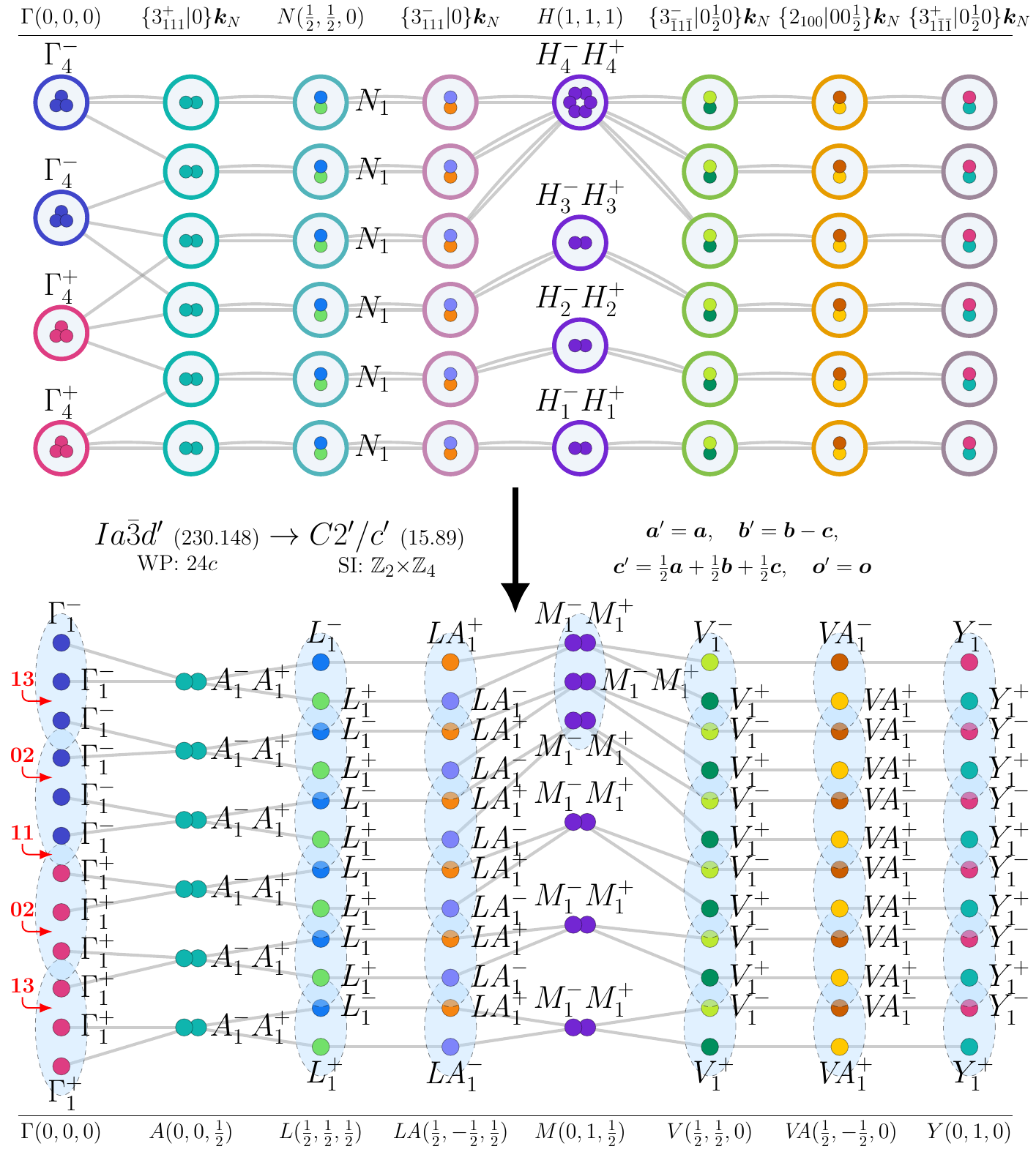}
\caption{Topological magnon bands in subgroup $C2'/c'~(15.89)$ for magnetic moments on Wyckoff position $24c$ of supergroup $Ia\bar{3}d'~(230.148)$.\label{fig_230.148_15.89_Bparallel100andstrainparallel111_24c}}
\end{figure}
\input{gap_tables_tex/230.148_15.89_Bparallel100andstrainparallel111_24c_table.tex}
\input{si_tables_tex/230.148_15.89_Bparallel100andstrainparallel111_24c_table.tex}
\subsubsection{Topological bands in subgroup $C2'/c'~(15.89)$}
\textbf{Perturbations:}
\begin{itemize}
\item B $\parallel$ [110],
\item B $\parallel$ [111] and strain $\parallel$ [110].
\end{itemize}
\begin{figure}[H]
\centering
\includegraphics[scale=0.6]{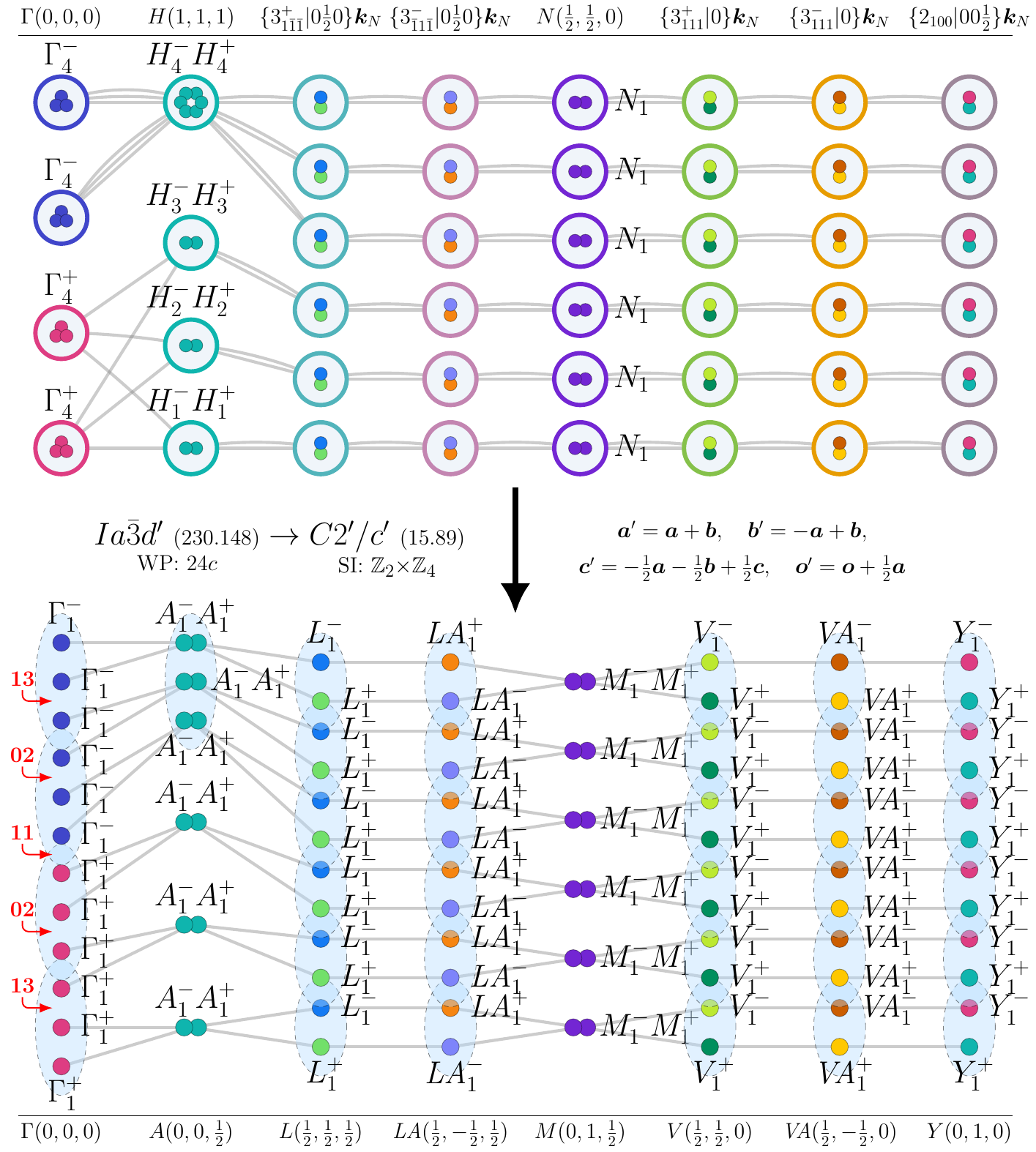}
\caption{Topological magnon bands in subgroup $C2'/c'~(15.89)$ for magnetic moments on Wyckoff position $24c$ of supergroup $Ia\bar{3}d'~(230.148)$.\label{fig_230.148_15.89_Bparallel110_24c}}
\end{figure}
\input{gap_tables_tex/230.148_15.89_Bparallel110_24c_table.tex}
\input{si_tables_tex/230.148_15.89_Bparallel110_24c_table.tex}
\subsubsection{Topological bands in subgroup $C2'/c'~(15.89)$}
\textbf{Perturbations:}
\begin{itemize}
\item strain $\perp$ [110],
\item B $\perp$ [110].
\end{itemize}
\begin{figure}[H]
\centering
\includegraphics[scale=0.6]{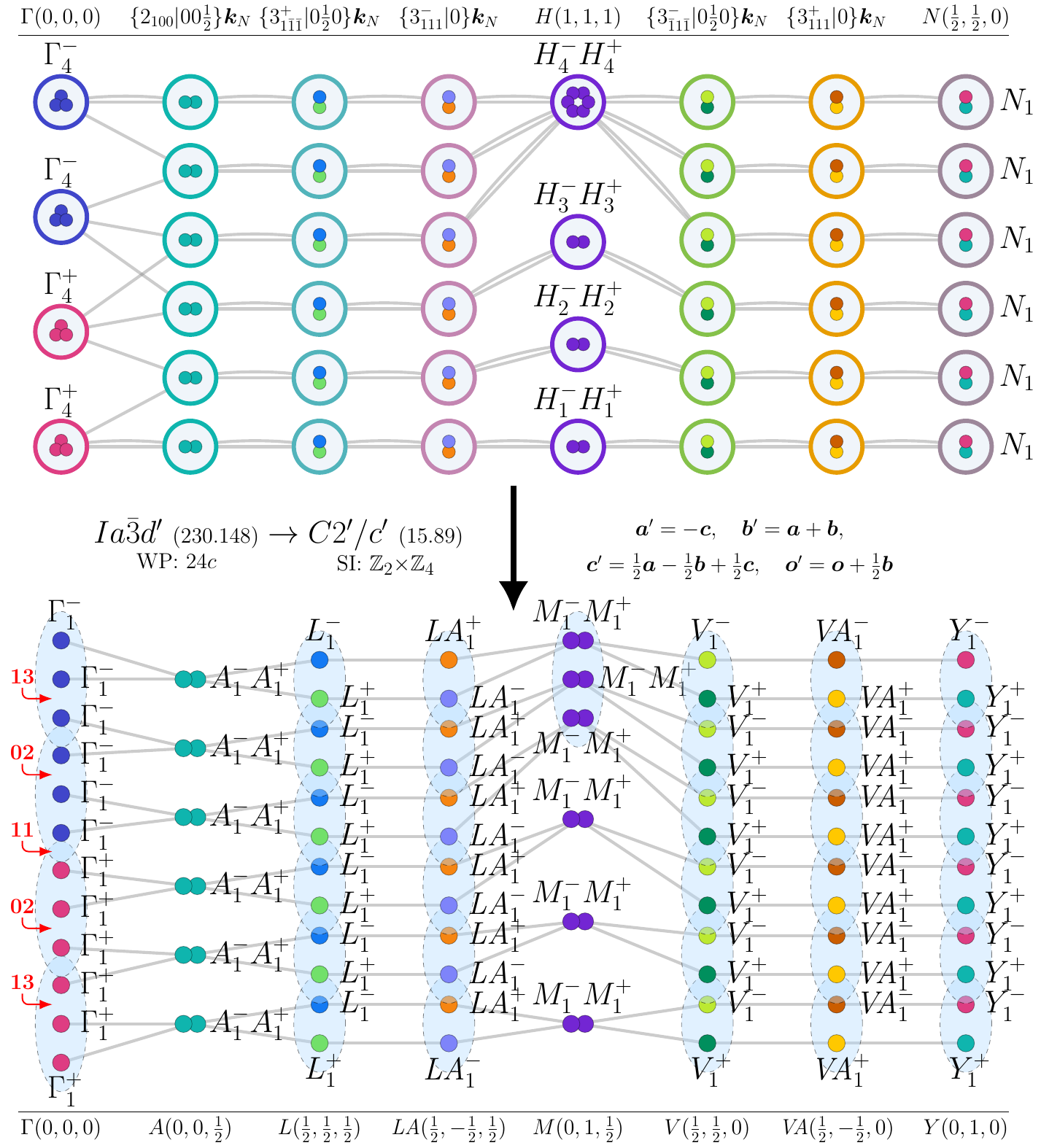}
\caption{Topological magnon bands in subgroup $C2'/c'~(15.89)$ for magnetic moments on Wyckoff position $24c$ of supergroup $Ia\bar{3}d'~(230.148)$.\label{fig_230.148_15.89_strainperp110_24c}}
\end{figure}
\input{gap_tables_tex/230.148_15.89_strainperp110_24c_table.tex}
\input{si_tables_tex/230.148_15.89_strainperp110_24c_table.tex}

\section{MSG $Pn'n'n~(48.260)$}
\textbf{Nontrivial-SI Subgroups:} $P\bar{1}~(2.4)$, $P2'/c'~(13.69)$, $P2'/c'~(13.69)$, $P2~(3.1)$, $P2/c~(13.65)$.\\

\textbf{Trivial-SI Subgroups:} $Pc'~(7.26)$, $Pc'~(7.26)$, $P2'~(3.3)$, $P2'~(3.3)$, $Pc~(7.24)$, $Pn'n2'~(34.158)$, $Pn'n2'~(34.158)$, $Pn'n'2~(34.159)$.\\

\subsection{WP: $2b+2c+2d+4f$}
\textbf{BCS Materials:} {CaCu\textsubscript{3}Fe\textsubscript{2}Sb\textsubscript{2}O\textsubscript{12}~(160 K)}\footnote{BCS web page: \texttt{\href{http://webbdcrista1.ehu.es/magndata/index.php?this\_label=0.672} {http://webbdcrista1.ehu.es/magndata/index.php?this\_label=0.672}}}.\\
\subsubsection{Topological bands in subgroup $P2'/c'~(13.69)$}
\textbf{Perturbations:}
\begin{itemize}
\item strain $\perp$ [010],
\item (B $\parallel$ [100] or B $\perp$ [010]).
\end{itemize}
\begin{figure}[H]
\centering
\includegraphics[scale=0.6]{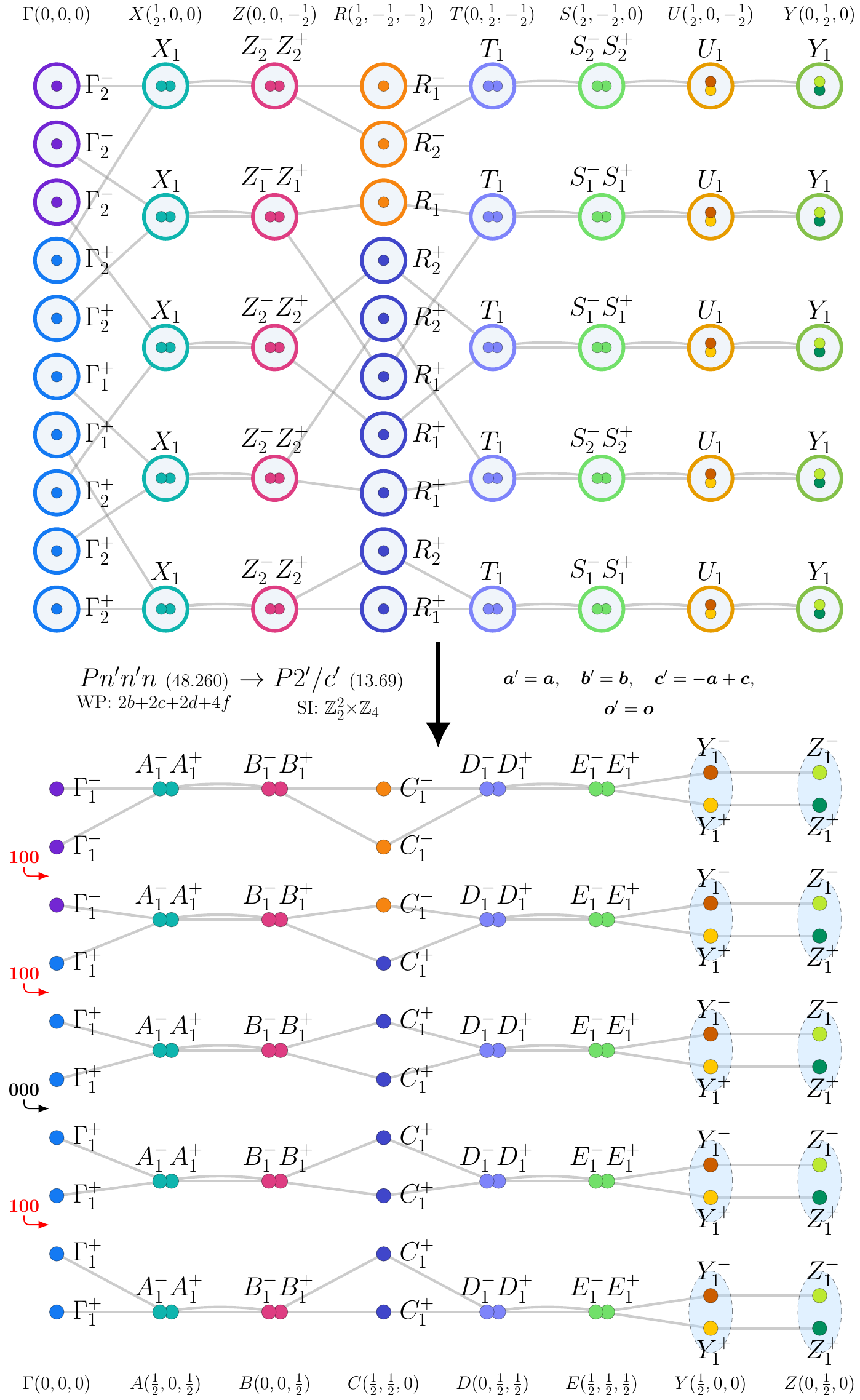}
\caption{Topological magnon bands in subgroup $P2'/c'~(13.69)$ for magnetic moments on Wyckoff positions $2b+2c+2d+4f$ of supergroup $Pn'n'n~(48.260)$.\label{fig_48.260_13.69_strainperp010_2b+2c+2d+4f}}
\end{figure}
\input{gap_tables_tex/48.260_13.69_strainperp010_2b+2c+2d+4f_table.tex}
\input{si_tables_tex/48.260_13.69_strainperp010_2b+2c+2d+4f_table.tex}
\subsubsection{Topological bands in subgroup $P2'/c'~(13.69)$}
\textbf{Perturbations:}
\begin{itemize}
\item strain $\perp$ [100],
\item (B $\parallel$ [010] or B $\perp$ [100]).
\end{itemize}
\begin{figure}[H]
\centering
\includegraphics[scale=0.6]{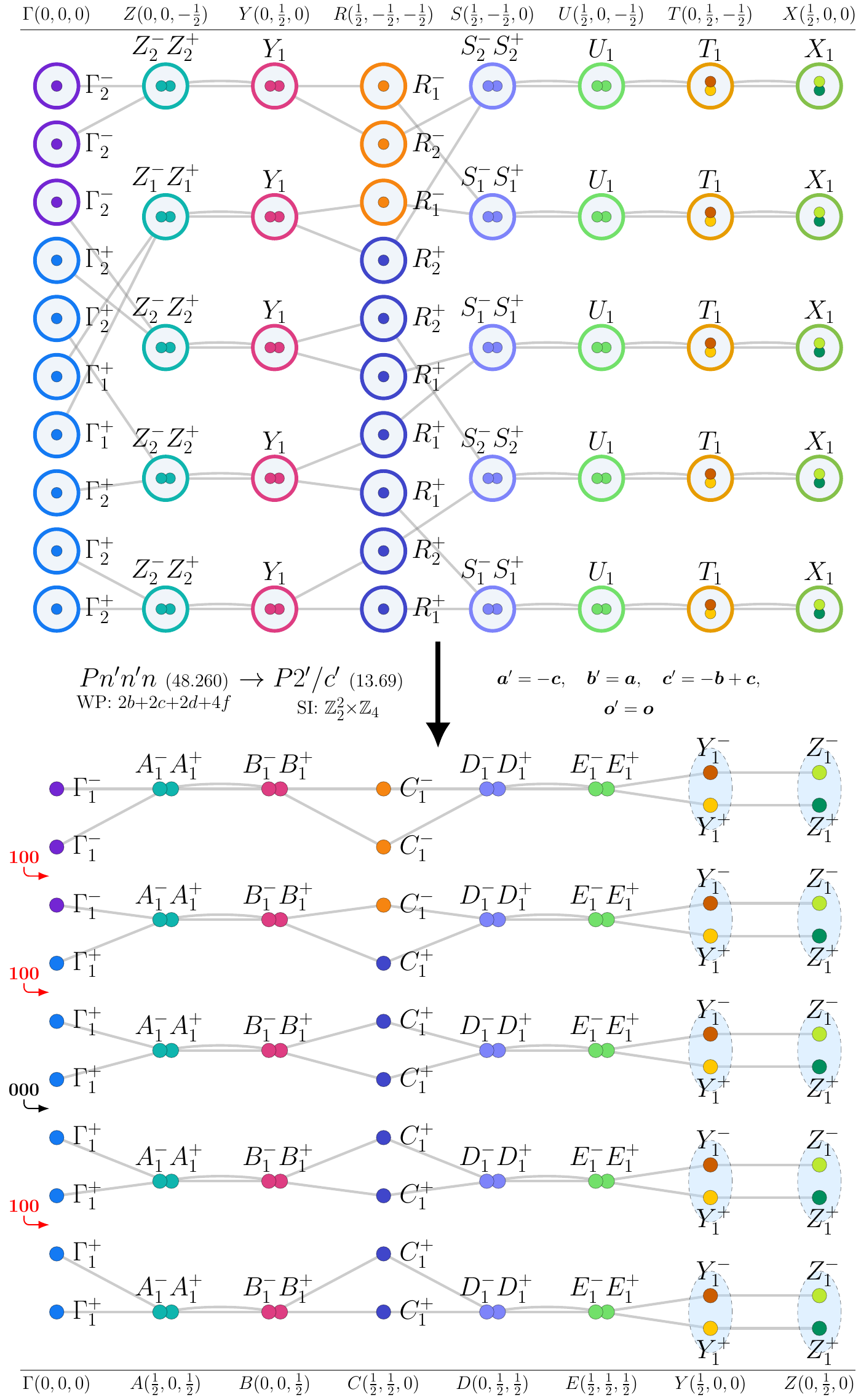}
\caption{Topological magnon bands in subgroup $P2'/c'~(13.69)$ for magnetic moments on Wyckoff positions $2b+2c+2d+4f$ of supergroup $Pn'n'n~(48.260)$.\label{fig_48.260_13.69_strainperp100_2b+2c+2d+4f}}
\end{figure}
\input{gap_tables_tex/48.260_13.69_strainperp100_2b+2c+2d+4f_table.tex}
\input{si_tables_tex/48.260_13.69_strainperp100_2b+2c+2d+4f_table.tex}

\section{MSG $Pb'an'~(50.282)$}
\textbf{Nontrivial-SI Subgroups:} $P\bar{1}~(2.4)$, $P2'/c'~(13.69)$, $P2'/c'~(13.69)$, $P2~(3.1)$, $P2/c~(13.65)$.\\

\textbf{Trivial-SI Subgroups:} $Pc'~(7.26)$, $Pc'~(7.26)$, $P2'~(3.3)$, $P2'~(3.3)$, $Pc~(7.24)$, $Pb'a2'~(32.137)$, $Pn'c2'~(30.113)$, $Pn'c'2~(30.115)$.\\

\subsection{WP: $4f+4f$}
\textbf{BCS Materials:} {CeMnCoGe\textsubscript{4}O\textsubscript{12}~(6 K)}\footnote{BCS web page: \texttt{\href{http://webbdcrista1.ehu.es/magndata/index.php?this\_label=0.190} {http://webbdcrista1.ehu.es/magndata/index.php?this\_label=0.190}}}.\\
\subsubsection{Topological bands in subgroup $P\bar{1}~(2.4)$}
\textbf{Perturbations:}
\begin{itemize}
\item strain in generic direction,
\item (B $\parallel$ [100] or B $\perp$ [001]) and strain $\perp$ [100],
\item (B $\parallel$ [100] or B $\perp$ [001]) and strain $\perp$ [010],
\item (B $\parallel$ [001] or B $\perp$ [100]) and strain $\perp$ [010],
\item (B $\parallel$ [001] or B $\perp$ [100]) and strain $\perp$ [001],
\item B in generic direction.
\end{itemize}
\begin{figure}[H]
\centering
\includegraphics[scale=0.6]{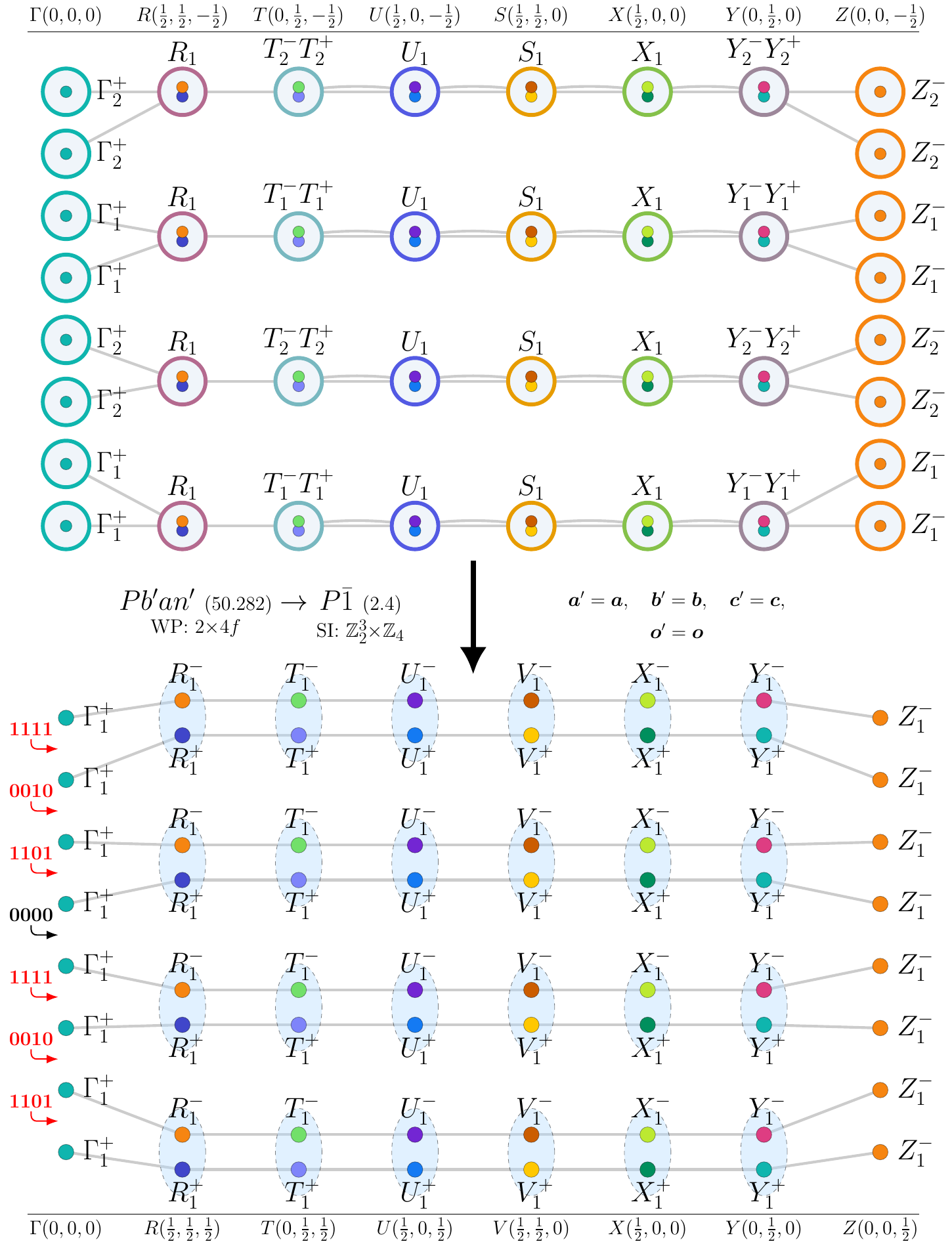}
\caption{Topological magnon bands in subgroup $P\bar{1}~(2.4)$ for magnetic moments on Wyckoff positions $4f+4f$ of supergroup $Pb'an'~(50.282)$.\label{fig_50.282_2.4_strainingenericdirection_4f+4f}}
\end{figure}
\input{gap_tables_tex/50.282_2.4_strainingenericdirection_4f+4f_table.tex}
\input{si_tables_tex/50.282_2.4_strainingenericdirection_4f+4f_table.tex}
\subsubsection{Topological bands in subgroup $P2'/c'~(13.69)$}
\textbf{Perturbations:}
\begin{itemize}
\item strain $\perp$ [001],
\item (B $\parallel$ [100] or B $\perp$ [001]).
\end{itemize}
\begin{figure}[H]
\centering
\includegraphics[scale=0.6]{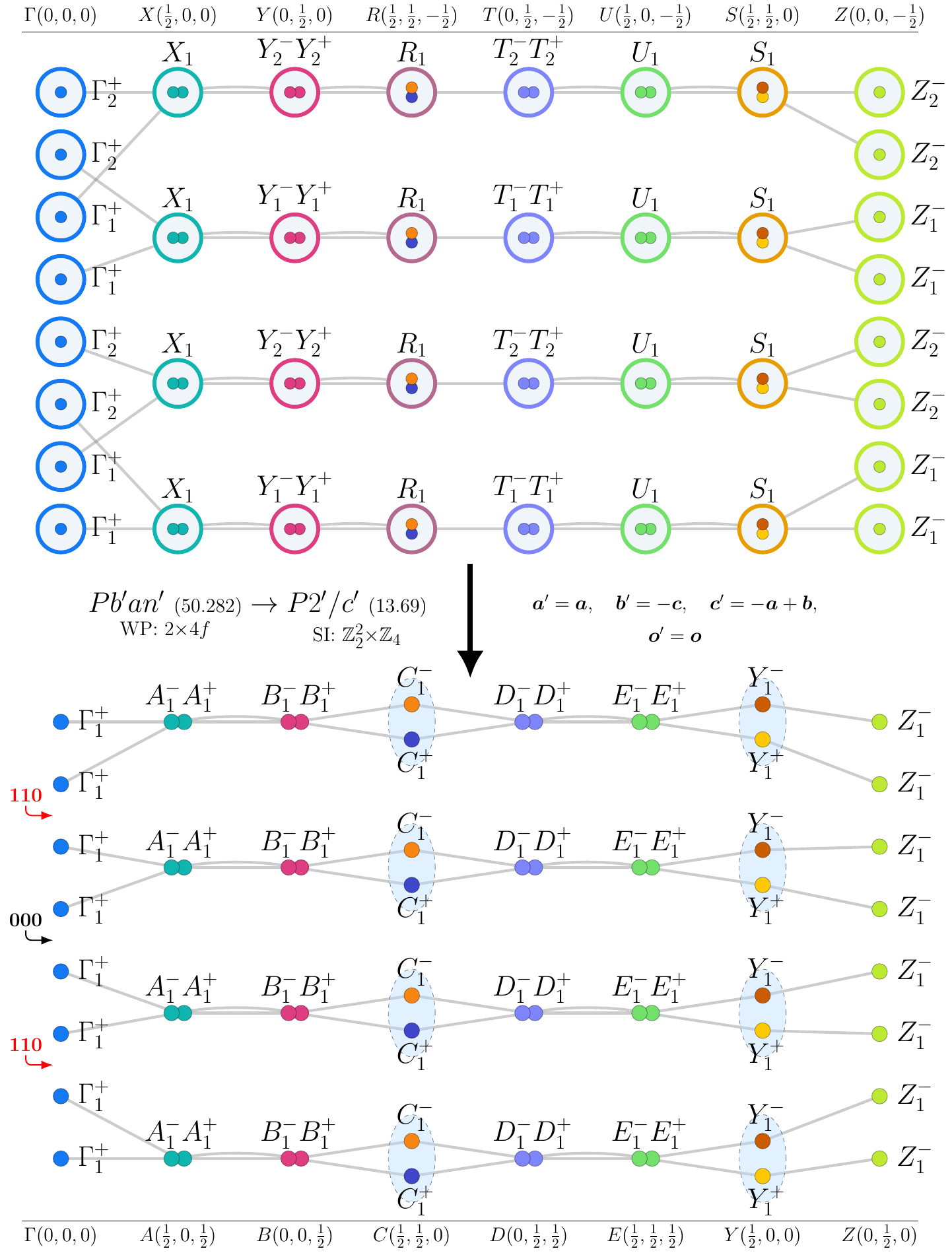}
\caption{Topological magnon bands in subgroup $P2'/c'~(13.69)$ for magnetic moments on Wyckoff positions $4f+4f$ of supergroup $Pb'an'~(50.282)$.\label{fig_50.282_13.69_strainperp001_4f+4f}}
\end{figure}
\input{gap_tables_tex/50.282_13.69_strainperp001_4f+4f_table.tex}
\input{si_tables_tex/50.282_13.69_strainperp001_4f+4f_table.tex}
\subsubsection{Topological bands in subgroup $P2'/c'~(13.69)$}
\textbf{Perturbations:}
\begin{itemize}
\item strain $\perp$ [100],
\item (B $\parallel$ [001] or B $\perp$ [100]).
\end{itemize}
\begin{figure}[H]
\centering
\includegraphics[scale=0.6]{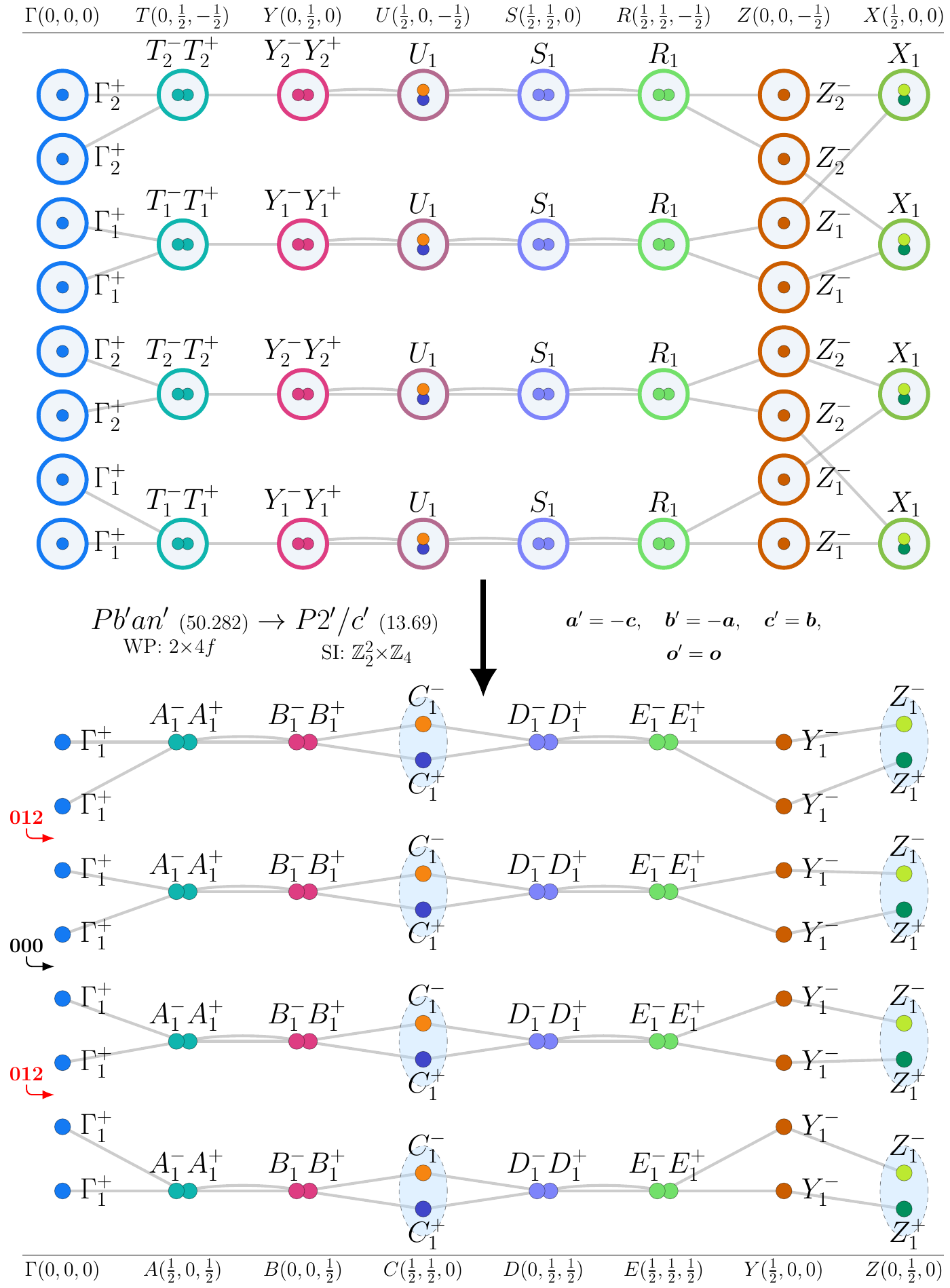}
\caption{Topological magnon bands in subgroup $P2'/c'~(13.69)$ for magnetic moments on Wyckoff positions $4f+4f$ of supergroup $Pb'an'~(50.282)$.\label{fig_50.282_13.69_strainperp100_4f+4f}}
\end{figure}
\input{gap_tables_tex/50.282_13.69_strainperp100_4f+4f_table.tex}
\input{si_tables_tex/50.282_13.69_strainperp100_4f+4f_table.tex}
\subsection{WP: $4f$}
\textbf{BCS Materials:} {ZrCo\textsubscript{2}Ge\textsubscript{4}O\textsubscript{12}~(3.5 K)}\footnote{BCS web page: \texttt{\href{http://webbdcrista1.ehu.es/magndata/index.php?this\_label=0.314} {http://webbdcrista1.ehu.es/magndata/index.php?this\_label=0.314}}}.\\
\subsubsection{Topological bands in subgroup $P\bar{1}~(2.4)$}
\textbf{Perturbations:}
\begin{itemize}
\item strain in generic direction,
\item (B $\parallel$ [100] or B $\perp$ [001]) and strain $\perp$ [100],
\item (B $\parallel$ [100] or B $\perp$ [001]) and strain $\perp$ [010],
\item (B $\parallel$ [001] or B $\perp$ [100]) and strain $\perp$ [010],
\item (B $\parallel$ [001] or B $\perp$ [100]) and strain $\perp$ [001],
\item B in generic direction.
\end{itemize}
\begin{figure}[H]
\centering
\includegraphics[scale=0.6]{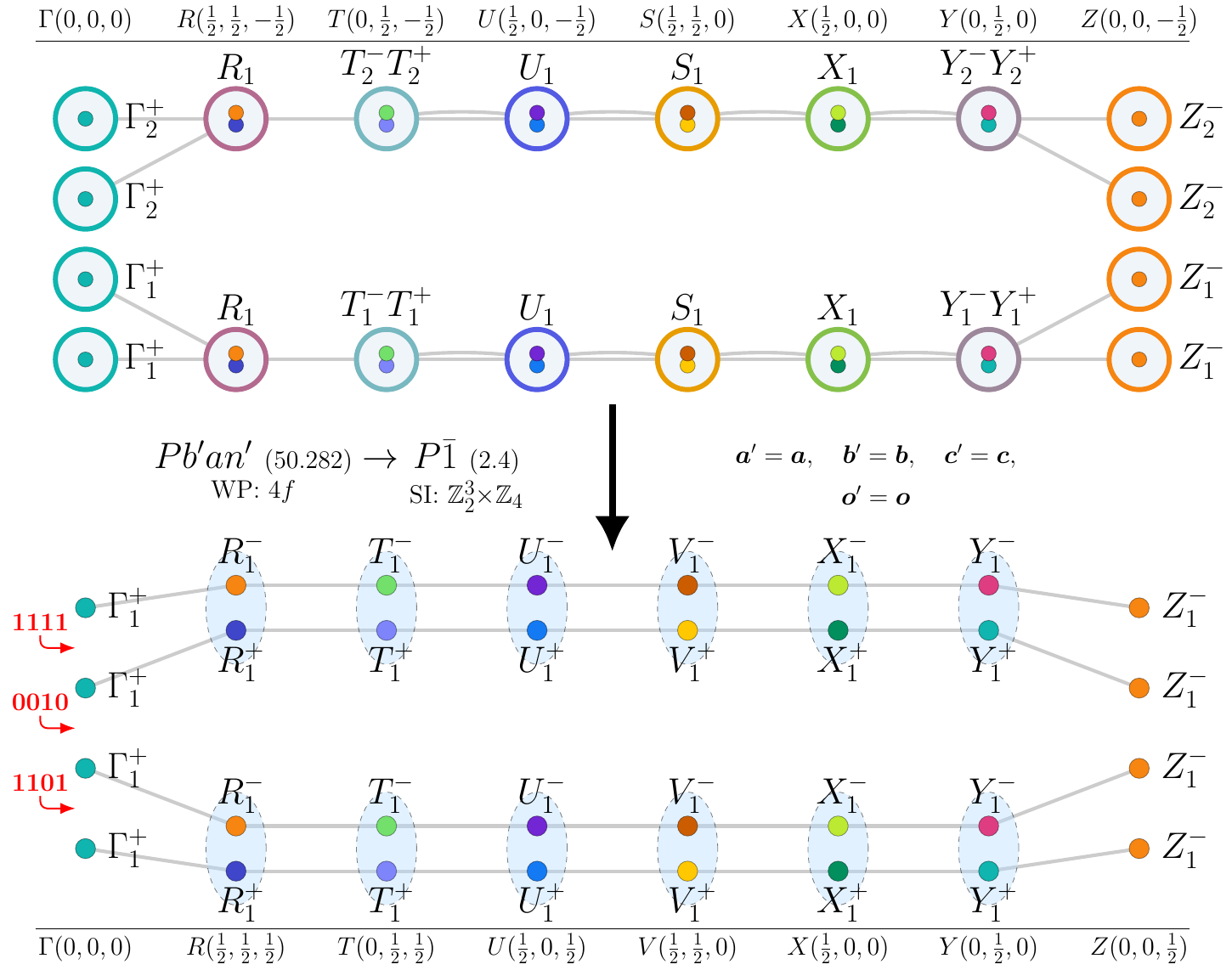}
\caption{Topological magnon bands in subgroup $P\bar{1}~(2.4)$ for magnetic moments on Wyckoff position $4f$ of supergroup $Pb'an'~(50.282)$.\label{fig_50.282_2.4_strainingenericdirection_4f}}
\end{figure}
\input{gap_tables_tex/50.282_2.4_strainingenericdirection_4f_table.tex}
\input{si_tables_tex/50.282_2.4_strainingenericdirection_4f_table.tex}
\subsubsection{Topological bands in subgroup $P2'/c'~(13.69)$}
\textbf{Perturbations:}
\begin{itemize}
\item strain $\perp$ [001],
\item (B $\parallel$ [100] or B $\perp$ [001]).
\end{itemize}
\begin{figure}[H]
\centering
\includegraphics[scale=0.6]{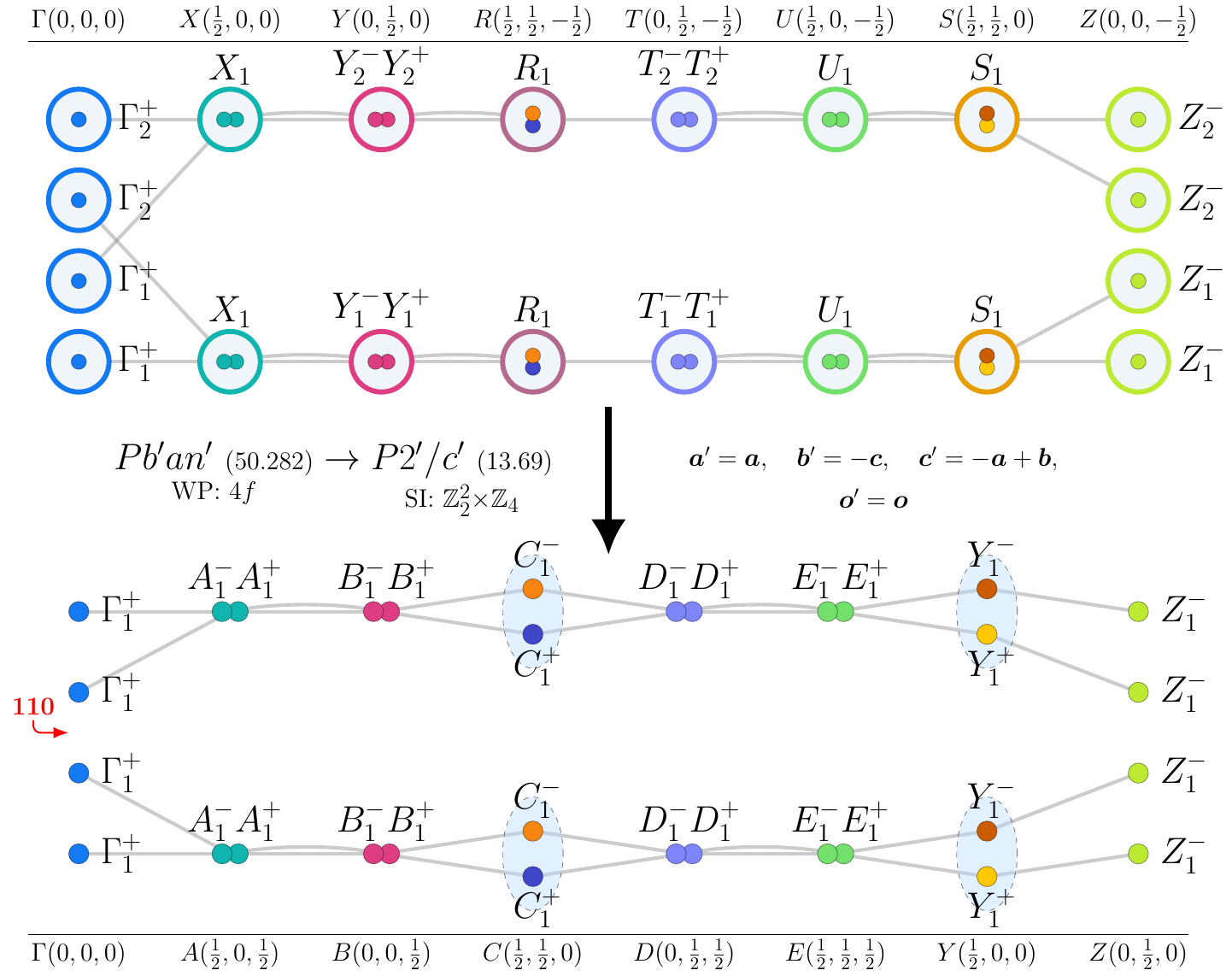}
\caption{Topological magnon bands in subgroup $P2'/c'~(13.69)$ for magnetic moments on Wyckoff position $4f$ of supergroup $Pb'an'~(50.282)$.\label{fig_50.282_13.69_strainperp001_4f}}
\end{figure}
\input{gap_tables_tex/50.282_13.69_strainperp001_4f_table.tex}
\input{si_tables_tex/50.282_13.69_strainperp001_4f_table.tex}
\subsubsection{Topological bands in subgroup $P2'/c'~(13.69)$}
\textbf{Perturbations:}
\begin{itemize}
\item strain $\perp$ [100],
\item (B $\parallel$ [001] or B $\perp$ [100]).
\end{itemize}
\begin{figure}[H]
\centering
\includegraphics[scale=0.6]{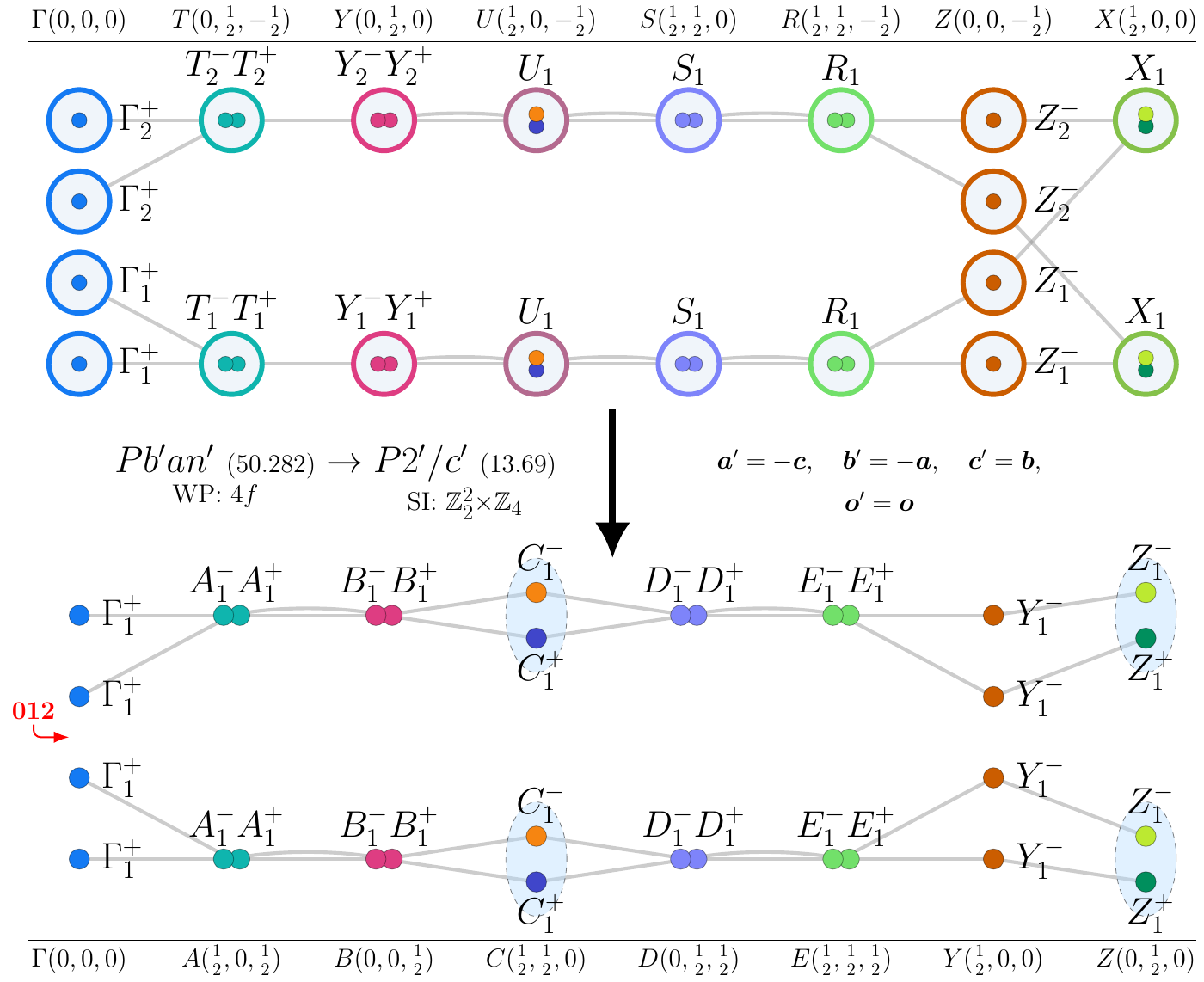}
\caption{Topological magnon bands in subgroup $P2'/c'~(13.69)$ for magnetic moments on Wyckoff position $4f$ of supergroup $Pb'an'~(50.282)$.\label{fig_50.282_13.69_strainperp100_4f}}
\end{figure}
\input{gap_tables_tex/50.282_13.69_strainperp100_4f_table.tex}
\input{si_tables_tex/50.282_13.69_strainperp100_4f_table.tex}

\section{MSG $Pn'n'a~(52.310)$}
\textbf{Nontrivial-SI Subgroups:} $P\bar{1}~(2.4)$, $P2_{1}'/c'~(14.79)$, $P2'/c'~(13.69)$, $P2~(3.1)$, $P2/c~(13.65)$.\\

\textbf{Trivial-SI Subgroups:} $Pc'~(7.26)$, $Pc'~(7.26)$, $P2_{1}'~(4.9)$, $P2'~(3.3)$, $Pc~(7.24)$, $Pn'a2_{1}'~(33.146)$, $Pn'c2'~(30.113)$, $Pn'n'2~(34.159)$.\\

\subsection{WP: $4d$}
\textbf{BCS Materials:} {[C(ND\textsubscript{2})\textsubscript{3}]Mn(DCOO)\textsubscript{3}~(8.73 K)}\footnote{BCS web page: \texttt{\href{http://webbdcrista1.ehu.es/magndata/index.php?this\_label=0.256} {http://webbdcrista1.ehu.es/magndata/index.php?this\_label=0.256}}}.\\
\subsubsection{Topological bands in subgroup $P\bar{1}~(2.4)$}
\textbf{Perturbations:}
\begin{itemize}
\item strain in generic direction,
\item (B $\parallel$ [100] or B $\perp$ [010]) and strain $\perp$ [100],
\item (B $\parallel$ [100] or B $\perp$ [010]) and strain $\perp$ [001],
\item (B $\parallel$ [010] or B $\perp$ [100]) and strain $\perp$ [010],
\item (B $\parallel$ [010] or B $\perp$ [100]) and strain $\perp$ [001],
\item B in generic direction.
\end{itemize}
\begin{figure}[H]
\centering
\includegraphics[scale=0.6]{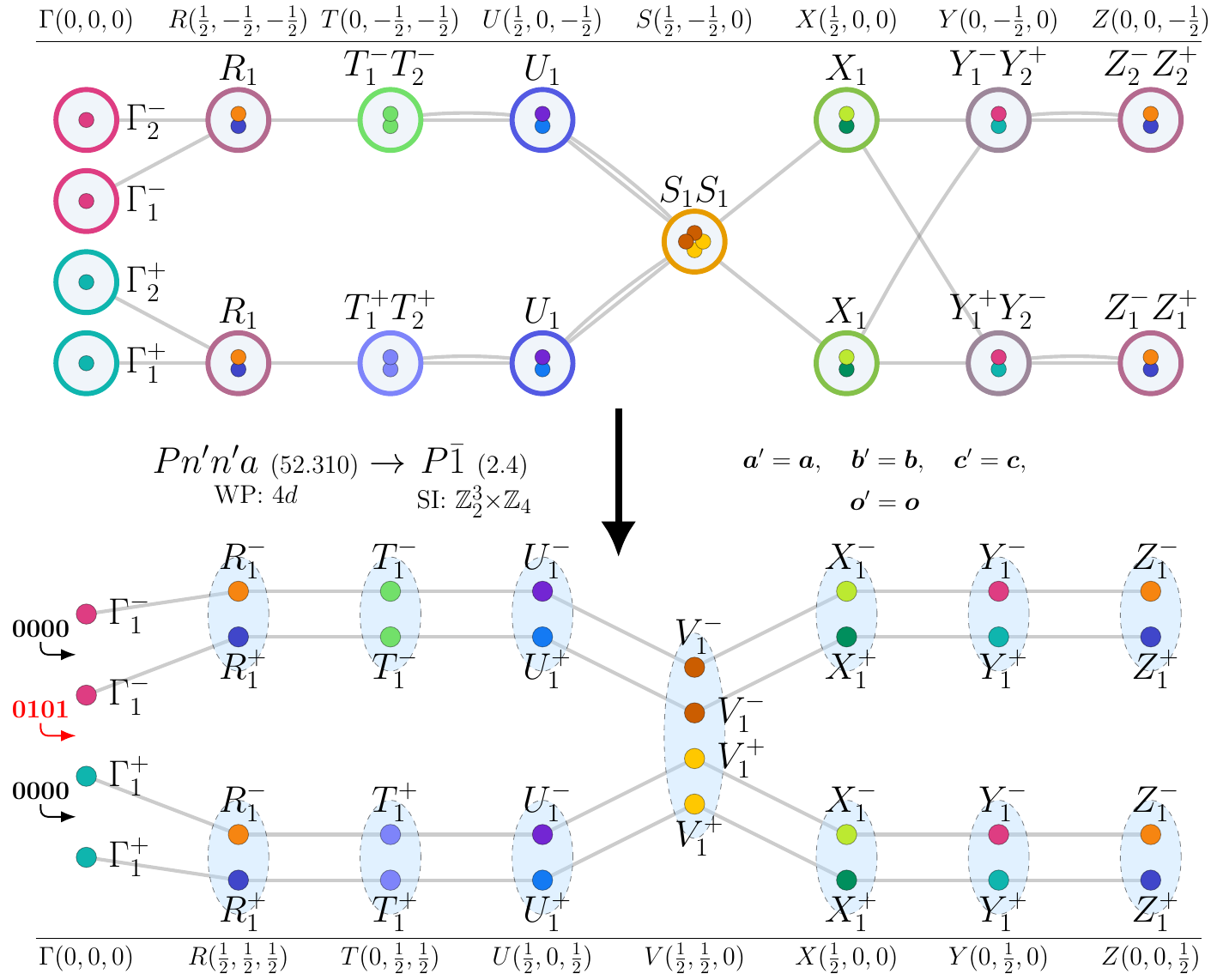}
\caption{Topological magnon bands in subgroup $P\bar{1}~(2.4)$ for magnetic moments on Wyckoff position $4d$ of supergroup $Pn'n'a~(52.310)$.\label{fig_52.310_2.4_strainingenericdirection_4d}}
\end{figure}
\input{gap_tables_tex/52.310_2.4_strainingenericdirection_4d_table.tex}
\input{si_tables_tex/52.310_2.4_strainingenericdirection_4d_table.tex}
\subsubsection{Topological bands in subgroup $P2_{1}'/c'~(14.79)$}
\textbf{Perturbations:}
\begin{itemize}
\item strain $\perp$ [010],
\item (B $\parallel$ [100] or B $\perp$ [010]).
\end{itemize}
\begin{figure}[H]
\centering
\includegraphics[scale=0.6]{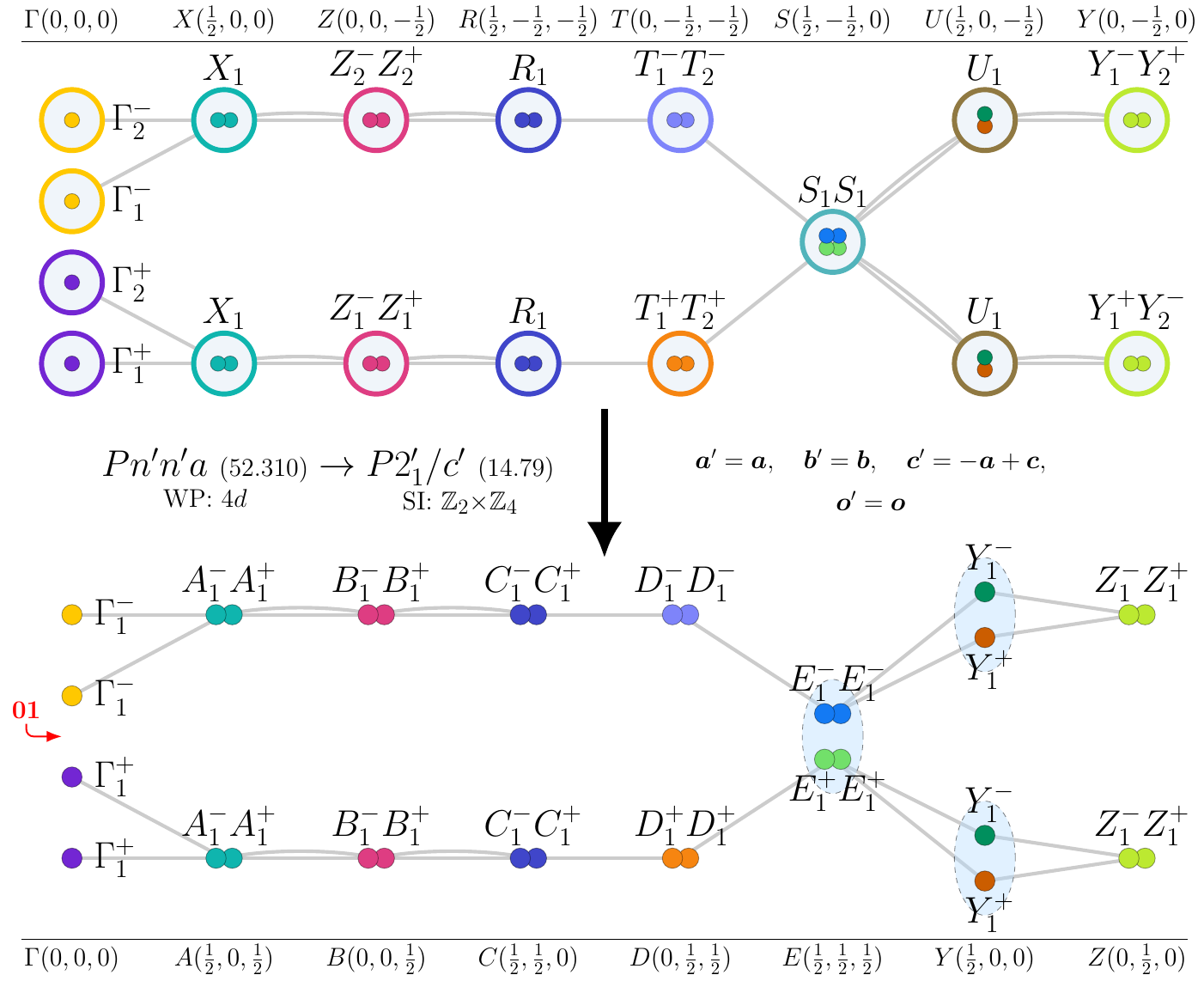}
\caption{Topological magnon bands in subgroup $P2_{1}'/c'~(14.79)$ for magnetic moments on Wyckoff position $4d$ of supergroup $Pn'n'a~(52.310)$.\label{fig_52.310_14.79_strainperp010_4d}}
\end{figure}
\input{gap_tables_tex/52.310_14.79_strainperp010_4d_table.tex}
\input{si_tables_tex/52.310_14.79_strainperp010_4d_table.tex}
\subsubsection{Topological bands in subgroup $P2'/c'~(13.69)$}
\textbf{Perturbations:}
\begin{itemize}
\item strain $\perp$ [100],
\item (B $\parallel$ [010] or B $\perp$ [100]).
\end{itemize}
\begin{figure}[H]
\centering
\includegraphics[scale=0.6]{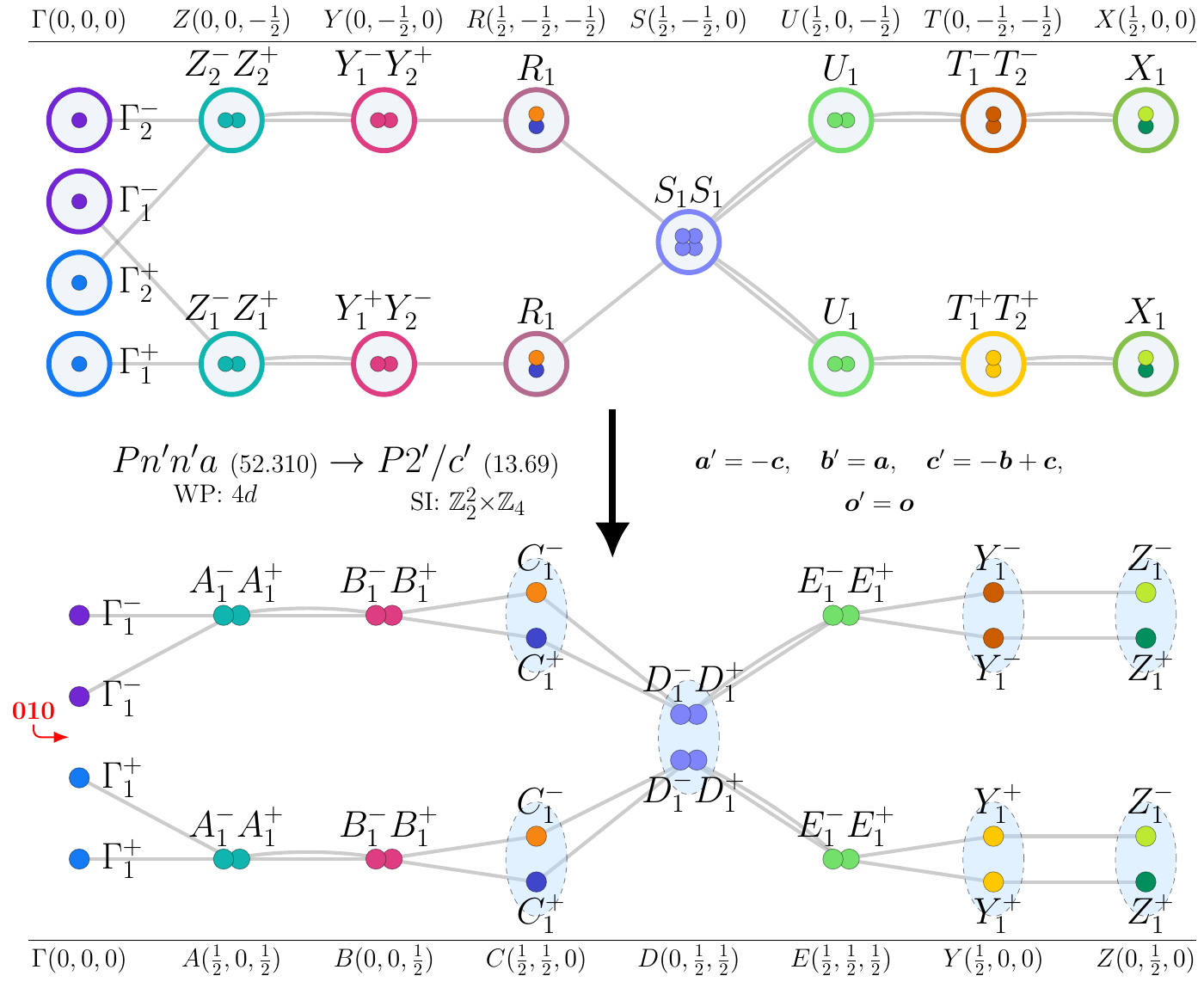}
\caption{Topological magnon bands in subgroup $P2'/c'~(13.69)$ for magnetic moments on Wyckoff position $4d$ of supergroup $Pn'n'a~(52.310)$.\label{fig_52.310_13.69_strainperp100_4d}}
\end{figure}
\input{gap_tables_tex/52.310_13.69_strainperp100_4d_table.tex}
\input{si_tables_tex/52.310_13.69_strainperp100_4d_table.tex}

\section{MSG $P_{b}nna~(52.315)$}
\textbf{Nontrivial-SI Subgroups:} $P\bar{1}~(2.4)$, $P2'/c'~(13.69)$, $P2'/c'~(13.69)$, $P2'/c'~(13.69)$, $P_{S}\bar{1}~(2.7)$, $P2~(3.1)$, $P_{a}2~(3.4)$, $P2/c~(13.65)$, $Pb'an'~(50.282)$, $P_{a}2/c~(13.70)$, $P2_{1}/c~(14.75)$, $Pn'na'~(52.312)$, $P_{b}2_{1}/c~(14.81)$, $P2~(3.1)$, $P_{a}2~(3.4)$, $P2/c~(13.65)$, $Pn'n'n~(48.260)$, $P_{a}2/c~(13.70)$.\\

\textbf{Trivial-SI Subgroups:} $Pc'~(7.26)$, $Pc'~(7.26)$, $Pc'~(7.26)$, $P2'~(3.3)$, $P2'~(3.3)$, $P2'~(3.3)$, $P_{S}1~(1.3)$, $Pc~(7.24)$, $Pb'a2'~(32.137)$, $Pn'c2'~(30.113)$, $P_{a}c~(7.27)$, $Pc~(7.24)$, $Pnc'2'~(30.114)$, $Pn'n2'~(34.158)$, $P_{b}c~(7.29)$, $Pc~(7.24)$, $Pn'n2'~(34.158)$, $Pn'n2'~(34.158)$, $P_{a}c~(7.27)$, $Pn'c'2~(30.115)$, $P_{a}nn2~(34.160)$, $P2_{1}~(4.7)$, $Pn'a'2_{1}~(33.148)$, $P_{b}2_{1}~(4.11)$, $P_{c}na2_{1}~(33.151)$, $Pn'n'2~(34.159)$, $P_{a}nc2~(30.116)$.\\

\subsection{WP: $4c+8f$}
\textbf{BCS Materials:} {GdFeZnGe\textsubscript{4}O\textsubscript{12}~(13.8 K)}\footnote{BCS web page: \texttt{\href{http://webbdcrista1.ehu.es/magndata/index.php?this\_label=1.313} {http://webbdcrista1.ehu.es/magndata/index.php?this\_label=1.313}}}.\\
\subsubsection{Topological bands in subgroup $P2'/c'~(13.69)$}
\textbf{Perturbations:}
\begin{itemize}
\item B $\parallel$ [100] and strain $\perp$ [001],
\item B $\parallel$ [010] and strain $\perp$ [001],
\item B $\perp$ [001].
\end{itemize}
\begin{figure}[H]
\centering
\includegraphics[scale=0.6]{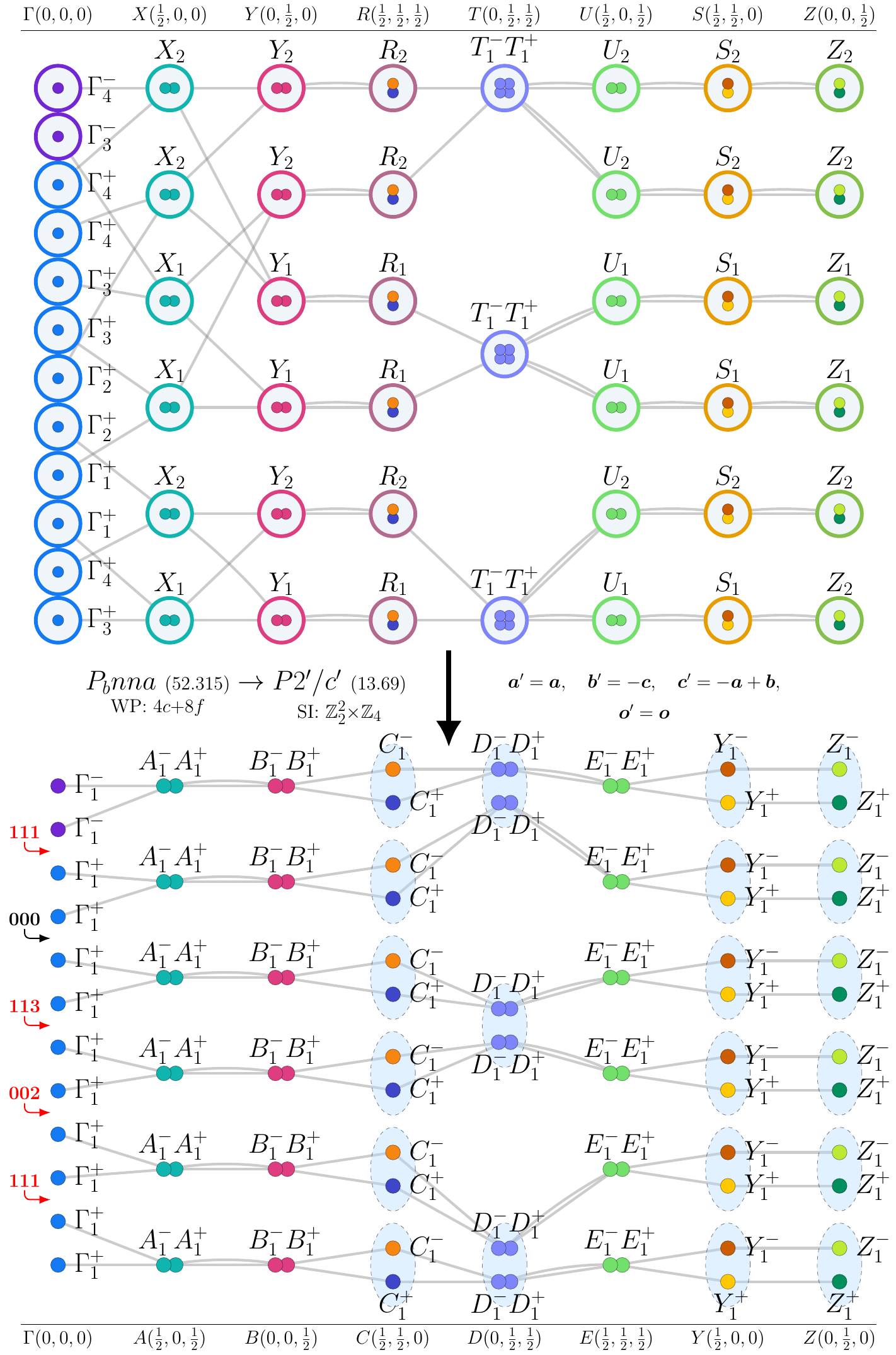}
\caption{Topological magnon bands in subgroup $P2'/c'~(13.69)$ for magnetic moments on Wyckoff positions $4c+8f$ of supergroup $P_{b}nna~(52.315)$.\label{fig_52.315_13.69_Bparallel100andstrainperp001_4c+8f}}
\end{figure}
\input{gap_tables_tex/52.315_13.69_Bparallel100andstrainperp001_4c+8f_table.tex}
\input{si_tables_tex/52.315_13.69_Bparallel100andstrainperp001_4c+8f_table.tex}
\subsubsection{Topological bands in subgroup $P2'/c'~(13.69)$}
\textbf{Perturbations:}
\begin{itemize}
\item B $\parallel$ [100] and strain $\perp$ [010],
\item B $\parallel$ [001] and strain $\perp$ [010],
\item B $\perp$ [010].
\end{itemize}
\begin{figure}[H]
\centering
\includegraphics[scale=0.6]{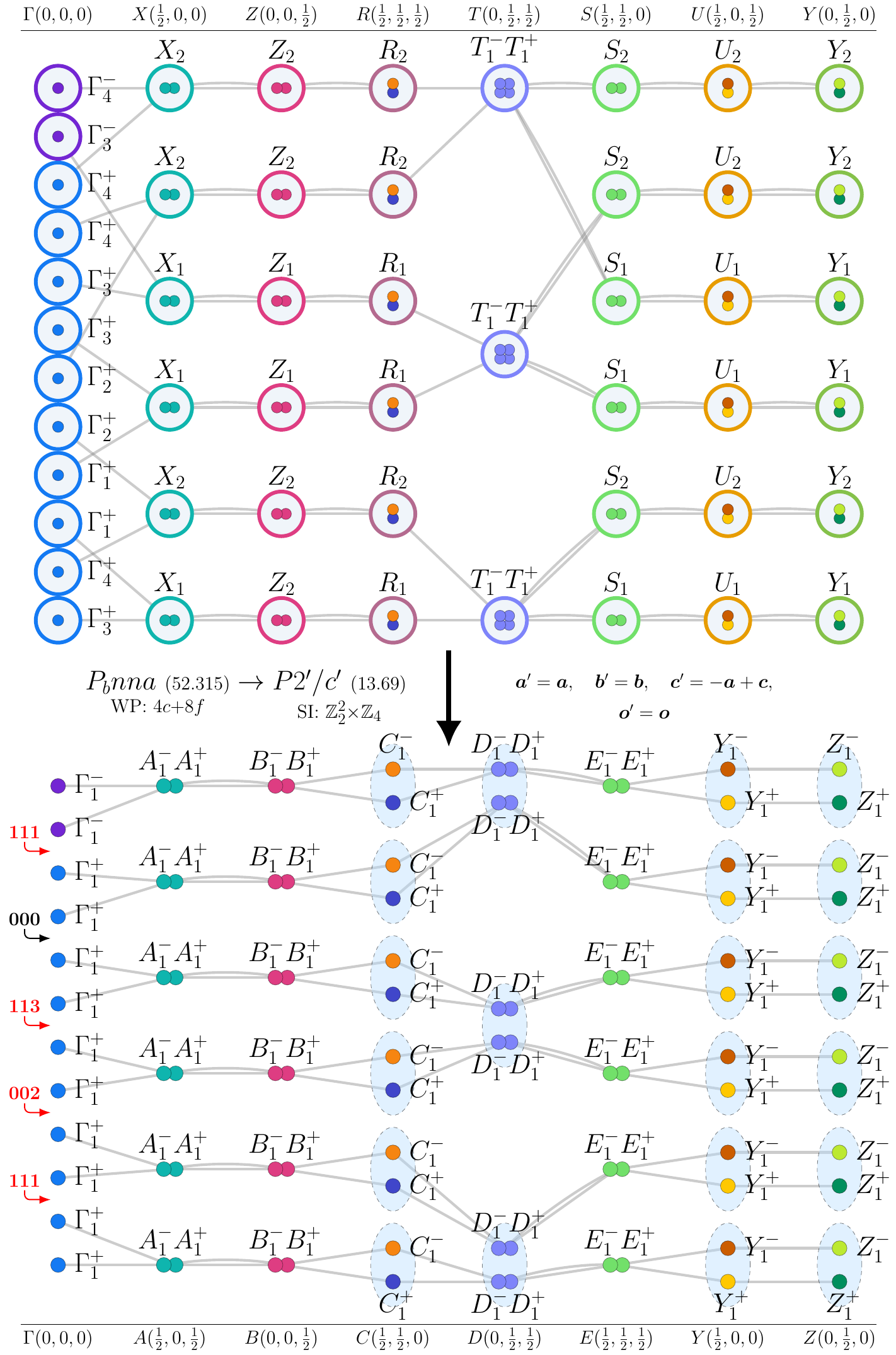}
\caption{Topological magnon bands in subgroup $P2'/c'~(13.69)$ for magnetic moments on Wyckoff positions $4c+8f$ of supergroup $P_{b}nna~(52.315)$.\label{fig_52.315_13.69_Bparallel100andstrainperp010_4c+8f}}
\end{figure}
\input{gap_tables_tex/52.315_13.69_Bparallel100andstrainperp010_4c+8f_table.tex}
\input{si_tables_tex/52.315_13.69_Bparallel100andstrainperp010_4c+8f_table.tex}
\subsubsection{Topological bands in subgroup $P2'/c'~(13.69)$}
\textbf{Perturbations:}
\begin{itemize}
\item B $\parallel$ [010] and strain $\perp$ [100],
\item B $\parallel$ [001] and strain $\perp$ [100],
\item B $\perp$ [100].
\end{itemize}
\begin{figure}[H]
\centering
\includegraphics[scale=0.6]{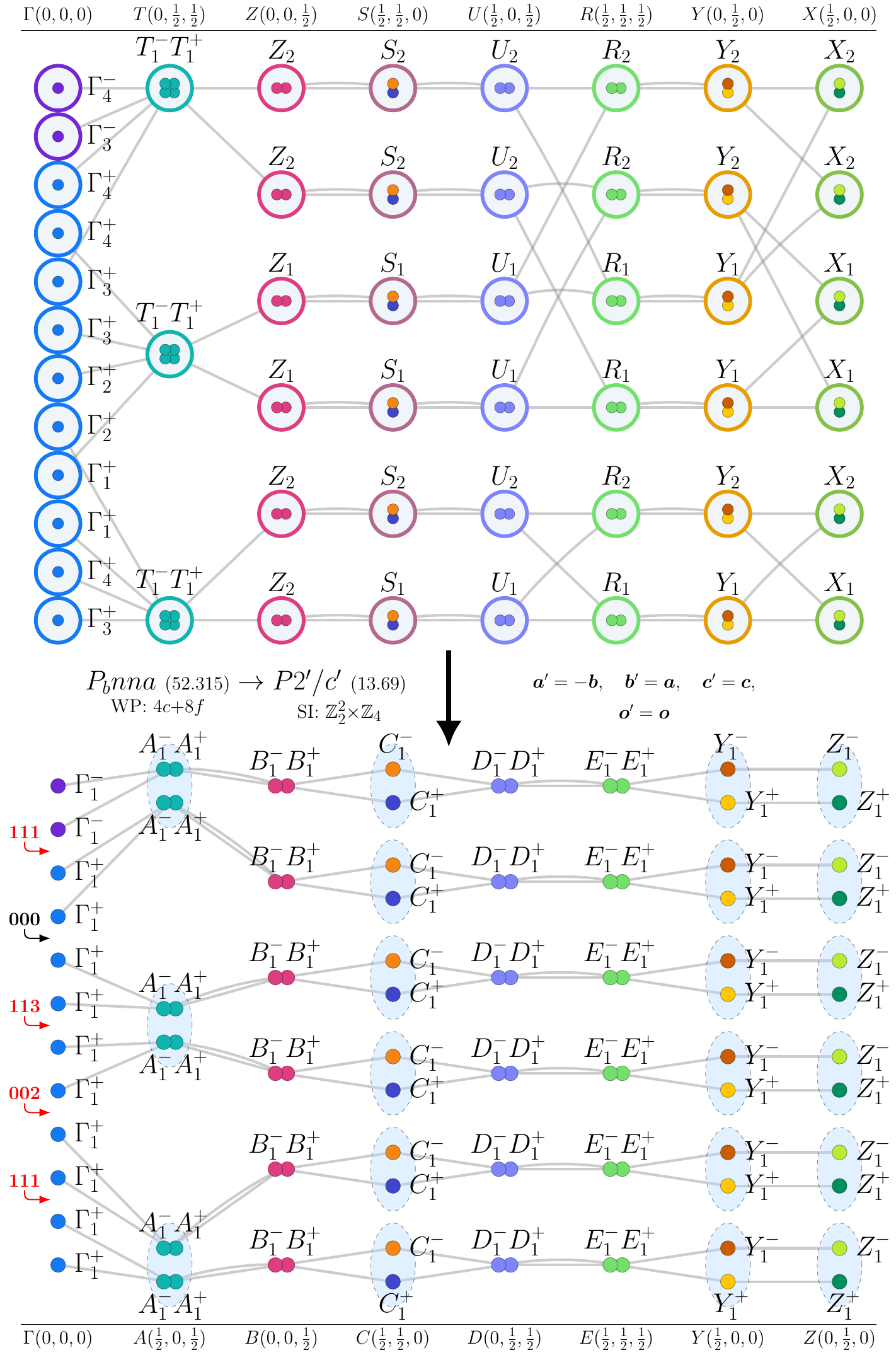}
\caption{Topological magnon bands in subgroup $P2'/c'~(13.69)$ for magnetic moments on Wyckoff positions $4c+8f$ of supergroup $P_{b}nna~(52.315)$.\label{fig_52.315_13.69_Bparallel010andstrainperp100_4c+8f}}
\end{figure}
\input{gap_tables_tex/52.315_13.69_Bparallel010andstrainperp100_4c+8f_table.tex}
\input{si_tables_tex/52.315_13.69_Bparallel010andstrainperp100_4c+8f_table.tex}
\subsubsection{Topological bands in subgroup $P_{S}\bar{1}~(2.7)$}
\textbf{Perturbation:}
\begin{itemize}
\item strain in generic direction.
\end{itemize}
\begin{figure}[H]
\centering
\includegraphics[scale=0.6]{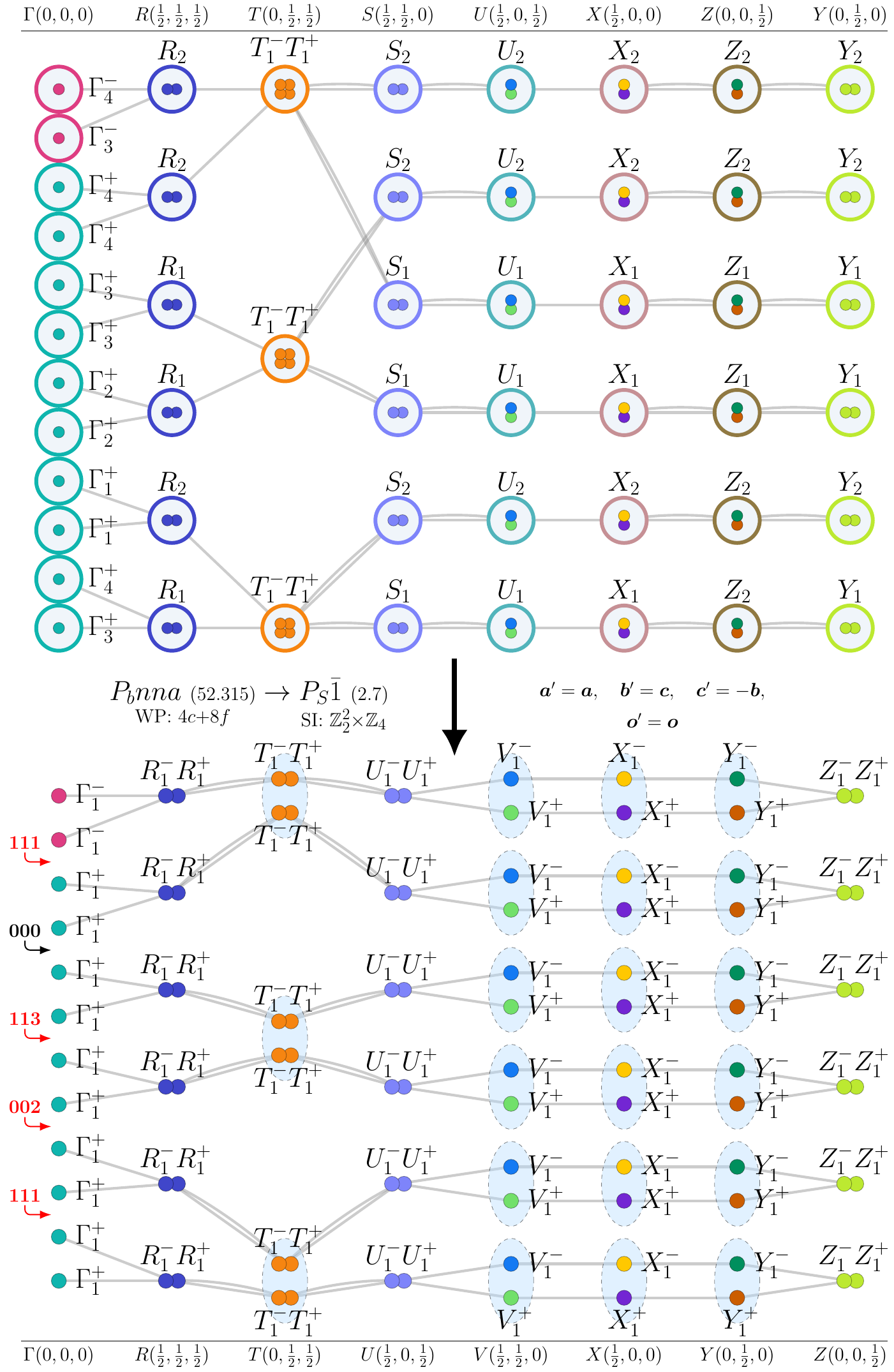}
\caption{Topological magnon bands in subgroup $P_{S}\bar{1}~(2.7)$ for magnetic moments on Wyckoff positions $4c+8f$ of supergroup $P_{b}nna~(52.315)$.\label{fig_52.315_2.7_strainingenericdirection_4c+8f}}
\end{figure}
\input{gap_tables_tex/52.315_2.7_strainingenericdirection_4c+8f_table.tex}
\input{si_tables_tex/52.315_2.7_strainingenericdirection_4c+8f_table.tex}
\subsection{WP: $8f$}
\textbf{BCS Materials:} {CeCo\textsubscript{2}Ge\textsubscript{4}O\textsubscript{12}~(4.5 K)}\footnote{BCS web page: \texttt{\href{http://webbdcrista1.ehu.es/magndata/index.php?this\_label=1.226} {http://webbdcrista1.ehu.es/magndata/index.php?this\_label=1.226}}}.\\
\subsubsection{Topological bands in subgroup $P\bar{1}~(2.4)$}
\textbf{Perturbations:}
\begin{itemize}
\item B $\parallel$ [100] and strain in generic direction,
\item B $\parallel$ [010] and strain in generic direction,
\item B $\parallel$ [001] and strain in generic direction,
\item B $\perp$ [100] and strain $\perp$ [010],
\item B $\perp$ [100] and strain $\perp$ [001],
\item B $\perp$ [100] and strain in generic direction,
\item B $\perp$ [010] and strain $\perp$ [100],
\item B $\perp$ [010] and strain $\perp$ [001],
\item B $\perp$ [010] and strain in generic direction,
\item B $\perp$ [001] and strain $\perp$ [100],
\item B $\perp$ [001] and strain $\perp$ [010],
\item B $\perp$ [001] and strain in generic direction,
\item B in generic direction.
\end{itemize}
\begin{figure}[H]
\centering
\includegraphics[scale=0.6]{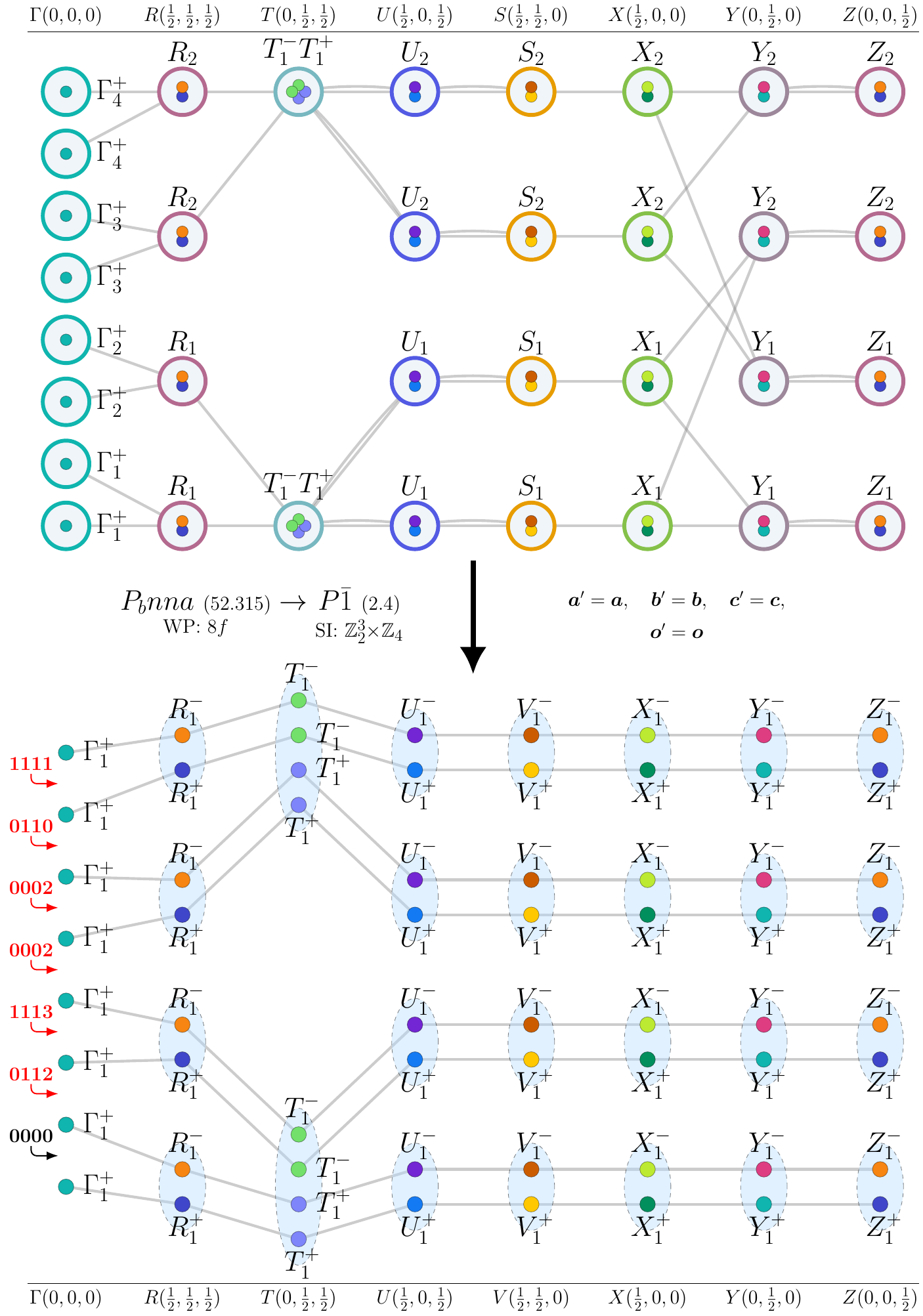}
\caption{Topological magnon bands in subgroup $P\bar{1}~(2.4)$ for magnetic moments on Wyckoff position $8f$ of supergroup $P_{b}nna~(52.315)$.\label{fig_52.315_2.4_Bparallel100andstrainingenericdirection_8f}}
\end{figure}
\input{gap_tables_tex/52.315_2.4_Bparallel100andstrainingenericdirection_8f_table.tex}
\input{si_tables_tex/52.315_2.4_Bparallel100andstrainingenericdirection_8f_table.tex}
\subsubsection{Topological bands in subgroup $P2'/c'~(13.69)$}
\textbf{Perturbations:}
\begin{itemize}
\item B $\parallel$ [100] and strain $\perp$ [001],
\item B $\parallel$ [010] and strain $\perp$ [001],
\item B $\perp$ [001].
\end{itemize}
\begin{figure}[H]
\centering
\includegraphics[scale=0.6]{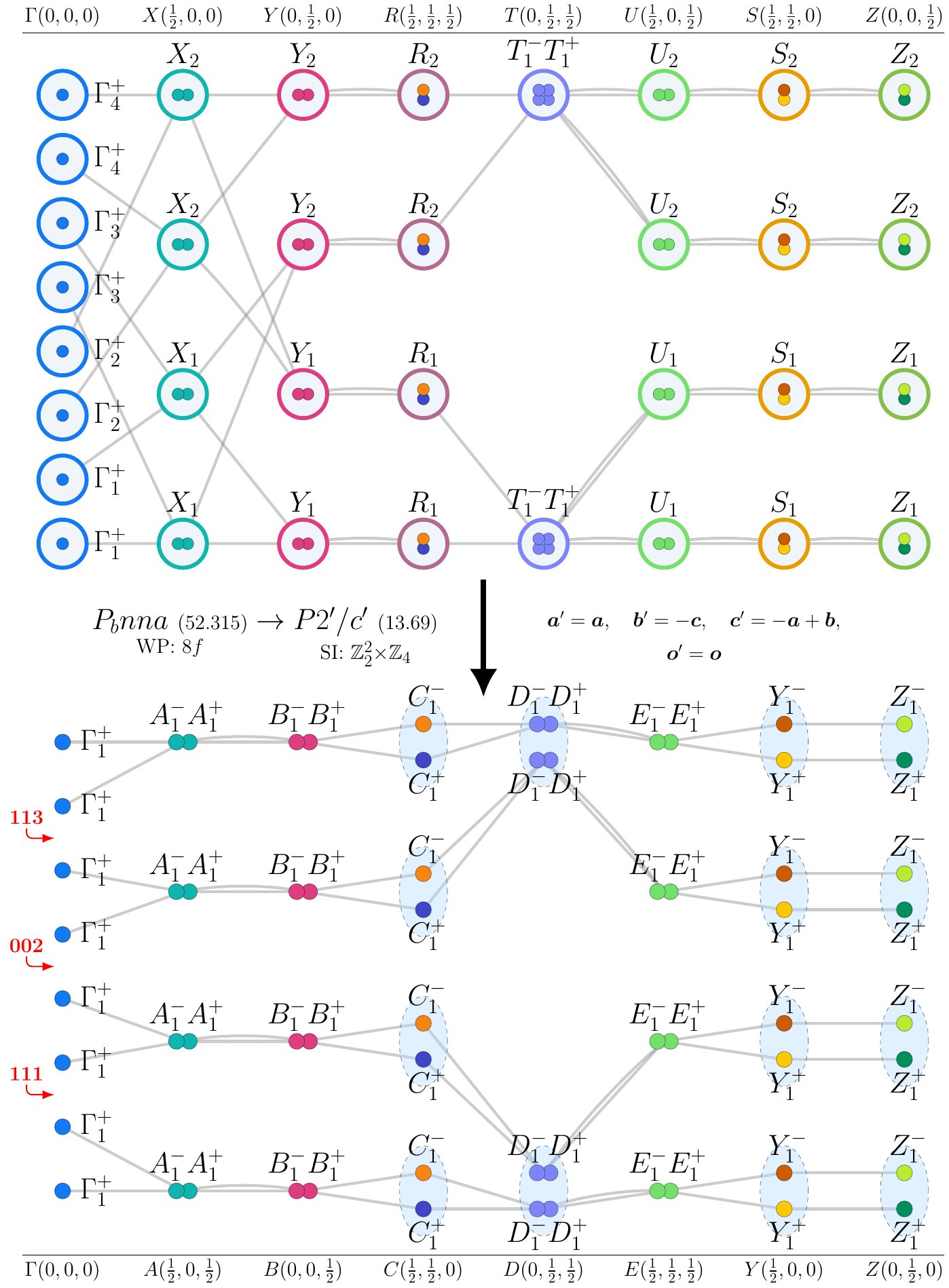}
\caption{Topological magnon bands in subgroup $P2'/c'~(13.69)$ for magnetic moments on Wyckoff position $8f$ of supergroup $P_{b}nna~(52.315)$.\label{fig_52.315_13.69_Bparallel100andstrainperp001_8f}}
\end{figure}
\input{gap_tables_tex/52.315_13.69_Bparallel100andstrainperp001_8f_table.tex}
\input{si_tables_tex/52.315_13.69_Bparallel100andstrainperp001_8f_table.tex}
\subsubsection{Topological bands in subgroup $P2'/c'~(13.69)$}
\textbf{Perturbations:}
\begin{itemize}
\item B $\parallel$ [100] and strain $\perp$ [010],
\item B $\parallel$ [001] and strain $\perp$ [010],
\item B $\perp$ [010].
\end{itemize}
\begin{figure}[H]
\centering
\includegraphics[scale=0.6]{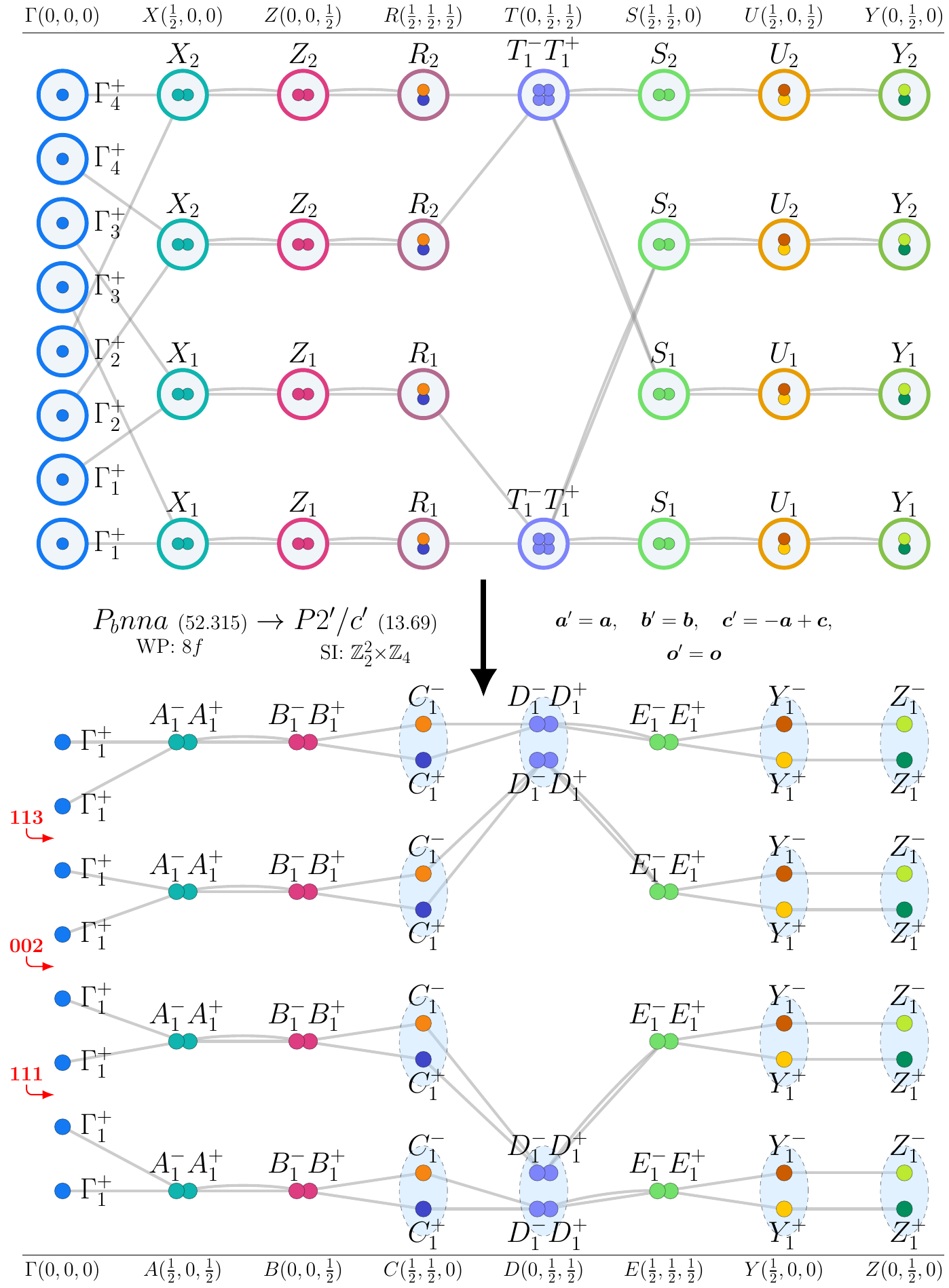}
\caption{Topological magnon bands in subgroup $P2'/c'~(13.69)$ for magnetic moments on Wyckoff position $8f$ of supergroup $P_{b}nna~(52.315)$.\label{fig_52.315_13.69_Bparallel100andstrainperp010_8f}}
\end{figure}
\input{gap_tables_tex/52.315_13.69_Bparallel100andstrainperp010_8f_table.tex}
\input{si_tables_tex/52.315_13.69_Bparallel100andstrainperp010_8f_table.tex}
\subsubsection{Topological bands in subgroup $P2'/c'~(13.69)$}
\textbf{Perturbations:}
\begin{itemize}
\item B $\parallel$ [010] and strain $\perp$ [100],
\item B $\parallel$ [001] and strain $\perp$ [100],
\item B $\perp$ [100].
\end{itemize}
\begin{figure}[H]
\centering
\includegraphics[scale=0.6]{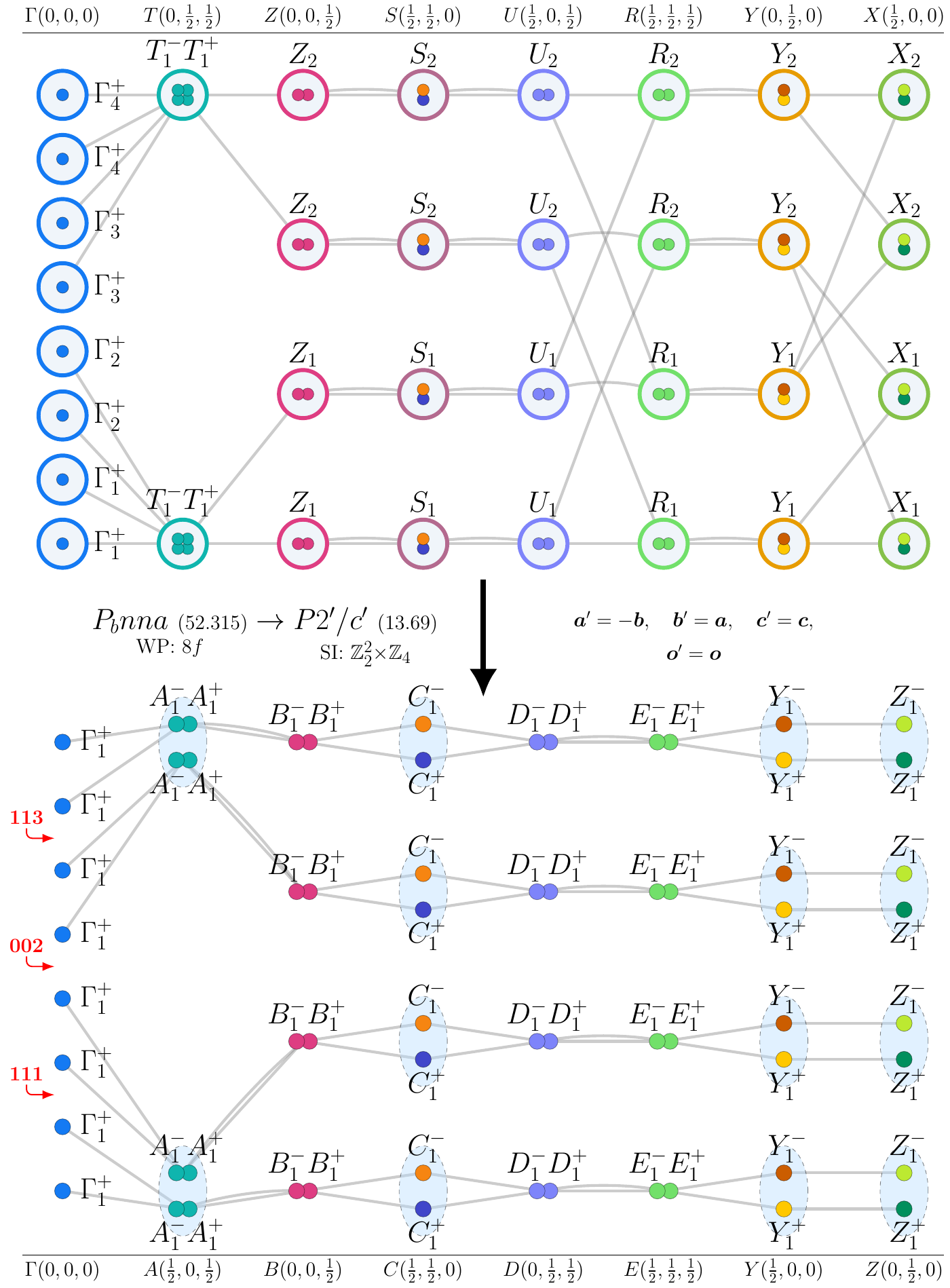}
\caption{Topological magnon bands in subgroup $P2'/c'~(13.69)$ for magnetic moments on Wyckoff position $8f$ of supergroup $P_{b}nna~(52.315)$.\label{fig_52.315_13.69_Bparallel010andstrainperp100_8f}}
\end{figure}
\input{gap_tables_tex/52.315_13.69_Bparallel010andstrainperp100_8f_table.tex}
\input{si_tables_tex/52.315_13.69_Bparallel010andstrainperp100_8f_table.tex}
\subsubsection{Topological bands in subgroup $P_{S}\bar{1}~(2.7)$}
\textbf{Perturbation:}
\begin{itemize}
\item strain in generic direction.
\end{itemize}
\begin{figure}[H]
\centering
\includegraphics[scale=0.6]{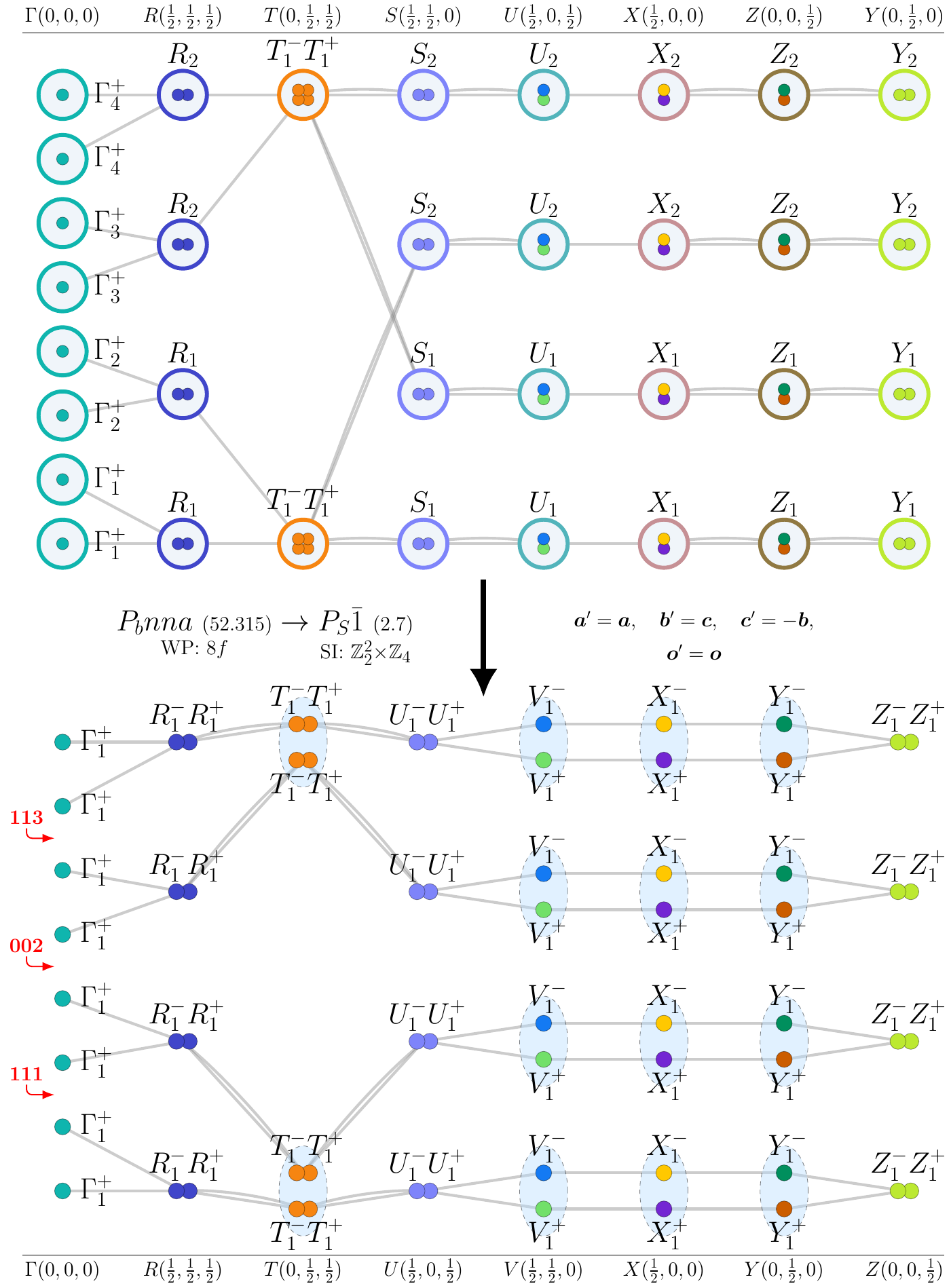}
\caption{Topological magnon bands in subgroup $P_{S}\bar{1}~(2.7)$ for magnetic moments on Wyckoff position $8f$ of supergroup $P_{b}nna~(52.315)$.\label{fig_52.315_2.7_strainingenericdirection_8f}}
\end{figure}
\input{gap_tables_tex/52.315_2.7_strainingenericdirection_8f_table.tex}
\input{si_tables_tex/52.315_2.7_strainingenericdirection_8f_table.tex}
\subsection{WP: $4b$}
\textbf{BCS Materials:} {CuCl(C\textsubscript{4}H\textsubscript{4}N\textsubscript{2})\textsubscript{2}(BF\textsubscript{4})~(3.9 K)}\footnote{BCS web page: \texttt{\href{http://webbdcrista1.ehu.es/magndata/index.php?this\_label=1.474} {http://webbdcrista1.ehu.es/magndata/index.php?this\_label=1.474}}}, {CuBr(C\textsubscript{4}H\textsubscript{4}N\textsubscript{2})\textsubscript{2}(BF\textsubscript{4})~(3.8 K)}\footnote{BCS web page: \texttt{\href{http://webbdcrista1.ehu.es/magndata/index.php?this\_label=1.473} {http://webbdcrista1.ehu.es/magndata/index.php?this\_label=1.473}}}.\\
\subsubsection{Topological bands in subgroup $P_{a}2~(3.4)$}
\textbf{Perturbation:}
\begin{itemize}
\item E $\parallel$ [100] and strain $\perp$ [100].
\end{itemize}
\begin{figure}[H]
\centering
\includegraphics[scale=0.6]{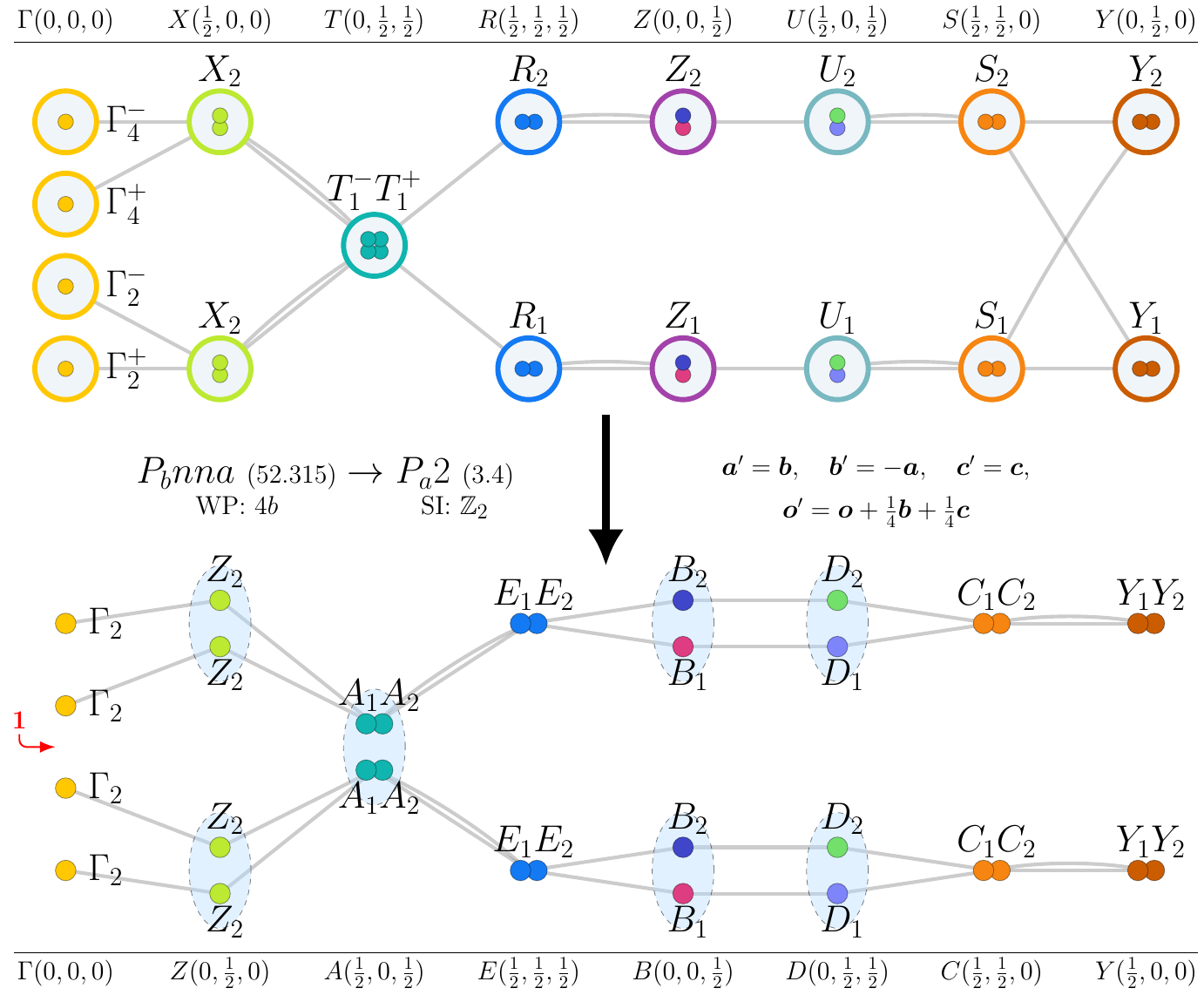}
\caption{Topological magnon bands in subgroup $P_{a}2~(3.4)$ for magnetic moments on Wyckoff position $4b$ of supergroup $P_{b}nna~(52.315)$.\label{fig_52.315_3.4_Eparallel100andstrainperp100_4b}}
\end{figure}
\input{gap_tables_tex/52.315_3.4_Eparallel100andstrainperp100_4b_table.tex}
\input{si_tables_tex/52.315_3.4_Eparallel100andstrainperp100_4b_table.tex}

\section{MSG $P_{B}nna~(52.318)$}
\textbf{Nontrivial-SI Subgroups:} $P\bar{1}~(2.4)$, $P2_{1}'/m'~(11.54)$, $P2_{1}'/m'~(11.54)$, $P2_{1}'/c'~(14.79)$, $P_{S}\bar{1}~(2.7)$, $P2~(3.1)$, $Pm'a'2~(28.91)$, $P2/c~(13.65)$, $Pbc'm'~(57.383)$, $P_{A}2/c~(13.73)$, $P2_{1}/c~(14.75)$, $Pnm'a'~(62.447)$, $P_{c}2_{1}/c~(14.82)$, $P2~(3.1)$, $Pm'm'2~(25.60)$, $P2/c~(13.65)$, $Pm'm'n~(59.409)$, $P_{C}2/c~(13.74)$.\\

\textbf{Trivial-SI Subgroups:} $Pm'~(6.20)$, $Pm'~(6.20)$, $Pc'~(7.26)$, $P2_{1}'~(4.9)$, $P2_{1}'~(4.9)$, $P2_{1}'~(4.9)$, $P_{S}1~(1.3)$, $Pc~(7.24)$, $Pc'a2_{1}'~(29.101)$, $Pm'c2_{1}'~(26.68)$, $P_{A}c~(7.31)$, $Pc~(7.24)$, $Pna'2_{1}'~(33.147)$, $Pm'n2_{1}'~(31.125)$, $P_{c}c~(7.28)$, $Pc~(7.24)$, $Pm'n2_{1}'~(31.125)$, $Pm'n2_{1}'~(31.125)$, $P_{C}c~(7.30)$, $P_{C}2~(3.6)$, $P_{A}nn2~(34.162)$, $P2_{1}~(4.7)$, $Pm'c'2_{1}~(26.70)$, $P_{a}2_{1}~(4.10)$, $P_{C}na2_{1}~(33.154)$, $P_{C}2~(3.6)$, $P_{A}nc2~(30.119)$.\\

\subsection{WP: $4c+4c$}
\textbf{BCS Materials:} {PrMnSi\textsubscript{2}~(34 K)}\footnote{BCS web page: \texttt{\href{http://webbdcrista1.ehu.es/magndata/index.php?this\_label=1.628} {http://webbdcrista1.ehu.es/magndata/index.php?this\_label=1.628}}}.\\
\subsubsection{Topological bands in subgroup $P2_{1}'/c'~(14.79)$}
\textbf{Perturbations:}
\begin{itemize}
\item B $\parallel$ [010] and strain $\perp$ [100],
\item B $\parallel$ [001] and strain $\perp$ [100],
\item B $\perp$ [100].
\end{itemize}
\begin{figure}[H]
\centering
\includegraphics[scale=0.6]{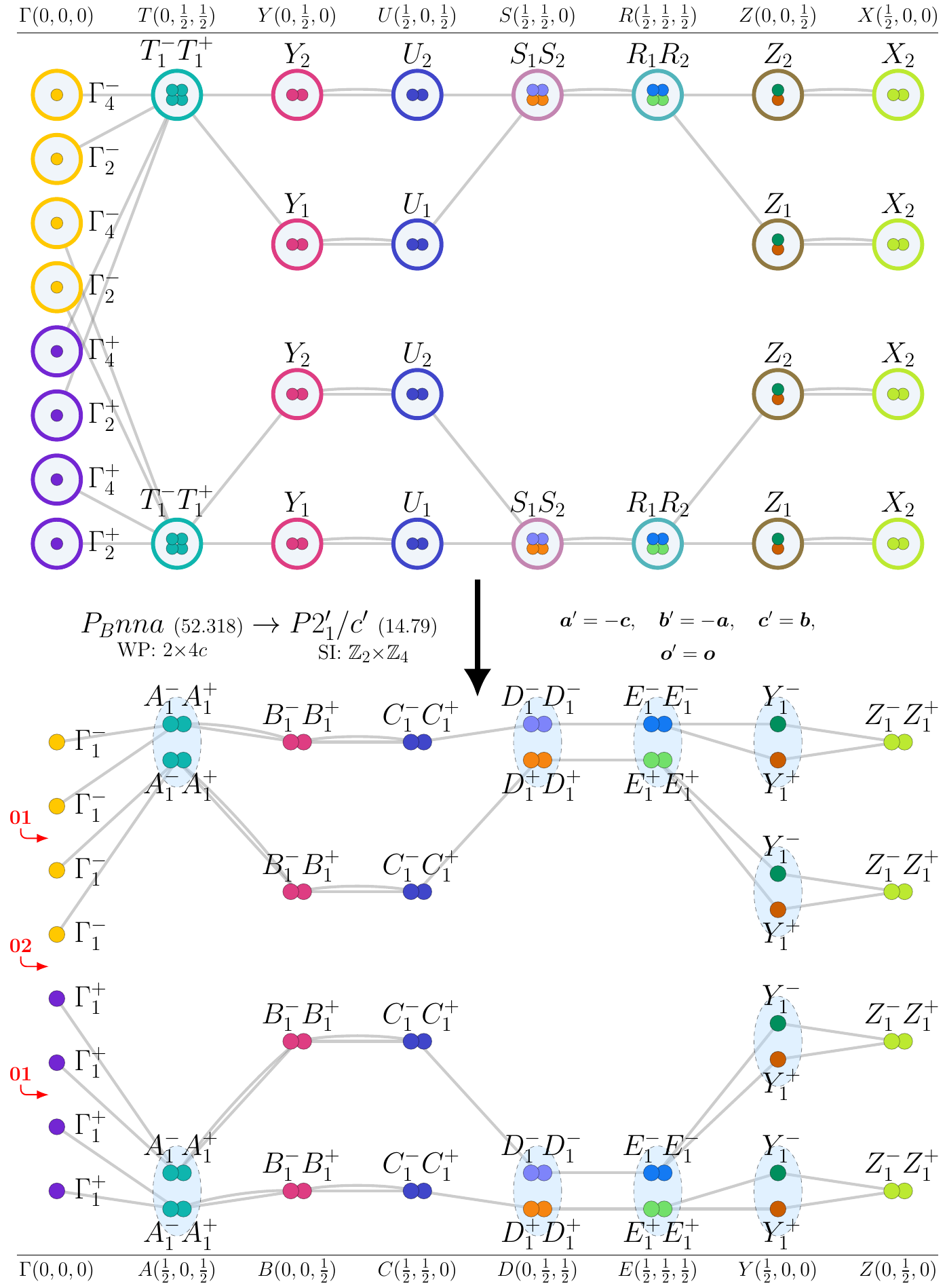}
\caption{Topological magnon bands in subgroup $P2_{1}'/c'~(14.79)$ for magnetic moments on Wyckoff positions $4c+4c$ of supergroup $P_{B}nna~(52.318)$.\label{fig_52.318_14.79_Bparallel010andstrainperp100_4c+4c}}
\end{figure}
\input{gap_tables_tex/52.318_14.79_Bparallel010andstrainperp100_4c+4c_table.tex}
\input{si_tables_tex/52.318_14.79_Bparallel010andstrainperp100_4c+4c_table.tex}
\subsection{WP: $8f$}
\textbf{BCS Materials:} {Ho\textsubscript{3}Ge\textsubscript{4}~(12 K)}\footnote{BCS web page: \texttt{\href{http://webbdcrista1.ehu.es/magndata/index.php?this\_label=1.356} {http://webbdcrista1.ehu.es/magndata/index.php?this\_label=1.356}}}.\\
\subsubsection{Topological bands in subgroup $P2_{1}'/c'~(14.79)$}
\textbf{Perturbations:}
\begin{itemize}
\item B $\parallel$ [010] and strain $\perp$ [100],
\item B $\parallel$ [001] and strain $\perp$ [100],
\item B $\perp$ [100].
\end{itemize}
\begin{figure}[H]
\centering
\includegraphics[scale=0.6]{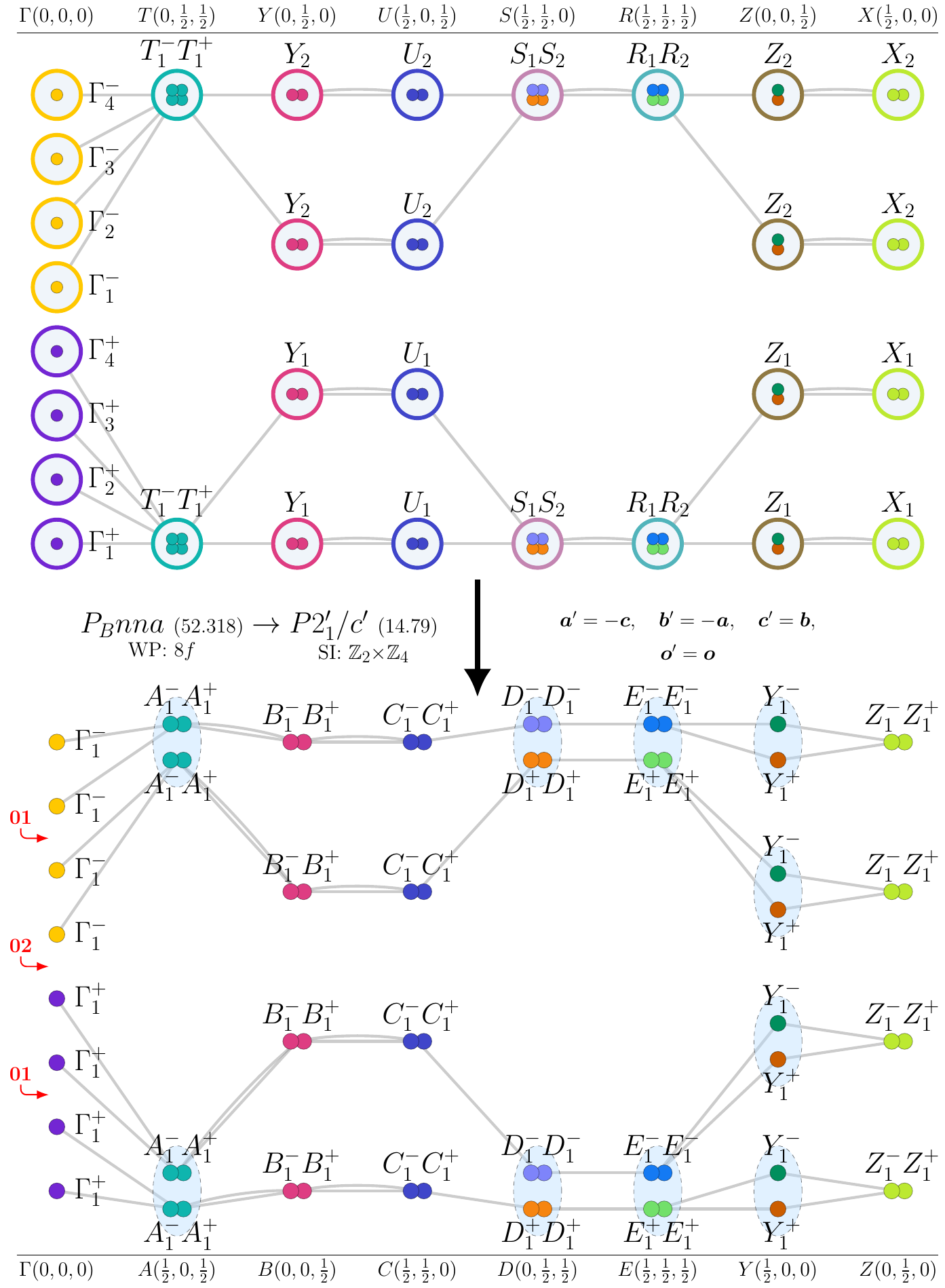}
\caption{Topological magnon bands in subgroup $P2_{1}'/c'~(14.79)$ for magnetic moments on Wyckoff position $8f$ of supergroup $P_{B}nna~(52.318)$.\label{fig_52.318_14.79_Bparallel010andstrainperp100_8f}}
\end{figure}
\input{gap_tables_tex/52.318_14.79_Bparallel010andstrainperp100_8f_table.tex}
\input{si_tables_tex/52.318_14.79_Bparallel010andstrainperp100_8f_table.tex}

\section{MSG $P_{A}mna~(53.333)$}
\textbf{Nontrivial-SI Subgroups:} $P\bar{1}~(2.4)$, $P2'/c'~(13.69)$, $P2_{1}'/c'~(14.79)$, $P2'/c'~(13.69)$, $P_{S}\bar{1}~(2.7)$, $P2_{1}/c~(14.75)$, $Pc'cn'~(56.370)$, $P_{C}2_{1}/c~(14.84)$, $P2~(3.1)$, $P2/c~(13.65)$, $Pn'n'n~(48.260)$, $P_{C}2/c~(13.74)$, $P2~(3.1)$, $P_{a}2~(3.4)$, $P2/m~(10.42)$, $Pmn'a'~(53.327)$, $P_{a}2/m~(10.47)$.\\

\textbf{Trivial-SI Subgroups:} $Pc'~(7.26)$, $Pc'~(7.26)$, $Pc'~(7.26)$, $P2'~(3.3)$, $P2_{1}'~(4.9)$, $P2'~(3.3)$, $P_{S}1~(1.3)$, $Pc~(7.24)$, $Pn'a2_{1}'~(33.146)$, $Pc'c2'~(27.80)$, $P_{C}c~(7.30)$, $Pc~(7.24)$, $Pn'n2'~(34.158)$, $Pn'n2'~(34.158)$, $P_{C}c~(7.30)$, $Pm~(6.18)$, $Pma'2'~(28.90)$, $Pmn'2_{1}'~(31.126)$, $P_{a}m~(6.21)$, $P2_{1}~(4.7)$, $Pn'a'2_{1}~(33.148)$, $P_{C}2_{1}~(4.12)$, $P_{A}mn2_{1}~(31.131)$, $Pn'n'2~(34.159)$, $P_{C}2~(3.6)$, $P_{A}ma2~(28.95)$, $Pn'c'2~(30.115)$, $P_{C}nc2~(30.121)$.\\

\subsection{WP: $4e$}
\textbf{BCS Materials:} {La\textsubscript{2}NiO\textsubscript{3}F\textsubscript{2}~(55 K)}\footnote{BCS web page: \texttt{\href{http://webbdcrista1.ehu.es/magndata/index.php?this\_label=1.388} {http://webbdcrista1.ehu.es/magndata/index.php?this\_label=1.388}}}.\\
\subsubsection{Topological bands in subgroup $P\bar{1}~(2.4)$}
\textbf{Perturbations:}
\begin{itemize}
\item B $\parallel$ [100] and strain in generic direction,
\item B $\parallel$ [010] and strain in generic direction,
\item B $\parallel$ [001] and strain in generic direction,
\item B $\perp$ [100] and strain $\perp$ [010],
\item B $\perp$ [100] and strain $\perp$ [001],
\item B $\perp$ [100] and strain in generic direction,
\item B $\perp$ [010] and strain $\perp$ [100],
\item B $\perp$ [010] and strain $\perp$ [001],
\item B $\perp$ [010] and strain in generic direction,
\item B $\perp$ [001] and strain $\perp$ [100],
\item B $\perp$ [001] and strain $\perp$ [010],
\item B $\perp$ [001] and strain in generic direction,
\item B in generic direction.
\end{itemize}
\begin{figure}[H]
\centering
\includegraphics[scale=0.6]{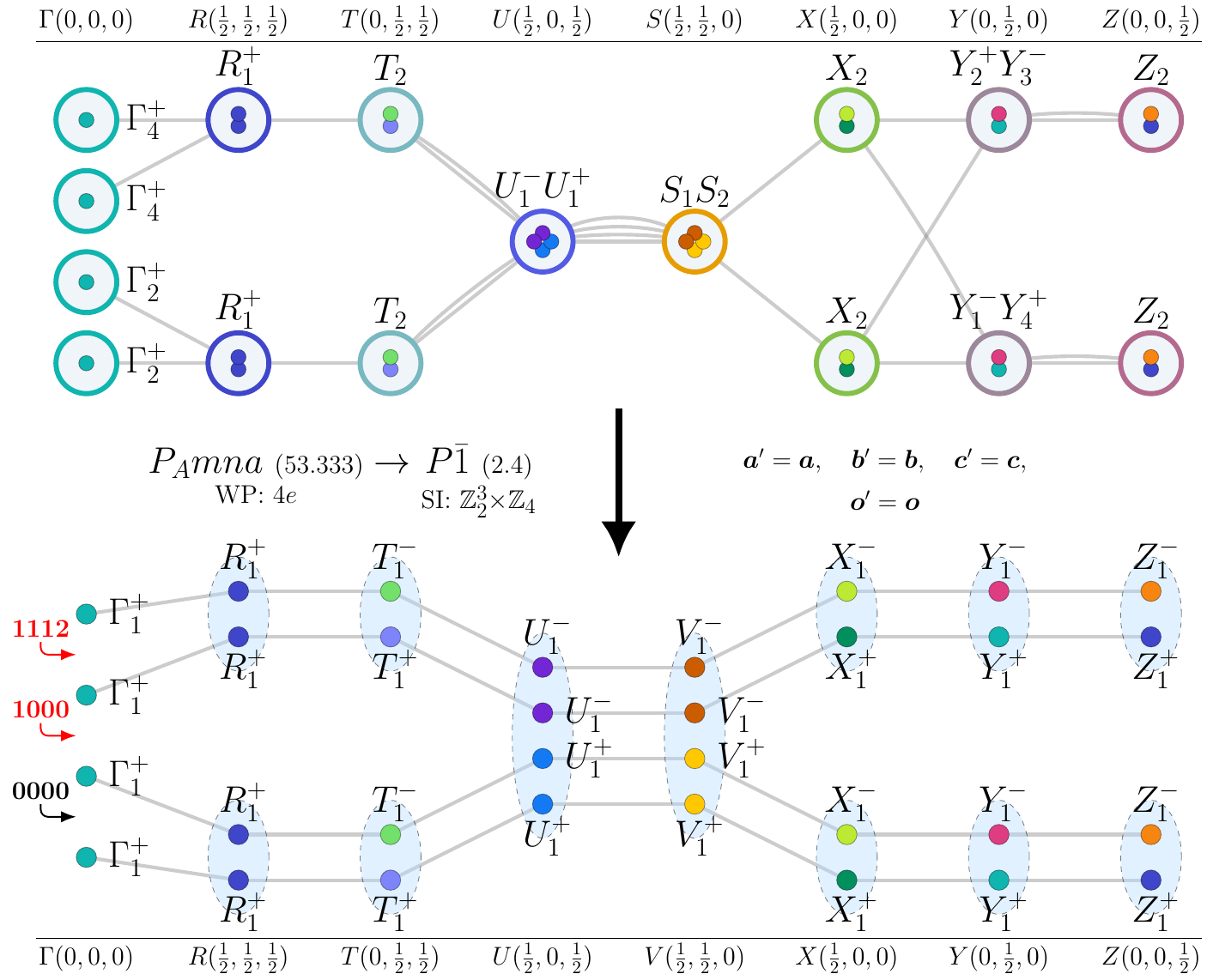}
\caption{Topological magnon bands in subgroup $P\bar{1}~(2.4)$ for magnetic moments on Wyckoff position $4e$ of supergroup $P_{A}mna~(53.333)$.\label{fig_53.333_2.4_Bparallel100andstrainingenericdirection_4e}}
\end{figure}
\input{gap_tables_tex/53.333_2.4_Bparallel100andstrainingenericdirection_4e_table.tex}
\input{si_tables_tex/53.333_2.4_Bparallel100andstrainingenericdirection_4e_table.tex}
\subsubsection{Topological bands in subgroup $P2'/c'~(13.69)$}
\textbf{Perturbations:}
\begin{itemize}
\item B $\parallel$ [100] and strain $\perp$ [001],
\item B $\parallel$ [010] and strain $\perp$ [001],
\item B $\perp$ [001].
\end{itemize}
\begin{figure}[H]
\centering
\includegraphics[scale=0.6]{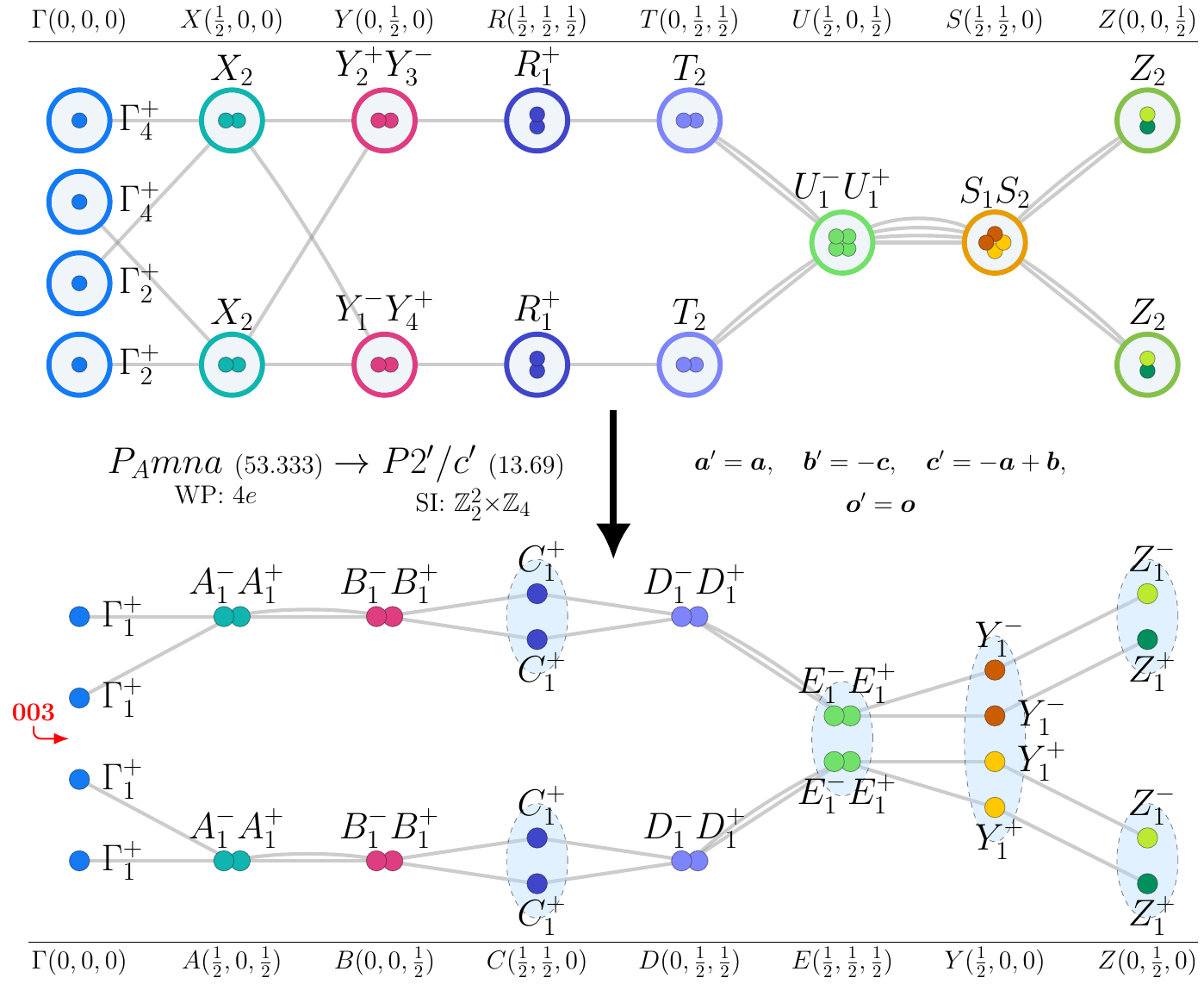}
\caption{Topological magnon bands in subgroup $P2'/c'~(13.69)$ for magnetic moments on Wyckoff position $4e$ of supergroup $P_{A}mna~(53.333)$.\label{fig_53.333_13.69_Bparallel100andstrainperp001_4e}}
\end{figure}
\input{gap_tables_tex/53.333_13.69_Bparallel100andstrainperp001_4e_table.tex}
\input{si_tables_tex/53.333_13.69_Bparallel100andstrainperp001_4e_table.tex}
\subsubsection{Topological bands in subgroup $P2_{1}'/c'~(14.79)$}
\textbf{Perturbations:}
\begin{itemize}
\item B $\parallel$ [100] and strain $\perp$ [010],
\item B $\parallel$ [001] and strain $\perp$ [010],
\item B $\perp$ [010].
\end{itemize}
\begin{figure}[H]
\centering
\includegraphics[scale=0.6]{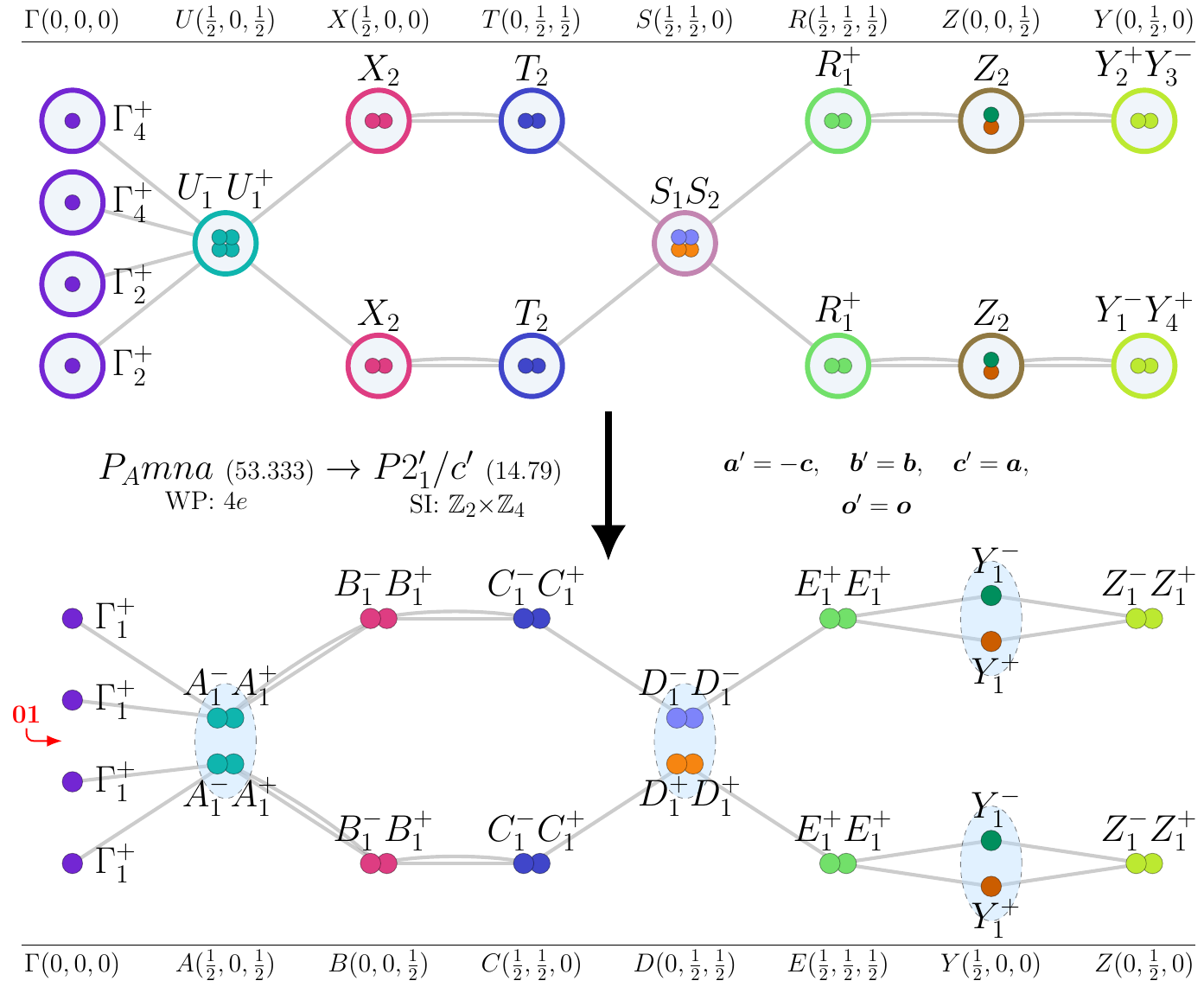}
\caption{Topological magnon bands in subgroup $P2_{1}'/c'~(14.79)$ for magnetic moments on Wyckoff position $4e$ of supergroup $P_{A}mna~(53.333)$.\label{fig_53.333_14.79_Bparallel100andstrainperp010_4e}}
\end{figure}
\input{gap_tables_tex/53.333_14.79_Bparallel100andstrainperp010_4e_table.tex}
\input{si_tables_tex/53.333_14.79_Bparallel100andstrainperp010_4e_table.tex}
\subsubsection{Topological bands in subgroup $P2'/c'~(13.69)$}
\textbf{Perturbations:}
\begin{itemize}
\item B $\parallel$ [010] and strain $\perp$ [100],
\item B $\parallel$ [001] and strain $\perp$ [100],
\item B $\perp$ [100].
\end{itemize}
\begin{figure}[H]
\centering
\includegraphics[scale=0.6]{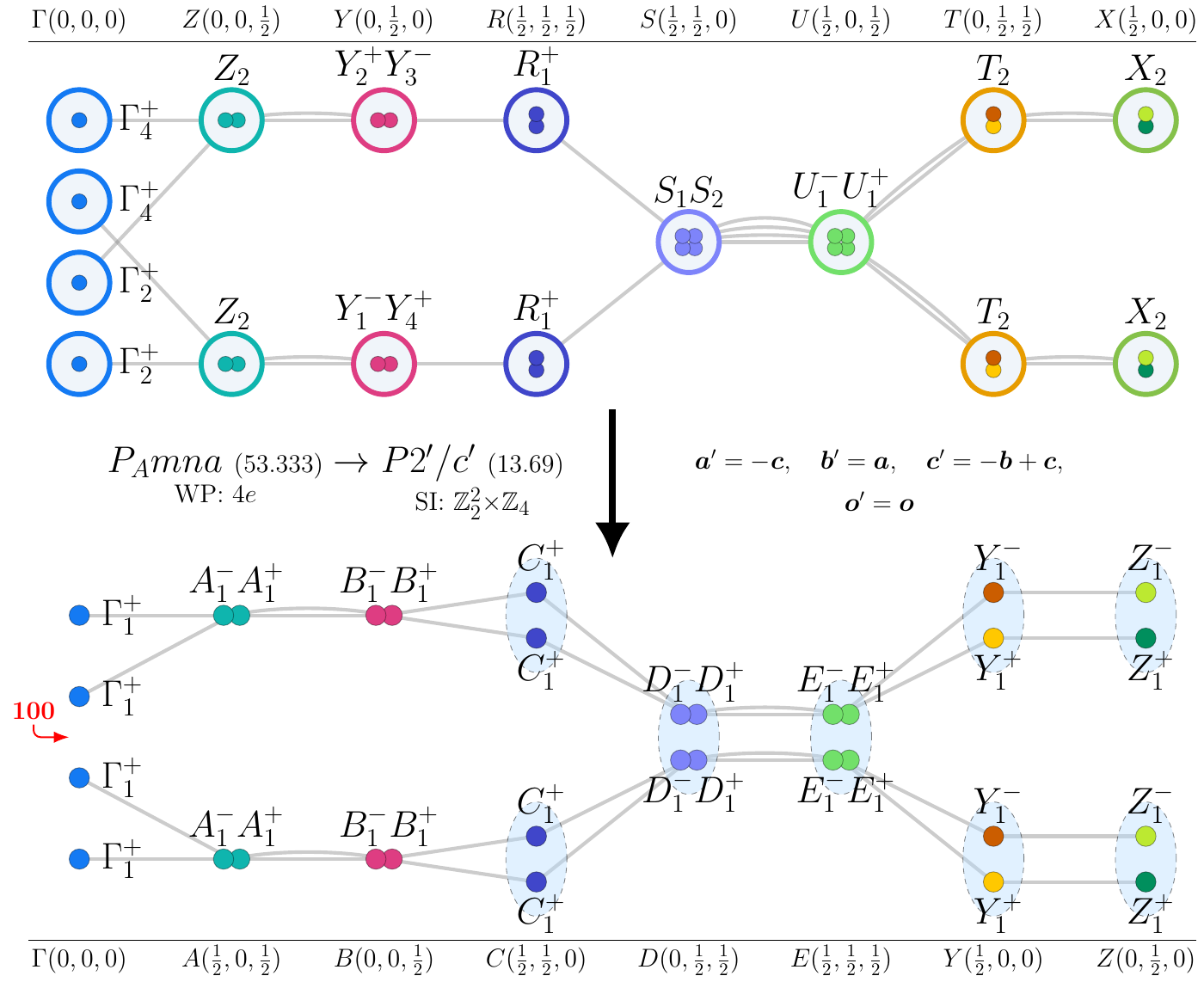}
\caption{Topological magnon bands in subgroup $P2'/c'~(13.69)$ for magnetic moments on Wyckoff position $4e$ of supergroup $P_{A}mna~(53.333)$.\label{fig_53.333_13.69_Bparallel010andstrainperp100_4e}}
\end{figure}
\input{gap_tables_tex/53.333_13.69_Bparallel010andstrainperp100_4e_table.tex}
\input{si_tables_tex/53.333_13.69_Bparallel010andstrainperp100_4e_table.tex}
\subsubsection{Topological bands in subgroup $P_{S}\bar{1}~(2.7)$}
\textbf{Perturbation:}
\begin{itemize}
\item strain in generic direction.
\end{itemize}
\begin{figure}[H]
\centering
\includegraphics[scale=0.6]{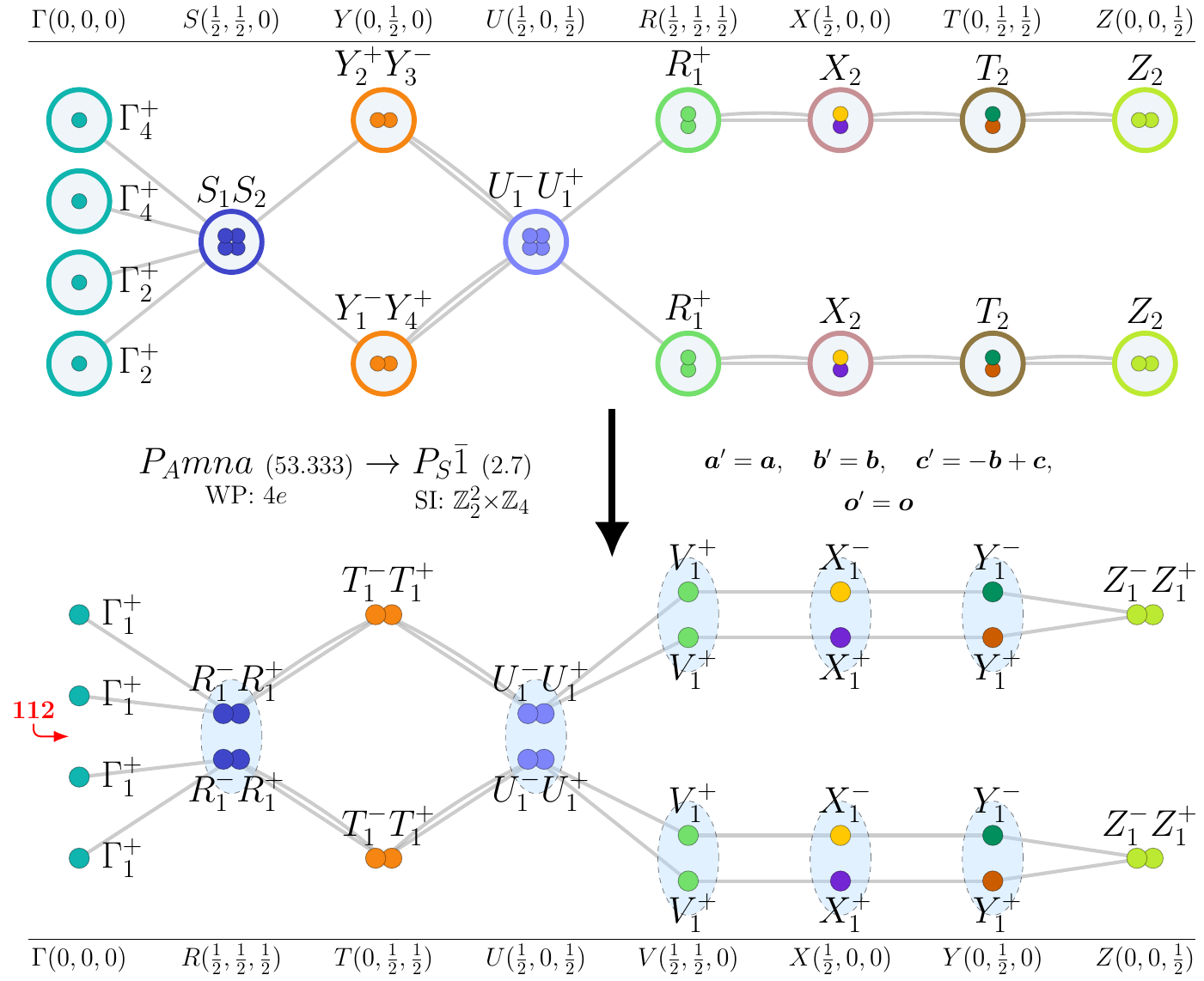}
\caption{Topological magnon bands in subgroup $P_{S}\bar{1}~(2.7)$ for magnetic moments on Wyckoff position $4e$ of supergroup $P_{A}mna~(53.333)$.\label{fig_53.333_2.7_strainingenericdirection_4e}}
\end{figure}
\input{gap_tables_tex/53.333_2.7_strainingenericdirection_4e_table.tex}
\input{si_tables_tex/53.333_2.7_strainingenericdirection_4e_table.tex}
\subsubsection{Topological bands in subgroup $P_{a}2~(3.4)$}
\textbf{Perturbation:}
\begin{itemize}
\item E $\parallel$ [100] and strain $\perp$ [100].
\end{itemize}
\begin{figure}[H]
\centering
\includegraphics[scale=0.6]{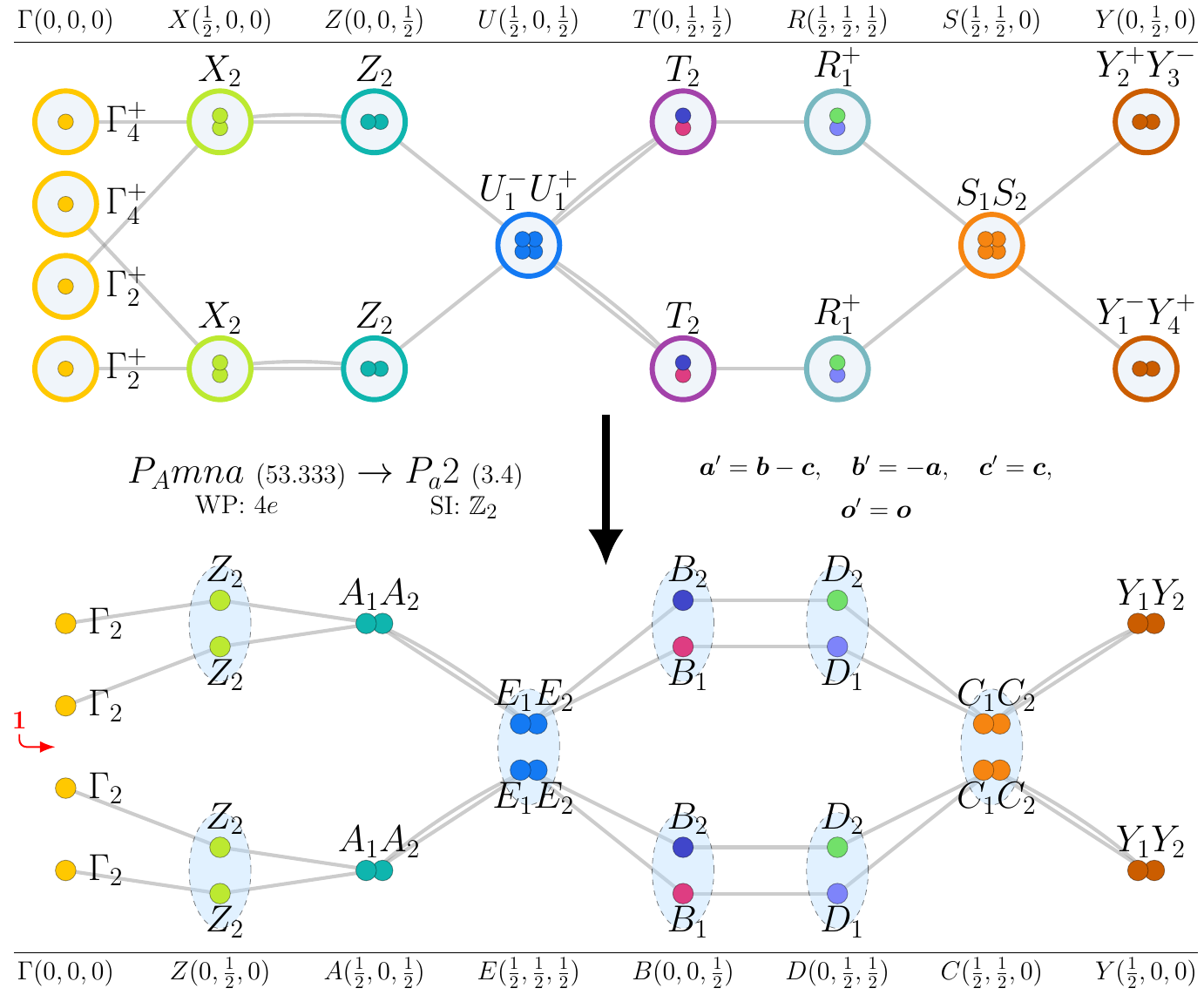}
\caption{Topological magnon bands in subgroup $P_{a}2~(3.4)$ for magnetic moments on Wyckoff position $4e$ of supergroup $P_{A}mna~(53.333)$.\label{fig_53.333_3.4_Eparallel100andstrainperp100_4e}}
\end{figure}
\input{gap_tables_tex/53.333_3.4_Eparallel100andstrainperp100_4e_table.tex}
\input{si_tables_tex/53.333_3.4_Eparallel100andstrainperp100_4e_table.tex}

\section{MSG $P_{C}mna~(53.335)$}
\textbf{Nontrivial-SI Subgroups:} $P\bar{1}~(2.4)$, $P2_{1}'/c'~(14.79)$, $P2_{1}'/c'~(14.79)$, $P2_{1}'/c'~(14.79)$, $P_{S}\bar{1}~(2.7)$, $P2_{1}/c~(14.75)$, $Pb'c'a~(61.436)$, $P_{a}2_{1}/c~(14.80)$, $P2~(3.1)$, $Pc'c'2~(27.81)$, $P2/c~(13.65)$, $Pc'c'n~(56.369)$, $P_{C}2/c~(13.74)$, $P2~(3.1)$, $Pb'a'2~(32.138)$, $P_{B}nc2~(30.120)$, $P2/m~(10.42)$, $Pb'a'm~(55.357)$, $P_{C}2/m~(10.49)$.\\

\textbf{Trivial-SI Subgroups:} $Pc'~(7.26)$, $Pc'~(7.26)$, $Pc'~(7.26)$, $P2_{1}'~(4.9)$, $P2_{1}'~(4.9)$, $P2_{1}'~(4.9)$, $P_{S}1~(1.3)$, $Pc~(7.24)$, $Pc'a2_{1}'~(29.101)$, $Pca'2_{1}'~(29.102)$, $P_{a}c~(7.27)$, $Pc~(7.24)$, $Pna'2_{1}'~(33.147)$, $Pna'2_{1}'~(33.147)$, $P_{C}c~(7.30)$, $Pm~(6.18)$, $Pmc'2_{1}'~(26.69)$, $Pmc'2_{1}'~(26.69)$, $P_{C}m~(6.23)$, $P2_{1}~(4.7)$, $Pc'a'2_{1}~(29.103)$, $P_{a}2_{1}~(4.10)$, $P_{C}mn2_{1}~(31.133)$, $P_{C}2~(3.6)$, $P_{B}ma2~(28.96)$, $P_{C}2~(3.6)$.\\

\subsection{WP: $4a$}
\textbf{BCS Materials:} {La\textsubscript{2}NiO\textsubscript{4}~(330 K)}\footnote{BCS web page: \texttt{\href{http://webbdcrista1.ehu.es/magndata/index.php?this\_label=1.42} {http://webbdcrista1.ehu.es/magndata/index.php?this\_label=1.42}}}, {Nd\textsubscript{2}NiO\textsubscript{4}~(320 K)}\footnote{BCS web page: \texttt{\href{http://webbdcrista1.ehu.es/magndata/index.php?this\_label=1.371} {http://webbdcrista1.ehu.es/magndata/index.php?this\_label=1.371}}}, {La\textsubscript{2}CoO\textsubscript{4}~(275 K)}\footnote{BCS web page: \texttt{\href{http://webbdcrista1.ehu.es/magndata/index.php?this\_label=1.403} {http://webbdcrista1.ehu.es/magndata/index.php?this\_label=1.403}}}.\\
\subsubsection{Topological bands in subgroup $P\bar{1}~(2.4)$}
\textbf{Perturbations:}
\begin{itemize}
\item B $\parallel$ [100] and strain in generic direction,
\item B $\parallel$ [010] and strain in generic direction,
\item B $\parallel$ [001] and strain in generic direction,
\item B $\perp$ [100] and strain $\perp$ [010],
\item B $\perp$ [100] and strain $\perp$ [001],
\item B $\perp$ [100] and strain in generic direction,
\item B $\perp$ [010] and strain $\perp$ [100],
\item B $\perp$ [010] and strain $\perp$ [001],
\item B $\perp$ [010] and strain in generic direction,
\item B $\perp$ [001] and strain $\perp$ [100],
\item B $\perp$ [001] and strain $\perp$ [010],
\item B $\perp$ [001] and strain in generic direction,
\item B in generic direction.
\end{itemize}
\begin{figure}[H]
\centering
\includegraphics[scale=0.6]{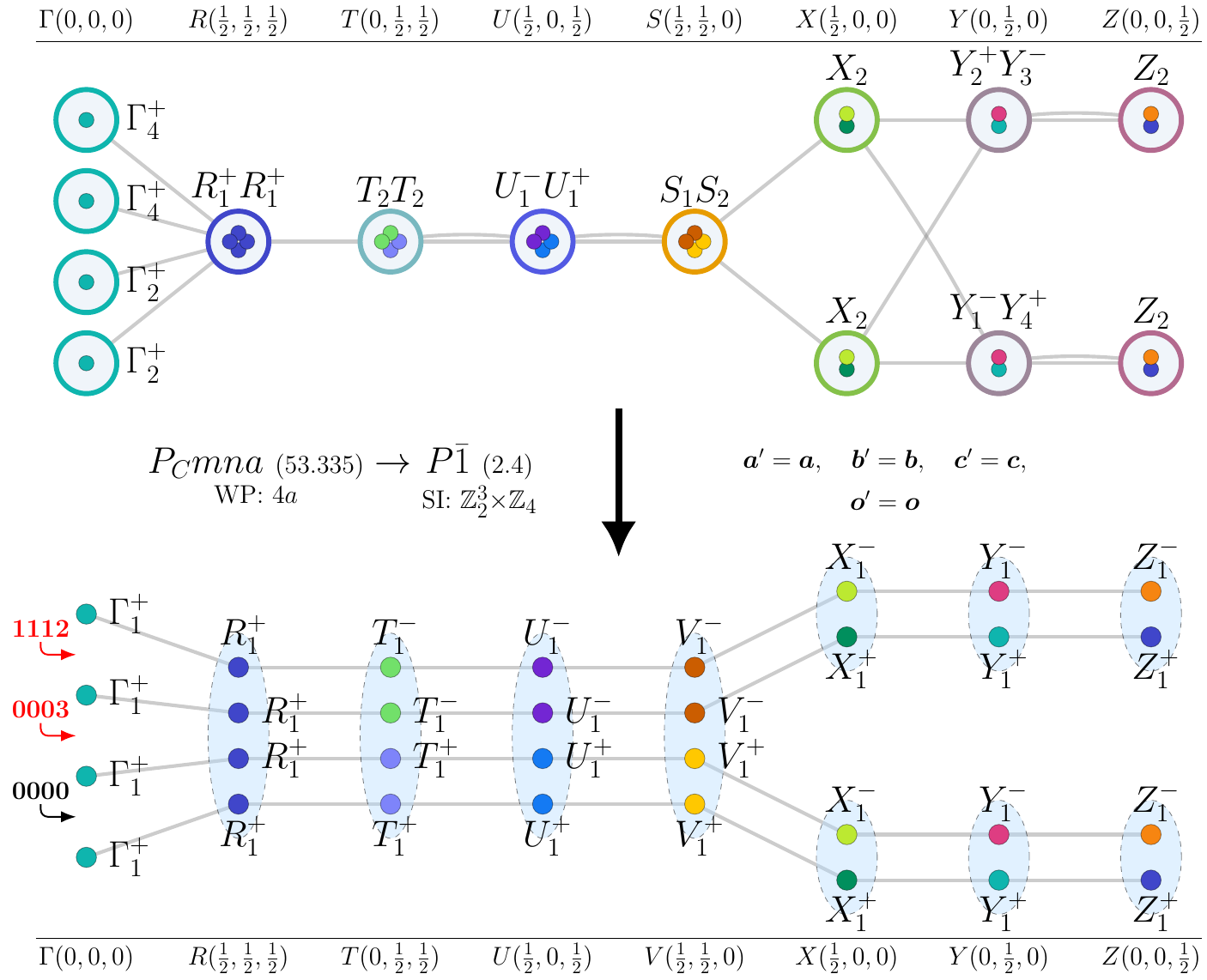}
\caption{Topological magnon bands in subgroup $P\bar{1}~(2.4)$ for magnetic moments on Wyckoff position $4a$ of supergroup $P_{C}mna~(53.335)$.\label{fig_53.335_2.4_Bparallel100andstrainingenericdirection_4a}}
\end{figure}
\input{gap_tables_tex/53.335_2.4_Bparallel100andstrainingenericdirection_4a_table.tex}
\input{si_tables_tex/53.335_2.4_Bparallel100andstrainingenericdirection_4a_table.tex}
\subsubsection{Topological bands in subgroup $P2_{1}'/c'~(14.79)$}
\textbf{Perturbations:}
\begin{itemize}
\item B $\parallel$ [100] and strain $\perp$ [001],
\item B $\parallel$ [010] and strain $\perp$ [001],
\item B $\perp$ [001].
\end{itemize}
\begin{figure}[H]
\centering
\includegraphics[scale=0.6]{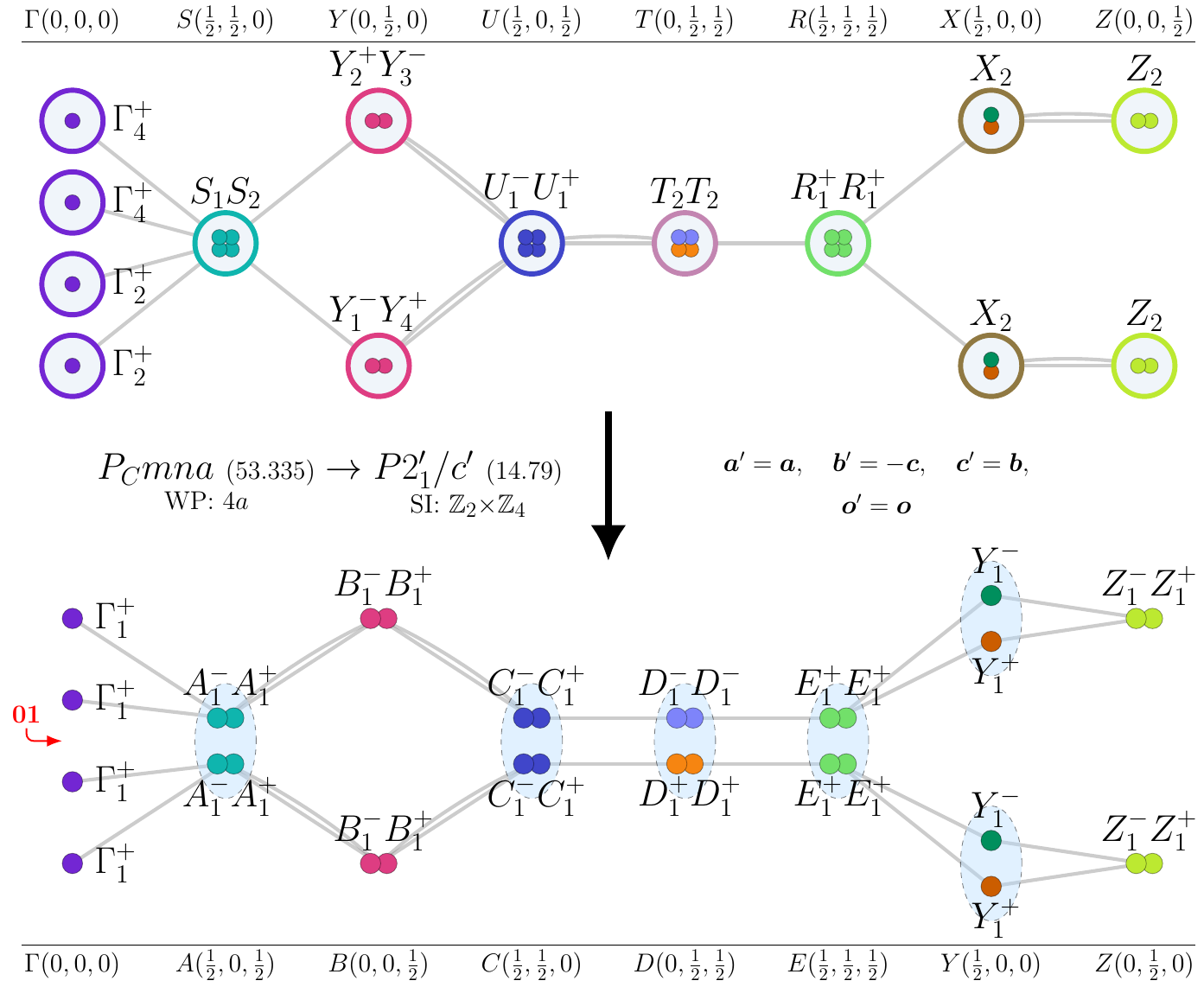}
\caption{Topological magnon bands in subgroup $P2_{1}'/c'~(14.79)$ for magnetic moments on Wyckoff position $4a$ of supergroup $P_{C}mna~(53.335)$.\label{fig_53.335_14.79_Bparallel100andstrainperp001_4a}}
\end{figure}
\input{gap_tables_tex/53.335_14.79_Bparallel100andstrainperp001_4a_table.tex}
\input{si_tables_tex/53.335_14.79_Bparallel100andstrainperp001_4a_table.tex}
\subsubsection{Topological bands in subgroup $P2_{1}'/c'~(14.79)$}
\textbf{Perturbations:}
\begin{itemize}
\item B $\parallel$ [100] and strain $\perp$ [010],
\item B $\parallel$ [001] and strain $\perp$ [010],
\item B $\perp$ [010].
\end{itemize}
\begin{figure}[H]
\centering
\includegraphics[scale=0.6]{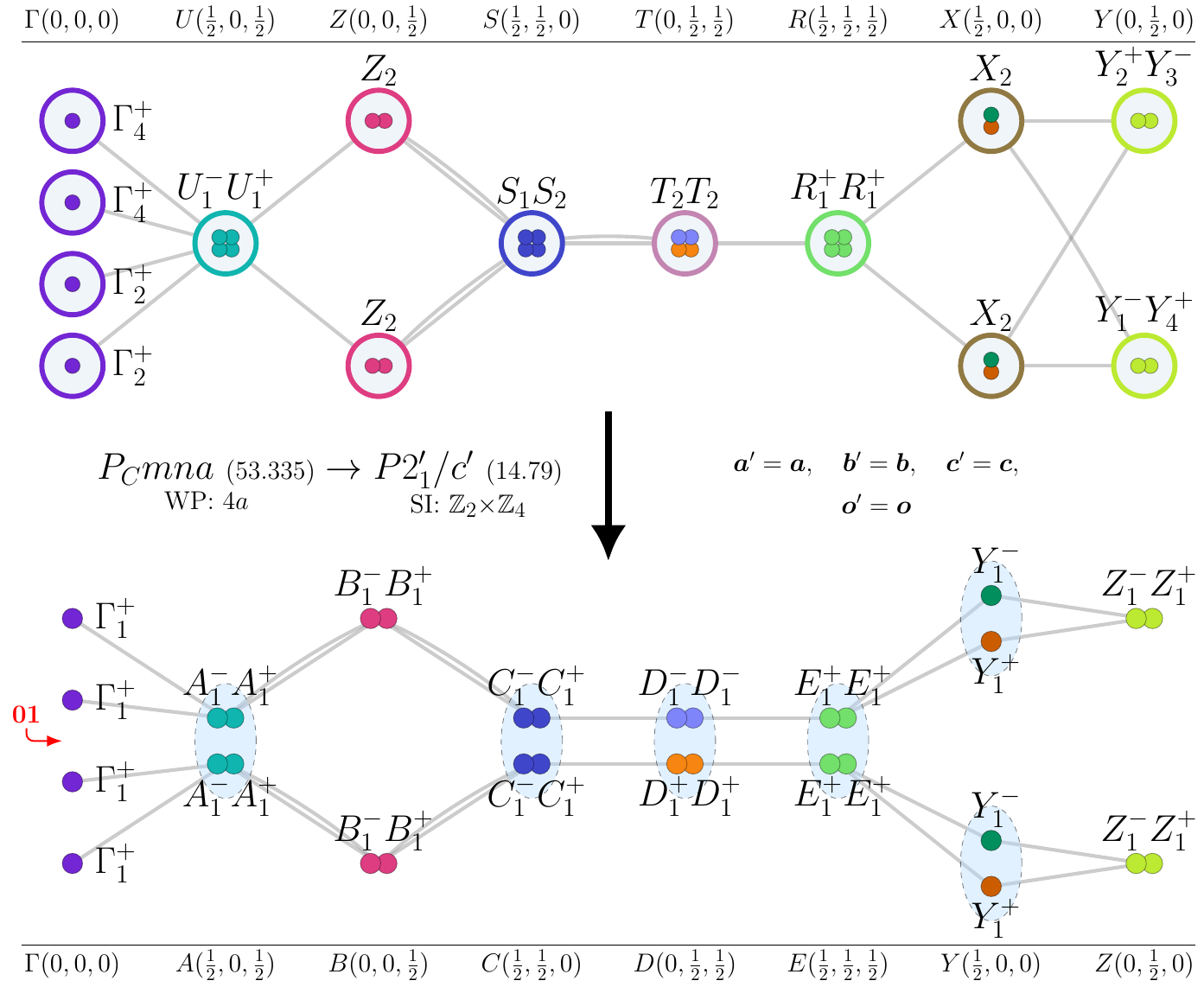}
\caption{Topological magnon bands in subgroup $P2_{1}'/c'~(14.79)$ for magnetic moments on Wyckoff position $4a$ of supergroup $P_{C}mna~(53.335)$.\label{fig_53.335_14.79_Bparallel100andstrainperp010_4a}}
\end{figure}
\input{gap_tables_tex/53.335_14.79_Bparallel100andstrainperp010_4a_table.tex}
\input{si_tables_tex/53.335_14.79_Bparallel100andstrainperp010_4a_table.tex}
\subsubsection{Topological bands in subgroup $P2_{1}'/c'~(14.79)$}
\textbf{Perturbations:}
\begin{itemize}
\item B $\parallel$ [010] and strain $\perp$ [100],
\item B $\parallel$ [001] and strain $\perp$ [100],
\item B $\perp$ [100].
\end{itemize}
\begin{figure}[H]
\centering
\includegraphics[scale=0.6]{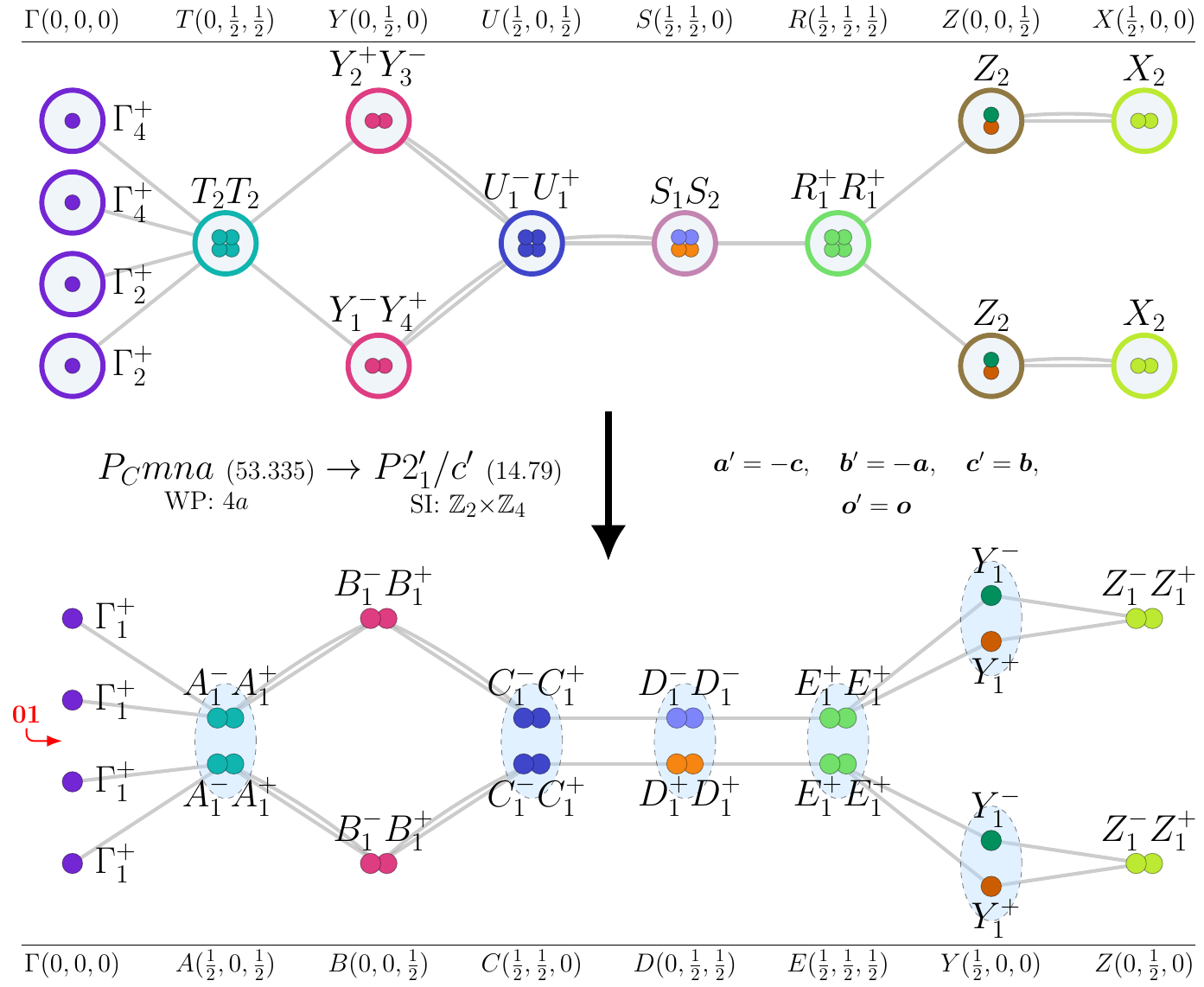}
\caption{Topological magnon bands in subgroup $P2_{1}'/c'~(14.79)$ for magnetic moments on Wyckoff position $4a$ of supergroup $P_{C}mna~(53.335)$.\label{fig_53.335_14.79_Bparallel010andstrainperp100_4a}}
\end{figure}
\input{gap_tables_tex/53.335_14.79_Bparallel010andstrainperp100_4a_table.tex}
\input{si_tables_tex/53.335_14.79_Bparallel010andstrainperp100_4a_table.tex}
\subsubsection{Topological bands in subgroup $P_{S}\bar{1}~(2.7)$}
\textbf{Perturbation:}
\begin{itemize}
\item strain in generic direction.
\end{itemize}
\begin{figure}[H]
\centering
\includegraphics[scale=0.6]{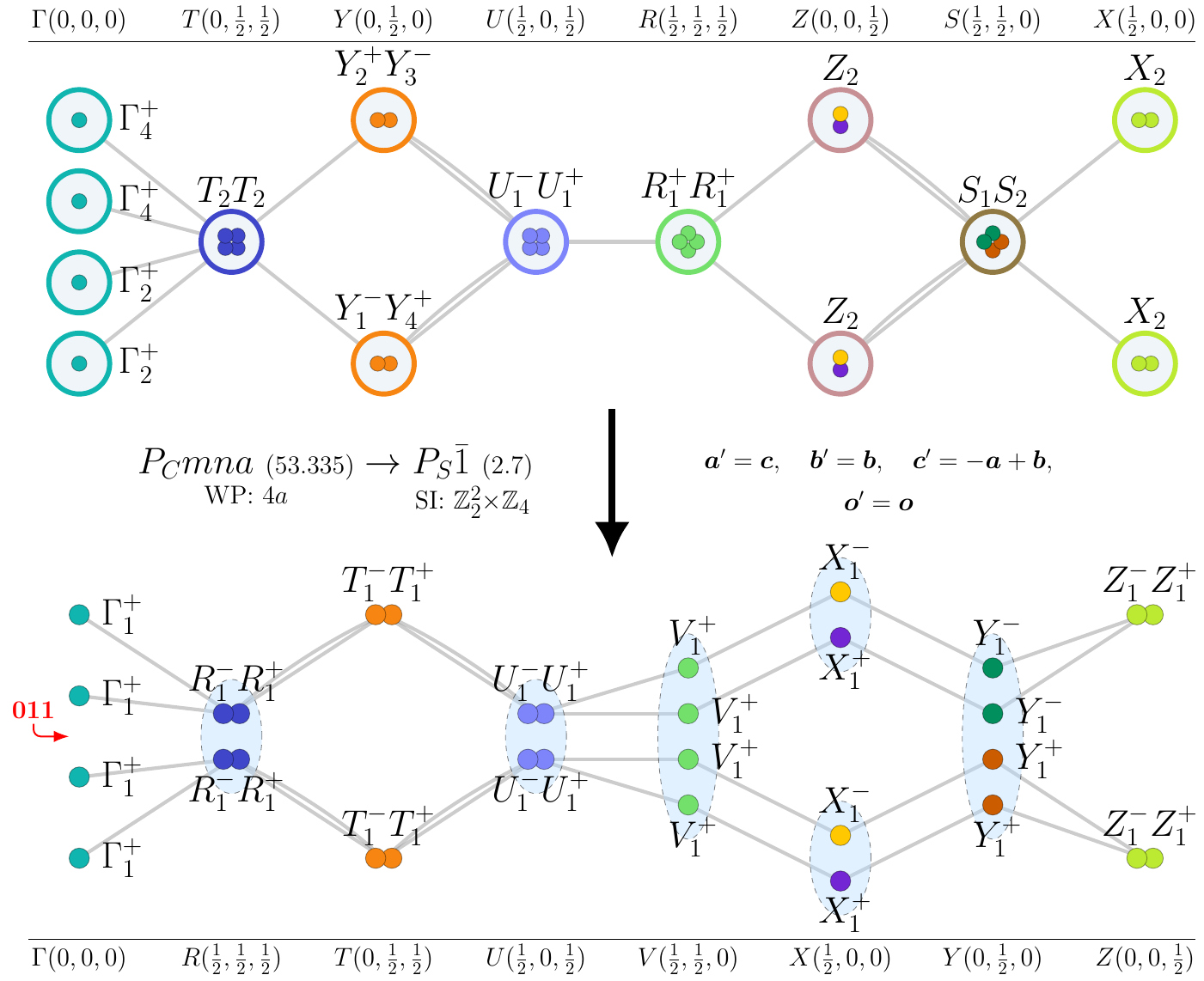}
\caption{Topological magnon bands in subgroup $P_{S}\bar{1}~(2.7)$ for magnetic moments on Wyckoff position $4a$ of supergroup $P_{C}mna~(53.335)$.\label{fig_53.335_2.7_strainingenericdirection_4a}}
\end{figure}
\input{gap_tables_tex/53.335_2.7_strainingenericdirection_4a_table.tex}
\input{si_tables_tex/53.335_2.7_strainingenericdirection_4a_table.tex}

\section{MSG $P_{I}mna~(53.336)$}
\textbf{Nontrivial-SI Subgroups:} $P\bar{1}~(2.4)$, $P2'/c'~(13.69)$, $P2_{1}'/m'~(11.54)$, $P2_{1}'/c'~(14.79)$, $P_{S}\bar{1}~(2.7)$, $P2_{1}/c~(14.75)$, $Pn'm'a~(62.446)$, $P_{C}2_{1}/c~(14.84)$, $P2~(3.1)$, $P2/c~(13.65)$, $Pnn'a'~(52.311)$, $P_{A}2/c~(13.73)$, $P2~(3.1)$, $Pm'a'2~(28.91)$, $P2/m~(10.42)$, $Pm'ma'~(51.296)$, $P_{C}2/m~(10.49)$.\\

\textbf{Trivial-SI Subgroups:} $Pc'~(7.26)$, $Pm'~(6.20)$, $Pc'~(7.26)$, $P2'~(3.3)$, $P2_{1}'~(4.9)$, $P2_{1}'~(4.9)$, $P_{S}1~(1.3)$, $Pc~(7.24)$, $Pn'a2_{1}'~(33.146)$, $Pm'c2_{1}'~(26.68)$, $P_{C}c~(7.30)$, $Pc~(7.24)$, $Pn'n2'~(34.158)$, $Pna'2_{1}'~(33.147)$, $P_{A}c~(7.31)$, $Pm~(6.18)$, $Pm'm2'~(25.59)$, $Pmc'2_{1}'~(26.69)$, $P_{C}m~(6.23)$, $P2_{1}~(4.7)$, $Pm'n'2_{1}~(31.127)$, $P_{C}2_{1}~(4.12)$, $P_{I}mn2_{1}~(31.134)$, $Pn'c'2~(30.115)$, $P_{C}2~(3.6)$, $P_{I}ma2~(28.98)$, $P_{C}2~(3.6)$, $P_{I}nc2~(30.122)$.\\

\subsection{WP: $4a+4a+8i$}
\textbf{BCS Materials:} {Sr\textsubscript{2}Fe\textsubscript{1.9}Cr\textsubscript{0.1}O\textsubscript{5}~(650 K)}\footnote{BCS web page: \texttt{\href{http://webbdcrista1.ehu.es/magndata/index.php?this\_label=1.410} {http://webbdcrista1.ehu.es/magndata/index.php?this\_label=1.410}}}.\\
\subsubsection{Topological bands in subgroup $P2'/c'~(13.69)$}
\textbf{Perturbations:}
\begin{itemize}
\item B $\parallel$ [100] and strain $\perp$ [001],
\item B $\parallel$ [010] and strain $\perp$ [001],
\item B $\perp$ [001].
\end{itemize}
\begin{figure}[H]
\centering
\includegraphics[scale=0.6]{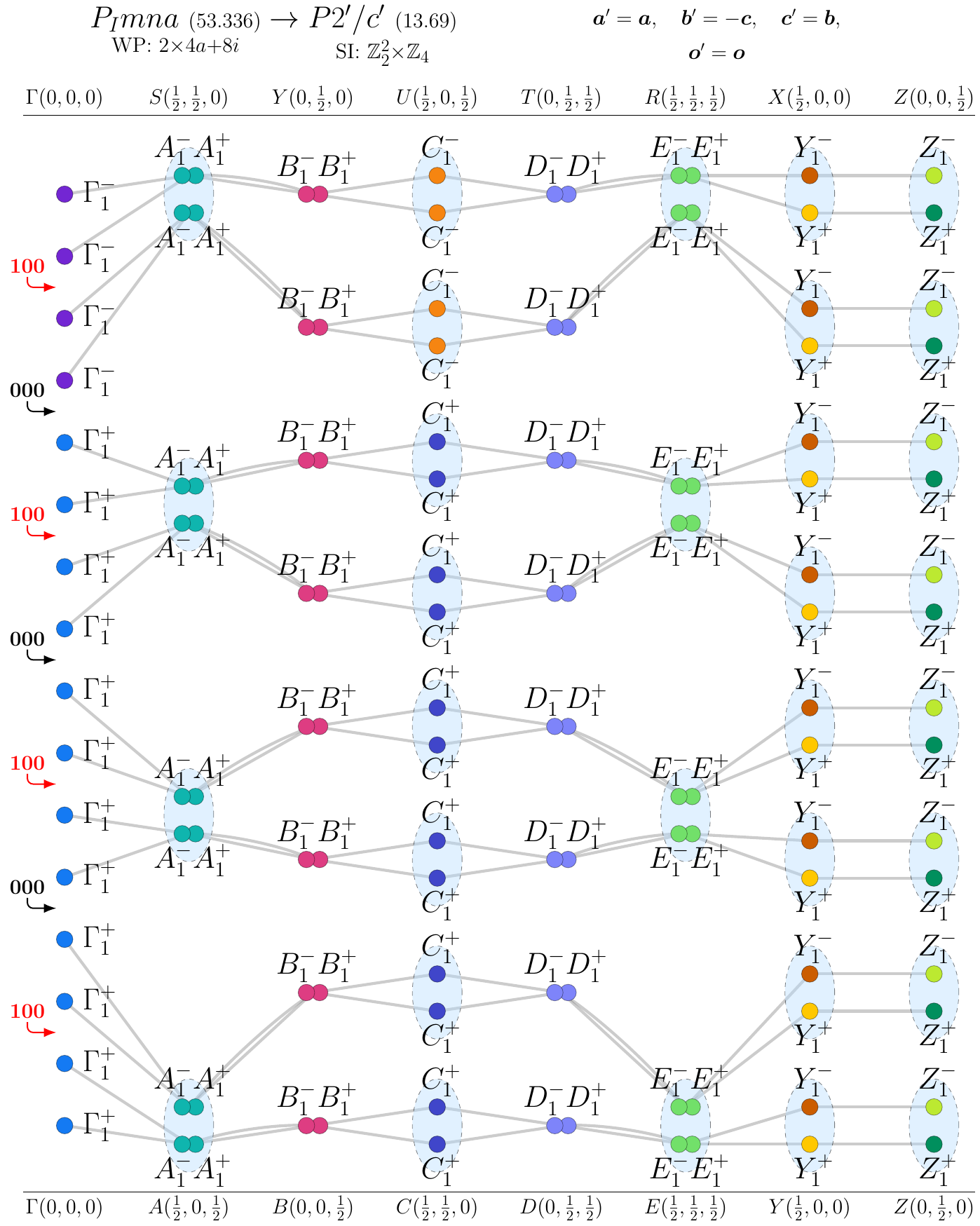}
\caption{Topological magnon bands in subgroup $P2'/c'~(13.69)$ for magnetic moments on Wyckoff positions $4a+4a+8i$ of supergroup $P_{I}mna~(53.336)$.\label{fig_53.336_13.69_Bparallel100andstrainperp001_4a+4a+8i}}
\end{figure}
\input{gap_tables_tex/53.336_13.69_Bparallel100andstrainperp001_4a+4a+8i_table.tex}
\input{si_tables_tex/53.336_13.69_Bparallel100andstrainperp001_4a+4a+8i_table.tex}
\subsubsection{Topological bands in subgroup $P2_{1}'/m'~(11.54)$}
\textbf{Perturbations:}
\begin{itemize}
\item B $\parallel$ [100] and strain $\perp$ [010],
\item B $\parallel$ [001] and strain $\perp$ [010],
\item B $\perp$ [010].
\end{itemize}
\begin{figure}[H]
\centering
\includegraphics[scale=0.6]{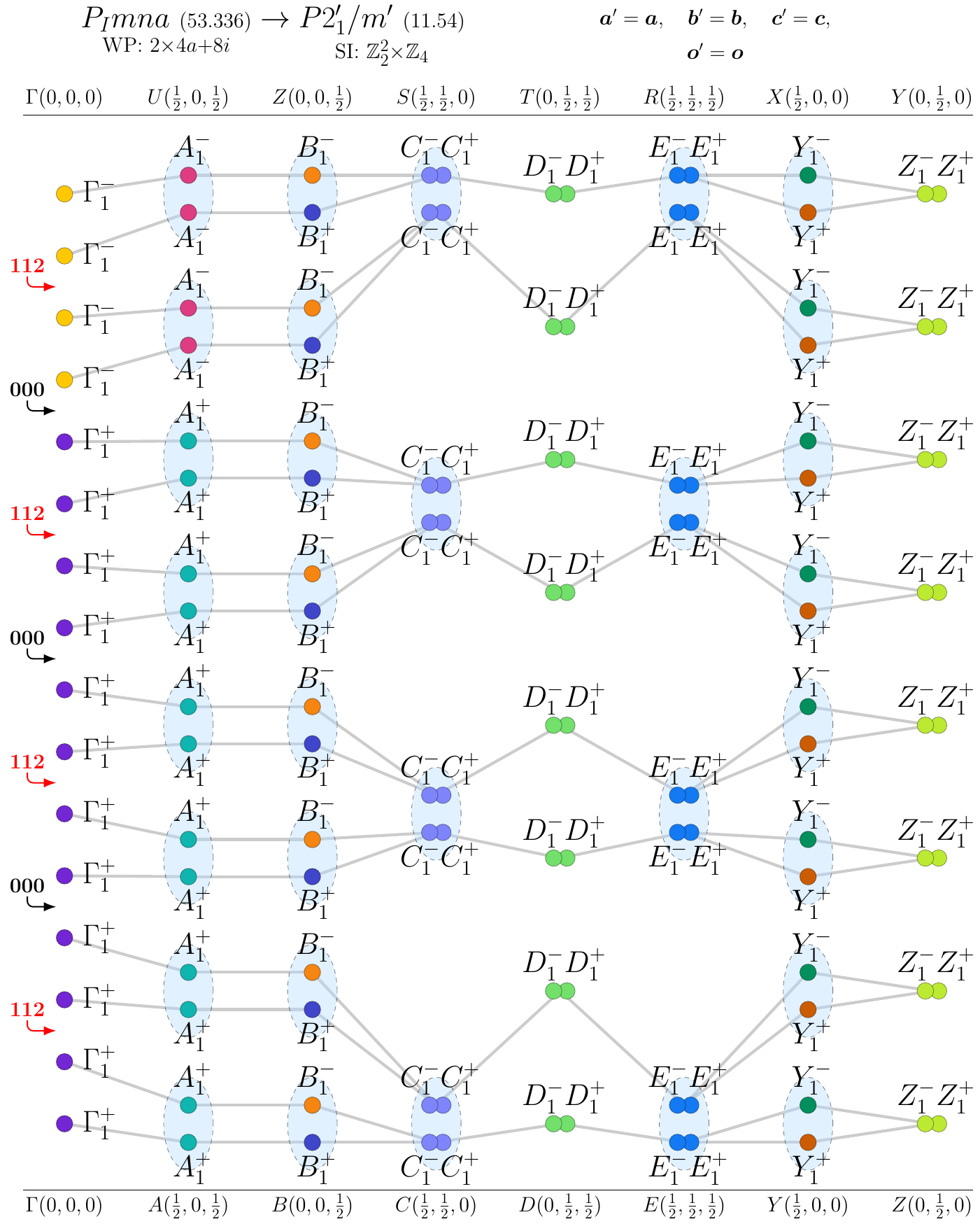}
\caption{Topological magnon bands in subgroup $P2_{1}'/m'~(11.54)$ for magnetic moments on Wyckoff positions $4a+4a+8i$ of supergroup $P_{I}mna~(53.336)$.\label{fig_53.336_11.54_Bparallel100andstrainperp010_4a+4a+8i}}
\end{figure}
\input{gap_tables_tex/53.336_11.54_Bparallel100andstrainperp010_4a+4a+8i_table.tex}
\input{si_tables_tex/53.336_11.54_Bparallel100andstrainperp010_4a+4a+8i_table.tex}
\subsubsection{Topological bands in subgroup $P2_{1}'/c'~(14.79)$}
\textbf{Perturbations:}
\begin{itemize}
\item B $\parallel$ [010] and strain $\perp$ [100],
\item B $\parallel$ [001] and strain $\perp$ [100],
\item B $\perp$ [100].
\end{itemize}
\begin{figure}[H]
\centering
\includegraphics[scale=0.6]{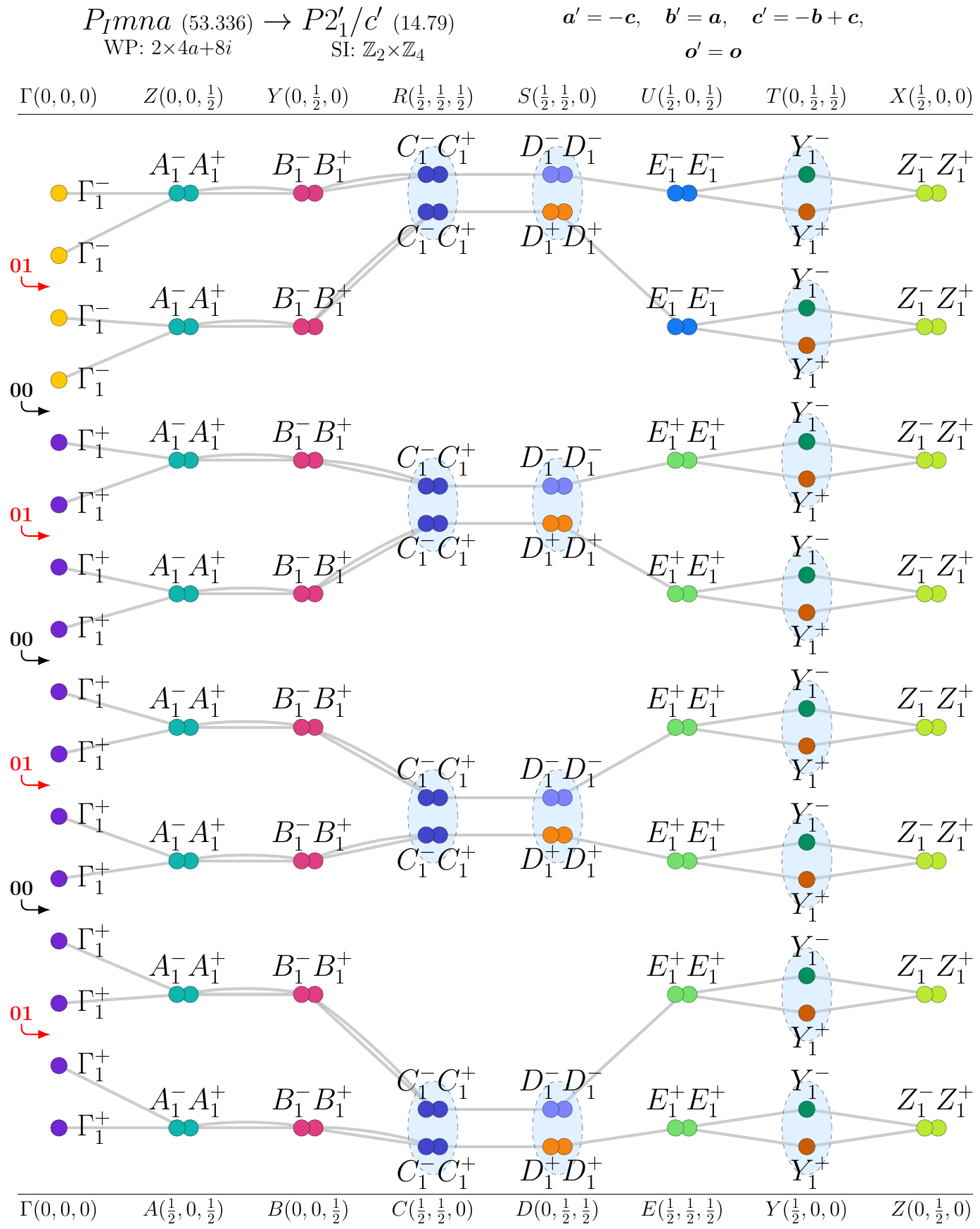}
\caption{Topological magnon bands in subgroup $P2_{1}'/c'~(14.79)$ for magnetic moments on Wyckoff positions $4a+4a+8i$ of supergroup $P_{I}mna~(53.336)$.\label{fig_53.336_14.79_Bparallel010andstrainperp100_4a+4a+8i}}
\end{figure}
\input{gap_tables_tex/53.336_14.79_Bparallel010andstrainperp100_4a+4a+8i_table.tex}
\input{si_tables_tex/53.336_14.79_Bparallel010andstrainperp100_4a+4a+8i_table.tex}

\section{MSG $P_{B}cca~(54.350)$}
\textbf{Nontrivial-SI Subgroups:} $P\bar{1}~(2.4)$, $P2_{1}'/m'~(11.54)$, $P2'/c'~(13.69)$, $P2'/m'~(10.46)$, $P_{S}\bar{1}~(2.7)$, $P2~(3.1)$, $Pm'a'2~(28.91)$, $P_{A}cc2~(27.85)$, $P2/c~(13.65)$, $Pc'cm'~(49.270)$, $P_{A}2/c~(13.73)$, $P2~(3.1)$, $Pm'm'2~(25.60)$, $P_{a}2~(3.4)$, $P2/c~(13.65)$, $Pm'm'a~(51.294)$, $P_{a}2/c~(13.70)$, $P2_{1}/c~(14.75)$, $Pb'cm'~(57.384)$, $P_{A}2_{1}/c~(14.83)$.\\

\textbf{Trivial-SI Subgroups:} $Pm'~(6.20)$, $Pc'~(7.26)$, $Pm'~(6.20)$, $P2_{1}'~(4.9)$, $P2'~(3.3)$, $P2'~(3.3)$, $P_{S}1~(1.3)$, $Pc~(7.24)$, $Pm'a2'~(28.89)$, $Pc'c2'~(27.80)$, $P_{A}c~(7.31)$, $Pc~(7.24)$, $Pm'c2_{1}'~(26.68)$, $Pm'a2'~(28.89)$, $P_{a}c~(7.27)$, $Pc~(7.24)$, $Pca'2_{1}'~(29.102)$, $Pm'a2'~(28.89)$, $P_{A}c~(7.31)$, $P_{C}2~(3.6)$, $P_{C}ba2~(32.141)$, $P2_{1}~(4.7)$, $Pm'c'2_{1}~(26.70)$, $P_{C}2_{1}~(4.12)$, $P_{B}ca2_{1}~(29.108)$.\\

\subsection{WP: $4g+4g$}
\textbf{BCS Materials:} {CeRh\textsubscript{2}Si\textsubscript{2}~(27 K)}\footnote{BCS web page: \texttt{\href{http://webbdcrista1.ehu.es/magndata/index.php?this\_label=2.30} {http://webbdcrista1.ehu.es/magndata/index.php?this\_label=2.30}}}.\\
\subsubsection{Topological bands in subgroup $P2~(3.1)$}
\textbf{Perturbation:}
\begin{itemize}
\item E $\parallel$ [010] and B $\parallel$ [010] and strain $\perp$ [010].
\end{itemize}
\begin{figure}[H]
\centering
\includegraphics[scale=0.6]{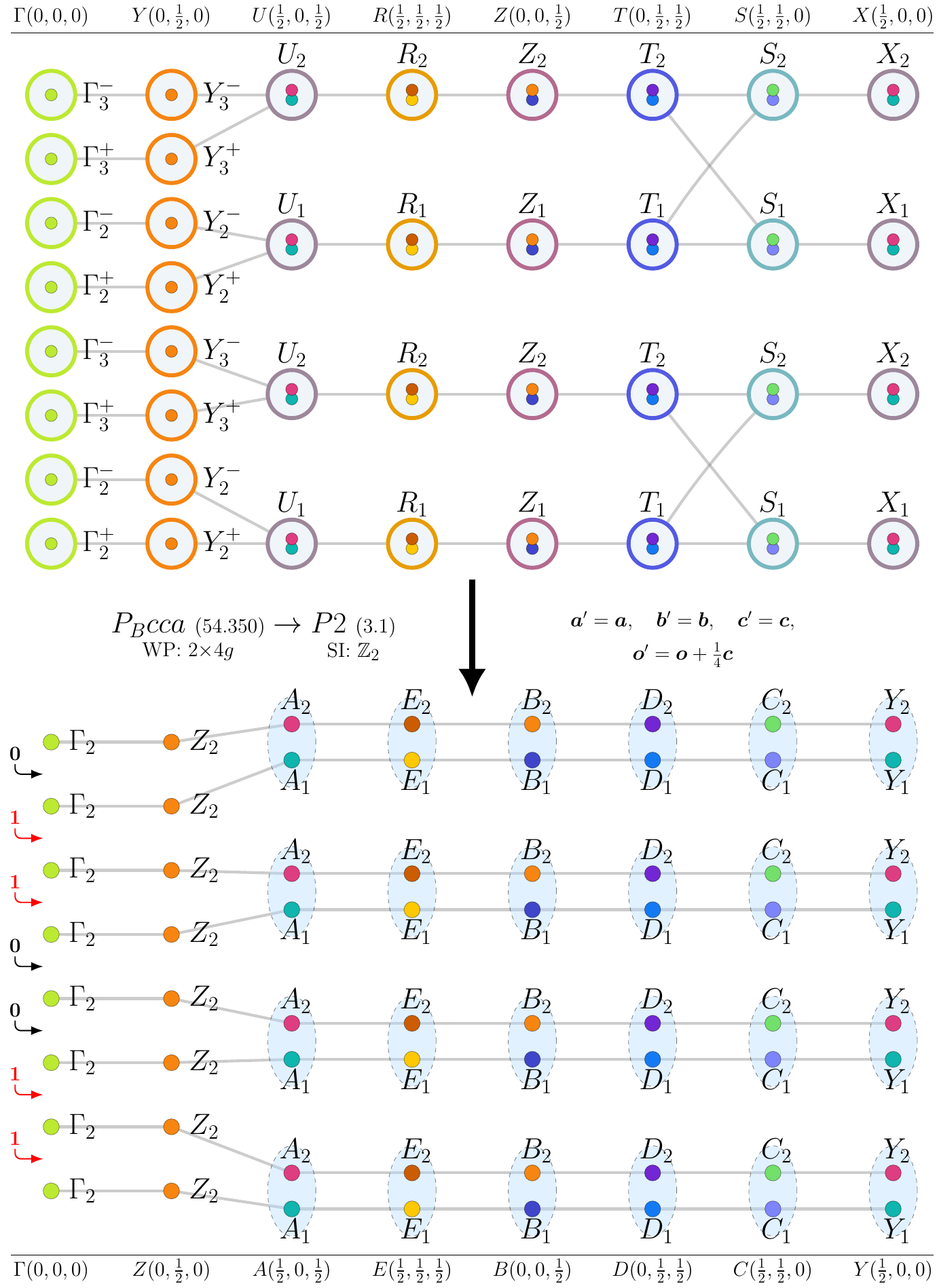}
\caption{Topological magnon bands in subgroup $P2~(3.1)$ for magnetic moments on Wyckoff positions $4g+4g$ of supergroup $P_{B}cca~(54.350)$.\label{fig_54.350_3.1_Eparallel010andBparallel010andstrainperp010_4g+4g}}
\end{figure}
\input{gap_tables_tex/54.350_3.1_Eparallel010andBparallel010andstrainperp010_4g+4g_table.tex}
\input{si_tables_tex/54.350_3.1_Eparallel010andBparallel010andstrainperp010_4g+4g_table.tex}
\subsubsection{Topological bands in subgroup $Pm'm'2~(25.60)$}
\textbf{Perturbation:}
\begin{itemize}
\item E $\parallel$ [010] and B $\parallel$ [010].
\end{itemize}
\begin{figure}[H]
\centering
\includegraphics[scale=0.6]{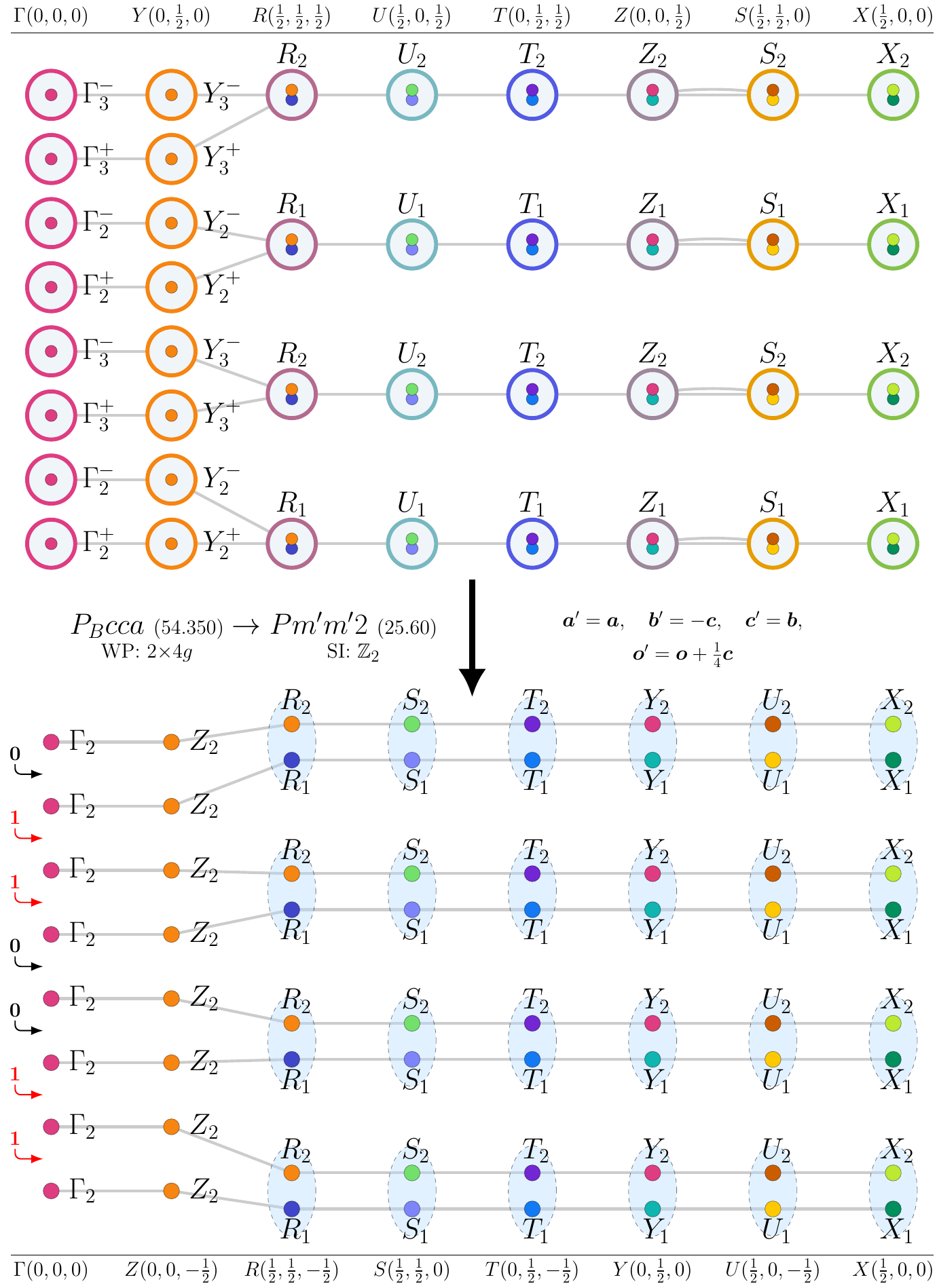}
\caption{Topological magnon bands in subgroup $Pm'm'2~(25.60)$ for magnetic moments on Wyckoff positions $4g+4g$ of supergroup $P_{B}cca~(54.350)$.\label{fig_54.350_25.60_Eparallel010andBparallel010_4g+4g}}
\end{figure}
\input{gap_tables_tex/54.350_25.60_Eparallel010andBparallel010_4g+4g_table.tex}
\input{si_tables_tex/54.350_25.60_Eparallel010andBparallel010_4g+4g_table.tex}
\subsubsection{Topological bands in subgroup $P_{a}2~(3.4)$}
\textbf{Perturbation:}
\begin{itemize}
\item E $\parallel$ [010] and strain $\perp$ [010].
\end{itemize}
\begin{figure}[H]
\centering
\includegraphics[scale=0.6]{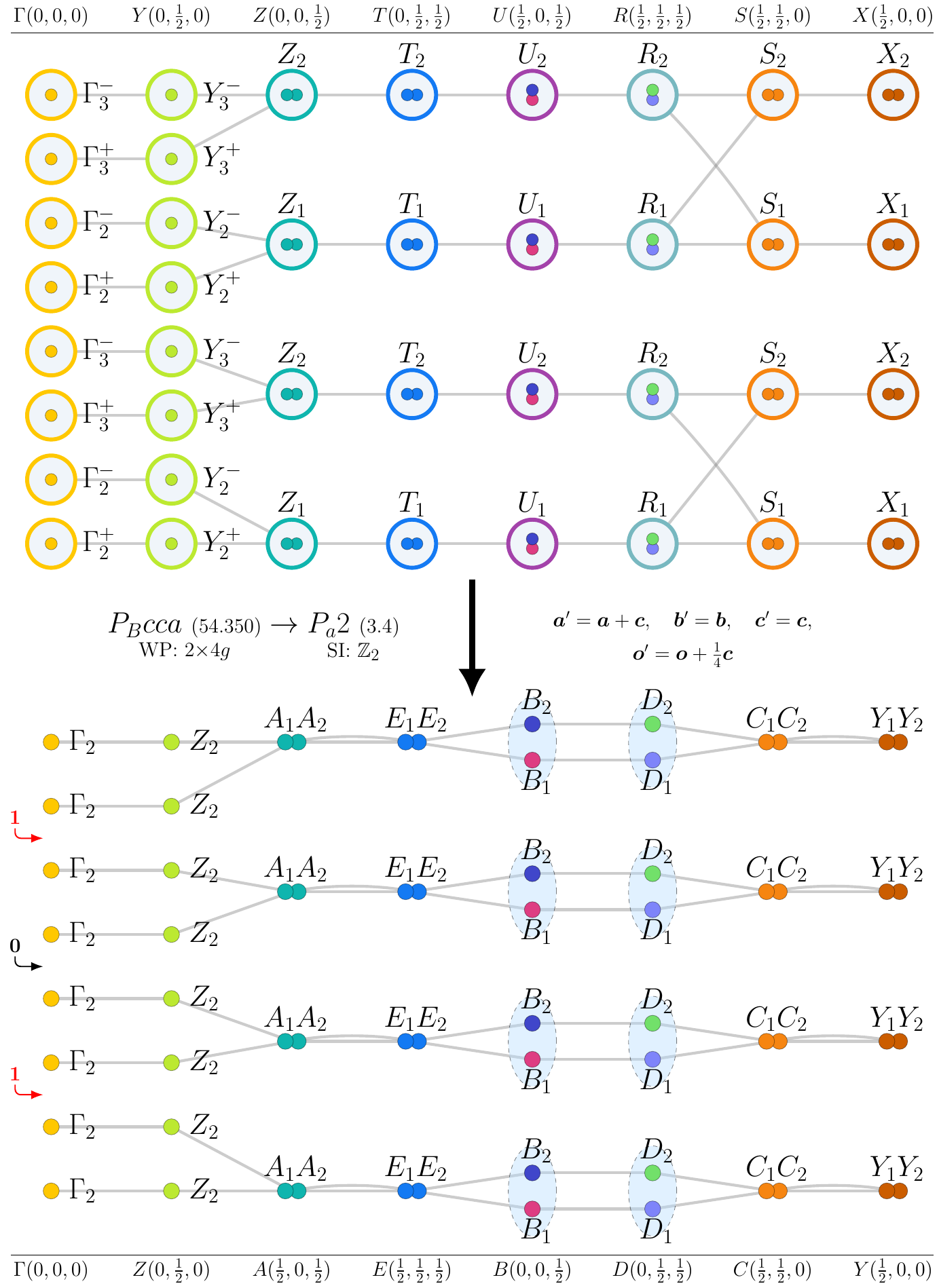}
\caption{Topological magnon bands in subgroup $P_{a}2~(3.4)$ for magnetic moments on Wyckoff positions $4g+4g$ of supergroup $P_{B}cca~(54.350)$.\label{fig_54.350_3.4_Eparallel010andstrainperp010_4g+4g}}
\end{figure}
\input{gap_tables_tex/54.350_3.4_Eparallel010andstrainperp010_4g+4g_table.tex}
\input{si_tables_tex/54.350_3.4_Eparallel010andstrainperp010_4g+4g_table.tex}
\subsection{WP: $4g$}
\textbf{BCS Materials:} {NdNiMg\textsubscript{15}~(9 K)}\footnote{BCS web page: \texttt{\href{http://webbdcrista1.ehu.es/magndata/index.php?this\_label=1.457} {http://webbdcrista1.ehu.es/magndata/index.php?this\_label=1.457}}}.\\
\subsubsection{Topological bands in subgroup $P2~(3.1)$}
\textbf{Perturbation:}
\begin{itemize}
\item E $\parallel$ [010] and B $\parallel$ [010] and strain $\perp$ [010].
\end{itemize}
\begin{figure}[H]
\centering
\includegraphics[scale=0.6]{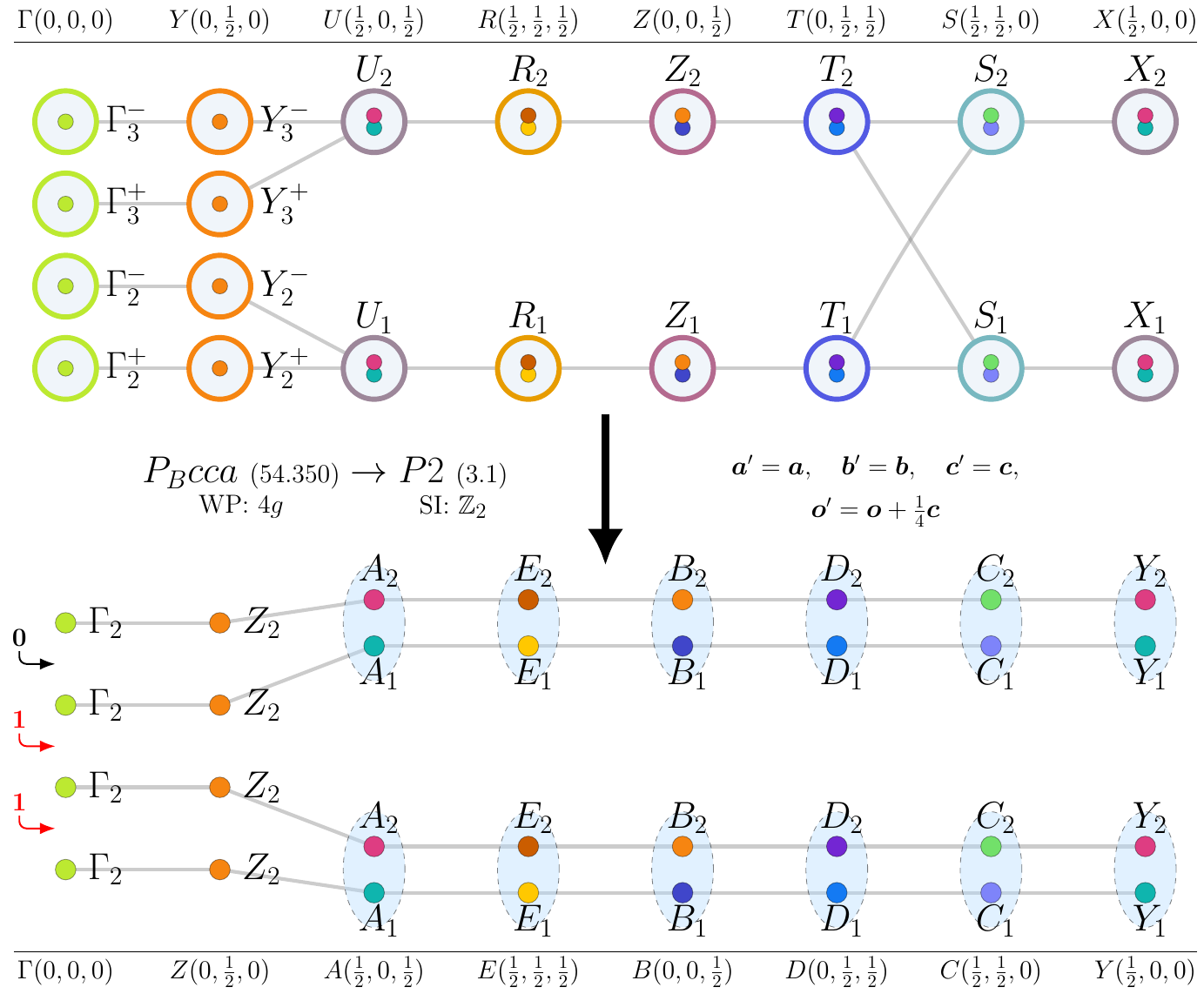}
\caption{Topological magnon bands in subgroup $P2~(3.1)$ for magnetic moments on Wyckoff position $4g$ of supergroup $P_{B}cca~(54.350)$.\label{fig_54.350_3.1_Eparallel010andBparallel010andstrainperp010_4g}}
\end{figure}
\input{gap_tables_tex/54.350_3.1_Eparallel010andBparallel010andstrainperp010_4g_table.tex}
\input{si_tables_tex/54.350_3.1_Eparallel010andBparallel010andstrainperp010_4g_table.tex}
\subsubsection{Topological bands in subgroup $Pm'm'2~(25.60)$}
\textbf{Perturbation:}
\begin{itemize}
\item E $\parallel$ [010] and B $\parallel$ [010].
\end{itemize}
\begin{figure}[H]
\centering
\includegraphics[scale=0.6]{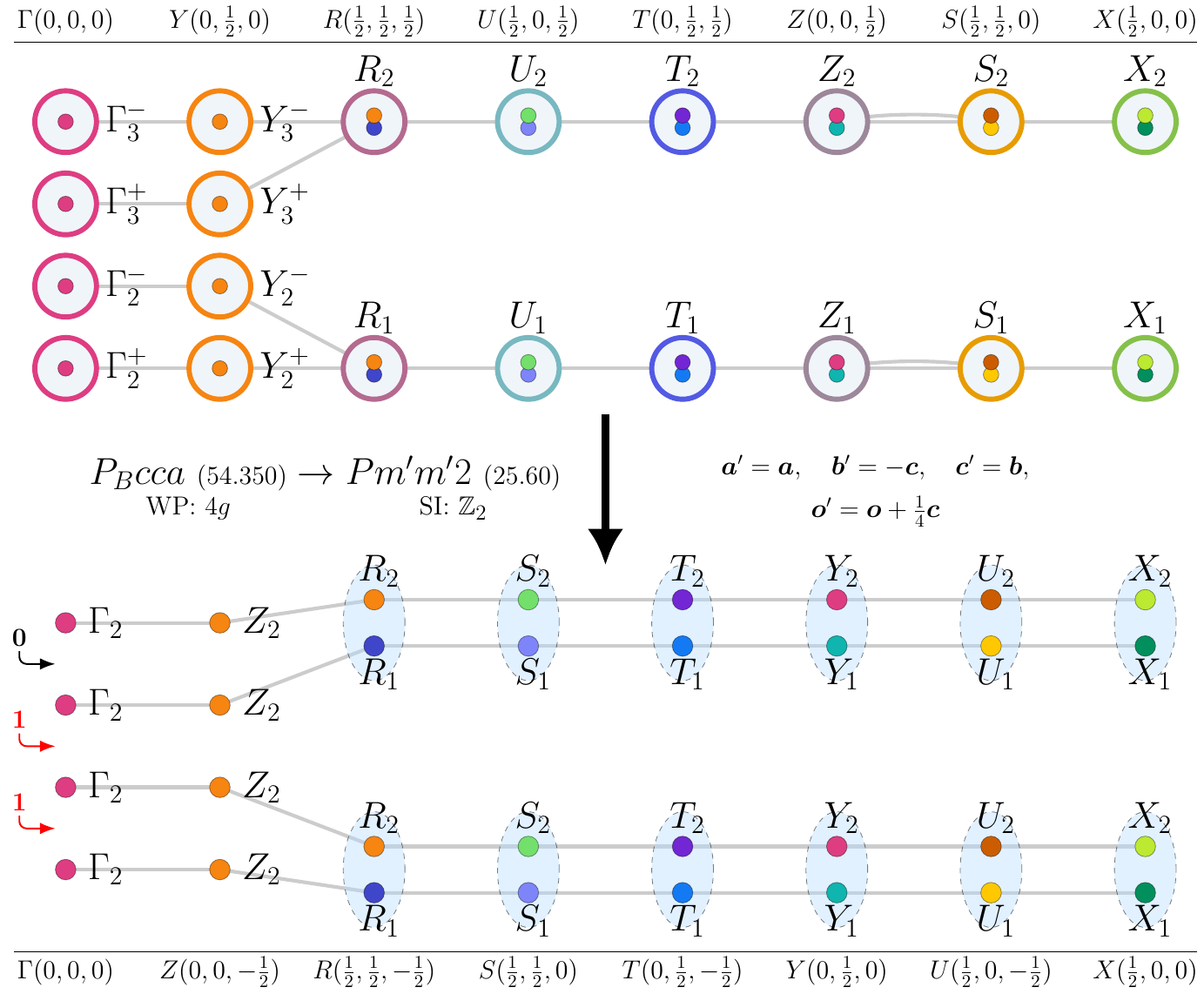}
\caption{Topological magnon bands in subgroup $Pm'm'2~(25.60)$ for magnetic moments on Wyckoff position $4g$ of supergroup $P_{B}cca~(54.350)$.\label{fig_54.350_25.60_Eparallel010andBparallel010_4g}}
\end{figure}
\input{gap_tables_tex/54.350_25.60_Eparallel010andBparallel010_4g_table.tex}
\input{si_tables_tex/54.350_25.60_Eparallel010andBparallel010_4g_table.tex}
\subsubsection{Topological bands in subgroup $P_{a}2~(3.4)$}
\textbf{Perturbation:}
\begin{itemize}
\item E $\parallel$ [010] and strain $\perp$ [010].
\end{itemize}
\begin{figure}[H]
\centering
\includegraphics[scale=0.6]{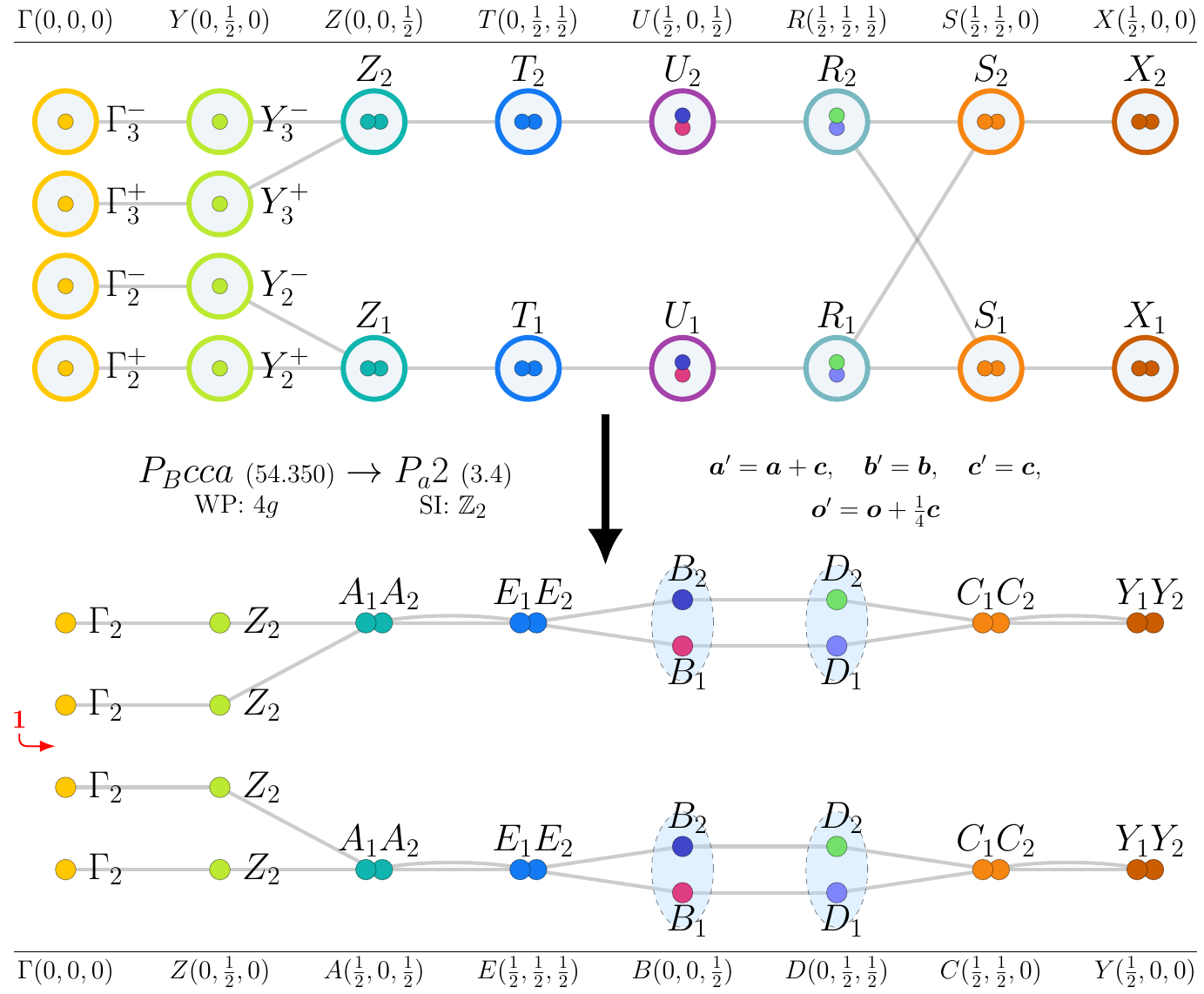}
\caption{Topological magnon bands in subgroup $P_{a}2~(3.4)$ for magnetic moments on Wyckoff position $4g$ of supergroup $P_{B}cca~(54.350)$.\label{fig_54.350_3.4_Eparallel010andstrainperp010_4g}}
\end{figure}
\input{gap_tables_tex/54.350_3.4_Eparallel010andstrainperp010_4g_table.tex}
\input{si_tables_tex/54.350_3.4_Eparallel010andstrainperp010_4g_table.tex}

\section{MSG $P_{I}cca~(54.352)$}
\textbf{Nontrivial-SI Subgroups:} $P\bar{1}~(2.4)$, $P2_{1}'/c'~(14.79)$, $P2_{1}'/c'~(14.79)$, $P2'/c'~(13.69)$, $P_{S}\bar{1}~(2.7)$, $P2~(3.1)$, $Pb'a'2~(32.138)$, $P_{I}cc2~(27.86)$, $P2/c~(13.65)$, $Pc'ca'~(54.344)$, $P_{C}2/c~(13.74)$, $P2~(3.1)$, $Pc'c'2~(27.81)$, $P2/c~(13.65)$, $Pc'c'a~(54.342)$, $P_{C}2/c~(13.74)$, $P2_{1}/c~(14.75)$, $Pb'c'a~(61.436)$, $P_{C}2_{1}/c~(14.84)$.\\

\textbf{Trivial-SI Subgroups:} $Pc'~(7.26)$, $Pc'~(7.26)$, $Pc'~(7.26)$, $P2_{1}'~(4.9)$, $P2_{1}'~(4.9)$, $P2'~(3.3)$, $P_{S}1~(1.3)$, $Pc~(7.24)$, $Pc'a2_{1}'~(29.101)$, $Pc'c2'~(27.80)$, $P_{C}c~(7.30)$, $Pc~(7.24)$, $Pca'2_{1}'~(29.102)$, $Pb'a2'~(32.137)$, $P_{C}c~(7.30)$, $Pc~(7.24)$, $Pca'2_{1}'~(29.102)$, $Pc'a2_{1}'~(29.101)$, $P_{C}c~(7.30)$, $P_{C}2~(3.6)$, $P_{C}2~(3.6)$, $P_{I}ba2~(32.143)$, $P2_{1}~(4.7)$, $Pc'a'2_{1}~(29.103)$, $P_{C}2_{1}~(4.12)$, $P_{I}ca2_{1}~(29.110)$.\\

\subsection{WP: $8c$}
\textbf{BCS Materials:} {Sr\textsubscript{2}IrO\textsubscript{4}~(240 K)}\footnote{BCS web page: \texttt{\href{http://webbdcrista1.ehu.es/magndata/index.php?this\_label=1.3} {http://webbdcrista1.ehu.es/magndata/index.php?this\_label=1.3}}}, {Sr\textsubscript{2}IrO\textsubscript{4}~(224 K)}\footnote{BCS web page: \texttt{\href{http://webbdcrista1.ehu.es/magndata/index.php?this\_label=1.77} {http://webbdcrista1.ehu.es/magndata/index.php?this\_label=1.77}}}.\\
\subsubsection{Topological bands in subgroup $P2_{1}'/c'~(14.79)$}
\textbf{Perturbations:}
\begin{itemize}
\item B $\parallel$ [100] and strain $\perp$ [001],
\item B $\parallel$ [010] and strain $\perp$ [001],
\item B $\perp$ [001].
\end{itemize}
\begin{figure}[H]
\centering
\includegraphics[scale=0.6]{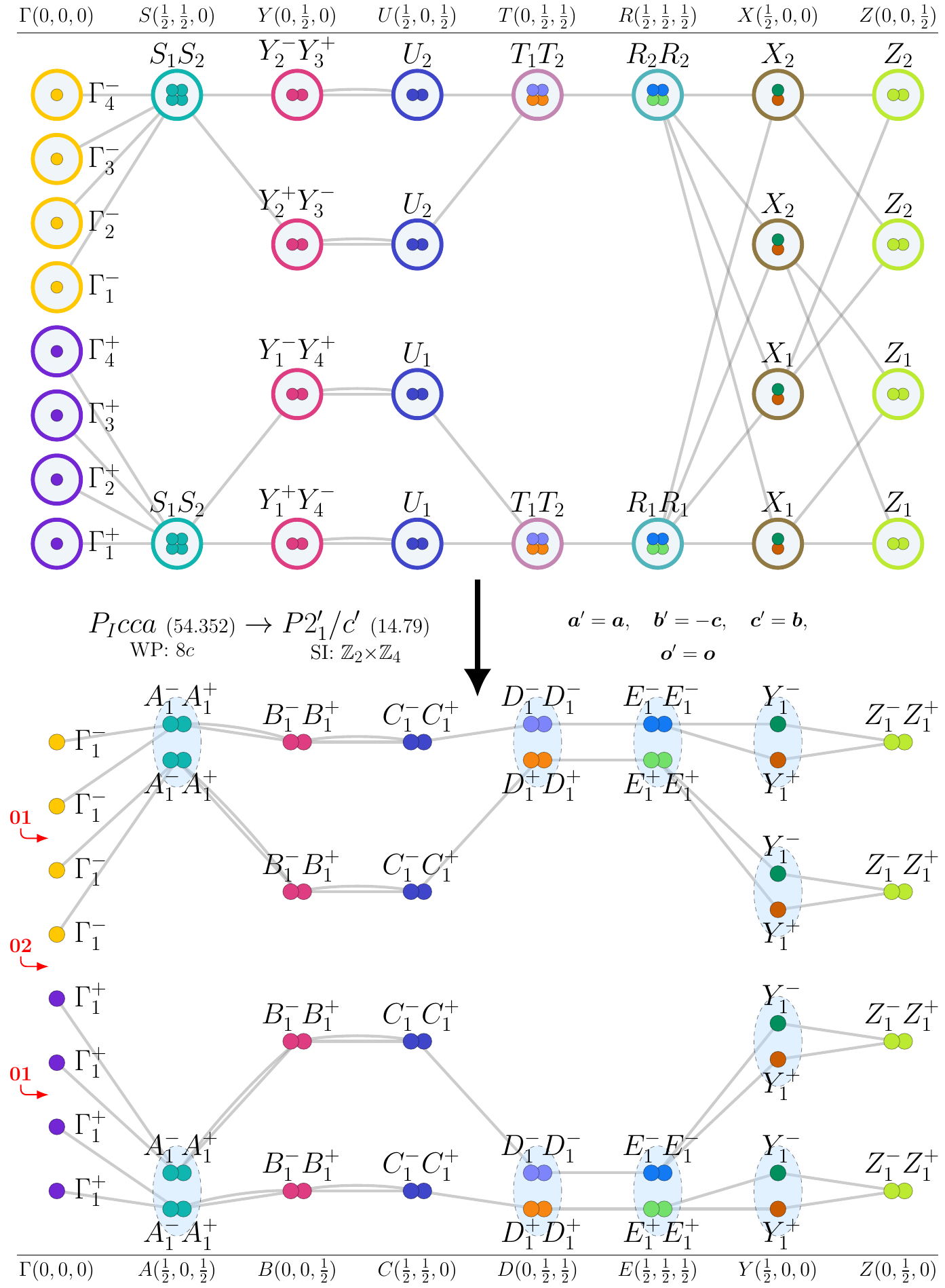}
\caption{Topological magnon bands in subgroup $P2_{1}'/c'~(14.79)$ for magnetic moments on Wyckoff position $8c$ of supergroup $P_{I}cca~(54.352)$.\label{fig_54.352_14.79_Bparallel100andstrainperp001_8c}}
\end{figure}
\input{gap_tables_tex/54.352_14.79_Bparallel100andstrainperp001_8c_table.tex}
\input{si_tables_tex/54.352_14.79_Bparallel100andstrainperp001_8c_table.tex}
\subsubsection{Topological bands in subgroup $P2_{1}'/c'~(14.79)$}
\textbf{Perturbations:}
\begin{itemize}
\item B $\parallel$ [100] and strain $\perp$ [010],
\item B $\parallel$ [001] and strain $\perp$ [010],
\item B $\perp$ [010].
\end{itemize}
\begin{figure}[H]
\centering
\includegraphics[scale=0.6]{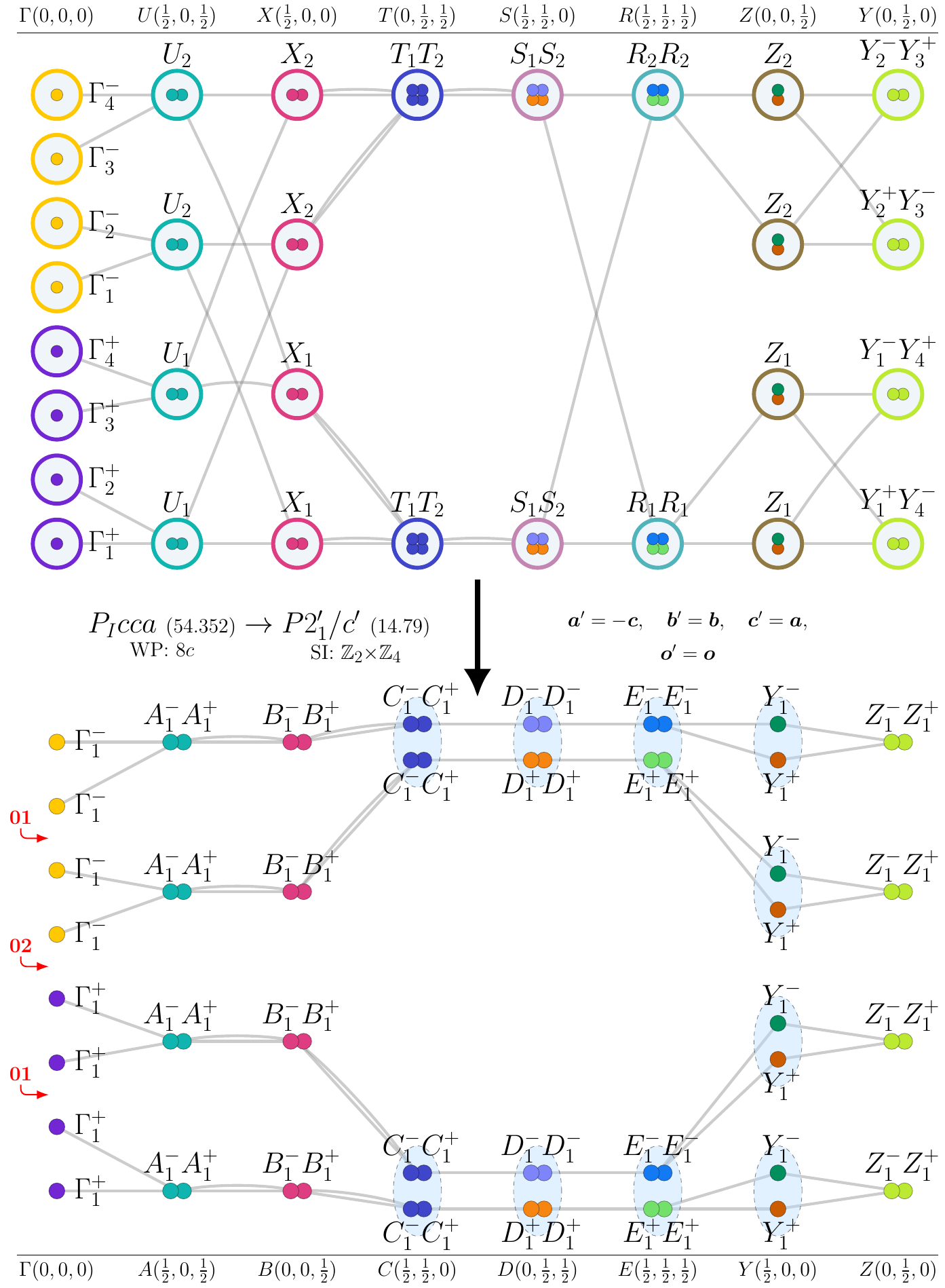}
\caption{Topological magnon bands in subgroup $P2_{1}'/c'~(14.79)$ for magnetic moments on Wyckoff position $8c$ of supergroup $P_{I}cca~(54.352)$.\label{fig_54.352_14.79_Bparallel100andstrainperp010_8c}}
\end{figure}
\input{gap_tables_tex/54.352_14.79_Bparallel100andstrainperp010_8c_table.tex}
\input{si_tables_tex/54.352_14.79_Bparallel100andstrainperp010_8c_table.tex}
\subsection{WP: $16f$}
\textbf{BCS Materials:} {BaCo\textsubscript{2}V\textsubscript{2}O\textsubscript{8}~(5 K)}\footnote{BCS web page: \texttt{\href{http://webbdcrista1.ehu.es/magndata/index.php?this\_label=1.30} {http://webbdcrista1.ehu.es/magndata/index.php?this\_label=1.30}}}.\\
\subsubsection{Topological bands in subgroup $P2_{1}'/c'~(14.79)$}
\textbf{Perturbations:}
\begin{itemize}
\item B $\parallel$ [100] and strain $\perp$ [001],
\item B $\parallel$ [010] and strain $\perp$ [001],
\item B $\perp$ [001].
\end{itemize}
\begin{figure}[H]
\centering
\includegraphics[scale=0.6]{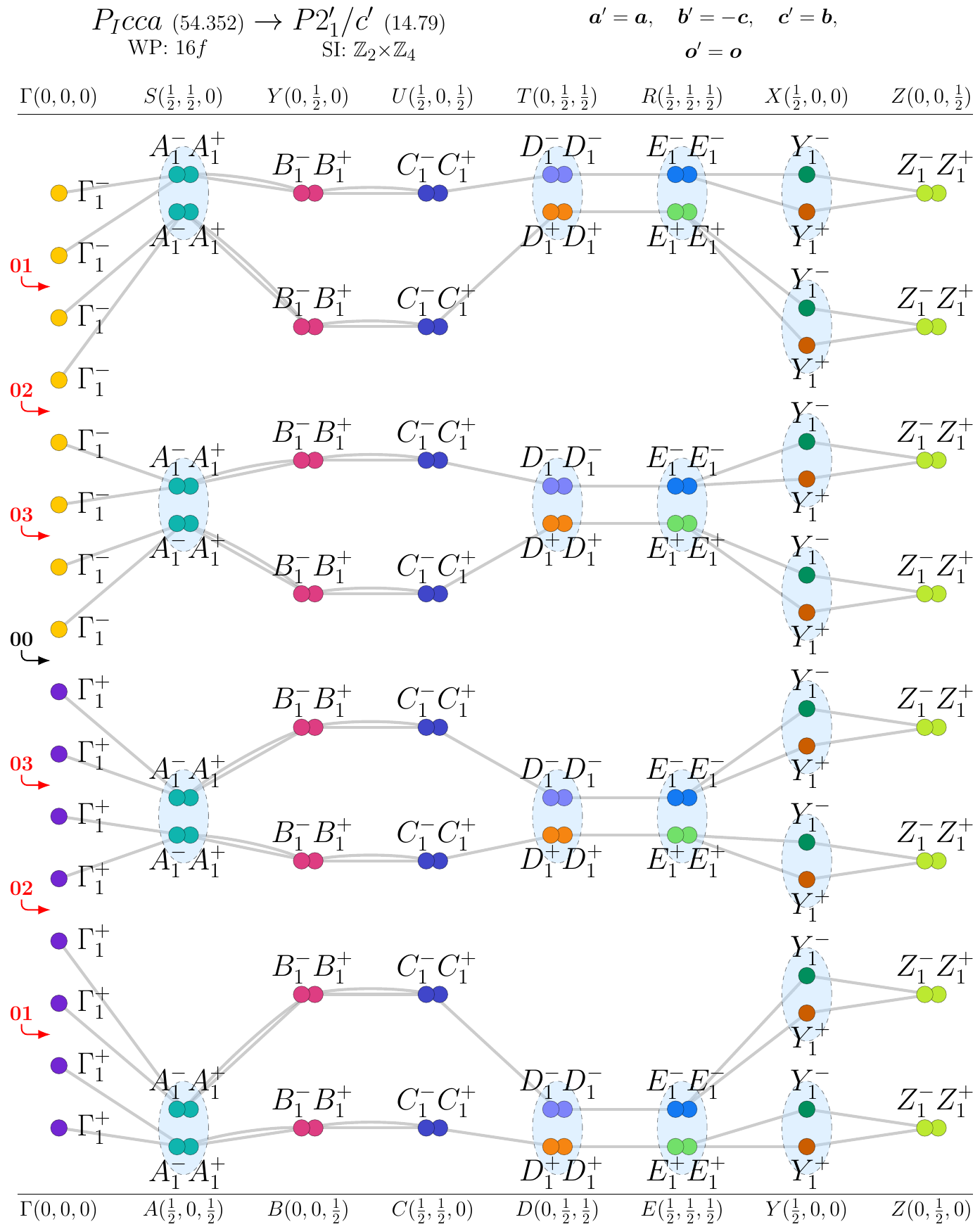}
\caption{Topological magnon bands in subgroup $P2_{1}'/c'~(14.79)$ for magnetic moments on Wyckoff position $16f$ of supergroup $P_{I}cca~(54.352)$.\label{fig_54.352_14.79_Bparallel100andstrainperp001_16f}}
\end{figure}
\input{gap_tables_tex/54.352_14.79_Bparallel100andstrainperp001_16f_table.tex}
\input{si_tables_tex/54.352_14.79_Bparallel100andstrainperp001_16f_table.tex}
\subsubsection{Topological bands in subgroup $P2_{1}'/c'~(14.79)$}
\textbf{Perturbations:}
\begin{itemize}
\item B $\parallel$ [100] and strain $\perp$ [010],
\item B $\parallel$ [001] and strain $\perp$ [010],
\item B $\perp$ [010].
\end{itemize}
\begin{figure}[H]
\centering
\includegraphics[scale=0.6]{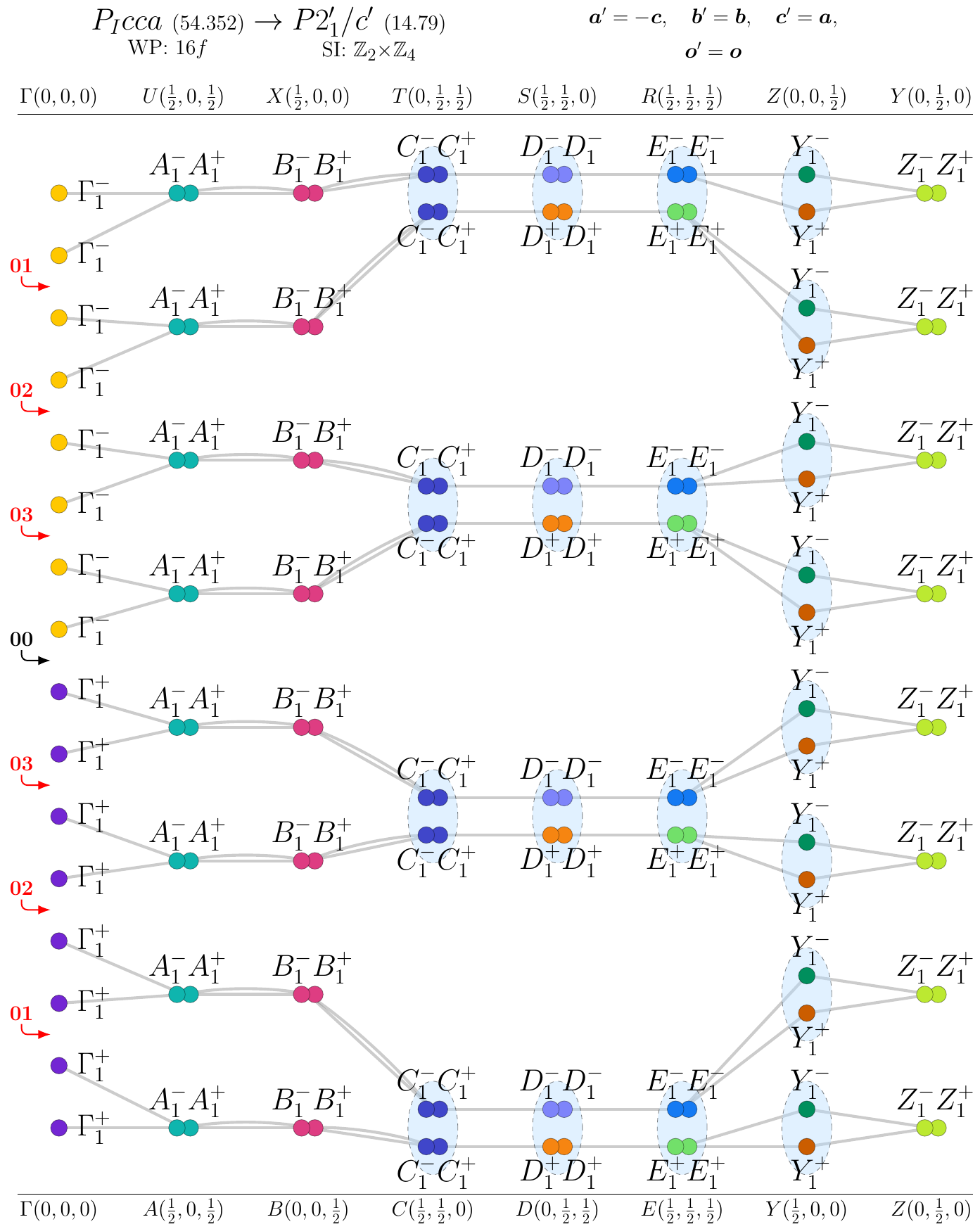}
\caption{Topological magnon bands in subgroup $P2_{1}'/c'~(14.79)$ for magnetic moments on Wyckoff position $16f$ of supergroup $P_{I}cca~(54.352)$.\label{fig_54.352_14.79_Bparallel100andstrainperp010_16f}}
\end{figure}
\input{gap_tables_tex/54.352_14.79_Bparallel100andstrainperp010_16f_table.tex}
\input{si_tables_tex/54.352_14.79_Bparallel100andstrainperp010_16f_table.tex}

\section{MSG $Pc'c'n~(56.369)$}
\textbf{Nontrivial-SI Subgroups:} $P\bar{1}~(2.4)$, $P2_{1}'/c'~(14.79)$, $P2_{1}'/c'~(14.79)$, $P2~(3.1)$, $Pc'c'2~(27.81)$, $P2/c~(13.65)$.\\

\textbf{Trivial-SI Subgroups:} $Pc'~(7.26)$, $Pc'~(7.26)$, $P2_{1}'~(4.9)$, $P2_{1}'~(4.9)$, $Pc~(7.24)$, $Pna'2_{1}'~(33.147)$, $Pna'2_{1}'~(33.147)$.\\

\subsection{WP: $4a$}
\textbf{BCS Materials:} {La\textsubscript{2}NiO\textsubscript{4}~(80 K)}\footnote{BCS web page: \texttt{\href{http://webbdcrista1.ehu.es/magndata/index.php?this\_label=0.45} {http://webbdcrista1.ehu.es/magndata/index.php?this\_label=0.45}}}.\\
\subsubsection{Topological bands in subgroup $P\bar{1}~(2.4)$}
\textbf{Perturbations:}
\begin{itemize}
\item strain in generic direction,
\item (B $\parallel$ [100] or B $\perp$ [010]) and strain $\perp$ [100],
\item (B $\parallel$ [100] or B $\perp$ [010]) and strain $\perp$ [001],
\item (B $\parallel$ [010] or B $\perp$ [100]) and strain $\perp$ [010],
\item (B $\parallel$ [010] or B $\perp$ [100]) and strain $\perp$ [001],
\item B in generic direction.
\end{itemize}
\begin{figure}[H]
\centering
\includegraphics[scale=0.6]{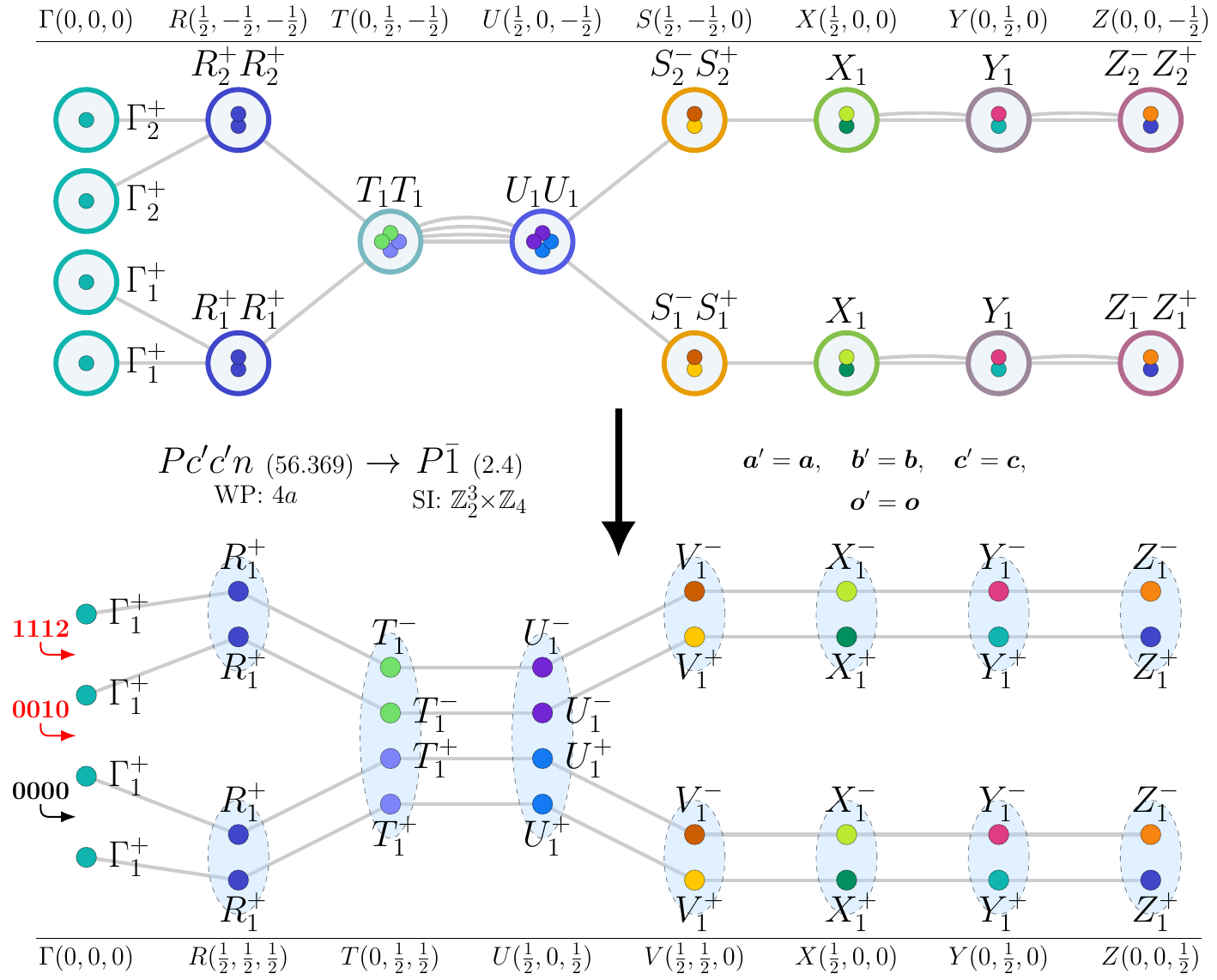}
\caption{Topological magnon bands in subgroup $P\bar{1}~(2.4)$ for magnetic moments on Wyckoff position $4a$ of supergroup $Pc'c'n~(56.369)$.\label{fig_56.369_2.4_strainingenericdirection_4a}}
\end{figure}
\input{gap_tables_tex/56.369_2.4_strainingenericdirection_4a_table.tex}
\input{si_tables_tex/56.369_2.4_strainingenericdirection_4a_table.tex}
\subsubsection{Topological bands in subgroup $P2_{1}'/c'~(14.79)$}
\textbf{Perturbations:}
\begin{itemize}
\item strain $\perp$ [010],
\item (B $\parallel$ [100] or B $\perp$ [010]).
\end{itemize}
\begin{figure}[H]
\centering
\includegraphics[scale=0.6]{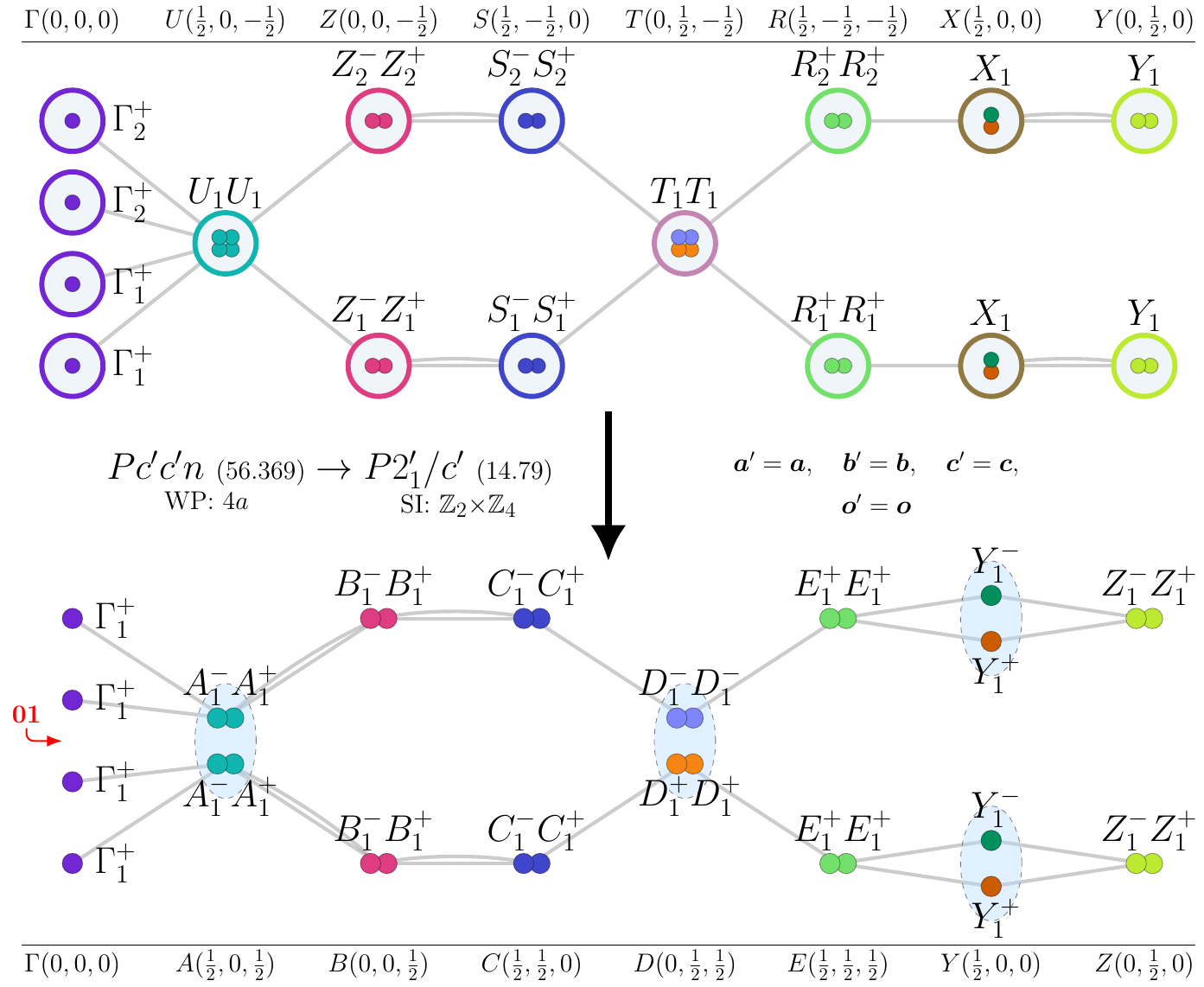}
\caption{Topological magnon bands in subgroup $P2_{1}'/c'~(14.79)$ for magnetic moments on Wyckoff position $4a$ of supergroup $Pc'c'n~(56.369)$.\label{fig_56.369_14.79_strainperp010_4a}}
\end{figure}
\input{gap_tables_tex/56.369_14.79_strainperp010_4a_table.tex}
\input{si_tables_tex/56.369_14.79_strainperp010_4a_table.tex}
\subsubsection{Topological bands in subgroup $P2_{1}'/c'~(14.79)$}
\textbf{Perturbations:}
\begin{itemize}
\item strain $\perp$ [100],
\item (B $\parallel$ [010] or B $\perp$ [100]).
\end{itemize}
\begin{figure}[H]
\centering
\includegraphics[scale=0.6]{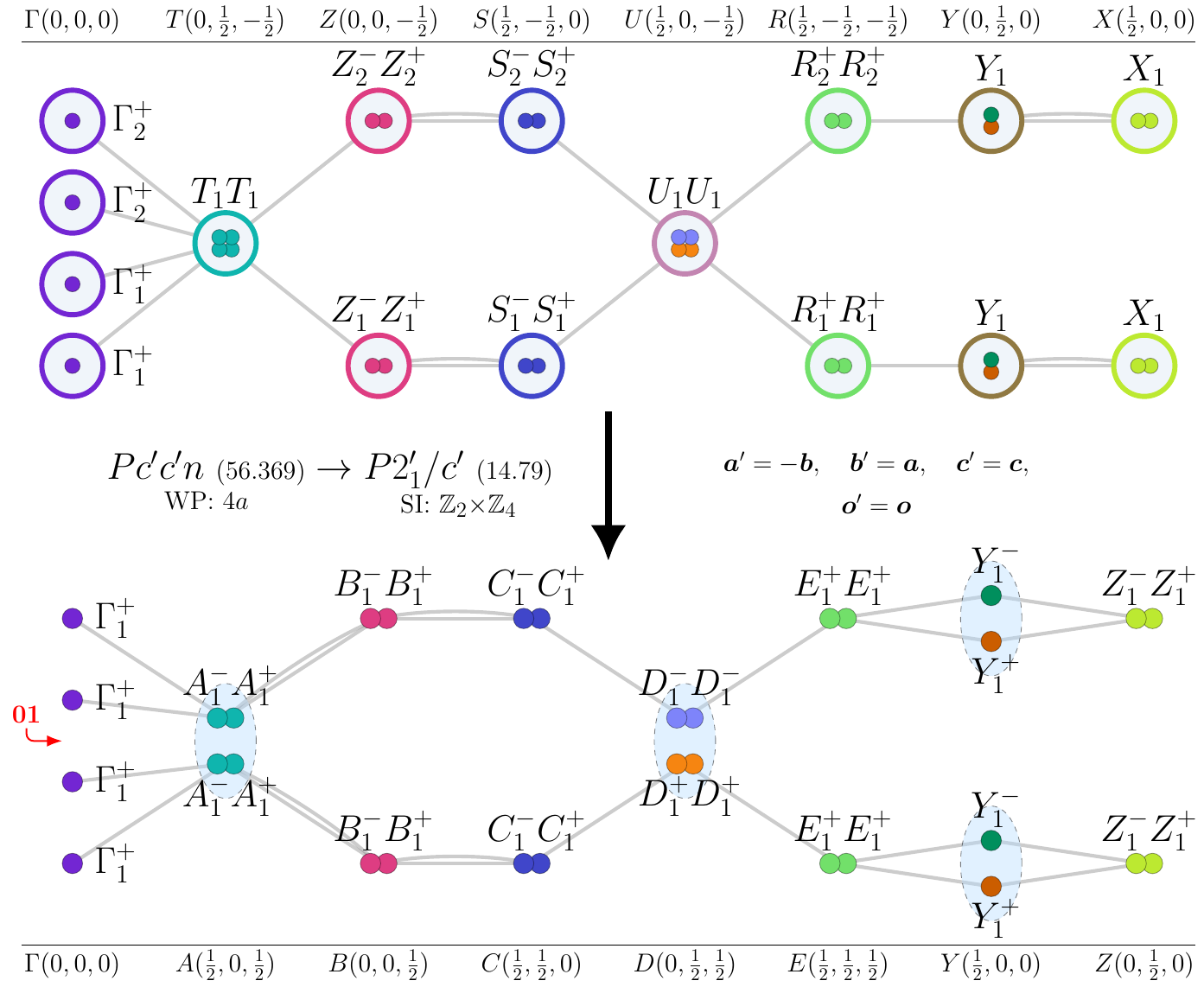}
\caption{Topological magnon bands in subgroup $P2_{1}'/c'~(14.79)$ for magnetic moments on Wyckoff position $4a$ of supergroup $Pc'c'n~(56.369)$.\label{fig_56.369_14.79_strainperp100_4a}}
\end{figure}
\input{gap_tables_tex/56.369_14.79_strainperp100_4a_table.tex}
\input{si_tables_tex/56.369_14.79_strainperp100_4a_table.tex}

\section{MSG $P_{b}ccn~(56.372)$}
\textbf{Nontrivial-SI Subgroups:} $P\bar{1}~(2.4)$, $P2'/c'~(13.69)$, $P2'/c'~(13.69)$, $P2_{1}'/c'~(14.79)$, $P_{S}\bar{1}~(2.7)$, $P2~(3.1)$, $P_{a}2~(3.4)$, $P_{a}cc2~(27.83)$, $P2/c~(13.65)$, $Pnn'a'~(52.311)$, $P_{a}2/c~(13.70)$, $P2_{1}/c~(14.75)$, $Pbc'n'~(60.423)$, $P_{b}2_{1}/c~(14.81)$, $P2_{1}/c~(14.75)$, $Pcc'a'~(54.343)$, $P_{a}2_{1}/c~(14.80)$.\\

\textbf{Trivial-SI Subgroups:} $Pc'~(7.26)$, $Pc'~(7.26)$, $Pc'~(7.26)$, $P2'~(3.3)$, $P2'~(3.3)$, $P2_{1}'~(4.9)$, $P_{S}1~(1.3)$, $Pc~(7.24)$, $Pn'n2'~(34.158)$, $Pna'2_{1}'~(33.147)$, $P_{a}c~(7.27)$, $Pc~(7.24)$, $Pn'c2'~(30.113)$, $Pc'a2_{1}'~(29.101)$, $P_{b}c~(7.29)$, $Pc~(7.24)$, $Pc'c2'~(27.80)$, $Pb'a2'~(32.137)$, $P_{a}c~(7.27)$, $Pn'c'2~(30.115)$, $P2_{1}~(4.7)$, $Pn'a'2_{1}~(33.148)$, $P_{b}2_{1}~(4.11)$, $P_{c}na2_{1}~(33.151)$, $P2_{1}~(4.7)$, $Pc'a'2_{1}~(29.103)$, $P_{a}2_{1}~(4.10)$, $P_{b}na2_{1}~(33.150)$.\\

\subsection{WP: $8c$}
\textbf{BCS Materials:} {CsCoCl\textsubscript{3}(D\textsubscript{2}O)\textsubscript{2}~(3.4 K)}\footnote{BCS web page: \texttt{\href{http://webbdcrista1.ehu.es/magndata/index.php?this\_label=1.99} {http://webbdcrista1.ehu.es/magndata/index.php?this\_label=1.99}}}.\\
\subsubsection{Topological bands in subgroup $P2_{1}'/c'~(14.79)$}
\textbf{Perturbations:}
\begin{itemize}
\item B $\parallel$ [010] and strain $\perp$ [100],
\item B $\parallel$ [001] and strain $\perp$ [100],
\item B $\perp$ [100].
\end{itemize}
\begin{figure}[H]
\centering
\includegraphics[scale=0.6]{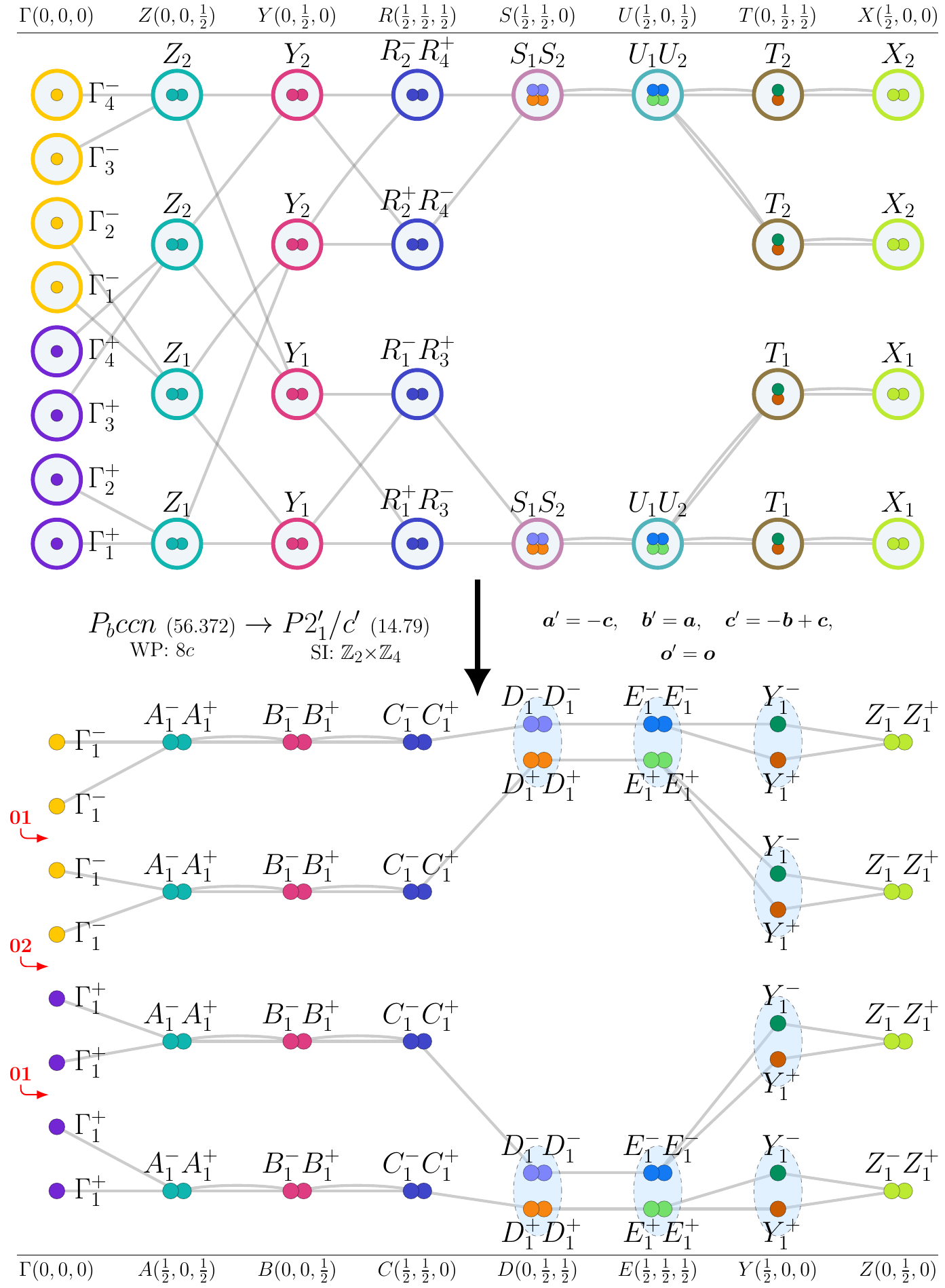}
\caption{Topological magnon bands in subgroup $P2_{1}'/c'~(14.79)$ for magnetic moments on Wyckoff position $8c$ of supergroup $P_{b}ccn~(56.372)$.\label{fig_56.372_14.79_Bparallel010andstrainperp100_8c}}
\end{figure}
\input{gap_tables_tex/56.372_14.79_Bparallel010andstrainperp100_8c_table.tex}
\input{si_tables_tex/56.372_14.79_Bparallel010andstrainperp100_8c_table.tex}

\section{MSG $P_{c}ccn~(56.373)$}
\textbf{Nontrivial-SI Subgroups:} $P\bar{1}~(2.4)$, $P2_{1}'/c'~(14.79)$, $P2_{1}'/m'~(11.54)$, $P2_{1}'/m'~(11.54)$, $P_{S}\bar{1}~(2.7)$, $P2~(3.1)$, $Pm'm'2~(25.60)$, $P_{b}2~(3.5)$, $P_{c}cc2~(27.82)$, $P2/c~(13.65)$, $Pm'm'n~(59.409)$, $P_{b}2/c~(13.71)$, $P2_{1}/c~(14.75)$, $Pn'm'a~(62.446)$, $P_{c}2_{1}/c~(14.82)$, $P2_{1}/c~(14.75)$, $Pn'm'a~(62.446)$, $P_{c}2_{1}/c~(14.82)$.\\

\textbf{Trivial-SI Subgroups:} $Pc'~(7.26)$, $Pm'~(6.20)$, $Pm'~(6.20)$, $P2_{1}'~(4.9)$, $P2_{1}'~(4.9)$, $P2_{1}'~(4.9)$, $P_{S}1~(1.3)$, $Pc~(7.24)$, $Pm'n2_{1}'~(31.125)$, $Pm'n2_{1}'~(31.125)$, $P_{b}c~(7.29)$, $Pc~(7.24)$, $Pm'c2_{1}'~(26.68)$, $Pn'a2_{1}'~(33.146)$, $P_{c}c~(7.28)$, $Pc~(7.24)$, $Pm'c2_{1}'~(26.68)$, $Pn'a2_{1}'~(33.146)$, $P_{c}c~(7.28)$, $P2_{1}~(4.7)$, $Pm'n'2_{1}~(31.127)$, $P_{a}2_{1}~(4.10)$, $P_{a}na2_{1}~(33.149)$, $P2_{1}~(4.7)$, $Pm'n'2_{1}~(31.127)$, $P_{a}2_{1}~(4.10)$, $P_{a}na2_{1}~(33.149)$.\\

\subsection{WP: $4a+8c$}
\textbf{BCS Materials:} {Cu\textsubscript{3}Y(SeO\textsubscript{3})\textsubscript{2}O\textsubscript{2}Cl~(35 K)}\footnote{BCS web page: \texttt{\href{http://webbdcrista1.ehu.es/magndata/index.php?this\_label=1.123} {http://webbdcrista1.ehu.es/magndata/index.php?this\_label=1.123}}}, {Cu\textsubscript{3}Bi(SeO\textsubscript{3})\textsubscript{2}O\textsubscript{2}Br~(27.4 K)}\footnote{BCS web page: \texttt{\href{http://webbdcrista1.ehu.es/magndata/index.php?this\_label=1.122} {http://webbdcrista1.ehu.es/magndata/index.php?this\_label=1.122}}}.\\
\subsubsection{Topological bands in subgroup $P2_{1}'/c'~(14.79)$}
\textbf{Perturbations:}
\begin{itemize}
\item B $\parallel$ [100] and strain $\perp$ [001],
\item B $\parallel$ [010] and strain $\perp$ [001],
\item B $\perp$ [001].
\end{itemize}
\begin{figure}[H]
\centering
\includegraphics[scale=0.6]{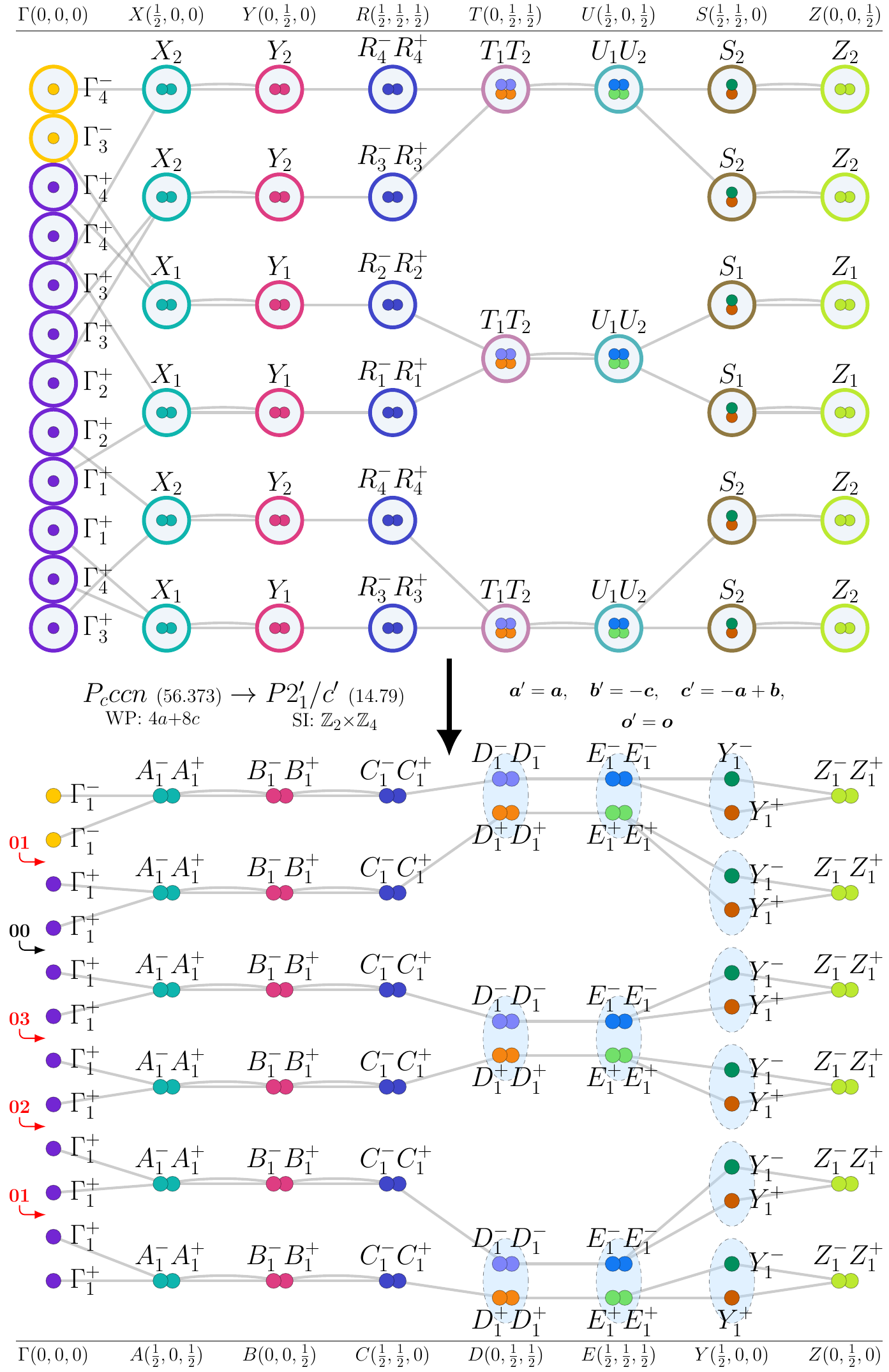}
\caption{Topological magnon bands in subgroup $P2_{1}'/c'~(14.79)$ for magnetic moments on Wyckoff positions $4a+8c$ of supergroup $P_{c}ccn~(56.373)$.\label{fig_56.373_14.79_Bparallel100andstrainperp001_4a+8c}}
\end{figure}
\input{gap_tables_tex/56.373_14.79_Bparallel100andstrainperp001_4a+8c_table.tex}
\input{si_tables_tex/56.373_14.79_Bparallel100andstrainperp001_4a+8c_table.tex}
\subsubsection{Topological bands in subgroup $P2_{1}'/m'~(11.54)$}
\textbf{Perturbations:}
\begin{itemize}
\item B $\parallel$ [100] and strain $\perp$ [010],
\item B $\parallel$ [001] and strain $\perp$ [010],
\item B $\perp$ [010].
\end{itemize}
\begin{figure}[H]
\centering
\includegraphics[scale=0.6]{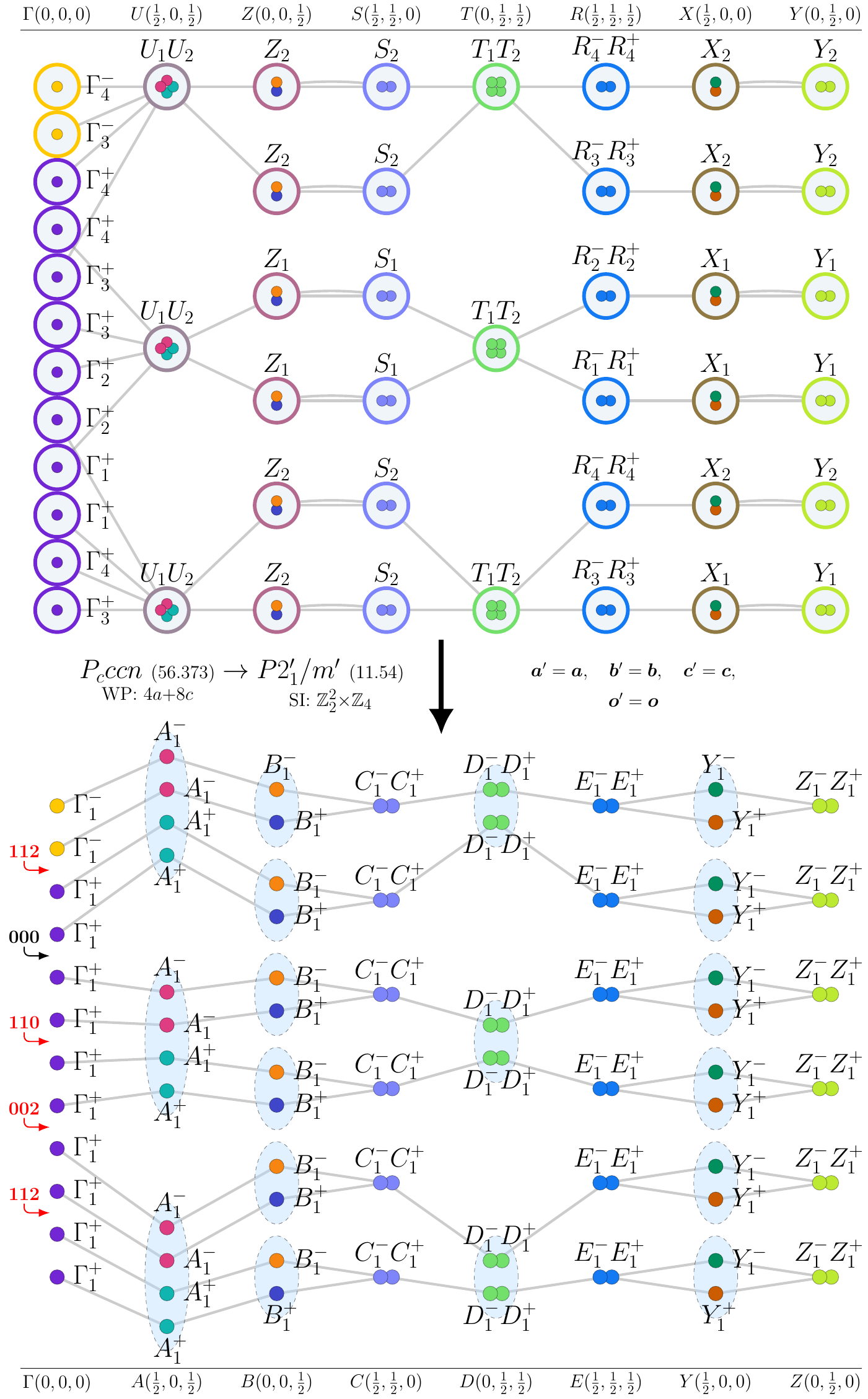}
\caption{Topological magnon bands in subgroup $P2_{1}'/m'~(11.54)$ for magnetic moments on Wyckoff positions $4a+8c$ of supergroup $P_{c}ccn~(56.373)$.\label{fig_56.373_11.54_Bparallel100andstrainperp010_4a+8c}}
\end{figure}
\input{gap_tables_tex/56.373_11.54_Bparallel100andstrainperp010_4a+8c_table.tex}
\input{si_tables_tex/56.373_11.54_Bparallel100andstrainperp010_4a+8c_table.tex}
\subsubsection{Topological bands in subgroup $P2_{1}'/m'~(11.54)$}
\textbf{Perturbations:}
\begin{itemize}
\item B $\parallel$ [010] and strain $\perp$ [100],
\item B $\parallel$ [001] and strain $\perp$ [100],
\item B $\perp$ [100].
\end{itemize}
\begin{figure}[H]
\centering
\includegraphics[scale=0.6]{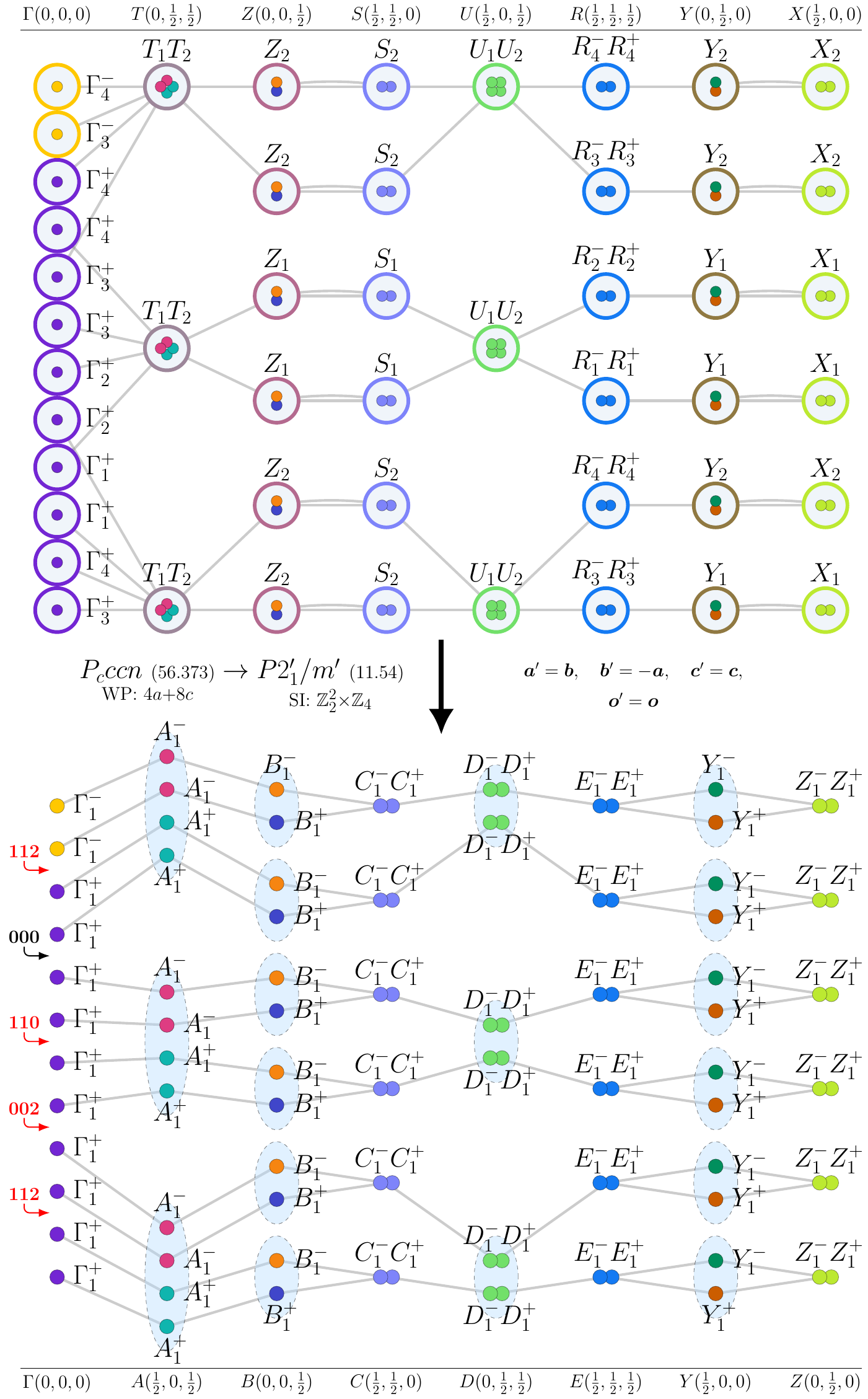}
\caption{Topological magnon bands in subgroup $P2_{1}'/m'~(11.54)$ for magnetic moments on Wyckoff positions $4a+8c$ of supergroup $P_{c}ccn~(56.373)$.\label{fig_56.373_11.54_Bparallel010andstrainperp100_4a+8c}}
\end{figure}
\input{gap_tables_tex/56.373_11.54_Bparallel010andstrainperp100_4a+8c_table.tex}
\input{si_tables_tex/56.373_11.54_Bparallel010andstrainperp100_4a+8c_table.tex}
\subsubsection{Topological bands in subgroup $P_{S}\bar{1}~(2.7)$}
\textbf{Perturbation:}
\begin{itemize}
\item strain in generic direction.
\end{itemize}
\begin{figure}[H]
\centering
\includegraphics[scale=0.6]{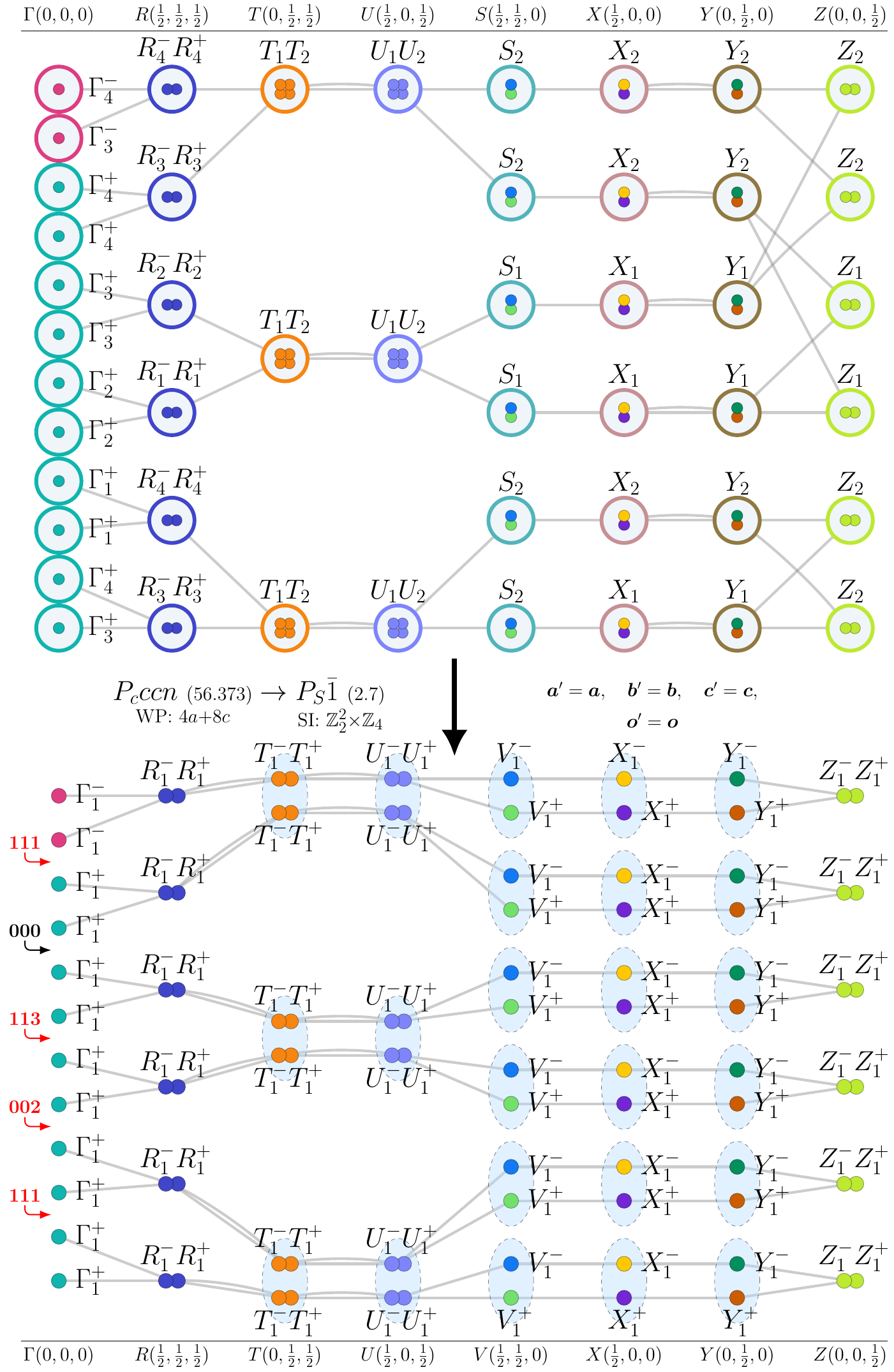}
\caption{Topological magnon bands in subgroup $P_{S}\bar{1}~(2.7)$ for magnetic moments on Wyckoff positions $4a+8c$ of supergroup $P_{c}ccn~(56.373)$.\label{fig_56.373_2.7_strainingenericdirection_4a+8c}}
\end{figure}
\input{gap_tables_tex/56.373_2.7_strainingenericdirection_4a+8c_table.tex}
\input{si_tables_tex/56.373_2.7_strainingenericdirection_4a+8c_table.tex}

\section{MSG $P_{A}ccn~(56.374)$}
\textbf{Nontrivial-SI Subgroups:} $P\bar{1}~(2.4)$, $P2_{1}'/c'~(14.79)$, $P2'/m'~(10.46)$, $P2_{1}'/c'~(14.79)$, $P_{S}\bar{1}~(2.7)$, $P2~(3.1)$, $Pm'a'2~(28.91)$, $P_{A}cc2~(27.85)$, $P2/c~(13.65)$, $Pm'na'~(53.328)$, $P_{C}2/c~(13.74)$, $P2_{1}/c~(14.75)$, $Pb'c'a~(61.436)$, $P_{A}2_{1}/c~(14.83)$, $P2_{1}/c~(14.75)$, $Pb'am'~(55.358)$, $P_{a}2_{1}/c~(14.80)$.\\

\textbf{Trivial-SI Subgroups:} $Pc'~(7.26)$, $Pm'~(6.20)$, $Pc'~(7.26)$, $P2_{1}'~(4.9)$, $P2'~(3.3)$, $P2_{1}'~(4.9)$, $P_{S}1~(1.3)$, $Pc~(7.24)$, $Pnc'2'~(30.114)$, $Pm'n2_{1}'~(31.125)$, $P_{C}c~(7.30)$, $Pc~(7.24)$, $Pca'2_{1}'~(29.102)$, $Pc'a2_{1}'~(29.101)$, $P_{A}c~(7.31)$, $Pc~(7.24)$, $Pm'c2_{1}'~(26.68)$, $Pb'a2'~(32.137)$, $P_{a}c~(7.27)$, $P_{C}2~(3.6)$, $P2_{1}~(4.7)$, $Pc'a'2_{1}~(29.103)$, $P_{C}2_{1}~(4.12)$, $P_{B}na2_{1}~(33.153)$, $P2_{1}~(4.7)$, $Pm'c'2_{1}~(26.70)$, $P_{a}2_{1}~(4.10)$, $P_{C}na2_{1}~(33.154)$.\\

\subsection{WP: $4a$}
\textbf{BCS Materials:} {La\textsubscript{2}CuO\textsubscript{4}~(316 K)}\footnote{BCS web page: \texttt{\href{http://webbdcrista1.ehu.es/magndata/index.php?this\_label=1.405} {http://webbdcrista1.ehu.es/magndata/index.php?this\_label=1.405}}}, {La\textsubscript{2}CuO\textsubscript{4}~(50 K)}\footnote{BCS web page: \texttt{\href{http://webbdcrista1.ehu.es/magndata/index.php?this\_label=1.23} {http://webbdcrista1.ehu.es/magndata/index.php?this\_label=1.23}}}, {Gd\textsubscript{2}CuO\textsubscript{4}~(18 K)}\footnote{BCS web page: \texttt{\href{http://webbdcrista1.ehu.es/magndata/index.php?this\_label=1.105} {http://webbdcrista1.ehu.es/magndata/index.php?this\_label=1.105}}}.\\
\subsubsection{Topological bands in subgroup $P\bar{1}~(2.4)$}
\textbf{Perturbations:}
\begin{itemize}
\item B $\parallel$ [100] and strain in generic direction,
\item B $\parallel$ [010] and strain in generic direction,
\item B $\parallel$ [001] and strain in generic direction,
\item B $\perp$ [100] and strain $\perp$ [010],
\item B $\perp$ [100] and strain $\perp$ [001],
\item B $\perp$ [100] and strain in generic direction,
\item B $\perp$ [010] and strain $\perp$ [100],
\item B $\perp$ [010] and strain $\perp$ [001],
\item B $\perp$ [010] and strain in generic direction,
\item B $\perp$ [001] and strain $\perp$ [100],
\item B $\perp$ [001] and strain $\perp$ [010],
\item B $\perp$ [001] and strain in generic direction,
\item B in generic direction.
\end{itemize}
\begin{figure}[H]
\centering
\includegraphics[scale=0.6]{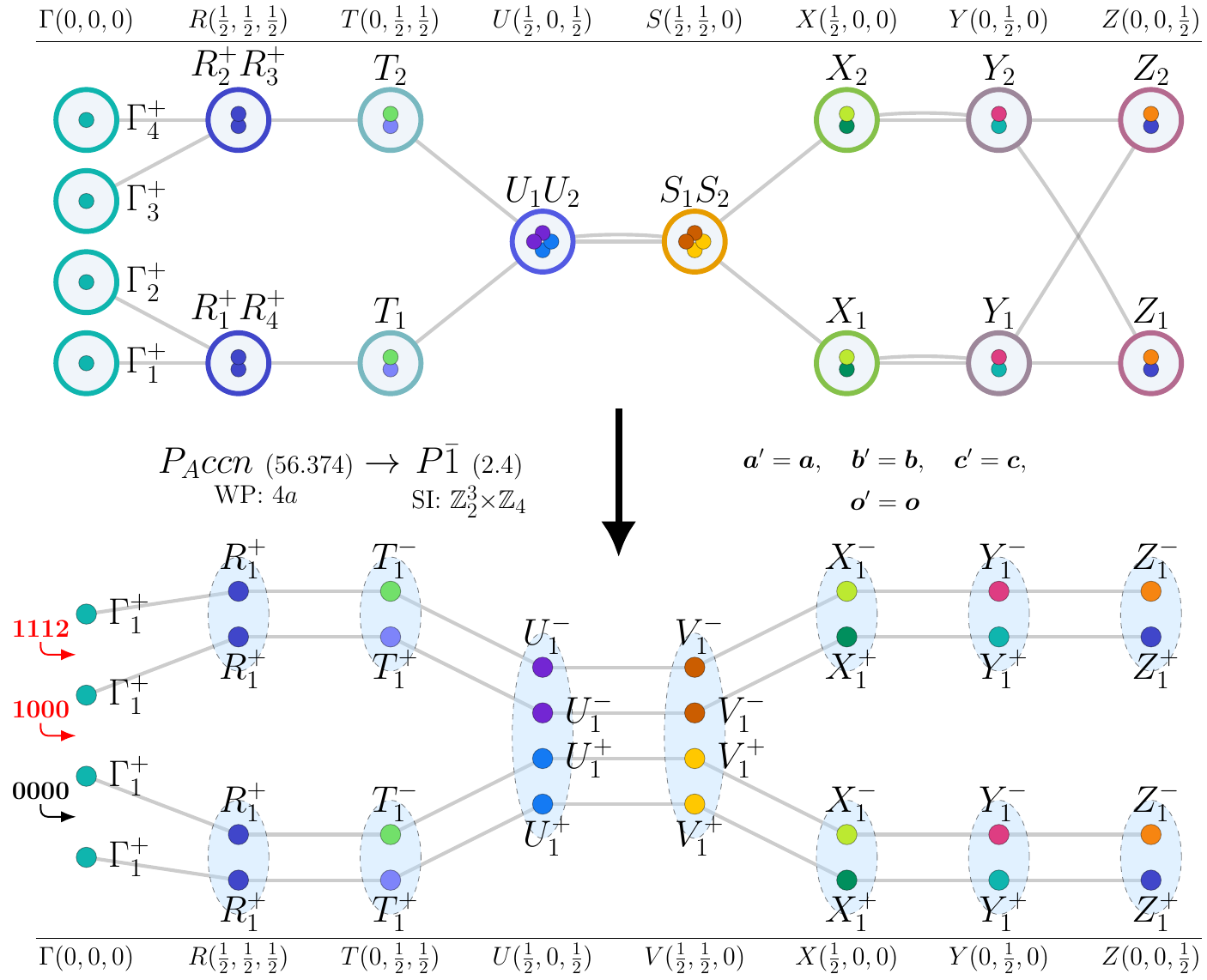}
\caption{Topological magnon bands in subgroup $P\bar{1}~(2.4)$ for magnetic moments on Wyckoff position $4a$ of supergroup $P_{A}ccn~(56.374)$.\label{fig_56.374_2.4_Bparallel100andstrainingenericdirection_4a}}
\end{figure}
\input{gap_tables_tex/56.374_2.4_Bparallel100andstrainingenericdirection_4a_table.tex}
\input{si_tables_tex/56.374_2.4_Bparallel100andstrainingenericdirection_4a_table.tex}
\subsubsection{Topological bands in subgroup $P2_{1}'/c'~(14.79)$}
\textbf{Perturbations:}
\begin{itemize}
\item B $\parallel$ [100] and strain $\perp$ [001],
\item B $\parallel$ [010] and strain $\perp$ [001],
\item B $\perp$ [001].
\end{itemize}
\begin{figure}[H]
\centering
\includegraphics[scale=0.6]{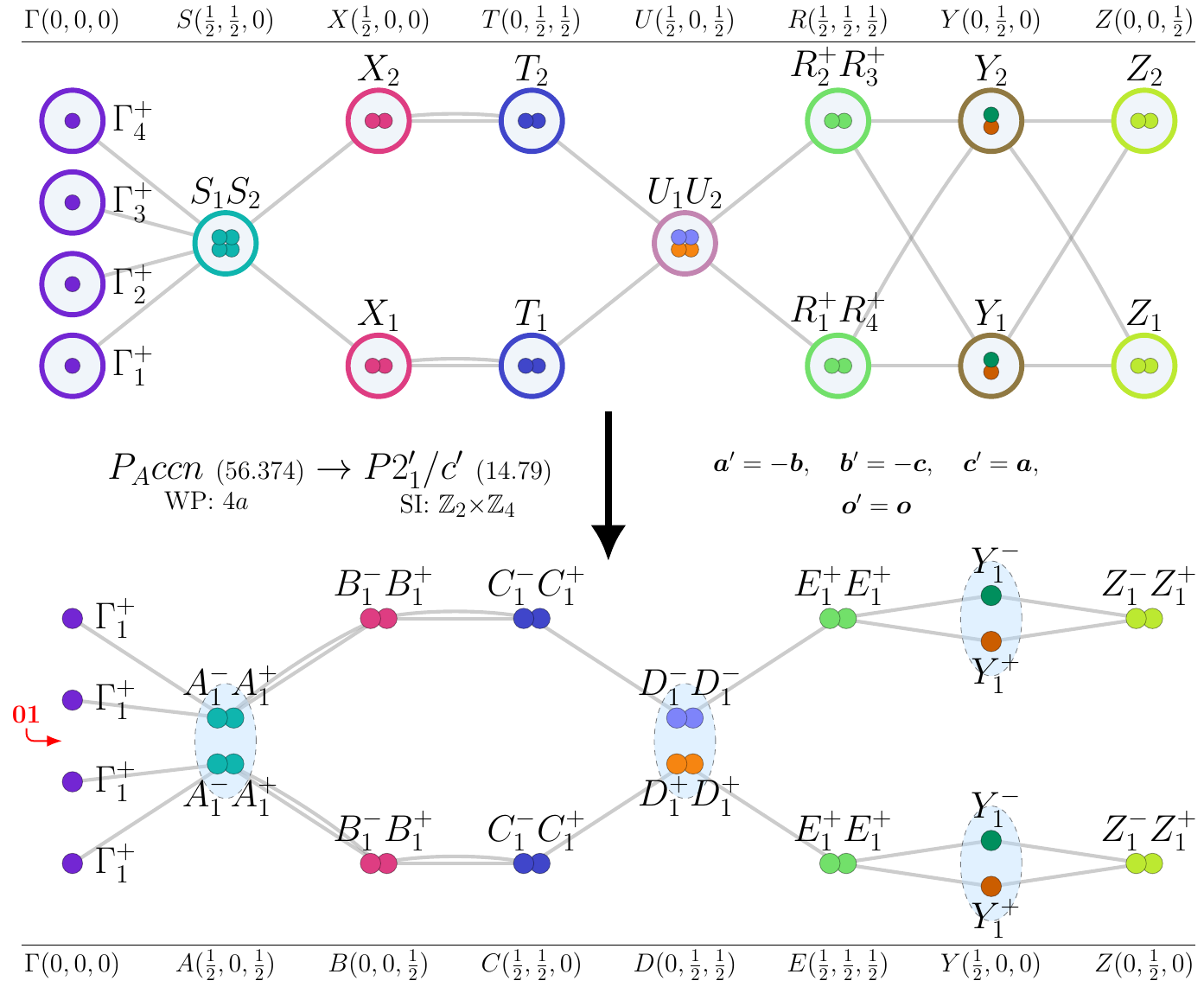}
\caption{Topological magnon bands in subgroup $P2_{1}'/c'~(14.79)$ for magnetic moments on Wyckoff position $4a$ of supergroup $P_{A}ccn~(56.374)$.\label{fig_56.374_14.79_Bparallel100andstrainperp001_4a}}
\end{figure}
\input{gap_tables_tex/56.374_14.79_Bparallel100andstrainperp001_4a_table.tex}
\input{si_tables_tex/56.374_14.79_Bparallel100andstrainperp001_4a_table.tex}
\subsubsection{Topological bands in subgroup $P2'/m'~(10.46)$}
\textbf{Perturbations:}
\begin{itemize}
\item B $\parallel$ [100] and strain $\perp$ [010],
\item B $\parallel$ [001] and strain $\perp$ [010],
\item B $\perp$ [010].
\end{itemize}
\begin{figure}[H]
\centering
\includegraphics[scale=0.6]{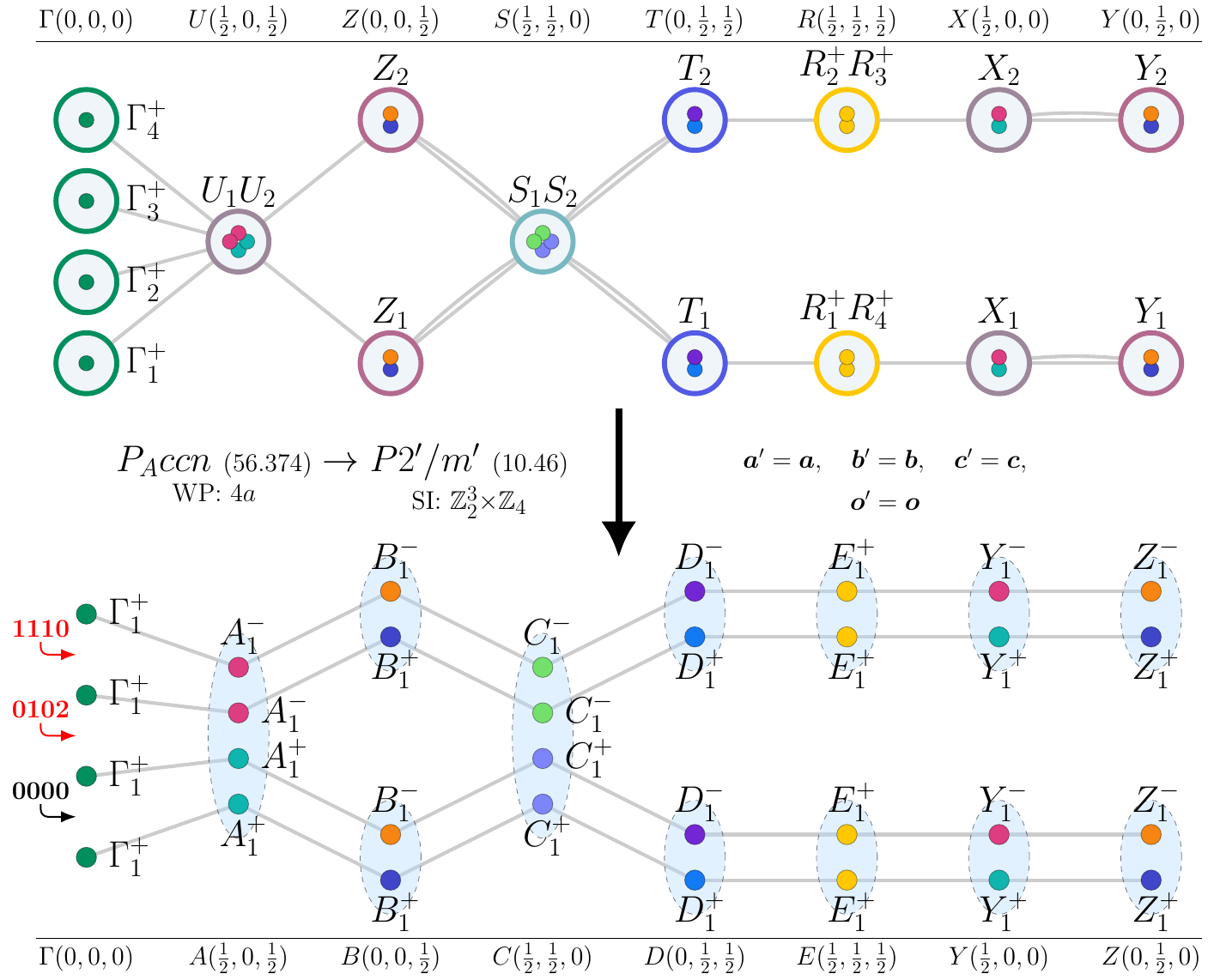}
\caption{Topological magnon bands in subgroup $P2'/m'~(10.46)$ for magnetic moments on Wyckoff position $4a$ of supergroup $P_{A}ccn~(56.374)$.\label{fig_56.374_10.46_Bparallel100andstrainperp010_4a}}
\end{figure}
\input{gap_tables_tex/56.374_10.46_Bparallel100andstrainperp010_4a_table.tex}
\input{si_tables_tex/56.374_10.46_Bparallel100andstrainperp010_4a_table.tex}
\subsubsection{Topological bands in subgroup $P2_{1}'/c'~(14.79)$}
\textbf{Perturbations:}
\begin{itemize}
\item B $\parallel$ [010] and strain $\perp$ [100],
\item B $\parallel$ [001] and strain $\perp$ [100],
\item B $\perp$ [100].
\end{itemize}
\begin{figure}[H]
\centering
\includegraphics[scale=0.6]{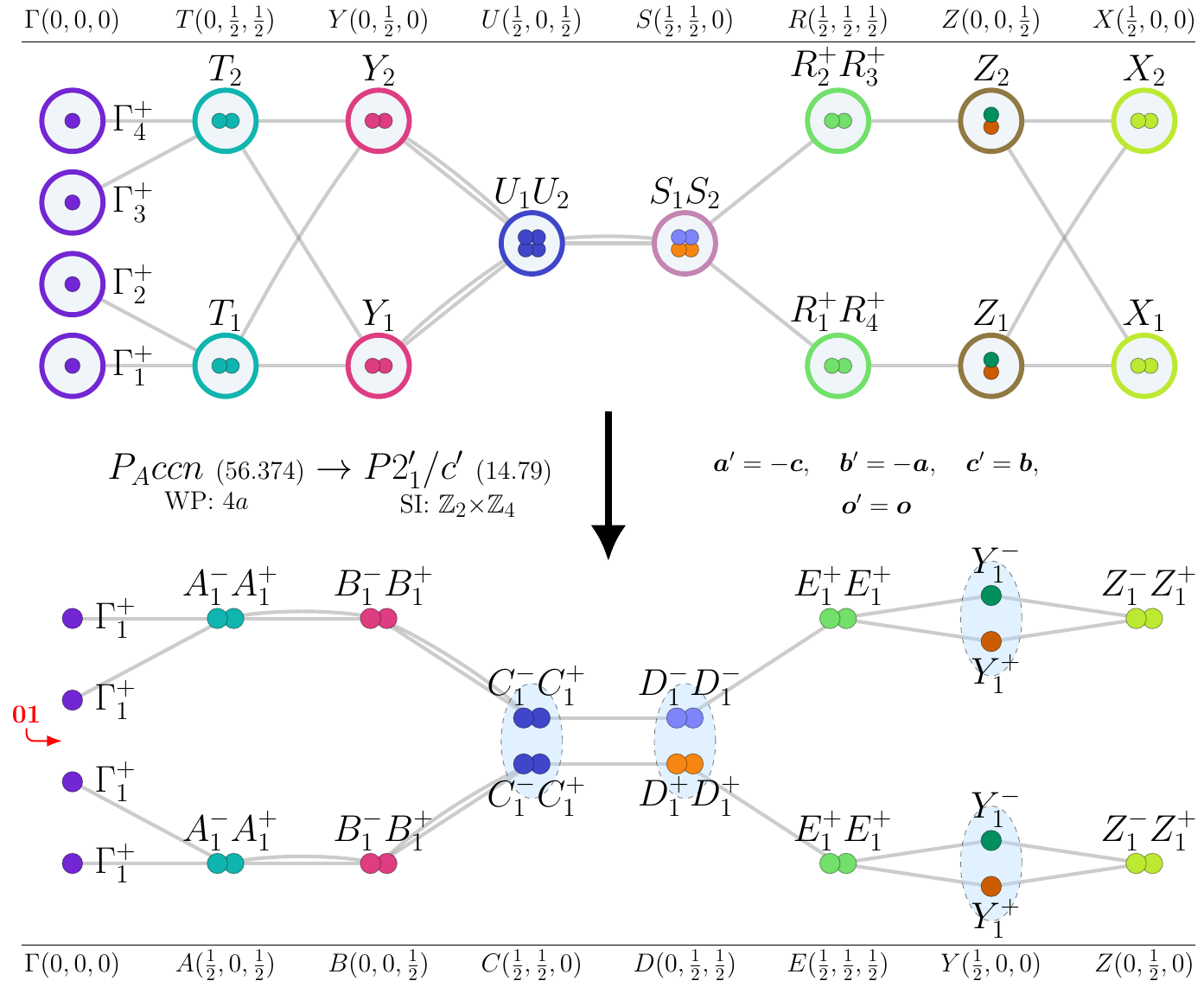}
\caption{Topological magnon bands in subgroup $P2_{1}'/c'~(14.79)$ for magnetic moments on Wyckoff position $4a$ of supergroup $P_{A}ccn~(56.374)$.\label{fig_56.374_14.79_Bparallel010andstrainperp100_4a}}
\end{figure}
\input{gap_tables_tex/56.374_14.79_Bparallel010andstrainperp100_4a_table.tex}
\input{si_tables_tex/56.374_14.79_Bparallel010andstrainperp100_4a_table.tex}
\subsubsection{Topological bands in subgroup $P_{S}\bar{1}~(2.7)$}
\textbf{Perturbation:}
\begin{itemize}
\item strain in generic direction.
\end{itemize}
\begin{figure}[H]
\centering
\includegraphics[scale=0.6]{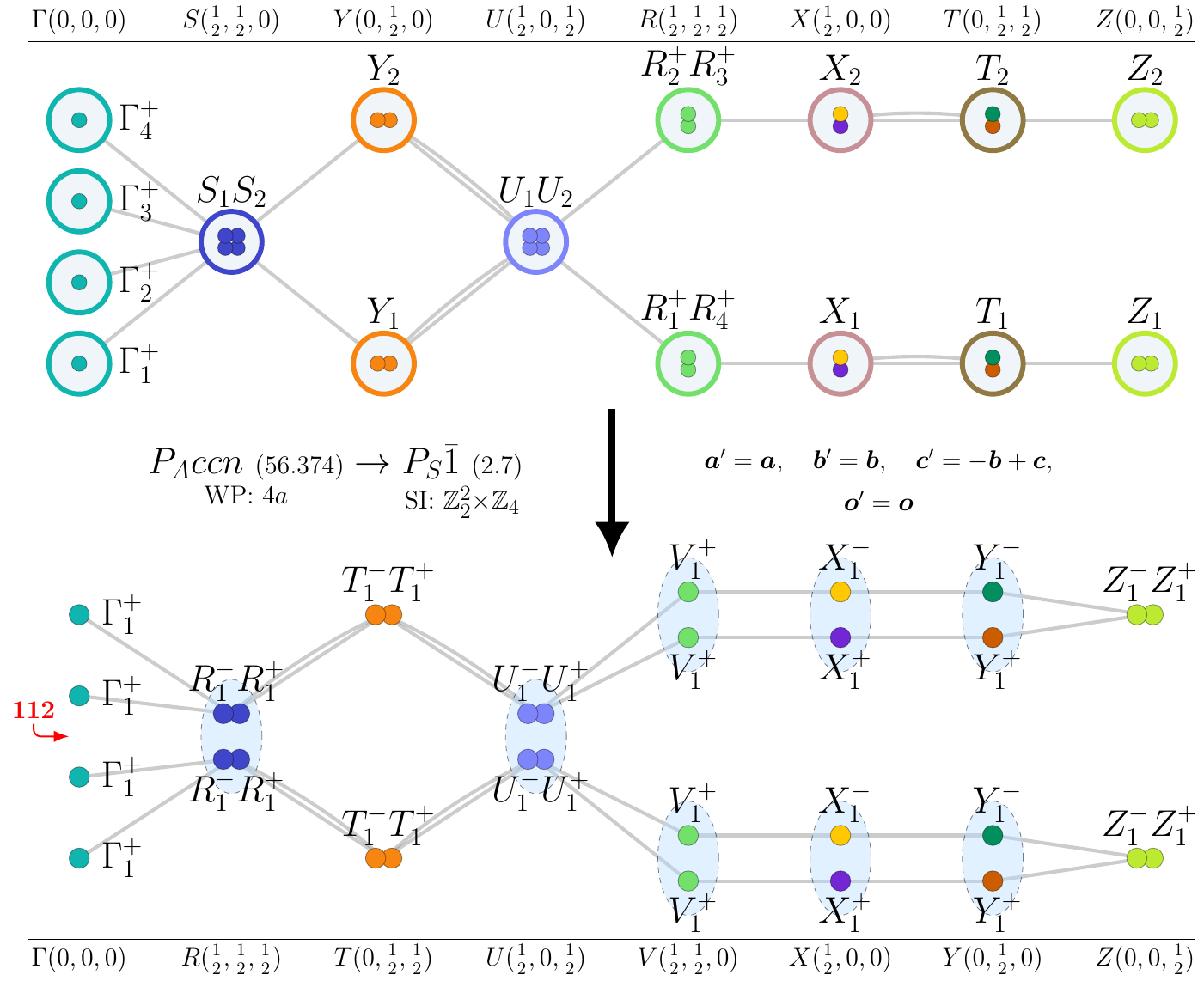}
\caption{Topological magnon bands in subgroup $P_{S}\bar{1}~(2.7)$ for magnetic moments on Wyckoff position $4a$ of supergroup $P_{A}ccn~(56.374)$.\label{fig_56.374_2.7_strainingenericdirection_4a}}
\end{figure}
\input{gap_tables_tex/56.374_2.7_strainingenericdirection_4a_table.tex}
\input{si_tables_tex/56.374_2.7_strainingenericdirection_4a_table.tex}

\section{MSG $P_{a}bcm~(57.386)$}
\textbf{Nontrivial-SI Subgroups:} $P\bar{1}~(2.4)$, $P2_{1}'/c'~(14.79)$, $P2_{1}'/c'~(14.79)$, $P2_{1}'/c'~(14.79)$, $P_{S}\bar{1}~(2.7)$, $P_{b}ca2_{1}~(29.105)$, $P2_{1}/m~(11.50)$, $Pn'ma'~(62.448)$, $P_{a}2_{1}/m~(11.55)$, $P2_{1}/c~(14.75)$, $Pb'c'a~(61.436)$, $P_{a}2_{1}/c~(14.80)$, $P2~(3.1)$, $P_{b}2~(3.5)$, $P2/c~(13.65)$, $Pb'cn'~(60.424)$, $P_{b}2/c~(13.71)$.\\

\textbf{Trivial-SI Subgroups:} $Pc'~(7.26)$, $Pc'~(7.26)$, $Pc'~(7.26)$, $P2_{1}'~(4.9)$, $P2_{1}'~(4.9)$, $P2_{1}'~(4.9)$, $P_{S}1~(1.3)$, $Pm~(6.18)$, $Pmc'2_{1}'~(26.69)$, $Pmn'2_{1}'~(31.126)$, $P_{a}m~(6.21)$, $Pc~(7.24)$, $Pca'2_{1}'~(29.102)$, $Pc'a2_{1}'~(29.101)$, $P_{a}c~(7.27)$, $Pc~(7.24)$, $Pn'a2_{1}'~(33.146)$, $Pca'2_{1}'~(29.102)$, $P_{b}c~(7.29)$, $P2_{1}~(4.7)$, $Pn'a'2_{1}~(33.148)$, $P_{a}2_{1}~(4.10)$, $P2_{1}~(4.7)$, $Pc'a'2_{1}~(29.103)$, $P_{a}2_{1}~(4.10)$, $P_{b}mc2_{1}~(26.72)$, $Pn'c'2~(30.115)$, $P_{c}ma2~(28.94)$.\\

\subsection{WP: $8d$}
\textbf{BCS Materials:} {BaCoSO~(220 K)}\footnote{BCS web page: \texttt{\href{http://webbdcrista1.ehu.es/magndata/index.php?this\_label=1.593} {http://webbdcrista1.ehu.es/magndata/index.php?this\_label=1.593}}}, {BaCoSO}\footnote{BCS web page: \texttt{\href{http://webbdcrista1.ehu.es/magndata/index.php?this\_label=1.594} {http://webbdcrista1.ehu.es/magndata/index.php?this\_label=1.594}}}.\\
\subsubsection{Topological bands in subgroup $P2_{1}'/c'~(14.79)$}
\textbf{Perturbations:}
\begin{itemize}
\item B $\parallel$ [100] and strain $\perp$ [001],
\item B $\parallel$ [010] and strain $\perp$ [001],
\item B $\perp$ [001].
\end{itemize}
\begin{figure}[H]
\centering
\includegraphics[scale=0.6]{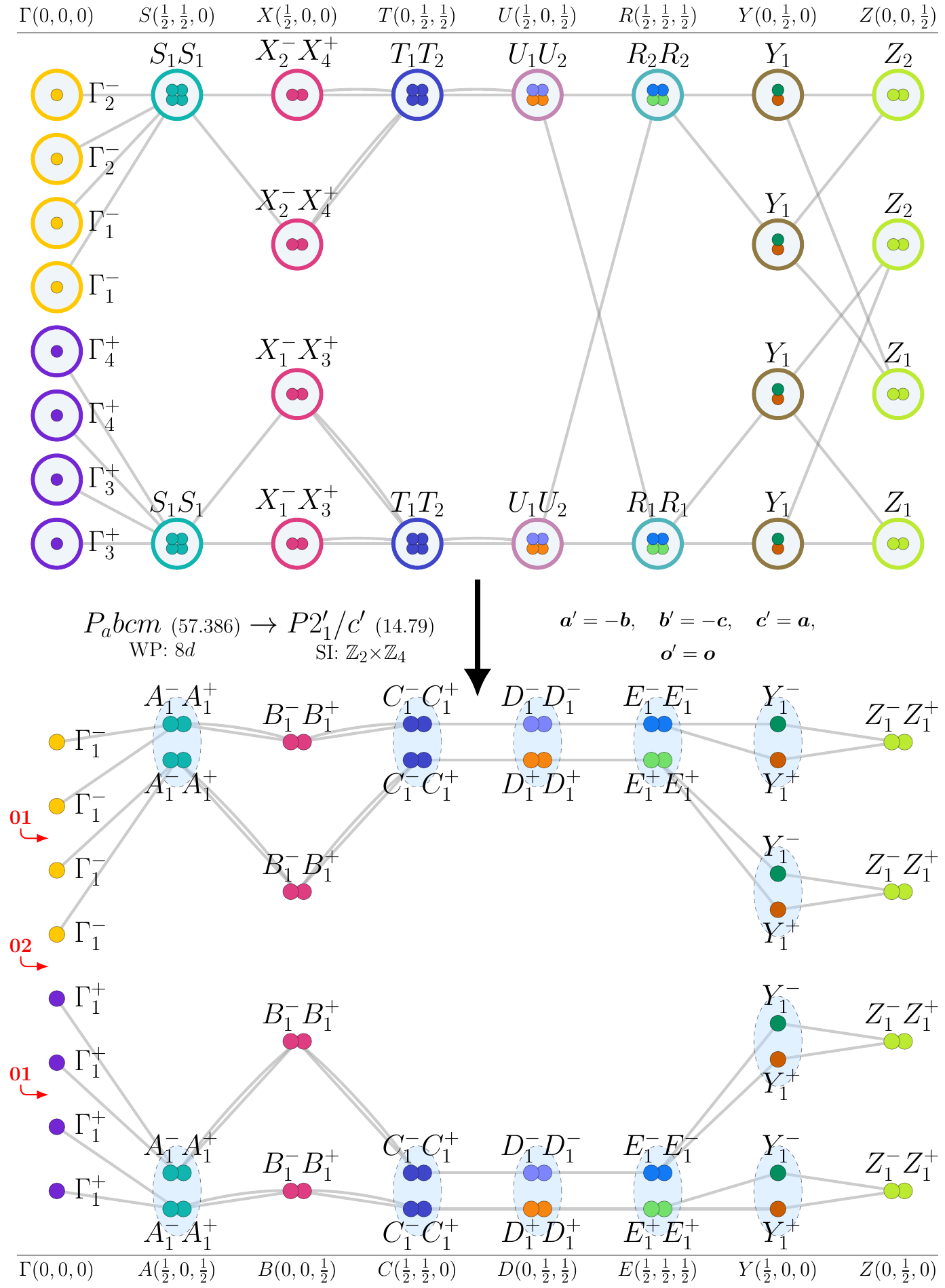}
\caption{Topological magnon bands in subgroup $P2_{1}'/c'~(14.79)$ for magnetic moments on Wyckoff position $8d$ of supergroup $P_{a}bcm~(57.386)$.\label{fig_57.386_14.79_Bparallel100andstrainperp001_8d}}
\end{figure}
\input{gap_tables_tex/57.386_14.79_Bparallel100andstrainperp001_8d_table.tex}
\input{si_tables_tex/57.386_14.79_Bparallel100andstrainperp001_8d_table.tex}
\subsubsection{Topological bands in subgroup $P2_{1}'/c'~(14.79)$}
\textbf{Perturbations:}
\begin{itemize}
\item B $\parallel$ [010] and strain $\perp$ [100],
\item B $\parallel$ [001] and strain $\perp$ [100],
\item B $\perp$ [100].
\end{itemize}
\begin{figure}[H]
\centering
\includegraphics[scale=0.6]{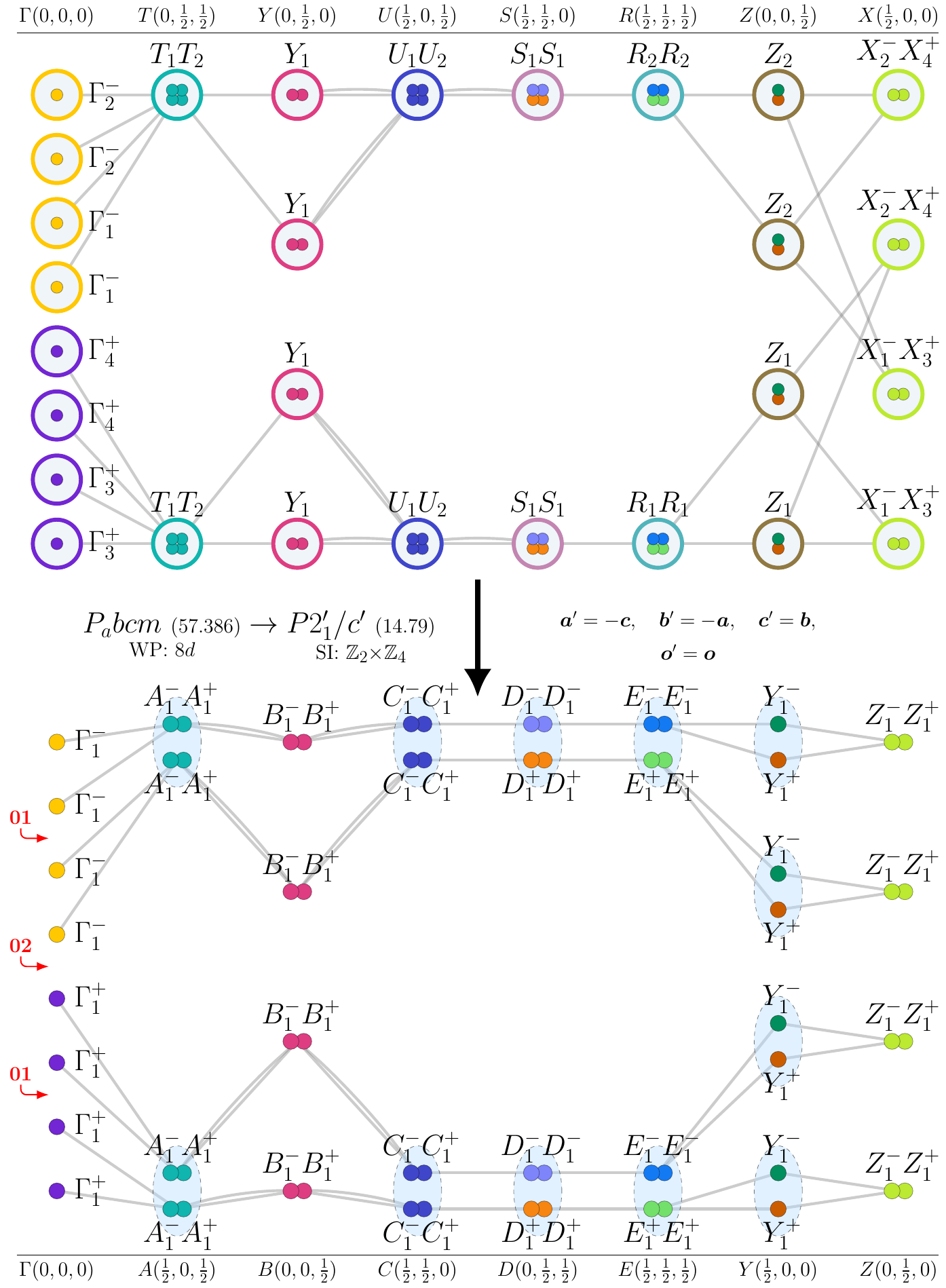}
\caption{Topological magnon bands in subgroup $P2_{1}'/c'~(14.79)$ for magnetic moments on Wyckoff position $8d$ of supergroup $P_{a}bcm~(57.386)$.\label{fig_57.386_14.79_Bparallel010andstrainperp100_8d}}
\end{figure}
\input{gap_tables_tex/57.386_14.79_Bparallel010andstrainperp100_8d_table.tex}
\input{si_tables_tex/57.386_14.79_Bparallel010andstrainperp100_8d_table.tex}

\section{MSG $P_{C}bcm~(57.391)$}
\textbf{Nontrivial-SI Subgroups:} $P\bar{1}~(2.4)$, $P2_{1}'/c'~(14.79)$, $P2'/c'~(13.69)$, $P2_{1}'/m'~(11.54)$, $P_{S}\bar{1}~(2.7)$, $P2_{1}/m~(11.50)$, $Pmm'n'~(59.410)$, $P_{a}2_{1}/m~(11.55)$, $P2_{1}/c~(14.75)$, $Pn'm'a~(62.446)$, $P_{C}2_{1}/c~(14.84)$, $P2~(3.1)$, $P2/c~(13.65)$, $Pn'n'a~(52.310)$, $P_{A}2/c~(13.73)$.\\

\textbf{Trivial-SI Subgroups:} $Pc'~(7.26)$, $Pc'~(7.26)$, $Pm'~(6.20)$, $P2_{1}'~(4.9)$, $P2'~(3.3)$, $P2_{1}'~(4.9)$, $P_{S}1~(1.3)$, $Pm~(6.18)$, $Pm'm2'~(25.59)$, $Pmn'2_{1}'~(31.126)$, $P_{a}m~(6.21)$, $Pc~(7.24)$, $Pm'c2_{1}'~(26.68)$, $Pn'a2_{1}'~(33.146)$, $P_{C}c~(7.30)$, $Pc~(7.24)$, $Pn'a2_{1}'~(33.146)$, $Pn'c2'~(30.113)$, $P_{A}c~(7.31)$, $P2_{1}~(4.7)$, $Pm'n'2_{1}~(31.127)$, $P_{a}2_{1}~(4.10)$, $P_{C}ca2_{1}~(29.109)$, $P2_{1}~(4.7)$, $Pm'n'2_{1}~(31.127)$, $P_{C}2_{1}~(4.12)$, $P_{A}mc2_{1}~(26.74)$, $Pn'n'2~(34.159)$, $P_{C}2~(3.6)$, $P_{A}ma2~(28.95)$.\\

\subsection{WP: $4c+8g$}
\textbf{BCS Materials:} {LaSr\textsubscript{3}Fe\textsubscript{3}O\textsubscript{9}~(350 K)}\footnote{BCS web page: \texttt{\href{http://webbdcrista1.ehu.es/magndata/index.php?this\_label=1.481} {http://webbdcrista1.ehu.es/magndata/index.php?this\_label=1.481}}}.\\
\subsubsection{Topological bands in subgroup $P2_{1}'/c'~(14.79)$}
\textbf{Perturbations:}
\begin{itemize}
\item B $\parallel$ [100] and strain $\perp$ [001],
\item B $\parallel$ [010] and strain $\perp$ [001],
\item B $\perp$ [001].
\end{itemize}
\begin{figure}[H]
\centering
\includegraphics[scale=0.6]{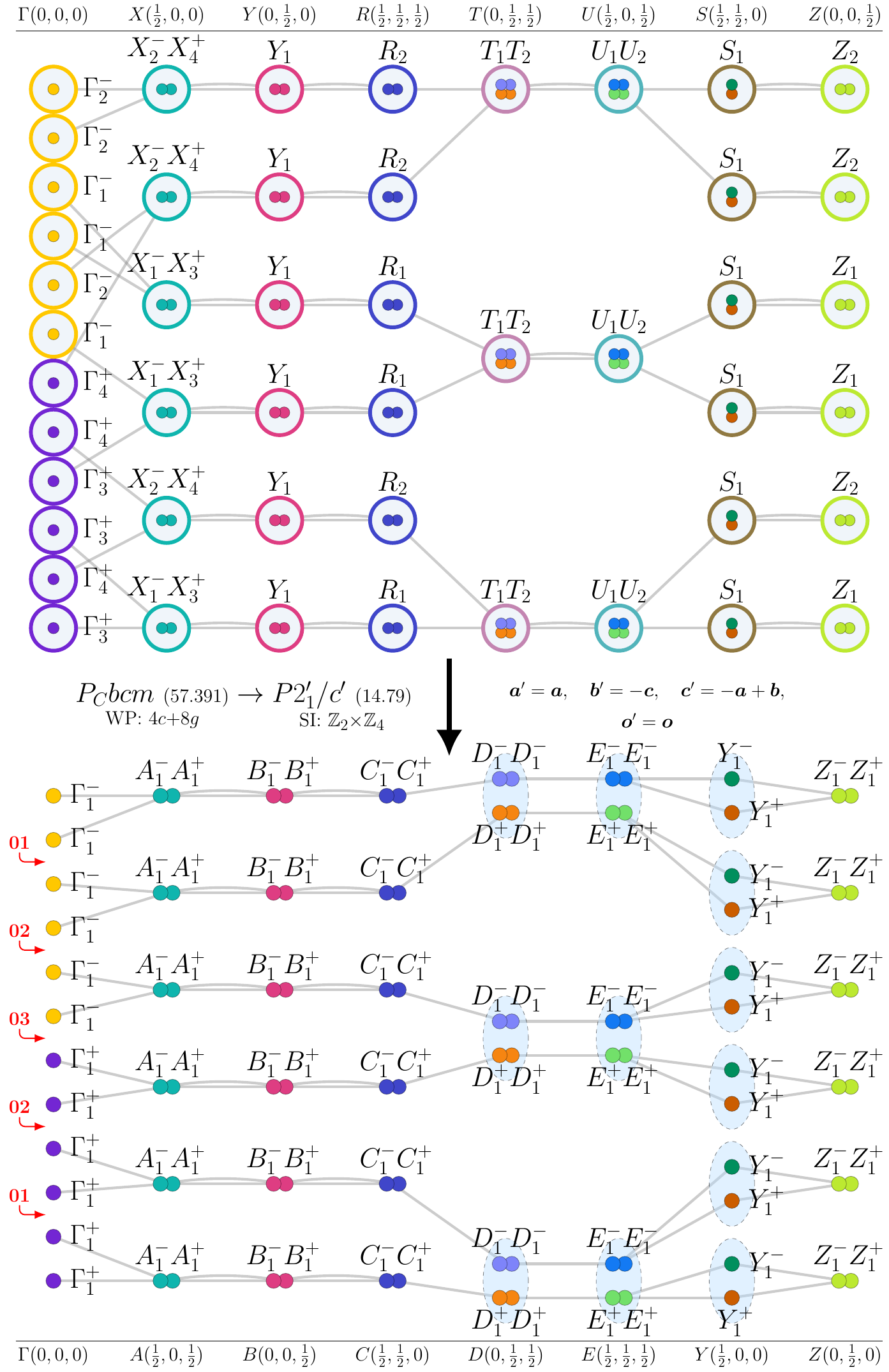}
\caption{Topological magnon bands in subgroup $P2_{1}'/c'~(14.79)$ for magnetic moments on Wyckoff positions $4c+8g$ of supergroup $P_{C}bcm~(57.391)$.\label{fig_57.391_14.79_Bparallel100andstrainperp001_4c+8g}}
\end{figure}
\input{gap_tables_tex/57.391_14.79_Bparallel100andstrainperp001_4c+8g_table.tex}
\input{si_tables_tex/57.391_14.79_Bparallel100andstrainperp001_4c+8g_table.tex}
\subsection{WP: $4c$}
\textbf{BCS Materials:} {CeRu\textsubscript{2}Al\textsubscript{10}~(27 K)}\footnote{BCS web page: \texttt{\href{http://webbdcrista1.ehu.es/magndata/index.php?this\_label=1.8} {http://webbdcrista1.ehu.es/magndata/index.php?this\_label=1.8}}}.\\
\subsubsection{Topological bands in subgroup $P2_{1}'/c'~(14.79)$}
\textbf{Perturbations:}
\begin{itemize}
\item B $\parallel$ [100] and strain $\perp$ [001],
\item B $\parallel$ [010] and strain $\perp$ [001],
\item B $\perp$ [001].
\end{itemize}
\begin{figure}[H]
\centering
\includegraphics[scale=0.6]{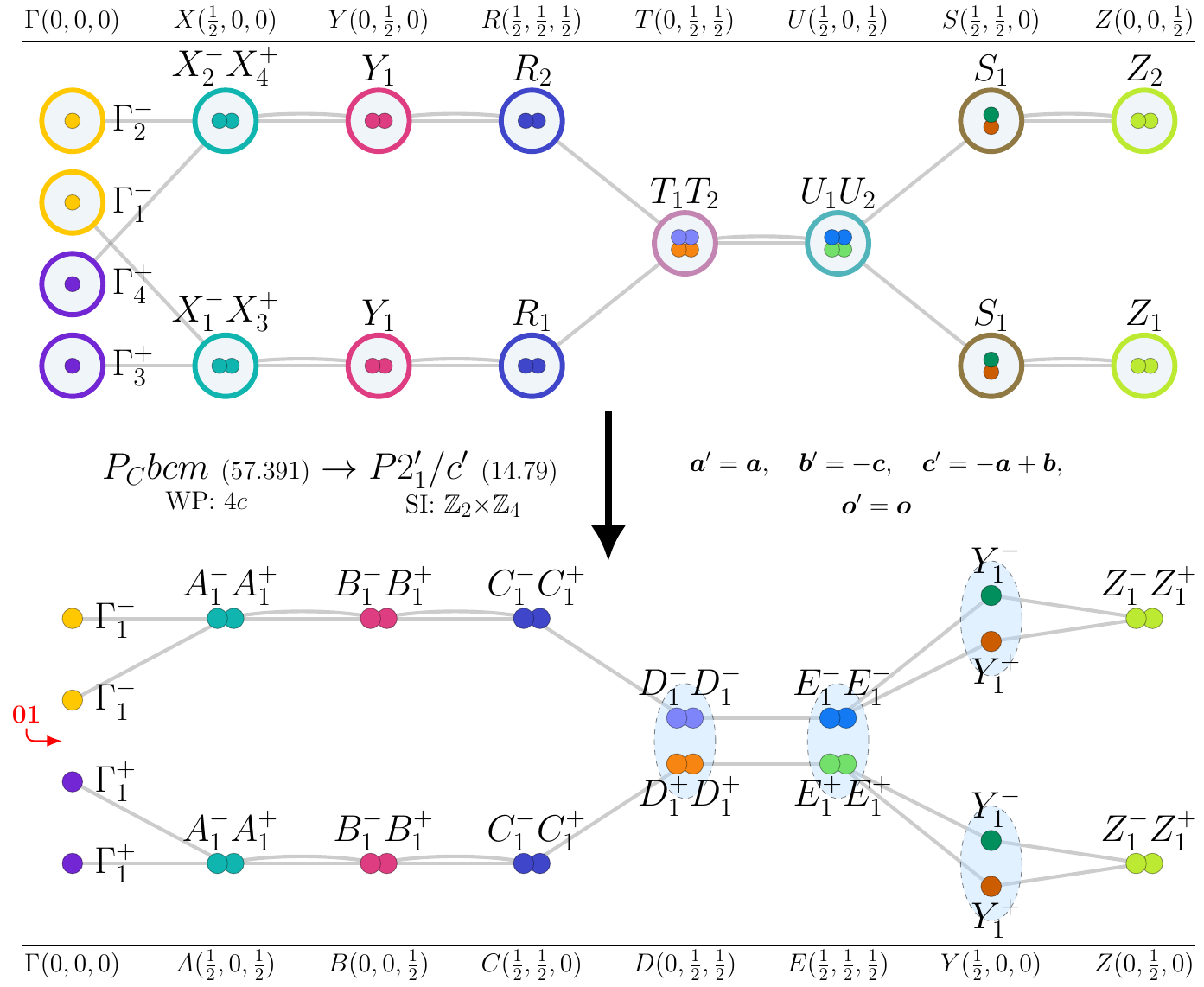}
\caption{Topological magnon bands in subgroup $P2_{1}'/c'~(14.79)$ for magnetic moments on Wyckoff position $4c$ of supergroup $P_{C}bcm~(57.391)$.\label{fig_57.391_14.79_Bparallel100andstrainperp001_4c}}
\end{figure}
\input{gap_tables_tex/57.391_14.79_Bparallel100andstrainperp001_4c_table.tex}
\input{si_tables_tex/57.391_14.79_Bparallel100andstrainperp001_4c_table.tex}
\subsection{WP: $4c+8f$}
\textbf{BCS Materials:} {Er\textsubscript{3}Ge\textsubscript{4}~(7.3 K)}\footnote{BCS web page: \texttt{\href{http://webbdcrista1.ehu.es/magndata/index.php?this\_label=1.362} {http://webbdcrista1.ehu.es/magndata/index.php?this\_label=1.362}}}.\\
\subsubsection{Topological bands in subgroup $P2_{1}'/c'~(14.79)$}
\textbf{Perturbations:}
\begin{itemize}
\item B $\parallel$ [100] and strain $\perp$ [001],
\item B $\parallel$ [010] and strain $\perp$ [001],
\item B $\perp$ [001].
\end{itemize}
\begin{figure}[H]
\centering
\includegraphics[scale=0.6]{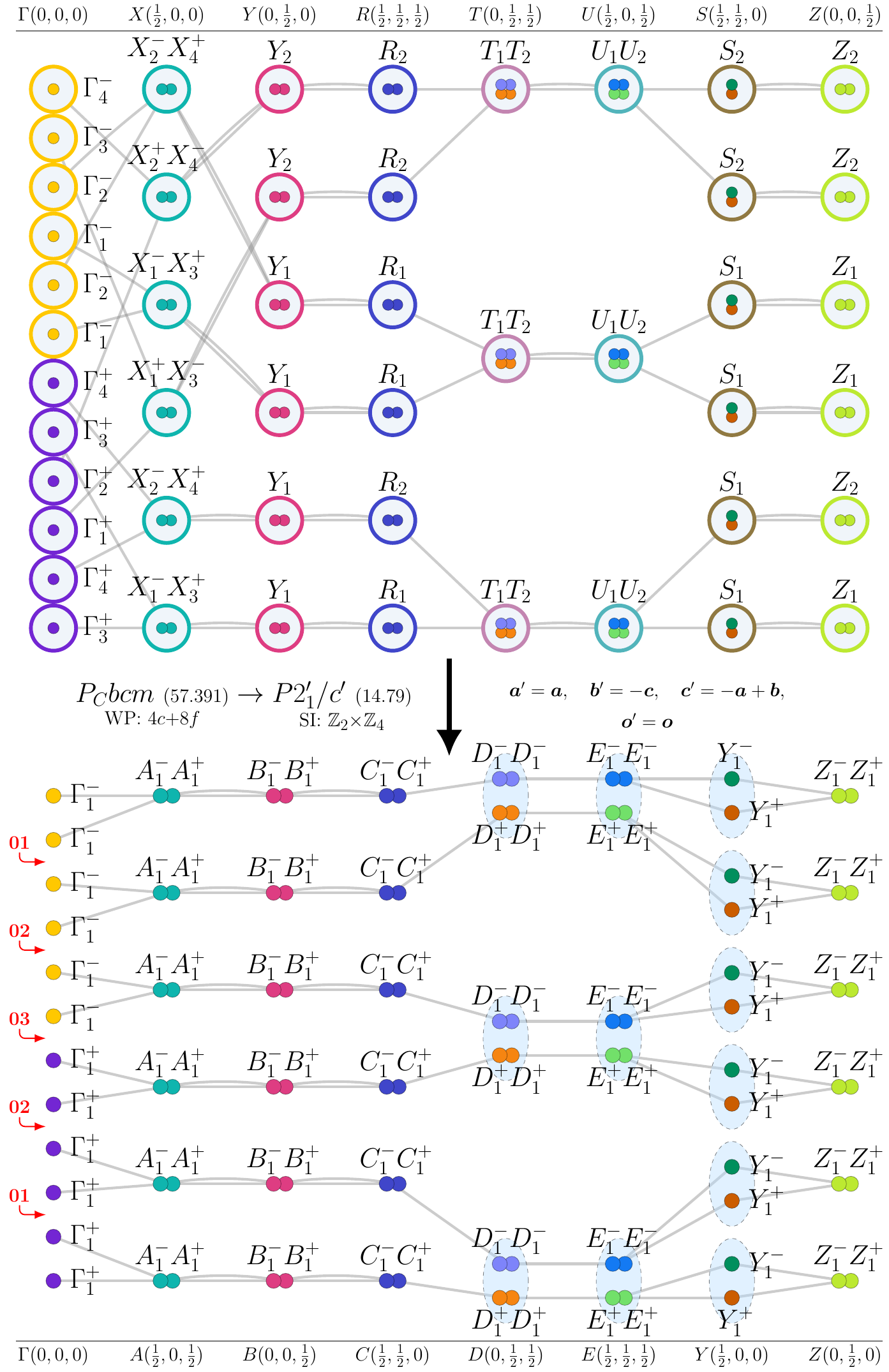}
\caption{Topological magnon bands in subgroup $P2_{1}'/c'~(14.79)$ for magnetic moments on Wyckoff positions $4c+8f$ of supergroup $P_{C}bcm~(57.391)$.\label{fig_57.391_14.79_Bparallel100andstrainperp001_4c+8f}}
\end{figure}
\input{gap_tables_tex/57.391_14.79_Bparallel100andstrainperp001_4c+8f_table.tex}
\input{si_tables_tex/57.391_14.79_Bparallel100andstrainperp001_4c+8f_table.tex}

\section{MSG $P_{B}nnm~(58.402)$}
\textbf{Nontrivial-SI Subgroups:} $P\bar{1}~(2.4)$, $P2_{1}'/c'~(14.79)$, $P2_{1}'/m'~(11.54)$, $P2'/c'~(13.69)$, $P_{S}\bar{1}~(2.7)$, $P2~(3.1)$, $Pm'a'2~(28.91)$, $P2/m~(10.42)$, $Pm'ma'~(51.296)$, $P_{C}2/m~(10.49)$, $P2_{1}/c~(14.75)$, $Pb'c'n~(60.422)$, $P_{c}2_{1}/c~(14.82)$, $P2_{1}/c~(14.75)$, $Pnm'a'~(62.447)$, $P_{C}2_{1}/c~(14.84)$.\\

\textbf{Trivial-SI Subgroups:} $Pc'~(7.26)$, $Pm'~(6.20)$, $Pc'~(7.26)$, $P2_{1}'~(4.9)$, $P2_{1}'~(4.9)$, $P2'~(3.3)$, $P_{S}1~(1.3)$, $Pm~(6.18)$, $Pmc'2_{1}'~(26.69)$, $Pm'm2'~(25.59)$, $P_{C}m~(6.23)$, $Pc~(7.24)$, $Pna'2_{1}'~(33.147)$, $Pnc'2'~(30.114)$, $P_{c}c~(7.28)$, $Pc~(7.24)$, $Pm'n2_{1}'~(31.125)$, $Pna'2_{1}'~(33.147)$, $P_{C}c~(7.30)$, $P_{C}2~(3.6)$, $P_{A}nn2~(34.162)$, $P2_{1}~(4.7)$, $Pc'a'2_{1}~(29.103)$, $P_{a}2_{1}~(4.10)$, $P_{C}mn2_{1}~(31.133)$, $P2_{1}~(4.7)$, $Pm'c'2_{1}~(26.70)$, $P_{C}2_{1}~(4.12)$, $P_{B}mn2_{1}~(31.132)$.\\

\subsection{WP: $4a$}
\textbf{BCS Materials:} {CsNiF\textsubscript{3}~(2.67 K)}\footnote{BCS web page: \texttt{\href{http://webbdcrista1.ehu.es/magndata/index.php?this\_label=1.527} {http://webbdcrista1.ehu.es/magndata/index.php?this\_label=1.527}}}.\\
\subsubsection{Topological bands in subgroup $P\bar{1}~(2.4)$}
\textbf{Perturbations:}
\begin{itemize}
\item B $\parallel$ [100] and strain in generic direction,
\item B $\parallel$ [010] and strain in generic direction,
\item B $\parallel$ [001] and strain in generic direction,
\item B $\perp$ [100] and strain $\perp$ [010],
\item B $\perp$ [100] and strain $\perp$ [001],
\item B $\perp$ [100] and strain in generic direction,
\item B $\perp$ [010] and strain $\perp$ [100],
\item B $\perp$ [010] and strain $\perp$ [001],
\item B $\perp$ [010] and strain in generic direction,
\item B $\perp$ [001] and strain $\perp$ [100],
\item B $\perp$ [001] and strain $\perp$ [010],
\item B $\perp$ [001] and strain in generic direction,
\item B in generic direction.
\end{itemize}
\begin{figure}[H]
\centering
\includegraphics[scale=0.6]{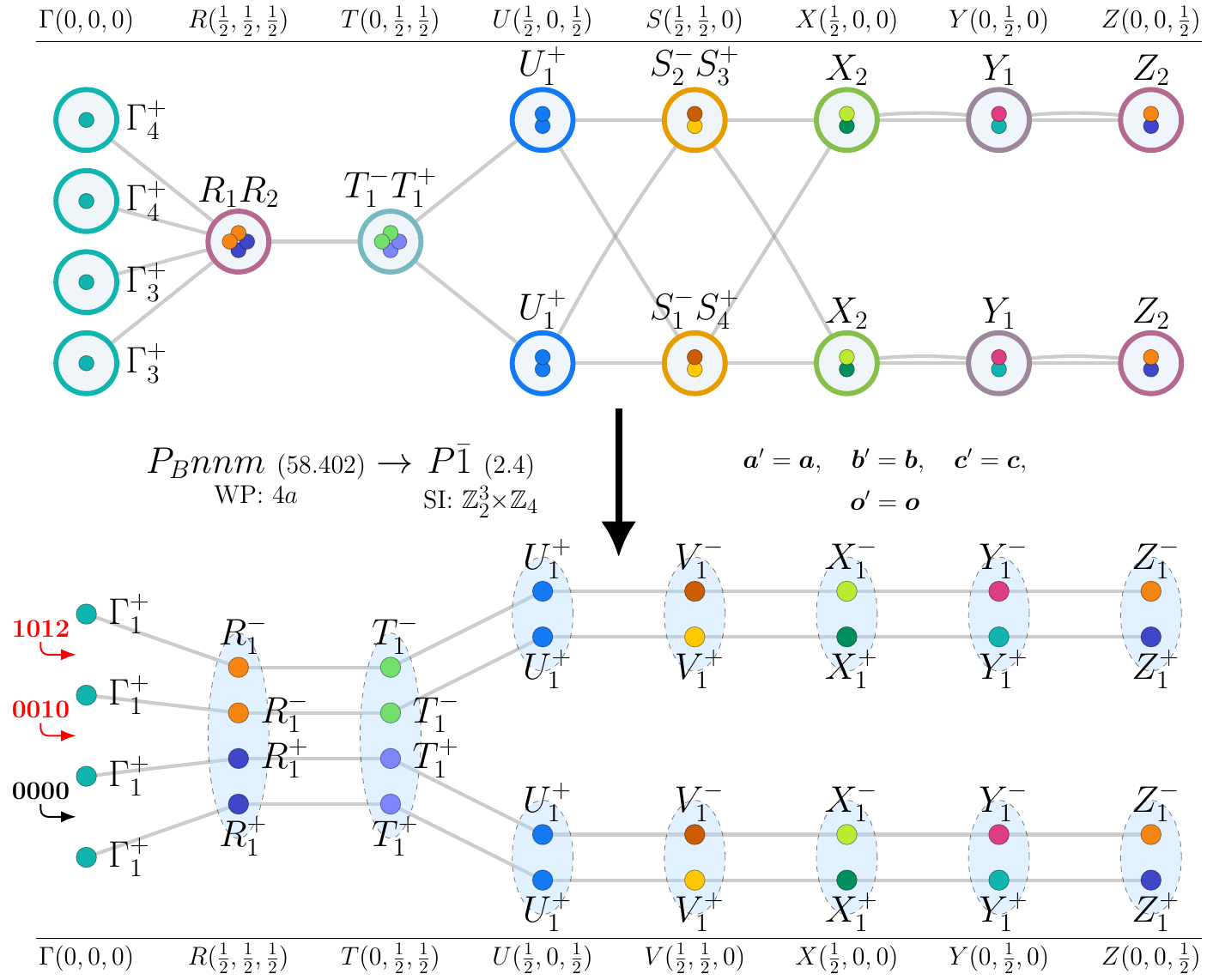}
\caption{Topological magnon bands in subgroup $P\bar{1}~(2.4)$ for magnetic moments on Wyckoff position $4a$ of supergroup $P_{B}nnm~(58.402)$.\label{fig_58.402_2.4_Bparallel100andstrainingenericdirection_4a}}
\end{figure}
\input{gap_tables_tex/58.402_2.4_Bparallel100andstrainingenericdirection_4a_table.tex}
\input{si_tables_tex/58.402_2.4_Bparallel100andstrainingenericdirection_4a_table.tex}
\subsubsection{Topological bands in subgroup $P2_{1}'/c'~(14.79)$}
\textbf{Perturbations:}
\begin{itemize}
\item B $\parallel$ [100] and strain $\perp$ [001],
\item B $\parallel$ [010] and strain $\perp$ [001],
\item B $\perp$ [001].
\end{itemize}
\begin{figure}[H]
\centering
\includegraphics[scale=0.6]{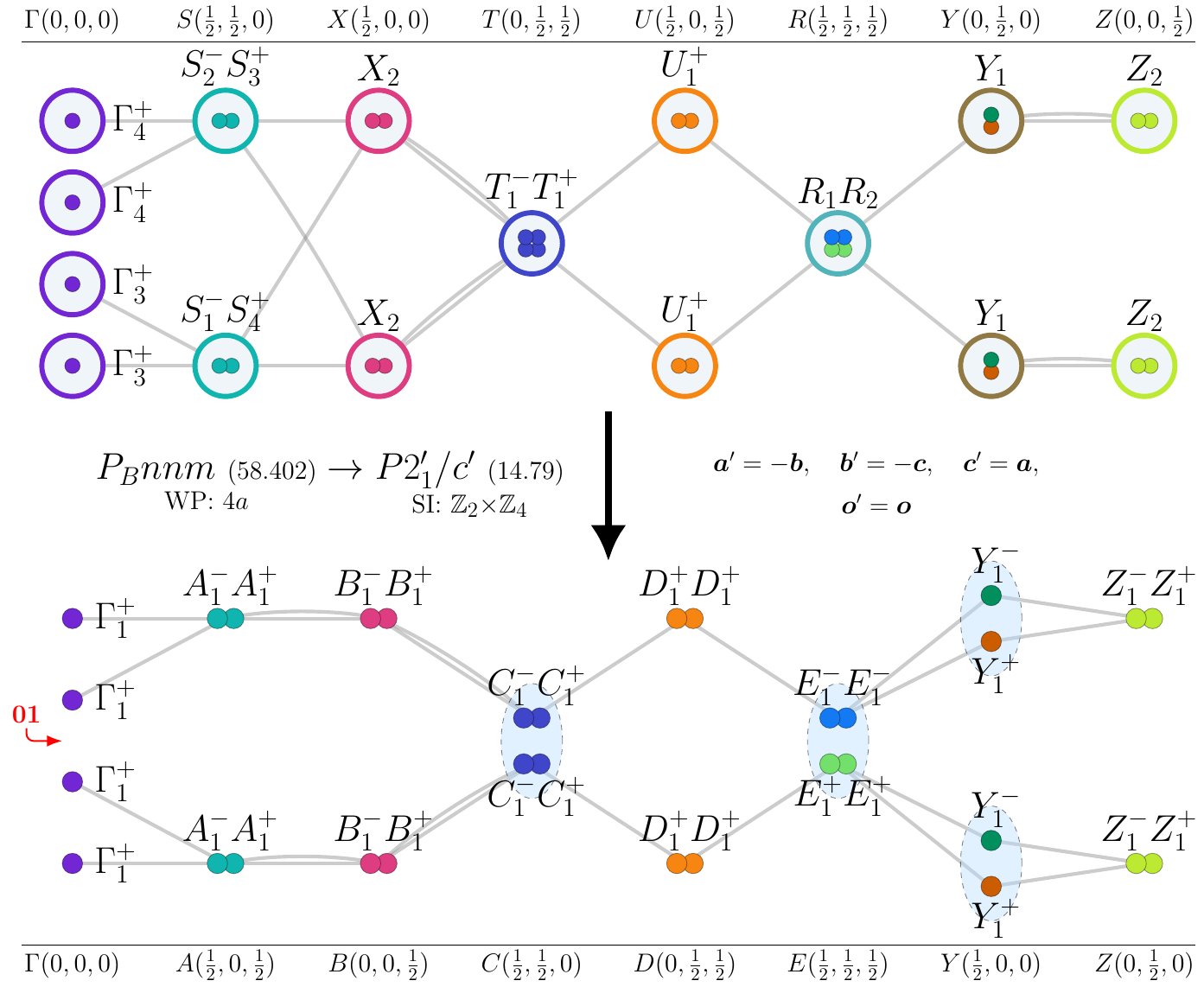}
\caption{Topological magnon bands in subgroup $P2_{1}'/c'~(14.79)$ for magnetic moments on Wyckoff position $4a$ of supergroup $P_{B}nnm~(58.402)$.\label{fig_58.402_14.79_Bparallel100andstrainperp001_4a}}
\end{figure}
\input{gap_tables_tex/58.402_14.79_Bparallel100andstrainperp001_4a_table.tex}
\input{si_tables_tex/58.402_14.79_Bparallel100andstrainperp001_4a_table.tex}
\subsubsection{Topological bands in subgroup $P2_{1}'/m'~(11.54)$}
\textbf{Perturbations:}
\begin{itemize}
\item B $\parallel$ [100] and strain $\perp$ [010],
\item B $\parallel$ [001] and strain $\perp$ [010],
\item B $\perp$ [010].
\end{itemize}
\begin{figure}[H]
\centering
\includegraphics[scale=0.6]{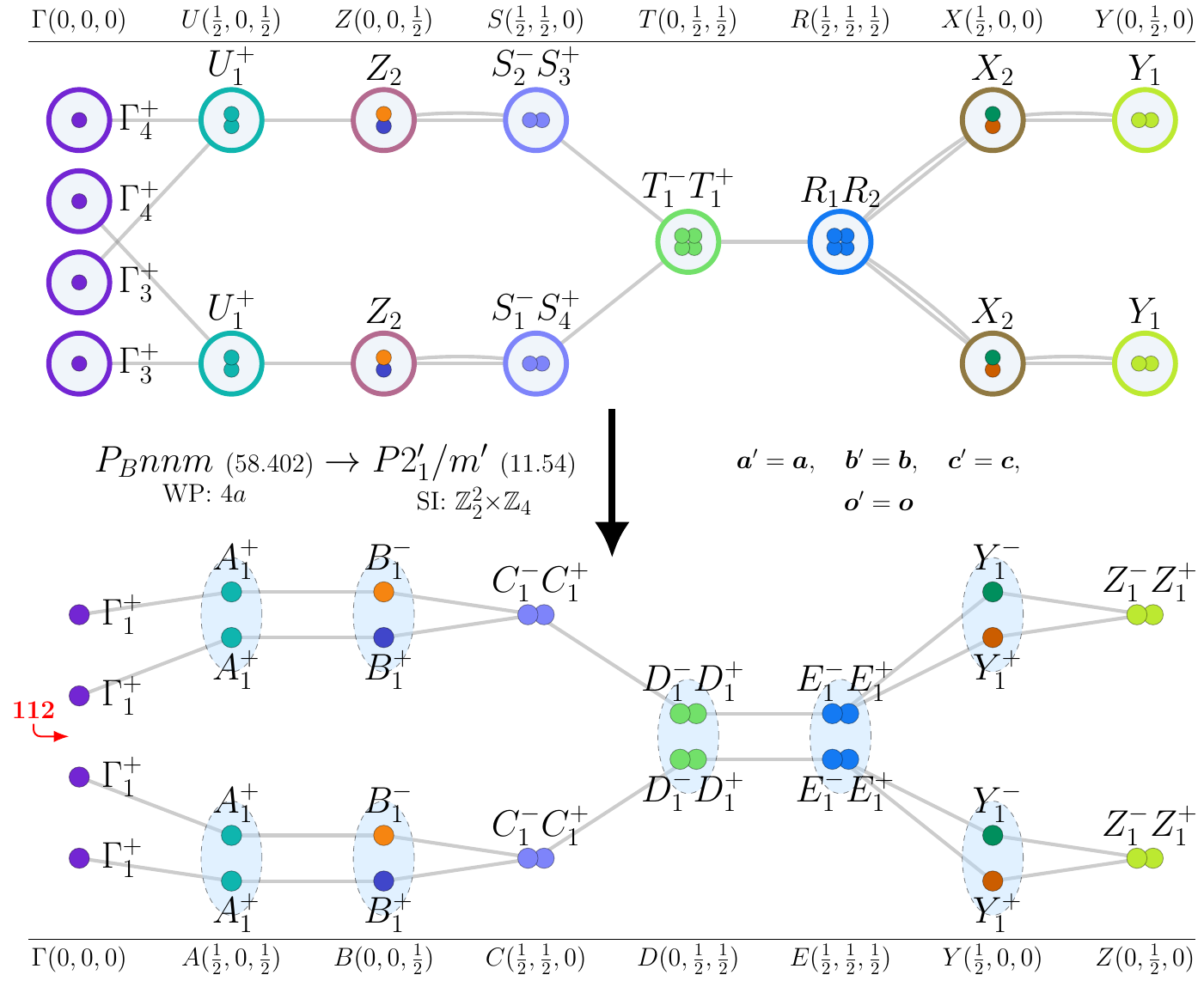}
\caption{Topological magnon bands in subgroup $P2_{1}'/m'~(11.54)$ for magnetic moments on Wyckoff position $4a$ of supergroup $P_{B}nnm~(58.402)$.\label{fig_58.402_11.54_Bparallel100andstrainperp010_4a}}
\end{figure}
\input{gap_tables_tex/58.402_11.54_Bparallel100andstrainperp010_4a_table.tex}
\input{si_tables_tex/58.402_11.54_Bparallel100andstrainperp010_4a_table.tex}
\subsubsection{Topological bands in subgroup $P2'/c'~(13.69)$}
\textbf{Perturbations:}
\begin{itemize}
\item B $\parallel$ [010] and strain $\perp$ [100],
\item B $\parallel$ [001] and strain $\perp$ [100],
\item B $\perp$ [100].
\end{itemize}
\begin{figure}[H]
\centering
\includegraphics[scale=0.6]{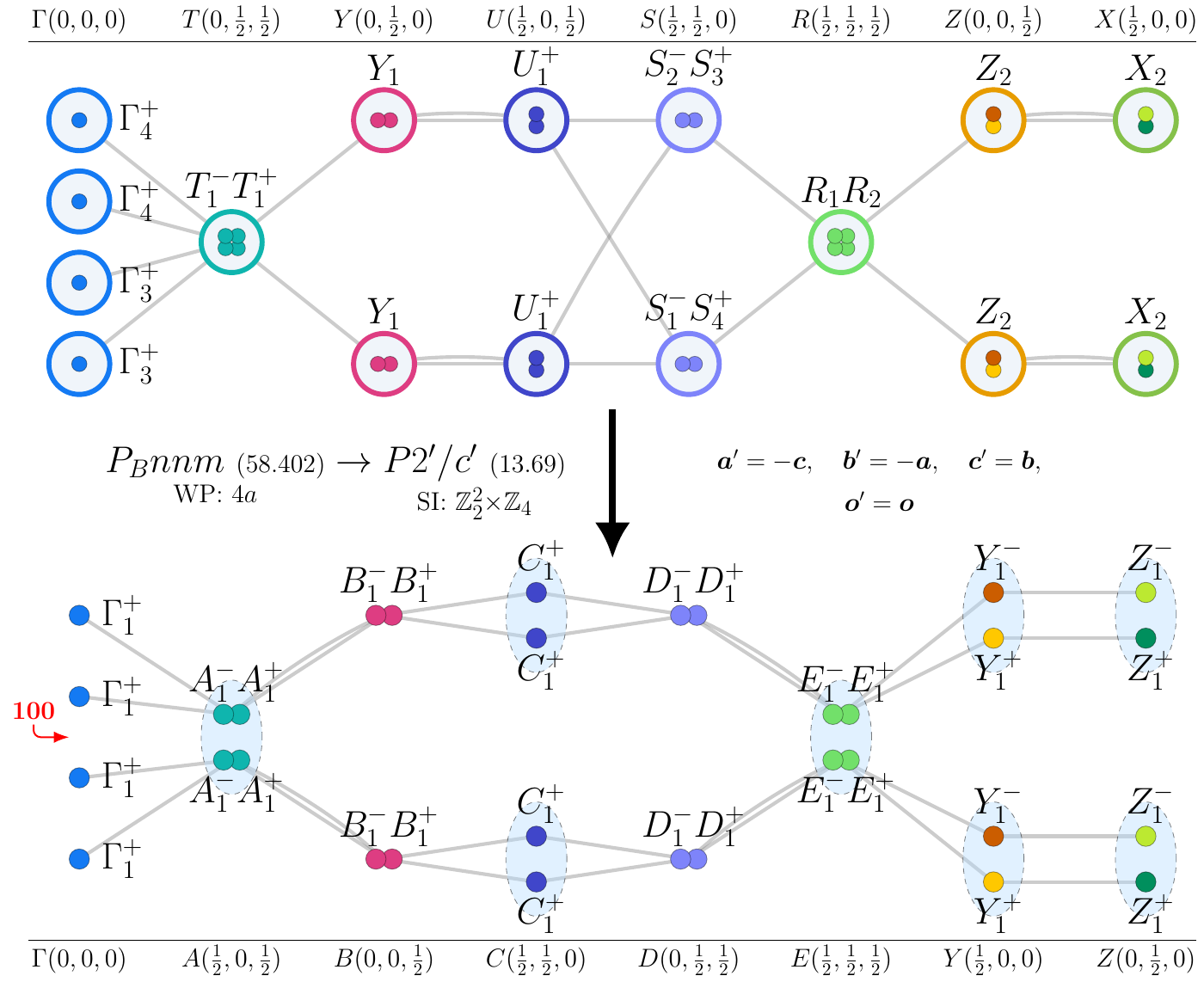}
\caption{Topological magnon bands in subgroup $P2'/c'~(13.69)$ for magnetic moments on Wyckoff position $4a$ of supergroup $P_{B}nnm~(58.402)$.\label{fig_58.402_13.69_Bparallel010andstrainperp100_4a}}
\end{figure}
\input{gap_tables_tex/58.402_13.69_Bparallel010andstrainperp100_4a_table.tex}
\input{si_tables_tex/58.402_13.69_Bparallel010andstrainperp100_4a_table.tex}
\subsubsection{Topological bands in subgroup $P_{S}\bar{1}~(2.7)$}
\textbf{Perturbation:}
\begin{itemize}
\item strain in generic direction.
\end{itemize}
\begin{figure}[H]
\centering
\includegraphics[scale=0.6]{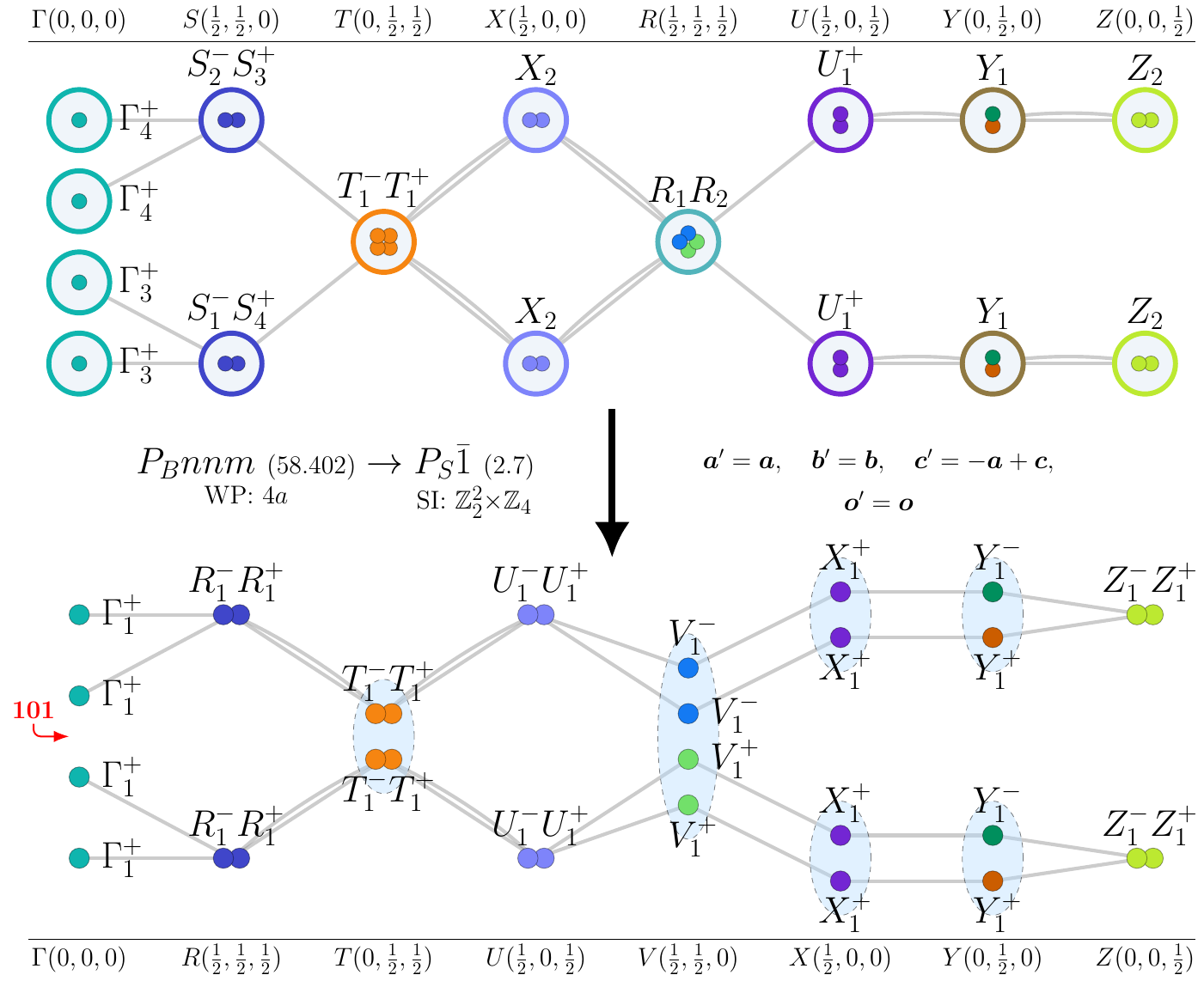}
\caption{Topological magnon bands in subgroup $P_{S}\bar{1}~(2.7)$ for magnetic moments on Wyckoff position $4a$ of supergroup $P_{B}nnm~(58.402)$.\label{fig_58.402_2.7_strainingenericdirection_4a}}
\end{figure}
\input{gap_tables_tex/58.402_2.7_strainingenericdirection_4a_table.tex}
\input{si_tables_tex/58.402_2.7_strainingenericdirection_4a_table.tex}
\subsubsection{Topological bands in subgroup $Pm'a'2~(28.91)$}
\textbf{Perturbation:}
\begin{itemize}
\item E $\parallel$ [001] and B $\parallel$ [001].
\end{itemize}
\begin{figure}[H]
\centering
\includegraphics[scale=0.6]{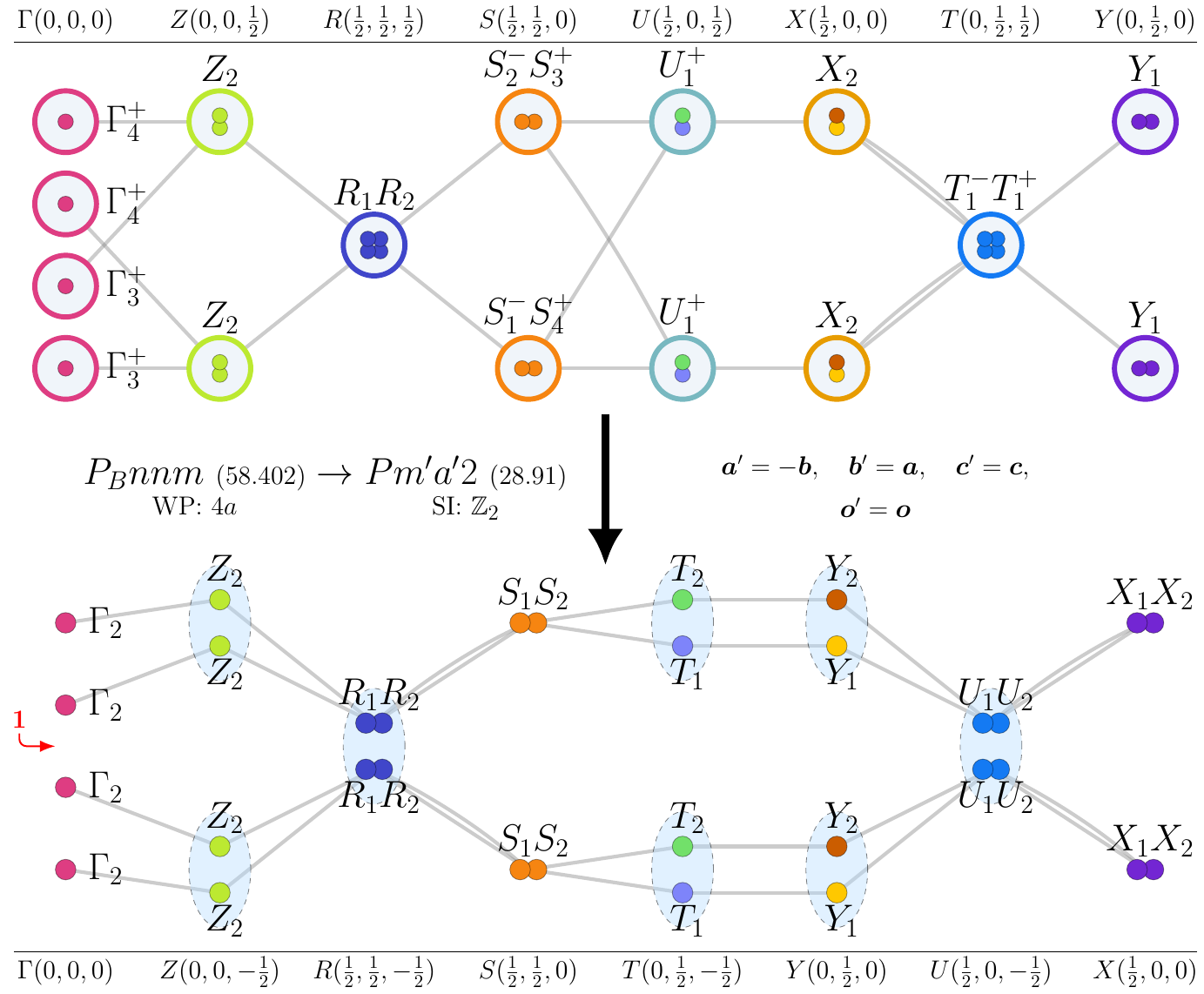}
\caption{Topological magnon bands in subgroup $Pm'a'2~(28.91)$ for magnetic moments on Wyckoff position $4a$ of supergroup $P_{B}nnm~(58.402)$.\label{fig_58.402_28.91_Eparallel001andBparallel001_4a}}
\end{figure}
\input{gap_tables_tex/58.402_28.91_Eparallel001andBparallel001_4a_table.tex}
\input{si_tables_tex/58.402_28.91_Eparallel001andBparallel001_4a_table.tex}

\section{MSG $Pm'm'n~(59.409)$}
\textbf{Nontrivial-SI Subgroups:} $P\bar{1}~(2.4)$, $P2_{1}'/m'~(11.54)$, $P2_{1}'/m'~(11.54)$, $P2~(3.1)$, $Pm'm'2~(25.60)$, $P2/c~(13.65)$.\\

\textbf{Trivial-SI Subgroups:} $Pm'~(6.20)$, $Pm'~(6.20)$, $P2_{1}'~(4.9)$, $P2_{1}'~(4.9)$, $Pc~(7.24)$, $Pm'n2_{1}'~(31.125)$, $Pm'n2_{1}'~(31.125)$.\\

\subsection{WP: $2b+2b+4c+4d$}
\textbf{BCS Materials:} {TmMn\textsubscript{3}O\textsubscript{6}~(74 K)}\footnote{BCS web page: \texttt{\href{http://webbdcrista1.ehu.es/magndata/index.php?this\_label=0.232} {http://webbdcrista1.ehu.es/magndata/index.php?this\_label=0.232}}}.\\
\subsubsection{Topological bands in subgroup $P2_{1}'/m'~(11.54)$}
\textbf{Perturbations:}
\begin{itemize}
\item strain $\perp$ [010],
\item (B $\parallel$ [100] or B $\perp$ [010]).
\end{itemize}
\begin{figure}[H]
\centering
\includegraphics[scale=0.6]{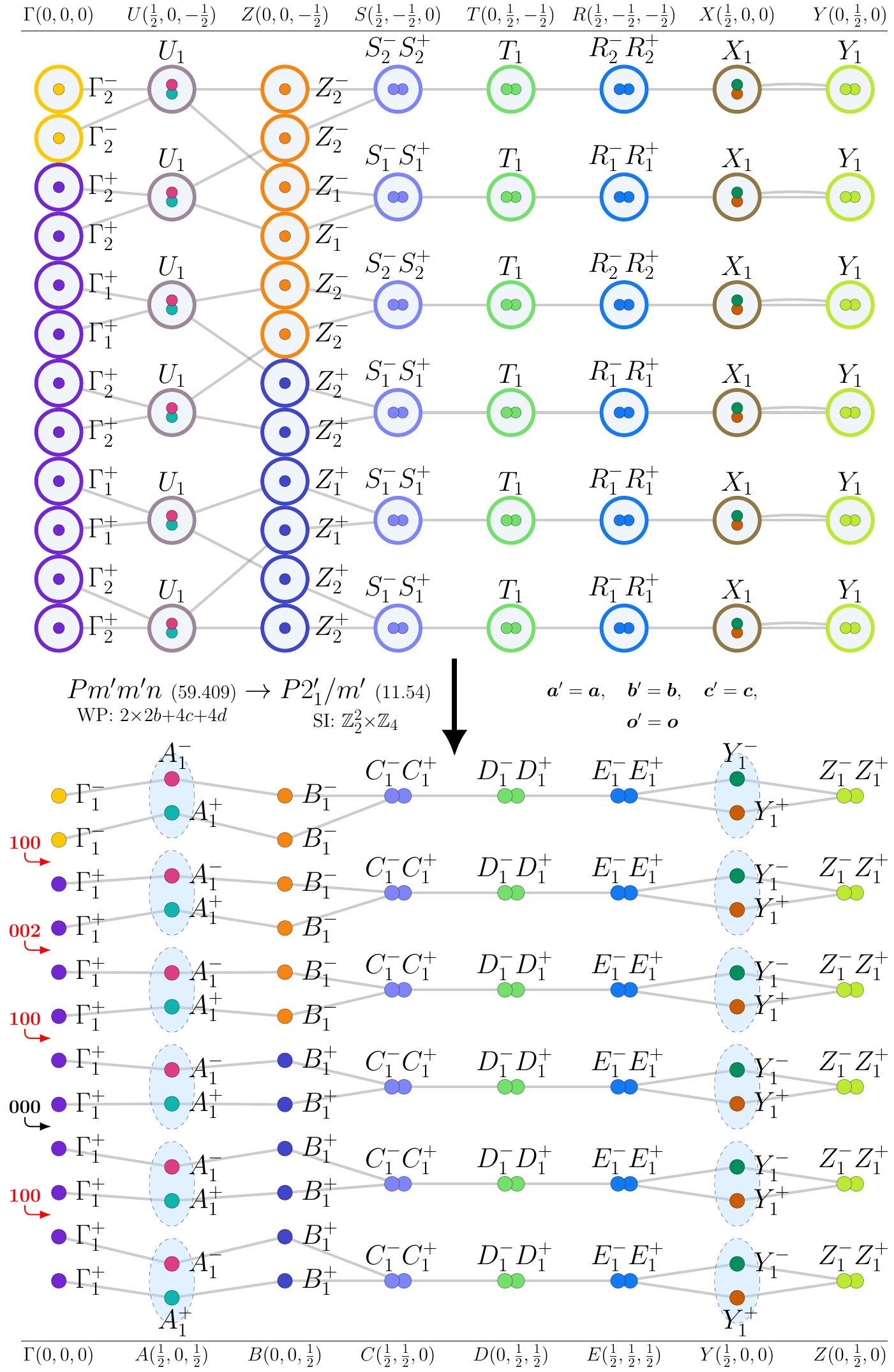}
\caption{Topological magnon bands in subgroup $P2_{1}'/m'~(11.54)$ for magnetic moments on Wyckoff positions $2b+2b+4c+4d$ of supergroup $Pm'm'n~(59.409)$.\label{fig_59.409_11.54_strainperp010_2b+2b+4c+4d}}
\end{figure}
\input{gap_tables_tex/59.409_11.54_strainperp010_2b+2b+4c+4d_table.tex}
\input{si_tables_tex/59.409_11.54_strainperp010_2b+2b+4c+4d_table.tex}
\subsubsection{Topological bands in subgroup $P2_{1}'/m'~(11.54)$}
\textbf{Perturbations:}
\begin{itemize}
\item strain $\perp$ [100],
\item (B $\parallel$ [010] or B $\perp$ [100]).
\end{itemize}
\begin{figure}[H]
\centering
\includegraphics[scale=0.6]{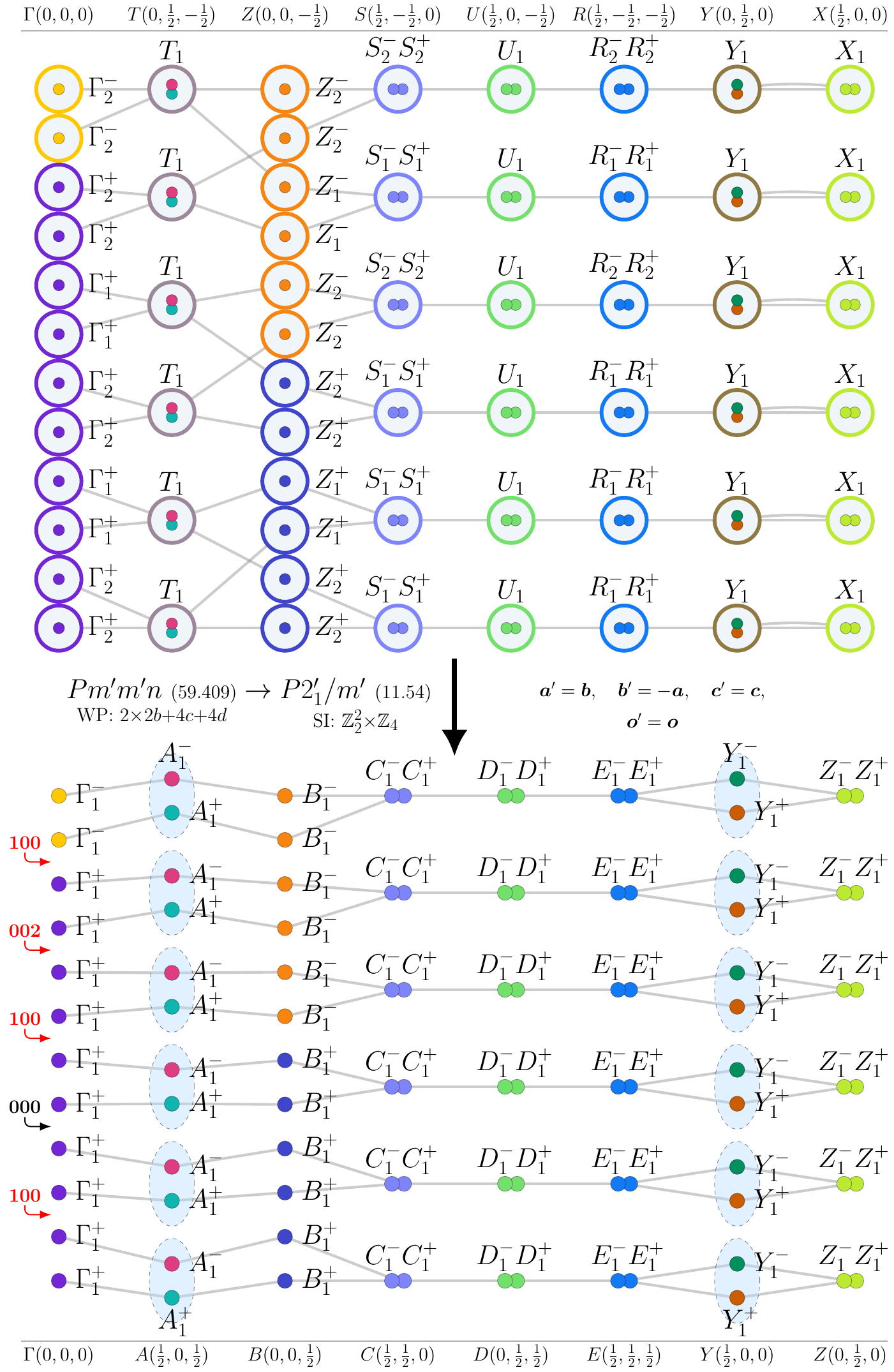}
\caption{Topological magnon bands in subgroup $P2_{1}'/m'~(11.54)$ for magnetic moments on Wyckoff positions $2b+2b+4c+4d$ of supergroup $Pm'm'n~(59.409)$.\label{fig_59.409_11.54_strainperp100_2b+2b+4c+4d}}
\end{figure}
\input{gap_tables_tex/59.409_11.54_strainperp100_2b+2b+4c+4d_table.tex}
\input{si_tables_tex/59.409_11.54_strainperp100_2b+2b+4c+4d_table.tex}

\section{MSG $Pmm'n'~(59.410)$}
\textbf{Nontrivial-SI Subgroups:} $P\bar{1}~(2.4)$, $P2'/c'~(13.69)$, $P2_{1}'/m'~(11.54)$, $P2_{1}/m~(11.50)$.\\

\textbf{Trivial-SI Subgroups:} $Pc'~(7.26)$, $Pm'~(6.20)$, $P2'~(3.3)$, $P2_{1}'~(4.9)$, $Pm~(6.18)$, $Pm'm2'~(25.59)$, $Pmn'2_{1}'~(31.126)$, $P2_{1}~(4.7)$, $Pm'n'2_{1}~(31.127)$.\\

\subsection{WP: $4c$}
\textbf{BCS Materials:} {CsMnF\textsubscript{4}~(18.9 K)}\footnote{BCS web page: \texttt{\href{http://webbdcrista1.ehu.es/magndata/index.php?this\_label=0.327} {http://webbdcrista1.ehu.es/magndata/index.php?this\_label=0.327}}}.\\
\subsubsection{Topological bands in subgroup $P\bar{1}~(2.4)$}
\textbf{Perturbations:}
\begin{itemize}
\item strain in generic direction,
\item (B $\parallel$ [010] or B $\perp$ [001]) and strain $\perp$ [100],
\item (B $\parallel$ [010] or B $\perp$ [001]) and strain $\perp$ [010],
\item (B $\parallel$ [001] or B $\perp$ [010]) and strain $\perp$ [100],
\item (B $\parallel$ [001] or B $\perp$ [010]) and strain $\perp$ [001],
\item B in generic direction.
\end{itemize}
\begin{figure}[H]
\centering
\includegraphics[scale=0.6]{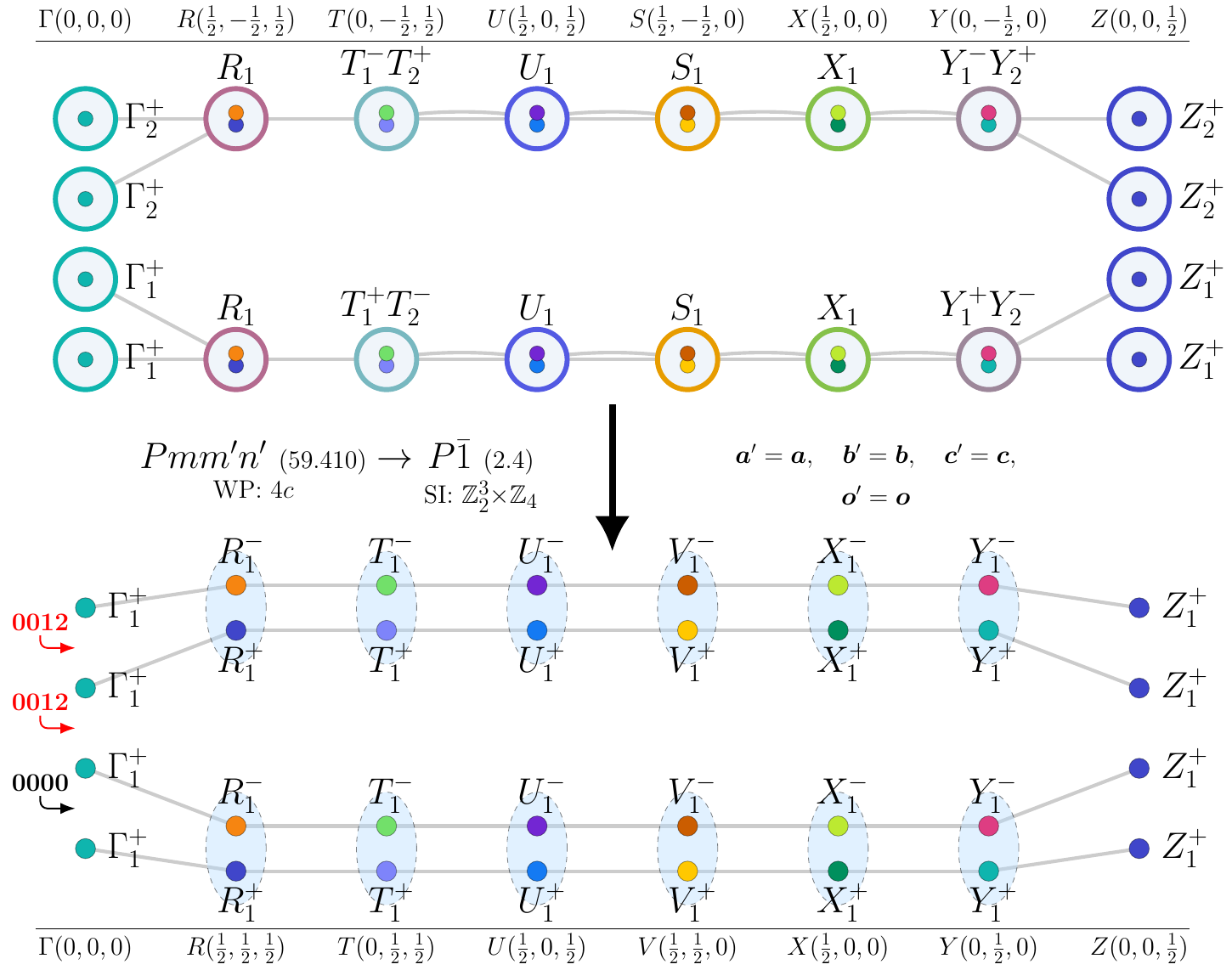}
\caption{Topological magnon bands in subgroup $P\bar{1}~(2.4)$ for magnetic moments on Wyckoff position $4c$ of supergroup $Pmm'n'~(59.410)$.\label{fig_59.410_2.4_strainingenericdirection_4c}}
\end{figure}
\input{gap_tables_tex/59.410_2.4_strainingenericdirection_4c_table.tex}
\input{si_tables_tex/59.410_2.4_strainingenericdirection_4c_table.tex}
\subsubsection{Topological bands in subgroup $P2'/c'~(13.69)$}
\textbf{Perturbations:}
\begin{itemize}
\item strain $\perp$ [001],
\item (B $\parallel$ [010] or B $\perp$ [001]).
\end{itemize}
\begin{figure}[H]
\centering
\includegraphics[scale=0.6]{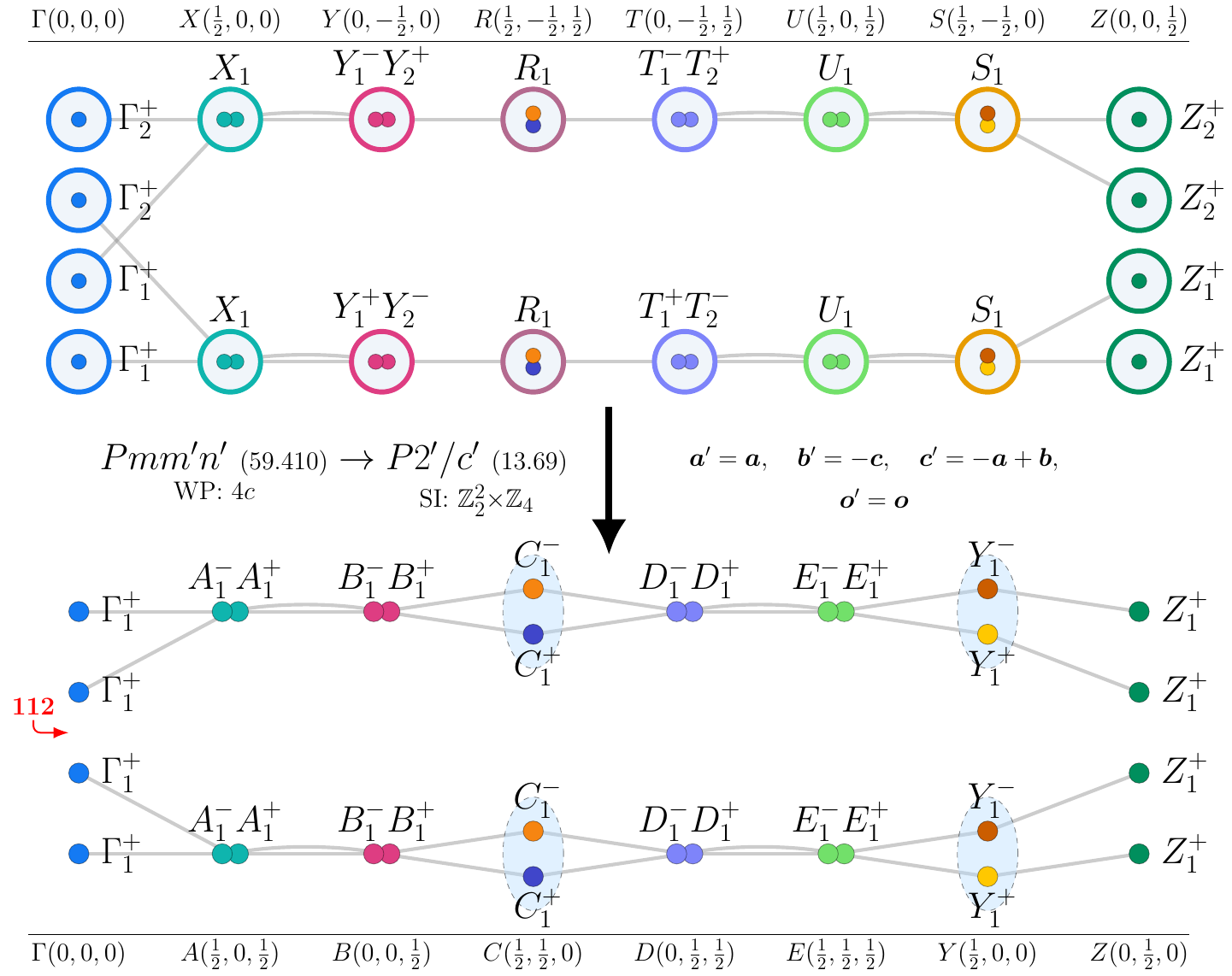}
\caption{Topological magnon bands in subgroup $P2'/c'~(13.69)$ for magnetic moments on Wyckoff position $4c$ of supergroup $Pmm'n'~(59.410)$.\label{fig_59.410_13.69_strainperp001_4c}}
\end{figure}
\input{gap_tables_tex/59.410_13.69_strainperp001_4c_table.tex}
\input{si_tables_tex/59.410_13.69_strainperp001_4c_table.tex}
\subsubsection{Topological bands in subgroup $P2_{1}'/m'~(11.54)$}
\textbf{Perturbations:}
\begin{itemize}
\item strain $\perp$ [010],
\item (B $\parallel$ [001] or B $\perp$ [010]).
\end{itemize}
\begin{figure}[H]
\centering
\includegraphics[scale=0.6]{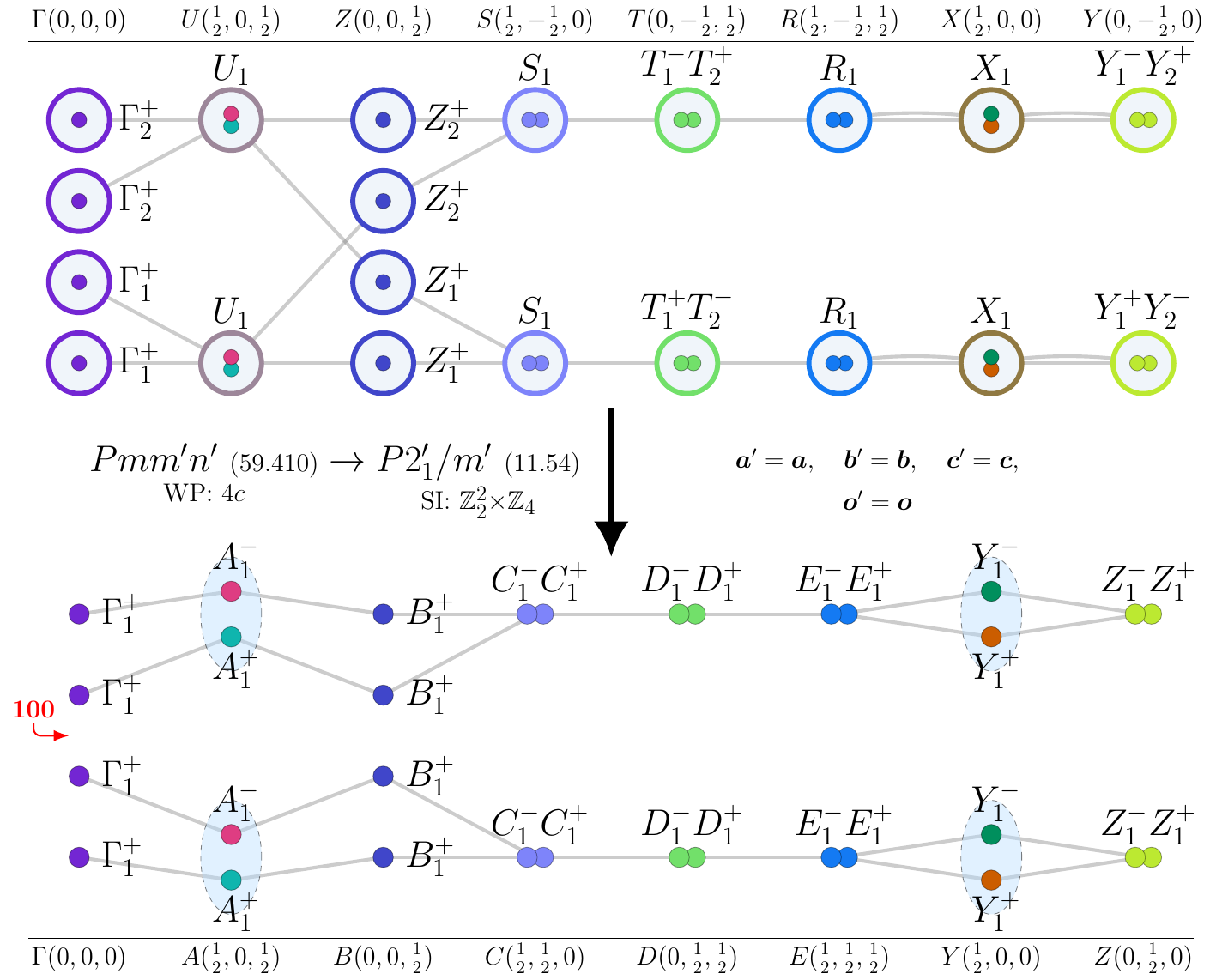}
\caption{Topological magnon bands in subgroup $P2_{1}'/m'~(11.54)$ for magnetic moments on Wyckoff position $4c$ of supergroup $Pmm'n'~(59.410)$.\label{fig_59.410_11.54_strainperp010_4c}}
\end{figure}
\input{gap_tables_tex/59.410_11.54_strainperp010_4c_table.tex}
\input{si_tables_tex/59.410_11.54_strainperp010_4c_table.tex}

\section{MSG $P_{C}bcn~(60.431)$}
\textbf{Nontrivial-SI Subgroups:} $P\bar{1}~(2.4)$, $P2_{1}'/m'~(11.54)$, $P2_{1}'/c'~(14.79)$, $P2'/m'~(10.46)$, $P_{S}\bar{1}~(2.7)$, $P2_{1}/c~(14.75)$, $Pnn'm'~(58.398)$, $P_{c}2_{1}/c~(14.82)$, $P2~(3.1)$, $Pm'm'2~(25.60)$, $P2/c~(13.65)$, $Pm'm'a~(51.294)$, $P_{C}2/c~(13.74)$, $P2_{1}/c~(14.75)$, $Pn'm'a~(62.446)$, $P_{A}2_{1}/c~(14.83)$.\\

\textbf{Trivial-SI Subgroups:} $Pm'~(6.20)$, $Pc'~(7.26)$, $Pm'~(6.20)$, $P2_{1}'~(4.9)$, $P2_{1}'~(4.9)$, $P2'~(3.3)$, $P_{S}1~(1.3)$, $Pc~(7.24)$, $Pm'n2_{1}'~(31.125)$, $Pn'n2'~(34.158)$, $P_{c}c~(7.28)$, $Pc~(7.24)$, $Pm'c2_{1}'~(26.68)$, $Pm'a2'~(28.89)$, $P_{C}c~(7.30)$, $Pc~(7.24)$, $Pn'a2_{1}'~(33.146)$, $Pm'c2_{1}'~(26.68)$, $P_{A}c~(7.31)$, $P2_{1}~(4.7)$, $Pm'n'2_{1}~(31.127)$, $P_{a}2_{1}~(4.10)$, $P_{C}ca2_{1}~(29.109)$, $P_{C}2~(3.6)$, $P_{A}nc2~(30.119)$, $P2_{1}~(4.7)$, $Pm'n'2_{1}~(31.127)$, $P_{C}2_{1}~(4.12)$, $P_{A}na2_{1}~(33.152)$.\\

\subsection{WP: $8g$}
\textbf{BCS Materials:} {Mn\textsubscript{5}Si\textsubscript{3}~(100 K)}\footnote{BCS web page: \texttt{\href{http://webbdcrista1.ehu.es/magndata/index.php?this\_label=1.88} {http://webbdcrista1.ehu.es/magndata/index.php?this\_label=1.88}}}, {Mn\textsubscript{5}Si\textsubscript{3}~(99 K)}\footnote{BCS web page: \texttt{\href{http://webbdcrista1.ehu.es/magndata/index.php?this\_label=1.305} {http://webbdcrista1.ehu.es/magndata/index.php?this\_label=1.305}}}.\\
\subsubsection{Topological bands in subgroup $P2_{1}'/m'~(11.54)$}
\textbf{Perturbations:}
\begin{itemize}
\item B $\parallel$ [100] and strain $\perp$ [001],
\item B $\parallel$ [010] and strain $\perp$ [001],
\item B $\perp$ [001].
\end{itemize}
\begin{figure}[H]
\centering
\includegraphics[scale=0.6]{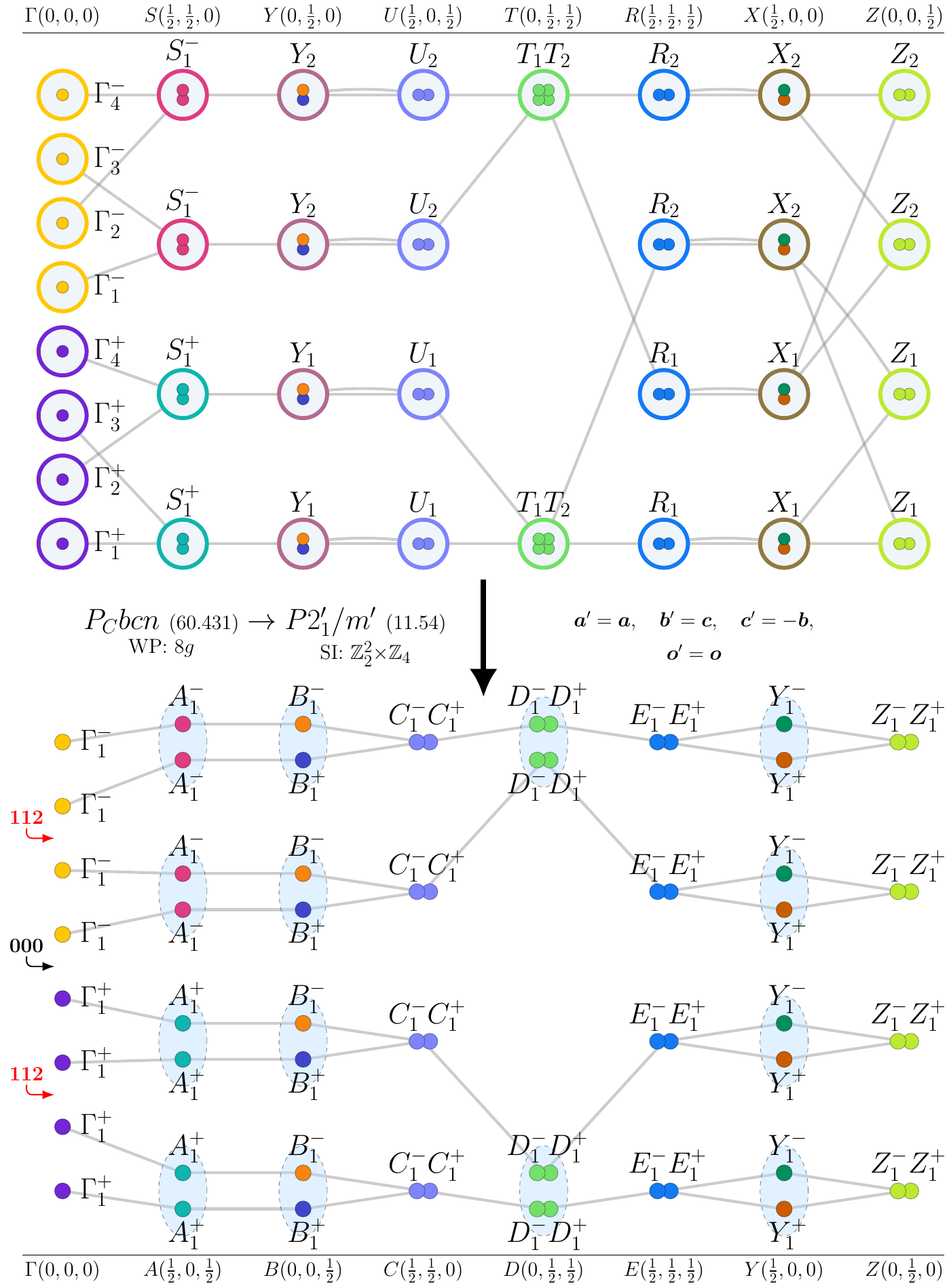}
\caption{Topological magnon bands in subgroup $P2_{1}'/m'~(11.54)$ for magnetic moments on Wyckoff position $8g$ of supergroup $P_{C}bcn~(60.431)$.\label{fig_60.431_11.54_Bparallel100andstrainperp001_8g}}
\end{figure}
\input{gap_tables_tex/60.431_11.54_Bparallel100andstrainperp001_8g_table.tex}
\input{si_tables_tex/60.431_11.54_Bparallel100andstrainperp001_8g_table.tex}
\subsubsection{Topological bands in subgroup $P2_{1}'/c'~(14.79)$}
\textbf{Perturbations:}
\begin{itemize}
\item B $\parallel$ [100] and strain $\perp$ [010],
\item B $\parallel$ [001] and strain $\perp$ [010],
\item B $\perp$ [010].
\end{itemize}
\begin{figure}[H]
\centering
\includegraphics[scale=0.6]{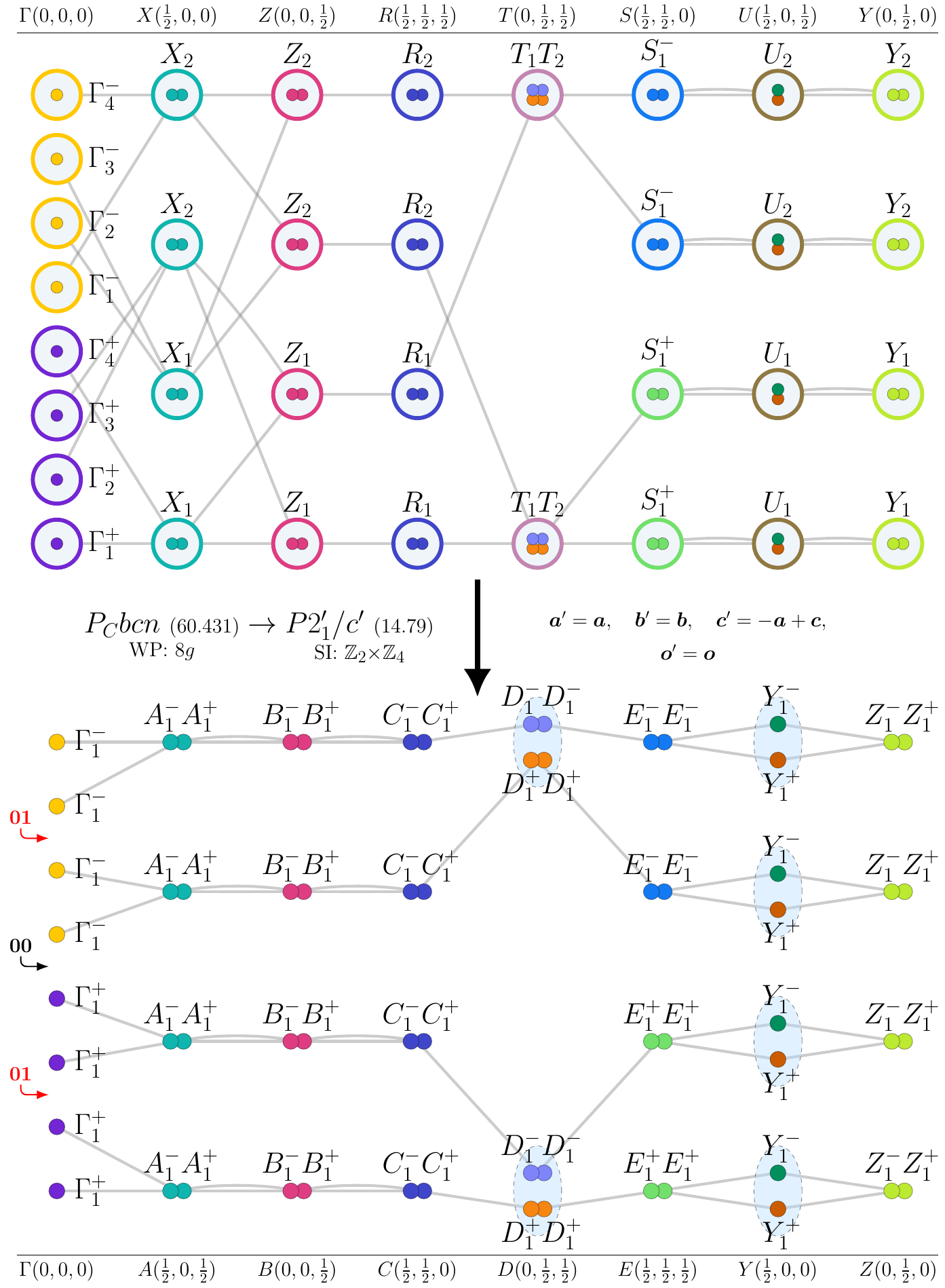}
\caption{Topological magnon bands in subgroup $P2_{1}'/c'~(14.79)$ for magnetic moments on Wyckoff position $8g$ of supergroup $P_{C}bcn~(60.431)$.\label{fig_60.431_14.79_Bparallel100andstrainperp010_8g}}
\end{figure}
\input{gap_tables_tex/60.431_14.79_Bparallel100andstrainperp010_8g_table.tex}
\input{si_tables_tex/60.431_14.79_Bparallel100andstrainperp010_8g_table.tex}
\subsubsection{Topological bands in subgroup $P_{S}\bar{1}~(2.7)$}
\textbf{Perturbation:}
\begin{itemize}
\item strain in generic direction.
\end{itemize}
\begin{figure}[H]
\centering
\includegraphics[scale=0.6]{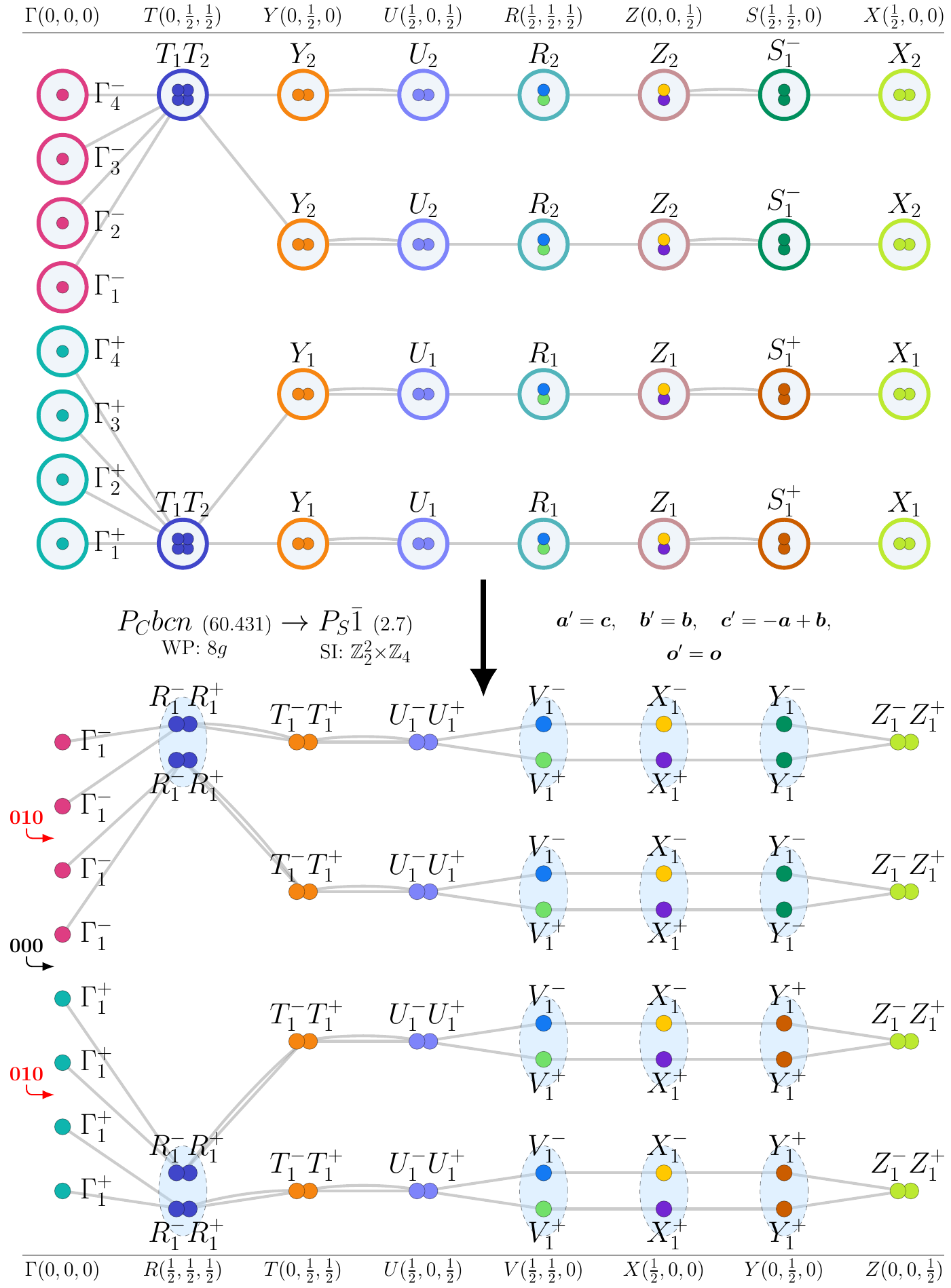}
\caption{Topological magnon bands in subgroup $P_{S}\bar{1}~(2.7)$ for magnetic moments on Wyckoff position $8g$ of supergroup $P_{C}bcn~(60.431)$.\label{fig_60.431_2.7_strainingenericdirection_8g}}
\end{figure}
\input{gap_tables_tex/60.431_2.7_strainingenericdirection_8g_table.tex}
\input{si_tables_tex/60.431_2.7_strainingenericdirection_8g_table.tex}
\subsection{WP: $8e$}
\textbf{BCS Materials:} {FeSn\textsubscript{2}~(93 K)}\footnote{BCS web page: \texttt{\href{http://webbdcrista1.ehu.es/magndata/index.php?this\_label=2.67} {http://webbdcrista1.ehu.es/magndata/index.php?this\_label=2.67}}}.\\
\subsubsection{Topological bands in subgroup $P2_{1}'/m'~(11.54)$}
\textbf{Perturbations:}
\begin{itemize}
\item B $\parallel$ [100] and strain $\perp$ [001],
\item B $\parallel$ [010] and strain $\perp$ [001],
\item B $\perp$ [001].
\end{itemize}
\begin{figure}[H]
\centering
\includegraphics[scale=0.6]{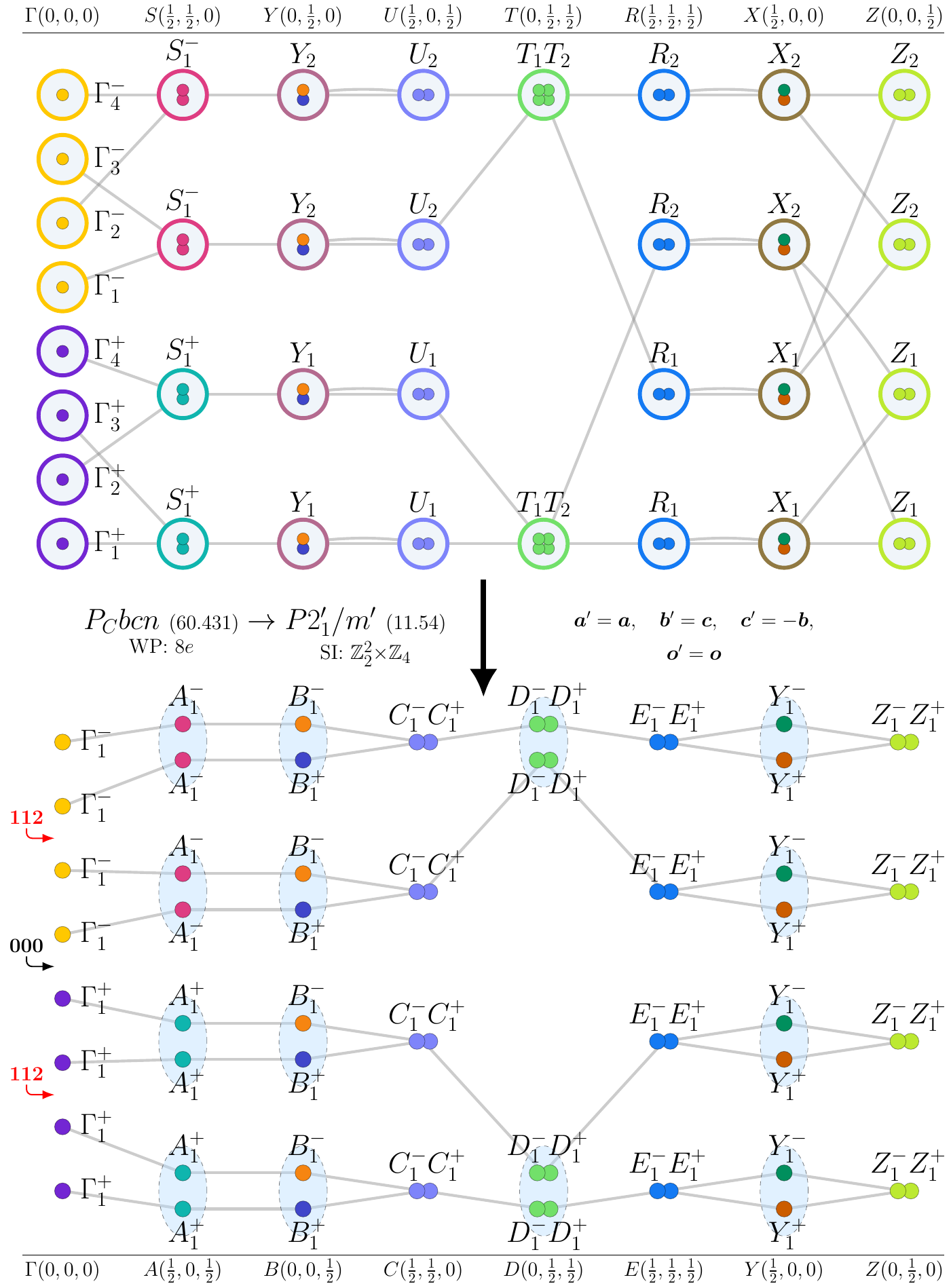}
\caption{Topological magnon bands in subgroup $P2_{1}'/m'~(11.54)$ for magnetic moments on Wyckoff position $8e$ of supergroup $P_{C}bcn~(60.431)$.\label{fig_60.431_11.54_Bparallel100andstrainperp001_8e}}
\end{figure}
\input{gap_tables_tex/60.431_11.54_Bparallel100andstrainperp001_8e_table.tex}
\input{si_tables_tex/60.431_11.54_Bparallel100andstrainperp001_8e_table.tex}
\subsubsection{Topological bands in subgroup $P2_{1}'/c'~(14.79)$}
\textbf{Perturbations:}
\begin{itemize}
\item B $\parallel$ [100] and strain $\perp$ [010],
\item B $\parallel$ [001] and strain $\perp$ [010],
\item B $\perp$ [010].
\end{itemize}
\begin{figure}[H]
\centering
\includegraphics[scale=0.6]{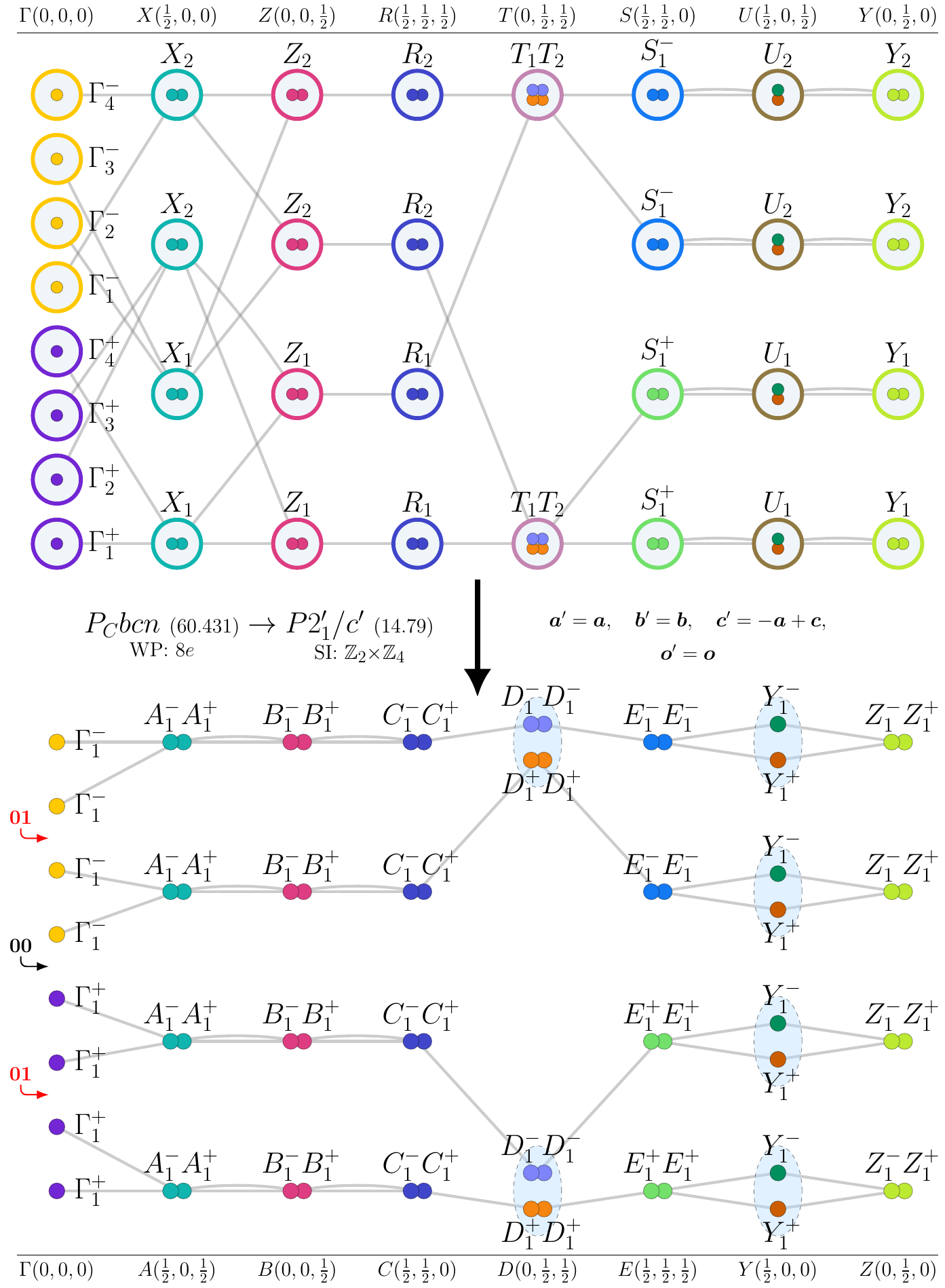}
\caption{Topological magnon bands in subgroup $P2_{1}'/c'~(14.79)$ for magnetic moments on Wyckoff position $8e$ of supergroup $P_{C}bcn~(60.431)$.\label{fig_60.431_14.79_Bparallel100andstrainperp010_8e}}
\end{figure}
\input{gap_tables_tex/60.431_14.79_Bparallel100andstrainperp010_8e_table.tex}
\input{si_tables_tex/60.431_14.79_Bparallel100andstrainperp010_8e_table.tex}
\subsubsection{Topological bands in subgroup $P_{S}\bar{1}~(2.7)$}
\textbf{Perturbation:}
\begin{itemize}
\item strain in generic direction.
\end{itemize}
\begin{figure}[H]
\centering
\includegraphics[scale=0.6]{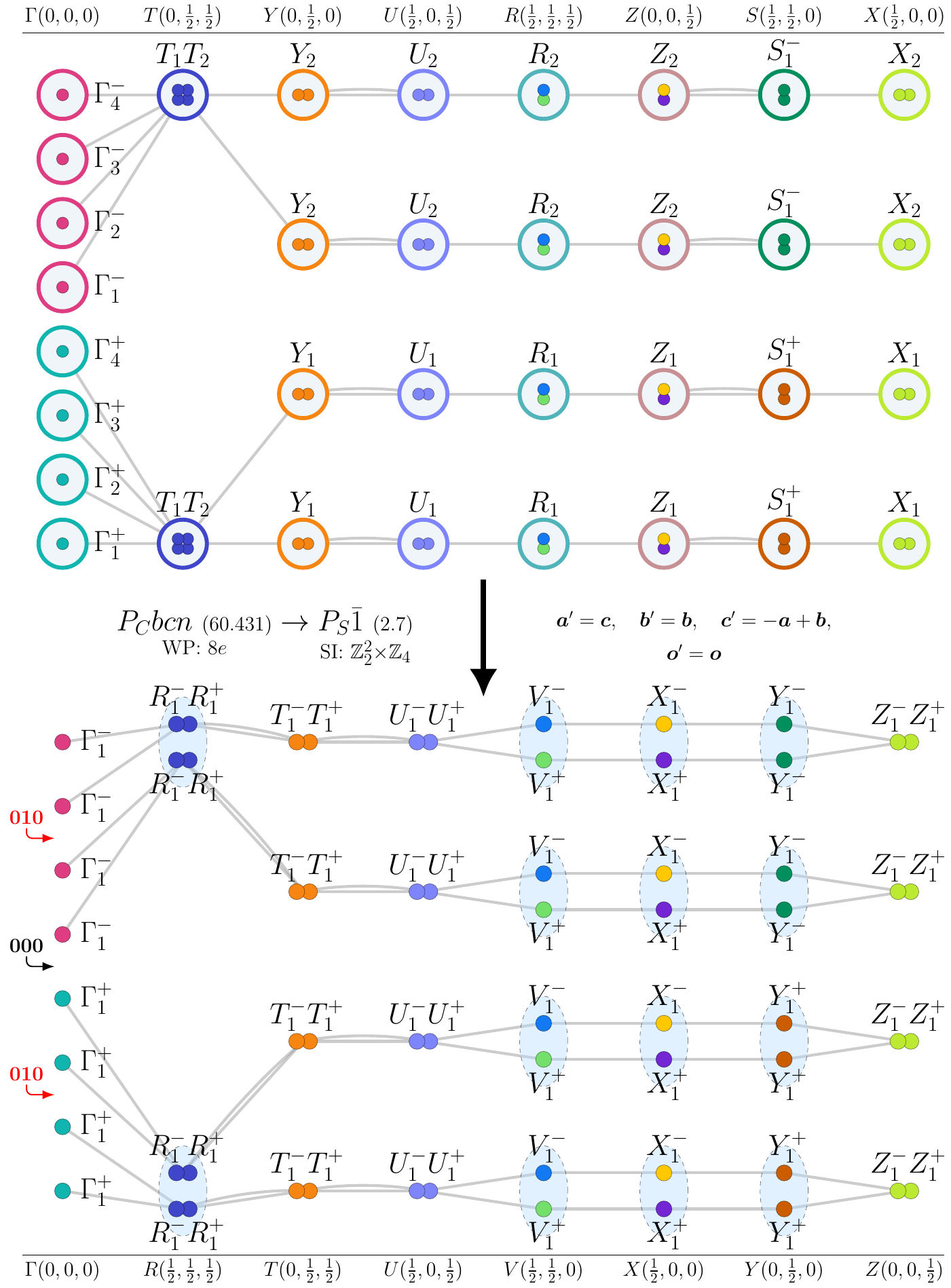}
\caption{Topological magnon bands in subgroup $P_{S}\bar{1}~(2.7)$ for magnetic moments on Wyckoff position $8e$ of supergroup $P_{C}bcn~(60.431)$.\label{fig_60.431_2.7_strainingenericdirection_8e}}
\end{figure}
\input{gap_tables_tex/60.431_2.7_strainingenericdirection_8e_table.tex}
\input{si_tables_tex/60.431_2.7_strainingenericdirection_8e_table.tex}
\subsection{WP: $4a$}
\textbf{BCS Materials:} {NiSO\textsubscript{4}~(37 K)}\footnote{BCS web page: \texttt{\href{http://webbdcrista1.ehu.es/magndata/index.php?this\_label=1.520} {http://webbdcrista1.ehu.es/magndata/index.php?this\_label=1.520}}}, {FeSO\textsubscript{4}~(21 K)}\footnote{BCS web page: \texttt{\href{http://webbdcrista1.ehu.es/magndata/index.php?this\_label=1.521} {http://webbdcrista1.ehu.es/magndata/index.php?this\_label=1.521}}}, {CoSO\textsubscript{4}~(15.5 K)}\footnote{BCS web page: \texttt{\href{http://webbdcrista1.ehu.es/magndata/index.php?this\_label=1.519} {http://webbdcrista1.ehu.es/magndata/index.php?this\_label=1.519}}}, {PrCuSi~(5.1 K)}\footnote{BCS web page: \texttt{\href{http://webbdcrista1.ehu.es/magndata/index.php?this\_label=1.460} {http://webbdcrista1.ehu.es/magndata/index.php?this\_label=1.460}}}.\\
\subsubsection{Topological bands in subgroup $P\bar{1}~(2.4)$}
\textbf{Perturbations:}
\begin{itemize}
\item B $\parallel$ [100] and strain in generic direction,
\item B $\parallel$ [010] and strain in generic direction,
\item B $\parallel$ [001] and strain in generic direction,
\item B $\perp$ [100] and strain $\perp$ [010],
\item B $\perp$ [100] and strain $\perp$ [001],
\item B $\perp$ [100] and strain in generic direction,
\item B $\perp$ [010] and strain $\perp$ [100],
\item B $\perp$ [010] and strain $\perp$ [001],
\item B $\perp$ [010] and strain in generic direction,
\item B $\perp$ [001] and strain $\perp$ [100],
\item B $\perp$ [001] and strain $\perp$ [010],
\item B $\perp$ [001] and strain in generic direction,
\item B in generic direction.
\end{itemize}
\begin{figure}[H]
\centering
\includegraphics[scale=0.6]{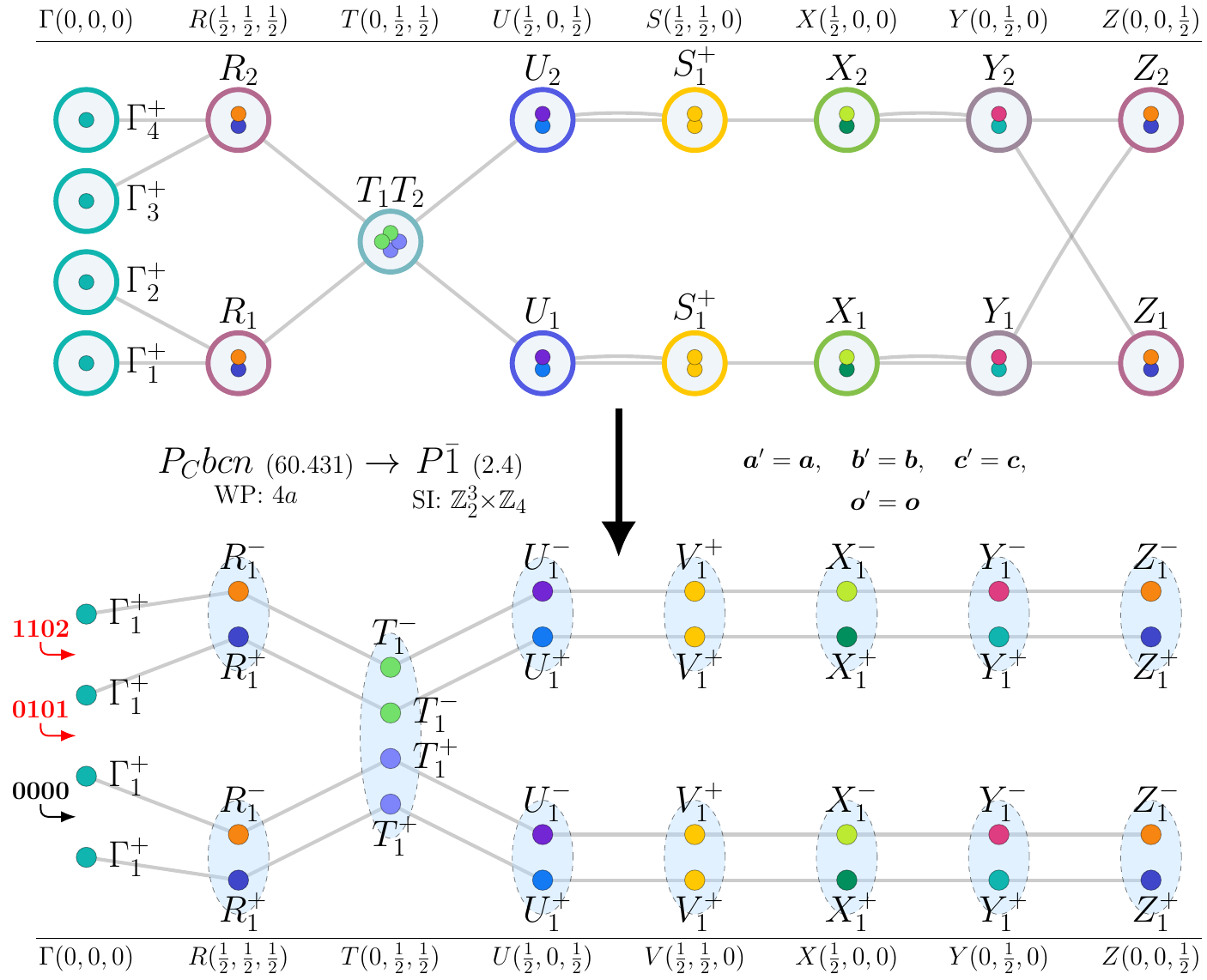}
\caption{Topological magnon bands in subgroup $P\bar{1}~(2.4)$ for magnetic moments on Wyckoff position $4a$ of supergroup $P_{C}bcn~(60.431)$.\label{fig_60.431_2.4_Bparallel100andstrainingenericdirection_4a}}
\end{figure}
\input{gap_tables_tex/60.431_2.4_Bparallel100andstrainingenericdirection_4a_table.tex}
\input{si_tables_tex/60.431_2.4_Bparallel100andstrainingenericdirection_4a_table.tex}
\subsubsection{Topological bands in subgroup $P2_{1}'/m'~(11.54)$}
\textbf{Perturbations:}
\begin{itemize}
\item B $\parallel$ [100] and strain $\perp$ [001],
\item B $\parallel$ [010] and strain $\perp$ [001],
\item B $\perp$ [001].
\end{itemize}
\begin{figure}[H]
\centering
\includegraphics[scale=0.6]{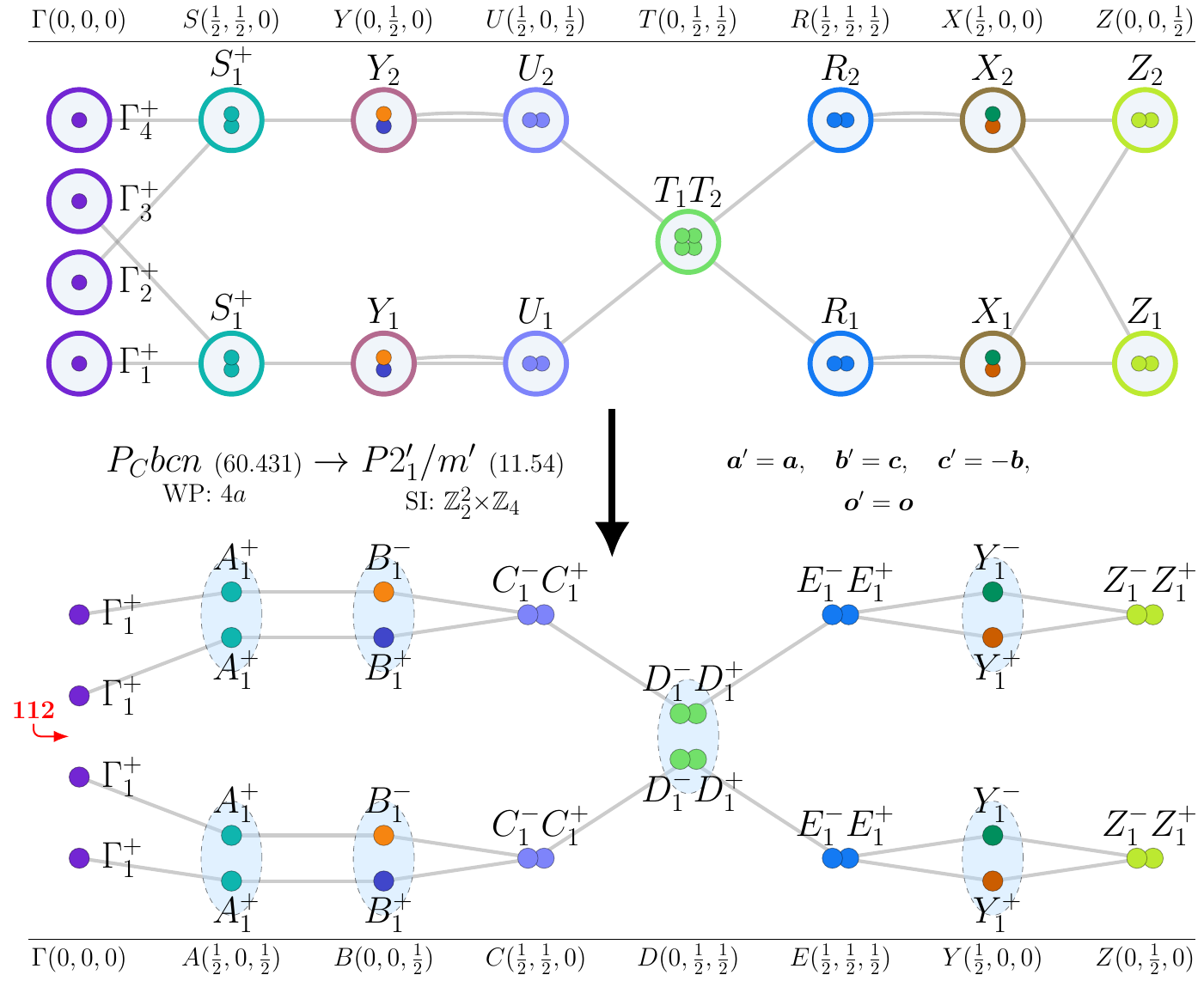}
\caption{Topological magnon bands in subgroup $P2_{1}'/m'~(11.54)$ for magnetic moments on Wyckoff position $4a$ of supergroup $P_{C}bcn~(60.431)$.\label{fig_60.431_11.54_Bparallel100andstrainperp001_4a}}
\end{figure}
\input{gap_tables_tex/60.431_11.54_Bparallel100andstrainperp001_4a_table.tex}
\input{si_tables_tex/60.431_11.54_Bparallel100andstrainperp001_4a_table.tex}
\subsubsection{Topological bands in subgroup $P2_{1}'/c'~(14.79)$}
\textbf{Perturbations:}
\begin{itemize}
\item B $\parallel$ [100] and strain $\perp$ [010],
\item B $\parallel$ [001] and strain $\perp$ [010],
\item B $\perp$ [010].
\end{itemize}
\begin{figure}[H]
\centering
\includegraphics[scale=0.6]{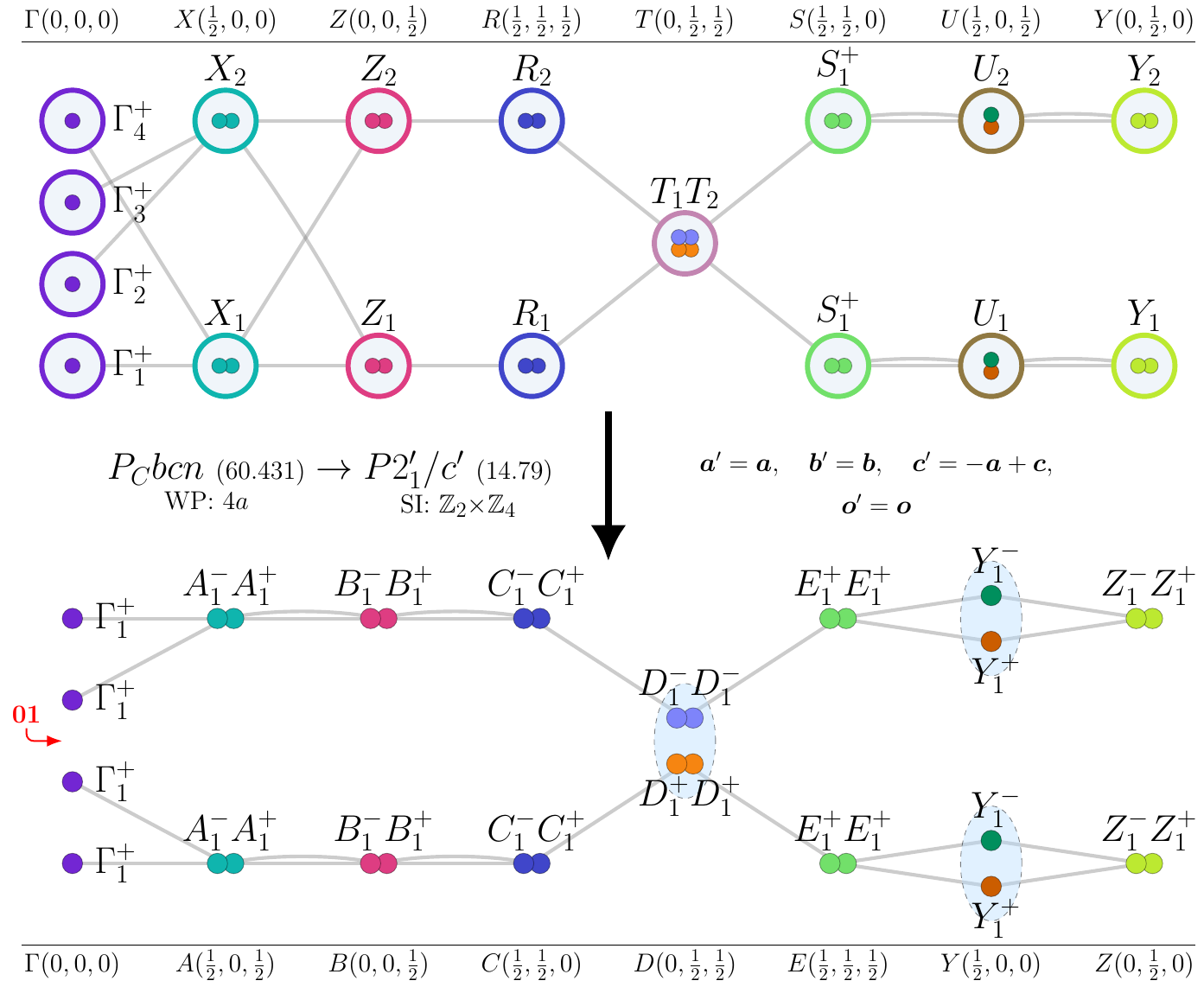}
\caption{Topological magnon bands in subgroup $P2_{1}'/c'~(14.79)$ for magnetic moments on Wyckoff position $4a$ of supergroup $P_{C}bcn~(60.431)$.\label{fig_60.431_14.79_Bparallel100andstrainperp010_4a}}
\end{figure}
\input{gap_tables_tex/60.431_14.79_Bparallel100andstrainperp010_4a_table.tex}
\input{si_tables_tex/60.431_14.79_Bparallel100andstrainperp010_4a_table.tex}
\subsubsection{Topological bands in subgroup $P2'/m'~(10.46)$}
\textbf{Perturbations:}
\begin{itemize}
\item B $\parallel$ [010] and strain $\perp$ [100],
\item B $\parallel$ [001] and strain $\perp$ [100],
\item B $\perp$ [100].
\end{itemize}
\begin{figure}[H]
\centering
\includegraphics[scale=0.6]{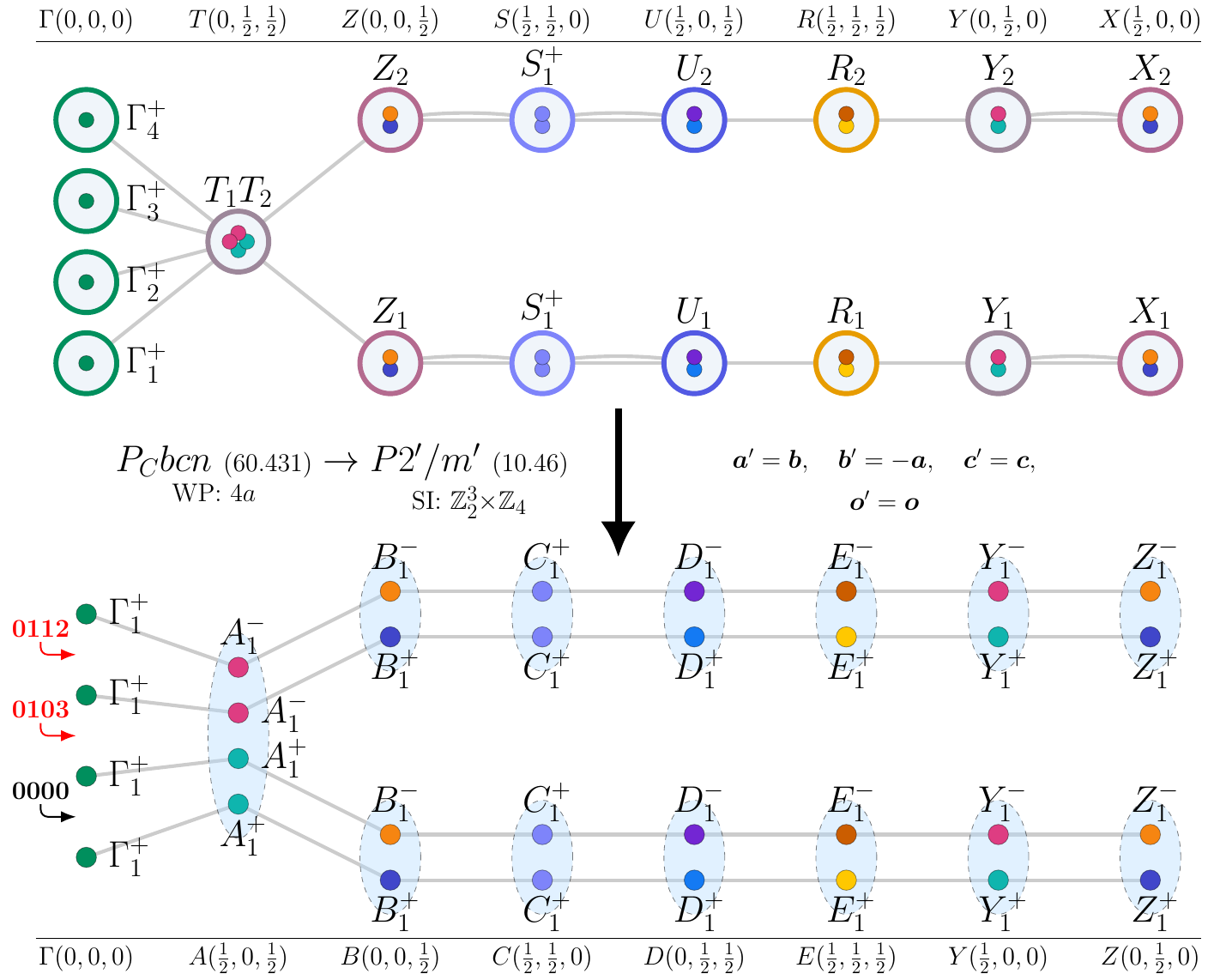}
\caption{Topological magnon bands in subgroup $P2'/m'~(10.46)$ for magnetic moments on Wyckoff position $4a$ of supergroup $P_{C}bcn~(60.431)$.\label{fig_60.431_10.46_Bparallel010andstrainperp100_4a}}
\end{figure}
\input{gap_tables_tex/60.431_10.46_Bparallel010andstrainperp100_4a_table.tex}
\input{si_tables_tex/60.431_10.46_Bparallel010andstrainperp100_4a_table.tex}
\subsubsection{Topological bands in subgroup $P_{S}\bar{1}~(2.7)$}
\textbf{Perturbation:}
\begin{itemize}
\item strain in generic direction.
\end{itemize}
\begin{figure}[H]
\centering
\includegraphics[scale=0.6]{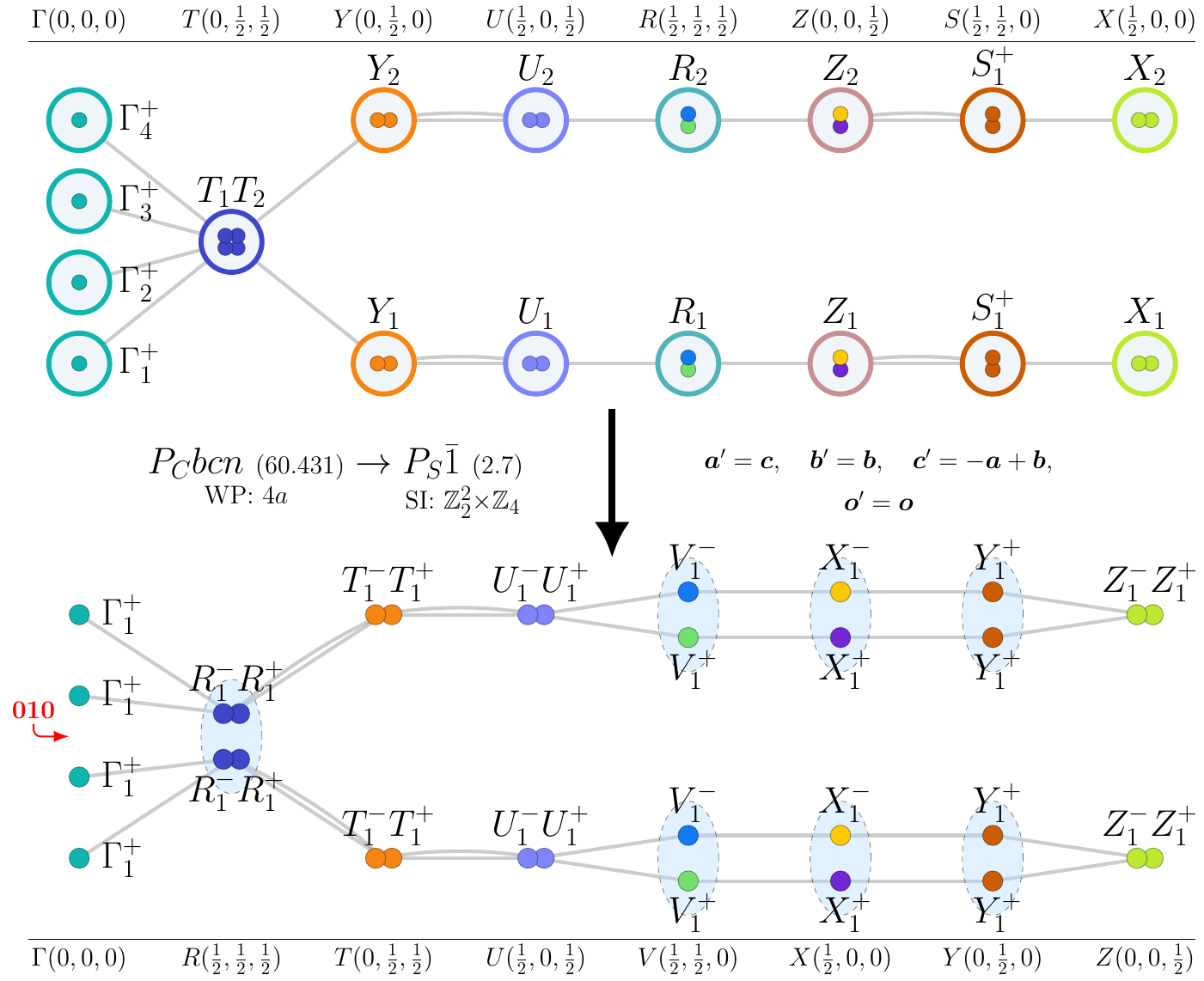}
\caption{Topological magnon bands in subgroup $P_{S}\bar{1}~(2.7)$ for magnetic moments on Wyckoff position $4a$ of supergroup $P_{C}bcn~(60.431)$.\label{fig_60.431_2.7_strainingenericdirection_4a}}
\end{figure}
\input{gap_tables_tex/60.431_2.7_strainingenericdirection_4a_table.tex}
\input{si_tables_tex/60.431_2.7_strainingenericdirection_4a_table.tex}

\section{MSG $P_{I}bcn~(60.432)$}
\textbf{Nontrivial-SI Subgroups:} $P\bar{1}~(2.4)$, $P2'/m'~(10.46)$, $P2_{1}'/c'~(14.79)$, $P2'/c'~(13.69)$, $P_{S}\bar{1}~(2.7)$, $P2_{1}/c~(14.75)$, $Pb'c'n~(60.422)$, $P_{A}2_{1}/c~(14.83)$, $P2~(3.1)$, $Pm'a'2~(28.91)$, $P2/c~(13.65)$, $Pc'cm'~(49.270)$, $P_{C}2/c~(13.74)$, $P2_{1}/c~(14.75)$, $Pb'am'~(55.358)$, $P_{C}2_{1}/c~(14.84)$.\\

\textbf{Trivial-SI Subgroups:} $Pm'~(6.20)$, $Pc'~(7.26)$, $Pc'~(7.26)$, $P2'~(3.3)$, $P2_{1}'~(4.9)$, $P2'~(3.3)$, $P_{S}1~(1.3)$, $Pc~(7.24)$, $Pna'2_{1}'~(33.147)$, $Pnc'2'~(30.114)$, $P_{A}c~(7.31)$, $Pc~(7.24)$, $Pc'c2'~(27.80)$, $Pm'a2'~(28.89)$, $P_{C}c~(7.30)$, $Pc~(7.24)$, $Pb'a2'~(32.137)$, $Pm'c2_{1}'~(26.68)$, $P_{C}c~(7.30)$, $P2_{1}~(4.7)$, $Pc'a'2_{1}~(29.103)$, $P_{C}2_{1}~(4.12)$, $P_{I}ca2_{1}~(29.110)$, $P_{C}2~(3.6)$, $P_{I}nc2~(30.122)$, $P2_{1}~(4.7)$, $Pm'c'2_{1}~(26.70)$, $P_{C}2_{1}~(4.12)$, $P_{I}na2_{1}~(33.155)$.\\

\subsection{WP: $4a$}
\textbf{BCS Materials:} {FeSn\textsubscript{2}~(378 K)}\footnote{BCS web page: \texttt{\href{http://webbdcrista1.ehu.es/magndata/index.php?this\_label=1.556} {http://webbdcrista1.ehu.es/magndata/index.php?this\_label=1.556}}}, {FeGe\textsubscript{2}~(315 K)}\footnote{BCS web page: \texttt{\href{http://webbdcrista1.ehu.es/magndata/index.php?this\_label=1.557} {http://webbdcrista1.ehu.es/magndata/index.php?this\_label=1.557}}}.\\
\subsubsection{Topological bands in subgroup $P\bar{1}~(2.4)$}
\textbf{Perturbations:}
\begin{itemize}
\item B $\parallel$ [100] and strain in generic direction,
\item B $\parallel$ [010] and strain in generic direction,
\item B $\parallel$ [001] and strain in generic direction,
\item B $\perp$ [100] and strain $\perp$ [010],
\item B $\perp$ [100] and strain $\perp$ [001],
\item B $\perp$ [100] and strain in generic direction,
\item B $\perp$ [010] and strain $\perp$ [100],
\item B $\perp$ [010] and strain $\perp$ [001],
\item B $\perp$ [010] and strain in generic direction,
\item B $\perp$ [001] and strain $\perp$ [100],
\item B $\perp$ [001] and strain $\perp$ [010],
\item B $\perp$ [001] and strain in generic direction,
\item B in generic direction.
\end{itemize}
\begin{figure}[H]
\centering
\includegraphics[scale=0.6]{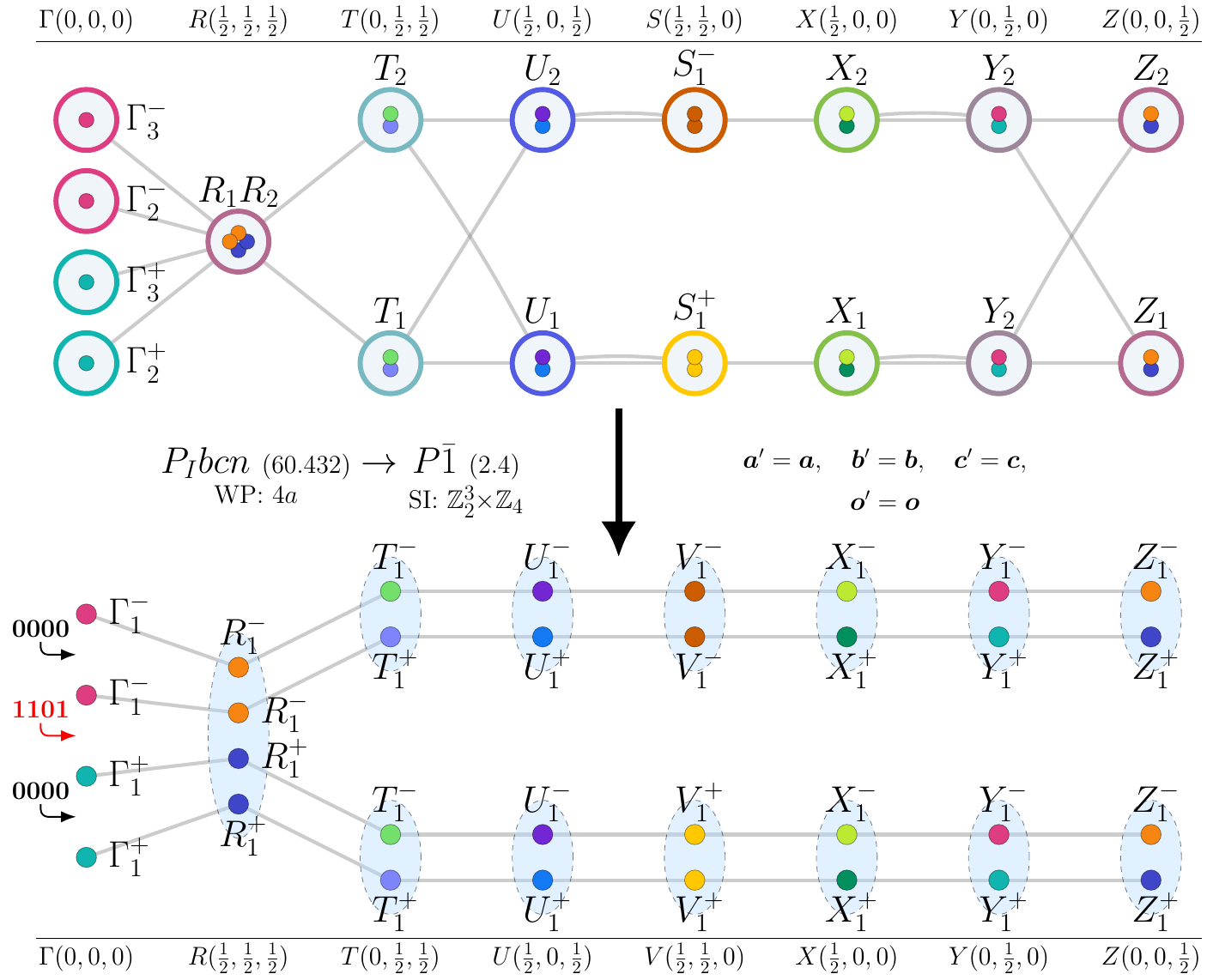}
\caption{Topological magnon bands in subgroup $P\bar{1}~(2.4)$ for magnetic moments on Wyckoff position $4a$ of supergroup $P_{I}bcn~(60.432)$.\label{fig_60.432_2.4_Bparallel100andstrainingenericdirection_4a}}
\end{figure}
\input{gap_tables_tex/60.432_2.4_Bparallel100andstrainingenericdirection_4a_table.tex}
\input{si_tables_tex/60.432_2.4_Bparallel100andstrainingenericdirection_4a_table.tex}
\subsubsection{Topological bands in subgroup $P2'/m'~(10.46)$}
\textbf{Perturbations:}
\begin{itemize}
\item B $\parallel$ [100] and strain $\perp$ [001],
\item B $\parallel$ [010] and strain $\perp$ [001],
\item B $\perp$ [001].
\end{itemize}
\begin{figure}[H]
\centering
\includegraphics[scale=0.6]{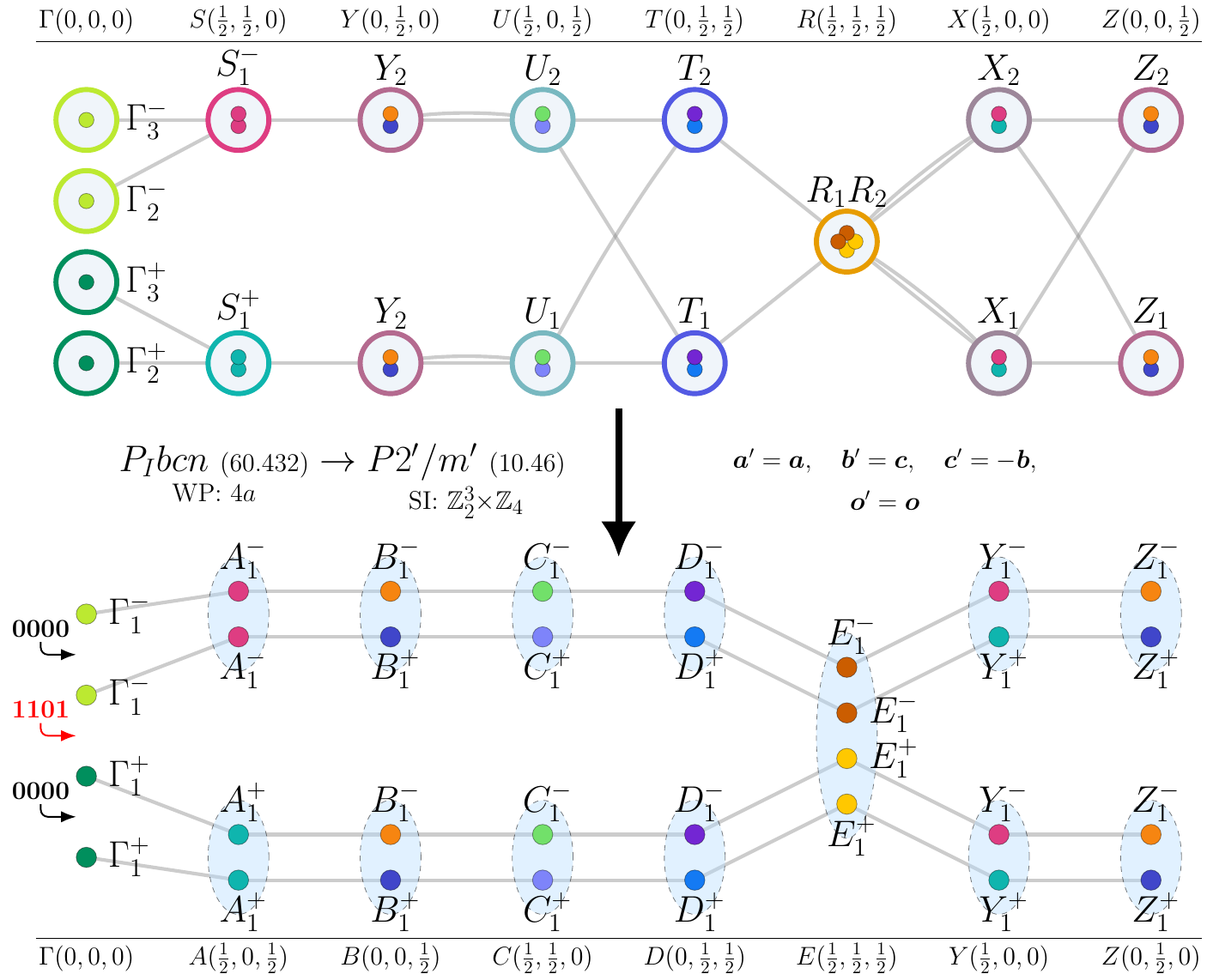}
\caption{Topological magnon bands in subgroup $P2'/m'~(10.46)$ for magnetic moments on Wyckoff position $4a$ of supergroup $P_{I}bcn~(60.432)$.\label{fig_60.432_10.46_Bparallel100andstrainperp001_4a}}
\end{figure}
\input{gap_tables_tex/60.432_10.46_Bparallel100andstrainperp001_4a_table.tex}
\input{si_tables_tex/60.432_10.46_Bparallel100andstrainperp001_4a_table.tex}
\subsubsection{Topological bands in subgroup $P2_{1}'/c'~(14.79)$}
\textbf{Perturbations:}
\begin{itemize}
\item B $\parallel$ [100] and strain $\perp$ [010],
\item B $\parallel$ [001] and strain $\perp$ [010],
\item B $\perp$ [010].
\end{itemize}
\begin{figure}[H]
\centering
\includegraphics[scale=0.6]{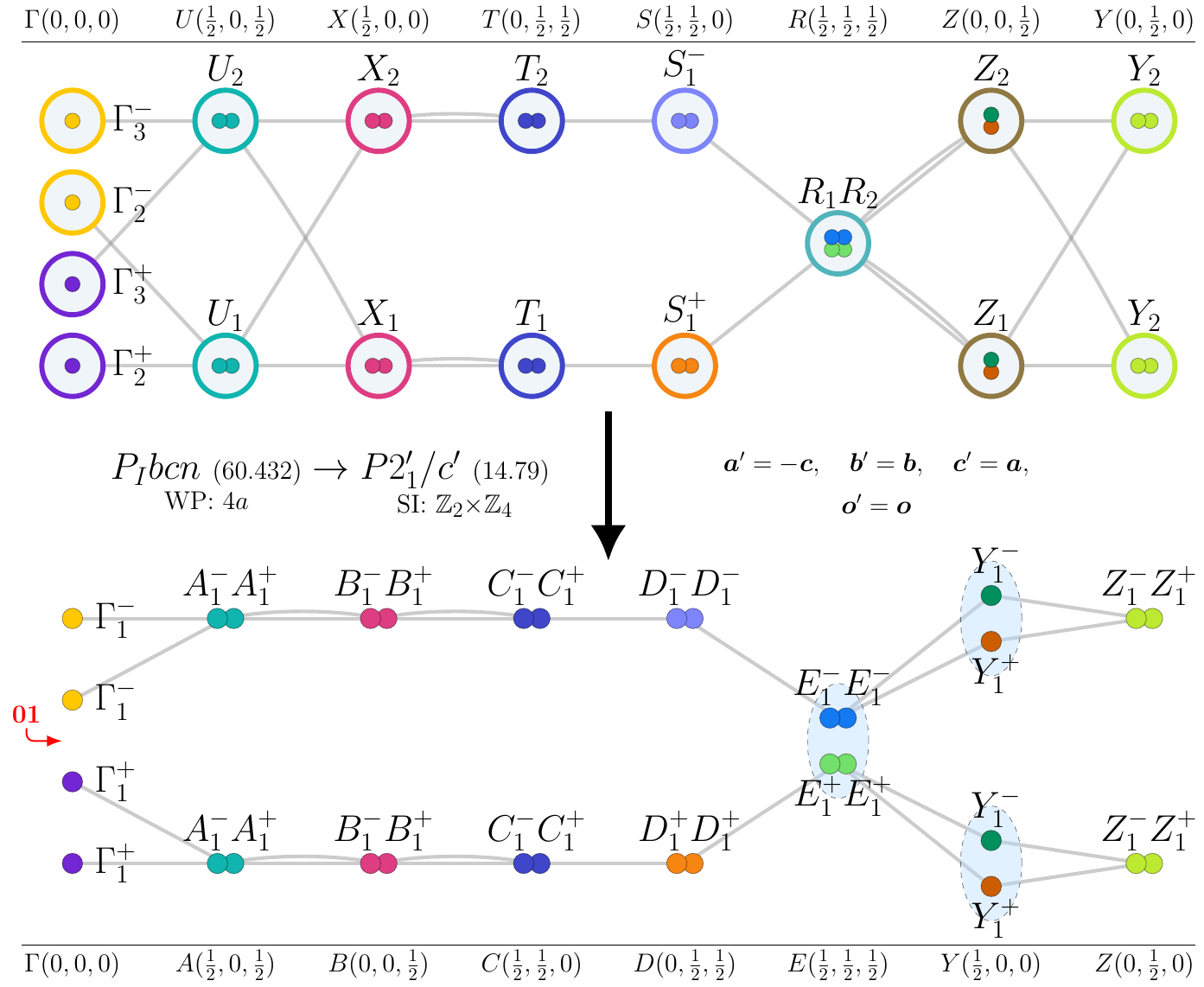}
\caption{Topological magnon bands in subgroup $P2_{1}'/c'~(14.79)$ for magnetic moments on Wyckoff position $4a$ of supergroup $P_{I}bcn~(60.432)$.\label{fig_60.432_14.79_Bparallel100andstrainperp010_4a}}
\end{figure}
\input{gap_tables_tex/60.432_14.79_Bparallel100andstrainperp010_4a_table.tex}
\input{si_tables_tex/60.432_14.79_Bparallel100andstrainperp010_4a_table.tex}
\subsubsection{Topological bands in subgroup $P2'/c'~(13.69)$}
\textbf{Perturbations:}
\begin{itemize}
\item B $\parallel$ [010] and strain $\perp$ [100],
\item B $\parallel$ [001] and strain $\perp$ [100],
\item B $\perp$ [100].
\end{itemize}
\begin{figure}[H]
\centering
\includegraphics[scale=0.6]{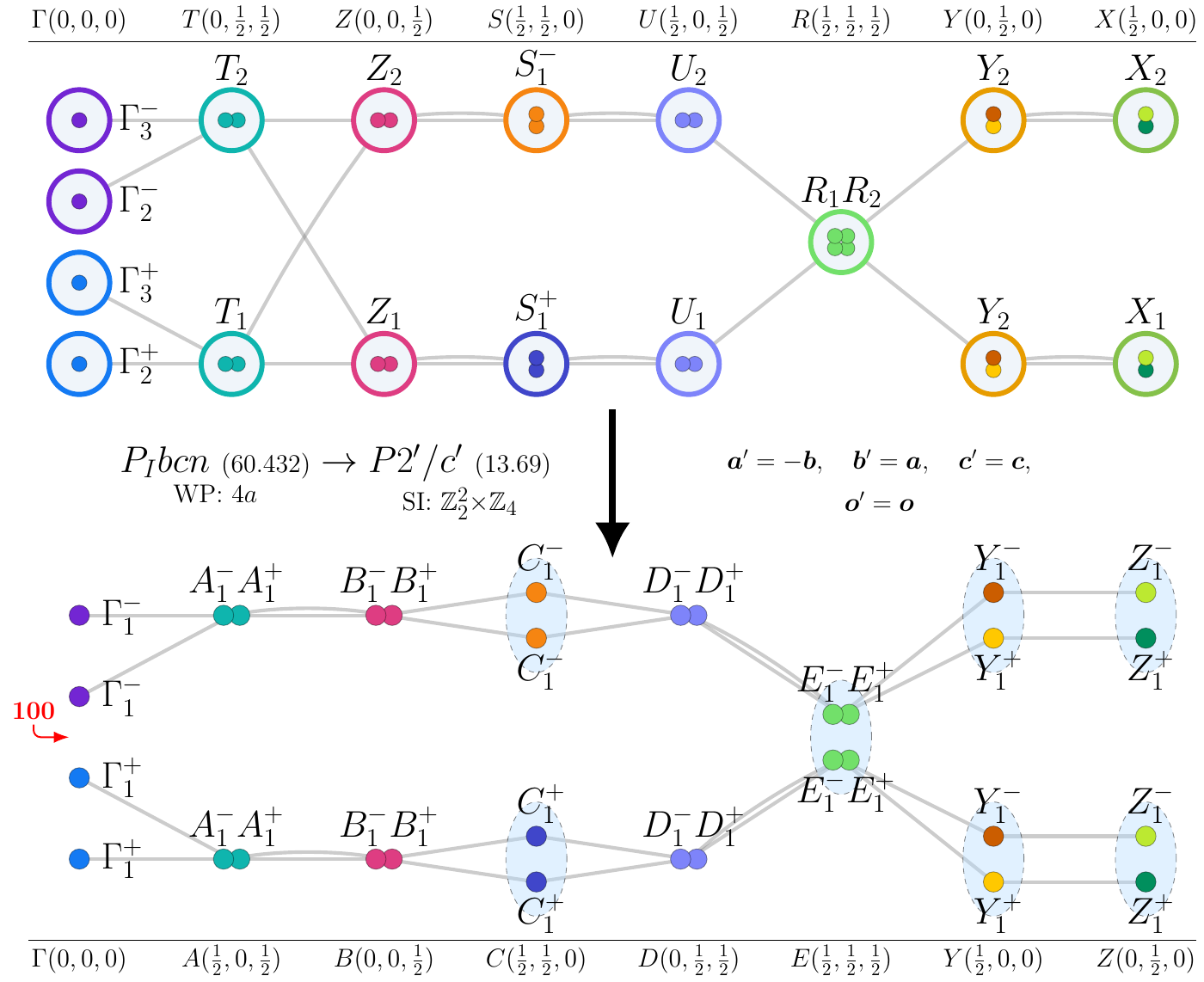}
\caption{Topological magnon bands in subgroup $P2'/c'~(13.69)$ for magnetic moments on Wyckoff position $4a$ of supergroup $P_{I}bcn~(60.432)$.\label{fig_60.432_13.69_Bparallel010andstrainperp100_4a}}
\end{figure}
\input{gap_tables_tex/60.432_13.69_Bparallel010andstrainperp100_4a_table.tex}
\input{si_tables_tex/60.432_13.69_Bparallel010andstrainperp100_4a_table.tex}
\subsubsection{Topological bands in subgroup $P_{S}\bar{1}~(2.7)$}
\textbf{Perturbation:}
\begin{itemize}
\item strain in generic direction.
\end{itemize}
\begin{figure}[H]
\centering
\includegraphics[scale=0.6]{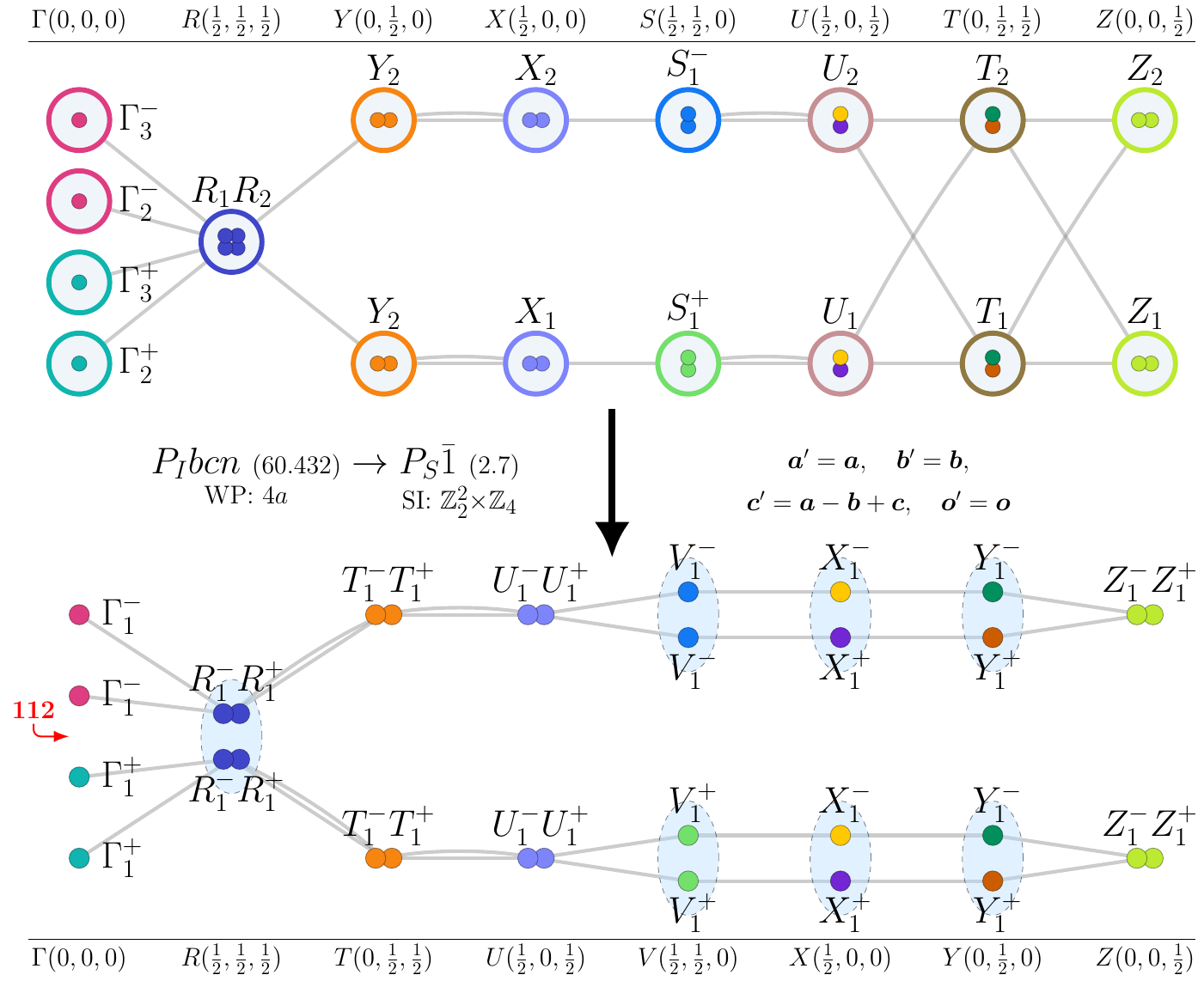}
\caption{Topological magnon bands in subgroup $P_{S}\bar{1}~(2.7)$ for magnetic moments on Wyckoff position $4a$ of supergroup $P_{I}bcn~(60.432)$.\label{fig_60.432_2.7_strainingenericdirection_4a}}
\end{figure}
\input{gap_tables_tex/60.432_2.7_strainingenericdirection_4a_table.tex}
\input{si_tables_tex/60.432_2.7_strainingenericdirection_4a_table.tex}
\subsubsection{Topological bands in subgroup $Pm'a'2~(28.91)$}
\textbf{Perturbation:}
\begin{itemize}
\item E $\parallel$ [010] and B $\parallel$ [010].
\end{itemize}
\begin{figure}[H]
\centering
\includegraphics[scale=0.6]{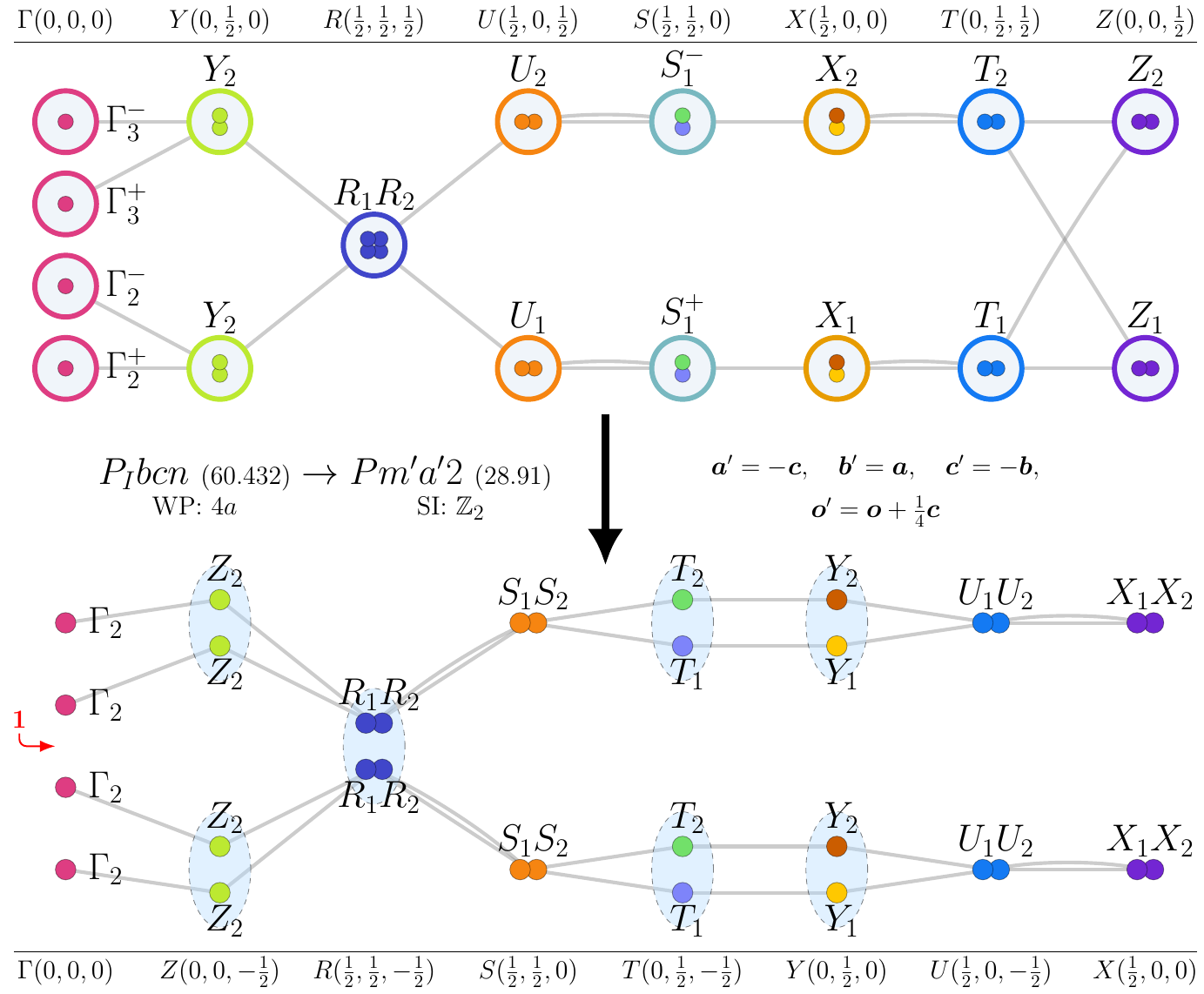}
\caption{Topological magnon bands in subgroup $Pm'a'2~(28.91)$ for magnetic moments on Wyckoff position $4a$ of supergroup $P_{I}bcn~(60.432)$.\label{fig_60.432_28.91_Eparallel010andBparallel010_4a}}
\end{figure}
\input{gap_tables_tex/60.432_28.91_Eparallel010andBparallel010_4a_table.tex}
\input{si_tables_tex/60.432_28.91_Eparallel010andBparallel010_4a_table.tex}
\subsection{WP: $4b+8j$}
\textbf{BCS Materials:} {Mn\textsubscript{3}ZnN~(140 K)}\footnote{BCS web page: \texttt{\href{http://webbdcrista1.ehu.es/magndata/index.php?this\_label=2.31} {http://webbdcrista1.ehu.es/magndata/index.php?this\_label=2.31}}}.\\
\subsubsection{Topological bands in subgroup $P2_{1}'/c'~(14.79)$}
\textbf{Perturbations:}
\begin{itemize}
\item B $\parallel$ [100] and strain $\perp$ [010],
\item B $\parallel$ [001] and strain $\perp$ [010],
\item B $\perp$ [010].
\end{itemize}
\begin{figure}[H]
\centering
\includegraphics[scale=0.6]{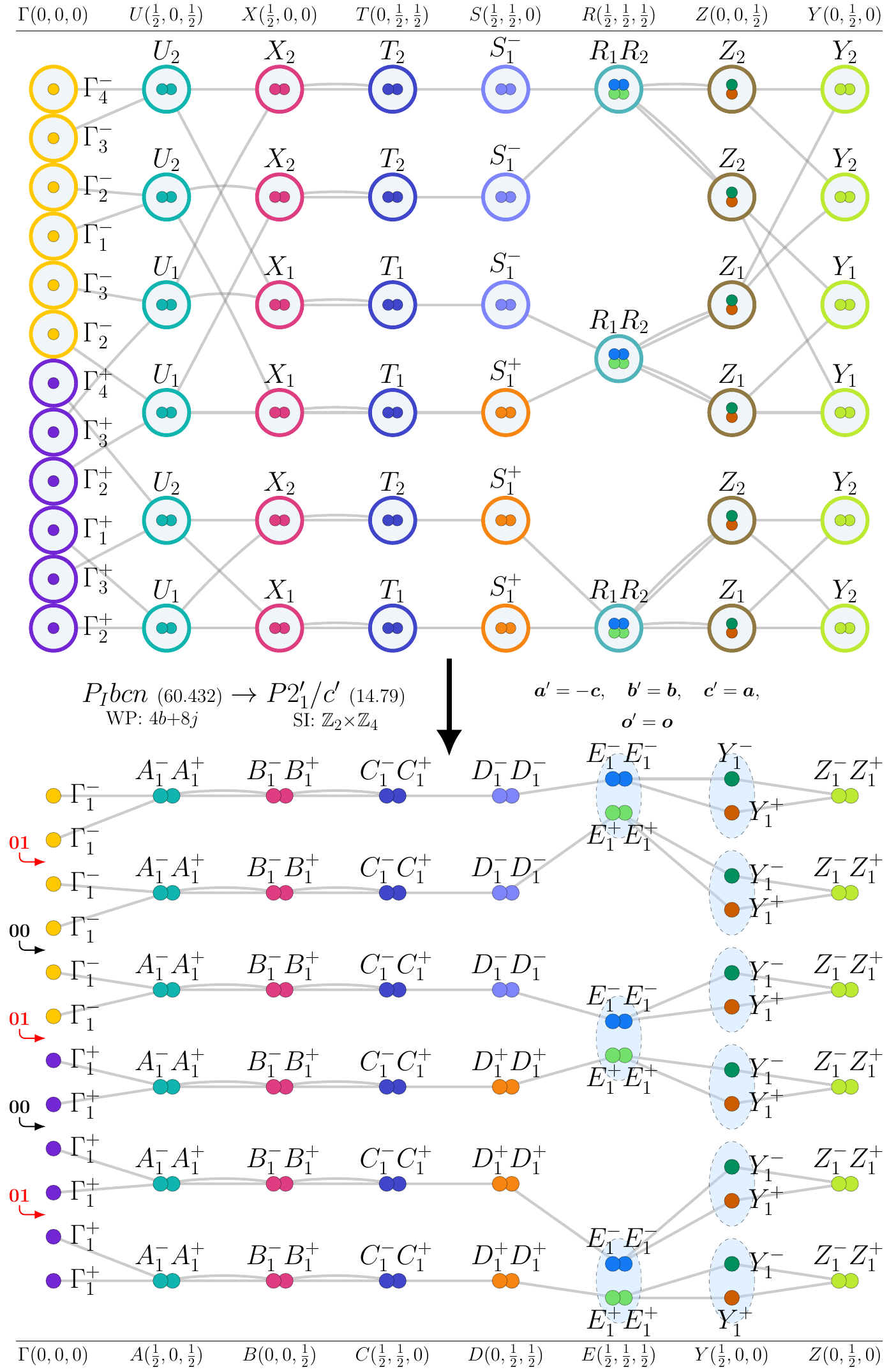}
\caption{Topological magnon bands in subgroup $P2_{1}'/c'~(14.79)$ for magnetic moments on Wyckoff positions $4b+8j$ of supergroup $P_{I}bcn~(60.432)$.\label{fig_60.432_14.79_Bparallel100andstrainperp010_4b+8j}}
\end{figure}
\input{gap_tables_tex/60.432_14.79_Bparallel100andstrainperp010_4b+8j_table.tex}
\input{si_tables_tex/60.432_14.79_Bparallel100andstrainperp010_4b+8j_table.tex}
\subsubsection{Topological bands in subgroup $P2'/c'~(13.69)$}
\textbf{Perturbations:}
\begin{itemize}
\item B $\parallel$ [010] and strain $\perp$ [100],
\item B $\parallel$ [001] and strain $\perp$ [100],
\item B $\perp$ [100].
\end{itemize}
\begin{figure}[H]
\centering
\includegraphics[scale=0.6]{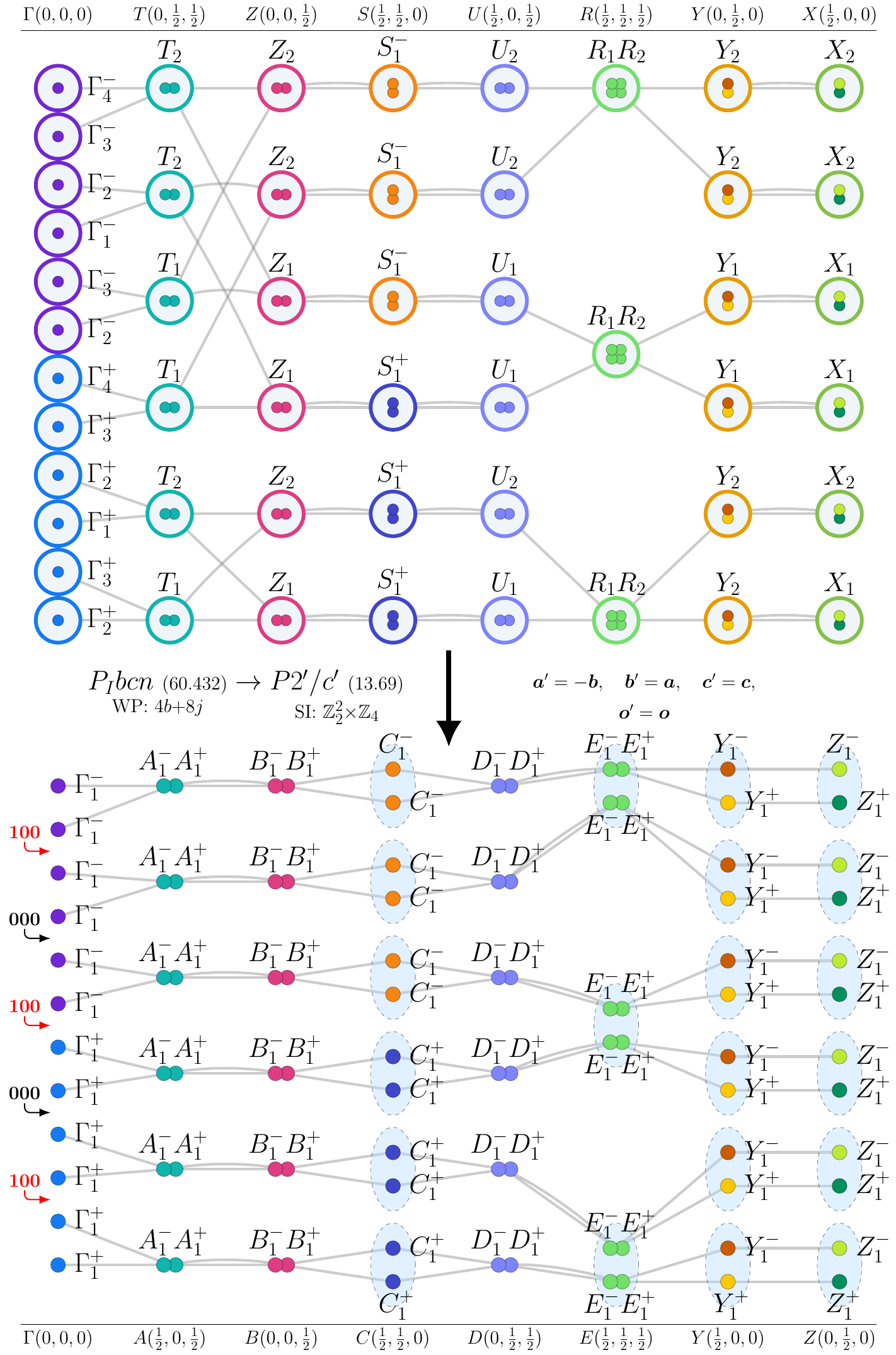}
\caption{Topological magnon bands in subgroup $P2'/c'~(13.69)$ for magnetic moments on Wyckoff positions $4b+8j$ of supergroup $P_{I}bcn~(60.432)$.\label{fig_60.432_13.69_Bparallel010andstrainperp100_4b+8j}}
\end{figure}
\input{gap_tables_tex/60.432_13.69_Bparallel010andstrainperp100_4b+8j_table.tex}
\input{si_tables_tex/60.432_13.69_Bparallel010andstrainperp100_4b+8j_table.tex}
\subsubsection{Topological bands in subgroup $P_{S}\bar{1}~(2.7)$}
\textbf{Perturbation:}
\begin{itemize}
\item strain in generic direction.
\end{itemize}
\begin{figure}[H]
\centering
\includegraphics[scale=0.6]{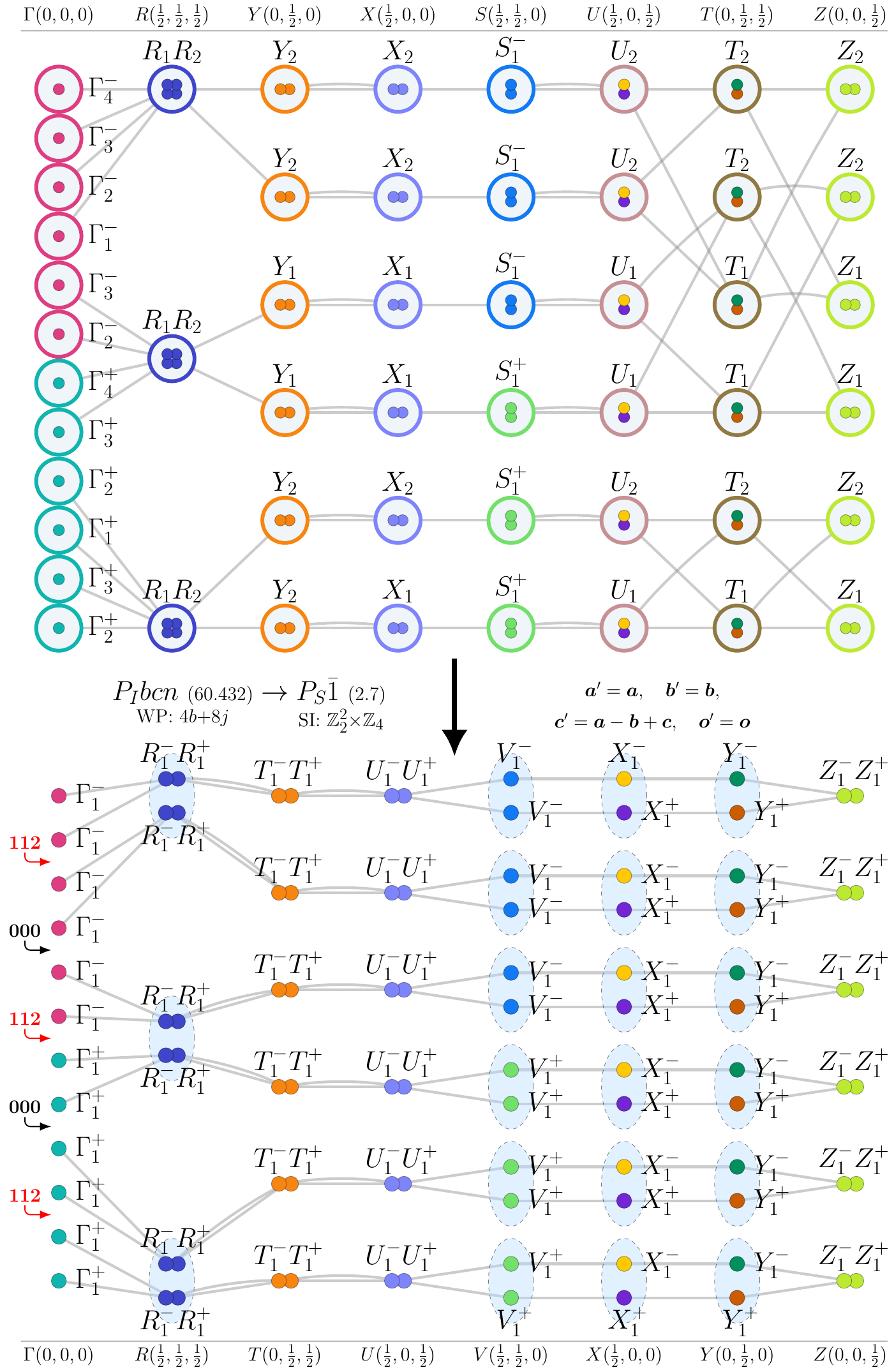}
\caption{Topological magnon bands in subgroup $P_{S}\bar{1}~(2.7)$ for magnetic moments on Wyckoff positions $4b+8j$ of supergroup $P_{I}bcn~(60.432)$.\label{fig_60.432_2.7_strainingenericdirection_4b+8j}}
\end{figure}
\input{gap_tables_tex/60.432_2.7_strainingenericdirection_4b+8j_table.tex}
\input{si_tables_tex/60.432_2.7_strainingenericdirection_4b+8j_table.tex}
\subsection{WP: $8j$}
\textbf{BCS Materials:} {Ce\textsubscript{2}Ni\textsubscript{3}Ge\textsubscript{5}~(4.8 K)}\footnote{BCS web page: \texttt{\href{http://webbdcrista1.ehu.es/magndata/index.php?this\_label=0.562} {http://webbdcrista1.ehu.es/magndata/index.php?this\_label=0.562}}}, {Ce\textsubscript{2}Ni\textsubscript{3}Ge\textsubscript{5}~(4.3 K)}\footnote{BCS web page: \texttt{\href{http://webbdcrista1.ehu.es/magndata/index.php?this\_label=0.565} {http://webbdcrista1.ehu.es/magndata/index.php?this\_label=0.565}}}.\\
\subsubsection{Topological bands in subgroup $P2_{1}'/c'~(14.79)$}
\textbf{Perturbations:}
\begin{itemize}
\item B $\parallel$ [100] and strain $\perp$ [010],
\item B $\parallel$ [001] and strain $\perp$ [010],
\item B $\perp$ [010].
\end{itemize}
\begin{figure}[H]
\centering
\includegraphics[scale=0.6]{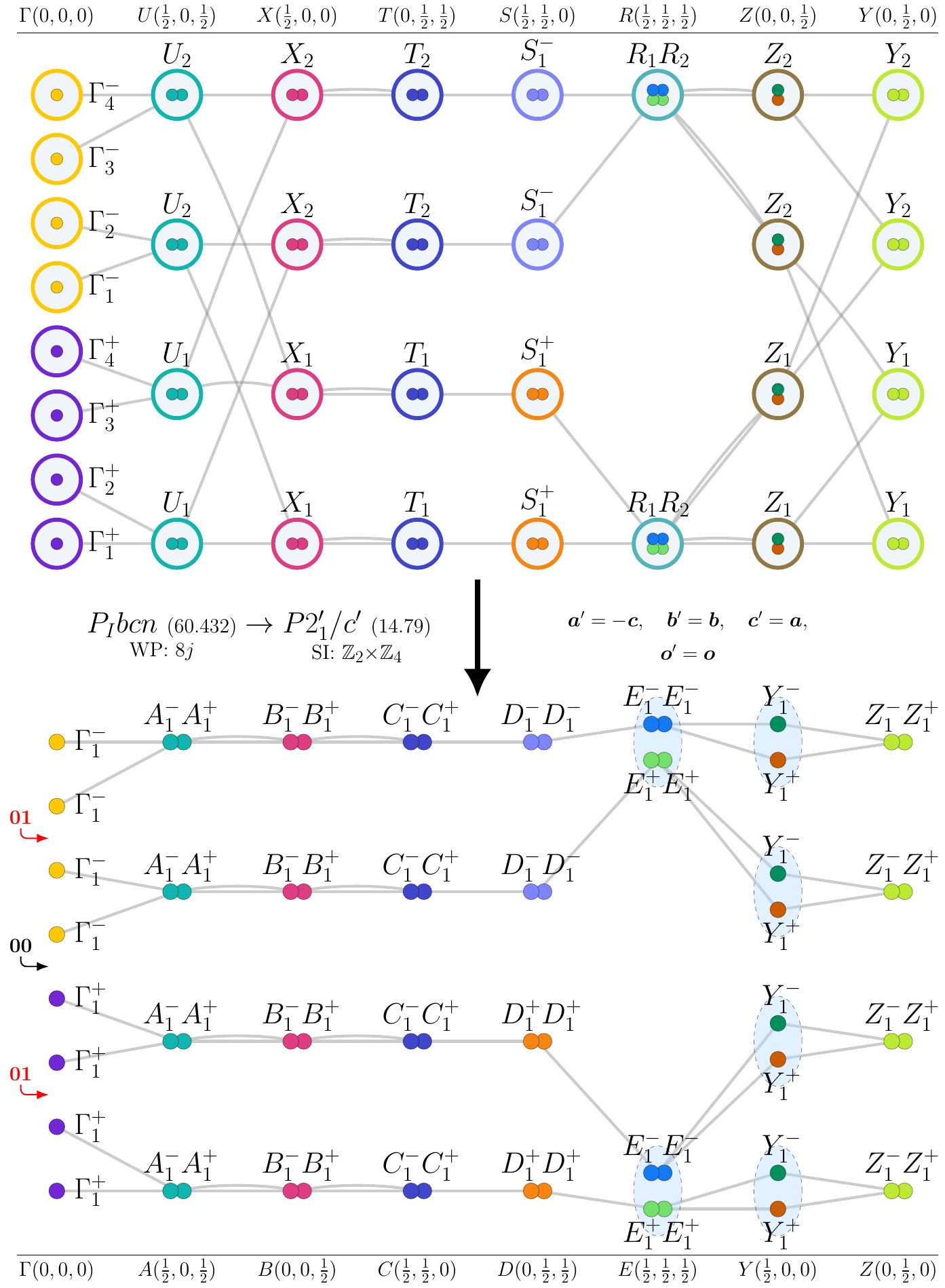}
\caption{Topological magnon bands in subgroup $P2_{1}'/c'~(14.79)$ for magnetic moments on Wyckoff position $8j$ of supergroup $P_{I}bcn~(60.432)$.\label{fig_60.432_14.79_Bparallel100andstrainperp010_8j}}
\end{figure}
\input{gap_tables_tex/60.432_14.79_Bparallel100andstrainperp010_8j_table.tex}
\input{si_tables_tex/60.432_14.79_Bparallel100andstrainperp010_8j_table.tex}
\subsubsection{Topological bands in subgroup $P2'/c'~(13.69)$}
\textbf{Perturbations:}
\begin{itemize}
\item B $\parallel$ [010] and strain $\perp$ [100],
\item B $\parallel$ [001] and strain $\perp$ [100],
\item B $\perp$ [100].
\end{itemize}
\begin{figure}[H]
\centering
\includegraphics[scale=0.6]{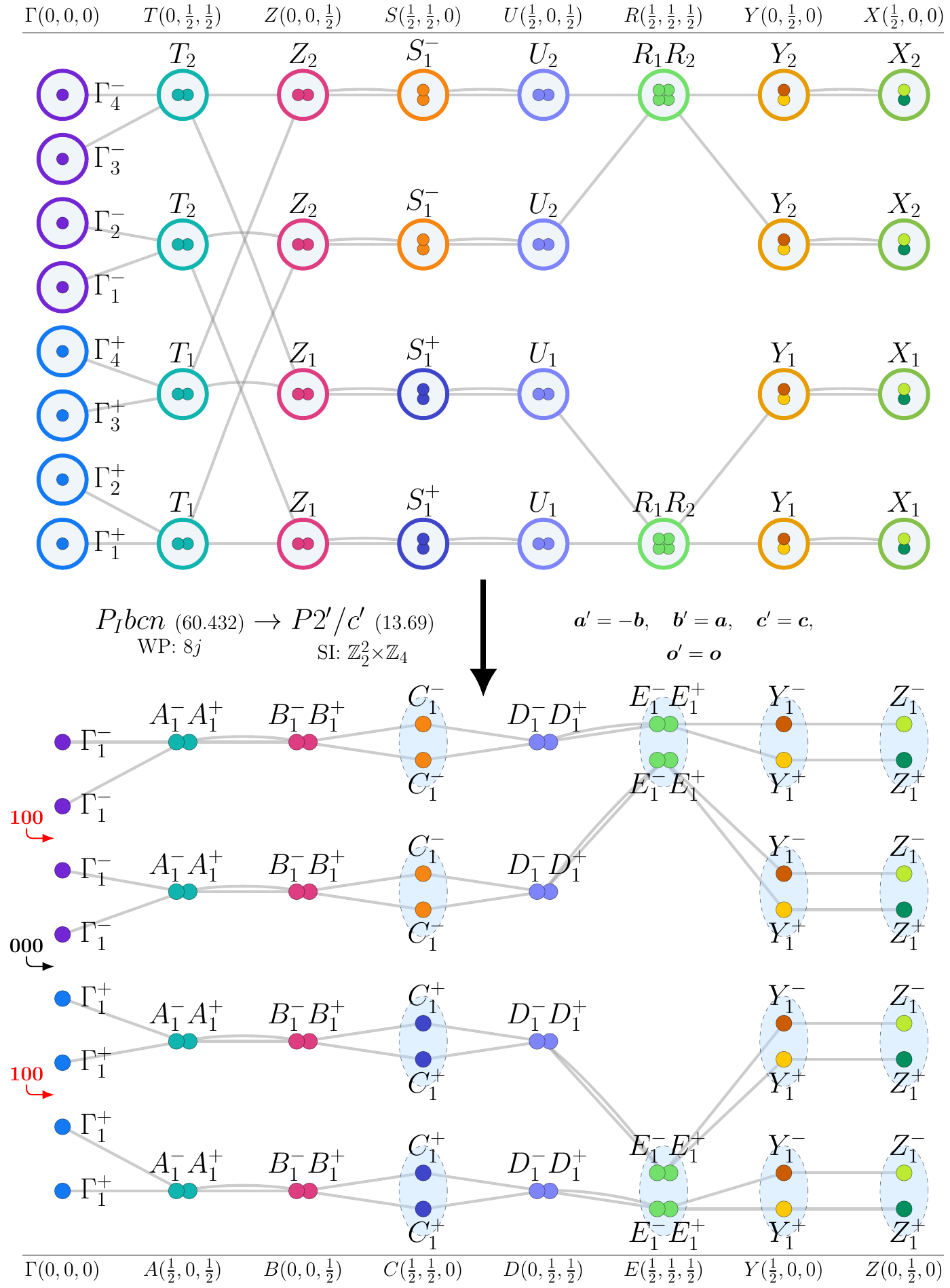}
\caption{Topological magnon bands in subgroup $P2'/c'~(13.69)$ for magnetic moments on Wyckoff position $8j$ of supergroup $P_{I}bcn~(60.432)$.\label{fig_60.432_13.69_Bparallel010andstrainperp100_8j}}
\end{figure}
\input{gap_tables_tex/60.432_13.69_Bparallel010andstrainperp100_8j_table.tex}
\input{si_tables_tex/60.432_13.69_Bparallel010andstrainperp100_8j_table.tex}
\subsubsection{Topological bands in subgroup $P_{S}\bar{1}~(2.7)$}
\textbf{Perturbation:}
\begin{itemize}
\item strain in generic direction.
\end{itemize}
\begin{figure}[H]
\centering
\includegraphics[scale=0.6]{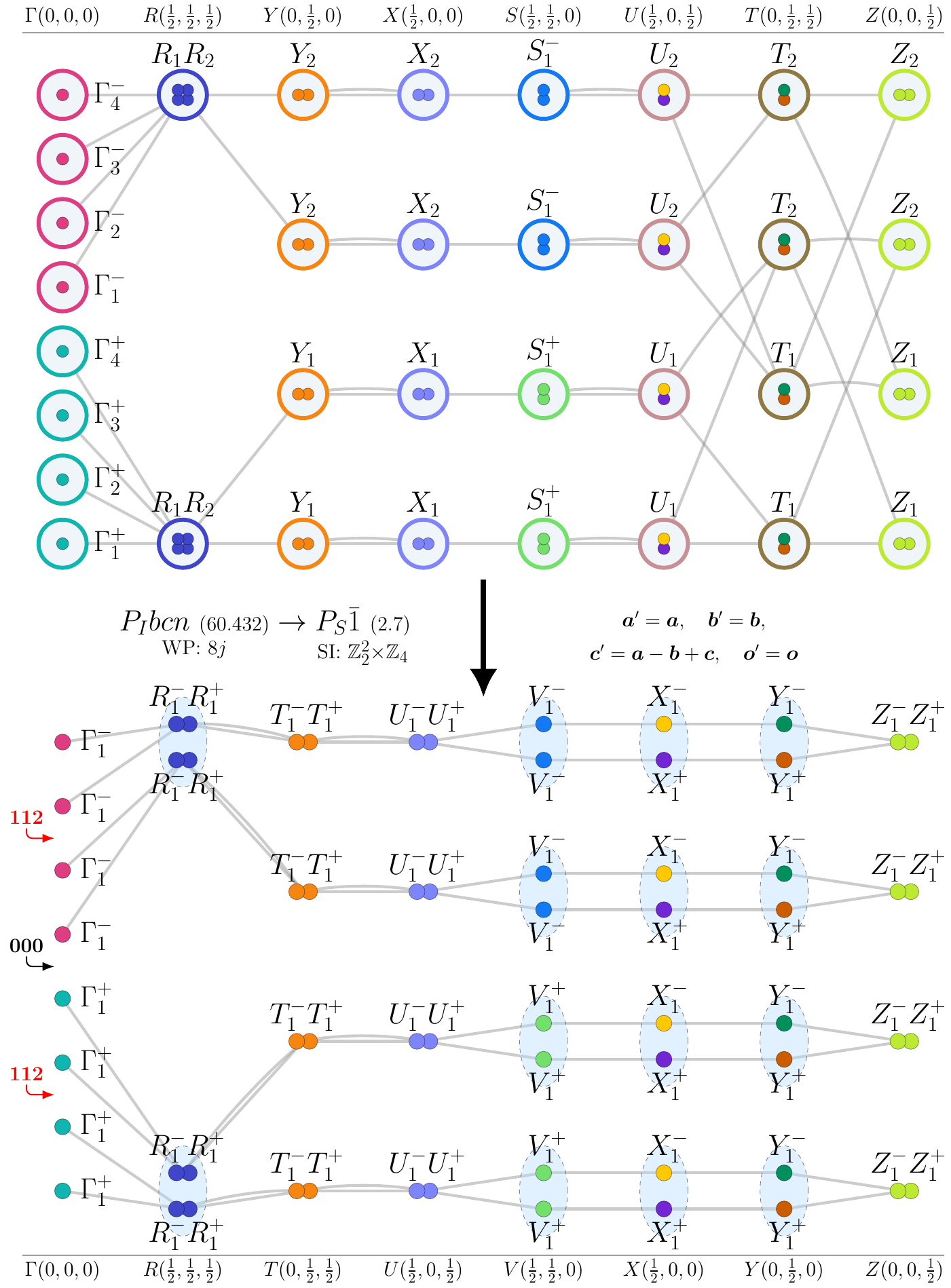}
\caption{Topological magnon bands in subgroup $P_{S}\bar{1}~(2.7)$ for magnetic moments on Wyckoff position $8j$ of supergroup $P_{I}bcn~(60.432)$.\label{fig_60.432_2.7_strainingenericdirection_8j}}
\end{figure}
\input{gap_tables_tex/60.432_2.7_strainingenericdirection_8j_table.tex}
\input{si_tables_tex/60.432_2.7_strainingenericdirection_8j_table.tex}

\section{MSG $Pbca~(61.433)$\label{Ca2RuO4SuppSection}}
\textbf{Nontrivial-SI Subgroups:} $P\bar{1}~(2.4)$, $P2_{1}/c~(14.75)$, $P2_{1}/c~(14.75)$, $P2_{1}/c~(14.75)$.\\

\textbf{Trivial-SI Subgroups:} $Pc~(7.24)$, $Pc~(7.24)$, $Pc~(7.24)$, $P2_{1}~(4.7)$, $Pca2_{1}~(29.99)$, $P2_{1}~(4.7)$, $Pca2_{1}~(29.99)$, $P2_{1}~(4.7)$, $Pca2_{1}~(29.99)$.\\

\subsection{WP: $4a$}
\textbf{BCS Materials:} {Ca\textsubscript{2}RuO\textsubscript{4}~(110 K)}\footnote{BCS web page: \texttt{\href{http://webbdcrista1.ehu.es/magndata/index.php?this\_label=0.398} {http://webbdcrista1.ehu.es/magndata/index.php?this\_label=0.398}}}.\\
\subsubsection{Topological bands in subgroup $P\bar{1}~(2.4)$}
\textbf{Perturbations:}
\begin{itemize}
\item strain in generic direction,
\item B $\parallel$ [100] and strain $\perp$ [010],
\item B $\parallel$ [100] and strain $\perp$ [001],
\item B $\parallel$ [010] and strain $\perp$ [100],
\item B $\parallel$ [010] and strain $\perp$ [001],
\item B $\parallel$ [001] and strain $\perp$ [100],
\item B $\parallel$ [001] and strain $\perp$ [010],
\item B in generic direction.
\end{itemize}
\begin{figure}[H]
\centering
\includegraphics[scale=0.6]{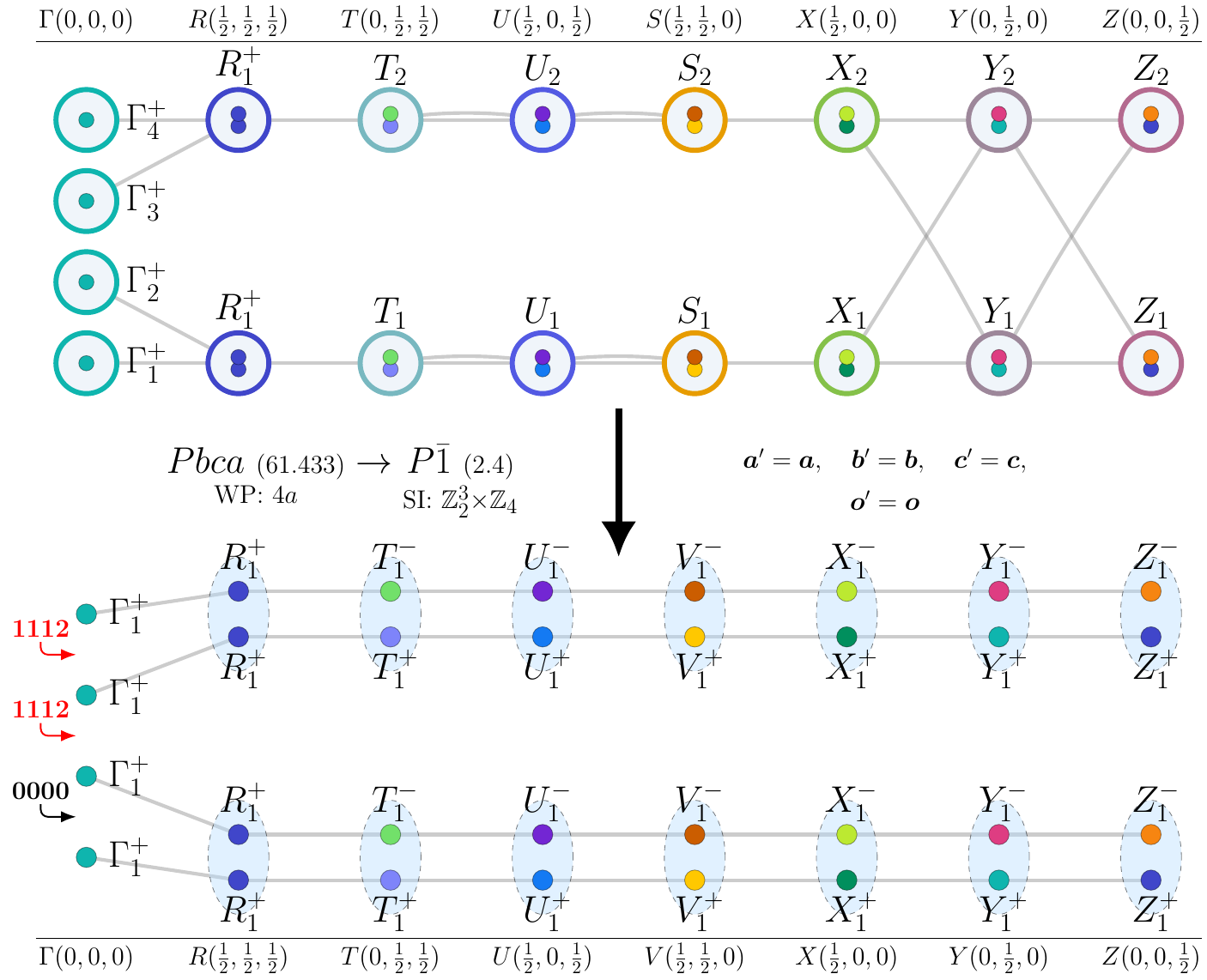}
\caption{Topological magnon bands in subgroup $P\bar{1}~(2.4)$ for magnetic moments on Wyckoff position $4a$ of supergroup $Pbca~(61.433)$.\label{fig_61.433_2.4_strainingenericdirection_4a}}
\end{figure}
\input{gap_tables_tex/61.433_2.4_strainingenericdirection_4a_table.tex}
\input{si_tables_tex/61.433_2.4_strainingenericdirection_4a_table.tex}

\section{MSG $P_{C}bca~(61.439)$}
\textbf{Nontrivial-SI Subgroups:} $P\bar{1}~(2.4)$, $P2_{1}'/c'~(14.79)$, $P2'/c'~(13.69)$, $P2'/m'~(10.46)$, $P_{S}\bar{1}~(2.7)$, $P2_{1}/c~(14.75)$, $Pm'n'a~(53.326)$, $P_{a}2_{1}/c~(14.80)$, $P2_{1}/c~(14.75)$, $Pb'am'~(55.358)$, $P_{C}2_{1}/c~(14.84)$, $P2_{1}/c~(14.75)$, $Pc'cn'~(56.370)$, $P_{A}2_{1}/c~(14.83)$.\\

\textbf{Trivial-SI Subgroups:} $Pc'~(7.26)$, $Pc'~(7.26)$, $Pm'~(6.20)$, $P2_{1}'~(4.9)$, $P2'~(3.3)$, $P2'~(3.3)$, $P_{S}1~(1.3)$, $Pc~(7.24)$, $Pm'a2'~(28.89)$, $Pn'c2'~(30.113)$, $P_{a}c~(7.27)$, $Pc~(7.24)$, $Pm'c2_{1}'~(26.68)$, $Pb'a2'~(32.137)$, $P_{C}c~(7.30)$, $Pc~(7.24)$, $Pn'a2_{1}'~(33.146)$, $Pc'c2'~(27.80)$, $P_{A}c~(7.31)$, $P2_{1}~(4.7)$, $Pm'n'2_{1}~(31.127)$, $P_{a}2_{1}~(4.10)$, $P_{C}ca2_{1}~(29.109)$, $P2_{1}~(4.7)$, $Pm'c'2_{1}~(26.70)$, $P_{C}2_{1}~(4.12)$, $P_{B}ca2_{1}~(29.108)$, $P2_{1}~(4.7)$, $Pn'a'2_{1}~(33.148)$, $P_{C}2_{1}~(4.12)$, $P_{A}ca2_{1}~(29.107)$.\\

\subsection{WP: $4a+8e$}
\textbf{BCS Materials:} {EuFe\textsubscript{2}As\textsubscript{2}~(19 K)}\footnote{BCS web page: \texttt{\href{http://webbdcrista1.ehu.es/magndata/index.php?this\_label=2.1} {http://webbdcrista1.ehu.es/magndata/index.php?this\_label=2.1}}}.\\
\subsubsection{Topological bands in subgroup $P2_{1}'/c'~(14.79)$}
\textbf{Perturbations:}
\begin{itemize}
\item B $\parallel$ [100] and strain $\perp$ [001],
\item B $\parallel$ [010] and strain $\perp$ [001],
\item B $\perp$ [001].
\end{itemize}
\begin{figure}[H]
\centering
\includegraphics[scale=0.6]{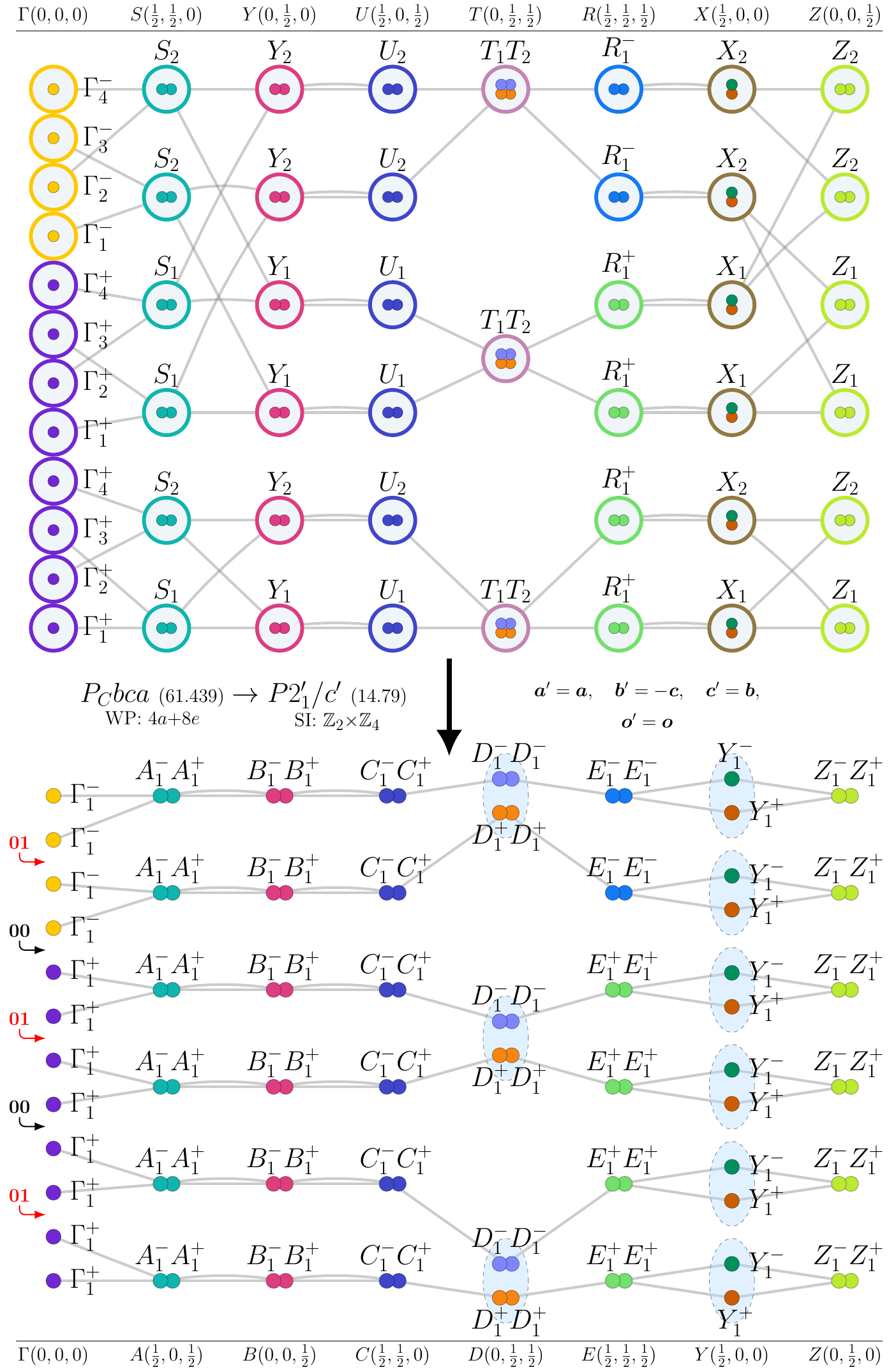}
\caption{Topological magnon bands in subgroup $P2_{1}'/c'~(14.79)$ for magnetic moments on Wyckoff positions $4a+8e$ of supergroup $P_{C}bca~(61.439)$.\label{fig_61.439_14.79_Bparallel100andstrainperp001_4a+8e}}
\end{figure}
\input{gap_tables_tex/61.439_14.79_Bparallel100andstrainperp001_4a+8e_table.tex}
\input{si_tables_tex/61.439_14.79_Bparallel100andstrainperp001_4a+8e_table.tex}
\subsubsection{Topological bands in subgroup $P2'/c'~(13.69)$}
\textbf{Perturbations:}
\begin{itemize}
\item B $\parallel$ [100] and strain $\perp$ [010],
\item B $\parallel$ [001] and strain $\perp$ [010],
\item B $\perp$ [010].
\end{itemize}
\begin{figure}[H]
\centering
\includegraphics[scale=0.6]{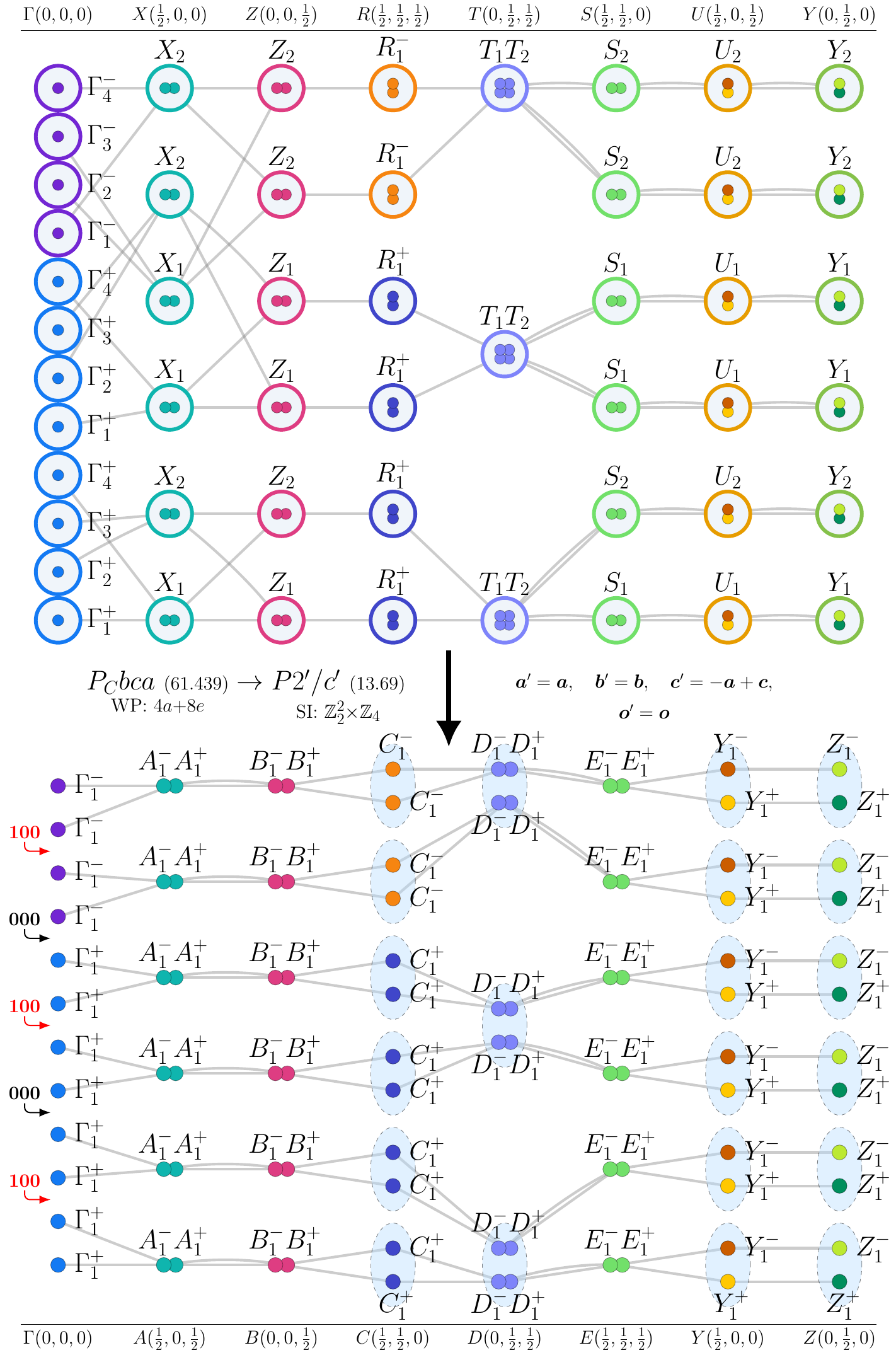}
\caption{Topological magnon bands in subgroup $P2'/c'~(13.69)$ for magnetic moments on Wyckoff positions $4a+8e$ of supergroup $P_{C}bca~(61.439)$.\label{fig_61.439_13.69_Bparallel100andstrainperp010_4a+8e}}
\end{figure}
\input{gap_tables_tex/61.439_13.69_Bparallel100andstrainperp010_4a+8e_table.tex}
\input{si_tables_tex/61.439_13.69_Bparallel100andstrainperp010_4a+8e_table.tex}
\subsubsection{Topological bands in subgroup $P_{S}\bar{1}~(2.7)$}
\textbf{Perturbation:}
\begin{itemize}
\item strain in generic direction.
\end{itemize}
\begin{figure}[H]
\centering
\includegraphics[scale=0.6]{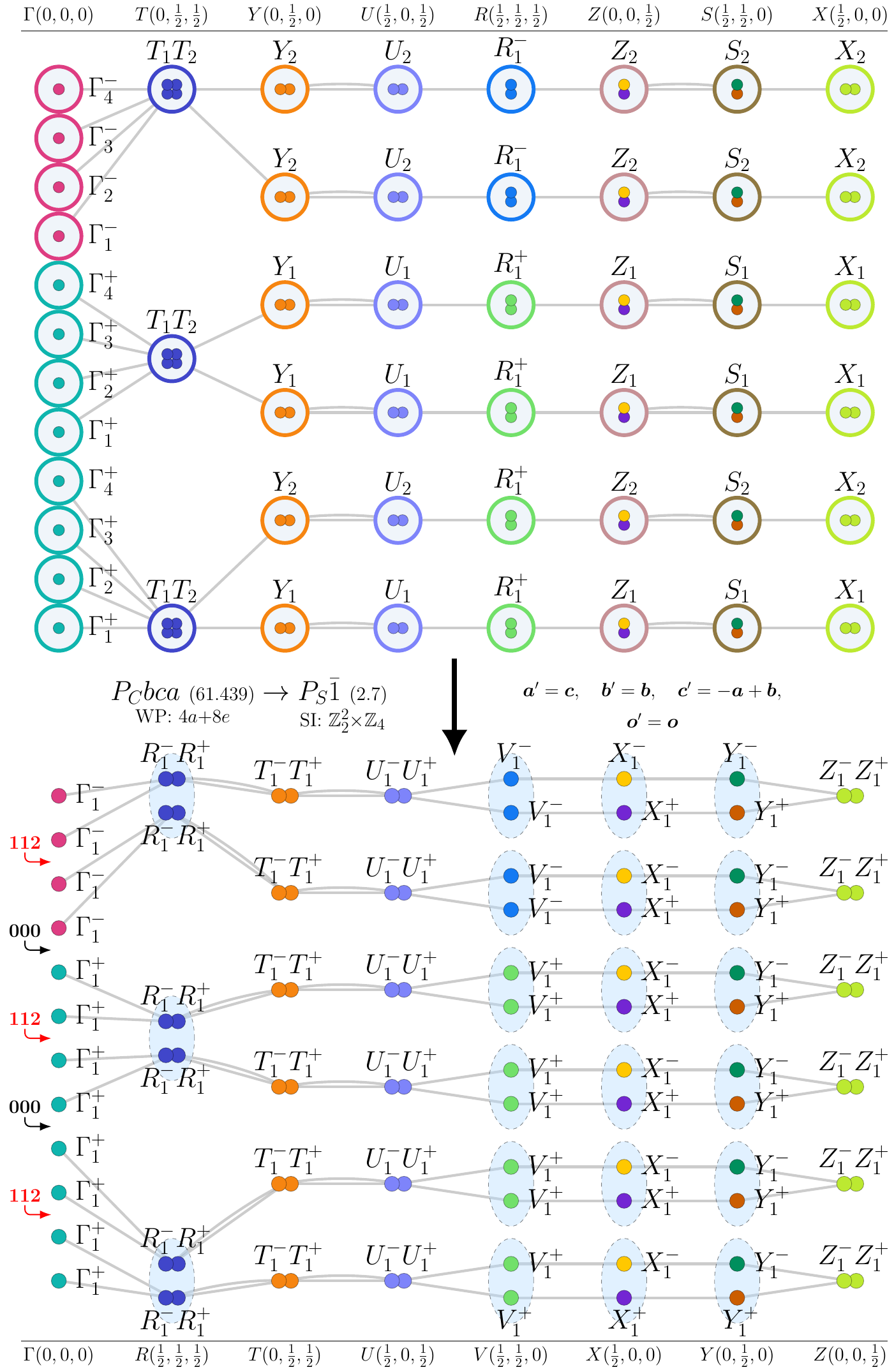}
\caption{Topological magnon bands in subgroup $P_{S}\bar{1}~(2.7)$ for magnetic moments on Wyckoff positions $4a+8e$ of supergroup $P_{C}bca~(61.439)$.\label{fig_61.439_2.7_strainingenericdirection_4a+8e}}
\end{figure}
\input{gap_tables_tex/61.439_2.7_strainingenericdirection_4a+8e_table.tex}
\input{si_tables_tex/61.439_2.7_strainingenericdirection_4a+8e_table.tex}

\section{MSG $Pnma~(62.441)$}
\textbf{Nontrivial-SI Subgroups:} $P\bar{1}~(2.4)$, $P2_{1}/c~(14.75)$, $P2_{1}/m~(11.50)$, $P2_{1}/c~(14.75)$.\\

\textbf{Trivial-SI Subgroups:} $Pc~(7.24)$, $Pm~(6.18)$, $Pc~(7.24)$, $P2_{1}~(4.7)$, $Pmn2_{1}~(31.123)$, $P2_{1}~(4.7)$, $Pna2_{1}~(33.144)$, $P2_{1}~(4.7)$, $Pmc2_{1}~(26.66)$.\\

\subsection{WP: $4b$}
\textbf{BCS Materials:} {La\textsubscript{0.75}Bi\textsubscript{0.25}Fe\textsubscript{0.5}Cr\textsubscript{0.5}O\textsubscript{3}~(350 K)}\footnote{BCS web page: \texttt{\href{http://webbdcrista1.ehu.es/magndata/index.php?this\_label=0.373} {http://webbdcrista1.ehu.es/magndata/index.php?this\_label=0.373}}}, {LaCrO\textsubscript{3}~(290 K)}\footnote{BCS web page: \texttt{\href{http://webbdcrista1.ehu.es/magndata/index.php?this\_label=0.323} {http://webbdcrista1.ehu.es/magndata/index.php?this\_label=0.323}}}, {CeFeO\textsubscript{3}~(220 K)}\footnote{BCS web page: \texttt{\href{http://webbdcrista1.ehu.es/magndata/index.php?this\_label=0.758} {http://webbdcrista1.ehu.es/magndata/index.php?this\_label=0.758}}}, {InCrO\textsubscript{3}~(93 K)}\footnote{BCS web page: \texttt{\href{http://webbdcrista1.ehu.es/magndata/index.php?this\_label=0.308} {http://webbdcrista1.ehu.es/magndata/index.php?this\_label=0.308}}}, {TlCrO\textsubscript{3}~(89 K)}\footnote{BCS web page: \texttt{\href{http://webbdcrista1.ehu.es/magndata/index.php?this\_label=0.309} {http://webbdcrista1.ehu.es/magndata/index.php?this\_label=0.309}}}, {ScCrO\textsubscript{3}~(73 K)}\footnote{BCS web page: \texttt{\href{http://webbdcrista1.ehu.es/magndata/index.php?this\_label=0.307} {http://webbdcrista1.ehu.es/magndata/index.php?this\_label=0.307}}}.\\
\subsubsection{Topological bands in subgroup $P\bar{1}~(2.4)$}
\textbf{Perturbations:}
\begin{itemize}
\item strain in generic direction,
\item B $\parallel$ [100] and strain $\perp$ [010],
\item B $\parallel$ [100] and strain $\perp$ [001],
\item B $\parallel$ [010] and strain $\perp$ [100],
\item B $\parallel$ [010] and strain $\perp$ [001],
\item B $\parallel$ [001] and strain $\perp$ [100],
\item B $\parallel$ [001] and strain $\perp$ [010],
\item B in generic direction.
\end{itemize}
\begin{figure}[H]
\centering
\includegraphics[scale=0.6]{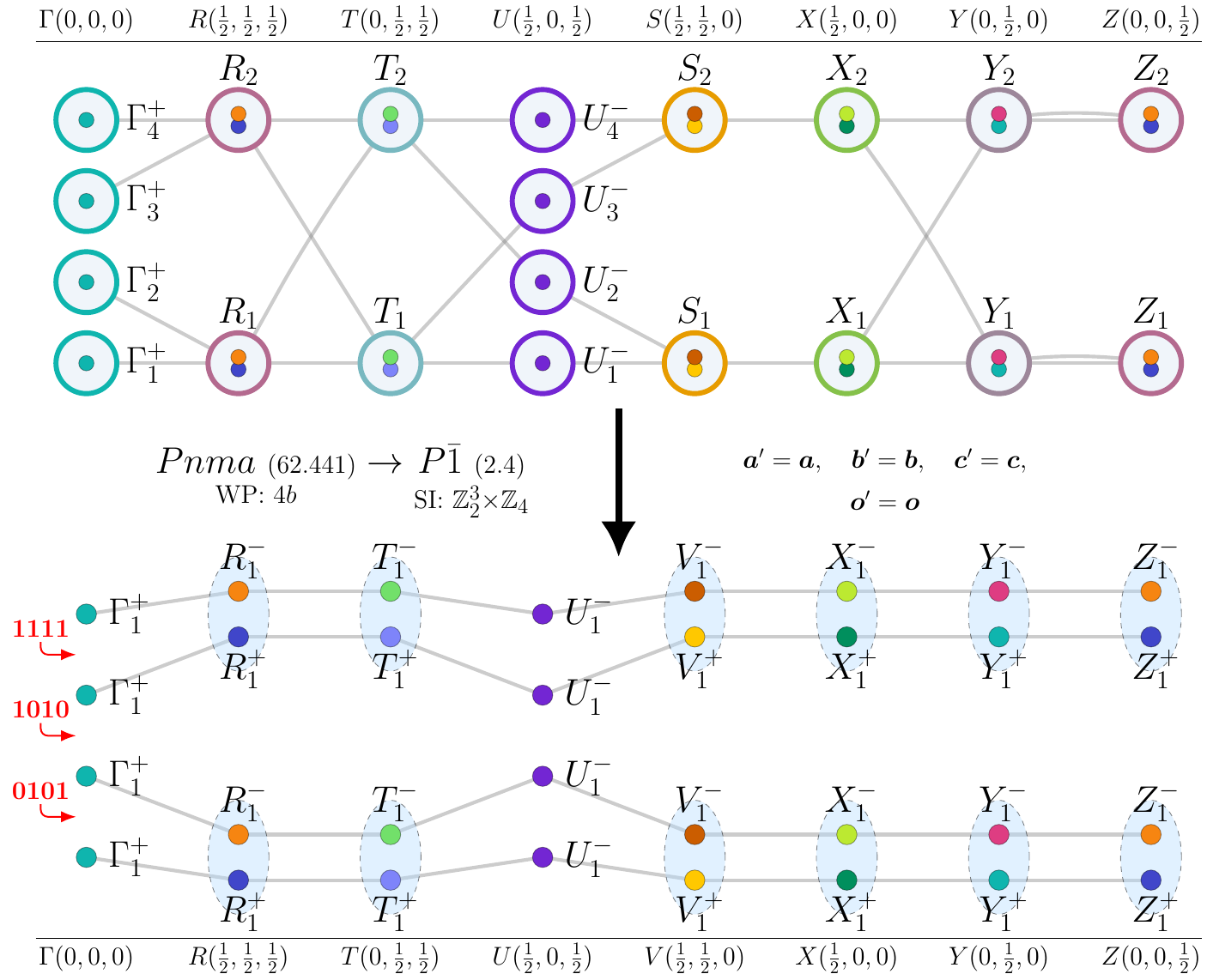}
\caption{Topological magnon bands in subgroup $P\bar{1}~(2.4)$ for magnetic moments on Wyckoff position $4b$ of supergroup $Pnma~(62.441)$.\label{fig_62.441_2.4_strainingenericdirection_4b}}
\end{figure}
\input{gap_tables_tex/62.441_2.4_strainingenericdirection_4b_table.tex}
\input{si_tables_tex/62.441_2.4_strainingenericdirection_4b_table.tex}
\subsection{WP: $4a+4c$}
\textbf{BCS Materials:} {Fe\textsubscript{2}PO\textsubscript{5}~(250 K)}\footnote{BCS web page: \texttt{\href{http://webbdcrista1.ehu.es/magndata/index.php?this\_label=0.263} {http://webbdcrista1.ehu.es/magndata/index.php?this\_label=0.263}}}, {CuFePO\textsubscript{5}~(195 K)}\footnote{BCS web page: \texttt{\href{http://webbdcrista1.ehu.es/magndata/index.php?this\_label=0.260} {http://webbdcrista1.ehu.es/magndata/index.php?this\_label=0.260}}}, {NiFePO\textsubscript{5}~(178 K)}\footnote{BCS web page: \texttt{\href{http://webbdcrista1.ehu.es/magndata/index.php?this\_label=0.261} {http://webbdcrista1.ehu.es/magndata/index.php?this\_label=0.261}}}, {Fe\textsubscript{2}SiO\textsubscript{4}~(65.3 K)}\footnote{BCS web page: \texttt{\href{http://webbdcrista1.ehu.es/magndata/index.php?this\_label=0.221} {http://webbdcrista1.ehu.es/magndata/index.php?this\_label=0.221}}}, {Co\textsubscript{2}SiO\textsubscript{4}~(49.5 K)}\footnote{BCS web page: \texttt{\href{http://webbdcrista1.ehu.es/magndata/index.php?this\_label=0.218} {http://webbdcrista1.ehu.es/magndata/index.php?this\_label=0.218}}}, {Co\textsubscript{2}SiO\textsubscript{4}~(49 K)}\footnote{BCS web page: \texttt{\href{http://webbdcrista1.ehu.es/magndata/index.php?this\_label=0.219} {http://webbdcrista1.ehu.es/magndata/index.php?this\_label=0.219}}}, {NH\textsubscript{4}Fe\textsubscript{2}F\textsubscript{6}~(19 K)}\footnote{BCS web page: \texttt{\href{http://webbdcrista1.ehu.es/magndata/index.php?this\_label=0.168} {http://webbdcrista1.ehu.es/magndata/index.php?this\_label=0.168}}}, {Mn\textsubscript{2}GeO\textsubscript{4}~(17 K)}\footnote{BCS web page: \texttt{\href{http://webbdcrista1.ehu.es/magndata/index.php?this\_label=0.102} {http://webbdcrista1.ehu.es/magndata/index.php?this\_label=0.102}}}, {RbFe\textsubscript{2}F\textsubscript{6}~(16 K)}\footnote{BCS web page: \texttt{\href{http://webbdcrista1.ehu.es/magndata/index.php?this\_label=0.192} {http://webbdcrista1.ehu.es/magndata/index.php?this\_label=0.192}}}.\\
\subsubsection{Topological bands in subgroup $P\bar{1}~(2.4)$}
\textbf{Perturbations:}
\begin{itemize}
\item strain in generic direction,
\item B $\parallel$ [100] and strain $\perp$ [010],
\item B $\parallel$ [100] and strain $\perp$ [001],
\item B $\parallel$ [010] and strain $\perp$ [100],
\item B $\parallel$ [010] and strain $\perp$ [001],
\item B $\parallel$ [001] and strain $\perp$ [100],
\item B $\parallel$ [001] and strain $\perp$ [010],
\item B in generic direction.
\end{itemize}
\begin{figure}[H]
\centering
\includegraphics[scale=0.6]{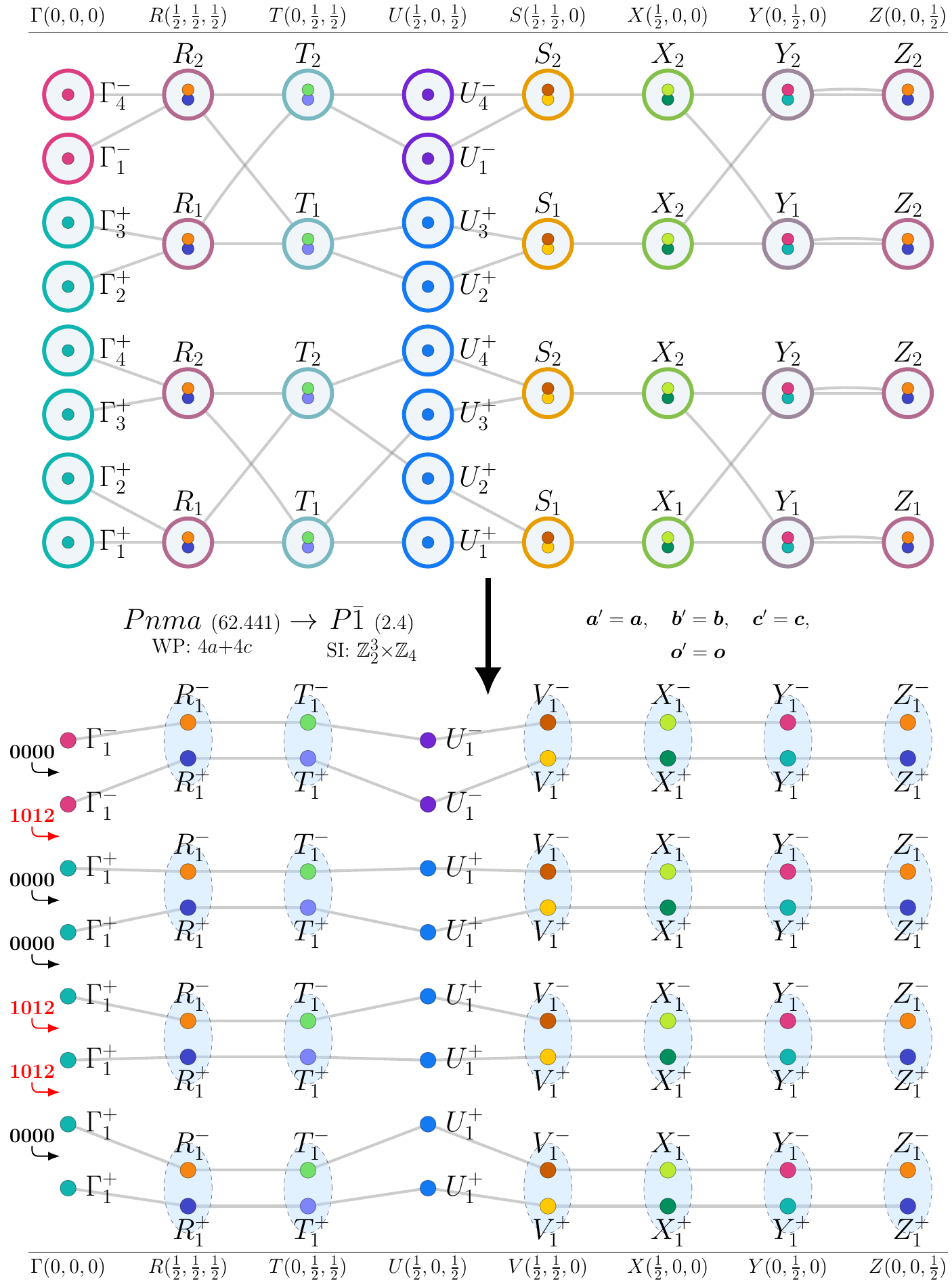}
\caption{Topological magnon bands in subgroup $P\bar{1}~(2.4)$ for magnetic moments on Wyckoff positions $4a+4c$ of supergroup $Pnma~(62.441)$.\label{fig_62.441_2.4_strainingenericdirection_4a+4c}}
\end{figure}
\input{gap_tables_tex/62.441_2.4_strainingenericdirection_4a+4c_table.tex}
\input{si_tables_tex/62.441_2.4_strainingenericdirection_4a+4c_table.tex}
\subsection{WP: $4b+4c$}
\textbf{BCS Materials:} {CeFeO\textsubscript{3}~(220 K)}\footnote{BCS web page: \texttt{\href{http://webbdcrista1.ehu.es/magndata/index.php?this\_label=0.759} {http://webbdcrista1.ehu.es/magndata/index.php?this\_label=0.759}}}, {Rb\textsubscript{2}Fe\textsubscript{2}O(AsO\textsubscript{4})\textsubscript{2}~(25 K)}\footnote{BCS web page: \texttt{\href{http://webbdcrista1.ehu.es/magndata/index.php?this\_label=0.90} {http://webbdcrista1.ehu.es/magndata/index.php?this\_label=0.90}}}, {SmCrO\textsubscript{3}~(10 K)}\footnote{BCS web page: \texttt{\href{http://webbdcrista1.ehu.es/magndata/index.php?this\_label=0.478} {http://webbdcrista1.ehu.es/magndata/index.php?this\_label=0.478}}}, {Nd\textsubscript{0.85}Sr\textsubscript{0.15}CrO\textsubscript{3}}\footnote{BCS web page: \texttt{\href{http://webbdcrista1.ehu.es/magndata/index.php?this\_label=0.678} {http://webbdcrista1.ehu.es/magndata/index.php?this\_label=0.678}}}, {Nd\textsubscript{0.9}Sr\textsubscript{0.1}CrO\textsubscript{3}}\footnote{BCS web page: \texttt{\href{http://webbdcrista1.ehu.es/magndata/index.php?this\_label=0.677} {http://webbdcrista1.ehu.es/magndata/index.php?this\_label=0.677}}}, {Nd\textsubscript{0.95}Sr\textsubscript{0.05}CrO\textsubscript{3}}\footnote{BCS web page: \texttt{\href{http://webbdcrista1.ehu.es/magndata/index.php?this\_label=0.676} {http://webbdcrista1.ehu.es/magndata/index.php?this\_label=0.676}}}.\\
\subsubsection{Topological bands in subgroup $P\bar{1}~(2.4)$}
\textbf{Perturbations:}
\begin{itemize}
\item strain in generic direction,
\item B $\parallel$ [100] and strain $\perp$ [010],
\item B $\parallel$ [100] and strain $\perp$ [001],
\item B $\parallel$ [010] and strain $\perp$ [100],
\item B $\parallel$ [010] and strain $\perp$ [001],
\item B $\parallel$ [001] and strain $\perp$ [100],
\item B $\parallel$ [001] and strain $\perp$ [010],
\item B in generic direction.
\end{itemize}
\begin{figure}[H]
\centering
\includegraphics[scale=0.6]{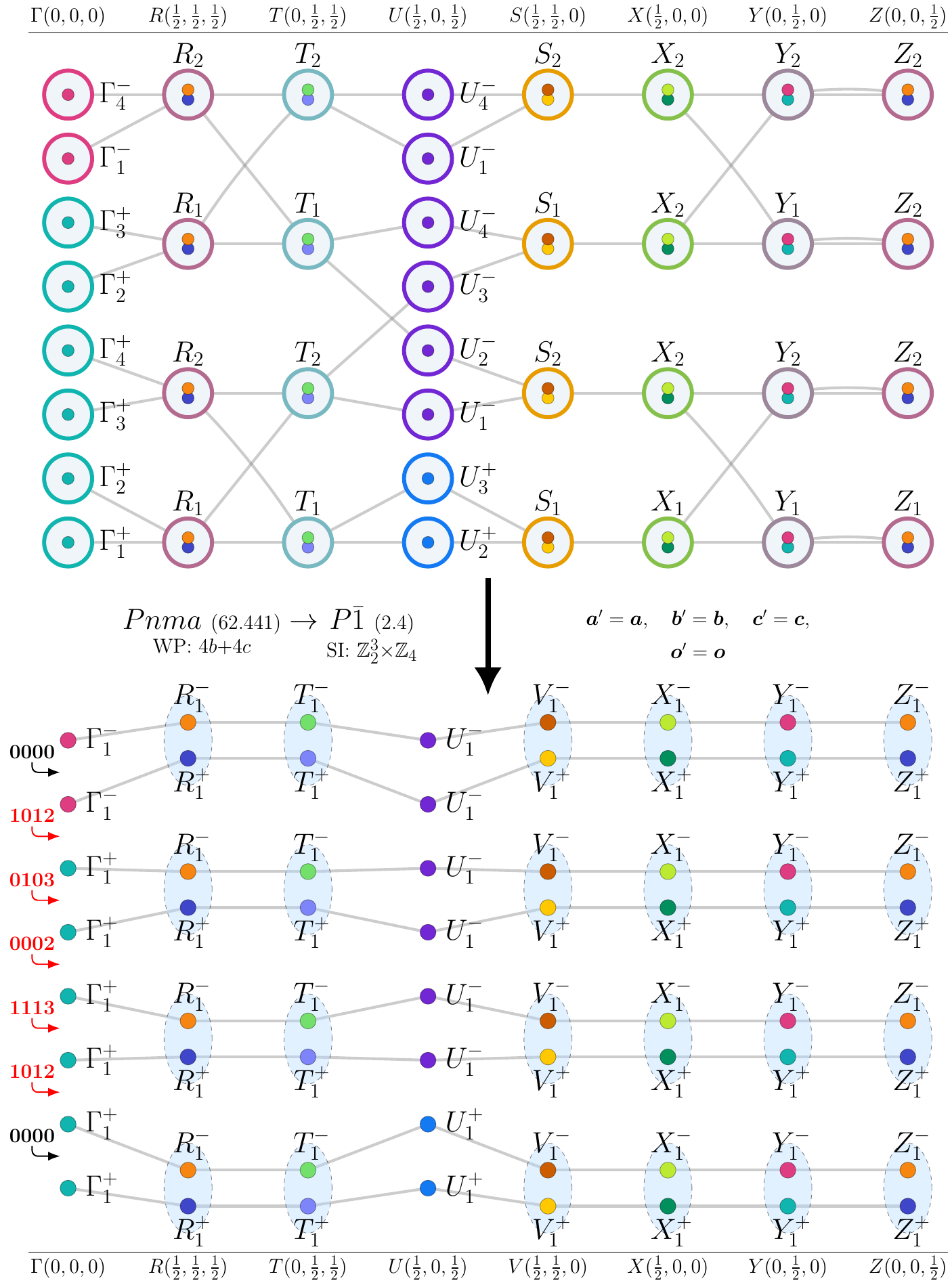}
\caption{Topological magnon bands in subgroup $P\bar{1}~(2.4)$ for magnetic moments on Wyckoff positions $4b+4c$ of supergroup $Pnma~(62.441)$.\label{fig_62.441_2.4_strainingenericdirection_4b+4c}}
\end{figure}
\input{gap_tables_tex/62.441_2.4_strainingenericdirection_4b+4c_table.tex}
\input{si_tables_tex/62.441_2.4_strainingenericdirection_4b+4c_table.tex}
\subsection{WP: $4a$}
\textbf{BCS Materials:} {CoSO\textsubscript{4}~(12 K)}\footnote{BCS web page: \texttt{\href{http://webbdcrista1.ehu.es/magndata/index.php?this\_label=0.571} {http://webbdcrista1.ehu.es/magndata/index.php?this\_label=0.571}}}, {CoSO\textsubscript{4}}\footnote{BCS web page: \texttt{\href{http://webbdcrista1.ehu.es/magndata/index.php?this\_label=0.96} {http://webbdcrista1.ehu.es/magndata/index.php?this\_label=0.96}}}.\\
\subsubsection{Topological bands in subgroup $P\bar{1}~(2.4)$}
\textbf{Perturbations:}
\begin{itemize}
\item strain in generic direction,
\item B $\parallel$ [100] and strain $\perp$ [010],
\item B $\parallel$ [100] and strain $\perp$ [001],
\item B $\parallel$ [010] and strain $\perp$ [100],
\item B $\parallel$ [010] and strain $\perp$ [001],
\item B $\parallel$ [001] and strain $\perp$ [100],
\item B $\parallel$ [001] and strain $\perp$ [010],
\item B in generic direction.
\end{itemize}
\begin{figure}[H]
\centering
\includegraphics[scale=0.6]{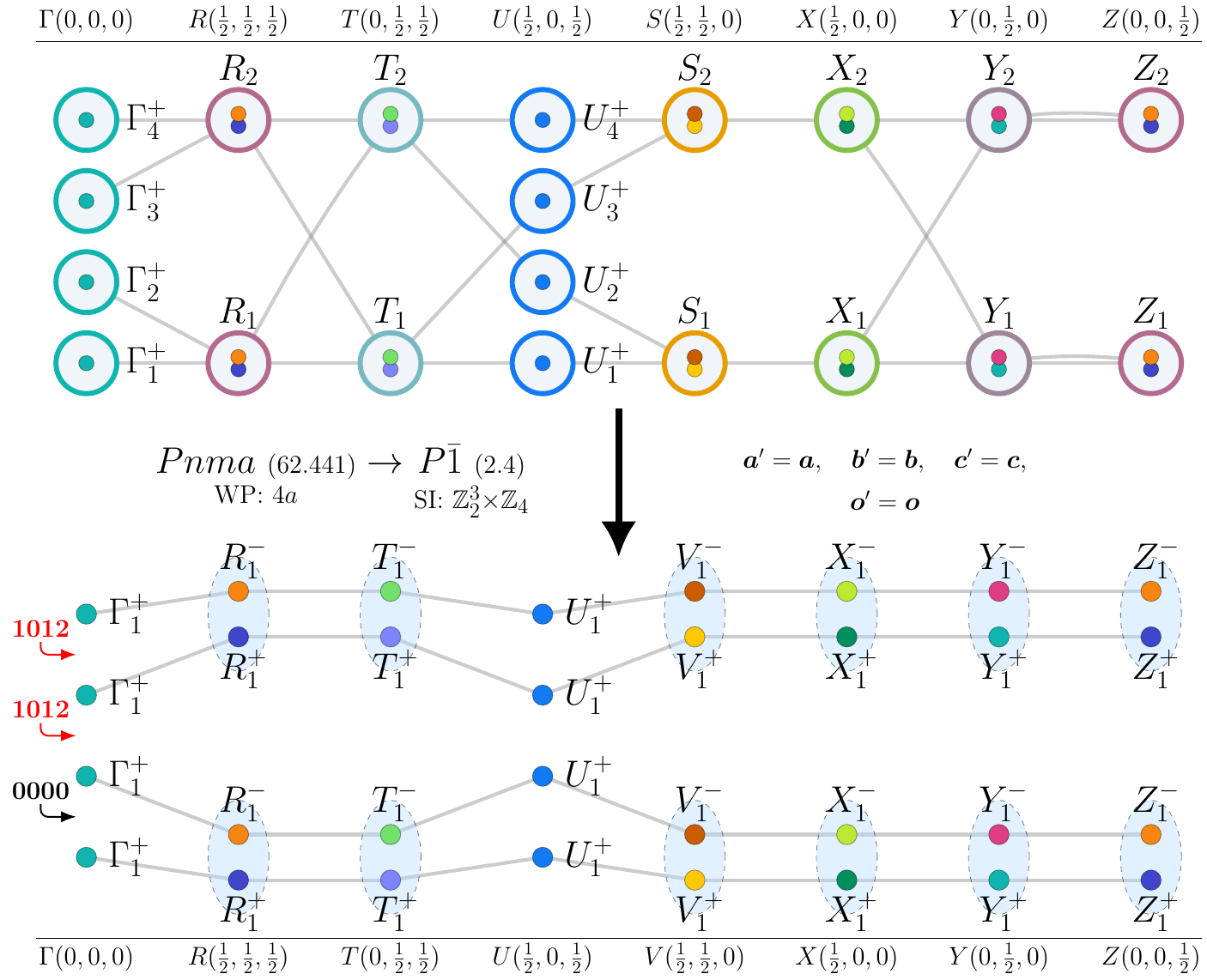}
\caption{Topological magnon bands in subgroup $P\bar{1}~(2.4)$ for magnetic moments on Wyckoff position $4a$ of supergroup $Pnma~(62.441)$.\label{fig_62.441_2.4_strainingenericdirection_4a}}
\end{figure}
\input{gap_tables_tex/62.441_2.4_strainingenericdirection_4a_table.tex}
\input{si_tables_tex/62.441_2.4_strainingenericdirection_4a_table.tex}

\section{MSG $Pn'm'a~(62.446)$}
\textbf{Nontrivial-SI Subgroups:} $P\bar{1}~(2.4)$, $P2_{1}'/m'~(11.54)$, $P2_{1}'/c'~(14.79)$, $P2_{1}/c~(14.75)$.\\

\textbf{Trivial-SI Subgroups:} $Pm'~(6.20)$, $Pc'~(7.26)$, $P2_{1}'~(4.9)$, $P2_{1}'~(4.9)$, $Pc~(7.24)$, $Pn'a2_{1}'~(33.146)$, $Pm'c2_{1}'~(26.68)$, $P2_{1}~(4.7)$, $Pm'n'2_{1}~(31.127)$.\\

\subsection{WP: $4b$}
\textbf{BCS Materials:} {SmFeO\textsubscript{3}~(480 K)}\footnote{BCS web page: \texttt{\href{http://webbdcrista1.ehu.es/magndata/index.php?this\_label=0.379} {http://webbdcrista1.ehu.es/magndata/index.php?this\_label=0.379}}}, {SrRuO\textsubscript{3}~(165 K)}\footnote{BCS web page: \texttt{\href{http://webbdcrista1.ehu.es/magndata/index.php?this\_label=0.732} {http://webbdcrista1.ehu.es/magndata/index.php?this\_label=0.732}}}, {TbCrO\textsubscript{3}~(158 K)}\footnote{BCS web page: \texttt{\href{http://webbdcrista1.ehu.es/magndata/index.php?this\_label=0.354} {http://webbdcrista1.ehu.es/magndata/index.php?this\_label=0.354}}}, {DyCrO\textsubscript{3}~(146 K)}\footnote{BCS web page: \texttt{\href{http://webbdcrista1.ehu.es/magndata/index.php?this\_label=0.592} {http://webbdcrista1.ehu.es/magndata/index.php?this\_label=0.592}}}, {TeNiO\textsubscript{3}~(130 K)}\footnote{BCS web page: \texttt{\href{http://webbdcrista1.ehu.es/magndata/index.php?this\_label=0.94} {http://webbdcrista1.ehu.es/magndata/index.php?this\_label=0.94}}}, {Ca\textsubscript{2}PrCr\textsubscript{2}TaO\textsubscript{9}~(130 K)}\footnote{BCS web page: \texttt{\href{http://webbdcrista1.ehu.es/magndata/index.php?this\_label=0.202} {http://webbdcrista1.ehu.es/magndata/index.php?this\_label=0.202}}}, {Ca\textsubscript{2}PrCr\textsubscript{2}NbO\textsubscript{9}~(110 K)}\footnote{BCS web page: \texttt{\href{http://webbdcrista1.ehu.es/magndata/index.php?this\_label=0.201} {http://webbdcrista1.ehu.es/magndata/index.php?this\_label=0.201}}}, {YVO\textsubscript{3}~(77 K)}\footnote{BCS web page: \texttt{\href{http://webbdcrista1.ehu.es/magndata/index.php?this\_label=0.787} {http://webbdcrista1.ehu.es/magndata/index.php?this\_label=0.787}}}.\\
\subsubsection{Topological bands in subgroup $P\bar{1}~(2.4)$}
\textbf{Perturbations:}
\begin{itemize}
\item strain in generic direction,
\item (B $\parallel$ [100] or B $\perp$ [010]) and strain $\perp$ [100],
\item (B $\parallel$ [100] or B $\perp$ [010]) and strain $\perp$ [001],
\item (B $\parallel$ [010] or B $\perp$ [100]) and strain $\perp$ [010],
\item (B $\parallel$ [010] or B $\perp$ [100]) and strain $\perp$ [001],
\item B in generic direction.
\end{itemize}
\begin{figure}[H]
\centering
\includegraphics[scale=0.6]{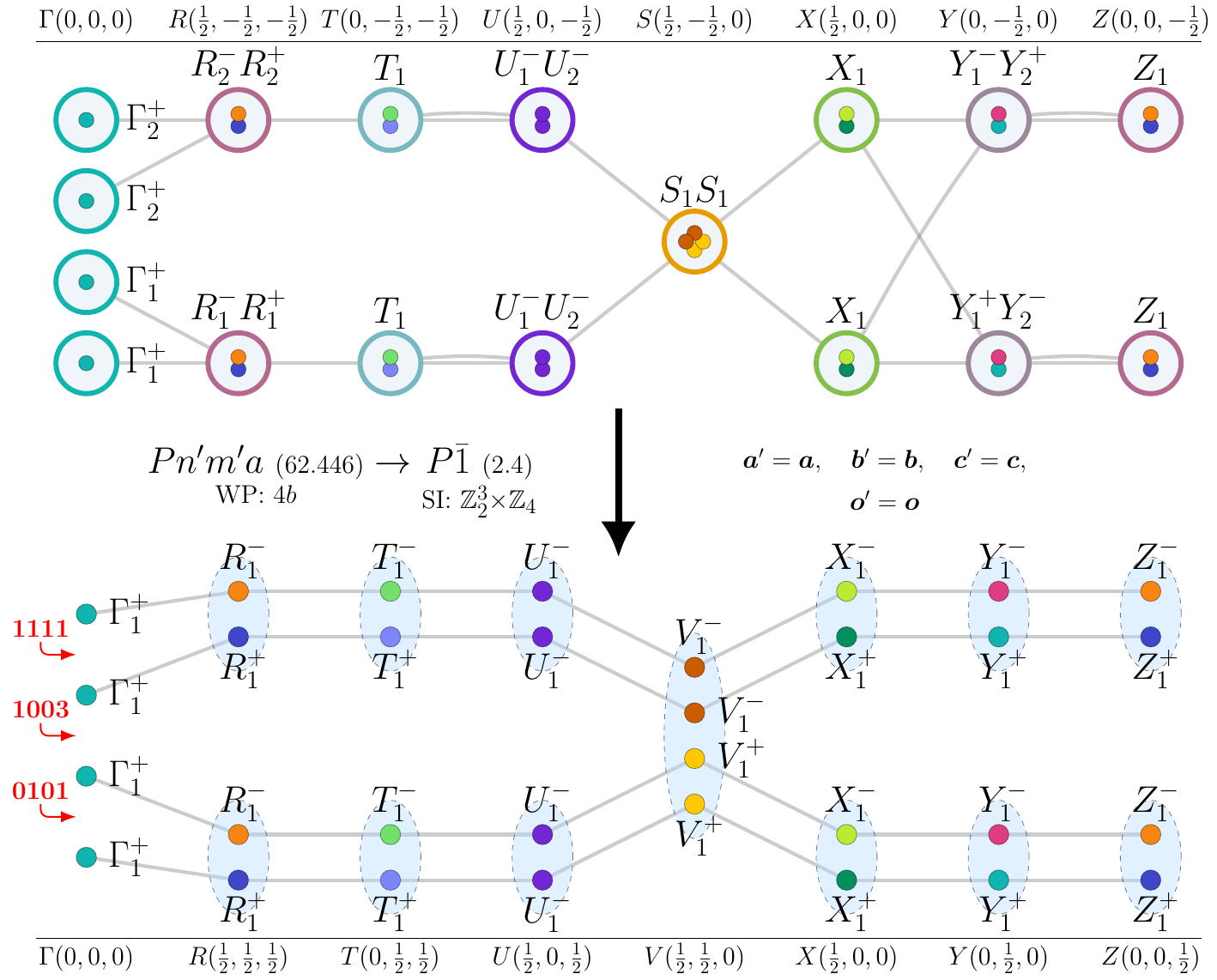}
\caption{Topological magnon bands in subgroup $P\bar{1}~(2.4)$ for magnetic moments on Wyckoff position $4b$ of supergroup $Pn'm'a~(62.446)$.\label{fig_62.446_2.4_strainingenericdirection_4b}}
\end{figure}
\input{gap_tables_tex/62.446_2.4_strainingenericdirection_4b_table.tex}
\input{si_tables_tex/62.446_2.4_strainingenericdirection_4b_table.tex}
\subsubsection{Topological bands in subgroup $P2_{1}'/m'~(11.54)$}
\textbf{Perturbations:}
\begin{itemize}
\item strain $\perp$ [010],
\item (B $\parallel$ [100] or B $\perp$ [010]).
\end{itemize}
\begin{figure}[H]
\centering
\includegraphics[scale=0.6]{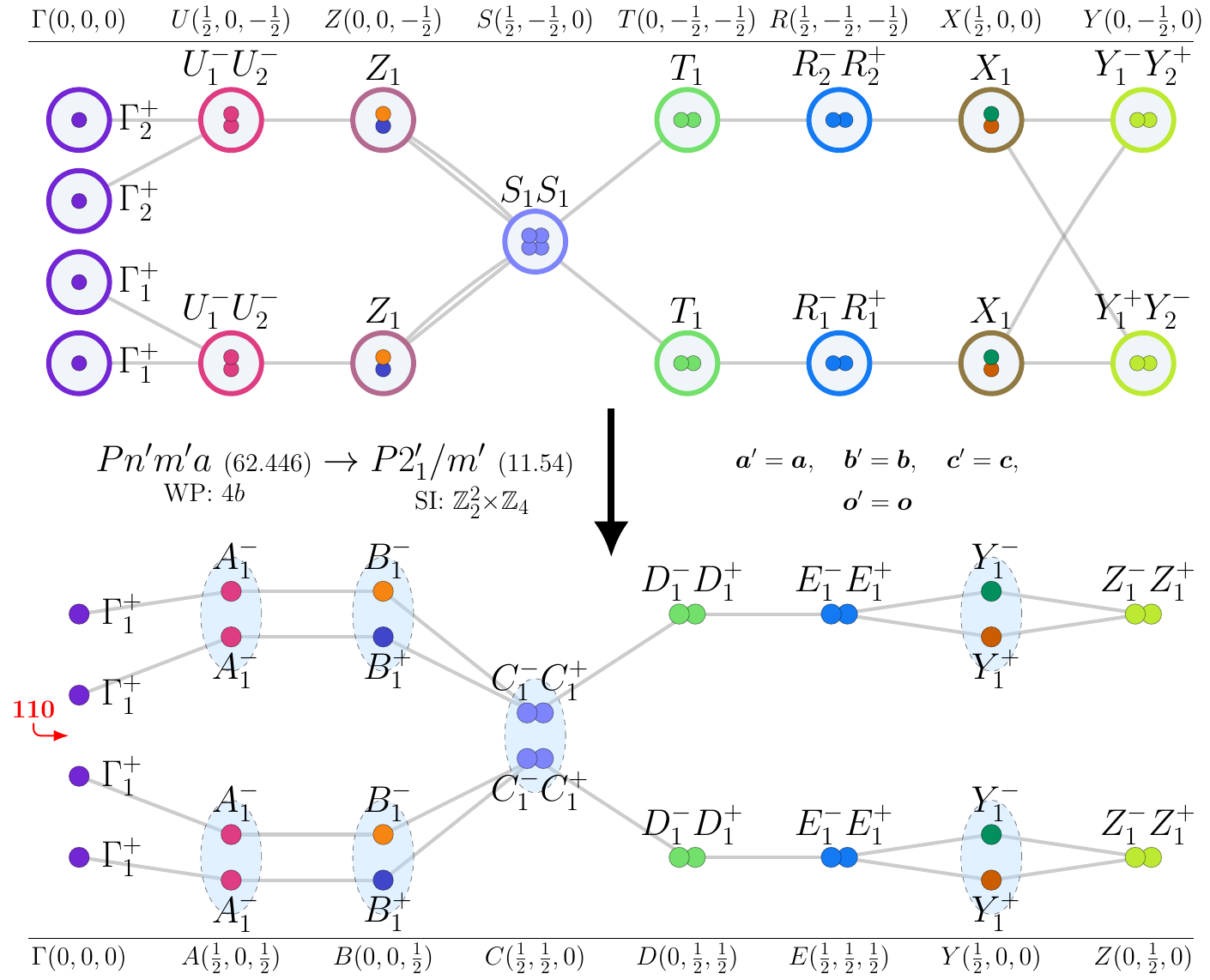}
\caption{Topological magnon bands in subgroup $P2_{1}'/m'~(11.54)$ for magnetic moments on Wyckoff position $4b$ of supergroup $Pn'm'a~(62.446)$.\label{fig_62.446_11.54_strainperp010_4b}}
\end{figure}
\input{gap_tables_tex/62.446_11.54_strainperp010_4b_table.tex}
\input{si_tables_tex/62.446_11.54_strainperp010_4b_table.tex}
\subsubsection{Topological bands in subgroup $P2_{1}'/c'~(14.79)$}
\textbf{Perturbations:}
\begin{itemize}
\item strain $\perp$ [100],
\item (B $\parallel$ [010] or B $\perp$ [100]).
\end{itemize}
\begin{figure}[H]
\centering
\includegraphics[scale=0.6]{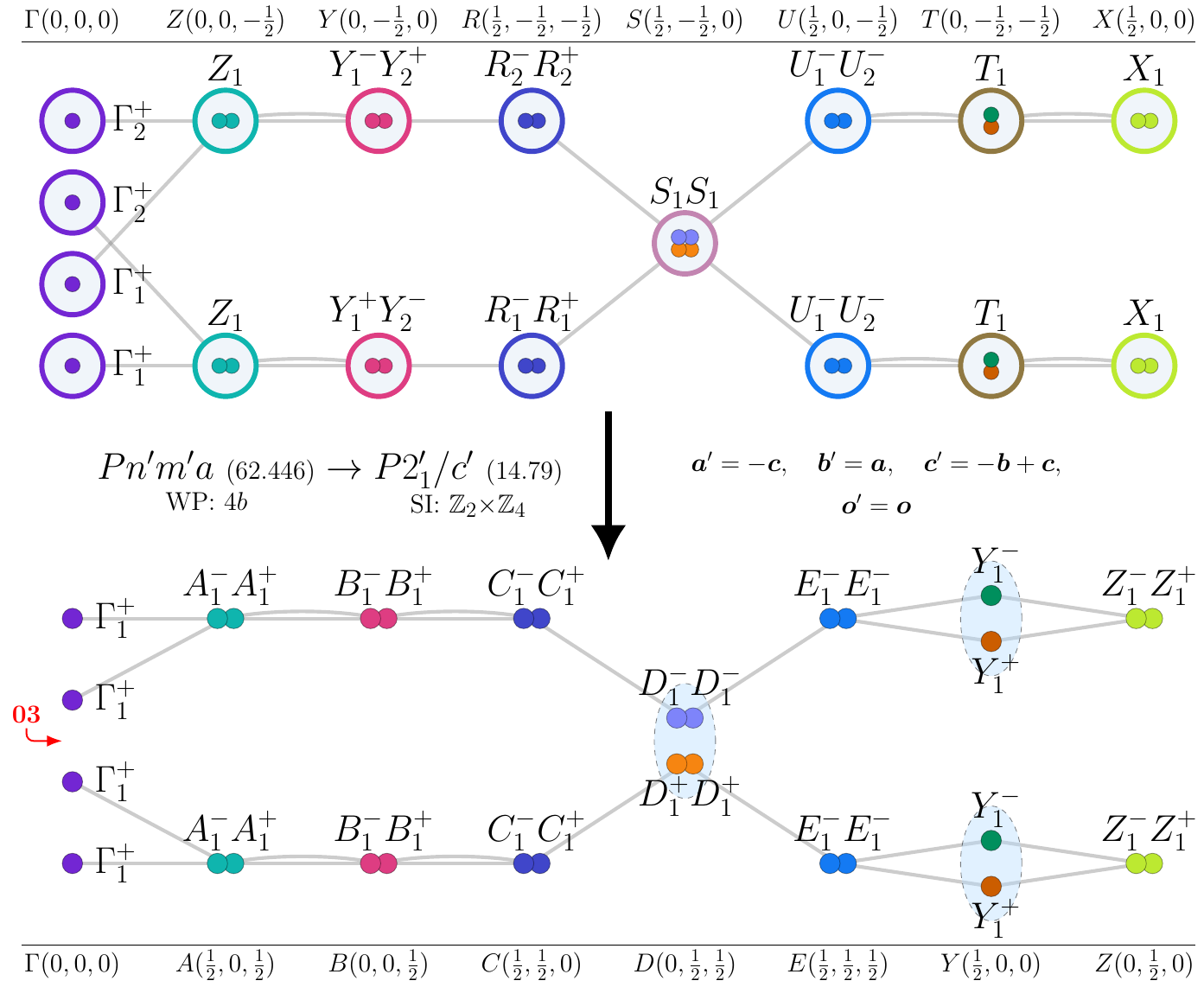}
\caption{Topological magnon bands in subgroup $P2_{1}'/c'~(14.79)$ for magnetic moments on Wyckoff position $4b$ of supergroup $Pn'm'a~(62.446)$.\label{fig_62.446_14.79_strainperp100_4b}}
\end{figure}
\input{gap_tables_tex/62.446_14.79_strainperp100_4b_table.tex}
\input{si_tables_tex/62.446_14.79_strainperp100_4b_table.tex}
\subsection{WP: $4c+4c$}
\textbf{BCS Materials:} {MnCoGe~(345 K)}\footnote{BCS web page: \texttt{\href{http://webbdcrista1.ehu.es/magndata/index.php?this\_label=0.445} {http://webbdcrista1.ehu.es/magndata/index.php?this\_label=0.445}}}.\\
\subsubsection{Topological bands in subgroup $P2_{1}'/m'~(11.54)$}
\textbf{Perturbations:}
\begin{itemize}
\item strain $\perp$ [010],
\item (B $\parallel$ [100] or B $\perp$ [010]).
\end{itemize}
\begin{figure}[H]
\centering
\includegraphics[scale=0.6]{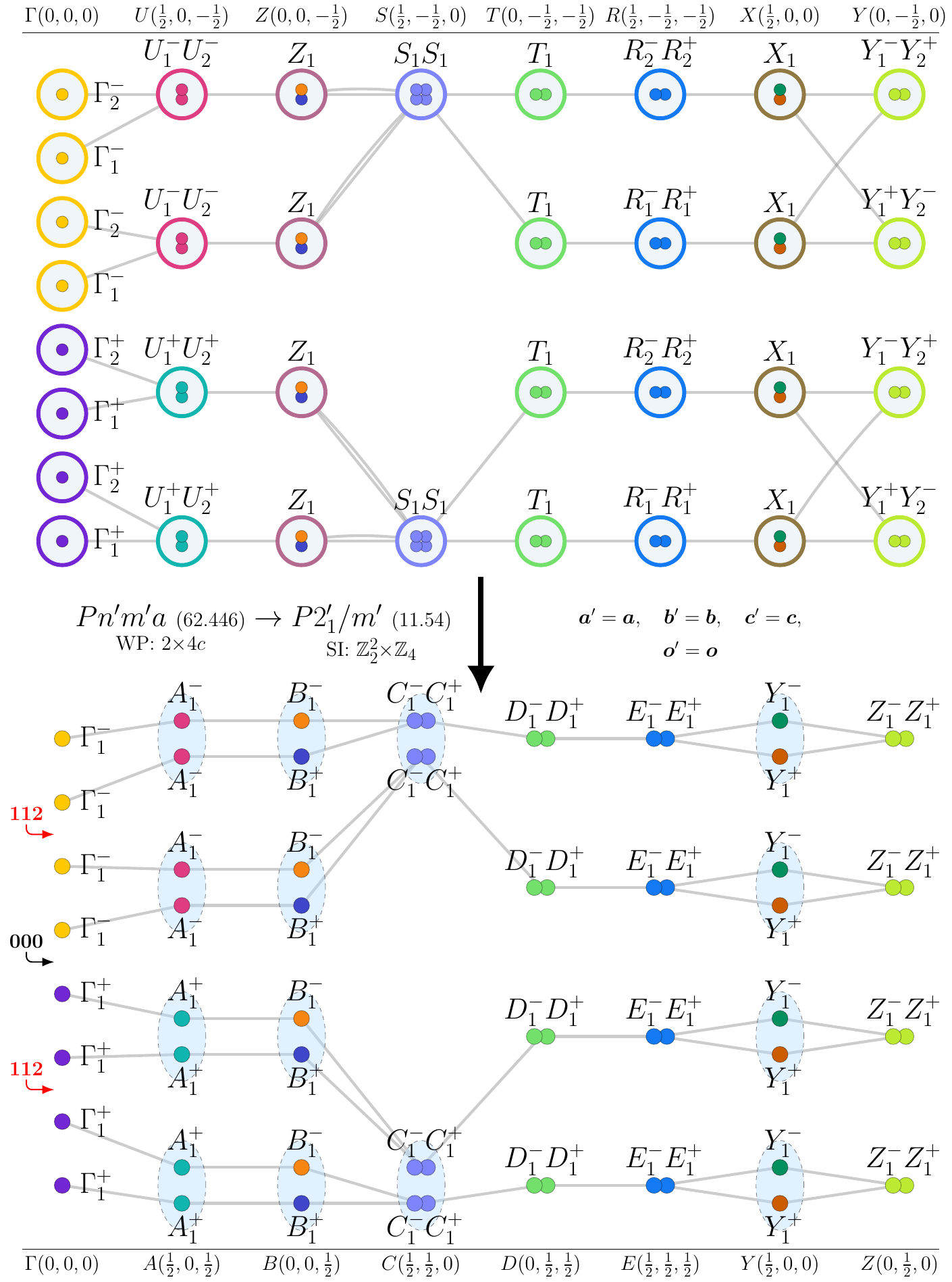}
\caption{Topological magnon bands in subgroup $P2_{1}'/m'~(11.54)$ for magnetic moments on Wyckoff positions $4c+4c$ of supergroup $Pn'm'a~(62.446)$.\label{fig_62.446_11.54_strainperp010_4c+4c}}
\end{figure}
\input{gap_tables_tex/62.446_11.54_strainperp010_4c+4c_table.tex}
\input{si_tables_tex/62.446_11.54_strainperp010_4c+4c_table.tex}
\subsubsection{Topological bands in subgroup $P2_{1}'/c'~(14.79)$}
\textbf{Perturbations:}
\begin{itemize}
\item strain $\perp$ [100],
\item (B $\parallel$ [010] or B $\perp$ [100]).
\end{itemize}
\begin{figure}[H]
\centering
\includegraphics[scale=0.6]{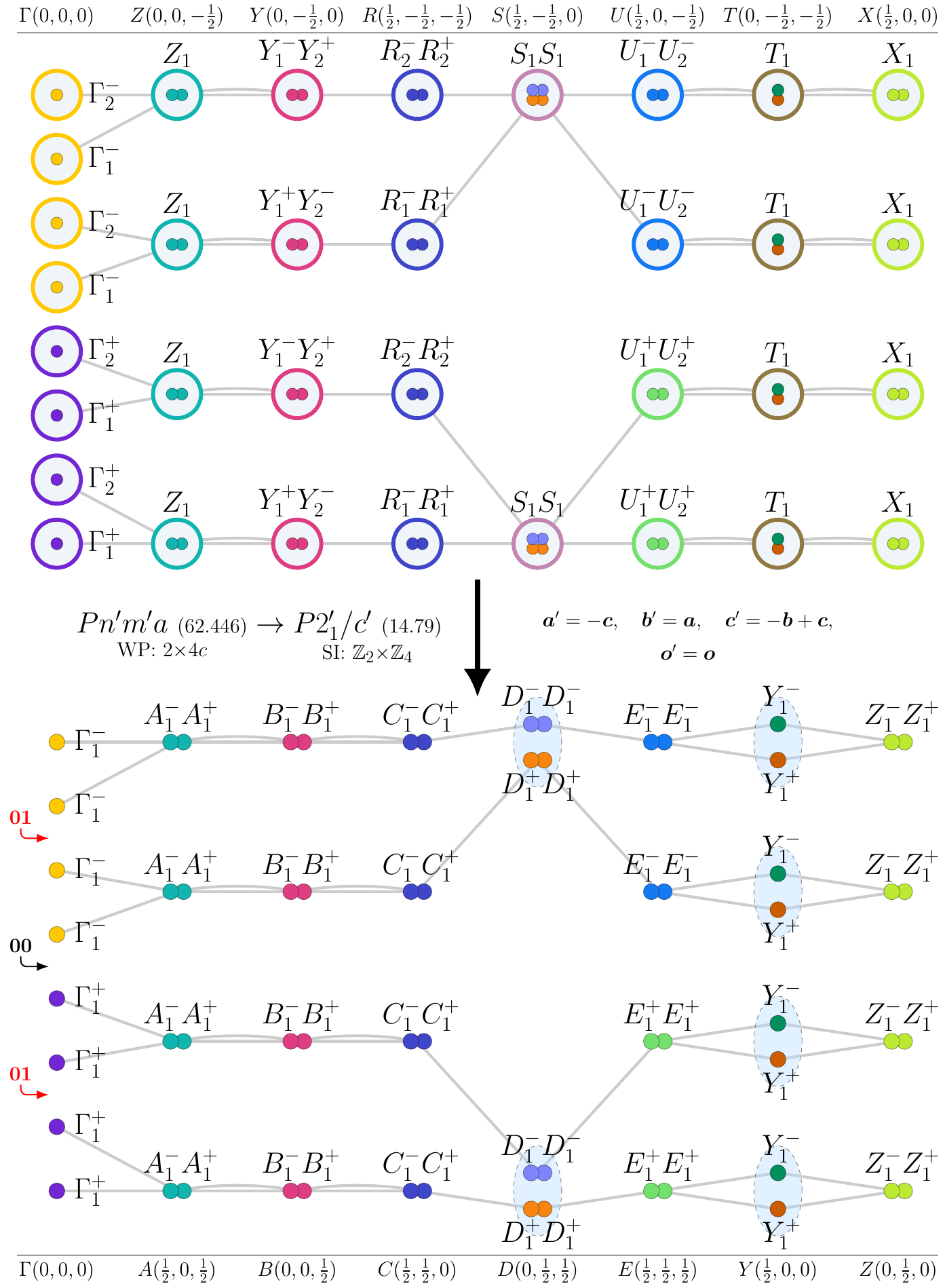}
\caption{Topological magnon bands in subgroup $P2_{1}'/c'~(14.79)$ for magnetic moments on Wyckoff positions $4c+4c$ of supergroup $Pn'm'a~(62.446)$.\label{fig_62.446_14.79_strainperp100_4c+4c}}
\end{figure}
\input{gap_tables_tex/62.446_14.79_strainperp100_4c+4c_table.tex}
\input{si_tables_tex/62.446_14.79_strainperp100_4c+4c_table.tex}
\subsection{WP: $4b+4b$}
\textbf{BCS Materials:} {Bi\textsubscript{0.85}Ca\textsubscript{0.15}Fe\textsubscript{0.55}Mn\textsubscript{0.45}O\textsubscript{3}~(268 K)}\footnote{BCS web page: \texttt{\href{http://webbdcrista1.ehu.es/magndata/index.php?this\_label=0.820} {http://webbdcrista1.ehu.es/magndata/index.php?this\_label=0.820}}}.\\
\subsubsection{Topological bands in subgroup $P\bar{1}~(2.4)$}
\textbf{Perturbations:}
\begin{itemize}
\item strain in generic direction,
\item (B $\parallel$ [100] or B $\perp$ [010]) and strain $\perp$ [100],
\item (B $\parallel$ [100] or B $\perp$ [010]) and strain $\perp$ [001],
\item (B $\parallel$ [010] or B $\perp$ [100]) and strain $\perp$ [010],
\item (B $\parallel$ [010] or B $\perp$ [100]) and strain $\perp$ [001],
\item B in generic direction.
\end{itemize}
\begin{figure}[H]
\centering
\includegraphics[scale=0.6]{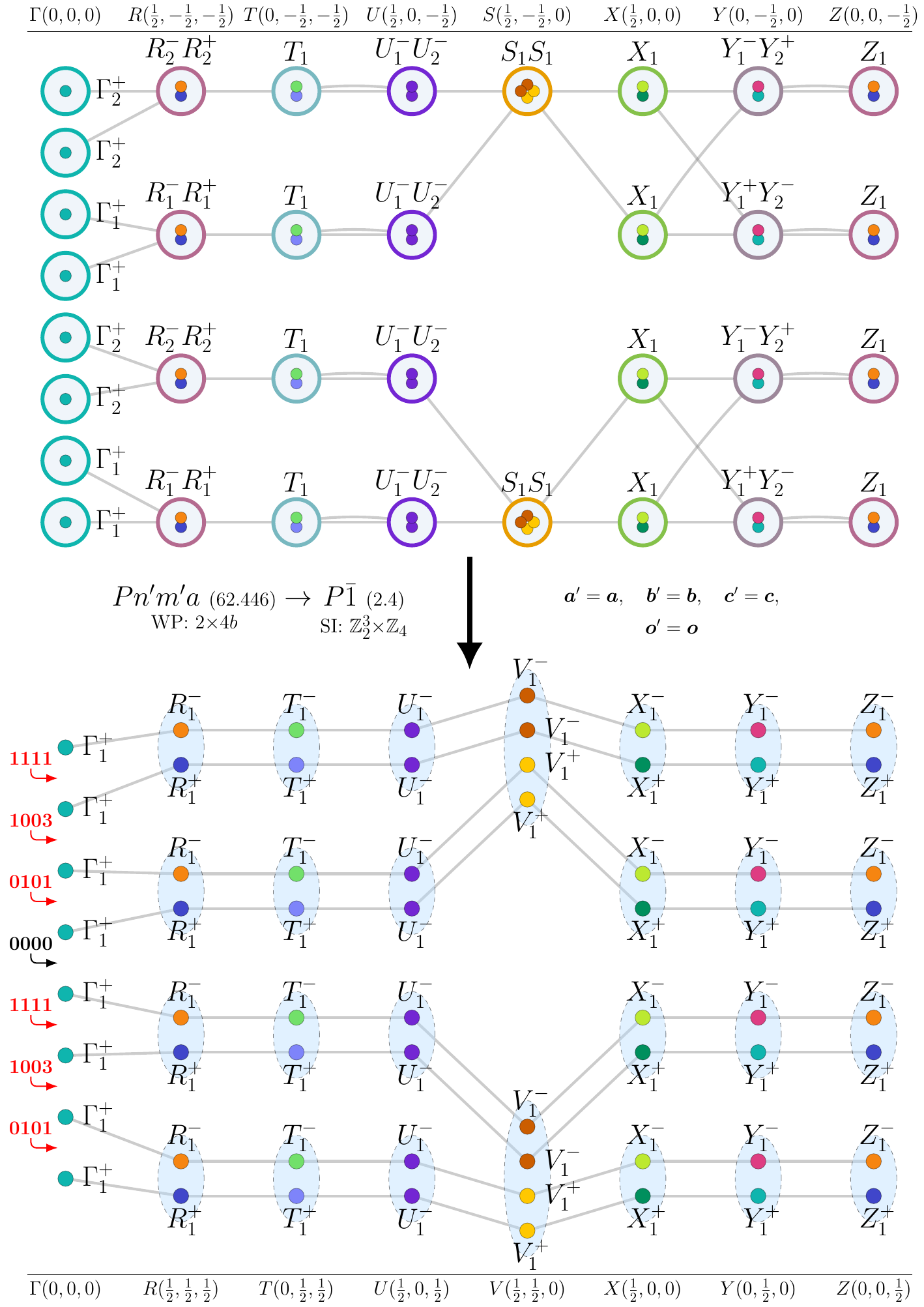}
\caption{Topological magnon bands in subgroup $P\bar{1}~(2.4)$ for magnetic moments on Wyckoff positions $4b+4b$ of supergroup $Pn'm'a~(62.446)$.\label{fig_62.446_2.4_strainingenericdirection_4b+4b}}
\end{figure}
\input{gap_tables_tex/62.446_2.4_strainingenericdirection_4b+4b_table.tex}
\input{si_tables_tex/62.446_2.4_strainingenericdirection_4b+4b_table.tex}
\subsubsection{Topological bands in subgroup $P2_{1}'/m'~(11.54)$}
\textbf{Perturbations:}
\begin{itemize}
\item strain $\perp$ [010],
\item (B $\parallel$ [100] or B $\perp$ [010]).
\end{itemize}
\begin{figure}[H]
\centering
\includegraphics[scale=0.6]{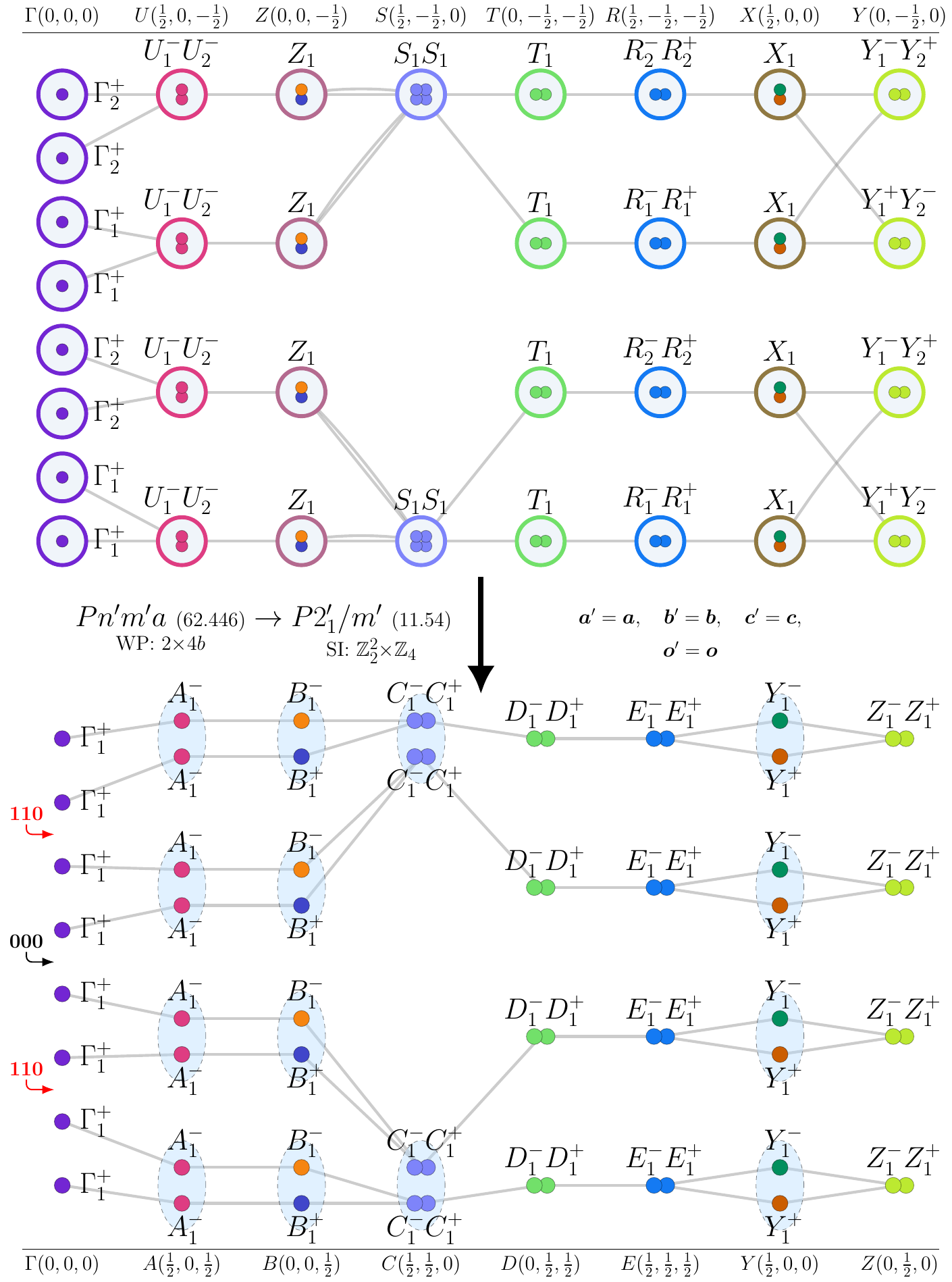}
\caption{Topological magnon bands in subgroup $P2_{1}'/m'~(11.54)$ for magnetic moments on Wyckoff positions $4b+4b$ of supergroup $Pn'm'a~(62.446)$.\label{fig_62.446_11.54_strainperp010_4b+4b}}
\end{figure}
\input{gap_tables_tex/62.446_11.54_strainperp010_4b+4b_table.tex}
\input{si_tables_tex/62.446_11.54_strainperp010_4b+4b_table.tex}
\subsubsection{Topological bands in subgroup $P2_{1}'/c'~(14.79)$}
\textbf{Perturbations:}
\begin{itemize}
\item strain $\perp$ [100],
\item (B $\parallel$ [010] or B $\perp$ [100]).
\end{itemize}
\begin{figure}[H]
\centering
\includegraphics[scale=0.6]{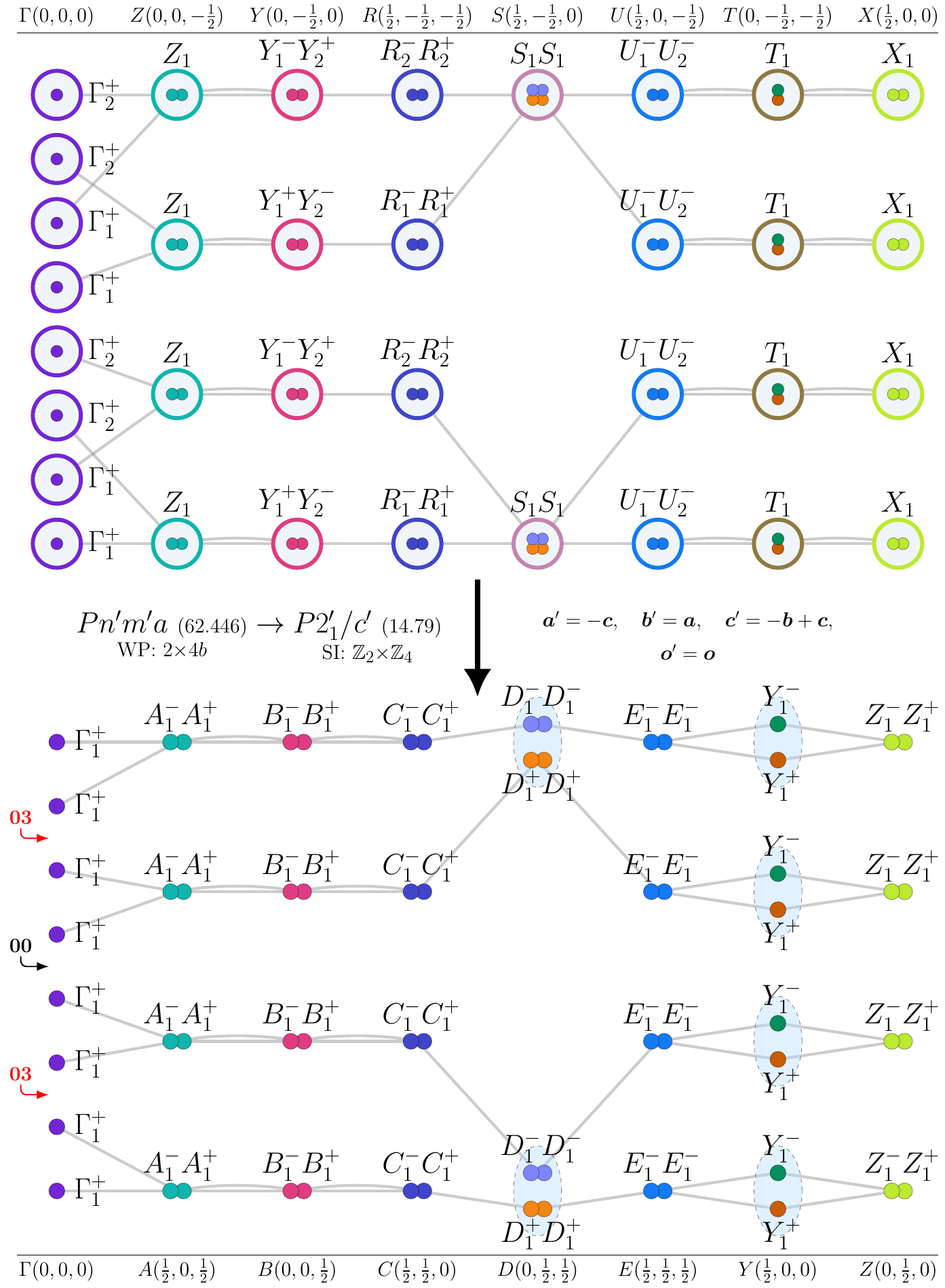}
\caption{Topological magnon bands in subgroup $P2_{1}'/c'~(14.79)$ for magnetic moments on Wyckoff positions $4b+4b$ of supergroup $Pn'm'a~(62.446)$.\label{fig_62.446_14.79_strainperp100_4b+4b}}
\end{figure}
\input{gap_tables_tex/62.446_14.79_strainperp100_4b+4b_table.tex}
\input{si_tables_tex/62.446_14.79_strainperp100_4b+4b_table.tex}
\subsection{WP: $4c$}
\textbf{BCS Materials:} {TbPt~(56 K)}\footnote{BCS web page: \texttt{\href{http://webbdcrista1.ehu.es/magndata/index.php?this\_label=0.684} {http://webbdcrista1.ehu.es/magndata/index.php?this\_label=0.684}}}, {TbPt\textsubscript{0.8}Cu\textsubscript{0.2}~(47 K)}\footnote{BCS web page: \texttt{\href{http://webbdcrista1.ehu.es/magndata/index.php?this\_label=0.248} {http://webbdcrista1.ehu.es/magndata/index.php?this\_label=0.248}}}, {NdSi~(46 K)}\footnote{BCS web page: \texttt{\href{http://webbdcrista1.ehu.es/magndata/index.php?this\_label=0.407} {http://webbdcrista1.ehu.es/magndata/index.php?this\_label=0.407}}}, {DyPt~(23 K)}\footnote{BCS web page: \texttt{\href{http://webbdcrista1.ehu.es/magndata/index.php?this\_label=0.687} {http://webbdcrista1.ehu.es/magndata/index.php?this\_label=0.687}}}, {TmNi~(7 K)}\footnote{BCS web page: \texttt{\href{http://webbdcrista1.ehu.es/magndata/index.php?this\_label=0.409} {http://webbdcrista1.ehu.es/magndata/index.php?this\_label=0.409}}}.\\
\subsubsection{Topological bands in subgroup $P\bar{1}~(2.4)$}
\textbf{Perturbations:}
\begin{itemize}
\item strain in generic direction,
\item (B $\parallel$ [100] or B $\perp$ [010]) and strain $\perp$ [100],
\item (B $\parallel$ [100] or B $\perp$ [010]) and strain $\perp$ [001],
\item (B $\parallel$ [010] or B $\perp$ [100]) and strain $\perp$ [010],
\item (B $\parallel$ [010] or B $\perp$ [100]) and strain $\perp$ [001],
\item B in generic direction.
\end{itemize}
\begin{figure}[H]
\centering
\includegraphics[scale=0.6]{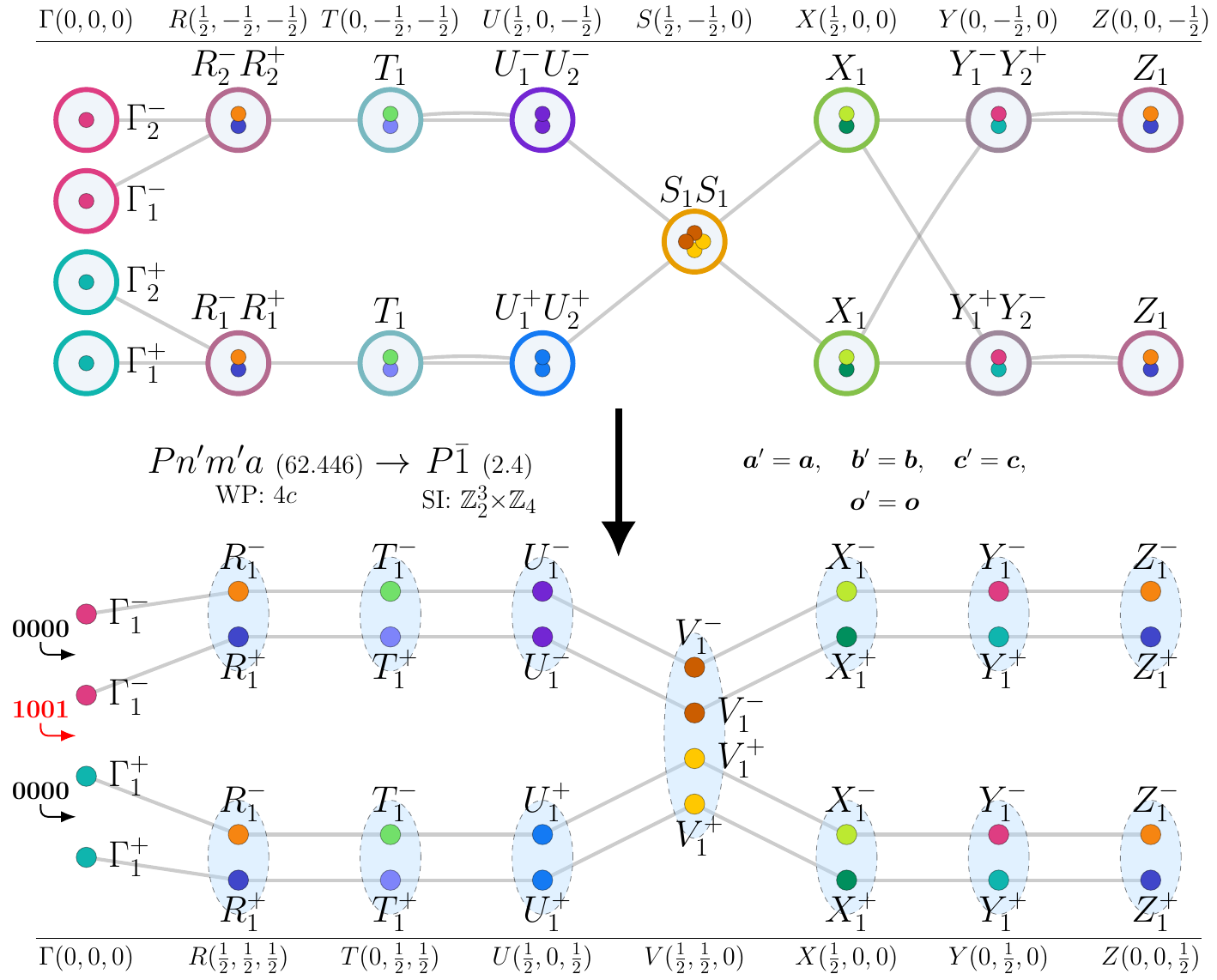}
\caption{Topological magnon bands in subgroup $P\bar{1}~(2.4)$ for magnetic moments on Wyckoff position $4c$ of supergroup $Pn'm'a~(62.446)$.\label{fig_62.446_2.4_strainingenericdirection_4c}}
\end{figure}
\input{gap_tables_tex/62.446_2.4_strainingenericdirection_4c_table.tex}
\input{si_tables_tex/62.446_2.4_strainingenericdirection_4c_table.tex}
\subsubsection{Topological bands in subgroup $P2_{1}'/m'~(11.54)$}
\textbf{Perturbations:}
\begin{itemize}
\item strain $\perp$ [010],
\item (B $\parallel$ [100] or B $\perp$ [010]).
\end{itemize}
\begin{figure}[H]
\centering
\includegraphics[scale=0.6]{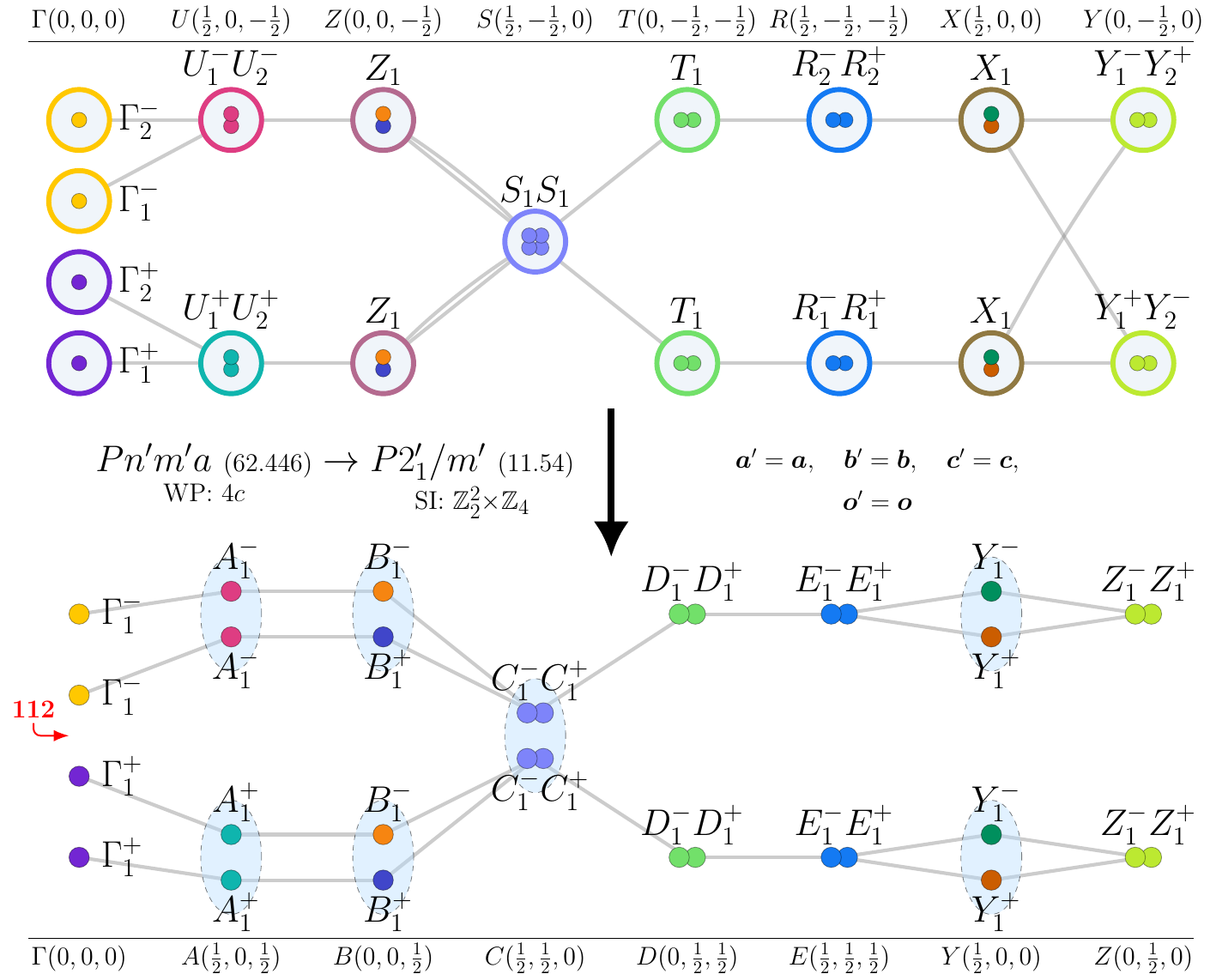}
\caption{Topological magnon bands in subgroup $P2_{1}'/m'~(11.54)$ for magnetic moments on Wyckoff position $4c$ of supergroup $Pn'm'a~(62.446)$.\label{fig_62.446_11.54_strainperp010_4c}}
\end{figure}
\input{gap_tables_tex/62.446_11.54_strainperp010_4c_table.tex}
\input{si_tables_tex/62.446_11.54_strainperp010_4c_table.tex}
\subsubsection{Topological bands in subgroup $P2_{1}'/c'~(14.79)$}
\textbf{Perturbations:}
\begin{itemize}
\item strain $\perp$ [100],
\item (B $\parallel$ [010] or B $\perp$ [100]).
\end{itemize}
\begin{figure}[H]
\centering
\includegraphics[scale=0.6]{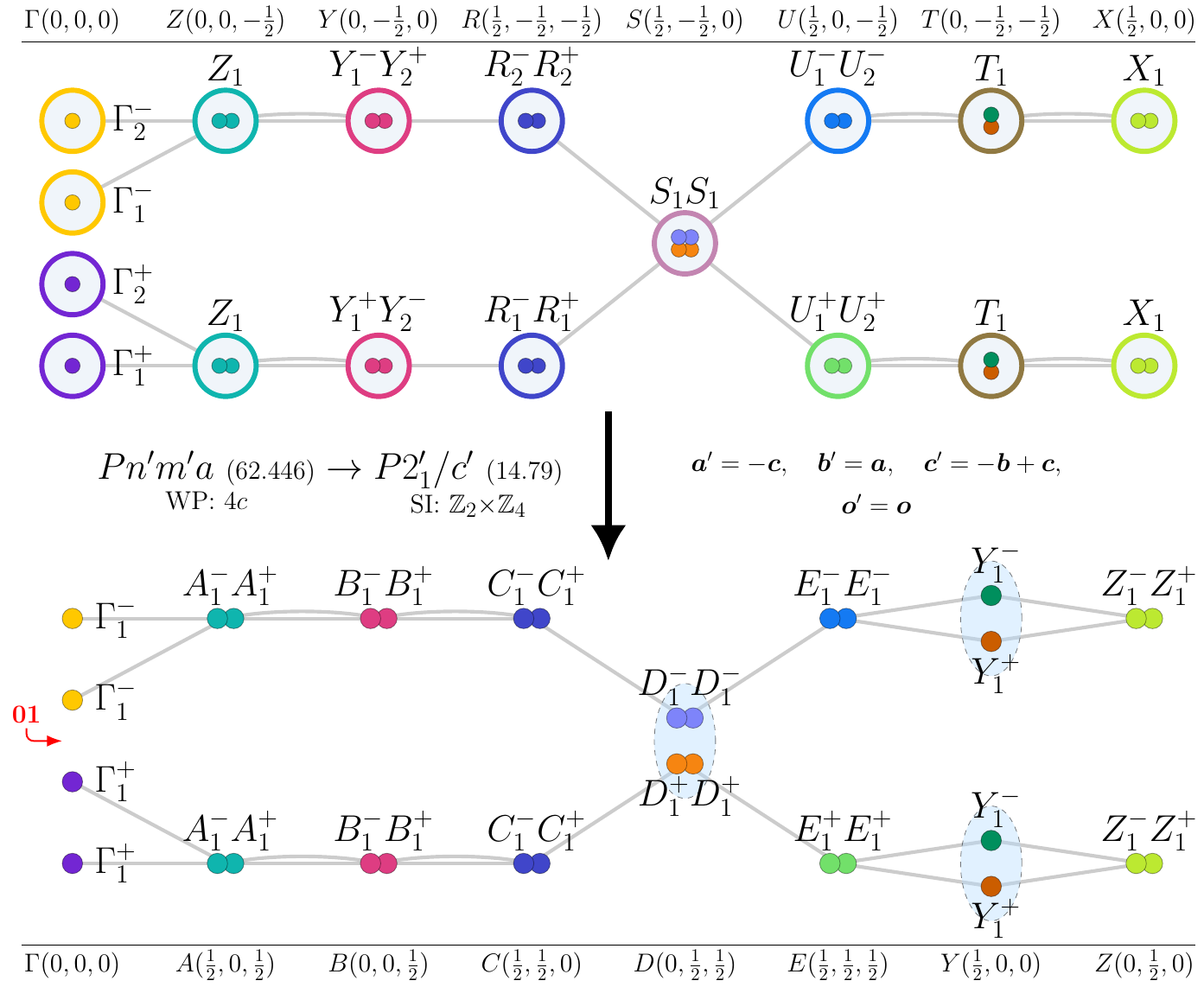}
\caption{Topological magnon bands in subgroup $P2_{1}'/c'~(14.79)$ for magnetic moments on Wyckoff position $4c$ of supergroup $Pn'm'a~(62.446)$.\label{fig_62.446_14.79_strainperp100_4c}}
\end{figure}
\input{gap_tables_tex/62.446_14.79_strainperp100_4c_table.tex}
\input{si_tables_tex/62.446_14.79_strainperp100_4c_table.tex}
\subsection{WP: $4a+4c$}
\textbf{BCS Materials:} {Mn\textsubscript{2}SiO\textsubscript{4}~(47.1 K)}\footnote{BCS web page: \texttt{\href{http://webbdcrista1.ehu.es/magndata/index.php?this\_label=0.220} {http://webbdcrista1.ehu.es/magndata/index.php?this\_label=0.220}}}, {Mn\textsubscript{2}GeO\textsubscript{4}~(47 K)}\footnote{BCS web page: \texttt{\href{http://webbdcrista1.ehu.es/magndata/index.php?this\_label=0.101} {http://webbdcrista1.ehu.es/magndata/index.php?this\_label=0.101}}}, {Ca\textsubscript{2}Fe\textsubscript{2}O\textsubscript{5}}\footnote{BCS web page: \texttt{\href{http://webbdcrista1.ehu.es/magndata/index.php?this\_label=0.93} {http://webbdcrista1.ehu.es/magndata/index.php?this\_label=0.93}}}.\\
\subsubsection{Topological bands in subgroup $P2_{1}'/m'~(11.54)$}
\textbf{Perturbations:}
\begin{itemize}
\item strain $\perp$ [010],
\item (B $\parallel$ [100] or B $\perp$ [010]).
\end{itemize}
\begin{figure}[H]
\centering
\includegraphics[scale=0.6]{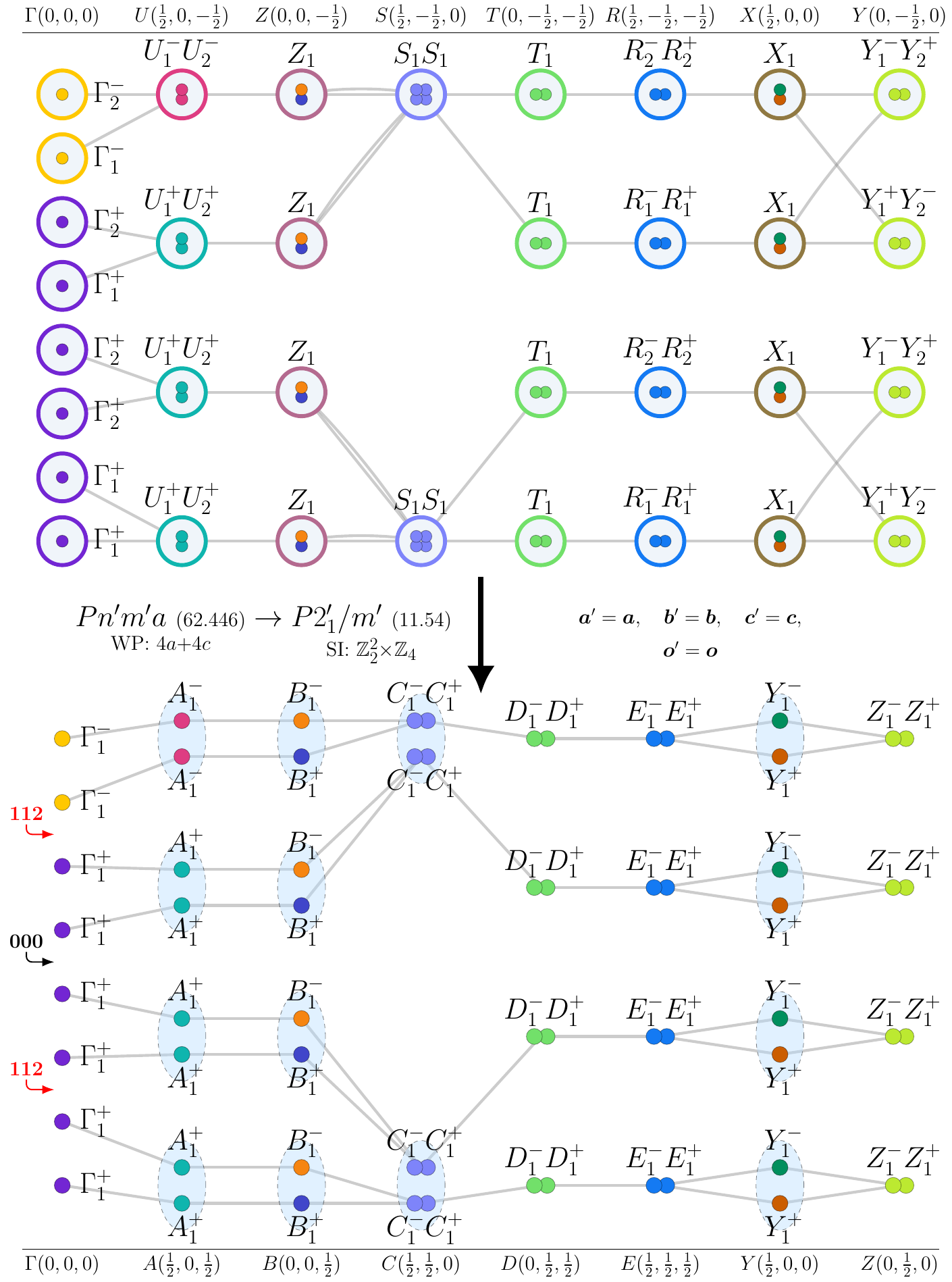}
\caption{Topological magnon bands in subgroup $P2_{1}'/m'~(11.54)$ for magnetic moments on Wyckoff positions $4a+4c$ of supergroup $Pn'm'a~(62.446)$.\label{fig_62.446_11.54_strainperp010_4a+4c}}
\end{figure}
\input{gap_tables_tex/62.446_11.54_strainperp010_4a+4c_table.tex}
\input{si_tables_tex/62.446_11.54_strainperp010_4a+4c_table.tex}
\subsubsection{Topological bands in subgroup $P2_{1}'/c'~(14.79)$}
\textbf{Perturbations:}
\begin{itemize}
\item strain $\perp$ [100],
\item (B $\parallel$ [010] or B $\perp$ [100]).
\end{itemize}
\begin{figure}[H]
\centering
\includegraphics[scale=0.6]{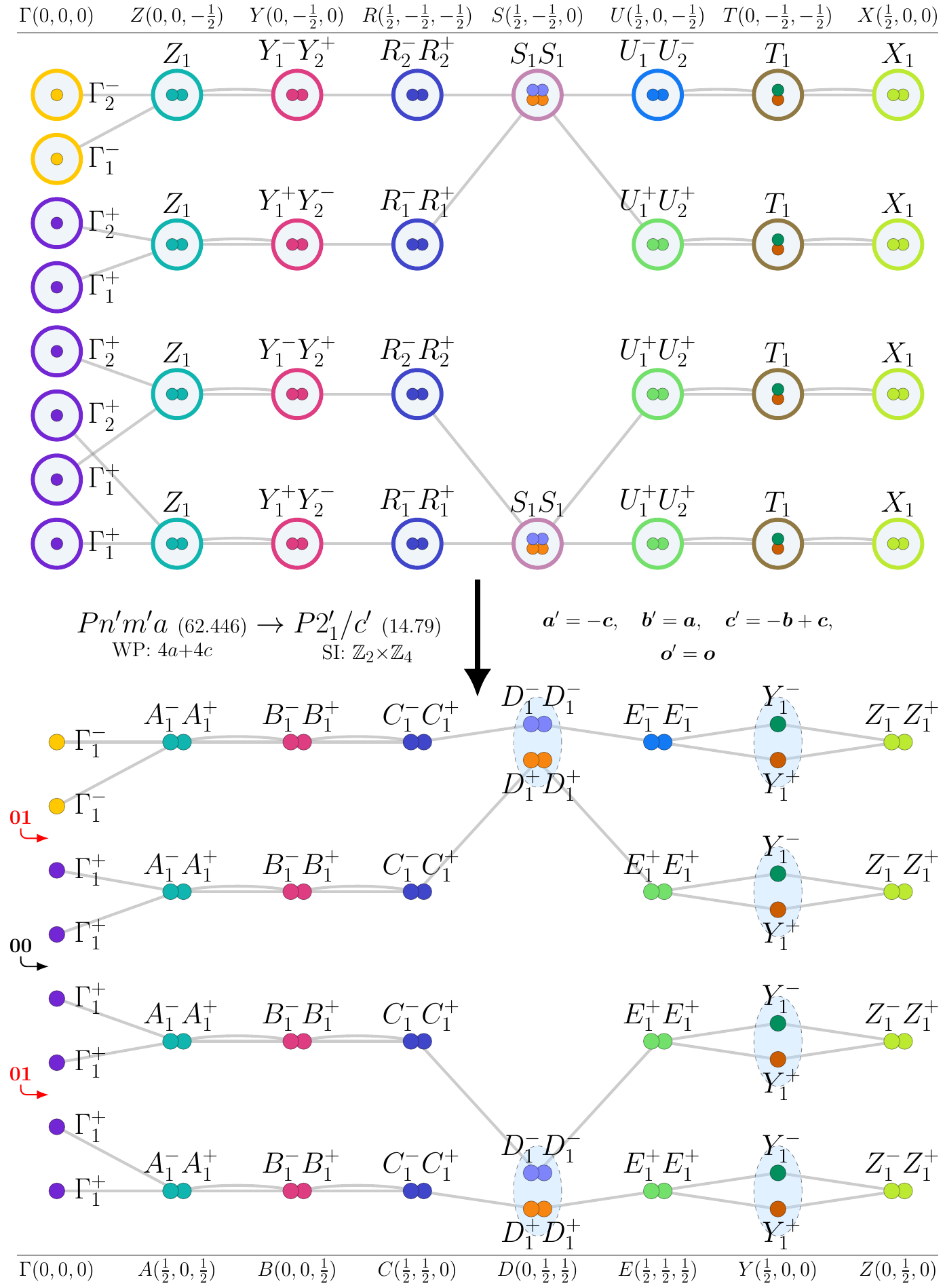}
\caption{Topological magnon bands in subgroup $P2_{1}'/c'~(14.79)$ for magnetic moments on Wyckoff positions $4a+4c$ of supergroup $Pn'm'a~(62.446)$.\label{fig_62.446_14.79_strainperp100_4a+4c}}
\end{figure}
\input{gap_tables_tex/62.446_14.79_strainperp100_4a+4c_table.tex}
\input{si_tables_tex/62.446_14.79_strainperp100_4a+4c_table.tex}
\subsection{WP: $4b+4c$}
\textbf{BCS Materials:} {SmCrO\textsubscript{3}~(37 K)}\footnote{BCS web page: \texttt{\href{http://webbdcrista1.ehu.es/magndata/index.php?this\_label=0.698} {http://webbdcrista1.ehu.es/magndata/index.php?this\_label=0.698}}}, {SmCrO\textsubscript{3}~(37 K)}\footnote{BCS web page: \texttt{\href{http://webbdcrista1.ehu.es/magndata/index.php?this\_label=0.697} {http://webbdcrista1.ehu.es/magndata/index.php?this\_label=0.697}}}, {TbFeO\textsubscript{3}~(8.4 K)}\footnote{BCS web page: \texttt{\href{http://webbdcrista1.ehu.es/magndata/index.php?this\_label=0.352} {http://webbdcrista1.ehu.es/magndata/index.php?this\_label=0.352}}}.\\
\subsubsection{Topological bands in subgroup $P2_{1}'/m'~(11.54)$}
\textbf{Perturbations:}
\begin{itemize}
\item strain $\perp$ [010],
\item (B $\parallel$ [100] or B $\perp$ [010]).
\end{itemize}
\begin{figure}[H]
\centering
\includegraphics[scale=0.6]{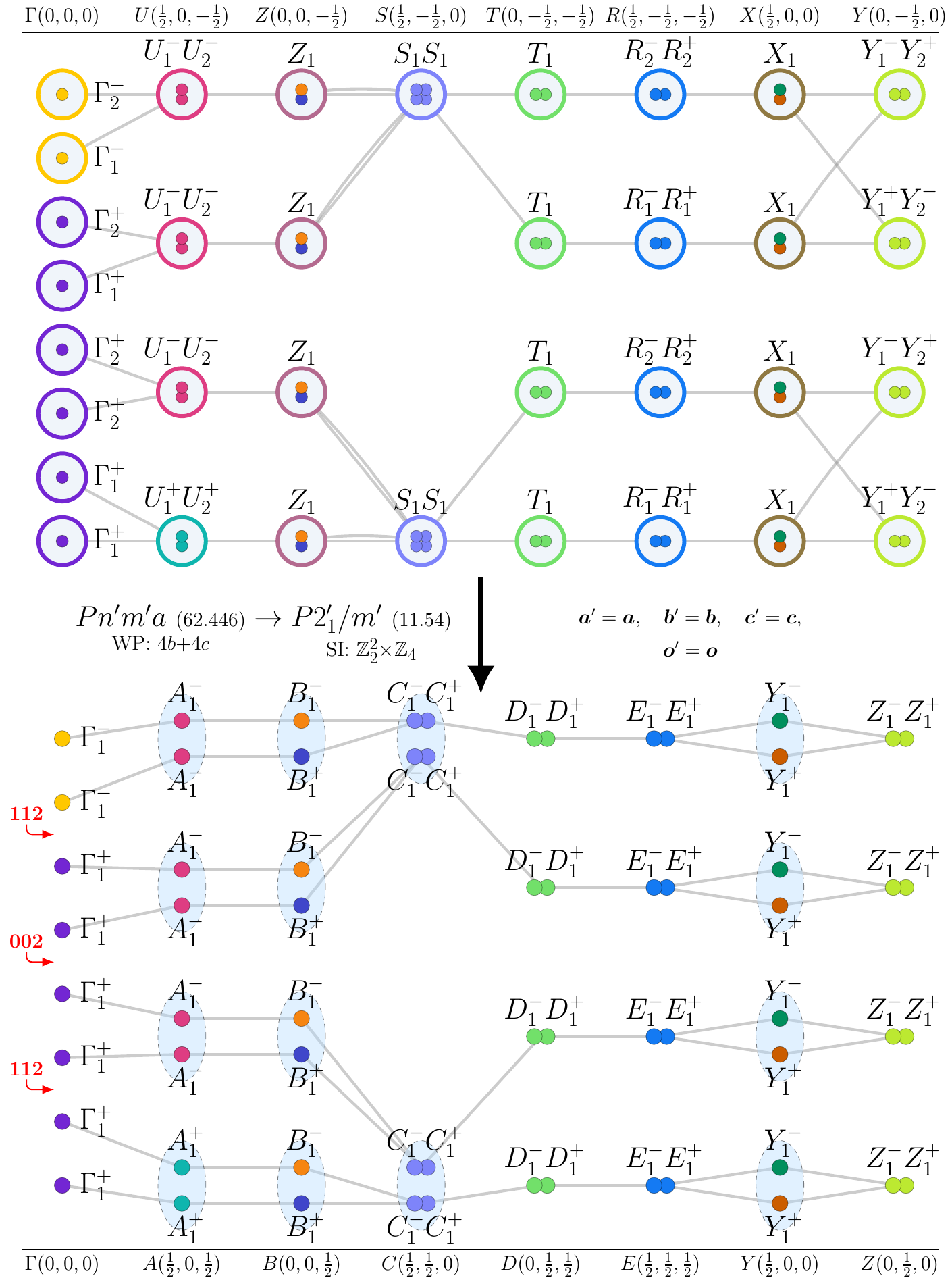}
\caption{Topological magnon bands in subgroup $P2_{1}'/m'~(11.54)$ for magnetic moments on Wyckoff positions $4b+4c$ of supergroup $Pn'm'a~(62.446)$.\label{fig_62.446_11.54_strainperp010_4b+4c}}
\end{figure}
\input{gap_tables_tex/62.446_11.54_strainperp010_4b+4c_table.tex}
\input{si_tables_tex/62.446_11.54_strainperp010_4b+4c_table.tex}
\subsubsection{Topological bands in subgroup $P2_{1}'/c'~(14.79)$}
\textbf{Perturbations:}
\begin{itemize}
\item strain $\perp$ [100],
\item (B $\parallel$ [010] or B $\perp$ [100]).
\end{itemize}
\begin{figure}[H]
\centering
\includegraphics[scale=0.6]{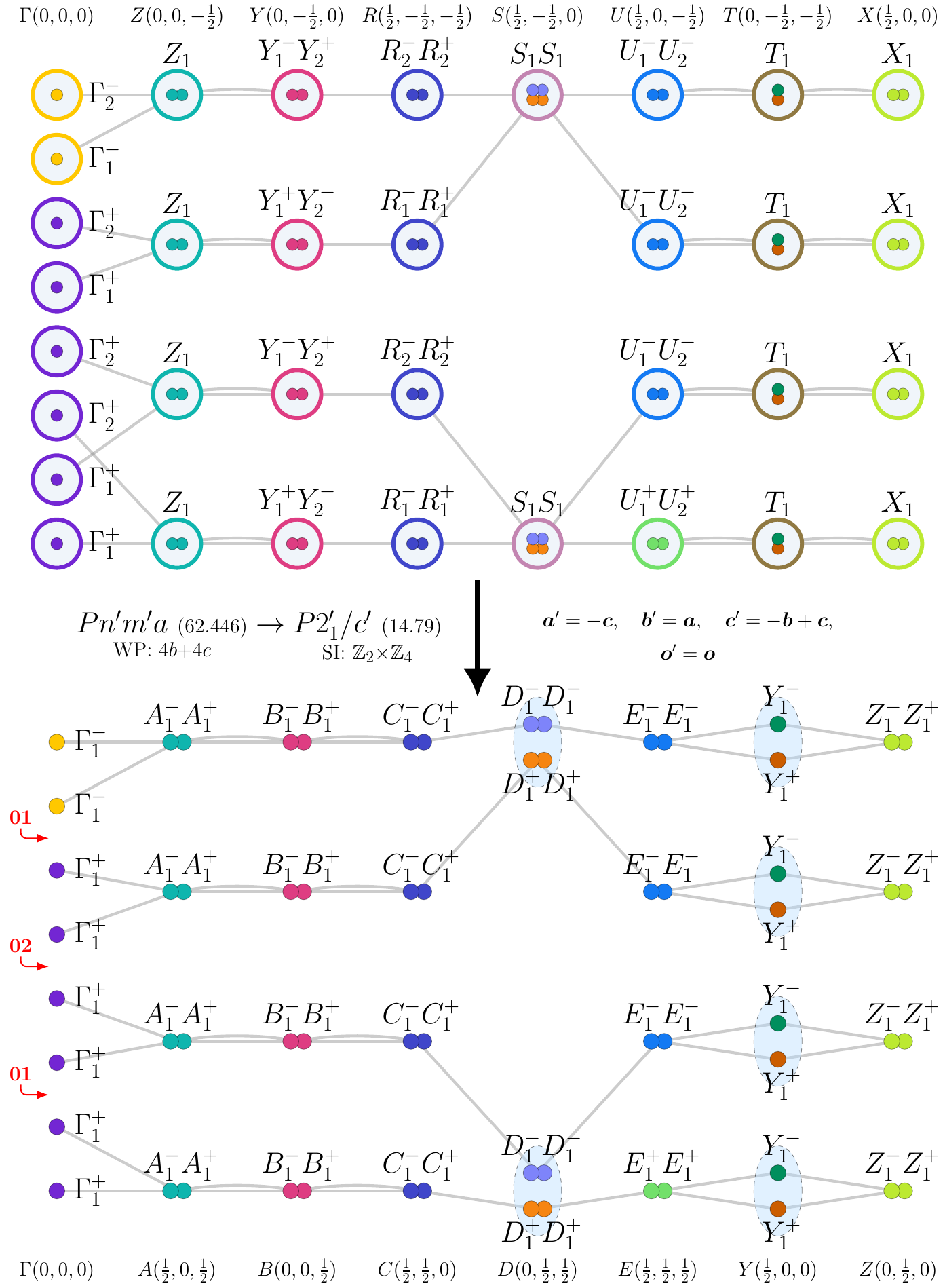}
\caption{Topological magnon bands in subgroup $P2_{1}'/c'~(14.79)$ for magnetic moments on Wyckoff positions $4b+4c$ of supergroup $Pn'm'a~(62.446)$.\label{fig_62.446_14.79_strainperp100_4b+4c}}
\end{figure}
\input{gap_tables_tex/62.446_14.79_strainperp100_4b+4c_table.tex}
\input{si_tables_tex/62.446_14.79_strainperp100_4b+4c_table.tex}
\subsection{WP: $4a+4a+4c$}
\textbf{BCS Materials:} {Ca\textsubscript{2}Fe\textsubscript{0.875}Cr\textsubscript{0.125}GaO\textsubscript{5}}\footnote{BCS web page: \texttt{\href{http://webbdcrista1.ehu.es/magndata/index.php?this\_label=0.206} {http://webbdcrista1.ehu.es/magndata/index.php?this\_label=0.206}}}.\\
\subsubsection{Topological bands in subgroup $P2_{1}'/m'~(11.54)$}
\textbf{Perturbations:}
\begin{itemize}
\item strain $\perp$ [010],
\item (B $\parallel$ [100] or B $\perp$ [010]).
\end{itemize}
\begin{figure}[H]
\centering
\includegraphics[scale=0.6]{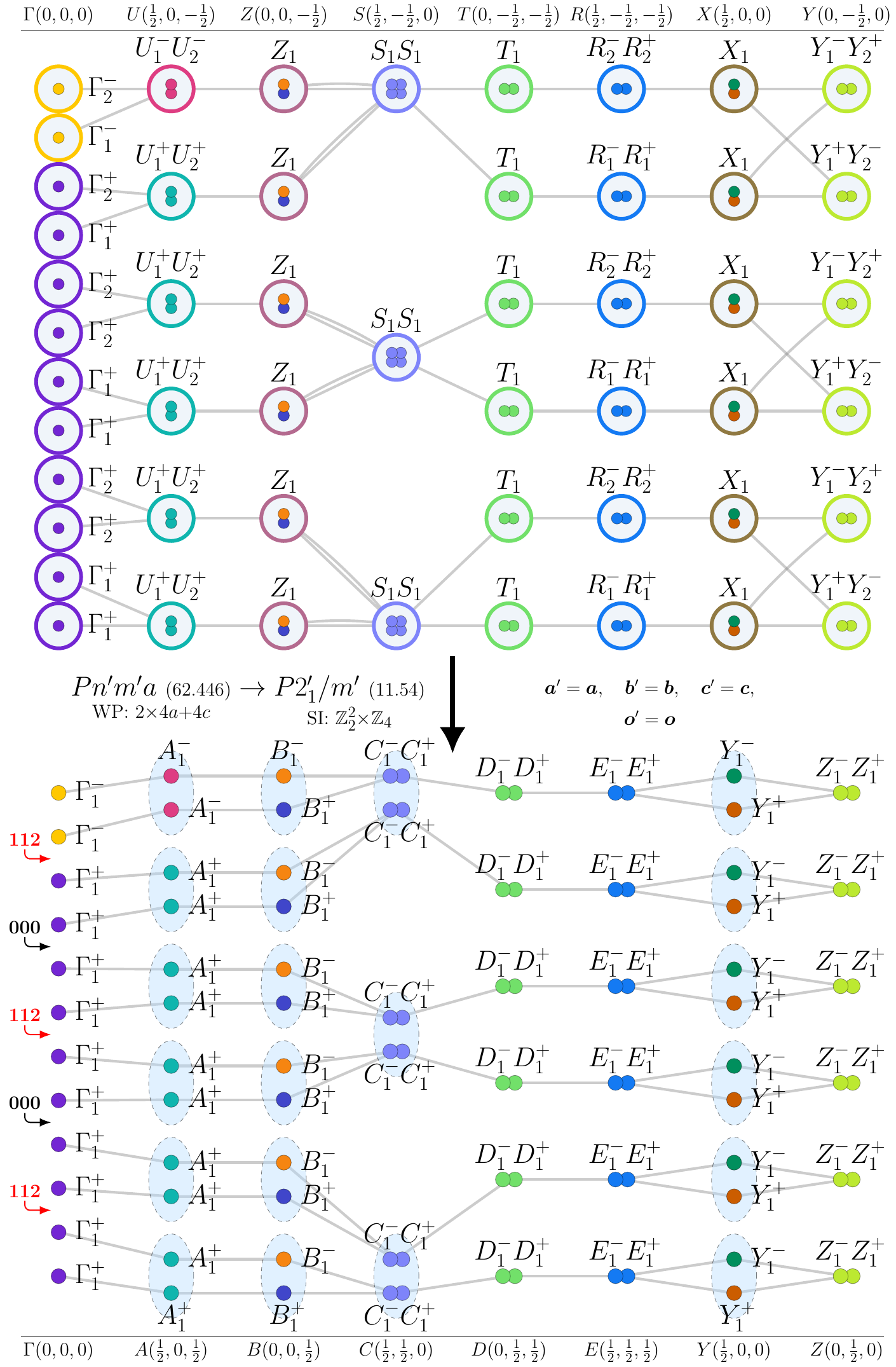}
\caption{Topological magnon bands in subgroup $P2_{1}'/m'~(11.54)$ for magnetic moments on Wyckoff positions $4a+4a+4c$ of supergroup $Pn'm'a~(62.446)$.\label{fig_62.446_11.54_strainperp010_4a+4a+4c}}
\end{figure}
\input{gap_tables_tex/62.446_11.54_strainperp010_4a+4a+4c_table.tex}
\input{si_tables_tex/62.446_11.54_strainperp010_4a+4a+4c_table.tex}
\subsubsection{Topological bands in subgroup $P2_{1}'/c'~(14.79)$}
\textbf{Perturbations:}
\begin{itemize}
\item strain $\perp$ [100],
\item (B $\parallel$ [010] or B $\perp$ [100]).
\end{itemize}
\begin{figure}[H]
\centering
\includegraphics[scale=0.6]{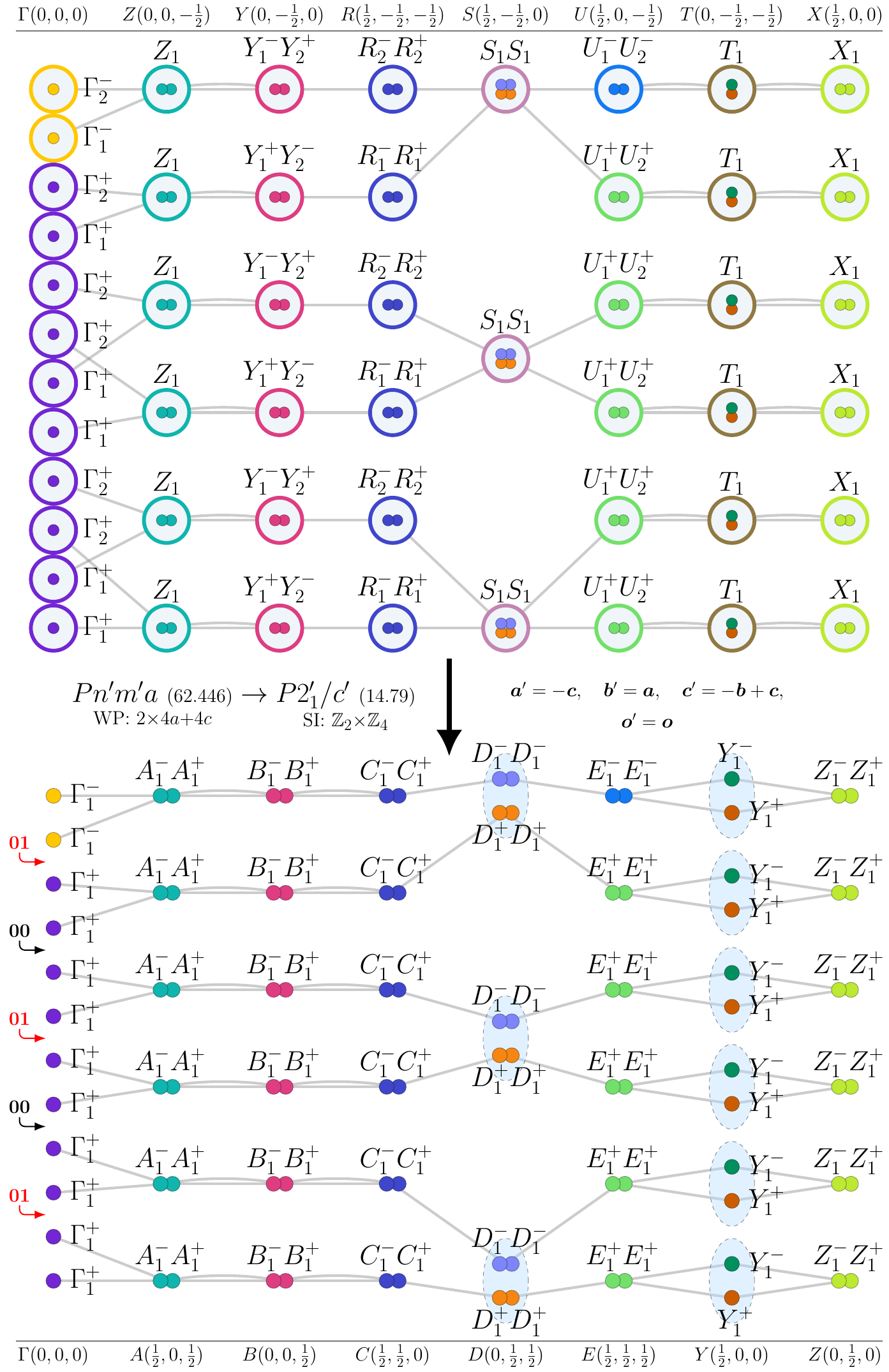}
\caption{Topological magnon bands in subgroup $P2_{1}'/c'~(14.79)$ for magnetic moments on Wyckoff positions $4a+4a+4c$ of supergroup $Pn'm'a~(62.446)$.\label{fig_62.446_14.79_strainperp100_4a+4a+4c}}
\end{figure}
\input{gap_tables_tex/62.446_14.79_strainperp100_4a+4a+4c_table.tex}
\input{si_tables_tex/62.446_14.79_strainperp100_4a+4a+4c_table.tex}

\section{MSG $Pnm'a'~(62.447)$}
\textbf{Nontrivial-SI Subgroups:} $P\bar{1}~(2.4)$, $P2_{1}'/c'~(14.79)$, $P2_{1}'/m'~(11.54)$, $P2_{1}/c~(14.75)$.\\

\textbf{Trivial-SI Subgroups:} $Pc'~(7.26)$, $Pm'~(6.20)$, $P2_{1}'~(4.9)$, $P2_{1}'~(4.9)$, $Pc~(7.24)$, $Pm'n2_{1}'~(31.125)$, $Pna'2_{1}'~(33.147)$, $P2_{1}~(4.7)$, $Pm'c'2_{1}~(26.70)$.\\

\subsection{WP: $4a+4c$}
\textbf{BCS Materials:} {CoFePO\textsubscript{5}~(175 K)}\footnote{BCS web page: \texttt{\href{http://webbdcrista1.ehu.es/magndata/index.php?this\_label=0.262} {http://webbdcrista1.ehu.es/magndata/index.php?this\_label=0.262}}}.\\
\subsubsection{Topological bands in subgroup $P2_{1}'/c'~(14.79)$}
\textbf{Perturbations:}
\begin{itemize}
\item strain $\perp$ [001],
\item (B $\parallel$ [010] or B $\perp$ [001]).
\end{itemize}
\begin{figure}[H]
\centering
\includegraphics[scale=0.6]{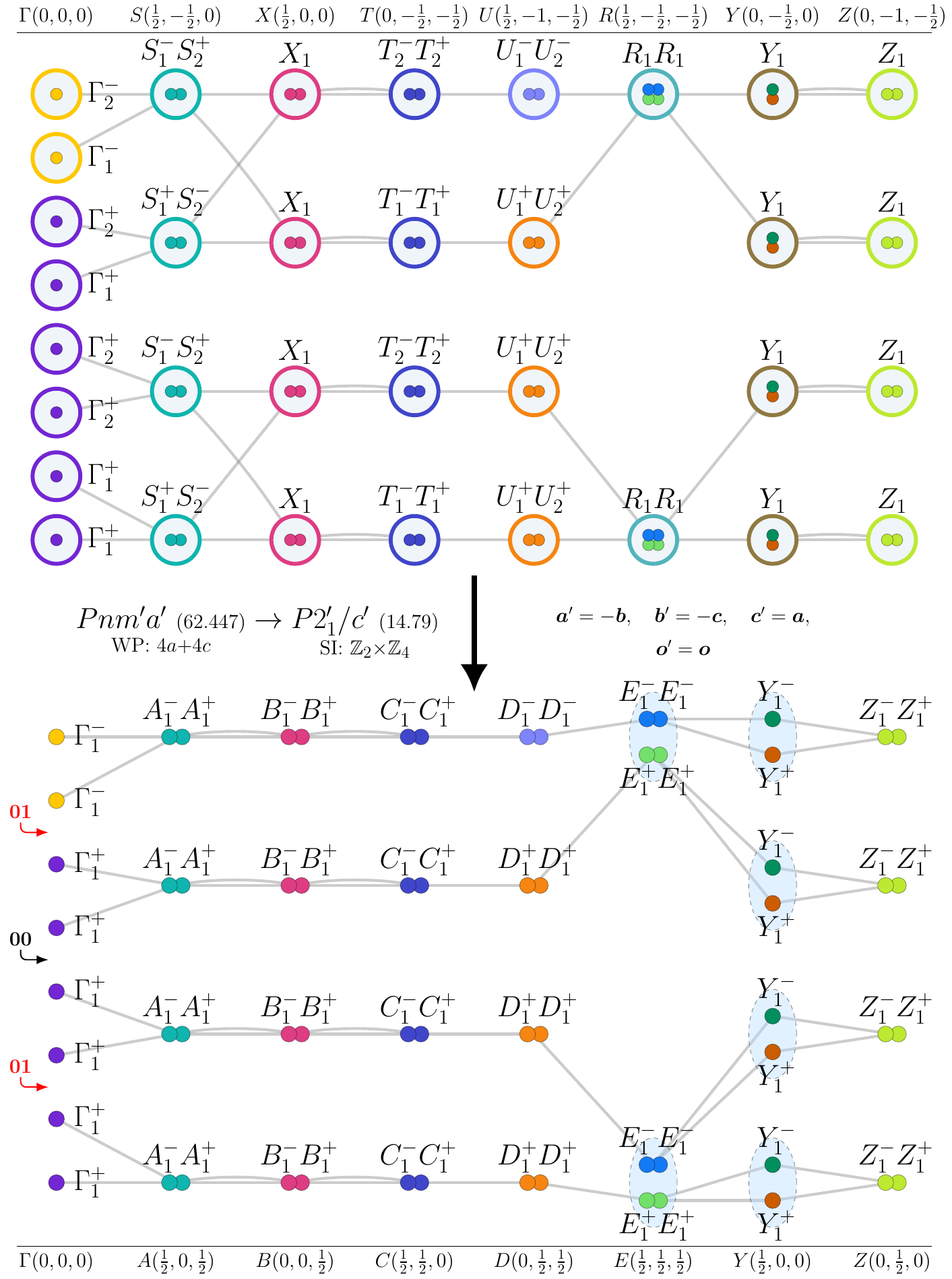}
\caption{Topological magnon bands in subgroup $P2_{1}'/c'~(14.79)$ for magnetic moments on Wyckoff positions $4a+4c$ of supergroup $Pnm'a'~(62.447)$.\label{fig_62.447_14.79_strainperp001_4a+4c}}
\end{figure}
\input{gap_tables_tex/62.447_14.79_strainperp001_4a+4c_table.tex}
\input{si_tables_tex/62.447_14.79_strainperp001_4a+4c_table.tex}
\subsubsection{Topological bands in subgroup $P2_{1}'/m'~(11.54)$}
\textbf{Perturbations:}
\begin{itemize}
\item strain $\perp$ [010],
\item (B $\parallel$ [001] or B $\perp$ [010]).
\end{itemize}
\begin{figure}[H]
\centering
\includegraphics[scale=0.6]{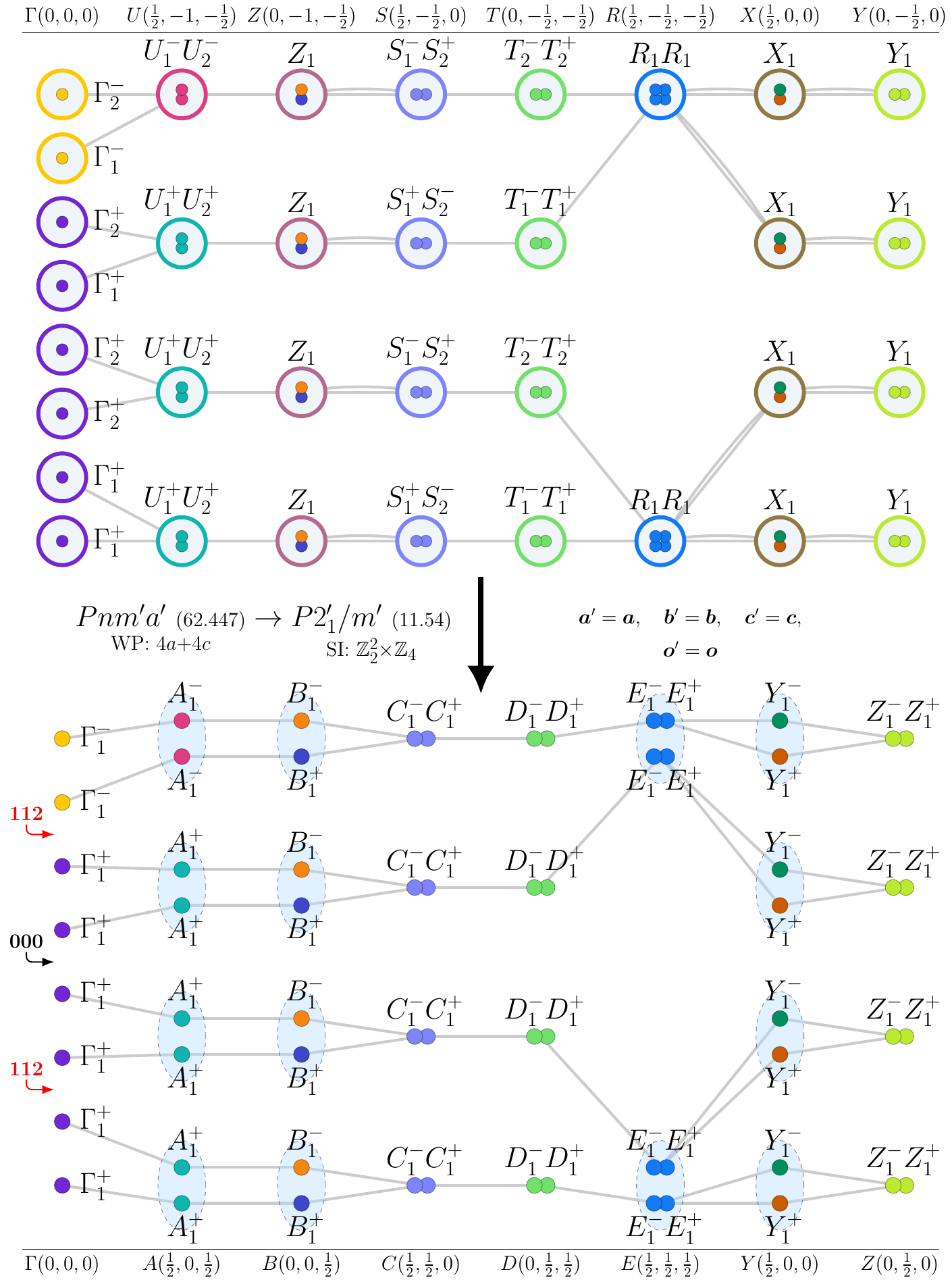}
\caption{Topological magnon bands in subgroup $P2_{1}'/m'~(11.54)$ for magnetic moments on Wyckoff positions $4a+4c$ of supergroup $Pnm'a'~(62.447)$.\label{fig_62.447_11.54_strainperp010_4a+4c}}
\end{figure}
\input{gap_tables_tex/62.447_11.54_strainperp010_4a+4c_table.tex}
\input{si_tables_tex/62.447_11.54_strainperp010_4a+4c_table.tex}
\subsection{WP: $4a$}
\textbf{BCS Materials:} {Ca\textsubscript{2}MnGaO\textsubscript{5}~(160 K)}\footnote{BCS web page: \texttt{\href{http://webbdcrista1.ehu.es/magndata/index.php?this\_label=0.825} {http://webbdcrista1.ehu.es/magndata/index.php?this\_label=0.825}}}.\\
\subsubsection{Topological bands in subgroup $P\bar{1}~(2.4)$}
\textbf{Perturbations:}
\begin{itemize}
\item strain in generic direction,
\item (B $\parallel$ [010] or B $\perp$ [001]) and strain $\perp$ [100],
\item (B $\parallel$ [010] or B $\perp$ [001]) and strain $\perp$ [010],
\item (B $\parallel$ [001] or B $\perp$ [010]) and strain $\perp$ [100],
\item (B $\parallel$ [001] or B $\perp$ [010]) and strain $\perp$ [001],
\item B in generic direction.
\end{itemize}
\begin{figure}[H]
\centering
\includegraphics[scale=0.6]{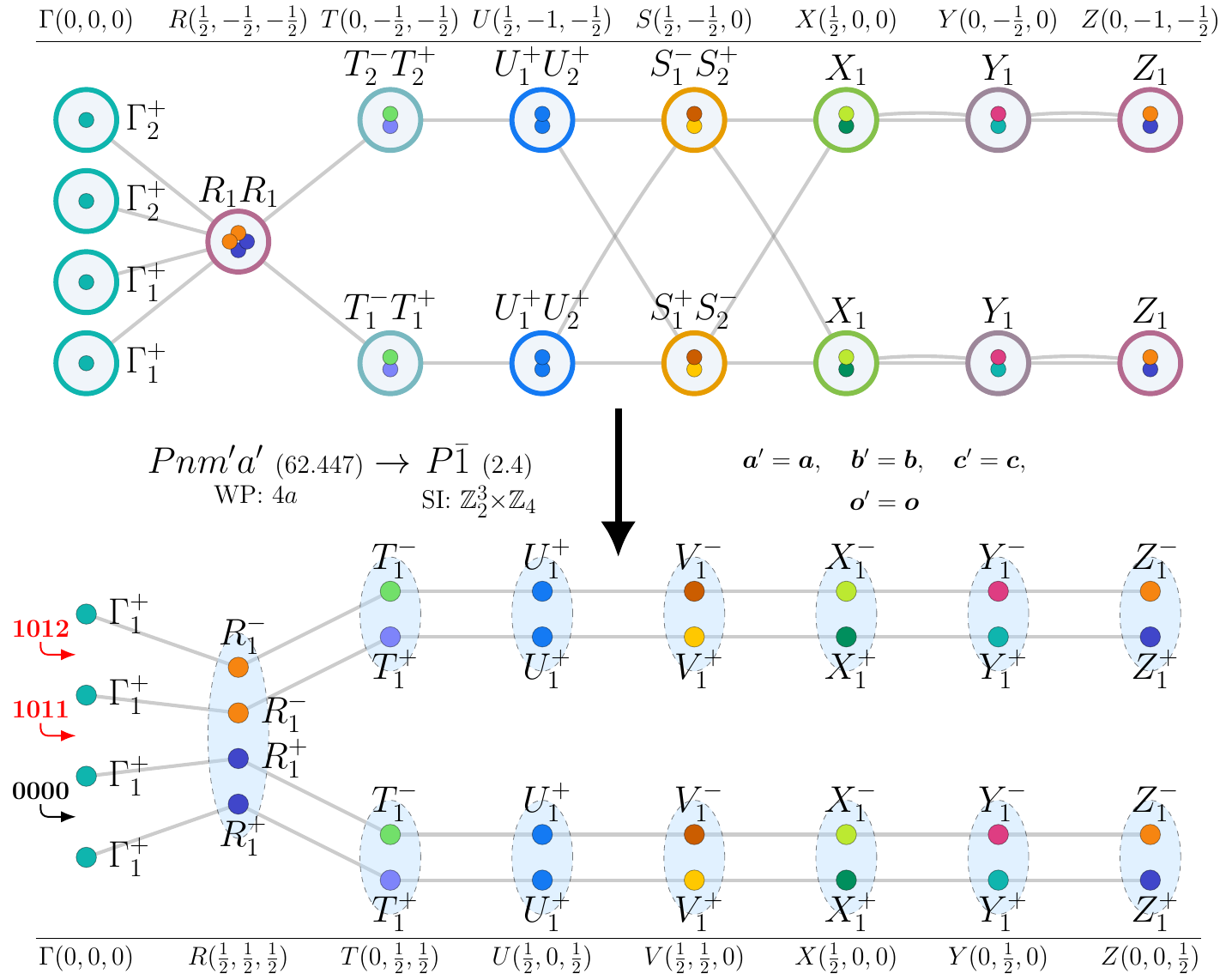}
\caption{Topological magnon bands in subgroup $P\bar{1}~(2.4)$ for magnetic moments on Wyckoff position $4a$ of supergroup $Pnm'a'~(62.447)$.\label{fig_62.447_2.4_strainingenericdirection_4a}}
\end{figure}
\input{gap_tables_tex/62.447_2.4_strainingenericdirection_4a_table.tex}
\input{si_tables_tex/62.447_2.4_strainingenericdirection_4a_table.tex}
\subsubsection{Topological bands in subgroup $P2_{1}'/c'~(14.79)$}
\textbf{Perturbations:}
\begin{itemize}
\item strain $\perp$ [001],
\item (B $\parallel$ [010] or B $\perp$ [001]).
\end{itemize}
\begin{figure}[H]
\centering
\includegraphics[scale=0.6]{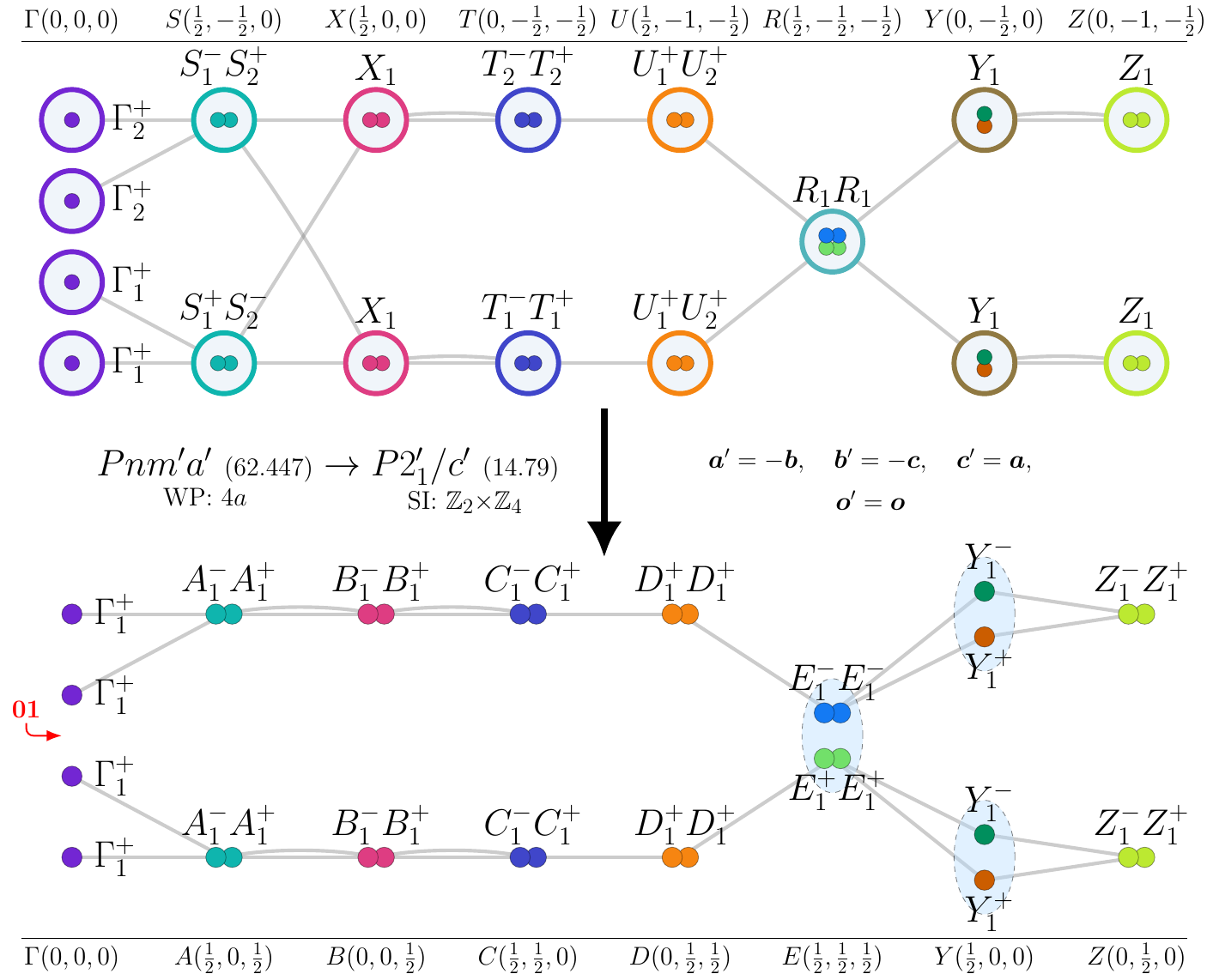}
\caption{Topological magnon bands in subgroup $P2_{1}'/c'~(14.79)$ for magnetic moments on Wyckoff position $4a$ of supergroup $Pnm'a'~(62.447)$.\label{fig_62.447_14.79_strainperp001_4a}}
\end{figure}
\input{gap_tables_tex/62.447_14.79_strainperp001_4a_table.tex}
\input{si_tables_tex/62.447_14.79_strainperp001_4a_table.tex}
\subsubsection{Topological bands in subgroup $P2_{1}'/m'~(11.54)$}
\textbf{Perturbations:}
\begin{itemize}
\item strain $\perp$ [010],
\item (B $\parallel$ [001] or B $\perp$ [010]).
\end{itemize}
\begin{figure}[H]
\centering
\includegraphics[scale=0.6]{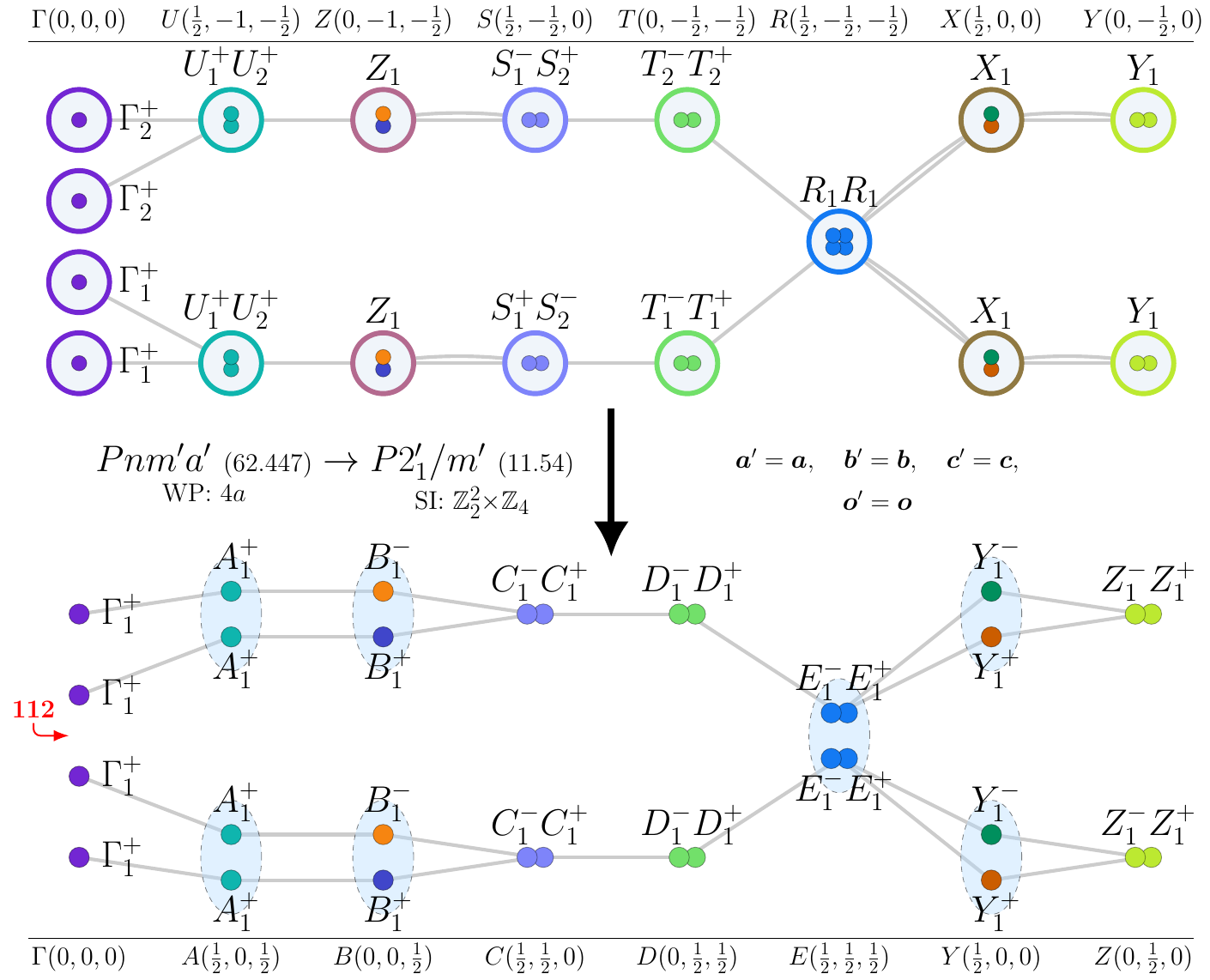}
\caption{Topological magnon bands in subgroup $P2_{1}'/m'~(11.54)$ for magnetic moments on Wyckoff position $4a$ of supergroup $Pnm'a'~(62.447)$.\label{fig_62.447_11.54_strainperp010_4a}}
\end{figure}
\input{gap_tables_tex/62.447_11.54_strainperp010_4a_table.tex}
\input{si_tables_tex/62.447_11.54_strainperp010_4a_table.tex}
\subsection{WP: $4b+4c+4c$}
\textbf{BCS Materials:} {(Tm\textsubscript{0.7}Mn\textsubscript{0.3})MnO\textsubscript{3}~(104 K)}\footnote{BCS web page: \texttt{\href{http://webbdcrista1.ehu.es/magndata/index.php?this\_label=0.293} {http://webbdcrista1.ehu.es/magndata/index.php?this\_label=0.293}}}, {(Ho\textsubscript{0.8}Mn\textsubscript{0.2})MnO\textsubscript{3}~(76 K)}\footnote{BCS web page: \texttt{\href{http://webbdcrista1.ehu.es/magndata/index.php?this\_label=0.648} {http://webbdcrista1.ehu.es/magndata/index.php?this\_label=0.648}}}, {(Ho\textsubscript{0.8}Mn\textsubscript{0.2})MnO\textsubscript{3}~(40 K)}\footnote{BCS web page: \texttt{\href{http://webbdcrista1.ehu.es/magndata/index.php?this\_label=0.649} {http://webbdcrista1.ehu.es/magndata/index.php?this\_label=0.649}}}.\\
\subsubsection{Topological bands in subgroup $P2_{1}'/c'~(14.79)$}
\textbf{Perturbations:}
\begin{itemize}
\item strain $\perp$ [001],
\item (B $\parallel$ [010] or B $\perp$ [001]).
\end{itemize}
\begin{figure}[H]
\centering
\includegraphics[scale=0.6]{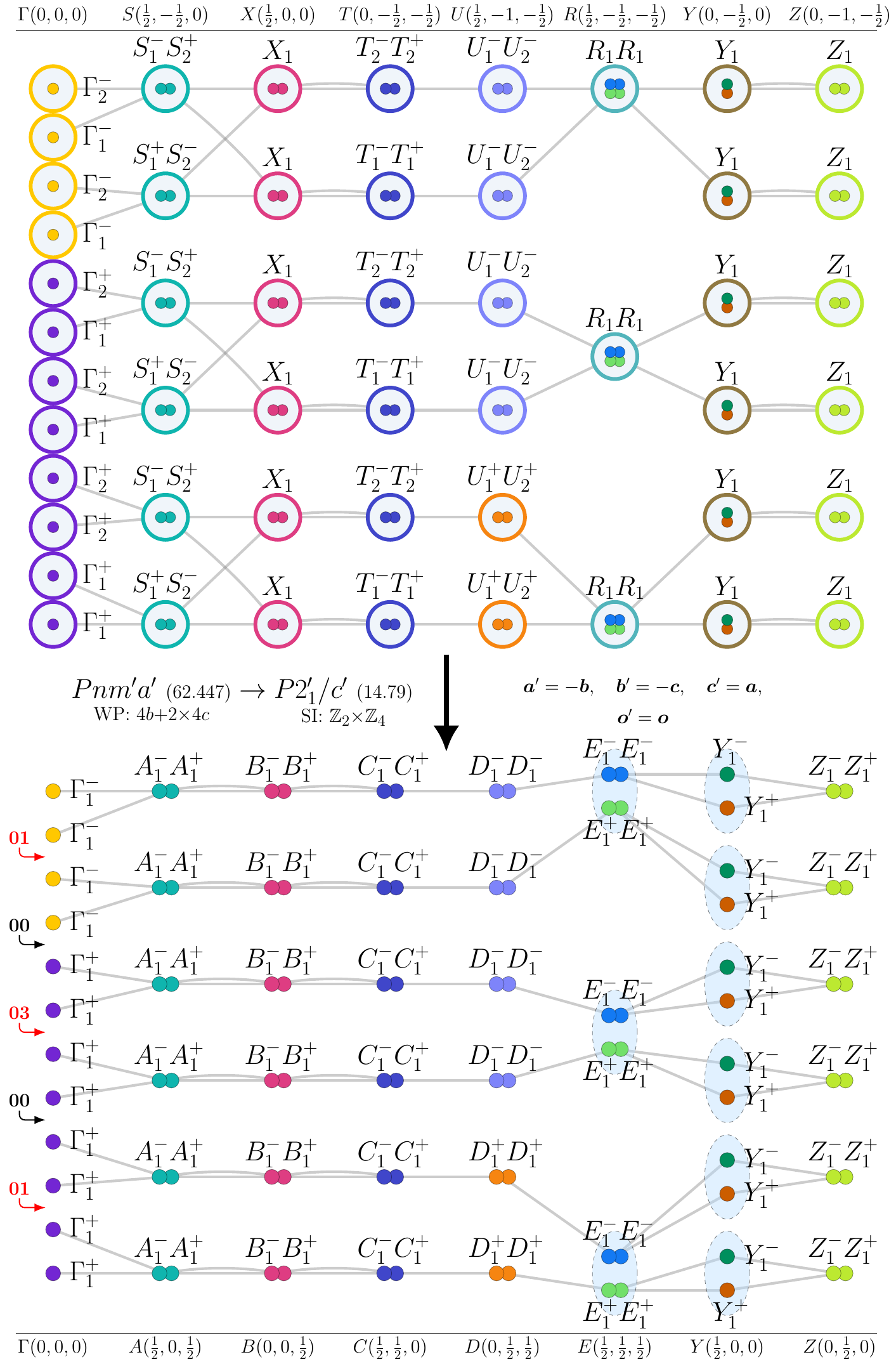}
\caption{Topological magnon bands in subgroup $P2_{1}'/c'~(14.79)$ for magnetic moments on Wyckoff positions $4b+4c+4c$ of supergroup $Pnm'a'~(62.447)$.\label{fig_62.447_14.79_strainperp001_4b+4c+4c}}
\end{figure}
\input{gap_tables_tex/62.447_14.79_strainperp001_4b+4c+4c_table.tex}
\input{si_tables_tex/62.447_14.79_strainperp001_4b+4c+4c_table.tex}
\subsubsection{Topological bands in subgroup $P2_{1}'/m'~(11.54)$}
\textbf{Perturbations:}
\begin{itemize}
\item strain $\perp$ [010],
\item (B $\parallel$ [001] or B $\perp$ [010]).
\end{itemize}
\begin{figure}[H]
\centering
\includegraphics[scale=0.6]{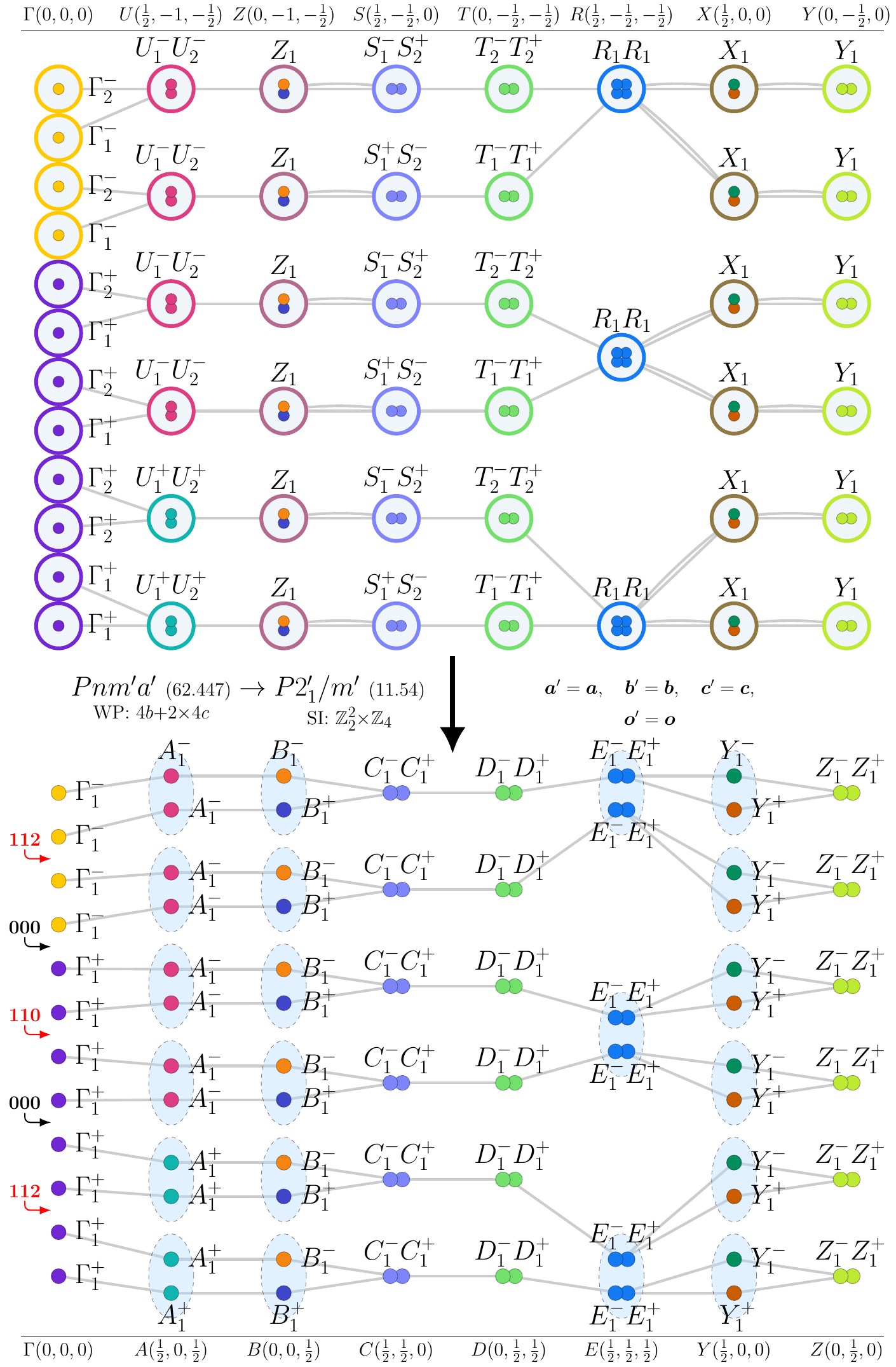}
\caption{Topological magnon bands in subgroup $P2_{1}'/m'~(11.54)$ for magnetic moments on Wyckoff positions $4b+4c+4c$ of supergroup $Pnm'a'~(62.447)$.\label{fig_62.447_11.54_strainperp010_4b+4c+4c}}
\end{figure}
\input{gap_tables_tex/62.447_11.54_strainperp010_4b+4c+4c_table.tex}
\input{si_tables_tex/62.447_11.54_strainperp010_4b+4c+4c_table.tex}
\subsection{WP: $4b+4c$}
\textbf{BCS Materials:} {(Lu\textsubscript{0.6}Mn\textsubscript{0.4})MnO\textsubscript{3}~(67 K)}\footnote{BCS web page: \texttt{\href{http://webbdcrista1.ehu.es/magndata/index.php?this\_label=0.661} {http://webbdcrista1.ehu.es/magndata/index.php?this\_label=0.661}}}, {CsMn\textsubscript{2}F\textsubscript{6}~(24.1 K)}\footnote{BCS web page: \texttt{\href{http://webbdcrista1.ehu.es/magndata/index.php?this\_label=0.726} {http://webbdcrista1.ehu.es/magndata/index.php?this\_label=0.726}}}, {Tb\textsubscript{0.55}Sr\textsubscript{0.45}MnO\textsubscript{3}~(8 K)}\footnote{BCS web page: \texttt{\href{http://webbdcrista1.ehu.es/magndata/index.php?this\_label=0.534} {http://webbdcrista1.ehu.es/magndata/index.php?this\_label=0.534}}}.\\
\subsubsection{Topological bands in subgroup $P2_{1}'/c'~(14.79)$}
\textbf{Perturbations:}
\begin{itemize}
\item strain $\perp$ [001],
\item (B $\parallel$ [010] or B $\perp$ [001]).
\end{itemize}
\begin{figure}[H]
\centering
\includegraphics[scale=0.6]{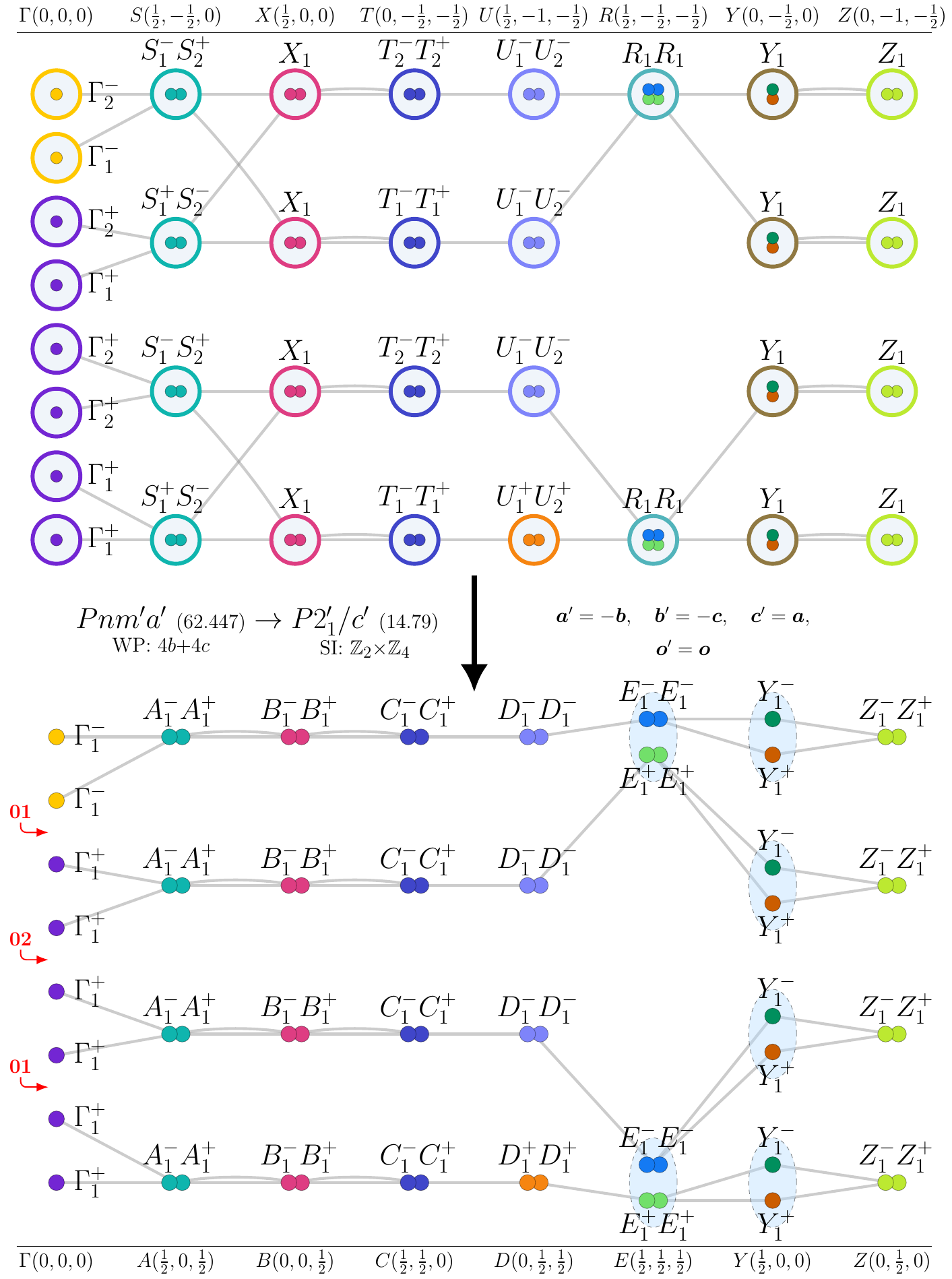}
\caption{Topological magnon bands in subgroup $P2_{1}'/c'~(14.79)$ for magnetic moments on Wyckoff positions $4b+4c$ of supergroup $Pnm'a'~(62.447)$.\label{fig_62.447_14.79_strainperp001_4b+4c}}
\end{figure}
\input{gap_tables_tex/62.447_14.79_strainperp001_4b+4c_table.tex}
\input{si_tables_tex/62.447_14.79_strainperp001_4b+4c_table.tex}
\subsubsection{Topological bands in subgroup $P2_{1}'/m'~(11.54)$}
\textbf{Perturbations:}
\begin{itemize}
\item strain $\perp$ [010],
\item (B $\parallel$ [001] or B $\perp$ [010]).
\end{itemize}
\begin{figure}[H]
\centering
\includegraphics[scale=0.6]{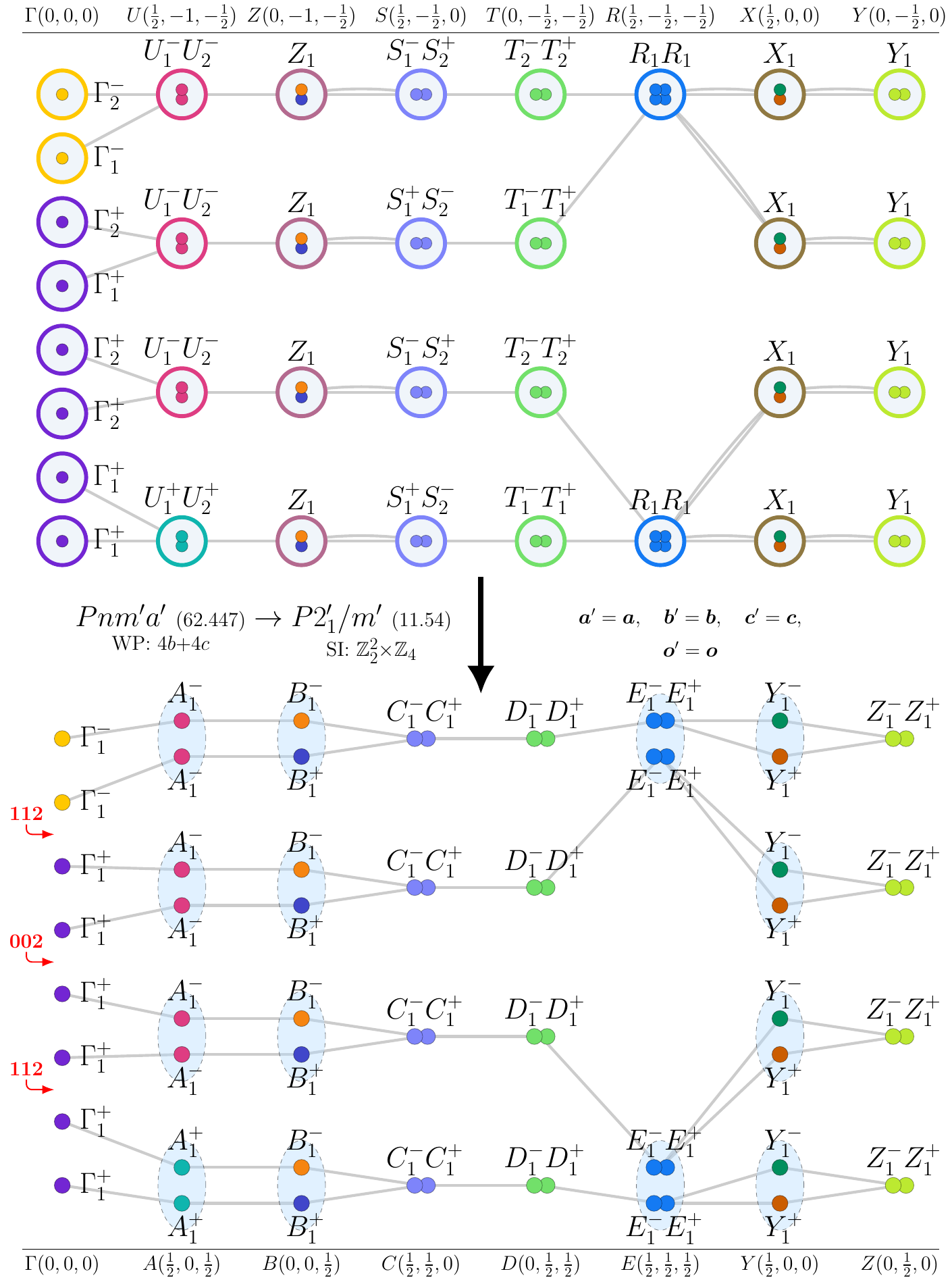}
\caption{Topological magnon bands in subgroup $P2_{1}'/m'~(11.54)$ for magnetic moments on Wyckoff positions $4b+4c$ of supergroup $Pnm'a'~(62.447)$.\label{fig_62.447_11.54_strainperp010_4b+4c}}
\end{figure}
\input{gap_tables_tex/62.447_11.54_strainperp010_4b+4c_table.tex}
\input{si_tables_tex/62.447_11.54_strainperp010_4b+4c_table.tex}
\subsection{WP: $4b$}
\textbf{BCS Materials:} {Tb\textsubscript{0.55}Sr\textsubscript{0.45}MnO\textsubscript{3}~(65 K)}\footnote{BCS web page: \texttt{\href{http://webbdcrista1.ehu.es/magndata/index.php?this\_label=0.536} {http://webbdcrista1.ehu.es/magndata/index.php?this\_label=0.536}}}, {Tb\textsubscript{0.55}Sr\textsubscript{0.45}MnO\textsubscript{3}~(65 K)}\footnote{BCS web page: \texttt{\href{http://webbdcrista1.ehu.es/magndata/index.php?this\_label=0.535} {http://webbdcrista1.ehu.es/magndata/index.php?this\_label=0.535}}}.\\
\subsubsection{Topological bands in subgroup $P\bar{1}~(2.4)$}
\textbf{Perturbations:}
\begin{itemize}
\item strain in generic direction,
\item (B $\parallel$ [010] or B $\perp$ [001]) and strain $\perp$ [100],
\item (B $\parallel$ [010] or B $\perp$ [001]) and strain $\perp$ [010],
\item (B $\parallel$ [001] or B $\perp$ [010]) and strain $\perp$ [100],
\item (B $\parallel$ [001] or B $\perp$ [010]) and strain $\perp$ [001],
\item B in generic direction.
\end{itemize}
\begin{figure}[H]
\centering
\includegraphics[scale=0.6]{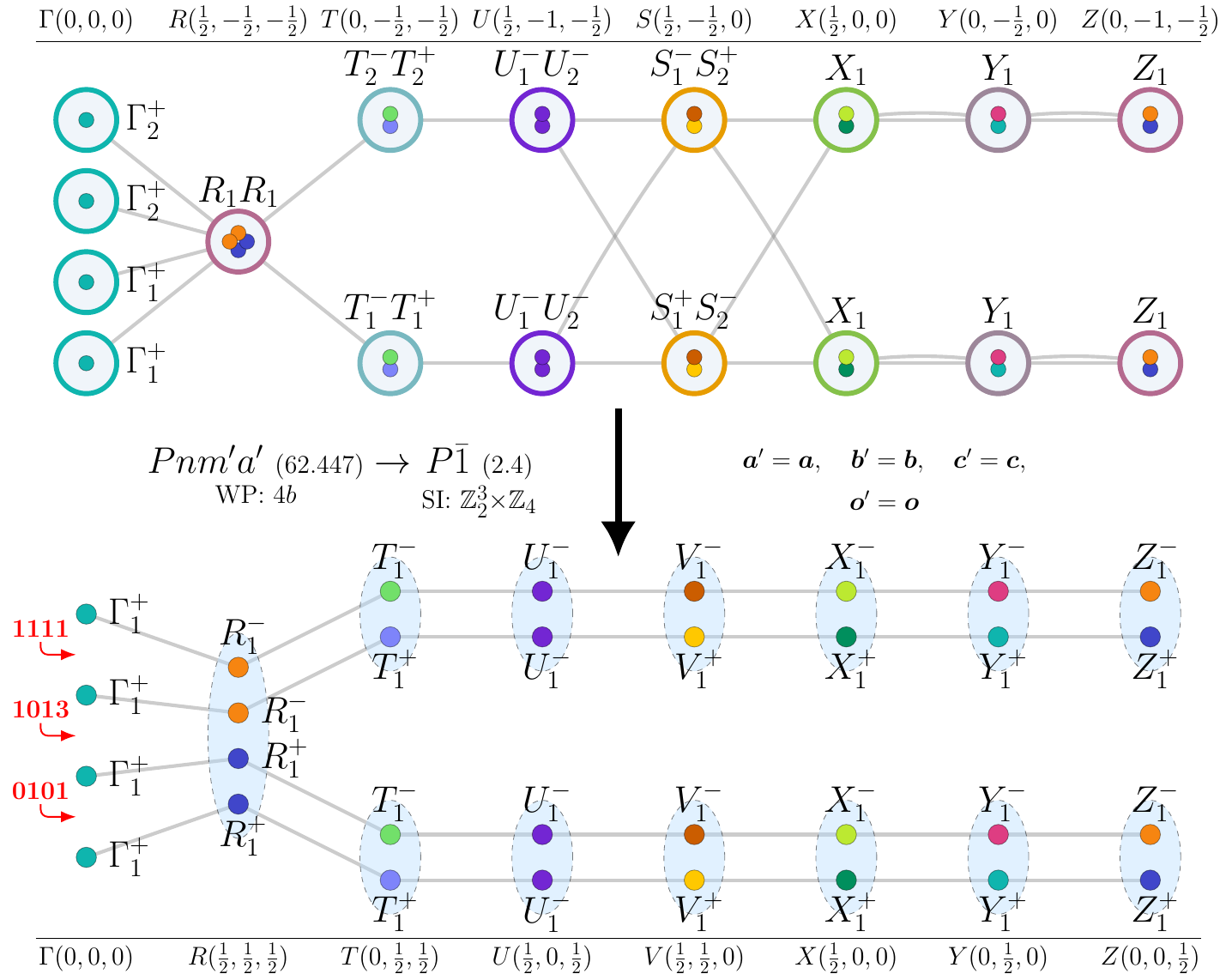}
\caption{Topological magnon bands in subgroup $P\bar{1}~(2.4)$ for magnetic moments on Wyckoff position $4b$ of supergroup $Pnm'a'~(62.447)$.\label{fig_62.447_2.4_strainingenericdirection_4b}}
\end{figure}
\input{gap_tables_tex/62.447_2.4_strainingenericdirection_4b_table.tex}
\input{si_tables_tex/62.447_2.4_strainingenericdirection_4b_table.tex}
\subsubsection{Topological bands in subgroup $P2_{1}'/c'~(14.79)$}
\textbf{Perturbations:}
\begin{itemize}
\item strain $\perp$ [001],
\item (B $\parallel$ [010] or B $\perp$ [001]).
\end{itemize}
\begin{figure}[H]
\centering
\includegraphics[scale=0.6]{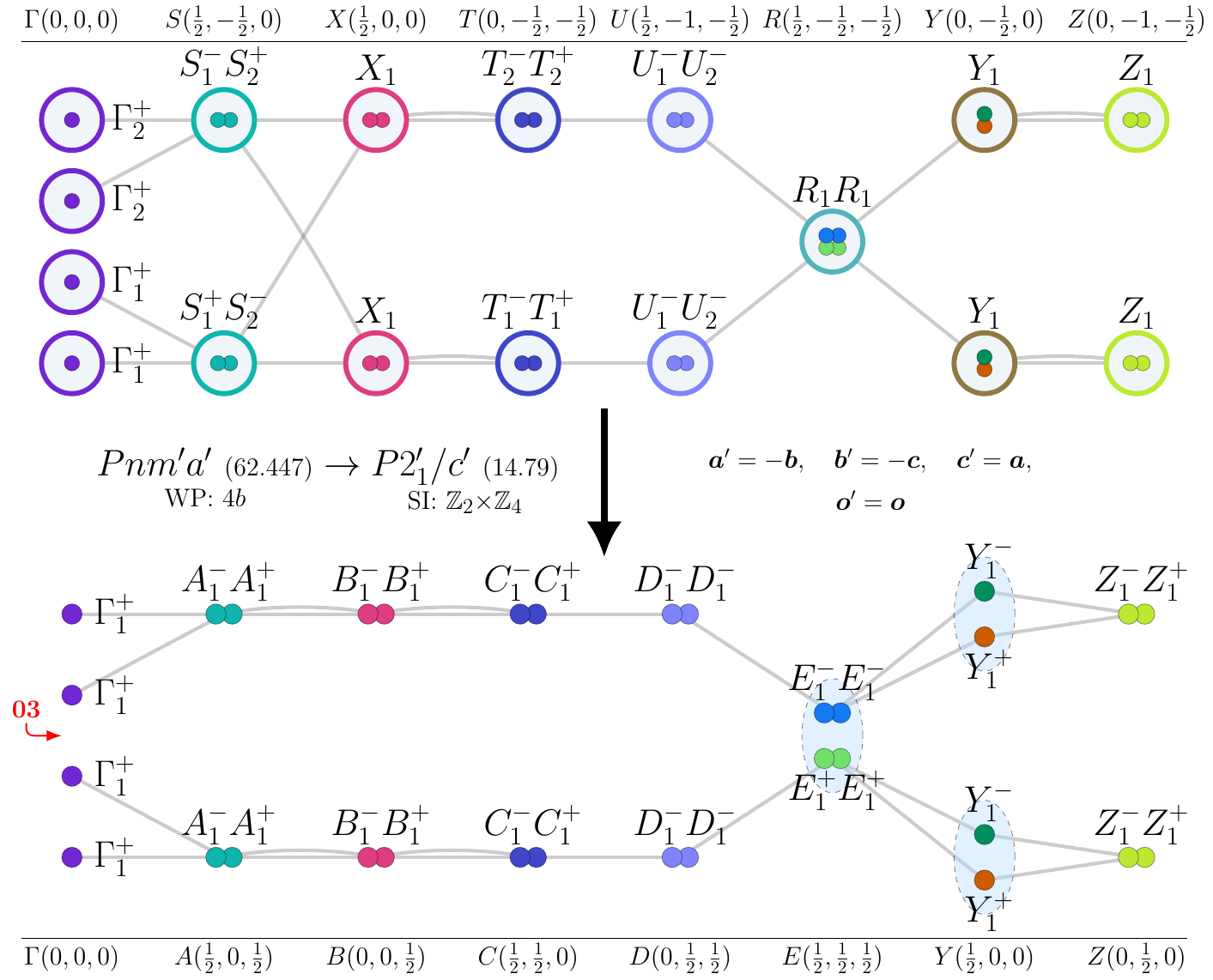}
\caption{Topological magnon bands in subgroup $P2_{1}'/c'~(14.79)$ for magnetic moments on Wyckoff position $4b$ of supergroup $Pnm'a'~(62.447)$.\label{fig_62.447_14.79_strainperp001_4b}}
\end{figure}
\input{gap_tables_tex/62.447_14.79_strainperp001_4b_table.tex}
\input{si_tables_tex/62.447_14.79_strainperp001_4b_table.tex}
\subsubsection{Topological bands in subgroup $P2_{1}'/m'~(11.54)$}
\textbf{Perturbations:}
\begin{itemize}
\item strain $\perp$ [010],
\item (B $\parallel$ [001] or B $\perp$ [010]).
\end{itemize}
\begin{figure}[H]
\centering
\includegraphics[scale=0.6]{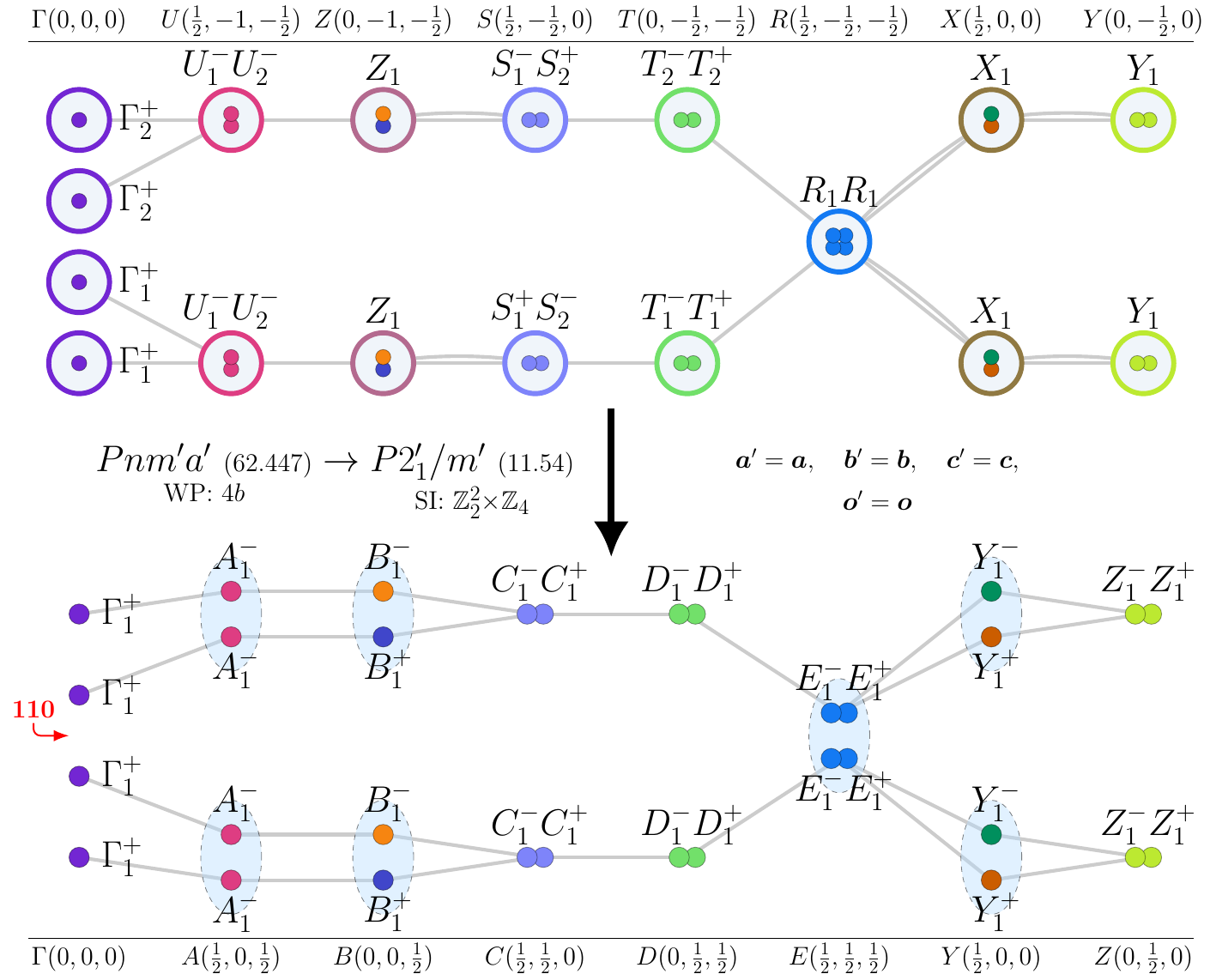}
\caption{Topological magnon bands in subgroup $P2_{1}'/m'~(11.54)$ for magnetic moments on Wyckoff position $4b$ of supergroup $Pnm'a'~(62.447)$.\label{fig_62.447_11.54_strainperp010_4b}}
\end{figure}
\input{gap_tables_tex/62.447_11.54_strainperp010_4b_table.tex}
\input{si_tables_tex/62.447_11.54_strainperp010_4b_table.tex}
\subsection{WP: $4c$}
\textbf{BCS Materials:} {PrSi~(54 K)}\footnote{BCS web page: \texttt{\href{http://webbdcrista1.ehu.es/magndata/index.php?this\_label=0.408} {http://webbdcrista1.ehu.es/magndata/index.php?this\_label=0.408}}}, {NdNi\textsubscript{0.6}Cu\textsubscript{0.4}~(37.5 K)}\footnote{BCS web page: \texttt{\href{http://webbdcrista1.ehu.es/magndata/index.php?this\_label=0.249} {http://webbdcrista1.ehu.es/magndata/index.php?this\_label=0.249}}}, {HoNi~(35.5 K)}\footnote{BCS web page: \texttt{\href{http://webbdcrista1.ehu.es/magndata/index.php?this\_label=0.480} {http://webbdcrista1.ehu.es/magndata/index.php?this\_label=0.480}}}, {HoPt~(16 K)}\footnote{BCS web page: \texttt{\href{http://webbdcrista1.ehu.es/magndata/index.php?this\_label=0.686} {http://webbdcrista1.ehu.es/magndata/index.php?this\_label=0.686}}}, {ErPt~(16 K)}\footnote{BCS web page: \texttt{\href{http://webbdcrista1.ehu.es/magndata/index.php?this\_label=0.685} {http://webbdcrista1.ehu.es/magndata/index.php?this\_label=0.685}}}, {TmPt~(6 K)}\footnote{BCS web page: \texttt{\href{http://webbdcrista1.ehu.es/magndata/index.php?this\_label=0.688} {http://webbdcrista1.ehu.es/magndata/index.php?this\_label=0.688}}}.\\
\subsubsection{Topological bands in subgroup $P\bar{1}~(2.4)$}
\textbf{Perturbations:}
\begin{itemize}
\item strain in generic direction,
\item (B $\parallel$ [010] or B $\perp$ [001]) and strain $\perp$ [100],
\item (B $\parallel$ [010] or B $\perp$ [001]) and strain $\perp$ [010],
\item (B $\parallel$ [001] or B $\perp$ [010]) and strain $\perp$ [100],
\item (B $\parallel$ [001] or B $\perp$ [010]) and strain $\perp$ [001],
\item B in generic direction.
\end{itemize}
\begin{figure}[H]
\centering
\includegraphics[scale=0.6]{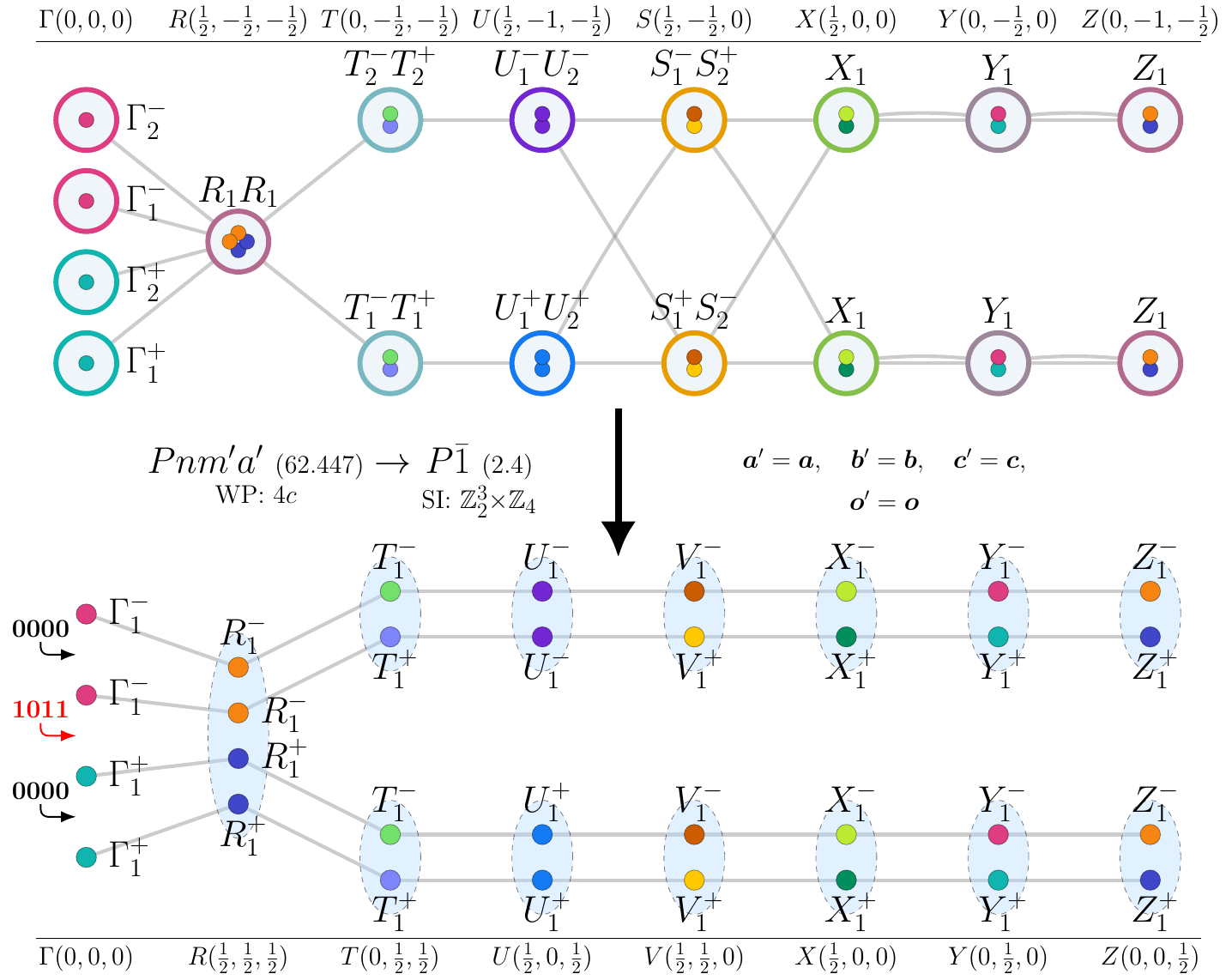}
\caption{Topological magnon bands in subgroup $P\bar{1}~(2.4)$ for magnetic moments on Wyckoff position $4c$ of supergroup $Pnm'a'~(62.447)$.\label{fig_62.447_2.4_strainingenericdirection_4c}}
\end{figure}
\input{gap_tables_tex/62.447_2.4_strainingenericdirection_4c_table.tex}
\input{si_tables_tex/62.447_2.4_strainingenericdirection_4c_table.tex}
\subsubsection{Topological bands in subgroup $P2_{1}'/c'~(14.79)$}
\textbf{Perturbations:}
\begin{itemize}
\item strain $\perp$ [001],
\item (B $\parallel$ [010] or B $\perp$ [001]).
\end{itemize}
\begin{figure}[H]
\centering
\includegraphics[scale=0.6]{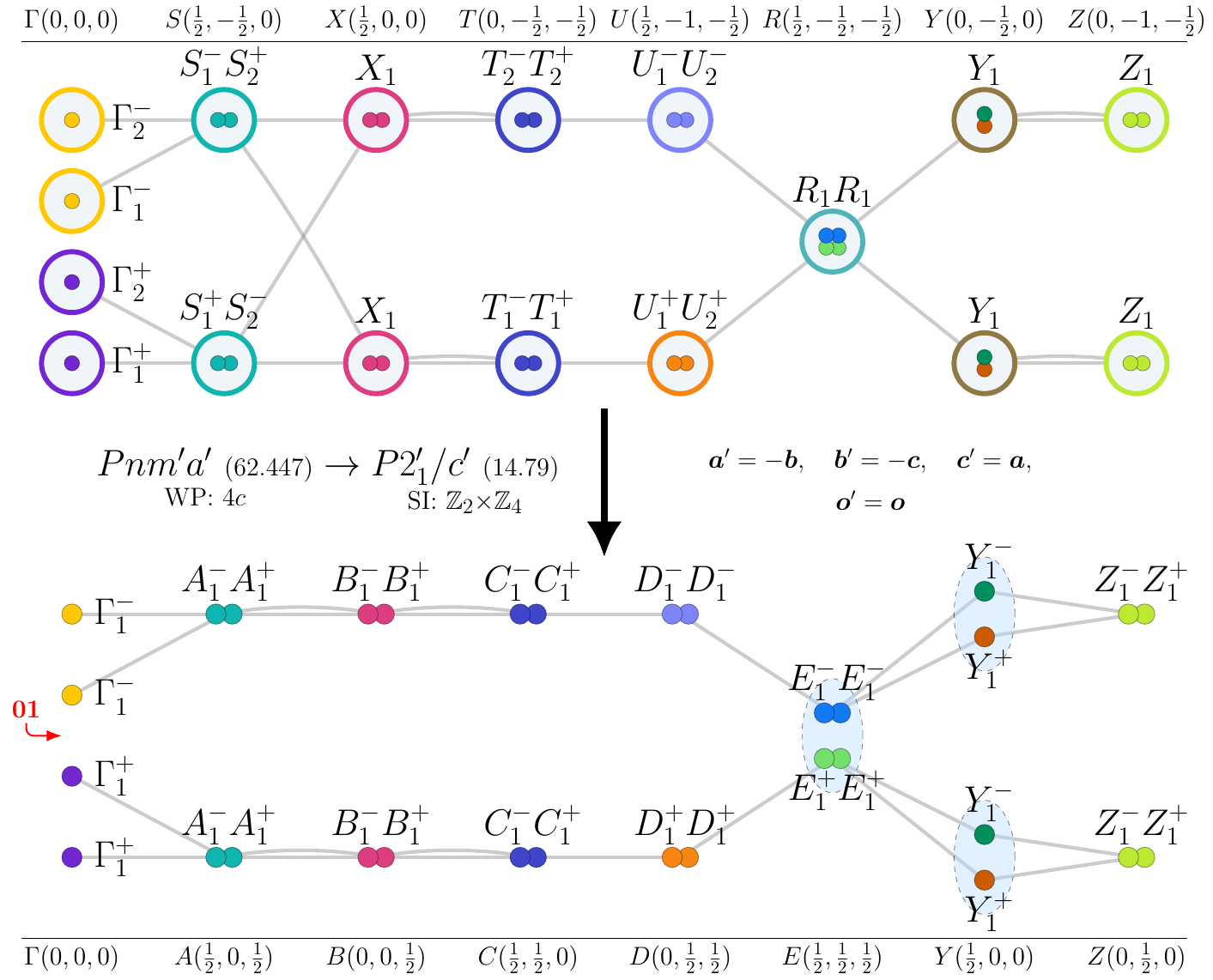}
\caption{Topological magnon bands in subgroup $P2_{1}'/c'~(14.79)$ for magnetic moments on Wyckoff position $4c$ of supergroup $Pnm'a'~(62.447)$.\label{fig_62.447_14.79_strainperp001_4c}}
\end{figure}
\input{gap_tables_tex/62.447_14.79_strainperp001_4c_table.tex}
\input{si_tables_tex/62.447_14.79_strainperp001_4c_table.tex}
\subsubsection{Topological bands in subgroup $P2_{1}'/m'~(11.54)$}
\textbf{Perturbations:}
\begin{itemize}
\item strain $\perp$ [010],
\item (B $\parallel$ [001] or B $\perp$ [010]).
\end{itemize}
\begin{figure}[H]
\centering
\includegraphics[scale=0.6]{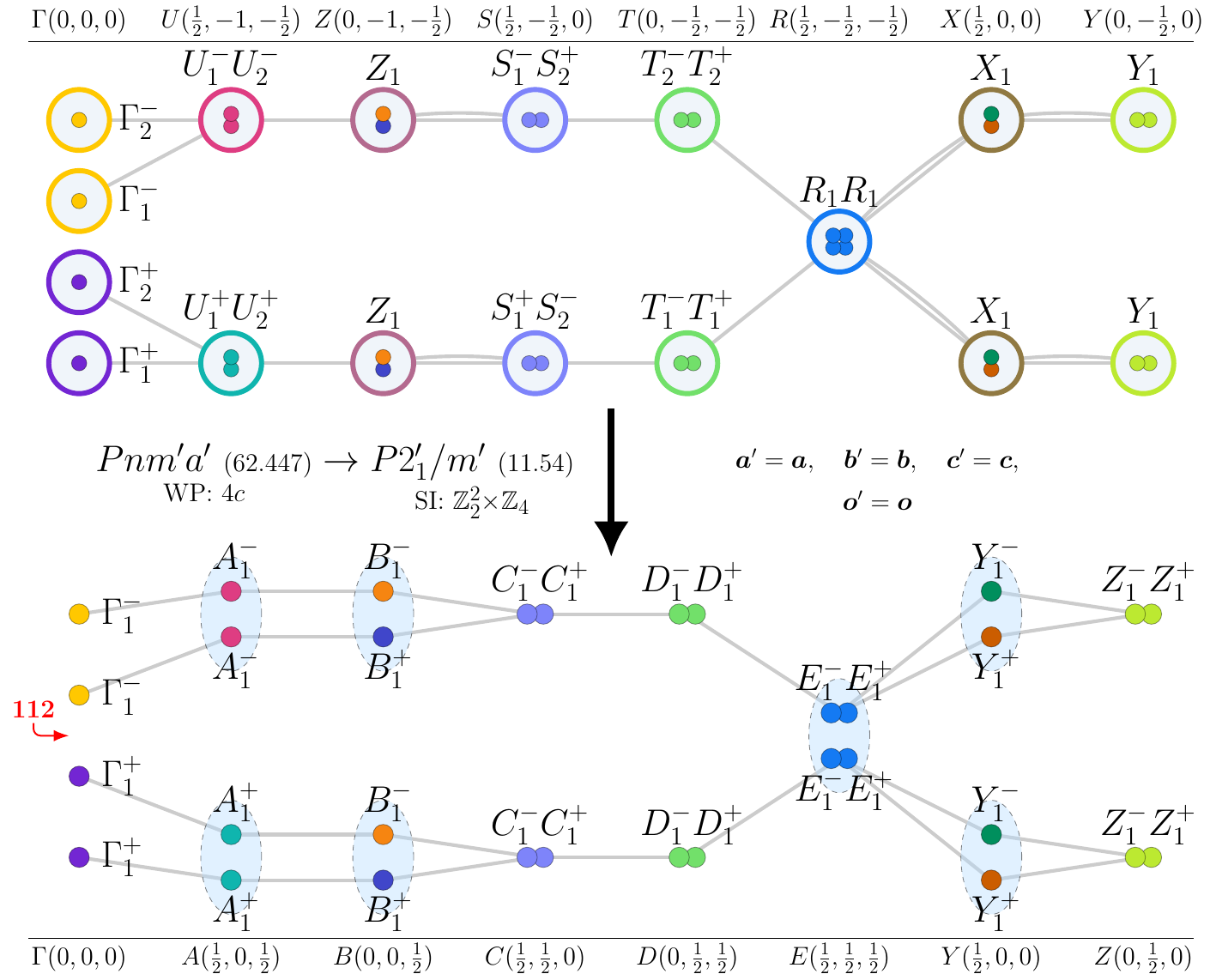}
\caption{Topological magnon bands in subgroup $P2_{1}'/m'~(11.54)$ for magnetic moments on Wyckoff position $4c$ of supergroup $Pnm'a'~(62.447)$.\label{fig_62.447_11.54_strainperp010_4c}}
\end{figure}
\input{gap_tables_tex/62.447_11.54_strainperp010_4c_table.tex}
\input{si_tables_tex/62.447_11.54_strainperp010_4c_table.tex}
\subsection{WP: $4c+4c+4c$}
\textbf{BCS Materials:} {Ho\textsubscript{3}NiGe\textsubscript{2}~(43 K)}\footnote{BCS web page: \texttt{\href{http://webbdcrista1.ehu.es/magndata/index.php?this\_label=0.437} {http://webbdcrista1.ehu.es/magndata/index.php?this\_label=0.437}}}, {Pr\textsubscript{3}CoGe\textsubscript{2}~(28 K)}\footnote{BCS web page: \texttt{\href{http://webbdcrista1.ehu.es/magndata/index.php?this\_label=0.438} {http://webbdcrista1.ehu.es/magndata/index.php?this\_label=0.438}}}.\\
\subsubsection{Topological bands in subgroup $P2_{1}'/c'~(14.79)$}
\textbf{Perturbations:}
\begin{itemize}
\item strain $\perp$ [001],
\item (B $\parallel$ [010] or B $\perp$ [001]).
\end{itemize}
\begin{figure}[H]
\centering
\includegraphics[scale=0.6]{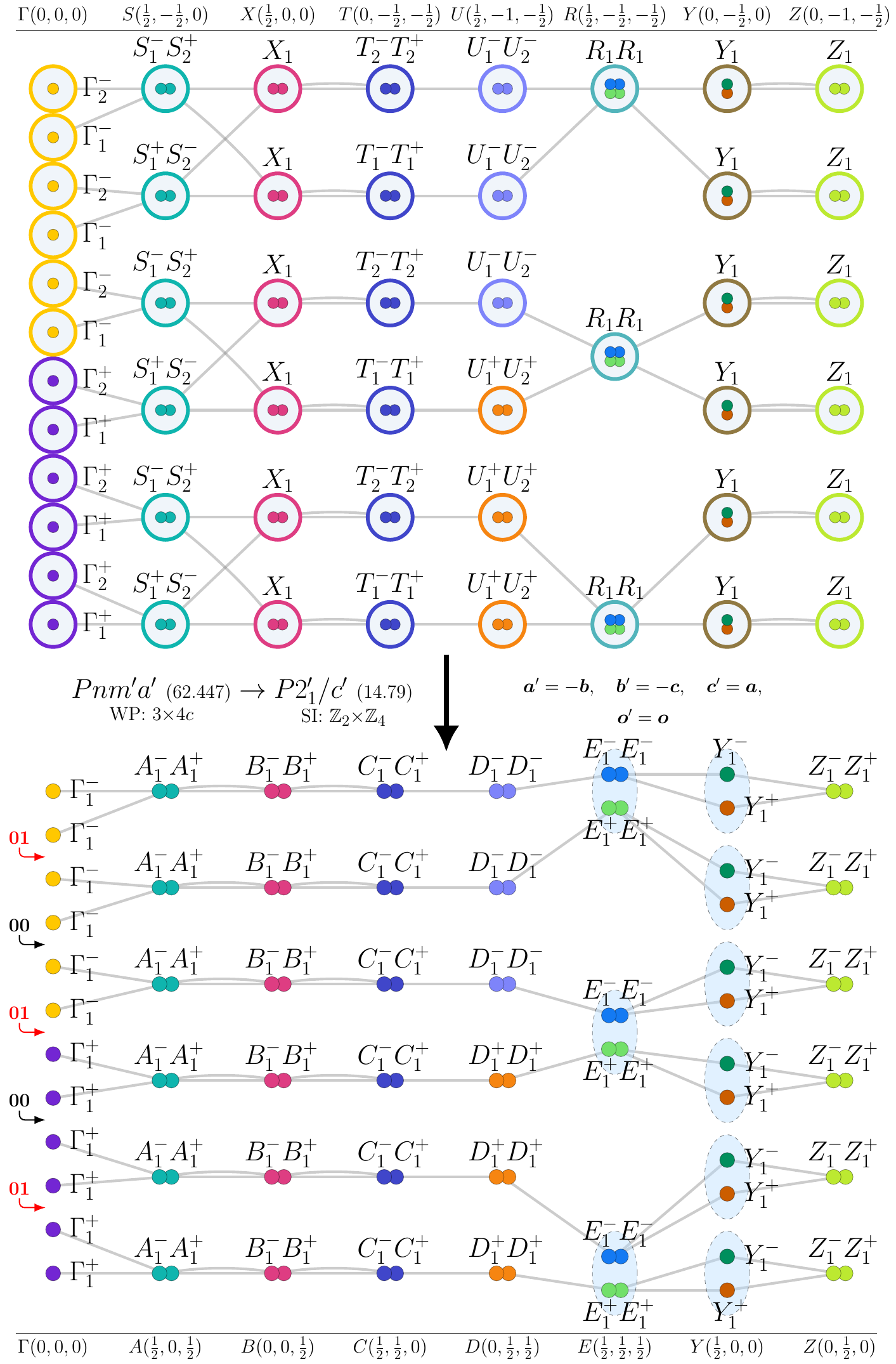}
\caption{Topological magnon bands in subgroup $P2_{1}'/c'~(14.79)$ for magnetic moments on Wyckoff positions $4c+4c+4c$ of supergroup $Pnm'a'~(62.447)$.\label{fig_62.447_14.79_strainperp001_4c+4c+4c}}
\end{figure}
\input{gap_tables_tex/62.447_14.79_strainperp001_4c+4c+4c_table.tex}
\input{si_tables_tex/62.447_14.79_strainperp001_4c+4c+4c_table.tex}
\subsubsection{Topological bands in subgroup $P2_{1}'/m'~(11.54)$}
\textbf{Perturbations:}
\begin{itemize}
\item strain $\perp$ [010],
\item (B $\parallel$ [001] or B $\perp$ [010]).
\end{itemize}
\begin{figure}[H]
\centering
\includegraphics[scale=0.6]{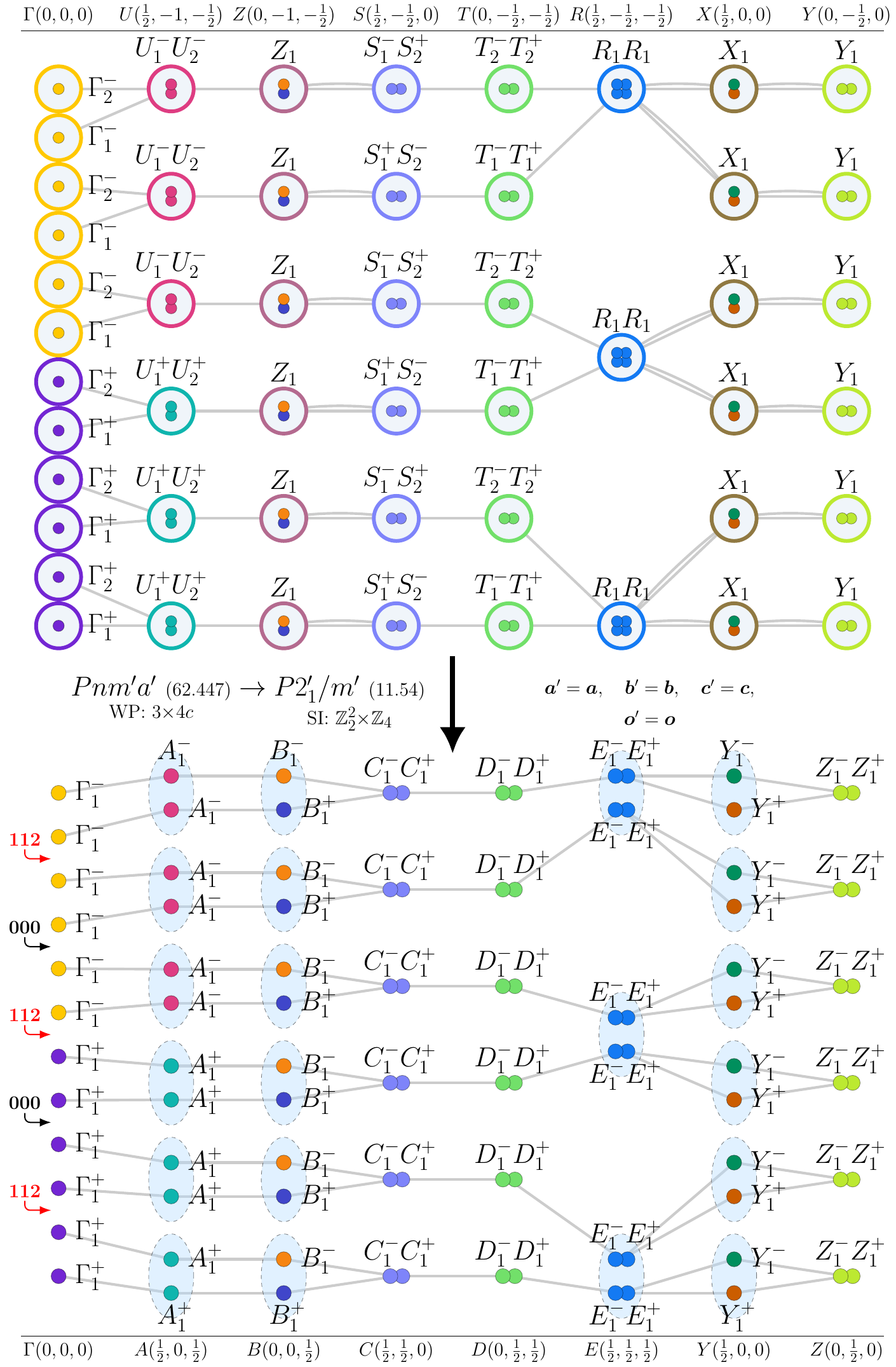}
\caption{Topological magnon bands in subgroup $P2_{1}'/m'~(11.54)$ for magnetic moments on Wyckoff positions $4c+4c+4c$ of supergroup $Pnm'a'~(62.447)$.\label{fig_62.447_11.54_strainperp010_4c+4c+4c}}
\end{figure}
\input{gap_tables_tex/62.447_11.54_strainperp010_4c+4c+4c_table.tex}
\input{si_tables_tex/62.447_11.54_strainperp010_4c+4c+4c_table.tex}

\section{MSG $Pn'ma'~(62.448)$}
\textbf{Nontrivial-SI Subgroups:} $P\bar{1}~(2.4)$, $P2_{1}'/c'~(14.79)$, $P2_{1}'/c'~(14.79)$, $P2_{1}/m~(11.50)$.\\

\textbf{Trivial-SI Subgroups:} $Pc'~(7.26)$, $Pc'~(7.26)$, $P2_{1}'~(4.9)$, $P2_{1}'~(4.9)$, $Pm~(6.18)$, $Pmn'2_{1}'~(31.126)$, $Pmc'2_{1}'~(26.69)$, $P2_{1}~(4.7)$, $Pn'a'2_{1}~(33.148)$.\\

\subsection{WP: $4b$\label{NdMnO3SuppSection}}
\textbf{BCS Materials:} {NdFeO\textsubscript{3}~(760 K)}\footnote{BCS web page: \texttt{\href{http://webbdcrista1.ehu.es/magndata/index.php?this\_label=0.336} {http://webbdcrista1.ehu.es/magndata/index.php?this\_label=0.336}}}, {CeFeO\textsubscript{3}~(720 K)}\footnote{BCS web page: \texttt{\href{http://webbdcrista1.ehu.es/magndata/index.php?this\_label=0.757} {http://webbdcrista1.ehu.es/magndata/index.php?this\_label=0.757}}}, {TbFeO\textsubscript{3}~(681 K)}\footnote{BCS web page: \texttt{\href{http://webbdcrista1.ehu.es/magndata/index.php?this\_label=0.351} {http://webbdcrista1.ehu.es/magndata/index.php?this\_label=0.351}}}, {SmFeO\textsubscript{3}~(670 K)}\footnote{BCS web page: \texttt{\href{http://webbdcrista1.ehu.es/magndata/index.php?this\_label=0.380} {http://webbdcrista1.ehu.es/magndata/index.php?this\_label=0.380}}}, {NaOsO\textsubscript{3}~(410 K)}\footnote{BCS web page: \texttt{\href{http://webbdcrista1.ehu.es/magndata/index.php?this\_label=0.25} {http://webbdcrista1.ehu.es/magndata/index.php?this\_label=0.25}}}, {LaCrO\textsubscript{3}~(310 K)}\footnote{BCS web page: \texttt{\href{http://webbdcrista1.ehu.es/magndata/index.php?this\_label=0.417} {http://webbdcrista1.ehu.es/magndata/index.php?this\_label=0.417}}}, {YCrO\textsubscript{3}~(141 K)}\footnote{BCS web page: \texttt{\href{http://webbdcrista1.ehu.es/magndata/index.php?this\_label=0.586} {http://webbdcrista1.ehu.es/magndata/index.php?this\_label=0.586}}}, {LaMnO\textsubscript{3}~(139.5 K)}\footnote{BCS web page: \texttt{\href{http://webbdcrista1.ehu.es/magndata/index.php?this\_label=0.1} {http://webbdcrista1.ehu.es/magndata/index.php?this\_label=0.1}}}, {ErCrO\textsubscript{3}~(133 K)}\footnote{BCS web page: \texttt{\href{http://webbdcrista1.ehu.es/magndata/index.php?this\_label=0.591} {http://webbdcrista1.ehu.es/magndata/index.php?this\_label=0.591}}}, {TmCrO\textsubscript{3}~(124 K)}\footnote{BCS web page: \texttt{\href{http://webbdcrista1.ehu.es/magndata/index.php?this\_label=0.587} {http://webbdcrista1.ehu.es/magndata/index.php?this\_label=0.587}}}, {YRuO\textsubscript{3}~(97 K)}\footnote{BCS web page: \texttt{\href{http://webbdcrista1.ehu.es/magndata/index.php?this\_label=0.513} {http://webbdcrista1.ehu.es/magndata/index.php?this\_label=0.513}}}, {PrMnO\textsubscript{3}~(91 K)}\footnote{BCS web page: \texttt{\href{http://webbdcrista1.ehu.es/magndata/index.php?this\_label=0.608} {http://webbdcrista1.ehu.es/magndata/index.php?this\_label=0.608}}}, {Pr\textsubscript{0.95}K\textsubscript{0.05}MnO\textsubscript{3}~(85 K)}\footnote{BCS web page: \texttt{\href{http://webbdcrista1.ehu.es/magndata/index.php?this\_label=0.610} {http://webbdcrista1.ehu.es/magndata/index.php?this\_label=0.610}}}, {NdMnO\textsubscript{3}~(78 K)}\footnote{BCS web page: \texttt{\href{http://webbdcrista1.ehu.es/magndata/index.php?this\_label=0.288} {http://webbdcrista1.ehu.es/magndata/index.php?this\_label=0.288}}}, {KMnF\textsubscript{3}~(77 K)}\footnote{BCS web page: \texttt{\href{http://webbdcrista1.ehu.es/magndata/index.php?this\_label=0.432} {http://webbdcrista1.ehu.es/magndata/index.php?this\_label=0.432}}}, {NdMnO\textsubscript{3}~(73 K)}\footnote{BCS web page: \texttt{\href{http://webbdcrista1.ehu.es/magndata/index.php?this\_label=0.370} {http://webbdcrista1.ehu.es/magndata/index.php?this\_label=0.370}}}, {(CH\textsubscript{3}NH\textsubscript{3})(Co(COOH)\textsubscript{3}~(15.9 K)}\footnote{BCS web page: \texttt{\href{http://webbdcrista1.ehu.es/magndata/index.php?this\_label=0.368} {http://webbdcrista1.ehu.es/magndata/index.php?this\_label=0.368}}}, {Ho\textsubscript{0.2}Bi\textsubscript{0.8}FeO\textsubscript{3}}\footnote{BCS web page: \texttt{\href{http://webbdcrista1.ehu.es/magndata/index.php?this\_label=0.560} {http://webbdcrista1.ehu.es/magndata/index.php?this\_label=0.560}}}, {Ho\textsubscript{0.15}Bi\textsubscript{0.85}FeO\textsubscript{3}}\footnote{BCS web page: \texttt{\href{http://webbdcrista1.ehu.es/magndata/index.php?this\_label=0.559} {http://webbdcrista1.ehu.es/magndata/index.php?this\_label=0.559}}}.\\
\subsubsection{Topological bands in subgroup $P\bar{1}~(2.4)$}
\textbf{Perturbations:}
\begin{itemize}
\item strain in generic direction,
\item (B $\parallel$ [100] or B $\perp$ [001]) and strain $\perp$ [100],
\item (B $\parallel$ [100] or B $\perp$ [001]) and strain $\perp$ [010],
\item (B $\parallel$ [001] or B $\perp$ [100]) and strain $\perp$ [010],
\item (B $\parallel$ [001] or B $\perp$ [100]) and strain $\perp$ [001],
\item B in generic direction.
\end{itemize}
\begin{figure}[H]
\centering
\includegraphics[scale=0.6]{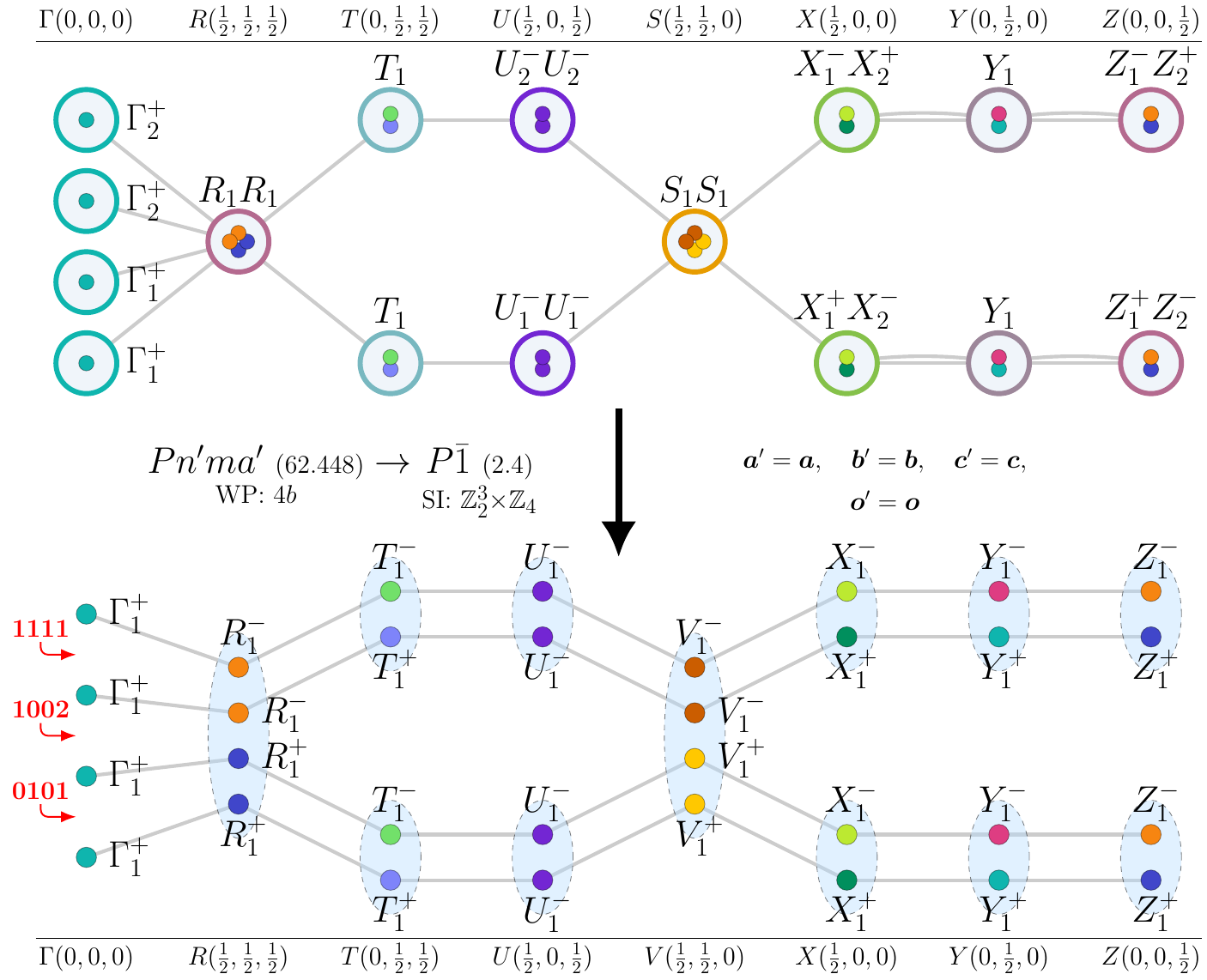}
\caption{Topological magnon bands in subgroup $P\bar{1}~(2.4)$ for magnetic moments on Wyckoff position $4b$ of supergroup $Pn'ma'~(62.448)$.\label{fig_62.448_2.4_strainingenericdirection_4b}}
\end{figure}
\input{gap_tables_tex/62.448_2.4_strainingenericdirection_4b_table.tex}
\input{si_tables_tex/62.448_2.4_strainingenericdirection_4b_table.tex}
\subsubsection{Topological bands in subgroup $P2_{1}'/c'~(14.79)$}
\textbf{Perturbations:}
\begin{itemize}
\item strain $\perp$ [001],
\item (B $\parallel$ [100] or B $\perp$ [001]).
\end{itemize}
\begin{figure}[H]
\centering
\includegraphics[scale=0.6]{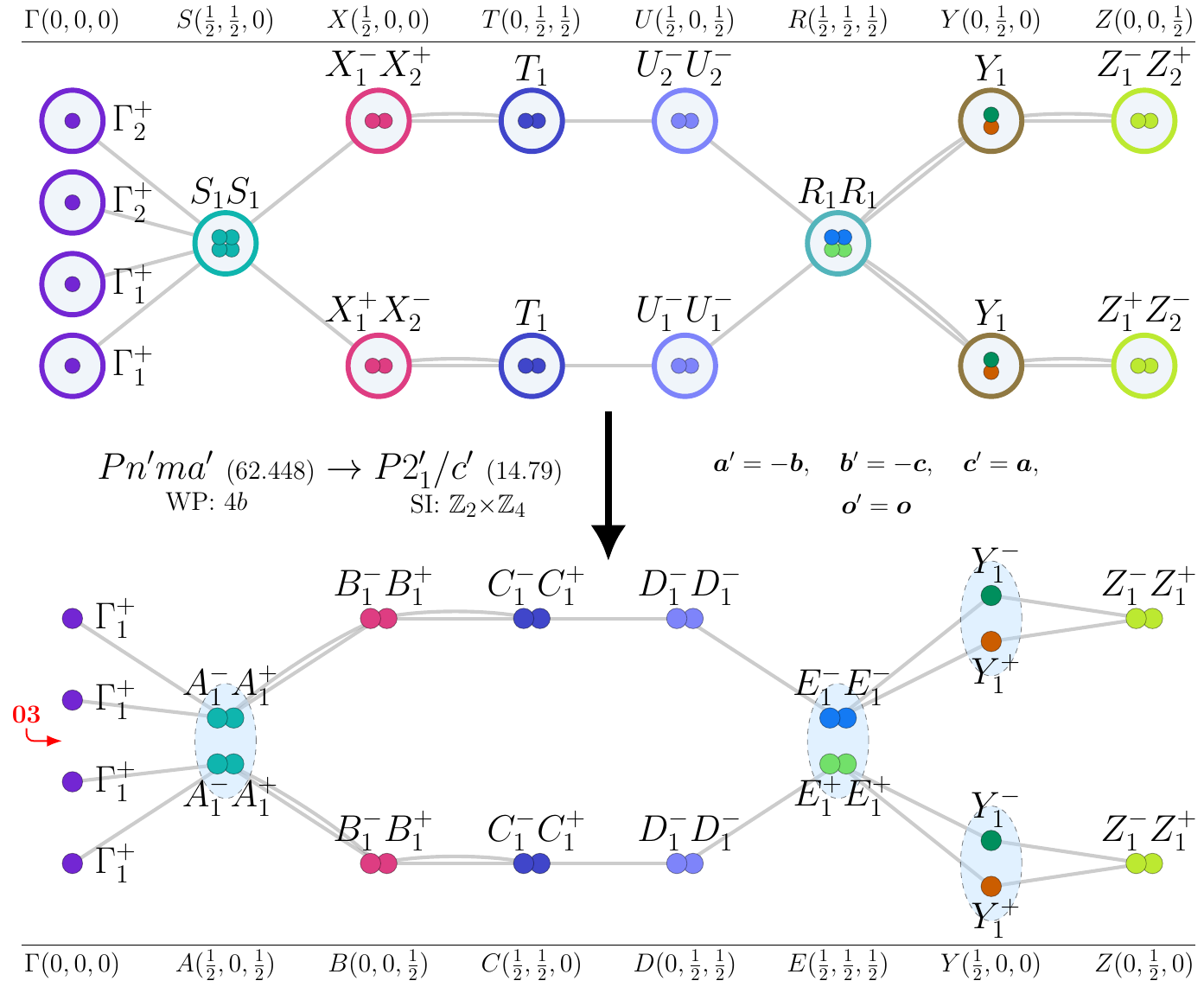}
\caption{Topological magnon bands in subgroup $P2_{1}'/c'~(14.79)$ for magnetic moments on Wyckoff position $4b$ of supergroup $Pn'ma'~(62.448)$.\label{fig_62.448_14.79_strainperp001_4b}}
\end{figure}
\input{gap_tables_tex/62.448_14.79_strainperp001_4b_table.tex}
\input{si_tables_tex/62.448_14.79_strainperp001_4b_table.tex}
\subsubsection{Topological bands in subgroup $P2_{1}'/c'~(14.79)$}
\textbf{Perturbations:}
\begin{itemize}
\item strain $\perp$ [100],
\item (B $\parallel$ [001] or B $\perp$ [100]).
\end{itemize}
\begin{figure}[H]
\centering
\includegraphics[scale=0.6]{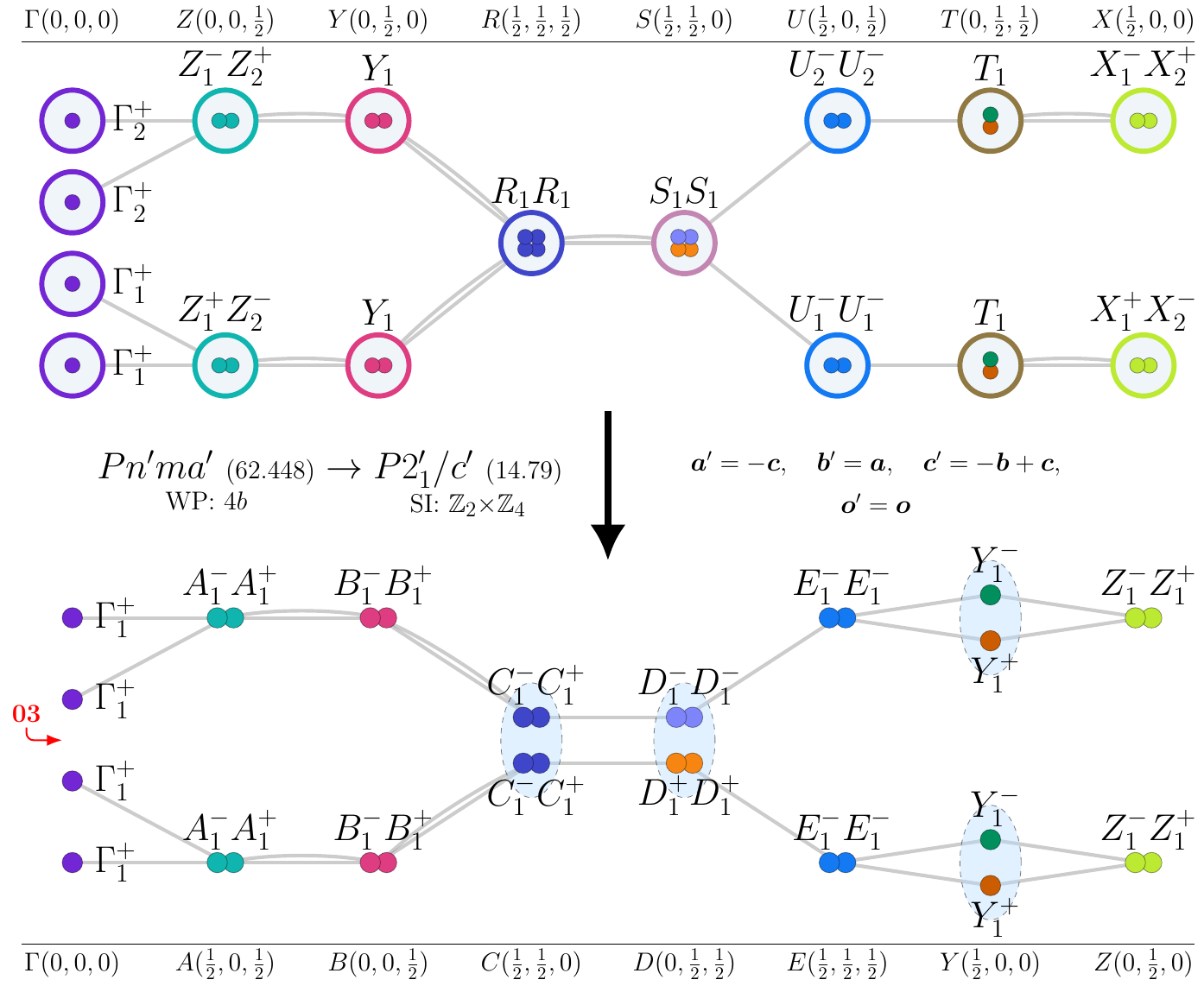}
\caption{Topological magnon bands in subgroup $P2_{1}'/c'~(14.79)$ for magnetic moments on Wyckoff position $4b$ of supergroup $Pn'ma'~(62.448)$.\label{fig_62.448_14.79_strainperp100_4b}}
\end{figure}
\input{gap_tables_tex/62.448_14.79_strainperp100_4b_table.tex}
\input{si_tables_tex/62.448_14.79_strainperp100_4b_table.tex}
\subsection{WP: $4a$}
\textbf{BCS Materials:} {La\textsubscript{0.5}Sr\textsubscript{0.5}FeO\textsubscript{2.5}F\textsubscript{0.5}~(700 K)}\footnote{BCS web page: \texttt{\href{http://webbdcrista1.ehu.es/magndata/index.php?this\_label=0.34} {http://webbdcrista1.ehu.es/magndata/index.php?this\_label=0.34}}}, {La\textsubscript{0.875}Ba\textsubscript{0.125}Mn\textsubscript{0.875}Ti\textsubscript{0.125}O\textsubscript{3}}\footnote{BCS web page: \texttt{\href{http://webbdcrista1.ehu.es/magndata/index.php?this\_label=0.647} {http://webbdcrista1.ehu.es/magndata/index.php?this\_label=0.647}}}, {La\textsubscript{0.90}Ba\textsubscript{0.10}Mn\textsubscript{0.90}Ti\textsubscript{0.10}O\textsubscript{3}}\footnote{BCS web page: \texttt{\href{http://webbdcrista1.ehu.es/magndata/index.php?this\_label=0.646} {http://webbdcrista1.ehu.es/magndata/index.php?this\_label=0.646}}}, {La\textsubscript{0.95}Ba\textsubscript{0.05}Mn\textsubscript{0.95}Ti\textsubscript{0.05}O\textsubscript{3}}\footnote{BCS web page: \texttt{\href{http://webbdcrista1.ehu.es/magndata/index.php?this\_label=0.645} {http://webbdcrista1.ehu.es/magndata/index.php?this\_label=0.645}}}, {La\textsubscript{0.95}Ba\textsubscript{0.05}MnO\textsubscript{3}}\footnote{BCS web page: \texttt{\href{http://webbdcrista1.ehu.es/magndata/index.php?this\_label=0.644} {http://webbdcrista1.ehu.es/magndata/index.php?this\_label=0.644}}}, {La\textsubscript{0.95}Ca\textsubscript{0.05}MnO\textsubscript{3}}\footnote{BCS web page: \texttt{\href{http://webbdcrista1.ehu.es/magndata/index.php?this\_label=0.643} {http://webbdcrista1.ehu.es/magndata/index.php?this\_label=0.643}}}, {LaMnO\textsubscript{3}}\footnote{BCS web page: \texttt{\href{http://webbdcrista1.ehu.es/magndata/index.php?this\_label=0.642} {http://webbdcrista1.ehu.es/magndata/index.php?this\_label=0.642}}}, {YCr\textsubscript{0.5}Mn\textsubscript{0.5}O\textsubscript{3}}\footnote{BCS web page: \texttt{\href{http://webbdcrista1.ehu.es/magndata/index.php?this\_label=0.100} {http://webbdcrista1.ehu.es/magndata/index.php?this\_label=0.100}}}.\\
\subsubsection{Topological bands in subgroup $P\bar{1}~(2.4)$}
\textbf{Perturbations:}
\begin{itemize}
\item strain in generic direction,
\item (B $\parallel$ [100] or B $\perp$ [001]) and strain $\perp$ [100],
\item (B $\parallel$ [100] or B $\perp$ [001]) and strain $\perp$ [010],
\item (B $\parallel$ [001] or B $\perp$ [100]) and strain $\perp$ [010],
\item (B $\parallel$ [001] or B $\perp$ [100]) and strain $\perp$ [001],
\item B in generic direction.
\end{itemize}
\begin{figure}[H]
\centering
\includegraphics[scale=0.6]{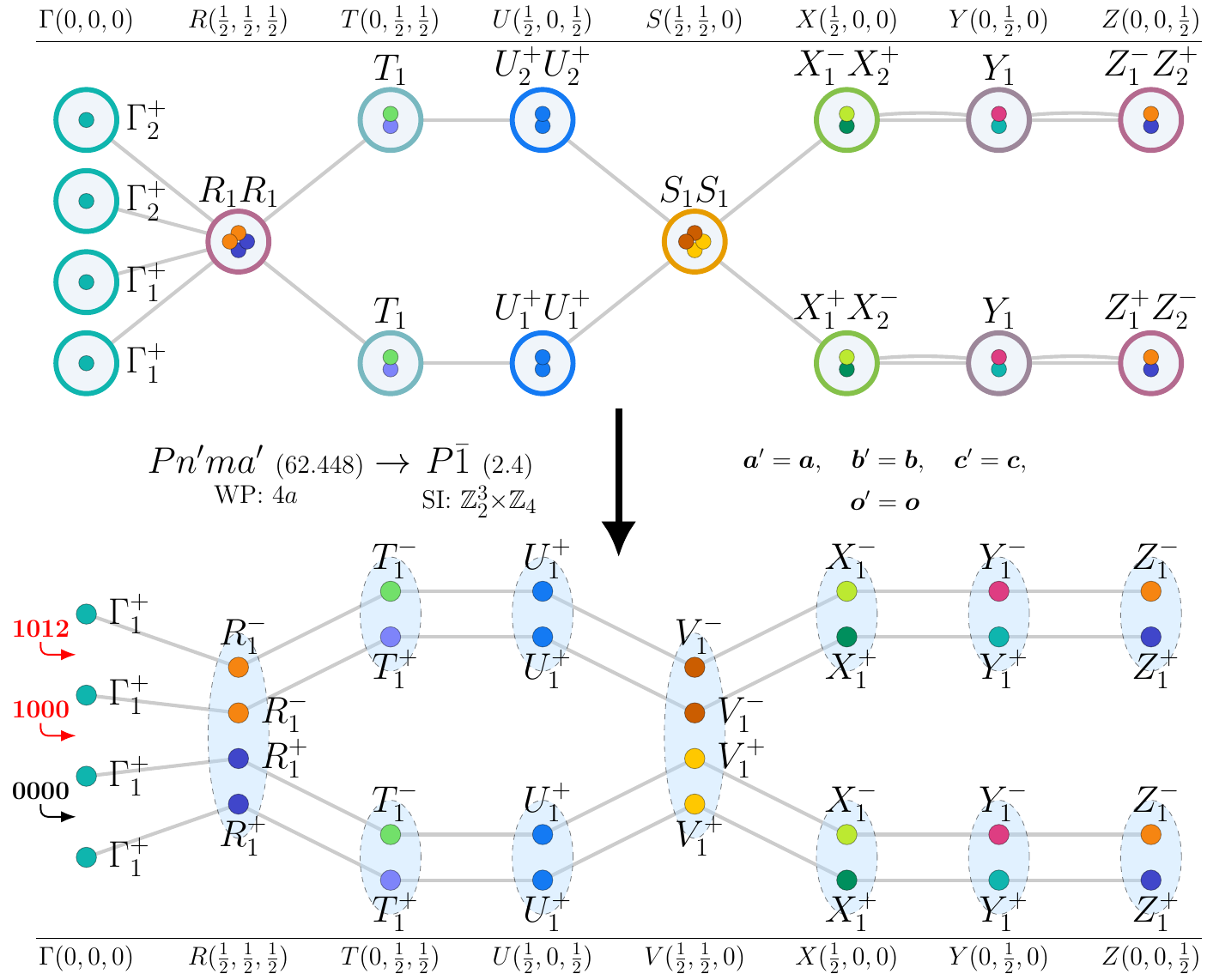}
\caption{Topological magnon bands in subgroup $P\bar{1}~(2.4)$ for magnetic moments on Wyckoff position $4a$ of supergroup $Pn'ma'~(62.448)$.\label{fig_62.448_2.4_strainingenericdirection_4a}}
\end{figure}
\input{gap_tables_tex/62.448_2.4_strainingenericdirection_4a_table.tex}
\input{si_tables_tex/62.448_2.4_strainingenericdirection_4a_table.tex}
\subsubsection{Topological bands in subgroup $P2_{1}'/c'~(14.79)$}
\textbf{Perturbations:}
\begin{itemize}
\item strain $\perp$ [001],
\item (B $\parallel$ [100] or B $\perp$ [001]).
\end{itemize}
\begin{figure}[H]
\centering
\includegraphics[scale=0.6]{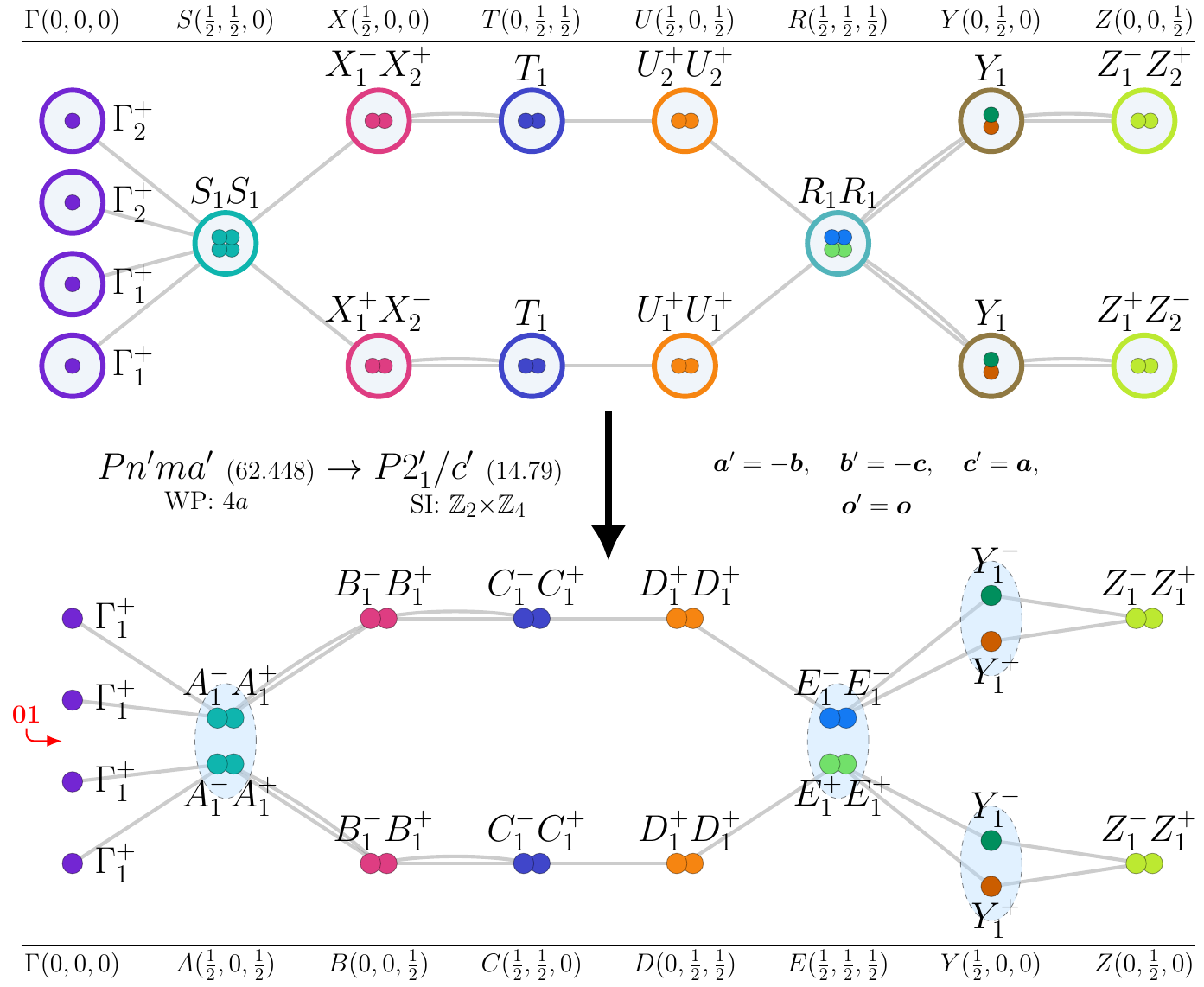}
\caption{Topological magnon bands in subgroup $P2_{1}'/c'~(14.79)$ for magnetic moments on Wyckoff position $4a$ of supergroup $Pn'ma'~(62.448)$.\label{fig_62.448_14.79_strainperp001_4a}}
\end{figure}
\input{gap_tables_tex/62.448_14.79_strainperp001_4a_table.tex}
\input{si_tables_tex/62.448_14.79_strainperp001_4a_table.tex}
\subsubsection{Topological bands in subgroup $P2_{1}'/c'~(14.79)$}
\textbf{Perturbations:}
\begin{itemize}
\item strain $\perp$ [100],
\item (B $\parallel$ [001] or B $\perp$ [100]).
\end{itemize}
\begin{figure}[H]
\centering
\includegraphics[scale=0.6]{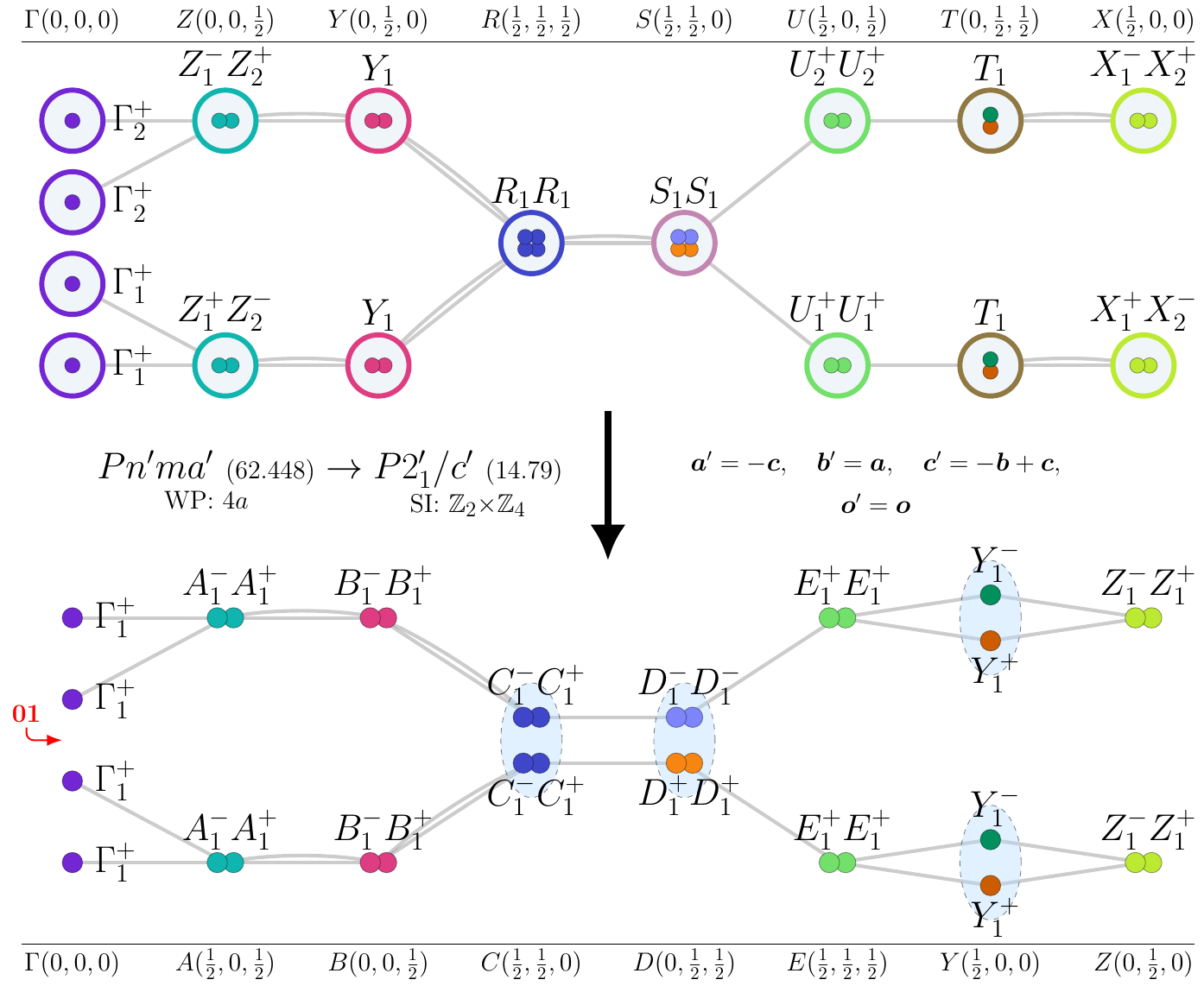}
\caption{Topological magnon bands in subgroup $P2_{1}'/c'~(14.79)$ for magnetic moments on Wyckoff position $4a$ of supergroup $Pn'ma'~(62.448)$.\label{fig_62.448_14.79_strainperp100_4a}}
\end{figure}
\input{gap_tables_tex/62.448_14.79_strainperp100_4a_table.tex}
\input{si_tables_tex/62.448_14.79_strainperp100_4a_table.tex}
\subsection{WP: $4c$}
\textbf{BCS Materials:} {SrMnSb\textsubscript{2}~(565 K)}\footnote{BCS web page: \texttt{\href{http://webbdcrista1.ehu.es/magndata/index.php?this\_label=0.767} {http://webbdcrista1.ehu.es/magndata/index.php?this\_label=0.767}}}.\\
\subsubsection{Topological bands in subgroup $P\bar{1}~(2.4)$}
\textbf{Perturbations:}
\begin{itemize}
\item strain in generic direction,
\item (B $\parallel$ [100] or B $\perp$ [001]) and strain $\perp$ [100],
\item (B $\parallel$ [100] or B $\perp$ [001]) and strain $\perp$ [010],
\item (B $\parallel$ [001] or B $\perp$ [100]) and strain $\perp$ [010],
\item (B $\parallel$ [001] or B $\perp$ [100]) and strain $\perp$ [001],
\item B in generic direction.
\end{itemize}
\begin{figure}[H]
\centering
\includegraphics[scale=0.6]{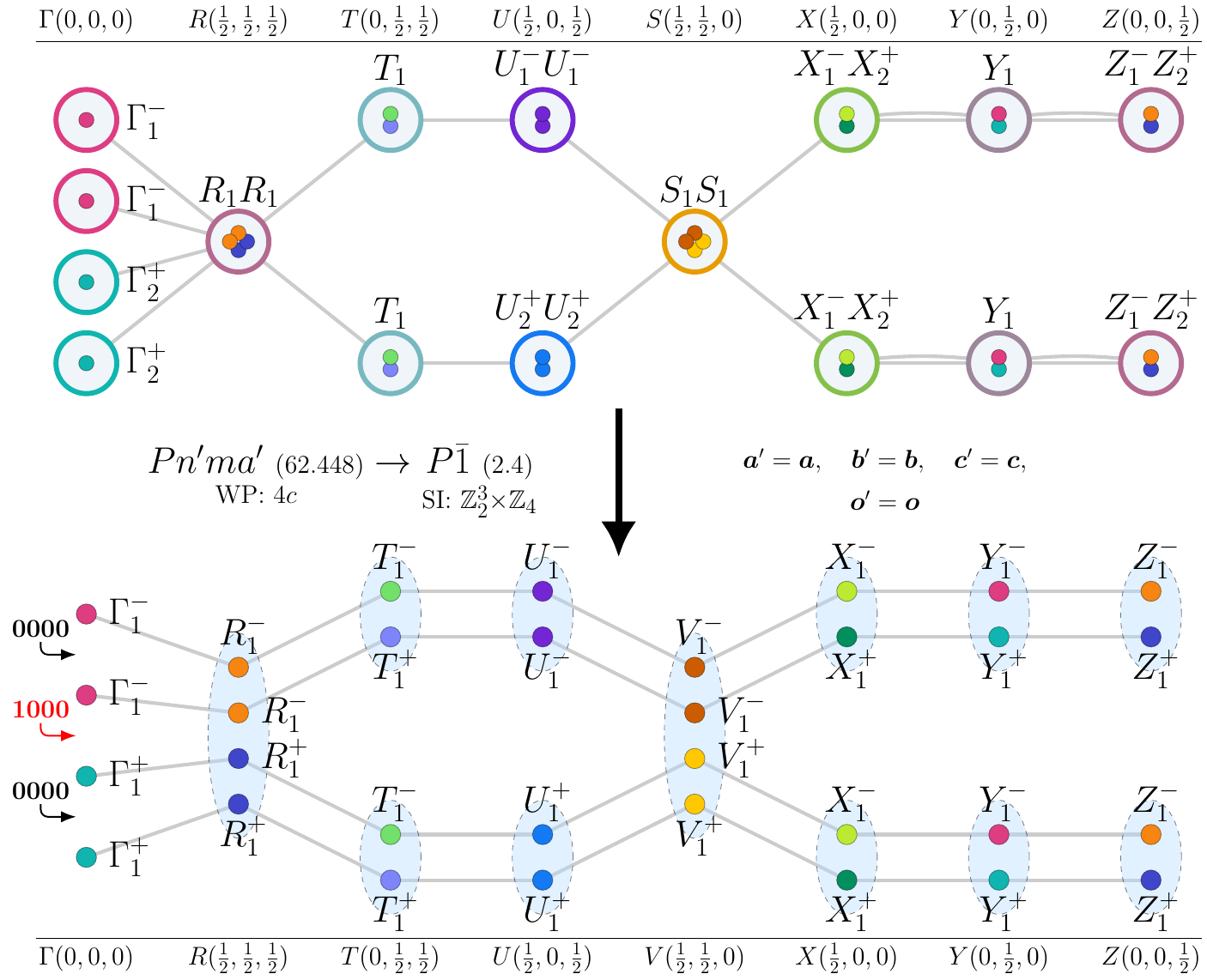}
\caption{Topological magnon bands in subgroup $P\bar{1}~(2.4)$ for magnetic moments on Wyckoff position $4c$ of supergroup $Pn'ma'~(62.448)$.\label{fig_62.448_2.4_strainingenericdirection_4c}}
\end{figure}
\input{gap_tables_tex/62.448_2.4_strainingenericdirection_4c_table.tex}
\input{si_tables_tex/62.448_2.4_strainingenericdirection_4c_table.tex}
\subsubsection{Topological bands in subgroup $P2_{1}'/c'~(14.79)$}
\textbf{Perturbations:}
\begin{itemize}
\item strain $\perp$ [001],
\item (B $\parallel$ [100] or B $\perp$ [001]).
\end{itemize}
\begin{figure}[H]
\centering
\includegraphics[scale=0.6]{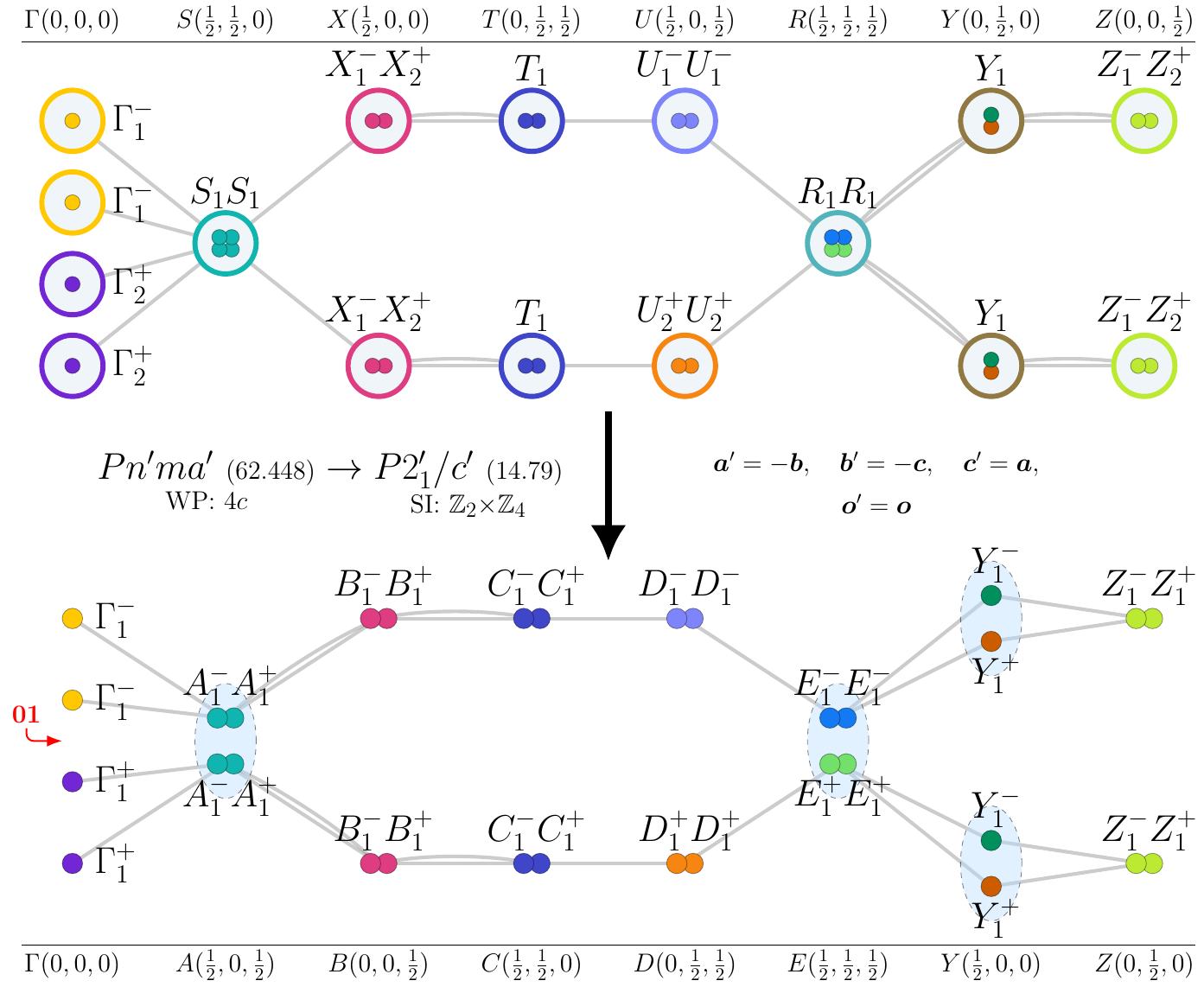}
\caption{Topological magnon bands in subgroup $P2_{1}'/c'~(14.79)$ for magnetic moments on Wyckoff position $4c$ of supergroup $Pn'ma'~(62.448)$.\label{fig_62.448_14.79_strainperp001_4c}}
\end{figure}
\input{gap_tables_tex/62.448_14.79_strainperp001_4c_table.tex}
\input{si_tables_tex/62.448_14.79_strainperp001_4c_table.tex}
\subsubsection{Topological bands in subgroup $P2_{1}'/c'~(14.79)$}
\textbf{Perturbations:}
\begin{itemize}
\item strain $\perp$ [100],
\item (B $\parallel$ [001] or B $\perp$ [100]).
\end{itemize}
\begin{figure}[H]
\centering
\includegraphics[scale=0.6]{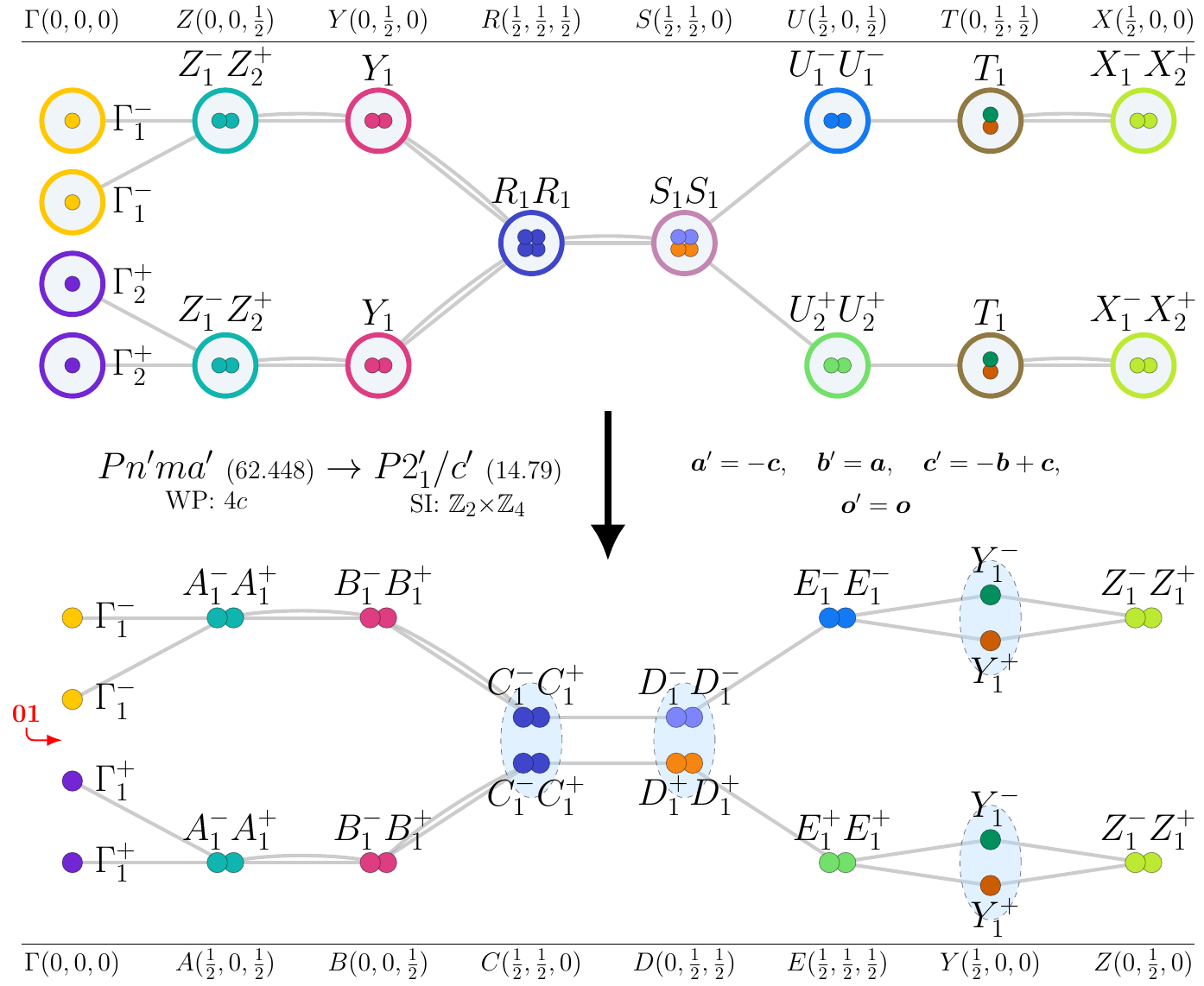}
\caption{Topological magnon bands in subgroup $P2_{1}'/c'~(14.79)$ for magnetic moments on Wyckoff position $4c$ of supergroup $Pn'ma'~(62.448)$.\label{fig_62.448_14.79_strainperp100_4c}}
\end{figure}
\input{gap_tables_tex/62.448_14.79_strainperp100_4c_table.tex}
\input{si_tables_tex/62.448_14.79_strainperp100_4c_table.tex}
\subsection{WP: $4c+8d$}
\textbf{BCS Materials:} {Mn\textsubscript{3}Sn\textsubscript{2}~(262 K)}\footnote{BCS web page: \texttt{\href{http://webbdcrista1.ehu.es/magndata/index.php?this\_label=0.662} {http://webbdcrista1.ehu.es/magndata/index.php?this\_label=0.662}}}, {Mn\textsubscript{3}Sn\textsubscript{2}~(227 K)}\footnote{BCS web page: \texttt{\href{http://webbdcrista1.ehu.es/magndata/index.php?this\_label=0.663} {http://webbdcrista1.ehu.es/magndata/index.php?this\_label=0.663}}}.\\
\subsubsection{Topological bands in subgroup $P2_{1}'/c'~(14.79)$}
\textbf{Perturbations:}
\begin{itemize}
\item strain $\perp$ [001],
\item (B $\parallel$ [100] or B $\perp$ [001]).
\end{itemize}
\begin{figure}[H]
\centering
\includegraphics[scale=0.6]{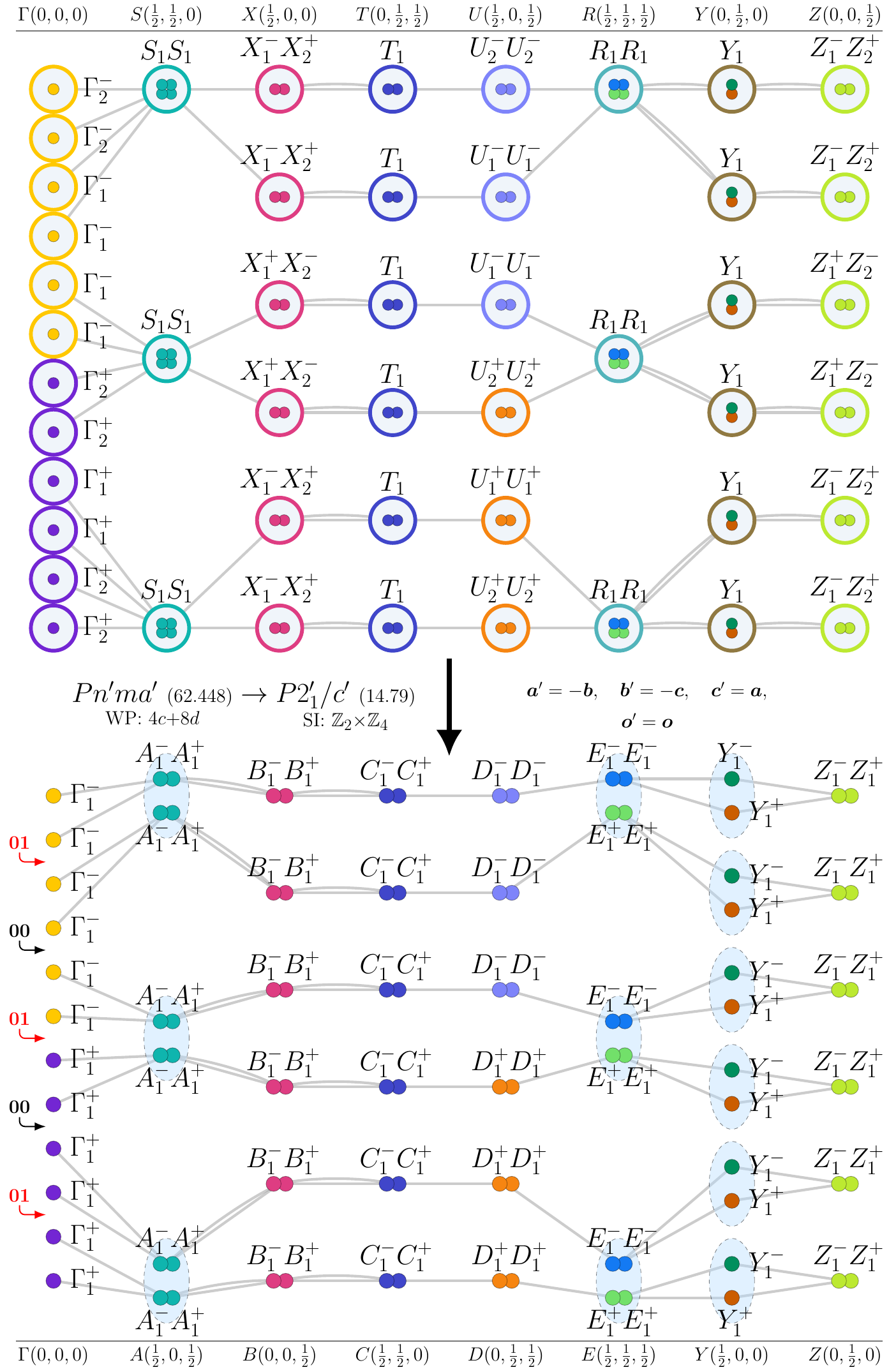}
\caption{Topological magnon bands in subgroup $P2_{1}'/c'~(14.79)$ for magnetic moments on Wyckoff positions $4c+8d$ of supergroup $Pn'ma'~(62.448)$.\label{fig_62.448_14.79_strainperp001_4c+8d}}
\end{figure}
\input{gap_tables_tex/62.448_14.79_strainperp001_4c+8d_table.tex}
\input{si_tables_tex/62.448_14.79_strainperp001_4c+8d_table.tex}
\subsubsection{Topological bands in subgroup $P2_{1}'/c'~(14.79)$}
\textbf{Perturbations:}
\begin{itemize}
\item strain $\perp$ [100],
\item (B $\parallel$ [001] or B $\perp$ [100]).
\end{itemize}
\begin{figure}[H]
\centering
\includegraphics[scale=0.6]{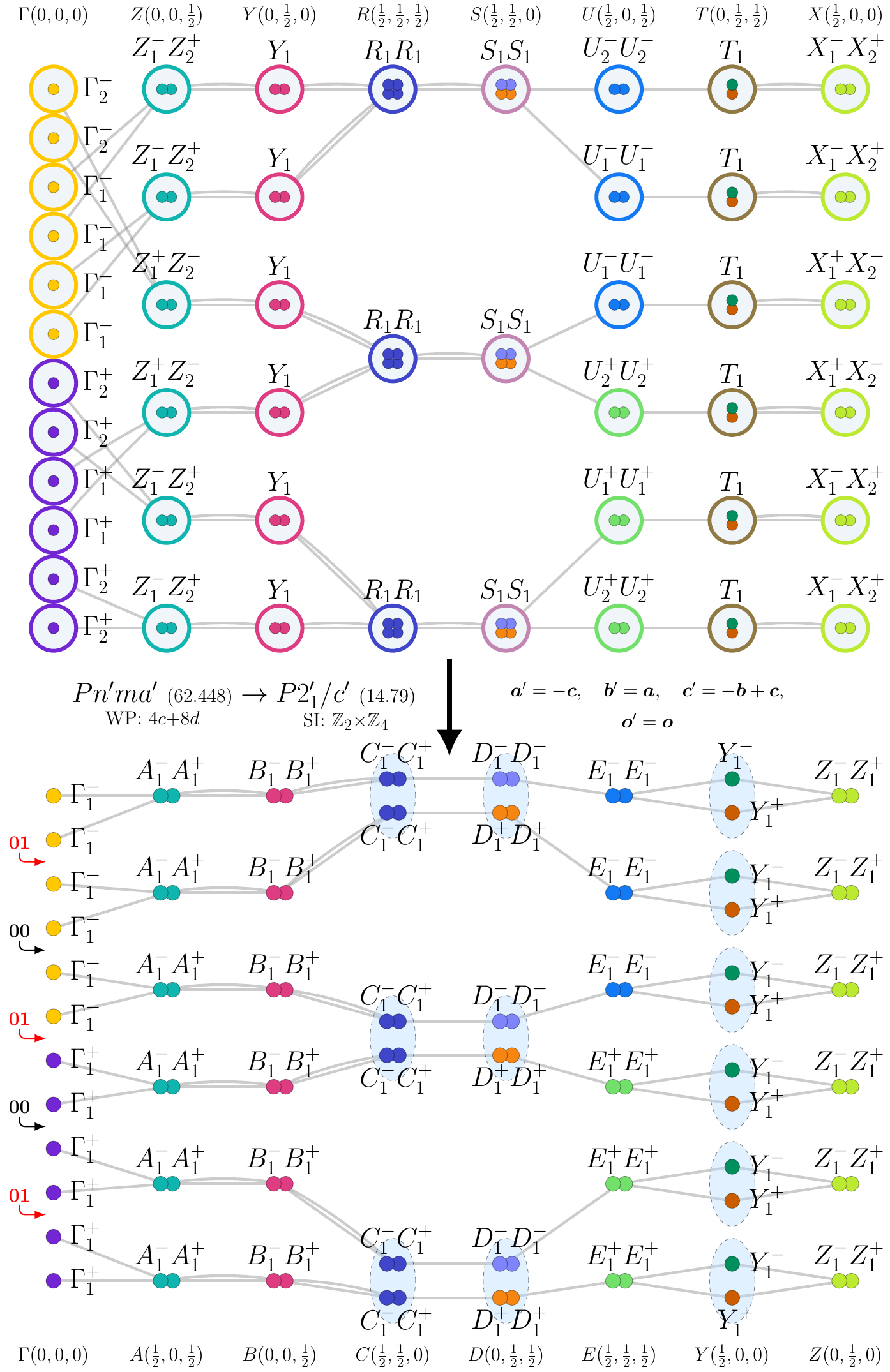}
\caption{Topological magnon bands in subgroup $P2_{1}'/c'~(14.79)$ for magnetic moments on Wyckoff positions $4c+8d$ of supergroup $Pn'ma'~(62.448)$.\label{fig_62.448_14.79_strainperp100_4c+8d}}
\end{figure}
\input{gap_tables_tex/62.448_14.79_strainperp100_4c+8d_table.tex}
\input{si_tables_tex/62.448_14.79_strainperp100_4c+8d_table.tex}
\subsection{WP: $4b+4c$}
\textbf{BCS Materials:} {PrCrO\textsubscript{3}~(239 K)}\footnote{BCS web page: \texttt{\href{http://webbdcrista1.ehu.es/magndata/index.php?this\_label=0.588} {http://webbdcrista1.ehu.es/magndata/index.php?this\_label=0.588}}}, {SmCrO\textsubscript{3}~(192 K)}\footnote{BCS web page: \texttt{\href{http://webbdcrista1.ehu.es/magndata/index.php?this\_label=0.696} {http://webbdcrista1.ehu.es/magndata/index.php?this\_label=0.696}}}, {SmCrO\textsubscript{3}~(191 K)}\footnote{BCS web page: \texttt{\href{http://webbdcrista1.ehu.es/magndata/index.php?this\_label=0.479} {http://webbdcrista1.ehu.es/magndata/index.php?this\_label=0.479}}}, {NdMn\textsubscript{0.8}Fe\textsubscript{0.2}O\textsubscript{3}~(59 K)}\footnote{BCS web page: \texttt{\href{http://webbdcrista1.ehu.es/magndata/index.php?this\_label=0.660} {http://webbdcrista1.ehu.es/magndata/index.php?this\_label=0.660}}}, {Rb\textsubscript{2}Fe\textsubscript{2}O(AsO\textsubscript{4})\textsubscript{2}~(25 K)}\footnote{BCS web page: \texttt{\href{http://webbdcrista1.ehu.es/magndata/index.php?this\_label=0.91} {http://webbdcrista1.ehu.es/magndata/index.php?this\_label=0.91}}}, {NdMnO\textsubscript{3}~(15 K)}\footnote{BCS web page: \texttt{\href{http://webbdcrista1.ehu.es/magndata/index.php?this\_label=0.371} {http://webbdcrista1.ehu.es/magndata/index.php?this\_label=0.371}}}, {NdMn\textsubscript{0.8}Fe\textsubscript{0.2}O\textsubscript{3}~(13 K)}\footnote{BCS web page: \texttt{\href{http://webbdcrista1.ehu.es/magndata/index.php?this\_label=0.659} {http://webbdcrista1.ehu.es/magndata/index.php?this\_label=0.659}}}, {NdMnO\textsubscript{3}~(13 K)}\footnote{BCS web page: \texttt{\href{http://webbdcrista1.ehu.es/magndata/index.php?this\_label=0.289} {http://webbdcrista1.ehu.es/magndata/index.php?this\_label=0.289}}}.\\
\subsubsection{Topological bands in subgroup $P2_{1}'/c'~(14.79)$}
\textbf{Perturbations:}
\begin{itemize}
\item strain $\perp$ [001],
\item (B $\parallel$ [100] or B $\perp$ [001]).
\end{itemize}
\begin{figure}[H]
\centering
\includegraphics[scale=0.6]{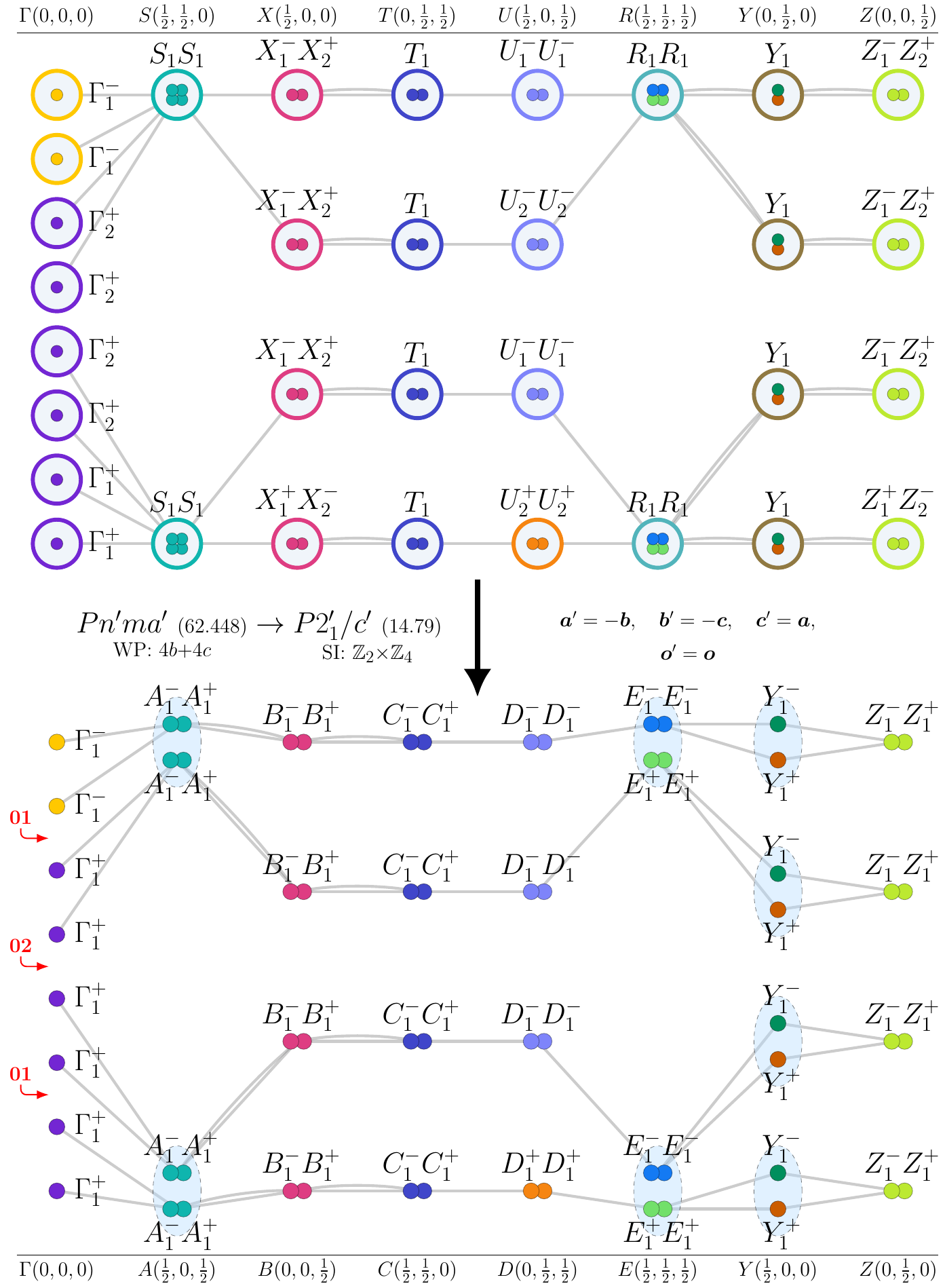}
\caption{Topological magnon bands in subgroup $P2_{1}'/c'~(14.79)$ for magnetic moments on Wyckoff positions $4b+4c$ of supergroup $Pn'ma'~(62.448)$.\label{fig_62.448_14.79_strainperp001_4b+4c}}
\end{figure}
\input{gap_tables_tex/62.448_14.79_strainperp001_4b+4c_table.tex}
\input{si_tables_tex/62.448_14.79_strainperp001_4b+4c_table.tex}
\subsubsection{Topological bands in subgroup $P2_{1}'/c'~(14.79)$}
\textbf{Perturbations:}
\begin{itemize}
\item strain $\perp$ [100],
\item (B $\parallel$ [001] or B $\perp$ [100]).
\end{itemize}
\begin{figure}[H]
\centering
\includegraphics[scale=0.6]{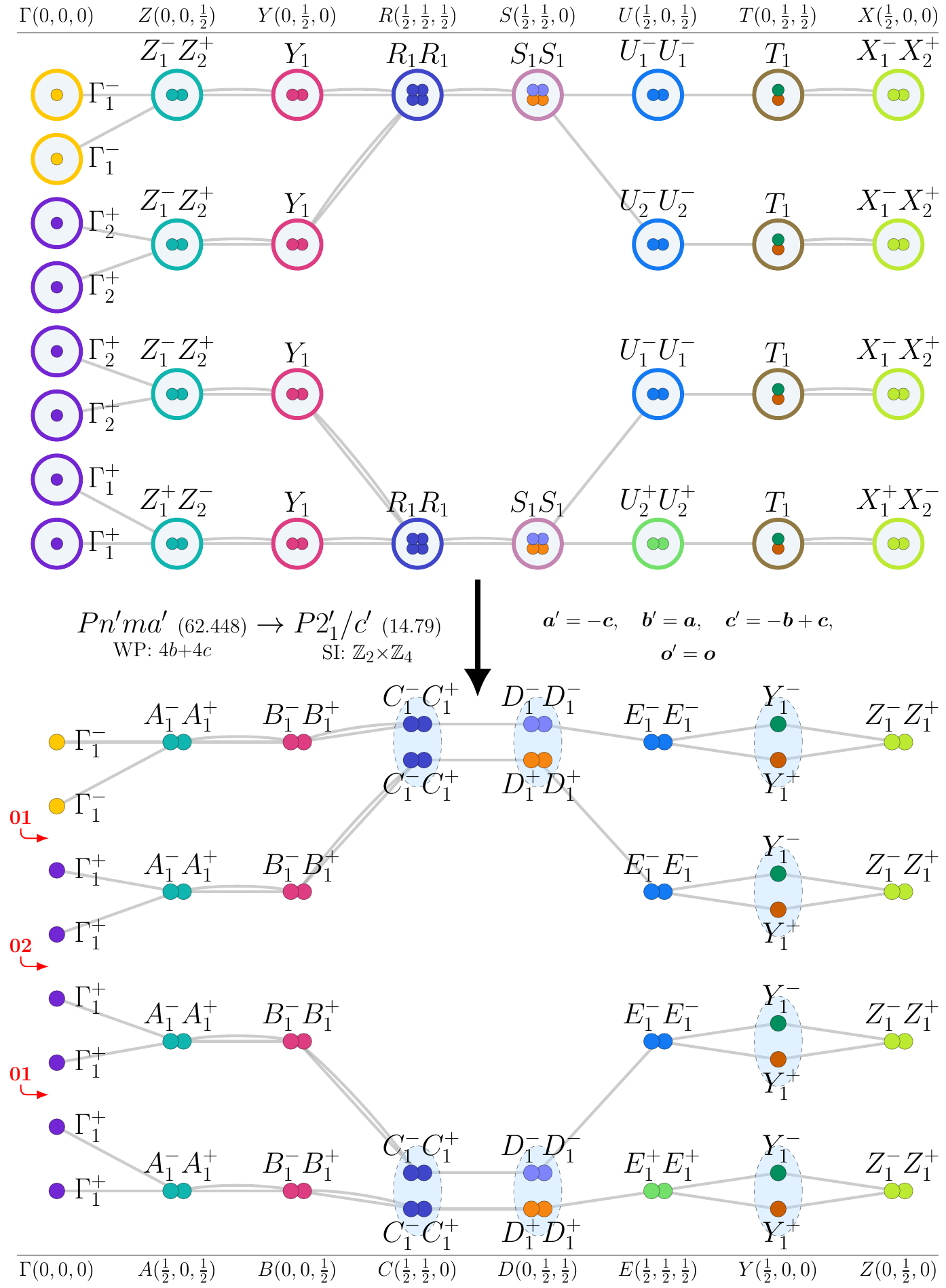}
\caption{Topological magnon bands in subgroup $P2_{1}'/c'~(14.79)$ for magnetic moments on Wyckoff positions $4b+4c$ of supergroup $Pn'ma'~(62.448)$.\label{fig_62.448_14.79_strainperp100_4b+4c}}
\end{figure}
\input{gap_tables_tex/62.448_14.79_strainperp100_4b+4c_table.tex}
\input{si_tables_tex/62.448_14.79_strainperp100_4b+4c_table.tex}
\subsection{WP: $4b+4b$}
\textbf{BCS Materials:} {TbCr\textsubscript{0.5}Mn\textsubscript{0.5}O\textsubscript{3}~(84 K)}\footnote{BCS web page: \texttt{\href{http://webbdcrista1.ehu.es/magndata/index.php?this\_label=0.679} {http://webbdcrista1.ehu.es/magndata/index.php?this\_label=0.679}}}.\\
\subsubsection{Topological bands in subgroup $P\bar{1}~(2.4)$}
\textbf{Perturbations:}
\begin{itemize}
\item strain in generic direction,
\item (B $\parallel$ [100] or B $\perp$ [001]) and strain $\perp$ [100],
\item (B $\parallel$ [100] or B $\perp$ [001]) and strain $\perp$ [010],
\item (B $\parallel$ [001] or B $\perp$ [100]) and strain $\perp$ [010],
\item (B $\parallel$ [001] or B $\perp$ [100]) and strain $\perp$ [001],
\item B in generic direction.
\end{itemize}
\begin{figure}[H]
\centering
\includegraphics[scale=0.6]{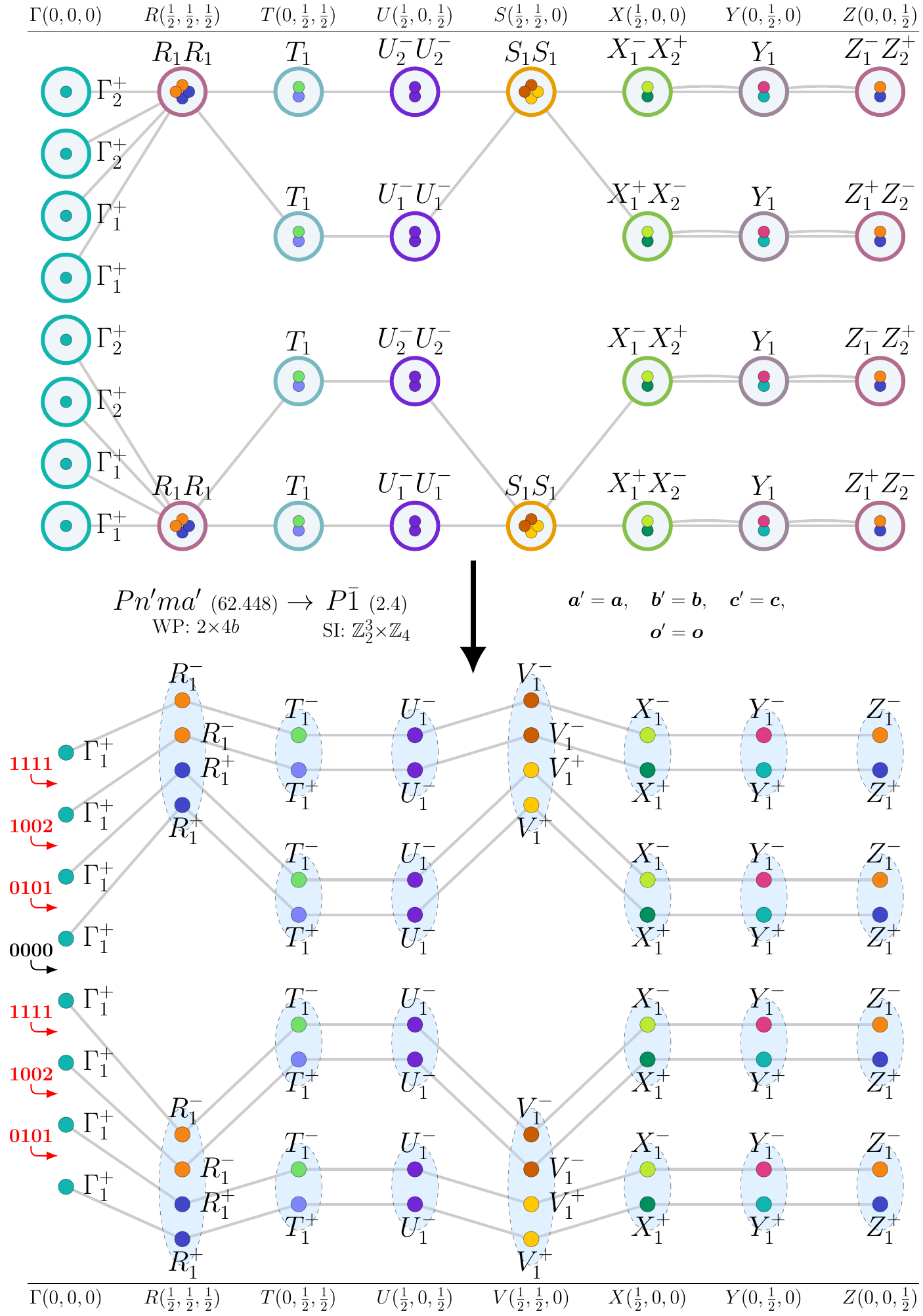}
\caption{Topological magnon bands in subgroup $P\bar{1}~(2.4)$ for magnetic moments on Wyckoff positions $4b+4b$ of supergroup $Pn'ma'~(62.448)$.\label{fig_62.448_2.4_strainingenericdirection_4b+4b}}
\end{figure}
\input{gap_tables_tex/62.448_2.4_strainingenericdirection_4b+4b_table.tex}
\input{si_tables_tex/62.448_2.4_strainingenericdirection_4b+4b_table.tex}
\subsubsection{Topological bands in subgroup $P2_{1}'/c'~(14.79)$}
\textbf{Perturbations:}
\begin{itemize}
\item strain $\perp$ [001],
\item (B $\parallel$ [100] or B $\perp$ [001]).
\end{itemize}
\begin{figure}[H]
\centering
\includegraphics[scale=0.6]{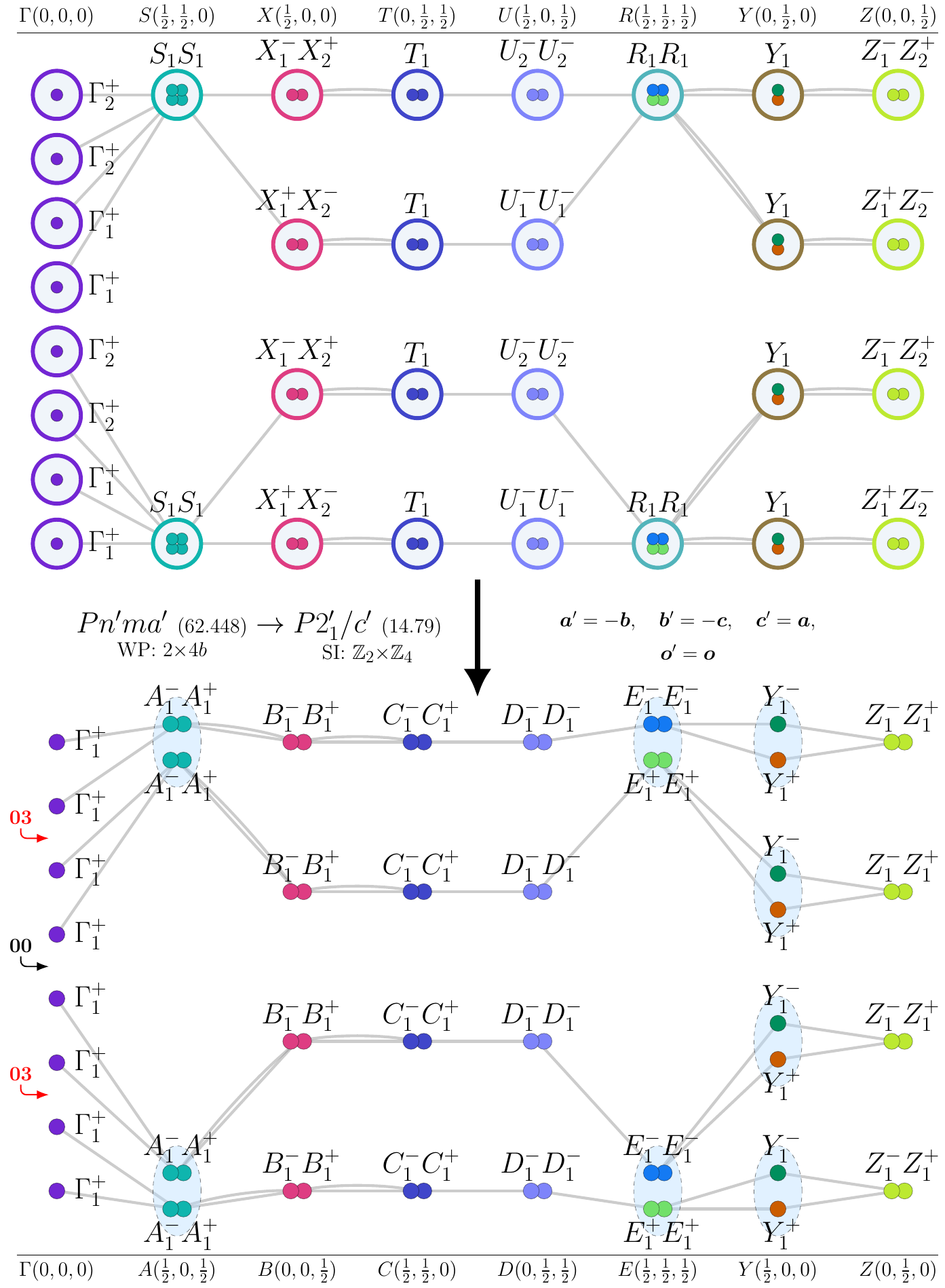}
\caption{Topological magnon bands in subgroup $P2_{1}'/c'~(14.79)$ for magnetic moments on Wyckoff positions $4b+4b$ of supergroup $Pn'ma'~(62.448)$.\label{fig_62.448_14.79_strainperp001_4b+4b}}
\end{figure}
\input{gap_tables_tex/62.448_14.79_strainperp001_4b+4b_table.tex}
\input{si_tables_tex/62.448_14.79_strainperp001_4b+4b_table.tex}
\subsubsection{Topological bands in subgroup $P2_{1}'/c'~(14.79)$}
\textbf{Perturbations:}
\begin{itemize}
\item strain $\perp$ [100],
\item (B $\parallel$ [001] or B $\perp$ [100]).
\end{itemize}
\begin{figure}[H]
\centering
\includegraphics[scale=0.6]{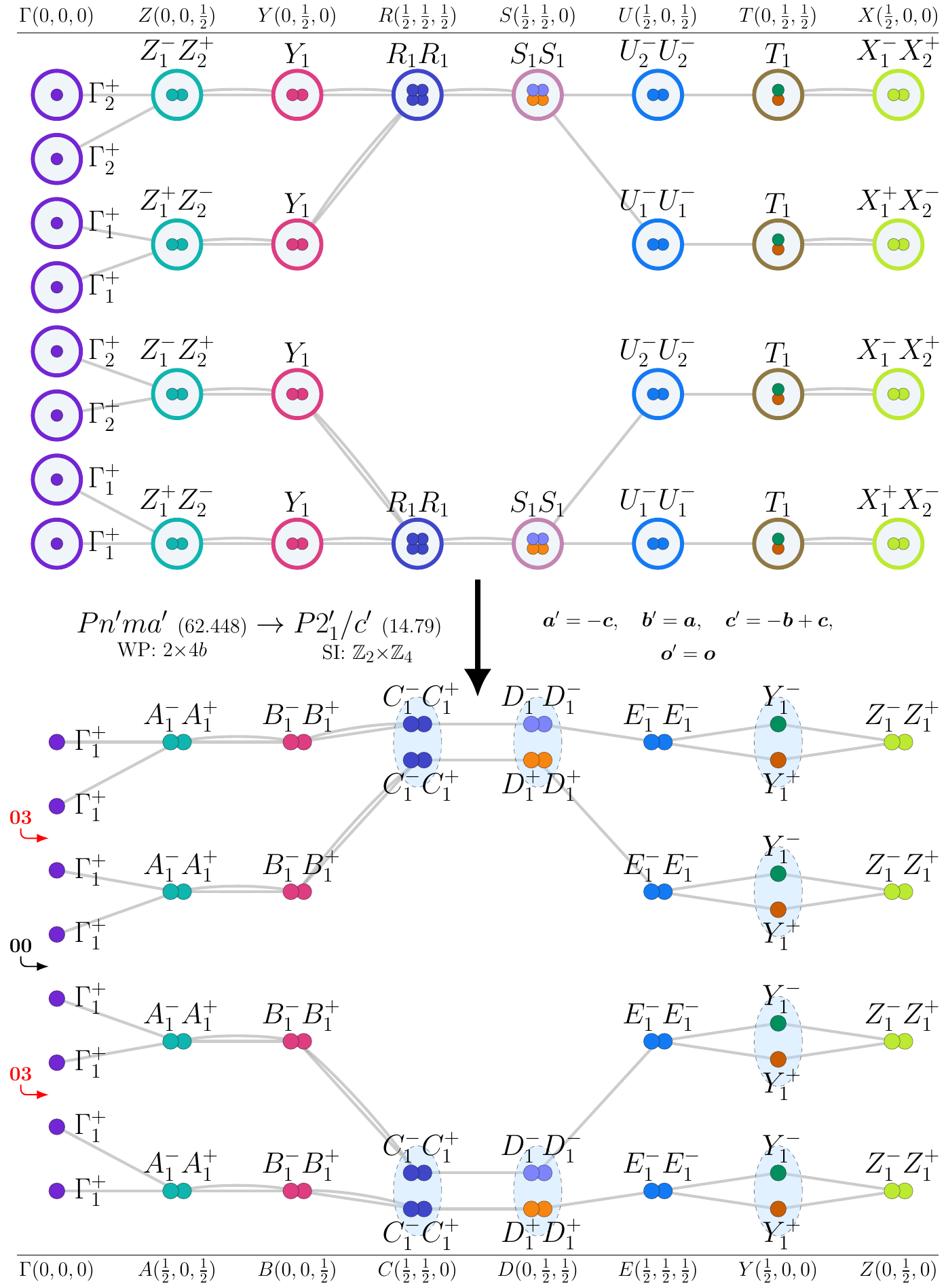}
\caption{Topological magnon bands in subgroup $P2_{1}'/c'~(14.79)$ for magnetic moments on Wyckoff positions $4b+4b$ of supergroup $Pn'ma'~(62.448)$.\label{fig_62.448_14.79_strainperp100_4b+4b}}
\end{figure}
\input{gap_tables_tex/62.448_14.79_strainperp100_4b+4b_table.tex}
\input{si_tables_tex/62.448_14.79_strainperp100_4b+4b_table.tex}
\subsection{WP: $8d$}
\textbf{BCS Materials:} {Mn\textsubscript{2}SeO\textsubscript{3}F\textsubscript{2}~(26 K)}\footnote{BCS web page: \texttt{\href{http://webbdcrista1.ehu.es/magndata/index.php?this\_label=0.755} {http://webbdcrista1.ehu.es/magndata/index.php?this\_label=0.755}}}.\\
\subsubsection{Topological bands in subgroup $P2_{1}'/c'~(14.79)$}
\textbf{Perturbations:}
\begin{itemize}
\item strain $\perp$ [001],
\item (B $\parallel$ [100] or B $\perp$ [001]).
\end{itemize}
\begin{figure}[H]
\centering
\includegraphics[scale=0.6]{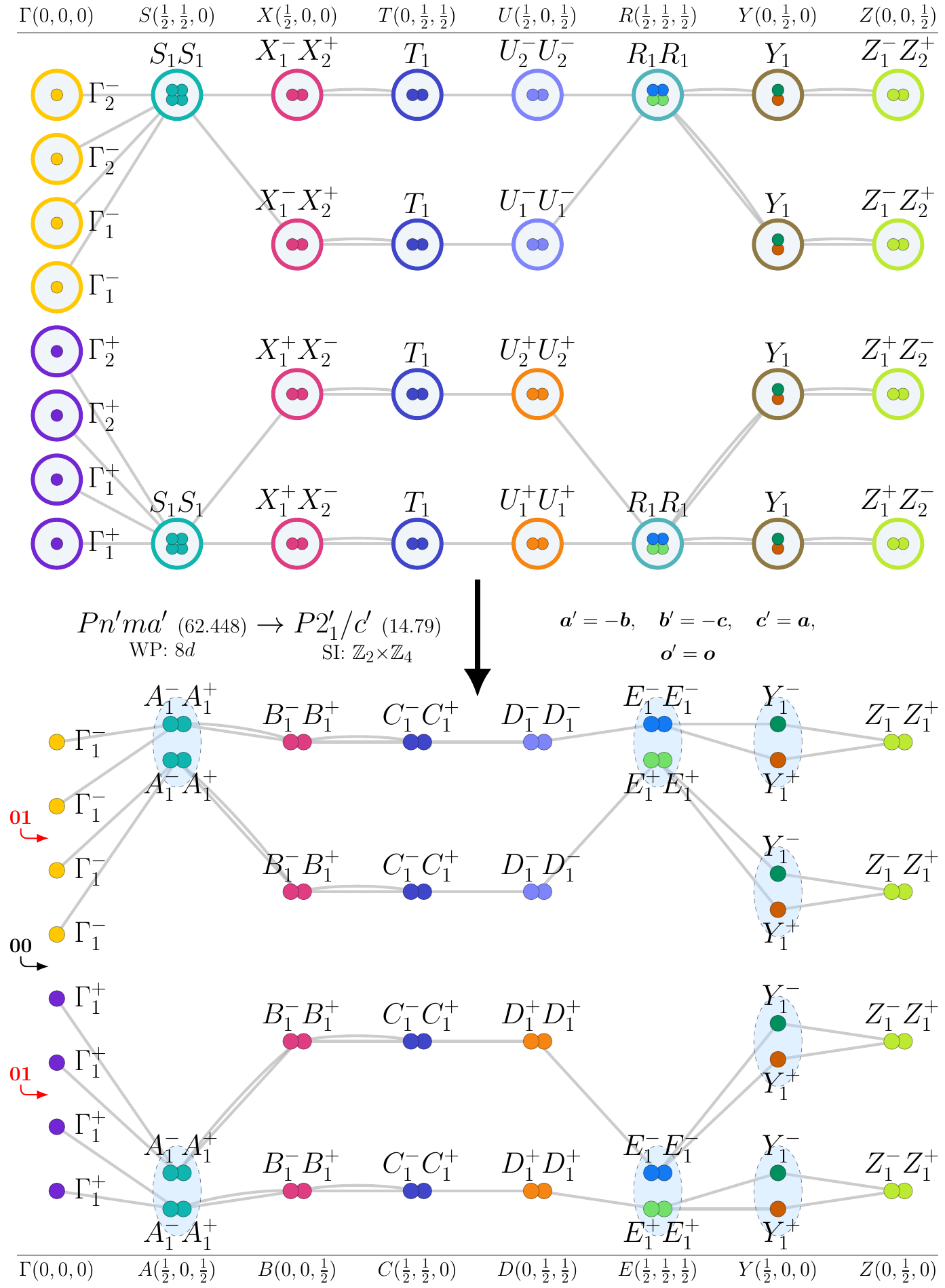}
\caption{Topological magnon bands in subgroup $P2_{1}'/c'~(14.79)$ for magnetic moments on Wyckoff position $8d$ of supergroup $Pn'ma'~(62.448)$.\label{fig_62.448_14.79_strainperp001_8d}}
\end{figure}
\input{gap_tables_tex/62.448_14.79_strainperp001_8d_table.tex}
\input{si_tables_tex/62.448_14.79_strainperp001_8d_table.tex}
\subsubsection{Topological bands in subgroup $P2_{1}'/c'~(14.79)$}
\textbf{Perturbations:}
\begin{itemize}
\item strain $\perp$ [100],
\item (B $\parallel$ [001] or B $\perp$ [100]).
\end{itemize}
\begin{figure}[H]
\centering
\includegraphics[scale=0.6]{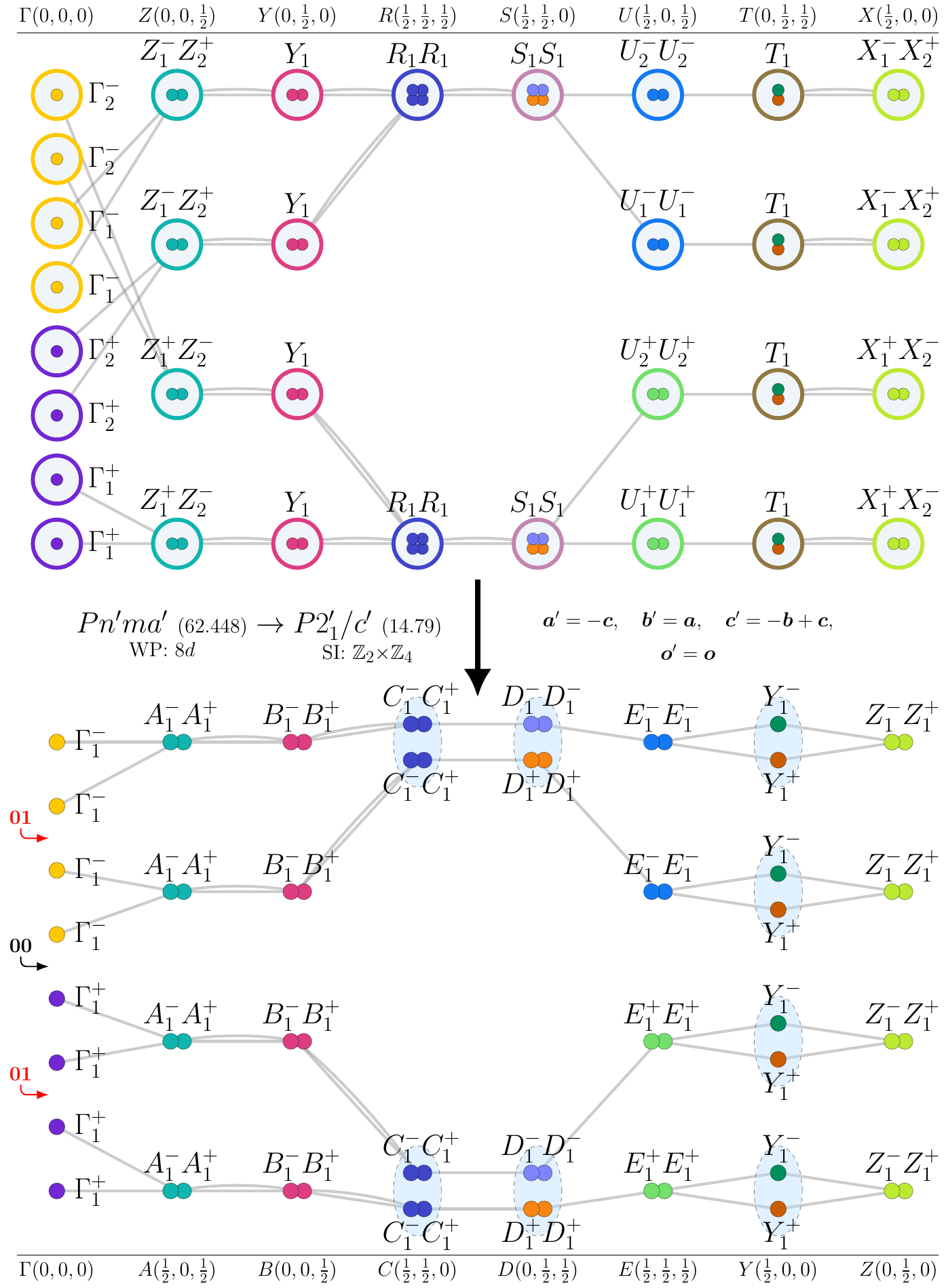}
\caption{Topological magnon bands in subgroup $P2_{1}'/c'~(14.79)$ for magnetic moments on Wyckoff position $8d$ of supergroup $Pn'ma'~(62.448)$.\label{fig_62.448_14.79_strainperp100_8d}}
\end{figure}
\input{gap_tables_tex/62.448_14.79_strainperp100_8d_table.tex}
\input{si_tables_tex/62.448_14.79_strainperp100_8d_table.tex}

\section{MSG $P_{a}nma~(62.450)$}
\textbf{Nontrivial-SI Subgroups:} $P\bar{1}~(2.4)$, $P2_{1}'/m'~(11.54)$, $P2_{1}'/c'~(14.79)$, $P2'/c'~(13.69)$, $P_{S}\bar{1}~(2.7)$, $P2_{1}/c~(14.75)$, $Pc'cn'~(56.370)$, $P_{c}2_{1}/c~(14.82)$, $P2_{1}/m~(11.50)$, $Pmm'n'~(59.410)$, $P_{a}2_{1}/m~(11.55)$, $P2_{1}/c~(14.75)$, $Pnm'a'~(62.447)$, $P_{b}2_{1}/c~(14.81)$.\\

\textbf{Trivial-SI Subgroups:} $Pm'~(6.20)$, $Pc'~(7.26)$, $Pc'~(7.26)$, $P2_{1}'~(4.9)$, $P2_{1}'~(4.9)$, $P2'~(3.3)$, $P_{S}1~(1.3)$, $Pc~(7.24)$, $Pn'a2_{1}'~(33.146)$, $Pc'c2'~(27.80)$, $P_{c}c~(7.28)$, $Pm~(6.18)$, $Pmn'2_{1}'~(31.126)$, $Pm'm2'~(25.59)$, $P_{a}m~(6.21)$, $Pc~(7.24)$, $Pna'2_{1}'~(33.147)$, $Pm'n2_{1}'~(31.125)$, $P_{b}c~(7.29)$, $P2_{1}~(4.7)$, $Pn'a'2_{1}~(33.148)$, $P_{a}2_{1}~(4.10)$, $P_{b}mn2_{1}~(31.129)$, $P2_{1}~(4.7)$, $Pm'n'2_{1}~(31.127)$, $P_{a}2_{1}~(4.10)$, $P_{a}na2_{1}~(33.149)$, $P2_{1}~(4.7)$, $Pm'c'2_{1}~(26.70)$, $P_{b}2_{1}~(4.11)$, $P_{c}mc2_{1}~(26.73)$.\\

\subsection{WP: $4a+4b$}
\textbf{BCS Materials:} {Mn\textsubscript{2}As~(573 K)}\footnote{BCS web page: \texttt{\href{http://webbdcrista1.ehu.es/magndata/index.php?this\_label=1.132} {http://webbdcrista1.ehu.es/magndata/index.php?this\_label=1.132}}}, {Cr\textsubscript{2}As~(393 K)}\footnote{BCS web page: \texttt{\href{http://webbdcrista1.ehu.es/magndata/index.php?this\_label=1.130} {http://webbdcrista1.ehu.es/magndata/index.php?this\_label=1.130}}}, {Fe\textsubscript{2}As~(353 K)}\footnote{BCS web page: \texttt{\href{http://webbdcrista1.ehu.es/magndata/index.php?this\_label=1.131} {http://webbdcrista1.ehu.es/magndata/index.php?this\_label=1.131}}}, {NdCoAsO~(14 K)}\footnote{BCS web page: \texttt{\href{http://webbdcrista1.ehu.es/magndata/index.php?this\_label=1.179} {http://webbdcrista1.ehu.es/magndata/index.php?this\_label=1.179}}}.\\
\subsubsection{Topological bands in subgroup $P2_{1}'/c'~(14.79)$}
\textbf{Perturbations:}
\begin{itemize}
\item B $\parallel$ [100] and strain $\perp$ [010],
\item B $\parallel$ [001] and strain $\perp$ [010],
\item B $\perp$ [010].
\end{itemize}
\begin{figure}[H]
\centering
\includegraphics[scale=0.6]{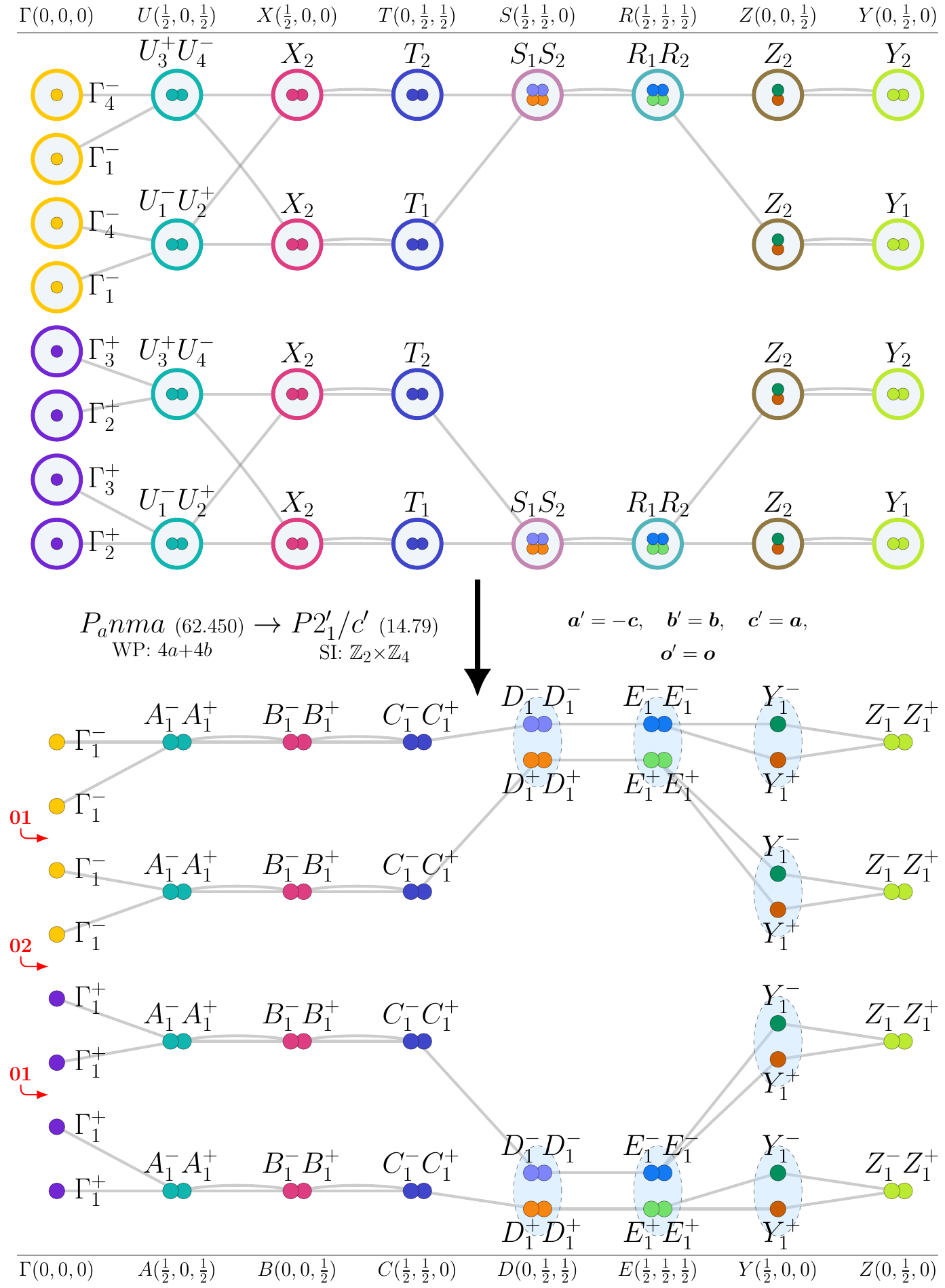}
\caption{Topological magnon bands in subgroup $P2_{1}'/c'~(14.79)$ for magnetic moments on Wyckoff positions $4a+4b$ of supergroup $P_{a}nma~(62.450)$.\label{fig_62.450_14.79_Bparallel100andstrainperp010_4a+4b}}
\end{figure}
\input{gap_tables_tex/62.450_14.79_Bparallel100andstrainperp010_4a+4b_table.tex}
\input{si_tables_tex/62.450_14.79_Bparallel100andstrainperp010_4a+4b_table.tex}
\subsection{WP: $4b$}
\textbf{BCS Materials:} {CrN~(273 K)}\footnote{BCS web page: \texttt{\href{http://webbdcrista1.ehu.es/magndata/index.php?this\_label=1.28} {http://webbdcrista1.ehu.es/magndata/index.php?this\_label=1.28}}}, {DyOCl~(10 K)}\footnote{BCS web page: \texttt{\href{http://webbdcrista1.ehu.es/magndata/index.php?this\_label=1.643} {http://webbdcrista1.ehu.es/magndata/index.php?this\_label=1.643}}}, {ErFe\textsubscript{2}Si\textsubscript{2}~(2.6 K)}\footnote{BCS web page: \texttt{\href{http://webbdcrista1.ehu.es/magndata/index.php?this\_label=1.635} {http://webbdcrista1.ehu.es/magndata/index.php?this\_label=1.635}}}.\\
\subsubsection{Topological bands in subgroup $P2_{1}'/c'~(14.79)$}
\textbf{Perturbations:}
\begin{itemize}
\item B $\parallel$ [100] and strain $\perp$ [010],
\item B $\parallel$ [001] and strain $\perp$ [010],
\item B $\perp$ [010].
\end{itemize}
\begin{figure}[H]
\centering
\includegraphics[scale=0.6]{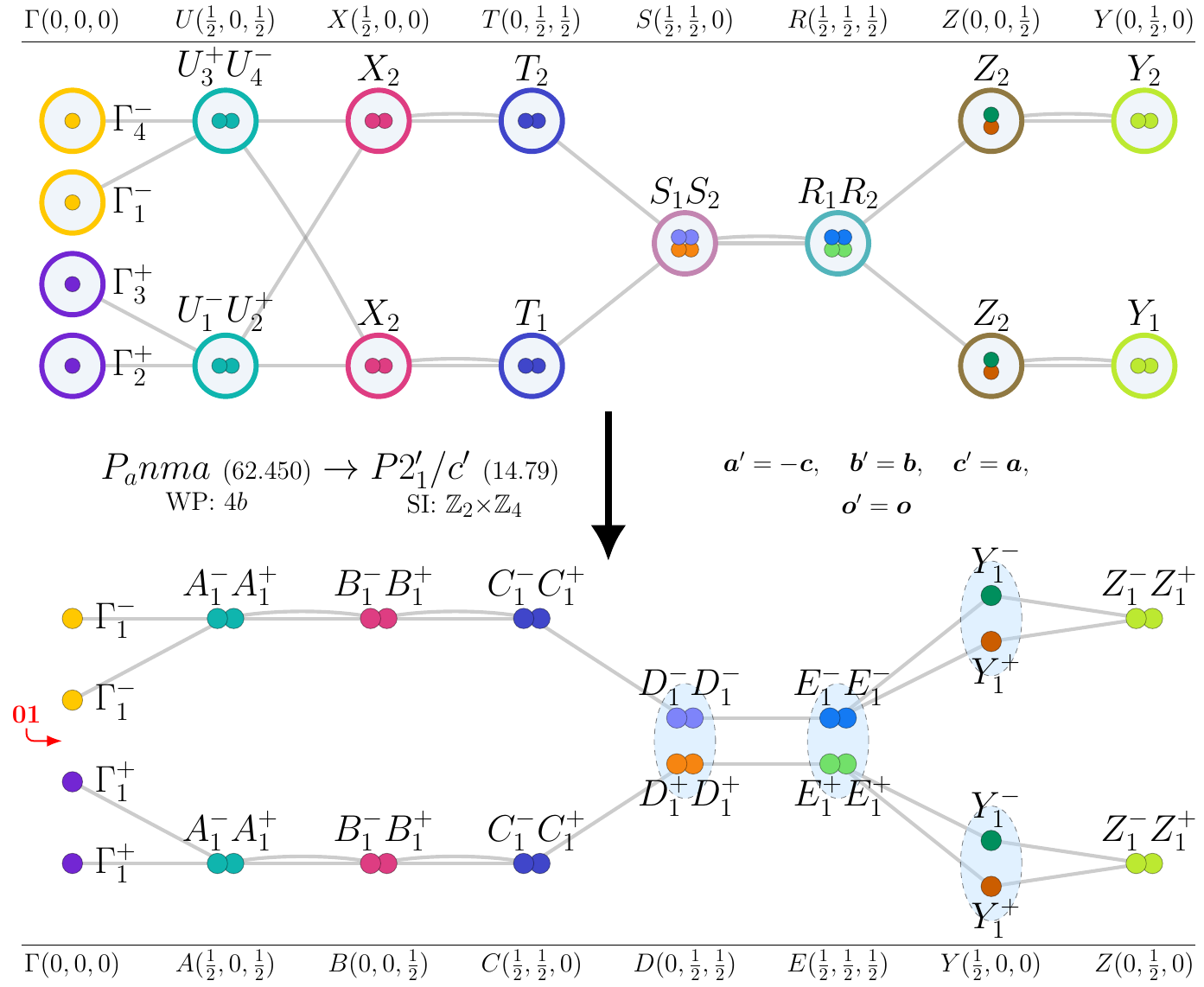}
\caption{Topological magnon bands in subgroup $P2_{1}'/c'~(14.79)$ for magnetic moments on Wyckoff position $4b$ of supergroup $P_{a}nma~(62.450)$.\label{fig_62.450_14.79_Bparallel100andstrainperp010_4b}}
\end{figure}
\input{gap_tables_tex/62.450_14.79_Bparallel100andstrainperp010_4b_table.tex}
\input{si_tables_tex/62.450_14.79_Bparallel100andstrainperp010_4b_table.tex}
\subsection{WP: $4a$}
\textbf{BCS Materials:} {NdPd\textsubscript{5}Al\textsubscript{2}~(1.3 K)}\footnote{BCS web page: \texttt{\href{http://webbdcrista1.ehu.es/magndata/index.php?this\_label=1.507} {http://webbdcrista1.ehu.es/magndata/index.php?this\_label=1.507}}}.\\
\subsubsection{Topological bands in subgroup $P2_{1}'/c'~(14.79)$}
\textbf{Perturbations:}
\begin{itemize}
\item B $\parallel$ [100] and strain $\perp$ [010],
\item B $\parallel$ [001] and strain $\perp$ [010],
\item B $\perp$ [010].
\end{itemize}
\begin{figure}[H]
\centering
\includegraphics[scale=0.6]{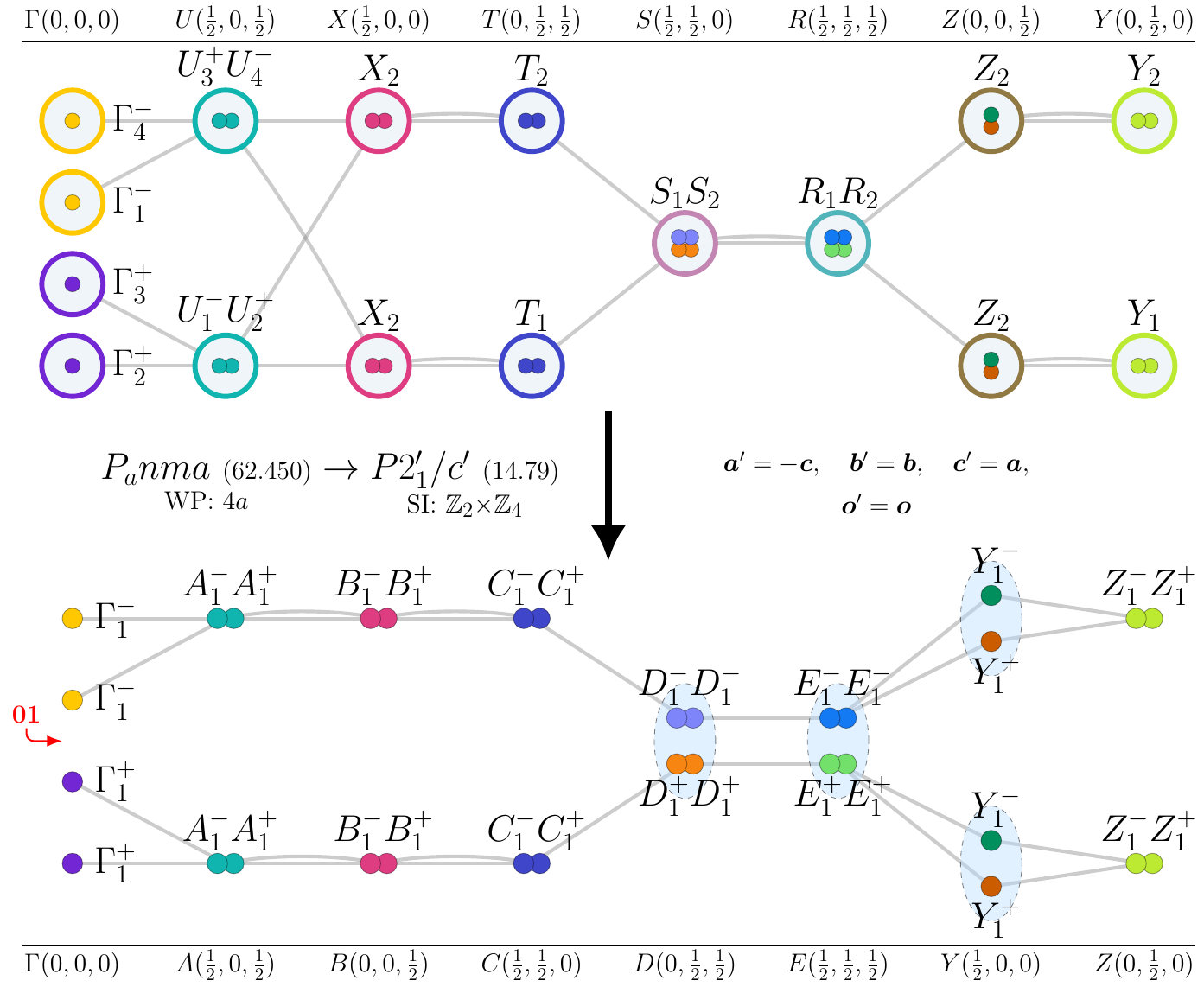}
\caption{Topological magnon bands in subgroup $P2_{1}'/c'~(14.79)$ for magnetic moments on Wyckoff position $4a$ of supergroup $P_{a}nma~(62.450)$.\label{fig_62.450_14.79_Bparallel100andstrainperp010_4a}}
\end{figure}
\input{gap_tables_tex/62.450_14.79_Bparallel100andstrainperp010_4a_table.tex}
\input{si_tables_tex/62.450_14.79_Bparallel100andstrainperp010_4a_table.tex}

\section{MSG $P_{b}nma~(62.451)$}
\textbf{Nontrivial-SI Subgroups:} $P\bar{1}~(2.4)$, $P2_{1}'/c'~(14.79)$, $P2'/m'~(10.46)$, $P2_{1}'/c'~(14.79)$, $P_{S}\bar{1}~(2.7)$, $P2_{1}/c~(14.75)$, $Pb'am'~(55.358)$, $P_{a}2_{1}/c~(14.80)$, $P2_{1}/m~(11.50)$, $Pn'ma'~(62.448)$, $P_{b}2_{1}/m~(11.56)$, $P2_{1}/c~(14.75)$, $Pnn'm'~(58.398)$, $P_{a}2_{1}/c~(14.80)$.\\

\textbf{Trivial-SI Subgroups:} $Pc'~(7.26)$, $Pm'~(6.20)$, $Pc'~(7.26)$, $P2_{1}'~(4.9)$, $P2'~(3.3)$, $P2_{1}'~(4.9)$, $P_{S}1~(1.3)$, $Pc~(7.24)$, $Pb'a2'~(32.137)$, $Pm'c2_{1}'~(26.68)$, $P_{a}c~(7.27)$, $Pm~(6.18)$, $Pmc'2_{1}'~(26.69)$, $Pmn'2_{1}'~(31.126)$, $P_{b}m~(6.22)$, $Pc~(7.24)$, $Pm'n2_{1}'~(31.125)$, $Pn'n2'~(34.158)$, $P_{a}c~(7.27)$, $P2_{1}~(4.7)$, $Pm'c'2_{1}~(26.70)$, $P_{a}2_{1}~(4.10)$, $P_{a}mn2_{1}~(31.128)$, $P2_{1}~(4.7)$, $Pn'a'2_{1}~(33.148)$, $P_{b}2_{1}~(4.11)$, $P_{c}na2_{1}~(33.151)$, $P2_{1}~(4.7)$, $Pm'n'2_{1}~(31.127)$, $P_{a}2_{1}~(4.10)$, $P_{a}mc2_{1}~(26.71)$.\\

\subsection{WP: $4d+8h$}
\textbf{BCS Materials:} {Fe\textsubscript{2}MnBO\textsubscript{5}~(90 K)}\footnote{BCS web page: \texttt{\href{http://webbdcrista1.ehu.es/magndata/index.php?this\_label=1.391} {http://webbdcrista1.ehu.es/magndata/index.php?this\_label=1.391}}}.\\
\subsubsection{Topological bands in subgroup $P2_{1}'/c'~(14.79)$}
\textbf{Perturbations:}
\begin{itemize}
\item B $\parallel$ [100] and strain $\perp$ [001],
\item B $\parallel$ [010] and strain $\perp$ [001],
\item B $\perp$ [001].
\end{itemize}
\begin{figure}[H]
\centering
\includegraphics[scale=0.6]{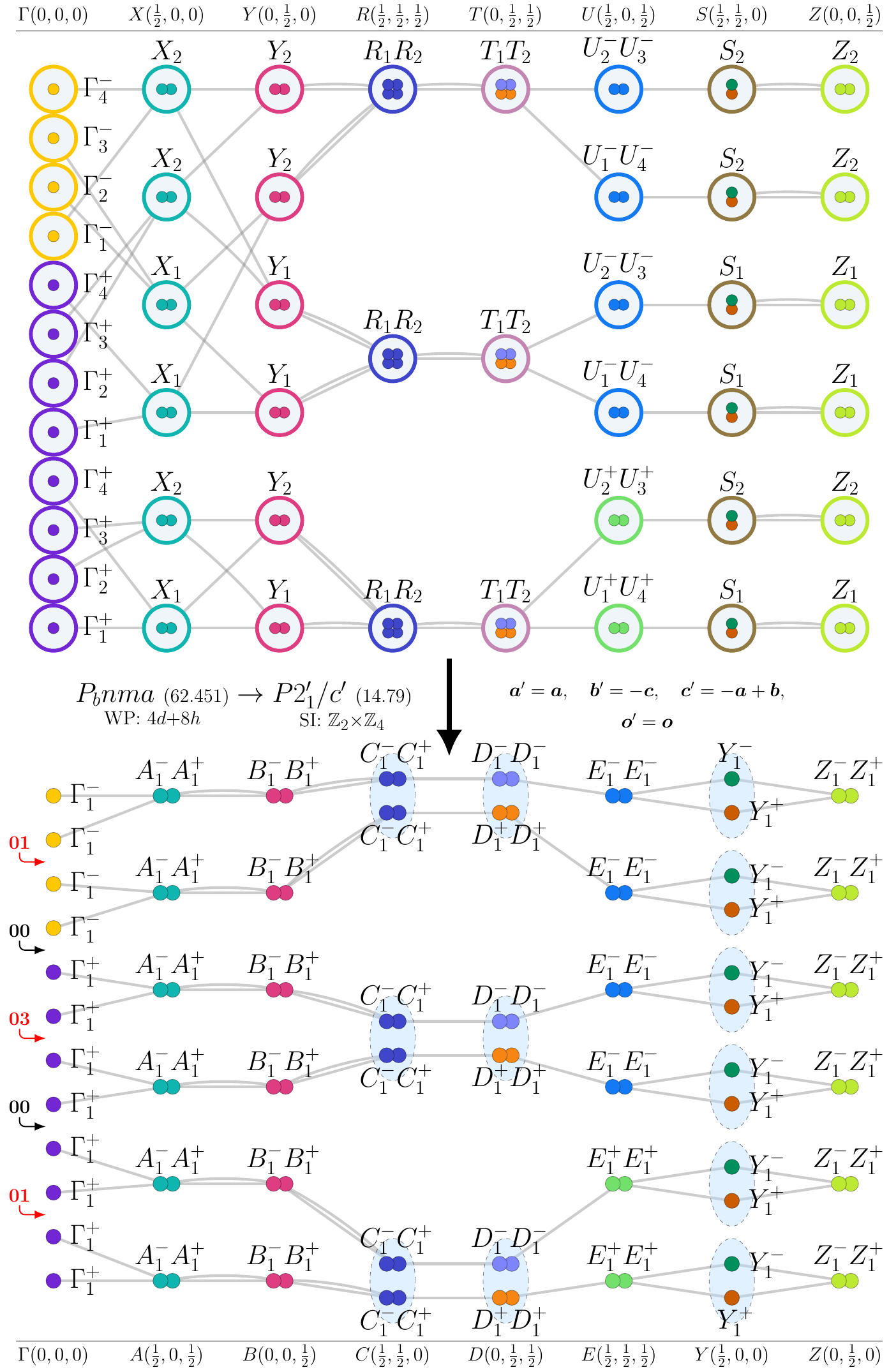}
\caption{Topological magnon bands in subgroup $P2_{1}'/c'~(14.79)$ for magnetic moments on Wyckoff positions $4d+8h$ of supergroup $P_{b}nma~(62.451)$.\label{fig_62.451_14.79_Bparallel100andstrainperp001_4d+8h}}
\end{figure}
\input{gap_tables_tex/62.451_14.79_Bparallel100andstrainperp001_4d+8h_table.tex}
\input{si_tables_tex/62.451_14.79_Bparallel100andstrainperp001_4d+8h_table.tex}
\subsubsection{Topological bands in subgroup $P2_{1}'/c'~(14.79)$}
\textbf{Perturbations:}
\begin{itemize}
\item B $\parallel$ [010] and strain $\perp$ [100],
\item B $\parallel$ [001] and strain $\perp$ [100],
\item B $\perp$ [100].
\end{itemize}
\begin{figure}[H]
\centering
\includegraphics[scale=0.6]{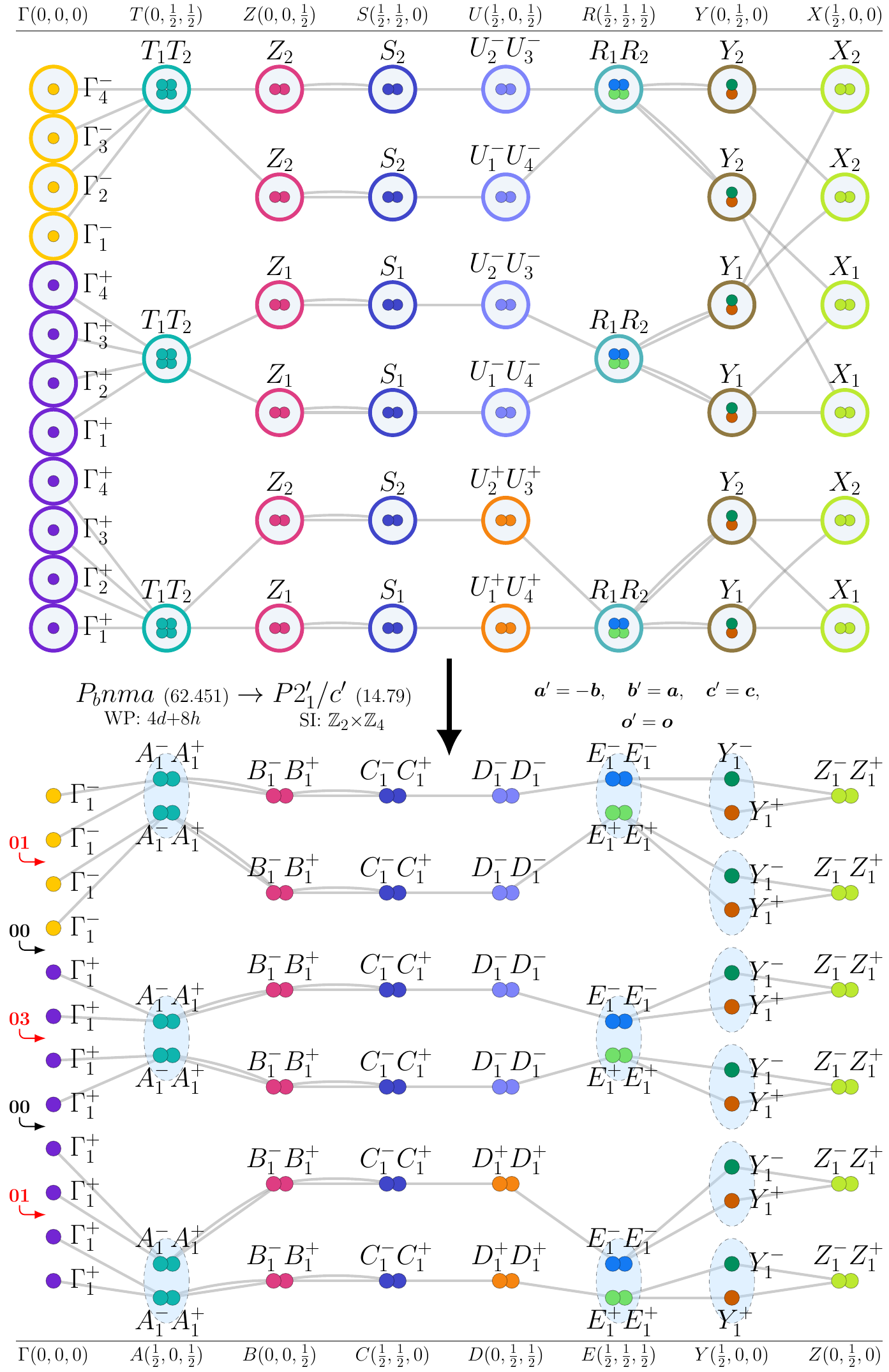}
\caption{Topological magnon bands in subgroup $P2_{1}'/c'~(14.79)$ for magnetic moments on Wyckoff positions $4d+8h$ of supergroup $P_{b}nma~(62.451)$.\label{fig_62.451_14.79_Bparallel010andstrainperp100_4d+8h}}
\end{figure}
\input{gap_tables_tex/62.451_14.79_Bparallel010andstrainperp100_4d+8h_table.tex}
\input{si_tables_tex/62.451_14.79_Bparallel010andstrainperp100_4d+8h_table.tex}
\subsubsection{Topological bands in subgroup $P_{S}\bar{1}~(2.7)$}
\textbf{Perturbation:}
\begin{itemize}
\item strain in generic direction.
\end{itemize}
\begin{figure}[H]
\centering
\includegraphics[scale=0.6]{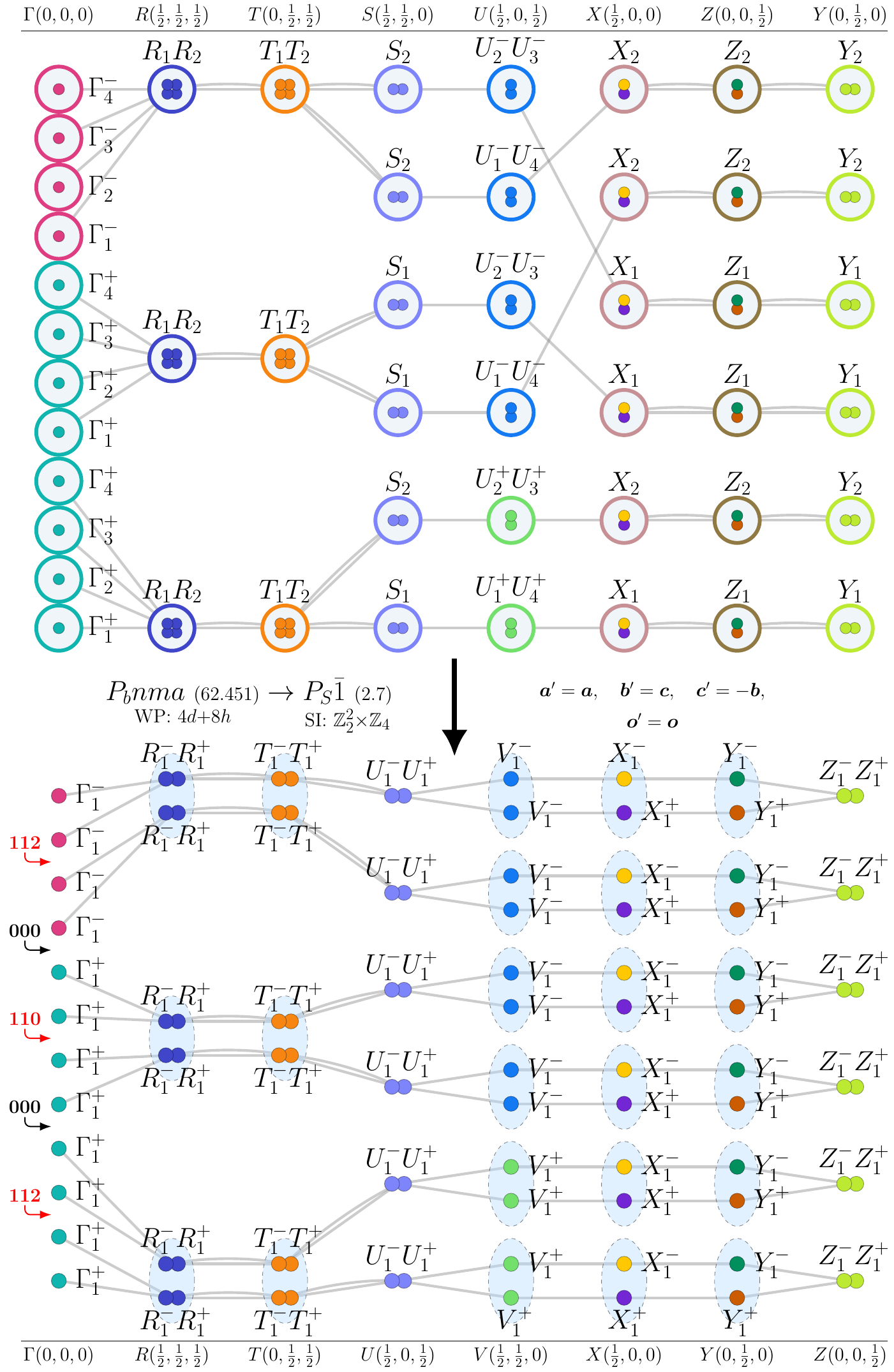}
\caption{Topological magnon bands in subgroup $P_{S}\bar{1}~(2.7)$ for magnetic moments on Wyckoff positions $4d+8h$ of supergroup $P_{b}nma~(62.451)$.\label{fig_62.451_2.7_strainingenericdirection_4d+8h}}
\end{figure}
\input{gap_tables_tex/62.451_2.7_strainingenericdirection_4d+8h_table.tex}
\input{si_tables_tex/62.451_2.7_strainingenericdirection_4d+8h_table.tex}
\subsection{WP: $8g$}
\textbf{BCS Materials:} {Sr\textsubscript{2}Fe\textsubscript{3}S\textsubscript{2}O\textsubscript{3}~(76 K)}\footnote{BCS web page: \texttt{\href{http://webbdcrista1.ehu.es/magndata/index.php?this\_label=1.625} {http://webbdcrista1.ehu.es/magndata/index.php?this\_label=1.625}}}, {Pr\textsubscript{2}Pd\textsubscript{2}In~(5 K)}\footnote{BCS web page: \texttt{\href{http://webbdcrista1.ehu.es/magndata/index.php?this\_label=1.334} {http://webbdcrista1.ehu.es/magndata/index.php?this\_label=1.334}}}.\\
\subsubsection{Topological bands in subgroup $P2_{1}'/c'~(14.79)$}
\textbf{Perturbations:}
\begin{itemize}
\item B $\parallel$ [100] and strain $\perp$ [001],
\item B $\parallel$ [010] and strain $\perp$ [001],
\item B $\perp$ [001].
\end{itemize}
\begin{figure}[H]
\centering
\includegraphics[scale=0.6]{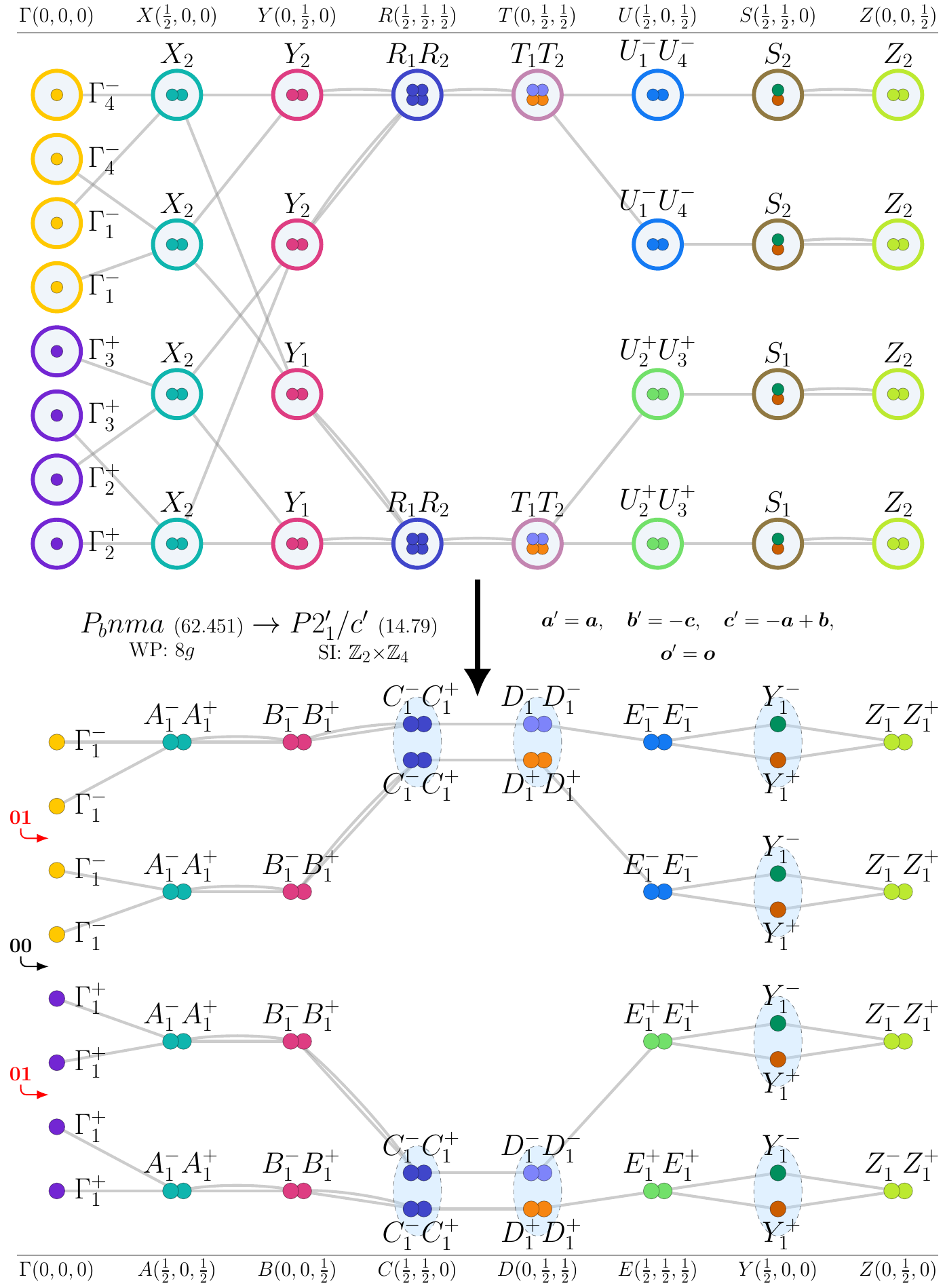}
\caption{Topological magnon bands in subgroup $P2_{1}'/c'~(14.79)$ for magnetic moments on Wyckoff position $8g$ of supergroup $P_{b}nma~(62.451)$.\label{fig_62.451_14.79_Bparallel100andstrainperp001_8g}}
\end{figure}
\input{gap_tables_tex/62.451_14.79_Bparallel100andstrainperp001_8g_table.tex}
\input{si_tables_tex/62.451_14.79_Bparallel100andstrainperp001_8g_table.tex}
\subsubsection{Topological bands in subgroup $P2_{1}'/c'~(14.79)$}
\textbf{Perturbations:}
\begin{itemize}
\item B $\parallel$ [010] and strain $\perp$ [100],
\item B $\parallel$ [001] and strain $\perp$ [100],
\item B $\perp$ [100].
\end{itemize}
\begin{figure}[H]
\centering
\includegraphics[scale=0.6]{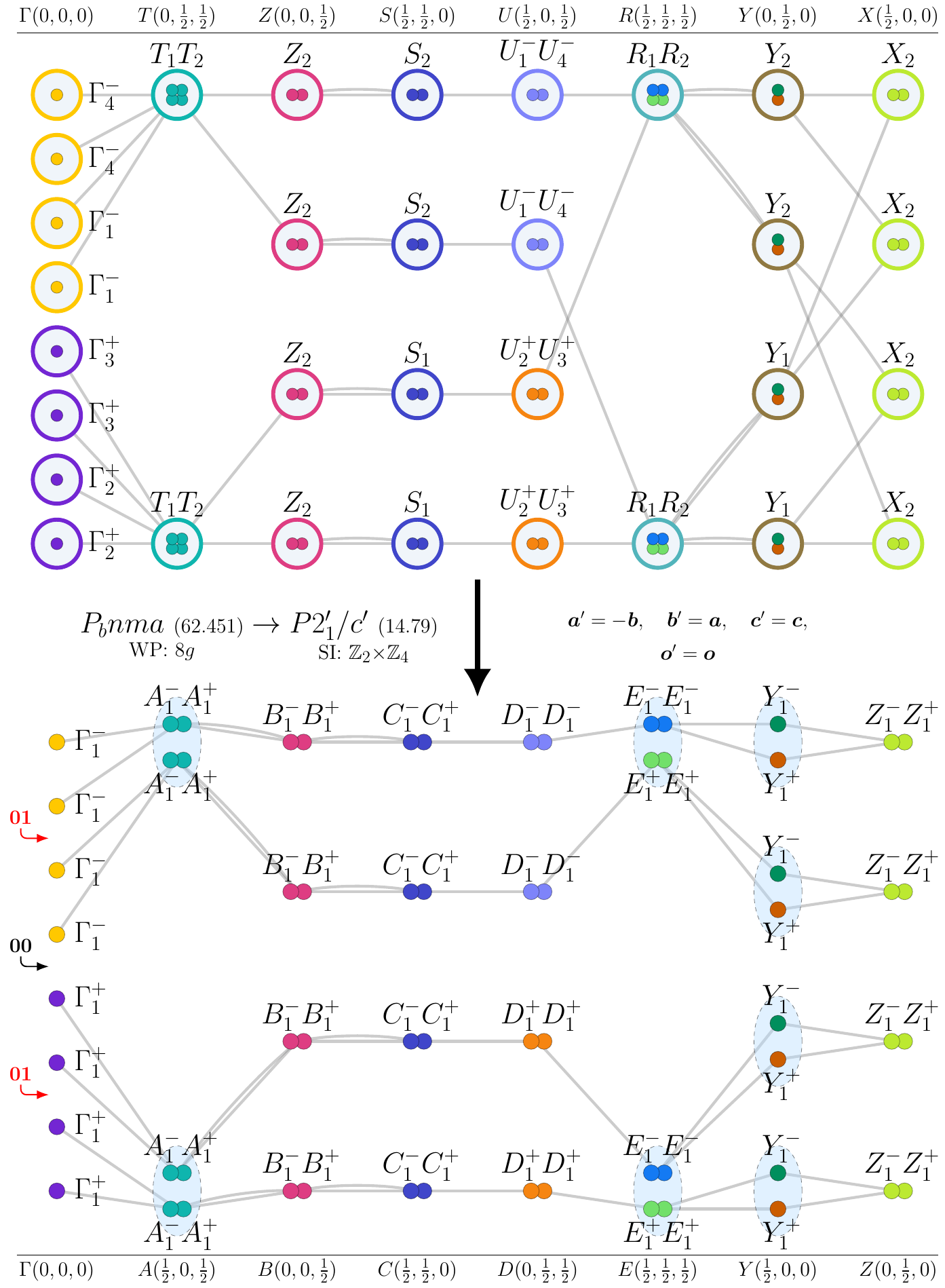}
\caption{Topological magnon bands in subgroup $P2_{1}'/c'~(14.79)$ for magnetic moments on Wyckoff position $8g$ of supergroup $P_{b}nma~(62.451)$.\label{fig_62.451_14.79_Bparallel010andstrainperp100_8g}}
\end{figure}
\input{gap_tables_tex/62.451_14.79_Bparallel010andstrainperp100_8g_table.tex}
\input{si_tables_tex/62.451_14.79_Bparallel010andstrainperp100_8g_table.tex}
\subsubsection{Topological bands in subgroup $P_{S}\bar{1}~(2.7)$}
\textbf{Perturbation:}
\begin{itemize}
\item strain in generic direction.
\end{itemize}
\begin{figure}[H]
\centering
\includegraphics[scale=0.6]{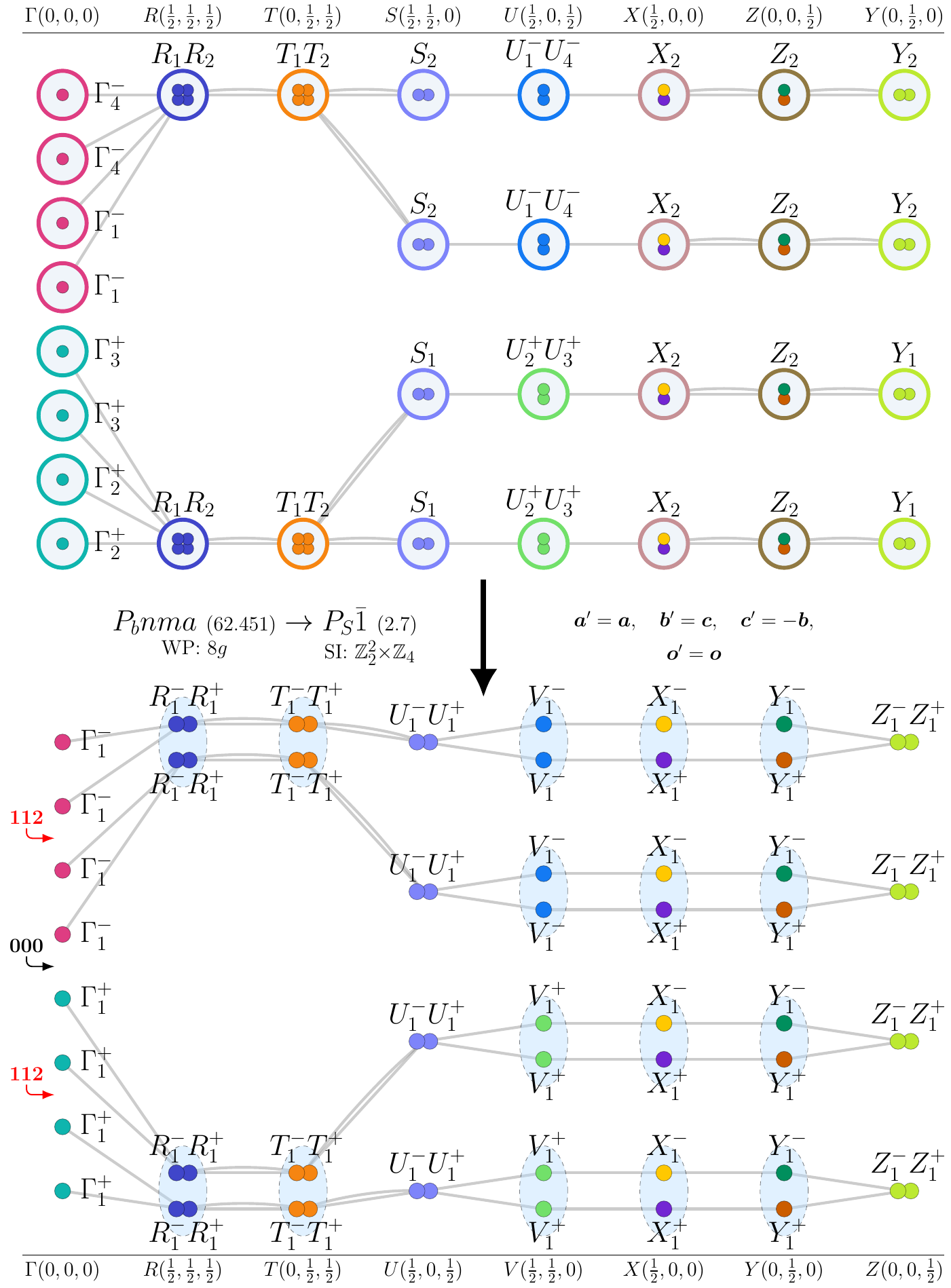}
\caption{Topological magnon bands in subgroup $P_{S}\bar{1}~(2.7)$ for magnetic moments on Wyckoff position $8g$ of supergroup $P_{b}nma~(62.451)$.\label{fig_62.451_2.7_strainingenericdirection_8g}}
\end{figure}
\input{gap_tables_tex/62.451_2.7_strainingenericdirection_8g_table.tex}
\input{si_tables_tex/62.451_2.7_strainingenericdirection_8g_table.tex}
\subsection{WP: $8h$}
\textbf{BCS Materials:} {Tb\textsubscript{2}Pd\textsubscript{2.05}Sn\textsubscript{0.95}~(20.8 K)}\footnote{BCS web page: \texttt{\href{http://webbdcrista1.ehu.es/magndata/index.php?this\_label=1.336} {http://webbdcrista1.ehu.es/magndata/index.php?this\_label=1.336}}}.\\
\subsubsection{Topological bands in subgroup $P2_{1}'/c'~(14.79)$}
\textbf{Perturbations:}
\begin{itemize}
\item B $\parallel$ [100] and strain $\perp$ [001],
\item B $\parallel$ [010] and strain $\perp$ [001],
\item B $\perp$ [001].
\end{itemize}
\begin{figure}[H]
\centering
\includegraphics[scale=0.6]{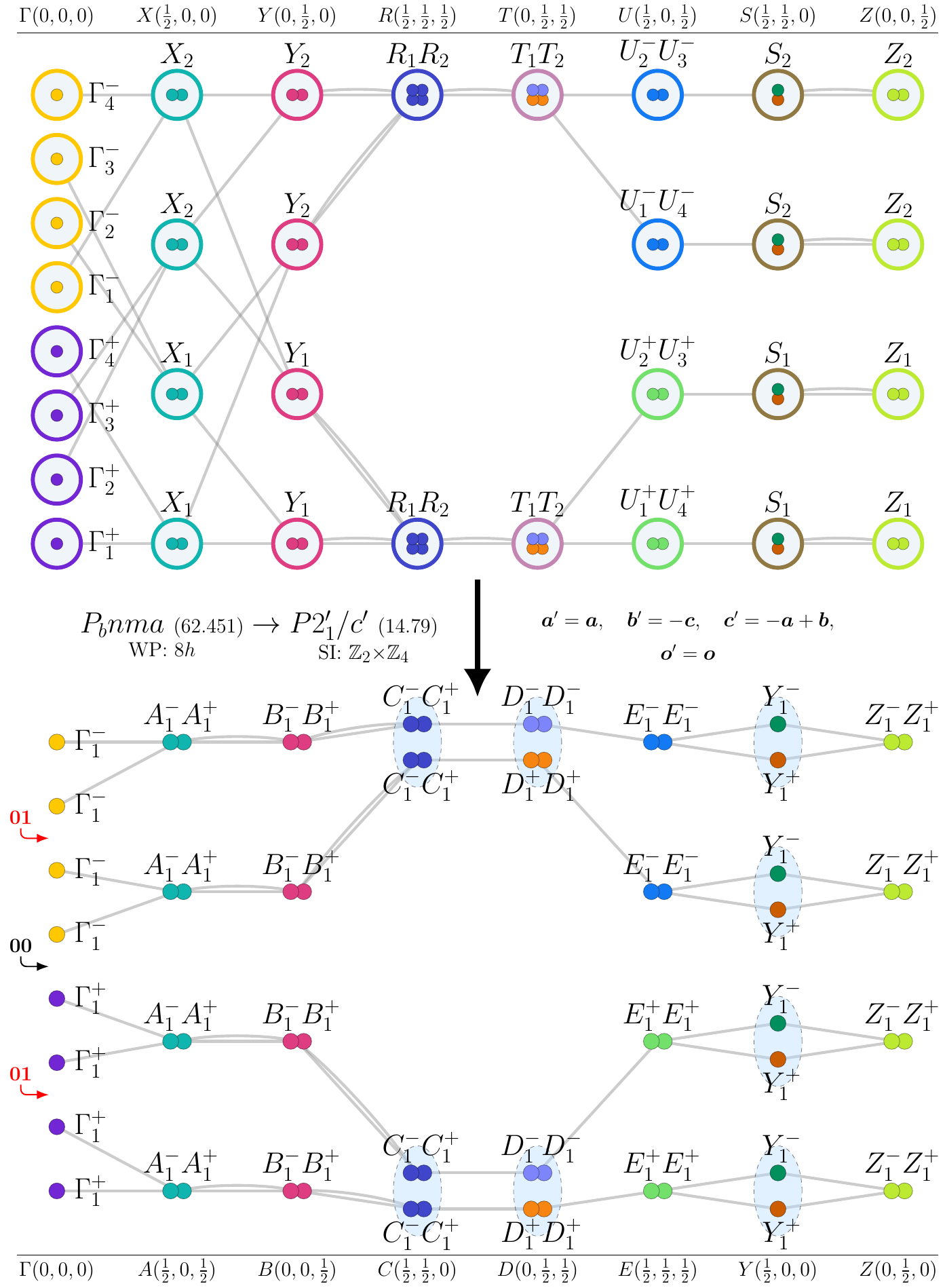}
\caption{Topological magnon bands in subgroup $P2_{1}'/c'~(14.79)$ for magnetic moments on Wyckoff position $8h$ of supergroup $P_{b}nma~(62.451)$.\label{fig_62.451_14.79_Bparallel100andstrainperp001_8h}}
\end{figure}
\input{gap_tables_tex/62.451_14.79_Bparallel100andstrainperp001_8h_table.tex}
\input{si_tables_tex/62.451_14.79_Bparallel100andstrainperp001_8h_table.tex}
\subsubsection{Topological bands in subgroup $P2_{1}'/c'~(14.79)$}
\textbf{Perturbations:}
\begin{itemize}
\item B $\parallel$ [010] and strain $\perp$ [100],
\item B $\parallel$ [001] and strain $\perp$ [100],
\item B $\perp$ [100].
\end{itemize}
\begin{figure}[H]
\centering
\includegraphics[scale=0.6]{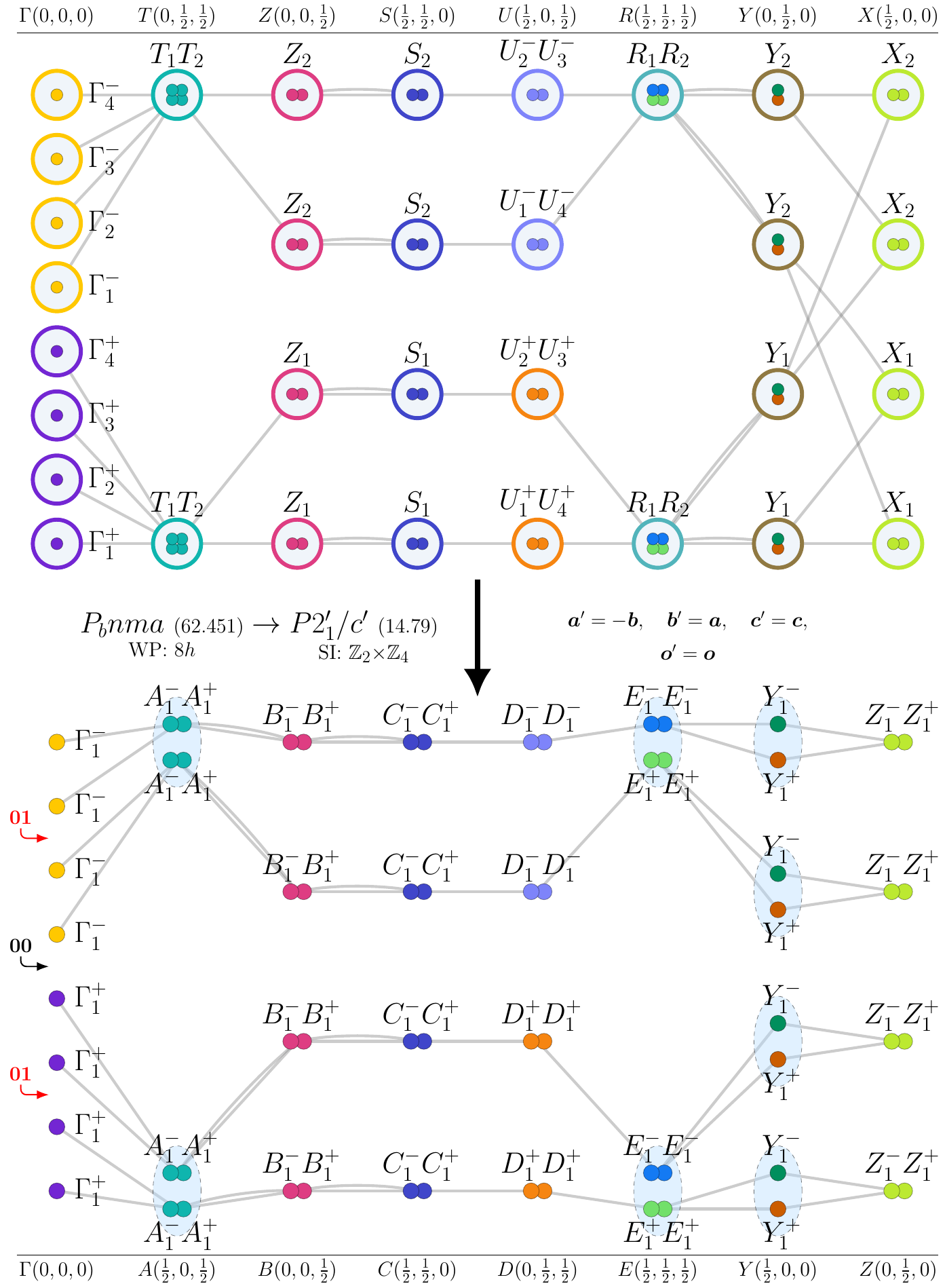}
\caption{Topological magnon bands in subgroup $P2_{1}'/c'~(14.79)$ for magnetic moments on Wyckoff position $8h$ of supergroup $P_{b}nma~(62.451)$.\label{fig_62.451_14.79_Bparallel010andstrainperp100_8h}}
\end{figure}
\input{gap_tables_tex/62.451_14.79_Bparallel010andstrainperp100_8h_table.tex}
\input{si_tables_tex/62.451_14.79_Bparallel010andstrainperp100_8h_table.tex}
\subsubsection{Topological bands in subgroup $P_{S}\bar{1}~(2.7)$}
\textbf{Perturbation:}
\begin{itemize}
\item strain in generic direction.
\end{itemize}
\begin{figure}[H]
\centering
\includegraphics[scale=0.6]{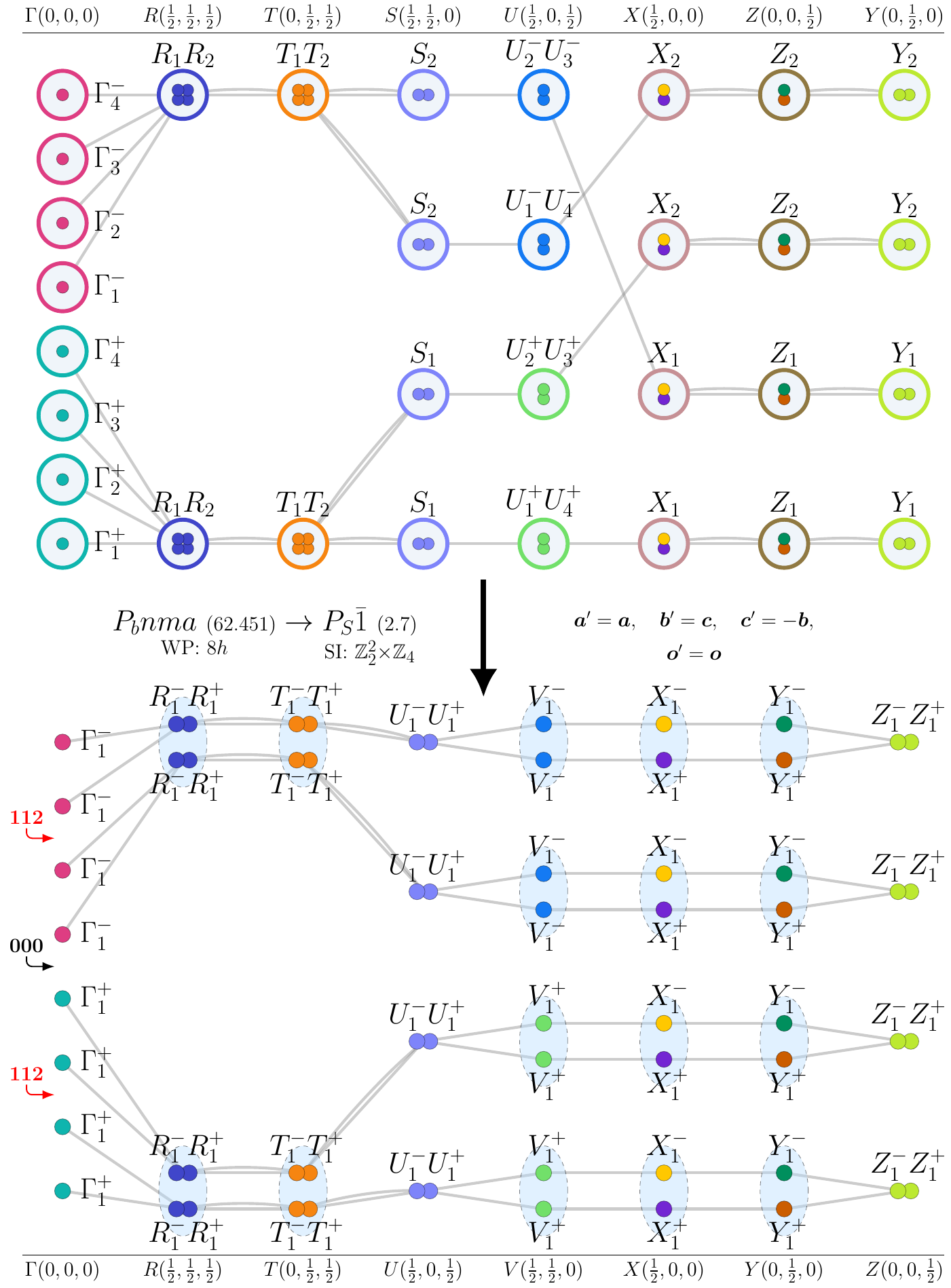}
\caption{Topological magnon bands in subgroup $P_{S}\bar{1}~(2.7)$ for magnetic moments on Wyckoff position $8h$ of supergroup $P_{b}nma~(62.451)$.\label{fig_62.451_2.7_strainingenericdirection_8h}}
\end{figure}
\input{gap_tables_tex/62.451_2.7_strainingenericdirection_8h_table.tex}
\input{si_tables_tex/62.451_2.7_strainingenericdirection_8h_table.tex}

\section{MSG $P_{c}nma~(62.452)$}
\textbf{Nontrivial-SI Subgroups:} $P\bar{1}~(2.4)$, $P2'/c'~(13.69)$, $P2_{1}'/c'~(14.79)$, $P2_{1}'/c'~(14.79)$, $P_{S}\bar{1}~(2.7)$, $P2_{1}/c~(14.75)$, $Pb'c'a~(61.436)$, $P_{b}2_{1}/c~(14.81)$, $P2_{1}/m~(11.50)$, $Pb'c'm~(57.382)$, $P_{a}2_{1}/m~(11.55)$, $P2_{1}/c~(14.75)$, $Pb'c'n~(60.422)$, $P_{a}2_{1}/c~(14.80)$.\\

\textbf{Trivial-SI Subgroups:} $Pc'~(7.26)$, $Pc'~(7.26)$, $Pc'~(7.26)$, $P2'~(3.3)$, $P2_{1}'~(4.9)$, $P2_{1}'~(4.9)$, $P_{S}1~(1.3)$, $Pc~(7.24)$, $Pc'a2_{1}'~(29.101)$, $Pca'2_{1}'~(29.102)$, $P_{b}c~(7.29)$, $Pm~(6.18)$, $Pma'2'~(28.90)$, $Pmc'2_{1}'~(26.69)$, $P_{a}m~(6.21)$, $Pc~(7.24)$, $Pnc'2'~(30.114)$, $Pna'2_{1}'~(33.147)$, $P_{a}c~(7.27)$, $P2_{1}~(4.7)$, $Pc'a'2_{1}~(29.103)$, $P_{b}2_{1}~(4.11)$, $P_{c}mn2_{1}~(31.130)$, $P2_{1}~(4.7)$, $Pc'a'2_{1}~(29.103)$, $P_{a}2_{1}~(4.10)$, $P_{b}na2_{1}~(33.150)$, $P2_{1}~(4.7)$, $Pc'a'2_{1}~(29.103)$, $P_{a}2_{1}~(4.10)$, $P_{b}mc2_{1}~(26.72)$.\\

\subsection{WP: $16e$}
\textbf{BCS Materials:} {Mn\textsubscript{3}O\textsubscript{4}~(210 K)}\footnote{BCS web page: \texttt{\href{http://webbdcrista1.ehu.es/magndata/index.php?this\_label=1.1} {http://webbdcrista1.ehu.es/magndata/index.php?this\_label=1.1}}}.\\
\subsubsection{Topological bands in subgroup $P2_{1}'/c'~(14.79)$}
\textbf{Perturbations:}
\begin{itemize}
\item B $\parallel$ [100] and strain $\perp$ [010],
\item B $\parallel$ [001] and strain $\perp$ [010],
\item B $\perp$ [010].
\end{itemize}
\begin{figure}[H]
\centering
\includegraphics[scale=0.6]{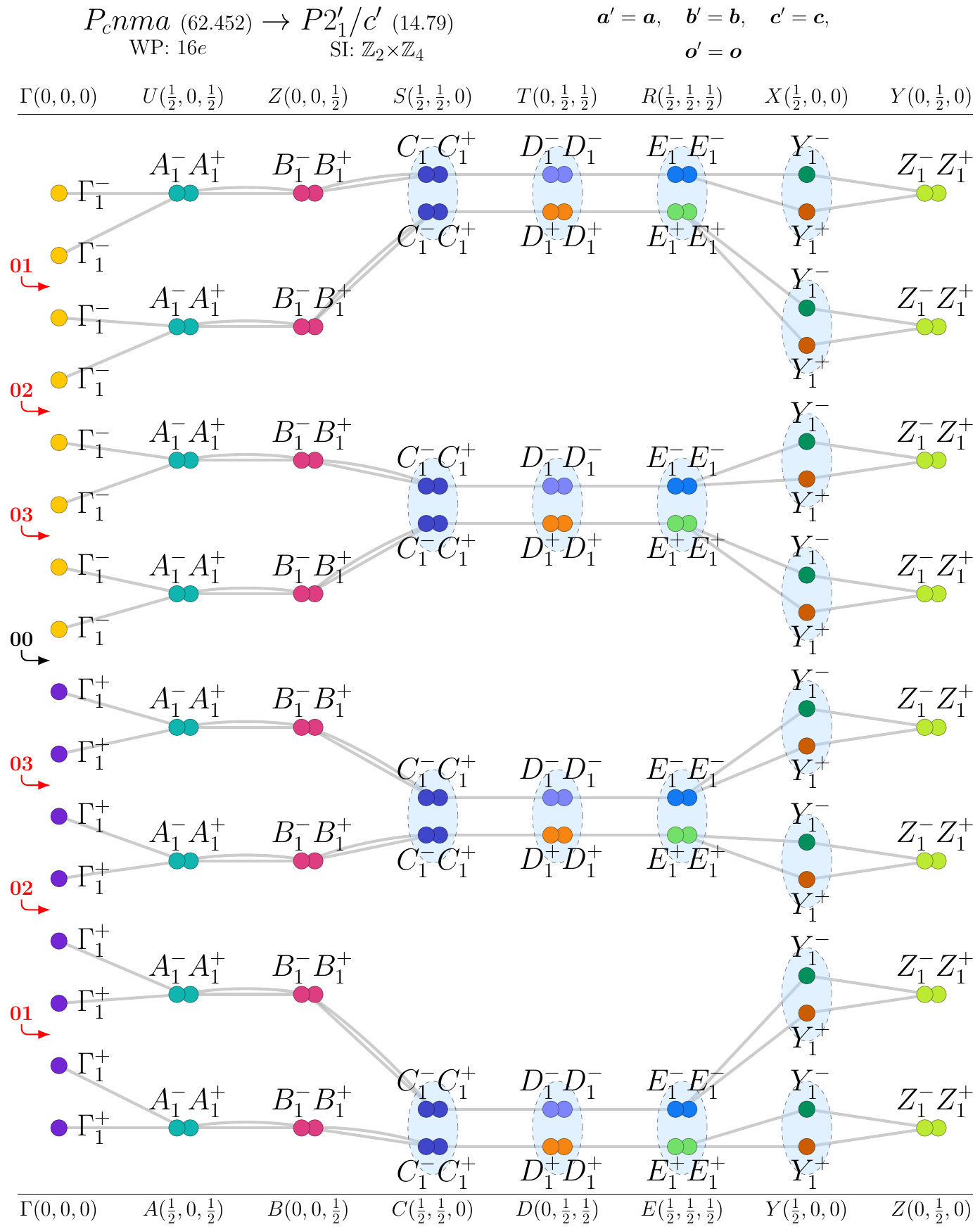}
\caption{Topological magnon bands in subgroup $P2_{1}'/c'~(14.79)$ for magnetic moments on Wyckoff position $16e$ of supergroup $P_{c}nma~(62.452)$.\label{fig_62.452_14.79_Bparallel100andstrainperp010_16e}}
\end{figure}
\input{gap_tables_tex/62.452_14.79_Bparallel100andstrainperp010_16e_table.tex}
\input{si_tables_tex/62.452_14.79_Bparallel100andstrainperp010_16e_table.tex}
\subsubsection{Topological bands in subgroup $P2_{1}'/c'~(14.79)$}
\textbf{Perturbations:}
\begin{itemize}
\item B $\parallel$ [010] and strain $\perp$ [100],
\item B $\parallel$ [001] and strain $\perp$ [100],
\item B $\perp$ [100].
\end{itemize}
\begin{figure}[H]
\centering
\includegraphics[scale=0.6]{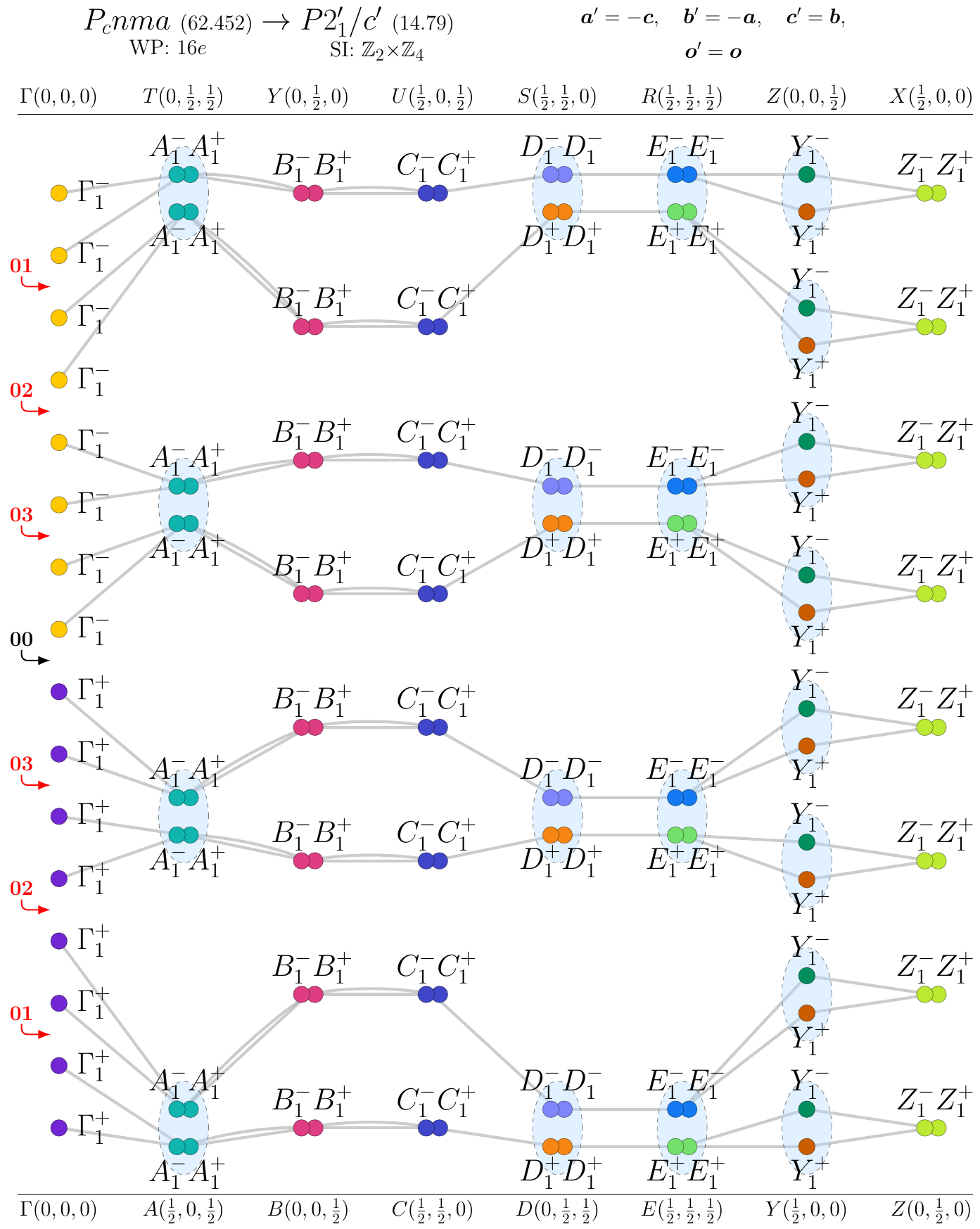}
\caption{Topological magnon bands in subgroup $P2_{1}'/c'~(14.79)$ for magnetic moments on Wyckoff position $16e$ of supergroup $P_{c}nma~(62.452)$.\label{fig_62.452_14.79_Bparallel010andstrainperp100_16e}}
\end{figure}
\input{gap_tables_tex/62.452_14.79_Bparallel010andstrainperp100_16e_table.tex}
\input{si_tables_tex/62.452_14.79_Bparallel010andstrainperp100_16e_table.tex}
\subsection{WP: $8c$}
\textbf{BCS Materials:} {VPO\textsubscript{4}~(10.3 K)}\footnote{BCS web page: \texttt{\href{http://webbdcrista1.ehu.es/magndata/index.php?this\_label=1.523} {http://webbdcrista1.ehu.es/magndata/index.php?this\_label=1.523}}}.\\
\subsubsection{Topological bands in subgroup $P2_{1}'/c'~(14.79)$}
\textbf{Perturbations:}
\begin{itemize}
\item B $\parallel$ [100] and strain $\perp$ [010],
\item B $\parallel$ [001] and strain $\perp$ [010],
\item B $\perp$ [010].
\end{itemize}
\begin{figure}[H]
\centering
\includegraphics[scale=0.6]{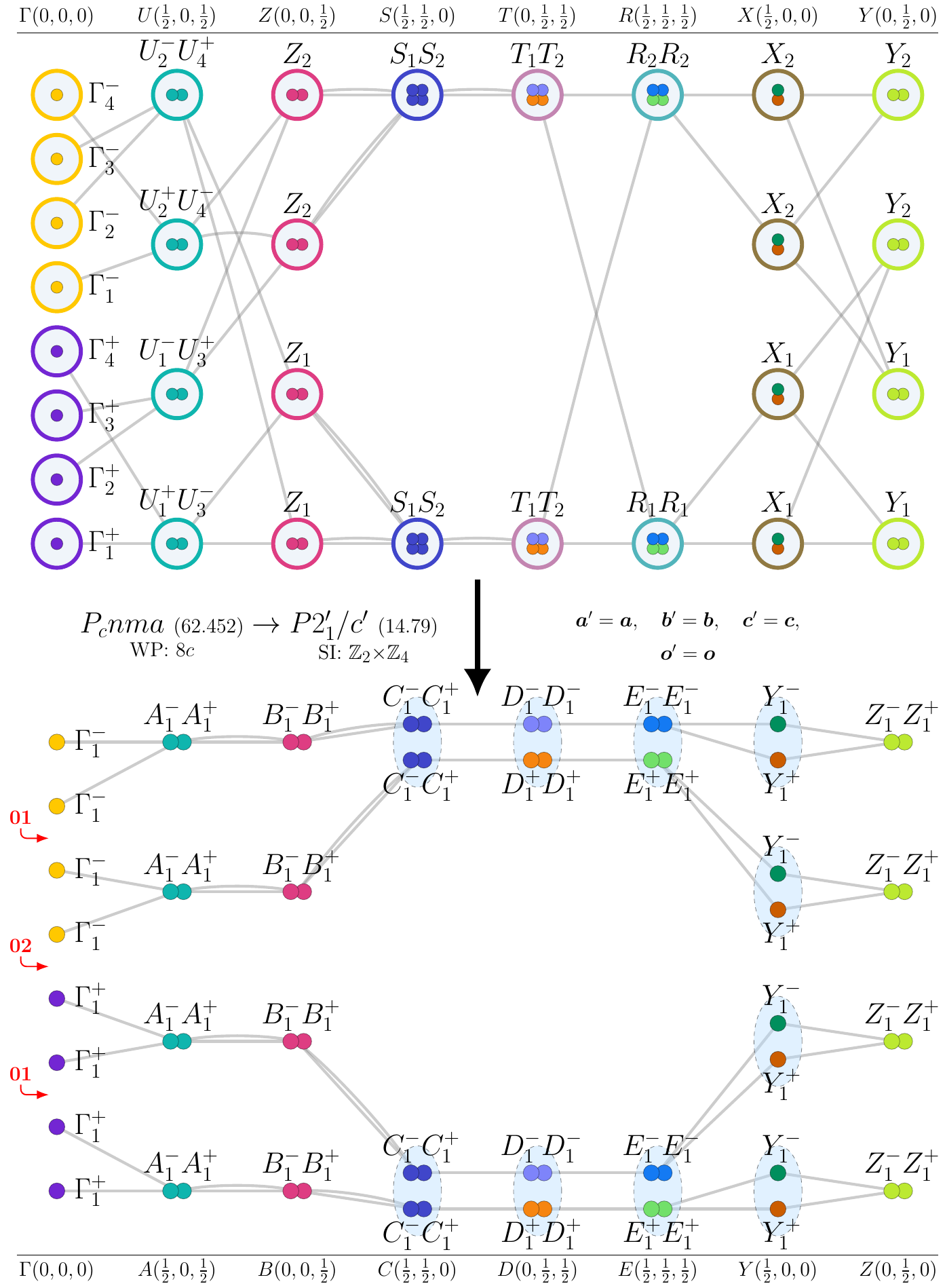}
\caption{Topological magnon bands in subgroup $P2_{1}'/c'~(14.79)$ for magnetic moments on Wyckoff position $8c$ of supergroup $P_{c}nma~(62.452)$.\label{fig_62.452_14.79_Bparallel100andstrainperp010_8c}}
\end{figure}
\input{gap_tables_tex/62.452_14.79_Bparallel100andstrainperp010_8c_table.tex}
\input{si_tables_tex/62.452_14.79_Bparallel100andstrainperp010_8c_table.tex}
\subsubsection{Topological bands in subgroup $P2_{1}'/c'~(14.79)$}
\textbf{Perturbations:}
\begin{itemize}
\item B $\parallel$ [010] and strain $\perp$ [100],
\item B $\parallel$ [001] and strain $\perp$ [100],
\item B $\perp$ [100].
\end{itemize}
\begin{figure}[H]
\centering
\includegraphics[scale=0.6]{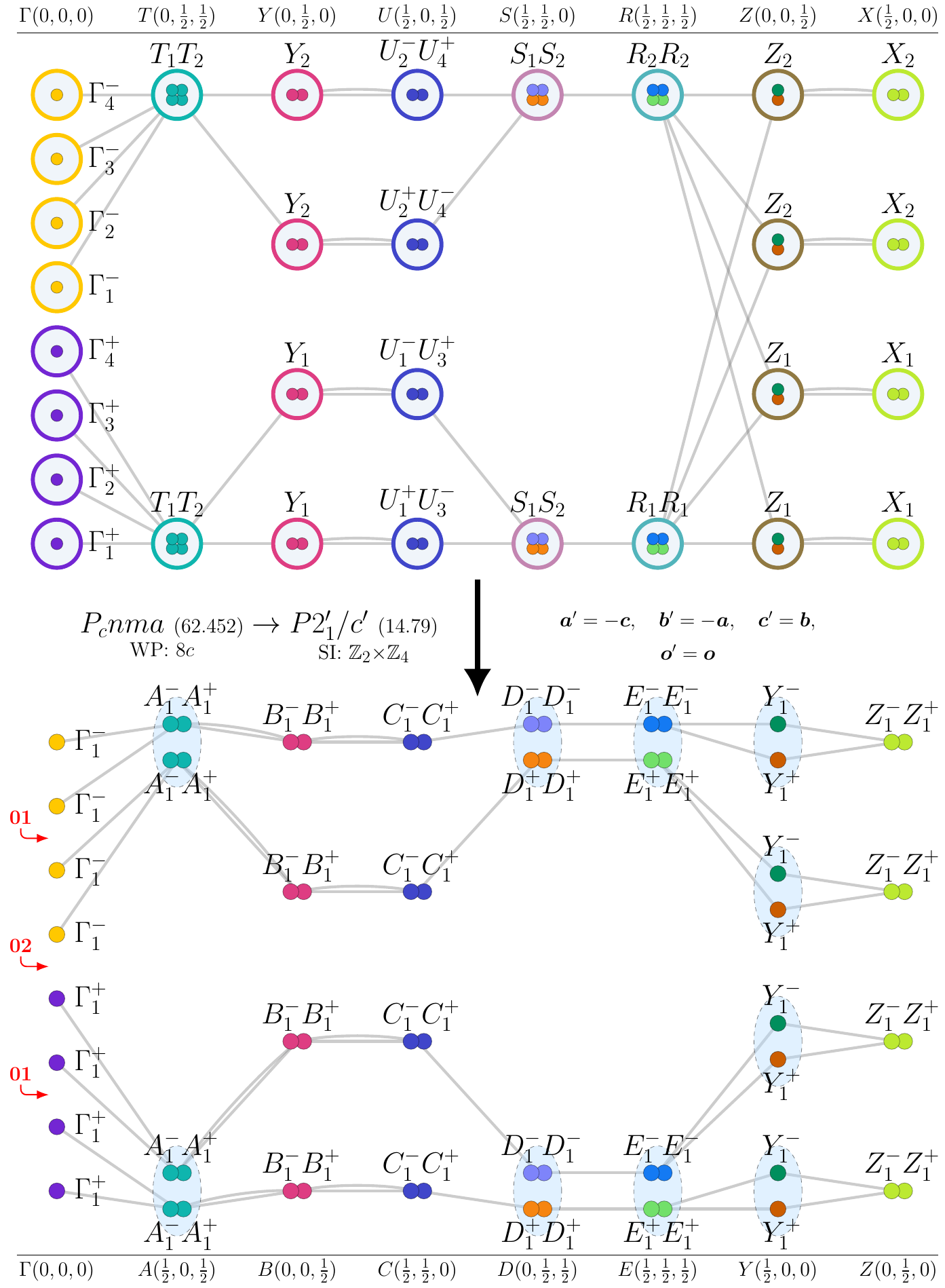}
\caption{Topological magnon bands in subgroup $P2_{1}'/c'~(14.79)$ for magnetic moments on Wyckoff position $8c$ of supergroup $P_{c}nma~(62.452)$.\label{fig_62.452_14.79_Bparallel010andstrainperp100_8c}}
\end{figure}
\input{gap_tables_tex/62.452_14.79_Bparallel010andstrainperp100_8c_table.tex}
\input{si_tables_tex/62.452_14.79_Bparallel010andstrainperp100_8c_table.tex}

\section{MSG $C_{c}mcm~(63.466)$}
\textbf{Nontrivial-SI Subgroups:} $P\bar{1}~(2.4)$, $P2'/m'~(10.46)$, $C2'/m'~(12.62)$, $C2'/c'~(15.89)$, $P_{S}\bar{1}~(2.7)$, $P2_{1}/m~(11.50)$, $Cm'c'm~(63.462)$, $P_{b}2_{1}/m~(11.56)$, $C2/c~(15.85)$, $Ccc'm'~(66.496)$, $C_{c}2/c~(15.90)$, $C2/m~(12.58)$, $Cmm'm'~(65.486)$, $C_{c}2/m~(12.63)$.\\

\textbf{Trivial-SI Subgroups:} $Pm'~(6.20)$, $Cm'~(8.34)$, $Cc'~(9.39)$, $P2'~(3.3)$, $C2'~(5.15)$, $C2'~(5.15)$, $P_{S}1~(1.3)$, $Pm~(6.18)$, $Ama'2'~(40.206)$, $Amm'2'~(38.190)$, $P_{b}m~(6.22)$, $Cc~(9.37)$, $Cc'c2'~(37.182)$, $Am'a2'~(40.205)$, $C_{c}c~(9.40)$, $Cm~(8.32)$, $Cm'm2'~(35.167)$, $Am'm2'~(38.189)$, $C_{c}m~(8.35)$, $P2_{1}~(4.7)$, $Cm'c'2_{1}~(36.176)$, $P_{b}2_{1}~(4.11)$, $C_{c}mc2_{1}~(36.177)$, $C2~(5.13)$, $Am'a'2~(40.207)$, $C_{c}2~(5.16)$, $A_{a}mm2~(38.192)$, $C2~(5.13)$, $Am'm'2~(38.191)$, $C_{c}2~(5.16)$, $A_{a}ma2~(40.208)$.\\

\subsection{WP: $4b+8e$}
\textbf{BCS Materials:} {FeSn~(373 K)}\footnote{BCS web page: \texttt{\href{http://webbdcrista1.ehu.es/magndata/index.php?this\_label=1.452} {http://webbdcrista1.ehu.es/magndata/index.php?this\_label=1.452}}}.\\
\subsubsection{Topological bands in subgroup $P\bar{1}~(2.4)$}
\textbf{Perturbations:}
\begin{itemize}
\item B $\parallel$ [100] and strain in generic direction,
\item B $\parallel$ [010] and strain in generic direction,
\item B $\parallel$ [001] and strain in generic direction,
\item B $\perp$ [100] and strain $\perp$ [010],
\item B $\perp$ [100] and strain $\perp$ [001],
\item B $\perp$ [100] and strain in generic direction,
\item B $\perp$ [010] and strain $\perp$ [100],
\item B $\perp$ [010] and strain $\perp$ [001],
\item B $\perp$ [010] and strain in generic direction,
\item B $\perp$ [001] and strain $\perp$ [100],
\item B $\perp$ [001] and strain $\perp$ [010],
\item B $\perp$ [001] and strain in generic direction,
\item B in generic direction.
\end{itemize}
\begin{figure}[H]
\centering
\includegraphics[scale=0.6]{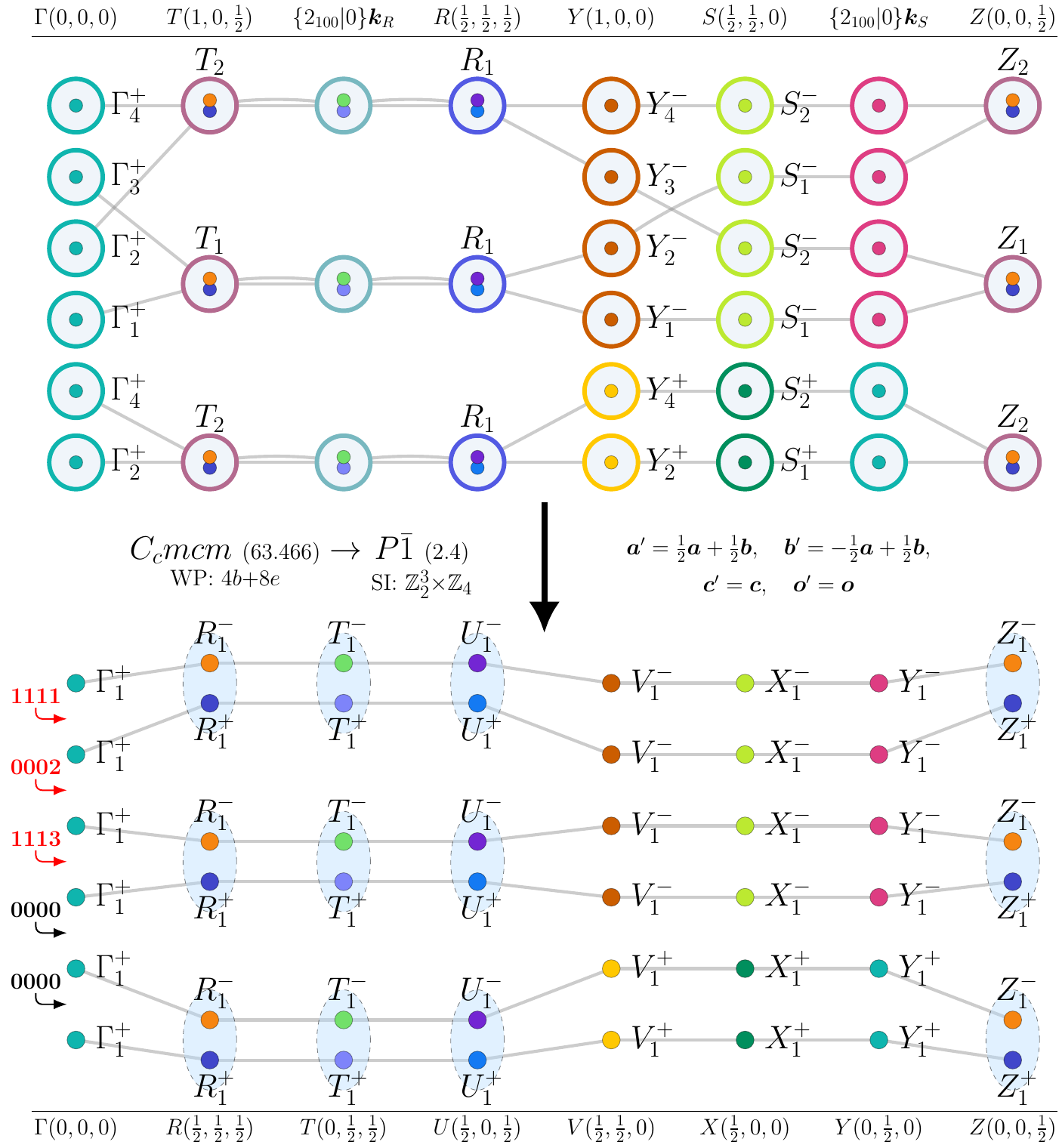}
\caption{Topological magnon bands in subgroup $P\bar{1}~(2.4)$ for magnetic moments on Wyckoff positions $4b+8e$ of supergroup $C_{c}mcm~(63.466)$.\label{fig_63.466_2.4_Bparallel100andstrainingenericdirection_4b+8e}}
\end{figure}
\input{gap_tables_tex/63.466_2.4_Bparallel100andstrainingenericdirection_4b+8e_table.tex}
\input{si_tables_tex/63.466_2.4_Bparallel100andstrainingenericdirection_4b+8e_table.tex}
\subsubsection{Topological bands in subgroup $P2'/m'~(10.46)$}
\textbf{Perturbations:}
\begin{itemize}
\item B $\parallel$ [100] and strain $\perp$ [001],
\item B $\parallel$ [010] and strain $\perp$ [001],
\item B $\perp$ [001].
\end{itemize}
\begin{figure}[H]
\centering
\includegraphics[scale=0.6]{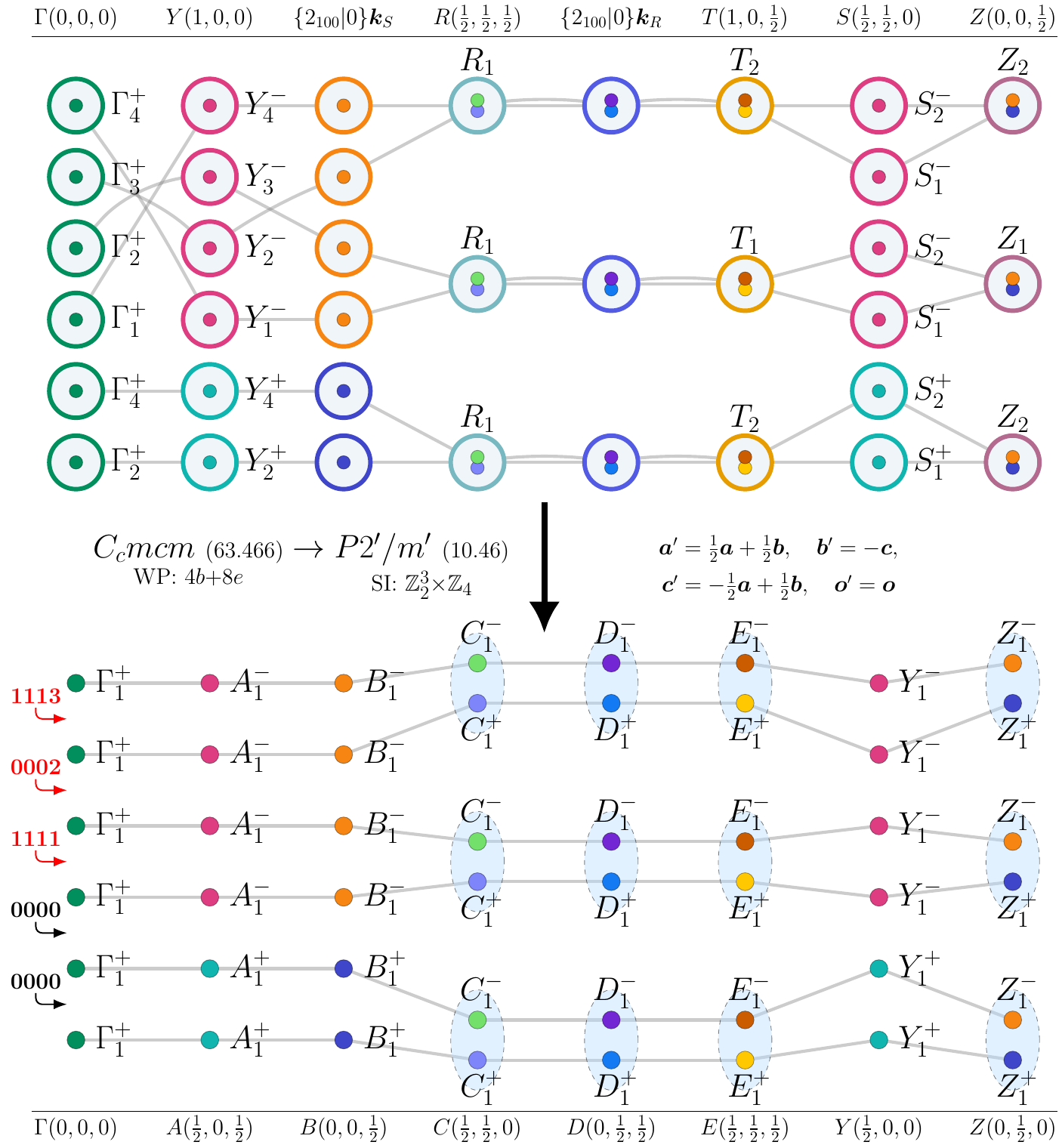}
\caption{Topological magnon bands in subgroup $P2'/m'~(10.46)$ for magnetic moments on Wyckoff positions $4b+8e$ of supergroup $C_{c}mcm~(63.466)$.\label{fig_63.466_10.46_Bparallel100andstrainperp001_4b+8e}}
\end{figure}
\input{gap_tables_tex/63.466_10.46_Bparallel100andstrainperp001_4b+8e_table.tex}
\input{si_tables_tex/63.466_10.46_Bparallel100andstrainperp001_4b+8e_table.tex}
\subsubsection{Topological bands in subgroup $C2'/m'~(12.62)$}
\textbf{Perturbations:}
\begin{itemize}
\item B $\parallel$ [100] and strain $\perp$ [010],
\item B $\parallel$ [001] and strain $\perp$ [010],
\item B $\perp$ [010].
\end{itemize}
\begin{figure}[H]
\centering
\includegraphics[scale=0.6]{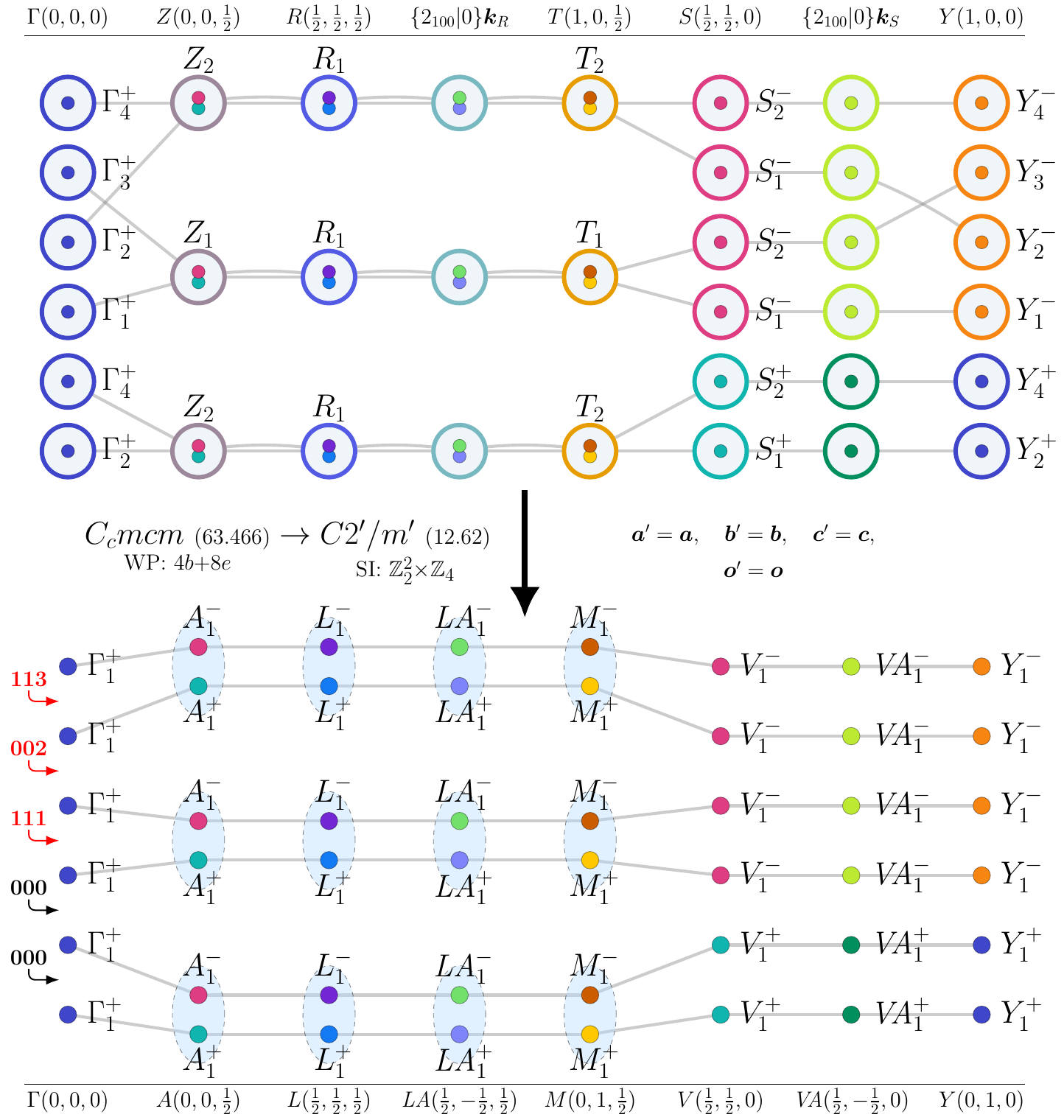}
\caption{Topological magnon bands in subgroup $C2'/m'~(12.62)$ for magnetic moments on Wyckoff positions $4b+8e$ of supergroup $C_{c}mcm~(63.466)$.\label{fig_63.466_12.62_Bparallel100andstrainperp010_4b+8e}}
\end{figure}
\input{gap_tables_tex/63.466_12.62_Bparallel100andstrainperp010_4b+8e_table.tex}
\input{si_tables_tex/63.466_12.62_Bparallel100andstrainperp010_4b+8e_table.tex}
\subsubsection{Topological bands in subgroup $C2'/c'~(15.89)$}
\textbf{Perturbations:}
\begin{itemize}
\item B $\parallel$ [010] and strain $\perp$ [100],
\item B $\parallel$ [001] and strain $\perp$ [100],
\item B $\perp$ [100].
\end{itemize}
\begin{figure}[H]
\centering
\includegraphics[scale=0.6]{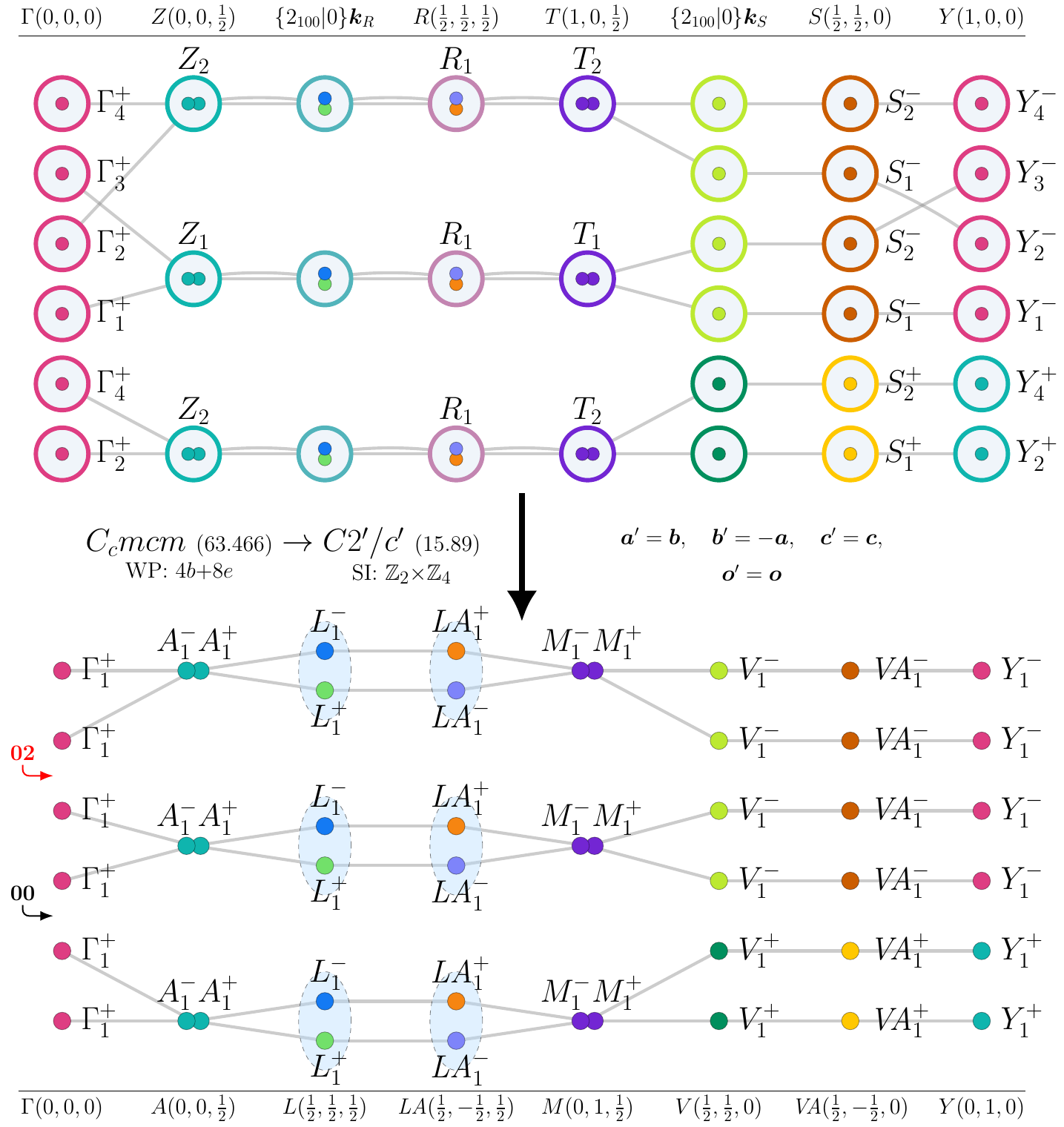}
\caption{Topological magnon bands in subgroup $C2'/c'~(15.89)$ for magnetic moments on Wyckoff positions $4b+8e$ of supergroup $C_{c}mcm~(63.466)$.\label{fig_63.466_15.89_Bparallel010andstrainperp100_4b+8e}}
\end{figure}
\input{gap_tables_tex/63.466_15.89_Bparallel010andstrainperp100_4b+8e_table.tex}
\input{si_tables_tex/63.466_15.89_Bparallel010andstrainperp100_4b+8e_table.tex}
\subsubsection{Topological bands in subgroup $P_{S}\bar{1}~(2.7)$}
\textbf{Perturbation:}
\begin{itemize}
\item strain in generic direction.
\end{itemize}
\begin{figure}[H]
\centering
\includegraphics[scale=0.6]{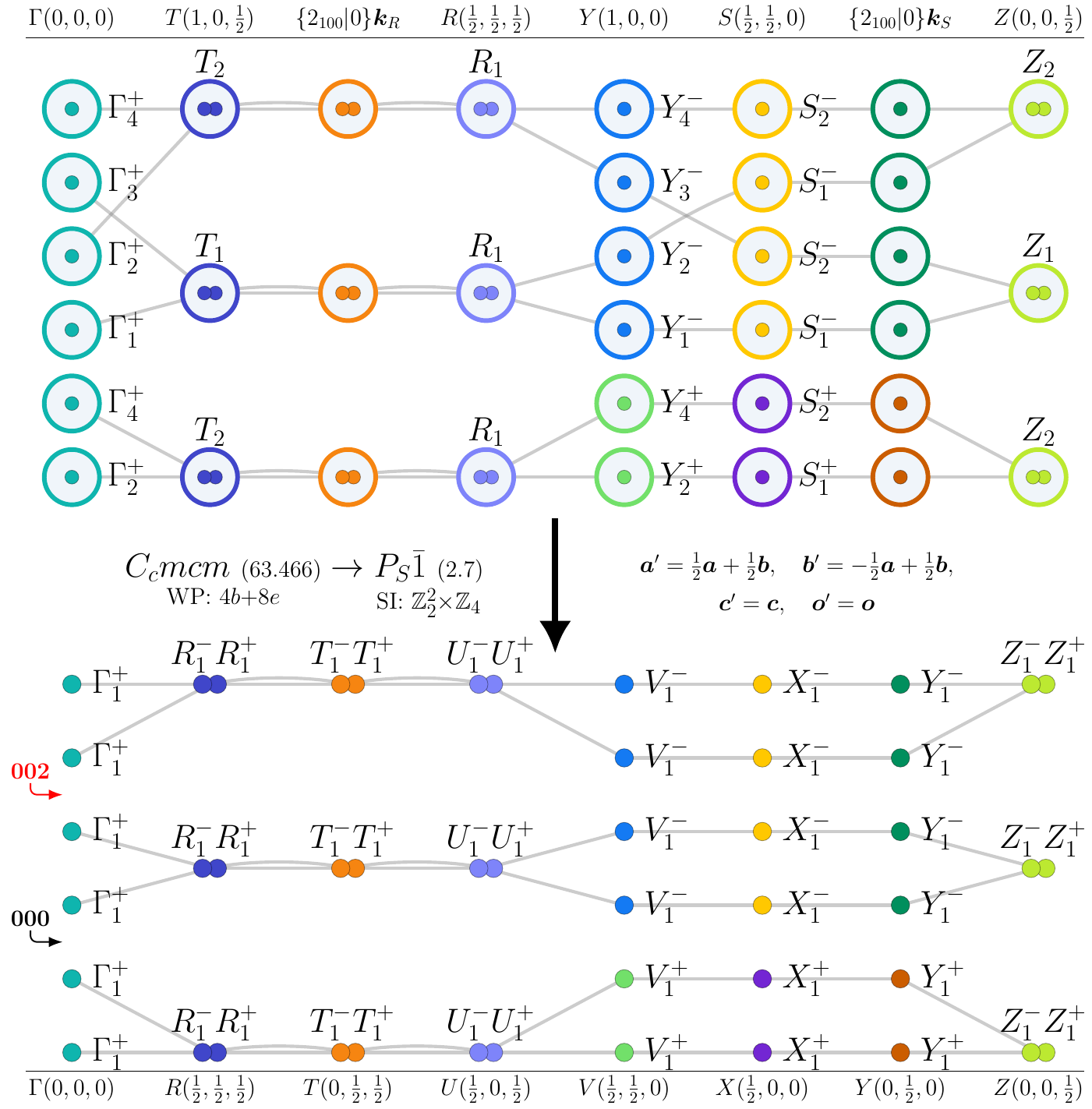}
\caption{Topological magnon bands in subgroup $P_{S}\bar{1}~(2.7)$ for magnetic moments on Wyckoff positions $4b+8e$ of supergroup $C_{c}mcm~(63.466)$.\label{fig_63.466_2.7_strainingenericdirection_4b+8e}}
\end{figure}
\input{gap_tables_tex/63.466_2.7_strainingenericdirection_4b+8e_table.tex}
\input{si_tables_tex/63.466_2.7_strainingenericdirection_4b+8e_table.tex}

\section{MSG $C_{A}mcm~(63.468)$}
\textbf{Nontrivial-SI Subgroups:} $P\bar{1}~(2.4)$, $P2'/c'~(13.69)$, $C2'/m'~(12.62)$, $C2'/c'~(15.89)$, $P_{S}\bar{1}~(2.7)$, $P2_{1}/m~(11.50)$, $Cm'c'm~(63.462)$, $P_{C}2_{1}/m~(11.57)$, $Ab'a'2~(41.215)$, $C2/c~(15.85)$, $Ccc'a'~(68.516)$, $C_{c}2/c~(15.90)$, $Ab'm'2~(39.199)$, $C2/m~(12.58)$, $Cmm'a'~(67.506)$, $C_{c}2/m~(12.63)$.\\

\textbf{Trivial-SI Subgroups:} $Pc'~(7.26)$, $Cm'~(8.34)$, $Cc'~(9.39)$, $P2'~(3.3)$, $C2'~(5.15)$, $C2'~(5.15)$, $P_{S}1~(1.3)$, $Pm~(6.18)$, $Ama'2'~(40.206)$, $Amm'2'~(38.190)$, $P_{C}m~(6.23)$, $Cc~(9.37)$, $Cc'c2'~(37.182)$, $Ab'a2'~(41.213)$, $C_{c}c~(9.40)$, $Cm~(8.32)$, $Cm'm2'~(35.167)$, $Ab'm2'~(39.197)$, $C_{c}m~(8.35)$, $P2_{1}~(4.7)$, $Cm'c'2_{1}~(36.176)$, $P_{C}2_{1}~(4.12)$, $C_{A}mc2_{1}~(36.179)$, $C2~(5.13)$, $C_{c}2~(5.16)$, $A_{B}mm2~(38.194)$, $C2~(5.13)$, $C_{c}2~(5.16)$, $A_{B}ma2~(40.210)$.\\

\subsection{WP: $8d$}
\textbf{BCS Materials:} {Pr\textsubscript{0.5}Sr\textsubscript{0.5}MnO\textsubscript{3}~(140 K)}\footnote{BCS web page: \texttt{\href{http://webbdcrista1.ehu.es/magndata/index.php?this\_label=1.273} {http://webbdcrista1.ehu.es/magndata/index.php?this\_label=1.273}}}.\\
\subsubsection{Topological bands in subgroup $P\bar{1}~(2.4)$}
\textbf{Perturbations:}
\begin{itemize}
\item B $\parallel$ [100] and strain in generic direction,
\item B $\parallel$ [010] and strain in generic direction,
\item B $\parallel$ [001] and strain in generic direction,
\item B $\perp$ [100] and strain $\perp$ [010],
\item B $\perp$ [100] and strain $\perp$ [001],
\item B $\perp$ [100] and strain in generic direction,
\item B $\perp$ [010] and strain $\perp$ [100],
\item B $\perp$ [010] and strain $\perp$ [001],
\item B $\perp$ [010] and strain in generic direction,
\item B $\perp$ [001] and strain $\perp$ [100],
\item B $\perp$ [001] and strain $\perp$ [010],
\item B $\perp$ [001] and strain in generic direction,
\item B in generic direction.
\end{itemize}
\begin{figure}[H]
\centering
\includegraphics[scale=0.6]{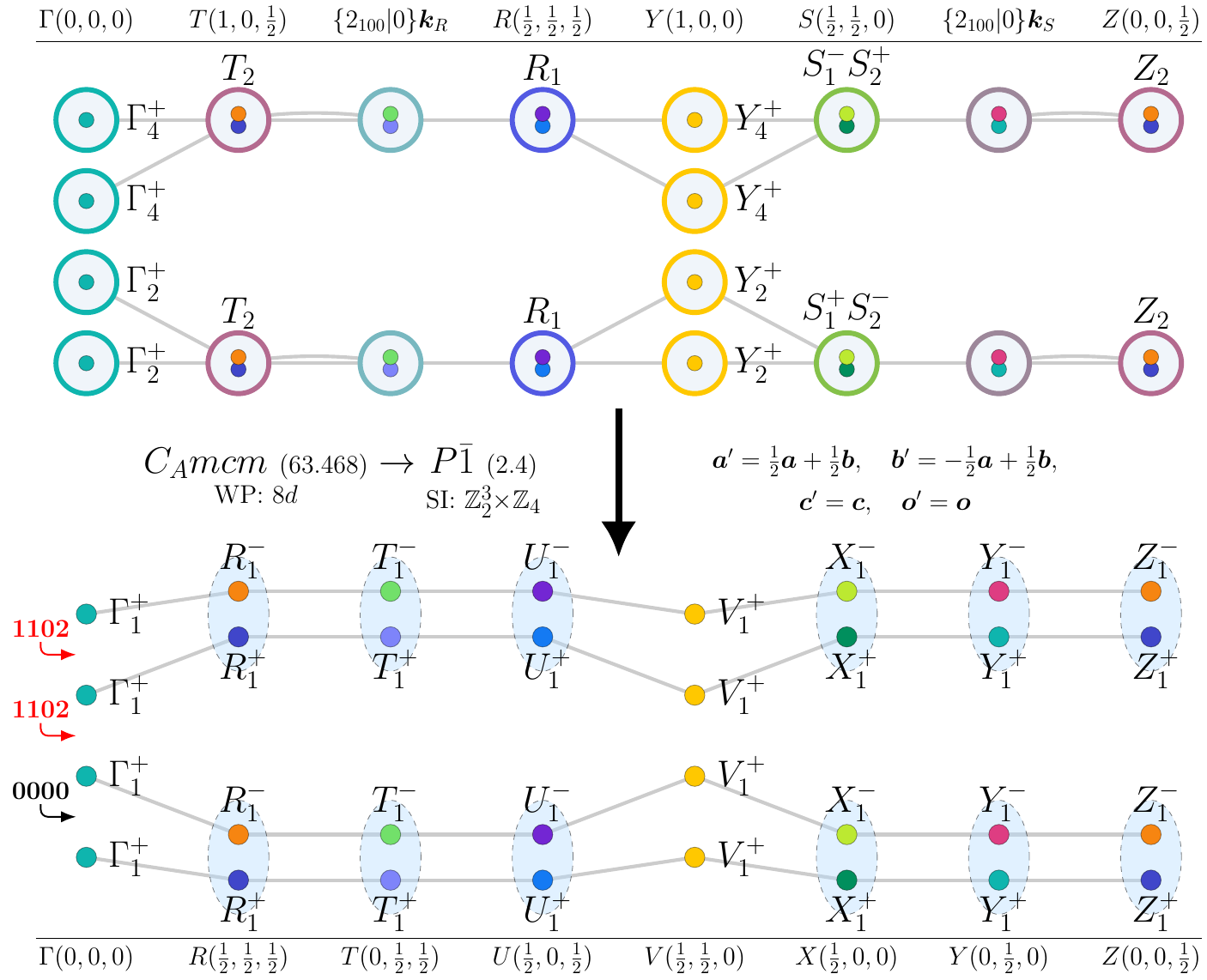}
\caption{Topological magnon bands in subgroup $P\bar{1}~(2.4)$ for magnetic moments on Wyckoff position $8d$ of supergroup $C_{A}mcm~(63.468)$.\label{fig_63.468_2.4_Bparallel100andstrainingenericdirection_8d}}
\end{figure}
\input{gap_tables_tex/63.468_2.4_Bparallel100andstrainingenericdirection_8d_table.tex}
\input{si_tables_tex/63.468_2.4_Bparallel100andstrainingenericdirection_8d_table.tex}
\subsubsection{Topological bands in subgroup $P2'/c'~(13.69)$}
\textbf{Perturbations:}
\begin{itemize}
\item B $\parallel$ [100] and strain $\perp$ [001],
\item B $\parallel$ [010] and strain $\perp$ [001],
\item B $\perp$ [001].
\end{itemize}
\begin{figure}[H]
\centering
\includegraphics[scale=0.6]{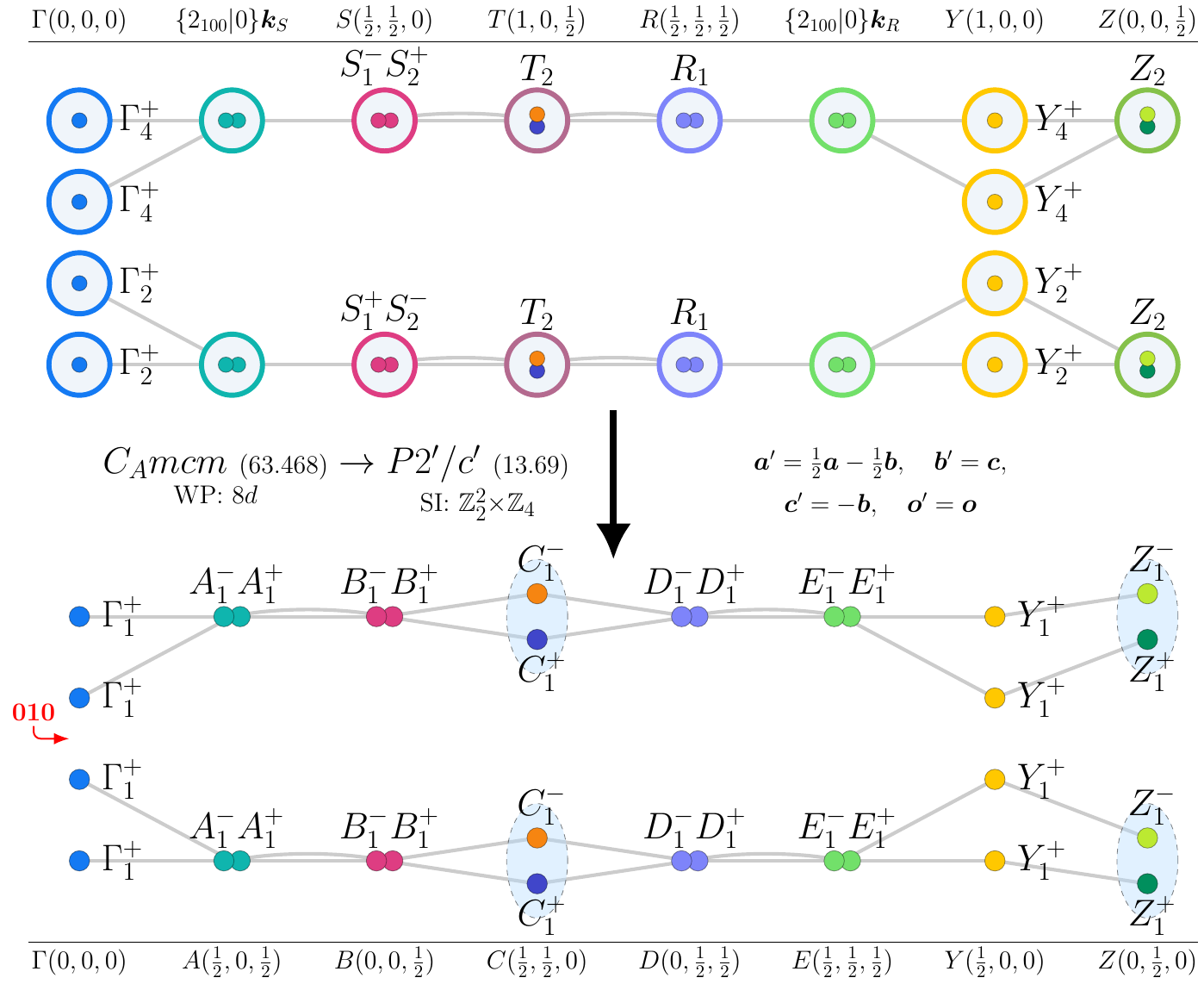}
\caption{Topological magnon bands in subgroup $P2'/c'~(13.69)$ for magnetic moments on Wyckoff position $8d$ of supergroup $C_{A}mcm~(63.468)$.\label{fig_63.468_13.69_Bparallel100andstrainperp001_8d}}
\end{figure}
\input{gap_tables_tex/63.468_13.69_Bparallel100andstrainperp001_8d_table.tex}
\input{si_tables_tex/63.468_13.69_Bparallel100andstrainperp001_8d_table.tex}
\subsubsection{Topological bands in subgroup $C2'/m'~(12.62)$}
\textbf{Perturbations:}
\begin{itemize}
\item B $\parallel$ [100] and strain $\perp$ [010],
\item B $\parallel$ [001] and strain $\perp$ [010],
\item B $\perp$ [010].
\end{itemize}
\begin{figure}[H]
\centering
\includegraphics[scale=0.6]{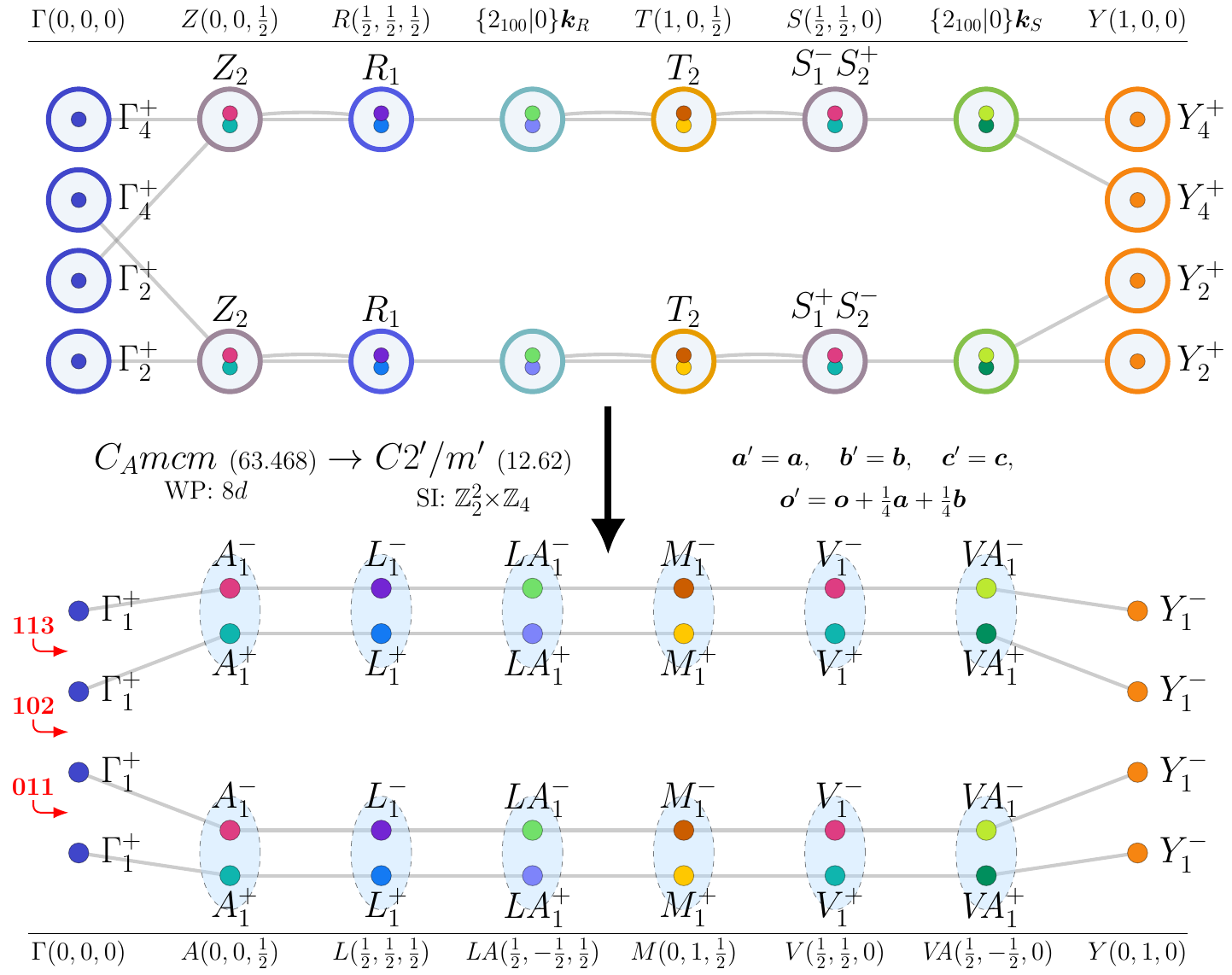}
\caption{Topological magnon bands in subgroup $C2'/m'~(12.62)$ for magnetic moments on Wyckoff position $8d$ of supergroup $C_{A}mcm~(63.468)$.\label{fig_63.468_12.62_Bparallel100andstrainperp010_8d}}
\end{figure}
\input{gap_tables_tex/63.468_12.62_Bparallel100andstrainperp010_8d_table.tex}
\input{si_tables_tex/63.468_12.62_Bparallel100andstrainperp010_8d_table.tex}
\subsubsection{Topological bands in subgroup $C2'/c'~(15.89)$}
\textbf{Perturbations:}
\begin{itemize}
\item B $\parallel$ [010] and strain $\perp$ [100],
\item B $\parallel$ [001] and strain $\perp$ [100],
\item B $\perp$ [100].
\end{itemize}
\begin{figure}[H]
\centering
\includegraphics[scale=0.6]{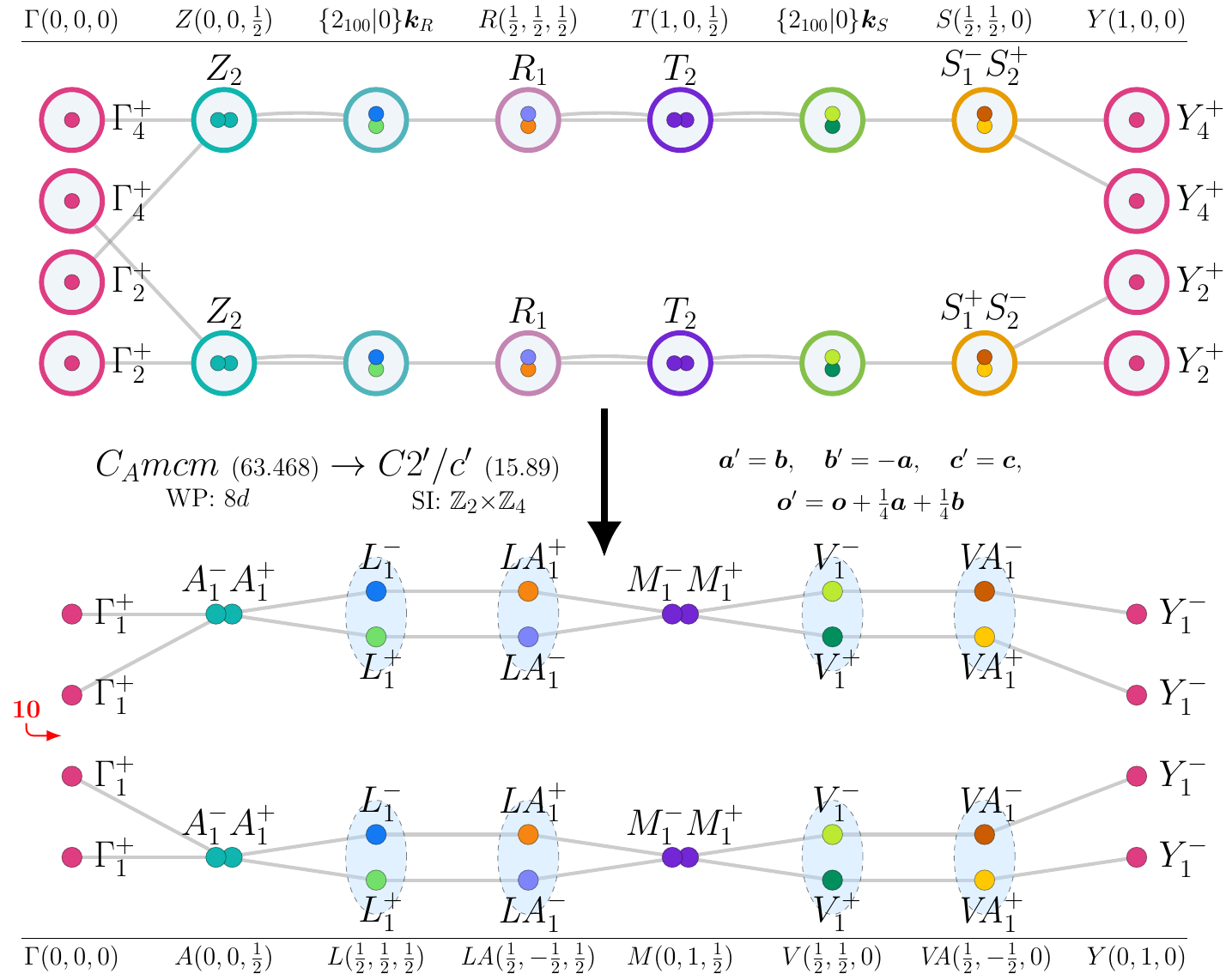}
\caption{Topological magnon bands in subgroup $C2'/c'~(15.89)$ for magnetic moments on Wyckoff position $8d$ of supergroup $C_{A}mcm~(63.468)$.\label{fig_63.468_15.89_Bparallel010andstrainperp100_8d}}
\end{figure}
\input{gap_tables_tex/63.468_15.89_Bparallel010andstrainperp100_8d_table.tex}
\input{si_tables_tex/63.468_15.89_Bparallel010andstrainperp100_8d_table.tex}
\subsubsection{Topological bands in subgroup $P_{S}\bar{1}~(2.7)$}
\textbf{Perturbation:}
\begin{itemize}
\item strain in generic direction.
\end{itemize}
\begin{figure}[H]
\centering
\includegraphics[scale=0.6]{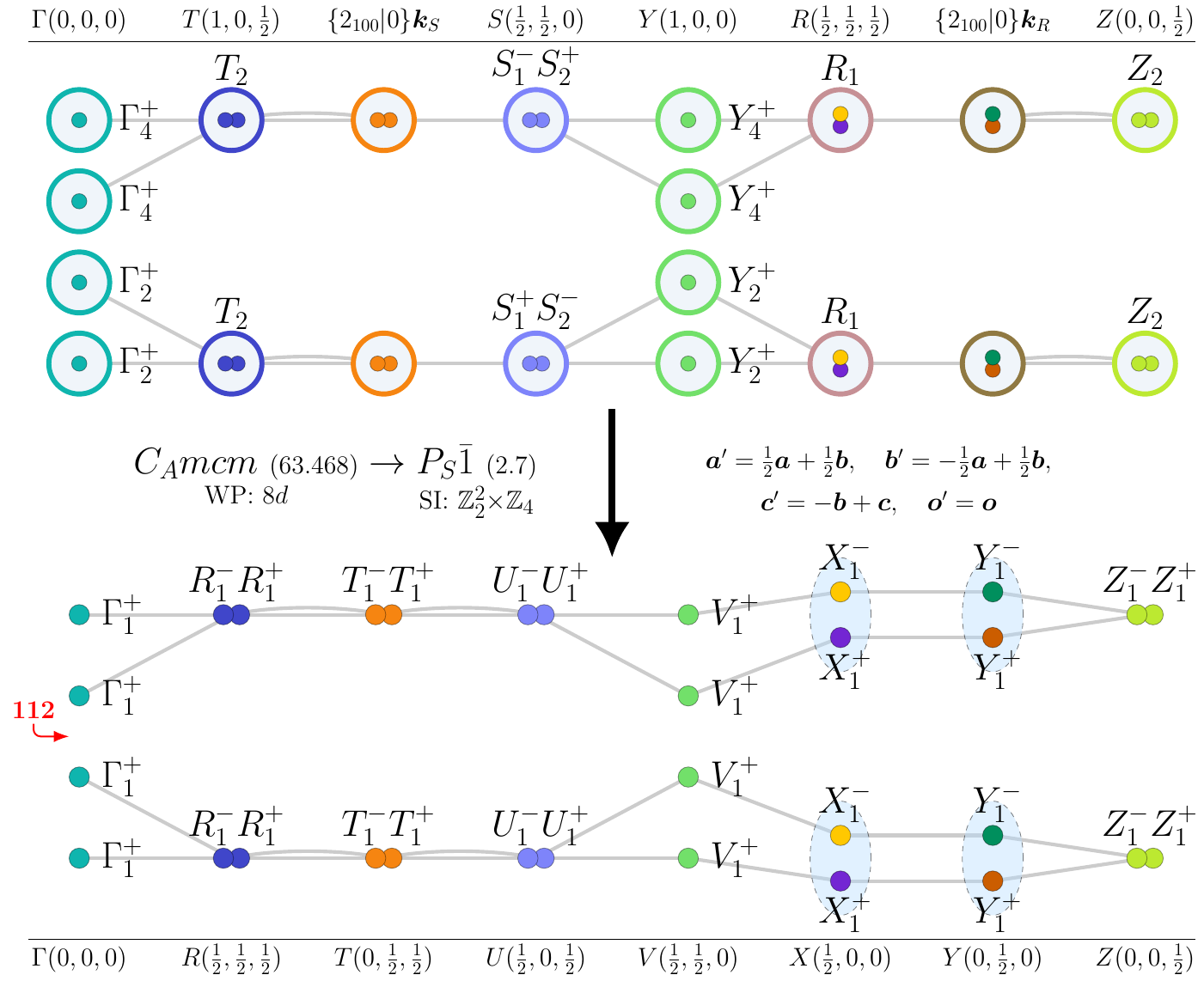}
\caption{Topological magnon bands in subgroup $P_{S}\bar{1}~(2.7)$ for magnetic moments on Wyckoff position $8d$ of supergroup $C_{A}mcm~(63.468)$.\label{fig_63.468_2.7_strainingenericdirection_8d}}
\end{figure}
\input{gap_tables_tex/63.468_2.7_strainingenericdirection_8d_table.tex}
\input{si_tables_tex/63.468_2.7_strainingenericdirection_8d_table.tex}
\subsubsection{Topological bands in subgroup $Ab'm'2~(39.199)$}
\textbf{Perturbation:}
\begin{itemize}
\item E $\parallel$ [100] and B $\parallel$ [100].
\end{itemize}
\begin{figure}[H]
\centering
\includegraphics[scale=0.6]{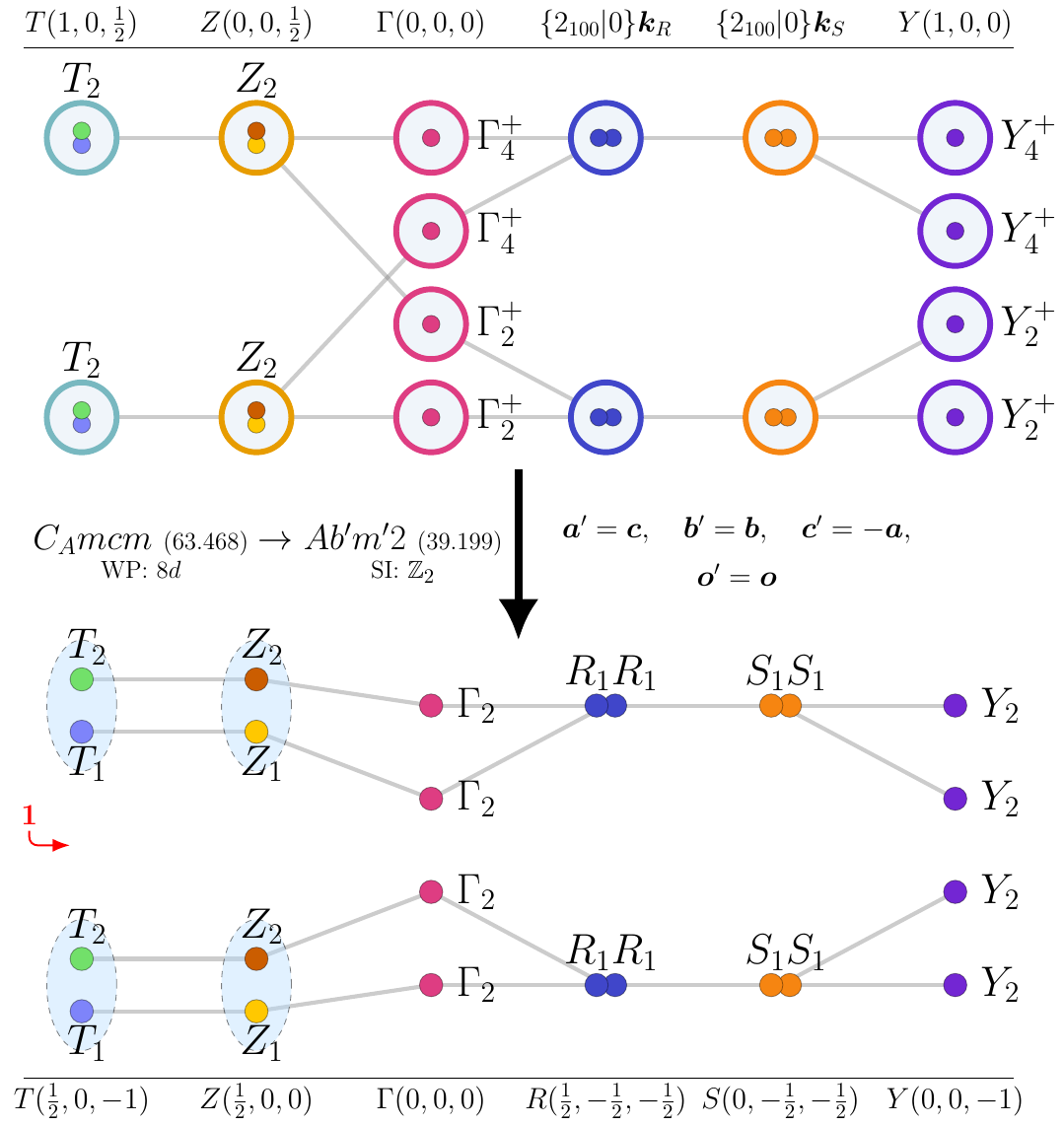}
\caption{Topological magnon bands in subgroup $Ab'm'2~(39.199)$ for magnetic moments on Wyckoff position $8d$ of supergroup $C_{A}mcm~(63.468)$.\label{fig_63.468_39.199_Eparallel100andBparallel100_8d}}
\end{figure}
\input{gap_tables_tex/63.468_39.199_Eparallel100andBparallel100_8d_table.tex}
\input{si_tables_tex/63.468_39.199_Eparallel100andBparallel100_8d_table.tex}
\subsubsection{Topological bands in subgroup $C2/m~(12.58)$}
\textbf{Perturbation:}
\begin{itemize}
\item B $\parallel$ [100] and strain $\perp$ [100].
\end{itemize}
\begin{figure}[H]
\centering
\includegraphics[scale=0.6]{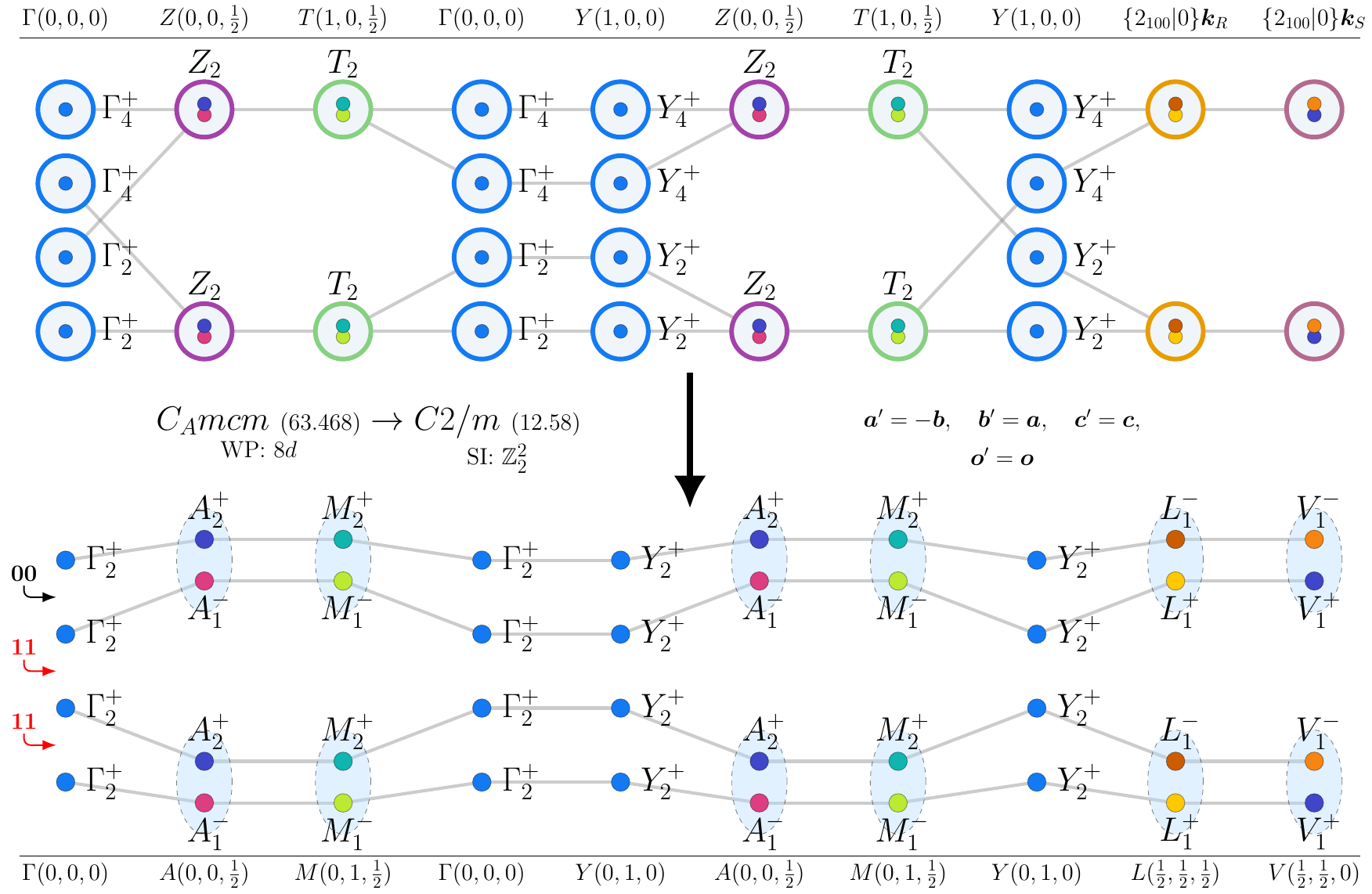}
\caption{Topological magnon bands in subgroup $C2/m~(12.58)$ for magnetic moments on Wyckoff position $8d$ of supergroup $C_{A}mcm~(63.468)$.\label{fig_63.468_12.58_Bparallel100andstrainperp100_8d}}
\end{figure}
\input{gap_tables_tex/63.468_12.58_Bparallel100andstrainperp100_8d_table.tex}
\input{si_tables_tex/63.468_12.58_Bparallel100andstrainperp100_8d_table.tex}
\subsubsection{Topological bands in subgroup $Cmm'a'~(67.506)$}
\textbf{Perturbation:}
\begin{itemize}
\item B $\parallel$ [100].
\end{itemize}
\begin{figure}[H]
\centering
\includegraphics[scale=0.6]{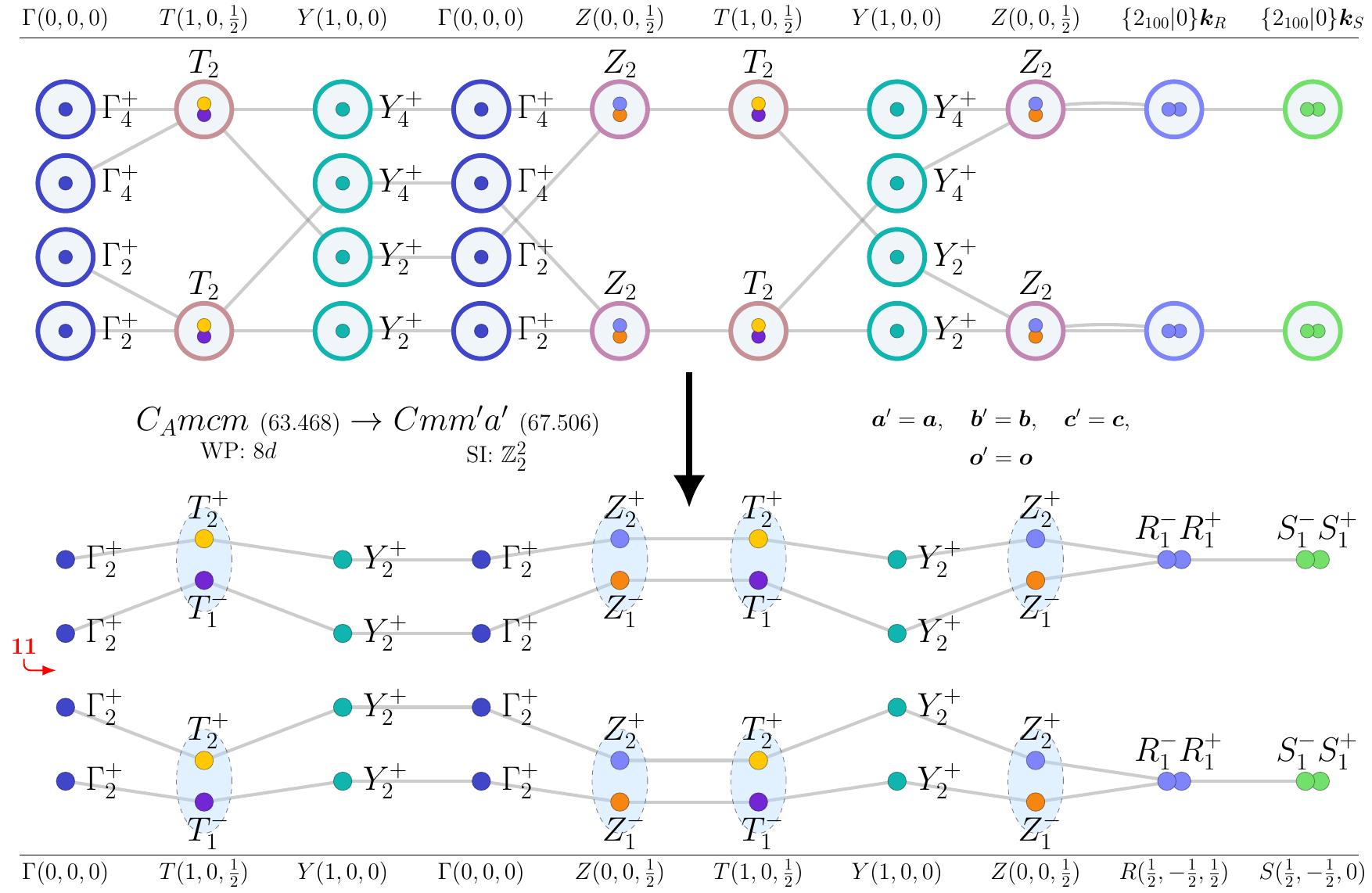}
\caption{Topological magnon bands in subgroup $Cmm'a'~(67.506)$ for magnetic moments on Wyckoff position $8d$ of supergroup $C_{A}mcm~(63.468)$.\label{fig_63.468_67.506_Bparallel100_8d}}
\end{figure}
\input{gap_tables_tex/63.468_67.506_Bparallel100_8d_table.tex}
\input{si_tables_tex/63.468_67.506_Bparallel100_8d_table.tex}
\subsubsection{Topological bands in subgroup $C_{c}2/m~(12.63)$}
\textbf{Perturbation:}
\begin{itemize}
\item strain $\perp$ [100].
\end{itemize}
\begin{figure}[H]
\centering
\includegraphics[scale=0.6]{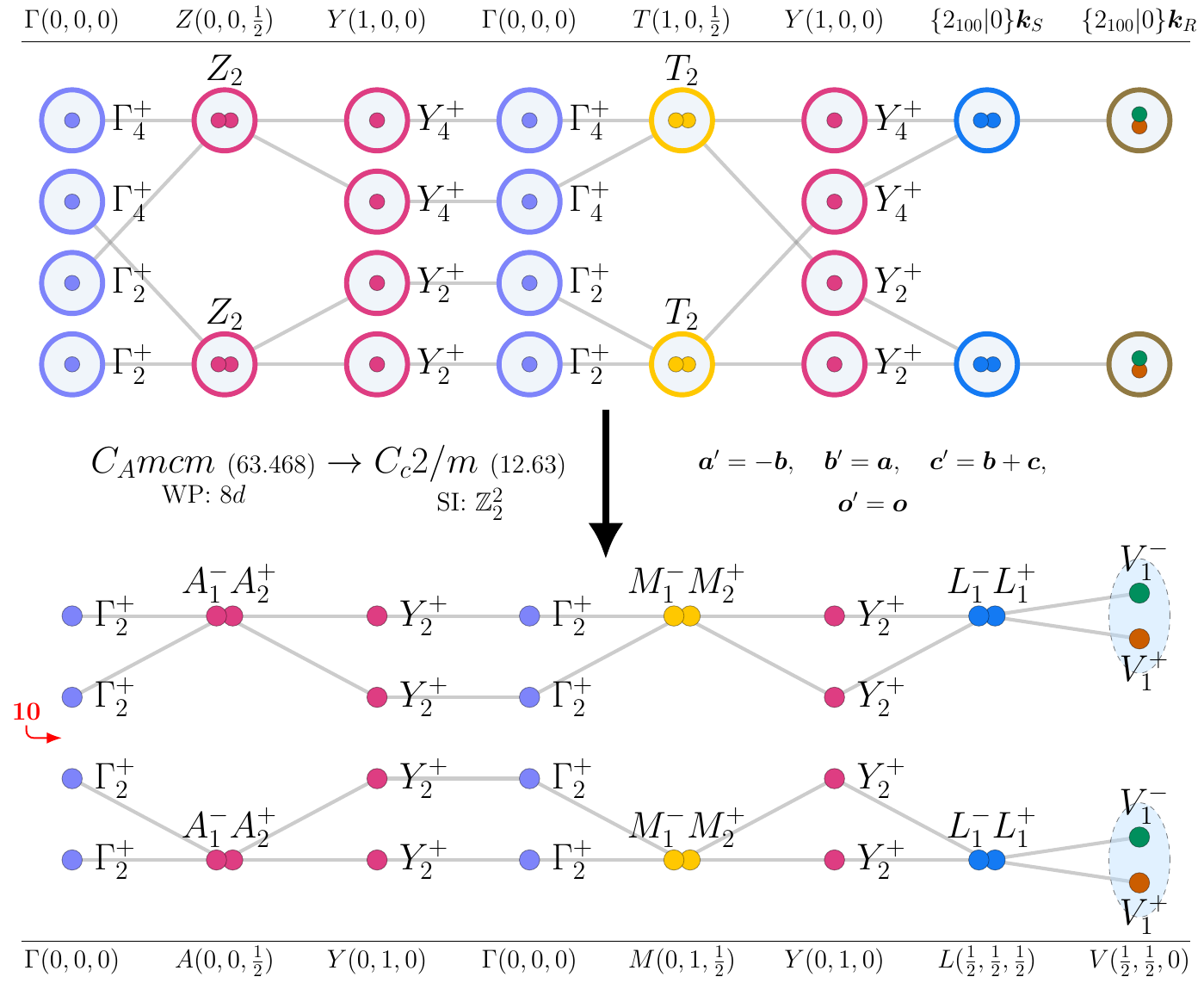}
\caption{Topological magnon bands in subgroup $C_{c}2/m~(12.63)$ for magnetic moments on Wyckoff position $8d$ of supergroup $C_{A}mcm~(63.468)$.\label{fig_63.468_12.63_strainperp100_8d}}
\end{figure}
\input{gap_tables_tex/63.468_12.63_strainperp100_8d_table.tex}
\input{si_tables_tex/63.468_12.63_strainperp100_8d_table.tex}
\subsection{WP: $8f$}
\textbf{BCS Materials:} {CsCo\textsubscript{2}Se\textsubscript{2}~(78 K)}\footnote{BCS web page: \texttt{\href{http://webbdcrista1.ehu.es/magndata/index.php?this\_label=1.458} {http://webbdcrista1.ehu.es/magndata/index.php?this\_label=1.458}}}.\\
\subsubsection{Topological bands in subgroup $Ab'm'2~(39.199)$}
\textbf{Perturbation:}
\begin{itemize}
\item E $\parallel$ [100] and B $\parallel$ [100].
\end{itemize}
\begin{figure}[H]
\centering
\includegraphics[scale=0.6]{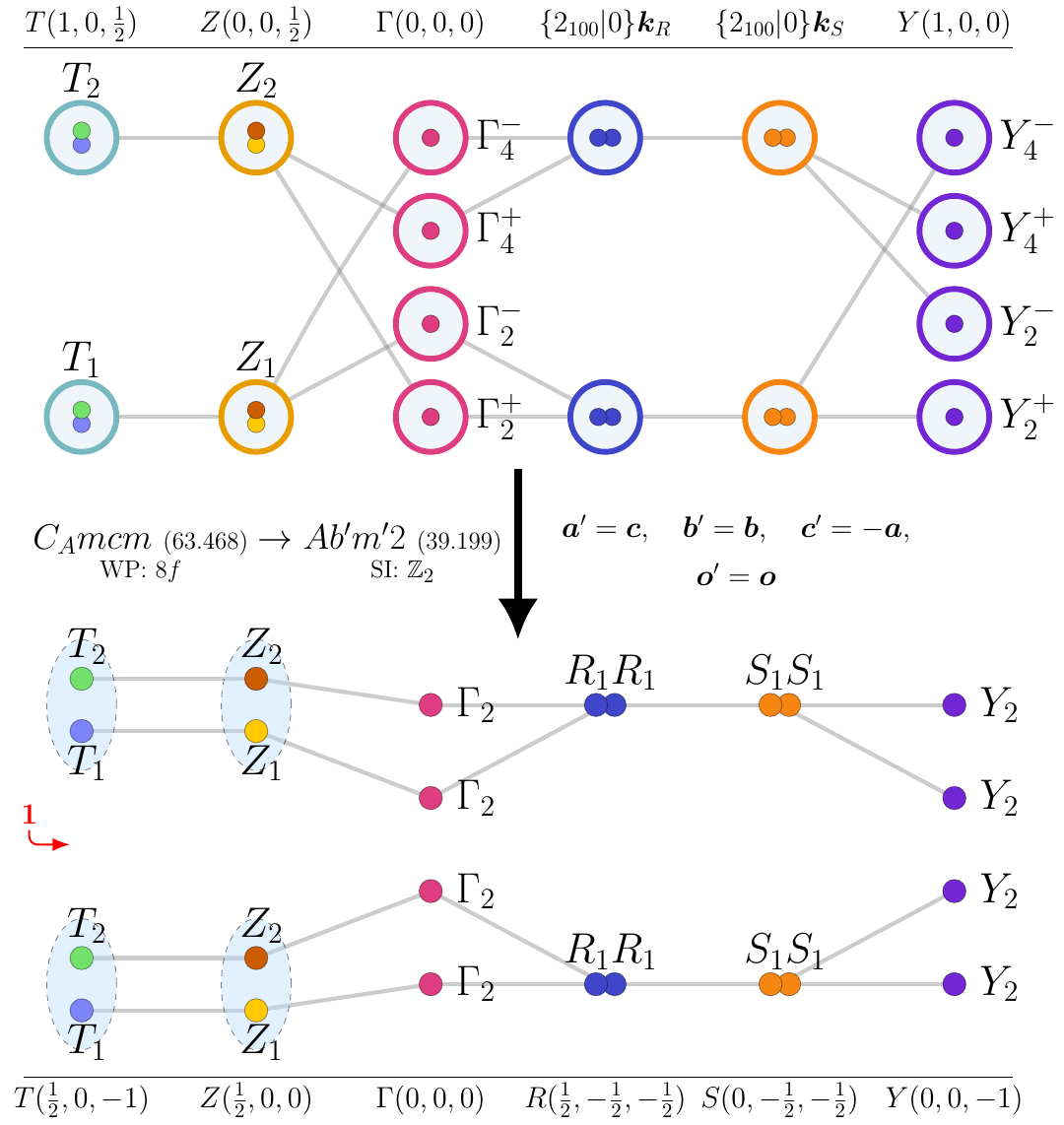}
\caption{Topological magnon bands in subgroup $Ab'm'2~(39.199)$ for magnetic moments on Wyckoff position $8f$ of supergroup $C_{A}mcm~(63.468)$.\label{fig_63.468_39.199_Eparallel100andBparallel100_8f}}
\end{figure}
\input{gap_tables_tex/63.468_39.199_Eparallel100andBparallel100_8f_table.tex}
\input{si_tables_tex/63.468_39.199_Eparallel100andBparallel100_8f_table.tex}

\section{MSG $C_{A}mmm~(65.490)$}
\textbf{Nontrivial-SI Subgroups:} $P\bar{1}~(2.4)$, $P2_{1}'/c'~(14.79)$, $C2'/c'~(15.89)$, $C2'/c'~(15.89)$, $P_{S}\bar{1}~(2.7)$, $P2~(3.1)$, $Cc'c'2~(37.183)$, $P2/m~(10.42)$, $Cc'c'm~(66.495)$, $P_{C}2/m~(10.49)$, $Ab'a'2~(41.215)$, $C2/m~(12.58)$, $Cmc'a'~(64.475)$, $C_{c}2/m~(12.63)$, $Ab'a'2~(41.215)$, $C2/m~(12.58)$, $Cmc'a'~(64.475)$, $C_{c}2/m~(12.63)$.\\

\textbf{Trivial-SI Subgroups:} $Pc'~(7.26)$, $Cc'~(9.39)$, $Cc'~(9.39)$, $P2_{1}'~(4.9)$, $C2'~(5.15)$, $C2'~(5.15)$, $P_{S}1~(1.3)$, $Pm~(6.18)$, $Ama'2'~(40.206)$, $Ama'2'~(40.206)$, $P_{C}m~(6.23)$, $Cm~(8.32)$, $Cmc'2_{1}'~(36.175)$, $Ab'm2'~(39.197)$, $C_{c}m~(8.35)$, $Cm~(8.32)$, $Cmc'2_{1}'~(36.175)$, $Ab'm2'~(39.197)$, $C_{c}m~(8.35)$, $P_{C}2~(3.6)$, $C_{A}mm2~(35.171)$, $C2~(5.13)$, $C_{c}2~(5.16)$, $A_{B}mm2~(38.194)$, $C2~(5.13)$, $C_{c}2~(5.16)$, $A_{B}mm2~(38.194)$.\\

\subsection{WP: $8f$}
\textbf{BCS Materials:} {KCuMnS\textsubscript{2}~(160 K)}\footnote{BCS web page: \texttt{\href{http://webbdcrista1.ehu.es/magndata/index.php?this\_label=1.392} {http://webbdcrista1.ehu.es/magndata/index.php?this\_label=1.392}}}.\\
\subsubsection{Topological bands in subgroup $Cc'c'2~(37.183)$}
\textbf{Perturbation:}
\begin{itemize}
\item E $\parallel$ [001] and B $\parallel$ [001].
\end{itemize}
\begin{figure}[H]
\centering
\includegraphics[scale=0.6]{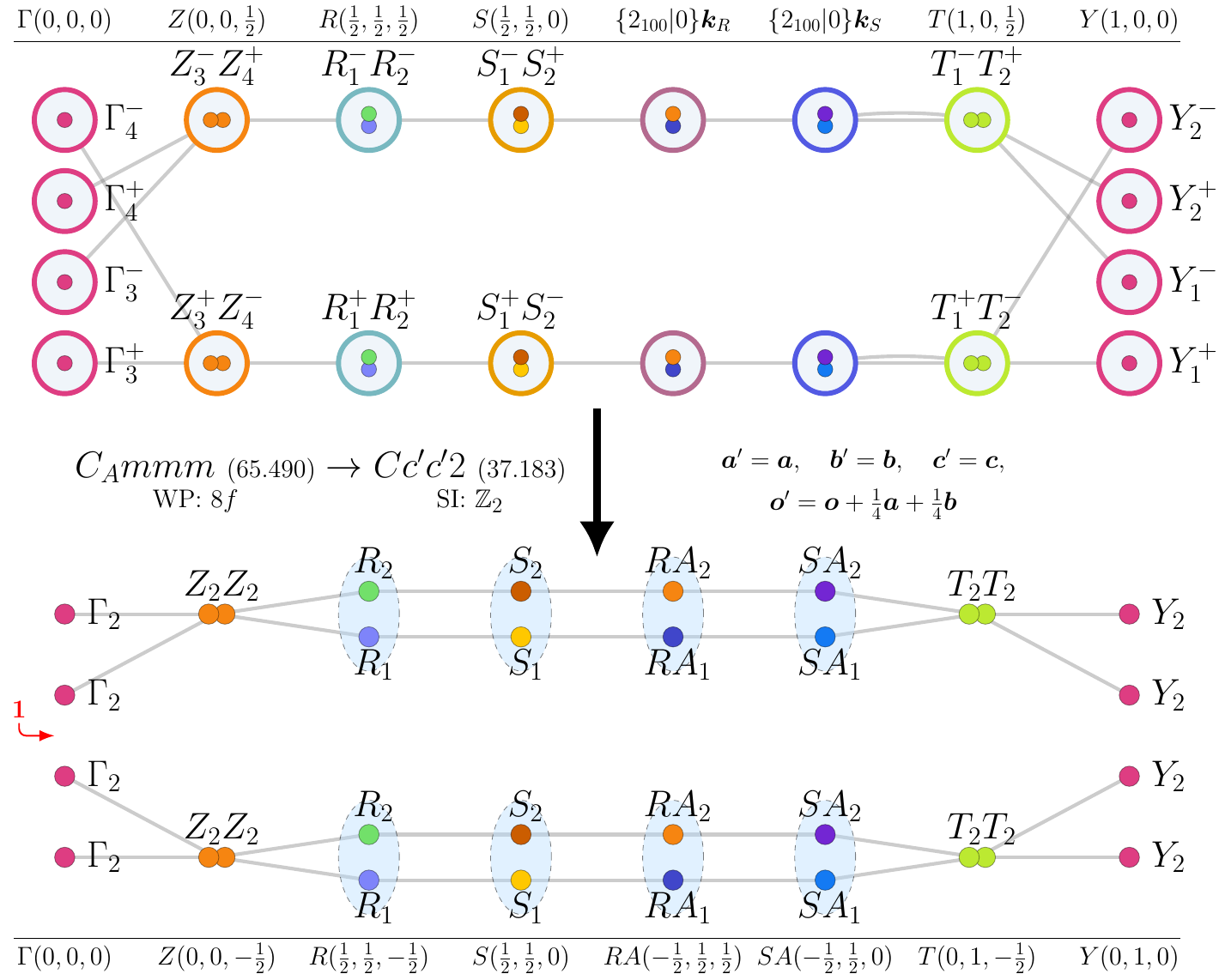}
\caption{Topological magnon bands in subgroup $Cc'c'2~(37.183)$ for magnetic moments on Wyckoff position $8f$ of supergroup $C_{A}mmm~(65.490)$.\label{fig_65.490_37.183_Eparallel001andBparallel001_8f}}
\end{figure}
\input{gap_tables_tex/65.490_37.183_Eparallel001andBparallel001_8f_table.tex}
\input{si_tables_tex/65.490_37.183_Eparallel001andBparallel001_8f_table.tex}

\section{MSG $C_{a}mma~(67.509)$}
\textbf{Nontrivial-SI Subgroups:} $P\bar{1}~(2.4)$, $P2'/m'~(10.46)$, $C2'/m'~(12.62)$, $C2'/m'~(12.62)$, $P_{S}\bar{1}~(2.7)$, $P2~(3.1)$, $Cm'm'2~(35.168)$, $P_{a}2~(3.4)$, $P2/c~(13.65)$, $Cm'm'a~(67.505)$, $P_{c}2/c~(13.72)$, $C_{a}2~(5.17)$, $C2/m~(12.58)$, $Cmm'm'~(65.486)$, $C_{a}2/m~(12.64)$, $C_{a}2~(5.17)$, $C2/m~(12.58)$, $Cmm'm'~(65.486)$, $C_{a}2/m~(12.64)$.\\

\textbf{Trivial-SI Subgroups:} $Pm'~(6.20)$, $Cm'~(8.34)$, $Cm'~(8.34)$, $P2'~(3.3)$, $C2'~(5.15)$, $C2'~(5.15)$, $P_{S}1~(1.3)$, $Pc~(7.24)$, $Abm'2'~(39.198)$, $Abm'2'~(39.198)$, $P_{c}c~(7.28)$, $Cm~(8.32)$, $Cm'm2'~(35.167)$, $Am'm2'~(38.189)$, $C_{a}m~(8.36)$, $Cm~(8.32)$, $Cm'm2'~(35.167)$, $Am'm2'~(38.189)$, $C_{a}m~(8.36)$, $C_{a}mm2~(35.170)$, $C2~(5.13)$, $Am'm'2~(38.191)$, $A_{b}bm2~(39.201)$, $C2~(5.13)$, $Am'm'2~(38.191)$, $A_{b}bm2~(39.201)$.\\

\subsection{WP: $4g$}
\textbf{BCS Materials:} {MnPt\textsubscript{0.5}Pd\textsubscript{0.5}~(866 K)}\footnote{BCS web page: \texttt{\href{http://webbdcrista1.ehu.es/magndata/index.php?this\_label=1.466} {http://webbdcrista1.ehu.es/magndata/index.php?this\_label=1.466}}}.\\
\subsubsection{Topological bands in subgroup $C2'/m'~(12.62)$}
\textbf{Perturbations:}
\begin{itemize}
\item B $\parallel$ [100] and strain $\perp$ [010],
\item B $\parallel$ [001] and strain $\perp$ [010],
\item B $\perp$ [010].
\end{itemize}
\begin{figure}[H]
\centering
\includegraphics[scale=0.6]{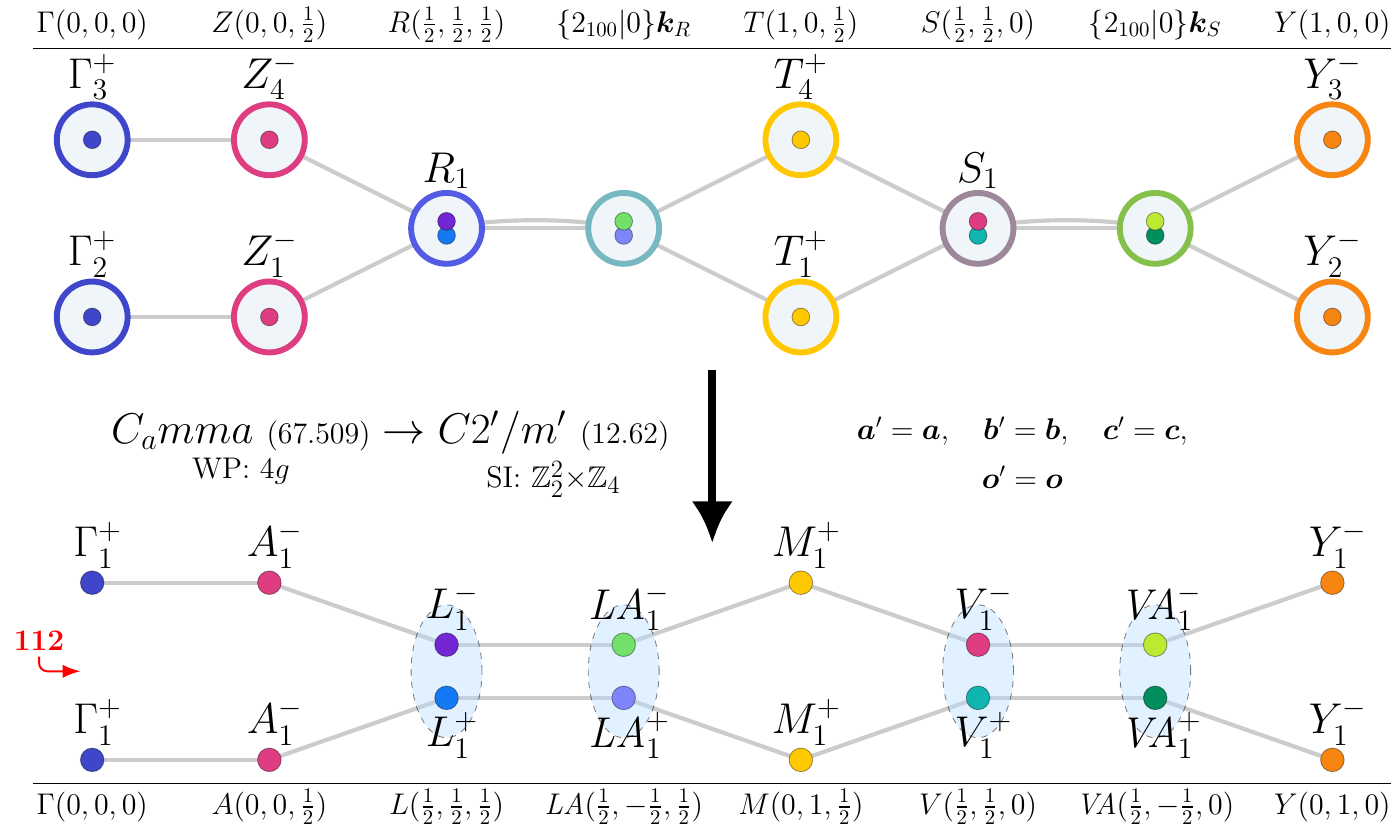}
\caption{Topological magnon bands in subgroup $C2'/m'~(12.62)$ for magnetic moments on Wyckoff position $4g$ of supergroup $C_{a}mma~(67.509)$.\label{fig_67.509_12.62_Bparallel100andstrainperp010_4g}}
\end{figure}
\input{gap_tables_tex/67.509_12.62_Bparallel100andstrainperp010_4g_table.tex}
\input{si_tables_tex/67.509_12.62_Bparallel100andstrainperp010_4g_table.tex}
\subsection{WP: $4e$}
\textbf{BCS Materials:} {Tb\textsubscript{0.6}Y\textsubscript{0.4}RhIn\textsubscript{5}~(37.6 K)}\footnote{BCS web page: \texttt{\href{http://webbdcrista1.ehu.es/magndata/index.php?this\_label=1.467} {http://webbdcrista1.ehu.es/magndata/index.php?this\_label=1.467}}}.\\
\subsubsection{Topological bands in subgroup $C2'/m'~(12.62)$}
\textbf{Perturbations:}
\begin{itemize}
\item B $\parallel$ [100] and strain $\perp$ [010],
\item B $\parallel$ [001] and strain $\perp$ [010],
\item B $\perp$ [010].
\end{itemize}
\begin{figure}[H]
\centering
\includegraphics[scale=0.6]{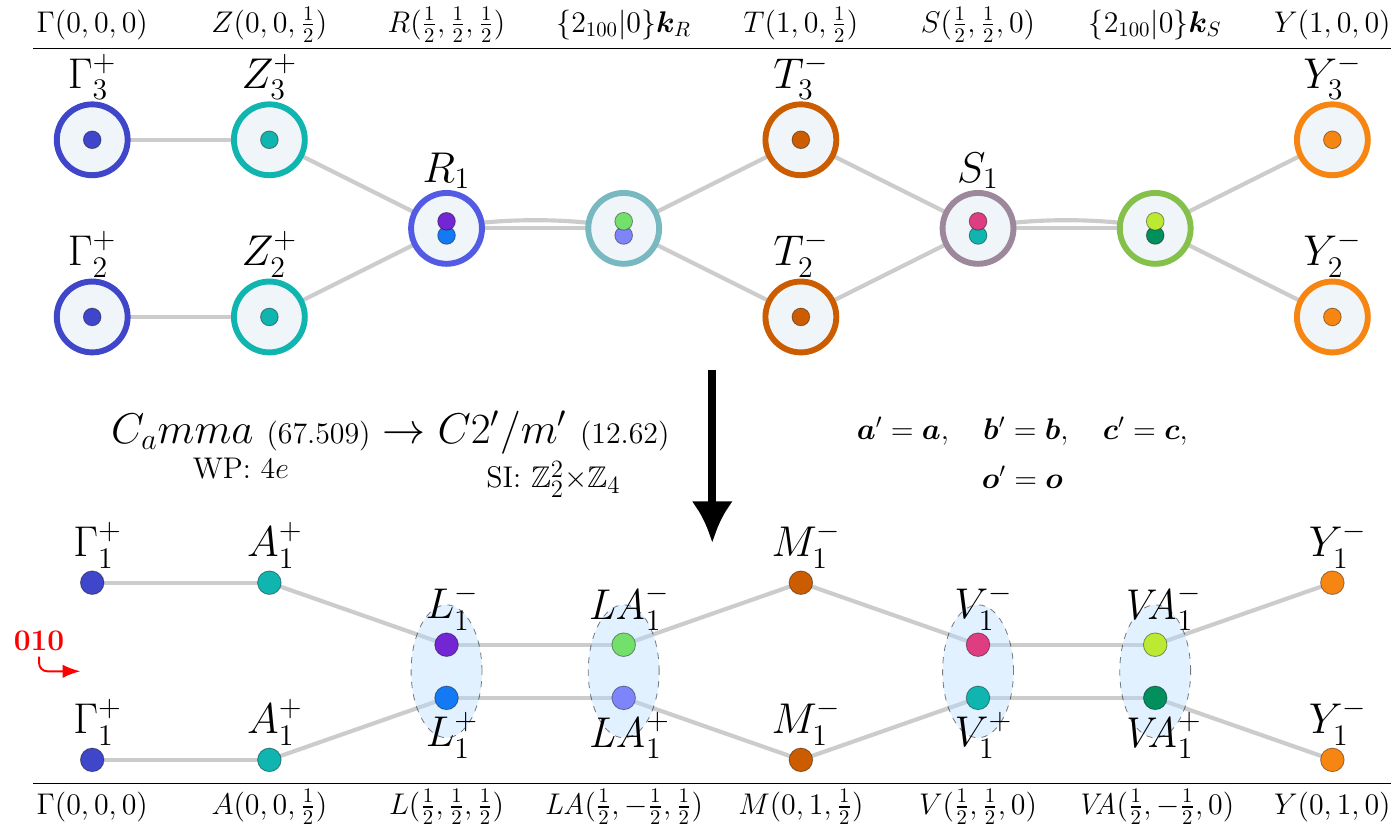}
\caption{Topological magnon bands in subgroup $C2'/m'~(12.62)$ for magnetic moments on Wyckoff position $4e$ of supergroup $C_{a}mma~(67.509)$.\label{fig_67.509_12.62_Bparallel100andstrainperp010_4e}}
\end{figure}
\input{gap_tables_tex/67.509_12.62_Bparallel100andstrainperp010_4e_table.tex}
\input{si_tables_tex/67.509_12.62_Bparallel100andstrainperp010_4e_table.tex}

\section{MSG $F_{S}mmm~(69.526)$}
\textbf{Nontrivial-SI Subgroups:} $P\bar{1}~(2.4)$, $C2'/m'~(12.62)$, $C2'/m'~(12.62)$, $C2'/m'~(12.62)$, $P_{S}\bar{1}~(2.7)$, $Fm'm'2~(42.222)$, $C_{a}2~(5.17)$, $C2/m~(12.58)$, $Fm'm'm~(69.524)$, $C_{a}2/m~(12.64)$, $Fm'm'2~(42.222)$, $C_{a}2~(5.17)$, $C2/m~(12.58)$, $Fm'm'm~(69.524)$, $C_{a}2/m~(12.64)$, $Fm'm'2~(42.222)$, $C_{a}2~(5.17)$, $C2/m~(12.58)$, $Fm'm'm~(69.524)$, $C_{a}2/m~(12.64)$.\\

\textbf{Trivial-SI Subgroups:} $Cm'~(8.34)$, $Cm'~(8.34)$, $Cm'~(8.34)$, $C2'~(5.15)$, $C2'~(5.15)$, $C2'~(5.15)$, $P_{S}1~(1.3)$, $Cm~(8.32)$, $Fm'm2'~(42.221)$, $Fm'm2'~(42.221)$, $C_{a}m~(8.36)$, $Cm~(8.32)$, $Fm'm2'~(42.221)$, $Fm'm2'~(42.221)$, $C_{a}m~(8.36)$, $Cm~(8.32)$, $Fm'm2'~(42.221)$, $Fm'm2'~(42.221)$, $C_{a}m~(8.36)$, $C2~(5.13)$, $F_{S}mm2~(42.223)$, $C2~(5.13)$, $F_{S}mm2~(42.223)$, $C2~(5.13)$, $F_{S}mm2~(42.223)$.\\

\subsection{WP: $8g$}
\textbf{BCS Materials:} {SrFeO\textsubscript{2}~(473 K)}\footnote{BCS web page: \texttt{\href{http://webbdcrista1.ehu.es/magndata/index.php?this\_label=1.65} {http://webbdcrista1.ehu.es/magndata/index.php?this\_label=1.65}}}.\\
\subsubsection{Topological bands in subgroup $C2'/m'~(12.62)$}
\textbf{Perturbations:}
\begin{itemize}
\item B $\parallel$ [010] and strain $\perp$ [100],
\item B $\parallel$ [001] and strain $\perp$ [100],
\item B $\perp$ [100].
\end{itemize}
\begin{figure}[H]
\centering
\includegraphics[scale=0.6]{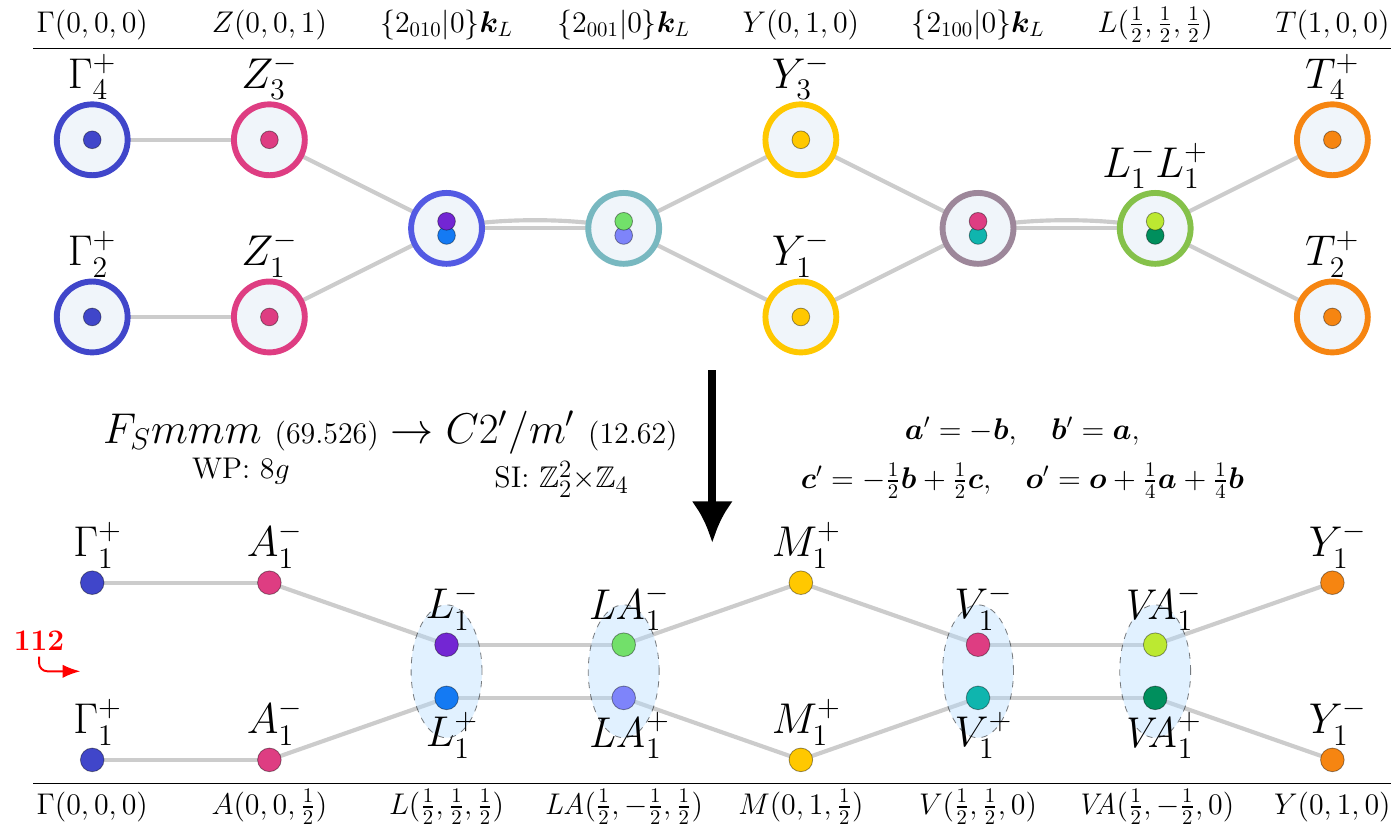}
\caption{Topological magnon bands in subgroup $C2'/m'~(12.62)$ for magnetic moments on Wyckoff position $8g$ of supergroup $F_{S}mmm~(69.526)$.\label{fig_69.526_12.62_Bparallel010andstrainperp100_8g}}
\end{figure}
\input{gap_tables_tex/69.526_12.62_Bparallel010andstrainperp100_8g_table.tex}
\input{si_tables_tex/69.526_12.62_Bparallel010andstrainperp100_8g_table.tex}

\section{MSG $Fd'd'd~(70.530)$}
\textbf{Nontrivial-SI Subgroups:} $P\bar{1}~(2.4)$, $C2'/c'~(15.89)$, $C2'/c'~(15.89)$, $C2/c~(15.85)$.\\

\textbf{Trivial-SI Subgroups:} $Cc'~(9.39)$, $Cc'~(9.39)$, $C2'~(5.15)$, $C2'~(5.15)$, $Cc~(9.37)$, $Fd'd2'~(43.226)$, $Fd'd2'~(43.226)$, $C2~(5.13)$, $Fd'd'2~(43.227)$.\\

\subsection{WP: $16c+8b$}
\textbf{BCS Materials:} {NiCr\textsubscript{2}O\textsubscript{4}~(74 K)}\footnote{BCS web page: \texttt{\href{http://webbdcrista1.ehu.es/magndata/index.php?this\_label=0.4} {http://webbdcrista1.ehu.es/magndata/index.php?this\_label=0.4}}}.\\
\subsubsection{Topological bands in subgroup $P\bar{1}~(2.4)$}
\textbf{Perturbations:}
\begin{itemize}
\item strain in generic direction,
\item (B $\parallel$ [100] or B $\perp$ [010]) and strain $\perp$ [100],
\item (B $\parallel$ [100] or B $\perp$ [010]) and strain $\perp$ [001],
\item (B $\parallel$ [010] or B $\perp$ [100]) and strain $\perp$ [010],
\item (B $\parallel$ [010] or B $\perp$ [100]) and strain $\perp$ [001],
\item B in generic direction.
\end{itemize}
\begin{figure}[H]
\centering
\includegraphics[scale=0.6]{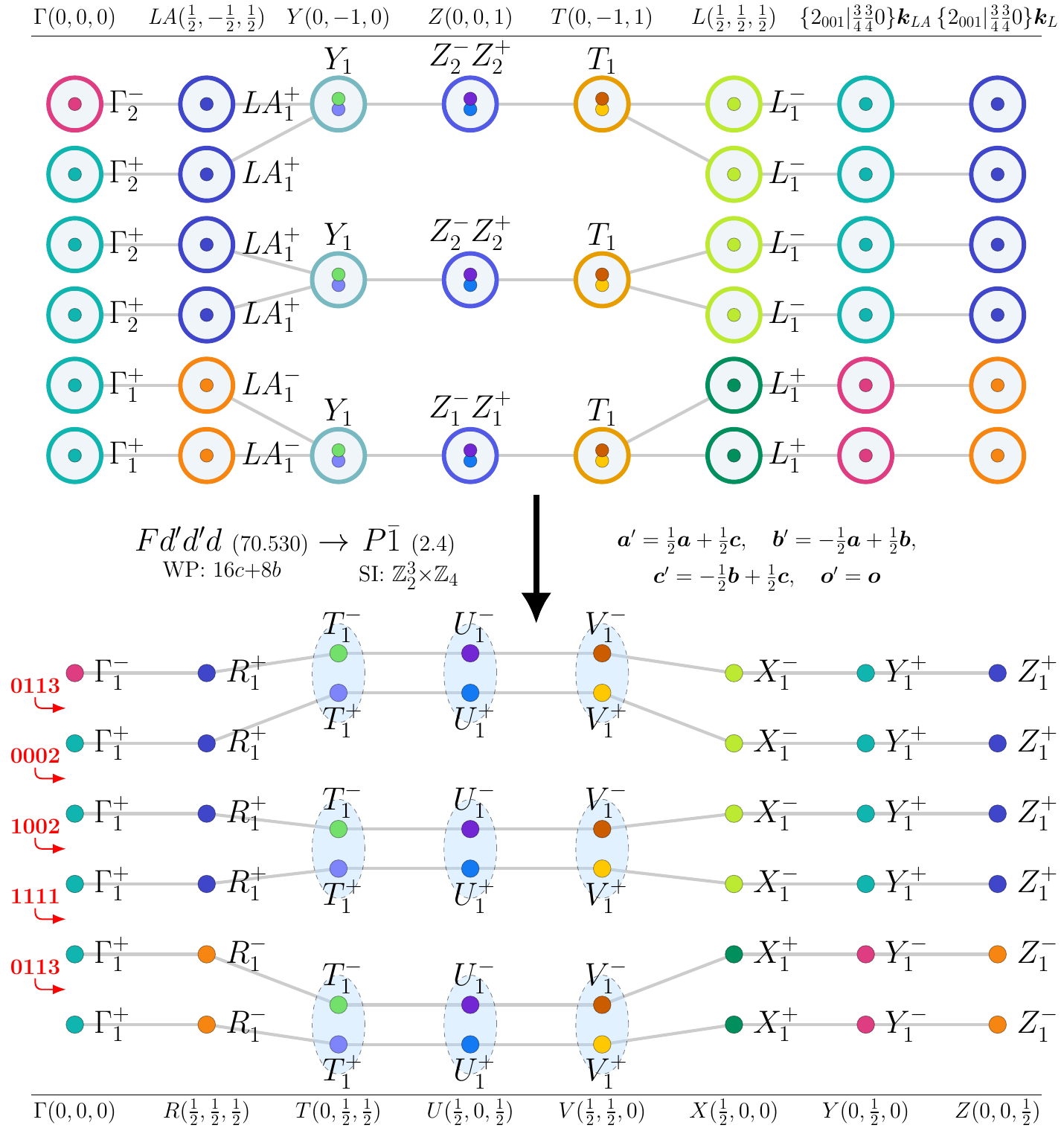}
\caption{Topological magnon bands in subgroup $P\bar{1}~(2.4)$ for magnetic moments on Wyckoff positions $16c+8b$ of supergroup $Fd'd'd~(70.530)$.\label{fig_70.530_2.4_strainingenericdirection_16c+8b}}
\end{figure}
\input{gap_tables_tex/70.530_2.4_strainingenericdirection_16c+8b_table.tex}
\input{si_tables_tex/70.530_2.4_strainingenericdirection_16c+8b_table.tex}
\subsubsection{Topological bands in subgroup $C2'/c'~(15.89)$}
\textbf{Perturbations:}
\begin{itemize}
\item strain $\perp$ [010],
\item (B $\parallel$ [100] or B $\perp$ [010]).
\end{itemize}
\begin{figure}[H]
\centering
\includegraphics[scale=0.6]{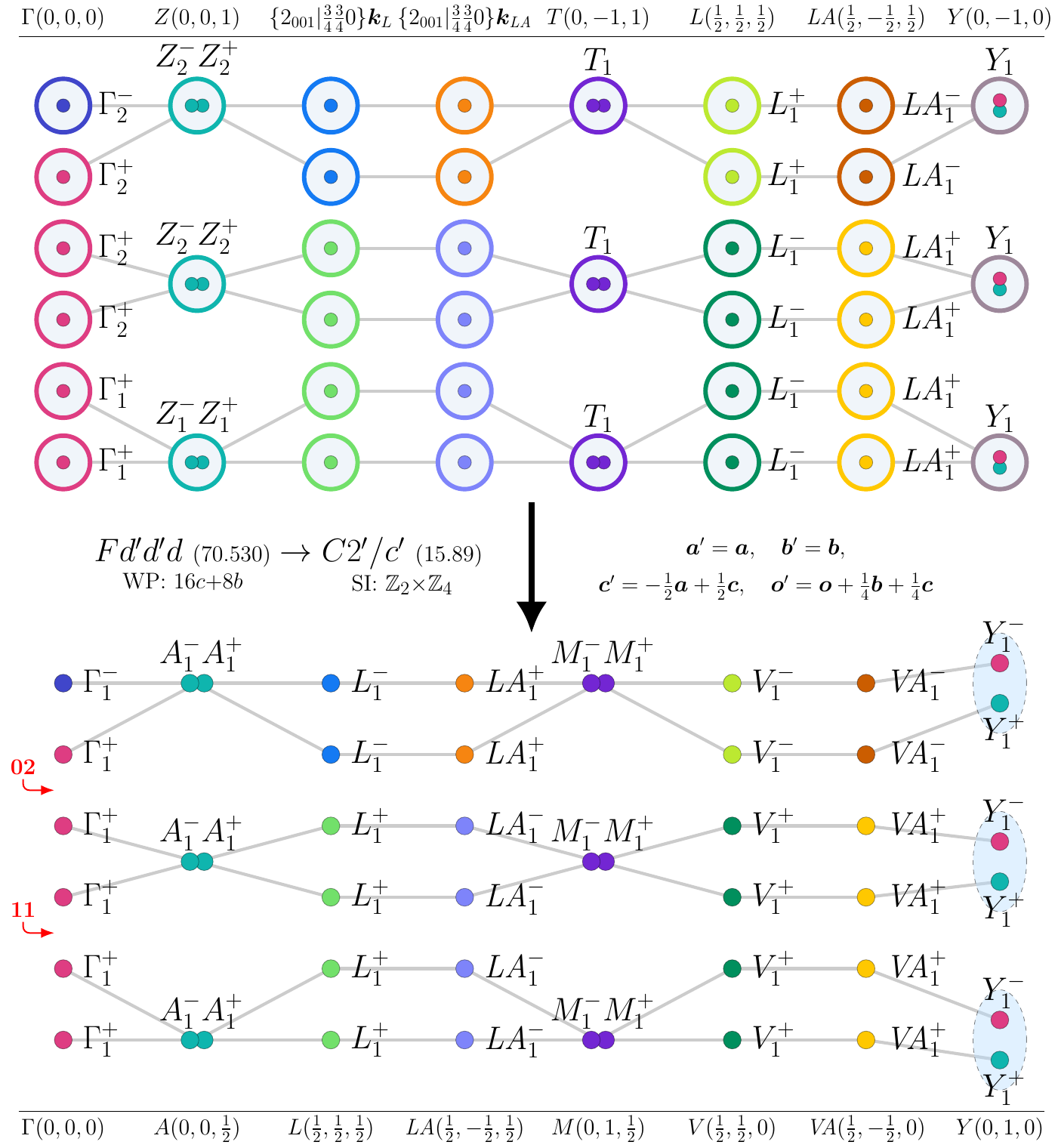}
\caption{Topological magnon bands in subgroup $C2'/c'~(15.89)$ for magnetic moments on Wyckoff positions $16c+8b$ of supergroup $Fd'd'd~(70.530)$.\label{fig_70.530_15.89_strainperp010_16c+8b}}
\end{figure}
\input{gap_tables_tex/70.530_15.89_strainperp010_16c+8b_table.tex}
\input{si_tables_tex/70.530_15.89_strainperp010_16c+8b_table.tex}
\subsubsection{Topological bands in subgroup $C2'/c'~(15.89)$}
\textbf{Perturbations:}
\begin{itemize}
\item strain $\perp$ [100],
\item (B $\parallel$ [010] or B $\perp$ [100]).
\end{itemize}
\begin{figure}[H]
\centering
\includegraphics[scale=0.6]{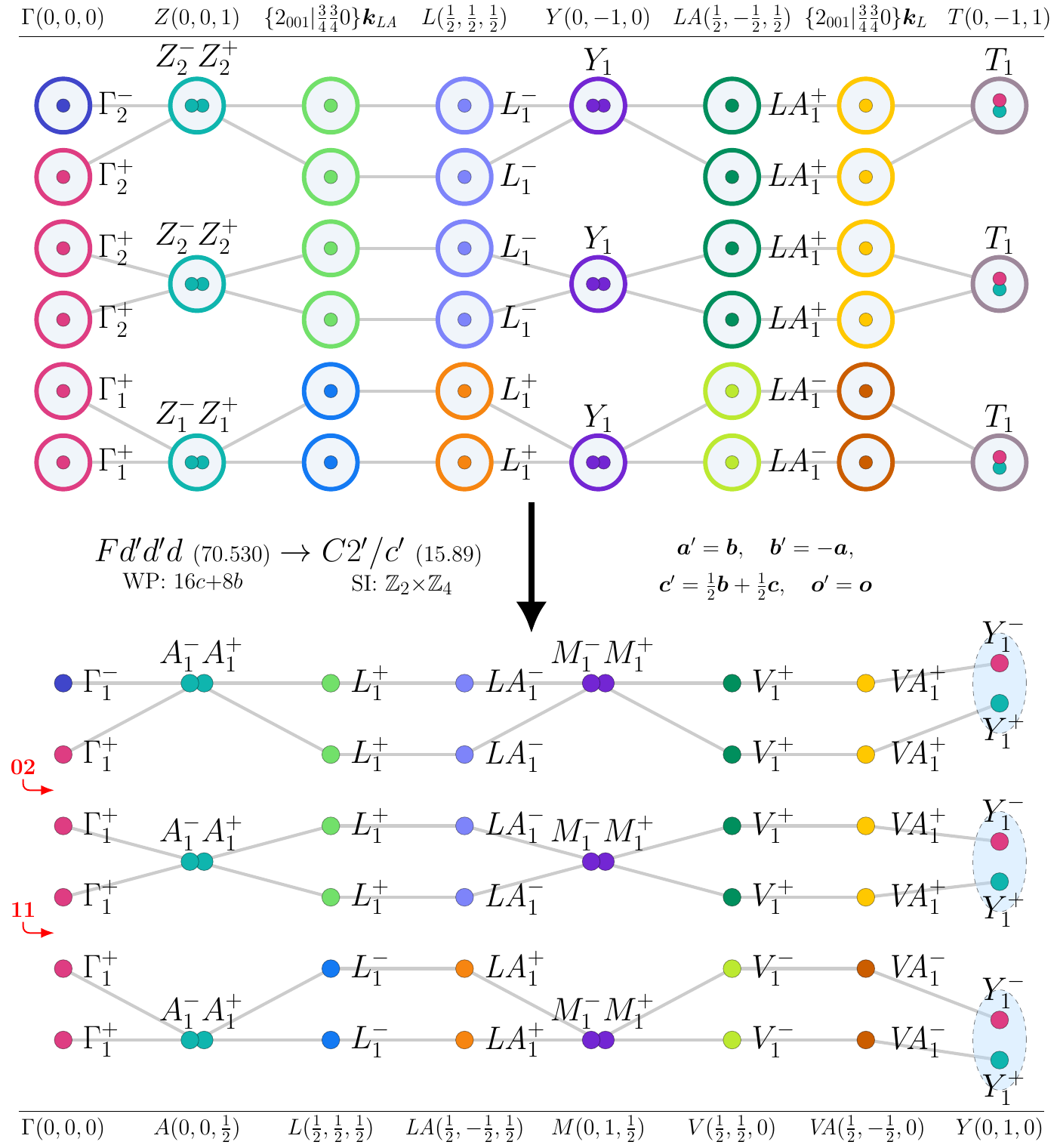}
\caption{Topological magnon bands in subgroup $C2'/c'~(15.89)$ for magnetic moments on Wyckoff positions $16c+8b$ of supergroup $Fd'd'd~(70.530)$.\label{fig_70.530_15.89_strainperp100_16c+8b}}
\end{figure}
\input{gap_tables_tex/70.530_15.89_strainperp100_16c+8b_table.tex}
\input{si_tables_tex/70.530_15.89_strainperp100_16c+8b_table.tex}
\subsection{WP: $16c$}
\textbf{BCS Materials:} {Bi\textsubscript{2}RuMnO\textsubscript{7}~(20 K)}\footnote{BCS web page: \texttt{\href{http://webbdcrista1.ehu.es/magndata/index.php?this\_label=0.153} {http://webbdcrista1.ehu.es/magndata/index.php?this\_label=0.153}}}.\\
\subsubsection{Topological bands in subgroup $P\bar{1}~(2.4)$}
\textbf{Perturbations:}
\begin{itemize}
\item strain in generic direction,
\item (B $\parallel$ [100] or B $\perp$ [010]) and strain $\perp$ [100],
\item (B $\parallel$ [100] or B $\perp$ [010]) and strain $\perp$ [001],
\item (B $\parallel$ [010] or B $\perp$ [100]) and strain $\perp$ [010],
\item (B $\parallel$ [010] or B $\perp$ [100]) and strain $\perp$ [001],
\item B in generic direction.
\end{itemize}
\begin{figure}[H]
\centering
\includegraphics[scale=0.6]{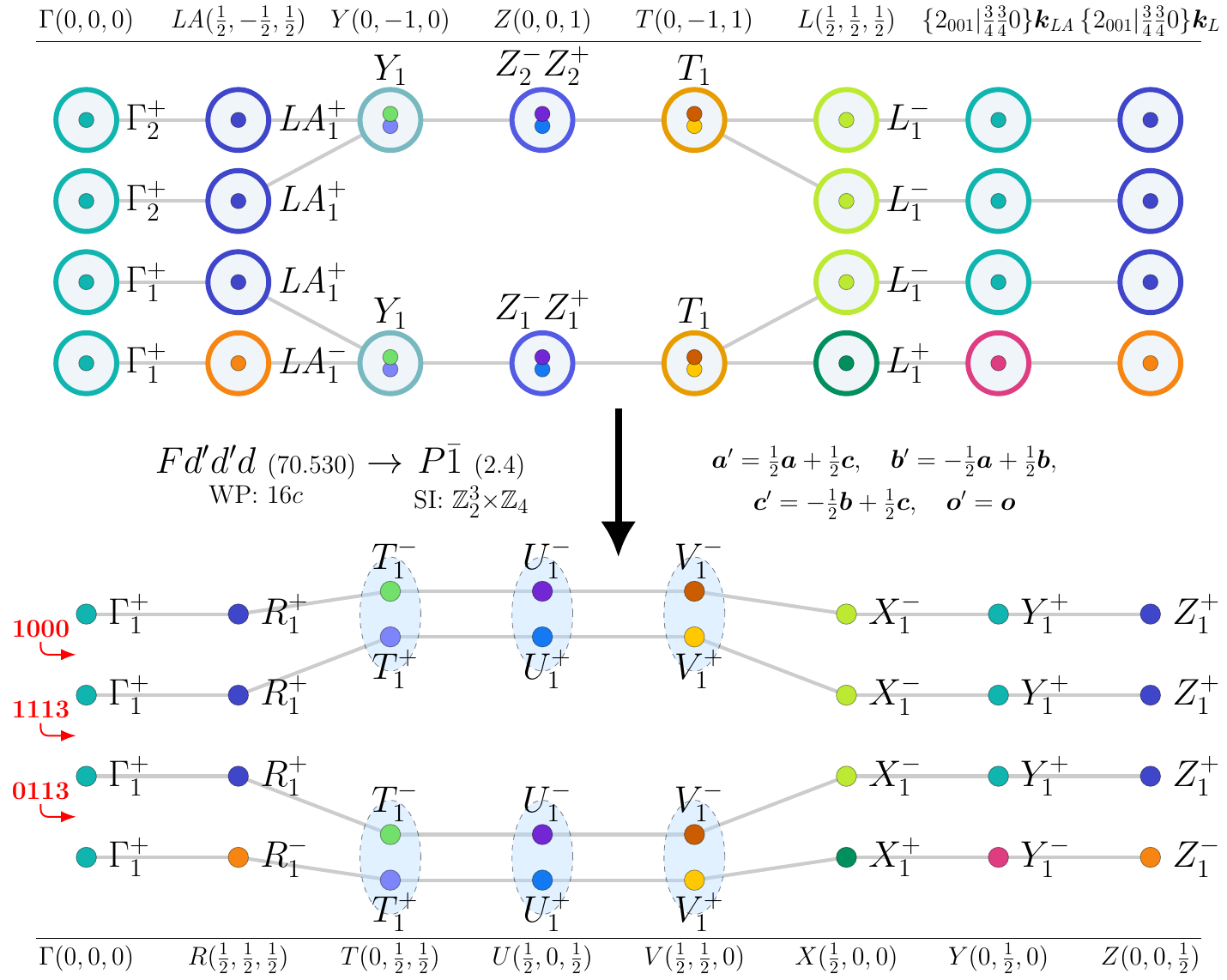}
\caption{Topological magnon bands in subgroup $P\bar{1}~(2.4)$ for magnetic moments on Wyckoff position $16c$ of supergroup $Fd'd'd~(70.530)$.\label{fig_70.530_2.4_strainingenericdirection_16c}}
\end{figure}
\input{gap_tables_tex/70.530_2.4_strainingenericdirection_16c_table.tex}
\input{si_tables_tex/70.530_2.4_strainingenericdirection_16c_table.tex}
\subsubsection{Topological bands in subgroup $C2'/c'~(15.89)$}
\textbf{Perturbations:}
\begin{itemize}
\item strain $\perp$ [010],
\item (B $\parallel$ [100] or B $\perp$ [010]).
\end{itemize}
\begin{figure}[H]
\centering
\includegraphics[scale=0.6]{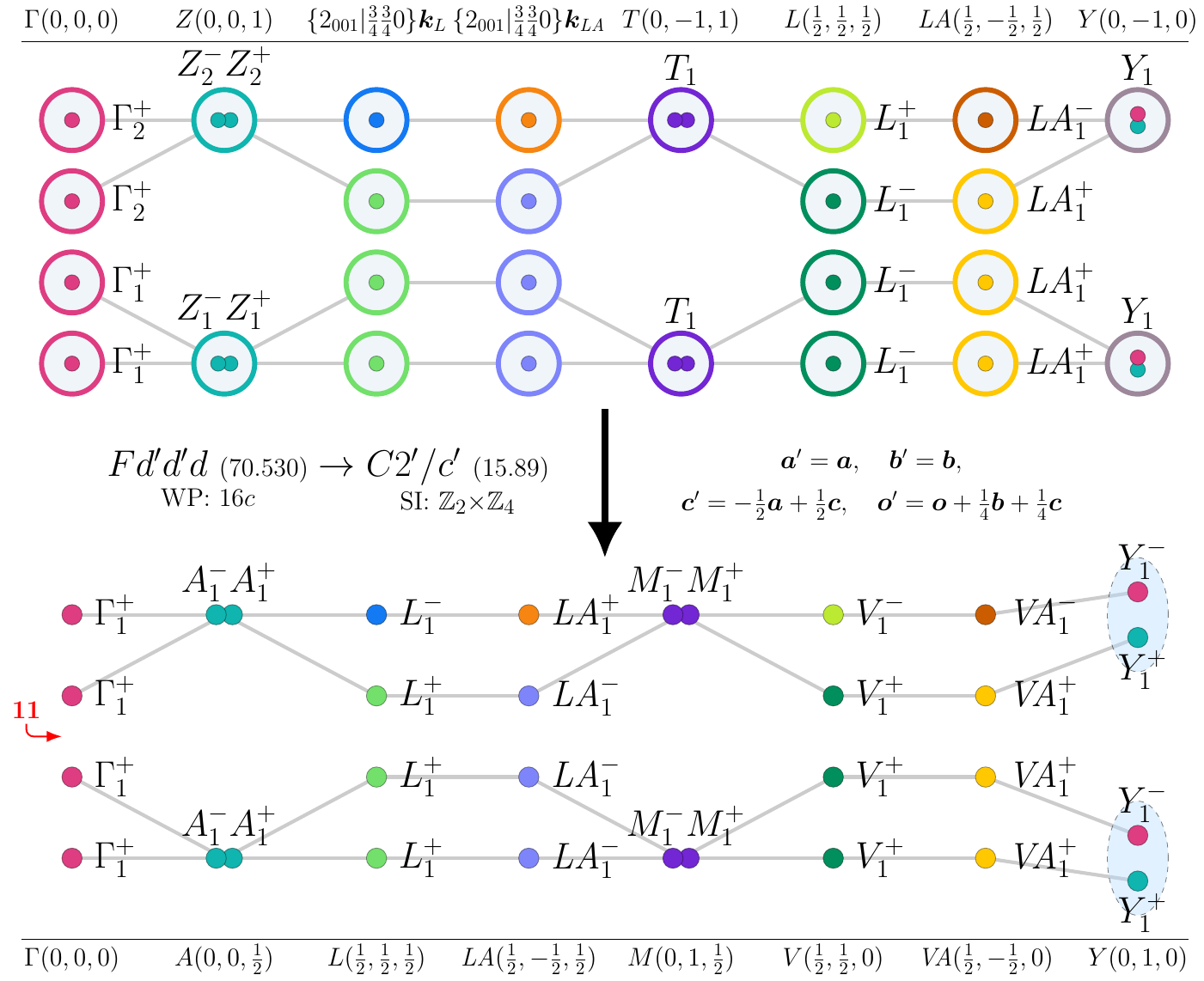}
\caption{Topological magnon bands in subgroup $C2'/c'~(15.89)$ for magnetic moments on Wyckoff position $16c$ of supergroup $Fd'd'd~(70.530)$.\label{fig_70.530_15.89_strainperp010_16c}}
\end{figure}
\input{gap_tables_tex/70.530_15.89_strainperp010_16c_table.tex}
\input{si_tables_tex/70.530_15.89_strainperp010_16c_table.tex}
\subsubsection{Topological bands in subgroup $C2'/c'~(15.89)$}
\textbf{Perturbations:}
\begin{itemize}
\item strain $\perp$ [100],
\item (B $\parallel$ [010] or B $\perp$ [100]).
\end{itemize}
\begin{figure}[H]
\centering
\includegraphics[scale=0.6]{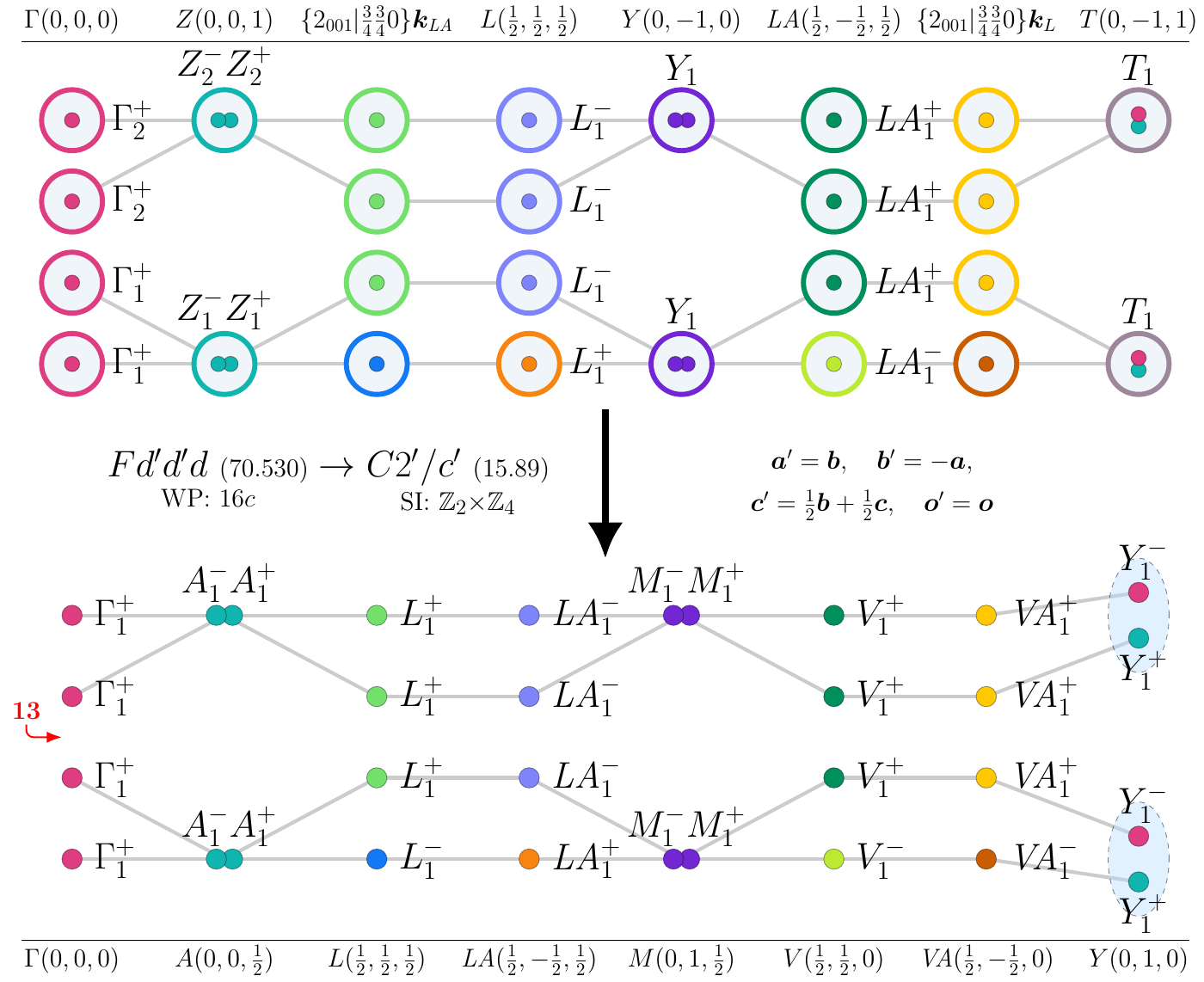}
\caption{Topological magnon bands in subgroup $C2'/c'~(15.89)$ for magnetic moments on Wyckoff position $16c$ of supergroup $Fd'd'd~(70.530)$.\label{fig_70.530_15.89_strainperp100_16c}}
\end{figure}
\input{gap_tables_tex/70.530_15.89_strainperp100_16c_table.tex}
\input{si_tables_tex/70.530_15.89_strainperp100_16c_table.tex}

\section{MSG $I_{c}bca~(73.553)$}
\textbf{Nontrivial-SI Subgroups:} $C_{a}c~(9.41)$, $P\bar{1}~(2.4)$, $C2'/c'~(15.89)$, $C2'/m'~(12.62)$, $C2'/m'~(12.62)$, $P_{S}\bar{1}~(2.7)$, $C_{a}2~(5.17)$, $I_{c}ba2~(45.239)$, $C2/c~(15.85)$, $Im'm'a~(74.558)$, $C_{a}2/c~(15.91)$, $I_{a}ba2~(45.240)$, $C2/c~(15.85)$, $Iba'm'~(72.544)$, $C_{c}2/c~(15.90)$, $I_{a}ba2~(45.240)$, $C2/c~(15.85)$, $Iba'm'~(72.544)$, $C_{c}2/c~(15.90)$.\\

\textbf{Trivial-SI Subgroups:} $Cc'~(9.39)$, $Cm'~(8.34)$, $Cm'~(8.34)$, $C2'~(5.15)$, $C2'~(5.15)$, $C2'~(5.15)$, $P_{S}1~(1.3)$, $Cc~(9.37)$, $Im'a2'~(46.243)$, $Im'a2'~(46.243)$, $Cc~(9.37)$, $Im'a2'~(46.243)$, $Ib'a2'~(45.237)$, $C_{c}c~(9.40)$, $Cc~(9.37)$, $Im'a2'~(46.243)$, $Ib'a2'~(45.237)$, $C_{c}c~(9.40)$, $C2~(5.13)$, $Im'm'2~(44.232)$, $C2~(5.13)$, $Im'a'2~(46.245)$, $C_{c}2~(5.16)$, $C2~(5.13)$, $Im'a'2~(46.245)$, $C_{c}2~(5.16)$.\\

\subsection{WP: $8g$}
\textbf{BCS Materials:} {YbCo\textsubscript{2}Si\textsubscript{2}~(0.9 K)}\footnote{BCS web page: \texttt{\href{http://webbdcrista1.ehu.es/magndata/index.php?this\_label=1.176} {http://webbdcrista1.ehu.es/magndata/index.php?this\_label=1.176}}}.\\
\subsubsection{Topological bands in subgroup $C_{a}2~(5.17)$}
\textbf{Perturbation:}
\begin{itemize}
\item E $\parallel$ [001] and strain $\perp$ [001].
\end{itemize}
\begin{figure}[H]
\centering
\includegraphics[scale=0.6]{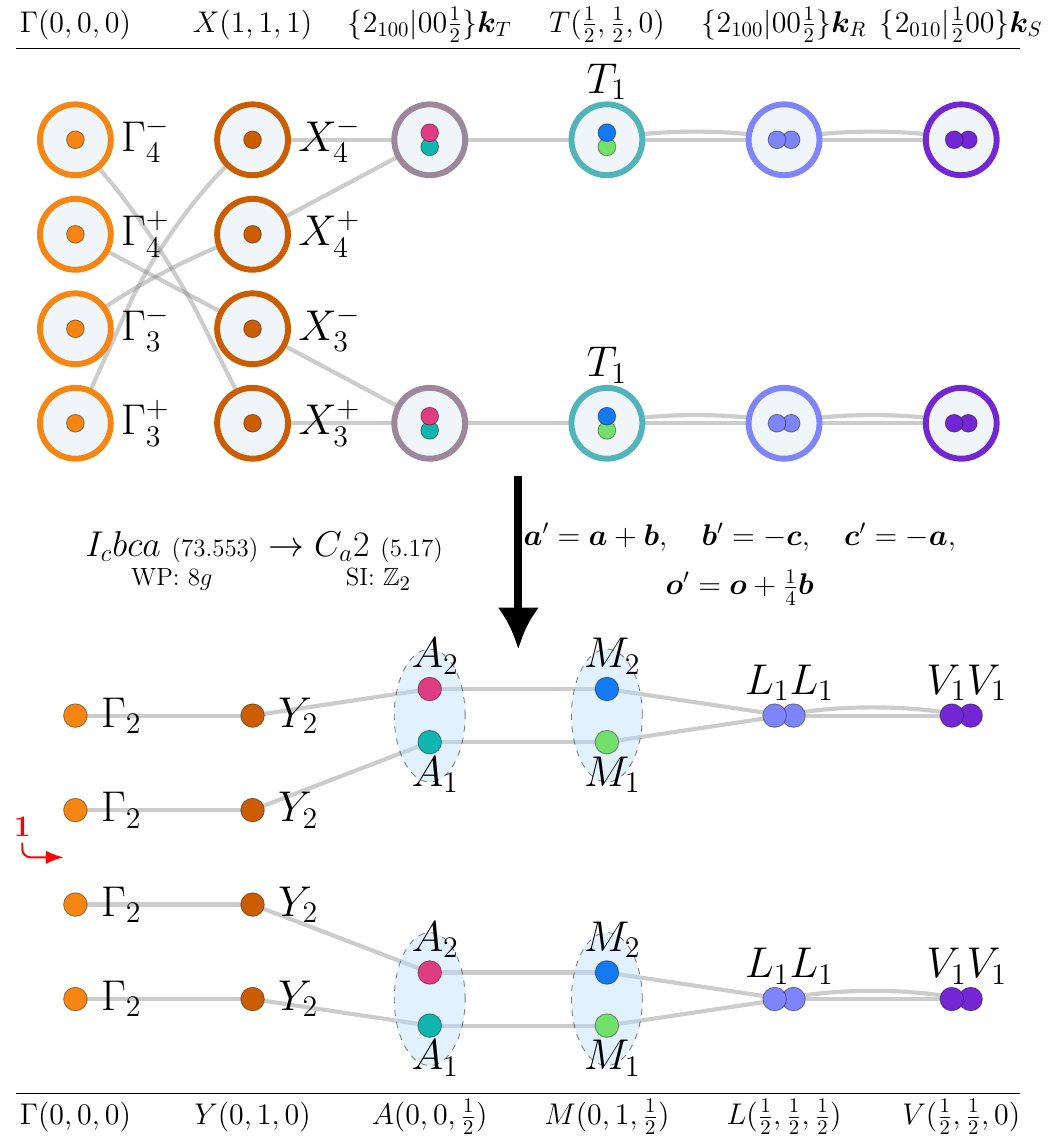}
\caption{Topological magnon bands in subgroup $C_{a}2~(5.17)$ for magnetic moments on Wyckoff position $8g$ of supergroup $I_{c}bca~(73.553)$.\label{fig_73.553_5.17_Eparallel001andstrainperp001_8g}}
\end{figure}
\input{gap_tables_tex/73.553_5.17_Eparallel001andstrainperp001_8g_table.tex}
\input{si_tables_tex/73.553_5.17_Eparallel001andstrainperp001_8g_table.tex}

\section{MSG $Im'm'a~(74.558)$}
\textbf{Nontrivial-SI Subgroups:} $P\bar{1}~(2.4)$, $C2'/m'~(12.62)$, $C2'/m'~(12.62)$, $C2/c~(15.85)$.\\

\textbf{Trivial-SI Subgroups:} $Cm'~(8.34)$, $Cm'~(8.34)$, $C2'~(5.15)$, $C2'~(5.15)$, $Cc~(9.37)$, $Im'a2'~(46.243)$, $Im'a2'~(46.243)$, $C2~(5.13)$, $Im'm'2~(44.232)$.\\

\subsection{WP: $4b$}
\textbf{BCS Materials:} {Pr\textsubscript{0.5}Sr\textsubscript{0.5}CoO\textsubscript{3}~(230 K)}\footnote{BCS web page: \texttt{\href{http://webbdcrista1.ehu.es/magndata/index.php?this\_label=0.304} {http://webbdcrista1.ehu.es/magndata/index.php?this\_label=0.304}}}, {Pr\textsubscript{0.5}Sr\textsubscript{0.4}Ba\textsubscript{0.1}CoO\textsubscript{3}~(226 K)}\footnote{BCS web page: \texttt{\href{http://webbdcrista1.ehu.es/magndata/index.php?this\_label=0.717} {http://webbdcrista1.ehu.es/magndata/index.php?this\_label=0.717}}}.\\
\subsubsection{Topological bands in subgroup $P\bar{1}~(2.4)$}
\textbf{Perturbations:}
\begin{itemize}
\item strain in generic direction,
\item (B $\parallel$ [100] or B $\perp$ [010]) and strain $\perp$ [100],
\item (B $\parallel$ [100] or B $\perp$ [010]) and strain $\perp$ [001],
\item (B $\parallel$ [010] or B $\perp$ [100]) and strain $\perp$ [010],
\item (B $\parallel$ [010] or B $\perp$ [100]) and strain $\perp$ [001],
\item B in generic direction.
\end{itemize}
\begin{figure}[H]
\centering
\includegraphics[scale=0.6]{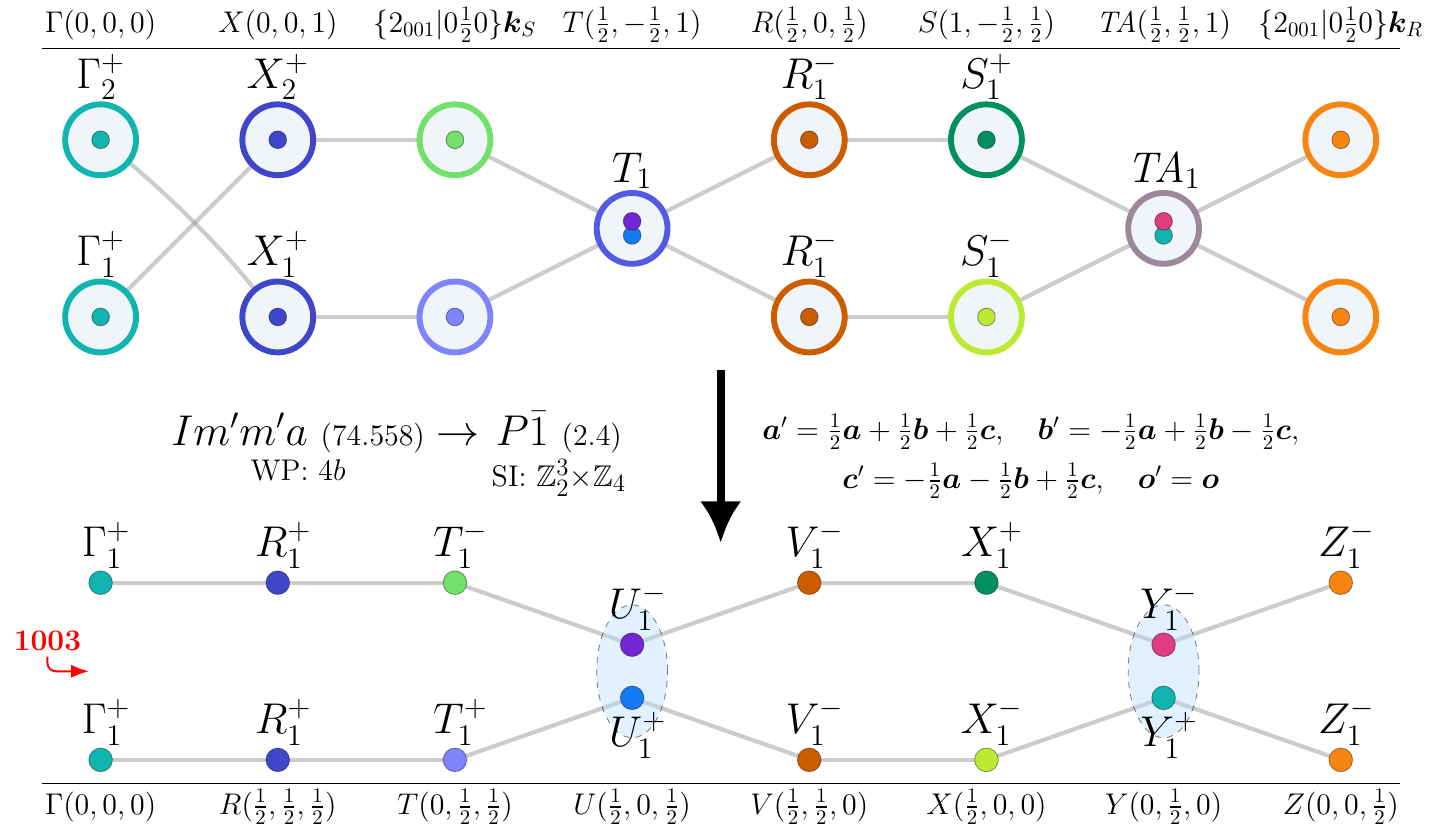}
\caption{Topological magnon bands in subgroup $P\bar{1}~(2.4)$ for magnetic moments on Wyckoff position $4b$ of supergroup $Im'm'a~(74.558)$.\label{fig_74.558_2.4_strainingenericdirection_4b}}
\end{figure}
\input{gap_tables_tex/74.558_2.4_strainingenericdirection_4b_table.tex}
\input{si_tables_tex/74.558_2.4_strainingenericdirection_4b_table.tex}
\subsubsection{Topological bands in subgroup $C2'/m'~(12.62)$}
\textbf{Perturbations:}
\begin{itemize}
\item strain $\perp$ [010],
\item (B $\parallel$ [100] or B $\perp$ [010]).
\end{itemize}
\begin{figure}[H]
\centering
\includegraphics[scale=0.6]{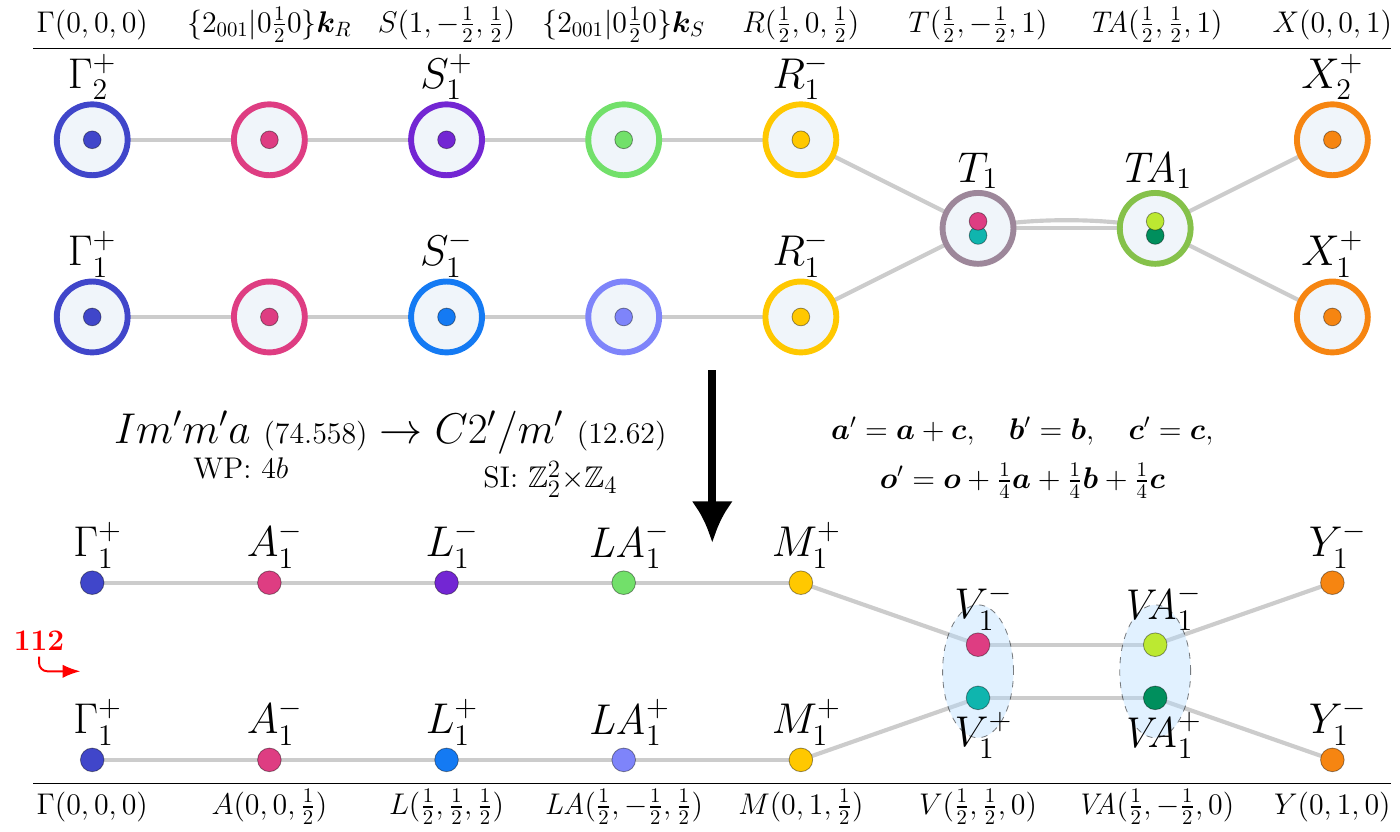}
\caption{Topological magnon bands in subgroup $C2'/m'~(12.62)$ for magnetic moments on Wyckoff position $4b$ of supergroup $Im'm'a~(74.558)$.\label{fig_74.558_12.62_strainperp010_4b}}
\end{figure}
\input{gap_tables_tex/74.558_12.62_strainperp010_4b_table.tex}
\input{si_tables_tex/74.558_12.62_strainperp010_4b_table.tex}
\subsubsection{Topological bands in subgroup $C2'/m'~(12.62)$}
\textbf{Perturbations:}
\begin{itemize}
\item strain $\perp$ [100],
\item (B $\parallel$ [010] or B $\perp$ [100]).
\end{itemize}
\begin{figure}[H]
\centering
\includegraphics[scale=0.6]{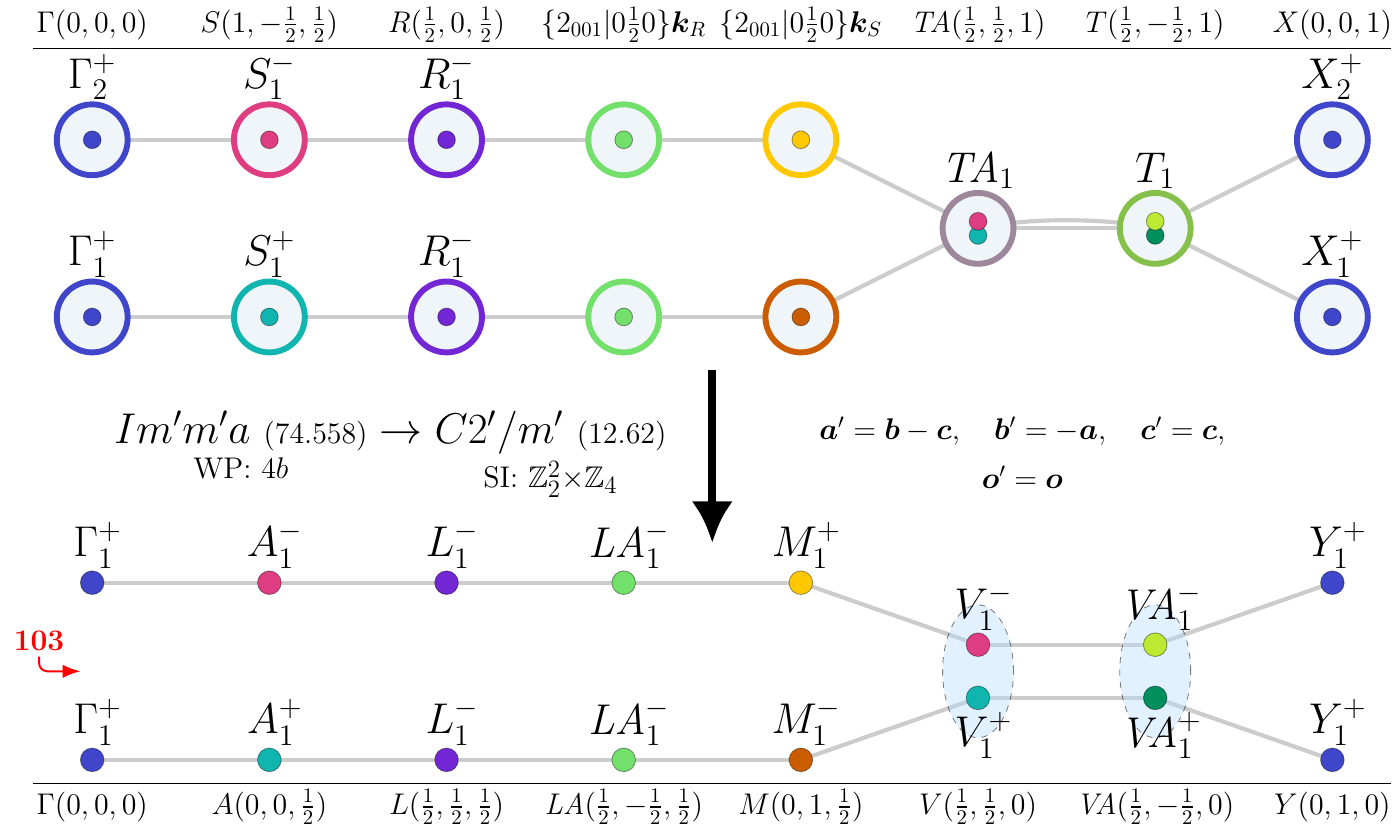}
\caption{Topological magnon bands in subgroup $C2'/m'~(12.62)$ for magnetic moments on Wyckoff position $4b$ of supergroup $Im'm'a~(74.558)$.\label{fig_74.558_12.62_strainperp100_4b}}
\end{figure}
\input{gap_tables_tex/74.558_12.62_strainperp100_4b_table.tex}
\input{si_tables_tex/74.558_12.62_strainperp100_4b_table.tex}
\subsection{WP: $4b+4c$}
\textbf{BCS Materials:} {Na\textsubscript{2}NiCrF\textsubscript{7}~(4 K)}\footnote{BCS web page: \texttt{\href{http://webbdcrista1.ehu.es/magndata/index.php?this\_label=0.573} {http://webbdcrista1.ehu.es/magndata/index.php?this\_label=0.573}}}, {Na\textsubscript{2}NiCrF\textsubscript{7}~(4 K)}\footnote{BCS web page: \texttt{\href{http://webbdcrista1.ehu.es/magndata/index.php?this\_label=0.572} {http://webbdcrista1.ehu.es/magndata/index.php?this\_label=0.572}}}.\\
\subsubsection{Topological bands in subgroup $P\bar{1}~(2.4)$}
\textbf{Perturbations:}
\begin{itemize}
\item strain in generic direction,
\item (B $\parallel$ [100] or B $\perp$ [010]) and strain $\perp$ [100],
\item (B $\parallel$ [100] or B $\perp$ [010]) and strain $\perp$ [001],
\item (B $\parallel$ [010] or B $\perp$ [100]) and strain $\perp$ [010],
\item (B $\parallel$ [010] or B $\perp$ [100]) and strain $\perp$ [001],
\item B in generic direction.
\end{itemize}
\begin{figure}[H]
\centering
\includegraphics[scale=0.6]{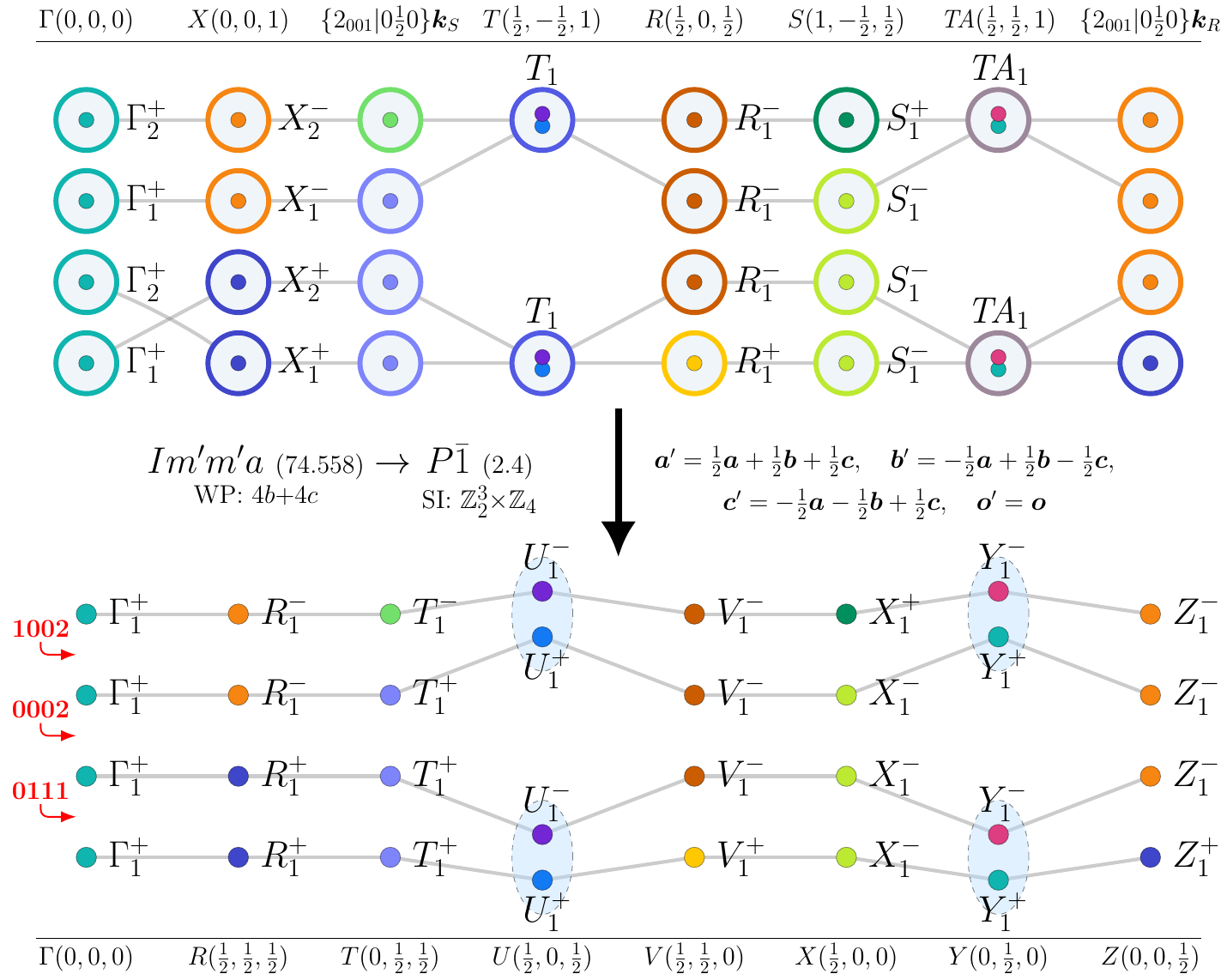}
\caption{Topological magnon bands in subgroup $P\bar{1}~(2.4)$ for magnetic moments on Wyckoff positions $4b+4c$ of supergroup $Im'm'a~(74.558)$.\label{fig_74.558_2.4_strainingenericdirection_4b+4c}}
\end{figure}
\input{gap_tables_tex/74.558_2.4_strainingenericdirection_4b+4c_table.tex}
\input{si_tables_tex/74.558_2.4_strainingenericdirection_4b+4c_table.tex}
\subsubsection{Topological bands in subgroup $C2'/m'~(12.62)$}
\textbf{Perturbations:}
\begin{itemize}
\item strain $\perp$ [010],
\item (B $\parallel$ [100] or B $\perp$ [010]).
\end{itemize}
\begin{figure}[H]
\centering
\includegraphics[scale=0.6]{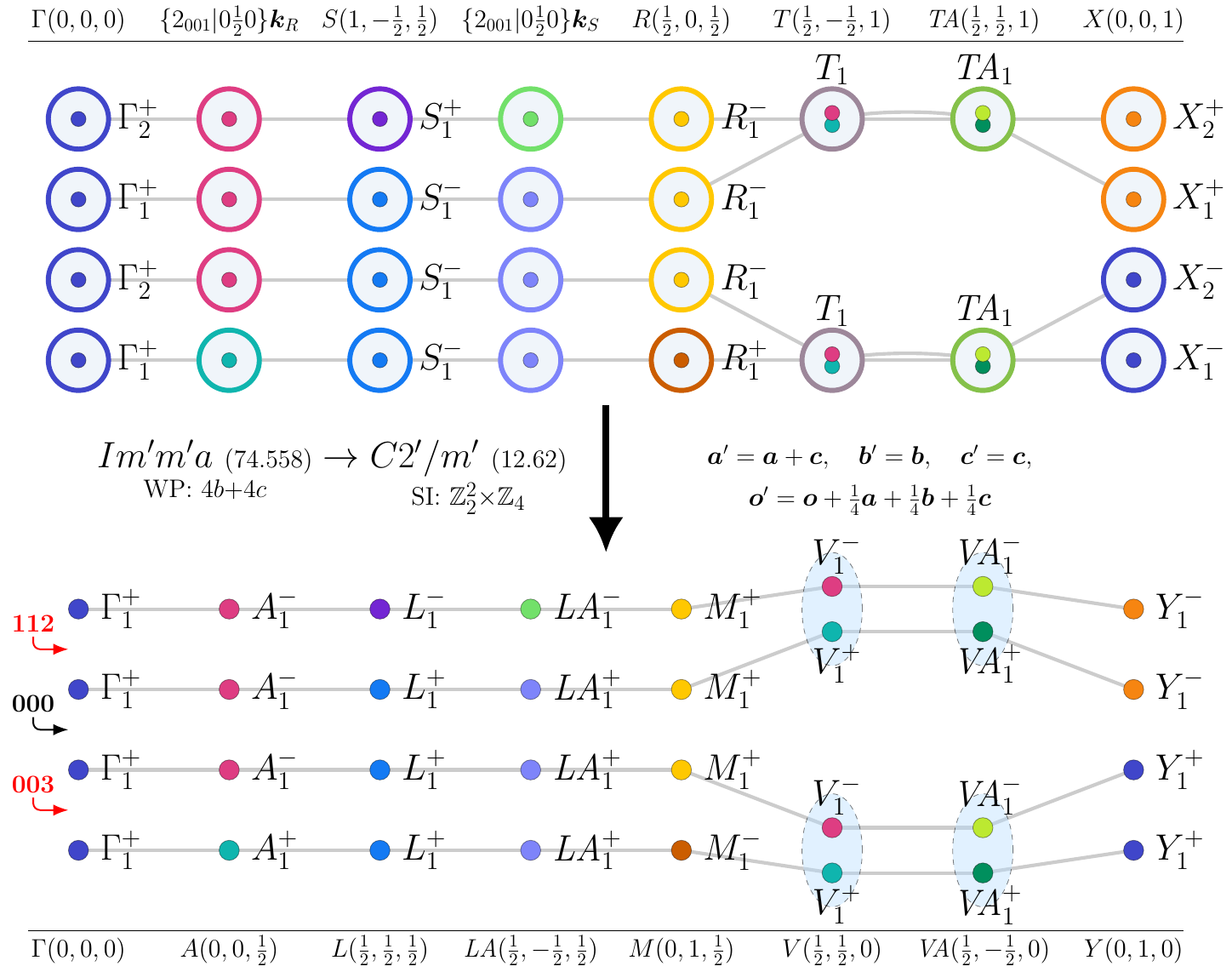}
\caption{Topological magnon bands in subgroup $C2'/m'~(12.62)$ for magnetic moments on Wyckoff positions $4b+4c$ of supergroup $Im'm'a~(74.558)$.\label{fig_74.558_12.62_strainperp010_4b+4c}}
\end{figure}
\input{gap_tables_tex/74.558_12.62_strainperp010_4b+4c_table.tex}
\input{si_tables_tex/74.558_12.62_strainperp010_4b+4c_table.tex}
\subsubsection{Topological bands in subgroup $C2'/m'~(12.62)$}
\textbf{Perturbations:}
\begin{itemize}
\item strain $\perp$ [100],
\item (B $\parallel$ [010] or B $\perp$ [100]).
\end{itemize}
\begin{figure}[H]
\centering
\includegraphics[scale=0.6]{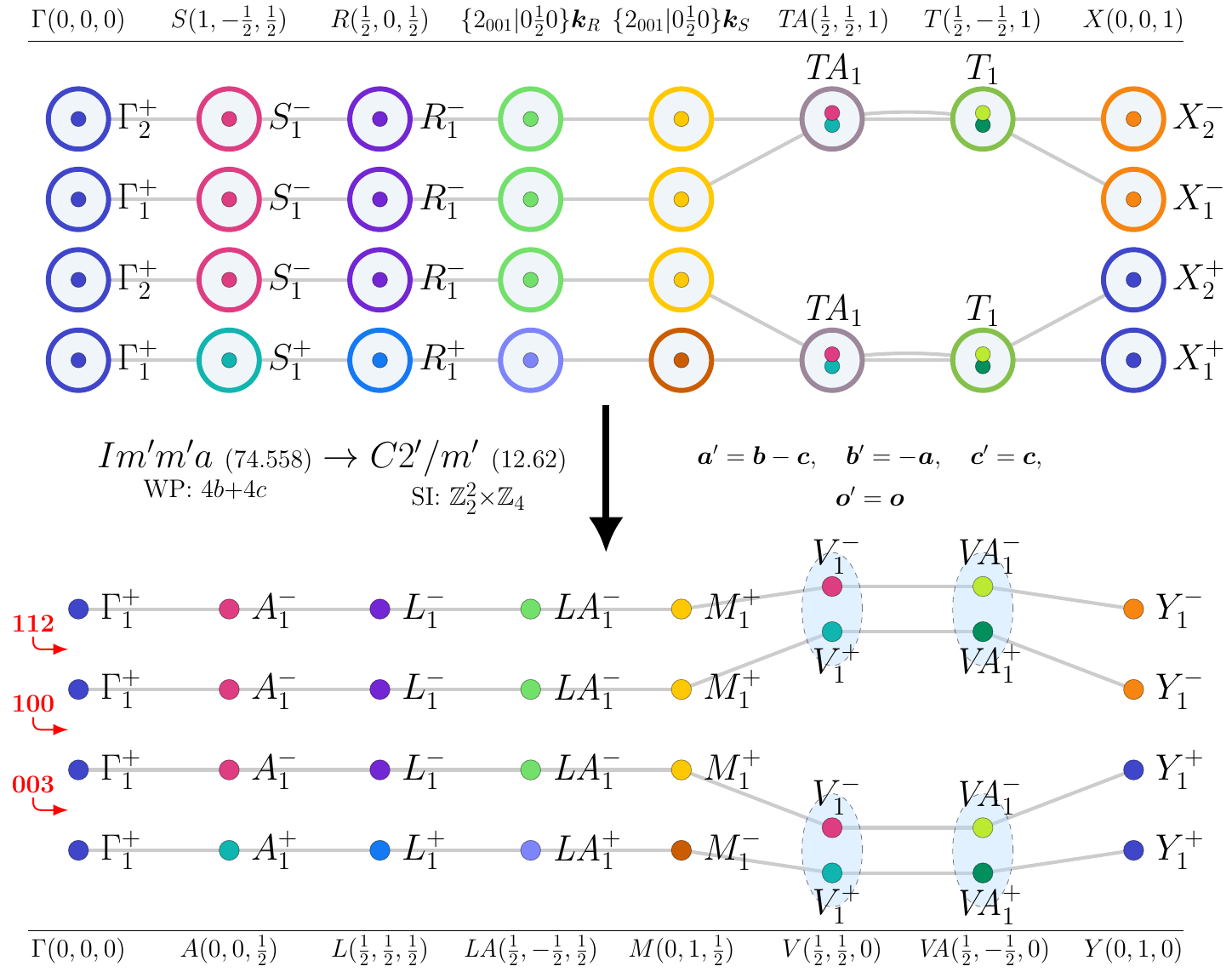}
\caption{Topological magnon bands in subgroup $C2'/m'~(12.62)$ for magnetic moments on Wyckoff positions $4b+4c$ of supergroup $Im'm'a~(74.558)$.\label{fig_74.558_12.62_strainperp100_4b+4c}}
\end{figure}
\input{gap_tables_tex/74.558_12.62_strainperp100_4b+4c_table.tex}
\input{si_tables_tex/74.558_12.62_strainperp100_4b+4c_table.tex}

\section{MSG $Imm'a'~(74.559)$}
\textbf{Nontrivial-SI Subgroups:} $P\bar{1}~(2.4)$, $C2'/c'~(15.89)$, $C2'/m'~(12.62)$, $C2/m~(12.58)$.\\

\textbf{Trivial-SI Subgroups:} $Cc'~(9.39)$, $Cm'~(8.34)$, $C2'~(5.15)$, $C2'~(5.15)$, $Cm~(8.32)$, $Im'm2'~(44.231)$, $Ima'2'~(46.244)$, $C2~(5.13)$, $Im'a'2~(46.245)$.\\

\subsection{WP: $4c+4c$}
\textbf{BCS Materials:} {Bi\textsubscript{0.8}La\textsubscript{0.2}Fe\textsubscript{0.5}Mn\textsubscript{0.5}O\textsubscript{3}~(240 K)}\footnote{BCS web page: \texttt{\href{http://webbdcrista1.ehu.es/magndata/index.php?this\_label=0.680} {http://webbdcrista1.ehu.es/magndata/index.php?this\_label=0.680}}}.\\
\subsubsection{Topological bands in subgroup $P\bar{1}~(2.4)$}
\textbf{Perturbations:}
\begin{itemize}
\item strain in generic direction,
\item (B $\parallel$ [010] or B $\perp$ [001]) and strain $\perp$ [100],
\item (B $\parallel$ [010] or B $\perp$ [001]) and strain $\perp$ [010],
\item (B $\parallel$ [001] or B $\perp$ [010]) and strain $\perp$ [100],
\item (B $\parallel$ [001] or B $\perp$ [010]) and strain $\perp$ [001],
\item B in generic direction.
\end{itemize}
\begin{figure}[H]
\centering
\includegraphics[scale=0.6]{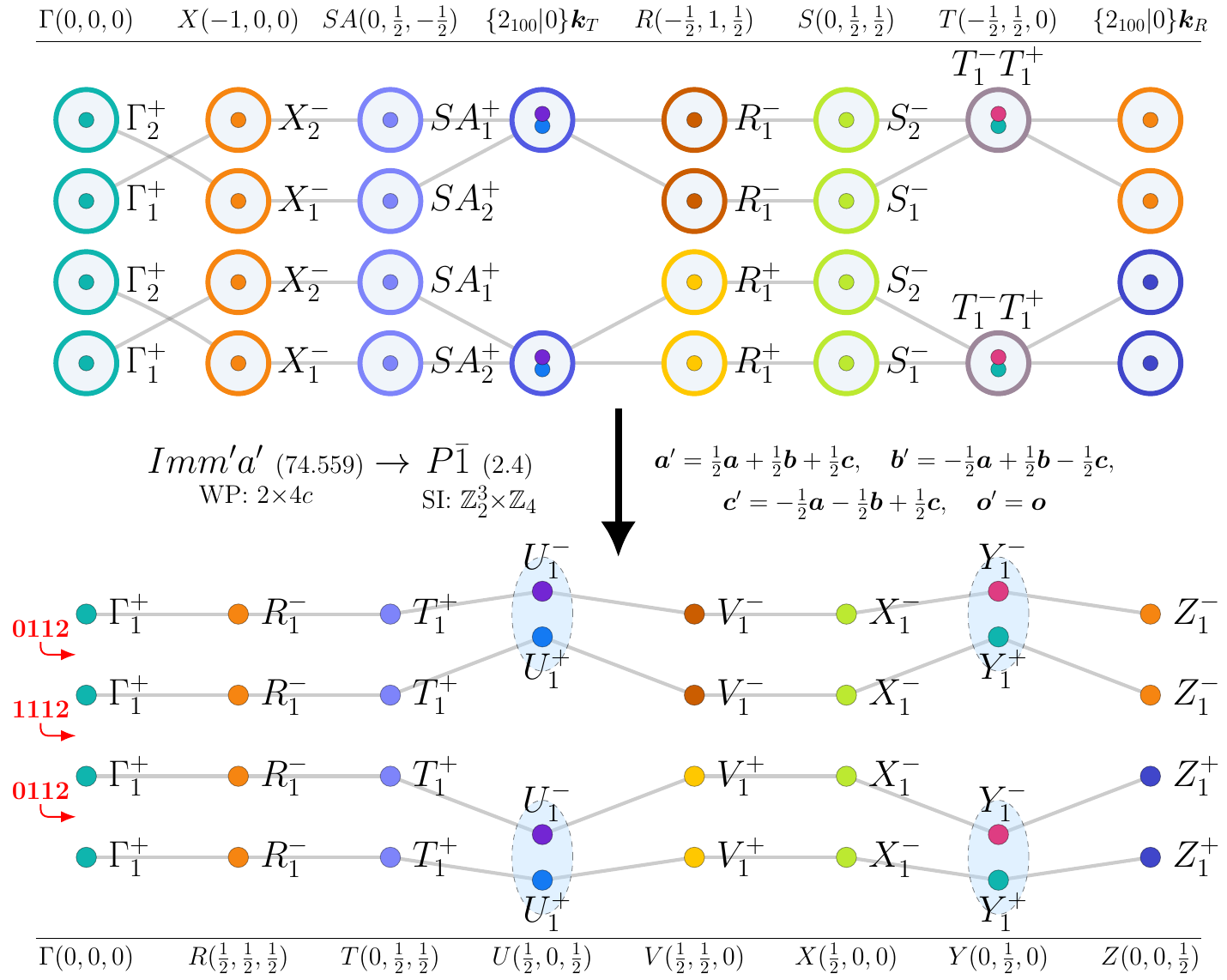}
\caption{Topological magnon bands in subgroup $P\bar{1}~(2.4)$ for magnetic moments on Wyckoff positions $4c+4c$ of supergroup $Imm'a'~(74.559)$.\label{fig_74.559_2.4_strainingenericdirection_4c+4c}}
\end{figure}
\input{gap_tables_tex/74.559_2.4_strainingenericdirection_4c+4c_table.tex}
\input{si_tables_tex/74.559_2.4_strainingenericdirection_4c+4c_table.tex}
\subsubsection{Topological bands in subgroup $C2'/m'~(12.62)$}
\textbf{Perturbations:}
\begin{itemize}
\item strain $\perp$ [010],
\item (B $\parallel$ [001] or B $\perp$ [010]).
\end{itemize}
\begin{figure}[H]
\centering
\includegraphics[scale=0.6]{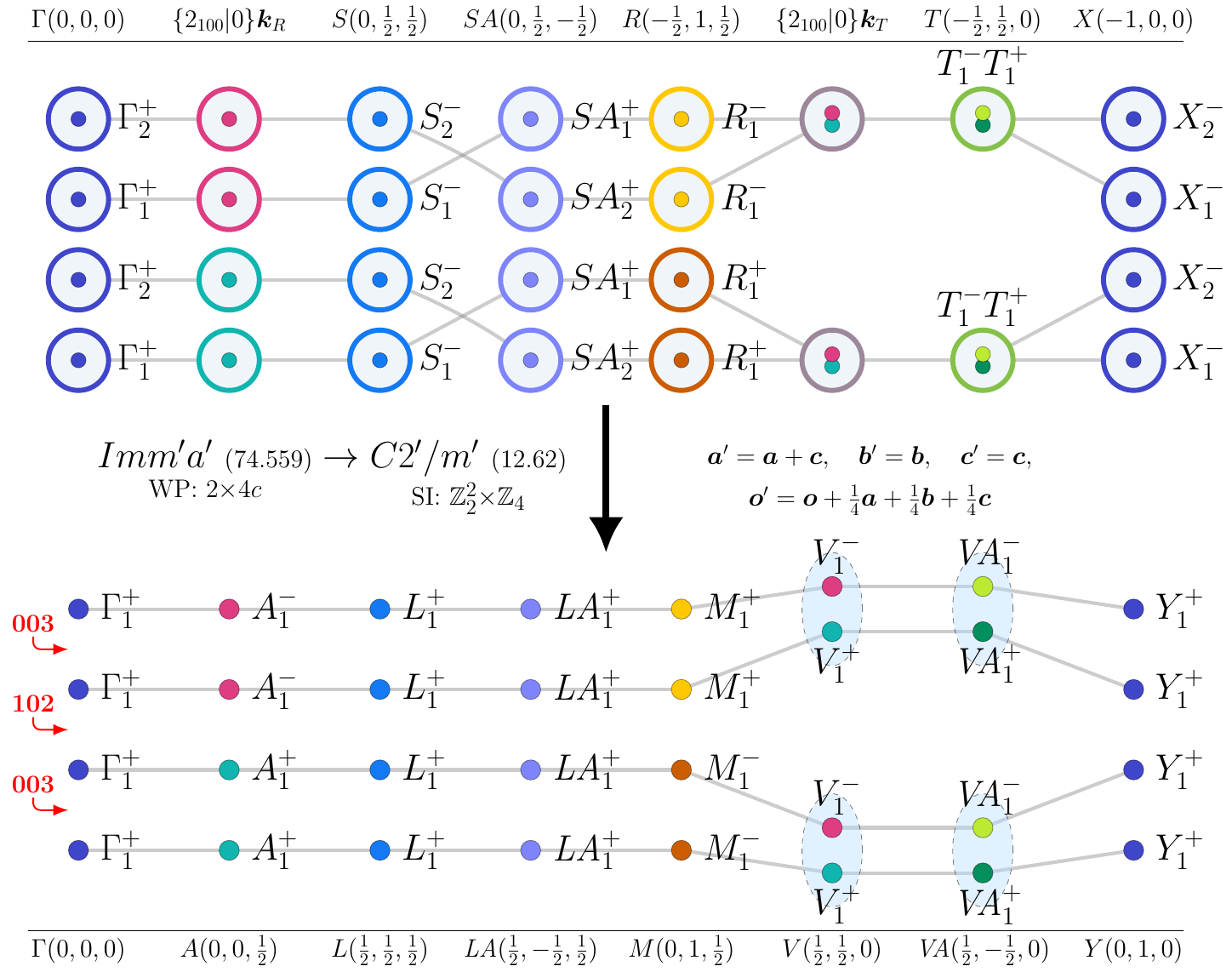}
\caption{Topological magnon bands in subgroup $C2'/m'~(12.62)$ for magnetic moments on Wyckoff positions $4c+4c$ of supergroup $Imm'a'~(74.559)$.\label{fig_74.559_12.62_strainperp010_4c+4c}}
\end{figure}
\input{gap_tables_tex/74.559_12.62_strainperp010_4c+4c_table.tex}
\input{si_tables_tex/74.559_12.62_strainperp010_4c+4c_table.tex}
\subsection{WP: $4b+4c$}
\textbf{BCS Materials:} {Na\textsubscript{2}NiFeF\textsubscript{7}~(88 K)}\footnote{BCS web page: \texttt{\href{http://webbdcrista1.ehu.es/magndata/index.php?this\_label=0.580} {http://webbdcrista1.ehu.es/magndata/index.php?this\_label=0.580}}}, {Na\textsubscript{2}NiFeF\textsubscript{7}~(88 K)}\footnote{BCS web page: \texttt{\href{http://webbdcrista1.ehu.es/magndata/index.php?this\_label=0.579} {http://webbdcrista1.ehu.es/magndata/index.php?this\_label=0.579}}}.\\
\subsubsection{Topological bands in subgroup $P\bar{1}~(2.4)$}
\textbf{Perturbations:}
\begin{itemize}
\item strain in generic direction,
\item (B $\parallel$ [010] or B $\perp$ [001]) and strain $\perp$ [100],
\item (B $\parallel$ [010] or B $\perp$ [001]) and strain $\perp$ [010],
\item (B $\parallel$ [001] or B $\perp$ [010]) and strain $\perp$ [100],
\item (B $\parallel$ [001] or B $\perp$ [010]) and strain $\perp$ [001],
\item B in generic direction.
\end{itemize}
\begin{figure}[H]
\centering
\includegraphics[scale=0.6]{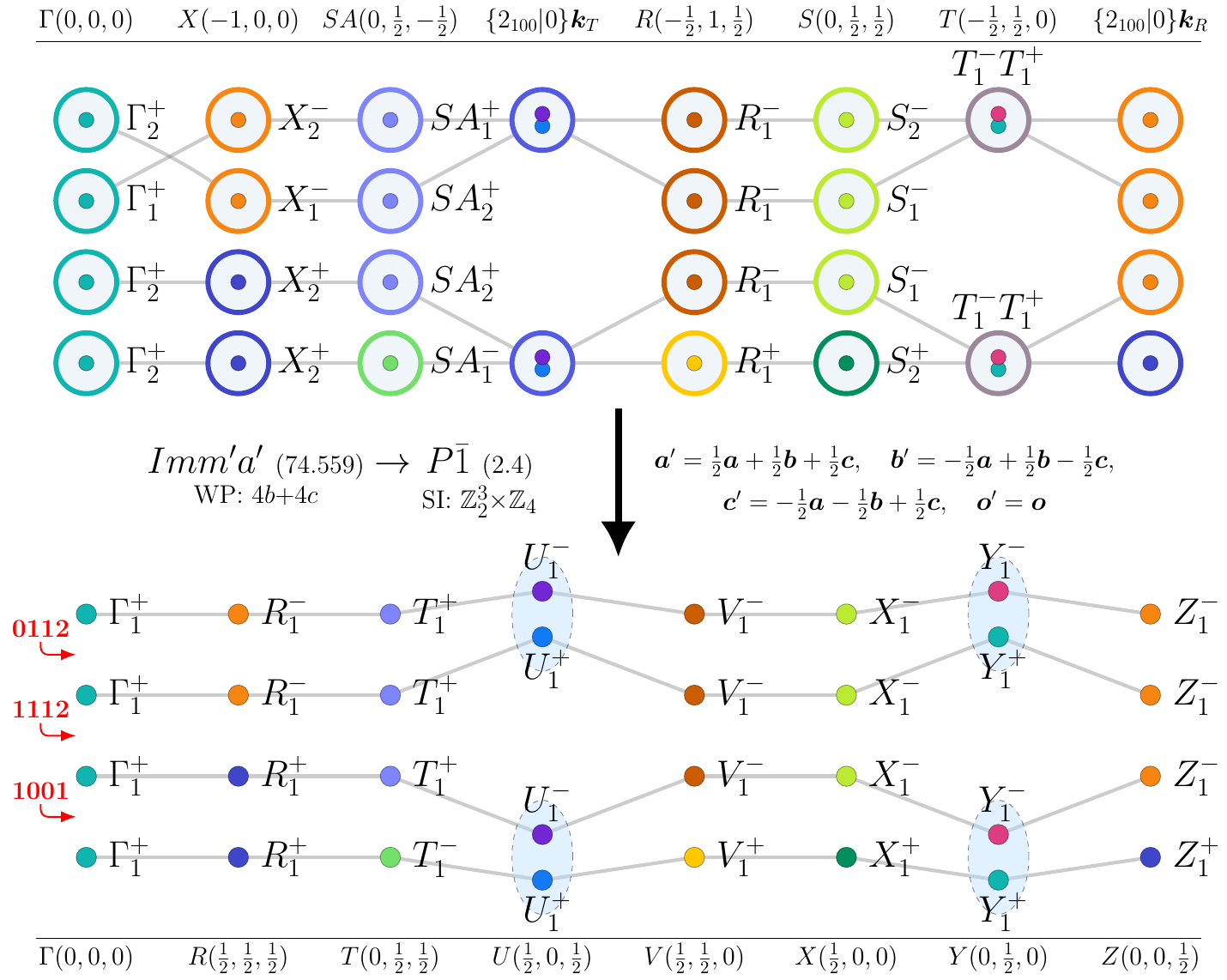}
\caption{Topological magnon bands in subgroup $P\bar{1}~(2.4)$ for magnetic moments on Wyckoff positions $4b+4c$ of supergroup $Imm'a'~(74.559)$.\label{fig_74.559_2.4_strainingenericdirection_4b+4c}}
\end{figure}
\input{gap_tables_tex/74.559_2.4_strainingenericdirection_4b+4c_table.tex}
\input{si_tables_tex/74.559_2.4_strainingenericdirection_4b+4c_table.tex}
\subsubsection{Topological bands in subgroup $C2'/m'~(12.62)$}
\textbf{Perturbations:}
\begin{itemize}
\item strain $\perp$ [010],
\item (B $\parallel$ [001] or B $\perp$ [010]).
\end{itemize}
\begin{figure}[H]
\centering
\includegraphics[scale=0.6]{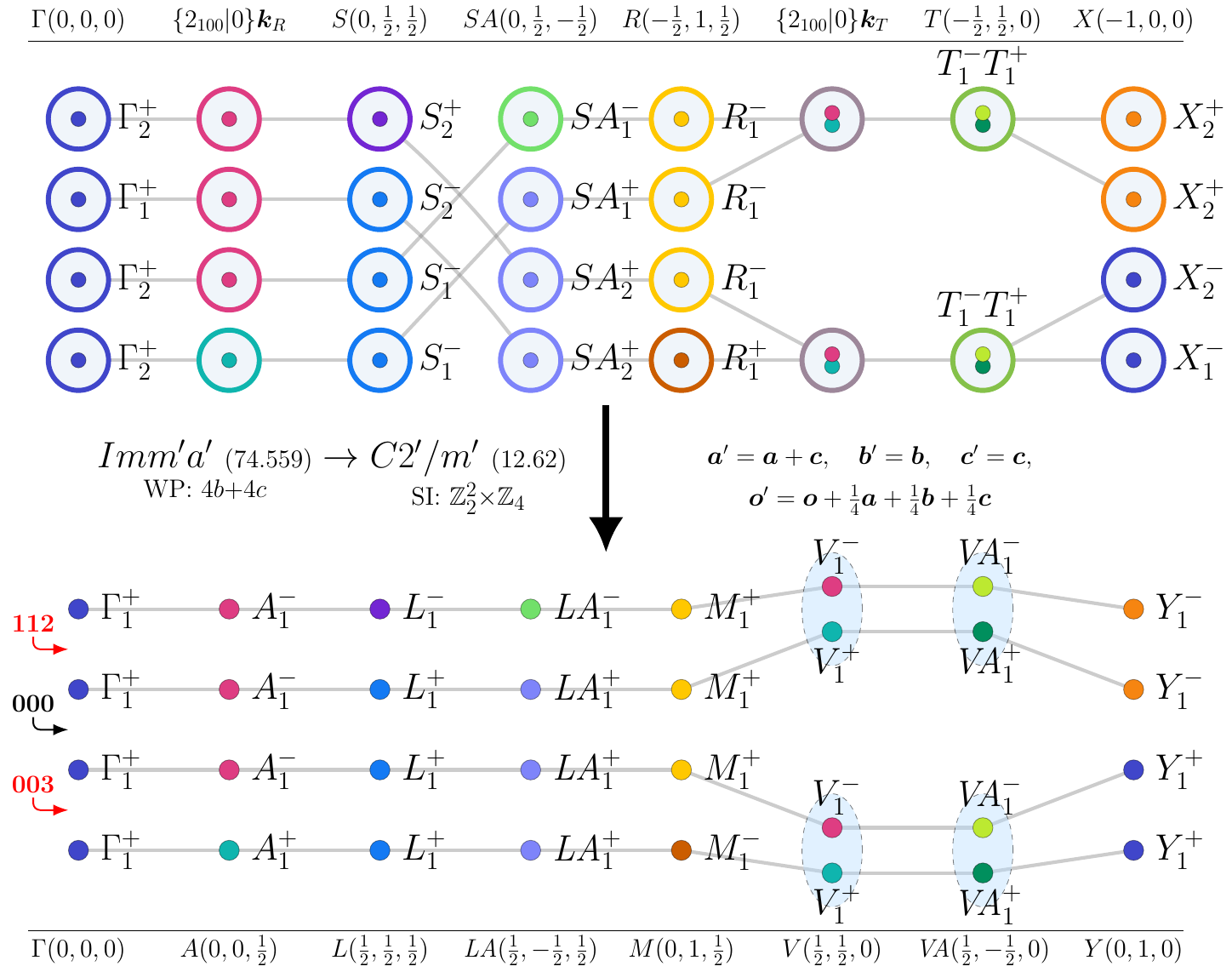}
\caption{Topological magnon bands in subgroup $C2'/m'~(12.62)$ for magnetic moments on Wyckoff positions $4b+4c$ of supergroup $Imm'a'~(74.559)$.\label{fig_74.559_12.62_strainperp010_4b+4c}}
\end{figure}
\input{gap_tables_tex/74.559_12.62_strainperp010_4b+4c_table.tex}
\input{si_tables_tex/74.559_12.62_strainperp010_4b+4c_table.tex}
\subsection{WP: $4a+4c$}
\textbf{BCS Materials:} {Fe\textsubscript{2}F\textsubscript{5}(H\textsubscript{2}O)\textsubscript{2}~(48 K)}\footnote{BCS web page: \texttt{\href{http://webbdcrista1.ehu.es/magndata/index.php?this\_label=0.583} {http://webbdcrista1.ehu.es/magndata/index.php?this\_label=0.583}}}.\\
\subsubsection{Topological bands in subgroup $P\bar{1}~(2.4)$}
\textbf{Perturbations:}
\begin{itemize}
\item strain in generic direction,
\item (B $\parallel$ [010] or B $\perp$ [001]) and strain $\perp$ [100],
\item (B $\parallel$ [010] or B $\perp$ [001]) and strain $\perp$ [010],
\item (B $\parallel$ [001] or B $\perp$ [010]) and strain $\perp$ [100],
\item (B $\parallel$ [001] or B $\perp$ [010]) and strain $\perp$ [001],
\item B in generic direction.
\end{itemize}
\begin{figure}[H]
\centering
\includegraphics[scale=0.6]{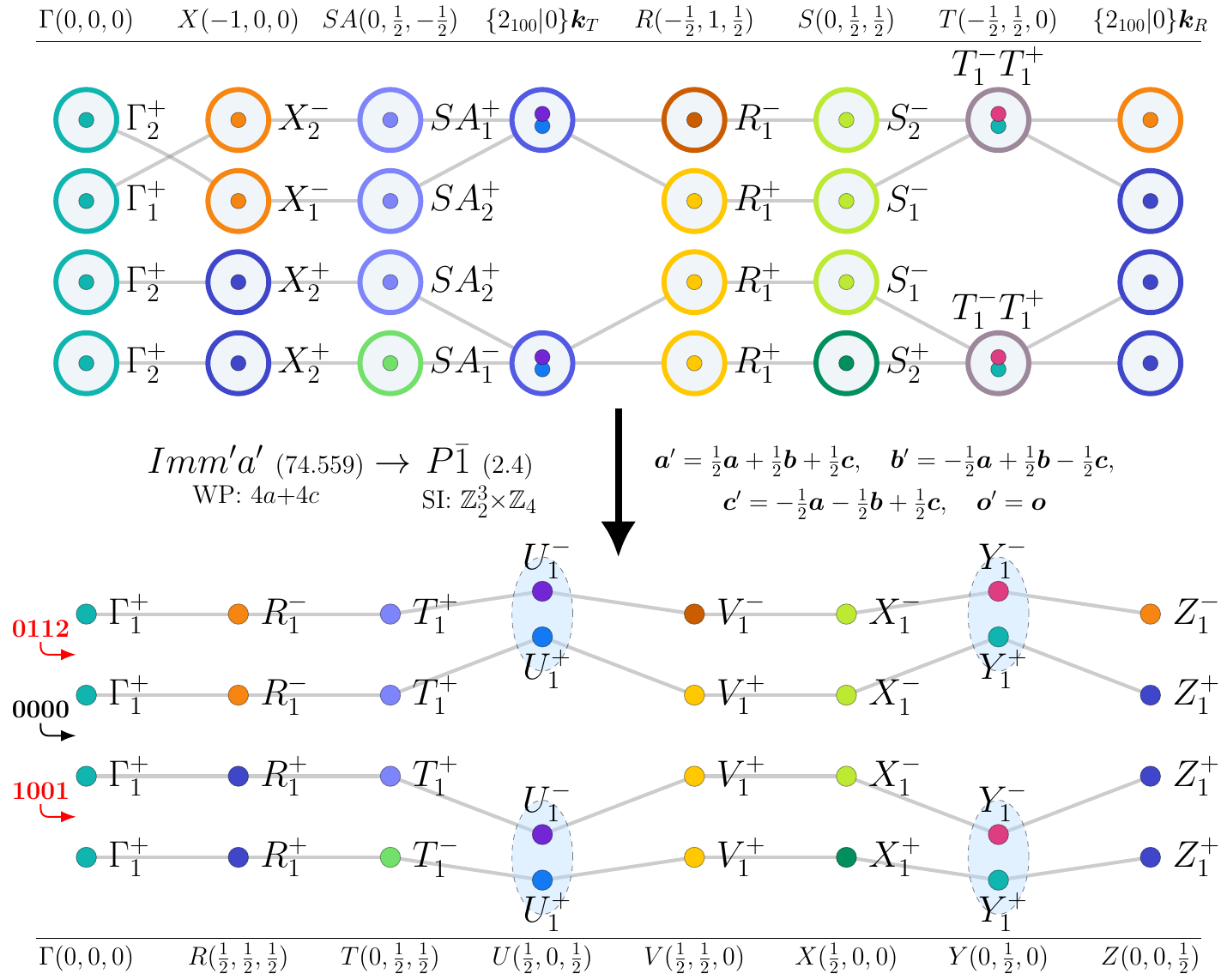}
\caption{Topological magnon bands in subgroup $P\bar{1}~(2.4)$ for magnetic moments on Wyckoff positions $4a+4c$ of supergroup $Imm'a'~(74.559)$.\label{fig_74.559_2.4_strainingenericdirection_4a+4c}}
\end{figure}
\input{gap_tables_tex/74.559_2.4_strainingenericdirection_4a+4c_table.tex}
\input{si_tables_tex/74.559_2.4_strainingenericdirection_4a+4c_table.tex}
\subsubsection{Topological bands in subgroup $C2'/m'~(12.62)$}
\textbf{Perturbations:}
\begin{itemize}
\item strain $\perp$ [010],
\item (B $\parallel$ [001] or B $\perp$ [010]).
\end{itemize}
\begin{figure}[H]
\centering
\includegraphics[scale=0.6]{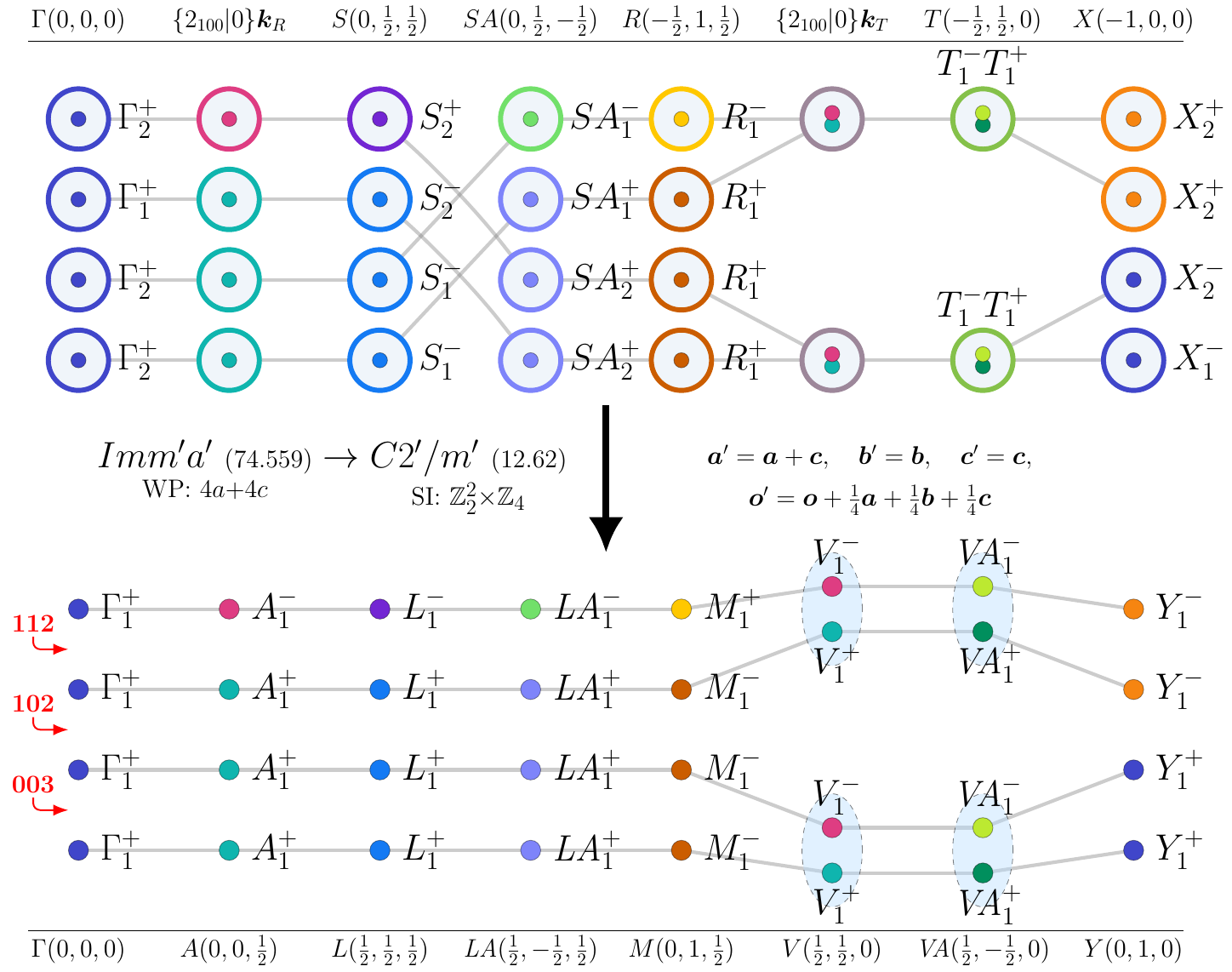}
\caption{Topological magnon bands in subgroup $C2'/m'~(12.62)$ for magnetic moments on Wyckoff positions $4a+4c$ of supergroup $Imm'a'~(74.559)$.\label{fig_74.559_12.62_strainperp010_4a+4c}}
\end{figure}
\input{gap_tables_tex/74.559_12.62_strainperp010_4a+4c_table.tex}
\input{si_tables_tex/74.559_12.62_strainperp010_4a+4c_table.tex}
\subsection{WP: $4b+4c+4e$}
\textbf{BCS Materials:} {NdCo\textsubscript{2}~(42 K)}\footnote{BCS web page: \texttt{\href{http://webbdcrista1.ehu.es/magndata/index.php?this\_label=0.403} {http://webbdcrista1.ehu.es/magndata/index.php?this\_label=0.403}}}.\\
\subsubsection{Topological bands in subgroup $P\bar{1}~(2.4)$}
\textbf{Perturbations:}
\begin{itemize}
\item strain in generic direction,
\item (B $\parallel$ [010] or B $\perp$ [001]) and strain $\perp$ [100],
\item (B $\parallel$ [010] or B $\perp$ [001]) and strain $\perp$ [010],
\item (B $\parallel$ [001] or B $\perp$ [010]) and strain $\perp$ [100],
\item (B $\parallel$ [001] or B $\perp$ [010]) and strain $\perp$ [001],
\item B in generic direction.
\end{itemize}
\begin{figure}[H]
\centering
\includegraphics[scale=0.6]{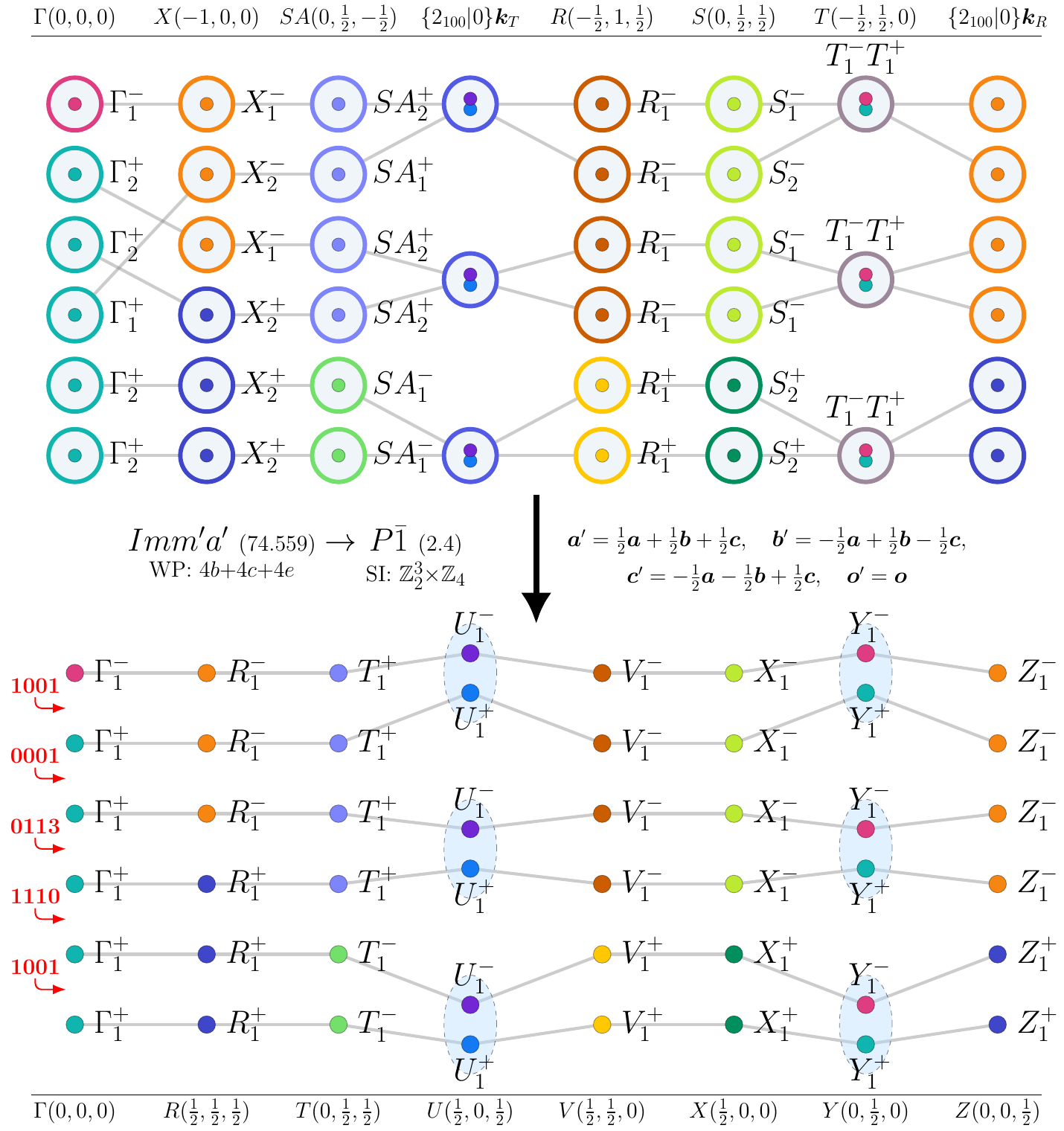}
\caption{Topological magnon bands in subgroup $P\bar{1}~(2.4)$ for magnetic moments on Wyckoff positions $4b+4c+4e$ of supergroup $Imm'a'~(74.559)$.\label{fig_74.559_2.4_strainingenericdirection_4b+4c+4e}}
\end{figure}
\input{gap_tables_tex/74.559_2.4_strainingenericdirection_4b+4c+4e_table.tex}
\input{si_tables_tex/74.559_2.4_strainingenericdirection_4b+4c+4e_table.tex}
\subsubsection{Topological bands in subgroup $C2'/m'~(12.62)$}
\textbf{Perturbations:}
\begin{itemize}
\item strain $\perp$ [010],
\item (B $\parallel$ [001] or B $\perp$ [010]).
\end{itemize}
\begin{figure}[H]
\centering
\includegraphics[scale=0.6]{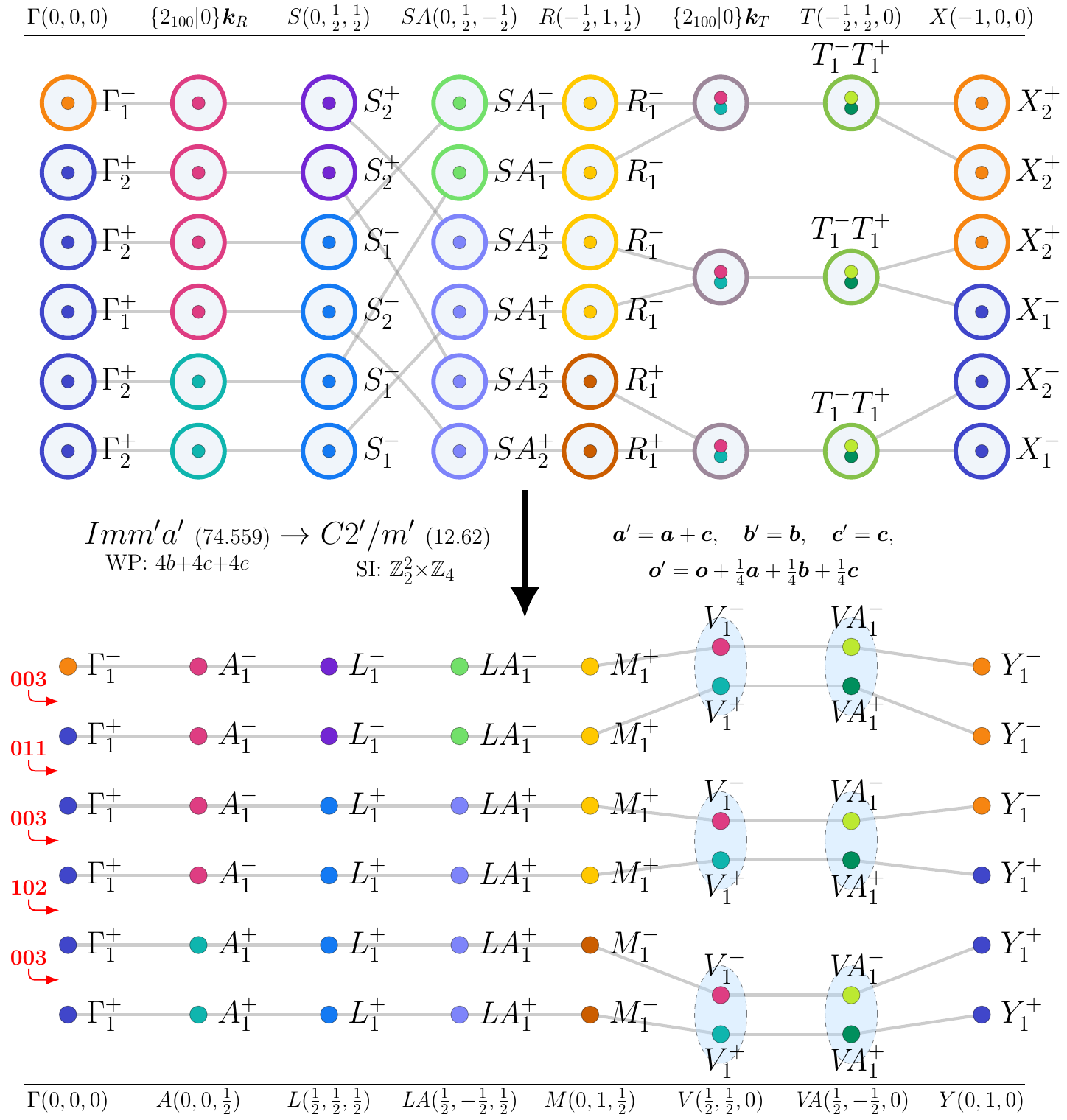}
\caption{Topological magnon bands in subgroup $C2'/m'~(12.62)$ for magnetic moments on Wyckoff positions $4b+4c+4e$ of supergroup $Imm'a'~(74.559)$.\label{fig_74.559_12.62_strainperp010_4b+4c+4e}}
\end{figure}
\input{gap_tables_tex/74.559_12.62_strainperp010_4b+4c+4e_table.tex}
\input{si_tables_tex/74.559_12.62_strainperp010_4b+4c+4e_table.tex}
\subsection{WP: $4d$}
\textbf{BCS Materials:} {NpNiGa\textsubscript{5}~(18 K)}\footnote{BCS web page: \texttt{\href{http://webbdcrista1.ehu.es/magndata/index.php?this\_label=2.28} {http://webbdcrista1.ehu.es/magndata/index.php?this\_label=2.28}}}.\\
\subsubsection{Topological bands in subgroup $P\bar{1}~(2.4)$}
\textbf{Perturbations:}
\begin{itemize}
\item strain in generic direction,
\item (B $\parallel$ [010] or B $\perp$ [001]) and strain $\perp$ [100],
\item (B $\parallel$ [010] or B $\perp$ [001]) and strain $\perp$ [010],
\item (B $\parallel$ [001] or B $\perp$ [010]) and strain $\perp$ [100],
\item (B $\parallel$ [001] or B $\perp$ [010]) and strain $\perp$ [001],
\item B in generic direction.
\end{itemize}
\begin{figure}[H]
\centering
\includegraphics[scale=0.6]{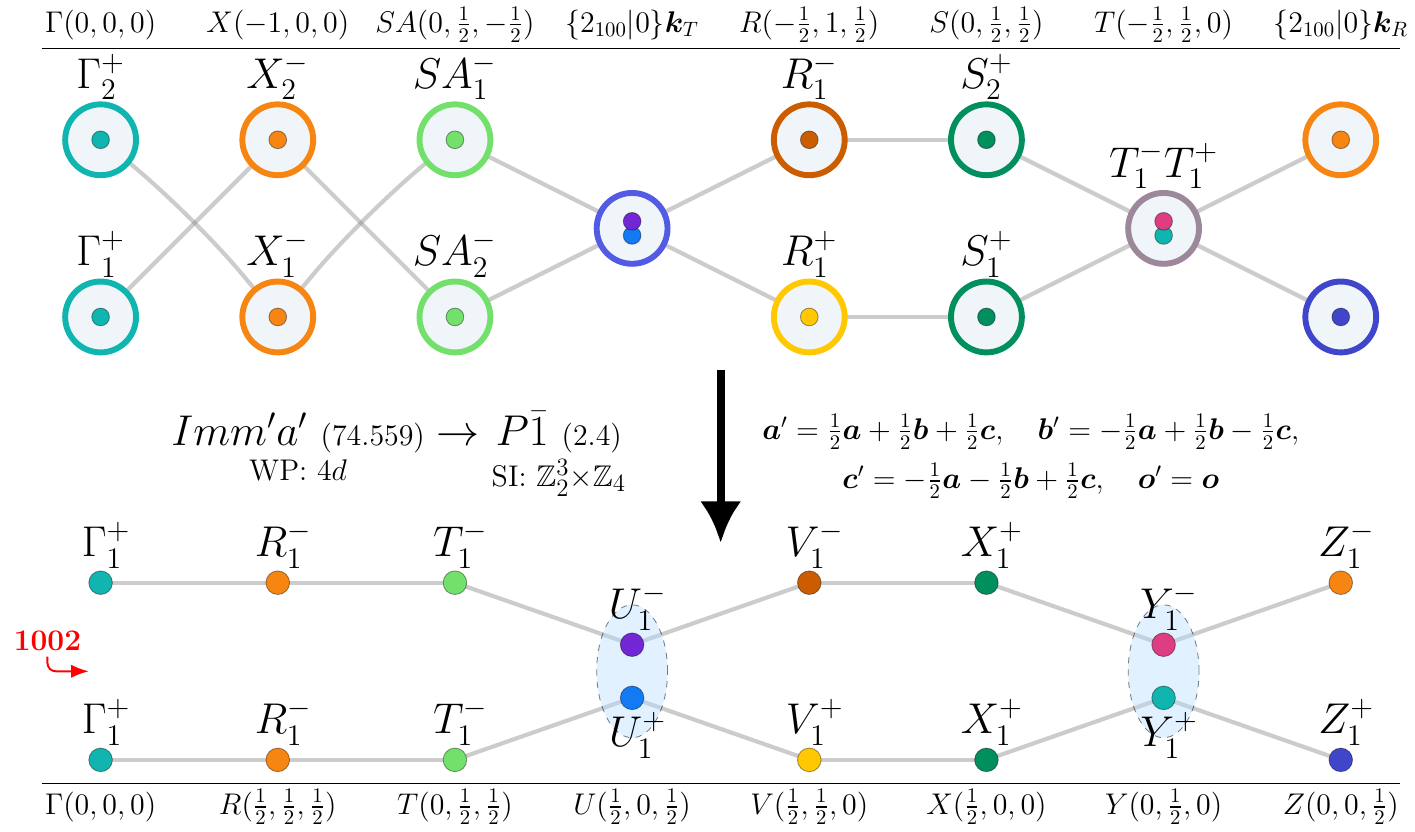}
\caption{Topological magnon bands in subgroup $P\bar{1}~(2.4)$ for magnetic moments on Wyckoff position $4d$ of supergroup $Imm'a'~(74.559)$.\label{fig_74.559_2.4_strainingenericdirection_4d}}
\end{figure}
\input{gap_tables_tex/74.559_2.4_strainingenericdirection_4d_table.tex}
\input{si_tables_tex/74.559_2.4_strainingenericdirection_4d_table.tex}
\subsubsection{Topological bands in subgroup $C2'/m'~(12.62)$}
\textbf{Perturbations:}
\begin{itemize}
\item strain $\perp$ [010],
\item (B $\parallel$ [001] or B $\perp$ [010]).
\end{itemize}
\begin{figure}[H]
\centering
\includegraphics[scale=0.6]{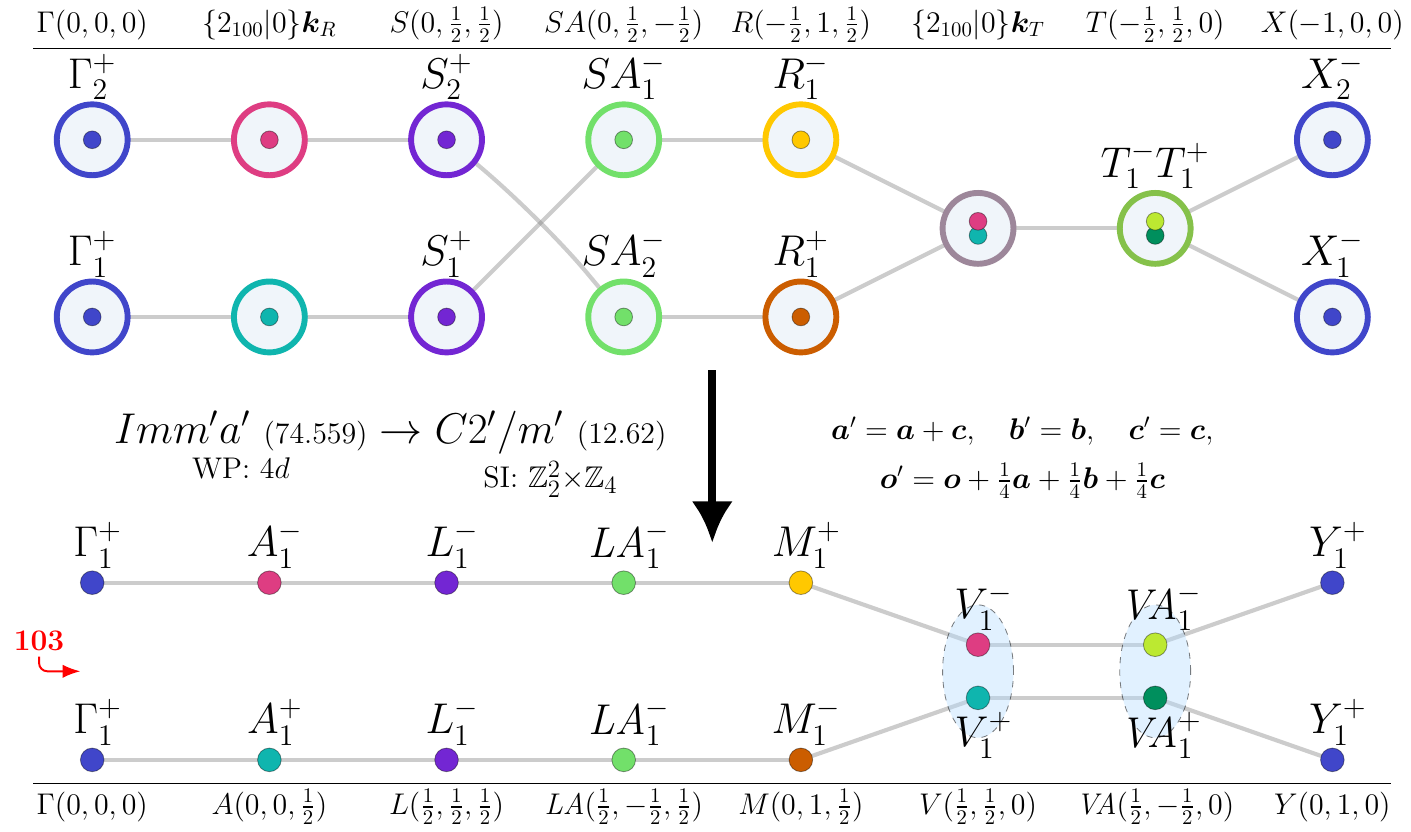}
\caption{Topological magnon bands in subgroup $C2'/m'~(12.62)$ for magnetic moments on Wyckoff position $4d$ of supergroup $Imm'a'~(74.559)$.\label{fig_74.559_12.62_strainperp010_4d}}
\end{figure}
\input{gap_tables_tex/74.559_12.62_strainperp010_4d_table.tex}
\input{si_tables_tex/74.559_12.62_strainperp010_4d_table.tex}

\section{MSG $P4/n~(85.59)$}
\textbf{Nontrivial-SI Subgroups:} $P\bar{1}~(2.4)$, $P2~(3.1)$, $P2/c~(13.65)$, $P4~(75.1)$.\\

\textbf{Trivial-SI Subgroups:} $Pc~(7.24)$.\\

\subsection{WP: $2c+4e$}
\textbf{BCS Materials:} {Mn\textsubscript{3}CuN~(143 K)}\footnote{BCS web page: \texttt{\href{http://webbdcrista1.ehu.es/magndata/index.php?this\_label=2.5} {http://webbdcrista1.ehu.es/magndata/index.php?this\_label=2.5}}}.\\
\subsubsection{Topological bands in subgroup $P\bar{1}~(2.4)$}
\textbf{Perturbations:}
\begin{itemize}
\item strain in generic direction,
\item B in generic direction.
\end{itemize}
\begin{figure}[H]
\centering
\includegraphics[scale=0.6]{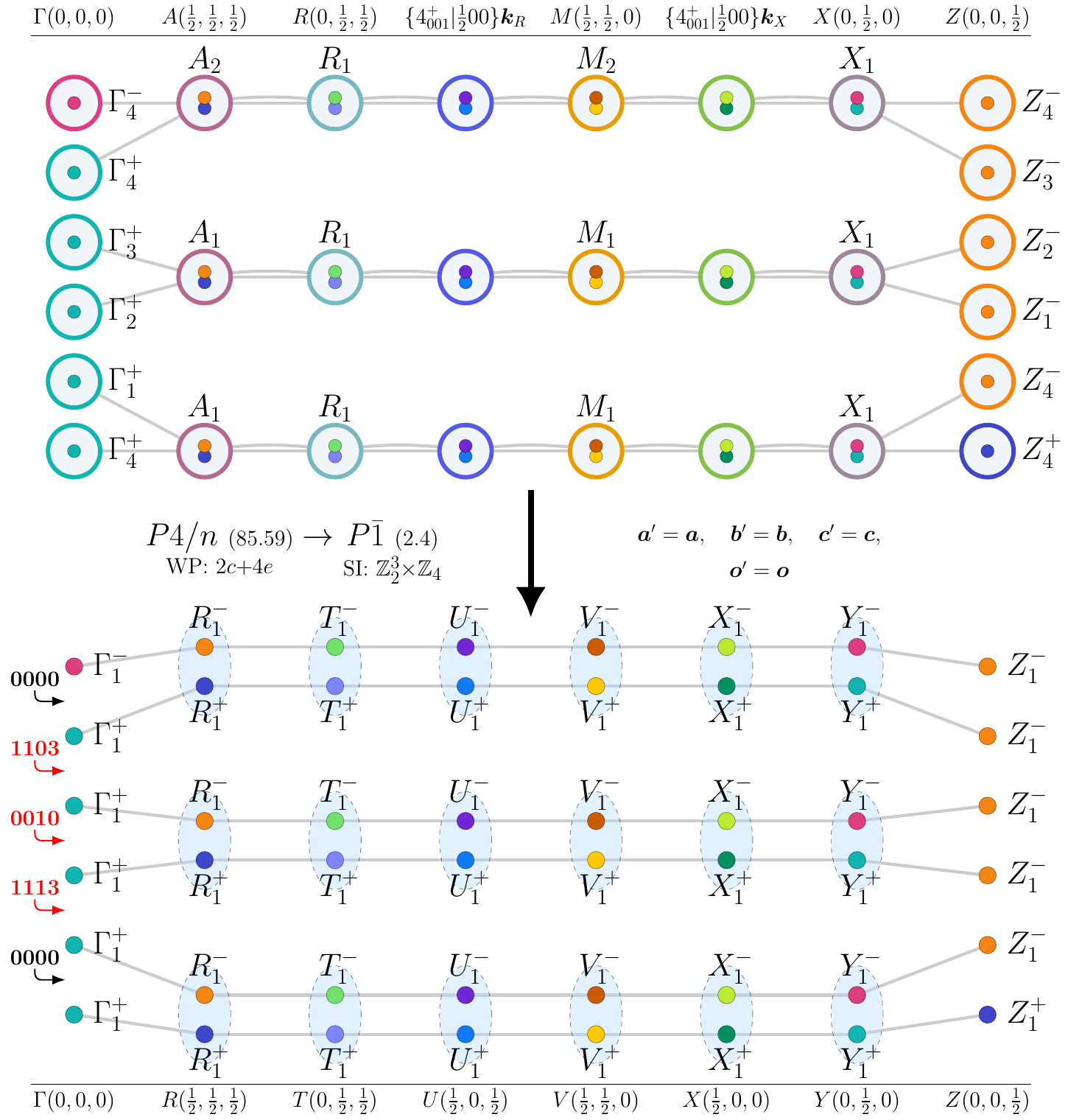}
\caption{Topological magnon bands in subgroup $P\bar{1}~(2.4)$ for magnetic moments on Wyckoff positions $2c+4e$ of supergroup $P4/n~(85.59)$.\label{fig_85.59_2.4_strainingenericdirection_2c+4e}}
\end{figure}
\input{gap_tables_tex/85.59_2.4_strainingenericdirection_2c+4e_table.tex}
\input{si_tables_tex/85.59_2.4_strainingenericdirection_2c+4e_table.tex}

\section{MSG $P_{c}4/n~(85.64)$}
\textbf{Nontrivial-SI Subgroups:} $P\bar{1}~(2.4)$, $P2_{1}'/c'~(14.79)$, $P_{S}\bar{1}~(2.7)$, $P2~(3.1)$, $P_{b}2~(3.5)$, $P2/c~(13.65)$, $P_{b}2/c~(13.71)$, $P4~(75.1)$, $P_{c}4~(75.4)$, $P4/n~(85.59)$.\\

\textbf{Trivial-SI Subgroups:} $Pc'~(7.26)$, $P2_{1}'~(4.9)$, $P_{S}1~(1.3)$, $Pc~(7.24)$, $P_{b}c~(7.29)$.\\

\subsection{WP: $4c+4c$}
\textbf{BCS Materials:} {Sr\textsubscript{2}FeOsO\textsubscript{6}~(67 K)}\footnote{BCS web page: \texttt{\href{http://webbdcrista1.ehu.es/magndata/index.php?this\_label=1.46} {http://webbdcrista1.ehu.es/magndata/index.php?this\_label=1.46}}}.\\
\subsubsection{Topological bands in subgroup $P2_{1}'/c'~(14.79)$}
\textbf{Perturbation:}
\begin{itemize}
\item (B $\parallel$ [100] or B $\parallel$ [110] or B $\perp$ [001]).
\end{itemize}
\begin{figure}[H]
\centering
\includegraphics[scale=0.6]{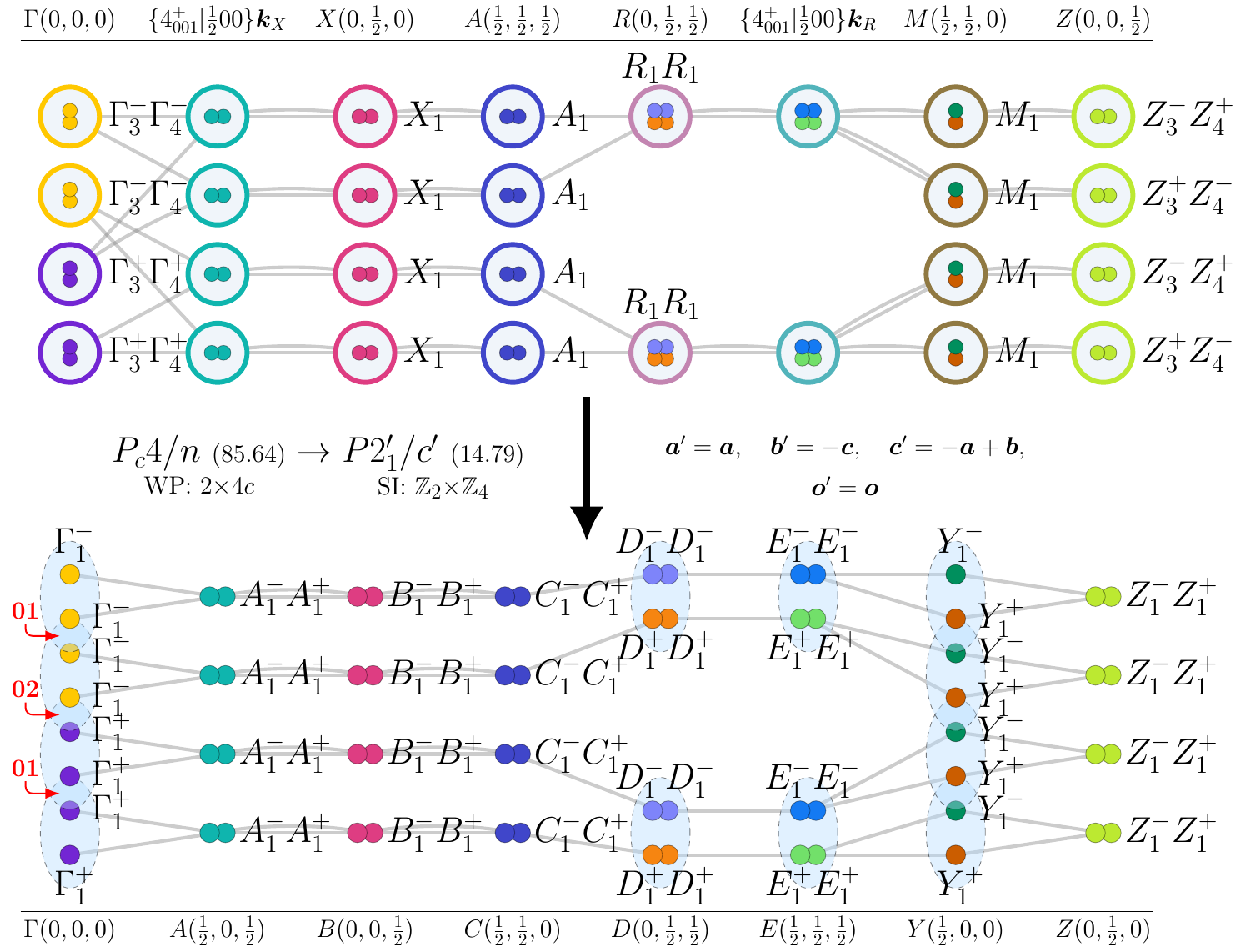}
\caption{Topological magnon bands in subgroup $P2_{1}'/c'~(14.79)$ for magnetic moments on Wyckoff positions $4c+4c$ of supergroup $P_{c}4/n~(85.64)$.\label{fig_85.64_14.79_Bparallel100orBparallel110orBperp001_4c+4c}}
\end{figure}
\input{gap_tables_tex/85.64_14.79_Bparallel100orBparallel110orBperp001_4c+4c_table.tex}
\input{si_tables_tex/85.64_14.79_Bparallel100orBparallel110orBperp001_4c+4c_table.tex}
\subsubsection{Topological bands in subgroup $P_{S}\bar{1}~(2.7)$}
\textbf{Perturbation:}
\begin{itemize}
\item strain in generic direction.
\end{itemize}
\begin{figure}[H]
\centering
\includegraphics[scale=0.6]{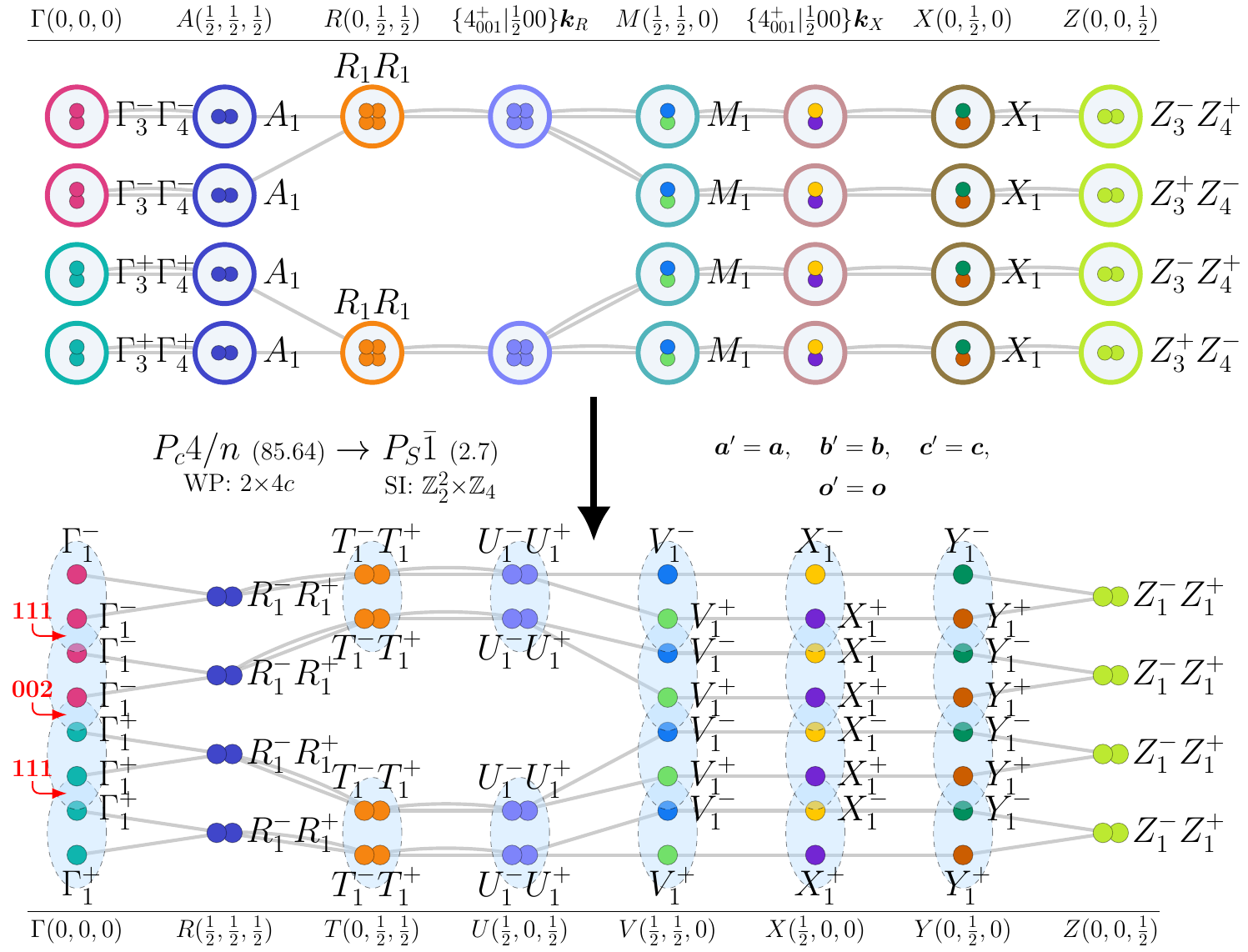}
\caption{Topological magnon bands in subgroup $P_{S}\bar{1}~(2.7)$ for magnetic moments on Wyckoff positions $4c+4c$ of supergroup $P_{c}4/n~(85.64)$.\label{fig_85.64_2.7_strainingenericdirection_4c+4c}}
\end{figure}
\input{gap_tables_tex/85.64_2.7_strainingenericdirection_4c+4c_table.tex}
\input{si_tables_tex/85.64_2.7_strainingenericdirection_4c+4c_table.tex}

\section{MSG $P_{C}4_{2}/n~(86.73)$}
\textbf{Nontrivial-SI Subgroups:} $P\bar{1}~(2.4)$, $P2'/m'~(10.46)$, $P_{S}\bar{1}~(2.7)$, $P2~(3.1)$, $P_{a}2~(3.4)$, $P2/c~(13.65)$, $P_{c}2/c~(13.72)$, $P4_{2}~(77.13)$, $P_{C}4_{2}~(77.17)$, $P4_{2}/n~(86.67)$.\\

\textbf{Trivial-SI Subgroups:} $Pm'~(6.20)$, $P2'~(3.3)$, $P_{S}1~(1.3)$, $Pc~(7.24)$, $P_{c}c~(7.28)$.\\

\subsection{WP: $4c$}
\textbf{BCS Materials:} {Sr\textsubscript{2}CoO\textsubscript{2}Ag\textsubscript{2}Se\textsubscript{2}~(190 K)}\footnote{BCS web page: \texttt{\href{http://webbdcrista1.ehu.es/magndata/index.php?this\_label=2.23} {http://webbdcrista1.ehu.es/magndata/index.php?this\_label=2.23}}}, {Ba\textsubscript{2}CoO\textsubscript{2}Ag\textsubscript{2}Se\textsubscript{2}~(180 K)}\footnote{BCS web page: \texttt{\href{http://webbdcrista1.ehu.es/magndata/index.php?this\_label=2.24} {http://webbdcrista1.ehu.es/magndata/index.php?this\_label=2.24}}}.\\
\subsubsection{Topological bands in subgroup $P\bar{1}~(2.4)$}
\textbf{Perturbations:}
\begin{itemize}
\item B $\parallel$ [001] and strain in generic direction,
\item (B $\parallel$ [100] or B $\parallel$ [110] or B $\perp$ [001]) and strain in generic direction,
\item B in generic direction.
\end{itemize}
\begin{figure}[H]
\centering
\includegraphics[scale=0.6]{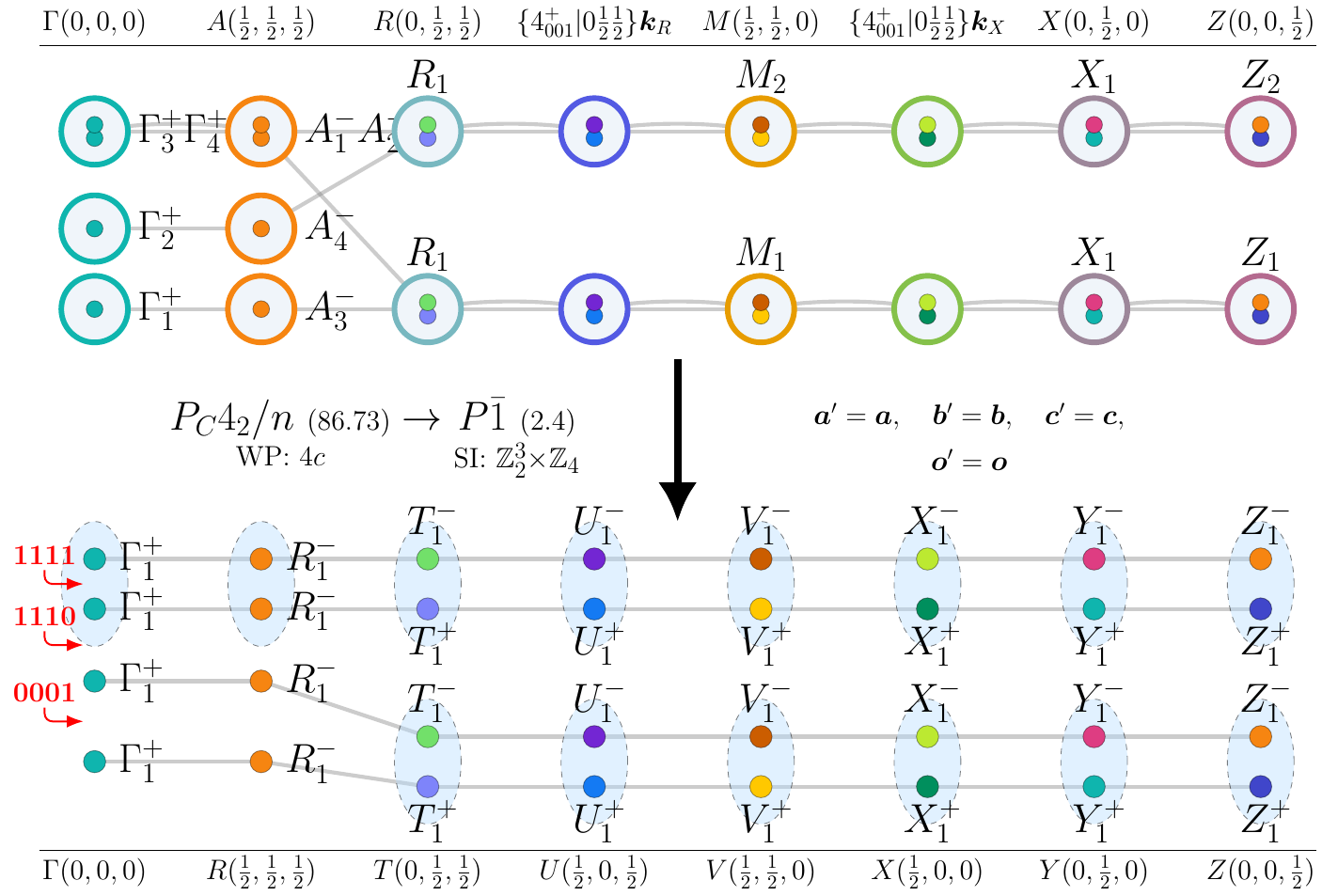}
\caption{Topological magnon bands in subgroup $P\bar{1}~(2.4)$ for magnetic moments on Wyckoff position $4c$ of supergroup $P_{C}4_{2}/n~(86.73)$.\label{fig_86.73_2.4_Bparallel001andstrainingenericdirection_4c}}
\end{figure}
\input{gap_tables_tex/86.73_2.4_Bparallel001andstrainingenericdirection_4c_table.tex}
\input{si_tables_tex/86.73_2.4_Bparallel001andstrainingenericdirection_4c_table.tex}
\subsubsection{Topological bands in subgroup $P2'/m'~(10.46)$}
\textbf{Perturbation:}
\begin{itemize}
\item (B $\parallel$ [100] or B $\parallel$ [110] or B $\perp$ [001]).
\end{itemize}
\begin{figure}[H]
\centering
\includegraphics[scale=0.6]{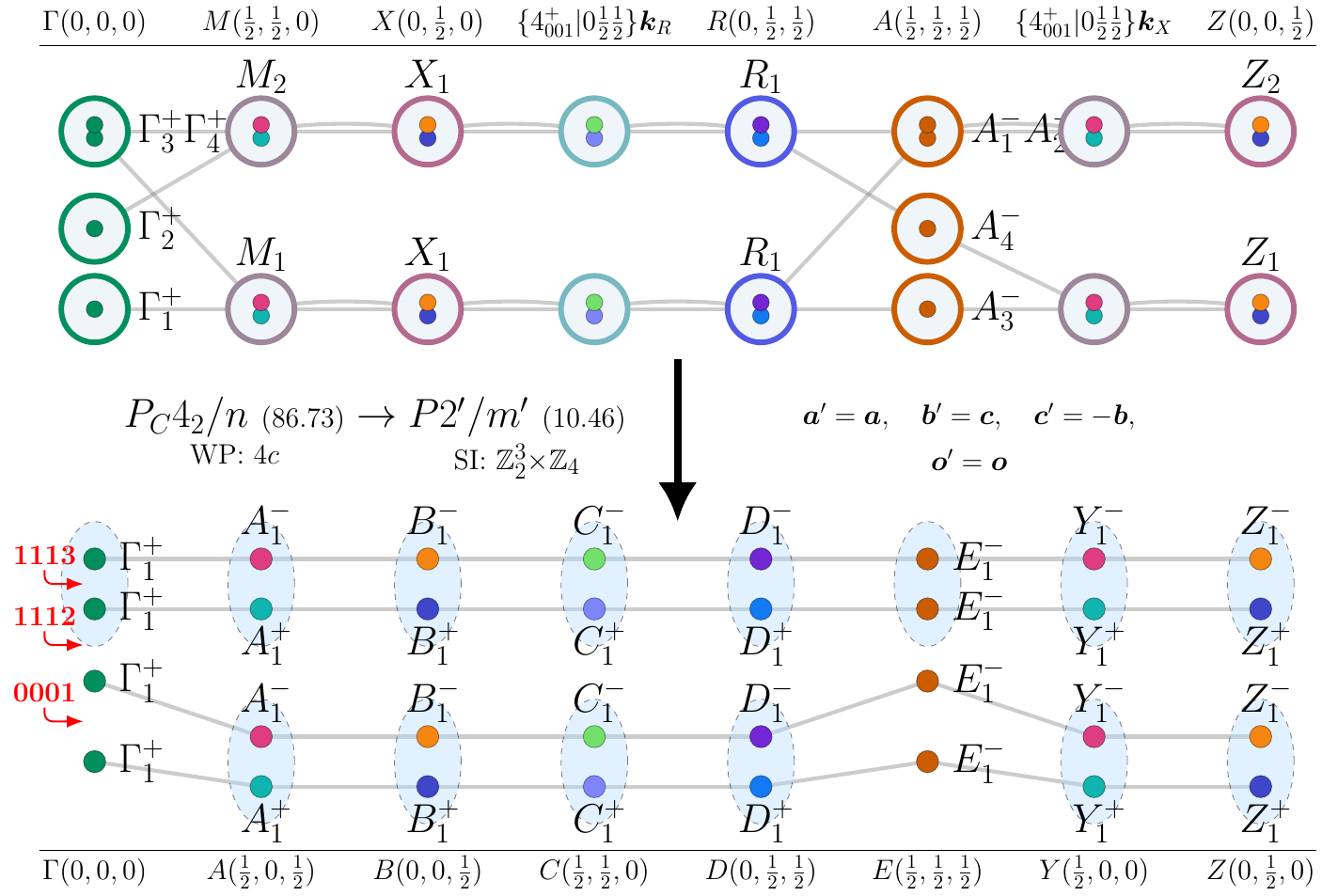}
\caption{Topological magnon bands in subgroup $P2'/m'~(10.46)$ for magnetic moments on Wyckoff position $4c$ of supergroup $P_{C}4_{2}/n~(86.73)$.\label{fig_86.73_10.46_Bparallel100orBparallel110orBperp001_4c}}
\end{figure}
\input{gap_tables_tex/86.73_10.46_Bparallel100orBparallel110orBperp001_4c_table.tex}
\input{si_tables_tex/86.73_10.46_Bparallel100orBparallel110orBperp001_4c_table.tex}
\subsubsection{Topological bands in subgroup $P_{S}\bar{1}~(2.7)$}
\textbf{Perturbation:}
\begin{itemize}
\item strain in generic direction.
\end{itemize}
\begin{figure}[H]
\centering
\includegraphics[scale=0.6]{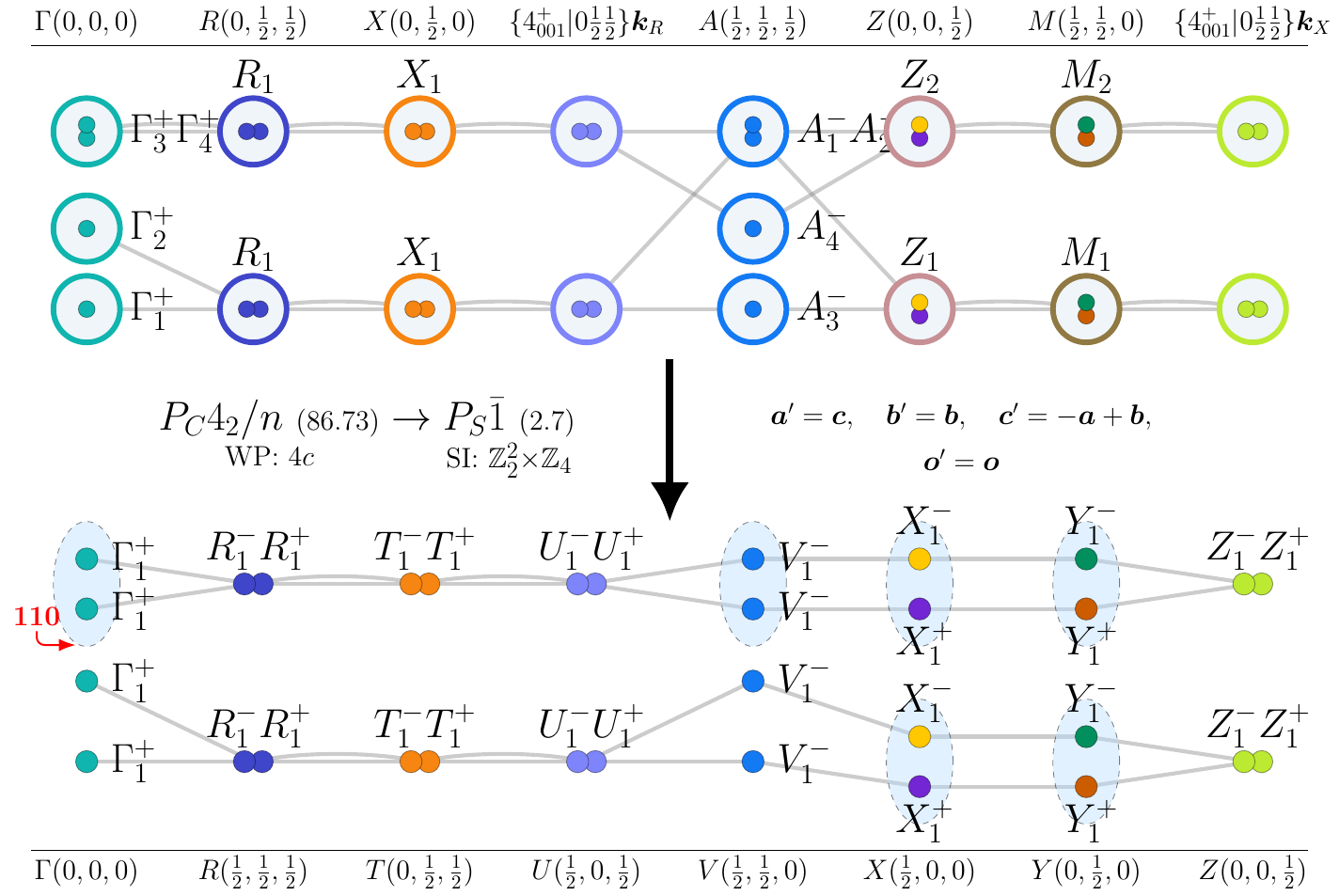}
\caption{Topological magnon bands in subgroup $P_{S}\bar{1}~(2.7)$ for magnetic moments on Wyckoff position $4c$ of supergroup $P_{C}4_{2}/n~(86.73)$.\label{fig_86.73_2.7_strainingenericdirection_4c}}
\end{figure}
\input{gap_tables_tex/86.73_2.7_strainingenericdirection_4c_table.tex}
\input{si_tables_tex/86.73_2.7_strainingenericdirection_4c_table.tex}

\section{MSG $I4_{1}/a~(88.81)$}
\textbf{Nontrivial-SI Subgroups:} $P\bar{1}~(2.4)$, $C2/c~(15.85)$.\\

\textbf{Trivial-SI Subgroups:} $Cc~(9.37)$, $C2~(5.13)$, $I4_{1}~(80.29)$.\\

\subsection{WP: $4a+8d$}
\textbf{BCS Materials:} {MnV\textsubscript{2}O\textsubscript{4}~(53 K)}\footnote{BCS web page: \texttt{\href{http://webbdcrista1.ehu.es/magndata/index.php?this\_label=0.64} {http://webbdcrista1.ehu.es/magndata/index.php?this\_label=0.64}}}.\\
\subsubsection{Topological bands in subgroup $P\bar{1}~(2.4)$}
\textbf{Perturbations:}
\begin{itemize}
\item strain in generic direction,
\item B in generic direction.
\end{itemize}
\begin{figure}[H]
\centering
\includegraphics[scale=0.6]{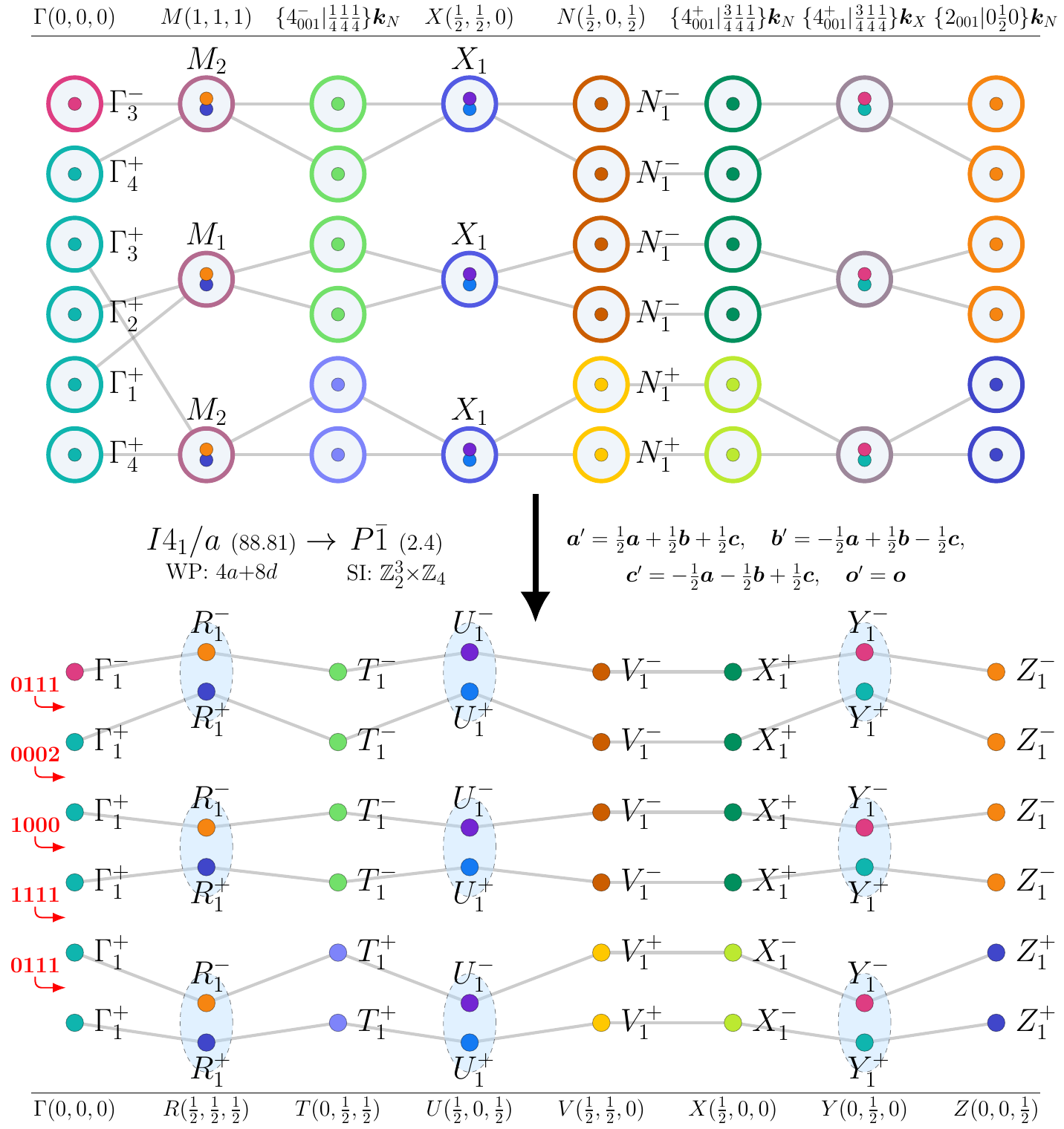}
\caption{Topological magnon bands in subgroup $P\bar{1}~(2.4)$ for magnetic moments on Wyckoff positions $4a+8d$ of supergroup $I4_{1}/a~(88.81)$.\label{fig_88.81_2.4_strainingenericdirection_4a+8d}}
\end{figure}
\input{gap_tables_tex/88.81_2.4_strainingenericdirection_4a+8d_table.tex}
\input{si_tables_tex/88.81_2.4_strainingenericdirection_4a+8d_table.tex}

\section{MSG $I_{c}4_{1}/a~(88.86)$}
\textbf{Nontrivial-SI Subgroups:} $C_{a}c~(9.41)$, $P\bar{1}~(2.4)$, $C2'/c'~(15.89)$, $P_{S}\bar{1}~(2.7)$, $C_{a}2~(5.17)$, $C2/c~(15.85)$, $C_{a}2/c~(15.91)$, $I_{c}4_{1}~(80.32)$, $I4_{1}/a~(88.81)$.\\

\textbf{Trivial-SI Subgroups:} $Cc'~(9.39)$, $C2'~(5.15)$, $P_{S}1~(1.3)$, $Cc~(9.37)$, $C2~(5.13)$, $I4_{1}~(80.29)$.\\

\subsection{WP: $16c$}
\textbf{BCS Materials:} {FeTa\textsubscript{2}O\textsubscript{6}~(13 K)}\footnote{BCS web page: \texttt{\href{http://webbdcrista1.ehu.es/magndata/index.php?this\_label=2.86} {http://webbdcrista1.ehu.es/magndata/index.php?this\_label=2.86}}}, {FeTa\textsubscript{2}O\textsubscript{6}~(8.5 K)}\footnote{BCS web page: \texttt{\href{http://webbdcrista1.ehu.es/magndata/index.php?this\_label=2.22} {http://webbdcrista1.ehu.es/magndata/index.php?this\_label=2.22}}}.\\
\subsubsection{Topological bands in subgroup $P\bar{1}~(2.4)$}
\textbf{Perturbations:}
\begin{itemize}
\item B $\parallel$ [001] and strain in generic direction,
\item (B $\parallel$ [100] or B $\parallel$ [110] or B $\perp$ [001]) and strain in generic direction,
\item B in generic direction.
\end{itemize}
\begin{figure}[H]
\centering
\includegraphics[scale=0.6]{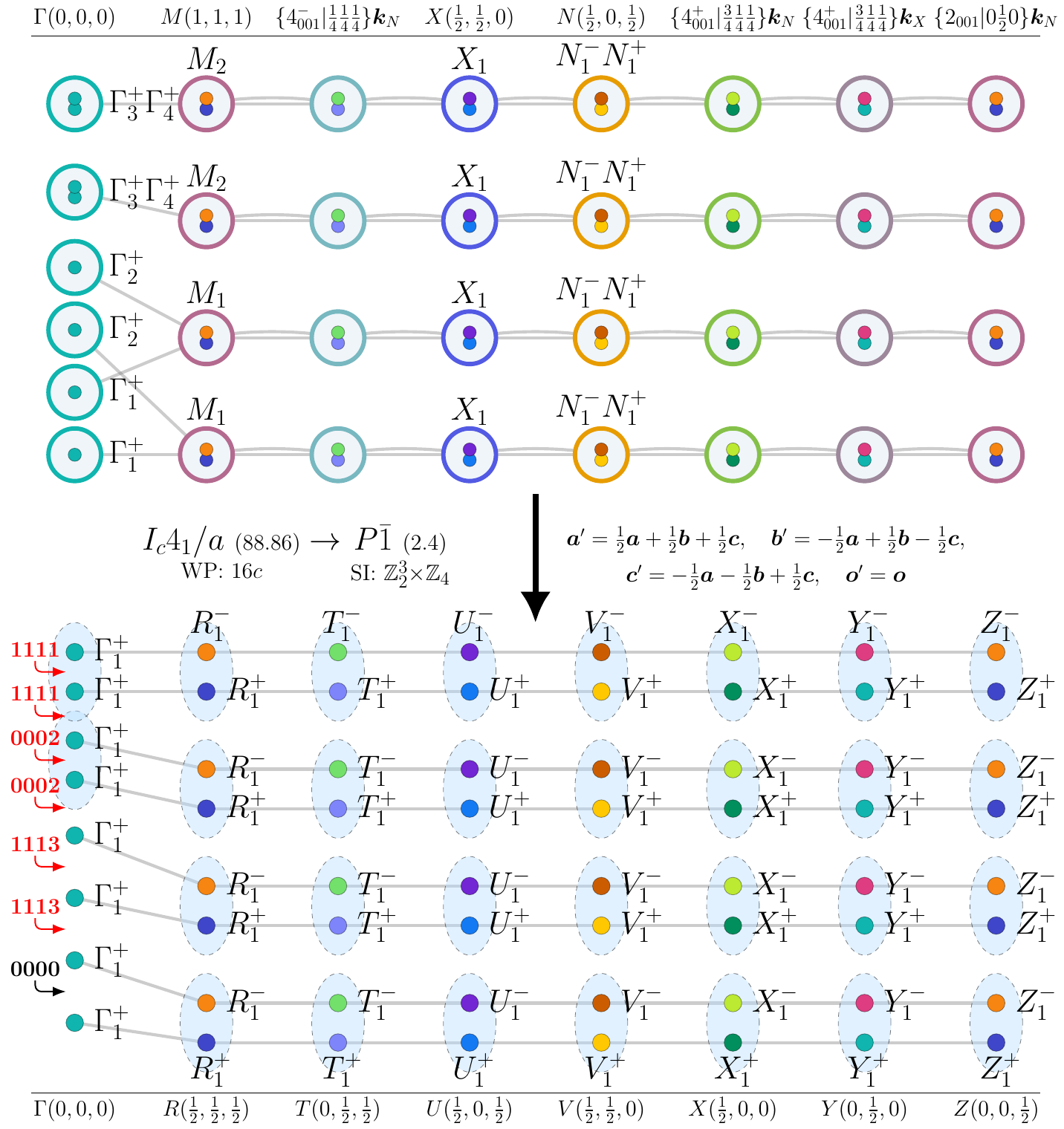}
\caption{Topological magnon bands in subgroup $P\bar{1}~(2.4)$ for magnetic moments on Wyckoff position $16c$ of supergroup $I_{c}4_{1}/a~(88.86)$.\label{fig_88.86_2.4_Bparallel001andstrainingenericdirection_16c}}
\end{figure}
\input{gap_tables_tex/88.86_2.4_Bparallel001andstrainingenericdirection_16c_table.tex}
\input{si_tables_tex/88.86_2.4_Bparallel001andstrainingenericdirection_16c_table.tex}
\subsubsection{Topological bands in subgroup $C2'/c'~(15.89)$}
\textbf{Perturbation:}
\begin{itemize}
\item (B $\parallel$ [100] or B $\parallel$ [110] or B $\perp$ [001]).
\end{itemize}
\begin{figure}[H]
\centering
\includegraphics[scale=0.6]{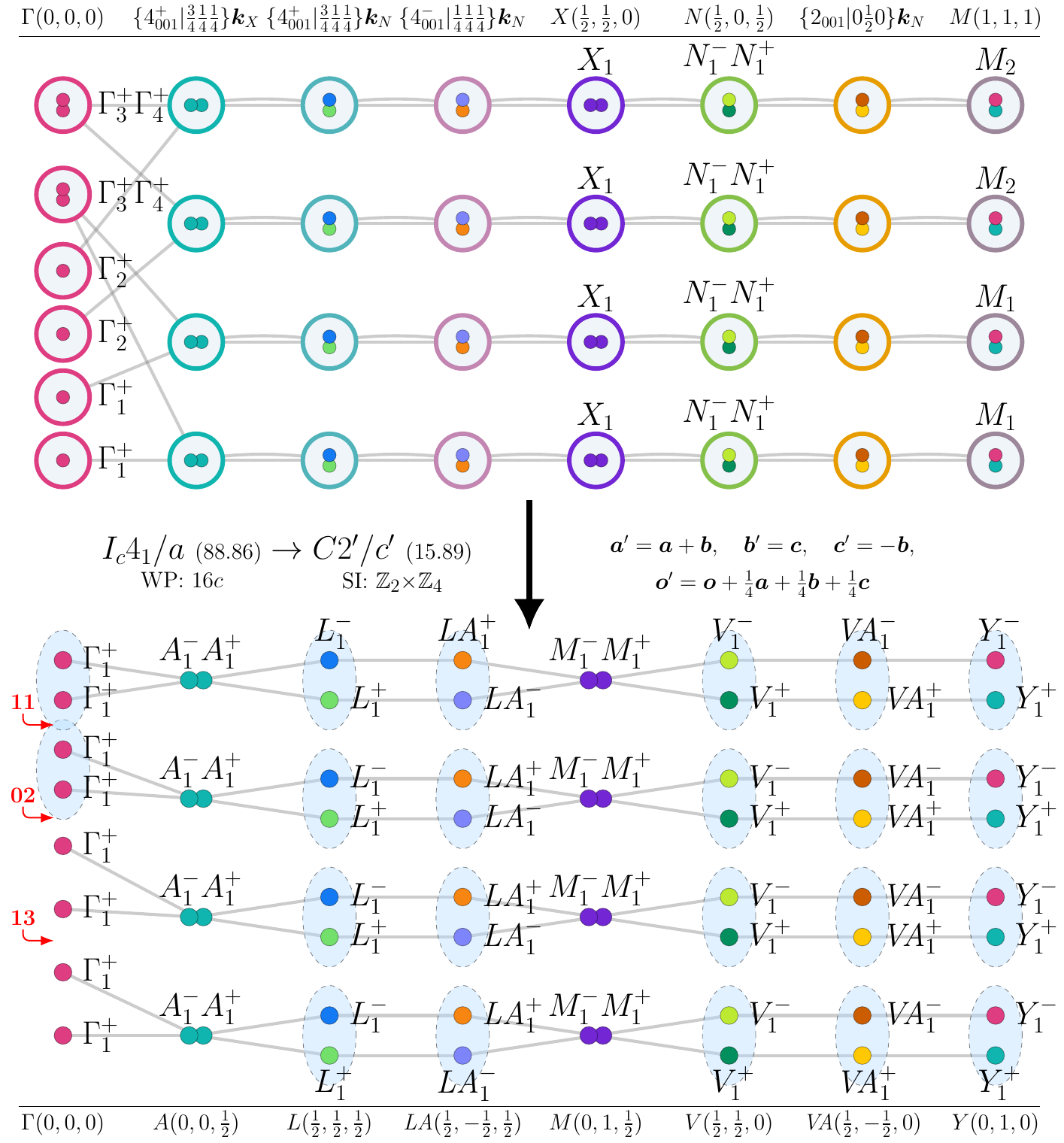}
\caption{Topological magnon bands in subgroup $C2'/c'~(15.89)$ for magnetic moments on Wyckoff position $16c$ of supergroup $I_{c}4_{1}/a~(88.86)$.\label{fig_88.86_15.89_Bparallel100orBparallel110orBperp001_16c}}
\end{figure}
\input{gap_tables_tex/88.86_15.89_Bparallel100orBparallel110orBperp001_16c_table.tex}
\input{si_tables_tex/88.86_15.89_Bparallel100orBparallel110orBperp001_16c_table.tex}
\subsubsection{Topological bands in subgroup $P_{S}\bar{1}~(2.7)$}
\textbf{Perturbation:}
\begin{itemize}
\item strain in generic direction.
\end{itemize}
\begin{figure}[H]
\centering
\includegraphics[scale=0.6]{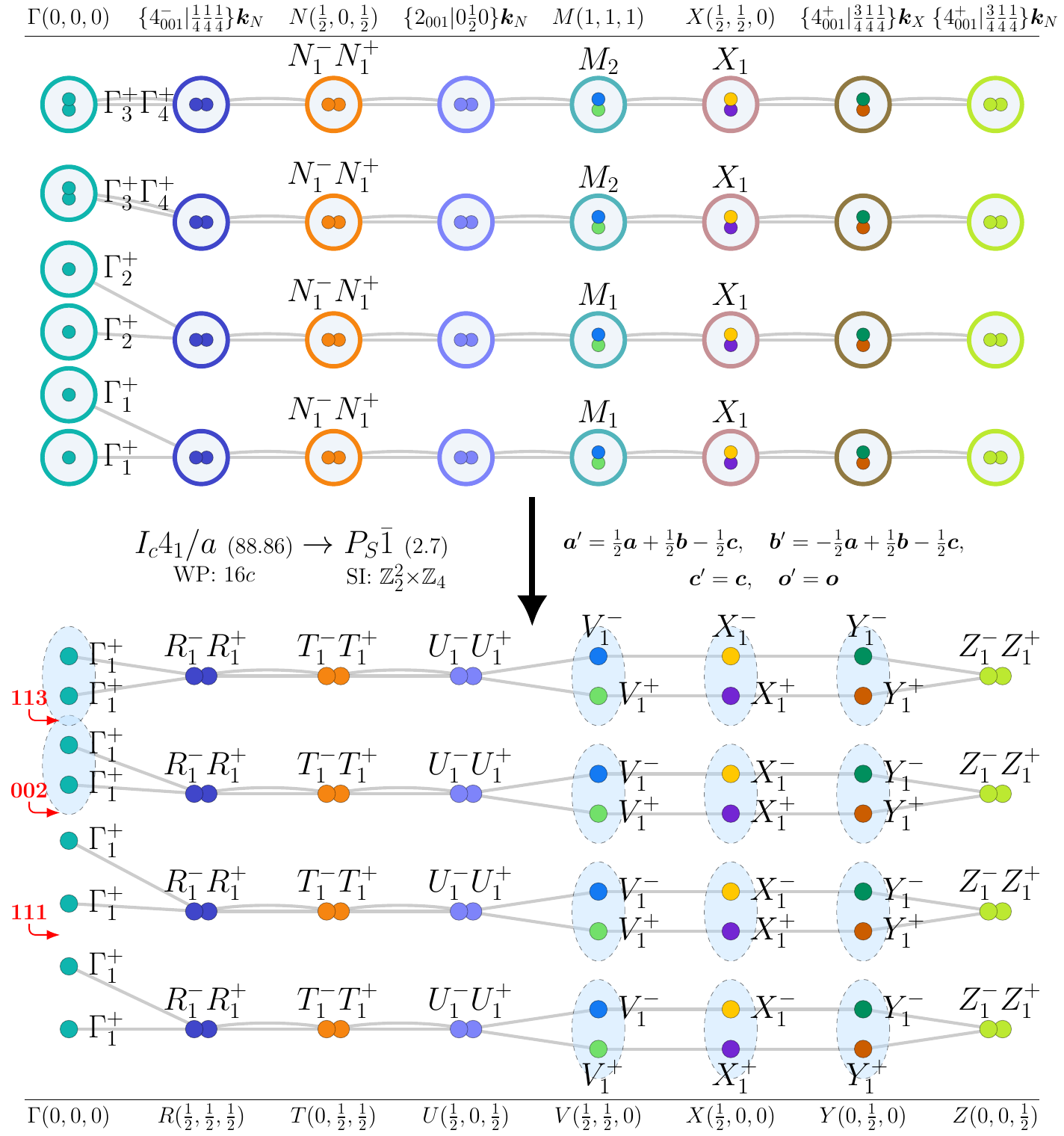}
\caption{Topological magnon bands in subgroup $P_{S}\bar{1}~(2.7)$ for magnetic moments on Wyckoff position $16c$ of supergroup $I_{c}4_{1}/a~(88.86)$.\label{fig_88.86_2.7_strainingenericdirection_16c}}
\end{figure}
\input{gap_tables_tex/88.86_2.7_strainingenericdirection_16c_table.tex}
\input{si_tables_tex/88.86_2.7_strainingenericdirection_16c_table.tex}

\section{SI formulas for subgroups}

\subsection{Subgroup $P2'/m'~(10.46)$}
\begin{align}
z_{1} &= n^{C_{1}^{+}}+n^{\Gamma_{1}^{+}}+n^{Y_{1}^{+}}+n^{Z_{1}^{+}} \bmod 2\\
z_{2} &= n^{A_{1}^{+}}+n^{A_{1}^{-}}+n^{B_{1}^{-}}+n^{D_{1}^{+}}+n^{\Gamma_{1}^{+}}+n^{Z_{1}^{+}} \bmod 2\\
z_{3} &= n^{A_{1}^{-}}+n^{B_{1}^{-}}+n^{\Gamma_{1}^{+}}+n^{Y_{1}^{+}} \bmod 2\\
z_{4} &= 3n^{A_{1}^{-}}+n^{B_{1}^{-}}+n^{C_{1}^{+}}+n^{D_{1}^{+}}+3n^{E_{1}^{+}}+n^{\Gamma_{1}^{+}}+3n^{Y_{1}^{+}}+3n^{Z_{1}^{+}} \bmod 4
\end{align}

\subsection{Subgroup $P_{I}4nc~(104.210)$}
\begin{align}
z_{1} &= n^{A_{5}}+n^{\Gamma_{1}}+n^{\Gamma_{2}} \bmod 2
\end{align}

\subsection{Subgroup $P2_{1}'/m'~(11.54)$}
\begin{align}
z_{1} &= n^{\Gamma_{1}^{+}}+n^{Y_{1}^{+}} \bmod 2\\
z_{2} &= n^{B_{1}^{+}}+n^{\Gamma_{1}^{+}} \bmod 2\\
z_{3} &= n^{A_{1}^{+}}+3n^{B_{1}^{+}}+n^{\Gamma_{1}^{+}}+3n^{Y_{1}^{+}} \bmod 4
\end{align}

\subsection{Subgroup $C2/m~(12.58)$}
\begin{align}
z_{1} &= n^{A_{2}^{+}}+n^{\Gamma_{2}^{+}}+n^{L_{1}^{+}}+n^{M_{1}^{+}}+n^{V_{1}^{+}}+n^{Y_{1}^{+}} \bmod 2\\
z_{2} &= n^{A_{2}^{+}}+n^{A_{2}^{-}}+n^{\Gamma_{2}^{+}}+n^{\Gamma_{2}^{-}}+n^{L_{1}^{+}}+n^{V_{1}^{+}} \bmod 2
\end{align}

\subsection{Subgroup $C2'/m'~(12.62)$}
\begin{align}
z_{1} &= n^{A_{1}^{+}}+n^{\Gamma_{1}^{+}}+n^{LA_{1}^{+}}+n^{VA_{1}^{+}} \bmod 2\\
z_{2} &= n^{\Gamma_{1}^{+}}+n^{Y_{1}^{+}} \bmod 2\\
z_{3} &= 3n^{A_{1}^{+}}+n^{\Gamma_{1}^{+}}+n^{M_{1}^{+}}+3n^{Y_{1}^{+}} \bmod 4
\end{align}

\subsection{Subgroup $C_{c}2/m~(12.63)$}
\begin{align}
z_{1} &= n^{\Gamma_{1}^{-}}+n^{V_{1}^{+}}+n^{Y_{1}^{+}} \bmod 2\\
z_{2} &= n^{\Gamma_{1}^{+}}+n^{Y_{1}^{+}} \bmod 2
\end{align}

\subsection{Subgroup $I\bar{4}2'd'~(122.337)$}
\begin{align}
z_{1} &= n^{\Gamma_{3}}+n^{M_{3}M_{4}}+n^{PA_{1}}+n^{PA_{3}}+n^{X_{1}X_{2}} \bmod 2\\
z_{2} &= n^{\Gamma_{2}}+n^{\Gamma_{3}}+n^{\Gamma_{4}}+n^{M_{3}M_{4}}+n^{PA_{1}}+n^{PA_{3}}+n^{X_{1}X_{2}} \bmod 2
\end{align}

\subsection{Subgroup $P2'/c'~(13.69)$}
\begin{align}
z_{1} &= n^{\Gamma_{1}^{+}}+n^{Y_{1}^{+}} \bmod 2\\
z_{2} &= n^{C_{1}^{+}}+n^{\Gamma_{1}^{+}} \bmod 2\\
z_{3} &= 3n^{C_{1}^{+}}+n^{\Gamma_{1}^{+}}+3n^{Y_{1}^{+}}+n^{Z_{1}^{+}} \bmod 4
\end{align}

\subsection{Subgroup $P2_{1}'/c'~(14.79)$}
\begin{align}
z_{1} &= n^{\Gamma_{1}^{+}} \bmod 2\\
z_{2} &= 2n^{D_{1}^{+}D_{1}^{+}}+2n^{E_{1}^{+}E_{1}^{+}}+n^{\Gamma_{1}^{+}}+3n^{Y_{1}^{+}} \bmod 4
\end{align}

\subsection{Subgroup $C2'/c'~(15.89)$}
\begin{align}
z_{1} &= n^{A_{1}^{-}A_{1}^{+}}+n^{\Gamma_{1}^{+}}+n^{LA_{1}^{+}}+n^{VA_{1}^{+}} \bmod 2\\
z_{2} &= 3n^{\Gamma_{1}^{+}}+2n^{VA_{1}^{+}}+3n^{Y_{1}^{+}} \bmod 4
\end{align}

\subsection{Subgroup $P22'2_{1}'~(17.10)$}
\begin{align}
z_{1} &= n^{\Gamma_{1}}+n^{S_{1}} \bmod 2
\end{align}

\subsection{Subgroup $P\bar{1}~(2.4)$}
\begin{align}
z_{1} &= n^{\Gamma_{1}^{-}}+n^{T_{1}^{-}}+n^{Y_{1}^{-}}+n^{Z_{1}^{-}} \bmod 2\\
z_{2} &= n^{\Gamma_{1}^{-}}+n^{U_{1}^{-}}+n^{X_{1}^{-}}+n^{Z_{1}^{-}} \bmod 2\\
z_{3} &= n^{\Gamma_{1}^{-}}+n^{V_{1}^{-}}+n^{X_{1}^{-}}+n^{Y_{1}^{-}} \bmod 2\\
z_{4} &= n^{\Gamma_{1}^{-}}+n^{R_{1}^{-}}+n^{T_{1}^{-}}+n^{U_{1}^{-}}+n^{V_{1}^{-}}+n^{X_{1}^{-}}+n^{Y_{1}^{-}}+n^{Z_{1}^{-}} \bmod 4
\end{align}

\subsection{Subgroup $P_{S}\bar{1}~(2.7)$}
\begin{align}
z_{1} &= n^{\Gamma_{1}^{+}}+n^{Y_{1}^{+}} \bmod 2\\
z_{2} &= n^{\Gamma_{1}^{+}}+n^{X_{1}^{+}} \bmod 2\\
z_{3} &= n^{\Gamma_{1}^{+}}+n^{V_{1}^{+}}+3n^{X_{1}^{+}}+3n^{Y_{1}^{+}} \bmod 4
\end{align}

\subsection{Subgroup $Pm'm'2~(25.60)$}
\begin{align}
z_{1} &= n^{\Gamma_{1}}+n^{R_{1}}+n^{T_{1}}+n^{U_{1}} \bmod 2
\end{align}

\subsection{Subgroup $Pm'a'2~(28.91)$}
\begin{align}
z_{1} &= n^{\Gamma_{1}}+n^{T_{1}} \bmod 2
\end{align}

\subsection{Subgroup $P2~(3.1)$}
\begin{align}
z_{1} &= n^{A_{2}}+n^{B_{2}}+n^{C_{1}}+n^{\Gamma_{1}} \bmod 2
\end{align}

\subsection{Subgroup $P_{a}2~(3.4)$}
\begin{align}
z_{1} &= n^{B_{1}}+n^{\Gamma_{1}} \bmod 2
\end{align}

\subsection{Subgroup $Cc'c'2~(37.183)$}
\begin{align}
z_{1} &= n^{RA_{1}}+n^{T_{1}T_{1}}+n^{Z_{1}Z_{1}} \bmod 2
\end{align}

\subsection{Subgroup $C_{A}cc2~(37.186)$}
\begin{align}
z_{1} &= n^{\Gamma_{3}}+n^{T_{1}T_{2}} \bmod 2
\end{align}

\subsection{Subgroup $Ab'm'2~(39.199)$}
\begin{align}
z_{1} &= n^{\Gamma_{1}}+n^{T_{1}} \bmod 2
\end{align}

\subsection{Subgroup $Ab'a'2~(41.215)$}
\begin{align}
z_{1} &= n^{\Gamma_{2}} \bmod 2
\end{align}

\subsection{Subgroup $C_{a}2~(5.17)$}
\begin{align}
z_{1} &= n^{A_{1}}+n^{\Gamma_{1}} \bmod 2
\end{align}

\subsection{Subgroup $Cmm'a'~(67.506)$}
\begin{align}
z_{1} &= n^{\Gamma_{1}^{+}}+n^{T_{1}^{+}}+n^{Y_{1}^{-}}+n^{Z_{1}^{-}} \bmod 2\\
z_{2} &= n^{\Gamma_{1}^{+}}+n^{\Gamma_{2}^{+}}+n^{T_{1}^{+}}+n^{T_{2}^{+}} \bmod 2
\end{align}

\stopcontents[sections]


\end{document}